\documentclass[12pt]{article}
\usepackage{jheppub}

\pdfoutput=1

\usepackage{amsmath,bbm,array,amsfonts,graphicx,wrapfig,lscape,float,mathtools,multirow,longtable}
\usepackage[dvipsnames]{xcolor}

\newcommand{\be}{\begin{equation}}
\newcommand{\ee}{\end{equation}}
\newcommand{\beq}{\begin{equation}}
\newcommand{\beql}[1]{\begin{equation}\label{#1}}
\newcommand{\eeq}{\end{equation}}
\newcommand{\ba}{\begin{array}}
\newcommand{\ea}{\end{array}}
\newcommand{\bea}{\begin{eqnarray}}
\newcommand{\beal}[1]{\begin{eqnarray}\label{#1}}
\newcommand{\eea}{\end{eqnarray}}
\newcommand{\ben}{\begin{enumerate}}
\newcommand{\een}{\end{enumerate}}
\newcommand{\bean}{\begin{eqnarray*}}
\newcommand{\eean}{\end{eqnarray*}}
\newcommand{\eref}[1]{(\ref{#1})}
\newcommand{\sref}[1]{\S\ref{#1}}
\newcommand{\tref}[1]{Table~\ref{#1}}
\newcommand{\nn}{\nonumber}

\newcommand{\fref}[1]{Figure \ref{#1}}
\newcommand{\btab}[1]{\begin{tabular}{#1}}
\newcommand{\etab}{\end{tabular}}

\def \Black {\pdfsetcolor{0 0 0 1}}

\newcommand{\comment}[1]{}

\newcommand{\ud}{\mathrm{d}}

\newcommand{\PL}{\mathrm{PL}}

\newcommand{\qed}{\nobreak \ifvmode \relax \else
      \ifdim\lastskip<1.5em \hskip-\lastskip
      \hskip1.5em plus0em minus0.5em \fi \nobreak
      \vrule height0.75em width0.5em depth0.25em\fi}

\definecolor{darkspringgreen}{rgb}{0.09, 0.45, 0.27}
\definecolor{forestgreen}{rgb}{0.13, 0.55, 0.13}

\usepackage{array}
\newcolumntype{C}[1]{>{\centering\let\newline\\\arraybackslash\hspace{0pt}}m{#1}}

\usepackage{breqn}

\title{Fano 3-Folds, Reflexive Polytopes \\ and Brane Brick Models} 

\author[a,b,c]{Sebasti\'an Franco} 
\author[d]{and Rak-Kyeong Seong}

\affiliation[a]{Physics Department, The City College of the CUNY\\
	160 Convent Avenue, New York, NY 10031, USA}
	
\affiliation[b]{Physics Program and \textsuperscript{$c$}Initiative for the Theoretical Sciences\\
	The Graduate School and University Center, The City University of New York\\
	365 Fifth Avenue, New York NY 10016, USA}
	
\affiliation[d]{
Department of Mathematical Sciences, 
Ulsan National Institute of Science and Technology,\\
50 UNIST-gil, Ulsan 44919, South Korea
}

\emailAdd{sfranco@ccny.cuny.edu}
\emailAdd{seong@unist.ac.kr}

\preprint{
\begin{flushright}
UNIST-MTH-22-RS-01 \\
\end{flushright}
}

\abstract{
Reflexive polytopes in $n$ dimensions have attracted much attention both in mathematics and theoretical physics due to their connection to Fano $n$-folds and mirror symmetry. 
This work focuses on the 18 regular reflexive polytopes corresponding to smooth Fano 3-folds.
For the first time, we show that all 18 regular reflexive polytopes have corresponding $2d$ $(0,2)$ gauge theories realized by brane brick models. 
These $2d$ gauge theories can be considered as the worldvolume theories of D1-branes probing the toric Calabi-Yau 4-singularities whose toric diagrams are given by the associated regular reflexive polytopes.
The generators of the mesonic moduli space of the brane brick models are shown to form a lattice of generators due to the charges under the rank 3 mesonic flavor symmetry.
It is shown that the lattice of generators is the exact polar dual reflexive polytope to the corresponding toric diagram of the brane brick model.
This duality not only highlights the close relationship between the geometry and $2d$ gauge theory, but also opens up pathways towards new discoveries in relation to reflexive polytopes and brane brick models.
}

\begin{document}

\maketitle

\section{Introduction}

The study of worldvolume theories of D-branes probing Calabi-Yau singularities has been immensely fruitful in the past \cite{Douglas:1996sw,Klebanov:1998hh,Acharya:1998db,Franco:2005rj}.
More recently, interest has grown in studying the setup of D1-branes probing Calabi-Yau 4-fold singularities as a pathway towards a better understanding of $2d$ $\mathcal{N}=(0,2)$ gauge theories \cite{Franco:2015tna,Franco:2015tya,Franco:2016tcm,Franco:2016qxh,Franco:2017cjj,Franco:2019pum,Franco:2021ixh}. 
When the Calabi-Yau 4-fold is toric, it was shown that the corresponding gauge theories are endowed with additional structural features. In this case, the map between the Calabi-Yau geometry and the gauge theory is considerably simplified by what we now call as brane brick models, which are Type IIA configurations that are connected to the D1-brane at the Calabi-Yau singularity via T-duality \cite{Franco:2015tna,Franco:2015tya}. One of the main virtues of these brane setups is that both the gauge theory and the underlying geometry can be easily determined from them.

There are various approaches for deriving the brane brick models associated to a given toric Calabi-Yau 4-fold, including partial resolution \cite{Franco:2015tna,Franco:2015tya}, mirror symmetry \cite{Franco:2016qxh} and the topological B-model \cite{Closset:2017yte,Closset:2018axq}. In addition, more efficient algorithms that apply to large classes of toric geometries, such as orbifold reduction, $3d$ printing and Calabi-Yau products, have been developed \cite{Franco:2016fxm,Franco:2018qsc,Franco:2020avj}.
Using these methods, brane brick models corresponding to a variety of toric Calabi-Yau 4-folds have been found.
Interestingly, some of these examples involve $2d$ $(0,2)$ gauge theories that exhibit novel gauge theory phenomena such as triality \cite{Gadde:2013lxa, Franco:2016nwv}, which through brane brick models obtained a brane picture interpretation as well as a geometrical interpretation through Calabi-Yau mirror symmetry \cite{Franco:2016tcm, Franco:2016qxh}.

A particular class of toric Calabi-Yau 4-folds however has not yet been systematically studied in the context of brane brick models and the corresponding $2d$ supersymmetric gauge theories. 
This class of toric Calabi-Yau 4-folds is realized as complex cones over Gorenstein Fano varieties that are constructed from a special set of lattice polytopes known as reflexive polytopes. 
This class of polytopes is special because they come in pairs that are related by a polar duality, which only exists between reflexive polytopes. 
Due to this property, reflexive polytopes have a long history in string theory, beginning with their introduction through mirror symmetry \cite{1997CMaPh.185..495K,Kreuzer:1998vb,Kreuzer:2000xy,Batyrev:1994hm,borisov1993towards} in the study of string theory compactifications on Calabi-Yau manifolds. 
Batyrev and Borisov \cite{Batyrev:1994hm,1998math1107B,Batyrev:1997tv,batyrev1994calabi,borisov1993towards} pioneered the systematic search for mirror paired Calabi-Yau manifolds that are realized as hypersurfaces in toric varieties given by pairs of dual reflexive polytopes. 

\begin{figure}[ht!!]
\begin{center}
\resizebox{0.8\hsize}{!}{
  \includegraphics[trim=0mm 10mm 0mm 0mm, width=8in]{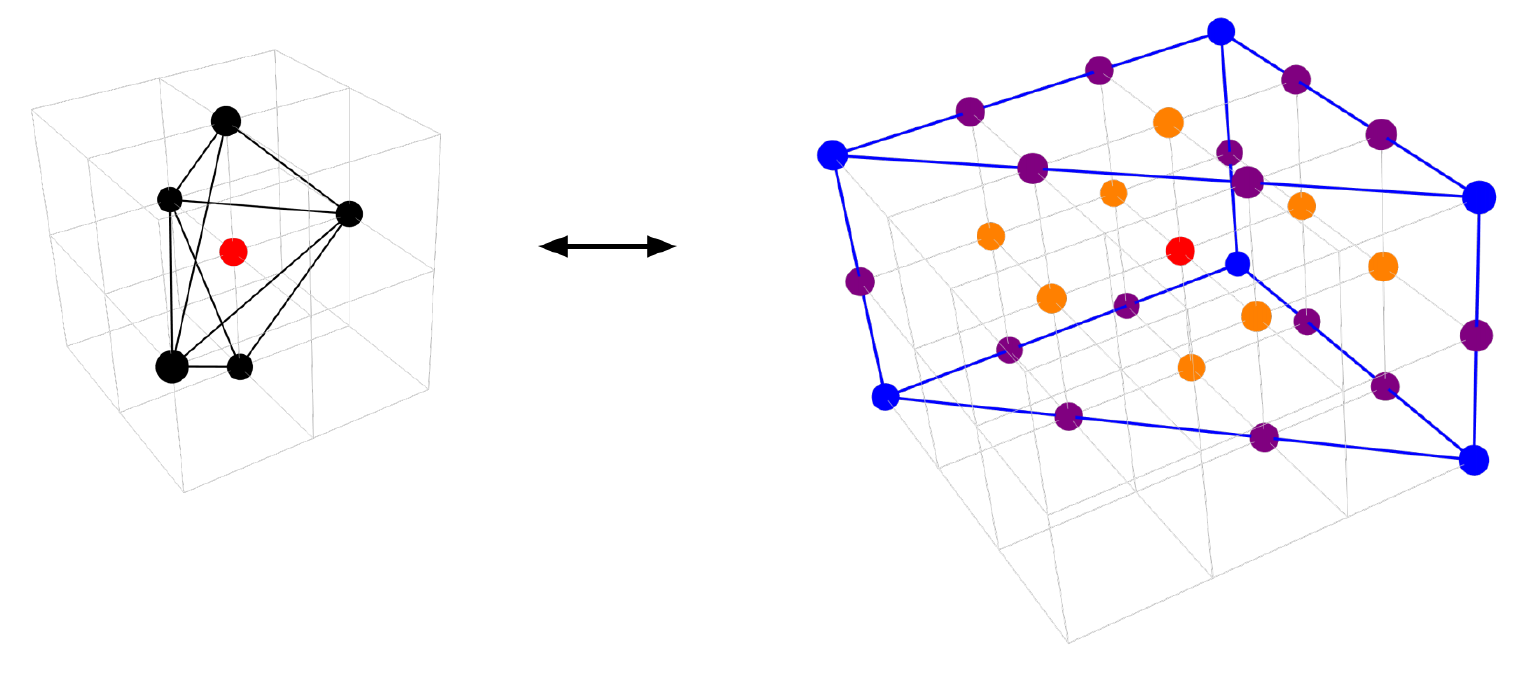}
}
\caption{
The toric diagram for $M^{3,2}$ ($\mathbb{P}^2 \times \mathbb{P}^1$) and the corresponding dual polytope.
Both are reflexive polytopes in dimension $3$ with the single interior point highlighted in red.
\label{f_alltoricdia}
}
\end{center}
\end{figure}

Thanks to the tour de force search for reflexive polytopes by Kreuzer and Skarke \cite{1997CMaPh.185..495K,Kreuzer:1998vb,Kreuzer:2000xy}, we know today how many of them exist up to lattice dimension 4.
Our interest lies in lattice dimension 3, where there are 4,319 reflexive polytopes up to $GL(3,\mathbb{Z})$ equivalence.
The corresponding toric varieties are known as Fano 3-folds.
In order to study brane brick models corresponding to these reflexive polytopes, 
we are interested in the associated complex cones -- the non-compact toric Calabi-Yau 4-folds.  

\begin{figure}[htt!!]
\begin{center}
\resizebox{0.85\hsize}{!}{
  \includegraphics[trim=0mm 0mm 0mm 5mm, width=8in]{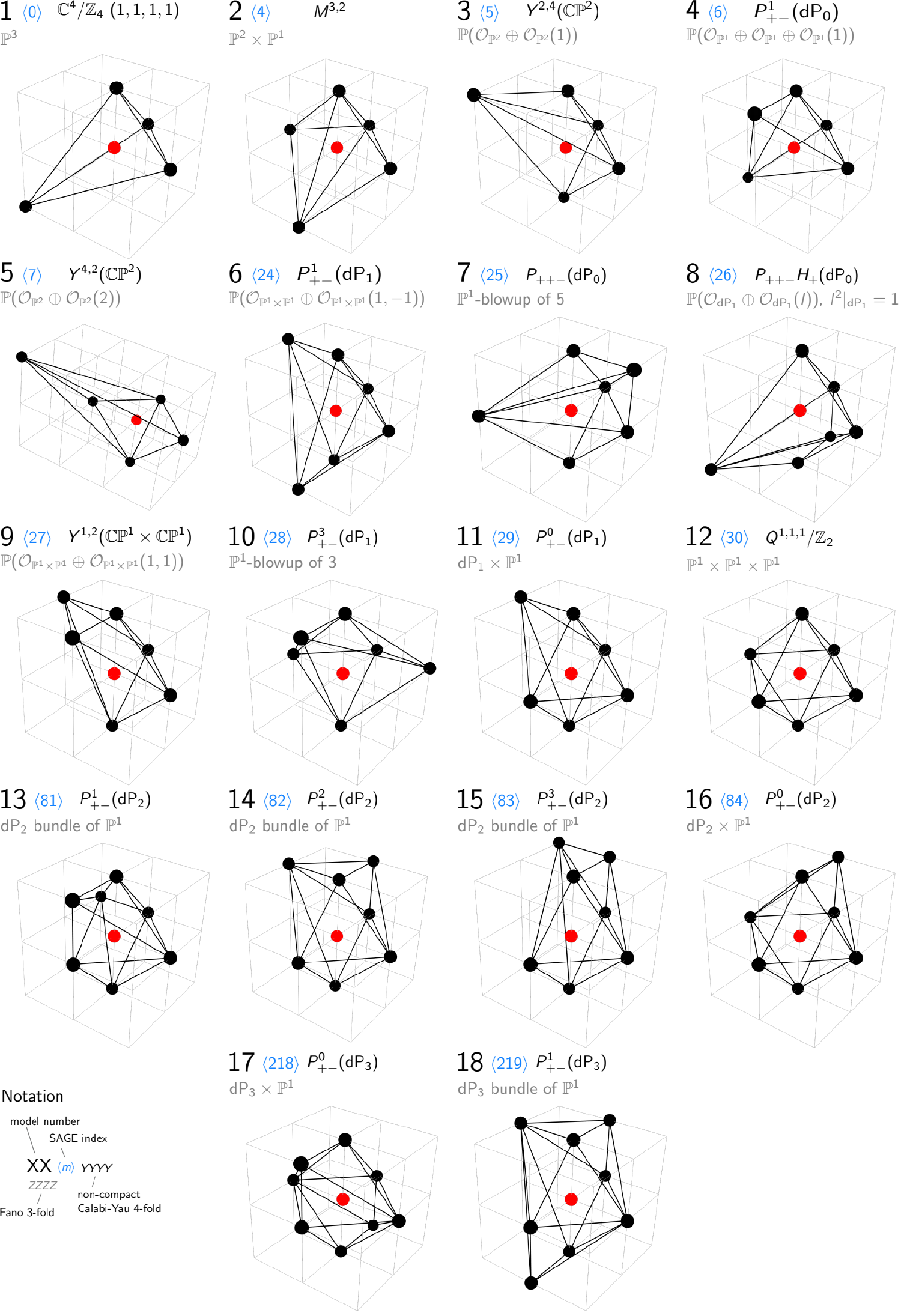}
}
\caption{
 The 18 regular reflexive polytopes in dimension 3 corresponding to toric non-compact Calabi-Yau 4-folds and corresponding smooth Fano 3-folds.
\label{f_alltoricdia}
}
\end{center}
\end{figure}

In this work, we focus on reflexive polytopes in dimension 3 that are also regular. A reflexive polytope is regular if every cone in the fan defined by the polytope has generators that form part of a $\mathbb{Z}$-basis.
Amongst the 4,319 reflexive polytopes in dimension 3, there are a manageable subset of 18 polytopes that are both reflexive and regular.
A property of regular reflexive polytopes is that their associated Gorenstein Fano varieties are smooth.
The fact that the polytopes are reflexive and regular has an interesting consequence for the corresponding brane brick models that we aim to explore as part of this work.

One of the main aims of this work is to find first the brane brick models and hence the corresponding $2d$ $(0,2)$ gauge theories for toric Calabi-Yau 4-folds whose toric diagrams are the 18 regular reflexive polytopes in dimension 3.
Because the toric diagrams are regular and reflexive, GLSM fields associated to brick matchings in the brane brick model either refer to the extremal corner points or the single interior point of the toric diagram.
Moreover, the toric Calabi-Yau 4-fold given by the regular reflexive polytopes is expected to be the mesonic moduli space of the $2d$ $(0,2)$ gauge theories represented by the brane brick models.
The spectrum of mesonic gauge invariant operators is therefore intricately lined to the GLSM fields corresponding to the points in the regular reflexive toric diagram.

In order to study this correspondence, we calculate the full spectrum of mesonic gauge invariant operators by obtaining the generating function known as the Hilbert series \cite{Benvenuti:2006qr,Hanany:2006uc,Butti:2007jv}.
Using plethystics \cite{Feng:2007ur,hanany2007counting}, we are able to identify the generating set of mesonic gauge invariant operators for each brane brick model associated to a regular reflexive polytope. 
As it is the case for any mesonic gauge invariant operator, the generators carry charges under the global symmetry of the gauge theory, which can be obtained from the isometry of the corresponding toric Calabi-Yau 4-fold.
The mesonic flavor symmetry, which is part of the global symmetry, assigns charges to the generators that can be scaled to be integer valued. 
This enables us to represent each set of mesonic flavor charges carried by a generator as a point on a $\mathbb{Z}^3$ lattice.
The convex hull of the set points associated to the set of generators of the mesonic moduli space forms a convex lattice polygon, which we call as the generator lattice.

One of the main results of this work is that for all 18 brane brick models corresponding to toric Calabi-Yau 4-folds whose toric diagram is a regular reflexive polygon, the corresponding mesonic moduli spaces have a generator lattice which is the polar dual to the toric diagram of the Calabi-Yau 4-fold. 
This correspondence between the toric diagrams and the generator lattices of the mesonic moduli space is exact because on both sides of the correspondence we are dealing with reflexive polytopes on a $\mathbb{Z}^3$ lattice. 
Our work illustrates this correspondence explicitly for all 18 regular reflexive polytopes corresponding to the 18 smooth Fano 3-folds for the first time, whilst making simultaneously a connection to $2d$ $(0,2)$ gauge theories realized by brane brick models.

Our work is structured as follows. In Section \sref{sec:background1}, we give a brief overview of reflexive polytopes and talk about the construction of Fano $n$-folds. The appearance of the corresponding non-compact toric Calabi-Yau 4-folds is discussed. Section \sref{sec:bbm} continues with a review on brane brick models and the realization of $2d$ $(0,2)$ gauge theories as worldvolume theories of D1-brane probing toric Calabi-Yau 4-folds. We discuss the characterization of the mesonic moduli spaces of brane brick models in terms of Hilbert series and the construction of generator lattices. The following sections then discuss the brane brick models for the 18 regular reflexive polytopes and illustrate how the toric diagrams are polar duals to the corresponding generator lattices. We conclude our work with a short summary of its main results.
\\

\section{18 Smooth Fano 3-Folds and Toric Calabi-Yau 4-Folds \label{sec:background1}}

Let us begin with a short review about the particular set of lattice polytopes known as reflexive polytopes and the role they play in our work.
\\

\subsection{Reflexive Polytopes  \label{sec:reflexive}}

\begin{table}[ht!!!]
\centering
\begin{tabular}{|c|c|c|}
\hline
$d$ & Number of Polytopes & Number of Regular Polytopes \\
\hline \hline
1 & 1  & 1\\
\hline
2 & 16 & 5\\
\hline
3 & 4,319 & 18 \\
\hline 
4 & 473,800,776 & 124 \\
\hline
\end{tabular}
\caption{
The number of distinct reflexive polytopes and distinct regular reflexive polytopes in dimension $d\leq 4$ \cite{1997CMaPh.185..495K,Kreuzer:1998vb,Kreuzer:2000xy}.
\label{t_counting}}
\end{table}

Reflexive polytopes form a special subset of \textit{lattice polytopes} $\Delta_n$ in $\mathbb{Z}^n$.
We call the convex hull of a finite number of points in $\mathbb{Z}^n$ a lattice polytope $\Delta_n$ in dimension $n$. 
The vertices of the lattice polytope form the set $\Delta_n \cap \mathbb{Z}^n$.
In this work, we call these convex lattice polytopes $\Delta_n$ also \textit{toric diagrams}.

A lattice polytope is called \textit{reflexive} if the dual polytope (sometimes also known as the \textit{polar polytope}) given by 
\beal{es00_1}
\Delta_n^{\circ} = \{ { \bf v} \in \mathbb{Z}^n ~|~ {\bf m}\cdot {\bf n} \geq -1 ~ \forall \, {\bf m} \in \Delta_n \} ~,~
\eea
is also a lattice polytope in $\mathbb{Z}^n$.
Reflexive polytopes and their duals have a unique interior lattice point at the origin $(0,\dots,0)\in \mathbb{Z}^n$.

For a given $n$, there is a finite number of such reflexive polytopes up to $GL(n,\mathbb{Z})$ equivalence and this makes the problem of classifying these reflexive polytopes an interesting combinatorial problem. 
For $n=2$, it is straightforward to identify the 16 reflexive polygons up to $GL(2,\mathbb{Z})$ equivalence.
For higher $n$, the problem of classifying all reflexive polytopes becomes significantly more complicated. 
Due to Kreuzer-Skarke \cite{1997CMaPh.185..495K,Kreuzer:1998vb,Kreuzer:2000xy}, the number of reflexive polytopes is known to be 4,319 in $n=3$  and 473,800 in $n=4$ as summarized in \tref{t_counting}.

A subset of reflexive polytopes in each dimension $n$ is known to be \textit{regular}. A polytope is called regular if every cone in the fan has generators that form part of a $\mathbb{Z}$-basis. That means, for instance in dimension $n=2$, the boundary edges of the polygons do not contain any internal points. For dimension $n=3$, the boundary faces of regular polytopes are all triangles that do not contain internal points on their boundary edges and in the interior of the triangles. There are exactly 18 regular reflexive polytopes in dimension $n=3$ as shown in \tref{t_counting}. 
Batyrev-Borisov \cite{Batyrev:1994hm,1998math1107B,Batyrev:1997tv,batyrev1994calabi,borisov1993towards} studied these reflexive polytopes $\Delta_n$ in $n$ dimensions in order construct new families of smooth Calabi-Yau hypersurfaces in toric varieties given by $\Delta_n$. To be more precise, if $\Delta_n$ is reflexive, it corresponds to a \textit{Gorenstein toric Fano variety} $X(\Delta_n)$.
We here also call $X(\Delta_n)$ a toric \textit{Fano $n$-fold}, which is smooth if $\Delta_n$ is regular in addition to being reflexive.
The duality between reflexive polytopes that relates dual families of smooth Fano $n$-folds $X(\Delta_n)$ has been shown to be \textit{mirror symmetry} \cite{1997CMaPh.185..495K,Kreuzer:1998vb,Kreuzer:2000xy,Batyrev:1994hm,borisov1993towards}.

Every reflexive polytope $\Delta_n$ is also associated with a non-compact toric Calabi-Yau $(n+1)$-fold, which is basically the affine cone over the base $X(\Delta_n)$. As discussed in \cite{He:2017gam}, many properties of the Fano $n$-folds can be related to geometrical properties of the corresponding toric Calabi-Yau $(n+1)$-folds.
The following section summarizes both constructions for given $\Delta_n$, their geometrical properties and the notation that we use to identify them. 
\\

\begin{table}[ht!!]
\centering
\begin{tabular}{|c|c|c|c|c|c|}
\hline
Model & Polytope  & $E$ &Calabi-Yau 4-fold& Fano 3-fold
\\
\hline\hline
1 & 
$\langle0\rangle$
&
4
&
$\mathbb{C}^4/\mathbb{Z}_4 ~(1,1,1,1)$
&
$\mathbb{P}^3$
\\
\hline
2 & 
$\langle4\rangle$
&
5
&
$M^{3,2}, Y^{2,3}(\mathbb{CP}^2), P_{+-}^{0}(\text{dP}_0)$
&
$\mathbb{P}^2\times\mathbb{P}^1$
\\
\hline
3 & 
$\langle5\rangle$
&
5
&
$Y^{2,4}(\mathbb{CP}^2)$
&
$\mathbb{P}(\mathcal{O}_{\mathbb{P}^2} \oplus \mathcal{O}_{\mathbb{P}^2}(1))$
\\
\hline
4 & 
$\langle6\rangle$
&
5
&
$P_{+-}^{1}(\text{dP}_0)$
&
$\mathbb{P}(\mathcal{O}_{\mathbb{P}^1} \oplus \mathcal{O}_{\mathbb{P}^1} \oplus \mathcal{O}_{\mathbb{P}^1} (1))$
\\
\hline
5 & 
$\langle7\rangle$
&
5
&
$Y^{2,5}(\mathbb{CP}^2)$
&
$\mathbb{P}(\mathcal{O}_{\mathbb{P}^2} \oplus \mathcal{O}_{\mathbb{P}^2}(2))$
\\
\hline
6 & 
$\langle24\rangle$
&
6
&
$P^{1}_{+-}(\text{dP}_1)$
&
$\mathbb{P}(\mathcal{O}_{\mathbb{P}^1\times \mathbb{P}^1} \oplus \mathcal{O}_{\mathbb{P}^1 \times\mathbb{P}^1}(1,-1))$
\\
\hline
7 & 
$\langle25\rangle$
&
6
&
$P_{++-}(\text{dP}_0)$
&
$\mathbb{P}^1$-blowup of $5$
\\
\hline
\end{tabular}
\caption{The names of smooth Fano 3-folds and toric Calabi-Yau 4-folds corresponding to the 18 regular reflexive polytopes in dimension 3. $E$ is the number of external points of the reflexive polytope. 
\label{t_names}}
\end{table}

\subsection{Fano $n$-Folds and Calabi-Yau $(n+1)$-Folds \label{sec:geometry}}

The following section gives a brief summary on Fano $n$-folds and non-compact toric Calabi-Yau $(n+1)$-folds that correspond to reflexive polytopes $\Delta_n$.

\paragraph{Fano $n$-folds.} 
Given a lattice polytope $\Delta_n$ in dimension $n$, one can construct a compact toric variety $X(\Delta_n)$ of complex dimension $n$ \cite{fulton,CLS}. 
From the lattice polytope, one can construct the normal fan $\Sigma(\Delta_n)$ as the positive hull of the $n$-cones over all the faces of $\Delta_n$. 
Given the fan $\Sigma$ from the polytope $\Delta_n$, the corresponding compact toric variety $X(\Sigma)$ follows by gluing in the standard way the affine varieties of each of the cones in $\Sigma$.
If the lattice polytope $\Delta_n$ is reflexive, $X(\Sigma)$ is known as a Gorenstein toric Fano variety or short as a Fano $n$-fold \cite{2004math......5448N}.
Furthermore, if $\Delta_n$ is regular, meaning every cone in the fan $\Sigma(\Delta_n)$ has generators that form part of a $\mathbb{Z}$-basis, $X(\Sigma)$ is also known to be smooth \cite{CLS}.

As discussed in Section \sref{sec:reflexive}, there are in dimension $n=3$ precisely 18 distinct reflexive polytopes that are regular. Each of these 18 polytopes corresponds to a smooth toric Fano $3$-fold $X(\Delta_3)$.
In this work, we concentrate on these 18 reflexive polytopes and the corresponding toric Fano $3$-folds. \fref{f_alltoricdia} shows the 18 reflexive polytopes with the corresponding names of toric Fano $3$-folds. 
\tref{t_names} also summarizes the names for the 18 toric Fano $3$-folds with the corresponding SAGE polytope index \cite{sage}.

\paragraph{Non-compact Calabi-Yau $(n+1)$-folds.}
A lattice polytope $\Delta_n$ can also be associated to a
non-compact toric Calabi-Yau $(n+1)$-fold. 
The complex cone over the toric variety $X(\Delta_n)$ associated to the lattice polytope $\Delta_n$ is an affine toric Calabi-Yau $(n+1)$-fold $\mathcal{X}_{n+1}$ of complex dimension $n+1$. 

Let us give a brief overview on the construction of $\mathcal{X}_{n+1}$.
First, let us explicitly define the polyhedral cones $\sigma$ that are generated by the vertices of $X(\Delta_n)$. 
Given that the vertices of $X(\Delta_n)$ are in $\mathbb{Z}^n$, the origin $N:= (0,\dots,0) \in \mathbb{Z}^{n+1}$ is taken to be the apex of the polytope in $\mathbb{Z}^n$. Then the vectors $\mathbf{u}_i$ from this apex to the vertices of $X(\Delta_n)$ generate the cone $\sigma$ as follows
\beal{es01_0}
\sigma = \left\{
\sum_{\mathbf{u}_i} \lambda_i \mathbf{u}_i ~|~ \lambda_{\mathbf{u}_i} \geq 0 
\right\}
\subset N_{\mathbb{R}}:= N \otimes_{\mathbb{Z}} \mathbb{R} ~.~
\eea
Like the polytope $\Delta_n$ having a dual following \eref{es00_1},
the cone also has a dual $\sigma^{\vee}$ that lives in $M_{\mathbb{R}}:= M \otimes_{\mathbb{Z}} \mathbb{R}$ where $M:= \text{hom}(N,\mathbb{Z})$.
The dual cone $\sigma^{\vee}$ is given by
\beal{es01_1}
\sigma^{\vee} 
=\left\{
\mathbf{v} \in M_{\mathbb{R}} ~|~
\mathbf{v} \cdot \mathbf{U} \geq 0 ~~
\forall \, \mathbf{u} \in \sigma
\right\} ~.~
\eea
Following the definition of the dual cone $\sigma^\vee$, $\mathcal{X}_{n+1}$ can be identified as the maximal spectrum of the group algebra generated by the lattice points covered by $\sigma^\vee$ in $M$.
This is given by
\beal{es01_2}
\mathcal{X}_{n+1} \simeq
\text{Spec} \left(
\mathbb{C}[\sigma^\vee \cap M]
\right) 
~.~
\eea
Note that the end-points of the vectors generating the cone are all co-hyperplanar. 
This guarantees that $\mathcal{X}_{n+1}$ is a Gorenstein singularity and hence relates to a Calabi-Yau manifold.

While $\mathcal{X}_{n+1}$ can be considered as the complex cone over the Gorenstein Fano variety $X(\Delta_n)$, 
as a toric non-compact Calabi-Yau $(n+1)$-fold it is also a real cone over a Sasaki-Einstein manifold $Y$ of real dimension $2n+1$.
The Sasaki-Einstein manifold $Y$ is a Riemannian manifold where the cone over $Y$ has a metric of the form
\beal{es01_3}
\ud s^2 (\mathcal{X}_{n+1}) = \ud r^2 + r^2 \ud s^2 (Y) ~,~
\eea
which is Ricci-flat and K\"ahler.
The metrics of large classes of Sasaki-Einstein manifolds were found in the past \cite{Gauntlett:2004zh,Gauntlett:2004yd,Gauntlett:2004hh}.
Moreover, these Sasaki-Einstein manifolds have a volume.
The volume has a minimum non-zero point that determines the Reeb vector field for the corresponding Sasaki-Einstein manifold \cite{Martelli:2006yb}. The volume functional that is used for the minimization process can be written in terms of the Reeb vector field. 
In \cite{He:2017gam}, it was shown that the minimum volume of Sasaki-Einstein manifolds corresponding to toric Calabi-Yau manifolds and obtained from reflexive polytopes up to dimension $n=4$ are related to topological quantities such as the Chern number and the Euler number of the toric varieties.

In this work, we concentrate on the 18 regular reflexive polytopes in dimension $n=3$ and the corresponding non-compact toric Calabi-Yau $4$-folds. 
The 18 polytopes with the names of the corresponding Calabi-Yau $4$-folds are shown in \fref{f_alltoricdia} and listed in \tref{t_names}.
Note that the toric Calabi-Yau $4$-folds corresponding to regular reflexive polytopes $\Delta_3$ and smooth Fano 3-folds are part 
of one or multiple large families of toric Calabi-Yau $4$-folds. Accordingly, the naming convention for the toric Calabi-Yau $4$-folds follows the naming conventions for these large families, which are as follows:
\begin{itemize}
\item \textbf{Abelian orbifolds of $\mathbb{C}^4$.}
Abelian orbifolds of $\mathbb{C}^4$ take the general form $\mathbb{C}^4/\mathbb{Z}_{n_1} \times \mathbb{Z}_{n_2} \times \mathbb{Z}_{n_3}$ where the order of the orbifold is $N=n_1 n_2 n_3$. For each $\mathbb{Z}_{n_i}$ factor, there is an integer 4 vector $(a_1,a_2,a_3,a_4)$ that determines the orbifold $\mathbb{Z}_{n_i}$-action on $\mathbb{C}^4$. 
Given that $\mathbb{C}^4$ has coordinates $z_k$, the orbifold $\mathbb{Z}_{n_i}$-action $(a_{i1},a_{i2},a_{i3},a_{i4})$ acts on the coordinates as follows
\beal{es01_05}
z_k \mapsto \omega_i^{a_{ik}} z_k ~,~ 
\eea
where $(\omega_i)^{n_i} = 1$.
The corresponding convex lattice polytopes $\Delta_3$ for Abelian orbifolds of $\mathbb{C}^4$ are 3-dimensional lattice tetrahedra with polytope volume $N$, where the volume of $\mathbb{C}^4$ is normalized to $1$.
Distinct Abelian orbifolds of $\mathbb{C}^4$ for a given order $N$ were counted and classified in \cite{Davey:2010px,Hanany:2010ne}.
Since all Abelian orbifolds of $\mathbb{C}^4$ are toric Calabi-Yau $4$-folds, we label the Calabi-Yau geometries with the orbifold name containing the orbifold action on $\mathbb{C}^4$.
In this work, the Abelian orbifold of the form $\mathbb{C}^4/\mathbb{Z}_4$ with action $(1,1,1,1)$ is the only example where the corresponding polytope $\Delta_4$ is a regular reflexive polytope. 
In fact, the toric diagram for $\mathbb{C}^4/\mathbb{Z}_4$ with action $(1,1,1,1)$ is the unique regular lattice tetrahedron with a single internal point as shown as Model 1 with polytope index $\langle 0 \rangle$ in \fref{f_alltoricdia} and \tref{t_names}.

\begin{figure}[ht!!]
\begin{center}
\resizebox{0.6\hsize}{!}{
  \includegraphics[trim=0mm 0mm 0mm 0mm, width=8in]{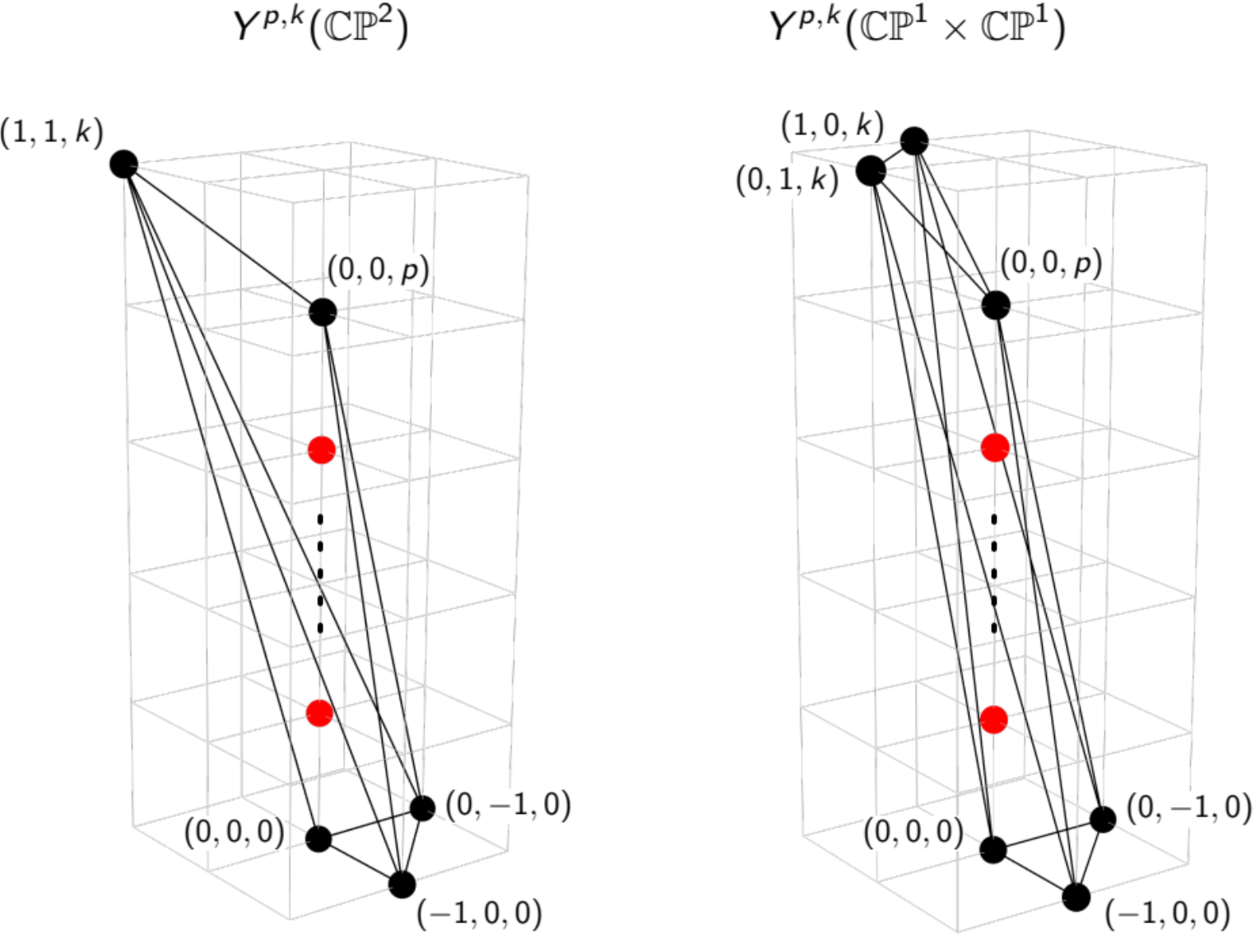}
}
\caption{
Toric diagrams corresponding to the $Y^{p,k}(\mathbb{CP}^2)$ and $Y^{p,k}(\mathbb{CP}^1\times \mathbb{CP}^1)$ families of toric Calabi-Yau 4-folds.
\label{f_toric_ypk}
}
\end{center}
\end{figure}

\item \textbf{$M^{3,2}$ and $Q^{1,1,1}/\mathbb{Z}_2$.}
$M^{3,2}$ and $Q^{1,1,1}$ are Sasaki-Einstein 7-manifolds \cite{1986PhR...130....1D} whose metric is explicitly known.
They are higher-dimensional generalizations of the famous $T^{1,1}$ Sasaki-Einstein 5-manifold and were some of the few known Sasaki-Einstein 7-manifolds before the discovery of some infinite families.
The cone over $M^{3,2}$ is a toric Calabi-Yau 4-fold which has a regular reflexive polytope as its toric diagram. In \fref{f_alltoricdia} and \tref{t_names}, this Calabi-Yau 4-fold is labeled as Model 2. In addition, even though $M^{3,2}$ refers to the Sasaki-Einstein 7-manifold, we use the name interchangeably also for the corresponding Calabi-Yau cone. 
The cone over $Q^{1,1,1}$ is a toric Calabi-Yau 4-fold whose toric diagram is the smallest lattice octahedron in $\mathbb{Z}^3$. The non-trivial $\mathbb{Z}_2$ orbifold on this Calabi-Yau $4$-fold has the regular lattice octahedron with a single internal point as its toric diagram as shown as Model 12 in \fref{f_alltoricdia} and \tref{t_names}. We refer to this Calabi-Yau $4$-fold as $Q^{1,1,1}/\mathbb{Z}_2$.

\item \textbf{$Y^{p,k}(\mathbb{CP}^2)$ and $Y^{p,k}(\mathbb{CP}^1\times \mathbb{CP}^1)$ families.}
It was shown in \cite{Gauntlett:2004zh, Gauntlett:2004yd} that in all odd dimensions there are infinite families of Sasaki-Einstein manifolds. In particular, \cite{Gauntlett:2004hh} shows that 
for any positive curvature K\"ahler-Einstein manifold $B_{2m}$
there is an infinite class of Sasaki-Einstein $(2m+3)$-manifolds. 
In the case for Sasaki-Einstein 7-manifolds, two infinite classes labeled by two integers $p$ and $k$ were found with $B_4$ being $\mathbb{CP}^2$ and $\mathbb{CP}^1\times \mathbb{CP}^1$. These are known as $Y^{p,k}(\mathbb{CP}^2)$ and $Y^{p,k}(\mathbb{CP}^1\times \mathbb{CP}^1)$ and the cone over these Sasaki-Einstein 7-manifolds is a toric Calabi-Yau 4-fold. 
The coordinates of the toric diagrams of these Calabi-Yau 4-folds can be written in terms of the integers $p$ and $k$ as illustrated in \fref{f_toric_ypk}.
In the particular cases of $Y^{2,3}(\mathbb{CP}^2)$, $Y^{2,4}(\mathbb{CP}^2)$, $Y^{2,5}(\mathbb{CP}^2)$ and $Y^{1,2}(\mathbb{CP}^1)$, the corresponding toric diagrams are regular reflexive polytopes as shown in \fref{f_alltoricdia} and \tref{t_names}. We refer to these as Models 2, 3, 5 and 9 respectively and also refer to the toric Calabi-Yau 4-folds by the names of the corresponding Sasaki-Einstein 7-manifolds.

\begin{figure}[ht!!]
\begin{center}
\resizebox{1\hsize}{!}{
  \includegraphics[trim=0mm 0mm 0mm 0mm, width=8in]{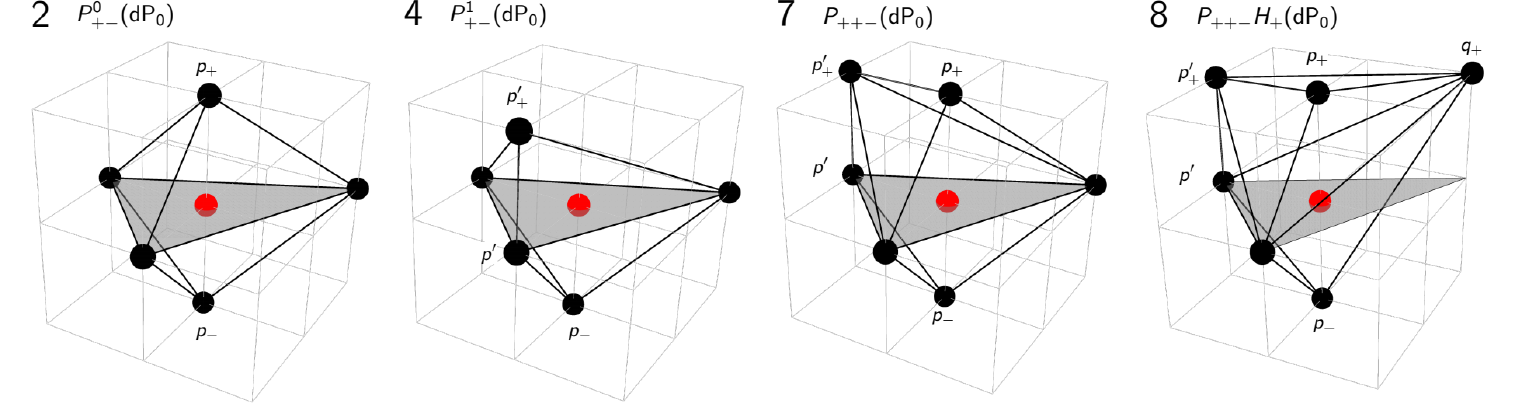}
}
\caption{
Naming convention for toric Calabi-Yau $4$-folds that are related to toric Calabi-Yau $3$-folds. The red lattice points indicate the unique internal point of the regular reflexive polytopes.
\label{f_toriclifting}
}
\end{center}
\end{figure}

\item \textbf{Calabi-Yau $4$-folds with a Calabi-Yau $3$-fold base.}
In this work, we introduce names for toric Calabi-Yau $4$-folds whose toric diagrams relate to toric diagrams of toric Calabi-Yau $3$-folds. 
Given the toric diagram of a toric Calabi-Yau 3-fold, it can be placed on the plane $z=0$ in a $\mathbb{Z}^3$ lattice. 
By adding additional lattice points above or below the $z=0$ plane, the convex hull of all the points forms a 3-dimensional convex lattice polytope corresponding to a toric Calabi-Yau 4-fold. 
For example, if one adds a single lattice point at height $z=1$ above the toric diagram of the Calabi-Yau 3-fold ($\text{CY}_3$) at height $z=0$, the resulting toric diagram will correspond to the toric Calabi-Yau 4-fold of the form $\mathbb{C}\times \text{CY}_3$.

\begin{figure}[ht!!]
\begin{center}
\resizebox{0.6\hsize}{!}{
  \includegraphics[trim=0mm 0mm 0mm 0mm, width=8in]{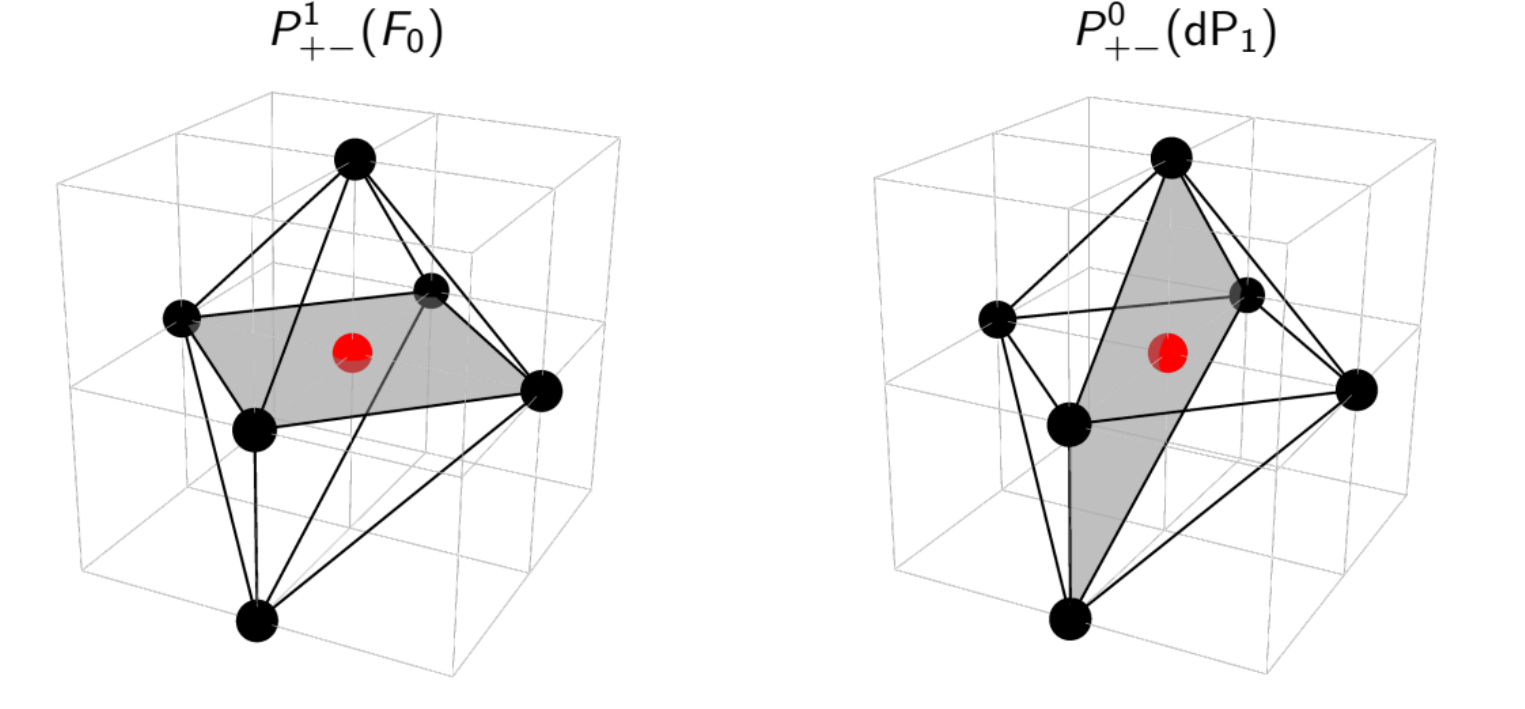}
}
\caption{
The toric diagram for $P^{1}_{+-}(F_0)$ is the same as $P^{0}_{+-}(\text{dP}_1)$.
\label{f_toricliftingf0dp1}
}
\end{center}
\end{figure}

In order to refer to toric Calabi-Yau 4-folds with reflexive polytopes $\Delta_3$ as their toric diagrams, we add two lattice points to a toric diagram $\Delta_2$ at height $z=0$. One point is added above $\Delta_2$ at height $z=1$ and another point is added below $\Delta_2$ at height $z=-1$. If $\Delta_2$ is a reflexive polygon in $2$-dimension then the resulting $\Delta_3$ from this construction is also a reflexive polygon. We denote such toric Calabi-Yau 4-folds as $P_{+-}^{i}(\text{CY}_3)$, where $i$ labels the $GL(3,\mathbb{Z})$-distinct combinations of adding a lattice point at height $z=1$ above and at height $z=-1$ below the toric diagram of $\text{CY}_3$. 
\fref{f_toriclifting} shows how Models 2 and 4 are respectively $P_{+-}^{0}(\text{dP}_0)$ and $P_{+-}^{1}(\text{dP}_0)$ and accordingly refer to examples of toric diagrams of
Calabi-Yau 4-folds obtained from the toric diagram of $\text{dP}_0$, which refers to both the $0$-th del Pezzo surface and the 3-dimensional Calabi-Yau cone over it. 

\fref{f_toriclifting} shows a generalization of this construction with Model 7, where starting from the toric diagram of $\text{dP}_0$, two lattice points are added at height $z=1$ and one at height $z=-1$, giving a regular reflexive polytope for a toric Calabi-Yau 4-fold that we call $P_{++-}(\text{dP}_0)$. The last example in \fref{f_toriclifting} is a case when starting with the toric diagram of $\text{dP}_0$, two lattice points are added at height $z=1$, one added at height $z=-1$ and one external point of the toric diagram of $\text{dP}_0$ lifted to height $z=1$. We refer to the toric Calabi-Yau 4-fold of the resulting regular reflexive polytope as $P_{++-}H_{+}(\text{dP}_0)$, where the added $H_{+}$ indicated the lifting of an external point of the toric diagram of $\text{dP}_0$. 

In this work, most toric Calabi-Yau 4-folds whose toric diagrams are regular reflexive lattice polytopes are of the form $P^{i}_{+-}(\text{CY}_3)$. For some cases, a given toric Calabi-Yau 4-fold $P^{i}_{+-}(\text{CY}_3)$ is the same as adding lattice points to another toric Calabi-Yau 3-fold $\text{CY}^\prime_3$, i.e. $P^{i}_{+-}(\text{CY}^\prime_3)$.
An example would be $P^{1}_{+-}(F_0)$, which is the same as $P^{0}_{+-}(\text{dP}_1)$ as illustrated in \fref{f_toricliftingf0dp1}. Here, $F_0$ refers to both the zeroth Hirzebruch surface and the Calabi-Yau cone over it similar to the first del Pezzo surface $\text{dP}_1$.

\end{itemize}

\section{Brane Brick Models and their Mesonic Moduli Space \label{sec:bbm}}

In the following section, we summarize the construction of brane brick models that correspond to $2d$ $(0,2)$ supersymmetric gauge theories corresponding to toric Calabi-Yau 4-folds. 
We discuss in particular the construction of the mesonic moduli space $\mathcal{M}^{mes}$ of these $2d$ $(0,2)$ supersymmetric gauge theories that exhibits interesting features for the case when the toric diagram for the corresponding Calabi-Yau 4-folds is a regular reflexive polytope.
For completeness, a brief review of brane brick models is presented here and the reader is referred to \cite{Franco:2015tna,Franco:2015tya} for more details. 
\\

\subsection{$2d$ $(0,2)$ theories and toric Calabi-Yau $4$-folds \label{sec:bbm1}}

\begin{figure}[ht!!]
\begin{center}
\resizebox{0.7\hsize}{!}{
\includegraphics[height=6cm]{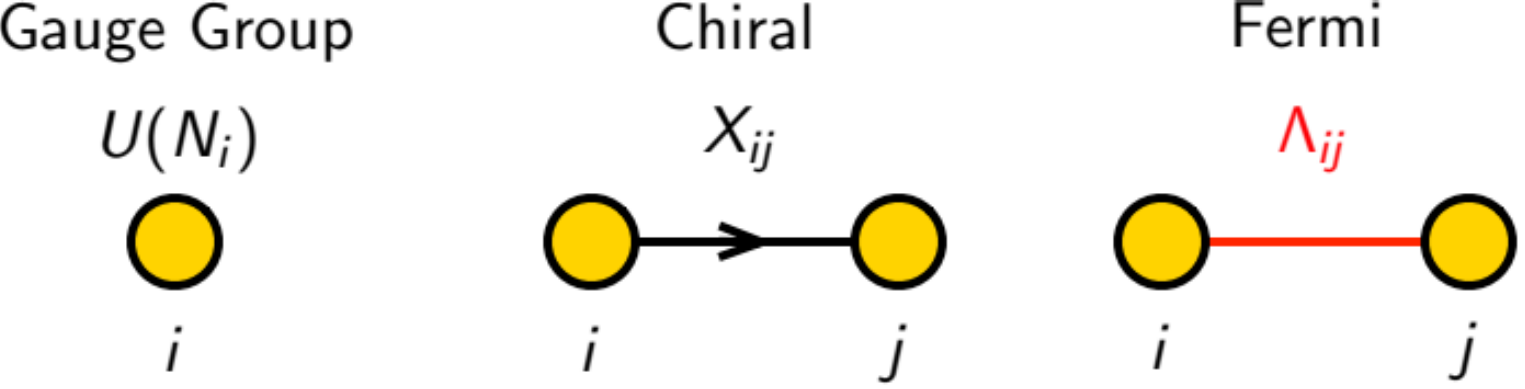} 
}
\caption{
Quiver representation of $U(N_i)$ gauge groups, chiral and Fermi fields. 
The subindices of field variables indicate the gauge nodes under which they transform. 
Fermi fields are not assigned an orientation due to the $\Lambda_{ij} \leftrightarrow \overline{\Lambda}_{ji}$ symmetry.
\label{f_quiverdic}}
 \end{center}
 \end{figure}

\begin{figure}[ht!!]
\begin{center}
\resizebox{0.5\hsize}{!}{
\includegraphics[height=6cm]{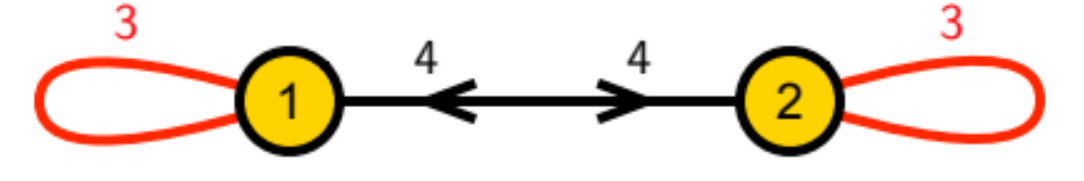} 
}
\caption{
Quiver for $\mathbb{C}^4/\mathbb{Z}_2~(1,1,1,1)$. The numbers next to edges correspond to the multiplicity of field between the adjacent nodes in the quiver diagram.
\label{f_quiversample}}
 \end{center}
 \end{figure}

\paragraph{Quiver.} The gauge symmetry and matter content of the $2d$ $(0,2)$ theories are encoded in a generalized quiver diagram.
The generalized quiver diagram contains two types of fields in the bifundamental or adjoint representation of the $U(N_i)$ gauge groups.
The fields are either chiral $X_{ij}$, represented by directed black arrows in the quiver diagram, or Fermi $\Lambda_{ij}$, represented by unoriented red lines in the quiver diagram. 
Fermi fields are not assigned an orientation, as illustrated in \fref{f_quiverdic}, due to the $\Lambda_{ij} \leftrightarrow \overline{\Lambda}_{ji}$ symmetry of $2d$ $(0,2)$ theories.
The nodes of the quiver correspond to the $U(N_i)$ gauge groups of the $2d$  $(0,2)$ theory.
\fref{f_quiversample} shows an example of a quiver diagram for the $2d$ $(0,2)$ theory corresponding to the Abelian orbifold of the form $\mathbb{C}^4/\mathbb{Z}_2$ $(1,1,1,1)$.

In this work, we focus on the case when all ranks of the gauge groups are equal, i.e. $N_i = N$. This simplifies the non-abelian $SU(N_i)^2$ anomaly cancellation conditions on the quiver \cite{Franco:2015tna,Franco:2015tya}.\footnote{For this class of $2d$ $(0,2)$ theories living on D1-branes probing Calabi-Yau singularities, Abelian gauge anomalies are cancelled by a generalized Green-Schwarz mechanism through interactions with bulk RR-field \cite{Mohri:1997ef}.}
It states that for each quiver node $i$, the number of connected arrows corresponding to chiral fields $n_i^\chi$ and Fermi fields $n_i^F$ need to satisfy the following condition
\beal{es02_01}
n_i^\chi - n_i^F = 2 ~.~
\eea
Adjoint chiral or Fermi fields contribute $2$ to $n_i^\chi$ and  $n_i^F$, respectively.

\paragraph{Binomial $J$- and $E$-terms.} 
Every Fermi field $\Lambda_{ij}$ corresponds to a pair of holomorphic functions of chiral fields, which we call $E_{ij}(X)$ and $J_{ji}(X)$. 
These functions are restricted to be binomial in the case when the probed Calabi-Yau 4-fold is toric. This is linked to the fact that ideals defining toric varieties are binomial prime ideals, which are precisely defined by the $J$- and $E$-terms of the $2d$ $(0,2)$ theory as we will see below in the discussion on moduli spaces.
For now, it is noted that this restriction was called the \textit{toric condition} in \cite{Franco:2015tna,Franco:2015tya} and implies the following general form of the $J$- and $E$-terms,
\beal{es02_02}
J_{ji} = J_{ji}^{+} - J_{ji}^{-} ~,~
E_{ij} = E_{ij}^{+} - E_{ij}^{-} ~,~
\eea 
where $J_{ji}^\pm$ and $E_{ij}^\pm$ are holomorphic monomials in chiral fields.
\\

As we will see in the following sections, this work will identify for each toric Calabi-Yau 4-fold whose toric diagram is a regular reflexive polygon a corresponding $2d$ $(0,2)$ theory, which is the worldvolume theory of the D1-brane probing the Calabi-Yau 4-fold. 
For each of the 18 regular reflexive polytopes, we will uniquely identify the corresponding $2d$ $(0,2)$ theory in terms of its $J$- and $E$-terms and its quiver diagram. 
\\

\subsection{Brane Brick Models \label{sec:bbm2}}

The $2d$ $(0,2)$ worldvolume theory of probe $D1$-branes on toric Calabi-Yau 4-folds can be represented in terms of a T-dual Type IIA brane configuration known as a \textit{brane brick model} \cite{Franco:2015tna,Franco:2015tya}.
Brane brick models are powerful tools to study $2d$ $(0,2)$ theories and corresponding toric Calabi-Yau 4-folds because they combine field theory information and information about the Calabi-Yau geometry in a single representation.
Accordingly, they play an analogous role for $2d$ $(0,2)$ theories and toric Calabi-Yau 4-folds as brane tilings do for $4d$ $\mathcal{N}=1$ theories and toric Calabi-Yau 3-folds \cite{Franco:2005rj,Hanany:2005ve,Hanany:2012hi}.

\begin{table}[ht!!]
\centering
\begin{tabular}{c|cccccccccc}
\; & 0 & 1 & 2 & 3 & 4 & 5 & 6 & 7 & 8 & 9\\
\hline
\text{D4} & $\times$ & $\times$ & $\times$ & $\cdot$ & $\times$ & $\cdot$ & $\times$ & $\cdot$ & $\cdot$ & $\cdot$ \\
\text{NS5} & $\times$ & $\times$ & \multicolumn{6}{c}{----------- \ $\Sigma$ \ ------------} & $\cdot$ & $\cdot$\\
\end{tabular}
\caption{
Brane brick models are Type IIA configurations where D4-branes are suspended from an NS5-brane that wraps a holomorphic surface $\Sigma$.
This configuration is T-dual to the D1-brane probing the toric Calabi-Yau 4-fold corresponding to $\Sigma$. 
}
\label{tbconfig}
\end{table}

\paragraph{Brane Configuration.} 
A brane brick model is a Type IIA brane configuration of D4-branes wrapping a 3-torus $T^3$ and suspended from a NS5-brane wrapping a holomorphic surface $\Sigma$. 
The holomorphic surface $\Sigma$ is the zero locus of the Newton polynomial corresponding to the toric Calabi-Yau 4-fold,
\beal{es02_10}
\sum_{(a,b,c)\in V} c_{(a,b,c)} x^a y^b z^c = 0 ~,~
\eea
where $c_{(a,b,c)}$ take values in $\mathbb{C}^*$ and $V$ is the set of lattice points in the toric diagram $\Delta_3$ of the toric Calabi-Yau 4-fold. 
The intersection between the holomorphic surface $\Sigma$ and the 3-torus $T^3$ is precisely where the D4-brane meets the NS5-brane as summarized in \tref{tbconfig}.

\begin{figure}[ht!!]
\begin{center}
\resizebox{0.9\hsize}{!}{
\includegraphics[height=6cm]{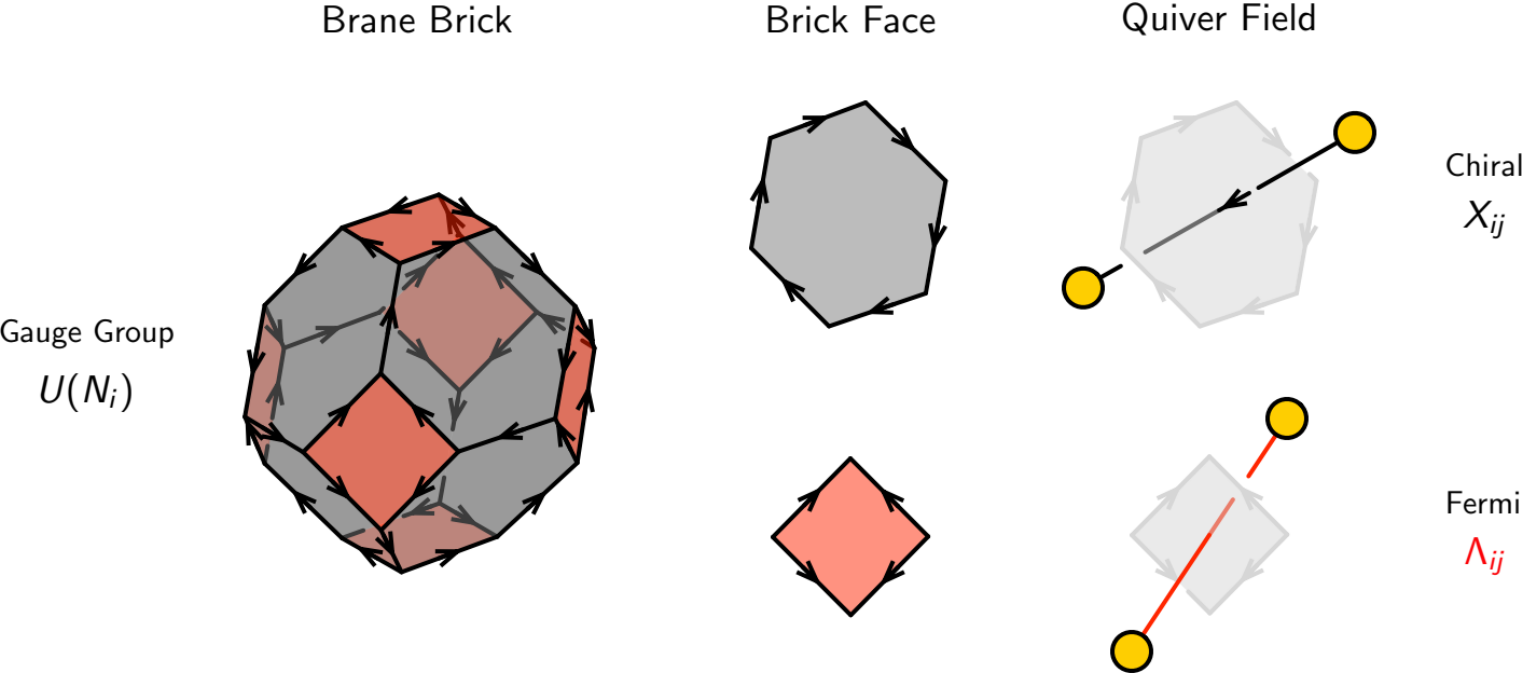} 
}
\caption{
$2d$ brick faces in the brane brick model can be systematically oriented along their boundary edges such that oriented even-sided faces correspond to chiral fields and unoriented 4-sided faces correspond to Fermi fields. 
\label{f_branebrickdic}}
 \end{center}
 \end{figure}

The intersections create a tessellation of the 3-torus $T^3$ which we call as the \textit{brane brick model}. 
For simplicity, we can replace $\Sigma$ by its simpler skeleton diagram that consists of $2d$ faces that indicate the locations of the NS5-brane wrapping $\Sigma$.
These $2d$ faces separate $T^3$ into $3d$ polytopes filled by D4-branes. 
We call these $3d$ polytopes as \textit{bricks}.

\paragraph{Dictionary.}
The brane brick model on $T^3$ consists of the following fundamental components:
\begin{itemize}
\item \textbf{Bricks.} Bricks are 3-dimensional polytopes that tessellate the 3-torus $T^3$. Each brick corresponds to a $U(N_i)$ gauge group of the $2d$ theory. As the 3-dimensional generalization of a brane interval, its interior indicates the location of the D4-branes suspended between the NS5-brane wrapping $\Sigma$.

\item \textbf{Faces.} The 2-dimensional brick faces are even-sided and can be oriented systematically along their boundary edges such they are either \textit{oriented} or \textit{unoriented}. Oriented and unoriented faces correspond to bifundamental (or adjoint) chiral and Fermi fields, respectively. Faces corresponding to Fermi fields are 4-sided. The two bricks adjacent to a given face correspond to the gauge groups under which the corresponding field transforms. \fref{f_branebrickdic} shows how the two different types of faces correspond to chiral and Fermi fields in the $2d$ theory. 

\begin{figure}[ht!!]
\begin{center}
\resizebox{0.58\hsize}{!}{
\includegraphics[height=6cm]{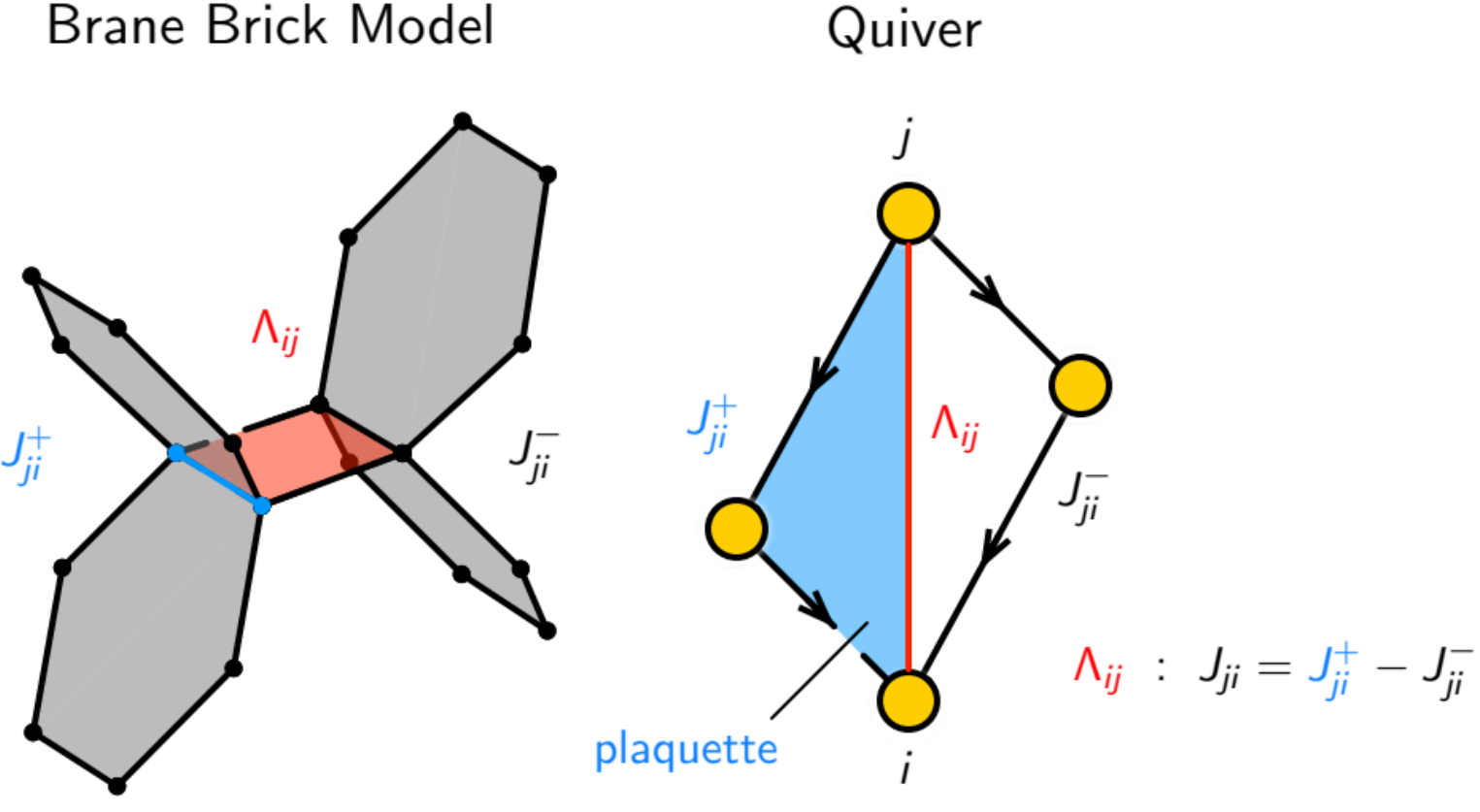} 
}
\caption{
Edges in the brane brick model correspond to monomials in the $J$- and $E$-terms of the $2d$ theory. 
The corresponding collection of chiral fields with the associated Fermi field forms what we call a \textit{plaquette} (blue). Each Fermi face with its four edges is associated with four plaquettes that form the $J$- and $E$-terms corresponding to the Fermi field.
\label{f_branebrickdic2}}
 \end{center}
 \end{figure}

\item \textbf{Edges.} Edges in a brane brick model are always adjacent to a Fermi face and a number of chiral faces. We call the collection of chiral fields and the Fermi field associated to the faces that coincide at a given edge a \textit{plaquette}, as illustrated in \fref{f_branebrickdic2}.
The chiral fields in a plaquette form one of the monomials $J_{ji}^\pm$ and $E_{ij}^\pm$ in the $J$- and $E$-terms of the $2d$ theory. 
Every Fermi face with its four edges is associated with four plaquettes that form the $J$- and $E$-terms corresponding to the Fermi field.
Opposite edges of a 4-sided Fermi face form a $J$- (or $E$-) term while the other pair of opposite edges form a $E$- (or $J$-) term, as illustrated in \fref{f_branebrickdic2}.

\end{itemize}
The complete dictionary between the $2d$ gauge theory and the brane brick model is summarized in \tref{tbrick}. For a thorough discussion of these constructions and generalizations beyond $T^3$ see \cite{Franco:2015tya,Franco:2021elb}.

\begin{table}[h]
\centering
\resizebox{0.9\hsize}{!}{
\begin{tabular}{|l|l|}
\hline
\ \ \ \ \ \ {\bf Brane Brick Model} \ \ \ \ \ & \ \ \ \ \ \ \ \ \ \ \ \ \ \ \ \ \ \ \ \ {\bf Gauge Theory} \ \ \ \ \ \ \ \ \ \ \ \ 
\\
\hline\hline
Brick  & Gauge group \\
\hline
Oriented face between bricks & Chiral field in the bifundamental representation \\
$i$ and $j$ & of nodes $i$ and $j$ (adjoint for $i=j$) \\
\hline
Unoriented square face between & Fermi field in the bifundamental representation \\
bricks $i$ and $j$ & of nodes $i$ and $j$ (adjoint for $i=j$) \\
\hline
Edge  & Plaquette encoding a monomial in a \\ 
& $J$- or $E$-term \\
\hline
\end{tabular}
}
\caption{
Dictionary between brane brick models and $2d$ gauge theories.
\label{tbrick}
}
\end{table}

In this work, we will identify one brane brick model representing a $2d$ $(0,2)$ theory for each of the toric Calabi-Yau 4-folds which have regular reflexive polytopes as their toric diagrams.\footnote{Generically, multiple brane brick models can be associated to a given toric Calabi-Yau 4-fold (see e.g. \cite{Franco:2016fxm}). Such brane brick models correspond to IR equivalent $2d$ $(0,2)$ gauge theories related by triality \cite{Franco:2016nwv}.  An exhaustive classification of the triality dual phases for each of the Calabi-Yau 4-folds we consider is beyond the scope of this paper.} 
We will identify the $J$- and $E$-terms as well as the quiver diagram which define the brane brick model as well as the corresponding $2d$ $(0,2)$ theory. 
\\

\subsection{The Mesonic Moduli Space and the Forward Algorithm \label{sec:bbm3}}

The Calabi-Yau 4-fold geometry that is probed by the D1-branes is related to the vacuum moduli space of the $2d$ gauge theory that lives on the worldvolume of the D1-branes. 
We call this particular vacuum moduli space the \textit{mesonic moduli space} $\mathcal{M}^{mes}$ of the brane brick model.
The mesonic moduli space is the toric Calabi-Yau 4-fold probed by the D1-brane when we consider the Abelian $2d$ theory with $U(1)$ gauge groups.
As usual, for $2d$ theories we have in mind the classical moduli spaces, which should be regarded, in the spirit of the Born-Oppenheimer approximation, as target spaces of non-linear sigma models.

\paragraph{Mesonic Moduli Space $\mathcal{M}^{mes}$}. The mesonic moduli space is determined by the $J$- and $E$-term constraints as well as the $D$-term constraints of the $2d$ $(0,2)$ theory. Let us summarize the properties and definition of the mesonic moduli space as follows:

\begin{itemize}
\item The $J$- and $E$-terms of the Abelian $2d$ $(0,2)$ theory form a binomial ideal of the following form
\beal{es03_00}
\mathcal{I}_{J=0, E=0} = \langle J_{ji}^{+} - J_{ji}^{-} = 0, E_{ij}^{+} - E_{ij}^{-} = 0 \rangle ~,~
\eea
where as discussed in Section \sref{sec:bbm1} $J_{ji}^\pm$ and $E_{ij}^\pm$ are monomials in chiral fields. 
We note that the quotient ring of the form 
\beal{es03_01}
R_{X} = \mathbb{C}[X_{ij}] / \mathcal{I}_{J=0, E=0}
\eea
captures the essence of a toric variety $X$ and we call it the coordinate ring of $X$.
We note that 
\beal{es03_02}
\mathcal{F}^\flat = \text{Spec}( \mathbb{C}[X_{ij}] / \mathcal{I}_{J=0, E=0} )~.~
\eea
We call $\mathcal{F}^\flat$ the \textit{master space} of the corresponding brane brick model.\footnote{The concept of master space was introduced in \cite{Forcella:2008eh,Forcella:2008bb} for $4d$ $\mathcal{N}=1$ gauge theories but, as mentioned above, can be naturally generalized to supersymmetric gauge theories in other dimensions \cite{Franco:2015tna}.}

Usually, the ideal $\mathcal{I}_{J=0, E=0}$ is reducible into irreducible components, which are known as primary ideals.
Taking the largest of these primary ideals in \eref{es03_02}, we obtain the \textit{coherent component} of the master space which we denote as ${}^{Irr}\mathcal{F}^\flat$.
In the following discussion, we will concentrate on the coherent component of the master space and for simplicity use ${}^{Irr}\mathcal{F}^\flat$ and $\mathcal{F}^\flat$ interchangeably for the Abelian $2d$ $(0,2)$ theories. 

\item The \textit{mesonic moduli space} of the one D1-brane theory is related to the master space ${}^{Irr}\mathcal{F}^\flat$.
It is obtained by quotienting out the $U(1)^G$ gauge charges, where $G$ is the number of gauge groups in the brane brick model. 
It is important to note that an overall $U(1)$ decouples, giving a total of $U(1)^{G-1}$ independent charges.
Accordingly, the mesonic moduli space takes the following form 
\beal{es03_01}
\mathcal{M}^{mes} = {}^{Irr}\mathcal{F}^\flat // U(1)^{G-1}~.~
\eea
The mesonic moduli space $\mathcal{M}^{mes}$ is a toric Calabi-Yau 4-fold for brane brick models. It is exactly the same Calabi-Yau that is probed by the D1-branes whose worldvolume theory is the $2d$ $(0,2)$ theory given by our brane brick model.

The dimension of the mesonic moduli space $\mathcal{M}^{mes}$ can be derived by starting with the number of chiral fields $n^\chi$. The $J$- and $E$-terms impose $n^F-3$ independent constraints where $n^F$ is the number of Fermi fields in the brane brick model. Further restrictions come from demanding invariance under the $G-1$ independent gauge charges and the anomaly cancellation condition from \eref{es02_01} that sets $n^\chi - n^F = G$. 
By combining all these constraints, one obtains the dimension of the mesonic moduli space $\mathcal{M}^{mes}$ to be $n^\chi - (n^F - 3) - (G-1) = 4$ as expected.

\end{itemize}

In the following work, we concentrate on brane brick models and corresponding $2d$ $(0,2)$ theories whose mesonic moduli space $\mathcal{M}^{mes}$ is a toric Calabi-Yau 4-fold that has a regular reflexive lattice polytope as its toric diagram.
In order to illustrate how the toric diagram can be systematically obtained from the brane brick model, a brief review on the \textit{forward algorithm} \cite{Franco:2015tya} which translates the $2d$ gauge theory information into toric data is given below.

\paragraph{Forward Algorithm.}
The forward algorithm for brane brick models involves the following steps:
\begin{itemize}
\item \textbf{$K$-matrix.} Let us denote by $X_m$ with $m=1,\dots, n^\chi$ the chiral fields of the brane brick model. 
The space of solutions of the $J$- and $E$-terms of the brane brick model can be expressed in terms of $G+3$ independent chiral fields, which we label by $v_k$.
Accordingly, with $G$ being the number of gauge groups, the chiral fields can be expressed as follows
\beal{es03_10}
X_m  = \prod_{k} v_k^{K_{mk}} ~,~
\eea
where $m=1,\dots, n^\chi$ and $k=1,\dots, G+3$. 
$K$ is a $n^\chi \times (G+3)$-dimensional matrix that encodes the relations from the vanishing $J$- and $E$-terms of the brane brick model. 

\item \textbf{$P$-matrix and Brick Matchings.} 
We note that the $K$-matrix can contain negative integers as entries meaning that chiral fields $X_m$ in the brane brick model sometimes are expressed in terms of negative powers of independent fields $v_k$.
In order to avoid negative entries in $K$,
we define a new matrix $T$ to be the space of vectors dual to $K$ as follows
\beal{es03_11}
K \cdot T \geq 0 ~.~
\eea
Using $T$, we can now express in terms of the independent chiral fields $v_k$ a new set of fields $p_\alpha$ as follows
\beal{es03_12}
v_k = \prod_\alpha p_{\alpha}^{T_{k \alpha}} ~,~
\eea
where $\alpha=1,\dots, c$.
The $p_\alpha$ are interpreted as GLSM fields \cite{Witten:1993yc} in the toric description of the mesonic moduli space $\mathcal{M}^{mes}$ of the brane brick model. 

Using \eref{es03_10} and \eref{es03_12}, all chiral fields $X_m$ of the brane brick model can be expressed in terms of GLSM fields as follows
\beal{es03_13}
X_m = \prod_\alpha p_\alpha^{P_{m\alpha}}
\eea
where the $(n^\chi \times c)$-dimensional $P$-matrix is given by 
\beal{es03_14}
P_{n^\chi \times c} = K_{n^\chi \times (G+3)} \cdot T_{(G+3) \times c} ~.~
\eea
The labels of the matrices above indicate their dimensions. We also note that the entries of the $P$-matrix are strictly greater or equal to zero. 

In the brane brick model, the GLSM fields encoded in the $P$-matrix in terms of chiral fields have an additional combinatorial meaning. The collection of chiral fields associated to a GLSM field is a special collection of fields that cover every plaquette in the brane brick model exactly once \cite{Franco:2015tya}. 
We call this collection of fields also a \textit{brick matching}.

\item \textbf{$Q_{JE}$-matrix.}
The relations between chiral fields of the brane brick model are given by the $J$- and $E$-terms.
These relations can be expressed in terms of a collection of $U(1)$-charges to the GLSM fields that form a new basis of fields parameterizing the $J$- and $E$-terms.
These charges are given by a $((c-(G+3))\times c)$-dimensional charge matrix called the $Q_{JE}$-matrix, which is the kernel of the $P$-matrix as follows
\beal{es03_15}
(Q_{JE})_{(c-(G+3))\times c} = \ker (P) ~.~
\eea

\item \textbf{$Q_{D}$-matrix.}
In order to obtain the mesonic moduli space of the abelian $2d$ theory corresponding to a brane brick model, we have to impose the $D$-terms.
Accordingly, we need to express the $U(1)$ gauge charges on chiral fields as $U(1)$ charges on the GLSM fields. 
The $U(1)$ gauge charges on chiral fields are given by the $(G\times n^\chi)$-dimensional quiver incidence matrix $d$, where $G$ is the number of nodes in the quiver. 
Because all chiral fields are either in the bifundamental or adjoint representation, the incidence matrix satisfies
\beal{es03_16}
\sum_a d_{ai} = 0 ~,~
\eea
where $a=1,\dots, G$ and $i=1,\dots, n^\chi$. 
Hence, only $G-1$ rows of the incidence matrix $d$ are independent allowing us to reduce it to a $((G-1)\times n^\chi)$-dimensional matrix $\overline{d}$. 
Using the reduced incidence matrix $\overline{d}$, the $U(1)$ charges on the GLSM fields $p_\alpha$ fields can be summarized in a $((G-1)\times c)$-dimensional matrix $Q_{D}$.
The $Q_D$-matrix can be obtained from the following relation
\beal{es03_17}
\overline{d}_{(G-1)\times n^\chi} = (Q_D)_{(G-1)\times c} \cdot P^t_{c\times n^\chi} ~.~
\eea

\item \textbf{Toric Diagram.}
The $Q_{JE}$- and $Q_D$- matrices contain the $U(1)$-charges on the GLSM fields which are associated with the vanishing $J$- and $E$-terms as well as the $D$-terms of the brane brick model, respectively. 
The charge matrices $Q_{JE}$ and $Q_D$ can be combined into the following total charge matrix
\beal{es03_20}
(Q_t)_{(c-4)\times c} = \left(
(Q_{JE})_{(c-(G+3))\times c} ~,~
(Q_D)_{(G-1)\times c}
\right)
~.~
\eea
The total charge matrix $Q_t$ is $((c-4)\times c)$-dimensional where $c$ is the number of GLSM fields.

The kernel of the total charge matrix $Q_t$ is $(4\times c)$-dimensional,
\beal{es03_21}
G_t = \ker(Q_t) ~,~
\eea
and encodes the toric diagram of the toric Calabi-Yau 4-fold that is the mesonic moduli space $\mathcal{M}^{mes}$ of the brane brick model. 
Every column of $G_t$ corresponds to a GLSM field and a brick matching of the brane brick model and determines the position of a point in the $\mathbb{Z}^4$-lattice.
The convex hull of the points forms the toric diagram $\Delta$ of the toric Calabi-Yau 4-fold. One can find a suitable $GL(4,\mathbb{Z})$ transformation of the coordinates of the points encoded in $G_t$ such that all points lie on the 3-dimensional hyperplane in $\mathbb{Z}^4$, giving us the 3-dimensional lattice polytope $\Delta$. Multiple GLSM fields may be mapped to the same point in the toric diagram.

\end{itemize}

One of the aims of our work is to identify for each toric diagram which is a regular reflexive lattice polytope a corresponding brane brick model. 
As a result, we expect to be able to identify 18 distinct brane brick models whose mesonic moduli space is a toric Calabi-Yau 4-fold with a toric diagram that is one of the 18 regular reflexive polytopes.

The definition of the mesonic moduli space $\mathcal{M}^{mes}$ in \eref{es03_01} and the formula for the toric diagram encoded in $G_t$ in \eref{es03_21} both refer to the same toric Calabi-Yau 4-fold associated to a brane brick model. 
Before we proceed, it is interesting to point out that for the $2d$ $(0,2)$ gauge theories associated to toric Calabi-Yau 4-folds, the forward algorithm often results in additional GLSM fields that we call as \textit{extra GLSM fields}. 

\paragraph{Extra GLSM fields.}
In some cases, the forward algorithm leads to extra GLSM fields in the $P$-matrix in \eref{es03_14}.
These in turn manifest themselves as additional points in the toric diagram given by the $G_t$-matrix in \eref{es03_21}.
These points lie outside the 3-dimensional hyperplane of the 3-dimensional toric diagram.

It is important to note that these extra GLSM fields are \textit{redundant} for the description of the mesonic moduli space $\mathcal{M}^{mes}$, meaning that the toric diagram without the points corresponding to the extra GLSM fields is the correct toric diagram for the toric Calabi-Yau 4-fold.
This is because these extra GLSM fields correspond to an over-parameterization of the mesonic moduli space $\mathcal{M}^{mes}$ \cite{Franco:2015tna}. 
While normally, the mesonic moduli space $\mathcal{M}^{mes}$ is parameterized by mesonic gauge invariant operators formed by the chiral fields, which form the spectrum of operators for the quotient in \eref{es03_01}, the presence or absence of the extra GLSM fields does not affect the spectrum of gauge invariant operators.
In fact, if we describe the mesonic moduli space $\mathcal{M}^{mes}$ in terms of mesonic gauge invariant operators that generate the entire spectrum of operators as well as their defining relations, then the extra GLSM fields do not affect the generators as well as the defining relations amongst them, leaving the mesonic moduli space $\mathcal{M}^{mes}$ unaffected. 

The \textit{Hilbert series} \cite{Benvenuti:2006qr,Hanany:2006uc,Butti:2007jv} is an important tool to characterize the spectrum of gauge invariant operators and hence the moduli space of a gauge theory. 
In the following sections, we will show whenever extra GLSM fields are present, that their removal does not affect the algebraic properties of the mesonic moduli space $\mathcal{M}^{mes}$ of a brane brick model using the Hilbert series.
As a result, the following toric diagrams that we calculate using the forward algorithm will only contain points corresponding to normal GLSM fields and have all points corresponding to extra GLSM fields removed. 
Let us in the following section discuss the calculation of Hilbert series for the mesonic moduli space $\mathcal{M}^{mes}$ and how it can be used to characterize the algebraic structure of $\mathcal{M}^{mes}$ for brane brick models.
\\

\subsection{Hilbert Series and Plethystics}

\paragraph{Hilbert Series.} 
The mesonic moduli space $\mathcal{M}^{mes}$ of a brane brick model is the space of gauge invariant operators under $J$- and $E$-term charges $Q_{JE}$ and $D$-term charges $Q_D$.
The \textit{Hilbert series} is a generating function that counts gauge invariant operators \cite{Benvenuti:2006qr,Hanany:2006uc,Butti:2007jv} of a moduli space. 
It contains information about the moduli space generators and the defining relations that they form amongst themselves. 
For charges $Q= (Q_{JE}, Q_D)$, the Hilbert series for the mesonic moduli space $\mathcal{M}=\mathcal{M}^{mes}$ is given by the \textit{Molien integral}
\beal{es04_01}
g_1(y_\alpha; \mathcal{M}) =
\prod_{i=1}^{|Q|} \oint_{|z_i|=1} \frac{\ud z_i}{2\pi i z_i} \prod_{\alpha=1}^c \frac{1}{1- y_\alpha \prod_{j=1}^{|Q|} z_j^{Q_{j\alpha}}}~,~
\eea
where $c$ is the number of brick matchings corresponding to the GLSM fields of the brane brick model and $|Q|$ is the number of rows in the total charge matrix $Q_t$. 

The fugacity $y_\alpha$ counts GLSM fields of the brane brick model and we set it to be $y_\alpha = t_i$ if the GLSM field corresponds to an extremal point $p_i$ of the toric diagram of the Calabi-Yau 4-fold. Furthermore, we set the fugacity to be $y_\alpha = y_{s_m}$ if it corresponds to brick matchings $s_m$ corresponding to the single internal point of the reflexive toric diagram. 
Finally, we set $y_\alpha = y_{o_k}$ if the GLSM field is an extra GLSM field that does not contribute to the algebraic structure of the moduli space. 
In fact, in the following discussion of brane brick models corresponding to regular reflexive polytopes, we see that setting the fugacities for extra GLSM fields $y_{o_k}=1$ does not change the algebraic structure of the mesonic moduli space captured by the Hilbert series. This will be indicated in the Hilbert series calculation when needed.

\paragraph{Plethystics.} 
The moduli space is specified by its generators and defining relations formed by the generators. In order to obtain information about the generators and relations amongst them, we make use of the \textit{plethystic logarithm} of the Hilbert series \cite{Feng:2007ur,hanany2007counting}.
The plethystic logarithm takes the form
\beal{es04_10}
\text{PL} \left[
g_1(y_\alpha; \mathcal{M}) 
\right]
= \sum_{k=1}^{\infty}
\frac{\mu(k)}{k}
\log\left[
g_1(y_\alpha^k; \mathcal{M}) 
\right]~,~
\eea
where $\mu(k)$ is the M\"obius function.
If the expansion of the plethystic logarithm is \textit{finite}, the corresponding moduli space is known to be a \textit{complete intersection} generated by a finite number of generators subject to a finite number of relations. 
The first positive terms of the finite expansion correspond to the counting of generators while the following negative terms correspond to the counting of the defining relations formed amongst them. 
On the other hand, if the expansion of the plethystic logarithm is infinite, 
the moduli space is known to be a \textit{non-complete intersection}. The first positive terms of the expansion again refer to generators of the moduli space while all higher order terms refer to relations amongst generators and relations amongst relations known as \textit{syzygies}. 

In the following work, the aim will be to identify the generators of the mesonic moduli space of brane brick models corresponding to toric Calabi-Yau 4-folds whose toric diagram is one of the 18 regular reflexive lattice polytopes in 3 dimensions. 
Using the plethystic logarithm of the Hilbert series of the mesonic moduli space, which is in terms of fugacities of the GLSM fields corresponding to brick matchings of the brane brick model, we are able to write the generators of the mesonic moduli space in terms of GLSM fields. 
Since the GLSM fields are themselves related to the chiral fields of the corresponding $2d$ gauge theory, which we know are subject to the $J$- and $E$-term constraints, we are not only able to write the generators of the mesonic moduli space in terms of chiral fields, but also reconstruct the different $J$- and $E$-term equivalent expressions of the generators in terms of chiral fields.
\\

\subsection{Duality between Generator Lattices and Toric Diagrams}

\begin{table}[ht!!]
\centering
\begin{tabular}{|c|c|}
\hline
Model & Global Symmetry
\\
\hline\hline
1 & 
$SU(4)\times U(1)$
\\
\hline
2 & 
$SU(3) \times SU(2) \times U(1)$
\\
\hline
3 & 
$SU(3) \times U(1) \times U(1)$
\\
\hline
4 & 
$SU(2)\times SU(2) \times U(1) \times U(1)$
\\
\hline
5 & 
$SU(3) \times U(1) \times U(1)$
\\
\hline
6 & 
$SU(2)\times SU(2)\times U(1) \times U(1)$
\\
\hline
7 & 
$SU(2)\times U(1) \times U(1)\times U(1)$
\\
\hline
\end{tabular}
\caption{
The global symmetries for the 18 models that can be obtained from the isometries of the toric Calabi-Yau 4-folds. 
\label{t_globalsym}}
\end{table}

\paragraph{Global Symmetries.}
From the isometry of the mesonic moduli space, which is a toric Calabi-Yau 4-fold, we can obtain the global symmetry of the brane brick model which contains $U(1)^4$. 
A linear combination of the four $U(1)$'s relates to the $R$-symmetry of $(0,2)$ supersymmetry. 
The non-R $U(1)^3$ symmetry is known as the \textit{mesonic flavor symmetry}.
For some mesonic moduli spaces, the mesonic flavor symmetry is enhanced to a non-abelian group.
In \tref{t_globalsym}, we summarize the global symmetries of the 18 brane brick models corresponding to regular reflexive polytopes.
The Hilbert series of the mesonic moduli space can be refined in terms of fugacities that count the global symmetry charges carried by each of the gauge invariant operators. 
This can be done by mapping the fugacities counting GLSM fields $y_\alpha$ into fugacities that count the global symmetry charges $(x_1,x_2,x_3,t)$.
Here, it is important to note that only the GLSM fields corresponding to extremal points in the toric diagram carry non-zero charges while all other GLSM fields carry no charges under the global symmetry. 

In the following discussion, without loss of generality, we restrict ourselves to fugacities in the Hilbert series that count the mesonic flavor charges $(x_1,x_2,x_3)$ and an additional fugacity $t_\alpha$ that instead of the $R$-symmetry charge counts the degree in GLSM fields $p_\alpha$ corresponding to the extremal toric points for each gauge invariant operator. 
This selection of independent fugacities $(x_1,x_2,x_3,t_\alpha)$ for each gauge invariant operator counted by the Hilbert series encodes the rank 4 global symmetry of the brane brick model.
Furthermore, this choice of fugacities enables us to identify for each gauge invariant operator of the mesonic moduli space their mesonic flavor charges through fugacities $(x_1,x_2,x_3)$ whilst still containing information about the composition of each gauge invariant operator in terms of GLSM fields through the fugacities $t_\alpha$. 

\paragraph{Lattice of Generators.}
The lattice of generators is formed by the mesonic flavor charges carried by the generators of the mesonic moduli space. 
The non-R mesonic flavor symmetry has rank 3 and the lattice that is formed by the charges is 3-dimensional.
By a suitable global scaling of the mesonic charges, the charges carried by the gauge invariant operators can be made to be in $\mathbb{Z}^3$.
There is only a finite number of generators for each mesonic moduli space of a brane brick model. 
Each integer 3-vector of mesonic flavor charges associated to a generator can be considered as the coordinates of a point on the $\mathbb{Z}^3$ lattice. 
The set of lattice points for the generators of the mesonic moduli space forms a convex polytope, which we refer from now on as the \textit{generator lattice} of the brane brick model.

\paragraph{Duality between Generator Lattices and Toric Diagrams.}
\begin{center}
\textit{The generator lattice of a brane brick model is the dual of the toric diagram.}
\end{center}
If the toric diagram is a reflexive polytope, then by duality of reflexive polytopes as stated in \eref{es00_1}, the generator lattice is also a convex lattice polytope that is reflexive. 
Accordingly, we are closing the circle of our discussion by stating that for brane brick models with toric diagrams that are reflexive polytopes, the corresponding mesonic moduli space has generator lattices which are the reflexive dual of the toric diagrams up to $GL(3,\mathbb{Z})$ isomorphism.

In the following sections, brane brick models corresponding to all 18 regular reflexive polytopes are classified. 
Furthermore, it is shown that the mesonic moduli spaces of the brane brick models have generator lattices that are reflexive polytopes which are reflexive dual to the toric diagrams of the brane brick models. 
In order to show this, we calculate the Hilbert series for each brane brick model and through plethystics write down the set of generators with their mesonic flavor charges and chiral field content. 
This is the first time \textit{all} Fano 3-folds and the associated Calabi-Yau 4-folds corresponding to the 18 regular reflexive polytopes have been systematically associated to quiver gauge theories realized in string theory. 
\\

\section{Model 1: $\mathbb{C}^4/\mathbb{Z}_4 ~(1,1,1,1)$~[$\mathbb{P}^3$,~$\langle0\rangle$] \label{smodel01}}
 
\begin{figure}[H]
\begin{center}
\resizebox{0.25\hsize}{!}{
\includegraphics[height=6cm]{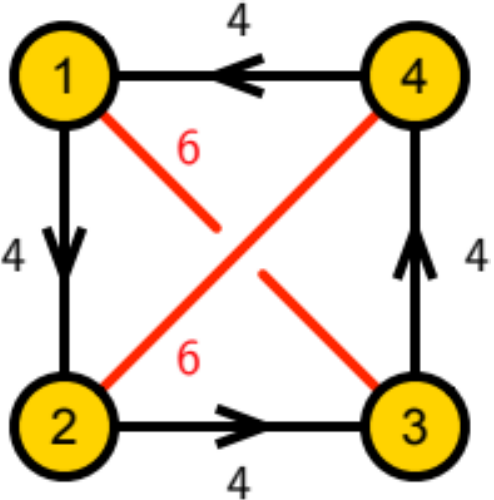} 
}
\caption{
Quiver for Model 1.
\label{f_quiver_01}}
 \end{center}
 \end{figure}
 
Model 1 corresponds to the Abelian orbifold of the form $\mathbb{C}^4/\mathbb{Z}_4 ~(1,1,1,1)$.
The corresponding brane brick model has the quiver in \fref{f_quiver_01} and the $J$- and $E$-terms are given as follows
  \beq
  {\footnotesize
\begin{array}{rrclrcl}
 &  & J &  & & E &  \\
\Lambda_{13}^{1}  : & Z_{34} Y_{41}  &-& Y_{34} Z_{41} &   P_{12} X_{23}  &-&  X_{12} P_{23} \\ 
\Lambda_{13}^{2}  : & X_{34} Z_{41}  &-&  Z_{34} X_{41}  &   P_{12} Y_{23}  &-&  Y_{12} P_{23} \\ 
\Lambda_{13}^{3}  : & Y_{34} X_{41}  &-& X_{34} Y_{41}  &   P_{12} Z_{23}  &-&  Z_{12} P_{23} \\ 
\Lambda_{24}^{1}  : & Z_{41} Y_{12}  &-& Y_{41} Z_{12}  &   P_{23} X_{34}  &-&  X_{23} P_{34} \\ 
\Lambda_{24}^{2}  : & X_{41} Z_{12}  &-&  Z_{41} X_{12}  &   P_{23} Y_{34}  &-&  Y_{23} P_{34} \\ 
\Lambda_{24}^{3}  : & Y_{41} X_{12}  &-& X_{41} Y_{12} &   P_{23} Z_{34}  &-&  Z_{23} P_{34} \\ 
\Lambda_{31}^{1}  : & Z_{12} Y_{23}  &-& Y_{12} Z_{23}  &   P_{34} X_{41}  &-&  X_{34} P_{41} \\ 
\Lambda_{31}^{2}  : & X_{12} Z_{23}  &-&  Z_{12} X_{23}  &   P_{34} Y_{41}  &-&  Y_{34} P_{41} \\ 
\Lambda_{31}^{3}  : & Y_{12} X_{23}   &-& X_{12} Y_{23} &   P_{34} Z_{41}  &-&  Z_{34} P_{41} \\ 
\Lambda_{42}^{1}  : & Z_{23} Y_{34}  &-& Y_{23} Z_{34}  &   P_{41} X_{12}  &-&  X_{41} P_{12} \\ 
\Lambda_{42}^{2}  : & X_{23} Z_{34}  &-&  Z_{23} X_{34}  &   P_{41} Y_{12}  &-&  Y_{41} P_{12} \\ 
\Lambda_{42}^{3}  : &  Y_{23} X_{34} &-& X_{23} Y_{34}  &   P_{41} Z_{12}  &-&  Z_{41} P_{12}
 \end{array} 
 }~.~
\label{E_J_C_+-}
 \eeq
 \\
 
Through the forward algorithm, the $P$-matrix of Model 1 can be calculated as follows
\beal{es0105}
{\footnotesize
 P=
\left(
\begin{array}{r|cccc|cccc}
\; & p_1 & p_2 & p_3 & p_4 & s_1 & s_2 & s_3 & s_4 \\
\hline
P_{12} & 1 & 0 & 0 & 0 & 1 & 0 & 0 & 0 \\
P_{23} & 1 & 0 & 0 & 0 & 0 & 1 & 0 & 0 \\
P_{34} & 1 & 0 & 0 & 0 & 0 & 0 & 1 & 0 \\
P_{41} & 1 & 0 & 0 & 0 & 0 & 0 & 0 & 1 \\
\hline
X_{12} & 0 & 1 & 0 & 0 & 1 & 0 & 0 & 0 \\
X_{23} & 0 & 1 & 0 & 0 & 0 & 1 & 0 & 0 \\
X_{34} & 0 & 1 & 0 & 0 & 0 & 0 & 1 & 0 \\
X_{41} & 0 & 1 & 0 & 0 & 0 & 0 & 0 & 1 \\
 \hline
Y_{12} & 0 & 0 & 1 & 0 & 1 & 0 & 0 & 0 \\
Y_{23} & 0 & 0 & 1 & 0 & 0 & 1 & 0 & 0 \\
Y_{34} & 0 & 0 & 1 & 0 & 0 & 0 & 1 & 0 \\
Y_{41} & 0 & 0 & 1 & 0 & 0 & 0 & 0 & 1 \\
 \hline
Z_{12} & 0 & 0 & 0 & 1 & 1 & 0 & 0 & 0 \\
Z_{23} & 0 & 0 & 0 & 1 & 0 & 1 & 0 & 0 \\
Z_{34} & 0 & 0 & 0 & 1 & 0 & 0 & 1 & 0 \\
Z_{41} & 0 & 0 & 0 & 1 & 0 & 0 & 0 & 1 \\
\end{array}
\right)
}~.~
\eea

The $J$- and $E$-term charges are given by
\beal{es0105}
{\footnotesize 
Q_{JE} =
\left(
\begin{array}{cccc|cccc}
p_1 & p_2 & p_3 & p_4 & s_1 & s_2 & s_3 & s_4 \\
\hline
 1 & 1 & 1 & 1 & -1 & -1 & -1 & -1 \\
\end{array}
\right)
}~,~
\eea
and the $D$-term charges are given by
\beal{es0106}
{\footnotesize
Q_D =
\left(
\begin{array}{cccc|cccc}
p_1 & p_2 & p_3 & p_4 & s_1 & s_2 & s_3 & s_4 \\
\hline
 0 & 0 & 0 & 0 & -1 & 1 & 0 & 0 \\
 0 & 0 & 0 & 0 & 0 & -1 & 1 & 0 \\
 0 & 0 & 0 & 0 & 0 & 0 & -1 & 1 \\
\end{array}
\right)
}~.~
\eea

\begin{figure}[H]
\begin{center}
\resizebox{0.4\hsize}{!}{
\includegraphics[trim=0mm 12mm 0mm 0mm, height=6cm]{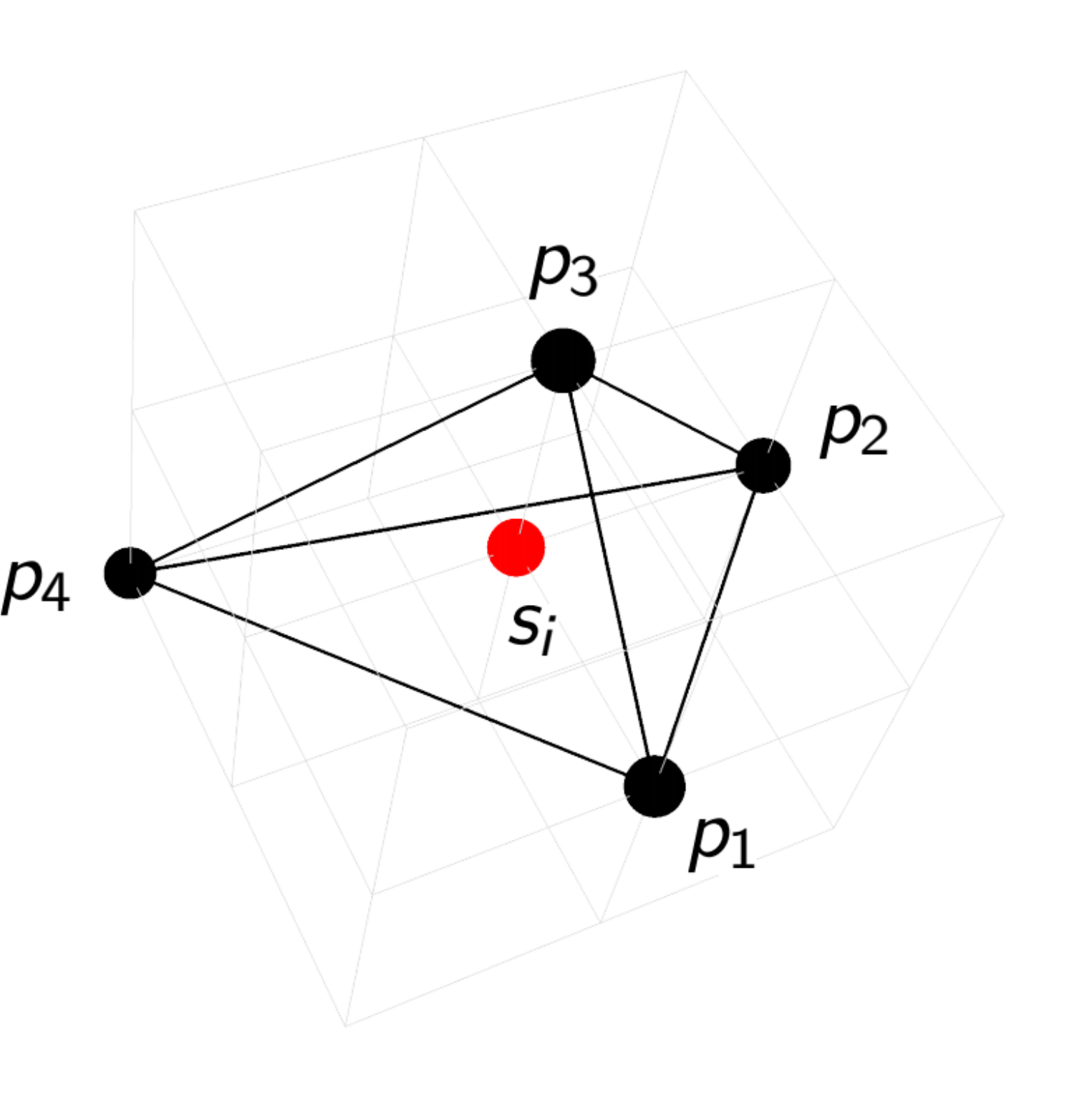} 
}
\caption{
Toric diagram for Model 1.
\label{f_toric_01}}
 \end{center}
 \end{figure}
 
The toric diagram of Model 1 can be found to be as follows, 
\beal{es0110}
{\footnotesize
G_t=
\left(
\begin{array}{cccc|cccc}
p_1 & p_2 & p_3 & p_4 & s_1 & s_2 & s_3 & s_4 \\
\hline
 1 & 1 & 1 & 1 & 1 & 1 & 1 & 1 \\
 1 & 0 & 0 & -1 & 0 & 0 & 0 & 0 \\
 0 & 1 & 0 & -1 & 0 & 0 & 0 & 0 \\
 0 & 0 & 1 & -1 & 0 & 0 & 0 & 0 \\
\end{array}
\right)
}~.~
\eea
The toric diagram with brick matching labels is shown in \fref{f_toric_01}.

Using the Molien integral formula, the Hilbert series of the mesonic moduli space of Model 1 is found as
\beal{es0120}
&&
g_1(t_i,y_s; \mathcal{M}_1) =
\frac{P(t_i,y_s; \mathcal{M}_1)}{(1 - y_s t_1^4) (1 - y_s t_2^4) (1 - y_s t_3^4) (1 - y_s t_4^4)}
~,~
\eea
where $t_i$ are the fugacities for the extremal brick matchings $p_i$.
$y_{s}$ counts the brick matching product $s_1 s_2 s_3 s_4$ corresponding to the single internal point of the toric diagram of Model 1. 
The explicit numerator $P(t_i,y_s; \mathcal{M}_1)$ of the Hilbert series is given in the Appendix Section \sref{app_num_01}.

\begin{table}[H]
\centering
\begin{tabular}{|c|c|c|c|}
\hline
\; & $SU(4)$ & $U(1)$ & \text{fugacity} \\
\hline
$p_1$ & (+1,0,0) & $r_1$ & $t_1$ \\
$p_2$ & (-1,+1,0) & $r_2$ & $t_2$ \\
$p_3$ & (0,-1,+1) & $r_3$ & $t_3$ \\
$p_4$ & (0,0,-1) & $r_4$ & $t_4$ \\
\hline
\end{tabular}
\caption{Global symmetry charges on the extremal brick matchings $p_i$ of Model 1.}
\label{t_pmcharges_01}
\end{table}

By setting $t_i=t$ for all $i$ and $y_s=1$, the unrefined Hilbert series takes the following form
\beal{es0121}
&&
g_1(t,1; \mathcal{M}_1) =
\frac{1 + 31 t^4 + 31 t^8 + t^{12}}{(1 - t^4)^4}
~,~
\eea
where the palindromic numerator indicates that the mesonic moduli space is Calabi-Yau. 

The global symmetry of Model 1 and the charges on the extremal brick matchings under the global symmetry are summarized in \tref{t_pmcharges_01}.
Using the following fugacity map,
\beal{es0122}
&&
t = t_1^{1/4} t_2^{1/4} t_3^{1/4} t_4^{1/4}~,~
x_1 = \frac{t_1^{3/4}}{t_2^{1/4} t_3^{1/4} t_4^{1/4}}~,~ 
x_2 = \frac{t_1^{1/2} t_2^{1/2}}{t_3^{1/2} t_4^{1/2}}~,~ 
x_3 = \frac{t_1^{1/4} t_2^{1/4} t_3^{1/4}}{t_4^{3/4}}~,~
\eea
\begin{table}[H]
\centering
\resizebox{.95\hsize}{!}{
\begin{minipage}[!b]{0.5\textwidth}
\begin{tabular}{|c|ccc|}
\hline
generator & \multicolumn{3}{c|}{$SU(4)_{(\tilde{x}_1,\tilde{x}_2,\tilde{x}_3)}$} \\
\hline
$p_1^4 ~s$  &  (3, & 1, & -1) \\
$p_1^3 p_2 ~s$  &  (2, & 1, & 0) \\
$p_1^2 p_2^2 ~s$  &  (1, & 1, & 1) \\
$p_1 p_2^3 ~s$  &  (0, & 1, & 2) \\
$p_2^4 ~s$  &  (-1, & 1, & 3) \\
$p_1^3 p_3 ~s$  &  (2, & 0, & -1) \\
$p_1^2 p_2 p_3 ~s$  &  (1, & 0, & 0) \\
$p_1 p_2^2 p_3 ~s$  &  (0, & 0, & 1) \\
$p_2^3 p_3 ~s$  &  (-1, & 0, & 2) \\
$p_1^2 p_3^2 ~s$  &  (1, & -1, & -1) \\
$p_1 p_2 p_3^2 ~s$  &  (0, & -1, & 0) \\
$p_2^2 p_3^2 ~s$  &  (-1, & -1, & 1) \\
$p_1 p_3^3 ~s$  &  (0, & -2, & -1) \\
$p_2 p_3^3 ~s$  &  (-1, & -2, & 0) \\
$p_3^4 ~s$  &  (-1, & -3, & -1) \\
$p_1^3 p_4 ~s$  &  (2, & 1, & -1) \\
$p_1^2 p_2 p_4 ~s$  &  (1, & 1, & 0) \\
$p_1 p_2^2 p_4 ~s$  &  (0, & 1, & 1) \\
$p_2^3 p_4 ~s$  &  (-1, & 1, & 2) \\
$p_1^2 p_3 p_4 ~s$  &  (1, & 0, & -1) \\
$p_1 p_2 p_3 p_4 ~s$  &  (0, & 0, & 0) \\
$p_2^2 p_3 p_4 ~s$  &  (-1, & 0, & 1) \\
$p_1 p_3^2 p_4 ~s$  &  (0, & -1, & -1) \\
$p_2 p_3^2 p_4 ~s$  &  (-1, & -1, & 0) \\
$p_3^3 p_4 ~s$  &  (-1, & -2, & -1) \\
$p_1^2 p_4^2 ~s$  &  (1, & 1, & -1) \\
$p_1 p_2 p_4^2 ~s$  &  (0, & 1, & 0) \\
$p_2^2 p_4^2 ~s$  &  (-1, & 1, & 1) \\
$p_1 p_3 p_4^2 ~s$  &  (0, & 0, & -1) \\
$p_2 p_3 p_4^2 ~s$  &  (-1, & 0, & 0) \\
$p_3^2 p_4^2 ~s$  &  (-1, & -1, & -1) \\
$p_1 p_4^3 ~s$  &  (0, & 1, & -1) \\
$p_2 p_4^3 ~s$  &  (-1, & 1, & 0) \\
$p_3 p_4^3 ~s$  &  (-1, & 0, & -1) \\
$p_4^4 ~s$  &  (-1, & 1, & -1) \\
\hline
\end{tabular}
\end{minipage}
\hspace{0cm}
\begin{minipage}[!b]{0.6\textwidth}
\includegraphics[height=9cm]{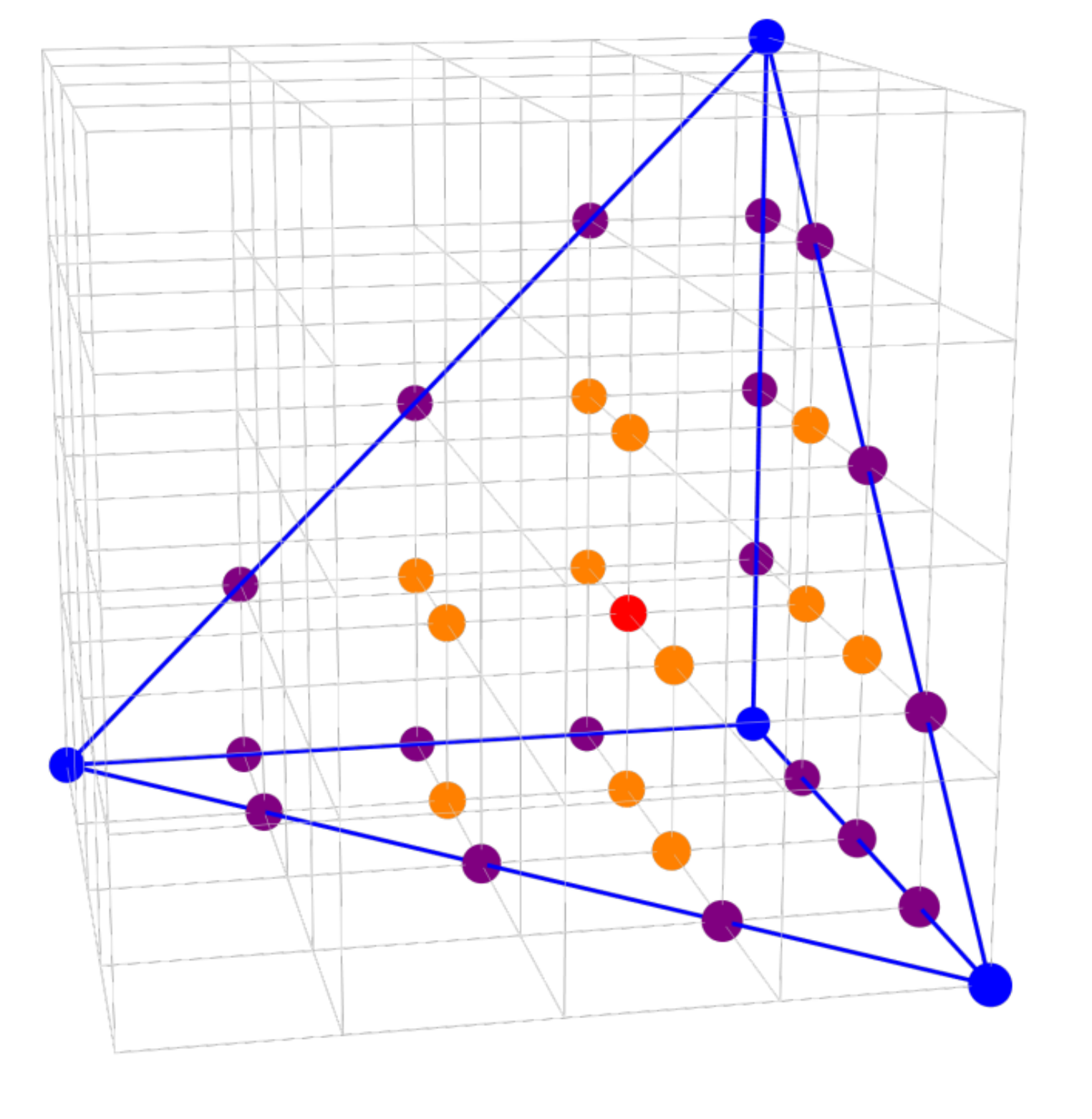} 
\end{minipage}
}
\caption{The generators and lattice of generators of the mesonic moduli space of Model 1 in terms of brick matchings with the corresponding flavor charges.
\label{f_genlattice_01}}
\end{table}

\noindent
the Hilbert series for Model 1 can be rewritten in terms of characters of irreducible representations of $SU(4)$, the mesonic flavor symmetry of Model 1, as follows 
\beal{es0125}
&&
g_1(t, x_i; \mathcal{M}_1) =
\sum_{n=0}^{\infty} [4n,0,0] t^{4n} ~.~
\eea

\begin{table}[H]
\centering
\resizebox{1\hsize}{!}{
\begin{tabular}{|l|ccc|}
\hline
generator & \multicolumn{3}{c|}{$SU(4)_{(\tilde{x}_1,\tilde{x}_2,\tilde{x}_3)}$} \\
\hline
$P_{12} P_{23} P_{34} P_{41}$  &  (3, & 1, & -1) \\
$X_{12} X_{23} X_{34} X_{41}$  &  (2, & 1, & 0) \\
$Y_{12} Y_{23} Y_{34} Y_{41}$  &  (1, & 1, & 1) \\
$Z_{12} Z_{23} Z_{34} Z_{41}$  &  (0, & 1, & 2) \\
$P_{12} P_{23} P_{34} X_{41}=  P_{12} P_{23} X_{34} P_{41}=  P_{12} X_{23} P_{41} P_{34}=  X_{12} P_{23} P_{34} P_{41}$  &  (-1, & 1, & 3) \\
$P_{12} P_{23} P_{34} Y_{41}=  P_{12} P_{23} Y_{34} P_{41}=  P_{12} Y_{23} P_{41} P_{34}=  Y_{12} P_{23} P_{34} P_{41}$  &  (2, & 0, & -1) \\
$P_{12} P_{23} P_{34} Z_{41}=  P_{12} P_{23} Z_{34} P_{41}=  P_{12} Z_{23} P_{41} P_{34}=  Z_{12} P_{23} P_{34} P_{41}$  &  (1, & 0, & 0) \\
\hline
\end{tabular}
}
\caption{The generators in terms of bifundamental chiral fields for Model 1.
\label{f_genfields_01}}
\end{table}

\noindent
Here, $[n_1, n_2, n_3] = [n_1, n_2, n_3]_{SU(4)}$ is the character of the irreducible representation of $SU(4)$ labeled by the highest weight $n_1,n_2,n_3$.
The corresponding plethystic logarithm is
\beal{es0126}
&&
\PL[g_1(t, x_i; \mathcal{M}_1)]=
[4,0,0] t^4 - ( [4,2,0] + [0,4,0] ) t^8 + \dots ~,~
\eea
where the mesonic moduli space is identified as a non-complete intersection.
The set of generators transform under the $[4,0,0]$ representation of the mesonic flavor symmetry.
Using the following fugacity map
\beal{es0127}
&&
\tilde{t} = t_1^{1/4} t_2^{1/4} t_3^{1/4} t_4^{1/4}~,~ 
\tilde{x}_1 = \frac{t_1}{t_4}~,~ 
\tilde{x}_2 = \frac{t_4}{t_3}~,~ 
\tilde{x}_3 = \frac{t_2}{t_4}~,~
\eea
the mesonic flavor charges on the gauge invariant operators become $\mathbb{Z}$-valued.
The generators in terms of brick matchings and their corresponding individual $\mathbb{Z}^3$-charges are summarized in \tref{f_genlattice_01}.
The generator lattice as shown in \tref{f_genlattice_01} is a convex lattice polytope, which is reflexive. It is the dual of the toric diagram of Model 1 in \fref{f_toric_01}.
For completeness, \tref{f_genfields_01} shows the generators of Model 1 in terms of chiral fields with the corresponding mesonic flavor charges. 
\\

\section{Model 2:  $M^{3,2}$~[$\mathbb{P}^2\times\mathbb{P}^1$,~$\langle4\rangle$] \label{smodel02}}

\begin{figure}[H]
\begin{center}
\resizebox{0.25\hsize}{!}{
\includegraphics[height=6cm]{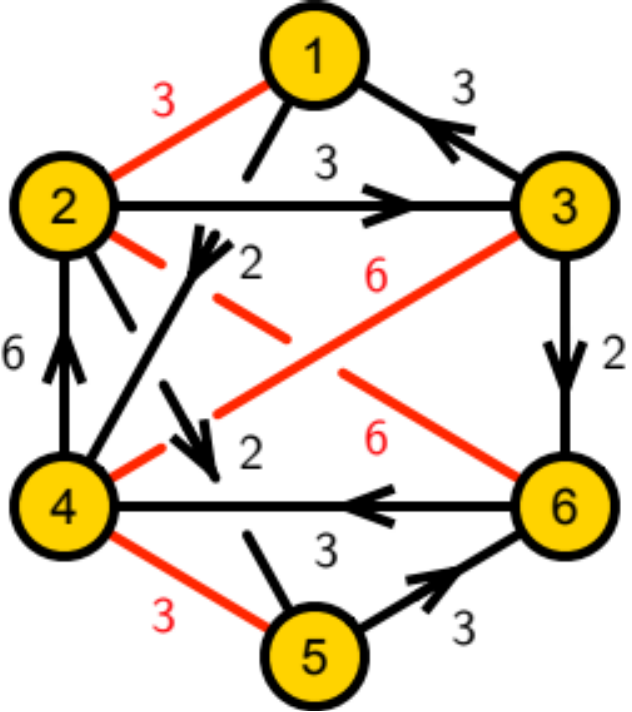} 
}
\caption{
Quiver for Model 2.
\label{f_quiver_02}}
 \end{center}
 \end{figure}
 
Model 2 corresponds to the Calabi-Yau cone over the $M^{3,2}$ surface.
The corresponding brane brick model has the quiver in \fref{f_quiver_02} and the $J$- and $E$-terms are given as follows
  \beq
  {\footnotesize
\begin{array}{rrclrcl}
 &  & J &  & & E &  \\
\Lambda_{12}^{1}  : & X_{23} X_{31}  &-&  Z_{23} Y_{31}  &   P_{14} X_{42}  &-&  Q_{14} R_{42} \\
\Lambda_{12}^{2}  : & Y_{23} Y_{31}  &-& X_{23} Z_{31}  &   P_{14} Y_{42}  &-&  Q_{14} S_{42} \\
\Lambda_{12}^{3}  : & Z_{23} Z_{31}  &-& Y_{23} X_{31}  &   P_{14} Z_{42}  &-&  Q_{14} T_{42} \\
\Lambda_{26}^{1}  : & X_{64} X_{42}  &-&  Z_{64} Y_{42}  &   P_{25} X_{56}  &-&  X_{23} P_{36} \\
\Lambda_{26}^{2}  : & Y_{64} Y_{42}   &-& X_{64} Z_{42}  &   P_{25} Y_{56}  &-&  Y_{23} P_{36} \\
\Lambda_{26}^{3}  : & Z_{64} Z_{42}  &-& Y_{64} X_{42} &   P_{25} Z_{56}  &-&  Z_{23} P_{36} \\
\Lambda_{26}^{4}  : & X_{64} R_{42}  &-&  Z_{64} S_{42}  &  X_{23} Q_{36}  &-& Q_{25} X_{56} \\
\Lambda_{26}^{5}  : & Y_{64} S_{42}  &-& X_{64} T_{42}  &  Y_{23} Q_{36}   &-& Q_{25} Y_{56} \\
\Lambda_{26}^{6}  : & Z_{64} T_{42}  &-& Y_{64} R_{42} &   Z_{23} Q_{36}  &-& Q_{25} Z_{56}\\
\Lambda_{34}^{1}  : & X_{42} X_{23}  &-&  Z_{42} Y_{23}  &   P_{36} X_{64}  &-&  X_{31} P_{14} \\
\Lambda_{34}^{2}  : & Y_{42} Y_{23}  &-& X_{42} Z_{23}  &   P_{36} Y_{64}  &-&  Y_{31} P_{14} \\
\Lambda_{34}^{3}  : & Z_{42} Z_{23}   &-& Y_{42} X_{23} &   P_{36} Z_{64}  &-&  Z_{31} P_{14} \\
\Lambda_{34}^{4}  : & R_{42} X_{23}  &-&  T_{42} Y_{23}  &  X_{31} Q_{14}  &-& Q_{36} X_{64}\\
\Lambda_{34}^{5}  : & S_{42} Y_{23}  &-& R_{42} Z_{23}  &   Y_{31} Q_{14}  &-& Q_{36} Y_{64}\\
\Lambda_{34}^{6}  : & T_{42} Z_{23}  &-& S_{42} X_{23}  &   Z_{31} Q_{14}  &-& Q_{36} Z_{64} \\
\Lambda_{45}^{1}  : & X_{56} X_{64}  &-&  Z_{56} Y_{64}  &   R_{42} Q_{25}  &-&  X_{42} P_{25} \\
\Lambda_{45}^{2}  : & Y_{56} Y_{64}   &-& X_{56} Z_{64} &   S_{42} Q_{25}  &-&  Y_{42} P_{25} \\
\Lambda_{45}^{3}  : & Z_{56} Z_{64}  &-& Y_{56} X_{64}   &   T_{42} Q_{25}  &-&  Z_{42} P_{25}
 \end{array} 
 }~.~
\label{E_J_C_+-}
 \eeq

The $P$-matrix of Model 2 takes the form
\beal{es0205}
 { \footnotesize
 P=
 \resizebox{0.45\textwidth}{!}{$
\left(
\begin{array}{r|ccccc|ccccccccc}
\; & p_1 & p_2 & p_3 & p_4 & p_5 & s_1 & s_2 & s_3 & s_4 & s_5 & s_6 & s_7 & s_8 & s_9\\
\hline
P_{14} & 1 & 0 & 0 & 0 & 0 & 1 & 1 & 0 & 0 & 1 & 0 & 0 & 0 & 0 \\
P_{25} & 1 & 0 & 0 & 0 & 0 & 1 & 0 & 1 & 0 & 0 & 1 & 0 & 0 & 0 \\
P_{36} & 1 & 0 & 0 & 0 & 0 & 1 & 0 & 0 & 1 & 1 & 1 & 0 & 0 & 0 \\
\hline
Q_{14} & 0 & 1 & 0 & 0 & 0 & 1 & 1 & 0 & 0 & 1 & 0 & 0 & 0 & 0 \\
Q_{25} & 0 & 1 & 0 & 0 & 0 & 1 & 0 & 1 & 0 & 0 & 1 & 0 & 0 & 0 \\
Q_{36} & 0 & 1 & 0 & 0 & 0 & 1 & 0 & 0 & 1 & 1 & 1 & 0 & 0 & 0 \\
\hline
R_{42} & 1 & 0 & 1 & 0 & 0 & 0 & 0 & 0 & 0 & 0 & 0 & 1 & 0 & 0 \\
S_{42} & 1 & 0 & 0 & 1 & 0 & 0 & 0 & 0 & 0 & 0 & 0 & 1 & 0 & 0 \\
T_{42} & 1 & 0 & 0 & 0 & 1 & 0 & 0 & 0 & 0 & 0 & 0 & 1 & 0 & 0 \\
\hline
X_{23} & 0 & 0 & 0 & 0 & 1 & 0 & 0 & 1 & 0 & 0 & 0 & 0 & 1 & 0 \\
X_{31} & 0 & 0 & 0 & 1 & 0 & 0 & 0 & 0 & 1 & 0 & 1 & 0 & 0 & 1 \\
X_{42} & 0 & 1 & 1 & 0 & 0 & 0 & 0 & 0 & 0 & 0 & 0 & 1 & 0 & 0 \\
X_{56} & 0 & 0 & 0 & 0 & 1 & 0 & 0 & 0 & 1 & 1 & 0 & 0 & 1 & 0 \\
X_{64} & 0 & 0 & 0 & 1 & 0 & 0 & 1 & 0 & 0 & 0 & 0 & 0 & 0 & 1 \\
\hline
Y_{23} & 0 & 0 & 1 & 0 & 0 & 0 & 0 & 1 & 0 & 0 & 0 & 0 & 1 & 0 \\
Y_{31} & 0 & 0 & 0 & 0 & 1 & 0 & 0 & 0 & 1 & 0 & 1 & 0 & 0 & 1 \\
Y_{42} & 0 & 1 & 0 & 1 & 0 & 0 & 0 & 0 & 0 & 0 & 0 & 1 & 0 & 0 \\
Y_{56} & 0 & 0 & 1 & 0 & 0 & 0 & 0 & 0 & 1 & 1 & 0 & 0 & 1 & 0 \\
Y_{64} & 0 & 0 & 0 & 0 & 1 & 0 & 1 & 0 & 0 & 0 & 0 & 0 & 0 & 1 \\
\hline
Z_{23} & 0 & 0 & 0 & 1 & 0 & 0 & 0 & 1 & 0 & 0 & 0 & 0 & 1 & 0 \\
Z_{31} & 0 & 0 & 1 & 0 & 0 & 0 & 0 & 0 & 1 & 0 & 1 & 0 & 0 & 1 \\
Z_{42} & 0 & 1 & 0 & 0 & 1 & 0 & 0 & 0 & 0 & 0 & 0 & 1 & 0 & 0 \\
Z_{56} & 0 & 0 & 0 & 1 & 0 & 0 & 0 & 0 & 1 & 1 & 0 & 0 & 1 & 0 \\
Z_{64} & 0 & 0 & 1 & 0 & 0 & 0 & 1 & 0 & 0 & 0 & 0 & 0 & 0 & 1 \\
\end{array}
\right)
$}
}~.~
\eea
 
The $J$- and $E$-term charges are given by
\beal{es0205}
{\footnotesize
Q_{JE}=
 \resizebox{0.42\textwidth}{!}{$
\left(
\begin{array}{ccccc|ccccccccc}
p_1 & p_2 & p_3 & p_4 & p_5 & s_1 & s_2 & s_3 & s_4 & s_5 & s_6 & s_7 & s_8 & s_9\\
\hline
 0 & 0 & 0 & 0 & 0 & 1 & -1 & -1 & -1 & 0 & 0 & 0 & 1 & 1 \\
 1 & 1 & -1 & -1 & -1 & 0 & -1 & -1 & -1 & 0 & 0 & 0 & 2 & 2 \\
 0 & 0 & 0 & 0 & 0 & 0 & -1 & 0 & -1 & 1 & 0 & 0 & 0 & 1 \\
 0 & 0 & 0 & 0 & 0 & 0 & 0 & -1 & -1 & 0 & 1 & 0 & 1 & 0 \\
 0 & 0 & -1 & -1 & -1 & 0 & 0 & 0 & 0 & 0 & 0 & 1 & 1 & 1 \\
\end{array}
\right)
$}
}~,~
\eea
and the $D$-term charges are given by
\beal{es0206}
{\footnotesize
Q_D =
 \resizebox{0.42\textwidth}{!}{$
\left(
\begin{array}{ccccc|ccccccccc}
p_1 & p_2 & p_3 & p_4 & p_5 & s_1 & s_2 & s_3 & s_4 & s_5 & s_6 & s_7 & s_8 & s_9\\
\hline
 0 & 0 & -1 & -1 & -1 & 0 & 0 & 1 & 0 & 0 & 0 & 0 & 1 & 1 \\
 0 & 0 & 0 & 0 & 0 & 0 & 0 & 0 & 1 & 0 & 0 & 0 & -1 & 0 \\
 0 & 0 & 1 & 1 & 1 & 0 & -1 & 0 & 0 & 0 & 0 & 0 & -1 & -1 \\
 0 & 0 & 0 & 0 & 0 & 0 & 0 & -1 & 0 & 0 & 0 & 0 & 1 & 0 \\
 0 & 0 & 0 & 0 & 0 & 0 & 0 & 0 & -1 & 0 & 0 & 0 & 0 & 1 \\
\end{array}
\right)
$}
}~.~
\eea

\begin{figure}[ht!!]
\begin{center}
\resizebox{0.4\hsize}{!}{
\includegraphics[height=6cm]{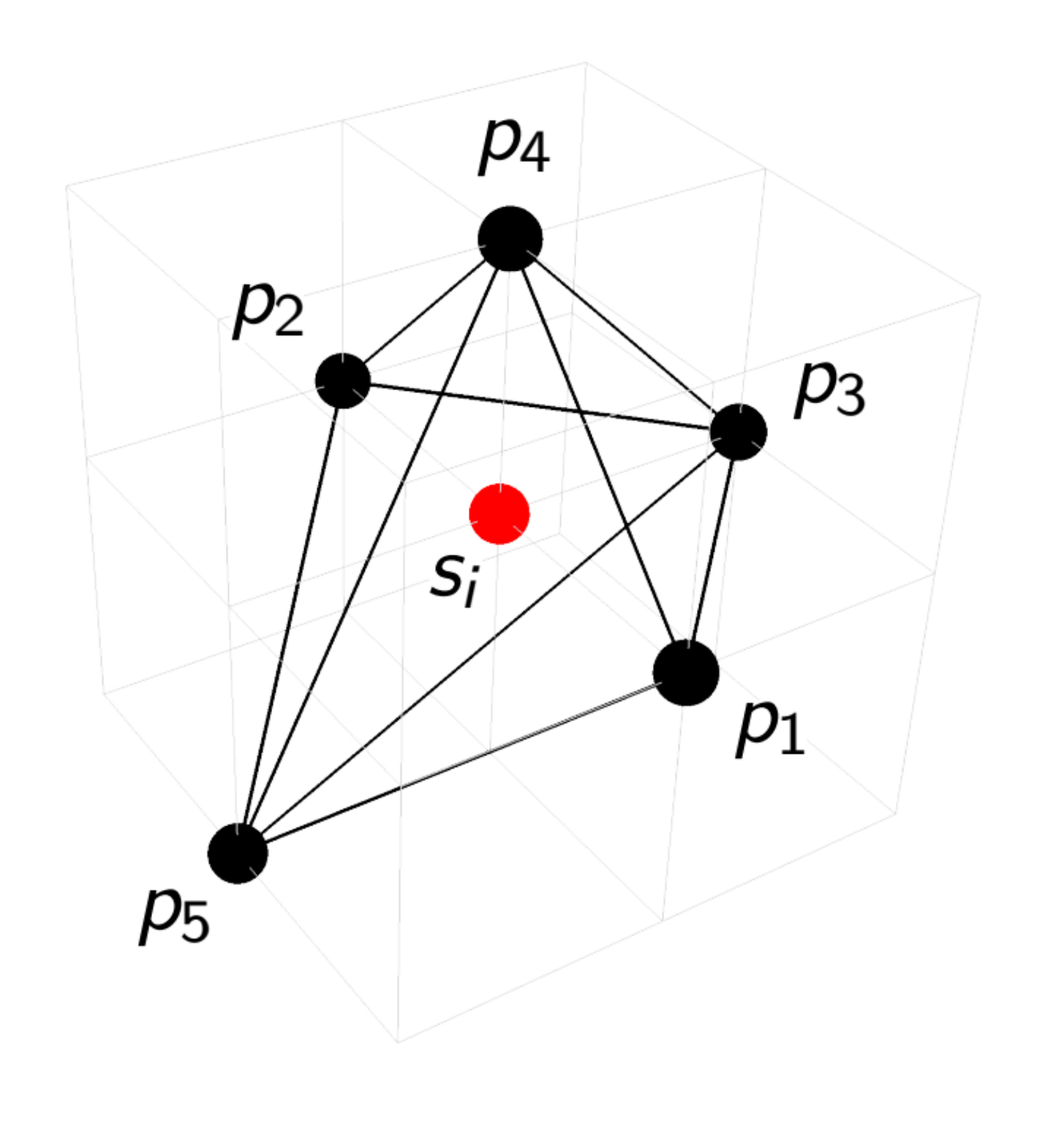} 
}
\caption{
Toric diagram for Model 2.
\label{f_toric_02}}
 \end{center}
 \end{figure}

The toric diagram of Model 2 is found to be as follows
\beal{es0210}
{\footnotesize
G_t=
\resizebox{0.42\textwidth}{!}{$
\left(
\begin{array}{ccccc|ccccccccc}
p_1 & p_2 & p_3 & p_4 & p_5 & s_1 & s_2 & s_3 & s_4 & s_5 & s_6 & s_7 & s_8 & s_9\\
\hline
 1 & 1 & 1 & 1 & 1 & 1 & 1 & 1 & 1 & 1 & 1 & 1 & 1 & 1 \\
 1 & -1 & 0 & 0 & 0 & 0 & 0 & 0 & 0 & 0 & 0 & 0 & 0 & 0 \\
 0 & 0 & 1 & 0 & -1 & 0 & 0 & 0 & 0 & 0 & 0 & 0 & 0 & 0 \\
 0 & 0 & 0 & 1 & -1 & 0 & 0 & 0 & 0 & 0 & 0 & 0 & 0 & 0 \\
\end{array}
\right)
$}
}~.~
\eea
The toric diagram with brick matching labels is shown in \fref{f_toric_02}.

Using the Molien integral formula, the Hilbert series of the mesonic moduli space of Model 2 is found to be as follows
\beal{es0220}
&&
g_1(t_i,y_s; \mathcal{M}_2) =
\frac{P(t_i, y_s; \mathcal{M}_2)}{(1 -  y_s t_1^2 t_3^3) (1 -  y_s t_2^2 t_3^3) (1 -  y_s t_1^2 t_4^3) (1 -   y_s t_2^2 t_4^3)} \times
\nn\\
&&
\hspace{1cm}
\frac{1}{(1 -  y_s t_1^2 t_5^3) (1 -  y_s t_2^2 t_5^3)} ~,~ 
\eea
where $t_i$ are the fugacities for the extremal brick matchings $p_i$.
$y_{s}$ counts the product of brick matchings $s_1 \dots s_9$ corresponding to the single internal point of the toric diagram of Model 2.
The explicit numerator $P(t_i, y_s; \mathcal{M}_2)$ of the Hilbert series is given in the Appendix Section \sref{app_num_02}.

\begin{table}[H]
\centering
\begin{tabular}{|c|cc|c|c|c|}
\hline
\; & \multicolumn{2}{c|}{$SU(3)_{(x_1,x_2)}$} & $SU(2)_{z}$ & $U(1)$ & \text{fugacity} \\
\hline
$p_1$ & (0,& 0)  & +1 & $r_1$ & $t_1$ \\
$p_2$ & (0,& 0)  & -1& $r_2$ & $t_2$ \\
$p_3$ & (+1,&0)  & 0 & $r_3$ & $t_3$ \\
$p_4$ & (-1, &+1) & 0 & $r_4$ & $t_4$ \\
$p_5$ & (0, &-1) & 0 & $r_5$ & $t_5$ \\
\hline
\end{tabular}
\caption{Global symmetry charges on the extremal brick matchings $p_i$ of Model 2.}
\label{t_pmcharges_02}
\end{table}

By setting $t_i=t$ for all $i$ and $y_s=1$, the unrefined Hilbert series takes the following form
\beal{es0221}
&&
g_1(t,1; \mathcal{M}_2) =
\frac{1 + 26 t^5 + 26 t^{10} + t^{15}}{(1 - t^5)^4} ~,~
\eea
where the palindromic numerator indicates that the mesonic moduli space is Calabi-Yau. 

The global symmetry of Model 2 and the charges on the extremal brick matchings under the global symmetry are summarized in \tref{t_pmcharges_02}.
Using the following fugacity map,
\beal{es0222}
&&
t = t_5 x_2 ~,~
x_1 = \frac{t_3^2 t_4 t_5}{t_1^2 t_2^2} ~,~
x_2 = \frac{t_3 t_4}{t_1 t_2} ~,~
z = \frac{t_5 x_2}{t_2} ~,~
\eea
the Hilbert series for Model 2 can be rewritten in terms of characters of irreducible representations of $SU(3)\times SU(2)$, the mesonic flavor symmetry of Model 2, as follows 
\beal{es0225}
&&
g_1(t, x_i, z; \mathcal{M}_2) =
\sum_{n=0}^{\infty} [3n,0;2n] t^{5n} ~,~
\eea
where $[m_1,m_2;n]=[m_1,m_2]_{SU(3)_{(x_1,x_2)}} [n]_{SU(2)_{z}}$.
The corresponding plethystic logarithm is
\beal{es0226}
&&
\PL[g_1(t, x_i, z; \mathcal{M}_2)]=
[3,0;2] t^5 
- ([6,0;0]+[4,1;2]+[2,2;4]+
\nn\\
&&
\hspace{1cm}
+ [2,2;0]+[0,3;2])t^{10}
+
\dots ~,~
\eea
where we see that the mesonic moduli space is a non-complete intersection.

\begin{table}[H]
\centering
\resizebox{.95\hsize}{!}{
\begin{minipage}[!b]{0.5\textwidth}
\begin{tabular}{|c|cc|c|}
\hline
generator & \multicolumn{2}{c|}{$SU(3)_{(\tilde{x}_1,\tilde{x}_2)}$} & $SU(2)_{\tilde{z}}$ \\
\hline
$p_1^2 p_3^3 s $ &   (1, &  1) &  1 \\
$p_1 p_2 p_3^3 s $ &   (1, &  1) &  0 \\
$p_2^2 p_3^3 s $ &   (1, &  1) &  -1 \\
$p_1^2 p_3^2 p_4 s $ &   (0, &  1) &  1 \\
$p_1 p_2 p_3^2 p_4 s $ &   (0, &  1) &  0 \\
$p_2^2 p_3^2 p_4 s $ &   (0, &  1) &  -1 \\
$p_1^2 p_3 p_4^2 s $ &   (-1, &  1) &  1 \\
$p_1 p_2 p_3 p_4^2 s $ &   (-1, &  1) &  0 \\
$p_2^2 p_3 p_4^2 s $ &   (-1, &  1) &  -1 \\
$p_1^2 p_4^3 s $ &   (-2, &  1) &  1 \\
$p_1 p_2 p_4^3 s $ &   (-2, &  1) &  0 \\
$p_2^2 p_4^3 s $ &   (-2, &  1) &  -1 \\
$p_1^2 p_3^2 p_5 s $ &   (1, &  0) &  1 \\
$p_1 p_2 p_3^2 p_5 s $ &   (1, &  0) &  0 \\
$p_2^2 p_3^2 p_5 s $ &   (1, &  0) &  -1 \\
$p_1^2 p_3 p_4 p_5 s $ &   (0, &  0) &  1 \\
$p_1 p_2 p_3 p_4 p_5 s $ &   (0, &  0) &  0 \\
$p_2^2 p_3 p_4 p_5 s $ &   (0, &  0) &  -1 \\
$p_1^2 p_4^2 p_5 s $ &   (-1, &  0) &  1 \\
$p_1 p_2 p_4^2 p_5 s $ &   (-1, &  0) &  0 \\
$p_2^2 p_4^2 p_5 s $ &   (-1, &  0) &  -1 \\
$p_1^2 p_3 p_5^2 s $ &   (1, &  -1) &  1 \\
$p_1 p_2 p_3 p_5^2 s $ &   (1, &  -1) &  0 \\
$p_2^2 p_3 p_5^2 s $ &   (1, &  -1) &  -1 \\
$p_1^2 p_4 p_5^2 s $ &   (0, &  -1) &  1 \\
$p_1 p_2 p_4 p_5^2 s $ &   (0, &  -1) &  0 \\
$p_2^2 p_4 p_5^2 s $ &   (0, &  -1) &  -1 \\
$p_1^2 p_5^3 s $ &   (1, &  -2) &  1 \\
$p_1 p_2 p_5^3 s $ &   (1, &  -2) &  0 \\
$p_2^2 p_5^3 s $ &   (1, &  -2) &  -1\\
\hline
\end{tabular}
\end{minipage}
\hspace{0cm}
\begin{minipage}[!b]{0.6\textwidth}
\includegraphics[height=6cm]{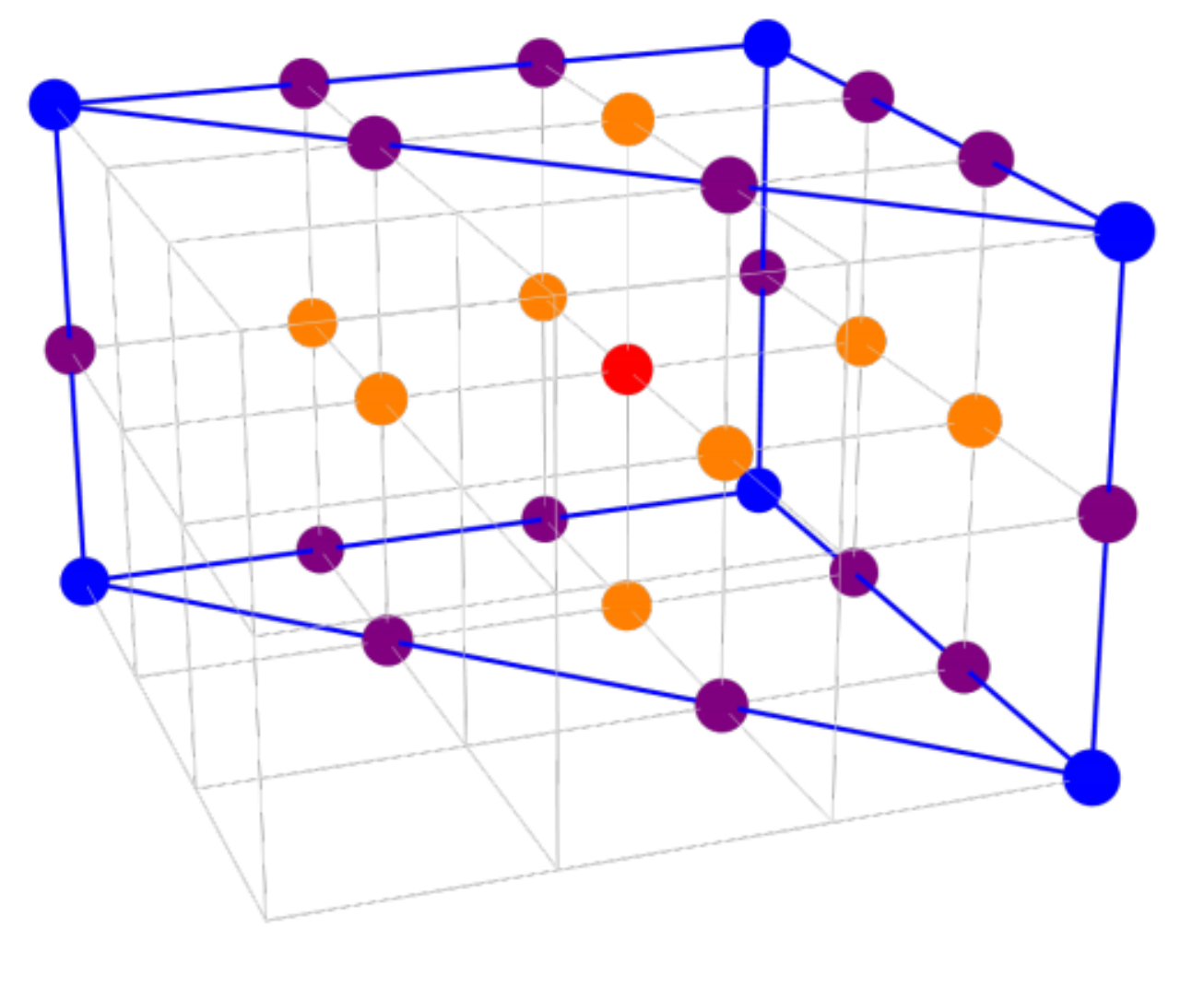} 
\end{minipage}
}
\caption{The generators and lattice of generators of the mesonic moduli space of Model 2 in terms of brick matchings with the corresponding flavor charges.
\label{f_genlattice_02}}
\end{table}

The set of generators transform under the $[3,0;2]$ representation of the mesonic flavor symmetry.
Using the following fugacity map
\beal{es0227}
&&
\tilde{t} = t_3^{1/3} t_4^{1/3} t_5^{1/3} ~,~ 
\tilde{x}_1 = \frac{t_3}{t_4}~,~ 
\tilde{x}_2 = \frac{t_3}{t_5}~,~ 
\tilde{z} = \frac{t_3^{2/3} t_4^{2/3} t_5^{2/3}}{t_2^2}
\eea
the mesonic flavor charges on the gauge invariant operators become $\mathbb{Z}$-valued.
The generators in terms of brick matchings and their corresponding rescaled mesonic flavor charges are summarized in \tref{f_genlattice_02}.

\begin{table}[H]
\centering
\resizebox{0.88\hsize}{!}{
\begin{tabular}{|l|cc|c|}
\hline
generator & \multicolumn{2}{c|}{$SU(3)_{(\tilde{x}_1,\tilde{x}_2)}$} & $SU(2)_{\tilde{z}}$ \\
\hline
$P_{14} R_{42} Y_{23} Z_{31}=  P_{25} Y_{56} Z_{64} R_{42}=  Y_{23} P_{36} Z_{64} R_{42}$ &   (1, &  1) &  1 \\
$P_{14} X_{42} Y_{23} Z_{31}=  P_{25} Y_{56} Z_{64} X_{42}=  Y_{23} P_{36} Z_{64} X_{42}=  Q_{14} R_{42} Y_{23} Z_{31}=  Q_{25} Y_{56} Z_{64} R_{42}=  Y_{23} Q_{36} Z_{64} R_{42}$ &   (1, &  1) &  0 \\
$Q_{14} X_{42} Y_{23} Z_{31}=  Q_{25} Y_{56} Z_{64} X_{42}=  Y_{23} Q_{36} Z_{64} X_{42}$ &   (1, &  1) &  -1 \\
$ P_{14} R_{42} Y_{23} X_{31}=  P_{14} R_{42} Z_{23} Z_{31}=  P_{14} S_{42} Y_{23} Z_{31}=  P_{25} Y_{56} X_{64} R_{42}=  P_{25} Z_{56} Z_{64} R_{42}=  P_{25} Y_{56} Z_{64} S_{42}$ &   (0, &  1) &  1 \\
$=  Y_{23} P_{36} X_{64} R_{42}=  Z_{23} P_{36} Z_{64} R_{42}=  Y_{23} P_{36} Z_{64} S_{42}$  & & & \\
$P_{14} X_{42} Y_{23} X_{31}=  P_{14} X_{42} Z_{23} Z_{31}=  P_{14} Y_{42} Y_{23} Z_{31}=  P_{25} Y_{56} X_{64} X_{42}=  P_{25} Z_{56} Z_{64} X_{42}=  P_{25} Y_{56} Z_{64} Y_{42}$ &   (0, &  1) &  0 \\
$=  Y_{23} P_{36} X_{64} X_{42}=  Z_{23} P_{36} Z_{64} X_{42}=  Y_{23} P_{36} Z_{64} Y_{42}=  Q_{14} R_{42} Y_{23} X_{31}=  Q_{14} R_{42} Z_{23} Z_{31}=  Q_{14} S_{42} Y_{23} Z_{31}$& & & \\
\hline
\end{tabular}
}
\caption{The generators in terms of bifundamental chiral fields for Model 2.
\label{f_genfields_02}}
\end{table}

The generator lattice as shown in \tref{f_genlattice_02} is a convex lattice polytope, which is reflexive. It is the dual of the toric diagram of Model 2 shown in \fref{f_toric_02}.
For completeness, \tref{f_genfields_02} shows the generators of Model 2 in terms of chiral fields with the corresponding mesonic flavor charges. 
\\

\section{Model 3: $Y^{2,4}(\mathbb{CP}^2)$~[$\mathbb{P}(\mathcal{O}_{\mathbb{P}^2} \oplus \mathcal{O}_{\mathbb{P}^2} (1) )$,~$\langle5\rangle$] \label{smodel03}}

\begin{figure}[H]
\begin{center}
\resizebox{0.25\hsize}{!}{
\includegraphics[height=6cm]{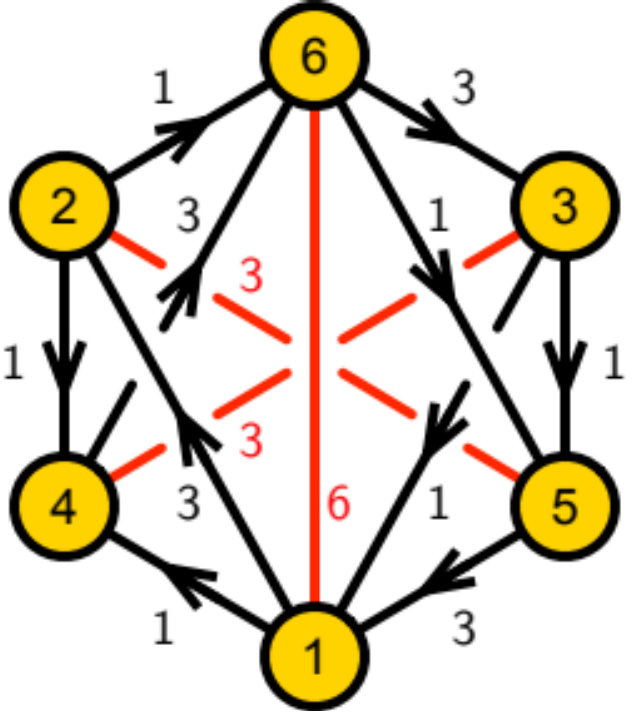} 
}
\caption{
Quiver for Model 3.
\label{f_quiver_03}}
 \end{center}
 \end{figure}
 
Model 3 corresponds to one of the $Y^{p,k}(\mathbb{CP}^2)$ models, $Y^{2,4}(\mathbb{CP}^2)$.
The corresponding brane brick model has the quiver in \fref{f_quiver_03} and the $J$- and $E$-terms are given as follows
\beq
{\footnotesize
\begin{array}{rrclrcl}
 &  & J &  & & E &  \\
\Lambda_{16}^{1}  : & X_{63} P_{35} Z_{51}  &-&  Z_{63} P_{35} X_{51}   &    P_{12} X_{26}  &-&  Q_{14} X_{46} \\ 
\Lambda_{16}^{2}  : & P_{63} P_{35} X_{51}  &-&  X_{63} P_{35} Y_{51}   &    X_{12} X_{26}  &-& Q_{14} Y_{46} \\ 
\Lambda_{16}^{3}  : & P_{63} P_{35} Z_{51}  &-&  Z_{63} P_{35} Y_{51}   &    Q_{14} Z_{46}  &-&  Y_{12} X_{26} \\ 
\Lambda_{43}^{1}  : & P_{35} Z_{51} Q_{14}  &-&  Q_{31} X_{12} P_{24}   &    X_{46} X_{63}  &-&  Z_{46} P_{63} \\ 
\Lambda_{43}^{2}  : & P_{35} X_{51} Q_{14}  &-&  Q_{31} Y_{12} P_{24}   &    Y_{46} P_{63}  &-& X_{46} Z_{63} \\ 
\Lambda_{43}^{3}  : & P_{35} Y_{51} Q_{14}  &-&  Q_{31} P_{12} P_{24}   &    Z_{46} Z_{63}  &-& Y_{46} X_{63} \\ 
\Lambda_{52}^{1}  : & P_{24} X_{46} Q_{65}  &-&  X_{26} P_{63} P_{35}   &    X_{51} X_{12}  &-&  Z_{51} Y_{12} \\ 
\Lambda_{52}^{2}  : & P_{24} Y_{46} Q_{65}  &-&  X_{26} Z_{63} P_{35}   &    Y_{51} Y_{12}  &-& X_{51} P_{12} \\ 
\Lambda_{52}^{3}  : & P_{24} Z_{46} Q_{65}  &-&  X_{26} X_{63} P_{35}   &    Z_{51} P_{12}  &-& Y_{51} X_{12} \\ 
\Lambda_{61}^{1}  : & X_{12} P_{24} X_{46}  &-& P_{12} P_{24} Y_{46}   &    X_{63} Q_{31}  &-& Q_{65} X_{51}  \\ 
\Lambda_{61}^{2}  : & Y_{12} P_{24} Y_{46}  &-& X_{12} P_{24} Z_{46}   &    P_{63} Q_{31}  &-&  Q_{65} Y_{51} \\ 
\Lambda_{61}^{3}  : & P_{12} P_{24} Z_{46}  &-&  Y_{12} P_{24} X_{46}   &    Z_{63} Q_{31}  &-& Q_{65} Z_{51} 
 \end{array} 
 }~.~
\label{es0301}
 \eeq
 
Using the forward algorithm, we can calculate the $P$-matrix of Model 3. It takes the form
\beal{es0305}
 {\footnotesize
 P=
 \resizebox{0.45\textwidth}{!}{$
\left(
\begin{array}{c|ccccc|cccccc|cc}
\; & p_1 & p_2 & p_3 & p_4 & p_5 & s_1 & s_2 & s_3 & s_4 & s_5 & s_6  & o_1 & o_2\\
\hline
P_{12} & 0 & 0 & 1 & 0 & 0 & 0 & 0 & 0 & 0 & 0 & 1 & 1 & 0 \\
P_{24} & 0 & 0 & 0 & 1 & 0 & 0 & 0 & 0 & 0 & 1 & 0 & 0 & 1 \\
P_{35} & 0 & 0 & 0 & 1 & 0 & 0 & 0 & 0 & 1 & 0 & 0 & 1 & 0 \\
P_{63} & 0 & 0 & 1 & 0 & 0 & 0 & 0 & 1 & 0 & 0 & 0 & 0 & 1 \\
\hline
Q_{14} & 0 & 0 & 0 & 0 & 1 & 0 & 0 & 0 & 0 & 1 & 1 & 0 & 0 \\
Q_{31} & 0 & 0 & 0 & 0 & 1 & 0 & 1 & 0 & 1 & 0 & 0 & 0 & 0 \\
Q_{65} & 0 & 0 & 0 & 0 & 1 & 0 & 0 & 1 & 1 & 0 & 0 & 0 & 0 \\
\hline
X_{12} & 0 & 1 & 0 & 0 & 0 & 0 & 0 & 0 & 0 & 0 & 1 & 1 & 0 \\
X_{26} & 0 & 0 & 0 & 0 & 1 & 1 & 0 & 0 & 0 & 1 & 0 & 0 & 0 \\
X_{46} & 0 & 0 & 1 & 0 & 0 & 1 & 0 & 0 & 0 & 0 & 0 & 1 & 0 \\
X_{51} & 1 & 0 & 0 & 0 & 0 & 0 & 1 & 0 & 0 & 0 & 0 & 0 & 1 \\
X_{63} & 1 & 0 & 0 & 0 & 0 & 0 & 0 & 1 & 0 & 0 & 0 & 0 & 1 \\
\hline
Y_{12} & 1 & 0 & 0 & 0 & 0 & 0 & 0 & 0 & 0 & 0 & 1 & 1 & 0 \\
Y_{46} & 0 & 1 & 0 & 0 & 0 & 1 & 0 & 0 & 0 & 0 & 0 & 1 & 0 \\
Y_{51} & 0 & 0 & 1 & 0 & 0 & 0 & 1 & 0 & 0 & 0 & 0 & 0 & 1 \\
\hline
Z_{4.6} & 1 & 0 & 0 & 0 & 0 & 1 & 0 & 0 & 0 & 0 & 0 & 1 & 0 \\
Z_{51} & 0 & 1 & 0 & 0 & 0 & 0 & 1 & 0 & 0 & 0 & 0 & 0 & 1 \\
Z_{63} & 0 & 1 & 0 & 0 & 0 & 0 & 0 & 1 & 0 & 0 & 0 & 0 & 1 \\
\end{array}
\right)
$}
~.~
}
 \eea
 
The $J$- and $E$-term charges are given by
\beal{es0305}
{\footnotesize
Q_{JE} =
 \resizebox{0.42\textwidth}{!}{$
\left(
\begin{array}{ccccc|cccccc|cc}
p_1 & p_2 & p_3 & p_4 & p_5 & s_1 & s_2 & s_3 & s_4 & s_5 & s_6  & o_1 & o_2\\
\hline
 0 & 0 & 0 & 0 & 1 & -1 & -1 & 0 & 0 & 0 & -1 & 1 & 0 \\
 1 & 1 & 1 & 0 & 1 & -1 & -1 & -1 & -1 & 0 & 0 & 0 & 0 \\
 0 & 0 & 0 & 1 & 1 & 0 & 0 & 0 & 0 & -1 & -1 & 0 & 0 \\
 0 & 0 & 0 & 0 & 1 & 0 & 0 & -1 & -1 & -1 & 0 & 0 & 1 \\
\end{array}
\right)
$}
}~,~
\eea
and the $Q_D$-matrix takes the form
\beal{es0306}
 {\footnotesize
Q_D =
 \resizebox{0.42\textwidth}{!}{$
\left(
\begin{array}{ccccc|cccccc|cc}
p_1 & p_2 & p_3 & p_4 & p_5 & s_1 & s_2 & s_3 & s_4 & s_5 & s_6  & o_1 & o_2\\
\hline
 0 & 0 & 0 & 0 & 0 & 0 & -1 & 0 & 0 & 1 & 0 & 0 & 0 \\
 0 & 0 & 0 & 0 & 0 & 0 & 0 & 0 & -1 & 0 & 1 & 0 & 0 \\
 0 & 0 & 0 & 0 & 0 & 1 & 0 & 0 & 0 & -1 & 0 & 0 & 0 \\
 0 & 0 & 0 & 0 & 0 & 0 & 0 & 1 & 0 & 0 & -1 & 0 & 0 \\
 0 & 0 & 0 & 0 & 0 & -1 & 0 & 0 & 1 & 0 & 0 & 0 & 0 \\
\end{array}
\right)
$}
}~.~
 \eea
 
\begin{figure}[ht!!]
\begin{center}
\resizebox{0.4\hsize}{!}{
\includegraphics[height=6cm]{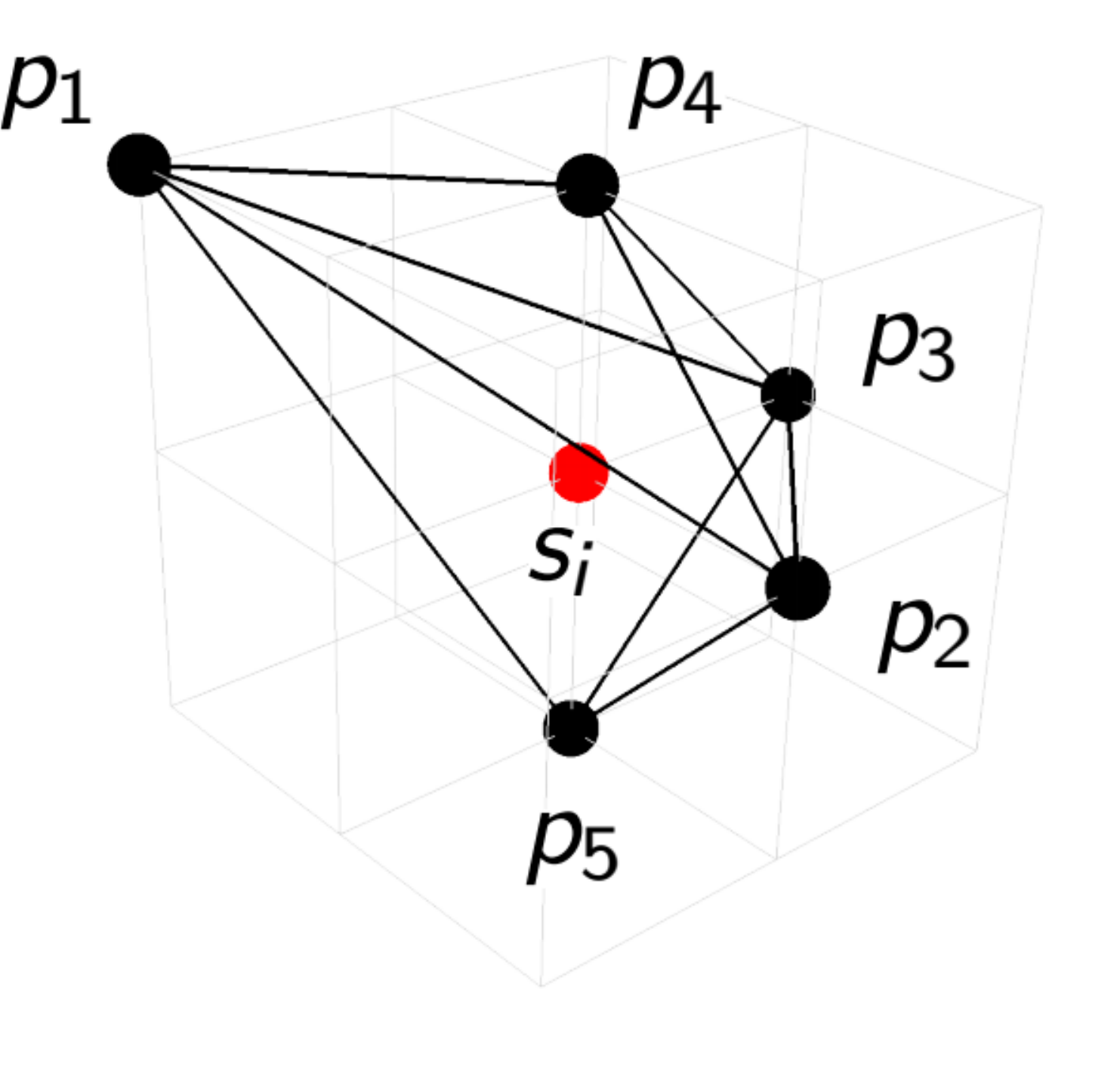} 
}
\caption{
Toric diagram for Model 3.
\label{f_toric_03}}
 \end{center}
 \end{figure}

The toric diagram of Model 3 is given by the $G_t$-matrix,
\beal{es0310}
{\footnotesize
G_t=
 \resizebox{0.42\textwidth}{!}{$
\left(
\begin{array}{ccccc|cccccc|cc}
p_1 & p_2 & p_3 & p_4 & p_5 & s_1 & s_2 & s_3 & s_4 & s_5 & s_6  & o_1 & o_2\\
\hline
 1 & 1 & 1 & 1 & 1 & 1 & 1 & 1 & 1 & 1 & 1 & 2 & 2 \\
 -1 & 1 & 0 & 0 & 0 & 0 & 0 & 0 & 0 & 0 & 0 & 0 & 0 \\
 -1 & 0 & 1 & 0 & 0 & 0 & 0 & 0 & 0 & 0 & 0 & 0 & 0 \\
 1 & 0 & 0 & 1 & -1 & 0 & 0 & 0 & 0 & 0 & 0 & 1 & 1 \\
\end{array}
\right)
$}
}~.~
 \eea
\fref{f_toric_03} shows the toric diagram of Model 3 with brick matching labels.

\begin{table}[H]
\centering
\begin{tabular}{|c|cc|c|c|c|}
\hline
\; & \multicolumn{2}{c|}{$SU(3)_{(x_1,x_2)}$} & $U(1)_{b}$ & $U(1)$ & \text{fugacity} \\
\hline
$p_1$ & (1,& 0)  & 0 & $r_1$ & $t_1$ \\
$p_2$ & (-1,& +1)  & 0& $r_2$ & $t_2$ \\
$p_3$ & (0,&-1)  & 0 & $r_3$ & $t_3$ \\
$p_4$ & (0, &0) & +1 & $r_4$ & $t_4$ \\
$p_5$ & (0, &0) & -1 & $r_5$ & $t_5$ \\
\hline
\end{tabular}
\caption{Global symmetry charges on the extremal brick matchings $p_i$ of Model 3.}
\label{t_pmcharges_03}
\end{table}

Using the Molien integral formula, the Hilbert series of the mesonic moduli space of Model 3 is found to be as follows
\beal{es0320}
&&
g_1(t_i,y_s; \mathcal{M}_3) =
\frac{P(t_i, y_s; \mathcal{M}_3)}{
(1 - y_{s} y_{o}^3 t_1^4 t_4^2) (1 - y_{s} y_{o}^3 t_2^4 t_4^2) (1 - y_{s} y_{o}^3 t_3^4 t_4^2) (1 - y_{s} y_{o} t_1^2 t_5^2) 
} 
\times
\nn\\
&&
\hspace{1cm}
\frac{1}{
(1 - y_{s} y_{o} t_2^2 t_5^2) (1 - y_{s} y_{o} t_3^2 t_5^2)
} ~,~ 
\eea
where $t_i$ are the fugacities for the extremal brick matchings $p_i$.
$y_{s}$ counts the product of brick matchings $s_1 \dots s_6$ corresponding to the single internal point of the toric diagram of Model 3.
Additionally, $y_o$ counts the product of extra GLSM fields $o_1 o_2$.
The explicit numerator of the Hilbert series $P(t_i, y_s; \mathcal{M}_3)$ is given in the Appendix Section \sref{app_num_03}.
We note that by setting $y_o=1$, the characterization of the mesonic moduli space by the Hilbert series does not change. This implies that the extra GLSM fields indeed are an over-parameterization of the moduli space as expected. 

By setting $t_i=t$ for all extremal brick matching fugacities, and $y_s=1$ and $y_o=1$ for all other fugacities, the unrefined Hilbert series takes the following form
\beal{es0321}
&&
g_1(t,1,1; \mathcal{M}_3) =
\frac{
1
}{
(1 - t^4)^3 (1 - t^6)^3
}
\times
\nn\\
&&
\hspace{1cm}
(1 + 3 t^4 + 10 t^5 + 12 t^6 - 9 t^9 - 26 t^{10} + 6 t^{11} + 3 t^{12} 
+ 3 t^{13} + 6 t^{14} - 26 t^{15} 
\nn\\
&&
\hspace{1cm}
- 9 t^{16} + 12 t^{19} + 10 t^{20} 
+ 3 t^{21} + t^{25})
~,~
\eea
where the palindromic numerator indicates that the mesonic moduli space is Calabi-Yau. 

\begin{table}[H]
\centering
\resizebox{.95\hsize}{!}{
\begin{minipage}[!b]{0.5\textwidth}
\begin{tabular}{|c|cc|c|}
\hline
generator & \multicolumn{2}{c|}{$SU(3)_{(\tilde{x}_1,\tilde{x}_2)}$} & $U(1)_{\tilde{b}}$ \\
\hline
$ p_1^2 p_5^2 s  o$ &    (1, &    1) &    -1 \\
$  p_1 p_2 p_5^2 s o$ &    (0, &    1) &    -1 \\
$  p_2^2 p_5^2 s o$ &    (-1, &    1) &    -1 \\
$  p_1 p_3 p_5^2 s o$ &    (1, &    0) &    -1 \\
$  p_2 p_3 p_5^2 s o$ &    (0, &    0) &    -1 \\
$  p_3^2 p_5^2 s o $ &    (1, &    -1) &    -1 \\
$  p_1^3 p_4 p_5 s o^2$ &    (1, &    1) &    0 \\
$  p_1^2 p_2 p_4 p_5 s o^2$ &    (0, &    1) &    0 \\
$  p_1 p_2^2 p_4 p_5 s o^2$ &    (-1, &    1) &    0 \\
$  p_2^3 p_4 p_5 s o^2$ &    (-2, &    1) &    0 \\
$  p_1^2 p_3 p_4 p_5 s o^2$ &    (1, &    0) &    0 \\
$  p_1 p_2 p_3 p_4 p_5 s o^2$ &    (0, &    0) &    0 \\
$  p_2^2 p_3 p_4 p_5 s o^2$ &    (-1, &    0) &    0 \\
$  p_1 p_3^2 p_4 p_5 s o^2$ &    (1, &    -1) &    0 \\
$  p_2 p_3^2 p_4 p_5 s o^2$ &    (0, &    -1) &    0 \\
$  p_3^3 p_4 p_5 s o^2$ &    (1, &    -2) &    0 \\
$  p_1^4 p_4^2 s o^3$ &    (1, &    1) &    1 \\
$  p_1^3 p_2 p_4^2 s o^3$ &    (0, &    1) &    1 \\
$  p_1^2 p_2^2 p_4^2 s o^3$ &    (-1, &    1) &    1 \\
$  p_1 p_2^3 p_4^2 s o^3$ &    (-2, &    1) &    1 \\
$  p_2^4 p_4^2 s o^3$ &    (-3, &    1) &    1 \\
$  p_1^3 p_3 p_4^2 s o^3$ &    (1, &    0) &    1 \\
$  p_1^2 p_2 p_3 p_4^2 s o^3$ &    (0, &    0) &    1 \\
$  p_1 p_2^2 p_3 p_4^2 s o^3$ &    (-1, &    0) &    1 \\
$  p_2^3 p_3 p_4^2 s o^3$ &    (-2, &    0) &    1 \\
$  p_1^2 p_3^2 p_4^2 s o^3$ &    (1, &    -1) &    1 \\
$  p_1 p_2 p_3^2 p_4^2 s o^3$ &    (0, &    -1) &    1 \\
$  p_2^2 p_3^2 p_4^2 s o^3$ &    (-1, &    -1) &    1 \\
$  p_1 p_3^3 p_4^2 s o^3$ &    (1, &    -2) &    1 \\
$  p_2 p_3^3 p_4^2 s o^3$ &    (0, &    -2) &    1 \\
$  p_3^4 p_4^2 s o^3$ &    (1, &    -3) &    1 \\
\hline
\end{tabular}
\end{minipage}
\hspace{0cm}
\begin{minipage}[!b]{0.6\textwidth}
\includegraphics[height=6cm]{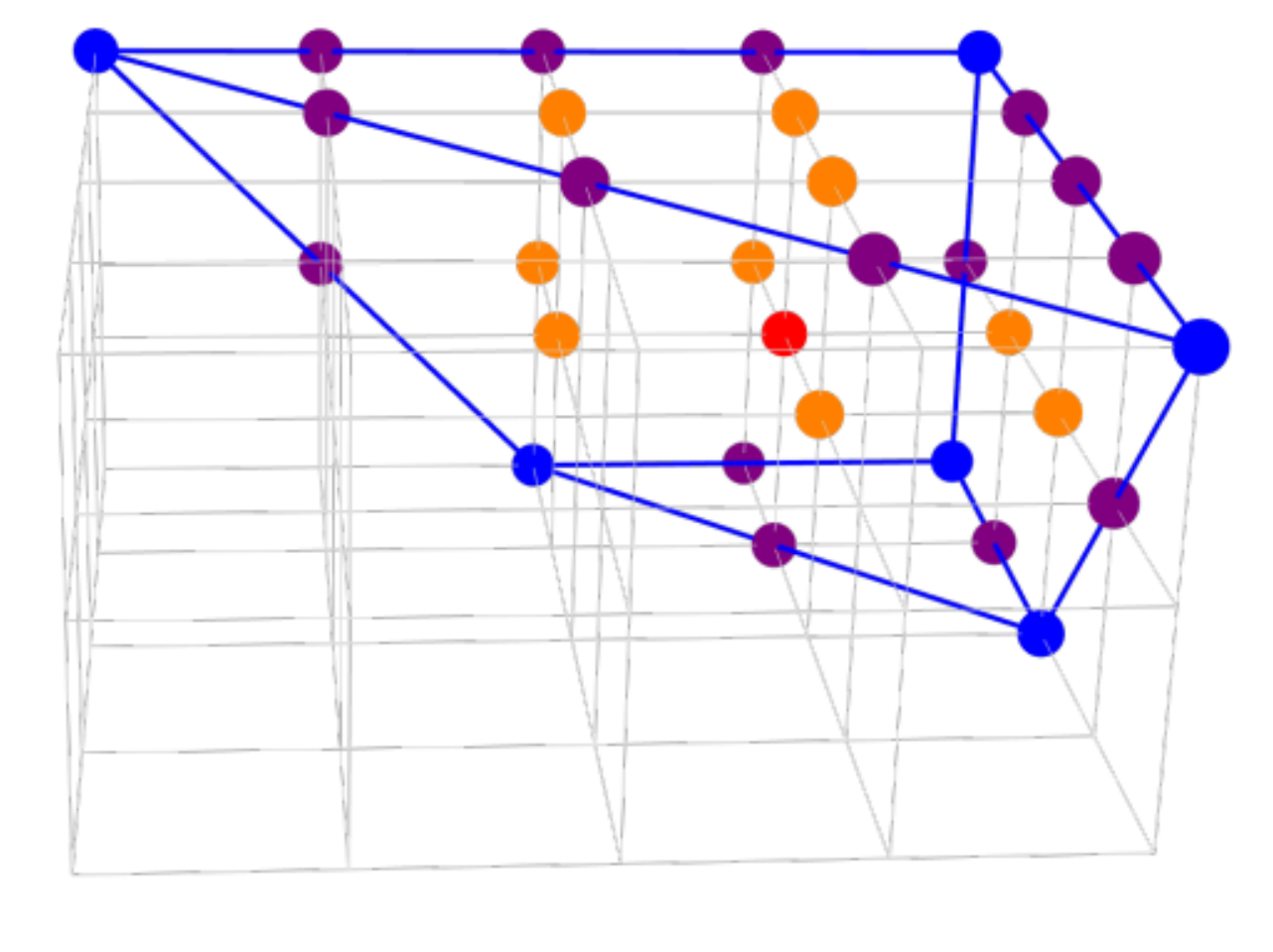} 
\end{minipage}
}
\caption{The generators and lattice of generators of the mesonic moduli space of Model 3 in terms of brick matchings with the corresponding flavor charges.
\label{f_genlattice_03}}
\end{table}

\begin{table}[H]
\centering
\resizebox{0.95\hsize}{!}{
\begin{tabular}{|l|cc|c|}
\hline
generator & \multicolumn{2}{c|}{$SU(3)_{(\tilde{x}_1,\tilde{x}_2)}$} & $U(1)_{\tilde{b}}$ \\
\hline
$Q_{14} Z_{46} X_{63} Q_{31}=  Q_{14} Z_{46} Q_{65} X_{51}=  Y_{12} X_{26} X_{63} Q_{31}=  Y_{12} X_{26} Q_{65} X_{51}$ &    (1, &    1) &    -1 \\
$Q_{14} Y_{46} X_{63} Q_{31}=  Q_{14} Z_{46} Z_{63} Q_{31}=  Q_{14} Y_{46} Q_{65} X_{51}=  Q_{14} Z_{46} Q_{65} Z_{51}=  X_{12} X_{26} X_{63} Q_{31}=  Y_{12} X_{26} Z_{63} Q_{31}$ &    (0, &    1) &    -1 \\
$=  X_{12} X_{26} Q_{65} X_{51}=  Y_{12} X_{26} Q_{65} Z_{51}$ & & & \\
$ Q_{14} Y_{46} Z_{63} Q_{31}=  Q_{14} Y_{46} Q_{65} Z_{51}=  X_{12} X_{26} Z_{63} Q_{31}=  X_{12} X_{26} Q_{65} Z_{51}$ &    (-1, &    1) &    -1 \\
$ P_{12} X_{26} X_{63} Q_{31}=  P_{12} X_{26} Q_{65} X_{51}=  Q_{14} Z_{46} P_{63} Q_{31}=  Y_{12} X_{26} P_{63} Q_{31}=  Q_{14} X_{46} X_{63} Q_{31}=  Q_{14} X_{46} Q_{65} X_{51}$ &    (1, &    0) &    -1 \\
$=  Q_{14} Z_{46} Q_{65} Y_{51}=  Y_{12} X_{26} Q_{65} Y_{51}$& & & \\
$ P_{12} X_{26} Z_{63} Q_{31}=  P_{12} X_{26} Q_{65} Z_{51}=  Q_{14} Y_{46} P_{63} Q_{31}=  X_{12} X_{26} P_{63} Q_{31}=  Q_{14} X_{46} Z_{63} Q_{31}=  Q_{14} X_{46} Q_{65} Z_{51}$&    (0, &    0) &    -1 \\
\hline
\end{tabular}
}
\caption{The generators in terms of bifundamental chiral fields for Model 3.
\label{f_genfields_03}}
\end{table}

The global symmetry of Model 3 and the charges on the extremal brick matchings under the global symmetry are summarized in \tref{t_pmcharges_03}.
Using the following fugacity map,
\beal{es0322}
&&
t = t_3 x_2 ~,~
x_1 = \frac{t_4 t_5}{t_2 t_3} ~,~
x_2 = \frac{t_2 x_1^2}{t_1} ~,~
b = \frac{t_4 x_1}{t_1} ~,~
\eea
the Hilbert series for Model 3 can be rewritten in terms of characters of irreducible representations of $SU(3)\times U(1)$, the mesonic flavor symmetry of Model 3, as follows 
\beal{es0325}
&&
g_1(t, x_i, b; \mathcal{M}_3) =
\sum_{n_1=0}^{\infty} \sum_{n_2=0}^{\infty}
\Big[
 [2n_1+3n_2,0] b^{-2n_1} t^{4n_1+5n_2}
\nn\\
&&
\hspace{4cm} 
+ [4n_1+3n_2+4,0] b^{2n_1+2} t^{6n_1+5n_2+6}
\Big]
 ~,~
\eea
where $ [m_1,m_2]= [m_1,m_2]_{SU(3)_{(x_1,x_2)}} $ and $b$ is the fugacity for the $U(1)$ factor of the mesonic flavor symmetry.
The corresponding plethystic logarithm is
\beal{es0326}
&&
\PL[g_1(t, x_i, b; \mathcal{M}_3)]=
[2,0] b^{-2} t^4
+[3,0] t^5
+[4,0] b^2 t^6
\nn\\
&&
\hspace{1cm}
- [0,2] b^{-4} t^8
- ( [3,1] b^{-2} + [1,2] b^{-2} ) t^9
- ( [6,0] + [4,1] + 2[2,2] ) t^{10}
\nn\\
&&
\hspace{1cm}
- ( [5,1] b^2 + [3,2] b^2 + [1,3] b^2 ) t^{11}
- ( [4,2] b^4 + [0,4] b^4) t^{12}
+
\dots ~,~
\eea
where we see that the mesonic moduli space is a non-complete intersection.
The generators form 3 sets which respectively transform as $[2,0] b^{-2}$, 
$[3,0]$ and $[4,0] b^2$ of the mesonic flavor symmetry. 
Using the following fugacity map
\beal{es0327}
&&
\tilde{t} = t_4^{1/2} t_5^{1/2} ~,~
\tilde{x}_1 = \frac{t_4^{3/2} t_5^{3/2}}{t_2^2 t_3} ~,~ 
\tilde{x}_2 = \frac{t_4^{3/2} t_5^{3/2}}{t_2 t_3^2} ~,~
\tilde{b} = \frac{t_4^2}{t_2 t_3} ~,~
\eea
the mesonic flavor charges on the gauge invariant operators become $\mathbb{Z}$-valued.
The generators in terms of brick matchings and their corresponding rescaled mesonic flavor charges are summarized in \tref{f_genlattice_03}.
The generator lattice as shown in \tref{f_genlattice_03} is a convex lattice polytope, which is reflexive. It is the dual of the toric diagram of Model 3 shown in \fref{f_toric_03}. Note that the 3 sets of generators transforming under $[2,0] b^{-2}$, 
$[3,0]$ and $[4,0] b^2$ of the mesonic flavor symmetry form the 3 layers of the generator lattice in \tref{f_genlattice_03}.
For completeness, \tref{f_genfields_03} shows the generators of Model 2 in terms of chiral fields with the corresponding mesonic flavor charges. 
\\

\section{Model 4: $P_{+-}^{1}(\text{dP}_0)$~[$\mathbb{P}(\mathcal{O}_{\mathbb{P}^1} \oplus \mathcal{O}_{\mathbb{P}^1} \oplus \mathcal{O}_{\mathbb{P}^1}  (1) )$,~$\langle6\rangle$] \label{smodel04}}
 
\begin{figure}[H]
\begin{center}
\resizebox{0.25\hsize}{!}{
\includegraphics[height=6cm]{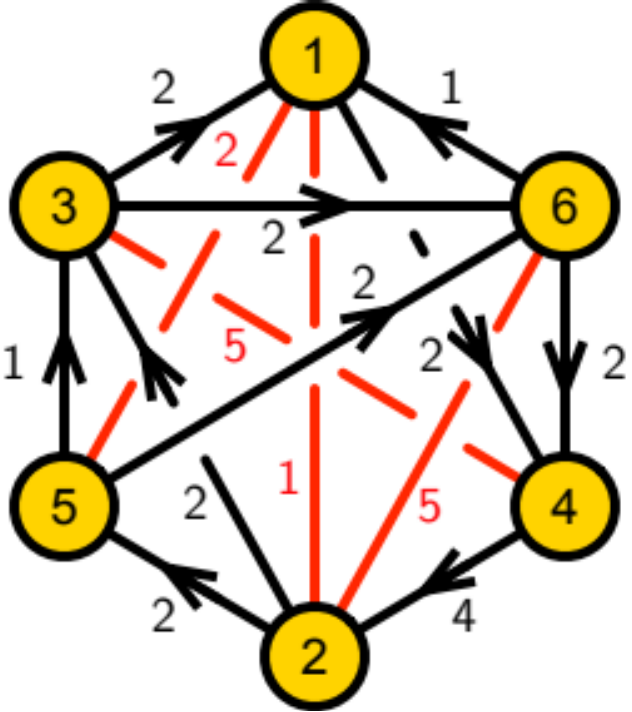} 
}
\caption{
Quiver for Model 4.
\label{f_quiver_04}}
 \end{center}
 \end{figure}
 
Model 4 corresponds to the Calabi-Yau 4-fold $P_{+-}^{1}(\text{dP}_0)$.
The corresponding brane brick model has the quiver in \fref{f_quiver_04} and the $J$- and $E$-terms are given as follows

  \beq
  {\footnotesize
\begin{array}{rrclrcl}
 &  & J &  & & E &  \\
\Lambda_{12}  : &  Z_{23} Z_{31}   &-& Y_{23} X_{31}  &   P_{14} Z_{42 }  &-&  Q_{14} T_{42} \\ 
\Lambda_{15}^{1}  : &  X_{53} X_{31}  &-&  Z_{56} Y_{61}  &   P_{14} X_{42} Q_{25 }  &-&  Q_{14} X_{42} P_{25} \\ 
\Lambda_{15}^{2}  : &  Y_{56} Y_{61}  &-& X_{53} Z_{31}  &   P_{14} Y_{42} Q_{25 }  &-&  Q_{14} Y_{42} P_{25} \\ 
\Lambda_{26}^{1}  : &  Y_{61} Q_{14} Y_{42}   &-& X_{64} Z_{42}  &   P_{25} Y_{56 }  &-&  Y_{23} P_{36} \\ 
\Lambda_{26}^{2}  : &  Z_{64} Z_{42 }  &-&  Y_{61} Q_{14} X_{42}  &   P_{25} Z_{56 }  &-&  Z_{23} P_{36} \\ 
\Lambda_{26}^{3}  : &   Y_{61} P_{14} Y_{42}  &-& X_{64} T_{42}   &   Y_{23} Q_{36}  &-& Q_{25} Y_{56} \\ 
\Lambda_{26}^{4}  : &  Z_{64} T_{42 }  &-&  Y_{61} P_{14} X_{42}  &    Z_{23} Q_{36}  &-& Q_{25} Z_{56}\\ 
\Lambda_{26}^{5}  : &  X_{64} X_{42 }  &-&  Z_{64} Y_{42}  &   P_{25} X_{53} Q_{36 }  &-&  Q_{25} X_{53} P_{36} \\ 
\Lambda_{34}^{1}  : &  X_{42} Q_{25} X_{53}   &-& Z_{42} Y_{23}  &   P_{36} X_{64 }  &-&  X_{31} P_{14} \\ 
\Lambda_{34}^{2}  : &  Z_{42} Z_{23 }  &-&  Y_{42} Q_{25} X_{53}  &   P_{36} Z_{64 }  &-&  Z_{31} P_{14} \\ 
\Lambda_{34}^{3}  : &  X_{42} P_{25} X_{53}  &-& T_{42} Y_{23}  &   X_{31} Q_{14}  &-& Q_{36} X_{64}  \\ 
\Lambda_{34}^{4}  : &  T_{42} Z_{23 }  &-&  Y_{42} P_{25} X_{53}  &   Z_{31} Q_{14}  &-& Q_{36} Z_{64} \\ 
\Lambda_{34}^{5}  : &  Y_{42} Y_{23}  &-& X_{42} Z_{23}  &   P_{36} Y_{61} Q_{14 }  &-&  Q_{36} Y_{61} P_{14} \\ 
\Lambda_{45}  : &  Z_{56} Z_{64}  &-& Y_{56} X_{64}  &   T_{42} Q_{25 }  &-&  Z_{42} P_{25}
 \end{array} 
 }~.~
\label{es0401}
 \eeq
 
Using the forward algorithm, we are able to obtain the brick matchings for Model 4. They are summarized in the $P$-matrix which takes the form
\beal{es0405}
{\footnotesize
 P=
\resizebox{0.45\textwidth}{!}{$
\left(
\begin{array}{c|ccccc|cccccc|cc}
\; & p_1 & p_2 & p_3 & p_4 & p_5 & s_1 & s_2 & s_3 & s_4 & s_5 & s_6  & o_1 & o_2\\
\hline
P_{14} & 1 & 0 & 0 & 0 & 0 & 1 & 0 & 0 & 0 & 0 & 0 & 1 & 0 \\
P_{25} & 1 & 0 & 0 & 0 & 0 & 0 & 1 & 0 & 0 & 0 & 0 & 0 & 1 \\
P_{36} & 1 & 0 & 0 & 0 & 0 & 0 & 0 & 1 & 0 & 0 & 0 & 1 & 1 \\
\hline
Q_{14} & 0 & 1 & 0 & 0 & 0 & 1 & 0 & 0 & 0 & 0 & 0 & 1 & 0 \\
Q_{25} & 0 & 1 & 0 & 0 & 0 & 0 & 1 & 0 & 0 & 0 & 0 & 0 & 1 \\
Q_{36} & 0 & 1 & 0 & 0 & 0 & 0 & 0 & 1 & 0 & 0 & 0 & 1 & 1 \\
\hline
T_{42} & 1 & 0 & 0 & 0 & 1 & 0 & 0 & 0 & 1 & 0 & 0 & 1 & 1 \\
\hline
X_{31} & 0 & 0 & 1 & 0 & 0 & 0 & 0 & 1 & 0 & 1 & 0 & 0 & 1 \\
X_{42} & 0 & 0 & 0 & 1 & 0 & 0 & 0 & 0 & 1 & 0 & 0 & 0 & 0 \\
X_{53} & 0 & 0 & 0 & 0 & 1 & 0 & 0 & 0 & 0 & 0 & 1 & 1 & 0 \\
X_{64} & 0 & 0 & 1 & 0 & 0 & 1 & 0 & 0 & 0 & 1 & 0 & 0 & 0 \\
\hline
Y_{23} & 0 & 0 & 0 & 1 & 0 & 0 & 1 & 0 & 0 & 0 & 1 & 0 & 0 \\
Y_{42} & 0 & 0 & 1 & 0 & 0 & 0 & 0 & 0 & 1 & 0 & 0 & 0 & 0 \\
Y_{56} & 0 & 0 & 0 & 1 & 0 & 0 & 0 & 1 & 0 & 0 & 1 & 1 & 0 \\
Y_{61} & 0 & 0 & 0 & 0 & 1 & 0 & 0 & 0 & 0 & 1 & 0 & 0 & 1 \\
\hline
Z_{23} & 0 & 0 & 1 & 0 & 0 & 0 & 1 & 0 & 0 & 0 & 1 & 0 & 0 \\
Z_{31} & 0 & 0 & 0 & 1 & 0 & 0 & 0 & 1 & 0 & 1 & 0 & 0 & 1 \\
Z_{42} & 0 & 1 & 0 & 0 & 1 & 0 & 0 & 0 & 1 & 0 & 0 & 1 & 1 \\
Z_{56} & 0 & 0 & 1 & 0 & 0 & 0 & 0 & 1 & 0 & 0 & 1 & 1 & 0 \\
Z_{64} & 0 & 0 & 0 & 1 & 0 & 1 & 0 & 0 & 0 & 1 & 0 & 0 & 0 \\
\end{array}
\right)
$}
}~.~
 \eea

The $J$- and $E$-term charges are given by
\beal{es0305}
{\footnotesize
Q_{JE} =
\resizebox{0.42\textwidth}{!}{$
\left(
\begin{array}{ccccc|cccccc|cc}
p_1 & p_2 & p_3 & p_4 & p_5 & s_1 & s_2 & s_3 & s_4 & s_5 & s_6  & o_1 & o_2\\
\hline
 1 & 1 & 0 & 0 & -1 & -1 & -1 & -1 & 0 & 1 & 1 & 0 & 0 \\
 0 & 0 & 0 & 0 & -1 & -1 & 0 & -1 & 0 & 1 & 0 & 1 & 0 \\
 0 & 0 & 0 & 0 & -1 & 0 & -1 & -1 & 0 & 0 & 1 & 0 & 1 \\
 0 & 0 & -1 & -1 & -1 & 0 & 0 & 0 & 1 & 1 & 1 & 0 & 0 \\
\end{array}
\right)
$}
}~,~
\eea
and the $D$-term charges are given by
\beal{e306}
{\footnotesize
Q_D =
\resizebox{0.42\textwidth}{!}{$
\left(
\begin{array}{ccccc|cccccc|cc}
p_1 & p_2 & p_3 & p_4 & p_5 & s_1 & s_2 & s_3 & s_4 & s_5 & s_6  & o_1 & o_2\\
\hline
 0 & 0 & -1 & -1 & -1 & 0 & 1 & 0 & 0 & 1 & 1 & 0 & 0 \\
 0 & 0 & 0 & 0 & 0 & 0 & 0 & 1 & 0 & 0 & -1 & 0 & 0 \\
 0 & 0 & 1 & 1 & 1 & -1 & 0 & 0 & 0 & -1 & -1 & 0 & 0 \\
 0 & 0 & 0 & 0 & 0 & 0 & -1 & 0 & 0 & 0 & 1 & 0 & 0 \\
 0 & 0 & 0 & 0 & 0 & 0 & 0 & -1 & 0 & 1 & 0 & 0 & 0 \\
\end{array}
\right)
$}
}~.~
\eea

\begin{figure}[H]
\begin{center}
\resizebox{0.4\hsize}{!}{
\includegraphics[height=6cm]{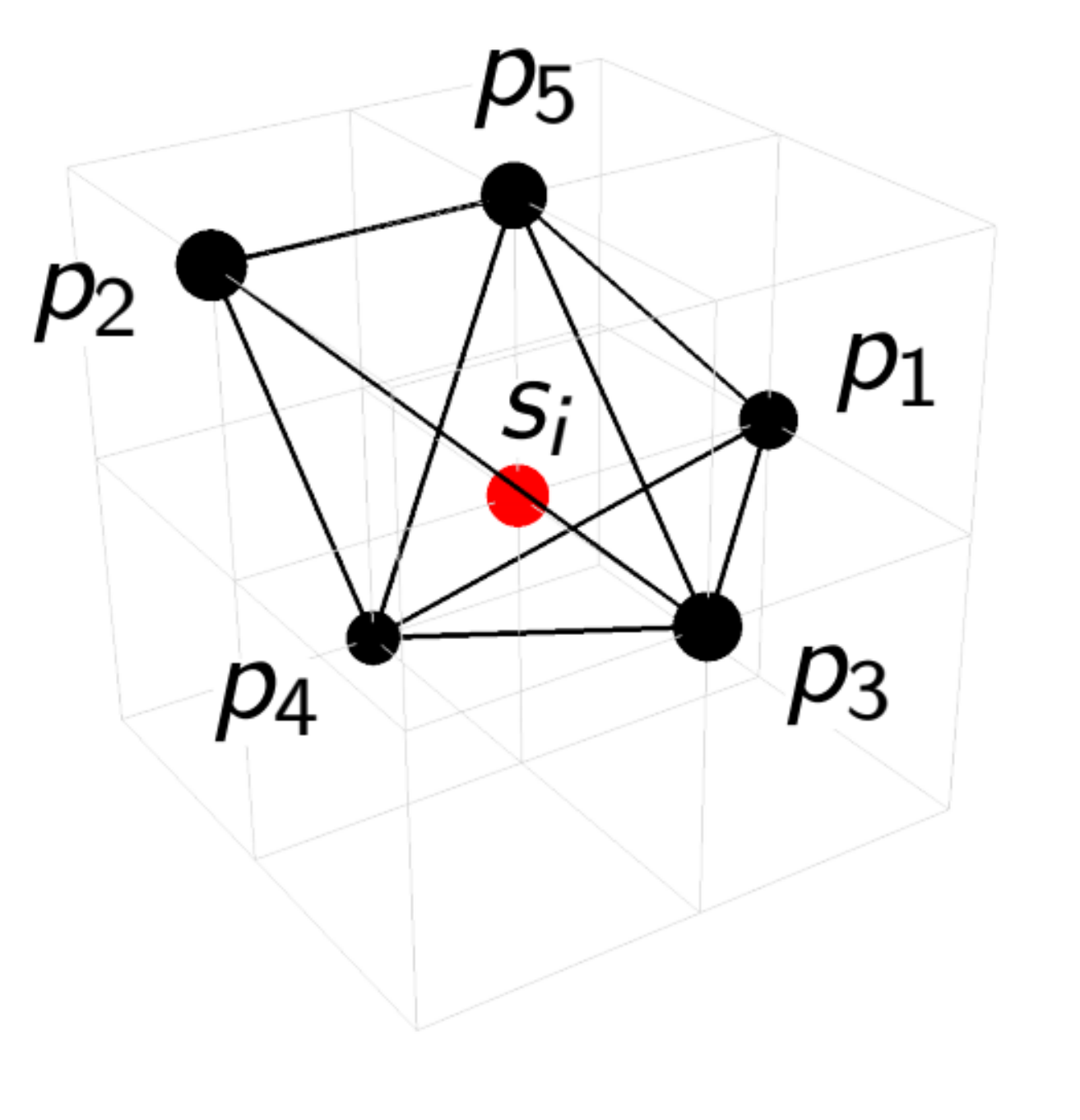} 
}
\caption{
Toric diagram for Model 4.
\label{f_toric_04}}
 \end{center}
 \end{figure}

The toric diagram of Model 4 is given by
\beal{es0410}
{\footnotesize
G_t=
\resizebox{0.42\textwidth}{!}{$
\left(
\begin{array}{ccccc|cccccc|cc}
p_1 & p_2 & p_3 & p_4 & p_5 & s_1 & s_2 & s_3 & s_4 & s_5 & s_6  & o_1 & o_2\\
\hline
 1 & 1 & 1 & 1 & 1 & 1 & 1 & 1 & 1 & 1 & 1 & 2 & 2 \\
 -1 & 1 & 0 & 0 & 0 & 0 & 0 & 0 & 0 & 0 & 0 & 0 & 0 \\
 -1 & 0 & 1 & 0 & 0 & 0 & 0 & 0 & 0 & 0 & 0 & 0 & 0 \\
 1 & 0 & 0 & 1 & -1 & 0 & 0 & 0 & 0 & 0 & 0 & 1 & 1 \\
\end{array}
\right)
$}
}~,~
\eea
where \fref{f_toric_04} shows the toric diagram with brick matching labels.
 
\begin{table}[H]
\centering
\begin{tabular}{|c|c|c|c|c|c|}
\hline
\; & $SU(2)_{x}$ & $SU(2)_{y}$ &  $U(1)_{b}$ & $U(1)$ & \text{fugacity} \\
\hline
$p_1$ & +1 & 0 & +1 & $r_1$ & $t_1$ \\
$p_2$ & -1 & 0 & +1 & $r_2$ & $t_2$ \\
$p_3$ & 0 & +1  & 0 & $r_3$ & $t_3$ \\
$p_4$ & 0 & -1 & 0 & $r_4$ & $t_4$ \\
$p_5$ & 0 &0 & -2 & $r_5$ & $t_5$ \\
\hline
\end{tabular}
\caption{Global symmetry charges on the extremal brick matchings $p_i$ of Model 4.}
\label{t_pmcharges_04}
\end{table}

\begin{table}[H]
\centering
\resizebox{.95\hsize}{!}{
\begin{minipage}[!b]{0.5\textwidth}
\begin{tabular}{|c|c|c|c|}
\hline
generator & $SU(2)_{\tilde{x}}$ & $SU(2)_{\tilde{y}}$ & $U(1)_{\tilde{b}}$ \\
\hline
$p_1 p_3^3 ~s o $ & 2 &  0 &  1  \\
$p_2 p_3^3 ~s o $ & 2 &  1 &  1  \\
$p_1 p_3^2 p_4 ~s o $ & 1 &  0 &  1  \\
$p_2 p_3^2 p_4 ~s o $ & 1 &  1 &  1  \\
$p_1 p_3 p_4^2 ~s o $ & 0 &  0 &  1  \\
$p_2 p_3 p_4^2 ~s o $ & 0 &  1 &  1  \\
$p_1 p_4^3 ~s o $ & -1 &  0 &  1  \\
$p_2 p_4^3 ~s o $ & -1 &  1 &  1  \\
$p_1^2 p_3^2 p_5 ~s o^2 $ & 1 &  -1 &  0  \\
$p_1 p_2 p_3^2 p_5 ~s o^2 $ & 1 &  0 &  0  \\
$p_2^2 p_3^2 p_5 ~s o^2 $ & 1 &  1 &  0  \\
$p_1^2 p_3 p_4 p_5 ~s o^2 $ & 0 &  -1 &  0  \\
$p_1 p_2 p_3 p_4 p_5 ~s o^2 $ & 0 &  0 &  0  \\
$p_2^2 p_3 p_4 p_5 ~s o^2 $ & 0 &  1 &  0  \\
$p_1^2 p_4^2 p_5 ~s o^2 $ & -1 &  -1 &  0  \\
$p_1 p_2 p_4^2 p_5 ~s o^2 $ & -1 &  0 &  0  \\
$p_2^2 p_4^2 p_5 ~s o^2 $ & -1 &  1 &  0  \\
$p_1^3 p_3 p_5^2 ~s o^3 $ & 0 &  -2 &  -1  \\
$p_1^2 p_2 p_3 p_5^2 ~s o^3 $ & 0 &  -1 &  -1  \\
$p_1 p_2^2 p_3 p_5^2 ~s o^3 $ & 0 &  0 &  -1  \\
$p_2^3 p_3 p_5^2 ~s o^3 $ & 0 &  1 &  -1  \\
$p_1^3 p_4 p_5^2 ~s o^3 $ & -1 &  -2 &  -1  \\
$p_1^2 p_2 p_4 p_5^2 ~s o^3 $ & -1 &  -1 &  -1  \\
$p_1 p_2^2 p_4 p_5^2 ~s o^3 $ & -1 &  0 &  -1  \\
$p_2^3 p_4 p_5^2 ~s o^3 $ & -1 &  1 &  -1  \\
$p_1^4 p_5^3 ~s o^4 $ & -1 &  -3 &  -2  \\
$p_1^3 p_2 p_5^3 ~s o^4 $ & -1 &  -2 &  -2  \\
$p_1^2 p_2^2 p_5^3 ~s o^4 $ & -1 &  -1 &  -2  \\
$p_1 p_2^3 p_5^3 ~s o^4 $ & -1 &  0 &  -2  \\
$p_2^4 p_5^3 ~s o^4 $ & -1 &  1 &  -2 \\
\hline
\end{tabular}
\end{minipage}
\hspace{0cm}
\begin{minipage}[!b]{0.6\textwidth}
\includegraphics[height=8cm]{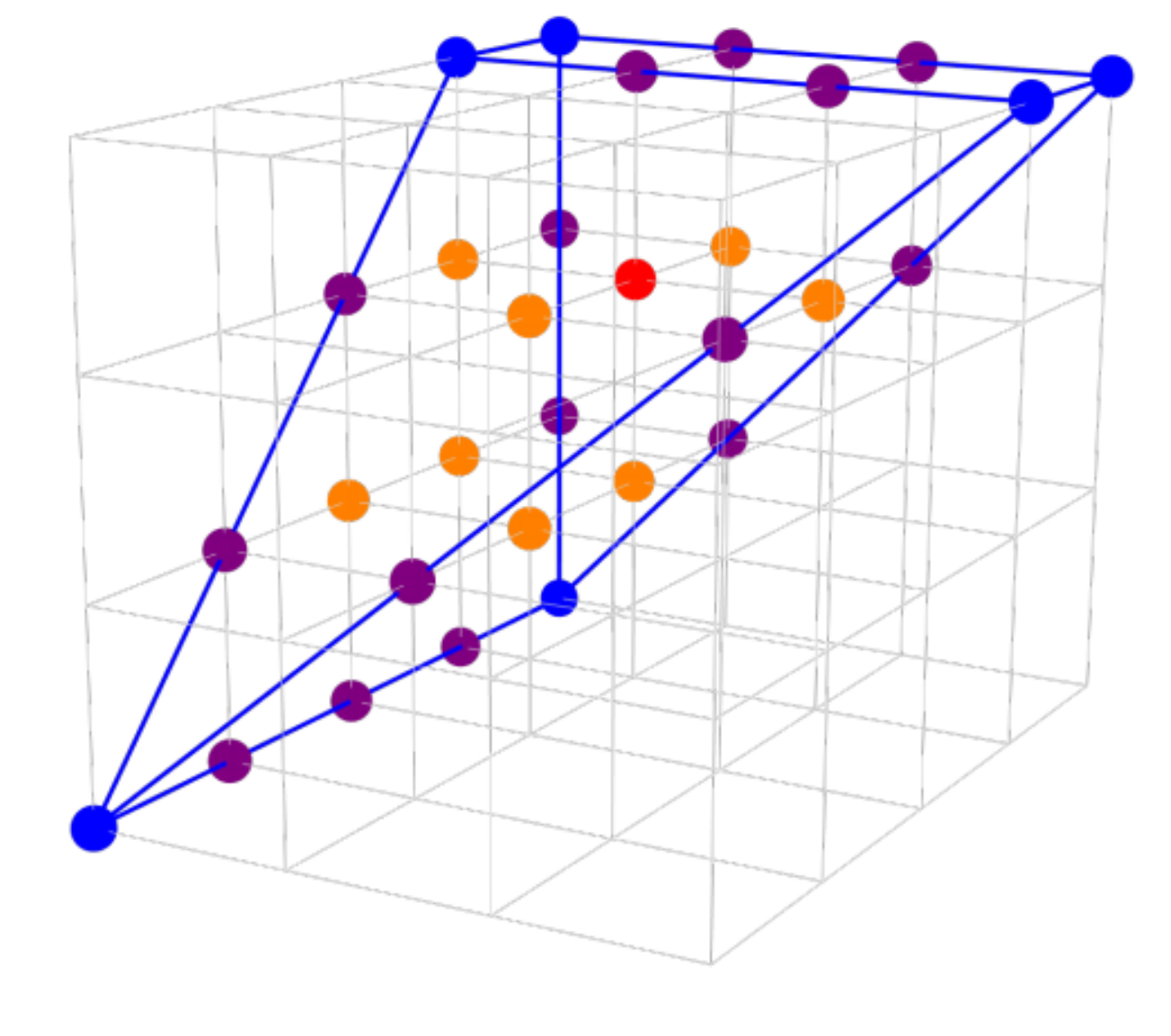} 
\end{minipage}
}
\caption{The generators and lattice of generators of the mesonic moduli space of Model 4 in terms of brick matchings with the corresponding flavor charges.
\label{f_genlattice_04}}
\end{table}

Using the Molien integral formula, the Hilbert series of the mesonic moduli space of Model 4 is found to be as follows
 \beal{es0420}
&&
g_1(t_i,y_s,y_o; \mathcal{M}_4) =
\frac{P(t_i,y_s,y_o; \mathcal{M}_4)}{
(1 - y_s y_o t_1 t_3^3) 
(1 - y_s y_o t_2 t_3^3) 
(1 - y_s y_o t_1 t_4^3) 
(1 - y_s y_o t_2 t_4^3) 
}
\nn\\
&&
\hspace{1cm} 
\times
\frac{1}{
(1 - y_s y_o^4 t_1^4 t_5^3) 
(1 - y_s y_o^4 t_2^4 t_5^3)
} ~,~
\eea
where $t_i$ are the fugacities for the extremal brick matchings $p_i$.
$y_{s}$ counts the product of brick matchings $s_1 \dots s_6$ corresponding to the single internal point of the toric diagram of Model 4.
Additionally, the fugacity $y_o$ corresponds to the product $o_1 o_2$ of extra GLSM fields.
The explicit numerator $P(t_i,y_s,y_o; \mathcal{M}_4)$ of the Hilbert series is given in the Appendix Section \sref{app_num_04}.
We note that setting the fugacity $y_o=1$ does not change the overall characterization of the mesonic moduli space by the Hilbert series, indicating that the extra GLSM fields, as expected, correspond to an over-parameterization of the moduli space. 

By setting $t_i=t$ for the fugacities of the extremal brick matchings, and all other fugacities to $y_s=1$ and $y_o=1$, the unrefined Hilbert series takes the following form
\beal{es0421}
&&
g_1(t,1,1; \mathcal{M}_4) =
\frac{
1
}{
(1 - t^4)^4 (1 -  t^7)^2
}
\times
\nn\\
&&
\hspace{1cm}
(
1 + 4 t^4 + 9 t^5 + 8 t^6 + 3 t^7 - 5 t^8 - 12 t^9 - 7 t^{10} 
- 4 t^{11} + 3 t^{12} + 3 t^{13} - 4 t^{14}
\nn\\
&&
\hspace{1cm}
 - 7 t^{15} - 12 t^{16} - 5 t^{17} 
+ 3 t^{18} + 8 t^{19} + 9 t^{20} + 4 t^{21} + t^{25}
)
~,~
\eea
where the palindromic numerator indicates that the mesonic moduli space is Calabi-Yau. 

The global symmetry of Model 4 and the charges on the extremal brick matchings under the global symmetry are summarized in \tref{t_pmcharges_04}.
Using the following fugacity map,
\beal{es0422}
&&
t = \frac{t_1 t_2 t_5}{t_3 t_4} ~,~
x = \frac{t_1^{1/2}}{t_2^{1/2}} ~,~
y = \frac{t_1 t_2 t_5}{t_3 t_4^2} ~,~
b = \frac{t_3 t_4}{t_1^{1/2} t_2^{1/2} t_5} ~,~
\eea
the Hilbert series for Model 4 can be rewritten in terms of characters of irreducible representations of $SU(2)\times SU(2)\times U(1)$.
In terms of fugacities of the mesonic flavor symmetry of Model 4, the Hilbert series becomes
\beal{es0525}
&&
g_1(t, x, y, b; \mathcal{M}_4) =
\sum_{n_1=0}^{\infty} \sum_{n_2=0}^{\infty}
\Big[
[n_1+2n_2; 3n_1+2n_2] b^{n_1} t^{4n_1+5n_2}
\nn\\
&&
\hspace{4.5cm} 
+ [3n_1+2n_2+3; n_1+2n_2+1] b^{-n_1-1} t^{6n_1+5n_2+6}
\nn\\
&&
\hspace{4.5cm} 
+ [3n_1+7n_2+7; n_1+n_2+1] b^{-n_1-3n_2-3} t^{6n_1+13n_2+13}
\nn\\
&&
\hspace{4.5cm} 
+ [4n_1+7n_2+4; n_2] b^{-2n_1-3n_2-2} t^{7n_1+13n_2+7}
\Big]
 ~,~
 \eea
 where $[m;n] = [m]_{SU(2)_{x}} [n]_{SU(2)_y}$. The fugacity $b$ counts charges under the $U(1)$ factor of the mesonic flavor symmetry. 
 
The corresponding plethystic logarithm is
\beal{es0426}
&&
\PL[g_1(t, x,y, b; \mathcal{M}_4)]=
[1;3] b t^4 + [2;2] t^5 + [3;1] b^{-1} t^6 + [4;0] b^{-2} t^7  
\nn\\
&&
\hspace{1cm}
- ([0;4] + [2;2]  + 1)  b^2 t^8
- ([1;5]  + [3;3] + [1;3] + [3;1] +[1;1] ) b t^9
\nn\\
&&
\hspace{1cm}
- ([4;4] + [2;4] + [4;2]+ [0;4] + 2[2;2] + [4;0] + 1) t^{10} 
\nn\\
&&
\hspace{1cm}
- ([5;3] + 2[3;3] + [5;1] + [1;3] + [3;1] + [1;1]) b^{-1} t^{11}
\nn\\
&&
\hspace{1cm}
- ([6;2] + [4;2] + 2[2;2] + [4;0] + 1) b^{-2} t^{12}
+
\dots ~,~
\eea
where we see that the mesonic moduli space is a non-complete intersection.
The generators form 4 sets that transform under 
$[1;3] b$, $[2;2]$, $[3;1] b^{-1}$ and $[4;0] b^{-2}$ of the mesonic flavor symmetry of Model 4, respectively.
Using the following fugacity map
\beal{es0427}
&&
\tilde{t} = t_3^{1/2} t_4^{1/2}~,~ 
\tilde{x} = \frac{t_3}{t_4}~,~ 
\tilde{y} =\frac{t_2^2 t_5}{t_3^{3/2}t_4^{3/2}}~,~ 
\tilde{b} = \frac{t_3^{1/2} t_4^{3/2}}{t_2 t_5}~,~
\eea
the mesonic flavor charges on the gauge invariant operators become $\mathbb{Z}$-valued.
The generators in terms of brick matchings and their corresponding rescaled mesonic flavor charges are summarized in \tref{f_genlattice_04}.
The generator lattice as shown in \tref{f_genlattice_04} is a convex lattice polytope, which is reflexive. It is the dual of the toric diagram of Model 4 shown in \fref{f_toric_04}.
We also note that the 4 layers of points in the generator lattice in \tref{f_genlattice_04} corresponds to the 4 sets of generators that transform respectively under $[1;3] b$, $[2;2]$, $[3;1] b^{-1}$ and $[4;0] b^{-2}$ of the mesonic flavor symmetry.
For completeness, 
\tref{f_genfields_04a} and \tref{f_genfields_04b} show the generators of Model 4 in terms of chiral fields with the corresponding mesonic flavor charges.
\\

\begin{table}[H]
\centering
\resizebox{0.95\hsize}{!}{
\begin{tabular}{|l|c|c|c|}
\hline
generator & $SU(2)_{\tilde{x}}$ & $SU(2)_{\tilde{y}}$ & $U(1)_{\tilde{b}}$ \\
\hline
$ P_{14} Y_{42} Z_{23} X_{31}=  P_{25} Z_{56} X_{64} Y_{42}=  Z_{23} P_{36} X_{64} Y_{42}$ & 2 &  0 &  1  \\
$Q_{14} Y_{42} Z_{23} X_{31}=  Q_{25} Z_{56} X_{64} Y_{42}=  Z_{23} Q_{36} X_{64} Y_{42}$ & 2 &  1 &  1  \\
$P_{14} X_{42} Z_{23} X_{31}=  P_{14} Y_{42} Y_{23} X_{31}=  P_{14} Y_{42} Z_{23} Z_{31}=  P_{25} Z_{56} X_{64} X_{42}=  P_{25} Y_{56} X_{64} Y_{42}=  P_{25} Z_{56} Z_{64} Y_{42}$ & 1 &  0 &  1  \\
$=  Z_{23} P_{36} X_{64} X_{42}=  Y_{23} P_{36} X_{64} Y_{42}=  Z_{23} P_{36} Z_{64} Y_{42}$ & & & \\
$Q_{14} X_{42} Z_{23} X_{31}=  Q_{14} Y_{42} Y_{23} X_{31}=  Q_{14} Y_{42} Z_{23} Z_{31}=  Q_{25} Z_{56} X_{64} X_{42}=  Q_{25} Y_{56} X_{64} Y_{42}=  Q_{25} Z_{56} Z_{64} Y_{42}$& 1 &  1 &  1  \\
$=  Z_{23} Q_{36} X_{64} X_{42}=  Y_{23} Q_{36} X_{64} Y_{42}=  Z_{23} Q_{36} Z_{64} Y_{42}$ & & & \\ 
$P_{14} X_{42} Y_{23} X_{31}=  P_{14} X_{42} Z_{23} Z_{31}=  P_{14} Y_{42} Y_{23} Z_{31}=  P_{25} Y_{56} X_{64} X_{42}=  P_{25} Z_{56} Z_{64} X_{42}=  P_{25} Y_{56} Z_{64} Y_{42}$& 0 &  0 &  1  \\
\hline
\end{tabular}
}
\caption{The generators in terms of bifundamental chiral fields for Model 4 \bf{(Part 1)}.
\label{f_genfields_04a}}
\end{table}

\begin{table}[H]
\centering
\resizebox{0.95\hsize}{!}{
\begin{tabular}{|l|c|c|c|}
\hline
generator & $SU(2)_{\tilde{x}}$ & $SU(2)_{\tilde{y}}$ & $U(1)_{\tilde{b}}$ \\
\hline
$P_{14} T_{42} Y_{23} X_{31}=  P_{14} T_{42} Z_{23} Z_{31}=  P_{25} Y_{56} X_{64} T_{42}=  P_{25} Z_{56} Z_{64} T_{42}=  Y_{23} P_{36} X_{64} T_{42}=  Z_{23} P_{36} Z_{64} T_{42}$& 0 &  -1 &  0  \\
$=  P_{14} X_{42} P_{25} X_{53} X_{31}=  P_{14} X_{42} P_{25} Z_{56} Y_{61}=  P_{14} Y_{42} P_{25} X_{53} Z_{31}=  P_{14} Y_{42} P_{25} Y_{56} Y_{61}=  P_{14} X_{42} Z_{23} P_{36} Y_{61}$& & & \\  
$=  P_{14} Y_{42} Y_{23} P_{36} Y_{61}=  P_{25} X_{53} P_{36} X_{64} X_{42}=  P_{25} X_{53} P_{36} Z_{64} Y_{42}$ & & & \\  
$P_{14} Z_{42} Y_{23} X_{31}=  P_{14} Z_{42} Z_{23} Z_{31}=  P_{25} Y_{56} X_{64} Z_{42}=  P_{25} Z_{56} Z_{64} Z_{42}=  Y_{23} P_{36} X_{64} Z_{42}=  Z_{23} P_{36} Z_{64} Z_{42}$ & 0 &  0 &  0  \\
$=  Q_{14} T_{42} Y_{23} X_{31}=  Q_{14} T_{42} Z_{23} Z_{31}=  Q_{25} Y_{56} X_{64} T_{42}=  Q_{25} Z_{56} Z_{64} T_{42}=  Y_{23} Q_{36} X_{64} T_{42}=  Z_{23} Q_{36} Z_{64} T_{42}$& & & \\  
$=  P_{14} X_{42} Q_{25} X_{53} X_{31}=  P_{14} X_{42} Q_{25} Z_{56} Y_{61}=  P_{14} Y_{42} Q_{25} X_{53} Z_{31}=  P_{14} Y_{42} Q_{25} Y_{56} Y_{61}=  P_{14} X_{42} Z_{23} Q_{36} Y_{61}$& & & \\  
$=  P_{14} Y_{42} Y_{23} Q_{36} Y_{61}=  Q_{14} X_{42} P_{25} X_{53} X_{31}=  Q_{14} X_{42} P_{25} Z_{56} Y_{61}=  Q_{14} Y_{42} P_{25} X_{53} Z_{31}=  Q_{14} Y_{42} P_{25} Y_{56} Y_{61}$& & & \\
\hline
\end{tabular}
}
\caption{The generators in terms of bifundamental chiral fields for Model 4 \bf{(Part 2)}.
\label{f_genfields_04b}}
\end{table}

\section{Model 5: $Y^{2,5}(\mathbb{CP}^2)$~[$\mathbb{P}(\mathcal{O}_{\mathbb{P}^2} \oplus \mathcal{O}_{\mathbb{P}^2}(2))$,~$\langle7\rangle$] \label{smodel05}}
 
\begin{figure}[H]
\begin{center}
\resizebox{0.25\hsize}{!}{
\includegraphics[height=6cm]{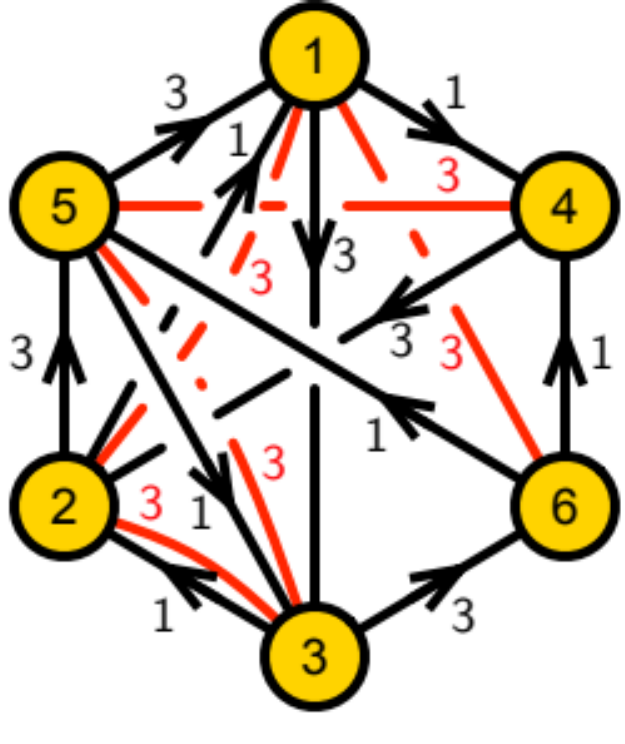} 
}
\caption{
Quiver for Model 5.
\label{f_quiver_05}}
 \end{center}
 \end{figure}
 
Model 5 corresponds to one of the $Y^{p,k}(\mathbb{CP}^2)$ models, $Y^{2,5}(\mathbb{CP}^2)$. 
The corresponding brane brick model has the quiver in \fref{f_quiver_05} and the $J$- and $E$-terms are given as follows
  \beq
  {\footnotesize
\begin{array}{rrclrcl}
 &  & J &  & & E &  \\
\Lambda_{12}^{1}  : &  X_{25} X_{51}  &-& Q_{25} Z_{51} &   P_{13} X_{32}  &-&  X_{14} P_{42} \\ 
\Lambda_{12}^{2}  : &  X_{25} R_{51}   &-& P_{25} Z_{51} &   X_{14} Q_{42}  &-& Q_{13} X_{32} \\ 
\Lambda_{16}^{1}  : &  X_{64} Q_{42} Y_{21}   &-& X_{65} X_{51}  &   P_{13} X_{36}  &-&  X_{13} P_{36} \\ 
\Lambda_{16}^{2}  : &  X_{65} Z_{51}  &-&  X_{64} X_{42} Y_{21}  &   P_{13} Q_{36}  &-&  Q_{13} P_{36} \\ 
\Lambda_{16}^{3}  : &  X_{64} P_{42} Y_{21}  &-& X_{65} R_{51}  &   X_{13} Q_{36}  &-& Q_{13} X_{36}  \\ 
\Lambda_{21}^{1}  : &  X_{14} X_{42}   &-& X_{13} X_{32} &   P_{25} X_{51}  &-&  Q_{25} R_{51} \\ 
\Lambda_{23}^{1}  : &  X_{36} X_{64} Q_{42}  &-& Q_{36} X_{64} X_{42}  &   P_{25} Y_{53}  &-&  Y_{21} P_{13} \\ 
\Lambda_{23}^{2}  : &  X_{36} X_{64} P_{42}  &-& P_{36} X_{64} X_{42}  &   Y_{21} Q_{13}  &-& Q_{25} Y_{53} \\ 
\Lambda_{32}^{1}  : &  Y_{21} X_{13}  &-& X_{25} Y_{53}  &   P_{36} X_{64} Q_{42}  &-&  Q_{36} X_{64} P_{42} \\ 
\Lambda_{35}^{1}  : &  Z_{51} Q_{13}   &-& X_{51} X_{13} &   P_{36} X_{65}  &-&  X_{32} P_{25} \\ 
\Lambda_{35}^{2}  : &  Z_{51} P_{13}   &-& R_{51} X_{13} &   X_{32} Q_{25}  &-& Q_{36} X_{65} \\ 
\Lambda_{45}^{1}  : &  X_{51} X_{14}  &-&  Y_{53} Q_{36} X_{64}  &   P_{42} X_{25}  &-&  X_{42} P_{25} \\ 
\Lambda_{45}^{2}  : &  Y_{53} X_{36} X_{64}  &-& Z_{51} X_{14}  &   P_{42} Q_{25}  &-&  Q_{42} P_{25} \\ 
\Lambda_{45}^{3}  : &  R_{51} X_{14}  &-&  Y_{53} P_{36} X_{64}  &   X_{42} Q_{25}  &-& Q_{42} X_{25} \\ 
\Lambda_{53}^{1}  : &  X_{32} X_{25}  &-&  X_{36} X_{65}  &   R_{51} Q_{13}  &-&  X_{51} P_{13}
 \end{array} 
 }~.~
\label{es0501}
 \eeq
 
Using the forward algorithm, we are able to obtain the brick matchings for Model 5. They are summarized in the $P$-matrix which takes the form
\beal{es0505}
{\footnotesize
 P=
 \resizebox{0.55\textwidth}{!}{$
\left(
\begin{array}{c|ccccc|cccccc|cccccc}
\; & p_1 & p_2 & p_3 & p_4 & p_5 & s_1 & s_2 & s_3 & s_4 & s_5 & s_6  & o_1 & o_2 & o_3 & o_4 & o_5 & o_6\\
\hline
P_{13} & 1 & 0 & 0 & 0 & 0 & 0 & 0 & 0 & 0 & 0 & 1 & 1 & 1 & 1 & 1 & 0 & 0 \\
P_{25} & 1 & 0 & 0 & 0 & 0 & 0 & 0 & 0 & 0 & 1 & 0 & 1 & 1 & 1 & 0 & 1 & 0 \\
P_{36} & 1 & 0 & 0 & 0 & 0 & 0 & 0 & 0 & 1 & 0 & 0 & 1 & 0 & 0 & 0 & 1 & 1 \\
P_{42} & 1 & 0 & 0 & 0 & 0 & 0 & 0 & 1 & 0 & 0 & 0 & 0 & 1 & 0 & 1 & 0 & 1 \\
\hline
Q_{13} & 0 & 1 & 0 & 0 & 0 & 0 & 0 & 0 & 0 & 0 & 1 & 1 & 1 & 1 & 1 & 0 & 0 \\
Q_{25} & 0 & 1 & 0 & 0 & 0 & 0 & 0 & 0 & 0 & 1 & 0 & 1 & 1 & 1 & 0 & 1 & 0 \\
Q_{36} & 0 & 1 & 0 & 0 & 0 & 0 & 0 & 0 & 1 & 0 & 0 & 1 & 0 & 0 & 0 & 1 & 1 \\
Q_{42} & 0 & 1 & 0 & 0 & 0 & 0 & 0 & 1 & 0 & 0 & 0 & 0 & 1 & 0 & 1 & 0 & 1 \\
\hline
R_{51} & 1 & 0 & 0 & 1 & 0 & 0 & 1 & 0 & 0 & 0 & 0 & 0 & 0 & 0 & 1 & 1 & 1 \\
\hline
X_{13} & 0 & 0 & 1 & 0 & 0 & 0 & 0 & 0 & 0 & 0 & 1 & 1 & 1 & 1 & 1 & 0 & 0 \\
X_{14} & 0 & 0 & 0 & 0 & 1 & 1 & 0 & 0 & 1 & 0 & 1 & 1 & 0 & 1 & 0 & 0 & 0 \\
X_{25} & 0 & 0 & 1 & 0 & 0 & 0 & 0 & 0 & 0 & 1 & 0 & 1 & 1 & 1 & 0 & 1 & 0 \\
X_{32} & 0 & 0 & 0 & 0 & 1 & 1 & 0 & 1 & 1 & 0 & 0 & 0 & 0 & 0 & 0 & 0 & 1 \\
X_{36} & 0 & 0 & 1 & 0 & 0 & 0 & 0 & 0 & 1 & 0 & 0 & 1 & 0 & 0 & 0 & 1 & 1 \\
X_{42} & 0 & 0 & 1 & 0 & 0 & 0 & 0 & 1 & 0 & 0 & 0 & 0 & 1 & 0 & 1 & 0 & 1 \\
X_{51} & 0 & 1 & 0 & 1 & 0 & 0 & 1 & 0 & 0 & 0 & 0 & 0 & 0 & 0 & 1 & 1 & 1 \\
X_{64} & 0 & 0 & 0 & 1 & 0 & 1 & 0 & 0 & 0 & 0 & 0 & 0 & 0 & 1 & 0 & 0 & 0 \\
X_{65} & 0 & 0 & 0 & 0 & 1 & 1 & 0 & 1 & 0 & 1 & 0 & 0 & 1 & 1 & 0 & 0 & 0 \\
\hline
Y_{21} & 0 & 0 & 0 & 0 & 1 & 0 & 1 & 0 & 0 & 1 & 0 & 0 & 0 & 0 & 0 & 1 & 0 \\
Y_{53} & 0 & 0 & 0 & 0 & 1 & 0 & 1 & 0 & 0 & 0 & 1 & 0 & 0 & 0 & 1 & 0 & 0 \\
\hline
Z_{51} & 0 & 0 & 1 & 1 & 0 & 0 & 1 & 0 & 0 & 0 & 0 & 0 & 0 & 0 & 1 & 1 & 1 \\
\end{array}
\right)
$}
}~.~
\eea

The $J$- and $E$-term charges are given by
\beal{es0505}
{\footnotesize
Q_{JE} =
 \resizebox{0.52\textwidth}{!}{$
\left(
\begin{array}{ccccc|cccccc|cccccc}
 p_1 & p_2 & p_3 & p_4 & p_5 & s_1 & s_2 & s_3 & s_4 & s_5 & s_6  & o_1 & o_2 & o_3 & o_4 & o_5 & o_6\\
\hline
 1 & 1 & 1 & 0 & 2 & 0 & -1 & -1 & -1 & -1 & -1 & 0 & 0 & 0 & 0 & 0 & 0 \\
 0 & 0 & 0 & 0 & 1 & 0 & 0 & 0 & -1 & -1 & -1 & 1 & 0 & 0 & 0 & 0 & 0 \\
 0 & 0 & 0 & 0 & 1 & 0 & 0 & -1 & 0 & -1 & -1 & 0 & 1 & 0 & 0 & 0 & 0 \\
 0 & 0 & 0 & 0 & 1 & -1 & 0 & 0 & 0 & -1 & -1 & 0 & 0 & 1 & 0 & 0 & 0 \\
 0 & 0 & 0 & 0 & 1 & 0 & -1 & -1 & 0 & 0 & -1 & 0 & 0 & 0 & 1 & 0 & 0 \\
 0 & 0 & 0 & 0 & 1 & 0 & -1 & 0 & -1 & -1 & 0 & 0 & 0 & 0 & 0 & 1 & 0 \\
 0 & 0 & 0 & 0 & 1 & 0 & -1 & -1 & -1 & 0 & 0 & 0 & 0 & 0 & 0 & 0 & 1 \\
 0 & 0 & 0 & 1 & 1 & -1 & -1 & 0 & 0 & 0 & 0 & 0 & 0 & 0 & 0 & 0 & 0 \\
\end{array}
\right)
$}
}~,~
\eea
and the $D$-term charges are given by
\beal{es0506}
{\footnotesize
Q_D =
 \resizebox{0.52\textwidth}{!}{$
\left(
\begin{array}{ccccc|cccccc|cccccc}
 p_1 & p_2 & p_3 & p_4 & p_5 & s_1 & s_2 & s_3 & s_4 & s_5 & s_6  & o_1 & o_2 & o_3 & o_4 & o_5 & o_6\\
\hline
 0 & 0 & 0 & 0 & 0 & 0 & 0 & -1 & 0 & 1 & 0 & 0 & 0 & 0 & 0 & 0 & 0 \\
 0 & 0 & 0 & 0 & 0 & 0 & 0 & 0 & 1 & 0 & -1 & 0 & 0 & 0 & 0 & 0 & 0 \\
 0 & 0 & 0 & 0 & 0 & -1 & 0 & 1 & 0 & 0 & 0 & 0 & 0 & 0 & 0 & 0 & 0 \\
 0 & 0 & 0 & 0 & 0 & 0 & 1 & 0 & 0 & -1 & 0 & 0 & 0 & 0 & 0 & 0 & 0 \\
 0 & 0 & 0 & 0 & 0 & 1 & 0 & 0 & -1 & 0 & 0 & 0 & 0 & 0 & 0 & 0 & 0 \\
\end{array}
\right)
$}
}~.~
\eea

\begin{figure}[H]
\begin{center}
\resizebox{0.4\hsize}{!}{
\includegraphics[height=6cm]{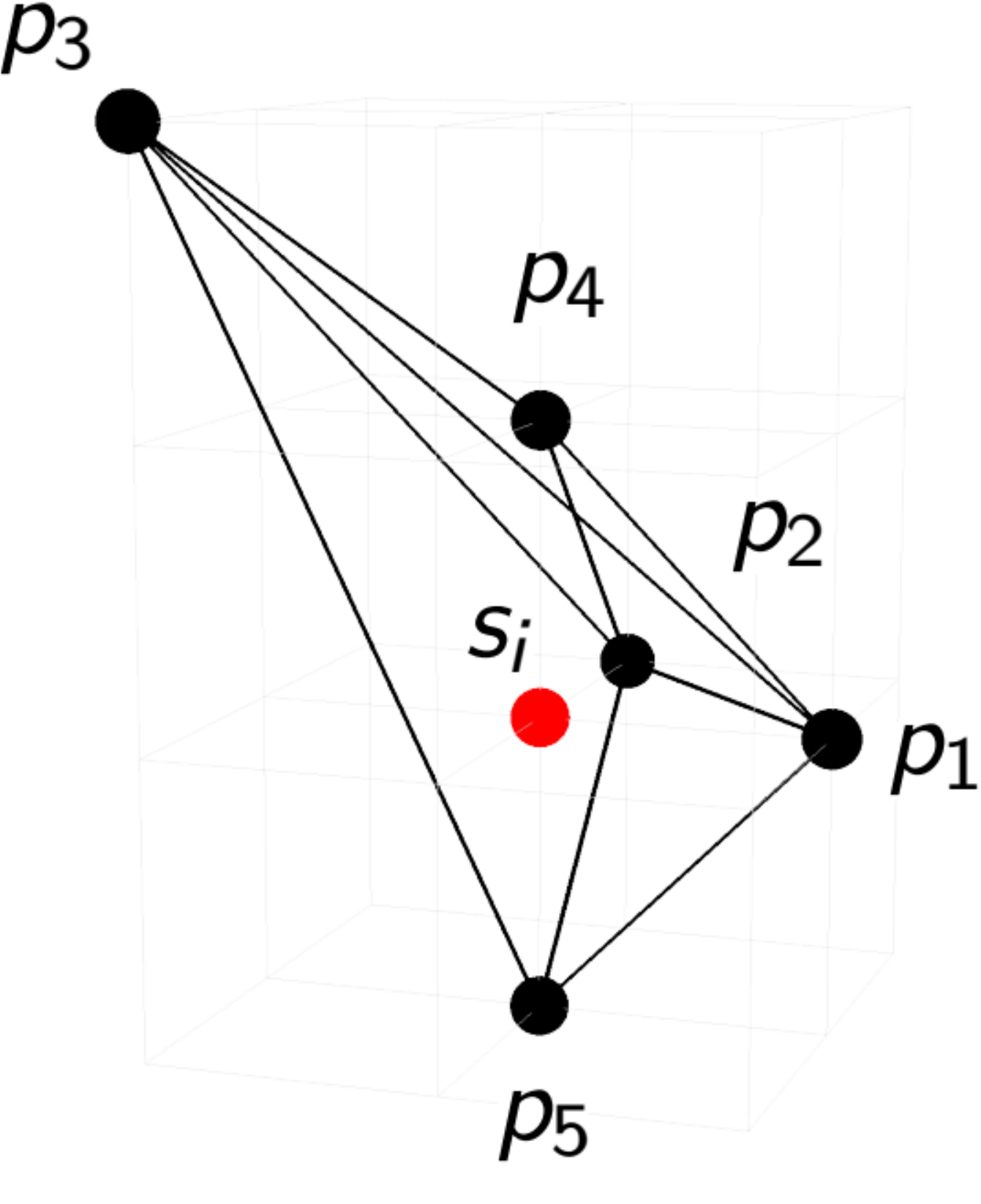} 
}
\caption{
Toric diagram for Model 5.
\label{f_toric_05}}
 \end{center}
 \end{figure}

The toric diagram of Model 5 is given by
\beal{es0510}
{\footnotesize
G_t=
\resizebox{0.52\textwidth}{!}{$
\left(
\begin{array}{ccccc|cccccc|cccccc}
 p_1 & p_2 & p_3 & p_4 & p_5 & s_1 & s_2 & s_3 & s_4 & s_5 & s_6  & o_1 & o_2 & o_3 & o_4 & o_5 & o_6\\
\hline
 1 & 1 & 1 & 1 & 1 & 1 & 1 & 1 & 1 & 1 & 1 & 2 & 2 & 2 & 2 & 2 & 2 \\
 1 & 0 & -1 & 0 & 0 & 0 & 0 & 0 & 0 & 0 & 0 & 0 & 0 & 0 & 0 & 0 & 0 \\
 0 & 1 & -1 & 0 & 0 & 0 & 0 & 0 & 0 & 0 & 0 & 0 & 0 & 0 & 0 & 0 & 0 \\
 0 & 0 & 2 & 1 & -1 & 0 & 0 & 0 & 0 & 0 & 0 & 1 & 1 & 1 & 1 & 1 & 1 \\
\end{array}
\right)
$}
}~,~
\eea
where \fref{f_toric_05} shows the toric diagram with brick matching labels.

The Hilbert series of the mesonic moduli space of Model 5 takes the form
\beal{es0520}
&&
g_1(t_i,y_s,y_o; \mathcal{M}_5) =
\frac{P(t_i,y_s,y_o; \mathcal{M}_5)}{
(1 -  y_s y_o^3 t_1^5 t_4^2 ) 
(1 -  y_s y_o^3 t_2^5 t_4^2 ) 
(1 -  y_s y_o^3 t_3^5 t_4^2 ) 
(1 -  y_s y_o t_1 t_5^2 ) 
}
\nn\\
&&
\hspace{1cm} 
\times
\frac{1}{
(1 -  y_s y_o t_2 t_5^2 ) 
(1 -  y_s y_o t_3 t_5^2 )
} ~,~
\eea
where $t_i$ are the fugacities for the extremal brick matchings $p_i$.
$y_{s}$ counts the product $s_1 \dots s_6$ and $y_{o}$ counts the product $o_1 \dots o_6$.
The explicit numerator $P(t_i,y_s,y_o; \mathcal{M}_5)$ of the Hilbert series is given in the Appendix Section \sref{app_num_05}.
We note that setting the fugacity $y_o=1$ does not change the overall characterization of the mesonic moduli space by the Hilbert series, indicating that the extra GLSM fields, as expected, correspond to an over-parameterization of the moduli space. 

By setting $t_i=t$ for the fugacities of the extremal brick matchings, and all other fugacities to $y_s=1$ and $y_o=1$, the unrefined Hilbert series takes the following form
\beal{es0521}
&&
g_1(t,1,1; \mathcal{M}_5) =
\frac{
1
}{
(1 - t^3)^3 (1 - t^7)^3
}
\times
\nn\\
&&
\hspace{1cm}
(
1 + 10 t^5 + 18 t^7 - 15 t^8 - 35 t^{10} + 6 t^{11} + 15 t^{12} + 15 t^{13} 
\nn\\
&&
\hspace{1cm}
+ 6 t^{14} - 35 t^{15} - 15 t^{17} + 18 t^{18} + 10 t^{20} + t^{25}
)
~,~
\eea
where the palindromic numerator indicates that the mesonic moduli space is Calabi-Yau. 

\begin{table}[H]
\centering
\begin{tabular}{|c|cc|c|c|c|}
\hline
\; & \multicolumn{2}{c|}{$SU(3)_{(x_1,x_2)}$} & $U(1)_{b}$ & $U(1)$ & \text{fugacity} \\
\hline
$p_1$ & (1,& 0)  & 0 & $r_1$ & $t_1$ \\
$p_2$ & (-1,& +1)  & 0& $r_2$ & $t_2$ \\
$p_3$ & (0,&-1)  & 0 & $r_3$ & $t_3$ \\
$p_4$ & (0, &0) & +1 & $r_4$ & $t_4$ \\
$p_5$ & (0, &0) & -1 & $r_5$ & $t_5$ \\
\hline
\end{tabular}
\caption{Global symmetry charges on the extremal brick matchings $p_i$ of Model 5.}
\label{t_pmcharges_05}
\end{table}

The global symmetry of Model 5 and the charges on the extremal brick matchings under the global symmetry are summarized in \tref{t_pmcharges_05}.
Using the following fugacity map,
\beal{es0522}
&&
t = \frac{t_4^2 t_5^2}{t_1 t_2 t_3} ~,~
x_1 = \frac{t_4 t_5}{t_2 t_3}~,~
x_2 = \frac{t_4^2 t_5^2}{t_1 t_2 t_3^2} ~,~
b = \frac{t_4^2 t_5}{t_1 t_2 t_3} ~,~
\eea
the Hilbert series for Model 5 can be rewritten in terms of characters of irreducible representations of $SU(3)\times U(1)$.
In terms of fugacities of the mesonic flavor symmetry of Model 5, the Hilbert series becomes
\beal{es0525}
&&
g_1(t, x_i, b; \mathcal{M}_5) =
\sum_{n_1=0}^{\infty} \sum_{n_2=0}^{\infty}
\Big[
 [n_1+3n_2,0] b^{-2n_1} t^{3 n_1+5n_2}
\nn\\
&&
\hspace{4cm} 
+ [5n_1+3n_2+5,0] b^{2n_1+2} t^{7n_1+5n_2+7}
\Big]
 ~,~
\eea
where $[m_1,m_2] = [m_1,m_2]_{SU(3)_{(x_1,x_2)}}$. The fugacity $b$ counts charges under the $U(1)$ factor of the mesonic flavor symmetry. 
The corresponding plethystic logarithm is
\beal{es0526}
&&
\PL[g_1(t, x_i, b; \mathcal{M}_5)]=
[1,0] b^{-2} t^3
+[3,0] t^5
+[5,0] b^2 t^7
\nn\\
&&
\hspace{1cm}
- [2,1] b^{-2} t^8
- ( [6,0]  + [4,1] +[2,2] ) t^{10}
+ [2,0] b^{-4} t^{11}
+
\dots ~,~
\eea
where we see that the mesonic moduli space is a non-complete intersection.

The generators form 3 sets that transform under 
$[1,0] b^{-2}$, $[3,0]$ and $[5,0] b^2$ of the mesonic flavor symmetry of Model 5, respectively.
Using the following fugacity map
\beal{es0527}
&&
\tilde{t} = t_4^{1/2} t_5^{1/2}~,~ 
\tilde{x}_1 =\frac{t_4^{3/2} t_5^{3/2}}{t_2^2 t_3}~,~ 
\tilde{x}_2 =\frac{t_4^{3/2} t_5^{3/2}}{t_2 t_3^2}~,~ 
\tilde{b} = \frac{t_4^3 t_5}{t_2^2 t_3^2}~,~
\eea
the mesonic flavor charges on the gauge invariant operators become $\mathbb{Z}$-valued.
The generators in terms of brick matchings and their corresponding rescaled mesonic flavor charges are summarized in \tref{f_genlattice_05}.

\begin{table}[H]
\centering
\resizebox{.95\hsize}{!}{
\begin{minipage}[!b]{0.5\textwidth}
\begin{tabular}{|c|cc|c|}
\hline
generator & \multicolumn{2}{c|}{$SU(3)_{(\tilde{x}_1,\tilde{x}_2)}$} & $U(1)_{\tilde{b}}$ \\
\hline
$p_1 p_5^2 ~s o$  &  (1, & 1) & -1 \\
$p_2 p_5^2 ~s o$  &  (0, & 1) & -1 \\
$p_3 p_5^2 ~s o$  &  (1, & 0) & -1 \\
$p_1^3 p_4 p_5 ~s o^2$  &  (1, & 1) & 0 \\
$p_1^2 p_2 p_4 p_5 ~s o^2$  &  (0, & 1) & 0 \\
$p_1 p_2^2 p_4 p_5 ~s o^2$  &  (-1, & 1) & 0 \\
$p_2^3 p_4 p_5 ~s o^2$  &  (-2, & 1) & 0 \\
$p_1^2 p_3 p_4 p_5 ~s o^2$  &  (1, & 0) & 0 \\
$p_1 p_2 p_3 p_4 p_5 ~s o^2$  &  (0, & 0) & 0 \\
$p_2^2 p_3 p_4 p_5 ~s o^2$  &  (-1, & 0) & 0 \\
$p_1 p_3^2 p_4 p_5 ~s o^2$  &  (1, & -1) & 0 \\
$p_2 p_3^2 p_4 p_5 ~s o^2$  &  (0, & -1) & 0 \\
$p_3^3 p_4 p_5 ~s o^2$  &  (1, & -2) & 0 \\
$p_1^5 p_4^2 ~s o^3$  &  (1, & 1) & 1 \\
$p_1^4 p_2 p_4^2 ~s o^3$  &  (0, & 1) & 1 \\
$p_1^3 p_2^2 p_4^2 ~s o^3$  &  (-1, & 1) & 1 \\
$p_1^2 p_2^3 p_4^2 ~s o^3$  &  (-2, & 1) & 1 \\
$p_1 p_2^4 p_4^2 ~s o^3$  &  (-3, & 1) & 1 \\
$p_2^5 p_4^2 ~s o^3$  &  (-4, & 1) & 1 \\
$p_1^4 p_3 p_4^2 ~s o^3$  &  (1, & 0) & 1 \\
$p_1^3 p_2 p_3 p_4^2 ~s o^3$  &  (0, & 0) & 1 \\
$p_1^2 p_2^2 p_3 p_4^2 ~s o^3$  &  (-1, & 0) & 1 \\
$p_1 p_2^3 p_3 p_4^2 ~s o^3$  &  (-2, & 0) & 1 \\
$p_2^4 p_3 p_4^2 ~s o^3$  &  (-3, & 0) & 1 \\
$p_1^3 p_3^2 p_4^2 ~s o^3$  &  (1, & -1) & 1 \\
$p_1^2 p_2 p_3^2 p_4^2 ~s o^3$  &  (0, & -1) & 1 \\
$p_1 p_2^2 p_3^2 p_4^2 ~s o^3$  &  (-1, & -1) & 1 \\
$p_2^3 p_3^2 p_4^2 ~s o^3$  &  (-2, & -1) & 1 \\
$p_1^2 p_3^3 p_4^2 ~s o^3$  &  (1, & -2) & 1 \\
$p_1 p_2 p_3^3 p_4^2 ~s o^3$  &  (0, & -2) & 1 \\
$p_2^2 p_3^3 p_4^2 ~s o^3$  &  (-1, & -2) & 1 \\
$p_1 p_3^4 p_4^2 ~s o^3$  &  (1, & -3) & 1 \\
$p_2 p_3^4 p_4^2 ~s o^3$  &  (0, & -3) & 1 \\
$p_3^5 p_4^2 ~s o^3$  &  (1, & -4) & 1 \\
\hline
\end{tabular}
\end{minipage}
\hspace{0cm}
\begin{minipage}[!b]{0.6\textwidth}
\includegraphics[height=9cm]{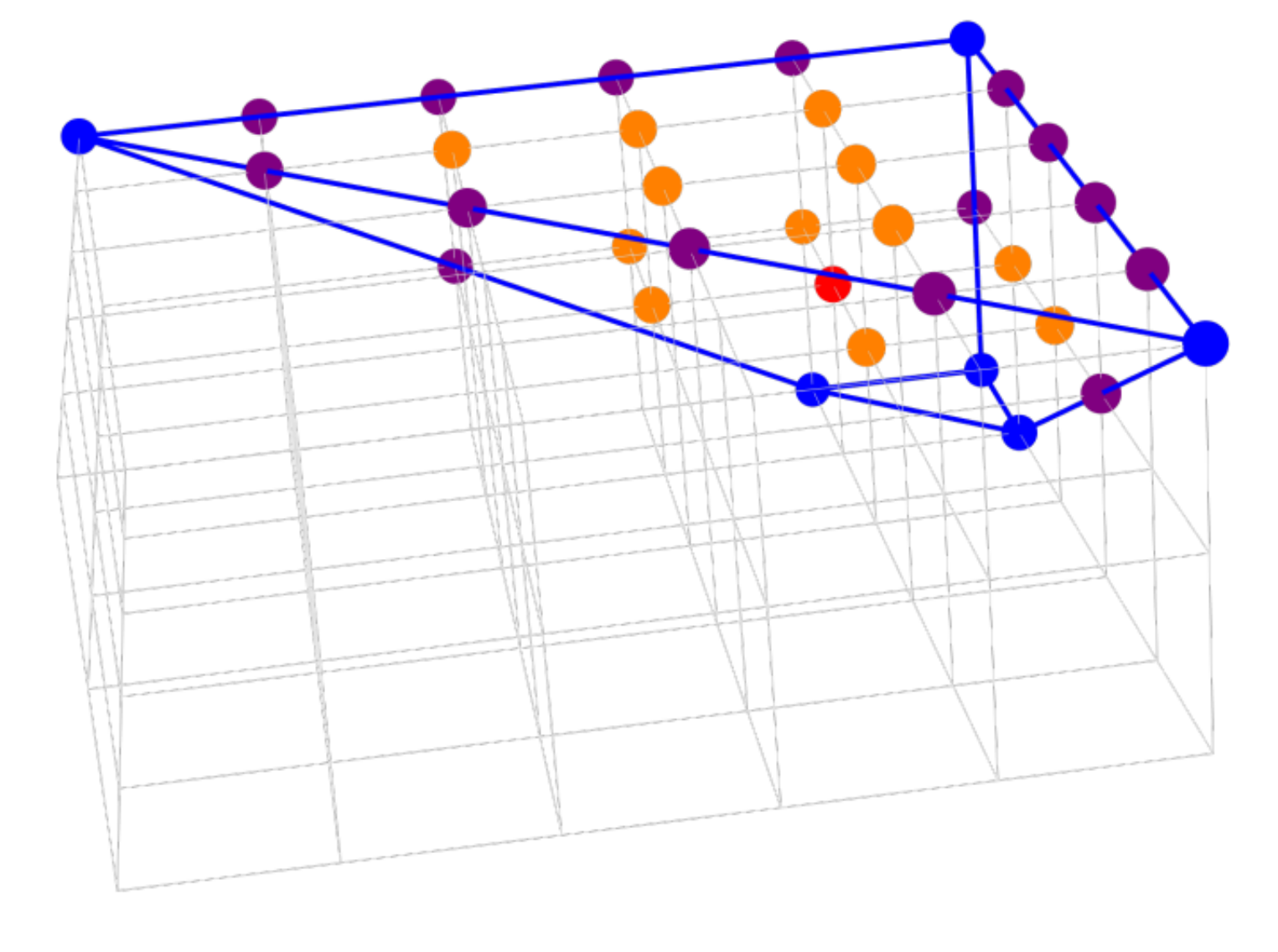} 
\end{minipage}
}
\caption{The generators and lattice of generators of the mesonic moduli space of Model 5 in terms of brick matchings with the corresponding flavor charges.
\label{f_genlattice_05}}
\end{table}

The generator lattice as shown in \tref{f_genlattice_05} is a convex lattice polytope, which is reflexive. It is the dual of the toric diagram of Model 5 shown in \fref{f_toric_05}.
We also note that the 3 layers of points in the generator lattice in \tref{f_genlattice_05} corresponds to the 3 sets that transform under 
$[1,0] b^{-2}$, $[3,0]$ and $[5,0] b^2$ of the mesonic flavor symmetry.
For completeness, \tref{f_genfields_05a} and \tref{f_genfields_05b} show the generators of Model 5 in terms of chiral fields with the corresponding mesonic flavor charges.
\\

\begin{table}[H]
\centering
\resizebox{0.95\hsize}{!}{
\begin{tabular}{|l|cc|c|}
\hline
generator & \multicolumn{2}{c|}{$SU(3)_{(\tilde{x}_1,\tilde{x}_2)}$} & $U(1)_{\tilde{b}}$ \\
\hline
$P_{13} X_{32} Y_{21}=  P_{25} Y_{53} X_{32}=  P_{36} X_{65} Y_{53}=  X_{14} P_{42} Y_{21}$  &  (1, & 1) & -1 \\
$ Q_{13} X_{32} Y_{21}=  Q_{25} Y_{53} X_{32}=  Q_{36} X_{65} Y_{53}=  X_{14} Q_{42} Y_{21}$  &  (0, & 1) & -1 \\
$ X_{13} X_{32} Y_{21}=  X_{14} X_{42} Y_{21}=  X_{25} Y_{53} X_{32}=  X_{36} X_{65} Y_{53}$  &  (1, & 0) & -1 \\
$P_{13} P_{36} X_{64} P_{42} Y_{21}=  P_{25} Y_{53} P_{36} X_{64} P_{42}=  P_{13} X_{32} P_{25} R_{51}=  P_{13} P_{36} X_{65} R_{51}=  X_{14} P_{42} P_{25} R_{51}$  &  (1, & 1) & 0 \\
$P_{13} P_{36} X_{64} Q_{42} Y_{21}=  P_{13} Q_{36} X_{64} P_{42} Y_{21}=  P_{25} Y_{53} P_{36} X_{64} Q_{42}=  P_{25} Y_{53} Q_{36} X_{64} P_{42}=  Q_{13} P_{36} X_{64} P_{42} Y_{21}$&  (0, & 1) & 0 \\
$=  Q_{25} Y_{53} P_{36} X_{64} P_{42}=  P_{13} X_{32} P_{25} X_{51}=  P_{13} P_{36} X_{65} X_{51}=  P_{13} X_{32} Q_{25} R_{51}=  P_{13} Q_{36} X_{65} R_{51}$& & & \\
$=  X_{14} P_{42} P_{25} X_{51}=  Q_{13} X_{32} P_{25} R_{51}=  X_{14} Q_{42} P_{25} R_{51}=  Q_{13} P_{36} X_{65} R_{51}=  X_{14} P_{42} Q_{25} R_{51}$  & & & \\
\hline
\end{tabular}
}
\caption{The generators in terms of bifundamental chiral fields for Model 5 \bf{(Part 1)}.
\label{f_genfields_05a}}
\end{table}

\begin{table}[H]
\centering
\resizebox{0.95\hsize}{!}{
\begin{tabular}{|l|cc|c|}
\hline
generator & \multicolumn{2}{c|}{$SU(3)_{(\tilde{x}_1,\tilde{x}_2)}$} & $U(1)_{\tilde{b}}$ \\
\hline
$ P_{13} Q_{36} X_{64} Q_{42} P_{25} Z_{51}=  P_{13} Q_{36} X_{64} X_{42} P_{25} X_{51}=  P_{13} X_{36} X_{64} Q_{42} P_{25} X_{51}=  P_{13} P_{36} X_{64} Q_{42} Q_{25} Z_{51}$ &  (-1, & 0) & 1 \\
$=  P_{13} P_{36} X_{64} X_{42} Q_{25} X_{51}=  P_{13} P_{36} X_{64} Q_{42} X_{25} X_{51}=  P_{13} Q_{36} X_{64} P_{42} Q_{25} Z_{51}=  P_{13} X_{36} X_{64} P_{42} Q_{25} X_{51}$  & & & \\
$=  P_{13} Q_{36} X_{64} P_{42} X_{25} X_{51}=  P_{13} Q_{36} X_{64} X_{42} Q_{25} R_{51}=  P_{13} X_{36} X_{64} Q_{42} Q_{25} R_{51}=  P_{13} Q_{36} X_{64} Q_{42} X_{25} R_{51}$ & & & \\
$=  Q_{13} P_{36} X_{64} Q_{42} P_{25} Z_{51}=  Q_{13} P_{36} X_{64} X_{42} P_{25} X_{51}=  X_{13} P_{36} X_{64} Q_{42} P_{25} X_{51}=  Q_{13} Q_{36} X_{64} P_{42} P_{25} Z_{51}$ & & & \\
$=  Q_{13} X_{36} X_{64} P_{42} P_{25} X_{51}=  X_{13} Q_{36} X_{64} P_{42} P_{25} X_{51}=  Q_{13} Q_{36} X_{64} X_{42} P_{25} R_{51}=  Q_{13} X_{36} X_{64} Q_{42} P_{25} R_{51}$ & & & \\
$=  X_{13} Q_{36} X_{64} Q_{42} P_{25} R_{51}=  Q_{13} P_{36} X_{64} P_{42} Q_{25} Z_{51}=  Q_{13} P_{36} X_{64} P_{42} X_{25} X_{51}=  X_{13} P_{36} X_{64} P_{42} Q_{25} X_{51}$ & & & \\
$=  Q_{13} P_{36} X_{64} X_{42} Q_{25} R_{51}=  Q_{13} P_{36} X_{64} Q_{42} X_{25} R_{51}=  X_{13} P_{36} X_{64} Q_{42} Q_{25} R_{51}=  Q_{13} X_{36} X_{64} P_{42} Q_{25} R_{51}$ & & & \\
\hline
\end{tabular}
}
\caption{The generators in terms of bifundamental chiral fields for Model 5 \bf{(Part 2)}.
\label{f_genfields_05b}}
\end{table}

\section{Model 6: $P^{1}_{+-}(\text{dP}_1)$~[$\mathbb{P}(\mathcal{O}_{\mathbb{P}^1\times \mathbb{P}^1} \oplus \mathcal{O}_{\mathbb{P}^1 \times\mathbb{P}^1}(1,-1))$,~$\langle24\rangle$] \label{smodel06}}

\begin{figure}[H]
\begin{center}
\resizebox{0.3\hsize}{!}{
\includegraphics[height=6cm]{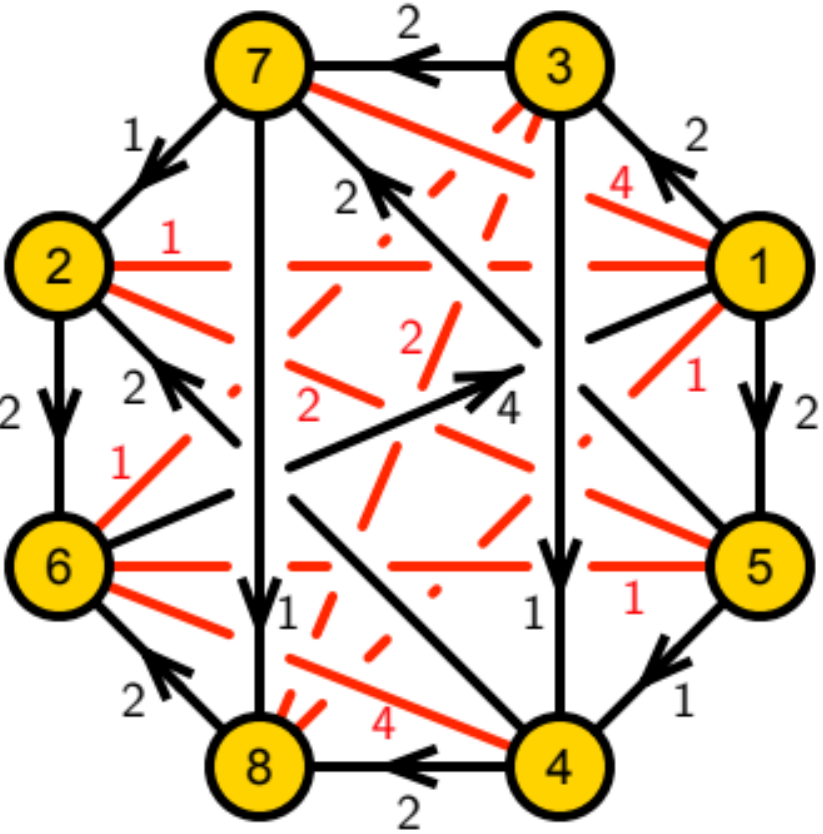} 
}
\caption{
Quiver for Model 6.
\label{f_quiver_06}}
 \end{center}
 \end{figure}

Model 6 corresponds to the Calabi-Yau 4-fold $P^{1}_{+-}(\text{dP}_1)$.
The corresponding brane brick model has the quiver in \fref{f_quiver_06} and the $J$- and $E$-terms are given as follows
\beq
{\footnotesize
\begin{array}{rrclrcl}
 &  & J &  & & E &  \\
\Lambda_{17}^{1} : &  X_{78} Y_{86} Y_{61}  &-& X_{72} Q_{26} X_{61} &   P_{15} X_{57}  &-&  X_{13} P_{37} \\ 
\Lambda_{17}^{2} : &  X_{72} Q_{26} Z_{61}  &-&  X_{78} X_{86} Y_{61}  &   P_{15} Y_{57}  &-&  Y_{13} P_{37} \\ 
\Lambda_{17}^{3} : &  X_{78} Y_{86} R_{61}  &-& X_{72} P_{26} X_{61}  &  X_{13} Q_{37}   &-& Q_{15} X_{57} \\ 
\Lambda_{17}^{4} : &  X_{72} P_{26} Z_{61}  &-&  X_{78} X_{86} R_{61}  &  Y_{13} Q_{37}  &-& Q_{15} Y_{57} \\ 
\Lambda_{18}^{1} : &  X_{86} X_{61}  &-&  Y_{86} Z_{61}  &   P_{15} X_{54} Q_{48}  &-&  Q_{15} X_{54} P_{48} \\ 
\Lambda_{21}^{1} : &  X_{13} X_{34} Y_{42}  &-&  Y_{13} X_{34} X_{42}  &   P_{26} Y_{61}  &-&  Q_{26} R_{61} \\ 
\Lambda_{25}^{1} : &  X_{54} X_{42}  &-&  X_{57} X_{72}  &   P_{26} X_{61} Q_{15}  &-&  Q_{26} X_{61} P_{15} \\ 
\Lambda_{25}^{2} : &  Y_{57} X_{72}  &-& X_{54} Y_{42}  &   P_{26} Z_{61} Q_{15}  &-&  Q_{26} Z_{61} P_{15} \\ 
\Lambda_{36}^{1} : &  Z_{61} Y_{13}  &-& X_{61} X_{13} &   P_{37} X_{72} Q_{26}  &-&  Q_{37} X_{72} P_{26} \\ 
\Lambda_{38}^{1} : &  Y_{86} Y_{61} X_{13}  &-& X_{86} Y_{61} Y_{13}  &   P_{37} X_{78}  &-&  X_{34} P_{48} \\ 
\Lambda_{38}^{2} : &  Y_{86} R_{61} X_{13}  &-& X_{86} R_{61} Y_{13}  &   X_{34} Q_{48}  &-& Q_{37} X_{78} \\ 
\Lambda_{46}^{1} : &  X_{61} Q_{15} X_{54}  &-&  Y_{61} Y_{13} X_{34}  &   P_{48} X_{86}  &-&  X_{42} P_{26} \\ 
\Lambda_{46}^{2} : &  Y_{61} X_{13} X_{34}  &-&  Z_{61} Q_{15} X_{54}  &   P_{48} Y_{86}  &-&  Y_{42} P_{26} \\ 
\Lambda_{46}^{3} : &  X_{61} P_{15} X_{54}  &-& R_{61} Y_{13} X_{34}  &  X_{42} Q_{26}   &-& Q_{48} X_{86}  \\ 
\Lambda_{46}^{4} : &  R_{61} X_{13} X_{34}  &-&  Z_{61} P_{15} X_{54}  &   Y_{42} Q_{26}  &-& Q_{48} Y_{86} \\ 
\Lambda_{65}^{1} : &  X_{57} X_{78} Y_{86}  &-&  Y_{57} X_{78} X_{86}  &   R_{61} Q_{15}  &-&  Y_{61} P_{15}
 \end{array} 
 }~.~
\label{es0601}
 \eeq

Following the forward algorithm, we obtain the brick matchings. They are summarized in 
the $P$-matrix, which takes the form

\beal{es0605}
{\footnotesize 
P=
\resizebox{0.7\textwidth}{!}{$
\left(
\begin{array}{c|cccccc|ccccccccccc|cccc}
\; & p_1 & p_2 & p_3 & p_4 & p_5 & p_6 & s_1 & s_2 & s_3 & s_4 & s_5 & s_6 & s_7 & s_8 & s_9 & s_{10} & s_{11} & o_1 & o_2 & o_3 & o_4 \\
\hline 
P_{15} & 1 & 0 & 0 & 0 & 0 & 0 & 0 & 0 & 0 & 0 & 0 & 0 & 0 & 0 & 0 & 1 & 1 & 0 & 0 & 0 & 1 \\
P_{26} & 1 & 0 & 0 & 0 & 0 & 0 & 0 & 0 & 0 & 1 & 0 & 0 & 0 & 0 & 1 & 0 & 0 & 0 & 0 & 1 & 0 \\
P_{37} & 1 & 0 & 0 & 0 & 0 & 0 & 0 & 0 & 0 & 0 & 1 & 0 & 1 & 0 & 0 & 0 & 1 & 0 & 1 & 1 & 1 \\
P_{48} & 1 & 0 & 0 & 0 & 0 & 0 & 0 & 0 & 0 & 0 & 0 & 1 & 1 & 0 & 1 & 0 & 0 & 1 & 0 & 1 & 1 \\
\hline
Q_{15} & 0 & 1 & 0 & 0 & 0 & 0 & 0 & 0 & 0 & 0 & 0 & 0 & 0 & 0 & 0 & 1 & 1 & 0 & 0 & 0 & 1 \\
Q_{26} & 0 & 1 & 0 & 0 & 0 & 0 & 0 & 0 & 0 & 1 & 0 & 0 & 0 & 0 & 1 & 0 & 0 & 0 & 0 & 1 & 0 \\
Q_{37} & 0 & 1 & 0 & 0 & 0 & 0 & 0 & 0 & 0 & 0 & 1 & 0 & 1 & 0 & 0 & 0 & 1 & 0 & 1 & 1 & 1 \\
\end{array}
\right)
$}
}~.~
\eea

The $J$- and $E$-term charges are given by
\beal{es0605}
{\footnotesize
Q_{JE} =
\resizebox{0.67\textwidth}{!}{$
\left(
\begin{array}{cccccc|ccccccccccc|cccc}
p_1 & p_2 & p_3 & p_4 & p_5 & p_6 & s_1 & s_2 & s_3 & s_4 & s_5 & s_6 & s_7 & s_8 & s_9 & s_{10} & s_{11} & o_1 & o_2 & o_3 & o_4 \\
\hline
 1 & 1 & 0 & 0 & 1 & 0 & 1 & 1 & 0 & -1 & -1 & -1 & 0 & -1 & 0 & -1 & 0 & 0 & 0 & 0 & 0
   \\
 0 & 0 & 0 & 0 & 1 & 0 & 0 & 1 & 0 & 0 & -1 & -1 & 0 & -1 & 0 & -1 & 0 & 0 & 0 & 0 & 1 \\
 0 & 0 & 0 & 0 & 0 & 0 & 0 & 1 & 0 & 0 & -1 & 0 & 0 & 0 & 0 & -1 & 1 & 0 & 0 & 0 & 0 \\
 0 & 0 & 0 & 0 & 1 & 0 & 1 & 0 & 0 & -1 & -1 & -1 & 0 & -1 & 0 & 0 & 0 & 0 & 0 & 1 & 0 \\
 0 & 0 & 0 & 0 & 0 & 0 & 1 & 0 & 0 & -1 & 0 & -1 & 0 & 0 & 1 & 0 & 0 & 0 & 0 & 0 & 0 \\
 0 & 0 & 0 & 0 & 0 & 0 & 0 & 0 & 1 & 0 & -1 & -1 & 1 & 0 & 0 & 0 & 0 & 0 & 0 & 0 & 0 \\
 0 & 0 & 0 & 0 & -1 & 0 & -1 & 0 & 1 & 0 & -1 & 0 & 0 & 0 & 0 & 0 & 0 & 0 & 1 & 0 & 0 \\
 0 & 0 & 0 & 0 & -1 & 0 & 0 & -1 & 1 & 0 & 0 & -1 & 0 & 0 & 0 & 0 & 0 & 1 & 0 & 0 & 0 \\
 0 & 0 & 0 & 0 & 1 & 1 & 0 & 0 & -1 & 0 & 0 & 0 & 0 & -1 & 0 & 0 & 0 & 0 & 0 & 0 & 0 \\
 0 & 0 & 1 & 1 & -1 & 0 & -1 & -1 & 1 & 0 & 0 & 0 & 0 & 0 & 0 & 0 & 0 & 0 & 0 & 0 & 0 \\
\end{array}
\right)
$}
}~,~
\eea
and the $D$-term charges are given by
\beal{es0606}
{\footnotesize
Q_D =
\resizebox{0.67\textwidth}{!}{$
\left(
\begin{array}{cccccc|ccccccccccc|cccc}
p_1 & p_2 & p_3 & p_4 & p_5 & p_6 & s_1 & s_2 & s_3 & s_4 & s_5 & s_6 & s_7 & s_8 & s_9 & s_{10} & s_{11} & o_1 & o_2 & o_3 & o_4 \\
\hline
 0 & 0 & 0 & 0 & 0 & 0 & -1 & 0 & 0 & 1 & 0 & 0 & 0 & 0 & 0 & 0 & 0 & 0 & 0 & 0 & 0 \\
 0 & 0 & 0 & 0 & 0 & 0 & 0 & -1 & 0 & 0 & 1 & 0 & 0 & 0 & 0 & 0 & 0 & 0 & 0 & 0 & 0 \\
 0 & 0 & 0 & 0 & 0 & 0 & 0 & 0 & -1 & 0 & 0 & 1 & 0 & 0 & 0 & 0 & 0 & 0 & 0 & 0 & 0 \\
 0 & 0 & 0 & 0 & 0 & 0 & 0 & 1 & 0 & 0 & 0 & 0 & 0 & 0 & 0 & -1 & 0 & 0 & 0 & 0 & 0 \\
 0 & 0 & 0 & 0 & 0 & 0 & 0 & 0 & 0 & -1 & 0 & 0 & 0 & 1 & 0 & 0 & 0 & 0 & 0 & 0 & 0 \\
 0 & 0 & 0 & 0 & 0 & 0 & 0 & 0 & 1 & 0 & -1 & 0 & 0 & 0 & 0 & 0 & 0 & 0 & 0 & 0 & 0 \\
 0 & 0 & 0 & 0 & 0 & 0 & 1 & 0 & 0 & 0 & 0 & -1 & 0 & 0 & 0 & 0 & 0 & 0 & 0 & 0 & 0 \\
\end{array}
\right)
$}
}~.~
\eea

\begin{figure}[H]
\begin{center}
\resizebox{0.4\hsize}{!}{
\includegraphics[height=6cm]{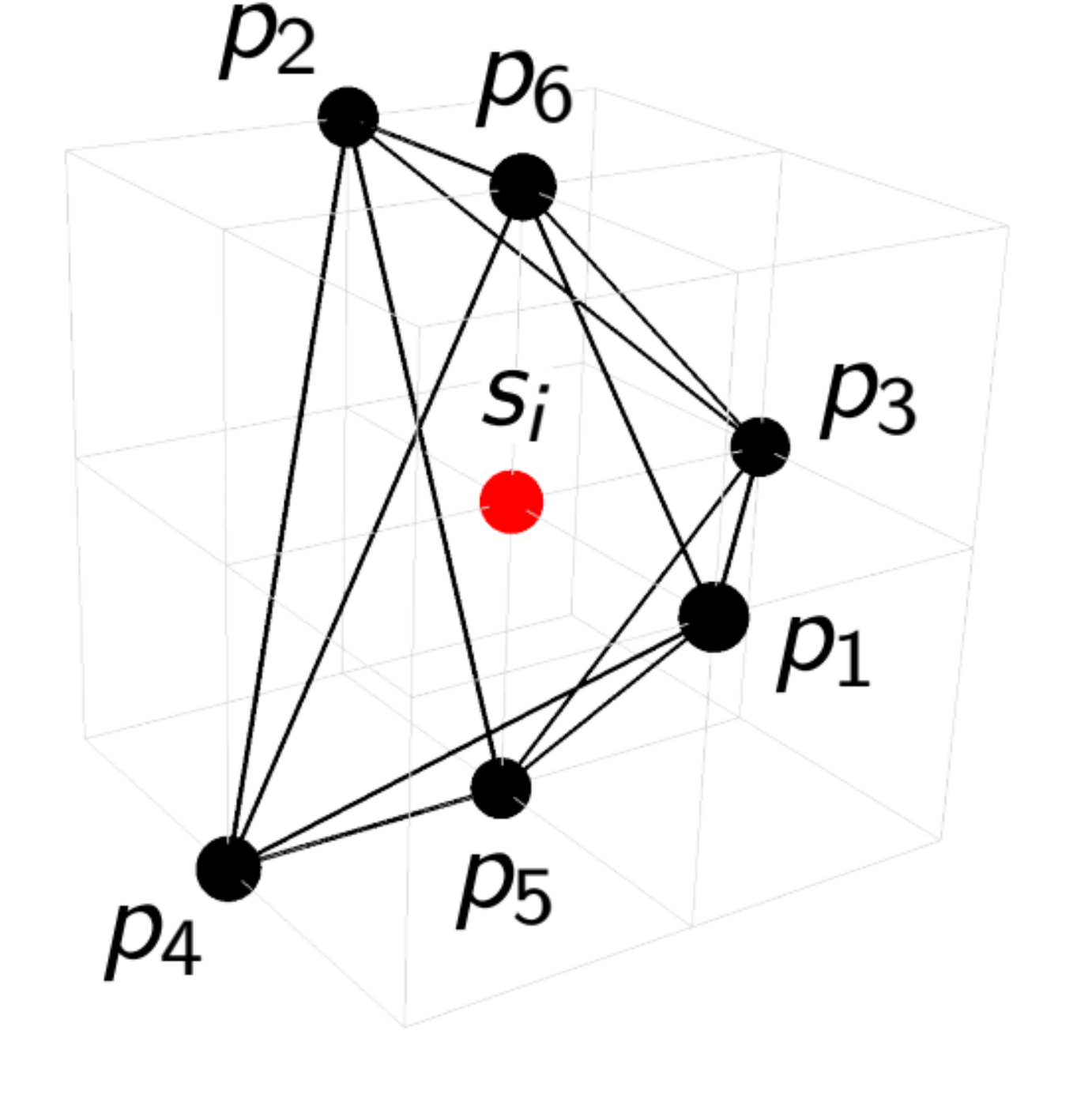} 
}
\caption{
Toric diagram for Model 6.
\label{f_toric_06}}
 \end{center}
 \end{figure}

The toric diagram of Model 6 is given by
\beal{es0610}
{\footnotesize
G_t=
\resizebox{0.67\textwidth}{!}{$
\left(
\begin{array}{cccccc|ccccccccccc|cccc}
p_1 & p_2 & p_3 & p_4 & p_5 & p_6 & s_1 & s_2 & s_3 & s_4 & s_5 & s_6 & s_7 & s_8 & s_9 & s_{10} & s_{11} & o_1 & o_2 & o_3 & o_4 \\
\hline
 1 & 1 & 1 & 1 & 1 & 1 & 1 & 1 & 1 & 1 & 1 & 1 & 1 & 1 & 1 & 1 & 1 & 2 & 2 & 2 & 2 \\
 1 & -1 & 0 & 0 & 0 & 0 & 0 & 0 & 0 & 0 & 0 & 0 & 0 & 0 & 0 & 0 & 0 & 0 & 0 & 0 & 0 \\
 0 & 0 & 1 & -1 & 0 & 0 & 0 & 0 & 0 & 0 & 0 & 0 & 0 & 0 & 0 & 0 & 0 & 0 & 0 & 0 & 0 \\
 0 & 1 & 0 & -1 & -1 & 1 & 0 & 0 & 0 & 0 & 0 & 0 & 0 & 0 & 0 & 0 & 0 & -1 & -1 & 1 & 1 \\
\end{array}
\right)
$}
}~,~
\eea
where \fref{f_toric_06} shows the toric diagram with brick matching labels.

The Hilbert series of the mesonic moduli space of Model 6 takes the form
\beal{es0620}
&&
g_1(t_i,y_s,y_{o_1},y_{o_2} ; \mathcal{M}_6) =
\frac{P(t_i,y_s,y_{o_1},y_{o_2}; \mathcal{M}_6)}{
(1 - y_{s} y_{o_1}^3 y_{o_2} t_1 t_3^3 t_5^2) (1 - y_{s} y_{o_1}^3 y_{o_2} t_2 t_3^3 t_5^2) (1 -   y_{s} y_{o_1}^3 y_{o_2} t_1 t_4^3 t_5^2) 
}
\nn\\
&&
\hspace{1cm} 
\times
\frac{1}{
(1 - y_{s} y_{o_1}^3 y_{o_2} t_2 t_4^3 t_5^2) (1 -   y_{s} y_{o_1} y_{o_2}^3 t_1^3 t_3 t_6^2) (1 - y_{s} y_{o_1} y_{o_2}^3 t_2^3 t_3 t_6^2) 
} 
\nn\\
&&
\hspace{1cm} 
\times
\frac{1}{
(1 -   y_{s} y_{o_1} y_{o_2}^3 t_1^3 t_4 t_6^2) (1 - y_{s} y_{o_1} y_{o_2}^3 t_2^3 t_4 t_6^2)
} ~,~
\eea
where $t_i$ are the fugacities for the extremal brick matchings $p_i$.
$y_{s}$ counts the brick matching product $s_1 \dots s_{11}$ corresponding to the single internal point of the toric diagram of Model 6.
Additionally, $y_{o_1}$ and $y_{o_2}$ count the products of extra GLSM fields $o_1 o_2$ and $o_3 o_4$, respectively.
The explicit numerator $P(t_i,y_s,y_{o_1},y_{o_2}; \mathcal{M}_6)$ of the Hilbert series is given in the Appendix Section \sref{app_num_06}.
We note that setting the fugacities $y_{o_1}=1$ and $y_{o_2}=1$ does not change the overall characterization of the mesonic moduli space by the Hilbert series, indicating that the extra GLSM fields, as expected, correspond to an over-parameterization of the moduli space. 

\begin{table}[H]
\centering
\begin{tabular}{|c|c|c|c|c|c|}
\hline
\; & $SU(2)_{x}$ & $SU(2)_{y}$ &  $U(1)_{b}$ & $U(1)$ & \text{fugacity} \\
\hline
$p_1$ & +1 & 0 & 0 & $r_1$ & $t_1$ \\
$p_2$ & -1 & 0 & 0 & $r_2$ & $t_2$ \\
$p_3$ & 0 & +1  & 0 & $r_3$ & $t_3$ \\
$p_4$ & 0 & -1 & 0 & $r_4$ & $t_4$ \\
$p_5$ & 0 &0 & +1 & $r_5$ & $t_5$ \\
$p_6$ & 0 &0 &  -1 & $r_6$ & $t_6$ \\
\hline
\end{tabular}
\caption{Global symmetry charges on the extremal brick matchings $p_i$ of Model 6.}
\label{t_pmcharges_06}
\end{table}

\begin{table}[H]
\centering
\resizebox{.95\hsize}{!}{
\begin{minipage}[!b]{0.5\textwidth}
\begin{tabular}{|c|c|c|c|}
\hline
generator & $SU(2)_{\tilde{x}}$ & $SU(2)_{\tilde{y}}$ & $U(1)_{\tilde{b}}$ \\
\hline
$p_1 p_3^3 p_5^2 ~s o_1^3 o_2 $ &  1 &  2 &  1 \\ 
$p_2 p_3^3 p_5^2 ~s o_1^3 o_2 $ &  0 &  2 &  1 \\ 
$p_1 p_3^2 p_4 p_5^2 ~s o_1^3 o_2 $ &  1 &  1 &  1 \\ 
$p_2 p_3^2 p_4 p_5^2 ~s o_1^3 o_2 $ &  0 &  1 &  1 \\ 
$p_1 p_3 p_4^2 p_5^2 ~s o_1^3 o_2 $ &  1 &  0 &  1 \\ 
$p_2 p_3 p_4^2 p_5^2 ~s o_1^3 o_2 $ &  0 &  0 &  1 \\ 
$p_1 p_4^3 p_5^2 ~s o_1^3 o_2 $ &  1 &  -1 &  1 \\ 
$p_2 p_4^3 p_5^2 ~s o_1^3 o_2 $ &  0 &  -1 &  1 \\ 
$p_1^2 p_3^2 p_5 p_6 ~s o_1^2 o_2^2 $ &  1 &  1 &  0 \\ 
$p_1 p_2 p_3^2 p_5 p_6 ~s o_1^2 o_2^2 $ &  0 &  1 &  0 \\ 
$p_2^2 p_3^2 p_5 p_6 ~s o_1^2 o_2^2 $ &  -1 &  1 &  0 \\ 
$p_1^2 p_3 p_4 p_5 p_6 ~s o_1^2 o_2^2 $ &  1 &  0 &  0 \\ 
$p_1 p_2 p_3 p_4 p_5 p_6 ~s o_1^2 o_2^2 $ &  0 &  0 &  0 \\ 
$p_2^2 p_3 p_4 p_5 p_6 ~s o_1^2 o_2^2 $ &  -1 &  0 &  0 \\ 
$p_1^2 p_4^2 p_5 p_6 ~s o_1^2 o_2^2 $ &  1 &  -1 &  0 \\ 
$p_1 p_2 p_4^2 p_5 p_6 ~s o_1^2 o_2^2 $ &  0 &  -1 &  0 \\ 
$p_2^2 p_4^2 p_5 p_6 ~s o_1^2 o_2^2 $ &  -1 &  -1 &  0 \\ 
$p_1^3 p_3 p_6^2 ~s o_1 o_2^3 $ &  1 &  0 &  -1 \\ 
$p_1^2 p_2 p_3 p_6^2 ~s o_1 o_2^3 $ &  0 &  0 &  -1 \\ 
$p_1 p_2^2 p_3 p_6^2 ~s o_1 o_2^3 $ &  -1 &  0 &  -1 \\ 
$p_2^3 p_3 p_6^2 ~s o_1 o_2^3 $ &  -2 &  0 &  -1 \\ 
$p_1^3 p_4 p_6^2 ~s o_1 o_2^3 $ &  1 &  -1 &  -1 \\ 
$p_1^2 p_2 p_4 p_6^2 ~s o_1 o_2^3 $ &  0 &  -1 &  -1 \\ 
$p_1 p_2^2 p_4 p_6^2 ~s o_1 o_2^3 $ &  -1 &  -1 &  -1 \\ 
$p_2^3 p_4 p_6^2 ~s o_1 o_2^3 $ &  -2 &  -1 &  -1\\
\hline
\end{tabular}
\end{minipage}
\hspace{2cm}
\begin{minipage}[!b]{0.4\textwidth}
\includegraphics[height=7cm]{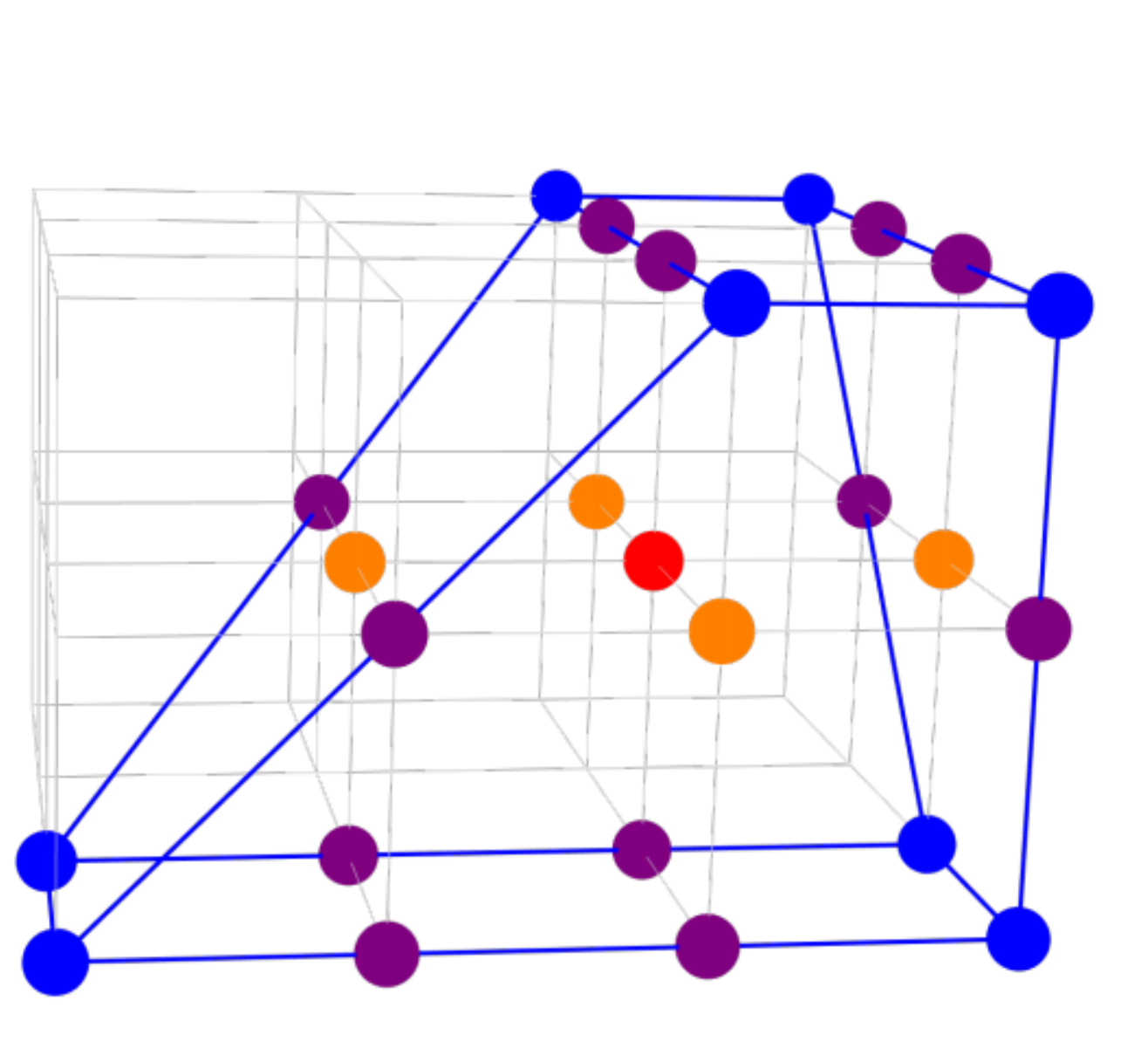} 
\end{minipage}
}
\caption{The generators and lattice of generators of the mesonic moduli space of Model 6 in terms of brick matchings with the corresponding flavor charges.
\label{f_genlattice_06}}
\end{table}

By setting $t_i=t$ for the fugacities of the extremal brick matchings, and all other fugacities to $y_s=1$, $y_{o_1}=1$ and $y_{o_2}=1$, the unrefined Hilbert series takes the following form
\beal{es0621}
&&
g_1(t,1,1,1; \mathcal{M}_6) =
\frac{
1 + 21 t^6 + 21 t^{12} + t^{18}
}{
(1 - t^6)^4
}
~,~
\eea
where the palindromic numerator indicates that the mesonic moduli space is Calabi-Yau. 

The global symmetry of Model 6 and the charges on the extremal brick matchings under the global symmetry are summarized in \tref{t_pmcharges_06}.
We can use the following fugacity map,
\beal{es0622}
&&
t = t_1^{1/2} t_2^{1/2} ~,~
x = \frac{t_1^{1/2}}{t_2^{1/2}} ~,~
y = \frac{t_3}{t_1^{1/2} t_2^{1/2}} ~,~
b = \frac{t_1^{1/2} t_2^{1/2}}{t_6} ~,~
\eea
where $t_3 = t_5 t_6 t_4^{-1}$ and $t_1 =t_5 t_6 t_2^{-1}$, in order to rewrite the 
Hilbert series for Model 6 in terms of characters of irreducible representations of $SU(2)\times SU(2)\times U(1)$.
In terms of fugacities of the mesonic flavor symmetry of Model 6, the Hilbert series becomes
\beal{es0625}
&&
g_1(t, x,y, b; \mathcal{M}_6) =
\sum_{n_1=0}^{\infty} \sum_{n_2=0}^{\infty}
\Big[
[3n_1+2n_2; n_1+2n_2] b^{-2n_1} t^{6n_1+6n_2}
\nn\\
&&
\hspace{1cm}
+[3n_1+n_2+1; 5n_1+3n_2+3] b^{2n_1+2n_2+2} t^{12n_1+6n_2+6}
\nn\\
&&
\hspace{1cm}
+[3n_1+5n_2+3; 5n_1+7n_2+5] b^{2n_1+2n_2+2} t^{12n_1+18n_2+12}
\nn\\
&&
\hspace{1cm}
+[7n_1+5n_2+5; 9n_1+7n_2+7] b^{2n_1+2n_2+2} t^{24n_1+18n_2+18}
\nn\\
&&
\hspace{1cm}
+[7n_1+2n_2+7; 9n_1+2n_2+9] b^{2n_1+2} t^{24n_1+6n_2+24}
\Big]
 ~,~
 \eea
 where $[m;n] = [m]_{SU(2)_{x}} [n]_{SU(2)_y}$. The fugacity $b$ counts charges under the $U(1)$ factor of the mesonic flavor symmetry. 
We can write the Hilbert series also in highest weight form, where each highest weight $n$ of a character is counted by its own fugacity $\mu$ to give
\beal{es0626}
h_1(t, \mu_1,\mu_2,b; \mathcal{M}_6) = \frac{
1+\mu_1^2 \mu_2^2 t^6
}{
(1-\mu_1^3 \mu_2 b^{-2} t^6) (1-\mu_1\mu_2^3 b^2 t^6)
}
~,~
\eea
where $\mu_1^{m} \mu_2^{n} \sim [m]_{SU(2)_{x}} [n]_{SU(2)_y}$.
The plethystic logarithm of the Hilbert series in \eref{es0625} is
\beal{es0626}
&&
\PL[g_1(t, x,y,b; \mathcal{M}_6)]=
( [1; 3] b^2 + [2; 2] + [3; 1] b^{-2} ) t^6
- 
(1 + b^{-4} + b^4 + [0; 4] 
\nn\\
&&
\hspace{1cm}
+  [0; 4] b^4 + [1; 1] b^{-2} +  [1; 1] b^2+ [1; 3]b^{-2} + [1; 3] b^2 +  [1; 5] b^2 + 2 [2; 2] + [2; 2] b^{-4} 
\nn\\
&&
\hspace{1cm}
+  [2; 2] b^4 + [2; 4] + [3; 1]b^{-2} + [3; 1] b^2 + [3; 3] b^{-2} +[3; 3]  b^2 + [4; 0] + [4; 0]b^{-4} 
\nn\\
&&
\hspace{1cm}
+ [4; 2] + [4; 4] + [5; 1] b^{-2}) t^{12}
+\dots ~,~
\eea
where $[m; n] = [m]_{SU(2)_{x}} [n]_{SU(2)_y}$. From the plethystic logarithm, we see that the mesonic moduli space is a non-complete intersection.

The generators form 3 sets that transform under 
$[1; 3] b^2$,  $[2; 2]$ and $[3; 1] b^{-2}$ of the mesonic flavor symmetry of Model 6, respectively.
Using the following fugacity map
 \beal{es0627}
&&
\tilde{t} = t_5^{1/2} t_6^{1/2}~,~ 
\tilde{x} = \frac{t_1}{t_2}~,~ 
\tilde{y} =\frac{t_3}{t_4}~,~ 
\tilde{b} = \frac{t_2 t_4}{t_6^2}
\eea
the mesonic flavor charges on the gauge invariant operators become $\mathbb{Z}$-valued.
The generators in terms of brick matchings and their corresponding rescaled mesonic flavor charges are summarized in \tref{f_genlattice_06}.
The generator lattice as shown in \tref{f_genlattice_06} is a convex lattice polytope, which is reflexive. It is the dual of the toric diagram of Model 6 shown in \fref{f_toric_06}.
We also note that the 3 layers of points in the generator lattice in \tref{f_genlattice_06} corresponds to the 3 sets that transform under 
$[1,0] b^{-2}$, $[3,0]$ and $[5,0] b^2$ of the mesonic flavor symmetry.
For completeness, \tref{f_genfields_06a} and \tref{f_genfields_06b} show the generators of Model 6 in terms of chiral fields with the corresponding mesonic flavor charges.
\\

\begin{table}[H]
\centering
\resizebox{0.95\hsize}{!}{
\begin{tabular}{|l|c|c|c|}
\hline
generator & $SU(2)_{\tilde{x}}$ & $SU(2)_{\tilde{y}}$ & $U(1)_{\tilde{b}}$ \\
\hline
$ P_{15} X_{57} X_{78} X_{86} Z_{61}=  X_{13} X_{34} X_{42} P_{26} Z_{61}=  X_{13} P_{37} X_{78} X_{86} Z_{61}=  X_{13} X_{34} P_{48} X_{86} Z_{61}$ &  1 &  2 &  1 \\ 
$ Q_{15} X_{57} X_{78} X_{86} Z_{61}=  X_{13} X_{34} X_{42} Q_{26} Z_{61}=  X_{13} Q_{37} X_{78} X_{86} Z_{61}=  X_{13} X_{34} Q_{48} X_{86} Z_{61} $ &  0 &  2 &  1 \\ 
$ P_{15} X_{57} X_{78} X_{86} X_{61}=  P_{15} X_{57} X_{78} Y_{86} Z_{61}=  P_{15} Y_{57} X_{78} X_{86} Z_{61}=  X_{13} X_{34} X_{42} P_{26} X_{61}=  X_{13} X_{34} Y_{42} P_{26} Z_{61}$&  1 &  1 &  1 \\ 
$=  Y_{13} X_{34} X_{42} P_{26} Z_{61}=  X_{13} P_{37} X_{78} X_{86} X_{61}=  X_{13} P_{37} X_{78} Y_{86} Z_{61}=  Y_{13} P_{37} X_{78} X_{86} Z_{61}=  X_{13} X_{34} P_{48} X_{86} X_{61}$ & & &\\
$=  X_{13} X_{34} P_{48} Y_{86} Z_{61}=  Y_{13} X_{34} P_{48} X_{86} Z_{61}$ & & &\\
$ Q_{15} X_{57} X_{78} X_{86} X_{61}=  Q_{15} X_{57} X_{78} Y_{86} Z_{61}=  Q_{15} Y_{57} X_{78} X_{86} Z_{61}=  X_{13} X_{34} X_{42} Q_{26} X_{61}=  X_{13} X_{34} Y_{42} Q_{26} Z_{61}$&  0 &  1 &  1 \\ 
$=  Y_{13} X_{34} X_{42} Q_{26} Z_{61}=  X_{13} Q_{37} X_{78} X_{86} X_{61}=  X_{13} Q_{37} X_{78} Y_{86} Z_{61}=  Y_{13} Q_{37} X_{78} X_{86} Z_{61}=  X_{13} X_{34} Q_{48} X_{86} X_{61}$& & &\\
\hline
\end{tabular}
}
\caption{The generators in terms of bifundamental chiral fields for Model 6 \bf{(Part 1)}.
\label{f_genfields_06a}}
\end{table}

\begin{table}[H]
\centering
\resizebox{0.95\hsize}{!}{
\begin{tabular}{|l|c|c|c|}
\hline
generator & $SU(2)_{\tilde{x}}$ & $SU(2)_{\tilde{y}}$ & $U(1)_{\tilde{b}}$ \\
\hline
$P_{15} X_{54} X_{42} Q_{26} X_{61}=  P_{15} X_{54} Y_{42} Q_{26} Z_{61}=  P_{15} X_{57} X_{72} Q_{26} X_{61}=  P_{15} Y_{57} X_{72} Q_{26} Z_{61}=  P_{15} X_{54} Q_{48} X_{86} X_{61}$ &  0 &  0 &  0 \\ 
$=  P_{15} X_{54} Q_{48} Y_{86} Z_{61}=  P_{15} X_{57} X_{78} Y_{86} Y_{61}=  P_{15} Y_{57} X_{78} X_{86} Y_{61}=  Q_{15} X_{54} X_{42} P_{26} X_{61}=  Q_{15} X_{54} Y_{42} P_{26} Z_{61}$& & &\\
$=  Q_{15} X_{57} X_{72} P_{26} X_{61}=  Q_{15} Y_{57} X_{72} P_{26} Z_{61}=  X_{13} Q_{37} X_{72} P_{26} X_{61}=  Y_{13} Q_{37} X_{72} P_{26} Z_{61}=  X_{13} X_{34} Y_{42} P_{26} Y_{61}$& & &\\
$=  Y_{13} X_{34} X_{42} P_{26} Y_{61}=  X_{13} P_{37} X_{72} Q_{26} X_{61}=  Y_{13} P_{37} X_{72} Q_{26} Z_{61}=  X_{13} P_{37} X_{78} Y_{86} Y_{61}=  Y_{13} P_{37} X_{78} X_{86} Y_{61}$& & &\\
$=  Q_{15} X_{54} P_{48} X_{86} X_{61}=  Q_{15} X_{54} P_{48} Y_{86} Z_{61}=  X_{13} X_{34} P_{48} Y_{86} Y_{61}=  Y_{13} X_{34} P_{48} X_{86} Y_{61}=  Q_{15} X_{57} X_{78} Y_{86} R_{61}$& & &\\
$=  Q_{15} Y_{57} X_{78} X_{86} R_{61}=  X_{13} X_{34} Y_{42} Q_{26} R_{61}=  Y_{13} X_{34} X_{42} Q_{26} R_{61}=  X_{13} Q_{37} X_{78} Y_{86} R_{61}=  Y_{13} Q_{37} X_{78} X_{86} R_{61}$& & &\\
$=  X_{13} X_{34} Q_{48} Y_{86} R_{61}=  Y_{13} X_{34} Q_{48} X_{86} R_{61}$ & & &\\
\hline
\end{tabular}
}
\caption{The generators in terms of bifundamental chiral fields for Model 6 \bf{(Part 2)}.
\label{f_genfields_06b}}
\end{table}

\section{Model 7: $P_{++-}(\text{dP}_0)$~[$\mathbb{P}^1$-blowup of $5$,~$\langle25\rangle$] \label{smodel07}}

\begin{figure}[H]
\begin{center}
\resizebox{0.3\hsize}{!}{
\includegraphics[height=6cm]{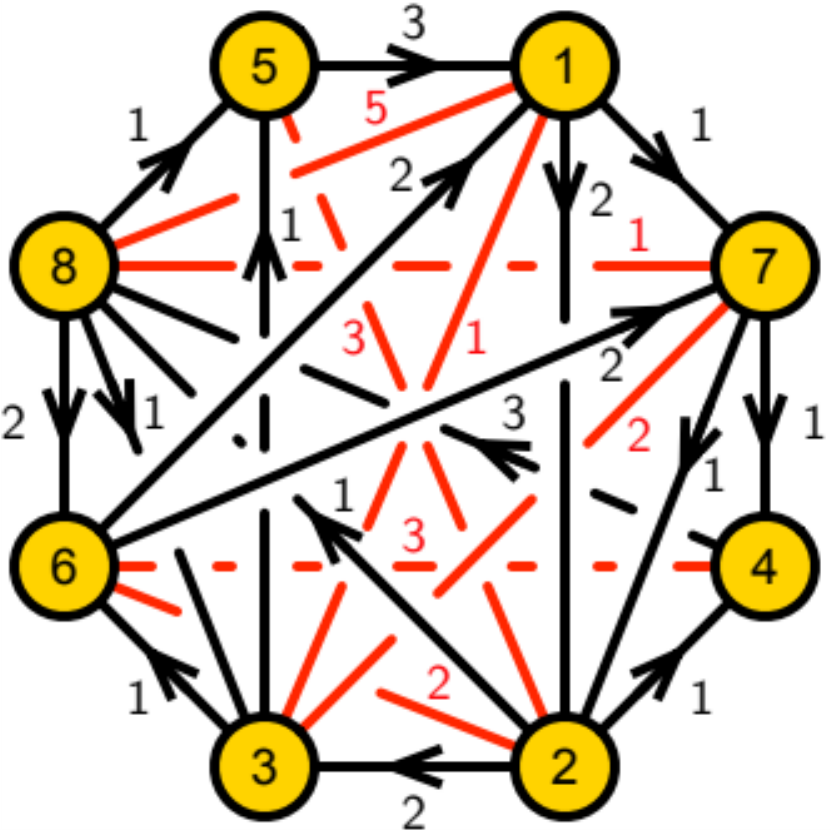} 
}
\caption{
Quiver for Model 7.
\label{f_quiver_07}}
 \end{center}
 \end{figure}

Model 7 corresponds to the Calabi-Yau 4-fold $P_{++-}(\text{dP}_0)$.
The corresponding brane brick model has the quiver in \fref{f_quiver_07} and the $J$- and $E$-terms are given as follows
\beq
{\footnotesize
\begin{array}{rrclrcl}
 &  & J &  & & E &  \\
\Lambda_{25}^{1} : &  X_{51} X_{12}  &-&  Z_{51} Y_{12} &  P_{24} X_{48} Q_{85}  &-&  X_{28} P_{83} P_{35} \\ 
\Lambda_{25}^{2} : &  Y_{51} Y_{12}  &-&  X_{51} Z_{17} P_{72} &  P_{24} Y_{48} Q_{85}  &-& Y_{23} P_{35} \\ 
\Lambda_{25}^{3} : &  Z_{51} Z_{17} P_{72}  &-& Y_{51} X_{12}  &  P_{24} Z_{48} Q_{85}  &-& Z_{23} P_{35} \\ 
\Lambda_{31}^{1} : &  Y_{12} Y_{23}  &-& X_{12} Z_{23} &  P_{35} Y_{51}  &-&  Y_{36} P_{61} \\ 
\Lambda_{62}^{1} : &  X_{28} X_{86}  &-&  Z_{23} Y_{36} &  P_{61} X_{12}  &-&  X_{67} P_{72} \\ 
\Lambda_{62}^{2} : &  Y_{23} Y_{36}  &-& X_{28} Z_{86}  &  P_{61} Y_{12}  &-&  Y_{67} P_{72} \\ 
\Lambda_{64}^{1} : &  X_{48} X_{86}  &-&  Z_{48} P_{83} Y_{36} &  X_{67} Q_{74}  &-&  Q_{61} X_{12} P_{24} \\ 
\Lambda_{64}^{2} : &  Y_{48} P_{83} Y_{36}  &-& X_{48} Z_{86} &  Y_{67} Q_{74}  &-&  Q_{61} Y_{12} P_{24} \\ 
\Lambda_{64}^{3} : &  Z_{48} Z_{86}  &-& Y_{48} X_{86}  &  P_{61} Z_{17} Q_{74}  &-&  Q_{61} Z_{17} P_{72} P_{24} \\ 
\Lambda_{73}^{1} : &  Y_{36} Y_{67}  &-&  P_{35} X_{51} Z_{17} &  P_{72} Y_{23}  &-&  Q_{74} Y_{48} P_{83} \\ 
\Lambda_{73}^{2} : &  P_{35} Z_{51} Z_{17}  &-& Y_{36} X_{67}  &  P_{72} Z_{23}  &-&  Q_{74} Z_{48} P_{83} \\ 
\Lambda_{78}^{1} : &  X_{86} X_{67}  &-&  Z_{86} Y_{67} &  P_{72} X_{28}  &-&  Q_{74} X_{48} \\ 
\Lambda_{81}^{1} : &  X_{12} P_{24} X_{48}  &-&  Z_{17} P_{72} P_{24} Y_{48} & X_{86} Q_{61}   &-& Q_{85} X_{51}\\ 
\Lambda_{81}^{2} : &  Y_{12} P_{24} Y_{48}  &-& X_{12} P_{24} Z_{48} &  P_{83} Y_{36} Q_{61}  &-& Q_{85} Y_{51} \\ 
\Lambda_{81}^{3} : &  Z_{17} P_{72} P_{24} Z_{48}  &-& Y_{12} P_{24} X_{48}  &  Z_{86} Q_{61}  &-& Q_{85} Z_{51} \\ 
\Lambda_{81}^{4} : &  X_{12} X_{28}  &-&  Z_{17} Q_{74} Y_{48} &  P_{83} P_{35} X_{51}  &-& X_{86} P_{61} \\ 
\Lambda_{81}^{5} : &  Z_{17} Q_{74} Z_{48}  &-& Y_{12} X_{28} &  P_{83} P_{35} Z_{51}  &-& Z_{86} P_{61} 
 \end{array} 
 }~.~
\label{E_J_C_+-}
\eeq
 
The $P$-matrix for Model 7 is 
\beal{es0705}
{\footnotesize
 P=
\resizebox{0.7\textwidth}{!}{$
\left(
\begin{array}{c|cccccc|ccccccccc|cccccccc}
\; & p_1 & p_2 & p_3 & p_4 & p_5 & p_6 & s_1 & s_2 & s_3 & s_4 & s_5 & s_6 & s_7 & s_8 & s_9 &  o_1 & o_2 & o_3 & o_4 & o_5 & o_6 & o_7 & o_8  \\
\hline
P_{24} & 0 & 0 & 1 & 0 & 0 & 0 & 0 & 1 & 0 & 0 & 0 & 0 & 0 & 0 & 0 & 1 & 1 & 0 & 0 & 0 & 0 & 1 & 0 \\
P_{35} & 0 & 0 & 1 & 0 & 0 & 0 & 1 & 0 & 1 & 0 & 0 & 0 & 0 & 0 & 0 & 1 & 0 & 1 & 1 & 0 & 0 & 0 & 1 \\
P_{61} & 0 & 0 & 1 & 1 & 0 & 0 & 1 & 0 & 0 & 1 & 0 & 0 & 0 & 0 & 0 & 1 & 1 & 1 & 1 & 1 & 1 & 1 & 1 \\
P_{72} & 0 & 0 & 0 & 1 & 0 & 0 & 0 & 0 & 0 & 0 & 1 & 0 & 0 & 0 & 0 & 1 & 0 & 1 & 1 & 1 & 0 & 0 & 1 \\
P_{83} & 0 & 0 & 0 & 1 & 0 & 0 & 0 & 0 & 0 & 0 & 0 & 1 & 0 & 0 & 0 & 0 & 1 & 0 & 0 & 0 & 1 & 1 & 0 \\
\hline
Q_{61} & 0 & 0 & 0 & 0 & 1 & 0 & 1 & 0 & 0 & 1 & 0 & 0 & 0 & 0 & 0 & 0 & 0 & 0 & 1 & 1 & 1 & 0 & 0 \\
Q_{74} & 0 & 0 & 0 & 0 & 1 & 0 & 0 & 1 & 0 & 0 & 1 & 0 & 0 & 0 & 0 & 1 & 0 & 0 & 1 & 1 & 0 & 0 & 0 \\
\end{array}
\right)
$}
}~.~
\eea

The $J$- and $E$-term charges are given by
\beal{es0705}
{\footnotesize
Q_{JE} =
\resizebox{0.67\textwidth}{!}{$
\left(
\begin{array}{cccccc|ccccccccc|cccccccc}
p_1 & p_2 & p_3 & p_4 & p_5 & p_6 & s_1 & s_2 & s_3 & s_4 & s_5 & s_6 & s_7 & s_8 & s_9 &  o_1 & o_2 & o_3 & o_4 & o_5 & o_6 & o_7 & o_8  \\
\hline
 1 & 0 & 0 & 0 & 1 & 1 & 0 & -1 & -1 & -1 & -1 & 0 & 0 & -1 & 0 & 1 & 0 & 0 & 0 & 0 & 0 & 0 & 0 \\
 0 & 0 & 1 & 0 & 1 & 0 & 0 & -1 & -1 & -1 & 0 & 0 & 0 & 0 & 1 & 0 & 0 & 0 & 0 & 0 & 0 & 0 & 0 \\
 0 & -1 & 0 & 0 & 1 & 0 & 0 & -1 & 0 & -1 & 0 & -1 & 0 & 0 & 1 & 0 & 1 & 0 & 0 & 0 & 0 & 0 & 0 \\
 0 & 0 & 0 & 0 & 1 & 0 & 0 & -1 & 0 & -1 & 0 & -1 & 0 & 0 & 0 & 0 & 0 & 0 & 0 & 0 & 0 & 1 & 0 \\
 1 & 0 & 0 & 0 & 0 & 1 & 0 & 0 & -1 & -1 & -1 & 0 & 0 & -1 & 0 & 0 & 0 & 0 & 1 & 0 & 0 & 0 & 0 \\
 0 & 0 & 0 & 0 & 1 & 0 & 0 & 0 & -1 & -1 & -1 & 0 & 0 & -1 & 1 & 0 & 0 & 0 & 0 & 0 & 0 & 0 & 1 \\
 1 & 0 & 0 & 0 & 1 & 1 & 0 & 0 & -1 & -1 & -1 & 0 & 0 & -2 & 0 & 0 & 0 & 1 & 0 & 0 & 0 & 0 & 0 \\
 0 & 0 & 0 & 0 & 0 & 0 & 1 & 0 & -1 & -1 & 0 & 0 & 0 & 0 & 1 & 0 & 0 & 0 & 0 & 0 & 0 & 0 & 0 \\
 1 & 0 & 0 & 1 & 1 & 1 & 0 & 0 & 0 & -1 & -1 & -1 & 0 & -1 & 0 & 0 & 0 & 0 & 0 & 0 & 0 & 0 & 0 \\
 1 & 0 & 0 & 0 & 0 & 1 & 0 & 0 & 0 & -1 & -1 & 0 & 0 & -1 & -1 & 0 & 0 & 0 & 0 & 1 & 0 & 0 & 0 \\
 0 & -1 & 0 & 0 & 0 & 0 & 0 & 0 & 0 & -1 & 0 & -1 & 0 & 0 & 1 & 0 & 0 & 0 & 0 & 0 & 1 & 0 & 0 \\
 -1 & -1 & 0 & 0 & 0 & -1 & 0 & 0 & 0 & 0 & 0 & 0 & 1 & 1 & 1 & 0 & 0 & 0 & 0 & 0 & 0 & 0 & 0 \\
\end{array}
\right)
$}
}~,~
\nn\\
\eea
and the $D$-term charges are given by
\beal{es0706}
{\footnotesize
Q_D =
\resizebox{0.67\textwidth}{!}{$
\left(
\begin{array}{cccccc|ccccccccc|cccccccc}
p_1 & p_2 & p_3 & p_4 & p_5 & p_6 & s_1 & s_2 & s_3 & s_4 & s_5 & s_6 & s_7 & s_8 & s_9 &  o_1 & o_2 & o_3 & o_4 & o_5 & o_6 & o_7 & o_8  \\
\hline
 0 & 0 & 0 & 0 & 0 & 0 & 0 & 1 & 0 & 0 & -1 & 0 & 0 & 0 & 0 & 0 & 0 & 0 & 0 & 0 & 0 & 0 & 0 \\
 0 & 0 & 0 & 0 & 0 & 0 & 0 & 0 & 1 & 0 & 0 & -1 & 0 & 0 & 0 & 0 & 0 & 0 & 0 & 0 & 0 & 0 & 0 \\
 0 & 0 & 0 & 0 & 0 & 0 & 0 & -1 & 0 & 0 & 0 & 0 & 0 & 1 & 0 & 0 & 0 & 0 & 0 & 0 & 0 & 0 & 0 \\
 0 & 0 & 0 & 0 & 0 & 0 & 0 & 0 & -1 & 0 & 0 & 0 & 0 & 0 & 1 & 0 & 0 & 0 & 0 & 0 & 0 & 0 & 0 \\
 0 & 0 & 0 & 0 & 0 & 0 & 0 & 0 & 0 & 1 & 0 & 0 & 0 & 0 & -1 & 0 & 0 & 0 & 0 & 0 & 0 & 0 & 0 \\
 -1 & -1 & 0 & 0 & 0 & -1 & 0 & 0 & 0 & 0 & 1 & 0 & 0 & 1 & 1 & 0 & 0 & 0 & 0 & 0 & 0 & 0 & 0 \\
 0 & 0 & 0 & 0 & 0 & 0 & 0 & 0 & 0 & 0 & 0 & 1 & 0 & -1 & 0 & 0 & 0 & 0 & 0 & 0 & 0 & 0 & 0 \\
\end{array}
\right)
$}
}~.~
\nn\\
\eea

\begin{figure}[H]
\begin{center}
\resizebox{0.4\hsize}{!}{
\includegraphics[height=6cm]{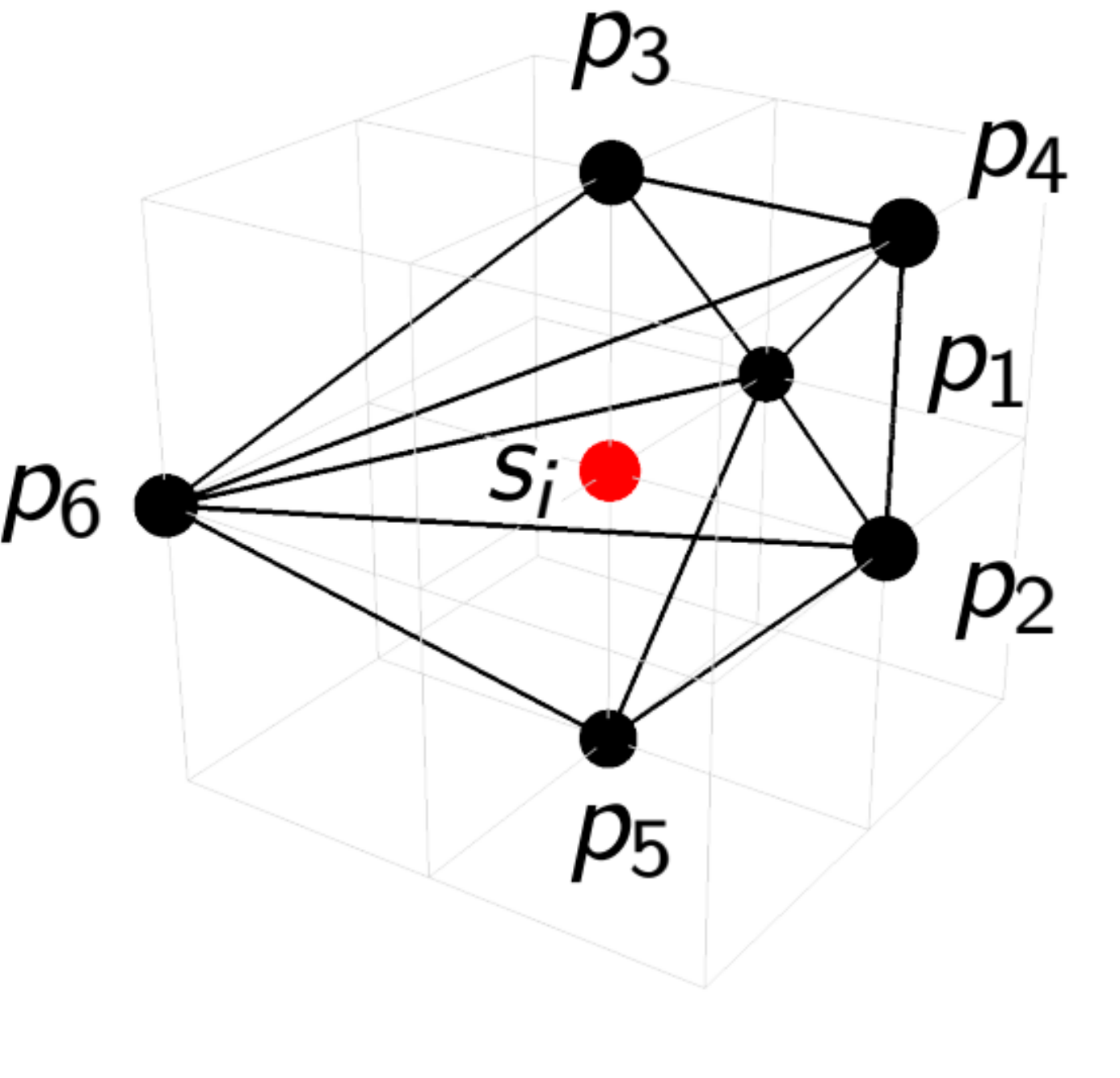} 
}
\caption{
Toric diagram for Model 7.
\label{f_toric_07}}
 \end{center}
 \end{figure}

The toric diagram of Model 7 is given by
\beal{es0710}
{\footnotesize
G_t=
\resizebox{0.67\textwidth}{!}{$
\left(
%
\right)
$}
}~,~
\eea
where \fref{f_toric_07} shows the toric diagram with brick matching labels.

The Hilbert series of the mesonic moduli space of Model 7 takes the form
 \beal{es0720}
&&
g_1(t_i,y_s,y_{o_1},y_{o_2},y_{o_3} ; \mathcal{M}_7) =
\frac{P(t_i,y_s,y_{o_1},y_{o_2},y_{o_3}; \mathcal{M}_7)}{
(1 - y_{s} y_{o_1}^2 y_{o_2} y_{o_3}^3 t_1^3 t_3^2 t_4) 
(1 - y_{s} y_{o_1}^5 y_{o_2}^4 y_{o_3}^3 t_2^3 t_3^2 t_4^4) 
}
\nn\\
&&
\hspace{1cm} 
\times
\frac{1}{
(1 - y_{s} y_{o_1} y_{o_2} y_{o_3}^2 t_1^3 t_3 t_5) 
(1 - y_{s} y_{o_1} y_{o_2}^2 y_{o_3} t_1^2 t_2 t_5^2) 
(1 - y_{s} y_{o_1}^3 y_{o_2}^4 y_{o_3} t_2^3 t_4^2 t_5^2) 
}
\nn\\
&&
\hspace{1cm} 
\times
\frac{1}{
(1 - y_{s} y_{o_1} y_{o_2}^2 y_{o_3} t_2 t_5^2 t_6^2) 
(1 - y_{s} y_{o_1}^2 y_{o_2} y_{o_3}^3 t_3^2 t_4 t_6^3) 
(1 - y_{s} y_{o_1} y_{o_2} y_{o_3}^2 t_3 t_5 t_6^3) 
} 
~,~
\nn\\
\eea
where $t_i$ are the fugacities for the extremal brick matchings $p_i$.
$y_{s}$ counts the brick matching product $s_1 \dots s_{9}$ corresponding to the single internal point of the toric diagram of Model 7.
Additionally, $y_{o_1}$, $y_{o_2}$ and $y_{o_3}$ count the products of extra GLSM fields $o_1 o_2 o_3$, $o_4 o_5 o_6$ and $o_7 o_8$, respectively.
The explicit numerator $P(t_i,y_s,y_{o_1},y_{o_2},y_{o_3}; \mathcal{M}_7)$ of the Hilbert series is given in the Appendix Section \sref{app_num_07}.
We note that setting the fugacities $y_{o_1}=1$, $y_{o_2}=1$ and $y_{o_3}=1$ does not change the overall characterization of the mesonic moduli space by the Hilbert series, indicating that the extra GLSM fields, as expected, correspond to an over-parameterization of the moduli space. 

\begin{table}[H]
\centering
\begin{tabular}{|c|c|c|c|c|c|}
\hline
\; & $SU(2)_{x}$ & $U(1)_{b_1}$ & $U(1)_{b_2}$ & $U(1)$ & \text{fugacity} \\
\hline
$p_1$ & +1 & 0 & 0 & $r_1$ & $t_1$ \\
$p_2$ & 0 & +1  & 0 & $r_2$ & $t_2$ \\
$p_3$ & 0 & -1  & 0 & $r_3$ & $t_3$ \\
$p_4$ & 0 & 0 & +1 & $r_4$ & $t_4$ \\
$p_5$ & 0 & 0 & -1 & $r_5$ & $t_5$ \\
$p_6$ & -1 & 0 & 0 & $r_6$ & $t_6$ \\
\hline
\end{tabular}
\caption{Global symmetry charges on the extremal brick matchings $p_i$ of Model 7.}
\label{t_pmcharges_07}
\end{table}

By setting $t_i=t$ for the fugacities of the extremal brick matchings, and all other fugacities to $y_s=1$, $y_{o_1}=1$, $y_{o_2}=1$ and $y_{o_3}=1$, the unrefined Hilbert series takes the following form
\beal{es0721}
&&
g_1(t,1,1,1,1; \mathcal{M}_7) =
\frac{
1
}{
(1 - t^5)^3 (1 - t^6)^2 (1 - t^7) (1 - t^9)
} 
\nn\\
&&
\hspace{0.5cm}
\times
(
1 + 4 t^5 + 7 t^6 + 5 t^7 + 3 t^8 - 13 t^{11} - 14 t^{12} - 8 t^{13} - 
 4 t^{14} - 2 t^{15} + t^{16} + 2 t^{17} 
 \nn\\
 &&
 \hspace{1cm}
+ 6 t^{18}  - 6 t^{19} - 2 t^{20} - t^{21} + 
 2 t^{22} + 4 t^{23} + 8 t^{24} + 14 t^{25} + 13 t^{26} - 3 t^{29} - 5 t^{30} 
 \nn\\
 &&
 \hspace{1cm}
 - 7 t^{31} - 4 t^{32} - t^{37}
 )
 ~,~
\eea
where the palindromic numerator indicates that the mesonic moduli space is Calabi-Yau. 

The global symmetry of Model 6 and the charges on the extremal brick matchings under the global symmetry are summarized in \tref{t_pmcharges_07}.
We can use the following fugacity map,
\beal{es0722}
&&
t = \frac{t_1 t_6}{t_4^{1/2} t_5^{1/2}} ~,~
x = \frac{t_1}{t_4^{1/2} t_5^{1/2}} ~,~
b_1 = \frac{t_2}{t_4^{1/2} t_5^{1/2}} ~,~
b_2 = \frac{t_4^{1/2} }{t_5^{1/2}} ~,~
\eea
where $t_2 t_3 = t_4 t_5$ and $t_1 t_6 = t_4 t_5$, in order to rewrite the 
Hilbert series for Model 6 in terms of characters of irreducible representations of $SU(2)\times U(1)\times U(1)$.
In terms of fugacities of the mesonic flavor symmetry of Model 7, the Hilbert series in highest weight form can be written as
\beal{es0725}
&&
h_1(t, \mu ,b_1, b_2; \mathcal{M}_7) =
\frac{1}{
(1 - b_1^3 t^7) 
(1 - b_1 b_2^4 t^9)
(1 - \mu^2 b_1 b_2^{-2} t^5) 
}
\nn\\
&&
\hspace{0.5cm}
\times 
\frac{1}{
(1 - \mu^3 b_1^{-1} b_2^{-1} t^5) 
(1 - \mu^3 b_1^{-2} b_2   t^6) 
}
\times 
(
1 + \mu b_1^2 b_2^{-1} t^6 + \mu^2 t^6 +\mu b_1 b_2  t^7 
\nn\\
&&
\hspace{1cm}
+ \mu^2 b_1^{-1} b_2^2 t^7 +  b_1^2 b_2^2 t^8 + \mu b_2^3  t^8 - \mu^4 b_1 b_2^{-2}  t^{11} -  \mu^5 b_1^{-1} b_2^{-1}  t^{11} -\mu^3  b_1^2 b_2^{-1}  t^{12}
\nn\\
&&
\hspace{1cm}
 - 2 \mu^4 t^{12} -  \mu^5 b_1^{-2} b_2  t^{12} -  \mu^2 b_1^3 t^{13} - 2\mu^3 b_1 b_2  t^{13}
 -  \mu^4 b_1^{-1} b_2^2 t^{13} -\mu^2  b_1^2 b_2^2  t^{14}
 \nn\\
 &&
 \hspace{1cm}
- \mu^3  b_2^3 t^{14} - \mu b_1^3 b_2^3  t^{15} + \mu^7 b_1^{-1} b_2^{-1}  t^{17}  
+ \mu^5 b_1^2 b_2^{-1}  t^{18} + \mu^6 t^{18} +  \mu^5 b_1 b_2  t^{19} 
 \nn\\
 &&
 \hspace{1cm}
+ \mu^7 b_1 b_2  t^{25} +  \mu^6 b_1^2 b_2^2 t^{26}
) ~,~
\eea
where $\mu^m \sim [m]_{SU(2)_x}$. Note that in highest weight form, the fugacity $\mu$ counts the highest weight of the irreducible representations of $SU(2)$. 
Additionally, the fugacities $b_1$ and $b_2$ count the charges under the $U(1)$ factors of the mesonic flavor symmetry.

\begin{table}[H]
\centering
\resizebox{.95\hsize}{!}{
\begin{minipage}[!b]{0.5\textwidth}
\begin{tabular}{|c|c|c|c|}
\hline
generator & $SU(2)_{\tilde{x}}$ & $U(1)_{\tilde{b_1}}$ & $U(1)_{\tilde{b_2}}$ \\
\hline
$p_1^3 p_3 p_5 ~s o_1 o_2 o_3^2  $ &2& 0& -1\\
$p_1^2 p_2 p_5^2 ~s o_1 o_2^2 o_3  $ &1& 1& 0\\
$p_1^2 p_3 p_5 p_6 ~s o_1 o_2 o_3^2  $ &1& 0& -1\\
$p_1 p_2 p_5^2 p_6 ~s o_1 o_2^2 o_3  $ &0& 1& 0\\
$p_1 p_3 p_5 p_6^2 ~s o_1 o_2 o_3^2  $ &0& 0& -1\\
$p_2 p_5^2 p_6^2 ~s o_1 o_2^2 o_3  $ &-1& 1& 0\\
$p_3 p_5 p_6^3 ~s o_1 o_2 o_3^2  $ &-1& 0& -1\\
$p_1^3 p_3^2 p_4 ~s o_1^2 o_2 o_3^3  $ &2& -1& -1\\
$p_1^2 p_2 p_3 p_4 p_5 ~s o_1^2 o_2^2 o_3^2  $ &1& 0& 0\\
$p_1 p_2^2 p_4 p_5^2 ~s o_1^2 o_2^3 o_3  $ &0& 1& 1\\
$p_1^2 p_3^2 p_4 p_6 ~s o_1^2 o_2 o_3^3  $ &1& -1& -1\\
$p_1 p_2 p_3 p_4 p_5 p_6 ~s o_1^2 o_2^2 o_3^2  $ &0& 0& 0\\
$p_2^2 p_4 p_5^2 p_6 ~s o_1^2 o_2^3 o_3  $ &-1& 1& 1\\
$p_1 p_3^2 p_4 p_6^2 ~s o_1^2 o_2 o_3^3  $ &0& -1& -1\\
$p_2 p_3 p_4 p_5 p_6^2 ~s o_1^2 o_2^2 o_3^2  $ &-1& 0& 0\\
$p_3^2 p_4 p_6^3 ~s o_1^2 o_2 o_3^3  $ &-1& -1& -1\\
$p_1^2 p_2 p_3^2 p_4^2 ~s o_1^3 o_2^2 o_3^3  $ &1& -1& 0\\
$p_1 p_2^2 p_3 p_4^2 p_5 ~s o_1^3 o_2^3 o_3^2  $ &0& 0& 1\\
$p_2^3 p_4^2 p_5^2 ~s o_1^3 o_2^4 o_3  $ &-1& 1& 2\\
$p_1 p_2 p_3^2 p_4^2 p_6 ~s o_1^3 o_2^2 o_3^3  $ &0& -1& 0\\
$p_2^2 p_3 p_4^2 p_5 p_6 ~s o_1^3 o_2^3 o_3^2  $ &-1& 0& 1\\
$p_2 p_3^2 p_4^2 p_6^2 ~s o_1^3 o_2^2 o_3^3  $ &-1& -1& 0\\
$p_1 p_2^2 p_3^2 p_4^3 ~s o_1^4 o_2^3 o_3^3  $ &0& -1& 1\\
$p_2^3 p_3 p_4^3 p_5 ~s o_1^4 o_2^4 o_3^2  $ &-1& 0& 2\\
$p_2^2 p_3^2 p_4^3 p_6 ~s o_1^4 o_2^3 o_3^3  $ &-1& -1& 1\\
$p_2^3 p_3^2 p_4^4 ~s o_1^5 o_2^4 o_3^3  $ &-1& -1& 2\\
\hline
\end{tabular}
\end{minipage}
\hspace{2cm}
\begin{minipage}[!b]{0.4\textwidth}
\includegraphics[height=8cm]{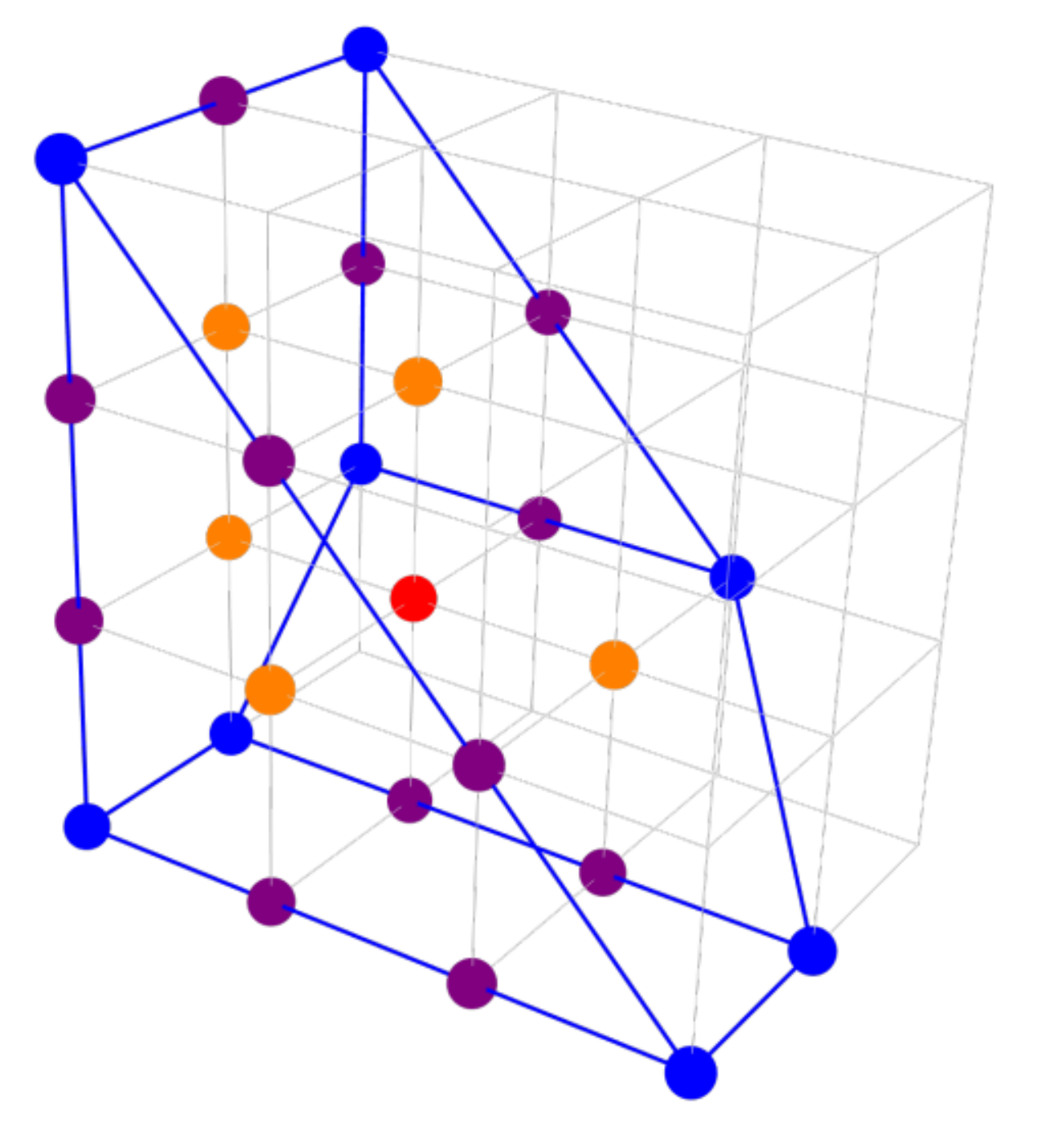} 
\end{minipage}
}
\caption{The generators and lattice of generators of the mesonic moduli space of Model 7 in terms of brick matchings with the corresponding flavor charges.
\label{f_genlattice_07}}
\end{table}

The plethystic logarithm of the Hilbert series takes the following form
\beal{es0726}
&&
\PL[g_1(t, x, b_1, b_2; \mathcal{M}_7)]=
([2] b_1 b_2^{-2} + [3] b_1^{-1} b_2^{-1}) t^5
+([1] b_1^2 b_2^{-1} + [2] + [3] b_1^{-2} b_2^{1}) t^6
\nn\\
&&
\hspace{1cm}
+(b_1^3 + [1] b_1 b_2 + [2] b_1^{-1} b_2^{2}) t^7
+(b_1^2 b_2^2 + [1]  b_2^3) t^8
+  b_1 b_2^4 t^9
\nn\\
&&
\hspace{1cm}
- ([3] b_2^{-3} + b_1^{2} b_2^{-4} + [1]  b_2^{-3} + [2] b_1^{-2} b_2^{-2}) t^{10}
- ([4] b_1 b_2^{-2} + [5] b_1^{-1} b_2^{-1} + [1] b_1^{3} b_2^{-3} 
\nn\\
&&
\hspace{1cm}
+ 2 [2] b_1 b_2^{-2} + 2 [3] b_1^{-1} b_2^{-1} + [4] b_1^{-3} + b_1 b_2^{-2} + 2 [1] b_1^{-1} b_2^{-1} + [2] b_1^{-3} + b_1^{-3} ) t^{11}
\nn\\
&&
\hspace{1cm}
- ([2] b_1^{4} b_2^{-2} + 2 [3] b_1^{2} b_2^{-1} + 3 [4] +  [5] b_1^{-2} b_2 + 2 [1] b_1^{2} b_2^{-1} + 3 [2] +  2 [3] b_1^{-2} b_2 + 2 
\nn\\
&& 
\hspace{1cm}
+ 2 [1] b_1^{-2} b_2 +  [2] b_1^{-4} b_2^{2}) t^{12}
- (2 [2]  b_1^{3} + 4 [3]  b_1 b_2 + 2 [4]  b_1^{-1} b_2^{2} + b_1^{3} + 3 [1]  b_1 b_2 
\nn\\
&& 
\hspace{1cm}
+ 3 [2]  b_1^{-1} b_2^{2} + [3]  b_1^{-3} b_2^{3} + b_1^{-1} b_2^{2} + [1]  b_1^{-3} b_2^{3}) t^{13}
- ([1] b_1^{4} b_2 + 4 [2] b_1^{2} b_2^{2} + 3 [3]  b_2^{3} 
\nn\\
&& 
\hspace{1cm}
+ [4] b_1^{-2} b_2^{4} + b_1^{2} b_2^{2} + 2 [1]  b_2^{3} + [2] b_1^{-2} b_2^{4} + b_1^{-2} b_2^{4}) t^{14}
+ \dots ~,~
\eea
where $[m] = [m]_{SU(2)_x}$. From the plethystic logarithm, we see that the mesonic moduli space is a non-complete intersection.

Using the following fugacity map
\beal{es0727}
&&
\tilde{t} = t_4^{1/2} t_5^{1/2}~,~ 
\tilde{x} = \frac{t_4 t_5}{t_6^2}~,~ 
\tilde{b_1} =\frac{t_5^{3/2}}{t_3 t_4^{1/2}}~,~ 
\tilde{b_2} = \frac{t_4^{3/2} t_5^{1/2}}{t_3 t_6}
~,~
\eea
where $t_1 t_6 = t_2 t_3 = t_4 t_5$, the mesonic flavor charges on the gauge invariant operators become $\mathbb{Z}$-valued.
The generators in terms of brick matchings and their corresponding rescaled mesonic flavor charges are summarized in \tref{f_genlattice_07}.
The generator lattice as shown in \tref{f_genlattice_07} is a convex lattice polytope, which is reflexive. It is the dual of the toric diagram of Model 7 shown in \fref{f_toric_07}.
For completeness, \tref{f_genfields_07a} and \tref{f_genfields_07b} show the generators of Model 7 in terms of chiral fields with the corresponding mesonic flavor charges.

\begin{table}[H]
\centering
\resizebox{0.95\hsize}{!}{

}
\caption{The generators in terms of bifundamental chiral fields for Model 7 \bf{(Part 1)}.
\label{f_genfields_07a}}
\end{table}

\begin{table}[H]
\centering
\resizebox{0.95\hsize}{!}{
%
}
\caption{The generators in terms of bifundamental chiral fields for Model 7 \bf{(Part 2)}.
\label{f_genfields_07b}}
\end{table}

\section{Model 8: $P_{++-}H_{+}(\text{dP}_0)$~[$\mathbb{P}(\mathcal{O}_{\text{dP}_1} \oplus \mathcal{O}_{\text{dP}_1}(l))$, $l^2|_{\text{dP}_1}=1$,~$\langle26\rangle$] \label{smodel08}}
 
\begin{figure}[H]
\begin{center}
\resizebox{0.3\hsize}{!}{
\includegraphics[height=6cm]{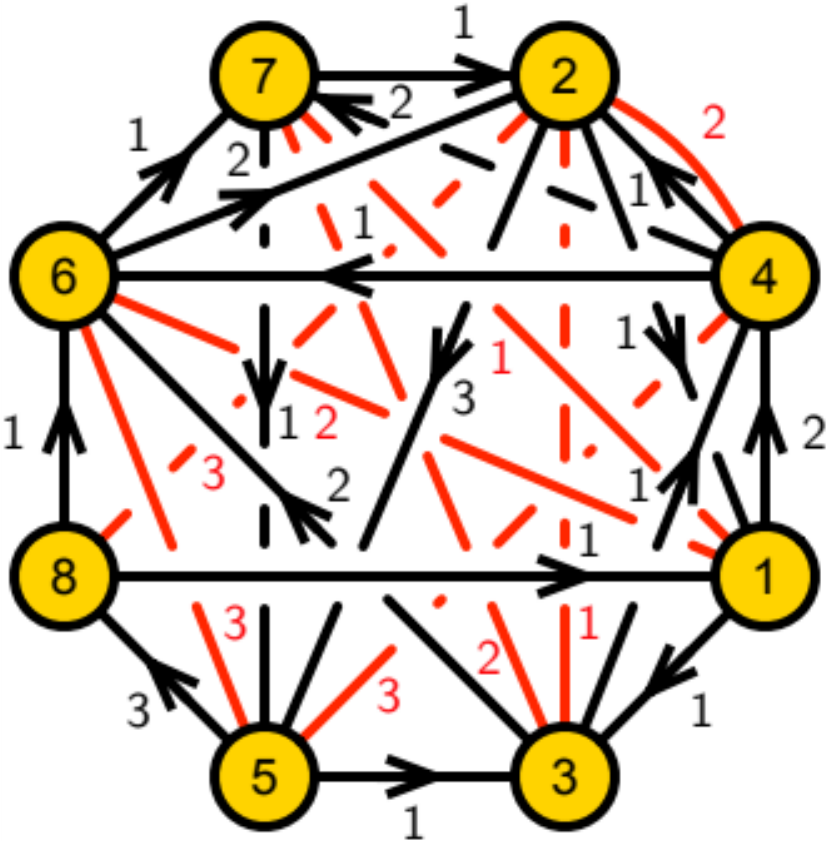} 
}
\caption{
Quiver for Model 8.
\label{f_quiver_08}}
 \end{center}
 \end{figure}

Model 8 corresponds to toric Calabi-Yau 4-fold $P_{++-}H_{+}(\text{dP}_0)$.
The corresponding brane brick model has the quiver in \fref{f_quiver_08} and the $J$- and $E$-terms are given as follows
\beq
{\footnotesize
 
 }~.~
\label{E_J_C_+-}
 \eeq
 
Following the forward algorithm, we obtain the brick matchings for Model 8. They are summarized in the $P$-matrix, which takes the form
\beal{es0805}
{\footnotesize
P=
\resizebox{0.87\textwidth}{!}{$
\left(
%
\right)
$}
}~.~
\nn\\
\eea

The $J$- and $E$-term charges are given by
\beal{es0806}
{\footnotesize
Q_{JE} =
\resizebox{0.84\hsize}{!}{$
\left(
%
\right)
$}
}~,~
\nn\\
\eea
and the $D$-term charges are given by
\beal{es0807}
{\footnotesize
Q_D =
\resizebox{0.84\hsize}{!}{$
\left(
%
\right)
$}
}~.~
\nn\\
\eea

\begin{figure}[H]
\begin{center}
\resizebox{0.4\hsize}{!}{
\includegraphics[height=6cm]{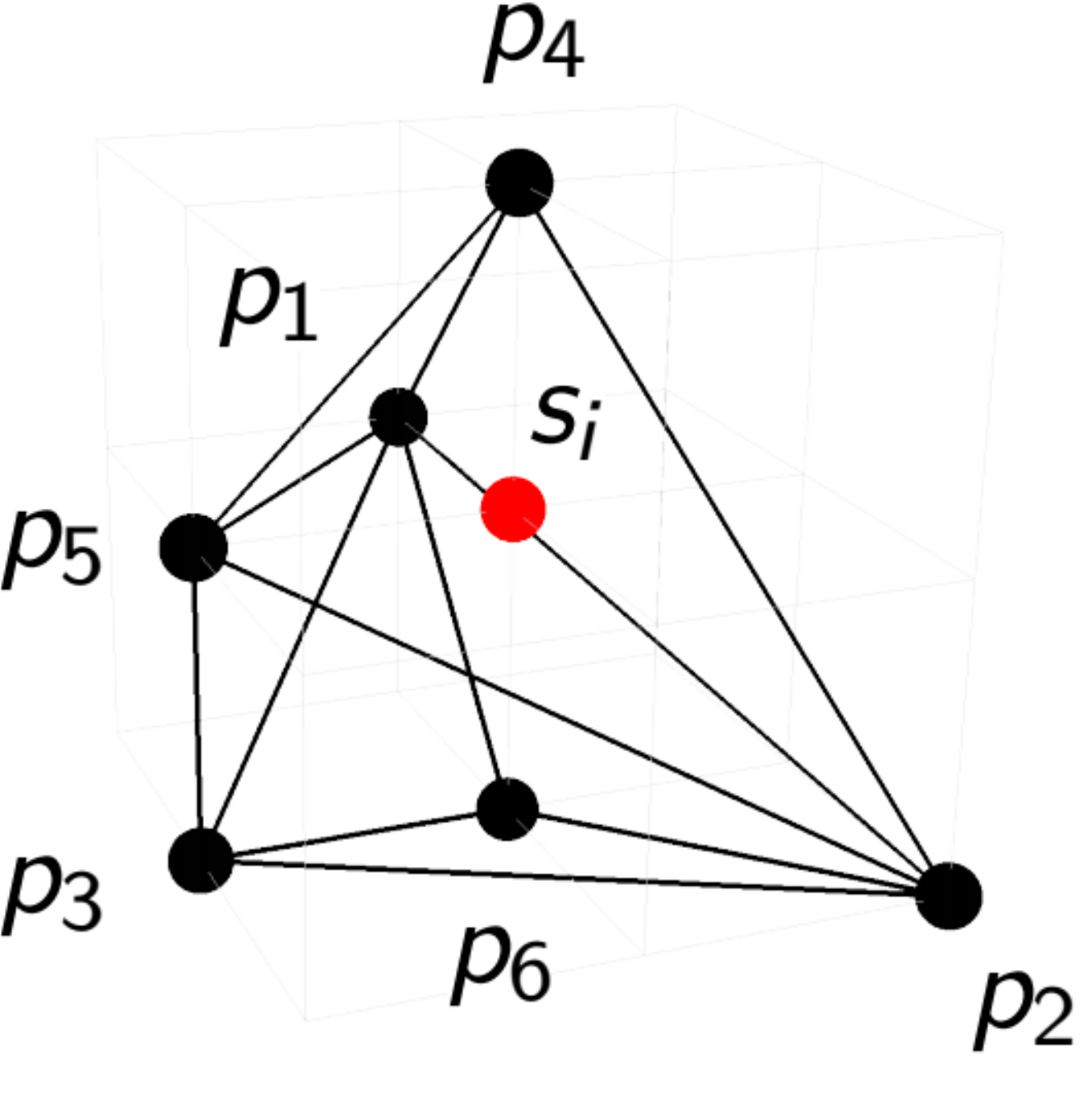} 
}
\caption{
Toric diagram for Model 8.
\label{f_toric_08}}
 \end{center}
 \end{figure}

The toric diagram of Model 8 is given by
\beal{es0810}
{\footnotesize
G_t=
\resizebox{0.84\hsize}{!}{$
\left(
%
\right)
$}
}~,~
\nn\\
\eea
where \fref{f_toric_08} shows the toric diagram with brick matching labels.

\begin{table}[H]
\centering
%
\caption{Global symmetry charges on the extremal brick matchings $p_i$ of Model 8.}
\label{t_pmcharges_08}
\end{table}

The Hilbert series of the mesonic moduli space of Model 8 takes the form
\beal{es0820}
&&
g_1(t_i,y_s,y_{o_1},y_{o_2},y_{o_3} ; \mathcal{M}_8) =
\frac{
P(t_i,y_s,y_{o_1},y_{o_2},y_{o_3}; \mathcal{M}_8)
}{
(1 - y_{s} y_{o_1}^2 y_{o_2} y_{o_3} t_1 t_4^2 t_5) 
(1 -    y_{s} y_{o_1}^2 y_{o_2} y_{o_3} t_2 t_4^2 t_5) 
}
\nn\\
&&
\hspace{1cm} 
\times
\frac{1}{
(1 -    y_{s} y_{o_1}^3 y_{o_2}^2 y_{o_3} t_3 t_4^2 t_5^2) 
(1 -    y_{s} y_{o_1} y_{o_2} y_{o_3}^2 t_1^3 t_4 t_6) 
(1 - y_{s} y_{o_1} y_{o_2} y_{o_3}^2 t_2^3 t_4 t_6) 
} 
\nn\\
&&
\hspace{1cm} 
\times
\frac{1}{
(1 -    y_{s} y_{o_1} y_{o_2}^2 y_{o_3}^3 t_1^4 t_3 t_6^2) 
(1 -    y_{s} y_{o_1} y_{o_2}^2 y_{o_3}^3 t_2^4 t_3 t_6^2) 
(1 -    y_{s} y_{o_1}^5 y_{o_2}^6 y_{o_3}^3 t_3^5 t_5^4 t_6^2)
} ~,~
\nn\\
\eea
where $t_i$ are the fugacities for the extremal brick matchings $p_i$.
$y_{s}$ counts the brick matching product $s_1 \dots s_{8}$ corresponding to the single internal point of the toric diagram of Model 8.
Additionally, $y_{o_1}$, $y_{o_2}$ and $y_{o_3}$ count the products of extra GLSM fields $o_1 o_2$, $o_3 o_4 o_5 o_6$ and $o_7 \dots o_{15}$, respectively.
The explicit numerator $P(t_i,y_s,y_{o_1},y_{o_2},y_{o_3}; \mathcal{M}_8)$ of the Hilbert series is given in the Appendix Section \sref{app_num_08}.
We note that setting the fugacities $y_{o_1}=1$, $y_{o_2}=1$ and $y_{o_3}=1$ does not change the overall characterization of the mesonic moduli space by the Hilbert series, indicating that the extra GLSM fields, as expected, correspond to an over-parameterization of the moduli space. 

By setting $t_i=t$ for the fugacities of the extremal brick matchings, and all other fugacities to $y_s=1$, $y_{o_1}=1$, $y_{o_2}=1$ and $y_{o_3}=1$, the unrefined Hilbert series takes the following form
\beal{es0821}
&&
g_1(t,1,1,1; \mathcal{M}_8) =
\frac{
(1 - t)^3
}{
(1- t^4)^2 (1- t^5)^2 (1- t^7)^2 (1- t^{11})
}
\nn\\
&&
\hspace{0.5cm}
\times
(
1 + 3 t + 6 t^{2} + 10 t^{3} + 15 t^{4} + 24 t^{5} + 40 t^{6} + 68 t^{7} +  113 t^{8} + 175 t^{9} + 253 t^{10}
\nn\\
&&
\hspace{1cm} 
+ 336 t^{11} + 409 t^{12} + 462 t^{13} +  490 t^{14} + 496 t^{15} + 493 t^{16} + 491 t^{17} + 493 t^{18} 
\nn\\
&&
\hspace{1cm} 
+ 496 t^{19} +  490 t^{20} + 462 t^{21} + 409 t^{22} + 336 t^{23} + 253 t^{24} + 175 t^{25} +  113 t^{26} 
\nn\\
&&
\hspace{1cm} 
+ 68 t^{27} + 40 t^{28} + 24 t^{29} + 15 t^{30} + 10 t^{31} +  6 t^{32} + 3 t^{33} + t^{34}
)
~,~
\eea
where the palindromic numerator shows that the mesonic moduli space is Calabi-Yau. 

\begin{table}[H]
\centering
\resizebox{.95\hsize}{!}{
\begin{minipage}[!b]{0.5\textwidth}
\begin{tabular}{|c|c|c|c|}
\hline
generator & $SU(2)_{\tilde{x}}$ & $U(1)_{\tilde{b_1}}$ & $U(1)_{\tilde{b_2}}$ \\
\hline
$ p_1 p_4^2 p_5  ~s o_1^2 o_2 o_3  $ & 1& 0& 1 \\ 
$ p_2 p_4^2 p_5  ~s o_1^2 o_2 o_3  $ & 0& 0& 1 \\ 
$ p_3 p_4^2 p_5^2  ~s o_1^3 o_2^2 o_3  $ & 1& 1& 1 \\ 
$ p_1^3 p_4 p_6  ~s o_1 o_2 o_3^2  $ & 1& -1& 0 \\ 
$ p_1^2 p_2 p_4 p_6  ~s o_1 o_2 o_3^2  $ & 0& -1& 0 \\ 
$ p_1 p_2^2 p_4 p_6  ~s o_1 o_2 o_3^2  $ & -1& -1& 0 \\ 
$ p_2^3 p_4 p_6  ~s o_1 o_2 o_3^2  $ & -2& -1& 0 \\ 
$ p_1^2 p_3 p_4 p_5 p_6  ~s o_1^2 o_2^2 o_3^2  $ & 1& 0& 0 \\ 
$ p_1 p_2 p_3 p_4 p_5 p_6  ~s o_1^2 o_2^2 o_3^2  $ & 0& 0& 0 \\ 
$ p_2^2 p_3 p_4 p_5 p_6  ~s o_1^2 o_2^2 o_3^2  $ & -1& 0& 0 \\ 
$ p_1 p_3^2 p_4 p_5^2 p_6  ~s o_1^3 o_2^3 o_3^2  $ & 1& 1& 0 \\ 
$ p_2 p_3^2 p_4 p_5^2 p_6  ~s o_1^3 o_2^3 o_3^2  $ & 0& 1& 0 \\ 
$ p_1^4 p_3 p_6^2  ~s o_1 o_2^2 o_3^3  $ & 1& -1& -1 \\ 
$ p_1^3 p_2 p_3 p_6^2  ~s o_1 o_2^2 o_3^3  $ & 0& -1& -1 \\ 
$ p_1^2 p_2^2 p_3 p_6^2  ~s o_1 o_2^2 o_3^3  $ & -1& -1& -1 \\ 
$ p_1 p_2^3 p_3 p_6^2  ~s o_1 o_2^2 o_3^3  $ & -2& -1& -1 \\ 
$ p_2^4 p_3 p_6^2  ~s o_1 o_2^2 o_3^3  $ & -3& -1& -1 \\ 
$ p_3^3 p_4 p_5^3 p_6  ~s o_1^4 o_2^4 o_3^2  $ & 1& 2& 0 \\ 
$ p_1^3 p_3^2 p_5 p_6^2  ~s o_1^2 o_2^3 o_3^3  $ & 1& 0& -1 \\ 
$ p_1^2 p_2 p_3^2 p_5 p_6^2  ~s o_1^2 o_2^3 o_3^3  $ & 0& 0& -1 \\ 
$ p_1 p_2^2 p_3^2 p_5 p_6^2  ~s o_1^2 o_2^3 o_3^3  $ & -1& 0& -1 \\ 
$ p_2^3 p_3^2 p_5 p_6^2  ~s o_1^2 o_2^3 o_3^3  $ & -2& 0& -1 \\ 
$ p_1^2 p_3^3 p_5^2 p_6^2  ~s o_1^3 o_2^4 o_3^3  $ & 1& 1& -1 \\ 
$ p_1 p_2 p_3^3 p_5^2 p_6^2  ~s o_1^3 o_2^4 o_3^3  $ & 0& 1& -1 \\ 
$ p_2^2 p_3^3 p_5^2 p_6^2  ~s o_1^3 o_2^4 o_3^3  $ & -1& 1& -1 \\ 
$ p_1 p_3^4 p_5^3 p_6^2  ~s o_1^4 o_2^5 o_3^3  $ & 1& 2& -1 \\ 
$ p_2 p_3^4 p_5^3 p_6^2  ~s o_1^4 o_2^5 o_3^3  $ & 0& 2& -1 \\ 
$ p_3^5 p_5^4 p_6^2  ~s o_1^5 o_2^6 o_3^3  $ & 1& 3& -1\\
\hline
\end{tabular}
\end{minipage}
\hspace{1.5cm}
\begin{minipage}[!b]{0.45\textwidth}
\includegraphics[height=6cm]{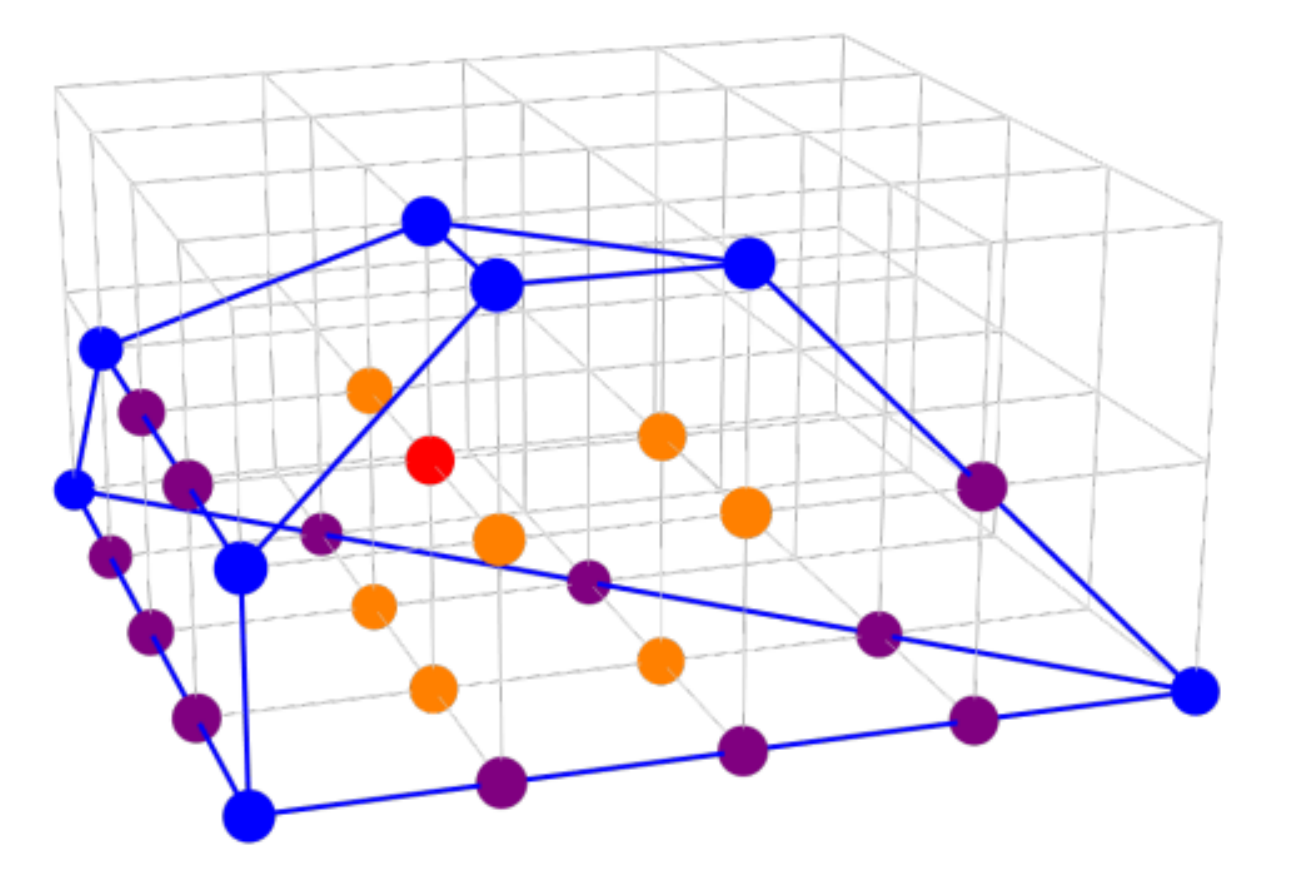} 
\end{minipage}
}
\caption{The generators and lattice of generators of the mesonic moduli space of Model 8 in terms of brick matchings with the corresponding flavor charges.
\label{f_genlattice_08}}
\end{table}

The global symmetry of Model 8 and the charges on the extremal brick matchings under the global symmetry are summarized in \tref{t_pmcharges_08}.
We can use the following fugacity map,
\beal{es0822}
&&
t = t_1^{1/2} t_2^{1/2} ~,~
x = \frac{t_1^{1/2}}{t_2^{1/2}} ~,~
b_1 = \frac{t_1^{1/2} t_2^{1/2}}{t_4^{1/2}} ~,~
b_2 = \frac{t_1^{1/2} t_2^{1/2} t_5}{t_3 t_4} ~,~
\eea
where $t_3 t_4 = t_5 t_6$ and $t_1 t_2 =t_5 t_6$, in order to rewrite the 
Hilbert series for Model 8 in terms of characters of irreducible representations of $SU(2)\times U(1) \times U(1)$.
In highest weight form, the Hilbert series of Model 8 can be written as
\beal{es0825}
&&
h_1(t, \mu,b_1, b_2; \mathcal{M}_8) =
\frac{1}{
(1 - \mu b_1^{-2} b_2 t^4) (1 - \mu^3 b_1^{-1} b_2^{-1} t^5) (1 - b_1^{-1} b_2^2 t^5)
}
\nn\\
&&
\hspace{0.5cm}
\times
\frac{1}{
(1 - \mu^4 b_1 b_2^{-2} t^7) (1 - b_1^5 b_2^2 t^{11})
}
\times 
(
1 + \mu^2 t^6 + \mu b_1 b_2 t^7 + \mu^3 b_1^2 b_2^{-1} t^8 + b_1^2 b_2^2 t^8 
\nn\\
&&
\hspace{1cm}
+ \mu^2 b_1^3 t^9 - \mu^3 b_1^{-2} b_2 t^{10}  + \mu b_1^4 b_2 t^{10} - \mu^5 b_1^{-1} b_2^{-1} t^{11} - \mu^2 b_1^{-1} b_2^2 t^{11} - 2 \mu^4 t^{12} 
\nn\\
&&
\hspace{1cm}
- \mu b_2^3 t^{12} - \mu^6 b_1 b_2^{-2} t^{13} - 2 \mu^3 b_1 b_2 t^{13} - \mu^5 b_1^2 b_2^{-1} t^{14} - \mu^2 b_1^2 b_2^2 t^{14} - \mu^4 b_1^3 t^{15} 
\nn\\
&&
\hspace{1cm}
- \mu b_1^3 b_2^3 t^{15} + \mu^5 b_1^{-2} b_2 t^{16} + \mu^7 b_1^{-1} b_2^{-1} t^{17} + \mu^4 b_1^{-1} b_2^2 t^{17} + \mu^6 t^{18} + \mu^5 b_1 b_2 t^{19} 
\nn\\
&&
\hspace{1cm}
- \mu^6 b_1^3 t^{21} + \mu^7 b_1 b_2 t^{25} + \mu^6 b_1^2 b_2^2 t^{26}
)~,~
\eea
where $\mu^m \sim [m]_{SU(2)_x}$. Here in highest weight form, the fugacity $\mu$ counts the highest weight of irreducible representations of $SU(2)$.

The plethystic logarithm of the Hilbert series takes the form
\beal{es0826}
&&
\PL[g_1(t, x, b_1, b_2; \mathcal{M}_8)]=
[1] b_1^{-2} b_2 t^4
+ ([3] b_1^{-1} b_2^{-1} + b_1^{-1} b_2^{2}) t^5
+ [2] t^6
\nn\\
&&
\hspace{1cm}
+ ([4] b_1 b_2^{-2} + [1] b_1 b_2) t^7
+ ([3] b_1^{2} b_2^{-1} + b_1^2 b_2^2) t^8
+ ([2] b_1^{3} + [2] b_1^{-3}) t^9
+ [1] b_1^4 b_2 t^{10}
\nn\\
&&
\hspace{1cm}
- ([3] b_1^{-2} b_2  + [2] b_1^{-2} b_2^{-2}  + [1] b_1^{-2} b_2) t^{10}
- (b_1^{5} b_2^{2}  + [5] b_1^{-1} b_2^{-1}  + 2 [3] b_1^{-1} b_2^{-1}  
\nn\\
&&
\hspace{1cm}
+ [2] b_1^{-1} b_2^{2}  + b [1] b_1^{-1} b_2^{-1}  + b_1^{-1} b_2^{2}  ) t^{11}
- ([5] b_2^{-3}  + 3 [4]  + [3] b_2^{-3}  + 2 [2]  + [1] b_2^{3}  
\nn\\
&&
\hspace{1cm}
+ [1] b_2^{-3}  + 1) t^{12}
- ([6] b_1 b_2^{-2}  + 2 [4] b_1 b_2^{-2}  +  3 [3] b_1 b_2  + 2 [2] b_1 b_2^{-2}  +  2 [1] b_1 b_2  
\nn\\
&&
\hspace{1cm}
+ b_1 b_2^{-2} ) t^{13}
- (2 [5] b_1^{2} b_2^{-1}  + [4] b_1^{2} b_2^{-4}  +  3 [3] b_1^{2} b_2^{-1}  + 3 [2] b_1^{2} b_2^{2}  +  2 [1] b_1^{2} b_2^{-1}  + b_1^2 b_2^2   
\nn\\
&&
\hspace{1cm}
+ b_1^2 b_2^{-4}  ) t^{14} 
+ \dots ~,~
\eea
where $[m]=[m]_{SU(2)_x}$. From the plethystic logarithm, we see that the mesonic moduli space is a non-complete intersection.

Using the following fugacity map
\beal{es0827}
&&
\tilde{t} = t_5^{1/2} t_6^{1/2}~,~ 
\tilde{x} = \frac{t_5 t_6}{t_2^2}~,~ 
\tilde{b_1} =\frac{t_2 t_5^{1/2}}{t_4 t_6^{1/2}}~,~ 
\tilde{b_2} = \frac{t_2 t_4^2}{t_5 t_6^2}
~,~
\eea
where $t_1 t_2 = t_3 t_4 = t_5 t_6$,
the mesonic flavor charges on the gauge invariant operators become $\mathbb{Z}$-valued.
The generators in terms of brick matchings and their corresponding rescaled mesonic flavor charges are summarized in \tref{f_genlattice_08}.
The generator lattice as shown in \tref{f_genlattice_08} is a convex lattice polytope, which is reflexive. It is the dual of the toric diagram of Model 8 shown in \fref{f_toric_08}.
For completeness, \tref{f_genfields_08a} and \tref{f_genfields_08b} show the generators of Model 8 in terms of chiral fields with the corresponding mesonic flavor charges.
\\

\begin{table}[H]
\centering
\resizebox{0.95\hsize}{!}{

}
\caption{The generators in terms of bifundamental chiral fields for Model 8 \bf{(Part 1)}.
\label{f_genfields_08a}}
\end{table}

\begin{table}[H]
\centering
\resizebox{0.95\hsize}{!}{
%
}
\caption{The generators in terms of bifundamental chiral fields for Model 8  \bf{(Part 2)}.
\label{f_genfields_08b}}
\end{table}

\vspace{1cm}

\section{Model 9: $Y^{1,2}(\mathbb{CP}^1\times\mathbb{CP}^1)$~[$\mathbb{P}(\mathcal{O}_{\mathbb{P}^1\times \mathbb{P}^1} \oplus \mathcal{O}_{\mathbb{P}^1 \times\mathbb{P}^1}(1,1))$,~$\langle27\rangle$] \label{smodel09}}
 
\begin{figure}[H]
\begin{center}
\resizebox{0.3\hsize}{!}{
\includegraphics[height=6cm]{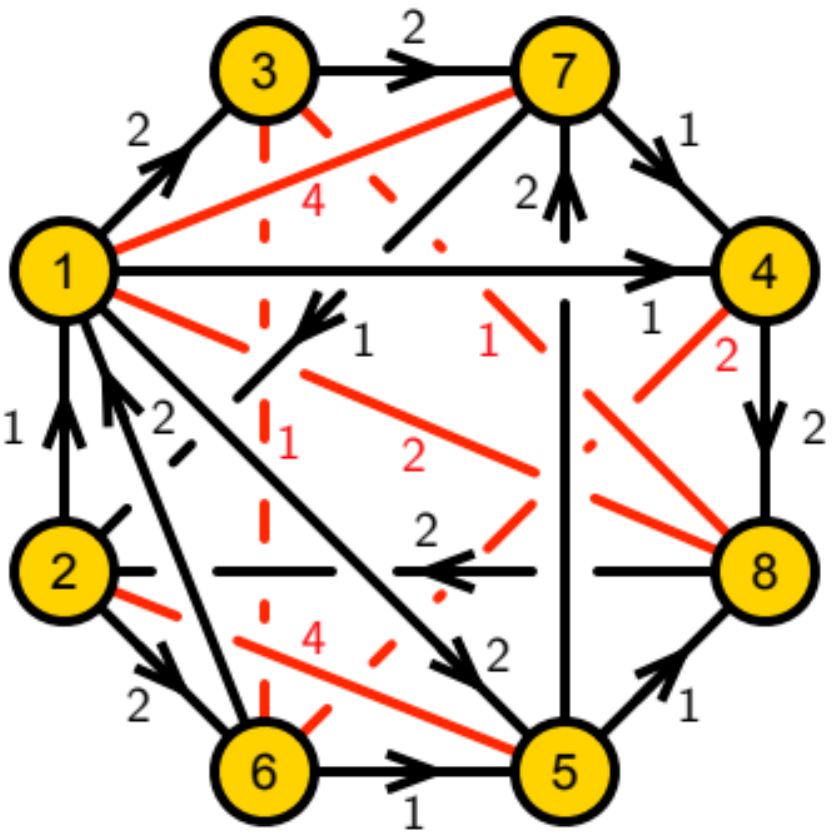} 
}
\caption{
Quiver for Model 9.
\label{f_quiver_09}}
 \end{center}
 \end{figure}
 
Model 9 corresponds to the $Y^{p,k}(\mathbb{CP}^1\times\mathbb{CP}^1))$ class of toric Calabi-Yau 4-folds. 
It is $Y^{1,2}(\mathbb{CP}^1\times\mathbb{CP}^1)$ and the corresponding brane brick model has the quiver in \fref{f_quiver_09}.
The $J$- and $E$-terms are 
\beq
{\footnotesize
 
 }~.~
\label{es0901}
 \eeq

Following the forward algorithm, we obtain the brick matchings for Model 9.
The brick matchings are summarized in the $P$-matrix, which takes the form
\beal{es0905}
{\footnotesize
 P=
\resizebox{0.75\hsize}{!}{$
\left( 
%
\right)
$}
}~.~
\eea

The $J$- and $E$-term charges are given by
\beal{es0905}
{\footnotesize
Q_{JE} =
\resizebox{0.72\hsize}{!}{$
\left(
%
\right)
$}
}~,~
\eea
and the $D$-term charges are given by
\beal{es0906}
{\footnotesize
Q_D =
\resizebox{0.72\hsize}{!}{$
\left(
%
\right)
$}
}~.~
\eea

\begin{figure}[H]
\begin{center}
\resizebox{0.4\hsize}{!}{
\includegraphics[trim=0mm 10mm 0mm 0mm, height=6cm]{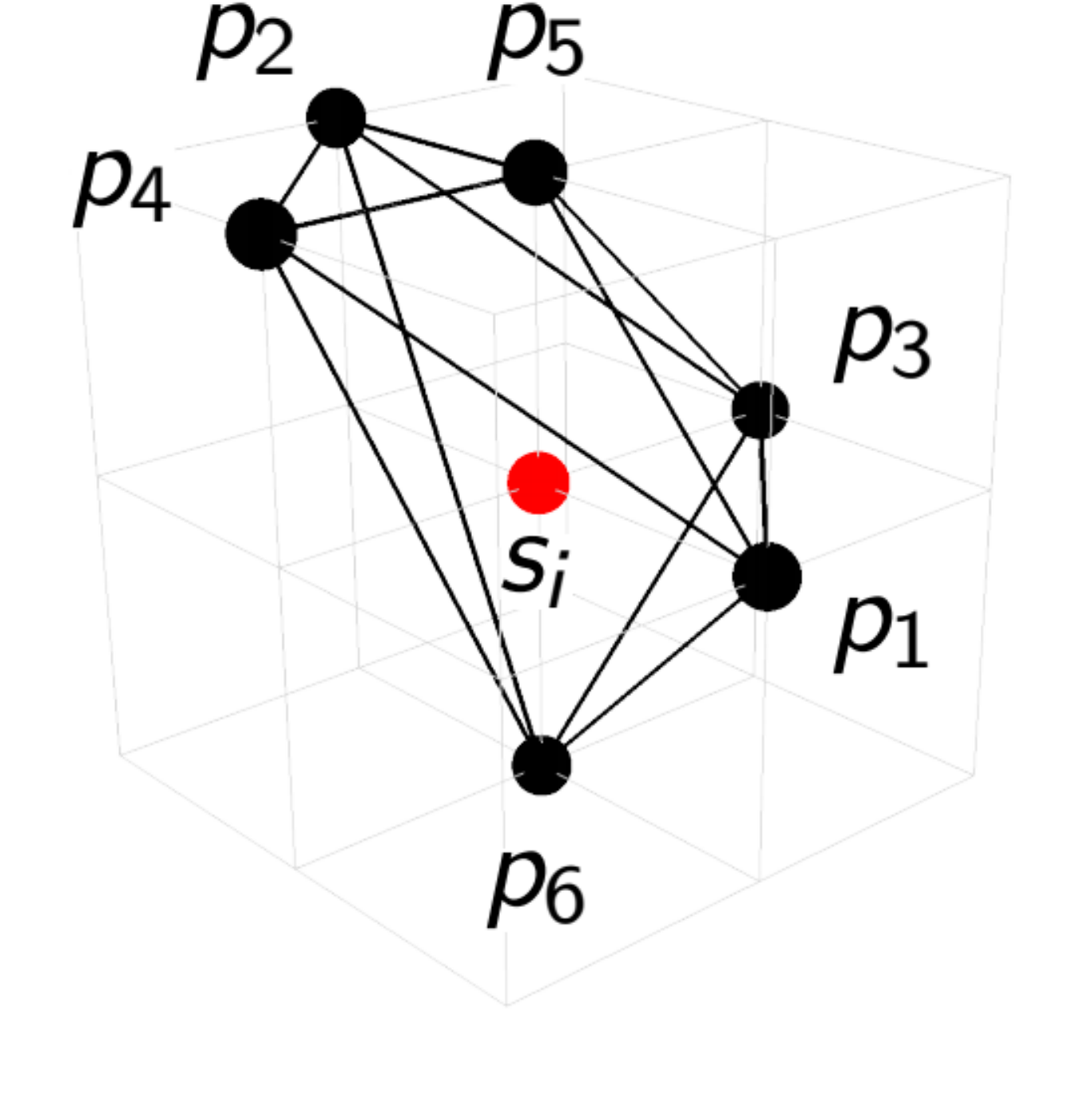} 
}
\caption{
Toric diagram for Model 9.
\label{f_toric_09}}
 \end{center}
 \end{figure}

\begin{table}[H]
\centering
%
\caption{Global symmetry charges on the extremal brick matchings $p_i$ of Model 9.}
\label{t_pmcharges_09}
\end{table}

The toric diagram of Model 9 is given by
\beal{es0910}
{\footnotesize 
G_t=
\resizebox{0.72\hsize}{!}{$
\left(
%
\right)
$}
}~,~
\eea
where \fref{f_toric_09} shows the toric diagram with brick matching labels.

The Hilbert series of the mesonic moduli space of Model 9 takes the form
\beal{es0920}
&&
g_1(t_i,y_s,y_{o_1},y_{o_2} ; \mathcal{M}_9) =
\frac{P(t_i,y_s,y_{o_1},y_{o_2}; \mathcal{M}_9)
}{
(1 - y_{s} y_{o_1}^3 y_{o_2}^4 t_1^3 t_3^3 t_5^2) 
(1 -    y_{s} y_{o_1}^3 y_{o_2}^4 t_2^3 t_3^3 t_5^2) 
}
\nn\\
&&
\hspace{1cm} 
\times
\frac{1}{
(1 -    y_{s} y_{o_1}^3 y_{o_2}^4 t_1^3 t_4^3 t_5^2) 
(1 -    y_{s} y_{o_1}^3 y_{o_2}^4 t_2^3 t_4^3 t_5^2) 
(1 - y_{s} y_{o_1} y_{o_2}^2 t_1 t_3 t_6^2) 
} 
\nn\\
&&
\hspace{1cm} 
\times
\frac{1}{
(1 -    y_{s} y_{o_1} y_{o_2}^2 t_2 t_3 t_6^2) 
(1 - y_{s} y_{o_1} y_{o_2}^2 t_1 t_4 t_6^2) 
(1 -    y_{s} y_{o_1} y_{o_2}^2 t_2 t_4 t_6^2)
} ~,~\nn\\
\eea
\begin{table}[H]
\centering
\resizebox{.95\hsize}{!}{
\begin{minipage}[!b]{0.5\textwidth}
\begin{tabular}{|c|c|c|c|}
\hline
generator & $SU(2)_{\tilde{x}}$ & $SU(2)_{\tilde{y}}$ & $U(1)_{\tilde{b}}$ \\
\hline
$ p_1 p_3 p_6^2 ~s o_1 o_2^2  $ & 1& 1& -1\\ 
$ p_2 p_3 p_6^2 ~s o_1 o_2^2  $ & 0& 1& -1\\ 
$ p_1 p_4 p_6^2 ~s o_1 o_2^2  $ & 1& 0& -1\\ 
$ p_2 p_4 p_6^2 ~s o_1 o_2^2  $ & 0& 0& -1\\ 
$ p_1^2 p_3^2 p_5 p_6 ~s o_1^2 o_2^3  $ & 1& 1& 0\\ 
$ p_1 p_2 p_3^2 p_5 p_6 ~s o_1^2 o_2^3  $ & 0& 1& 0\\ 
$ p_2^2 p_3^2 p_5 p_6 ~s o_1^2 o_2^3  $ & -1& 1& 0\\ 
$ p_1^2 p_3 p_4 p_5 p_6 ~s o_1^2 o_2^3  $ & 1& 0& 0\\ 
$ p_1 p_2 p_3 p_4 p_5 p_6 ~s o_1^2 o_2^3  $ & 0& 0& 0\\ 
$ p_2^2 p_3 p_4 p_5 p_6 ~s o_1^2 o_2^3  $ & -1& 0& 0\\ 
$ p_1^2 p_4^2 p_5 p_6 ~s o_1^2 o_2^3  $ & 1& -1& 0\\ 
$ p_1 p_2 p_4^2 p_5 p_6 ~s o_1^2 o_2^3  $ & 0& -1& 0\\ 
$ p_2^2 p_4^2 p_5 p_6 ~s o_1^2 o_2^3  $ & -1& -1& 0\\ 
$ p_1^3 p_3^3 p_5^2 ~s o_1^3 o_2^4  $ & 1& 1& 1\\ 
$ p_1^2 p_2 p_3^3 p_5^2 ~s o_1^3 o_2^4  $ & 0& 1& 1\\ 
$ p_1 p_2^2 p_3^3 p_5^2 ~s o_1^3 o_2^4  $ & -1& 1& 1\\ 
$ p_2^3 p_3^3 p_5^2 ~s o_1^3 o_2^4  $ & -2& 1& 1\\ 
$ p_1^3 p_3^2 p_4 p_5^2 ~s o_1^3 o_2^4  $ & 1& 0& 1\\ 
$ p_1^2 p_2 p_3^2 p_4 p_5^2 ~s o_1^3 o_2^4  $ & 0& 0& 1\\ 
$ p_1 p_2^2 p_3^2 p_4 p_5^2 ~s o_1^3 o_2^4  $ & -1& 0& 1\\ 
$ p_2^3 p_3^2 p_4 p_5^2 ~s o_1^3 o_2^4  $ & -2& 0& 1\\ 
$ p_1^3 p_3 p_4^2 p_5^2 ~s o_1^3 o_2^4  $ & 1& -1& 1\\ 
$ p_1^2 p_2 p_3 p_4^2 p_5^2 ~s o_1^3 o_2^4  $ & 0& -1& 1\\ 
$ p_1 p_2^2 p_3 p_4^2 p_5^2 ~s o_1^3 o_2^4  $ & -1& -1& 1\\ 
$ p_2^3 p_3 p_4^2 p_5^2 ~s o_1^3 o_2^4  $ & -2& -1& 1\\ 
$ p_1^3 p_4^3 p_5^2 ~s o_1^3 o_2^4  $ & 1& -2& 1\\ 
$ p_1^2 p_2 p_4^3 p_5^2 ~s o_1^3 o_2^4  $ & 0& -2& 1\\ 
$ p_1 p_2^2 p_4^3 p_5^2 ~s o_1^3 o_2^4  $ & -1& -2& 1\\ 
$ p_2^3 p_4^3 p_5^2 ~s o_1^3 o_2^4  $ & -2& -2& 1\\ 
\hline
\end{tabular}
\end{minipage}
\hspace{1.5cm}
\begin{minipage}[!b]{0.4\textwidth}
\includegraphics[height=5.5cm]{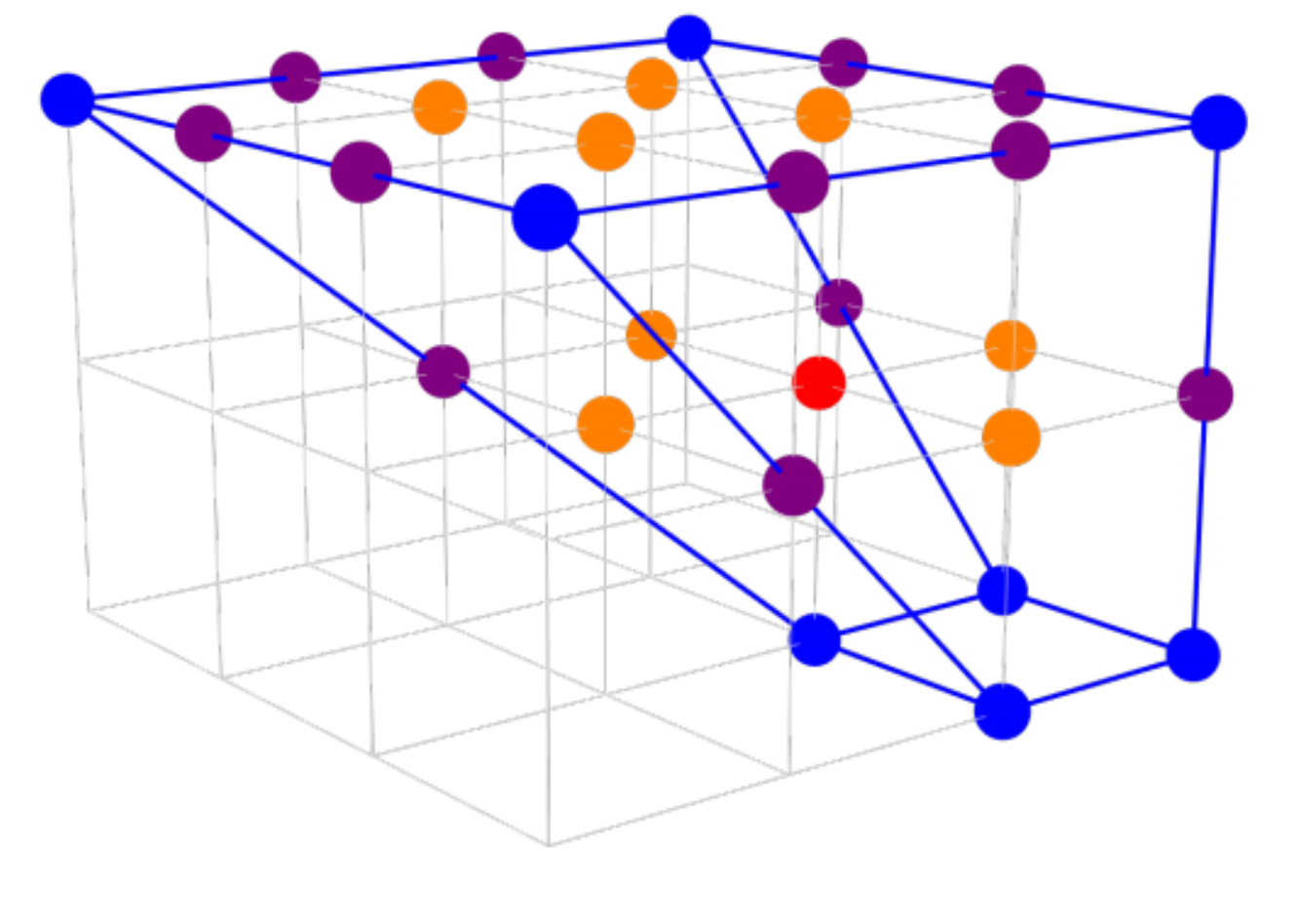} 
\end{minipage}
}
\caption{The generators and lattice of generators of the mesonic moduli space of Model 9 in terms of brick matchings with the corresponding flavor charges.
\label{f_genlattice_09}}
\end{table}
\noindent
where $t_i$ are the fugacities for the extremal brick matchings $p_i$.
$y_{s}$ counts the brick matching product $s_1 \dots s_{9}$ corresponding to the single internal point of the toric diagram of Model 9.
Additionally, $y_{o_1}$ and $y_{o_2}$ count the products of extra GLSM fields $o_1 \dots o_9$ and $o_9 o_{10}$, respectively.
The explicit numerator $P(t_i,y_s,y_{o_1},y_{o_2}; \mathcal{M}_9)$ of the Hilbert series is given in the Appendix Section \sref{app_num_09}.
We note that setting the fugacities $y_{o_1}=1$ and $y_{o_2}=1$ does not change the overall characterization of the mesonic moduli space by the Hilbert series, indicating that the extra GLSM fields, as expected, correspond to an over-parameterization of the moduli space. 

By setting $t_i=t$ for the fugacities of the extremal brick matchings, and all other fugacities to $y_s=1$, $y_{o_1}=1$ and $y_{o_2}=1$, the unrefined Hilbert series takes the following form
\beal{es0921}
&&
g_1(t,1,1,1; \mathcal{M}_9) =
\frac{
1
}{
(1 - t^2) (1 - t^8)^3
}
\times
(
1 - t^2 + 4 t^4 + 5 t^6 + 13 t^8 - 6 t^{10} + 13 t^{12} 
\nn\\
&&
\hspace{1cm}
+ 5 t^{14} + 4 t^{16} - t^{18} + t^{20}
 )
~,~
\eea
where the palindromic numerator indicates that the mesonic moduli space is Calabi-Yau. 

The global symmetry of Model 9 and the charges on the extremal brick matchings under the global symmetry are summarized in \tref{t_pmcharges_09}.
We can use the following fugacity map,
\beal{es0922}
&&
t = t_1^{1/2} t_2^{1/2} ~,~
x = \frac{t_1^{1/2}}{t_2^{1/2}} ~,~
y = \frac{t_3}{t_1^{1/2} t_2^{1/2}} ~,~
b = \frac{t_1^{1/2} t_2^{1/2}}{t_6} ~,~
\eea
where $t_3 t_4 = t_5 t_6 $ and $t_1 t_2 =t_5 t_6$, in order to rewrite the 
Hilbert series for Model 8 in terms of characters of irreducible representations of $SU(2)\times SU(2) \times U(1)$.
In highest weight form, the Hilbert series of Model 9 can be written as
\beal{es0925}
&&
h_1(t, \mu_1,\mu_2, b; \mathcal{M}_9) =
\frac{
1 + \mu_1^2  \mu_2^2 t^6
}{
(1 - \mu_1 \mu_2 b^{-2} t^4) (1 - \mu_1^3 \mu_2^3 b^2 t^8)
} ~,~
\eea
where $\mu_1^m \mu_2^n \sim [m]_{SU(2)_x} [n]_{SU(2)_y}$. Here in highest weight form, the fugacities $\mu_1$ and $\mu_2$ count the highest weight of irreducible representations of $SU(2)_x \times SU(2)_y$.

The plethystic logarithm of the Hilbert series takes the form
\beal{es0925}
&&
g_1(t, x,y, b; \mathcal{M}_9) =
[1; 1] b^{-2} t^4
+ [2; 2] t^6
+ [3; 3] b^2  t^8
- b^{-4} t^8
\nn\\
&&
\hspace{1cm}
- ([1; 3] b^{-2} + [3; 1] b^{-2} + [1; 1] b^{-2}) t^{10}
- ([4; 4] + [2; 4] + [4; 2] + [0; 4] 
\nn\\
&&
\hspace{1cm}
+ [4; 0] + 2 [2; 2] + 1) t^{12}
- [3; 5] b^2 t^{14}
+ \dots ~,~
\eea
where $[m; n] = [m]_{SU(2)_x} [n]_{SU(2)_y}$.
From the plethystic logarithm, we see that the mesonic moduli space is a non-complete intersection.

The generators form 3 sets that transform under 
$[1; 1] b^{-2}$, 
$[2; 2]$ and $[3; 3] b^2$ of the mesonic flavor symmetry of Model 9, respectively.
Using the following fugacity map
\beal{es0927}
&&
\tilde{t} = t_5^{1/2} t_6^{1/2}~,~ 
\tilde{x} = \frac{t_1}{t_2}~,~ 
\tilde{y} =\frac{t_3}{t_4}~,~ 
\tilde{b} = \frac{t_5^2}{t_2 t_4}
~,~
\eea
where $t_1 t_2 = t_3 t_4 = t_5 t_6$,
the mesonic flavor charges on the gauge invariant operators become $\mathbb{Z}$-valued.
The generators in terms of brick matchings and their corresponding rescaled mesonic flavor charges are summarized in \tref{f_genlattice_09}.
The generator lattice as shown in \tref{f_genlattice_09} is a convex lattice polytope, which is reflexive. It is the dual of the toric diagram of Model 9 shown in \fref{f_toric_09}.
We also note that the 3 layers of points in the generator lattice in \tref{f_genlattice_09} corresponds to the 3 sets that transform under 
$[1; 1] b^{-2}$, $[2; 2]$ and $[3; 3] b^2$ of the mesonic flavor symmetry.
For completeness, \tref{f_genfields_09a} and \tref{f_genfields_09b} show the generators of Model 9 in terms of chiral fields with the corresponding mesonic flavor charges.
\\

\begin{table}[H]
\centering
\resizebox{0.9\hsize}{!}{

}
\caption{The generators in terms of bifundamental chiral fields for Model 9 \bf{(Part 1)}.
\label{f_genfields_09a}}
\end{table}

\begin{table}[H]
\centering
\resizebox{0.9\hsize}{!}{
%
}
\caption{The generators in terms of bifundamental chiral fields for Model 9 \bf{(Part 2)}.
\label{f_genfields_09b}}
\end{table}
\vspace{-0.2cm}

\section{Model 10: $P^{3}_{+-}(\text{dP}_1)$~[$\mathbb{P}^1$-blowup of $3$,~$\langle28\rangle$] \label{smodel10}}
 
\begin{figure}[H]
\begin{center}
\resizebox{0.3\hsize}{!}{
\includegraphics[height=6cm]{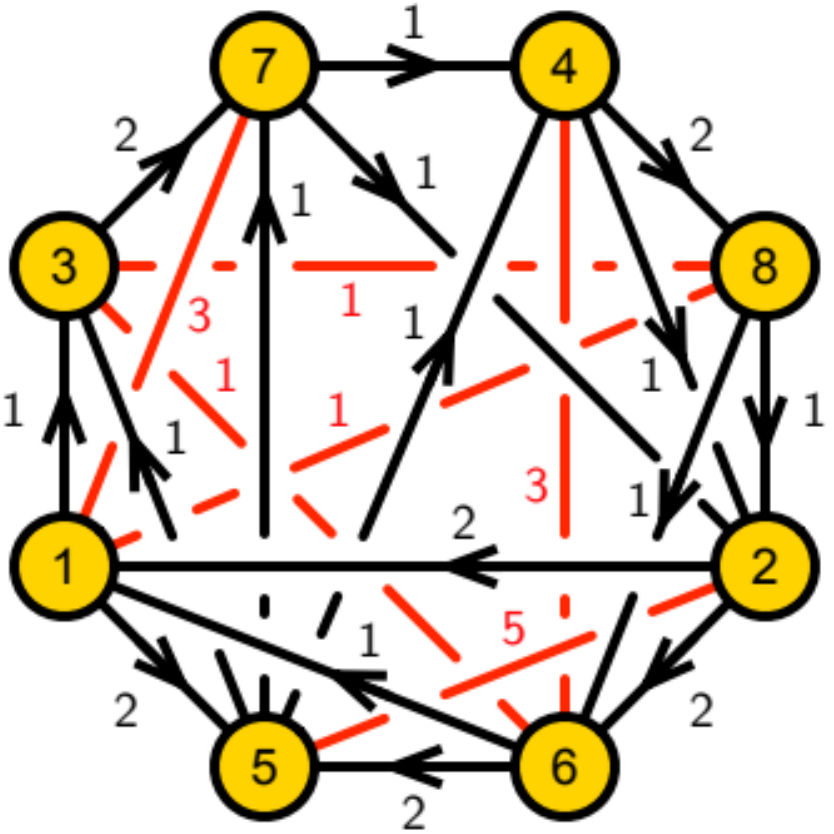} 
}
\caption{
Quiver for Model 10.
\label{f_quiver_10}}
 \end{center}
 \end{figure}
 
Model 10 corresponds to the toric Calabi-Yau 4-fold $P^{3}_{+-}(\text{dP}_1)$. 
The corresponding brane brick model has the quiver in \fref{f_quiver_10} and the $J$- and $E$-terms are 

\beq
{\footnotesize
 
 }~.~
\label{E_J_C_+-}
 \eeq

Following the forward algorithm, we obtain the brick matchings for Model 10.
The brick matchings are summarized in the $P$-matrix, which takes the form
\beal{es1005}
{\footnotesize 
P=
\resizebox{0.7\textwidth}{!}{$
\left( 
%
\right)
$}
}~.~
\eea

The $J$- and $E$-term charges are given by
\beal{es1005}
{\footnotesize
Q_{JE} =
\resizebox{0.67\textwidth}{!}{$
\left(
%
\right)
$}
}~,~
\eea
and the $D$-term charges are given by
\beal{es1005}
{\footnotesize
Q_D =
\resizebox{0.67\textwidth}{!}{$
\left(
%
\right)
$}
}~.~
\eea

\begin{figure}[H]
\begin{center}
\resizebox{0.4\hsize}{!}{
\includegraphics[height=6cm]{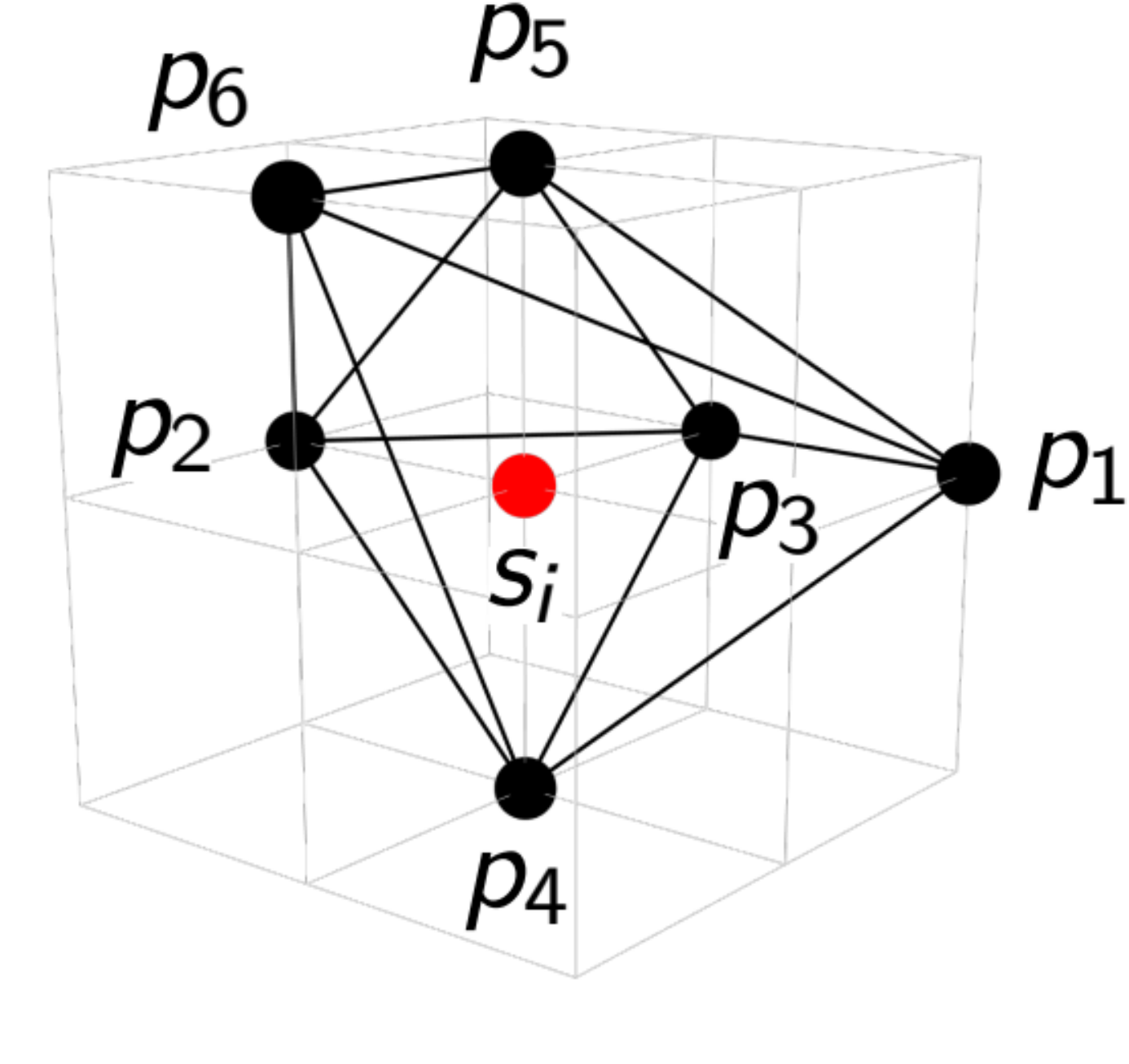} 
}
\caption{
Toric diagram for Model 10.
\label{f_toric_10}}
 \end{center}
 \end{figure}

The toric diagram of Model 10 is given by
\beal{es1010}
{\footnotesize 
G_t=
\resizebox{0.67\textwidth}{!}{$
\left(
%
\right)
$}
}~,~
\eea
where \fref{f_toric_10} shows the toric diagram with brick matching labels.

\begin{table}[H]
\centering
%
\caption{Global symmetry charges on the extremal brick matchings $p_i$ of Model 10.}
\label{t_pmcharges_10}
\end{table}

The Hilbert series of the mesonic moduli space of Model 10 takes the form
\beal{es1020}
&&
g_1(t_i,y_s,y_{o_1},y_{o_2} ; \mathcal{M}_{10}) =
\frac{P(t_i,y_s,y_{o_1},y_{o_2}; \mathcal{M}_{10})
}{
(1 - y_{s} y_{o_1}^2 y_{o_2} t_1^2 t_3 t_4^2) 
(1 - y_{s} y_{o_1}^2 y_{o_2} t_2^2 t_3 t_4^2) 
}
\nn\\
&&
\hspace{1cm} 
\times
\frac{1}{
(1 -  y_{s} y_{o_1}^4 y_{o_2}^3 t_1^4 t_3^3 t_5^2) 
(1 -  y_{s} y_{o_1}^4 y_{o_2}^3 t_2^4 t_3^3 t_5^2) 
(1 - y_{s} y_{o_1} y_{o_2} t_1 t_4^2 t_6) 
} 
\nn\\
&&
\hspace{1cm} 
\times
\frac{1}{
(1 -  y_{s} y_{o_1} y_{o_2} t_2 t_4^2 t_6) 
(1 - y_{s} y_{o_1} y_{o_2}^3 t_1 t_5^2 t_6^3) 
(1 -  y_{s} y_{o_1} y_{o_2}^3 t_2 t_5^2 t_6^3)
} ~,~
\eea
where $t_i$ are the fugacities for the extremal brick matchings $p_i$.
$y_{s}$ counts the brick matching product $s_1 \dots s_{8}$ corresponding to the single internal point of the toric diagram of Model 10.
Additionally, $y_{o_1}$ and $y_{o_2}$ count the products of extra GLSM fields $o_1 \dots o_5$ and $o_6\dots o_{9}$, respectively.
The explicit numerator $P(t_i,y_s,y_{o_1},y_{o_2}; \mathcal{M}_{10})$ of the Hilbert series is given in the Appendix Section \sref{app_num_10}.

By setting $t_i=t$ for the fugacities of the extremal brick matchings, and all other fugacities to $y_s=1$, $y_{o_1}=1$ and $y_{o_2}=1$, the unrefined Hilbert series takes the following form
\beal{es1021}
&&
g_1(t,1,1,1; \mathcal{M}_{10}) =
\frac{
(1 - t^2)^2 (1 - t^3)^2
}{
(1 - t^4)^2 (1 - t^5)^2 (1 - t^6)^2 (1 - t^9)^2
}
\times
(
1 + 2 t^{2} + 2 t^{3} + 3 t^{4} 
\nn\\
&&
\hspace{1cm}
+ 7 t^{5} + 10 t^{6} + 19 t^{7} + 27 t^{8} + 41 t^{9} + 51 t^{10} + 62 t^{11} + 76 t^{12} + 73 t^{13} + 85 t^{14} 
\nn\\
&&
\hspace{1cm}
+ 85 t^{15} + 81 t^{16} + 85 t^{17} + 85 t^{18} + 73 t^{19} + 76 t^{20} + 62 t^{21} + 51 t^{22} + 41 t^{23} 
\nn\\
&&
\hspace{1cm}
+ 27 t^{24} + 19 t^{25} + 10 t^{26} + 7 t^{27} + 3 t^{28} + 2 t^{29} + 2 t^{30} + t^{32}
)
~,~
\eea
where the palindromic numerator indicates that the mesonic moduli space is Calabi-Yau. 

The global symmetry of Model 10 and the charges on the extremal brick matchings under the global symmetry are summarized in \tref{t_pmcharges_10}.
We can use the following fugacity map,
\beal{es1022}
&&
t = t_1^{1/2} t_2^{1/2} ~,~
x = \frac{t_1^{1/2}}{t_2^{1/2}} ~,~
y = \frac{t_1^{1/2} t_2^{1/2}}{t_4} ~,~
b = \frac{t_1^{1/2} t_2^{1/2}}{t_6} ~,~
\eea
where $t_3 t_4 = t_5 t_6$ and $t_1 t_2 =t_5 t_6$, in order to rewrite the 
Hilbert series for Model 10 in terms of characters of irreducible representations of $SU(2)\times U(1) \times U(1)$.

\begin{table}[H]
\centering
\resizebox{.95\hsize}{!}{
\begin{minipage}[!b]{0.5\textwidth}
\begin{tabular}{|c|c|c|c|}
\hline
generator & $SU(2)_{\tilde{x}}$ & $U(1)_{\tilde{b_1}}$ & $U(1)_{\tilde{b_2}}$ \\
\hline
$ p_2^2 p_3 p_4^2  ~s o_1^2 o_2  $ & -1& -1& 0 \\ 
$ p_2 p_4^2 p_6  ~s o_1 o_2  $ & 0& -1& -1 \\ 
$ p_2^3 p_3^2 p_4 p_5  ~s o_1^3 o_2^2  $ & -2& 0& 1 \\ 
$ p_2^2 p_3 p_4 p_5 p_6  ~s o_1^2 o_2^2  $ & -1& 0& 0 \\ 
$ p_2 p_4 p_5 p_6^2  ~s o_1 o_2^2  $ & 0& 0& -1 \\ 
$ p_1 p_2 p_3 p_4^2  ~s o_1^2 o_2  $ & 0& -1& 0 \\ 
$ p_1 p_4^2 p_6  ~s o_1 o_2  $ & 1& -1& -1 \\ 
$ p_2^4 p_3^3 p_5^2  ~s o_1^4 o_2^3  $ & -3& 1& 2 \\ 
$ p_2^3 p_3^2 p_5^2 p_6  ~s o_1^3 o_2^3  $ & -2& 1& 1 \\ 
$ p_2^2 p_3 p_5^2 p_6^2  ~s o_1^2 o_2^3  $ & -1& 1& 0 \\ 
$ p_1 p_2^2 p_3^2 p_4 p_5  ~s o_1^3 o_2^2  $ & -1& 0& 1 \\ 
$ p_2 p_5^2 p_6^3  ~s o_1 o_2^3  $ & 0& 1& -1 \\ 
$ p_1 p_2 p_3 p_4 p_5 p_6  ~s o_1^2 o_2^2  $ & 0& 0& 0 \\ 
$ p_1 p_4 p_5 p_6^2  ~s o_1 o_2^2  $ & 1& 0& -1 \\ 
$ p_1^2 p_3 p_4^2  ~s o_1^2 o_2  $ & 1& -1& 0 \\ 
$ p_1 p_2^3 p_3^3 p_5^2  ~s o_1^4 o_2^3  $ & -2& 1& 2 \\ 
$ p_1 p_2^2 p_3^2 p_5^2 p_6  ~s o_1^3 o_2^3  $ & -1& 1& 1 \\ 
$ p_1 p_2 p_3 p_5^2 p_6^2  ~s o_1^2 o_2^3  $ & 0& 1& 0 \\ 
$ p_1^2 p_2 p_3^2 p_4 p_5  ~s o_1^3 o_2^2  $ & 0& 0& 1 \\ 
$ p_1 p_5^2 p_6^3  ~s o_1 o_2^3  $ & 1& 1& -1 \\ 
$ p_1^2 p_3 p_4 p_5 p_6  ~s o_1^2 o_2^2  $ & 1& 0& 0 \\ 
$ p_1^2 p_2^2 p_3^3 p_5^2  ~s o_1^4 o_2^3  $ & -1& 1& 2 \\ 
$ p_1^2 p_2 p_3^2 p_5^2 p_6  ~s o_1^3 o_2^3  $ & 0& 1& 1 \\ 
$ p_1^2 p_3 p_5^2 p_6^2  ~s o_1^2 o_2^3  $ & 1& 1& 0 \\ 
$ p_1^3 p_3^2 p_4 p_5  ~s o_1^3 o_2^2  $ & 1& 0& 1 \\ 
$ p_1^3 p_2 p_3^3 p_5^2  ~s o_1^4 o_2^3  $ & 0& 1& 2 \\ 
$ p_1^3 p_3^2 p_5^2 p_6  ~s o_1^3 o_2^3  $ & 1& 1& 1 \\ 
$ p_1^4 p_3^3 p_5^2  ~s o_1^4 o_2^3  $ & 1& 1& 2 \\
\hline
\end{tabular}
\end{minipage}
\hspace{1.5cm}
\begin{minipage}[!b]{0.4\textwidth}
\includegraphics[height=7cm]{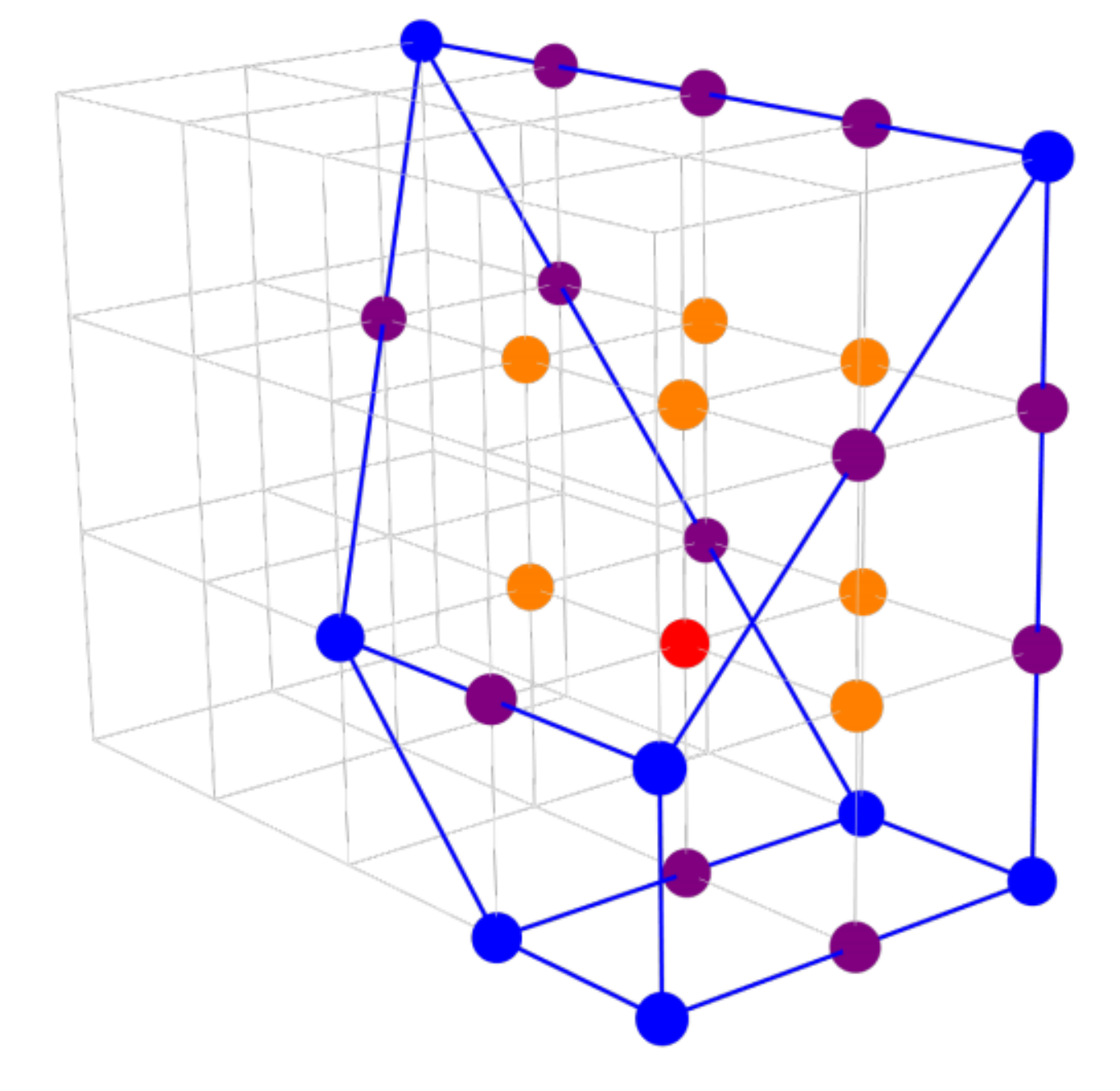} 
\end{minipage}
}
\caption{The generators and lattice of generators of the mesonic moduli space of Model 10 in terms of brick matchings with the corresponding flavor charges.
\label{f_genlattice_10}}
\end{table}

In highest weight form, the Hilbert series of Model 10 can be written as
\beal{es1025}
&&
h_1(t, \mu_1 ,\mu_2, b; \mathcal{M}_{10}) =
\frac{1}{
(1 - \mu b_1^{-2} b_2^{-1} t^4) (1 - \mu^2 b_1^{-1} t^5) (1 - \mu b_2^{-1} t^6) (1 - \mu^4 b_1^3 b_2^2 t^9)
}
\nn\\
&&
\hspace{0.5cm}
\times
(
1 + \mu b_1^{-1} b_2^{-1} t^5 + \mu^2 t^6 + \mu^2 b_1 t^7 + \mu^3 b_1 b_2 t^7 +  \mu^3 b_1^2 b_2 t^8 - \mu^3 b_1^{-2} b_2^{-1} t^{10} - \mu^4 b_1^{-1} t^{11} 
\nn\\
&&
\hspace{1cm}
- \mu^3 b_1^{-1} b_2^{-1} t^{11} - \mu^4 t^{12} - \mu^5 b_1 b_2 t^{13} - \mu^6 t^{18}
)~,~
\eea
where $\mu^m \sim [m]_{SU(2)_x}$. Here in highest weight form, the fugacity $\mu$ counts the highest weight of irreducible representations of $SU(2)_x$.

The plethystic logarithm of the Hilbert series takes the form
\beal{es1026}
&&
\PL[g_1(t, x, b_1, b_2; \mathcal{M}_{10})]=
[1] b_1^{-2} b_2^{-1} t^4
+ ([2] b_1^{-1} + [1] b_1^{-1} b_2^{-1}) t^5
+ ([2] + [1] b_2^{-1} )t^6
\nn\\
&&
\hspace{1cm}
+ ([3] b_1 b_2 + [2] b_1 )t^7
+ [3] b_1^{2} b_2 t^8
+ [4] b_1^{3} b_2^{2} t^9 
\nn\\
&&
\hspace{1cm}
- ([1] b_1^{-3} b_2^{-1}  + b_1^{-3} b_2^{-2}) t^9
- ([3] b_1^{-2} b_2^{-1}  + [2] b_1^{-2} b_2^{-2}  + 2 [1] b_1^{-2} b_2^{-1}  + b_1^{-2} b_2^{-2}  
\nn\\
&&
\hspace{1cm}
+ b_1^{-2} ) t^{10}
- ([4] b_1^{-1}  + 2 [3] b_1^{-1} b_2^{-1}  +  2 [2] b_1^{-1}  + 3 [1] b_1^{-1} b_2^{-1}  +  b_1^{-1} b_2^{-2}  + b_1^{-1} ) t^{11}
\nn\\
&&
\hspace{1cm}
- (3 [4]  + [3] b2  + [3] b2^{-1}  +  3 [2]  + [1] b2  + 2 [1] b2^{-1}  +  2 ) t^{12}
- (2 [5] b_1 b_2  + 2 [4] b_1  
\nn\\
&&
\hspace{1cm}
+  3 [3] b_1 b_2  + 3 [2] b_1  +  2 [1] b_1 b_2  + [1] b_1 b_2^{-1}  +  b_1 ) t^{13}
- ([6] b_1^{2} b_2^{2}  + 2 [5] b_1^{2} b_2  +  [4] b_1^{2} b_2^{2}  
\nn\\
&&
\hspace{1cm}
+ [4] b_1^{2}  +  3 [3] b_1^{2} b_2  + 2 [2] b_1^{2} b_2^{2}  +  [2] b_1^{2}  + 2 [1] b_1^{2} b_2  + b_1^{2} ) t^{14}
+ \dots ~,~
\eea
where $[m] = [m]_{SU(2)}$.
From the plethystic logarithm, we see that the mesonic moduli space is a non-complete intersection.

Using the following fugacity map
\beal{es1027}
&&
\tilde{t} = t_5^{1/2} t_6^{1/2}~,~ 
\tilde{x} = \frac{t_1}{t_2}~,~ 
\tilde{b_1} =\frac{t_5^{1/2} t_6^{1/2}}{t_4}~,~ 
\tilde{b_2} = \frac{t_5^{3/2} t_6^{1/2}}{t_2 t_4}
~,~
\eea
where $t_1 t_2 = t_3 t_4 = t_5 t_6$,
the mesonic flavor charges on the gauge invariant operators become $\mathbb{Z}$-valued.
The generators in terms of brick matchings and their corresponding rescaled mesonic flavor charges are summarized in \tref{f_genlattice_10}.
The generator lattice as shown in \tref{f_genlattice_10} is a convex lattice polytope, which is reflexive. It is the dual of the toric diagram of Model 10 shown in \fref{f_toric_10}.
For completeness, \tref{f_genfields_10a} and \tref{f_genfields_10b} show the generators of Model 10 in terms of chiral fields with the corresponding mesonic flavor charges.

\begin{table}[H]
\centering
\resizebox{0.95\hsize}{!}{

}
\caption{The generators in terms of bifundamental chiral fields for Model 10 \bf{(Part 1)}.
\label{f_genfields_10a}}
\end{table}

\begin{table}[H]
\centering
\resizebox{0.95\hsize}{!}{
%
}
\caption{The generators in terms of bifundamental chiral fields for Model 10 \bf{(Part 2)}.
\label{f_genfields_10b}}
\end{table}

\section{Model 11: $P^{0}_{+-}(\text{dP}_1)$~[$\text{dP}_1 \times \mathbb{P}^1$,~$\langle29\rangle$] \label{smodel11}}

\begin{figure}[H]
\begin{center}
\resizebox{0.3\hsize}{!}{
\includegraphics[height=6cm]{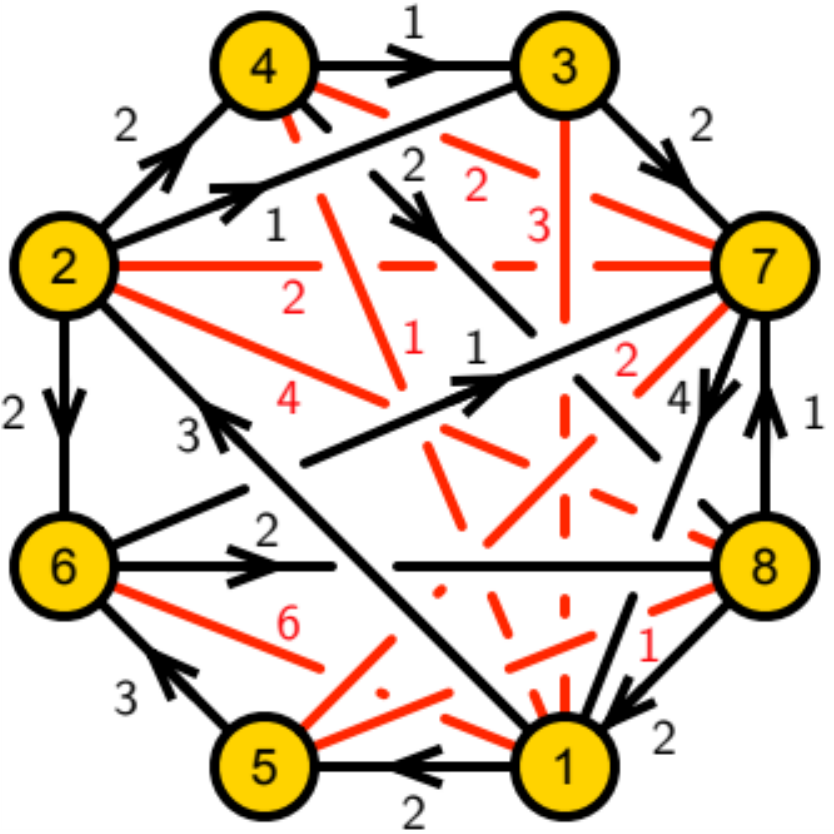} 
}
\caption{
Quiver for Model 11.
\label{f_quiver_11}}
 \end{center}
 \end{figure}

Model 11 corresponds to the toric Calabi-Yau 4-fold $P^{0}_{+-}(\text{dP}_1)$. The corresponding brane brick model has the quiver in \fref{f_quiver_11} and the $J$- and $E$-terms are 
\beq
{\footnotesize
 
 }
\label{E_J_C_+-}
 \eeq
 
Following the forward algorithm, we obtain the brick matchings for Model 11.
The brick matchings are summarized in the $P$-matrix, which takes the form
\beal{es1105}
{\footnotesize 
 P=
 \resizebox{0.77\hsize}{!}{$
\left( 
%
\right)
$}
}~.~
 \eea
 
The $J$- and $E$-term charges are given by
\beal{es1105}
{\footnotesize
Q_{JE} =
\resizebox{0.74\hsize}{!}{$
\left(
%
\right)
$}
}~,~
\nn\\
\eea
and the $D$-term charges are given by
\beal{es1105}
{\footnotesize
Q_D =
\resizebox{0.74\hsize}{!}{$
\left(
%
\right)
$}
}~.~
\nn\\
\eea

\begin{figure}[H]
\begin{center}
\resizebox{0.4\hsize}{!}{
\includegraphics[height=6cm]{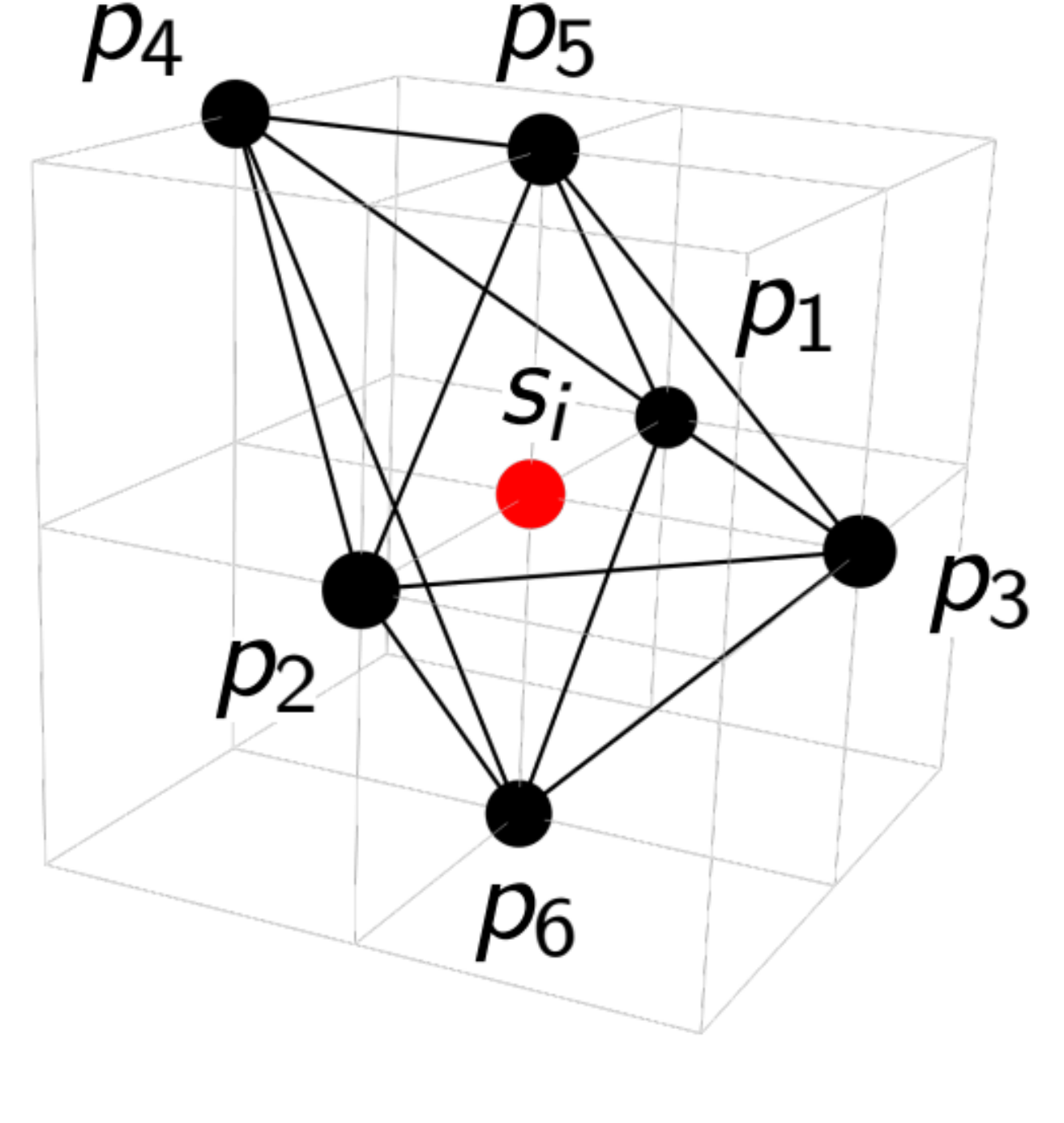} 
}
\caption{
Toric diagram for Model 11.
\label{f_toric_11}}
 \end{center}
 \end{figure}

The toric diagram of Model 11 is given by
\beal{es1110}
{\footnotesize 
G_t=
\resizebox{0.74\hsize}{!}{$
\left(
%
\right)
$}
}~,~
\nn\\
\eea
where \fref{f_toric_11} shows the toric diagram with brick matching labels.

The Hilbert series of the mesonic moduli space of Model 11 takes the form
\beal{es1120}
&&
g_1(t_i,y_s,y_{o} ; \mathcal{M}_{11}) =
\frac{P(t_i,y_s,y_{o}; \mathcal{M}_{11})
}{
(1 - y_{s} y_{o}^3 t_1^2 t_3^3 t_5^2) 
(1 - y_{s} y_{o}^3 t_2^2 t_3^3 t_5^2) 
(1 -  y_{s} y_{o}^3 t_1^2 t_4^3 t_5^2) 
(1 - y_{s} y_{o}^3 t_2^2 t_4^3 t_5^2) 
}
\nn\\
&&
\hspace{1cm} 
\times
\frac{1}{
(1 -  y_{s} y_{o} t_1^2 t_3 t_6^2) 
(1 - y_{s} y_{o} t_2^2 t_3 t_6^2) 
(1 -  y_{s} y_{o} t_1^2 t_4 t_6^2) 
(1 - y_{s} y_{o} t_2^2 t_4 t_6^2)
} 
~,~
\eea
where $t_i$ are the fugacities for the extremal brick matchings $p_i$.
$y_{s}$ counts the brick matching product $s_1 \dots s_{14}$ corresponding to the single internal point of the toric diagram of Model 11.
Additionally, $y_{o}$ counts the product of extra GLSM fields $o_1 \dots o_4$.
The explicit numerator $P(t_i,y_s,y_{o}; \mathcal{M}_{11})$ of the Hilbert series is given in the Appendix Section \sref{app_num_11}.
We note that setting the fugacity $y_{o}=1$ does not change the overall characterization of the mesonic moduli space by the Hilbert series, indicating that the extra GLSM fields, as expected, correspond to an over-parameterization of the moduli space. 

\begin{table}[H]
\centering
\begin{tabular}{|c|c|c|c|c|c|}
\hline
\; & $SU(2)_{x}$ & $SU(2)_{y}$ &  $U(1)_{b}$ & $U(1)$ & \text{fugacity} \\
\hline
$p_1$ & +1 & 0 & 0 & $r_1$ & $t_1$ \\
$p_2$ & -1 & 0 & 0 & $r_2$ & $t_2$ \\
$p_3$ & 0 & +1  & 0 & $r_3$ & $t_3$ \\
$p_4$ & 0 & -1 & 0 & $r_4$ & $t_4$ \\
$p_5$ & 0 &0 & +1 & $r_5$ & $t_5$ \\
$p_6$ & 0 &0 &  -1 & $r_6$ & $t_6$ \\
\hline
\end{tabular}
\caption{Global symmetry charges on the extremal brick matchings $p_i$ of Model 11.}
\label{t_pmcharges_11}
\end{table}

By setting $t_i=t$ for the fugacities of the extremal brick matchings, and all other fugacities to $y_s=1$ and $y_{o}=1$, the unrefined Hilbert series takes the following form
\beal{es1121}
&&
g_1(t,1,1,1; \mathcal{M}_{11}) =
\frac{
(1 - t)^2
}{
(1 - t^5)^3 (1 - t^7)^3
}
\times 
(
1 + 2 t + 3 t^{2} + 4 t^{3} + 5 t^{4} + 9 t^{5} + 22 t^{6} 
\nn\\
&&
\hspace{1cm}
+ 44 t^{7}  + 66 t^{8} +  88 t^{9} + 110 t^{10} + 125 t^{11} + 120 t^{12} + 118 t^{13} + 118 t^{14} +  118 t^{15} 
\nn\\
&&
\hspace{1cm}
+ 120 t^{16} + 125 t^{17} + 110 t^{18} + 88 t^{19} + 66 t^{20} +  44 t^{21} + 22 t^{22} + 9 t^{23} + 5 t^{24} 
\nn\\
&&
\hspace{1cm}
+ 4 t^{25} + 3 t^{26} + 2 t^{27} + t^{28}
)
~,~
\eea
where the palindromic numerator indicates that the mesonic moduli space is Calabi-Yau. 

The global symmetry of Model 11 and the charges on the extremal brick matchings under the global symmetry are summarized in \tref{t_pmcharges_11}.
We can use the following fugacity map,
\beal{es1122}
&&
t = t_1^{1/2} t_2^{1/2} ~,~
x = \frac{t_1^{1/2}}{t_2^{1/2}} ~,~
y = \frac{t_3^{1/2}}{t_4^{1/2}} ~,~
b = \frac{t_1^{1/2} t_2^{1/2}}{t_6} ~,~
\eea
where $t_3 t_4 = t_5 t_6$ and $t_1 t_2=t_5 t_6$, in order to rewrite the 
Hilbert series for Model 11 in terms of characters of irreducible representations of $SU(2)\times SU(2) \times U(1)$.
The character expansion of the Hilbert series is
\beal{es1124}
&&
g_1(t, x, y, b; \mathcal{M}_{11}) =
\sum_{n_1=0}^{\infty}
\sum_{n_2=0}^{\infty}
\Big[
[2n_1+ 2n_2; n_1+2n_2] b^{-2n_1} t^{5n_1 +6n_2}
\nn\\
&& 
\hspace{3cm}
+ [2n_1+ 2n_2+2; 2n_1+3n_2+3] b^{2n_1+2} t^{6n_1 +7n_2+7}
\Big]
~,~
\eea
where $[m;n]=[m]_{SU(2)_x} [n]_{SU(2)_y}$.

\begin{table}[H]
\centering
\resizebox{.9\hsize}{!}{
\begin{minipage}[!b]{0.5\textwidth}
\begin{tabular}{|c|c|c|c|}
\hline
generator & $SU(2)_{\tilde{x}}$ & $SU(2)_{\tilde{y}}$ & $U(1)_{\tilde{b}}$ \\
\hline
$p_1^2 p_3 p_6^2 ~s o  $ &1& 0& -1\\
$p_1 p_2 p_3 p_6^2 ~s o  $ &0& 0& -1\\
$p_2^2 p_3 p_6^2 ~s o  $ &-1& 0& -1\\
$p_1^2 p_4 p_6^2 ~s o  $ &1& -1& -1\\
$p_1 p_2 p_4 p_6^2 ~s o  $ &0& -1& -1\\
$p_2^2 p_4 p_6^2 ~s o  $ &-1& -1& -1\\
$p_1^2 p_3^2 p_5 p_6 ~s o ^2 $ &1& 1& 0\\
$p_1 p_2 p_3^2 p_5 p_6 ~s o ^2 $ &0& 1& 0\\
$p_2^2 p_3^2 p_5 p_6 ~s o ^2 $ &-1& 1& 0\\
$p_1^2 p_3 p_4 p_5 p_6 ~s o ^2 $ &1& 0& 0\\
$p_1 p_2 p_3 p_4 p_5 p_6 ~s o ^2 $ &0& 0& 0\\
$p_2^2 p_3 p_4 p_5 p_6 ~s o ^2 $ &-1& 0& 0\\
$p_1^2 p_4^2 p_5 p_6 ~s o ^2 $ &1& -1& 0\\
$p_1 p_2 p_4^2 p_5 p_6 ~s o ^2 $ &0& -1& 0\\
$p_2^2 p_4^2 p_5 p_6 ~s o ^2 $ &-1& -1& 0\\
$p_1^2 p_3^3 p_5^2 ~s o ^3 $ &1& 2& 1\\
$p_1 p_2 p_3^3 p_5^2 ~s o ^3 $ &0& 2& 1\\
$p_2^2 p_3^3 p_5^2 ~s o ^3 $ &-1& 2& 1\\
$p_1^2 p_3^2 p_4 p_5^2 ~s o ^3 $ &1& 1& 1\\
$p_1 p_2 p_3^2 p_4 p_5^2 ~s o ^3 $ &0& 1& 1\\
$p_2^2 p_3^2 p_4 p_5^2 ~s o ^3 $ &-1& 1& 1\\
$p_1^2 p_3 p_4^2 p_5^2 ~s o ^3 $ &1& 0& 1\\
$p_1 p_2 p_3 p_4^2 p_5^2 ~s o ^3 $ &0& 0& 1\\
$p_2^2 p_3 p_4^2 p_5^2 ~s o ^3 $ &-1& 0& 1\\
$p_1^2 p_4^3 p_5^2 ~s o ^3 $ &1& -1& 1\\
$p_1 p_2 p_4^3 p_5^2 ~s o ^3 $ &0& -1& 1\\
$p_2^2 p_4^3 p_5^2 ~s o ^3 $ &-1& -1& 1\\
\hline
\end{tabular}
\end{minipage}
\hspace{2cm}
\begin{minipage}[!b]{0.4\textwidth}
\includegraphics[height=6cm]{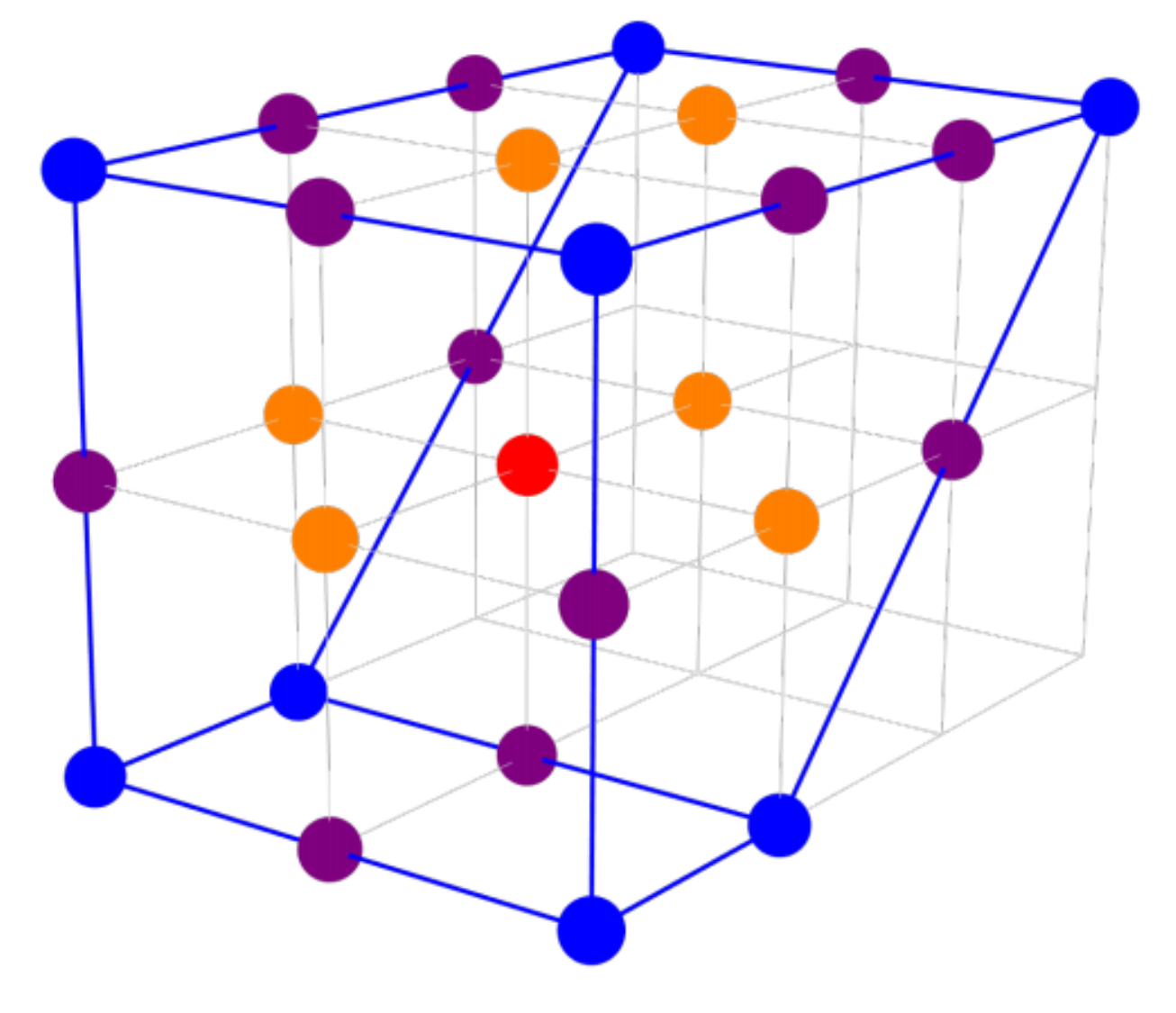} 
\end{minipage}
}
\caption{The generators and lattice of generators of the mesonic moduli space of Model 11 in terms of brick matchings with the corresponding flavor charges.
\label{f_genlattice_11}}
\end{table}

In highest weight form, the Hilbert series of Model 11 can be written as
\beal{es1125}
&&
h_1(t, \mu_1,\mu_2, b; \mathcal{M}_{11}) =
\frac{
(1 + \mu_1^2 \mu_2^2 t^6)
}{
(1 - \mu_1^2 \mu_2 b^{-2} t^5) (1 -  \mu_1^2 \mu_2^3 b^2 t^7)
}
~,~
\eea
where $\mu_1^m \sim [m]_{SU(2)_x}$ and $\mu_2^n \sim [n]_{SU(2)_y}$. Here in highest weight form, the fugacities $\mu_1$ and $\mu_2$ count the highest weight of irreducible representations of $SU(2)_x \times SU(2)_y$.

The plethystic logarithm of the Hilbert series takes the form
\beal{es1126}
&&
\PL[g_1(t, x, y, b; \mathcal{M}_{11})]=
[2; 1]b^{-2} t^5
+ [2; 2] t^6
+ [2; 3] b^2 t^7
\nn\\
&&
\hspace{1cm}
- ( [0; 2]b^{-4} + [2; 0]b^{-4}) t^{10}
- ( [0; 1]b^{-2} +  [0; 3]b^{-2} +  [2; 1]b^{-2} +  [2; 3]b^{-2} 
\nn\\
&&
\hspace{1cm}
+  [4; 1]b^{-2}) t^{11}
+ (1 + [0; 2] + 2[0; 4] + 2[2; 2] + [2; 4] +  [4; 0] + [4; 2] 
\nn\\
&&
\hspace{1cm}
+ [4; 4]) t^{12}
- (  [0; 1] b^2+  [0; 3]b^2  +   [0; 5] b^2 +  [2; 1]b^2  +   [2; 3]  b^2 +  [2; 5] b^2  
\nn\\
&&
\hspace{1cm}
+   [4; 1]  b^2 +  [4; 3] b^2 ) t^{13}
- ( [0; 2] b^4 +  [0; 6] b^4 +   [2; 0] b^4 +  [2; 4] b^4 +   [4; 2] b^4 ) t^{14}
\nn\\
&&
\hspace{1cm}
+ \dots ~,~
\eea
where $[m; n]= [m]_{SU(2)_x}  [n]_{SU(2)_y}$.
From the plethystic logarithm, we see that the mesonic moduli space is a non-complete intersection.

The generators form 3 sets that transform under 
$[2; 1]b^{-2}$, $[2; 2]$ and $[2; 3] b^2$ of the mesonic flavor symmetry of Model 11, respectively.
Using the following fugacity map
\beal{es1127}
&&
\tilde{t} = t_5^{1/2} t_6^{1/2}~,~ 
\tilde{x} = \frac{t_1}{t_2}~,~ 
\tilde{y} =\frac{t_3}{t_4}~,~ 
\tilde{b} = \frac{t_4 t_5^{1/2}}{t_6^{3/2}}
~,~
\eea
where $t_1 t_2 = t_3 t_4 = t_5 t_6$,
the mesonic flavor charges on the gauge invariant operators become $\mathbb{Z}$-valued.
The generators in terms of brick matchings and their corresponding rescaled mesonic flavor charges are summarized in \tref{f_genlattice_11}.
The generator lattice as shown in \tref{f_genlattice_11} is a convex lattice polytope, which is reflexive. It is the dual of the toric diagram of Model 11 shown in \fref{f_toric_11}.
We also note that the 3 layers of points in the generator lattice in \tref{f_genlattice_11} corresponds to the 3 sets that transform under 
$[2; 1]b^{-2}$, $[2; 2]$ and $[2; 3] b^2$ of the mesonic flavor symmetry.
For completeness, \tref{f_genfields_11a} and \tref{f_genfields_11b} show the generators of Model 11 in terms of chiral fields with the corresponding mesonic flavor charges.

\begin{table}[H]
\centering
\resizebox{0.95\hsize}{!}{

}
\caption{The generators in terms of bifundamental chiral fields for Model 11 \bf{(Part 1)}.
\label{f_genfields_11a}}
\end{table}

\begin{table}[H]
\centering
\resizebox{0.95\hsize}{!}{
%
}
\caption{The generators in terms of bifundamental chiral fields for Model 11 \bf{(Part 2)}.
\label{f_genfields_11b}}
\end{table}

\section{Model 12: $Q^{1,1,1}/\mathbb{Z}_2$~[$\mathbb{P}^1\times\mathbb{P}^1\times\mathbb{P}^1$,~$\langle30\rangle$] \label{smodel12}}

\begin{figure}[H]
\begin{center}
\resizebox{0.3\hsize}{!}{
\includegraphics[height=6cm]{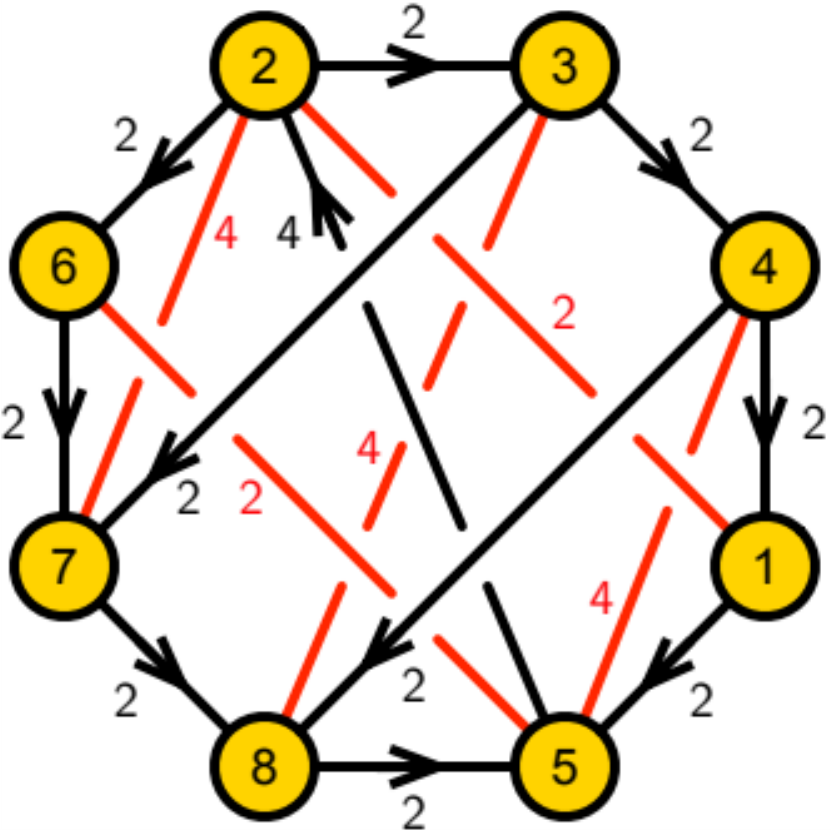} 
}
\caption{
Quiver for Model 12.
\label{f_quiver_12}}
 \end{center}
 \end{figure}

Model 12 corresponds to the toric Calabi-Yau 4-fold $Q^{1,1,1}/\mathbb{Z}_2$. 
The corresponding brane brick model has the quiver in \fref{f_quiver_11} and the $J$- and $E$-terms are 
\beq
{\footnotesize
 
 }
\label{E_J_C_+-}
\eeq
 
Using the forward algorithm, we can find the brick matchings for Model 12. 
The resulting brick matchings are summarized in the $P$-matrix, which takes the following form

\beal{es1205}
{\footnotesize
 P=
 \resizebox{0.7\hsize}{!}{$
\left( 
%
\right)
$}
}~.~
\eea

The $J$- and $E$-term charges are given by
\beal{es1205}
{\footnotesize
Q_{JE} =
\resizebox{0.67\hsize}{!}{$
\left(
%
\right)
$}
}~,~
\eea
and the $D$-term charges are given by
\beal{es1205}
{\footnotesize
Q_D =
\resizebox{0.67\hsize}{!}{$
\left(
%
\right)
$}
}~.~
\eea

\begin{figure}[H]
\begin{center}
\resizebox{0.4\hsize}{!}{
\includegraphics[height=6cm]{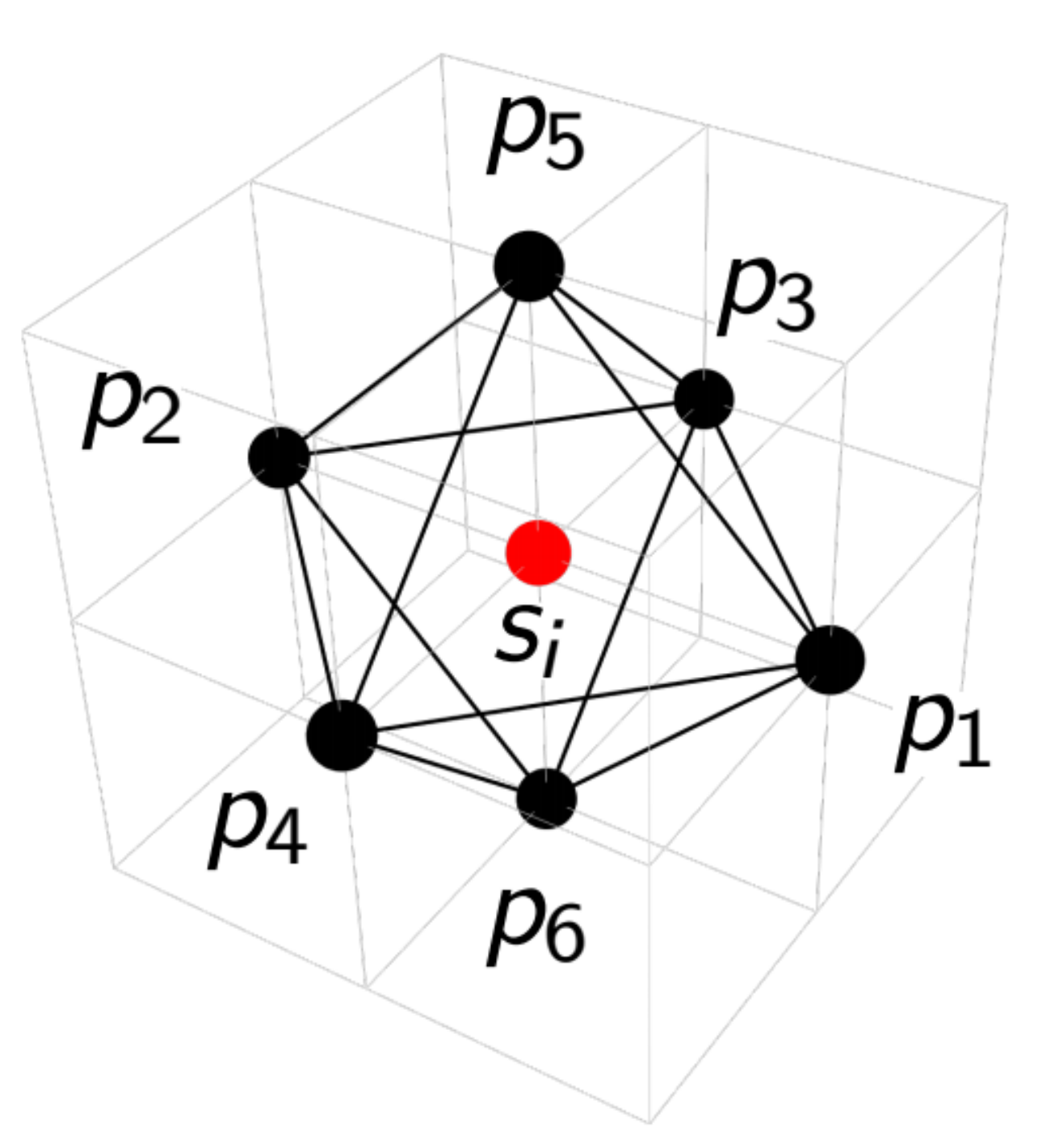} 
}
\caption{
Toric diagram for Model 12.
\label{f_toric_12}}
 \end{center}
 \end{figure}

The toric diagram of Model 12 is given by
\beal{es1210}
{\footnotesize
G_t=
\resizebox{0.67\hsize}{!}{$
\left(
%
\right)
$}
}~,~
\eea
where \fref{f_toric_12} shows the toric diagram with brick matching labels.

\begin{table}[H]
\centering
%
\caption{Global symmetry charges on the extremal brick matchings $p_i$ of Model 12.}
\label{t_pmcharges_12}
\end{table}

The Hilbert series of the mesonic moduli space of Model 12 takes the form
\beal{es1220}
&&
g_1(t_i,y_s ; \mathcal{M}_{12}) =
\frac{P(t_i,y_s; \mathcal{M}_{12})}{
(1 - y_{s} t_1^2 t_3^2 t_5^2) 
(1 - y_{s} t_2^2 t_3^2 t_5^2) 
(1 - y_{s} t_1^2 t_4^2 t_5^2) 
(1 - y_{s} t_2^2 t_4^2 t_5^2) 
}
\nn\\
&&
\hspace{1cm} 
\times
\frac{1}{
(1 - y_{s} t_1^2 t_3^2 t_6^2) 
(1 - y_{s} t_2^2 t_3^2 t_6^2) 
(1 - y_{s} t_1^2 t_4^2 t_6^2) 
(1 - y_{s} t_2^2 t_4^2 t_6^2)
} ~,~
\eea
where $t_i$ are the fugacities for the extremal brick matchings $p_i$.
$y_{s}$ counts the brick matching product $s_1 \dots s_{14}$ corresponding to the single internal point of the toric diagram of Model 12.
The explicit numerator $P(t_i,y_s; \mathcal{M}_{12})$ of the Hilbert series is given in the Appendix Section \sref{app_num_12}.

\begin{table}[H]
\centering
\resizebox{.95\hsize}{!}{
\begin{minipage}[!b]{0.5\textwidth}
\begin{tabular}{|c|c|c|c|}
\hline
generator & $SU(2)_{\tilde{x}}$ & $SU(2)_{\tilde{y}}$ & $SU(2)_{\tilde{z}}$ \\
\hline
$ p_1^2 p_3^2 p_5^2  ~s  $ &1& 1& 1\\
$ p_1 p_2 p_3^2 p_5^2  ~s  $ &0& 1& 1\\
$ p_2^2 p_3^2 p_5^2  ~s  $ &-1& 1& 1\\
$ p_1^2 p_3 p_4 p_5^2  ~s  $ &1& 0& 1\\
$ p_1 p_2 p_3 p_4 p_5^2  ~s  $ &0& 0& 1\\
$ p_2^2 p_3 p_4 p_5^2  ~s  $ &-1& 0& 1\\
$ p_1^2 p_4^2 p_5^2  ~s  $ &1& -1& 1\\
$ p_1 p_2 p_4^2 p_5^2  ~s  $ &0& -1& 1\\
$ p_2^2 p_4^2 p_5^2  ~s  $ &-1& -1& 1\\
$ p_1^2 p_3^2 p_5 p_6  ~s  $ &1& 1& 0\\
$ p_1 p_2 p_3^2 p_5 p_6  ~s  $ &0& 1& 0\\
$ p_2^2 p_3^2 p_5 p_6  ~s  $ &-1& 1& 0\\
$ p_1^2 p_3 p_4 p_5 p_6  ~s  $ &1& 0& 0\\
$ p_1 p_2 p_3 p_4 p_5 p_6  ~s  $ &0& 0& 0\\
$ p_2^2 p_3 p_4 p_5 p_6  ~s  $ &-1& 0& 0\\
$ p_1^2 p_4^2 p_5 p_6  ~s  $ &1& -1& 0\\
$ p_1 p_2 p_4^2 p_5 p_6  ~s  $ &0& -1& 0\\
$ p_2^2 p_4^2 p_5 p_6  ~s  $ &-1& -1& 0\\
$ p_1^2 p_3^2 p_6^2  ~s  $ &1& 1& -1\\
$ p_1 p_2 p_3^2 p_6^2  ~s  $ &0& 1& -1\\
$ p_2^2 p_3^2 p_6^2  ~s  $ &-1& 1& -1\\
$ p_1^2 p_3 p_4 p_6^2  ~s  $ &1& 0& -1\\
$ p_1 p_2 p_3 p_4 p_6^2  ~s  $ &0& 0& -1\\
$ p_2^2 p_3 p_4 p_6^2  ~s  $ &-1& 0& -1\\
$ p_1^2 p_4^2 p_6^2  ~s  $ &1& -1& -1\\
$ p_1 p_2 p_4^2 p_6^2  ~s  $ &0& -1& -1\\
$ p_2^2 p_4^2 p_6^2  ~s  $ &-1& -1& -1\\
\hline
\end{tabular}
\end{minipage}
\hspace{1.5cm}
\begin{minipage}[!b]{0.4\textwidth}
\includegraphics[height=7cm]{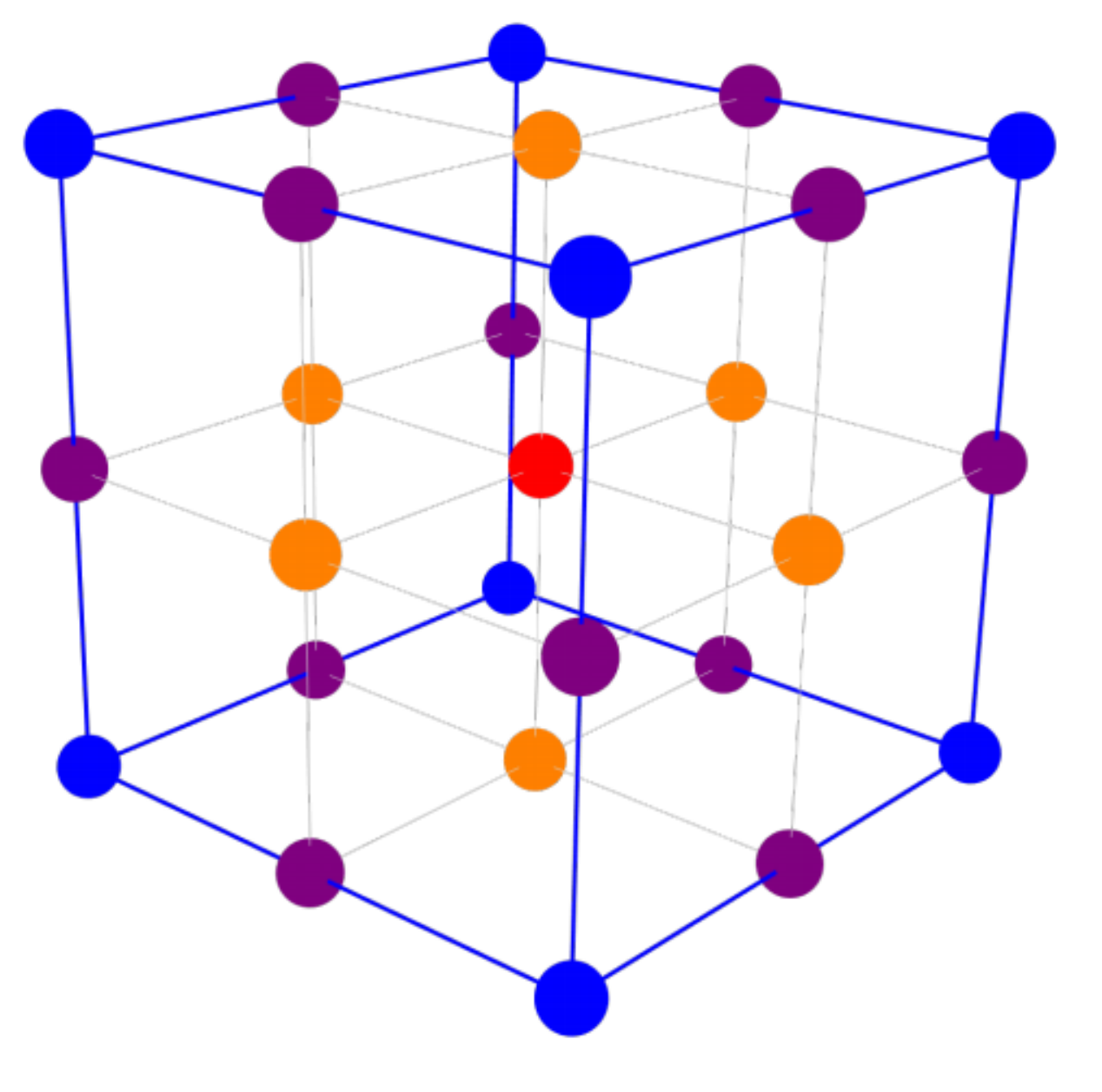} 
\end{minipage}
}
\caption{The generators and lattice of generators of the mesonic moduli space of Model 12 in terms of brick matchings with the corresponding flavor charges.
\label{f_genlattice_12}}
\end{table}

By setting $t_i=t$ and $y_s=1$, the unrefined Hilbert series takes the following form
\beal{es1221}
&&
g_1(t,1; \mathcal{M}_{12}) =
\frac{
1 + 19 t^6 - 63 t^{12} + 43 t^{18} + 43 t^{24} - 63 t^{30} + 19 t^{36} + t^{42}
}{
(1 - t^6)^8
}
~,~
\eea
where the palindromic numerator indicates that the mesonic moduli space is Calabi-Yau. 

The global symmetry of Model 12 and the charges on the extremal brick matchings under the global symmetry are summarized in \tref{t_pmcharges_12}.
We can use the following fugacity map,
\beal{es1222}
&&
t = t_1^{1/2} t_2^{1/2} ~,~
x = \frac{t_1^{1/2}}{t_2^{1/2}} ~,~
y = \frac{t_3^{1/2}}{t_4^{1/2}} ~,~
z = \frac{t_5^{1/2}}{t_6^{1/2}} ~,~
\eea
where $t_1 t_2 =t_3 t_4 =t_5 t_6 $, in order to rewrite the 
Hilbert series for Model 12 in terms of characters of irreducible representations of $SU(2)\times SU(2) \times SU(2)$.
The character expansion of the Hilbert series is
\beal{es1225}
&&
g_1(t, x,y, z; \mathcal{M}_{12}) = 
\sum_{n=0}^{\infty} [2n; 2n; 2n] t^{6n} ~,~
\eea
where $[m_1; m_2; m_3] = [m_1]_{SU(2)_x} [m_2]_{SU(2)_y} [m_3]_{SU(2)_z}$.
In highest weight form, the Hilbert series of Model 12 can be written as
\beal{es1226}
&&
h_1(t, \mu_1,\mu_2, \mu_3; \mathcal{M}_{12}) = 
\frac{1}{
(1 - \mu_1^2 \mu_2^2 \mu_3^2 t^6)
}~,~
\eea
where $\mu_1^{m_1} \mu_2^{m_2} \mu_3^{m_3} \sim [m_1]_{SU(2)_x} [m_2]_{SU(2)_y} [m_3]_{SU(2)_z}$.
Here in highest weight form, the fugacities $\mu_1$, $\mu_2$ and $\mu_3$ count the highest weights of irreducible representations of $SU(2)_x \times SU(2)_y \times SU(2)_z$.

The plethystic logarithm of the Hilbert series takes the form
\beal{es1226}
&&
\PL[g_1(t, x,y,z; \mathcal{M}_{12})]=
 [2; 2; 2] t^6
 - (1 +  [0; 0; 4] +  [0; 2; 2] +   [0; 4; 0] 
 \nn\\
 &&
 \hspace{1cm}
+  [0; 4; 4]  +   [2; 0; 2] +  [2; 2; 0] +   [2; 2; 4] +  [2; 4; 2] +   [4; 0; 0] +  [4; 0; 4] 
 \nn\\
 &&
 \hspace{1cm}
+   [4; 2; 2] +  [4; 4; 0]) t^{12}
+ \dots ~,~
\eea
where $[k;m;n]=[k]_{SU(2)_{x}} [m]_{SU(2)_{y}} [n]_{SU(2)_{z}}$.
From the plethystic logarithm, we see that the mesonic moduli space is a non-complete intersection.

We note that all generators transform together in the adjoint representation of each of the $SU(2)$ factors of the mesonic flavor symmetry of Model 12. 
Using the following fugacity map
\beal{es1227}
&&
\tilde{t} = t_5^{1/2} t_6^{1/2}~,~ 
\tilde{x} = \frac{t_1}{t_2}~,~ 
\tilde{y} =\frac{t_3}{t_4}~,~ 
\tilde{z} = \frac{t_5}{t_6}
~,~
\eea
where $t_1 t_2 = t_3 t_4 = t_5 t_6$,
the mesonic flavor charges on the gauge invariant operators become $\mathbb{Z}$-valued.
The generators in terms of brick matchings and their corresponding rescaled mesonic flavor charges are summarized in \tref{f_genlattice_12}.
The generator lattice as shown in \tref{f_genlattice_12} is a convex lattice polytope, which is reflexive. It is the dual of the toric diagram of Model 12 shown in \fref{f_toric_12}.
It is also interesting to note that the generator lattice forms layers in any of the 3 planar directions of $\mathbb{Z}^3$.
Each layer of points in the generator lattice forms a subset of generators that transforms under a the adjoint representations of a pair of $SU(2)$ factors of the mesonic flavor symmetry of Model 12.
For completeness, we also refer to \tref{f_genfields_12a} and \tref{f_genfields_12b} that show the generators of Model 12 in terms of chiral fields with the corresponding mesonic flavor charges.
\\

\begin{table}[H]
\centering
\resizebox{0.95\hsize}{!}{

}
\caption{The generators in terms of bifundamental chiral fields for Model 12 \bf{(Part 1)}.
\label{f_genfields_12a}}
\end{table}

\begin{table}[H]
\centering
\resizebox{0.95\hsize}{!}{
%
}
\caption{The generators in terms of bifundamental chiral fields for Model 12 \bf{(Part 2)}.
\label{f_genfields_12b}}
\end{table}
\vspace{1cm}

\section{Model 13: $P^{1}_{+-}(\text{dP}_2)$~[$\text{dP}_2$ bundle of $\mathbb{P}^1$,~$\langle81\rangle$] \label{smodel13}}
 
\begin{figure}[H]
\begin{center}
\resizebox{0.35\hsize}{!}{
\includegraphics[height=6cm]{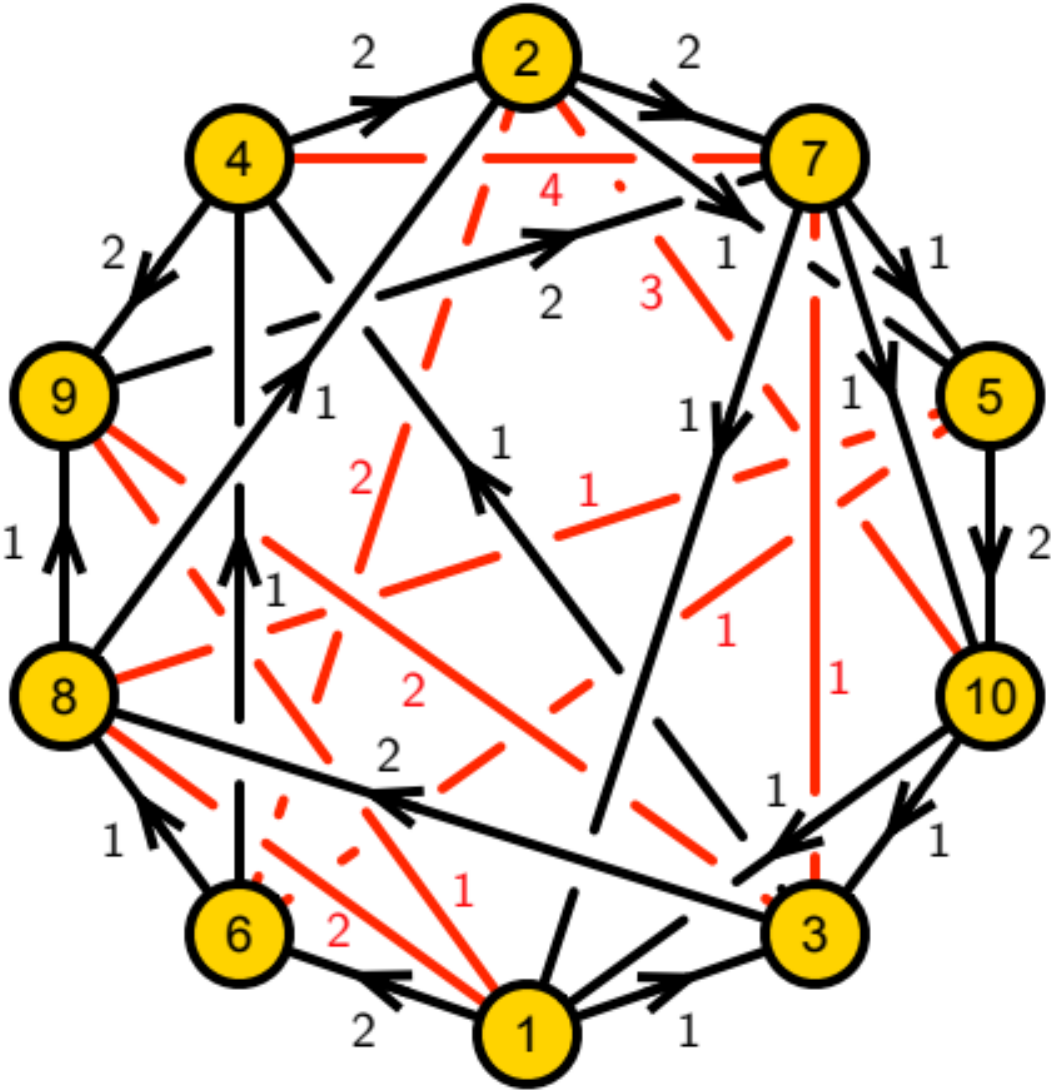} 
}
\caption{
Quiver for Model 13.
\label{f_quiver_13}}
 \end{center}
 \end{figure}
 
Model 13 corresponds to the toric Calabi-Yau 4-fold $P^{1}_{+-}(\text{dP}_2)$. 
The corresponding brane brick model has the quiver in \fref{f_quiver_13} and the $J$- and $E$-terms are 
\beq
 {\footnotesize
 
 }~.~
\label{E_J_C_+-}
 \eeq
 
Using the forward algorithm, we can obtain the brick matchings for Model 13. 
The brick matchings are summarized in the $P$-matrix, which takes the form

 \beal{es1305}
 {\footnotesize
 P=
\resizebox{0.9\textwidth}{!}{$
\left( 
%
\right)
$ }
}~.~
\nn\\
 \eea

The $J$- and $E$-term charges are given by
\beal{es1305}
{\footnotesize
Q_{JE} =
\resizebox{0.87\textwidth}{!}{$
\left(
%
\right)
$}
}~,~
\nn\\
\eea
and the $D$-term charges are given by
\beal{es1305}
{\footnotesize
Q_D =
\resizebox{0.87\textwidth}{!}{$
\left(
%
\right)
$}
}~.~
\nn\\
\eea

\begin{figure}[ht!!]
\begin{center}
\resizebox{0.4\hsize}{!}{
\includegraphics[height=6cm]{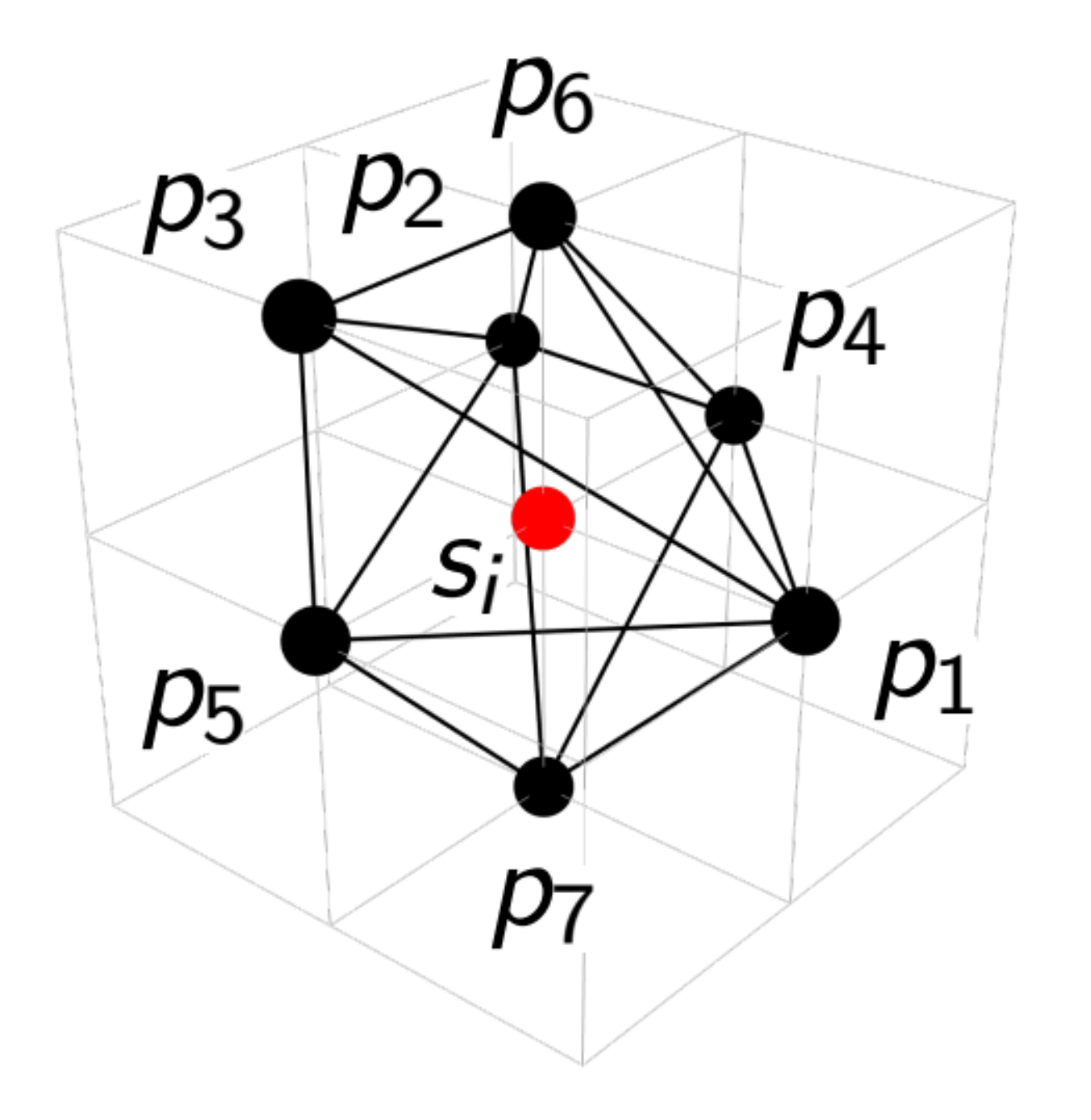} 
}
\caption{
Toric diagram for Model 13.
\label{f_toric_13}}
 \end{center}
 \end{figure}

The toric diagram of Model 13 is given by
\beal{es1310}
{\footnotesize
G_t=
\resizebox{0.87\textwidth}{!}{$
\left(
%
\right)
$}
}~,~
\nn\\
\eea
where \fref{f_toric_13} shows the toric diagram with brick matching labels.

\begin{table}[H]
\centering
%
\caption{Global symmetry charges on the extremal brick matchings $p_i$ of Model 13.}
\label{t_pmcharges_13}
\end{table}

The Hilbert series of the mesonic moduli space of Model 13 takes the form
\beal{es1320}
&&
g_1(t_i,y_s,y_{o_1},y_{o_2},y_{o_3},y_{o_4} ; \mathcal{M}_{13}) =
\frac{P(t_i,y_s,y_{o_1},y_{o_2},y_{o_3},y_{o_4}; \mathcal{M}_{13})
}{
(1 - y_{s} y_{o_1}^2 y_{o_2} y_{o_3}^3 y_{o_4}^3 t_1^3 t_3 t_4^2 t_6^2) 
(1 - y_{s} y_{o_1}^2 y_{o_2} y_{o_3}^3 y_{o_4}^3 t_2^3 t_3 t_4^2 t_6^2) 
}
\nn\\
&&
\hspace{1cm} 
\times
\frac{1}{
(1 - y_{s} y_{o_1}^4 y_{o_2}^3 y_{o_3}^3 y_{o_4} t_1 t_3^3 t_5^2 t_6^2) 
(1 - y_{s} y_{o_1}^4 y_{o_2}^3 y_{o_3}^3 y_{o_4} t_2 t_3^3 t_5^2 t_6^2) 
(1 - y_{s} y_{o_1} y_{o_2} y_{o_3}^2 y_{o_4}^3 t_1^3 t_4^2 t_6 t_7) 
} 
\nn\\
&&
\hspace{1cm} 
\times
\frac{1}{
(1 - y_{s} y_{o_1} y_{o_2} y_{o_3}^2 y_{o_4}^3 t_2^3 t_4^2 t_6 t_7) 
(1 - y_{s} y_{o_1} y_{o_2}^2 y_{o_3} y_{o_4}^2 t_1^2 t_4 t_5 t_7^2) 
(1 - y_{s} y_{o_1} y_{o_2}^2 y_{o_3} y_{o_4}^2 t_2^2 t_4 t_5 t_7^2) 
} 
\nn\\
&&
\hspace{1cm} 
\times
\frac{1}{
(1 - y_{s} y_{o_1}^2 y_{o_2}^3 y_{o_3} y_{o_4} t_1 t_3 t_5^2 t_7^2) 
(1 - y_{s} y_{o_1}^2 y_{o_2}^3 y_{o_3} y_{o_4} t_2 t_3 t_5^2 t_7^2)
} 
~,~
\eea
where $t_i$ are the fugacities for the extremal brick matchings $p_i$.
$y_{s}$ counts the brick matching product $s_1 \dots s_{13}$ corresponding to the single internal point of the toric diagram of Model 13.
Additionally, $y_{o_1}$, $y_{o_2}$, $y_{o_3}$ and $y_{o_4}$ count the products of extra GLSM fields $o_1 o_2$, $o_3 o_4$ and $o_5  \dots o_8$ and $o_9 \dots o_{14}$, respectively.
The explicit numerator $P(t_i,y_s,y_{o_1},y_{o_2},y_{o_3},y_{o_4}; \mathcal{M}_{13})$ of the Hilbert series is given in the Appendix Section \sref{app_num_13}.
We note that setting the fugacities $y_{o_1}=1, \dots, y_{o_4}=1$ does not change the overall characterization of the mesonic moduli space by the Hilbert series, indicating that the extra GLSM fields, as expected, correspond to an over-parameterization of the moduli space.

By setting $t_i=t$ for the fugacities of the extremal brick matchings, and all other fugacities to $1$, the unrefined Hilbert series takes the following form
\beal{es1321}
&&
g_1(t,1,1,1,1,1; \mathcal{M}_{13}) =
\frac{
(1 - t) (1 - t^2)^3
}{
(1 - t^6)^3 (1 - t^7)^2 (1 - t^8)^3
}
\times 
(
1 + t + 4 t^{2} + 4 t^{3} + 10 t^{4} 
\nn\\
&&
\hspace{1cm}
+ 10 t^{5} + 22 t^{6} + 29 t^{7} + 56 t^{8} + 77 t^{9} + 128 t^{10} + 170 t^{11} + 254 t^{12} + 311 t^{13} 
\nn\\
&&
\hspace{1cm}
+ 412 t^{14} + 464 t^{15} + 567 t^{16} + 594 t^{17} + 684 t^{18} + 669 t^{19} + 742 t^{20} + 694 t^{21} 
\nn\\
&&
\hspace{1cm}
+ 742 t^{22} + 669 t^{23} + 684 t^{24} + 594 t^{25} + 567 t^{26} + 464 t^{27} + 412 t^{28} + 311 t^{29} 
\nn\\
&&
\hspace{1cm}
+ 254 t^{30} + 170 t^{31} + 128 t^{32} + 77 t^{33} + 56 t^{34} + 29 t^{35} + 22 t^{36} + 10 t^{37} + 10 t^{38} 
\nn\\
&&
\hspace{1cm}
+ 4 t^{39} + 4 t^{40} + t^{41} + t^{42}
)
~,~
\eea
where the palindromic numerator indicates that the mesonic moduli space is Calabi-Yau. 

The global symmetry of Model 13 and the charges on the extremal brick matchings under the global symmetry are summarized in \tref{t_pmcharges_13}.
We can use the following fugacity map,
\beal{es1322}
&&
t = t_7 ~,~
x = \frac{t_7}{t_2} ~,~
b_1 = \frac{t_7}{t_4} ~,~
b_2 = \frac{t_5}{t_7} ~,~
\eea
where $t_1 t_2 = t_3 t_4 = t_5 t_6 = t_7^2$, in order to rewrite the 
Hilbert series for Model 13 in terms of characters of irreducible representations of $SU(2)\times U(1) \times U(1)$.

The highest weight form of the Hilbert series of Model 13 is
\beal{es1325}
&&
h_1(t, \mu, b_1, b_2; \mathcal{M}_{13}) =
\frac{
1
}{
(1 - \mu^2 b_1^{-1} b_2 t^6) (1 - \mu b_1 b_2^2 t^6) (1 - \mu^3 b_1^{-2} b_2^{-1} t^7) (1 - \mu^3 b_1^{-1} b_2^{-2} t^8) 
}
\nn\\
&&
\hspace{0.5cm}
\frac{1}{
(1 - \mu b_1^3 t^8)
}
\times (
1 + \mu^2 t^7 + \mu b_1^2 b_2 t^7 + \mu^2 b_1 b_2^{-1} t^8 - \mu^4 b_1^{-1} b_2 t^{13} -  \mu^3 b_1 b_2^2 t^{13} - 2 \mu^4 t^{14} 
\nn\\
&&
\hspace{0.5cm}
- \mu^5 b_1^{-2} b_2^{-1} t^{14} -  \mu^3 b_1^2 b_2 t^{14} - \mu^5 b_1^{-1} b_2^{-2} t^{15} - \mu^4 b_1 b_2^{-1} t^{15} 
+ \mu^6 b_1^{-1} b_2 t^{20} + \mu^6 t^{21} 
\nn\\
&&
\hspace{0.5cm}
+ \mu^7 b_1^{-2} b_2^{-1} t^{21} + \mu^8 t^{28}
)~,~
\eea
where $\mu^m \sim [m]_{SU(2)_x}$.
Here in highest weight form, the fugacity $\mu$ counts the highest weight of irreducible representations of $SU(2)_x$.
The fugacities $b_1$ and $b_2$ count charges under the two $U(1)$ factors of the mesonic flavor symmetry.

\begin{table}[H]
\centering
\resizebox{.95\hsize}{!}{
\begin{minipage}[!b]{0.5\textwidth}
\begin{tabular}{|c|c|c|c|}
\hline
generator & $SU(2)_{\tilde{x}}$ & $U(1)_{\tilde{b_1}}$ & $U(1)_{\tilde{b_2}}$ \\
\hline
$ p_1^2 p_4 p_5 p_7^2  ~s o_1 o_2^2 o_3 o_4^2  $ &0& -1& 1\\
$ p_1 p_2 p_4 p_5 p_7^2  ~s o_1 o_2^2 o_3 o_4^2  $ &0& 0& 1\\
$ p_2^2 p_4 p_5 p_7^2  ~s o_1 o_2^2 o_3 o_4^2  $ &0& 1& 1\\
$ p_1 p_3 p_5^2 p_7^2  ~s o_1^2 o_2^3 o_3 o_4  $ &1& 0& 1\\
$ p_2 p_3 p_5^2 p_7^2  ~s o_1^2 o_2^3 o_3 o_4  $ &1& 1& 1\\
$ p_1^3 p_4^2 p_6 p_7  ~s o_1 o_2 o_3^2 o_4^3  $ &-1& -2& 0\\
$ p_1^2 p_2 p_4^2 p_6 p_7  ~s o_1 o_2 o_3^2 o_4^3  $ &-1& -1& 0\\
$ p_1 p_2^2 p_4^2 p_6 p_7  ~s o_1 o_2 o_3^2 o_4^3  $ &-1& 0& 0\\
$ p_2^3 p_4^2 p_6 p_7  ~s o_1 o_2 o_3^2 o_4^3  $ &-1& 1& 0\\
$ p_1^2 p_3 p_4 p_5 p_6 p_7  ~s o_1^2 o_2^2 o_3^2 o_4^2  $ &0& -1& 0\\
$ p_1 p_2 p_3 p_4 p_5 p_6 p_7  ~s o_1^2 o_2^2 o_3^2 o_4^2  $ &0& 0& 0\\
$ p_2^2 p_3 p_4 p_5 p_6 p_7  ~s o_1^2 o_2^2 o_3^2 o_4^2  $ &0& 1& 0\\
$ p_1 p_3^2 p_5^2 p_6 p_7  ~s o_1^3 o_2^3 o_3^2 o_4  $ &1& 0& 0\\
$ p_2 p_3^2 p_5^2 p_6 p_7  ~s o_1^3 o_2^3 o_3^2 o_4  $ &1& 1& 0\\
$ p_1^3 p_3 p_4^2 p_6^2  ~s o_1^2 o_2 o_3^3 o_4^3  $ &-1& -2& -1\\
$ p_1^2 p_2 p_3 p_4^2 p_6^2  ~s o_1^2 o_2 o_3^3 o_4^3  $ &-1& -1& -1\\
$ p_1 p_2^2 p_3 p_4^2 p_6^2  ~s o_1^2 o_2 o_3^3 o_4^3  $ &-1& 0& -1\\
$ p_2^3 p_3 p_4^2 p_6^2  ~s o_1^2 o_2 o_3^3 o_4^3  $ &-1& 1& -1\\
$ p_1^2 p_3^2 p_4 p_5 p_6^2  ~s o_1^3 o_2^2 o_3^3 o_4^2  $ &0& -1& -1\\
$ p_1 p_2 p_3^2 p_4 p_5 p_6^2  ~s o_1^3 o_2^2 o_3^3 o_4^2  $ &0& 0& -1\\
$ p_2^2 p_3^2 p_4 p_5 p_6^2  ~s o_1^3 o_2^2 o_3^3 o_4^2  $ &0& 1& -1\\
$ p_1 p_3^3 p_5^2 p_6^2  ~s o_1^4 o_2^3 o_3^3 o_4  $ &1& 0& -1\\
$ p_2 p_3^3 p_5^2 p_6^2  ~s o_1^4 o_2^3 o_3^3 o_4  $ &1& 1& -1\\
\hline
\end{tabular}
\end{minipage}
\hspace{2cm}
\begin{minipage}[!b]{0.4\textwidth}
\includegraphics[height=6cm]{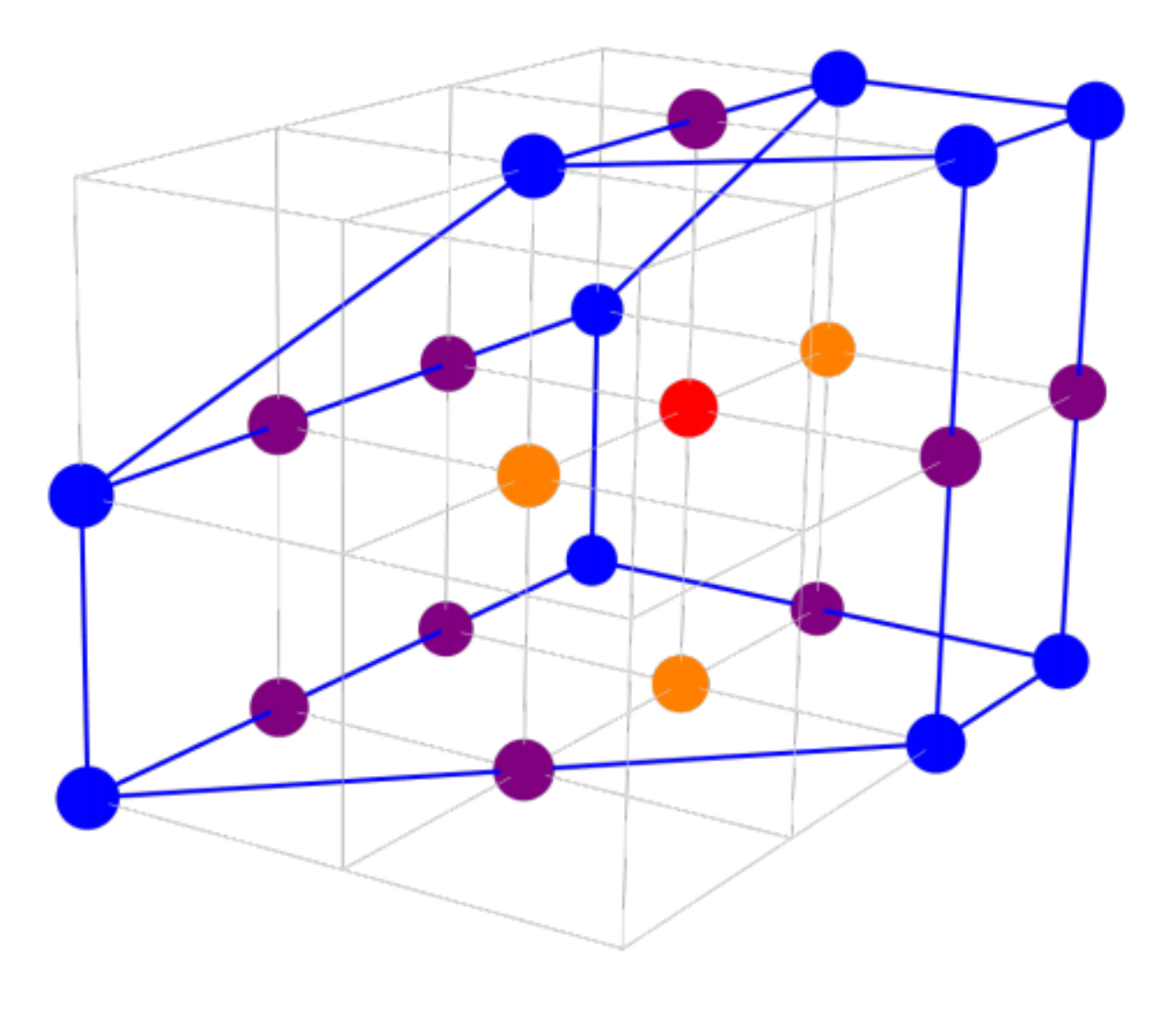} 
\end{minipage}
}
\caption{The generators and lattice of generators of the mesonic moduli space of Model 13 in terms of brick matchings with the corresponding flavor charges.
\label{f_genlattice_13}}
\end{table}

The plethystic logarithm of the Hilbert series takes the form
\beal{es1326}
&&
\PL[g_1(t, x,b_1,b_2; \mathcal{M}_{13})]=
(  [1] b_1 b_2^2  +  [2] )t^6
+ ( [1] b_1^2 b_2  +  [2] b_1 b_2^{-1}  +  [3] b_2^{-3} )t^7
\nn\\
&&
\hspace{1cm}
+ ( [1] b_1^3  +  [2]  b_1^2 b_2^{-2} + [3] b_1 b_2^{-4})t^8
\nn\\
&&
\hspace{1cm}
- (1 + [1] b_1 b_2^2) t^{12}
- (b_1 b_2^{-1} + b_1^3 b_2^3 + [1] b_2^{-3} +  2 [1] b_1^2 b_2 + 2 [2] b_1 b_2^{-1}  + [3] b_2^{-3} 
\nn\\
&&
\hspace{1cm}
+ [3] b_1^2 b_2 + [4]  b_1  b_2^{-1} ) t^{13}
- (2 b_1^2 b_2^{-2} + b_1^4 b_2^2  + 3 [1] b_1^3   + 2  [1] b_1  b_2^{-4} +  [2] b_2^{-6} 
\nn\\
&&
\hspace{1cm}
+ 3 [2] b_1^2 b_2^{-2} + [2] b_1^4 b_2^2 +  2 [3] b_1^3 + 2  [3] b_1  b_2^{-4} +  3  [4] b_1^2  b_2^{-2} + [5] b_1 b_2^{-4}) t^{14}
\nn\\
&&
\hspace{1cm}
- (b_1 b_2^{-7} + b_1^3 b_2^{-3} + b_1^5 b_2  + 2 [1]  b_1^2 b_2^{-5} + 2 [1] b_1^4  b_2^{-1} +   [2]  b_1 b_2^{-7} + 3 [2] b_1^3  b_2^{-3} 
\nn\\
&&
\hspace{1cm}
+ 2[3] b_1^2  b_2^{-5} +  [3] b_1^4  b_2^{-1} +  [4]  b_1  b_2^{-7} + 2  [4] b_1^3  b_2^{-3} +  [5]  b_1^2 b_2^{-5}) t^{15}
- (b_1^4 b_2^{-4} 
\nn\\
&&
\hspace{1cm}
+ [1] b_1^3 b_2^{-6} + [1]  b_1^5 b_2^{-2} + [2] b_1^2 b_2^{-8} + [2] b_1^4  b_2^{-4} + [3] b_1^3 b_2^{-6} + [4] b_1^4 b_2^{-4}) t^{16}
+ \dots ~,~
\nn\\
\eea
where $[m] = [m]_{SU(2)_x}$.
From the plethystic logarithm, we see that the mesonic moduli space is a non-complete intersection.

By using the following fugacity map
\beal{es1327}
&&
\tilde{t} = t_5^{1/2} t_6^{1/2}~,~ 
\tilde{x} = \frac{t_5^3}{t_2 t_7^2}~,~ 
\tilde{b_1} =\frac{t_2^2 t_7}{t_5^{3/2} t_6^{3/2}}~,~ 
\tilde{b_2} = \frac{t_7^2}{t_5 t_6}
~,~
\eea
where $t_1^2 t_2^2 t_7^2 = t_5^3 t_6^3$, $t_4^2 t_5 = t_6 t_7^2$ and $t_3^2 t_7^2 = t_5^3 t_6$,
the mesonic flavor charges on the gauge invariant operators become $\mathbb{Z}$-valued.
The generators in terms of brick matchings and their corresponding rescaled mesonic flavor charges are summarized in \tref{f_genlattice_13}.
The generator lattice as shown in \tref{f_genlattice_13} is a convex lattice polytope, which is reflexive. It is the dual of the toric diagram of Model 13 shown in \fref{f_toric_13}.
For completeness, \tref{f_genfields_13a} and \tref{f_genfields_13b} show the generators of Model 13 in terms of chiral fields with the corresponding mesonic flavor charges.

\begin{table}[H]
\centering
\resizebox{0.9\hsize}{!}{

}
\caption{The generators in terms of bifundamental chiral fields for Model 13 \bf{(Part 1)}.
\label{f_genfields_13a}}
\end{table}

\begin{table}[H]
\centering
\resizebox{0.9\hsize}{!}{
%
}
\caption{The generators in terms of bifundamental chiral fields for Model 13 \bf{(Part 2)}.
\label{f_genfields_13b}}
\end{table}

\section{Model 14: $P^{2}_{+-}(\text{dP}_2)$~[$\text{dP}_2$ bundle of $\mathbb{P}^1$,~$\langle82\rangle$] \label{smodel14}}

\begin{figure}[H]
\begin{center}
\resizebox{0.35\hsize}{!}{
\includegraphics[height=6cm]{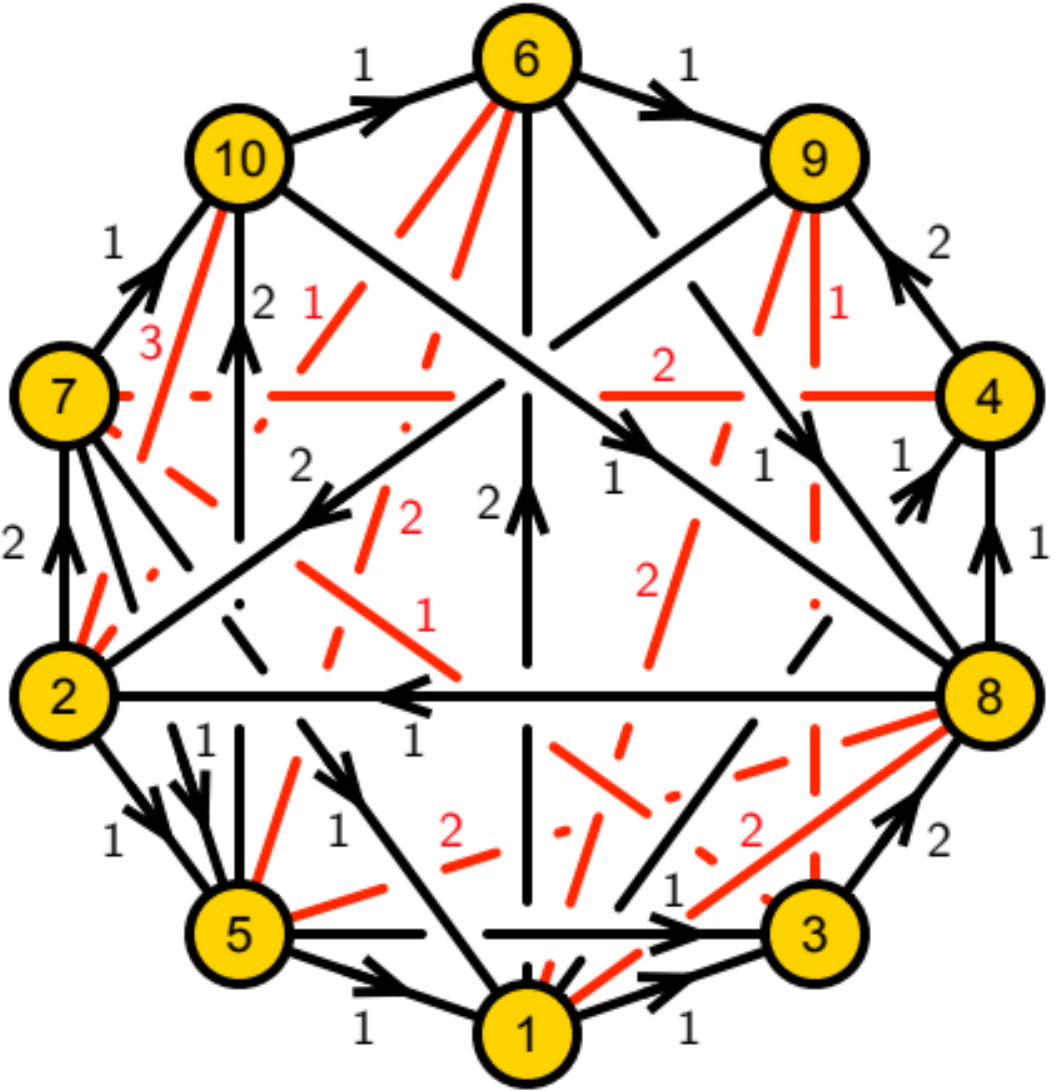} 
}
\caption{
Quiver for Model 14.
\label{f_quiver_14}}
 \end{center}
 \end{figure}

Model 14 corresponds to the toric Calabi-Yau 4-fold $P^{2}_{+-}(\text{dP}_2)$. 
The corresponding brane brick model has the quiver in \fref{f_quiver_14} and the $J$- and $E$-terms are 

 \beq
 {\footnotesize
 
 }~.~
\label{E_J_C_+-}
 \eeq
 
The brick matchings are summarized in the $P$-matrix, which takes the form
\beal{es1405}
{\footnotesize 
P=
\resizebox{0.90\textwidth}{!}{$
\left( 
%
\right)
$ }
}~.~
\nn\\
\eea

The $J$- and $E$-term charges are given by
\beal{es1405}
{\footnotesize
Q_{JE} =
\resizebox{0.87\textwidth}{!}{$
\left(
%
\right)
$}
}~,~
\nn\\
\eea
and the $D$-term charges are given by
\beal{es1405}
{\footnotesize
Q_D =
\resizebox{0.87\textwidth}{!}{$
\left(
%
\right)
$}
}~.~
\nn\\
\eea

\begin{figure}[H]
\begin{center}
\resizebox{0.4\hsize}{!}{
\includegraphics[trim=0mm 10mm 0mm 0mm, height=6cm]{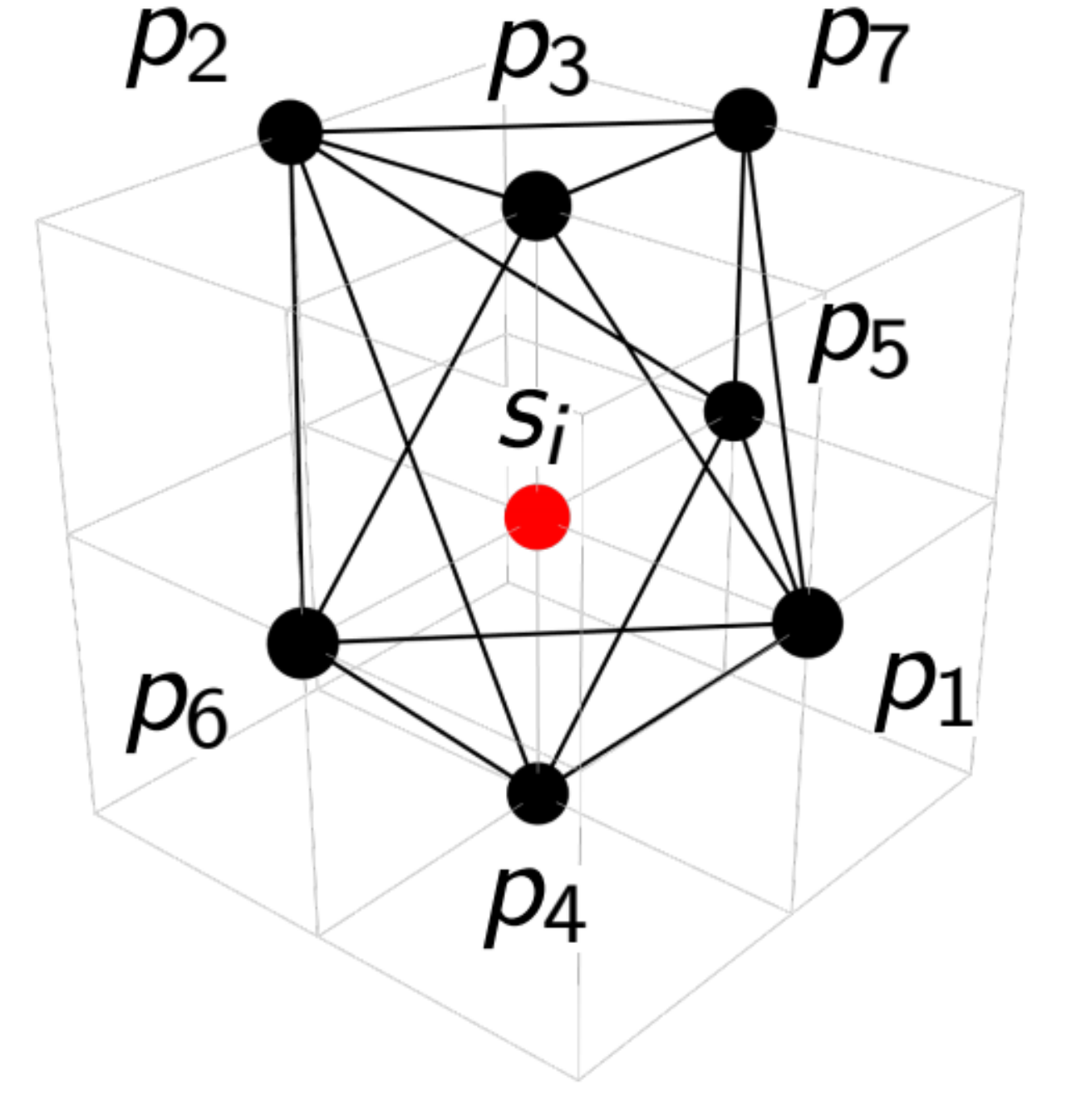} 
}
\caption{
Toric diagram for Model 14.
\label{f_toric_14}}
 \end{center}
 \end{figure}
 
The toric diagram of Model 14 is given by
\beal{es1410}
{\footnotesize
G_t=
\resizebox{0.87\textwidth}{!}{$
\left(
%
\right)
$}
}~,~
\nn\\
\eea
where \fref{f_toric_14} shows the toric diagram with brick matching labels.

\begin{table}[H]
\centering
%
\caption{Global symmetry charges on the extremal brick matchings $p_i$ of Model 14.}
\label{t_pmcharges_14}
\end{table}

The Hilbert series of the mesonic moduli space of Model 14 takes the form
\beal{es1420}
&&
g_1(t_i,y_s,y_{o_1},y_{o_2},y_{o_3},y_{o_4} ; \mathcal{M}_{14}) =
\frac{P(t_i,y_s,y_{o_1},y_{o_2},y_{o_3},y_{o_4}; \mathcal{M}_{14})
}{
(1 - y_{s} y_{o_1} y_{o_2}^2 y_{o_3} y_{o_4}^2 t_1 t_4^2 t_5 t_6) 
(1 - y_{s} y_{o_1} y_{o_2}^2 y_{o_3} y_{o_4}^2 t_2 t_4^2 t_5 t_6) 
}
\nn\\
&&
\hspace{1cm} 
\times
\frac{1}{
(1 - y_{s} y_{o_1} y_{o_2} y_{o_3}^2 y_{o_4}^3 t_1^2 t_3 t_4 t_6^2) 
(1 - y_{s} y_{o_1} y_{o_2} y_{o_3}^2 y_{o_4}^3 t_2^2 t_3 t_4 t_6^2) 
(1 - y_{s} y_{o_1}^2 y_{o_2}^3 y_{o_3} y_{o_4}^2 t_1 t_4^2 t_5^2 t_7) 
} 
\nn\\
&&
\hspace{1cm} 
\times
\frac{1}{
(1 - y_{s} y_{o_1}^2 y_{o_2}^3 y_{o_3} y_{o_4}^2 t_2 t_4^2 t_5^2 t_7) 
(1 - y_{s} y_{o_1}^2 y_{o_2} y_{o_3}^3 y_{o_4}^4 t_1^3 t_3^2 t_6^2 t_7) 
(1 - y_{s} y_{o_1}^2 y_{o_2} y_{o_3}^3 y_{o_4}^4 t_2^3 t_3^2 t_6^2 t_7) 
} 
\nn\\
&&
\hspace{1cm} 
\times
\frac{1}{
(1 - y_{s} y_{o_1}^4 y_{o_2}^3 y_{o_3}^3 y_{o_4}^4 t_1^3 t_3^2 t_5^2 t_7^3) 
(1 - y_{s} y_{o_1}^4 y_{o_2}^3 y_{o_3}^3 y_{o_4}^4 t_2^3 t_3^2 t_5^2 t_7^3)
} 
~,~
\eea
where $t_i$ are the fugacities for the extremal brick matchings $p_i$.
$y_{s}$ counts the brick matching product $s_1 \dots s_{13}$ corresponding to the single internal point of the toric diagram of Model 14.
Additionally, $y_{o_1}$ counts $o_1 \dots o_6$, $y_{o_2}$ counts $o_7 \dots o_9$, $y_{o_3}$ counts $o_{10} \dots o_{20}$ and $y_{o_4}$ counts $o_{21} \dots o_{24}$.
The explicit numerator $P(t_i,y_s,y_{o_1},y_{o_2},y_{o_3},y_{o_4}; \mathcal{M}_{14})$ of the Hilbert series is given in the Appendix Section \sref{app_num_14}.
We note that setting the fugacities $y_{o_1}=1, \dots, y_{o_4}=1$ does not change the overall characterization of the mesonic moduli space by the Hilbert series, indicating that the extra GLSM fields, as expected, correspond to an over-parameterization of the moduli space. 

By setting $t_i=t$ for the fugacities of the extremal brick matchings, and all other fugacities to $1$, the unrefined Hilbert series takes the following form
\beal{es1421}
&&
g_1(t,1,1,1,1,1; \mathcal{M}_{14}) =
\frac{
(1 - t^2)^4
}{
(1 - t) (1 - t^6)^2 (-1 + t^8)^2 (1 - t^{10})^2
}
\times
(
1 - t + 3 t^{2} 
\nn\\
&&
\hspace{1cm} 
- 3 t^{3}  + 6 t^{4} - 4 t^{5} + 11 t^{6} - 4 t^{7} + 20 t^{8} -  4 t^{9} + 34 t^{10} - 6 t^{11} + 53 t^{12} - 14 t^{13} 
\nn\\
&&
\hspace{1cm} 
+ 73 t^{14} - 26 t^{15} +  91 t^{16} - 34 t^{17} + 97 t^{18} - 34 t^{19} + 91 t^{20} - 26 t^{21} +  73 t^{22} 
\nn\\
&&
\hspace{1cm} 
- 14 t^{23}  + 53 t^{24} - 6 t^{25}+ 34 t^{26} - 4 t^{27} + 20 t^{28} -  4 t^{29} + 11 t^{30} - 4 t^{31} + 6 t^{32} 
\nn\\
&&
\hspace{1cm} 
- 3 t^{33}  + 3 t^{34} - t^{35} + t^{36}
) ~,~
\eea
where the palindromic numerator indicates that the mesonic moduli space is Calabi-Yau. 

The global symmetry of Model 14 and the charges on the extremal brick matchings under the global symmetry are summarized in \tref{t_pmcharges_14}.
We can use the following fugacity map,
\beal{es1422}
&&
t = t_7 ~,~
x = \frac{t_7}{t_2} ~,~
b_1 = \frac{t_7}{t_4} ~,~
b_2 = \frac{t_5}{t_7} ~,~
\eea
where $t_1 t_2 = t_3 t_4 = t_5 t_6 = t_7^2$, in order to rewrite the 
Hilbert series for Model 14 in terms of characters of irreducible representations of $SU(2)\times U(1) \times U(1)$.

The character expansion of the Hilbert series of Model 14 takes the following form
\beal{es1425}
&&
g_1(t, x, b_1, b_2; \mathcal{M}_{14}) =
1+  [1] b_1^{-2} t^5
+ ([1] b_1^{-2} b_2^2 + [2] b_2^{-2}) t^6
+ [2] t^7
+ ([2] b_2^2 + [3] b_1^2 b_2^{-2}) t^{8}
\nn\\
&& 
\hspace{1cm}
+[3] b_1^2 t^9
+ ([2] b_1^{-4} + [3] b_1^2 b_2^2) t^{10}
+ ([2] b_1^{-4} b_2^{2} + [3] b_1^{-2} b_2^{-2}) t^{11}
+ ([2] b_1^{-4} b_2^{4} + [3] b_1^{-2} 
\nn\\
&&
\hspace{1cm}
+ [4] b_2^{-4}) t^{12}
+ ([3] b_1^{-2} b_2^2 + [4] b_2^{-2}) t^{13}
+ ([3] b_1^{-2} b_2^4 + [4] + [5] b_1^2 b_2^{-4}) t^{14}
+ ([3] b_1^{-6} 
\nn\\
&& 
\hspace{1cm}
+ [4] b_2^2 + [5] b_1^2 b_2^{-2}) t^{15}
+ ([3] b_1^{-6} b_2^2 + [4] b_1^{-4} b_2^{-2} + [4] b_2^4 + [5] b_1^2 + [6] b_1^4 b_2^{-4}) t^{16}
\nn\\
&& 
\hspace{1cm}
+ ([3] b_1^{-6} b_2^4 + [4] b_1^{-4} + [5] b_1^{-2} b_2^{-4} + [5] b_1^2 b_2^2 + [6] b_1^4 b_2^{-2} ) t^{17}
+ ([3] b_1^{-6} b_2^6 + [4] b_1^{-4} b_2^2 
\nn\\
&& 
\hspace{1cm}
+ [5] b_1^{-2} b_2^{-2} + [5] b_1^2 b_2^4 + [6] b_1^4 + [6] b_2^{-6} ) t^{18}
+ ( [4] b_1^{-4} b_2^4 + [5] b_1^{-2} + [6] b_2^{-4} 
\nn\\
&& 
\hspace{1cm}
+ [6] b_1^4 b_2^2) t^{19}
+ ([4] b_1^{-8} + [4] b_1^{-4} b_2^6 + [5] b_1^{-2} b_2^2 + [6] b_2^{-2} + [6] b_1^4 b_2^4 + [7] b_1^2 b_2^{-6}) t^{20}
\nn\\
&& 
\hspace{1cm}
+ \dots 
~,~
\eea
where $[m] = [m]_{SU(2)_{x}}$.
The fugacities $b_1$ and $b_2$ count charges under the two $U(1)$ factors of the mesonic flavor symmetry.

\begin{table}[H]
\centering
\resizebox{.95\hsize}{!}{
\begin{minipage}[!b]{0.5\textwidth}
\begin{tabular}{|c|c|c|c|}
\hline
generator & $SU(2)_{\tilde{x}}$ & $U(1)_{\tilde{b_1}}$ & $U(1)_{\tilde{b_2}}$ \\
\hline
$ p_1 p_4^2 p_5 p_6  ~s o_1 o_2^2 o_3 o_4^2   $ &0& -1& 0\\
$ p_2 p_4^2 p_5 p_6  ~s o_1 o_2^2 o_3 o_4^2   $ &-1& -1& 0\\
$ p_1^2 p_3 p_4 p_6^2  ~s o_1 o_2 o_3^2 o_4^3   $ &1& 0& -1\\
$ p_1 p_2 p_3 p_4 p_6^2  ~s o_1 o_2 o_3^2 o_4^3   $ &0& 0& -1\\
$ p_2^2 p_3 p_4 p_6^2  ~s o_1 o_2 o_3^2 o_4^3   $ &-1& 0& -1\\
$ p_1 p_4^2 p_5^2 p_7  ~s o_1^2 o_2^3 o_3 o_4^2   $ &0& -1& 1\\
$ p_2 p_4^2 p_5^2 p_7  ~s o_1^2 o_2^3 o_3 o_4^2   $ &-1& -1& 1\\
$ p_1^2 p_3 p_4 p_5 p_6 p_7  ~s o_1^2 o_2^2 o_3^2 o_4^3   $ &1& 0& 0\\
$ p_1 p_2 p_3 p_4 p_5 p_6 p_7  ~s o_1^2 o_2^2 o_3^2 o_4^3   $ &0& 0& 0\\
$ p_2^2 p_3 p_4 p_5 p_6 p_7  ~s o_1^2 o_2^2 o_3^2 o_4^3   $ &-1& 0& 0\\
$ p_1^3 p_3^2 p_6^2 p_7  ~s o_1^2 o_2 o_3^3 o_4^4   $ &2& 1& -1\\
$ p_1^2 p_2 p_3^2 p_6^2 p_7  ~s o_1^2 o_2 o_3^3 o_4^4   $ &1& 1& -1\\
$ p_1 p_2^2 p_3^2 p_6^2 p_7  ~s o_1^2 o_2 o_3^3 o_4^4   $ &0& 1& -1\\
$ p_2^3 p_3^2 p_6^2 p_7  ~s o_1^2 o_2 o_3^3 o_4^4   $ &-1& 1& -1\\
$ p_1^2 p_3 p_4 p_5^2 p_7^2  ~s o_1^3 o_2^3 o_3^2 o_4^3   $ &1& 0& 1\\
$ p_1 p_2 p_3 p_4 p_5^2 p_7^2  ~s o_1^3 o_2^3 o_3^2 o_4^3   $ &0& 0& 1\\
$ p_2^2 p_3 p_4 p_5^2 p_7^2  ~s o_1^3 o_2^3 o_3^2 o_4^3   $ &-1& 0& 1\\
$ p_1^3 p_3^2 p_5 p_6 p_7^2  ~s o_1^3 o_2^2 o_3^3 o_4^4   $ &2& 1& 0\\
$ p_1^2 p_2 p_3^2 p_5 p_6 p_7^2  ~s o_1^3 o_2^2 o_3^3 o_4^4   $ &1& 1& 0\\
$ p_1 p_2^2 p_3^2 p_5 p_6 p_7^2  ~s o_1^3 o_2^2 o_3^3 o_4^4   $ &0& 1& 0\\
$ p_2^3 p_3^2 p_5 p_6 p_7^2  ~s o_1^3 o_2^2 o_3^3 o_4^4   $ &-1& 1& 0\\
$ p_1^3 p_3^2 p_5^2 p_7^3  ~s o_1^4 o_2^3 o_3^3 o_4^4   $ &2& 1& 1\\
$ p_1^2 p_2 p_3^2 p_5^2 p_7^3  ~s o_1^4 o_2^3 o_3^3 o_4^4   $ &1& 1& 1\\
$ p_1 p_2^2 p_3^2 p_5^2 p_7^3  ~s o_1^4 o_2^3 o_3^3 o_4^4   $ &0& 1& 1\\
$ p_2^3 p_3^2 p_5^2 p_7^3  ~s o_1^4 o_2^3 o_3^3 o_4^4   $ &-1& 1& 1\\
\hline
\end{tabular}
\end{minipage}
\hspace{2.5cm}
\begin{minipage}[!b]{0.4\textwidth}
\includegraphics[height=6cm]{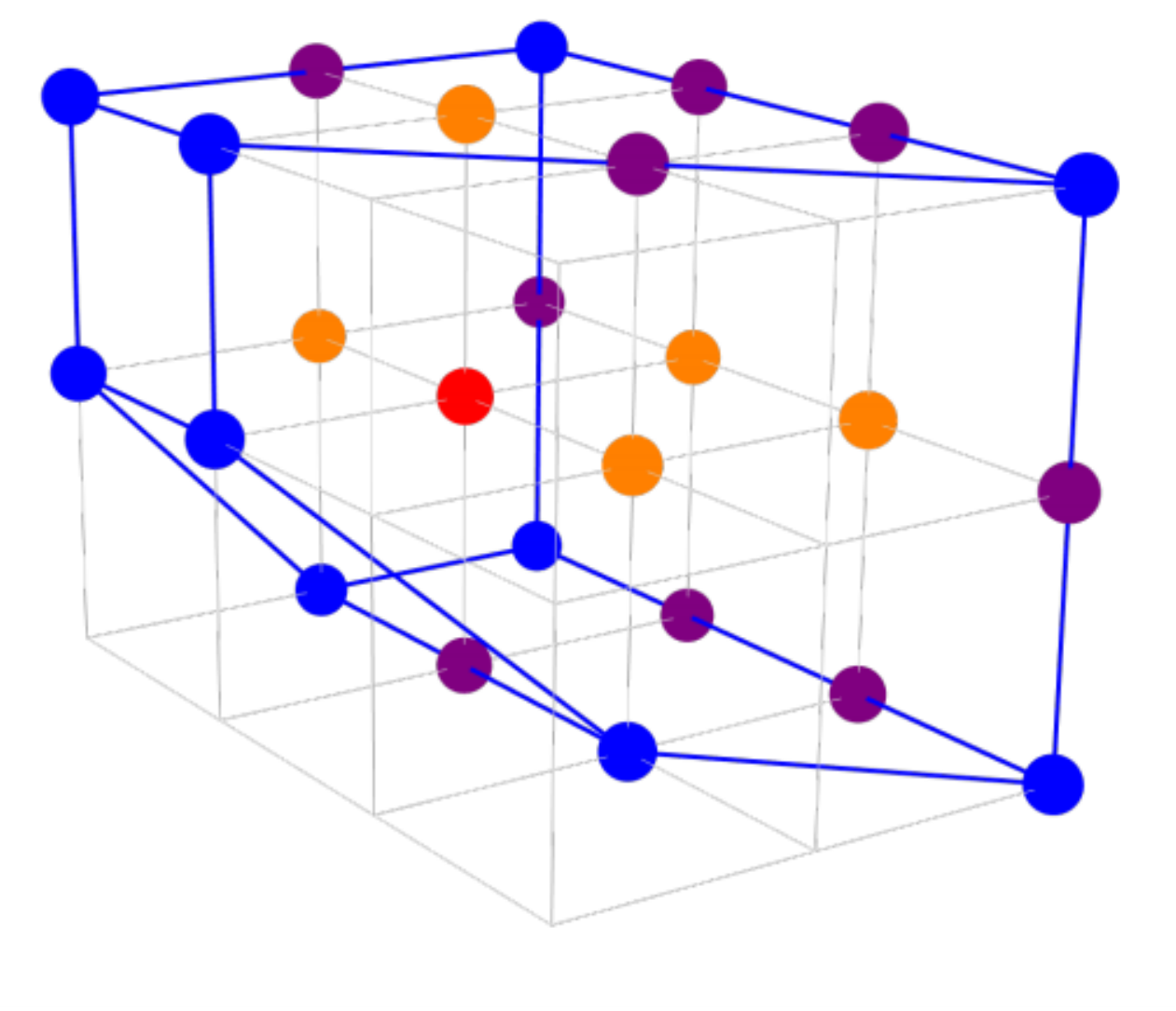} 
\end{minipage}
}
\caption{The generators and lattice of generators of the mesonic moduli space of Model 14 in terms of brick matchings with the corresponding flavor charges.
\label{f_genlattice_14}}
\end{table}

The plethystic logarithm of the Hilbert series takes the form
\beal{es1426}
&&
\PL[g_1(t, x,b_1,b_2; \mathcal{M}_{14})]=
[1] b_1^{-2} t^5 
+ ([1] b_1^{-2} b_2^2 + [2] b_2^{-2}) t^6
+ [2] t^7
\nn\\
&&
\hspace{1cm}
+ ( [2] b_2^2  + [3] b_1^2  b_2^{-2}) t^8
+ [3] b_1^2 t^9 
+ [3] b_1^2 b_2^2 t^{10}
\nn\\
&&
\hspace{1cm}
- (b_1^{-4} b_2^2 + [1] b_1^{-2} b_2^{-2}) t^{11}
- (b_2^{-4} + 2 [1] b_1^{-2} + [3] b_1^{-2}) t^{12}
- (b_2^{-2} + 2 [1] b_1^{-2} b_2^2 
\nn\\
&&
\hspace{1cm}
+ 2  [2]b_2^{-2} +   [3]b_1^{-2}b_2^2 + [4]b_2^{-2}) t^{13}
- (2  + [1] b_1^2 b_2^{-4} + [1] b_1^{-2} b_2^4 + 3  [2] + [3] b_1^2 b_2^{-4} 
\nn\\
&&
\hspace{1cm}
+ 3  [4]) t^{14}
- (b_2^2  + [1] 2 b_1^2 b_2^{-2} +  3  [2] b_2^2  + 2  [3] b_1^2  b_2^{-2} +  2  [4] b_2^2  +  [5] b_1^2  b_2^{-2} ) t^{15}
\nn\\
&&
\hspace{1cm}
- (b_2^4  + 3 [1] b_1^2  +   [2] b_1^4 b_2^{-4} +  [2] b_2^4   + 3  [3] b_1^2 +  [4] b_2^4 +  2  [5] b_1^2 ) t^{16}
\nn\\
&&
\hspace{1cm}
- (b_1^4 b_2^{-2} + 2  [1] b_1^2 b_2^2  +   [2] b_1^4 b_2^{-2} + 2 [3]  b_1^2 b_2^2  +   [4]  b_1^4 b_2^{-2} +   [5] b_1^2 b_2^2) t^{17}
+ \dots
~,~ 
\nn\\
\eea
where $[m] = [m]_{SU(2)_x}$.
From the plethystic logarithm, we see that the mesonic moduli space is a non-complete intersection.

By using the following fugacity map
\beal{es1427}
&&
\tilde{t} = t_7~,~ 
\tilde{x} = \frac{t_7^2}{t_2^2}~,~ 
\tilde{b_1} =\frac{t_2 t_7}{t_4^2}~,~ 
\tilde{b_2} = \frac{t_7^2}{t_6^2}~,~
~,~
\eea
where $t_1 t_2 = t_3 t_4 = t_5 t_6 = t_7^3$,
the mesonic flavor charges on the gauge invariant operators become $\mathbb{Z}$-valued.
The generators in terms of brick matchings and their corresponding rescaled mesonic flavor charges are summarized in \tref{f_genlattice_14}.
The generator lattice as shown in \tref{f_genlattice_14} is a convex lattice polytope, which is reflexive. It is the dual of the toric diagram of Model 14 shown in \fref{f_toric_14}.
For completeness, \tref{f_genfields_14a} and \tref{f_genfields_14b} show the generators of Model 14 in terms of chiral fields with the corresponding mesonic flavor charges.
\\

\begin{table}[H]
\centering
\resizebox{0.95\hsize}{!}{

}
\caption{The generators in terms of bifundamental chiral fields for Model 14 \bf{(Part 1)}.
\label{f_genfields_14a}}
\end{table}

\begin{table}[H]
\centering
\resizebox{0.95\hsize}{!}{
%
}
\caption{The generators in terms of bifundamental chiral fields for Model 14 \bf{(Part 2)}.
\label{f_genfields_14b}}
\end{table}

\section{Model 15: $P^{3}_{+-}(\text{dP}_2)$~[$\text{dP}_2$ bundle of $\mathbb{P}^1$,~$\langle83\rangle$] \label{smodel15}}
 
\begin{figure}[H]
\begin{center}
\resizebox{0.35\hsize}{!}{
\includegraphics[height=6cm]{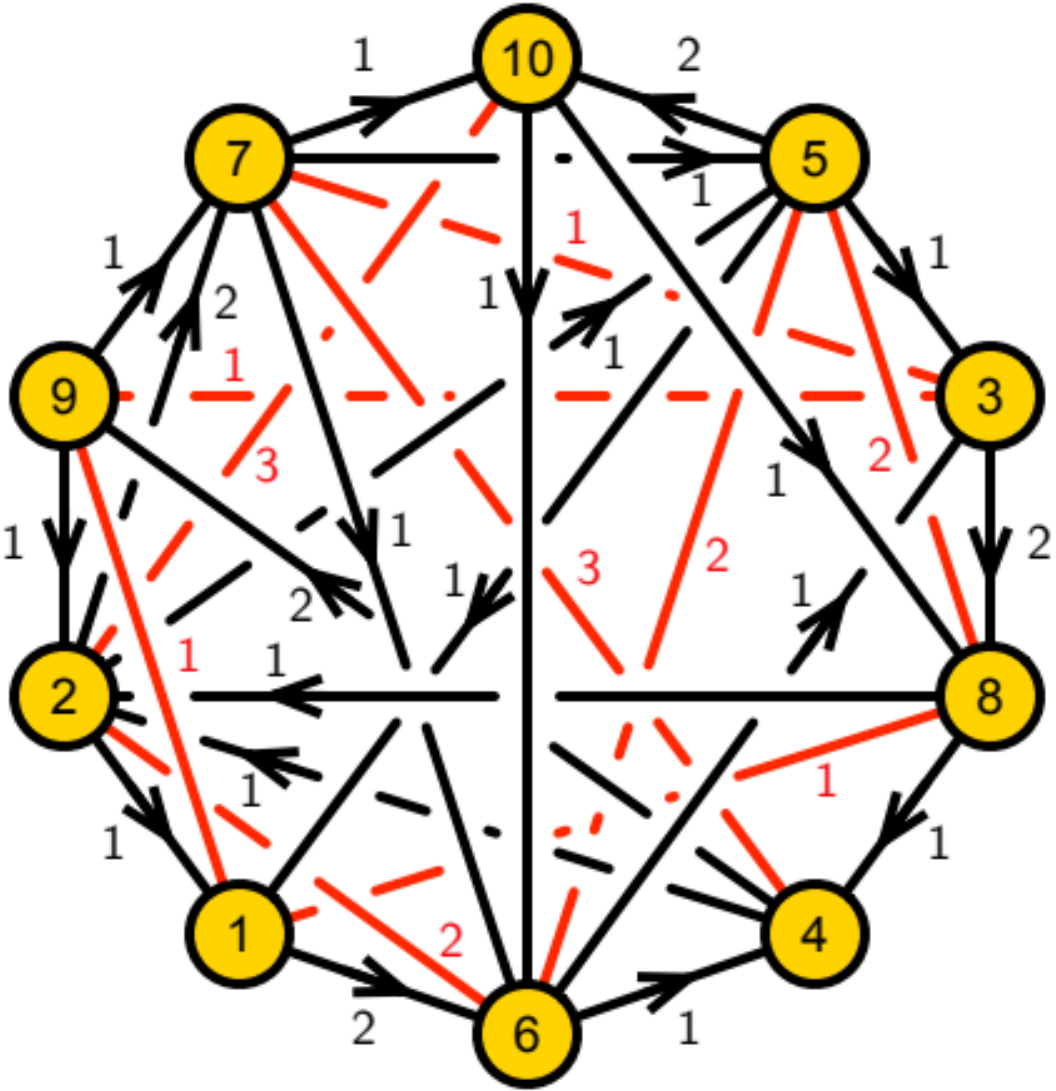} 
}
\caption{
Quiver for Model 15.
\label{f_quiver_15}}
 \end{center}
 \end{figure}

Model 15 corresponds to the toric Calabi-Yau 4-fold $P^{3}_{+-}(\text{dP}_2)$. 
The corresponding brane brick model has the quiver in \fref{f_quiver_15} and the $J$- and $E$-terms are 
\beq
{\footnotesize
 
 }~.~
\label{E_J_C_+-}
 \eeq
 
Using the forward algorithm, we are able to obtain the brick matchings for Model 15.
The brick matchings are summarized in the $P$-matrix, which takes the form 

\beal{es1505}
{\footnotesize
P=
\resizebox{0.9\textwidth}{!}{$
\left(
%
\right)
$ }
}~.~
\nn\\
\eea

The $J$- and $E$-term charges are given by
\beal{es1505}
{\footnotesize
Q_{JE} =
\resizebox{0.87\textwidth}{!}{$
\left(
%
\right)
$}
}~,~
\nn\\
\eea
and the $D$-term charges are given by
\beal{es1505}
{\footnotesize
Q_D =
\resizebox{0.87\textwidth}{!}{$
\left(
%
\right)
$}
}~.~
\nn\\
\eea

\begin{figure}[ht!!]
\begin{center}
\resizebox{0.4\hsize}{!}{
\includegraphics[height=6cm]{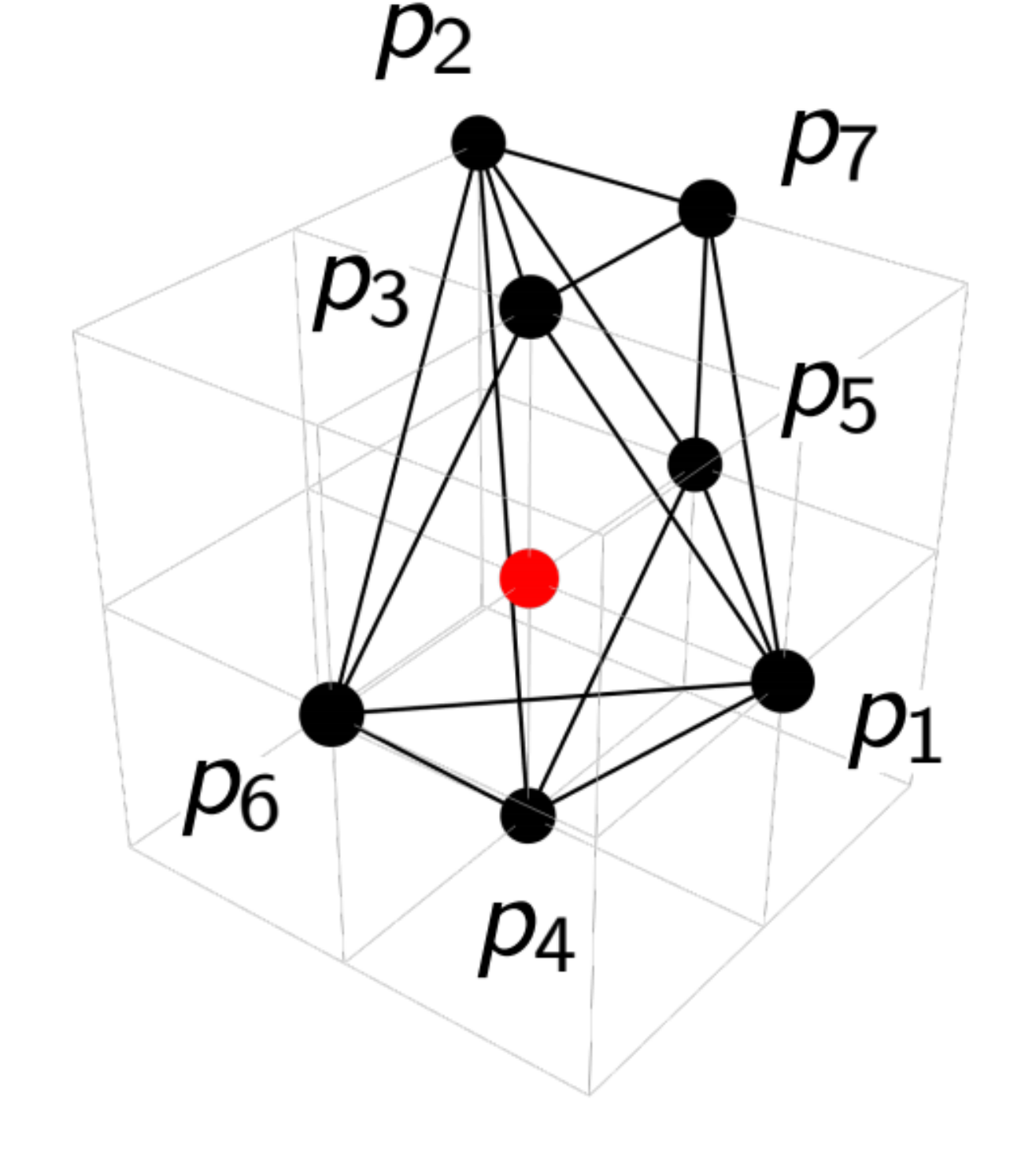} 
}
\caption{
Toric diagram for Model 15.
\label{f_toric_15}}
 \end{center}
 \end{figure}

The toric diagram of Model 15 is given by
\beal{es1510}
{\footnotesize
G_t=
\resizebox{0.87\textwidth}{!}{$
\left(
%
\right)
$}
}~,~
\nn\\
\eea
where \fref{f_toric_15} shows the toric diagram with brick matching labels.

The Hilbert series of the mesonic moduli space of Model 15 is given by
\beal{es1520}
&&
g_1(t_i,y_s,y_{o_1},y_{o_2},y_{o_3},y_{o_4} ; \mathcal{M}_{15}) =
\frac{P(t_i,y_s,y_{o_1},y_{o_2},y_{o_3},y_{o_4}; \mathcal{M}_{15})
}{
(1 - y_{s} y_{o_1} y_{o_2} y_{o_3}^2 y_{o_4}^2 t_1 t_4^2 t_5 t_6) 
}
\nn\\
&&
\hspace{0.5cm} 
\times
\frac{1}{
(1 - y_{s} y_{o_1} y_{o_2} y_{o_3}^2 y_{o_4}^2 t_2 t_4^2 t_5 t_6) 
(1 - y_{s} y_{o_1} y_{o_2}^2 y_{o_3} y_{o_4}^2 t_1 t_3 t_4 t_6^2) 
(1 - y_{s} y_{o_1} y_{o_2}^2 y_{o_3} y_{o_4}^2 t_2 t_3 t_4 t_6^2) 
} 
\nn\\
&&
\hspace{0.5cm} 
\times
\frac{1}{
(1 - y_{s} y_{o_1}^2 y_{o_2} y_{o_3}^3 y_{o_4}^3 t_1^2 t_4^2 t_5^2 t_7) 
(1 - y_{s} y_{o_1}^2 y_{o_2} y_{o_3}^3 y_{o_4}^3 t_2^2 t_4^2 t_5^2 t_7) 
(1 - y_{s} y_{o_1}^2 y_{o_2}^3 y_{o_3} y_{o_4}^3 t_1^2 t_3^2 t_6^2 t_7) 
} 
\nn\\
&&
\hspace{0.5cm} 
\times
\frac{1}{
(1 - y_{s} y_{o_1}^2 y_{o_2}^3 y_{o_3} y_{o_4}^3 t_2^2 t_3^2 t_6^2 t_7) 
(1 - y_{s} y_{o_1}^4 y_{o_2}^3 y_{o_3}^3 y_{o_4}^5 t_1^4 t_3^2 t_5^2 t_7^3) 
(1 - y_{s} y_{o_1}^4 y_{o_2}^3 y_{o_3}^3 y_{o_4}^5 t_2^4 t_3^2 t_5^2 t_7^3)
} 
~,~
\nn\\
\eea
where $t_i$ are the fugacities for the extremal brick matchings $p_i$.
Additionally, $y_{s}$ counts $s_1 \dots s_{13}$, $y_{o_1}$ counts $o_1\dots o_{13}$, $y_{o_2}$ counts $o_{14} \dots o_{19}$, $y_{o_3}$ counts $o_{20}\dots o_{24}$ and $y_{o_4}$ counts $o_{25}\dots o_{29}$.
The explicit numerator $P(t_i,y_s,y_{o_1},y_{o_2},y_{o_3},y_{o_4}; \mathcal{M}_{15})$ of the Hilbert series is given in the Appendix Section \sref{app_num_15}.
We note that setting the fugacities $y_{o_1}=1, \dots, y_{o_4}=1$ does not change the overall characterization of the mesonic moduli space by the Hilbert series, indicating that the extra GLSM fields, as expected, correspond to an over-parameterization of the moduli space. 

By setting $t_i=t$ for the fugacities of the extremal brick matchings, and all other fugacities to $1$, the unrefined Hilbert series takes the following form
\beal{es1521}
&&
g_1(t,1,1,1,1,1; \mathcal{M}_{15}) =
\frac{
(1 - t) (1 - t^2) (1 - t^3)
}{
(1 - t^5)^3 (1 - t^7)^2 (1 - t^{11})^2
}
\times
(
1 + t + 2 t^{2} + 3 t^{3} + 4 t^{4} 
\nn\\
&&
\hspace{1cm}
+ 6 t^{5} + 8 t^{6} + 17 t^{7} + 20 t^{8} + 38 t^{9} + 48 t^{10} + 70 t^{11} + 76 t^{12} + 105 t^{13} + 99 t^{14} 
\nn\\
&&
\hspace{1cm}
+ 124 t^{15} + 116 t^{16} + 133 t^{17} + 118 t^{18} + 141 t^{19} + 118 t^{20} + 133 t^{21} + 116 t^{22} 
\nn\\
&&
\hspace{1cm}
+ 124 t^{23} + 99 t^{24} + 105 t^{25} + 76 t^{26} + 70 t^{27} + 48 t^{28} + 38 t^{29} + 20 t^{30} + 17 t^{31} 
\nn\\
&&
\hspace{1cm}
+ 8 t^{32} + 6 t^{33} + 4 t^{34} + 3 t^{35} + 2 t^{36} + t^{37} + t^{38}
) ~,~
\eea
where the palindromic numerator indicates that the mesonic moduli space is Calabi-Yau. 

\begin{table}[H]
\centering
\begin{tabular}{|c|c|c|c|c|c|}
\hline
\; & $SU(2)_{x}$ & $U(1)_{b_1}$ &  $U(1)_{b_2}$ & $U(1)$ & \text{fugacity} \\
\hline
$p_1$ & +1 & 0 & 0 & $r_1$ & $t_1$ \\
$p_2$ & -1 & 0 & 0 & $r_2$ & $t_2$ \\
$p_3$ & 0 & +1  & 0 & $r_3$ & $t_3$ \\
$p_4$ & 0 & -1 & 0 & $r_4$ & $t_4$ \\
$p_5$ & 0 &0 & +1 & $r_5$ & $t_5$ \\
$p_6$ & 0 &0 &  -1 & $r_6$ & $t_6$ \\
$p_7$ & 0 &0 & 0 & $r_7$ & $t_7$ \\
\hline
\end{tabular}
\caption{Global symmetry charges on the extremal brick matchings $p_i$ of Model 15.}
\label{t_pmcharges_15}
\end{table}

The global symmetry of Model 15 and the charges on the extremal brick matchings under the global symmetry are summarized in \tref{t_pmcharges_15}.
We can use the following fugacity map,
\beal{es1522}
&&
t = t_7 ~,~
x = \frac{t_7}{t_2} ~,~
b_1 = \frac{t_7}{t_4} ~,~
b_2 = \frac{t_5}{t_7} ~,~
\eea
where $t_1 t_2 = t_3 t_4 = t_5 t_6 = t_7^2$, in order to rewrite the 
Hilbert series for Model 15 in terms of characters of irreducible representations of $SU(2)\times U(1) \times U(1)$.
\\

\begin{table}[H]
\centering
\resizebox{.95\hsize}{!}{
\begin{minipage}[!b]{0.5\textwidth}
\begin{tabular}{|c|c|c|c|}
\hline
generator & $SU(2)_{\tilde{x}}$ & $U(1)_{\tilde{b_1}}$ & $U(1)_{\tilde{b_2}}$ \\
\hline
$ p_1 p_4^2 p_5 p_6  ~s o_1 o_2 o_3^2 o_4^2   $ &0& -1& 0\\
$ p_2 p_4^2 p_5 p_6  ~s o_1 o_2 o_3^2 o_4^2   $ &-1& -1& 0\\
$ p_1 p_3 p_4 p_6^2  ~s o_1 o_2^2 o_3 o_4^2   $ &0& 0& -1\\
$ p_2 p_3 p_4 p_6^2  ~s o_1 o_2^2 o_3 o_4^2   $ &-1& 0& -1\\
$ p_1^2 p_4^2 p_5^2 p_7  ~s o_1^2 o_2 o_3^3 o_4^3   $ &1& -1& 1\\
$ p_1 p_2 p_4^2 p_5^2 p_7  ~s o_1^2 o_2 o_3^3 o_4^3   $ &0& -1& 1\\
$ p_2^2 p_4^2 p_5^2 p_7  ~s o_1^2 o_2 o_3^3 o_4^3   $ &-1& -1& 1\\
$ p_1^2 p_3 p_4 p_5 p_6 p_7  ~s o_1^2 o_2^2 o_3^2 o_4^3   $ &1& 0& 0\\
$ p_1 p_2 p_3 p_4 p_5 p_6 p_7  ~s o_1^2 o_2^2 o_3^2 o_4^3   $ &0& 0& 0\\
$ p_2^2 p_3 p_4 p_5 p_6 p_7  ~s o_1^2 o_2^2 o_3^2 o_4^3   $ &-1& 0& 0\\
$ p_1^2 p_3^2 p_6^2 p_7  ~s o_1^2 o_2^3 o_3 o_4^3   $ &1& 1& -1\\
$ p_1 p_2 p_3^2 p_6^2 p_7  ~s o_1^2 o_2^3 o_3 o_4^3   $ &0& 1& -1\\
$ p_2^2 p_3^2 p_6^2 p_7  ~s o_1^2 o_2^3 o_3 o_4^3   $ &-1& 1& -1\\
$ p_1^3 p_3 p_4 p_5^2 p_7^2  ~s o_1^3 o_2^2 o_3^3 o_4^4   $ &2& 0& 1\\
$ p_1^2 p_2 p_3 p_4 p_5^2 p_7^2  ~s o_1^3 o_2^2 o_3^3 o_4^4   $ &1& 0& 1\\
$ p_1 p_2^2 p_3 p_4 p_5^2 p_7^2  ~s o_1^3 o_2^2 o_3^3 o_4^4   $ &0& 0& 1\\
$ p_2^3 p_3 p_4 p_5^2 p_7^2  ~s o_1^3 o_2^2 o_3^3 o_4^4   $ &-1& 0& 1\\
$ p_1^3 p_3^2 p_5 p_6 p_7^2  ~s o_1^3 o_2^3 o_3^2 o_4^4   $ &2& 1& 0\\
$ p_1^2 p_2 p_3^2 p_5 p_6 p_7^2  ~s o_1^3 o_2^3 o_3^2 o_4^4   $ &1& 1& 0\\
$ p_1 p_2^2 p_3^2 p_5 p_6 p_7^2  ~s o_1^3 o_2^3 o_3^2 o_4^4   $ &0& 1& 0\\
$ p_2^3 p_3^2 p_5 p_6 p_7^2  ~s o_1^3 o_2^3 o_3^2 o_4^4   $ &-1& 1& 0\\
$ p_1^4 p_3^2 p_5^2 p_7^3  ~s o_1^4 o_2^3 o_3^3 o_4^5   $ &3& 1& 1\\
$ p_1^3 p_2 p_3^2 p_5^2 p_7^3  ~s o_1^4 o_2^3 o_3^3 o_4^5   $ &2& 1& 1\\
$ p_1^2 p_2^2 p_3^2 p_5^2 p_7^3  ~s o_1^4 o_2^3 o_3^3 o_4^5   $ &1& 1& 1\\
$ p_1 p_2^3 p_3^2 p_5^2 p_7^3  ~s o_1^4 o_2^3 o_3^3 o_4^5   $ &0& 1& 1\\
$ p_2^4 p_3^2 p_5^2 p_7^3  ~s o_1^4 o_2^3 o_3^3 o_4^5   $ &-1& 1& 1\\
\hline
\end{tabular}
\end{minipage}
\hspace{2.5cm}
\begin{minipage}[!b]{0.4\textwidth}
\includegraphics[height=6cm]{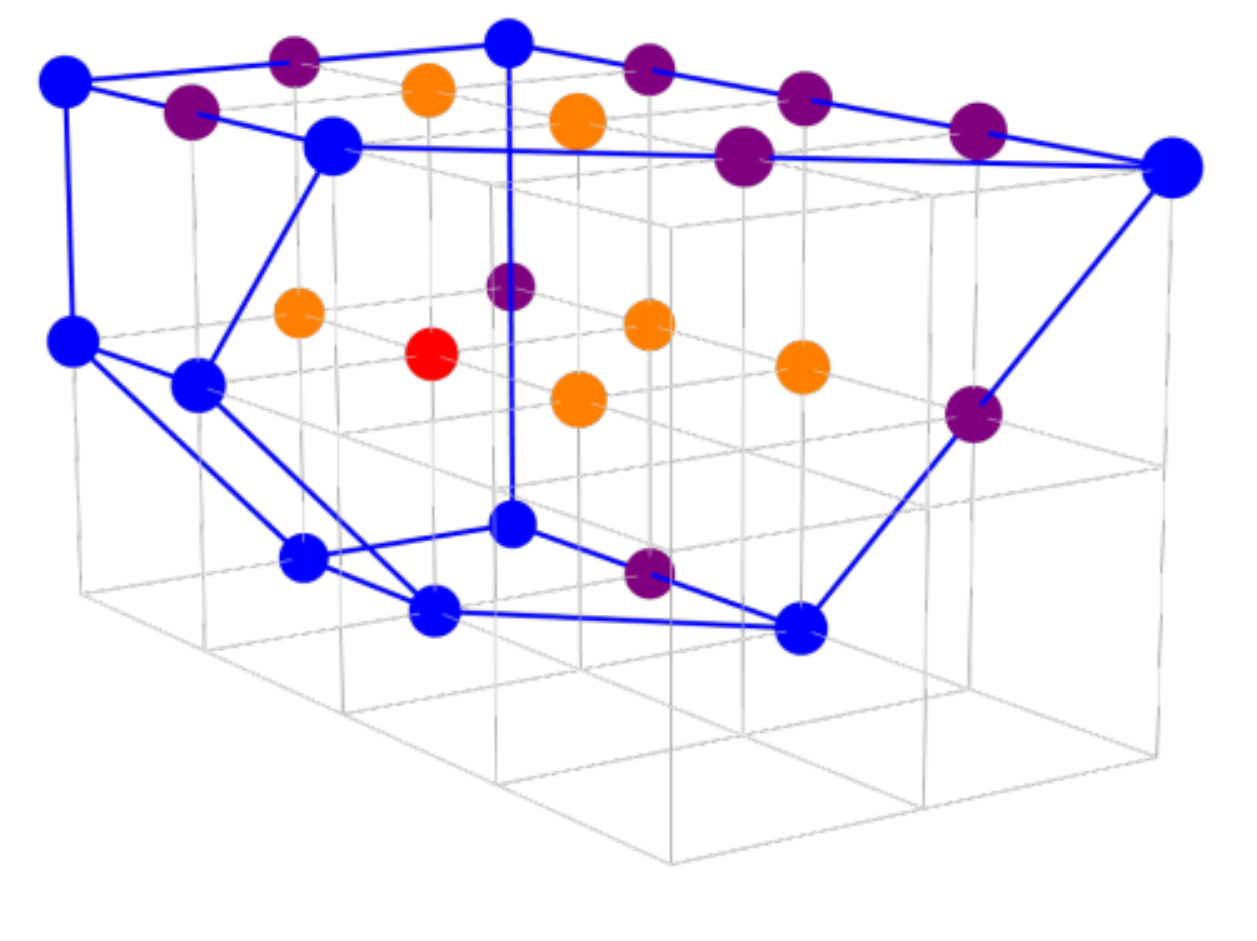} 
\end{minipage}
}
\caption{The generators and lattice of generators of the mesonic moduli space of Model 15 in terms of brick matchings with the corresponding flavor charges.
\label{f_genlattice_15}}
\end{table}

The highest weight form of the Hilbert series of Model 15 is
\beal{es1525}
&&
h_1(t, \mu, b_1, b_2; \mathcal{M}_{15}) =
\frac{1}{
(1 - \mu b_1^{-2} t^5) (1 - \mu b_2^{-2} t^5) (1 - \mu^2 b_1^2 b_2^{-2} t^7) (1 - \mu^2 b_1^{-2} b_2^2 t^7) 
}
\nn\\
&&
\hspace{0.5cm}
\times
\frac{1}{(1 - \mu^4 b_1^2 b_2^2 t^{11})}
\times
(
1 + \mu^2 t^7 + \mu^3 b_1^2 t^9 + \mu^3 b_2^2 t^9 - \mu^3 b_1^{-2} t^{12} - \mu^3 b_2^{-2} t^{12} - 2 \mu^4 t^{14} 
\nn\\
&&
\hspace{1cm}
- \mu^4 b_1^2 b_2^{-2} t^{14} - \mu^4 b_1^{-2} b_2^2 t^{14} - \mu^5 b_1^2 t^{16} - \mu^5 b_2^2 t^{16} + \mu^5 b_1^{-2} t^{19} 
+ \mu^5 b_2^{-2} t^{19} + \mu^6 t^{21}
\nn\\
&&
\hspace{1cm}
 + \mu^8 t^{28}
)
~,~
\eea
where $\mu^m \sim [m]_{SU(2)_x}$.
Here in highest weight form, the fugacity $\mu$ counts the highest weight of irreducible representations of $SU(2)_x$.
The fugacities $b_1$ and $b_2$ count charges under the two $U(1)$ factors of the mesonic flavor symmetry.

The plethystic logarithm of the Hilbert series takes the form
\beal{es1526}
&&
\PL[g_1(t, x,b_1,b_2; \mathcal{M}_{15})]=
([1] b_2^{-2}  + [1] b_1^{-2} )t^5
+ ([2] b_1^{2} b_2^{-2} + [2] + [2] b_1^{-2} b_2^{2} )t^7
\nn\\
&&
\hspace{1cm}
+ b_1^{-2} b_2^{-2} t^{10}
- [4] b_1^{2} b_2^{2} t^{11}
- ([3] b_2^{-2}  + [1] b_1^{2} b_2^{-4}  + [3] b_1^{-2}  + 2 [1] b_2^{-2}  
\nn\\
&&
\hspace{1cm}
+  2 [1] b_1^{-2}  + [1] b_1^{-4} b_2^{2} ) t^{12}
- ([4] b_1^{2} b_2^{-2}  + 3 [4]  +  2 [2] b_1^{2} b_2^{-2}  + b_1^{4} b_2^{-4}  +  [4] b_1^{-2} b_2^{2}  
\nn\\
&&
\hspace{1cm}
+ 3 [2]  +  b_1^{2} b_2^{-2}  + 2 [2] b_1^{-2} b_2^{2}  + 2  +  b_1^{-2} b_2^{2}  + b_1^{-4} b_2^{4} ) t^{14}
- (2 [5] b_1^{2}  + [3] b_1^{4} b_2^{-2}  
\nn\\
&&
\hspace{1cm}
+  2 [5] b_2^{2}  + 3 [3] b_1^{2}  +  [1] b_1^{4} b_2^{-2}  + 3 [3] b_2^{2}  +  2 [1] b_1^{2}  + [3] b_1^{-2} b_2^{4}  +  2 [1] b_2^{2}  
\nn\\
&&
\hspace{1cm}
+ [1] b_1^{-2} b_2^{4} )t^{16}
+ \dots ~,~ 
\eea
where $[m] = [m]_{SU(2)_x}$.
From the plethystic logarithm, we see that the mesonic moduli space is a non-complete intersection.

By using the following fugacity map
\beal{es1527}
&&
\tilde{t} = t_7~,~ 
\tilde{x} = \frac{t_7^2}{t_2^2}~,~ 
\tilde{b_1} =\frac{t_2 t_7}{t_4^2}~,~ 
\tilde{b_2} = \frac{t_2 t_7}{t_6^2}
~,~
\eea
where $t_1 t_2 = t_3 t_4 = t_5 t_6 = t_7^2$, 
the mesonic flavor charges on the gauge invariant operators become $\mathbb{Z}$-valued.
The generators in terms of brick matchings and their corresponding rescaled mesonic flavor charges are summarized in \tref{f_genlattice_15}.
The generator lattice as shown in \tref{f_genlattice_15} is a convex lattice polytope, which is reflexive. It is the dual of the toric diagram of Model 15 shown in \fref{f_toric_15}.
For completeness, \tref{f_genfields_15a} and \tref{f_genfields_15b} show the generators of Model 15 in terms of chiral fields with the corresponding mesonic flavor charges.

\begin{table}[H]
\centering
\resizebox{0.95\hsize}{!}{

}
\caption{The generators in terms of bifundamental chiral fields for Model 15 \bf{(Part 1)}.
\label{f_genfields_15a}}
\end{table}

\begin{table}[H]
\centering
\resizebox{0.95\hsize}{!}{
%
}
\caption{The generators in terms of bifundamental chiral fields for Model 15 \bf{(Part 2)}.
\label{f_genfields_15b}}
\end{table}

\section{Model 16: $P^{0}_{+-}(\text{dP}_2)$~[$\text{dP}_2 \times \mathbb{P}^1$,~$\langle84\rangle$] \label{smodel16}}

\begin{figure}[H]
\begin{center}
\resizebox{0.35\hsize}{!}{
\includegraphics[height=6cm]{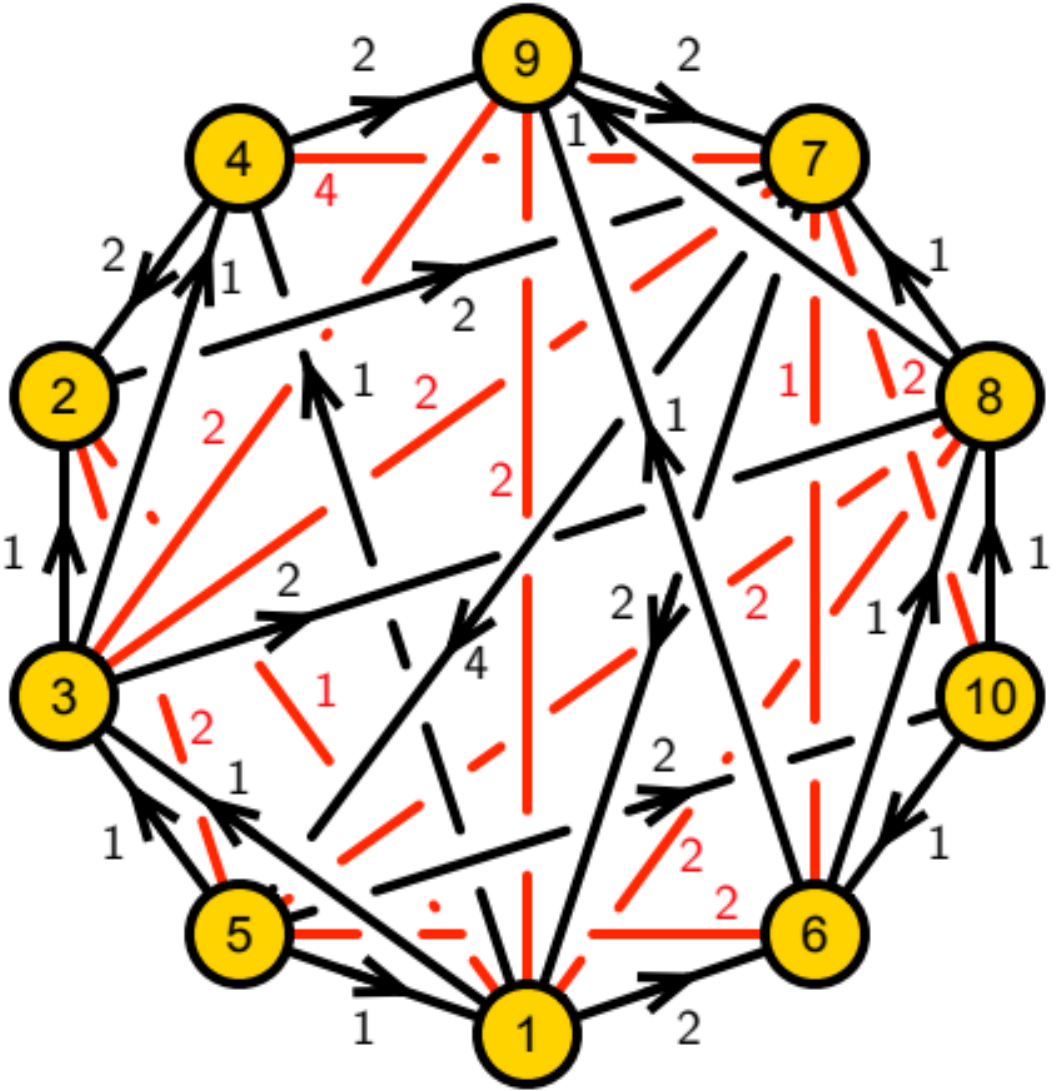} 
}
\caption{
Quiver for Model 16.
\label{f_quiver_16}}
 \end{center}
 \end{figure}

Model 16 corresponds to the toric Calabi-Yau 4-fold $P^{0}_{+-}(\text{dP}_2)$. 
The corresponding brane brick model has the quiver in \fref{f_quiver_16} and the $J$- and $E$-terms are 
\beq
{\footnotesize
 
 }~.~
\label{E_J_C_+-}
\eeq

The brick matchings are summarized in the $P$-matrix, which takes the form 
\beal{es1605}
{\footnotesize
P=
\resizebox{0.9\textwidth}{!}{$
\left(
%
\right)
$ }
}~.~
\nn\\
\eea

The $J$- and $E$-term charges are given by
\beal{es1605}
{\footnotesize
Q_{JE} =
\resizebox{0.87\textwidth}{!}{$
\left(
%
\right)
$}
}~,~
\nn\\
\eea
and the $D$-term charges are given by
\beal{es1605}
{\footnotesize
Q_D =
\resizebox{0.87\textwidth}{!}{$
\left(
%
\right)
$}
}~.~
\nn\\
\eea
 
\begin{figure}[ht!!]
\begin{center}
\resizebox{0.4\hsize}{!}{
\includegraphics[height=6cm]{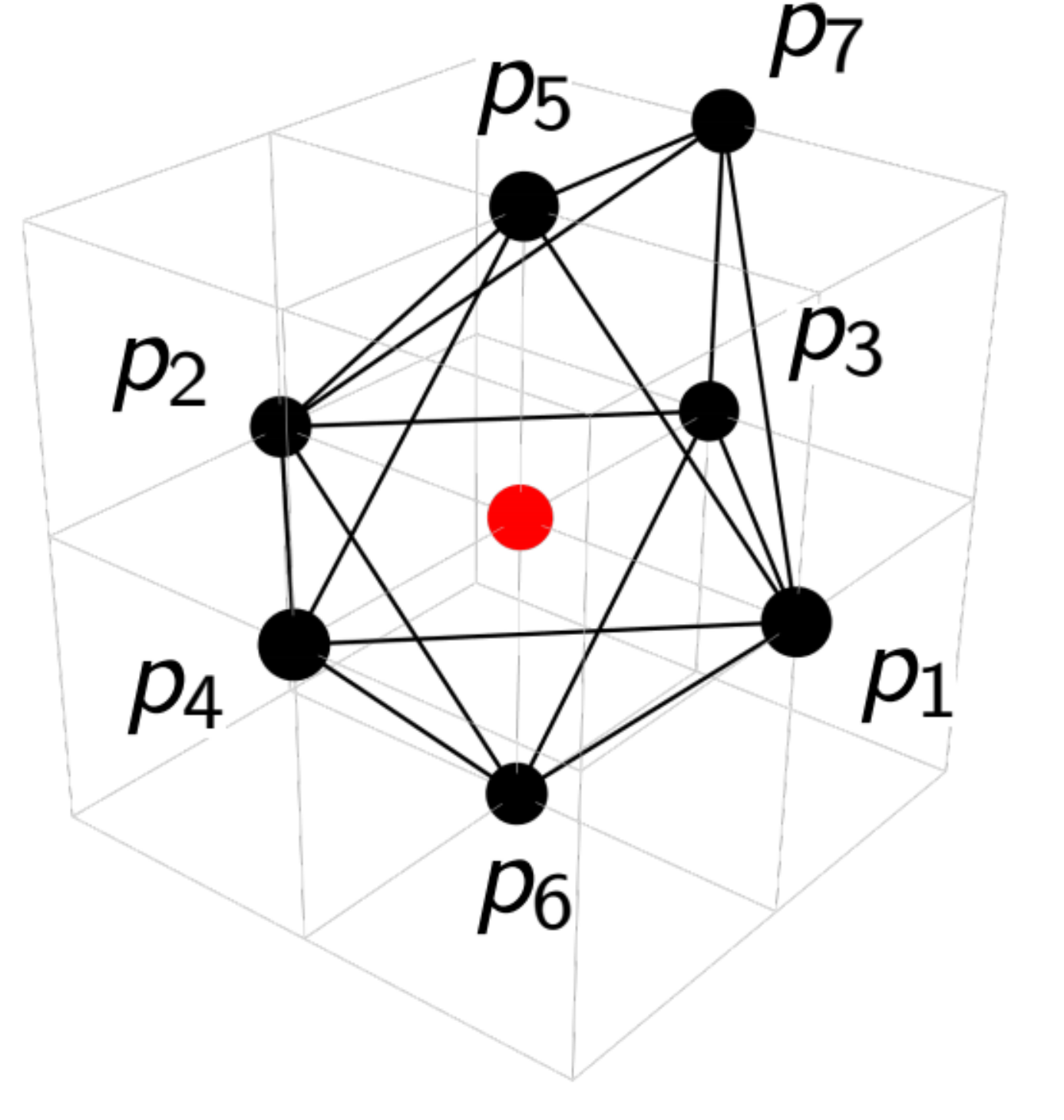} 
}
\caption{
Toric diagram for Model 16.
\label{f_toric_16}}
 \end{center}
 \end{figure}

The toric diagram of Model 16 is given by
\beal{es1610}
{\footnotesize
G_t=
\resizebox{0.87\textwidth}{!}{$
\left(
%
\right)
$}
}~,~
\nn\\
\eea
where \fref{f_toric_16} shows the toric diagram with brick matching labels.
 
The Hilbert series of the mesonic moduli space of Model 16 is
\beal{es1620}
&&
g_1(t_i,y_s,y_{o_1},y_{o_2},y_{o_3} ; \mathcal{M}_{16}) =
\frac{P(t_i,y_s,y_{o_1},y_{o_2},y_{o_3}; \mathcal{M}_{16})
}{
(1 - y_{s} y_{o_1} y_{o_2} y_{o_3}^2 t_1^2 t_4^2 t_5 t_6) 
(1 - y_{s} y_{o_1} y_{o_2} y_{o_3}^2 t_2^2 t_4^2 t_5 t_6) 
}
\nn\\
&&
\hspace{1cm} 
\times
\frac{1}{
(1 - y_{s} y_{o_1}^2 y_{o_2} y_{o_3} t_1^2 t_3 t_4 t_6^2) 
(1 - y_{s} y_{o_1}^2 y_{o_2} y_{o_3} t_2^2 t_3 t_4 t_6^2) 
(1 - y_{s} y_{o_1} y_{o_2}^2 y_{o_3}^3 t_1^2 t_4^2 t_5^2 t_7) 
} 
\nn\\
&&
\hspace{1cm} 
\times
\frac{1}{
(1 - y_{s} y_{o_1} y_{o_2}^2 y_{o_3}^3 t_2^2 t_4^2 t_5^2 t_7) 
(1 - y_{s} y_{o_1}^3 y_{o_2}^2 y_{o_3} t_1^2 t_3^2 t_6^2 t_7) 
(1 - y_{s} y_{o_1}^3 y_{o_2}^2 y_{o_3} t_2^2 t_3^2 t_6^2 t_7) 
} 
\nn\\
&&
\hspace{1cm} 
\times
\frac{1}{
(1 - y_{s} y_{o_1}^3 y_{o_2}^4 y_{o_3}^3 t_1^2 t_3^2 t_5^2 t_7^3) 
(1 - y_{s} y_{o_1}^3 y_{o_2}^4 y_{o_3}^3 t_2^2 t_3^2 t_5^2 t_7^3)
} 
~,~
\eea
where $t_i$ are the fugacities for the extremal brick matchings $p_i$.
Furthermore, $y_{s}$ counts $s_1 \dots s_{20}$, $y_{o_1}$ counts $o_1 \dots o_5$, $y_{o_2}$ counts $o_6 \dots o_{10}$ and $y_{o_3}$ counts $o_{11} \dots o_{15}$.
The explicit numerator $P(t_i,y_s,y_{o_1},y_{o_2},y_{o_3}; \mathcal{M}_{16})$ of the Hilbert series is given in the Appendix Section \sref{app_num_16}.
We note that setting the fugacities $y_{o_1}=1, \dots, y_{o_3}=1$ does not change the overall characterization of the mesonic moduli space by the Hilbert series, indicating that the extra GLSM fields, as expected, correspond to an over-parameterization of the moduli space. 

\begin{table}[H]
\centering
\begin{tabular}{|c|c|c|c|c|c|}
\hline
\; & $SU(2)_{x}$ & $U(1)_{b_1}$ &  $U(1)_{b_2}$ & $U(1)$ & \text{fugacity} \\
\hline
$p_1$ & +1 & 0 & 0 & $r_1$ & $t_1$ \\
$p_2$ & -1 & 0 & 0 & $r_2$ & $t_2$ \\
$p_3$ & 0 & +1  & 0 & $r_3$ & $t_3$ \\
$p_4$ & 0 & -1 & 0 & $r_4$ & $t_4$ \\
$p_5$ & 0 &0 & +1 & $r_5$ & $t_5$ \\
$p_6$ & 0 &0 &  -1 & $r_6$ & $t_6$ \\
$p_7$ & 0 &0 & 0 & $r_7$ & $t_7$ \\
\hline
\end{tabular}
\caption{Global symmetry charges on the extremal brick matchings $p_i$ of Model 16.}
\label{t_pmcharges_16}
\end{table}

By setting $t_i=t$ for the fugacities of the extremal brick matchings, and all other fugacities to $1$, the unrefined Hilbert series takes the following form
\beal{es1621}
&&
g_1(t,1,1,1,1,1; \mathcal{M}_{16}) =
\frac{
(1 - t)^2 (1 - t^3)^2
}{
(1 - t^6)^3 (1 - t^7)^2 (1 - t^9)^2
}
\times
(
1 + t + t^{2} + 3 t^{3} + 3 t^{4} + 3 t^{5} 
\nn\\
&&
\hspace{1cm}
+ 9 t^{6} + 16 t^{7} + 22 t^{8} + 33 t^{9} + 47 t^{10} + 59 t^{11} + 75 t^{12} + 83 t^{13} + 91 t^{14} + 105 t^{15} 
\nn\\
&&
\hspace{1cm}
+ 102 t^{16} + 104 t^{17} + 116 t^{18} + 104 t^{19} + 102 t^{20} + 105 t^{21} + 91 t^{22} + 83 t^{23} + 75 t^{24} 
\nn\\
&&
\hspace{1cm}
+ 59 t^{25} + 47 t^{26} + 33 t^{27} + 22 t^{28} + 16 t^{29} + 9 t^{30} + 3 t^{31} + 3 t^{32} + 3 t^{33} + t^{34} + t^{35} 
\nn\\
&&
\hspace{1cm}
+ t^{36}
)~,~
\eea
where the palindromic numerator indicates that the mesonic moduli space is Calabi-Yau. 

The global symmetry of Model 16 and the charges on the extremal brick matchings under the global symmetry are summarized in \tref{t_pmcharges_16}.
We can use the following fugacity map,
\beal{es1622}
&&
t = t_7 ~,~
x = \frac{t_7}{t_2} ~,~
b_1 = \frac{t_7}{t_4} ~,~
b_2 = \frac{t_5}{t_7} ~,~
\eea
where $t_1 t_2 = t_3 t_4 = t_5 t_6 = t_7^2$, in order to rewrite the 
Hilbert series for Model 16 in terms of characters of irreducible representations of $SU(2)\times U(1) \times U(1)$.

The highest weight form of the Hilbert series of Model 16 is
\beal{es1625}
&&
h_1(t, \mu, b_1, b_2; \mathcal{M}_{16}) =
\frac{1}{
(1 - \mu^2 b_1^{-2} t^6) (1 - \mu^2 b_2^{-2} t^6) (1 - \mu^2 b_1^2 b_2^{-2} t^7) (1 - \mu^2 b_1^{-2} b_2^2 t^7) 
}
\nn\\
&&
\hspace{0.5cm}
\times
\frac{1}{(1 - \mu^2 b_1^2 b_2^2 t^9)}
\times
(
1 + \mu^2 t^7 + \mu^2 b_1^2 t^8 + \mu^2 b_2^2 t^8 - \mu^4 b_1^{-2} t^{13} - \mu^4 b_2^{-2} t^{13} - 2 \mu^4 t^{14} 
\nn\\
&&
\hspace{1cm}
- \mu^4 b_1^2 b_2^{-2} t^{14} - \mu^4 b_1^{-2} b_2^2 t^{14} - \mu^4 b_1^2 t^{15} - \mu^4 b_2^2 t^{15} 
+ \mu^6 b_1^{-2} t^{20} + \mu^6 b_2^{-2} t^{20} + \mu^6 t^{21} 
\nn\\
&&
\hspace{1cm}
+ \mu^8 t^{28}
)~,~
\eea
where $\mu^m \sim [m]_{SU(2)_x}$.
Here in highest weight form, the fugacity $\mu$ counts the highest weight of irreducible representations of $SU(2)_x$.
The fugacities $b_1$ and $b_2$ count charges under the two $U(1)$ factors of the mesonic flavor symmetry.

The plethystic logarithm of the Hilbert series takes the form
\beal{es1626}
&&
\PL[g_1(t, x,b_1,b_2; \mathcal{M}_{16})]=
([2] b_2^{-2} + [2] b_1^{-2}) t^6
+([2] b_1^{2} b_2^{-2} + [2] + [2] b_1^{-2} b_2^{2}) t^7
\nn\\
&&
\hspace{1cm}
+ ([2] b_1^{2} + [2] b_2^{2} )t^8
+[2] b_1^{2} b_2^{2} t^9
- ([2] b_1^{-2} b_2^{-2}  + b_2^{-4}  + b_1^{-2} b_2^{-2}  + b_1^{-4} ) t^{12}
\nn\\
&&
\hspace{1cm}
- ([4] b_2^{-2}  + [2] b_1^{2} b_2^{-4}  +  [4] b_1^{-2}  + 2 [2] b_2^{-2}  +  b_1^{2} b_2^{-4}  + 2 [2] b_1^{-2}  + 2 b_2^{-2}  +  [2] b_1^{-4} b_2^{2}  
\nn\\
&&
\hspace{1cm}
+ 2 b_1^{-2}  + b_1^{-4} b_2^{2} ) t^{13}
- ([4] b_1^{2} b_2^{-2}  + 3 [4]  +  2 [2] b_1^{2} b_2^{-2}  + b_1^{4} b_2^{-4}  +  [4] b_1^{-2} b_2^{2}  + 3 [2]  
\nn\\
&&
\hspace{1cm}
+  2 b_1^{2} b_2^{-2}  + 2 [2] b_1^{-2} b_2^{2}  + 4  +  2 b_1^{-2} b_2^{2}  + b_1^{-4} b_2^{4} ) t^{14}
-(2 [4] b_1^{2}  + [2] b_1^{4} b_2^{-2}  +  2 [4] b_2^{2}  
\nn\\
&&
\hspace{1cm}
+ 3 [2] b_1^{2}  +  b_1^{4} b_2^{-2}  + 3 [2] b_2^{2}  + 3 b_1^{2}  +  [2] b_1^{-2} b_2^{4}  + 3 b_2^{2}  + b_1^{-2} b_2^{4} ) t^{15}
-([4] b_1^{4}  + [4] b_1^{2} b_2^{2}  
\nn\\
&&
\hspace{1cm}
+  [2] b_1^{4}  + [4] b_2^{4}  +  2 [2] b_1^{2} b_2^{2}  + 2 b_1^{4}  +  [2] b_2^{4}  + 2 b_1^{2} b_2^{2}  + 2 b_2^{4} ) t^{16}
- ([2] b_1^{4} b_2^{2}  + [2] b_1^{2} b_2^{4}  
\nn\\
&&
\hspace{1cm}
+ b_1^{4} b_2^{2}  + b_1^{2} b_2^{4} ) t^{17}
- ([2] b_1^{-2} b_2^{-4}  + [2] b_1^{-4} b_2^{-2}  + b_1^{-2} b_2^{-4}  + b_1^{-4} b_2^{-2} ) t^{18}
+ \dots ~,~
\nn\\
\eea
where $[m] = [m]_{SU(2)_x}$.
From the plethystic logarithm, we see that the mesonic moduli space is a non-complete intersection.

By using the following fugacity map
\beal{es1627}
&&
\tilde{t} = t_7~,~ 
\tilde{x} = \frac{t_7^2}{t_2^2}~,~ 
\tilde{b_1} =\frac{t_7^2}{t_4^2}~,~ 
\tilde{b_2} = \frac{t_7^2}{t_6^2}
~,~
\eea
where $t_1 t_2 = t_3 t_4 = t_5 t_6 = t_7^2$, 
the mesonic flavor charges on the gauge invariant operators become $\mathbb{Z}$-valued.
The generators in terms of brick matchings and their corresponding rescaled mesonic flavor charges are summarized in \tref{f_genlattice_16}.
The generator lattice as shown in \tref{f_genlattice_16} is a convex lattice polytope, which is reflexive. It is the dual of the toric diagram of Model 16 shown in \fref{f_toric_16}.
For completeness, \tref{f_genfields_16a} and \tref{f_genfields_16b} show the generators of Model 16 in terms of chiral fields with the corresponding mesonic flavor charges.

\begin{table}[H]
\centering
\resizebox{.95\hsize}{!}{
\begin{minipage}[!b]{0.5\textwidth}

\end{minipage}
\hspace{2cm}
\begin{minipage}[!b]{0.4\textwidth}
\includegraphics[height=7cm]{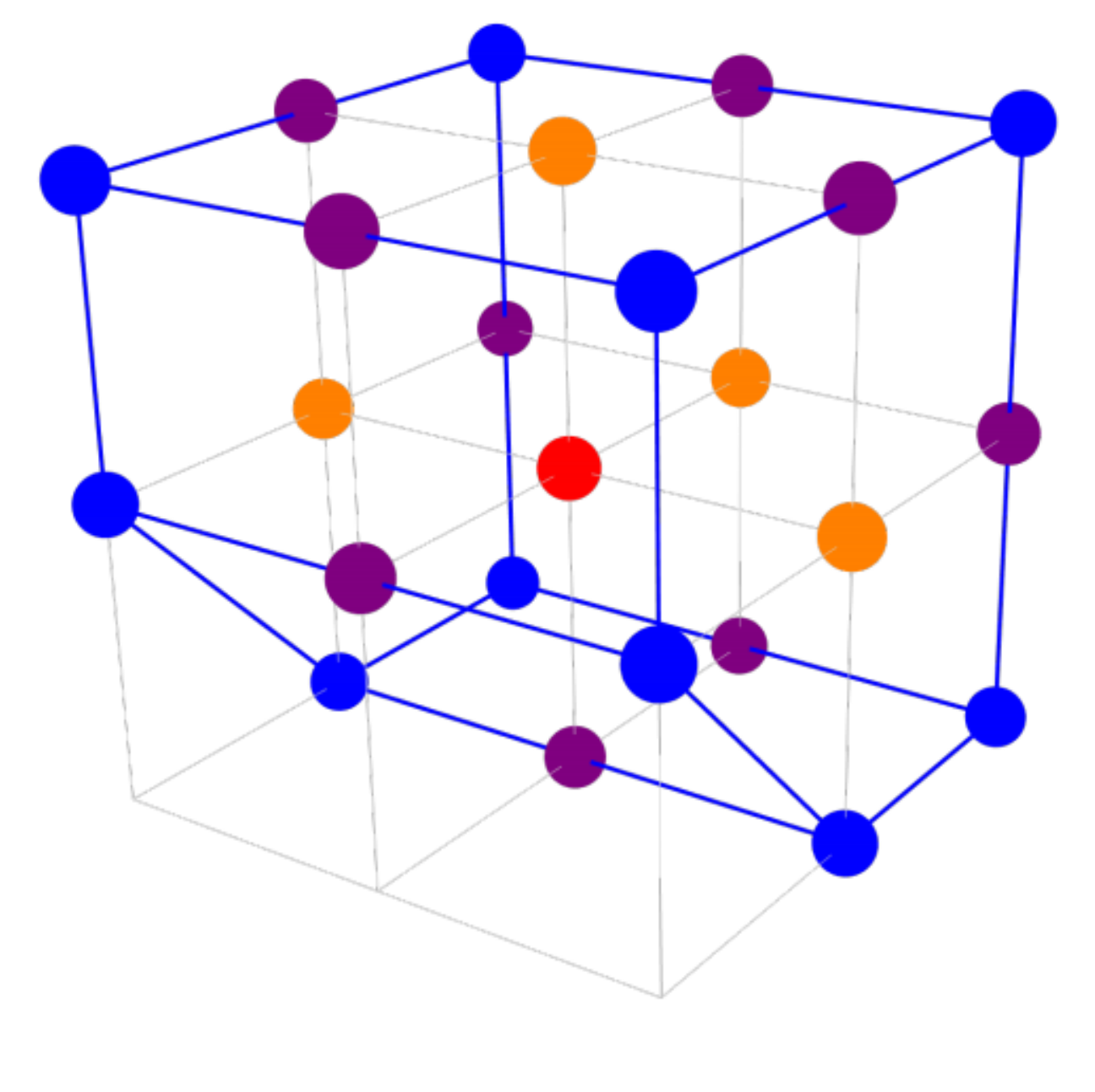} 
\end{minipage}
}
\caption{The generators and lattice of generators of the mesonic moduli space of Model 16 in terms of brick matchings with the corresponding flavor charges.
\label{f_genlattice_16}}
\end{table}

\begin{table}[H]
\centering
\resizebox{0.95\hsize}{!}{
%
}
\caption{The generators in terms of bifundamental chiral fields for Model 16 \bf{(Part 1)}.
\label{f_genfields_16a}}
\end{table}

\begin{table}[H]
\centering
\resizebox{0.95\hsize}{!}{
%
}
\caption{The generators in terms of bifundamental chiral fields for Model 16 \bf{(Part 2)}.
\label{f_genfields_16b}}
\end{table}

\section{Model 17: $P^{0}_{+-}(\text{dP}_3)$~[$\text{dP}_3 \times \mathbb{P}^1$,~$\langle218\rangle$] \label{smodel17}}

\begin{figure}[H]
\begin{center}
\resizebox{0.40\hsize}{!}{
\includegraphics[height=6cm]{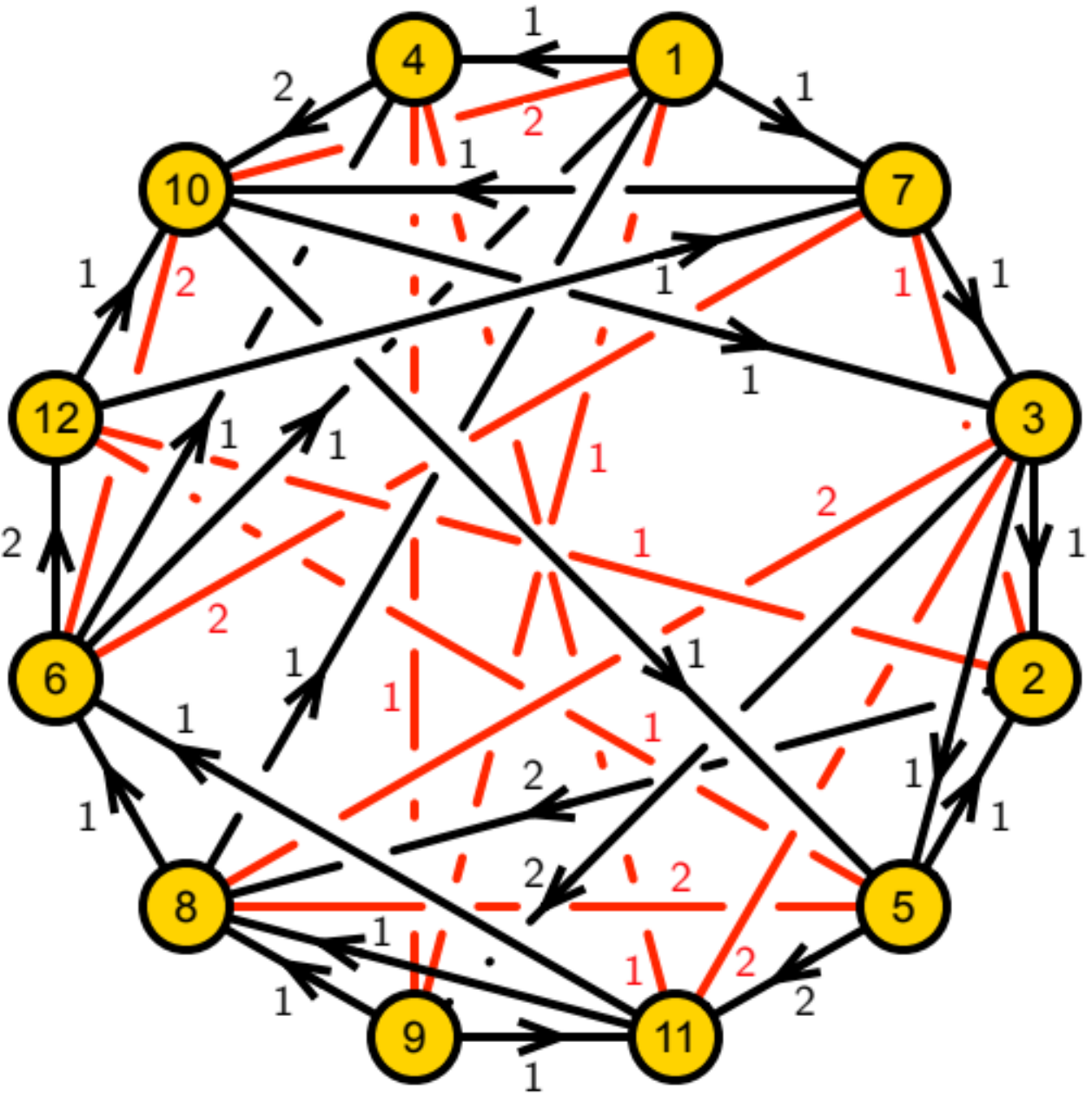} 
}
\caption{
Quiver for Model 17.
\label{f_quiver_17}}
 \end{center}
 \end{figure}

Model 17 corresponds to the toric Calabi-Yau 4-fold $P^{0}_{+-}(\text{dP}_3)$. 
The corresponding brane brick model has the quiver in \fref{f_quiver_17} and the $J$- and $E$-terms are 
\beq
{\footnotesize
 
}~.~
\label{E_J_C_+-}
\eeq

We can use the forward algorithm to obtain the brick matchings for Model 17.
The brick matchings are summarized in the $P$-matrix, which takes the form 

\beal{es1705}
{\footnotesize
P=
\resizebox{0.9\textwidth}{!}{$
\left(
%
\right)
$ }
}~.~
\nn\\
\eea

The $J$- and $E$-term charges are given by
\beal{es1705}
{\footnotesize
Q_{JE} =
\resizebox{0.87\textwidth}{!}{$
\left(
%
\right)
$}
}~,~
\nn\\
\eea
and the $D$-term charges are given by
\beal{es1705}
{\footnotesize
Q_D =
\resizebox{0.87\textwidth}{!}{$
\left(
%
\right)
$}
}~.~
\nn\\
\eea

\begin{figure}[ht!!]
\begin{center}
\resizebox{0.4\hsize}{!}{
\includegraphics[trim=0 0 0 0, height=6cm]{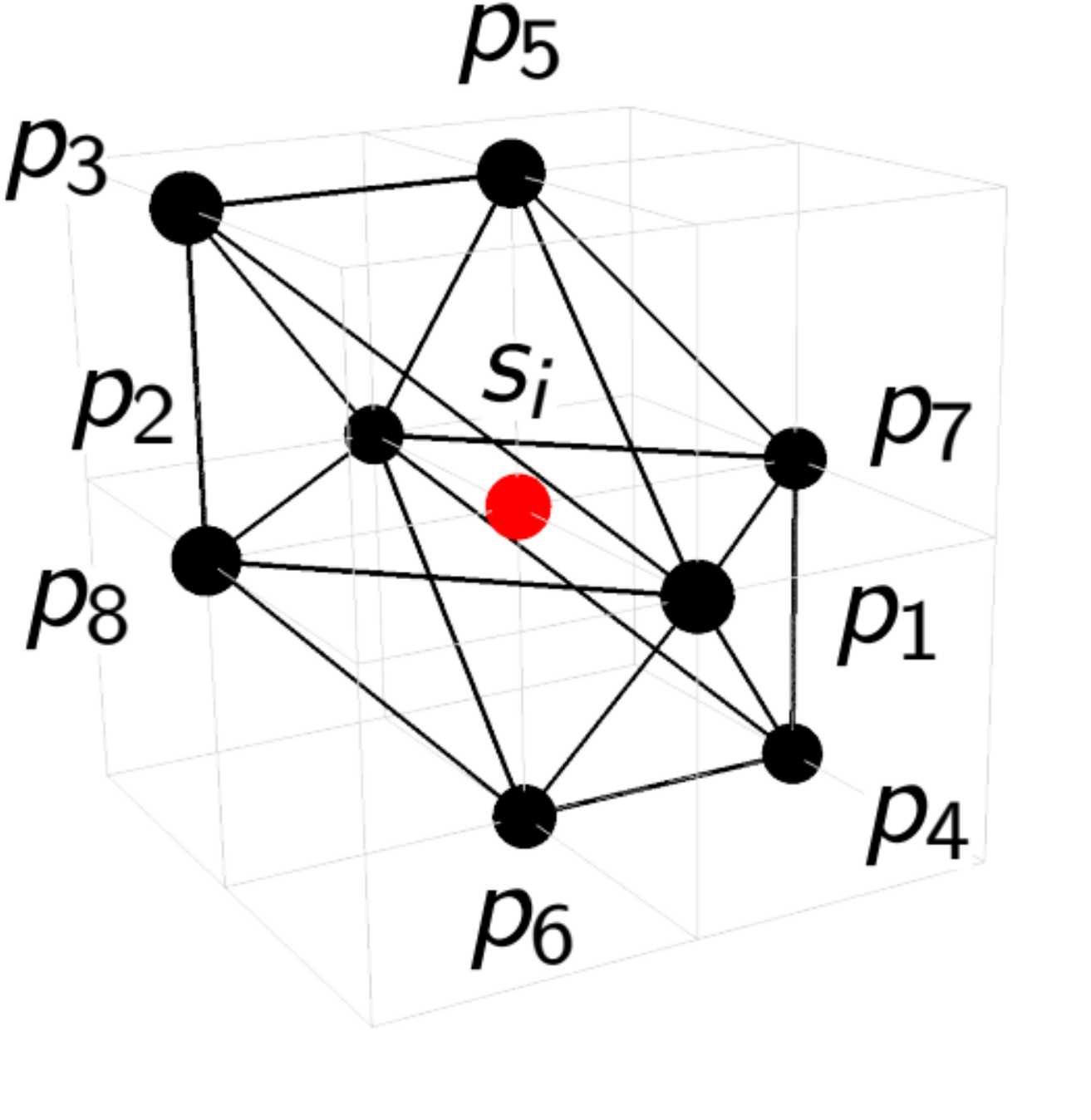} 
}
\caption{
Toric diagram for Model 17.
\label{f_toric_17}}
 \end{center}
 \end{figure}

The toric diagram of Model 17 is given by 
\beal{es1710}
{\footnotesize
G_t=
\resizebox{0.87\textwidth}{!}{$
\left(
%
\right)
$}
}~,~
\nn\\
\eea
where \fref{f_toric_17} shows the toric diagram with brick matching labels.

The Hilbert series of the mesonic moduli space of Model 17 is
\beal{es1320}
&&
g_1(t_i,y_s,y_{o_1},y_{o_2},y_{o_3},y_{o_4},y_{o_5},y_{o_6} ; \mathcal{M}_{17}) =
\frac{P(t_i,y_s,y_{o_1},y_{o_2},y_{o_3},y_{o_4},y_{o_5},y_{o_6}; \mathcal{M}_{17})
}{
(1 - y_{s} y_{o_1}^3 y_{o_2} y_{o_3}^3 y_{o_4} y_{o_5}^2 y_{o_6}^2 t_1^2 t_3 t_4 t_5^2 t_7^2) 
}
\nn\\
&&
\hspace{0.5cm} 
\times
\frac{1}{
(1 - y_{s} y_{o_1}^3 y_{o_2} y_{o_3}^3 y_{o_4} y_{o_5}^2 y_{o_6}^2 t_2^2 t_3 t_4 t_5^2 t_7^2) 
(1 - y_{s} y_{o_1}^3 y_{o_2} y_{o_3}^2 y_{o_4}^2 y_{o_5}^3 y_{o_6} t_1^2 t_4^2 t_5 t_6 t_7^2) 
} 
\nn\\
&&
\hspace{0.5cm} 
\times
\frac{1}{
(1 - y_{s} y_{o_1}^3 y_{o_2} y_{o_3}^2 y_{o_4}^2 y_{o_5}^3 y_{o_6} t_2^2 t_4^2 t_5 t_6 t_7^2) 
(1 - y_{s} y_{o_1}^2 y_{o_2}^2 y_{o_3}^3 y_{o_4} y_{o_5} y_{o_6}^3 t_1^2 t_3^2 t_5^2 t_7 t_8) 
} 
\nn\\
&&
\hspace{0.5cm} 
\times
\frac{1}{
(1 - y_{s} y_{o_1}^2 y_{o_2}^2 y_{o_3}^3 y_{o_4} y_{o_5} y_{o_6}^3 t_2^2 t_3^2 t_5^2 t_7 t_8) 
(1 - y_{s} y_{o_1}^2 y_{o_2}^2 y_{o_3} y_{o_4}^3 y_{o_5}^3 y_{o_6} t_1^2 t_4^2 t_6^2 t_7 t_8) 
} 
\nn\\
&&
\hspace{0.5cm} 
\times
\frac{1}{
(1 - y_{s} y_{o_1}^2 y_{o_2}^2 y_{o_3} y_{o_4}^3 y_{o_5}^3 y_{o_6} t_2^2 t_4^2 t_6^2 t_7 t_8) 
(1 - y_{s} y_{o_1} y_{o_2}^3 y_{o_3}^2 y_{o_4}^2 y_{o_5} y_{o_6}^3 t_1^2 t_3^2 t_5 t_6 t_8^2) 
}
\nn\\
&&
\hspace{0.5cm} 
\times
\frac{1}{
(1 - y_{s} y_{o_1} y_{o_2}^3 y_{o_3}^2 y_{o_4}^2 y_{o_5} y_{o_6}^3 t_2^2 t_3^2 t_5 t_6 t_8^2) 
(1 - y_{s} y_{o_1} y_{o_2}^3 y_{o_3} y_{o_4}^3 y_{o_5}^2 y_{o_6}^2 t_1^2 t_3 t_4 t_6^2 t_8^2) 
}
\nn\\
&&
\hspace{0.5cm} 
\times
\frac{1}{
(1 - y_{s} y_{o_1} y_{o_2}^3 y_{o_3} y_{o_4}^3 y_{o_5}^2 y_{o_6}^2 t_2^2 t_3 t_4 t_6^2 t_8^2)
}
~,~
\eea
where $t_i$ are the fugacities for the extremal brick matchings $p_i$.
Additionally, $y_{s}$ counts $s_1 \dots s_{18}$, $y_{o_1}$ counts $o_1 o_2 o_3$, $y_{o_2}$ counts $o_4 o_5 o_6$, $y_{o_3}$ counts $o_7 o_8 o_9$,$y_{o_4}$ counts $o_{10} o_{11} o_{12}$, $y_{o_5}$ counts $o_{13} o_{14} o_{15}$ and $y_{o_6}$ counts $o_{16} o_{17} o_{18}$.
The explicit numerator $P(t_i,y_s,y_{o_1},\dots,y_{o_6}; \mathcal{M}_{17})$ of the Hilbert series is given in the Appendix Section \sref{app_num_17}.
We note that setting the fugacities $y_{o_1}=1, \dots, y_{o_6}=1$ does not change the overall characterization of the mesonic moduli space by the Hilbert series, indicating that the extra GLSM fields, as expected, correspond to an over-parameterization of the moduli space. 

By setting $t_i=t$ for the fugacities of the extremal brick matchings, and all other fugacities to $1$, the unrefined Hilbert series takes the following form
\beal{es1721}
&&
g_1(t,1,1,1,1,1,1,1; \mathcal{M}_{17}) =
\frac{1 + 17 t^8 + 17 t^{16} + t^{24}}{(1 - t^8)^4} ~,~
\eea
where the palindromic numerator indicates that the mesonic moduli space is Calabi-Yau. 

\begin{table}[H]
\centering
\begin{tabular}{|c|c|c|c|c|c|}
\hline
\; & $SU(2)_{x}$ & $U(1)_{b_1}$ &  $U(1)_{b_2}$ & $U(1)$ & \text{fugacity} \\
\hline
$p_1$ & +1 & 0 & 0 & $r_1$ & $t_1$ \\
$p_2$ & -1 & 0 & 0 & $r_2$ & $t_2$ \\
$p_3$ & 0 & +1  & 0 & $r_3$ & $t_3$ \\
$p_4$ & 0 & -1 & 0 & $r_4$ & $t_4$ \\
$p_5$ & 0 &0 & +1 & $r_5$ & $t_5$ \\
$p_6$ & 0 &0 &  -1 & $r_6$ & $t_6$ \\
$p_7$ & 0 &0 & 0 & $r_7$ & $t_7$ \\
$p_8$ & 0 &0 & 0 & $r_8$ & $t_8$ \\
\hline
\end{tabular}
\caption{Global symmetry charges on the extremal brick matchings $p_i$ of Model 17.}
\label{t_pmcharges_17}
\end{table}

The global symmetry of Model 17 and the charges on the extremal brick matchings under the global symmetry are summarized in \tref{t_pmcharges_17}.
We can use the following fugacity map,
\beal{es1722}
&&
t = t_7 = t_8 ~,~
x = \frac{t_8}{t_2} ~,~
b_1 = \frac{t_8}{t_4} ~,~
b_2 = \frac{t_5}{t_8} ~,~
\eea
where $t_1 t_2 = t_3 t_4 = t_5 t_6 = t_8^2$, in order to rewrite the 
Hilbert series for Model 17 in terms of characters of irreducible representations of $SU(2)\times U(1) \times U(1)$.

\begin{table}[H]
\centering
\resizebox{.95\hsize}{!}{
\begin{minipage}[!b]{0.5\textwidth}
\begin{tabular}{|c|c|c|c|}
\hline
generator & $SU(2)_{\tilde{x}}$ & $U(1)_{\tilde{b_1}}$ & $U(1)_{\tilde{b_2}}$ \\
\hline
$ p_1^2 p_3 p_4 p_5^2 p_7^2  ~s o_1^3 o_2 o_3^3 o_4 o_5^2 o_6^2   $ &1& 0& 1\\
$ p_1 p_2 p_3 p_4 p_5^2 p_7^2  ~s o_1^3 o_2 o_3^3 o_4 o_5^2 o_6^2   $ &0& 0& 1\\
$ p_2^2 p_3 p_4 p_5^2 p_7^2  ~s o_1^3 o_2 o_3^3 o_4 o_5^2 o_6^2   $ &-1& 0& 1\\
$ p_1^2 p_4^2 p_5 p_6 p_7^2  ~s o_1^3 o_2 o_3^2 o_4^2 o_5^3 o_6   $ &1& -1& 0\\
$ p_1 p_2 p_4^2 p_5 p_6 p_7^2  ~s o_1^3 o_2 o_3^2 o_4^2 o_5^3 o_6   $ &0& -1& 0\\
$ p_2^2 p_4^2 p_5 p_6 p_7^2  ~s o_1^3 o_2 o_3^2 o_4^2 o_5^3 o_6   $ &-1& -1& 0\\
$ p_1^2 p_3^2 p_5^2 p_7 p_8  ~s o_1^2 o_2^2 o_3^3 o_4 o_5 o_6^3   $ &1& 1& 1\\
$ p_1 p_2 p_3^2 p_5^2 p_7 p_8  ~s o_1^2 o_2^2 o_3^3 o_4 o_5 o_6^3   $ &0& 1& 1\\
$ p_2^2 p_3^2 p_5^2 p_7 p_8  ~s o_1^2 o_2^2 o_3^3 o_4 o_5 o_6^3   $ &-1& 1& 1\\
$ p_1^2 p_3 p_4 p_5 p_6 p_7 p_8  ~s o_1^2 o_2^2 o_3^2 o_4^2 o_5^2 o_6^2   $ &1& 0& 0\\
$ p_1 p_2 p_3 p_4 p_5 p_6 p_7 p_8  ~s o_1^2 o_2^2 o_3^2 o_4^2 o_5^2 o_6^2   $ &0& 0& 0\\
$ p_2^2 p_3 p_4 p_5 p_6 p_7 p_8  ~s o_1^2 o_2^2 o_3^2 o_4^2 o_5^2 o_6^2   $ &-1& 0& 0\\
$ p_1^2 p_4^2 p_6^2 p_7 p_8  ~s o_1^2 o_2^2 o_3 o_4^3 o_5^3 o_6   $ &1& -1& -1\\
$ p_1 p_2 p_4^2 p_6^2 p_7 p_8  ~s o_1^2 o_2^2 o_3 o_4^3 o_5^3 o_6   $ &0& -1& -1\\
$ p_2^2 p_4^2 p_6^2 p_7 p_8  ~s o_1^2 o_2^2 o_3 o_4^3 o_5^3 o_6   $ &-1& -1& -1\\
$ p_1^2 p_3^2 p_5 p_6 p_8^2  ~s o_1 o_2^3 o_3^2 o_4^2 o_5 o_6^3   $ &1& 1& 0\\
$ p_1 p_2 p_3^2 p_5 p_6 p_8^2  ~s o_1 o_2^3 o_3^2 o_4^2 o_5 o_6^3   $ &0& 1& 0\\
$ p_2^2 p_3^2 p_5 p_6 p_8^2  ~s o_1 o_2^3 o_3^2 o_4^2 o_5 o_6^3   $ &-1& 1& 0\\
$ p_1^2 p_3 p_4 p_6^2 p_8^2  ~s o_1 o_2^3 o_3 o_4^3 o_5^2 o_6^2   $ &1& 0& -1\\
$ p_1 p_2 p_3 p_4 p_6^2 p_8^2  ~s o_1 o_2^3 o_3 o_4^3 o_5^2 o_6^2   $ &0& 0& -1\\
$ p_2^2 p_3 p_4 p_6^2 p_8^2  ~s o_1 o_2^3 o_3 o_4^3 o_5^2 o_6^2   $ &-1& 0& -1\\
\hline
\end{tabular}
\end{minipage}
\hspace{3cm}
\begin{minipage}[!b]{0.4\textwidth}
\includegraphics[height=7cm]{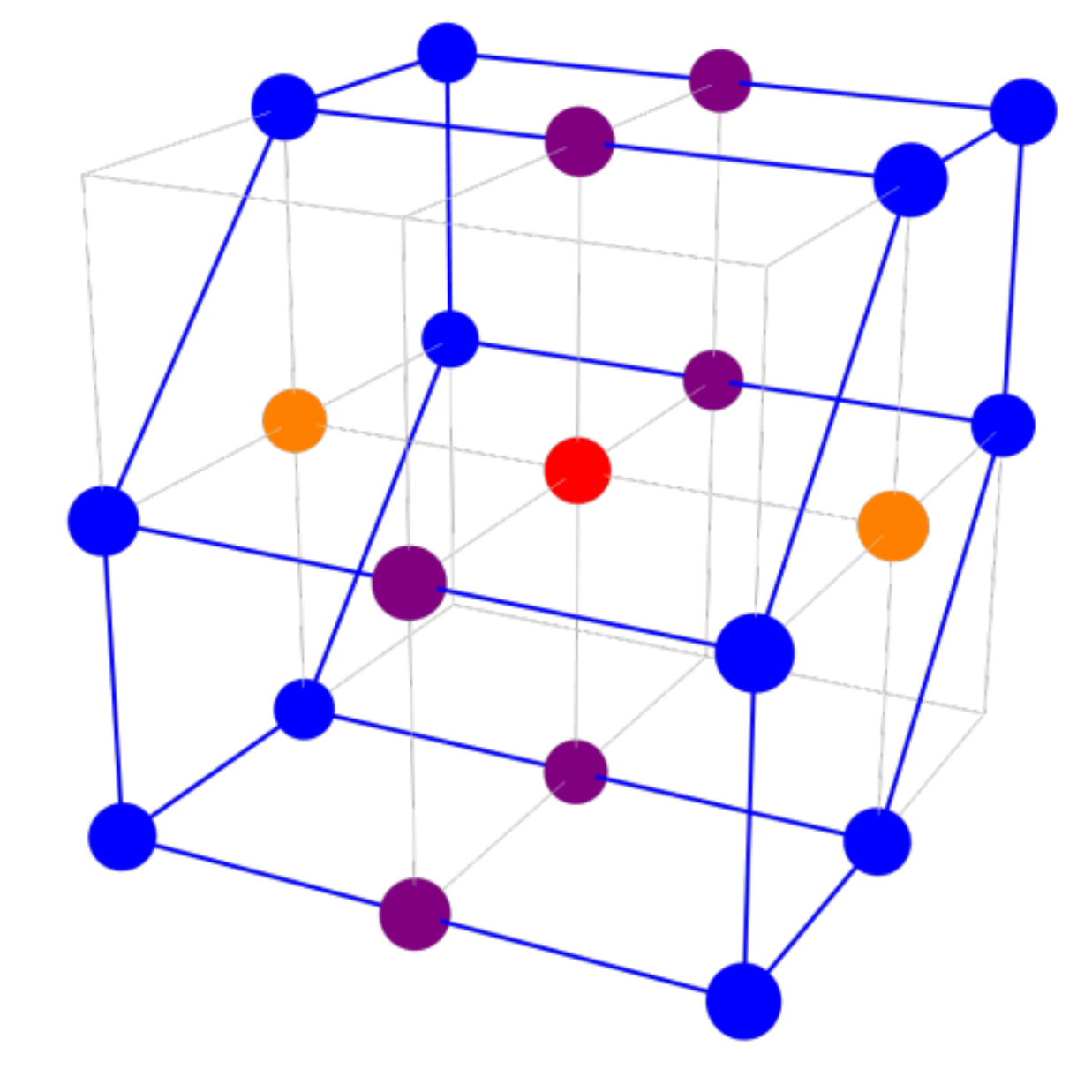} 
\end{minipage}
}
\caption{The generators and lattice of generators of the mesonic moduli space of Model 17 in terms of brick matchings with the corresponding flavor charges.
\label{f_genlattice_17}}
\end{table}

The highest weight form of the Hilbert series of Model 17 is
\beal{es1725}
&&
h_1(t, \mu, b_1, b_2; \mathcal{M}_{17}) =
\frac{1}{
(1 - \mu^2 b_1^{-2} t^8) (1 - \mu^2 b_2^{-2} t^8) (1 - \mu^2 b_1^{-2} b_2^{-2} t^8) (1 - b_1^2 \mu^2 t^8) 
}
\nn\\
&&
\hspace{0.5cm}
\times
\frac{1}{
(1 - b_2^2 \mu^2 t^8) (1 - b_1^2 b_2^2 \mu^2 t^8)
}
\times
(
1 + \mu^2 t^8 - 2 \mu^4 t^{16} - \mu^4 b_1^{-2} t^{16} - \mu^4 b_1^2 t^{16} - \mu^4 b_2^{-2} t^{16} 
\nn\\
&&
\hspace{1cm}
- \mu^4 b_1^{-2} b_2^{-2} t^{16} - \mu^4 b_2^2 t^{16} - \mu^4 b_1^2 b_2^2 t^{16} + 2 \mu^6 t^{24} + \mu^6 b_1^{-2} t^{24} + \mu^6 b_1^2 t^{24} + \mu^6 b_2^{-2} t^{24} 
\nn\\
&&
\hspace{1cm}
+ \mu^6 b_1^{-2} b_2^{-2} t^{24} + \mu^6 b_2^2 t^{24} + \mu^6 b_1^2 b_2^2 t^{24} - \mu^8 t^{32} - \mu^{10} t^{40}
)
~,~
\eea
where $\mu^m \sim [m]_{SU(2)_x}$.
Here in highest weight form, the fugacity $\mu$ counts the highest weight of irreducible representations of $SU(2)_x$.
The fugacities $b_1$ and $b_2$ count charges under the two $U(1)$ factors of the mesonic flavor symmetry.

The plethystic logarithm of the Hilbert series is
\beal{es1726}
&&
\PL[g_1(t, x, b_1, b_2; \mathcal{M}_{17})]=
([2] + [2] b_1^{-2} + [2] b_1^2 + [2] b_2^{-2} + [2] b_1^{-2} b_2^{-2} + [2] b_2^2 + [2] b_1^2 b_2^2 ) t^8
\nn\\
&&
\hspace{1cm}
- (b_1^{-4} + 4  + 2 b_1^{-2} + 2 b_1^2  +  b_1^4  + b_2^{-4} + b_1^{-4} b_2^{-4} +  b_1^{-2} b_2^{-4} + 2b_2^{-2} + b_1^{-4} b_2^{-2} + 2 b_1^{-2} b_2^{-2} 
\nn\\
&&
\hspace{1cm}
+ b_1^2 b_2^{-2} + 2 b_2^2  + b_1^{-2} b_2^2 +  2 b_1^2 b_2^2  + b_1^4 b_2^2  + b_2^4  +  b_1^2 b_2^4  + b_1^4 b_2^4  +  3 [2] + 2 [2] b_1^{-2} + 2 [2] b_1^2  
\nn\\
&&
\hspace{1cm}
+  [2] b_1^{-2} b_2^{-4} + 2 [2] b_2^{-2} +  [2]b_1^{-4} b_2^{-2} + 2 [2]b_1^{-2} b_2^{-2} + [2] b_1^2    b_2^{-2} + 2 [2]  b_2^2+ [2] b_1^{-2} b_2^2  
\nn\\
&&
\hspace{1cm}
+  2 [2] b_1^2 b_2^2  + [2] b_1^4 b_2^2  + [2] b_1^2 b_2^4  +  3 [4] + [4] b_1^{-2} + [4] b_1^2  + [4] b_2^{-2} +  [4]b_1^{-2} b_2^{-2} + [4] b_2^2  
\nn\\
&&
\hspace{1cm}
+  [4] b_1^2 b_2^2) t^{16}
+ \dots ~,~
\eea
where $[m]=[m]_{SU(2)_x}$.
From the plethystic logarithm, we see that the mesonic moduli space is a non-complete intersection.

By using the following fugacity map
\beal{es1727}
&&
\tilde{t} = t_7 = t_8~,~ 
\tilde{x} = \frac{t_8^2}{t_2^2}~,~ 
\tilde{b_1} =\frac{t_8^2}{t_4^2}~,~ 
\tilde{b_2} = \frac{t_8^2}{t_6^2}
~,~
\eea
where $t_1 t_2 = t_3 t_4 = t_5 t_6 = t_8^2$,
the mesonic flavor charges on the gauge invariant operators become $\mathbb{Z}$-valued.
The generators in terms of brick matchings and their corresponding rescaled mesonic flavor charges are summarized in \tref{f_genlattice_17}.
The generator lattice as shown in \tref{f_genlattice_17} is a convex lattice polytope, which is reflexive. It is the dual of the toric diagram of Model 17 shown in \fref{f_toric_17}.
For completeness, \tref{f_genfields_17a} and \tref{f_genfields_17b} show the generators of Model 17 in terms of chiral fields with the corresponding mesonic flavor charges.

\begin{table}[H]
\centering
\resizebox{0.9\hsize}{!}{

}
\caption{The generators in terms of bifundamental chiral fields for Model 17 \bf{(Part 1)}.
\label{f_genfields_17a}}
\end{table}

\begin{table}[H]
\centering
\resizebox{0.9\hsize}{!}{
%
}
\caption{The generators in terms of bifundamental chiral fields for Model 17 \bf{(Part 2)}.
\label{f_genfields_17b}}
\end{table}

\section{Model 18: $P^{1}_{+-}(\text{dP}_3)$~[$\text{dP}_3$ bundle of $\mathbb{P}^1$,~$\langle219\rangle$] \label{smodel18}}
 
\begin{figure}[H]
\begin{center}
\resizebox{0.40\hsize}{!}{
\includegraphics[height=6cm]{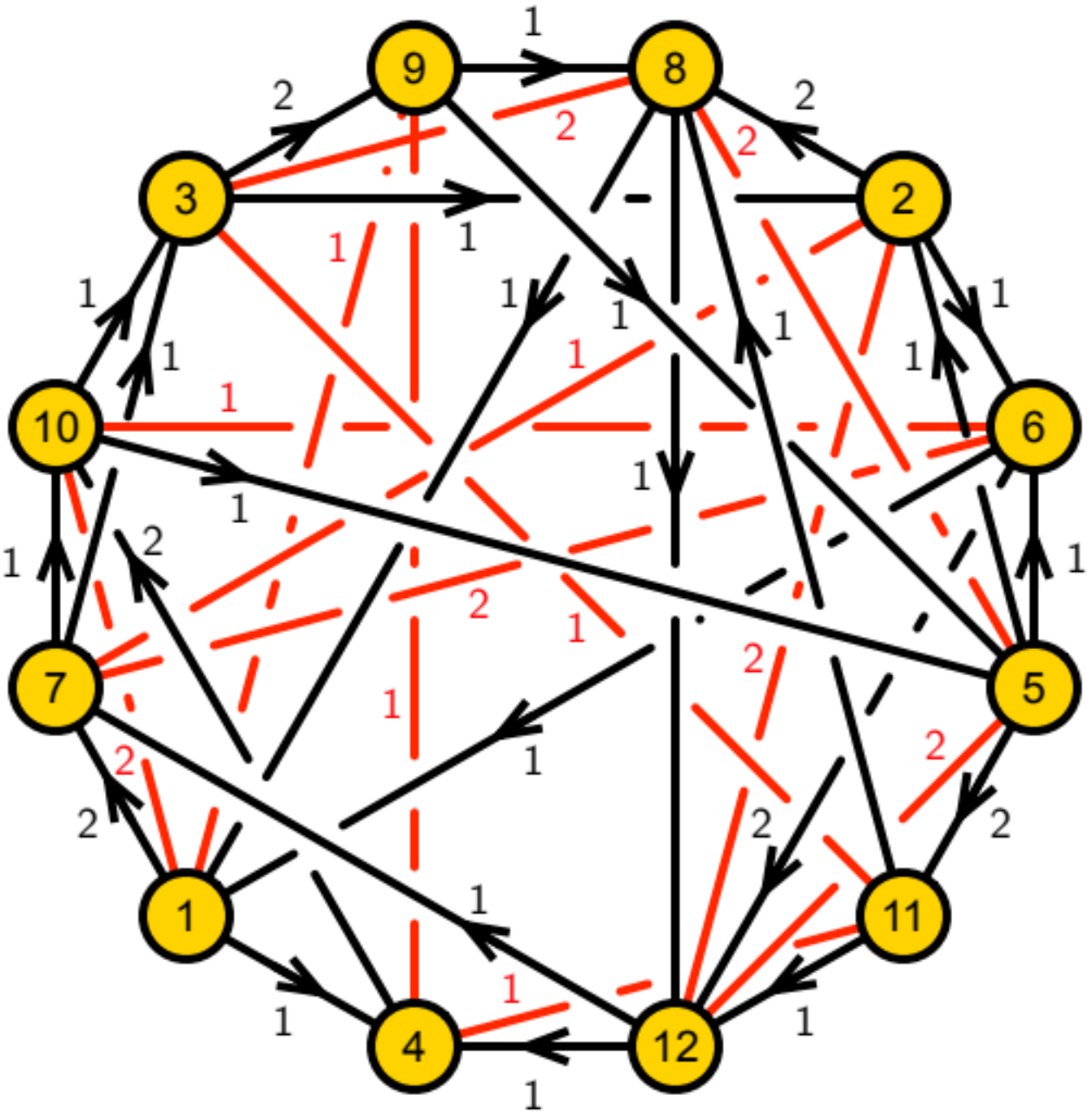} 
}
\caption{
Quiver for Model 18.
\label{f_quiver_18}}
 \end{center}
 \end{figure}

Model 18 corresponds to the toric Calabi-Yau 4-fold $P^{1}_{+-}(\text{dP}_3)$. 
The corresponding brane brick model has the quiver in \fref{f_quiver_18} and the $J$- and $E$-terms are 
\beq
{\footnotesize
 
}~.~
\label{E_J_C_+-}
\eeq

We can use the forward algorithm to obtain the brick matchings for Model 18.
The brick matchings are summarized in the $P$-matrix, which takes the form 

\beal{es1805}
{\footnotesize
P=
\resizebox{0.9\textwidth}{!}{$
\left(
%
\right)
$ }
}~.~
\nn\\
\eea

The $J$- and $E$-term charges are given by
\beal{es1805}
{\footnotesize
Q_{JE} =
\resizebox{0.87\textwidth}{!}{$
\left(
%
\right)
$}
}~,~
\nn\\
\eea
and the $D$-term charges are given by
\beal{es1805}
{\footnotesize
Q_D =
\resizebox{0.87\textwidth}{!}{$
\left(
%
\right)
$}
}~.~
\nn\\
\eea

\begin{figure}[ht!!]
\begin{center}
\resizebox{0.4\hsize}{!}{
\includegraphics[height=6cm]{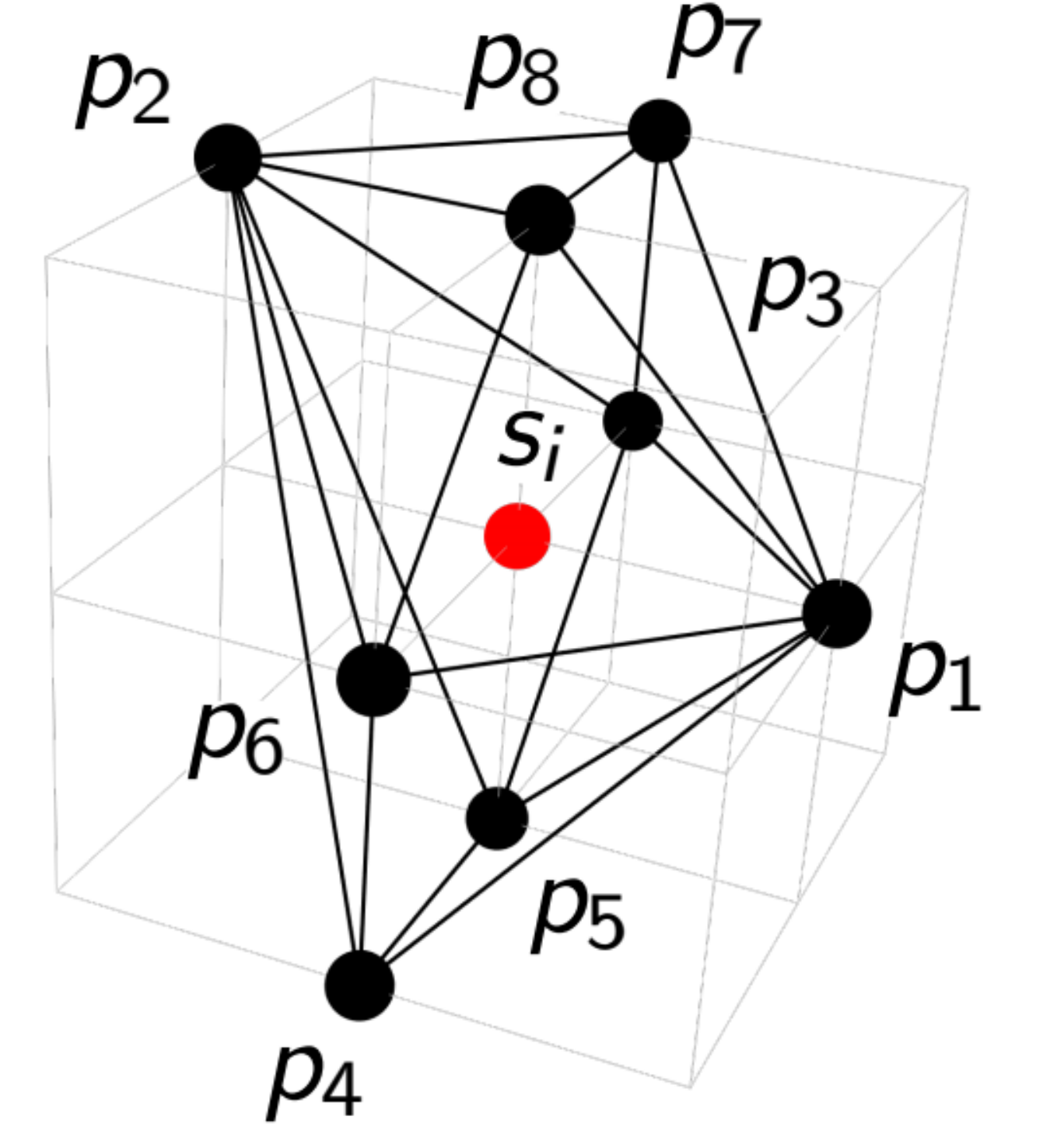} 
}
\caption{
Toric diagram for Model 18.
\label{f_toric_18}}
 \end{center}
 \end{figure}

The toric diagram of Model 18 is given by 
\beal{es1810}
{\footnotesize
G_t=
\resizebox{0.87\textwidth}{!}{$
\left(
%
\right)
$}
}~,~
\nn\\
\eea
where \fref{f_toric_18} shows the toric diagram with brick matching labels.
 
The Hilbert series of the mesonic moduli space of Model 18 is
\beal{es1820}
&&
g_1(t_i,y_s,y_{o_1},y_{o_2},y_{o_3},y_{o_4},y_{o_5},y_{o_6},y_{o_7} ; \mathcal{M}_{18}) =
\frac{P(t_i,y_s,y_{o_1},y_{o_2},y_{o_3},y_{o_4},y_{o_5},y_{o_6},y_{o_7}; \mathcal{M}_{18} )
}{
(1 - y_{s} y_{o_1} y_{o_2}^3 y_{o_3}^2 y_{o_4}^2 y_{o_5} y_{o_6}^3 y_{o_7}^2 t_1 t_3 t_4^2 t_5^2 t_6)
}
\nn\\
&&
\hspace{0.5cm} 
\times
\frac{1}{
(1 - y_{s} y_{o_1} y_{o_2}^3 y_{o_3}^2 y_{o_4}^2 y_{o_5} y_{o_6}^3 y_{o_7}^2 t_2 t_3 t_4^2 t_5^2 t_6)
(1 - y_{s} y_{o_1}^2 y_{o_2}^2 y_{o_3}^3 y_{o_4} y_{o_5} y_{o_6}^3 y_{o_7}^2 t_1 t_3^2 t_4 t_5^2 t_7)
} 
\nn\\
&&
\hspace{0.5cm} 
\times
\frac{1}{
(1 - y_{s} y_{o_1}^2 y_{o_2}^2 y_{o_3}^3 y_{o_4} y_{o_5} y_{o_6}^3 y_{o_7}^2 t_2 t_3^2 t_4 t_5^2 t_7)
(1 - y_{s} y_{o_1} y_{o_2}^3 y_{o_3} y_{o_4}^3 y_{o_5}^2 y_{o_6}^2 y_{o_7}^3 t_1^2 t_4^2 t_5 t_6^2 t_8)
} 
\nn\\
&&
\hspace{0.5cm} 
\times
\frac{1}{
(1 - y_{s} y_{o_1} y_{o_2}^3 y_{o_3} y_{o_4}^3 y_{o_5}^2 y_{o_6}^2 y_{o_7}^3 t_2^2 t_4^2 t_5 t_6^2 t_8)
(1 - y_{s} y_{o_1}^3 y_{o_2} y_{o_3}^3 y_{o_4} y_{o_5}^2 y_{o_6}^2 y_{o_7}^3 t_1^2 t_3^2 t_5 t_7^2 t_8)
} 
\nn\\
&&
\hspace{0.5cm} 
\times
\frac{1}{
(1 - y_{s} y_{o_1}^3 y_{o_2} y_{o_3}^3 y_{o_4} y_{o_5}^2 y_{o_6}^2 y_{o_7}^3 t_2^2 t_3^2 t_5 t_7^2 t_8)
(1 - y_{s} y_{o_1}^2 y_{o_2}^2 y_{o_3} y_{o_4}^3 y_{o_5}^3 y_{o_6} y_{o_7}^4 t_1^3 t_4 t_6^2 t_7 t_8^2)
}
\nn\\
&&
\hspace{0.5cm} 
\times
\frac{1}{
(1 - y_{s} y_{o_1}^2 y_{o_2}^2 y_{o_3} y_{o_4}^3 y_{o_5}^3 y_{o_6} y_{o_7}^4 t_2^3 t_4 t_6^2 t_7 t_8^2)
(1 - y_{s} y_{o_1}^3 y_{o_2} y_{o_3}^2 y_{o_4}^2 y_{o_5}^3 y_{o_6} y_{o_7}^4 t_1^3 t_3 t_6 t_7^2 t_8^2)
}
\nn\\
&&
\hspace{0.5cm} 
\times
\frac{1}{
(1 - y_{s} y_{o_1}^3 y_{o_2} y_{o_3}^2 y_{o_4}^2 y_{o_5}^3 y_{o_6} y_{o_7}^4 t_2^3 t_3 t_6 t_7^2 t_8^2)
}
~,~
\eea
where $t_i$ are the fugacities for the extremal brick matchings $p_i$.
Furthermore, we have the fugacity $y_s$ that counts the product brick matchings of the internal toric point $s_1 \dots s_{17}$.
The fugacities $y_{o_1}, \dots, y_{o_6}$, 
count products of extra GLSM fields, which are 
$o_1 \dots o_5$,
$o_6 \dots o_8$,
$o_9 \dots o_{10}$,
$o_{11} \dots o_{17}$,
$o_{18}\dots o_{34}$
and $o_{35}\dots o_{39}$, respectively. 
We note that setting the fugacities $y_{o_1}=1, \dots, y_{o_6}=1$ does not change the overall characterization of the mesonic moduli space by the Hilbert series, indicating that the extra GLSM fields, as expected, correspond to an over-parameterization of the moduli space. 

\begin{table}[H]
\centering
\begin{tabular}{|c|c|c|c|c|c|}
\hline
\; & $SU(2)_{x}$ & $U(1)_{b_1}$ &  $U(1)_{b_2}$ & $U(1)$ & \text{fugacity} \\
\hline
$p_1$ & +1 & 0 & 0 & $r_1$ & $t_1$ \\
$p_2$ & -1 & 0 & 0 & $r_2$ & $t_2$ \\
$p_3$ & 0 & +1  & 0 & $r_3$ & $t_3$ \\
$p_4$ & 0 & -1 & 0 & $r_4$ & $t_4$ \\
$p_5$ & 0 &0 & +1 & $r_5$ & $t_5$ \\
$p_6$ & 0 &0 &  -1 & $r_6$ & $t_6$ \\
$p_7$ & 0 &0 & 0 & $r_7$ & $t_7$ \\
$p_8$ & 0 &0 & 0 & $r_8$ & $t_8$ \\
\hline
\end{tabular}
\caption{Global symmetry charges on the extremal brick matchings $p_i$ of Model 18.}
\label{t_pmcharges_18}
\end{table}

By setting $t_i=t$ for the fugacities of the extremal brick matchings, and all other fugacities to $1$, the unrefined Hilbert series takes the following form
\beal{es1821}
&&
g_1(t,1,1,1,1,1,1,1,1; \mathcal{M}_{18}) =
\frac{
(1 - t)^4
}{
(1 - t^7)^3 (1 - t^8)^2 (1 - t^9)^3
}
\times
(
1 + 4 t + 10 t^{2} + 20 t^{3} 
\nn\\
&&
\hspace{1cm}
+ 35 t^{4} + 56 t^{5} + 84 t^{6} + 121 t^{7} + 176 t^{8} + 263 t^{9} + 396 t^{10} + 589 t^{11} + 856 t^{12} + 1211 t^{13} 
\nn\\
&&
\hspace{1cm}
+ 1668 t^{14} + 2228 t^{15} + 2873 t^{16} + 3572 t^{17} + 4294 t^{18} + 5008 t^{19} + 5683 t^{20} + 6288 t^{21} 
\nn\\
&&
\hspace{1cm}
+ 6796 t^{22} + 7188 t^{23} + 7468 t^{24} + 7636 t^{25} + 7692 t^{26} + 7636 t^{27} + 7468 t^{28} + 7188 t^{29} 
\nn\\
&&
\hspace{1cm}
+ 6796 t^{30} + 6288 t^{31} + 5683 t^{32} + 5008 t^{33} + 4294 t^{34} + 3572 t^{35} + 2873 t^{36} + 2228 t^{37} 
\nn\\
&&
\hspace{1cm}
+ 1668 t^{38} + 1211 t^{39} + 856 t^{40} + 589 t^{41} + 396 t^{42} + 263 t^{43} + 176 t^{44} + 121 t^{45} + 84 t^{46} 
\nn\\
&&
\hspace{1cm}
+ 56 t^{47} + 35 t^{48} + 20 t^{49} + 10 t^{50} + 4 t^{51} + t^{52}
)~,~
\eea
where the palindromic numerator indicates that the mesonic moduli space is Calabi-Yau. 

The global symmetry of Model 18 and the charges on the extremal brick matchings under the global symmetry are summarized in \tref{t_pmcharges_18}.
We can use the following fugacity map,
\beal{es1822}
&&
t = t_7 = t_8 ~,~
x = \frac{t_8}{t_2} ~,~
b_1 = \frac{t_8}{t_4} ~,~
b_2 = \frac{t_5}{t_8} ~,~
\eea
where $t_1 t_2 = t_3 t_4 = t_5 t_6 = t_8^2$, in order to rewrite the 
Hilbert series for Model 18 in terms of characters of irreducible representations of $SU(2)\times U(1) \times U(1)$.

The highest weight form of the Hilbert series of Model 18 is
\beal{es1825}
&&
h_1(t, \mu, b_1, b_2; \mathcal{M}_{18}) =
\frac{
1 - \mu^2 t^8
}{
(1 - \mu^2 b_1^{-2} t^8) (1 - \mu^2 b_2^{-2} t^8) (1 - \mu^2 b_1^{-2} b_2^{-2} t^8) 
(1 - \mu^2 b_1^2 t^8) 
}
\nn\\
&&
\hspace{0.5cm}
\times
\frac{1}{
(1 - \mu^2 b_2^2 t^8) (1 - \mu^2 b_1^2 b_2^2 t^8)
}
\times
(
1 + 2 \mu^2 t^8 - \mu^4 b_1^{-2} t^{16} - \mu^4 b_1^2 t^{16} - \mu^4 b_2^{-2} t^{16} 
\nn\\
&&
\hspace{1cm}
- \mu^4 b_1^{-2} b_2^{-2} t^{16} - \mu^4 b_2^2 t^{16} - \mu^4 b_1^2 b_2^2 t^{16} + 2 \mu^6 t^{24} + \mu^8 t^{32}
) ~,~
\eea
where $\mu^m \sim [m]_{SU(2)_x}$.
Here in highest weight form, the fugacity $\mu$ counts the highest weight of irreducible representations of $SU(2)_x$.
The fugacities $b_1$ and $b_2$ count charges under the two $U(1)$ factors of the mesonic flavor symmetry.

The plethystic logarithm of the Hilbert series is
\beal{es1826}
&&
\PL[g_1(t, x, b_1, b_2; \mathcal{M}_{18})]=
([1] b_1^{-1}b_2  + [1]b_1 b_2^2) t^7 
+([2] + [2] b_1^{-2} b_2^{-1} +  [2] b_1^2 b_2) t^8
\nn\\
&&
\hspace{1cm}
+([3] b_1^{-1} b_2^{-2} +  [3] b_1 b_2^{-1}) t^9
- b_2^3 t^{14}
- ([1] b_1^{-3} + 2 [1] b_1^{-1}b_2  + 2 [1] b_1 b_2^2
\nn\\
&&
\hspace{1cm}
+  [1] b_1^3 b_2^3  + [3]b_1^{-1} b_2  + [3] b_1 b_2^2) t^{15}
- (2 + b_1^{-4} b_2^{-2} + b_1^{-2} b_2^{-1} + b_1^2 b_2 + b_1^4 b_2^2 
\nn\\
&&
\hspace{1cm}
+ 3 [2] + 2 [2]b_1^{-2} b_2^{-1} + 2 [2] b_1^2 b_2  + 3 [4] + [4] b_1^{-2} b_2^{-1} + [4] b_1^2 b_2) t^{16}
\nn\\
&&
\hspace{1cm}
-( [1] b_1^3  +  [1] b_1^{-3} b_2^{-3} + 2  [1] b_1^{-1} b_2^{-2} + 2 [1] b_1  b_2^{-1} +   [3] b_1^3  +  [3] b_1^{-3} b_2^{-3} 
\nn\\
&&
\hspace{1cm}
+ 2  [3] b_1^{-1} b_2^{-2} + 2 [3] b_1    b_2^{-1} + [5] b_1^{-1} b_2^{-2} +   [5] b_1 b_2^{-1}) t^{17}
- (b_2^{-3} + [2] b_1^{-2} b_2^{-4} 
\nn\\
&&
\hspace{1cm}
+ [2] b_2^{-3} +   [2] b_1^2 b_2^{-2} +  [4] b_2^{-3}) t^{18}
\dots ~,~
\eea
where $[m]=[m]_{SU(2)_x}$.
From the plethystic logarithm, we see that the mesonic moduli space is a non-complete intersection.

By using the following fugacity map
\beal{es1827}
&&
\tilde{t} = t_7 = t_8 ~,~ 
\tilde{x} = \frac{t_8^2}{t_2^2}~,~ 
\tilde{b_1} =\frac{t_6 t_8}{t_2 t_4}~,~ 
\tilde{b_2} = \frac{t_2 t_8^2}{t_4 t_6^2}
~,~
\eea
where $t_1 t_2 = t_3 t_4 = t_5 t_6 = t_8^2$,
the mesonic flavor charges on the gauge invariant operators become $\mathbb{Z}$-valued.
The generators in terms of brick matchings and their corresponding rescaled mesonic flavor charges are summarized in \tref{f_genlattice_18}.
The generator lattice as shown in \tref{f_genlattice_18} is a convex lattice polytope, which is reflexive. It is the dual of the toric diagram of Model 18 shown in \fref{f_toric_18}.
For completeness, \tref{f_genfields_18a} and \tref{f_genfields_18b} show the generators of Model 18 in terms of chiral fields with the corresponding mesonic flavor charges.

\begin{table}[H]
\centering
\resizebox{.95\hsize}{!}{
\begin{minipage}[!b]{0.5\textwidth}

\end{minipage}
\hspace{3.8cm}
\begin{minipage}[!b]{0.4\textwidth}
\includegraphics[height=6cm]{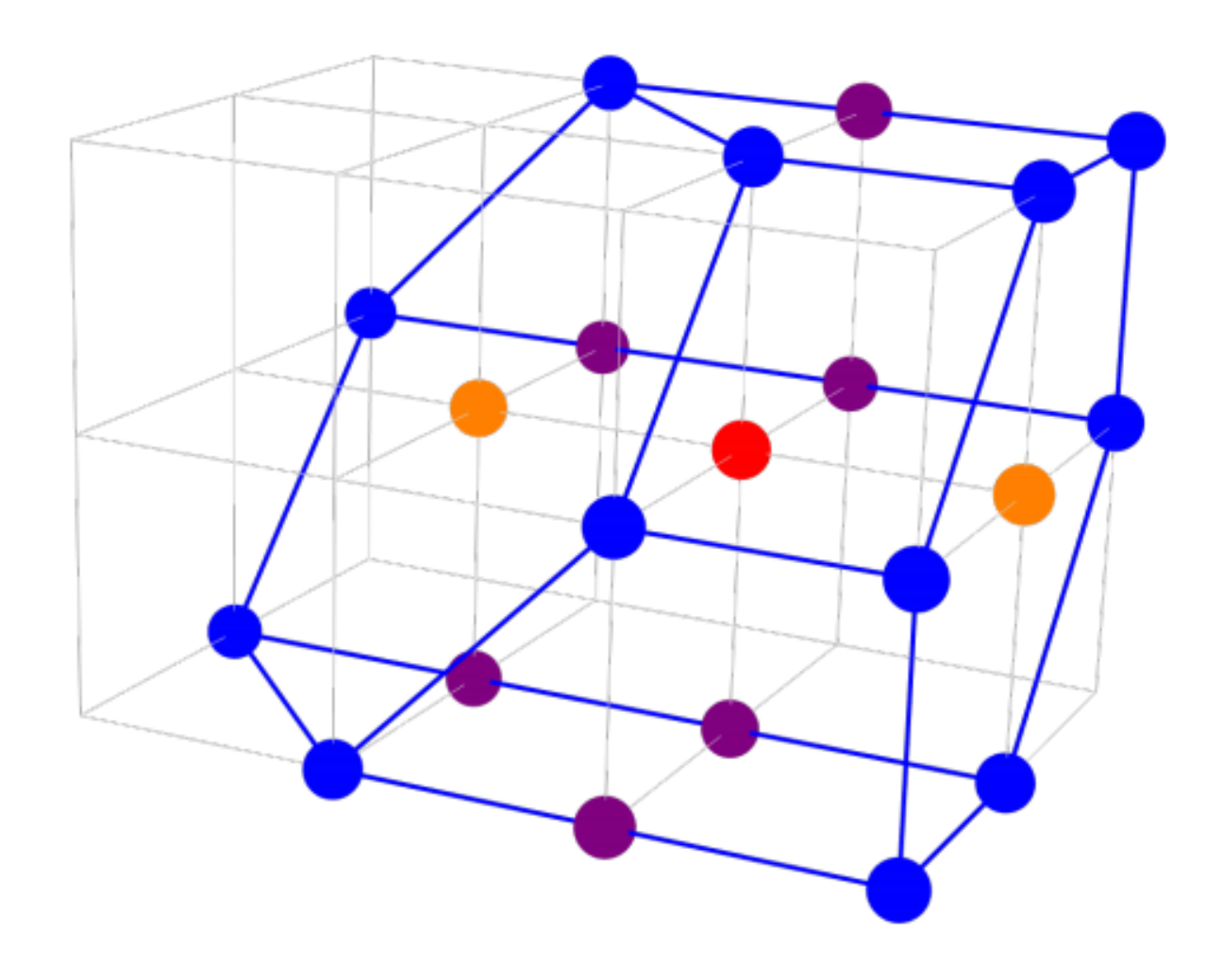} 
\end{minipage}
}
\caption{The generators and lattice of generators of the mesonic moduli space of Model 18 in terms of brick matchings with the corresponding flavor charges.
\label{f_genlattice_18}}
\end{table}

\begin{table}[H]
\centering
\resizebox{0.95\hsize}{!}{
%
}
\caption{The generators in terms of bifundamental chiral fields for Model 18 \bf{(Part 1)}.
\label{f_genfields_18a}}
\end{table}

\begin{table}[H]
\centering
\resizebox{0.95\hsize}{!}{
%
}
\caption{The generators in terms of bifundamental chiral fields for Model 18 \bf{(Part 2)}.
\label{f_genfields_18b}}
\end{table}

\section{Conclusions}

In this work, we have studied the 18 regular reflexive polytopes in dimension $3$ as toric diagrams of Calabi-Yau 4-folds corresponding to $2d$ $(0,2)$ supersymmetric gauge theories. These $2d$ quiver gauge theories are realized by brane brick models. It is natural to search for brane brick models for each of the Calabi-Yau 4-folds associated to smooth Fano 3-folds and to investigate what detailed information about these gauge theories can be extracted from the underlying geometries. Guided by these questions, the following comprehensive results have been presented in this work:
\begin{itemize}
\item
\textbf{Gauge Theory Identification.}
For each of the 18 regular reflexive polytopes in $3$ dimensions corresponding to toric Calabi-Yau 4-folds and smooth Fano 3-folds, we constructed a brane brick model realizing a $2d$ $(0,2)$ supersymmetric gauge theory.
These $2d$ theories are worldvolume theories of D1-branes probing the Calabi-Yau singularities. 

\item 
\textbf{Moduli Space Computation.}
The toric Calabi-Yau 4-folds are the mesonic moduli spaces of the $2d$ $(0,2)$ supersymmetric gauge theories realized by brane brick models.
For each model, the generating function of mesonic gauge invariant operators, the Hilbert series, was calculated using the Molien integral formula.
The fugacities of the Hilbert series can be chosen to count brick matchings of the brane brick model that correspond to GLSM fields as well as to points in the toric diagram of the Calabi-Yau 4-fold. Furthermore, the fugacities also refer to charges under the global symmetry of the $2d$ gauge theory.

\item
\textbf{Moduli Space Characterization.}
For the 18 brane brick models under consideration, we expressed the generators of the mesonic moduli space both in terms of chiral fields of the $2d$ gauge theory as well as brick matchings. This is the first time the geometry of all toric Calabi-Yau 4-folds and smooth Fano 3-folds corresponding to the 18 regular reflexive polytopes has been associated to the moduli space generators of supersymmetric gauge theories.  

\item 
\textbf{Reflexive Polytope Duality.}
The generators of the mesonic moduli space carry mesonic flavor symmetry charges.
The mesonic charges carried by the generators can be represented as points on a $\mathbb{Z}^3$ lattice, similar to the points of the toric diagram $\Delta_3$ of the Calabi-Yau 4-folds.
The convex hull of all such points forms a convex polytope, which we call as the generator lattice of the mesonic moduli space.
It turns out that for the brane brick models corresponding to the 18 toric Calabi-Yau 4-folds with regular reflexive polytopes as their toric diagrams, 
the generator lattice of the corresponding mesonic moduli spaces is another reflexive polytope.
In fact, this work verified for all 18 brane brick models that the generator lattice of the corresponding mesonic moduli spaces is the polar reflexive dual of the toric diagram $\Delta_3$.

\end{itemize} 

Overall, this work has considerably expanded our understanding of the correspondence between $2d$ $(0,2)$ supersymmetric gauge theories, brane brick models and toric Calabi-Yau 4-folds. At the same time, it opens several avenues of future inquiry. We expect to be able to report on further related developments in the near future. 
\\

\section*{Acknowledgements}

R.K.-S. would like to thank Amihay Hanany, Yang-Hui He and Shing-Tung Yau for collaborations and discussions on related work that gave the inspiration to start this project. 
S.F. and R.K.-S. are also grateful to Dongwook Ghim for related ongoing collaborations. S.F. is supported by the U.S. National Science Foundation grants PHY-1820721, PHY-2112729 and DMS-1854179. R.K.-S.  is supported by the New Faculty Start-up Research Grant (1.210139.01) of the Ulsan National Institute of Science and Technology. He is also supported by the BK21 Program (``Next Generation Education Program for Mathematical Sciences'', 4299990414089) funded by the Ministry of Education (MOE) in Korea and the National Research Foundation of Korea (NRF).
\\

\addtocontents{toc}{\protect\setcounter{tocdepth}{1}}
\appendix

\section{Numerators for the fully refined Hilbert Series}

\subsection{Model 1 \label{app_num_01}}

\begingroup\makeatletter\def\f@size{7}\check@mathfonts
\begin{quote}\raggedright
$
P(t_i,y_s; \mathcal{M}_1) =
1 + y_s t_1^3 t_2 + y_s t_1^2 t_2^2 
+ y_s t_1 t_2^3  + y_s t_1^3 t_3 + y_s t_1^2 t_2 t_3 + y_s t_1 t_2^2 t_3 + y_s t_2^3 t_3 + y_s t_1^2 t_3^2 +  y_s t_1 t_2 t_3^2 + y_s t_2^2 t_3^2 
+ y_s^2 t_1^3 t_2^3 t_3^2 + y_s t_1 t_3^3 +  y_s t_2 t_3^3 + y_s^2 t_1^3 t_2^2 t_3^3 + y_s^2 t_1^2 t_2^3 t_3^3 + y_s t_1^3 t_4 +  y_s t_1^2 t_2 t_4 
+ y_s t_1 t_2^2 t_4 + y_s t_2^3 t_4 + y_s t_1^2 t_3 t_4 +  y_s t_1 t_2 t_3 t_4 + y_s t_2^2 t_3 t_4 + y_s^2 t_1^3 t_2^3 t_3 t_4 +  y_s t_1 t_3^2 t_4 
+ y_s t_2 t_3^2 t_4 + y_s^2 t_1^3 t_2^2 t_3^2 t_4 +  y_s^2 t_1^2 t_2^3 t_3^2 t_4 + y_s t_3^3 t_4 + y_s^2 t_1^3 t_2 t_3^3 t_4 +  y_s^2 t_1^2 t_2^2 t_3^3 t_4 
+ y_s^2 t_1 t_2^3 t_3^3 t_4 + y_s t_1^2 t_4^2 +  y_s t_1 t_2 t_4^2 + y_s t_2^2 t_4^2 + y_s^2 t_1^3 t_2^3 t_4^2 + y_s t_1 t_3 t_4^2 +  y_s t_2 t_3 t_4^2 
+ y_s^2 t_1^3 t_2^2 t_3 t_4^2 + y_s^2 t_1^2 t_2^3 t_3 t_4^2 +  y_s t_3^2 t_4^2 + y_s^2 t_1^3 t_2 t_3^2 t_4^2 + y_s^2 t_1^2 t_2^2 t_3^2 t_4^2 
+  y_s^2 t_1 t_2^3 t_3^2 t_4^2 + y_s^2 t_1^3 t_3^3 t_4^2 +  y_s^2 t_1^2 t_2 t_3^3 t_4^2 + y_s^2 t_1 t_2^2 t_3^3 t_4^2 +  y_s^2 t_2^3 t_3^3 t_4^2 
+ y_s t_1 t_4^3 + y_s t_2 t_4^3 + y_s^2 t_1^3 t_2^2 t_4^3 +  y_s^2 t_1^2 t_2^3 t_4^3 + y_s t_3 t_4^3 + y_s^2 t_1^3 t_2 t_3 t_4^3 +  y_s^2 t_1^2 t_2^2 t_3 t_4^3 
+ y_s^2 t_1 t_2^3 t_3 t_4^3 + y_s^2 t_1^3 t_3^2 t_4^3 +  y_s^2 t_1^2 t_2 t_3^2 t_4^3 + y_s^2 t_1 t_2^2 t_3^2 t_4^3 +  y_s^2 t_2^3 t_3^2 t_4^3 
+ y_s^2 t_1^2 t_3^3 t_4^3 + y_s^2 t_1 t_2 t_3^3 t_4^3 +  y_s^2 t_2^2 t_3^3 t_4^3 + y_s^3 t_1^3 t_2^3 t_3^3 t_4^3
~,~
$
\end{quote}
\endgroup

\subsection{Model 2 \label{app_num_02}}

\begingroup\makeatletter\def\f@size{7}\check@mathfonts
\begin{quote}\raggedright
$
P(t_i, y_s; \mathcal{M}_2) =
 1 + y_s t_1 t_2 t_3^3 + y_s t_1^2 t_3^2 t_4 + y_s t_1 t_2 t_3^2 t_4 +  y_s t_2^2 t_3^2 t_4 - y_s^2 t_1^2 t_2^2 t_3^5 t_4 + y_s t_1^2 t_3 t_4^2+  y_s t_1 t_2 t_3 t_4^2 + y_s t_2^2 t_3 t_4^2 - y_s^2 t_1^2 t_2^2 t_3^4 t_4^2 +  y_s t_1 t_2 t_4^3 - y_s^2 t_1^3 t_2 t_3^3 t_4^3 - y_s^2 t_1^2 t_2^2 t_3^3 t_4^3 -  y_s^2 t_1 t_2^3 t_3^3 t_4^3  - y_s^2 t_1^2 t_2^2 t_3^2 t_4^4 -  y_s^3 t_1^3 t_2^3 t_3^5 t_4^4 - y_s^2 t_1^2 t_2^2 t_3 t_4^5 -  y_s^3 t_1^3 t_2^3 t_3^4 t_4^5 + y_s t_1^2 t_3^2 t_5 + y_s t_1 t_2 t_3^2 t_5+  y_s t_2^2 t_3^2 t_5 - y_s^2 t_1^2 t_2^2 t_3^5 t_5 + y_s t_1^2 t_3 t_4 t_5 +  y_s t_1 t_2 t_3 t_4 t_5 + y_s t_2^2 t_3 t_4 t_5 - y_s^2 t_1^2 t_2^2 t_3^4 t_4 t_5 +  y_s t_1^2 t_4^2 t_5 + y_s t_1 t_2 t_4^2 t_5 + y_s t_2^2 t_4^2 t_5 -  y_s^2 t_1^2 t_2^2 t_3^3 t_4^2 t_5 - y_s^2 t_1^2 t_2^2 t_3^2 t_4^3 t_5 -  y_s^3 t_1^3 t_2^3 t_3^5 t_4^3 t_5 - y_s^2 t_1^2 t_2^2 t_3 t_4^4 t_5  -  y_s^3 t_1^3 t_2^3 t_3^4 t_4^4 t_5 - y_s^2 t_1^2 t_2^2 t_4^5 t_5 -  y_s^3 t_1^3 t_2^3 t_3^3 t_4^5 t_5 + y_s t_1^2 t_3 t_5^2 + y_s t_1 t_2 t_3 t_5^2 +  y_s t_2^2 t_3 t_5^2 - y_s^2 t_1^2 t_2^2 t_3^4 t_5^2 + y_s t_1^2 t_4 t_5^2 +  y_s t_1 t_2 t_4 t_5^2 + y_s t_2^2 t_4 t_5^2 - y_s^2 t_1^2 t_2^2 t_3^3 t_4 t_5^2 +  y_s^2 t_1^4 t_3^2 t_4^2 t_5^2 + y_s^2 t_1^3 t_2 t_3^2 t_4^2 t_5^2 +  y_s^2 t_1^2 t_2^2 t_3^2 t_4^2 t_5^2 + y_s^2 t_1 t_2^3 t_3^2 t_4^2 t_5^2 +  y_s^2 t_2^4 t_3^2 t_4^2 t_5^2 - y_s^3 t_1^4 t_2^2 t_3^5 t_4^2 t_5^2 -  y_s^3 t_1^3 t_2^3 t_3^5 t_4^2 t_5^2 - y_s^3 t_1^2 t_2^4 t_3^5 t_4^2 t_5^2 -  y_s^2 t_1^2 t_2^2 t_3 t_4^3 t_5^2 - y_s^3 t_1^3 t_2^3 t_3^4 t_4^3 t_5^2 -  y_s^2 t_1^2 t_2^2 t_4^4 t_5^2  - y_s^3 t_1^3 t_2^3 t_3^3 t_4^4 t_5^2 -  y_s^3 t_1^4 t_2^2 t_3^2 t_4^5 t_5^2 - y_s^3 t_1^3 t_2^3 t_3^2 t_4^5 t_5^2 -  y_s^3 t_1^2 t_2^4 t_3^2 t_4^5 t_5^2 + y_s^4 t_1^4 t_2^4 t_3^5 t_4^5 t_5^2 +  y_s t_1 t_2 t_5^3 - y_s^2 t_1^3 t_2 t_3^3 t_5^3 - y_s^2 t_1^2 t_2^2 t_3^3 t_5^3 -  y_s^2 t_1 t_2^3 t_3^3 t_5^3 - y_s^2 t_1^2 t_2^2 t_3^2 t_4 t_5^3 - y_s^3 t_1^3 t_2^3 t_3^5 t_4 t_5^3 - y_s^2 t_1^2 t_2^2 t_3 t_4^2 t_5^3 -  y_s^3 t_1^3 t_2^3 t_3^4 t_4^2 t_5^3 - y_s^2 t_1^3 t_2 t_4^3 t_5^3- y_s^2 t_1^2 t_2^2 t_4^3 t_5^3 - y_s^2 t_1 t_2^3 t_4^3 t_5^3 +  y_s^3 t_1^5 t_2 t_3^3 t_4^3 t_5^3 + y_s^3 t_1^4 t_2^2 t_3^3 t_4^3 t_5^3 +  y_s^3 t_1^3 t_2^3 t_3^3 t_4^3 t_5^3 + y_s^3 t_1^2 t_2^4 t_3^3 t_4^3 t_5^3 +  y_s^3 t_1 t_2^5 t_3^3 t_4^3 t_5^3 - y_s^3 t_1^3 t_2^3 t_3^2 t_4^4 t_5^3 +  y_s^4 t_1^5 t_2^3 t_3^5 t_4^4 t_5^3 + y_s^4 t_1^4 t_2^4 t_3^5 t_4^4 t_5^3 +  y_s^4 t_1^3 t_2^5 t_3^5 t_4^4 t_5^3 - y_s^3 t_1^3 t_2^3 t_3 t_4^5 t_5^3 +  y_s^4 t_1^5 t_2^3 t_3^4 t_4^5 t_5^3 + y_s^4 t_1^4 t_2^4 t_3^4 t_4^5 t_5^3 +  y_s^4 t_1^3 t_2^5 t_3^4 t_4^5 t_5^3 - y_s^2 t_1^2 t_2^2 t_3^2 t_5^4 -  y_s^3 t_1^3 t_2^3 t_3^5 t_5^4 - y_s^2 t_1^2 t_2^2 t_3 t_4 t_5^4 -  y_s^3 t_1^3 t_2^3 t_3^4 t_4 t_5^4 - y_s^2 t_1^2 t_2^2 t_4^2 t_5^4 -  y_s^3 t_1^3 t_2^3 t_3^3 t_4^2 t_5^4 - y_s^3 t_1^3 t_2^3 t_3^2 t_4^3 t_5^4 +  y_s^4 t_1^5 t_2^3 t_3^5 t_4^3 t_5^4 + y_s^4 t_1^4 t_2^4 t_3^5 t_4^3 t_5^4 +  y_s^4 t_1^3 t_2^5 t_3^5 t_4^3 t_5^4 - y_s^3 t_1^3 t_2^3 t_3 t_4^4 t_5^4 +  y_s^4 t_1^5 t_2^3 t_3^4 t_4^4 t_5^4 + y_s^4 t_1^4 t_2^4 t_3^4 t_4^4 t_5^4 +  y_s^4 t_1^3 t_2^5 t_3^4 t_4^4 t_5^4 - y_s^3 t_1^3 t_2^3 t_4^5 t_5^4+  y_s^4 t_1^5 t_2^3 t_3^3 t_4^5 t_5^4 + y_s^4 t_1^4 t_2^4 t_3^3 t_4^5 t_5^4 +  y_s^4 t_1^3 t_2^5 t_3^3 t_4^5 t_5^4 - y_s^2 t_1^2 t_2^2 t_3 t_5^5 -  y_s^3 t_1^3 t_2^3 t_3^4 t_5^5- y_s^2 t_1^2 t_2^2 t_4 t_5^5-  y_s^3 t_1^3 t_2^3 t_3^3 t_4 t_5^5 - y_s^3 t_1^4 t_2^2 t_3^2 t_4^2 t_5^5 -  y_s^3 t_1^3 t_2^3 t_3^2 t_4^2 t_5^5 - y_s^3 t_1^2 t_2^4 t_3^2 t_4^2 t_5^5 +  y_s^4 t_1^4 t_2^4 t_3^5 t_4^2 t_5^5 - y_s^3 t_1^3 t_2^3 t_3 t_4^3 t_5^5 +  y_s^4 t_1^5 t_2^3 t_3^4 t_4^3 t_5^5 + y_s^4 t_1^4 t_2^4 t_3^4 t_4^3 t_5^5 +  y_s^4 t_1^3 t_2^5 t_3^4 t_4^3 t_5^5 - y_s^3 t_1^3 t_2^3 t_4^4 t_5^5 +  y_s^4 t_1^5 t_2^3 t_3^3 t_4^4 t_5^5 + y_s^4 t_1^4 t_2^4 t_3^3 t_4^4 t_5^5 +  y_s^4 t_1^3 t_2^5 t_3^3 t_4^4 t_5^5 + y_s^4 t_1^4 t_2^4 t_3^2 t_4^5 t_5^5 +  y_s^5 t_1^5 t_2^5 t_3^5 t_4^5 t_5^5
~,~
$
\end{quote}
\endgroup

\subsection{Model 3 \label{app_num_03}}

\begingroup\makeatletter\def\f@size{7}\check@mathfonts
\begin{quote}\raggedright
$ P(t_i,y_s,y_o; \mathcal{M}_3)= 1 + y_{s} y_{o}^3 t_1^3 t_2 t_4^2 + y_{s} y_{o}^3 t_1^2 t_2^2 t_4^2 +  y_{s} y_{o}^3 t_1 t_2^3 t_4^2 + y_{s} y_{o}^3 t_1^3 t_3 t_4^2 + y_{s} y_{o}^3 t_1^2 t_2 t_3 t_4^2 +  y_{s} y_{o}^3 t_1 t_2^2 t_3 t_4^2 + y_{s} y_{o}^3 t_2^3 t_3 t_4^2 +  y_{s} y_{o}^3 t_1^2 t_3^2 t_4^2 + y_{s} y_{o}^3 t_1 t_2 t_3^2 t_4^2 +  y_{s} y_{o}^3 t_2^2 t_3^2 t_4^2 + y_{s} y_{o}^3 t_1 t_3^3 t_4^2 + y_{s} y_{o}^3 t_2 t_3^3 t_4^2 +  y_{s}^2 y_{o}^6 t_1^3 t_2^3 t_3^2 t_4^4 + y_{s}^2 y_{o}^6 t_1^3 t_2^2 t_3^3 t_4^4 +  y_{s}^2 y_{o}^6 t_1^2 t_2^3 t_3^3 t_4^4 + y_{s} y_{o}^2 t_1^3 t_4 t_5 +  y_{s} y_{o}^2 t_1^2 t_2 t_4 t_5 + y_{s} y_{o}^2 t_1 t_2^2 t_4 t_5 + y_{s} y_{o}^2 t_2^3 t_4 t_5 +  y_{s} y_{o}^2 t_1^2 t_3 t_4 t_5 + y_{s} y_{o}^2 t_1 t_2 t_3 t_4 t_5 + y_{s} y_{o}^2 t_2^2 t_3 t_4 t_5 +  y_{s} y_{o}^2 t_1 t_3^2 t_4 t_5 + y_{s} y_{o}^2 t_2 t_3^2 t_4 t_5 + y_{s} y_{o}^2 t_3^3 t_4 t_5 +  y_{s}^2 y_{o}^5 t_1^3 t_2^3 t_3 t_4^3 t_5 + y_{s}^2 y_{o}^5 t_1^3 t_2^2 t_3^2 t_4^3 t_5 +  y_{s}^2 y_{o}^5 t_1^2 t_2^3 t_3^2 t_4^3 t_5 + y_{s}^2 y_{o}^5 t_1^3 t_2 t_3^3 t_4^3 t_5 +  y_{s}^2 y_{o}^5 t_1^2 t_2^2 t_3^3 t_4^3 t_5 + y_{s}^2 y_{o}^5 t_1 t_2^3 t_3^3 t_4^3 t_5 +  y_{s} y_{o} t_1 t_2 t_5^2 + y_{s} y_{o} t_1 t_3 t_5^2 + y_{s} y_{o} t_2 t_3 t_5^2 -  y_{s}^2 y_{o}^4 t_1^5 t_2 t_4^2 t_5^2 - y_{s}^2 y_{o}^4 t_1^4 t_2^2 t_4^2 t_5^2 -  y_{s}^2 y_{o}^4 t_1^3 t_2^3 t_4^2 t_5^2 - y_{s}^2 y_{o}^4 t_1^2 t_2^4 t_4^2 t_5^2 -  y_{s}^2 y_{o}^4 t_1 t_2^5 t_4^2 t_5^2 - y_{s}^2 y_{o}^4 t_1^5 t_3 t_4^2 t_5^2 -  y_{s}^2 y_{o}^4 t_1^4 t_2 t_3 t_4^2 t_5^2 - y_{s}^2 y_{o}^4 t_1^3 t_2^2 t_3 t_4^2 t_5^2 -  y_{s}^2 y_{o}^4 t_1^2 t_2^3 t_3 t_4^2 t_5^2 - y_{s}^2 y_{o}^4 t_1 t_2^4 t_3 t_4^2 t_5^2 -  y_{s}^2 y_{o}^4 t_2^5 t_3 t_4^2 t_5^2 - y_{s}^2 y_{o}^4 t_1^4 t_3^2 t_4^2 t_5^2 -  y_{s}^2 y_{o}^4 t_1^3 t_2 t_3^2 t_4^2 t_5^2 -  2 y_{s}^2 y_{o}^4 t_1^2 t_2^2 t_3^2 t_4^2 t_5^2 -  y_{s}^2 y_{o}^4 t_1 t_2^3 t_3^2 t_4^2 t_5^2 - y_{s}^2 y_{o}^4 t_2^4 t_3^2 t_4^2 t_5^2 -  y_{s}^2 y_{o}^4 t_1^3 t_3^3 t_4^2 t_5^2 - y_{s}^2 y_{o}^4 t_1^2 t_2 t_3^3 t_4^2 t_5^2 -  y_{s}^2 y_{o}^4 t_1 t_2^2 t_3^3 t_4^2 t_5^2 - y_{s}^2 y_{o}^4 t_2^3 t_3^3 t_4^2 t_5^2 -  y_{s}^2 y_{o}^4 t_1^2 t_3^4 t_4^2 t_5^2 - y_{s}^2 y_{o}^4 t_1 t_2 t_3^4 t_4^2 t_5^2 -  y_{s}^2 y_{o}^4 t_2^2 t_3^4 t_4^2 t_5^2 - y_{s}^2 y_{o}^4 t_1 t_3^5 t_4^2 t_5^2 -  y_{s}^2 y_{o}^4 t_2 t_3^5 t_4^2 t_5^2 - y_{s}^3 y_{o}^7 t_1^5 t_2^3 t_3^2 t_4^4 t_5^2 -  y_{s}^3 y_{o}^7 t_1^3 t_2^5 t_3^2 t_4^4 t_5^2 -  y_{s}^3 y_{o}^7 t_1^5 t_2^2 t_3^3 t_4^4 t_5^2 -  y_{s}^3 y_{o}^7 t_1^4 t_2^3 t_3^3 t_4^4 t_5^2 -  y_{s}^3 y_{o}^7 t_1^3 t_2^4 t_3^3 t_4^4 t_5^2 -  y_{s}^3 y_{o}^7 t_1^2 t_2^5 t_3^3 t_4^4 t_5^2 -  y_{s}^3 y_{o}^7 t_1^3 t_2^3 t_3^4 t_4^4 t_5^2 -  y_{s}^3 y_{o}^7 t_1^3 t_2^2 t_3^5 t_4^4 t_5^2 -  y_{s}^3 y_{o}^7 t_1^2 t_2^3 t_3^5 t_4^4 t_5^2 - y_{s}^2 y_{o}^3 t_1^3 t_2^2 t_4 t_5^3 -  y_{s}^2 y_{o}^3 t_1^2 t_2^3 t_4 t_5^3 - y_{s}^2 y_{o}^3 t_1^2 t_2^2 t_3 t_4 t_5^3 -  y_{s}^2 y_{o}^3 t_1^3 t_3^2 t_4 t_5^3 - y_{s}^2 y_{o}^3 t_1^2 t_2 t_3^2 t_4 t_5^3 -  y_{s}^2 y_{o}^3 t_1 t_2^2 t_3^2 t_4 t_5^3 - y_{s}^2 y_{o}^3 t_2^3 t_3^2 t_4 t_5^3 -  y_{s}^2 y_{o}^3 t_1^2 t_3^3 t_4 t_5^3 - y_{s}^2 y_{o}^3 t_2^2 t_3^3 t_4 t_5^3 -  y_{s}^3 y_{o}^6 t_1^5 t_2^4 t_4^3 t_5^3 - y_{s}^3 y_{o}^6 t_1^4 t_2^5 t_4^3 t_5^3 -  y_{s}^3 y_{o}^6 t_1^5 t_2^3 t_3 t_4^3 t_5^3 - y_{s}^3 y_{o}^6 t_1^4 t_2^4 t_3 t_4^3 t_5^3 -  y_{s}^3 y_{o}^6 t_1^3 t_2^5 t_3 t_4^3 t_5^3 - y_{s}^3 y_{o}^6 t_1^5 t_2^2 t_3^2 t_4^3 t_5^3 -  y_{s}^3 y_{o}^6 t_1^4 t_2^3 t_3^2 t_4^3 t_5^3 -  y_{s}^3 y_{o}^6 t_1^3 t_2^4 t_3^2 t_4^3 t_5^3 -  y_{s}^3 y_{o}^6 t_1^2 t_2^5 t_3^2 t_4^3 t_5^3 - y_{s}^3 y_{o}^6 t_1^5 t_2 t_3^3 t_4^3 t_5^3 -  y_{s}^3 y_{o}^6 t_1^4 t_2^2 t_3^3 t_4^3 t_5^3 -  2 y_{s}^3 y_{o}^6 t_1^3 t_2^3 t_3^3 t_4^3 t_5^3 -  y_{s}^3 y_{o}^6 t_1^2 t_2^4 t_3^3 t_4^3 t_5^3 - y_{s}^3 y_{o}^6 t_1 t_2^5 t_3^3 t_4^3 t_5^3 -  y_{s}^3 y_{o}^6 t_1^5 t_3^4 t_4^3 t_5^3 - y_{s}^3 y_{o}^6 t_1^4 t_2 t_3^4 t_4^3 t_5^3 -  y_{s}^3 y_{o}^6 t_1^3 t_2^2 t_3^4 t_4^3 t_5^3 -  y_{s}^3 y_{o}^6 t_1^2 t_2^3 t_3^4 t_4^3 t_5^3 - y_{s}^3 y_{o}^6 t_1 t_2^4 t_3^4 t_4^3 t_5^3 -  y_{s}^3 y_{o}^6 t_2^5 t_3^4 t_4^3 t_5^3 - y_{s}^3 y_{o}^6 t_1^4 t_3^5 t_4^3 t_5^3 -  y_{s}^3 y_{o}^6 t_1^3 t_2 t_3^5 t_4^3 t_5^3 - y_{s}^3 y_{o}^6 t_1^2 t_2^2 t_3^5 t_4^3 t_5^3 -  y_{s}^3 y_{o}^6 t_1 t_2^3 t_3^5 t_4^3 t_5^3 - y_{s}^3 y_{o}^6 t_2^4 t_3^5 t_4^3 t_5^3 +  y_{s}^4 y_{o}^9 t_1^5 t_2^4 t_3^4 t_4^5 t_5^3 +  y_{s}^4 y_{o}^9 t_1^4 t_2^5 t_3^4 t_4^5 t_5^3 +  y_{s}^4 y_{o}^9 t_1^4 t_2^4 t_3^5 t_4^5 t_5^3 +  y_{s}^3 y_{o}^5 t_1^4 t_2^2 t_3^2 t_4^2 t_5^4 +  y_{s}^3 y_{o}^5 t_1^3 t_2^3 t_3^2 t_4^2 t_5^4 +  y_{s}^3 y_{o}^5 t_1^2 t_2^4 t_3^2 t_4^2 t_5^4 +  y_{s}^3 y_{o}^5 t_1^3 t_2^2 t_3^3 t_4^2 t_5^4 +  y_{s}^3 y_{o}^5 t_1^2 t_2^3 t_3^3 t_4^2 t_5^4 +  y_{s}^3 y_{o}^5 t_1^2 t_2^2 t_3^4 t_4^2 t_5^4 +  y_{s}^4 y_{o}^8 t_1^5 t_2^5 t_3^2 t_4^4 t_5^4 +  y_{s}^4 y_{o}^8 t_1^5 t_2^4 t_3^3 t_4^4 t_5^4 +  y_{s}^4 y_{o}^8 t_1^4 t_2^5 t_3^3 t_4^4 t_5^4 +  y_{s}^4 y_{o}^8 t_1^5 t_2^3 t_3^4 t_4^4 t_5^4 +  y_{s}^4 y_{o}^8 t_1^4 t_2^4 t_3^4 t_4^4 t_5^4 +  y_{s}^4 y_{o}^8 t_1^3 t_2^5 t_3^4 t_4^4 t_5^4 +  y_{s}^4 y_{o}^8 t_1^5 t_2^2 t_3^5 t_4^4 t_5^4 +  y_{s}^4 y_{o}^8 t_1^4 t_2^3 t_3^5 t_4^4 t_5^4 +  y_{s}^4 y_{o}^8 t_1^3 t_2^4 t_3^5 t_4^4 t_5^4 +  y_{s}^4 y_{o}^8 t_1^2 t_2^5 t_3^5 t_4^4 t_5^4 + y_{s}^3 y_{o}^4 t_1^3 t_2^2 t_3^2 t_4 t_5^5 +  y_{s}^3 y_{o}^4 t_1^2 t_2^3 t_3^2 t_4 t_5^5 + y_{s}^3 y_{o}^4 t_1^2 t_2^2 t_3^3 t_4 t_5^5 +  y_{s}^4 y_{o}^7 t_1^5 t_2^4 t_3^2 t_4^3 t_5^5 +  y_{s}^4 y_{o}^7 t_1^4 t_2^5 t_3^2 t_4^3 t_5^5 +  y_{s}^4 y_{o}^7 t_1^5 t_2^3 t_3^3 t_4^3 t_5^5 +  y_{s}^4 y_{o}^7 t_1^4 t_2^4 t_3^3 t_4^3 t_5^5 +  y_{s}^4 y_{o}^7 t_1^3 t_2^5 t_3^3 t_4^3 t_5^5 +  y_{s}^4 y_{o}^7 t_1^5 t_2^2 t_3^4 t_4^3 t_5^5 +  y_{s}^4 y_{o}^7 t_1^4 t_2^3 t_3^4 t_4^3 t_5^5 +  y_{s}^4 y_{o}^7 t_1^3 t_2^4 t_3^4 t_4^3 t_5^5 +  y_{s}^4 y_{o}^7 t_1^2 t_2^5 t_3^4 t_4^3 t_5^5 +  y_{s}^4 y_{o}^7 t_1^4 t_2^2 t_3^5 t_4^3 t_5^5 +  y_{s}^4 y_{o}^7 t_1^3 t_2^3 t_3^5 t_4^3 t_5^5 +  y_{s}^4 y_{o}^7 t_1^2 t_2^4 t_3^5 t_4^3 t_5^5 +  y_{s}^5 y_{o}^10 t_1^5 t_2^5 t_3^5 t_4^5 t_5^5
~,~
$
\end{quote}
\endgroup

\subsection{Model 4 \label{app_num_04}}

\begingroup\makeatletter\def\f@size{7}\check@mathfonts
\begin{quote}\raggedright
$
P(t_i,y_s,y_o; \mathcal{M}_4) =1 + y_s y_o t_1 t_3^2 t_4 + y_s y_o t_2 t_3^2 t_4 -  y_s^2 y_o^2 t_1 t_2 t_3^5 t_4 + y_s y_o t_1 t_3 t_4^2 + y_s y_o t_2 t_3 t_4^2-  y_s^2 y_o^2 t_1 t_2 t_3^4 t_4^2 - y_s^2 y_o^2 t_1 t_2 t_3^3 t_4^3 -  y_s^2 y_o^2 t_1 t_2 t_3^2 t_4^4 - y_s^2 y_o^2 t_1 t_2 t_3 t_4^5 +  y_s y_o^2 t_1^2 t_3^2 t_5 + y_s y_o^2 t_1 t_2 t_3^2 t_5 +  y_s y_o^2 t_2^2 t_3^2 t_5 - y_s^2 y_o^3 t_1^2 t_2 t_3^5 t_5 -  y_s^2 y_o^3 t_1 t_2^2 t_3^5 t_5 + y_s y_o^2 t_1^2 t_3 t_4 t_5 +  y_s y_o^2 t_1 t_2 t_3 t_4 t_5 + y_s y_o^2 t_2^2 t_3 t_4 t_5 -  y_s^2 y_o^3 t_1^2 t_2 t_3^4 t_4 t_5 - y_s^2 y_o^3 t_1 t_2^2 t_3^4 t_4 t_5 +  y_s y_o^2 t_1^2 t_4^2 t_5 + y_s y_o^2 t_1 t_2 t_4^2 t_5 +  y_s y_o^2 t_2^2 t_4^2 t_5 - y_s^2 y_o^3 t_1^2 t_2 t_3^3 t_4^2 t_5 -  y_s^2 y_o^3 t_1 t_2^2 t_3^3 t_4^2 t_5 - y_s^2 y_o^3 t_1^2 t_2 t_3^2 t_4^3 t_5 -  y_s^2 y_o^3 t_1 t_2^2 t_3^2 t_4^3 t_5 +  y_s^3 y_o^4 t_1^2 t_2^2 t_3^5 t_4^3 t_5 - y_s^2 y_o^3 t_1^2 t_2 t_3 t_4^4 t_5 -  y_s^2 y_o^3 t_1 t_2^2 t_3 t_4^4 t_5 + y_s^3 y_o^4 t_1^2 t_2^2 t_3^4 t_4^4 t_5 -  y_s^2 y_o^3 t_1^2 t_2 t_4^5 t_5 - y_s^2 y_o^3 t_1 t_2^2 t_4^5 t_5 +  y_s^3 y_o^4 t_1^2 t_2^2 t_3^3 t_4^5 t_5 + y_s y_o^3 t_1^3 t_3 t_5^2 +  y_s y_o^3 t_1^2 t_2 t_3 t_5^2 + y_s y_o^3 t_1 t_2^2 t_3 t_5^2 +  y_s y_o^3 t_2^3 t_3 t_5^2 - y_s^2 y_o^4 t_1^3 t_2 t_3^4 t_5^2 -  y_s^2 y_o^4 t_1^2 t_2^2 t_3^4 t_5^2 - y_s^2 y_o^4 t_1 t_2^3 t_3^4 t_5^2 +  y_s y_o^3 t_1^3 t_4 t_5^2 + y_s y_o^3 t_1^2 t_2 t_4 t_5^2 +  y_s y_o^3 t_1 t_2^2 t_4 t_5^2 + y_s y_o^3 t_2^3 t_4 t_5^2 -  y_s^2 y_o^4 t_1^3 t_2 t_3^3 t_4 t_5^2 -  y_s^2 y_o^4 t_1^2 t_2^2 t_3^3 t_4 t_5^2 -  y_s^2 y_o^4 t_1 t_2^3 t_3^3 t_4 t_5^2 + y_s^2 y_o^4 t_1^4 t_3^2 t_4^2 t_5^2 +  y_s^2 y_o^4 t_1^3 t_2 t_3^2 t_4^2 t_5^2 +  y_s^2 y_o^4 t_1^2 t_2^2 t_3^2 t_4^2 t_5^2 +  y_s^2 y_o^4 t_1 t_2^3 t_3^2 t_4^2 t_5^2 + y_s^2 y_o^4 t_2^4 t_3^2 t_4^2 t_5^2 -  y_s^3 y_o^5 t_1^4 t_2 t_3^5 t_4^2 t_5^2 -  y_s^3 y_o^5 t_1^3 t_2^2 t_3^5 t_4^2 t_5^2 -  y_s^3 y_o^5 t_1^2 t_2^3 t_3^5 t_4^2 t_5^2 -  y_s^3 y_o^5 t_1 t_2^4 t_3^5 t_4^2 t_5^2 -  y_s^2 y_o^4 t_1^3 t_2 t_3 t_4^3 t_5^2 -  y_s^2 y_o^4 t_1^2 t_2^2 t_3 t_4^3 t_5^2 -  y_s^2 y_o^4 t_1 t_2^3 t_3 t_4^3 t_5^2 +  y_s^3 y_o^5 t_1^3 t_2^2 t_3^4 t_4^3 t_5^2 +  y_s^3 y_o^5 t_1^2 t_2^3 t_3^4 t_4^3 t_5^2 - y_s^2 y_o^4 t_1^3 t_2 t_4^4 t_5^2 -  y_s^2 y_o^4 t_1^2 t_2^2 t_4^4 t_5^2 - y_s^2 y_o^4 t_1 t_2^3 t_4^4 t_5^2 +  y_s^3 y_o^5 t_1^3 t_2^2 t_3^3 t_4^4 t_5^2 +  y_s^3 y_o^5 t_1^2 t_2^3 t_3^3 t_4^4 t_5^2 -  y_s^3 y_o^5 t_1^4 t_2 t_3^2 t_4^5 t_5^2 -  y_s^3 y_o^5 t_1^3 t_2^2 t_3^2 t_4^5 t_5^2 -  y_s^3 y_o^5 t_1^2 t_2^3 t_3^2 t_4^5 t_5^2 -  y_s^3 y_o^5 t_1 t_2^4 t_3^2 t_4^5 t_5^2 +  y_s^4 y_o^6 t_1^4 t_2^2 t_3^5 t_4^5 t_5^2 +  y_s^4 y_o^6 t_1^3 t_2^3 t_3^5 t_4^5 t_5^2 +  y_s^4 y_o^6 t_1^2 t_2^4 t_3^5 t_4^5 t_5^2 + y_s y_o^4 t_1^3 t_2 t_5^3 +  y_s y_o^4 t_1^2 t_2^2 t_5^3 + y_s y_o^4 t_1 t_2^3 t_5^3 -  y_s^2 y_o^5 t_1^4 t_2 t_3^3 t_5^3 - y_s^2 y_o^5 t_1^3 t_2^2 t_3^3 t_5^3 -  y_s^2 y_o^5 t_1^2 t_2^3 t_3^3 t_5^3 - y_s^2 y_o^5 t_1 t_2^4 t_3^3 t_5^3 +  y_s^2 y_o^5 t_1^3 t_2^2 t_3^2 t_4 t_5^3 +  y_s^2 y_o^5 t_1^2 t_2^3 t_3^2 t_4 t_5^3 -  y_s^3 y_o^6 t_1^4 t_2^2 t_3^5 t_4 t_5^3 -  y_s^3 y_o^6 t_1^3 t_2^3 t_3^5 t_4 t_5^3 -  y_s^3 y_o^6 t_1^2 t_2^4 t_3^5 t_4 t_5^3 +  y_s^2 y_o^5 t_1^3 t_2^2 t_3 t_4^2 t_5^3 +  y_s^2 y_o^5 t_1^2 t_2^3 t_3 t_4^2 t_5^3 -  y_s^3 y_o^6 t_1^4 t_2^2 t_3^4 t_4^2 t_5^3 -  y_s^3 y_o^6 t_1^3 t_2^3 t_3^4 t_4^2 t_5^3 -  y_s^3 y_o^6 t_1^2 t_2^4 t_3^4 t_4^2 t_5^3 - y_s^2 y_o^5 t_1^4 t_2 t_4^3 t_5^3 -  y_s^2 y_o^5 t_1^3 t_2^2 t_4^3 t_5^3 - y_s^2 y_o^5 t_1^2 t_2^3 t_4^3 t_5^3 -  y_s^2 y_o^5 t_1 t_2^4 t_4^3 t_5^3 + y_s^3 y_o^6 t_1^5 t_2 t_3^3 t_4^3 t_5^3 +  y_s^3 y_o^6 t_1^4 t_2^2 t_3^3 t_4^3 t_5^3 +  y_s^3 y_o^6 t_1^3 t_2^3 t_3^3 t_4^3 t_5^3 +  y_s^3 y_o^6 t_1^2 t_2^4 t_3^3 t_4^3 t_5^3 +  y_s^3 y_o^6 t_1 t_2^5 t_3^3 t_4^3 t_5^3 -  y_s^3 y_o^6 t_1^4 t_2^2 t_3^2 t_4^4 t_5^3 -  y_s^3 y_o^6 t_1^3 t_2^3 t_3^2 t_4^4 t_5^3 -  y_s^3 y_o^6 t_1^2 t_2^4 t_3^2 t_4^4 t_5^3 +  y_s^4 y_o^7 t_1^5 t_2^2 t_3^5 t_4^4 t_5^3 +  y_s^4 y_o^7 t_1^4 t_2^3 t_3^5 t_4^4 t_5^3 +  y_s^4 y_o^7 t_1^3 t_2^4 t_3^5 t_4^4 t_5^3 +  y_s^4 y_o^7 t_1^2 t_2^5 t_3^5 t_4^4 t_5^3 -  y_s^3 y_o^6 t_1^4 t_2^2 t_3 t_4^5 t_5^3 -  y_s^3 y_o^6 t_1^3 t_2^3 t_3 t_4^5 t_5^3 -  y_s^3 y_o^6 t_1^2 t_2^4 t_3 t_4^5 t_5^3 +  y_s^4 y_o^7 t_1^5 t_2^2 t_3^4 t_4^5 t_5^3 +  y_s^4 y_o^7 t_1^4 t_2^3 t_3^4 t_4^5 t_5^3 +  y_s^4 y_o^7 t_1^3 t_2^4 t_3^4 t_4^5 t_5^3 +  y_s^4 y_o^7 t_1^2 t_2^5 t_3^4 t_4^5 t_5^3 +  y_s^2 y_o^6 t_1^3 t_2^3 t_3^2 t_5^4 - y_s^3 y_o^7 t_1^4 t_2^3 t_3^5 t_5^4 -  y_s^3 y_o^7 t_1^3 t_2^4 t_3^5 t_5^4 + y_s^2 y_o^6 t_1^3 t_2^3 t_3 t_4 t_5^4 -  y_s^3 y_o^7 t_1^4 t_2^3 t_3^4 t_4 t_5^4 -  y_s^3 y_o^7 t_1^3 t_2^4 t_3^4 t_4 t_5^4 + y_s^2 y_o^6 t_1^3 t_2^3 t_4^2 t_5^4 -  y_s^3 y_o^7 t_1^4 t_2^3 t_3^3 t_4^2 t_5^4 -  y_s^3 y_o^7 t_1^3 t_2^4 t_3^3 t_4^2 t_5^4 -  y_s^3 y_o^7 t_1^4 t_2^3 t_3^2 t_4^3 t_5^4 -  y_s^3 y_o^7 t_1^3 t_2^4 t_3^2 t_4^3 t_5^4 +  y_s^4 y_o^8 t_1^5 t_2^3 t_3^5 t_4^3 t_5^4 +  y_s^4 y_o^8 t_1^4 t_2^4 t_3^5 t_4^3 t_5^4 +  y_s^4 y_o^8 t_1^3 t_2^5 t_3^5 t_4^3 t_5^4 -  y_s^3 y_o^7 t_1^4 t_2^3 t_3 t_4^4 t_5^4 -  y_s^3 y_o^7 t_1^3 t_2^4 t_3 t_4^4 t_5^4 +  y_s^4 y_o^8 t_1^5 t_2^3 t_3^4 t_4^4 t_5^4 +  y_s^4 y_o^8 t_1^4 t_2^4 t_3^4 t_4^4 t_5^4 +  y_s^4 y_o^8 t_1^3 t_2^5 t_3^4 t_4^4 t_5^4 -  y_s^3 y_o^7 t_1^4 t_2^3 t_4^5 t_5^4 - y_s^3 y_o^7 t_1^3 t_2^4 t_4^5 t_5^4 +  y_s^4 y_o^8 t_1^5 t_2^3 t_3^3 t_4^5 t_5^4 +  y_s^4 y_o^8 t_1^4 t_2^4 t_3^3 t_4^5 t_5^4 +  y_s^4 y_o^8 t_1^3 t_2^5 t_3^3 t_4^5 t_5^4 -  y_s^3 y_o^8 t_1^4 t_2^4 t_3^4 t_5^5 - y_s^3 y_o^8 t_1^4 t_2^4 t_3^3 t_4 t_5^5 -  y_s^3 y_o^8 t_1^4 t_2^4 t_3^2 t_4^2 t_5^5 -  y_s^3 y_o^8 t_1^4 t_2^4 t_3 t_4^3 t_5^5 +  y_s^4 y_o^9 t_1^5 t_2^4 t_3^4 t_4^3 t_5^5 +  y_s^4 y_o^9 t_1^4 t_2^5 t_3^4 t_4^3 t_5^5 -  y_s^3 y_o^8 t_1^4 t_2^4 t_4^4 t_5^5 +  y_s^4 y_o^9 t_1^5 t_2^4 t_3^3 t_4^4 t_5^5 +  y_s^4 y_o^9 t_1^4 t_2^5 t_3^3 t_4^4 t_5^5 +  y_s^5 y_o^{10} t_1^5 t_2^5 t_3^5 t_4^5 t_5^5
~,~
$
\end{quote}
\endgroup

\subsection{Model 5 \label{app_num_05}}

\begingroup\makeatletter\def\f@size{7}\check@mathfonts
\begin{quote}\raggedright
$
P(t_i,y_s,y_o; \mathcal{M}_5) =
1 + y_s y_o^2 t_1^3 t_4 t_5 +y_s y_o^2  t_1^2 t_2 t_4 t_5 + y_s y_o^2  t_1 t_2^2 t_4 t_5 + y_s y_o^2 t_2^3 t_4 t_5 +y_s y_o^2  t_1^2 t_3 t_4 t_5 +y_s y_o^2  t_1 t_2 t_3 t_4 t_5 +y_s y_o^2  t_2^2 t_3 t_4 t_5 +y_s y_o^2  t_1 t_3^2 t_4 t_5 +y_s y_o^2  t_2 t_3^2 t_4 t_5 +y_s y_o^2 t_3^3 t_4 t_5 +y_s y_o^3  t_1^4 t_2 t_4^2 +y_s y_o^3  t_1^3 t_2^2 t_4^2 +y_s y_o^3  t_1^2 t_2^3 t_4^2 +y_s y_o^3  t_1 t_2^4 t_4^2 +y_s y_o^3  t_1^4 t_3 t_4^2 +y_s y_o^3  t_1^3 t_2 t_3 t_4^2 + y_s y_o^3 t_1^2 t_2^2 t_3 t_4^2 + y_s y_o^3 t_1 t_2^3 t_3 t_4^2 + y_s y_o^3 t_2^4 t_3 t_4^2 + y_s y_o^3 t_1^3 t_3^2 t_4^2 + y_s y_o^3 t_1^2 t_2 t_3^2 t_4^2 + y_s y_o^3 t_1 t_2^2 t_3^2 t_4^2 + y_s y_o^3 t_2^3 t_3^2 t_4^2 + y_s y_o^3 t_1^2 t_3^3 t_4^2 + y_s y_o^3 t_1 t_2 t_3^3 t_4^2 + y_s y_o^3 t_2^2 t_3^3 t_4^2 + y_s y_o^3 t_1 t_3^4 t_4^2 + y_s y_o^3 t_2 t_3^4 t_4^2 - y_s^2 y_o^3 t_1^3 t_2 t_4 t_5^3 - y_s^2 y_o^3 t_1^2 t_2^2 t_4 t_5^3 - y_s^2 y_o^3 t_1 t_2^3 t_4 t_5^3 - y_s^2 y_o^3 t_1^3 t_3 t_4 t_5^3 - y_s^2 y_o^3 2 t_1^2 t_2 t_3 t_4 t_5^3 - y_s^2 y_o^3 2 t_1 t_2^2 t_3 t_4 t_5^3 - y_s^2 y_o^3 t_2^3 t_3 t_4 t_5^3 - y_s^2 y_o^3 t_1^2 t_3^2 t_4 t_5^3 - y_s^2 y_o^3 2 t_1 t_2 t_3^2 t_4 t_5^3 - y_s^2 y_o^3 t_2^2 t_3^2 t_4 t_5^3 - y_s^2 y_o^3 t_1 t_3^3 t_4 t_5^3 - y_s^2 y_o^3 t_2 t_3^3 t_4 t_5^3 - y_s^2 y_o^4 t_1^5 t_2 t_4^2 t_5^2 - y_s^2 y_o^4 t_1^4 t_2^2 t_4^2 t_5^2 - y_s^2 y_o^4 t_1^3 t_2^3 t_4^2 t_5^2 - y_s^2 y_o^4 t_1^2 t_2^4 t_4^2 t_5^2 - y_s^2 y_o^4 t_1 t_2^5 t_4^2 t_5^2 - y_s^2 y_o^4 t_1^5 t_3 t_4^2 t_5^2 - 2 y_s^2 y_o^4 t_1^4 t_2 t_3 t_4^2 t_5^2 - 2 y_s^2 y_o^4  t_1^3 t_2^2 t_3 t_4^2 t_5^2 - 2y_s^2 y_o^4  t_1^2 t_2^3 t_3 t_4^2 t_5^2 - 2y_s^2 y_o^4  t_1 t_2^4 t_3 t_4^2 t_5^2 - y_s^2 y_o^4 t_2^5 t_3 t_4^2 t_5^2 - y_s^2 y_o^4 t_1^4 t_3^2 t_4^2 t_5^2 - 2y_s^2 y_o^4  t_1^3 t_2 t_3^2 t_4^2 t_5^2 - 2y_s^2 y_o^4  t_1^2 t_2^2 t_3^2 t_4^2 t_5^2 - 2y_s^2 y_o^4  t_1 t_2^3 t_3^2 t_4^2 t_5^2 - y_s^2 y_o^4 t_2^4 t_3^2 t_4^2 t_5^2 - y_s^2 y_o^4 t_1^3 t_3^3 t_4^2 t_5^2 - 2y_s^2 y_o^4  t_1^2 t_2 t_3^3 t_4^2 t_5^2 - 2y_s^2 y_o^4  t_1 t_2^2 t_3^3 t_4^2 t_5^2 - y_s^2 y_o^4 t_2^3 t_3^3 t_4^2 t_5^2 - y_s^2 y_o^4 t_1^2 t_3^4 t_4^2 t_5^2 - 2y_s^2 y_o^4  t_1 t_2 t_3^4 t_4^2 t_5^2 - y_s^2 y_o^4 t_2^2 t_3^4 t_4^2 t_5^2 - y_s^2 y_o^4 t_1 t_3^5 t_4^2 t_5^2 - y_s^2 y_o^4 t_2 t_3^5 t_4^2 t_5^2 + y_s^3 y_o^4 t_1^3 t_2 t_3 t_4 t_5^5 + y_s^3 y_o^4 t_1^2 t_2^2 t_3 t_4 t_5^5 + y_s^3 y_o^4 t_1 t_2^3 t_3 t_4 t_5^5 + y_s^3 y_o^4 t_1^2 t_2 t_3^2 t_4 t_5^5 + y_s^3 y_o^4 t_1 t_2^2 t_3^2 t_4 t_5^5 + y_s^3 y_o^4 t_1 t_2 t_3^3 t_4 t_5^5 + y_s^2 y_o^5 t_1^4 t_2^4 t_4^3 t_5 + y_s^2 y_o^5 t_1^4 t_2^3 t_3 t_4^3 t_5 + y_s^2 y_o^5 t_1^3 t_2^4 t_3 t_4^3 t_5 + y_s^2 y_o^5 t_1^4 t_2^2 t_3^2 t_4^3 t_5 + y_s^2 y_o^5 t_1^3 t_2^3 t_3^2 t_4^3 t_5 + y_s^2 y_o^5 t_1^2 t_2^4 t_3^2 t_4^3 t_5 + y_s^2 y_o^5 t_1^4 t_2 t_3^3 t_4^3 t_5 + y_s^2 y_o^5 t_1^3 t_2^2 t_3^3 t_4^3 t_5 + y_s^2 y_o^5 t_1^2 t_2^3 t_3^3 t_4^3 t_5 + y_s^2 y_o^5 t_1 t_2^4 t_3^3 t_4^3 t_5 + y_s^2 y_o^5 t_1^4 t_3^4 t_4^3 t_5 + y_s^2 y_o^5 t_1^3 t_2 t_3^4 t_4^3 t_5 + y_s^2 y_o^5 t_1^2 t_2^2 t_3^4 t_4^3 t_5 + y_s^2 y_o^5 t_1 t_2^3 t_3^4 t_4^3 t_5 + y_s^2 y_o^5 t_2^4 t_3^4 t_4^3 t_5 + y_s^3 y_o^5 t_1^5 t_2 t_3 t_4^2 t_5^4 + y_s^3 y_o^5 t_1^4 t_2^2 t_3 t_4^2 t_5^4 + y_s^3 y_o^5 t_1^3 t_2^3 t_3 t_4^2 t_5^4 + y_s^3 y_o^5 t_1^2 t_2^4 t_3 t_4^2 t_5^4 + y_s^3 y_o^5 t_1 t_2^5 t_3 t_4^2 t_5^4 + y_s^3 y_o^5 t_1^4 t_2 t_3^2 t_4^2 t_5^4 + y_s^3 y_o^5 t_1^3 t_2^2 t_3^2 t_4^2 t_5^4 + y_s^3 y_o^5 t_1^2 t_2^3 t_3^2 t_4^2 t_5^4 + y_s^3 y_o^5 t_1 t_2^4 t_3^2 t_4^2 t_5^4 + y_s^3 y_o^5 t_1^3 t_2 t_3^3 t_4^2 t_5^4 + y_s^3 y_o^5 t_1^2 t_2^2 t_3^3 t_4^2 t_5^4 + y_s^3 y_o^5 t_1 t_2^3 t_3^3 t_4^2 t_5^4 + y_s^3 y_o^5 t_1^2 t_2 t_3^4 t_4^2 t_5^4 + y_s^3 y_o^5 t_1 t_2^2 t_3^4 t_4^2 t_5^4 + y_s^3 y_o^5 t_1 t_2 t_3^5 t_4^2 t_5^4 + y_s^2 y_o^6 t_1^4 t_2^4 t_3^2 t_4^4 + y_s^2 y_o^6 t_1^4 t_2^3 t_3^3 t_4^4 +  y_s^2 y_o^6 t_1^3 t_2^4 t_3^3 t_4^4+ y_s^2 y_o^6 t_1^4 t_2^2 t_3^4 t_4^4 + y_s^2 y_o^6 t_1^3 t_2^3 t_3^4 t_4^4 + y_s^2 y_o^6 t_1^2 t_2^4 t_3^4 t_4^4 - y_s^3 y_o^6 t_1^5 t_2^4 t_4^3 t_5^3  - y_s^3 y_o^6 t_1^4 t_2^5 t_4^3 t_5^3 - y_s^3 y_o^6t_1^5 t_2^3 t_3 t_4^3 t_5^3  - 2y_s^3 y_o^6 t_1^4 t_2^4 t_3 t_4^3 t_5^3  - y_s^3 y_o^6t_1^3 t_2^5 t_3 t_4^3 t_5^3  - y_s^3 y_o^6t_1^5 t_2^2 t_3^2 t_4^3 t_5^3  - 2y_s^3 y_o^6  t_1^4 t_2^3 t_3^2 t_4^3 t_5^3 - 2y_s^3 y_o^6 t_1^3 t_2^4 t_3^2 t_4^3 t_5^3  - y_s^3 y_o^6 t_1^2 t_2^5 t_3^2 t_4^3 t_5^3  - y_s^3 y_o^6 t_1^5 t_2 t_3^3 t_4^3 t_5^3  - 2y_s^3 y_o^6 t_1^4 t_2^2 t_3^3 t_4^3 t_5^3  - 2y_s^3 y_o^6  t_1^3 t_2^3 t_3^3 t_4^3 t_5^3 - 2y_s^3 y_o^6  t_1^2 t_2^4 t_3^3 t_4^3 t_5^3 - y_s^3 y_o^6 t_1 t_2^5 t_3^3 t_4^3 t_5^3 - y_s^3 y_o^6 t_1^5 t_3^4 t_4^3 t_5^3  - 2y_s^3 y_o^6 t_1^4 t_2 t_3^4 t_4^3 t_5^3  - 2y_s^3 y_o^6 t_1^3 t_2^2 t_3^4 t_4^3 t_5^3  - 2y_s^3 y_o^6 t_1^2 t_2^3 t_3^4 t_4^3 t_5^3  - 2y_s^3 y_o^6 t_1 t_2^4 t_3^4 t_4^3 t_5^3  - y_s^3 y_o^6 t_2^5 t_3^4 t_4^3 t_5^3  - y_s^3 y_o^6 t_1^4 t_3^5 t_4^3 t_5^3  - y_s^3 y_o^6 t_1^3 t_2 t_3^5 t_4^3 t_5^3  -y_s^3 y_o^6  t_1^2 t_2^2 t_3^5 t_4^3 t_5^3   - y_s^3 y_o^6 t_1 t_2^3 t_3^5 t_4^3 t_5^3   - y_s^3 y_o^6 t_2^4 t_3^5 t_4^3 t_5^3   - y_s^3 y_o^7 t_1^5 t_2^4 t_3^2 t_4^4 t_5^2  - y_s^3 y_o^7 t_1^4 t_2^5 t_3^2 t_4^4 t_5^2  - y_s^3 y_o^7 t_1^5 t_2^3 t_3^3 t_4^4 t_5^2  - 2y_s^3 y_o^7 t_1^4 t_2^4 t_3^3 t_4^4 t_5^2  - y_s^3 y_o^7 t_1^3 t_2^5 t_3^3 t_4^4 t_5^2  - y_s^3 y_o^7 t_1^5 t_2^2 t_3^4 t_4^4 t_5^2  - 2y_s^3 y_o^7 t_1^4 t_2^3 t_3^4 t_4^4 t_5^2  - 2y_s^3 y_o^7 t_1^3 t_2^4 t_3^4 t_4^4 t_5^2  - y_s^3 y_o^7 t_1^2 t_2^5 t_3^4 t_4^4 t_5^2  - y_s^3 y_o^7 t_1^4 t_2^2 t_3^5 t_4^4 t_5^2  - y_s^3 y_o^7 t_1^3 t_2^3 t_3^5 t_4^4 t_5^2  - y_s^3 y_o^7 t_1^2 t_2^4 t_3^5 t_4^4 t_5^2  + y_s^4 y_o^7 t_1^5 t_2^4 t_3 t_4^3 t_5^5  + y_s^4 y_o^7 t_1^4 t_2^5 t_3 t_4^3 t_5^5  + y_s^4 y_o^7 t_1^5 t_2^3 t_3^2 t_4^3 t_5^5  + y_s^4 y_o^7 t_1^4 t_2^4 t_3^2 t_4^3 t_5^5  + y_s^4 y_o^7 t_1^3 t_2^5 t_3^2 t_4^3 t_5^5  + y_s^4 y_o^7 t_1^5 t_2^2 t_3^3 t_4^3 t_5^5  + y_s^4 y_o^7 t_1^4 t_2^3 t_3^3 t_4^3 t_5^5  + y_s^4 y_o^7 t_1^3 t_2^4 t_3^3 t_4^3 t_5^5  + y_s^4 y_o^7 t_1^2 t_2^5 t_3^3 t_4^3 t_5^5  + y_s^4 y_o^7 t_1^5 t_2 t_3^4 t_4^3 t_5^5  + y_s^4 y_o^7 t_1^4 t_2^2 t_3^4 t_4^3 t_5^5  + y_s^4 y_o^7 t_1^3 t_2^3 t_3^4 t_4^3 t_5^5  + y_s^4 y_o^7 t_1^2 t_2^4 t_3^4 t_4^3 t_5^5  + y_s^4 y_o^7 t_1 t_2^5 t_3^4 t_4^3 t_5^5  + y_s^4 y_o^7 t_1^4 t_2 t_3^5 t_4^3 t_5^5  + y_s^4 y_o^7 t_1^3 t_2^2 t_3^5 t_4^3 t_5^5  + y_s^4 y_o^7 t_1^2 t_2^3 t_3^5 t_4^3 t_5^5  + y_s^4 y_o^7 t_1 t_2^4 t_3^5 t_4^3 t_5^5  + y_s^4 y_o^8 t_1^5 t_2^5 t_3^2 t_4^4 t_5^4  +y_s^4 y_o^8  t_1^5 t_2^4 t_3^3 t_4^4 t_5^4 +y_s^4 y_o^8  t_1^4 t_2^5 t_3^3 t_4^4 t_5^4 +y_s^4 y_o^8  t_1^5 t_2^3 t_3^4 t_4^4 t_5^4 +y_s^4 y_o^8  t_1^4 t_2^4 t_3^4 t_4^4 t_5^4 +y_s^4 y_o^8  t_1^3 t_2^5 t_3^4 t_4^4 t_5^4 +y_s^4 y_o^8  t_1^5 t_2^2 t_3^5 t_4^4 t_5^4 +y_s^4 y_o^8  t_1^4 t_2^3 t_3^5 t_4^4 t_5^4 +y_s^4 y_o^8  t_1^3 t_2^4 t_3^5 t_4^4 t_5^4 +y_s^4 y_o^8  t_1^2 t_2^5 t_3^5 t_4^4 t_5^4 +y_s^5 y_o^{10} t_1^5 t_2^5 t_3^5 t_4^5 t_5^5 
~,~$
\end{quote}
\endgroup

\subsection{Model 6 \label{app_num_06}}

%
\begingroup\makeatletter\def\f@size{7}\check@mathfonts
\begin{quote}\raggedright
$P(t_i,y_s,y_{o_1},y_{o_2}; \mathcal{M}_6) = 1 + y_{s} y_{o_1}^3 y_{o_2} t_1 t_3^2 t_4 t_5^2 + y_{s} y_{o_1}^3 y_{o_2} t_2 t_3^2 t_4 t_5^2 +  y_{s} y_{o_1}^3 y_{o_2} t_1 t_3 t_4^2 t_5^2 + y_{s} y_{o_1}^3 y_{o_2} t_2 t_3 t_4^2 t_5^2 -  y_{s}^2 y_{o_1}^6 y_{o_2}^2 t_1 t_2 t_3^5 t_4 t_5^4 -  y_{s}^2 y_{o_1}^6 y_{o_2}^2 t_1 t_2 t_3^4 t_4^2 t_5^4 -  y_{s}^2 y_{o_1}^6 y_{o_2}^2 t_1 t_2 t_3^3 t_4^3 t_5^4 -  y_{s}^2 y_{o_1}^6 y_{o_2}^2 t_1 t_2 t_3^2 t_4^4 t_5^4 -  y_{s}^2 y_{o_1}^6 y_{o_2}^2 t_1 t_2 t_3 t_4^5 t_5^4 + y_{s} y_{o_1}^2 y_{o_2}^2 t_1^2 t_3^2 t_5 t_6 +  y_{s} y_{o_1}^2 y_{o_2}^2 t_1 t_2 t_3^2 t_5 t_6 + y_{s} y_{o_1}^2 y_{o_2}^2 t_2^2 t_3^2 t_5 t_6 +  y_{s} y_{o_1}^2 y_{o_2}^2 t_1^2 t_3 t_4 t_5 t_6 + y_{s} y_{o_1}^2 y_{o_2}^2 t_1 t_2 t_3 t_4 t_5 t_6 +  y_{s} y_{o_1}^2 y_{o_2}^2 t_2^2 t_3 t_4 t_5 t_6 + y_{s} y_{o_1}^2 y_{o_2}^2 t_1^2 t_4^2 t_5 t_6 +  y_{s} y_{o_1}^2 y_{o_2}^2 t_1 t_2 t_4^2 t_5 t_6 + y_{s} y_{o_1}^2 y_{o_2}^2 t_2^2 t_4^2 t_5 t_6 -  y_{s}^2 y_{o_1}^5 y_{o_2}^3 t_1^2 t_2 t_3^5 t_5^3 t_6 -  y_{s}^2 y_{o_1}^5 y_{o_2}^3 t_1 t_2^2 t_3^5 t_5^3 t_6 -  y_{s}^2 y_{o_1}^5 y_{o_2}^3 t_1^2 t_2 t_3^4 t_4 t_5^3 t_6 -  y_{s}^2 y_{o_1}^5 y_{o_2}^3 t_1 t_2^2 t_3^4 t_4 t_5^3 t_6 -  y_{s}^2 y_{o_1}^5 y_{o_2}^3 t_1^2 t_2 t_3^3 t_4^2 t_5^3 t_6 -  y_{s}^2 y_{o_1}^5 y_{o_2}^3 t_1 t_2^2 t_3^3 t_4^2 t_5^3 t_6 -  y_{s}^2 y_{o_1}^5 y_{o_2}^3 t_1^2 t_2 t_3^2 t_4^3 t_5^3 t_6 -  y_{s}^2 y_{o_1}^5 y_{o_2}^3 t_1 t_2^2 t_3^2 t_4^3 t_5^3 t_6 -  y_{s}^2 y_{o_1}^5 y_{o_2}^3 t_1^2 t_2 t_3 t_4^4 t_5^3 t_6 -  y_{s}^2 y_{o_1}^5 y_{o_2}^3 t_1 t_2^2 t_3 t_4^4 t_5^3 t_6 -  y_{s}^2 y_{o_1}^5 y_{o_2}^3 t_1^2 t_2 t_4^5 t_5^3 t_6 -  y_{s}^2 y_{o_1}^5 y_{o_2}^3 t_1 t_2^2 t_4^5 t_5^3 t_6 +  y_{s}^3 y_{o_1}^8 y_{o_2}^4 t_1^2 t_2^2 t_3^5 t_4^3 t_5^5 t_6 +  y_{s}^3 y_{o_1}^8 y_{o_2}^4 t_1^2 t_2^2 t_3^4 t_4^4 t_5^5 t_6 +  y_{s}^3 y_{o_1}^8 y_{o_2}^4 t_1^2 t_2^2 t_3^3 t_4^5 t_5^5 t_6 +  y_{s} y_{o_1} y_{o_2}^3 t_1^2 t_2 t_3 t_6^2 + y_{s} y_{o_1} y_{o_2}^3 t_1 t_2^2 t_3 t_6^2 +  y_{s} y_{o_1} y_{o_2}^3 t_1^2 t_2 t_4 t_6^2 + y_{s} y_{o_1} y_{o_2}^3 t_1 t_2^2 t_4 t_6^2 -  y_{s}^2 y_{o_1}^4 y_{o_2}^4 t_1^3 t_2 t_3^4 t_5^2 t_6^2 -  y_{s}^2 y_{o_1}^4 y_{o_2}^4 t_1^2 t_2^2 t_3^4 t_5^2 t_6^2 -  y_{s}^2 y_{o_1}^4 y_{o_2}^4 t_1 t_2^3 t_3^4 t_5^2 t_6^2 -  y_{s}^2 y_{o_1}^4 y_{o_2}^4 t_1^4 t_3^3 t_4 t_5^2 t_6^2 -  2 y_{s}^2 y_{o_1}^4 y_{o_2}^4 t_1^3 t_2 t_3^3 t_4 t_5^2 t_6^2 -  y_{s}^2 y_{o_1}^4 y_{o_2}^4 t_1^2 t_2^2 t_3^3 t_4 t_5^2 t_6^2 -  2 y_{s}^2 y_{o_1}^4 y_{o_2}^4 t_1 t_2^3 t_3^3 t_4 t_5^2 t_6^2 -  y_{s}^2 y_{o_1}^4 y_{o_2}^4 t_2^4 t_3^3 t_4 t_5^2 t_6^2 -  y_{s}^2 y_{o_1}^4 y_{o_2}^4 t_1^4 t_3^2 t_4^2 t_5^2 t_6^2 -  y_{s}^2 y_{o_1}^4 y_{o_2}^4 t_1^3 t_2 t_3^2 t_4^2 t_5^2 t_6^2 +  y_{s}^2 y_{o_1}^4 y_{o_2}^4 t_1^2 t_2^2 t_3^2 t_4^2 t_5^2 t_6^2 -  y_{s}^2 y_{o_1}^4 y_{o_2}^4 t_1 t_2^3 t_3^2 t_4^2 t_5^2 t_6^2 -  y_{s}^2 y_{o_1}^4 y_{o_2}^4 t_2^4 t_3^2 t_4^2 t_5^2 t_6^2 -  y_{s}^2 y_{o_1}^4 y_{o_2}^4 t_1^4 t_3 t_4^3 t_5^2 t_6^2 -  2 y_{s}^2 y_{o_1}^4 y_{o_2}^4 t_1^3 t_2 t_3 t_4^3 t_5^2 t_6^2 -  y_{s}^2 y_{o_1}^4 y_{o_2}^4 t_1^2 t_2^2 t_3 t_4^3 t_5^2 t_6^2 -  2 y_{s}^2 y_{o_1}^4 y_{o_2}^4 t_1 t_2^3 t_3 t_4^3 t_5^2 t_6^2 -  y_{s}^2 y_{o_1}^4 y_{o_2}^4 t_2^4 t_3 t_4^3 t_5^2 t_6^2 -  y_{s}^2 y_{o_1}^4 y_{o_2}^4 t_1^3 t_2 t_4^4 t_5^2 t_6^2 -  y_{s}^2 y_{o_1}^4 y_{o_2}^4 t_1^2 t_2^2 t_4^4 t_5^2 t_6^2 -  y_{s}^2 y_{o_1}^4 y_{o_2}^4 t_1 t_2^3 t_4^4 t_5^2 t_6^2 +  y_{s}^3 y_{o_1}^7 y_{o_2}^5 t_1^4 t_2 t_3^6 t_4 t_5^4 t_6^2 +  y_{s}^3 y_{o_1}^7 y_{o_2}^5 t_1 t_2^4 t_3^6 t_4 t_5^4 t_6^2 +  y_{s}^3 y_{o_1}^7 y_{o_2}^5 t_1^4 t_2 t_3^5 t_4^2 t_5^4 t_6^2 -  y_{s}^3 y_{o_1}^7 y_{o_2}^5 t_1^3 t_2^2 t_3^5 t_4^2 t_5^4 t_6^2 -  y_{s}^3 y_{o_1}^7 y_{o_2}^5 t_1^2 t_2^3 t_3^5 t_4^2 t_5^4 t_6^2 +  y_{s}^3 y_{o_1}^7 y_{o_2}^5 t_1 t_2^4 t_3^5 t_4^2 t_5^4 t_6^2 +  2 y_{s}^3 y_{o_1}^7 y_{o_2}^5 t_1^4 t_2 t_3^4 t_4^3 t_5^4 t_6^2 +  y_{s}^3 y_{o_1}^7 y_{o_2}^5 t_1^3 t_2^2 t_3^4 t_4^3 t_5^4 t_6^2 +  y_{s}^3 y_{o_1}^7 y_{o_2}^5 t_1^2 t_2^3 t_3^4 t_4^3 t_5^4 t_6^2 +  2 y_{s}^3 y_{o_1}^7 y_{o_2}^5 t_1 t_2^4 t_3^4 t_4^3 t_5^4 t_6^2 +  2 y_{s}^3 y_{o_1}^7 y_{o_2}^5 t_1^4 t_2 t_3^3 t_4^4 t_5^4 t_6^2 +  y_{s}^3 y_{o_1}^7 y_{o_2}^5 t_1^3 t_2^2 t_3^3 t_4^4 t_5^4 t_6^2 +  y_{s}^3 y_{o_1}^7 y_{o_2}^5 t_1^2 t_2^3 t_3^3 t_4^4 t_5^4 t_6^2 +  2 y_{s}^3 y_{o_1}^7 y_{o_2}^5 t_1 t_2^4 t_3^3 t_4^4 t_5^4 t_6^2 +  y_{s}^3 y_{o_1}^7 y_{o_2}^5 t_1^4 t_2 t_3^2 t_4^5 t_5^4 t_6^2 -  y_{s}^3 y_{o_1}^7 y_{o_2}^5 t_1^3 t_2^2 t_3^2 t_4^5 t_5^4 t_6^2 -  y_{s}^3 y_{o_1}^7 y_{o_2}^5 t_1^2 t_2^3 t_3^2 t_4^5 t_5^4 t_6^2 +  y_{s}^3 y_{o_1}^7 y_{o_2}^5 t_1 t_2^4 t_3^2 t_4^5 t_5^4 t_6^2 +  y_{s}^3 y_{o_1}^7 y_{o_2}^5 t_1^4 t_2 t_3 t_4^6 t_5^4 t_6^2 +  y_{s}^3 y_{o_1}^7 y_{o_2}^5 t_1 t_2^4 t_3 t_4^6 t_5^4 t_6^2 +  y_{s}^4 y_{o_1}^{10} y_{o_2}^6 t_1^4 t_2^2 t_3^5 t_4^5 t_5^6 t_6^2 +  y_{s}^4 y_{o_1}^{10} y_{o_2}^6 t_1^3 t_2^3 t_3^5 t_4^5 t_5^6 t_6^2 +  y_{s}^4 y_{o_1}^{10} y_{o_2}^6 t_1^2 t_2^4 t_3^5 t_4^5 t_5^6 t_6^2 -  y_{s}^2 y_{o_1}^3 y_{o_2}^5 t_1^5 t_3^2 t_4 t_5 t_6^3 -  y_{s}^2 y_{o_1}^3 y_{o_2}^5 t_1^4 t_2 t_3^2 t_4 t_5 t_6^3 -  y_{s}^2 y_{o_1}^3 y_{o_2}^5 t_1^3 t_2^2 t_3^2 t_4 t_5 t_6^3 -  y_{s}^2 y_{o_1}^3 y_{o_2}^5 t_1^2 t_2^3 t_3^2 t_4 t_5 t_6^3 -  y_{s}^2 y_{o_1}^3 y_{o_2}^5 t_1 t_2^4 t_3^2 t_4 t_5 t_6^3 -  y_{s}^2 y_{o_1}^3 y_{o_2}^5 t_2^5 t_3^2 t_4 t_5 t_6^3 -  y_{s}^2 y_{o_1}^3 y_{o_2}^5 t_1^5 t_3 t_4^2 t_5 t_6^3 -  y_{s}^2 y_{o_1}^3 y_{o_2}^5 t_1^4 t_2 t_3 t_4^2 t_5 t_6^3 -  y_{s}^2 y_{o_1}^3 y_{o_2}^5 t_1^3 t_2^2 t_3 t_4^2 t_5 t_6^3 -  y_{s}^2 y_{o_1}^3 y_{o_2}^5 t_1^2 t_2^3 t_3 t_4^2 t_5 t_6^3 -  y_{s}^2 y_{o_1}^3 y_{o_2}^5 t_1 t_2^4 t_3 t_4^2 t_5 t_6^3 -  y_{s}^2 y_{o_1}^3 y_{o_2}^5 t_2^5 t_3 t_4^2 t_5 t_6^3 -  y_{s}^3 y_{o_1}^6 y_{o_2}^6 t_1^3 t_2^3 t_3^6 t_5^3 t_6^3 +  y_{s}^3 y_{o_1}^6 y_{o_2}^6 t_1^5 t_2 t_3^5 t_4 t_5^3 t_6^3 +  y_{s}^3 y_{o_1}^6 y_{o_2}^6 t_1^4 t_2^2 t_3^5 t_4 t_5^3 t_6^3 -  y_{s}^3 y_{o_1}^6 y_{o_2}^6 t_1^3 t_2^3 t_3^5 t_4 t_5^3 t_6^3 +  y_{s}^3 y_{o_1}^6 y_{o_2}^6 t_1^2 t_2^4 t_3^5 t_4 t_5^3 t_6^3 +  y_{s}^3 y_{o_1}^6 y_{o_2}^6 t_1 t_2^5 t_3^5 t_4 t_5^3 t_6^3 +  y_{s}^3 y_{o_1}^6 y_{o_2}^6 t_1^5 t_2 t_3^4 t_4^2 t_5^3 t_6^3 +  y_{s}^3 y_{o_1}^6 y_{o_2}^6 t_1^4 t_2^2 t_3^4 t_4^2 t_5^3 t_6^3 -  y_{s}^3 y_{o_1}^6 y_{o_2}^6 t_1^3 t_2^3 t_3^4 t_4^2 t_5^3 t_6^3 +  y_{s}^3 y_{o_1}^6 y_{o_2}^6 t_1^2 t_2^4 t_3^4 t_4^2 t_5^3 t_6^3 +  y_{s}^3 y_{o_1}^6 y_{o_2}^6 t_1 t_2^5 t_3^4 t_4^2 t_5^3 t_6^3 -  y_{s}^3 y_{o_1}^6 y_{o_2}^6 t_1^6 t_3^3 t_4^3 t_5^3 t_6^3 -  y_{s}^3 y_{o_1}^6 y_{o_2}^6 t_1^5 t_2 t_3^3 t_4^3 t_5^3 t_6^3 -  y_{s}^3 y_{o_1}^6 y_{o_2}^6 t_1^4 t_2^2 t_3^3 t_4^3 t_5^3 t_6^3 -  3 y_{s}^3 y_{o_1}^6 y_{o_2}^6 t_1^3 t_2^3 t_3^3 t_4^3 t_5^3 t_6^3 -  y_{s}^3 y_{o_1}^6 y_{o_2}^6 t_1^2 t_2^4 t_3^3 t_4^3 t_5^3 t_6^3 -  y_{s}^3 y_{o_1}^6 y_{o_2}^6 t_1 t_2^5 t_3^3 t_4^3 t_5^3 t_6^3 -  y_{s}^3 y_{o_1}^6 y_{o_2}^6 t_2^6 t_3^3 t_4^3 t_5^3 t_6^3 +  y_{s}^3 y_{o_1}^6 y_{o_2}^6 t_1^5 t_2 t_3^2 t_4^4 t_5^3 t_6^3 +  y_{s}^3 y_{o_1}^6 y_{o_2}^6 t_1^4 t_2^2 t_3^2 t_4^4 t_5^3 t_6^3 -  y_{s}^3 y_{o_1}^6 y_{o_2}^6 t_1^3 t_2^3 t_3^2 t_4^4 t_5^3 t_6^3 +  y_{s}^3 y_{o_1}^6 y_{o_2}^6 t_1^2 t_2^4 t_3^2 t_4^4 t_5^3 t_6^3 +  y_{s}^3 y_{o_1}^6 y_{o_2}^6 t_1 t_2^5 t_3^2 t_4^4 t_5^3 t_6^3 +  y_{s}^3 y_{o_1}^6 y_{o_2}^6 t_1^5 t_2 t_3 t_4^5 t_5^3 t_6^3 +  y_{s}^3 y_{o_1}^6 y_{o_2}^6 t_1^4 t_2^2 t_3 t_4^5 t_5^3 t_6^3 -  y_{s}^3 y_{o_1}^6 y_{o_2}^6 t_1^3 t_2^3 t_3 t_4^5 t_5^3 t_6^3 +  y_{s}^3 y_{o_1}^6 y_{o_2}^6 t_1^2 t_2^4 t_3 t_4^5 t_5^3 t_6^3 +  y_{s}^3 y_{o_1}^6 y_{o_2}^6 t_1 t_2^5 t_3 t_4^5 t_5^3 t_6^3 -  y_{s}^3 y_{o_1}^6 y_{o_2}^6 t_1^3 t_2^3 t_4^6 t_5^3 t_6^3 +  y_{s}^4 y_{o_1}^9 y_{o_2}^7 t_1^6 t_2 t_3^6 t_4^3 t_5^5 t_6^3 +  y_{s}^4 y_{o_1}^9 y_{o_2}^7 t_1^5 t_2^2 t_3^6 t_4^3 t_5^5 t_6^3 +  2 y_{s}^4 y_{o_1}^9 y_{o_2}^7 t_1^4 t_2^3 t_3^6 t_4^3 t_5^5 t_6^3 +  2 y_{s}^4 y_{o_1}^9 y_{o_2}^7 t_1^3 t_2^4 t_3^6 t_4^3 t_5^5 t_6^3 +  y_{s}^4 y_{o_1}^9 y_{o_2}^7 t_1^2 t_2^5 t_3^6 t_4^3 t_5^5 t_6^3 +  y_{s}^4 y_{o_1}^9 y_{o_2}^7 t_1 t_2^6 t_3^6 t_4^3 t_5^5 t_6^3 -  y_{s}^4 y_{o_1}^9 y_{o_2}^7 t_1^5 t_2^2 t_3^5 t_4^4 t_5^5 t_6^3 +  y_{s}^4 y_{o_1}^9 y_{o_2}^7 t_1^4 t_2^3 t_3^5 t_4^4 t_5^5 t_6^3 +  y_{s}^4 y_{o_1}^9 y_{o_2}^7 t_1^3 t_2^4 t_3^5 t_4^4 t_5^5 t_6^3 -  y_{s}^4 y_{o_1}^9 y_{o_2}^7 t_1^2 t_2^5 t_3^5 t_4^4 t_5^5 t_6^3 -  y_{s}^4 y_{o_1}^9 y_{o_2}^7 t_1^5 t_2^2 t_3^4 t_4^5 t_5^5 t_6^3 +  y_{s}^4 y_{o_1}^9 y_{o_2}^7 t_1^4 t_2^3 t_3^4 t_4^5 t_5^5 t_6^3 +  y_{s}^4 y_{o_1}^9 y_{o_2}^7 t_1^3 t_2^4 t_3^4 t_4^5 t_5^5 t_6^3 -  y_{s}^4 y_{o_1}^9 y_{o_2}^7 t_1^2 t_2^5 t_3^4 t_4^5 t_5^5 t_6^3 +  y_{s}^4 y_{o_1}^9 y_{o_2}^7 t_1^6 t_2 t_3^3 t_4^6 t_5^5 t_6^3 +  y_{s}^4 y_{o_1}^9 y_{o_2}^7 t_1^5 t_2^2 t_3^3 t_4^6 t_5^5 t_6^3 +  2 y_{s}^4 y_{o_1}^9 y_{o_2}^7 t_1^4 t_2^3 t_3^3 t_4^6 t_5^5 t_6^3 +  2 y_{s}^4 y_{o_1}^9 y_{o_2}^7 t_1^3 t_2^4 t_3^3 t_4^6 t_5^5 t_6^3 +  y_{s}^4 y_{o_1}^9 y_{o_2}^7 t_1^2 t_2^5 t_3^3 t_4^6 t_5^5 t_6^3 +  y_{s}^4 y_{o_1}^9 y_{o_2}^7 t_1 t_2^6 t_3^3 t_4^6 t_5^5 t_6^3 -  y_{s}^5 y_{o_1}^{12} y_{o_2}^8 t_1^6 t_2^2 t_3^6 t_4^6 t_5^7 t_6^3 -  y_{s}^5 y_{o_1}^{12} y_{o_2}^8 t_1^5 t_2^3 t_3^6 t_4^6 t_5^7 t_6^3 -  y_{s}^5 y_{o_1}^{12} y_{o_2}^8 t_1^4 t_2^4 t_3^6 t_4^6 t_5^7 t_6^3 -  y_{s}^5 y_{o_1}^{12} y_{o_2}^8 t_1^3 t_2^5 t_3^6 t_4^6 t_5^7 t_6^3 -  y_{s}^5 y_{o_1}^{12} y_{o_2}^8 t_1^2 t_2^6 t_3^6 t_4^6 t_5^7 t_6^3 -  y_{s}^2 y_{o_1}^2 y_{o_2}^6 t_1^5 t_2 t_3 t_4 t_6^4 -  y_{s}^2 y_{o_1}^2 y_{o_2}^6 t_1^4 t_2^2 t_3 t_4 t_6^4 -  y_{s}^2 y_{o_1}^2 y_{o_2}^6 t_1^3 t_2^3 t_3 t_4 t_6^4 -  y_{s}^2 y_{o_1}^2 y_{o_2}^6 t_1^2 t_2^4 t_3 t_4 t_6^4 -  y_{s}^2 y_{o_1}^2 y_{o_2}^6 t_1 t_2^5 t_3 t_4 t_6^4 +  y_{s}^3 y_{o_1}^5 y_{o_2}^7 t_1^6 t_2 t_3^4 t_4 t_5^2 t_6^4 +  y_{s}^3 y_{o_1}^5 y_{o_2}^7 t_1^5 t_2^2 t_3^4 t_4 t_5^2 t_6^4 +  2 y_{s}^3 y_{o_1}^5 y_{o_2}^7 t_1^4 t_2^3 t_3^4 t_4 t_5^2 t_6^4 +  2 y_{s}^3 y_{o_1}^5 y_{o_2}^7 t_1^3 t_2^4 t_3^4 t_4 t_5^2 t_6^4 +  y_{s}^3 y_{o_1}^5 y_{o_2}^7 t_1^2 t_2^5 t_3^4 t_4 t_5^2 t_6^4 +  y_{s}^3 y_{o_1}^5 y_{o_2}^7 t_1 t_2^6 t_3^4 t_4 t_5^2 t_6^4 -  y_{s}^3 y_{o_1}^5 y_{o_2}^7 t_1^5 t_2^2 t_3^3 t_4^2 t_5^2 t_6^4 +  y_{s}^3 y_{o_1}^5 y_{o_2}^7 t_1^4 t_2^3 t_3^3 t_4^2 t_5^2 t_6^4 +  y_{s}^3 y_{o_1}^5 y_{o_2}^7 t_1^3 t_2^4 t_3^3 t_4^2 t_5^2 t_6^4 -  y_{s}^3 y_{o_1}^5 y_{o_2}^7 t_1^2 t_2^5 t_3^3 t_4^2 t_5^2 t_6^4 -  y_{s}^3 y_{o_1}^5 y_{o_2}^7 t_1^5 t_2^2 t_3^2 t_4^3 t_5^2 t_6^4 +  y_{s}^3 y_{o_1}^5 y_{o_2}^7 t_1^4 t_2^3 t_3^2 t_4^3 t_5^2 t_6^4 +  y_{s}^3 y_{o_1}^5 y_{o_2}^7 t_1^3 t_2^4 t_3^2 t_4^3 t_5^2 t_6^4 -  y_{s}^3 y_{o_1}^5 y_{o_2}^7 t_1^2 t_2^5 t_3^2 t_4^3 t_5^2 t_6^4 +  y_{s}^3 y_{o_1}^5 y_{o_2}^7 t_1^6 t_2 t_3 t_4^4 t_5^2 t_6^4 +  y_{s}^3 y_{o_1}^5 y_{o_2}^7 t_1^5 t_2^2 t_3 t_4^4 t_5^2 t_6^4 +  2 y_{s}^3 y_{o_1}^5 y_{o_2}^7 t_1^4 t_2^3 t_3 t_4^4 t_5^2 t_6^4 +  2 y_{s}^3 y_{o_1}^5 y_{o_2}^7 t_1^3 t_2^4 t_3 t_4^4 t_5^2 t_6^4 +  y_{s}^3 y_{o_1}^5 y_{o_2}^7 t_1^2 t_2^5 t_3 t_4^4 t_5^2 t_6^4 +  y_{s}^3 y_{o_1}^5 y_{o_2}^7 t_1 t_2^6 t_3 t_4^4 t_5^2 t_6^4 -  y_{s}^4 y_{o_1}^8 y_{o_2}^8 t_1^4 t_2^4 t_3^7 t_4 t_5^4 t_6^4 +  y_{s}^4 y_{o_1}^8 y_{o_2}^8 t_1^6 t_2^2 t_3^6 t_4^2 t_5^4 t_6^4 +  y_{s}^4 y_{o_1}^8 y_{o_2}^8 t_1^5 t_2^3 t_3^6 t_4^2 t_5^4 t_6^4 -  y_{s}^4 y_{o_1}^8 y_{o_2}^8 t_1^4 t_2^4 t_3^6 t_4^2 t_5^4 t_6^4 +  y_{s}^4 y_{o_1}^8 y_{o_2}^8 t_1^3 t_2^5 t_3^6 t_4^2 t_5^4 t_6^4 +  y_{s}^4 y_{o_1}^8 y_{o_2}^8 t_1^2 t_2^6 t_3^6 t_4^2 t_5^4 t_6^4 +  y_{s}^4 y_{o_1}^8 y_{o_2}^8 t_1^6 t_2^2 t_3^5 t_4^3 t_5^4 t_6^4 +  y_{s}^4 y_{o_1}^8 y_{o_2}^8 t_1^5 t_2^3 t_3^5 t_4^3 t_5^4 t_6^4 -  y_{s}^4 y_{o_1}^8 y_{o_2}^8 t_1^4 t_2^4 t_3^5 t_4^3 t_5^4 t_6^4 +  y_{s}^4 y_{o_1}^8 y_{o_2}^8 t_1^3 t_2^5 t_3^5 t_4^3 t_5^4 t_6^4 +  y_{s}^4 y_{o_1}^8 y_{o_2}^8 t_1^2 t_2^6 t_3^5 t_4^3 t_5^4 t_6^4 -  y_{s}^4 y_{o_1}^8 y_{o_2}^8 t_1^7 t_2 t_3^4 t_4^4 t_5^4 t_6^4 -  y_{s}^4 y_{o_1}^8 y_{o_2}^8 t_1^6 t_2^2 t_3^4 t_4^4 t_5^4 t_6^4 -  y_{s}^4 y_{o_1}^8 y_{o_2}^8 t_1^5 t_2^3 t_3^4 t_4^4 t_5^4 t_6^4 -  3 y_{s}^4 y_{o_1}^8 y_{o_2}^8 t_1^4 t_2^4 t_3^4 t_4^4 t_5^4 t_6^4 -  y_{s}^4 y_{o_1}^8 y_{o_2}^8 t_1^3 t_2^5 t_3^4 t_4^4 t_5^4 t_6^4 -  y_{s}^4 y_{o_1}^8 y_{o_2}^8 t_1^2 t_2^6 t_3^4 t_4^4 t_5^4 t_6^4 -  y_{s}^4 y_{o_1}^8 y_{o_2}^8 t_1 t_2^7 t_3^4 t_4^4 t_5^4 t_6^4 +  y_{s}^4 y_{o_1}^8 y_{o_2}^8 t_1^6 t_2^2 t_3^3 t_4^5 t_5^4 t_6^4 +  y_{s}^4 y_{o_1}^8 y_{o_2}^8 t_1^5 t_2^3 t_3^3 t_4^5 t_5^4 t_6^4 -  y_{s}^4 y_{o_1}^8 y_{o_2}^8 t_1^4 t_2^4 t_3^3 t_4^5 t_5^4 t_6^4 +  y_{s}^4 y_{o_1}^8 y_{o_2}^8 t_1^3 t_2^5 t_3^3 t_4^5 t_5^4 t_6^4 +  y_{s}^4 y_{o_1}^8 y_{o_2}^8 t_1^2 t_2^6 t_3^3 t_4^5 t_5^4 t_6^4 +  y_{s}^4 y_{o_1}^8 y_{o_2}^8 t_1^6 t_2^2 t_3^2 t_4^6 t_5^4 t_6^4 +  y_{s}^4 y_{o_1}^8 y_{o_2}^8 t_1^5 t_2^3 t_3^2 t_4^6 t_5^4 t_6^4 -  y_{s}^4 y_{o_1}^8 y_{o_2}^8 t_1^4 t_2^4 t_3^2 t_4^6 t_5^4 t_6^4 +  y_{s}^4 y_{o_1}^8 y_{o_2}^8 t_1^3 t_2^5 t_3^2 t_4^6 t_5^4 t_6^4 +  y_{s}^4 y_{o_1}^8 y_{o_2}^8 t_1^2 t_2^6 t_3^2 t_4^6 t_5^4 t_6^4 -  y_{s}^4 y_{o_1}^8 y_{o_2}^8 t_1^4 t_2^4 t_3 t_4^7 t_5^4 t_6^4 -  y_{s}^5 y_{o_1}^{11} y_{o_2}^9 t_1^7 t_2^2 t_3^6 t_4^5 t_5^6 t_6^4 -  y_{s}^5 y_{o_1}^{11} y_{o_2}^9 t_1^6 t_2^3 t_3^6 t_4^5 t_5^6 t_6^4 -  y_{s}^5 y_{o_1}^{11} y_{o_2}^9 t_1^5 t_2^4 t_3^6 t_4^5 t_5^6 t_6^4 -  y_{s}^5 y_{o_1}^{11} y_{o_2}^9 t_1^4 t_2^5 t_3^6 t_4^5 t_5^6 t_6^4 -  y_{s}^5 y_{o_1}^{11} y_{o_2}^9 t_1^3 t_2^6 t_3^6 t_4^5 t_5^6 t_6^4 -  y_{s}^5 y_{o_1}^{11} y_{o_2}^9 t_1^2 t_2^7 t_3^6 t_4^5 t_5^6 t_6^4 -  y_{s}^5 y_{o_1}^{11} y_{o_2}^9 t_1^7 t_2^2 t_3^5 t_4^6 t_5^6 t_6^4 -  y_{s}^5 y_{o_1}^{11} y_{o_2}^9 t_1^6 t_2^3 t_3^5 t_4^6 t_5^6 t_6^4 -  y_{s}^5 y_{o_1}^{11} y_{o_2}^9 t_1^5 t_2^4 t_3^5 t_4^6 t_5^6 t_6^4 -  y_{s}^5 y_{o_1}^{11} y_{o_2}^9 t_1^4 t_2^5 t_3^5 t_4^6 t_5^6 t_6^4 -  y_{s}^5 y_{o_1}^{11} y_{o_2}^9 t_1^3 t_2^6 t_3^5 t_4^6 t_5^6 t_6^4 -  y_{s}^5 y_{o_1}^{11} y_{o_2}^9 t_1^2 t_2^7 t_3^5 t_4^6 t_5^6 t_6^4 +  y_{s}^3 y_{o_1}^4 y_{o_2}^8 t_1^5 t_2^3 t_3^2 t_4^2 t_5 t_6^5 +  y_{s}^3 y_{o_1}^4 y_{o_2}^8 t_1^4 t_2^4 t_3^2 t_4^2 t_5 t_6^5 +  y_{s}^3 y_{o_1}^4 y_{o_2}^8 t_1^3 t_2^5 t_3^2 t_4^2 t_5 t_6^5 +  y_{s}^4 y_{o_1}^7 y_{o_2}^9 t_1^6 t_2^3 t_3^6 t_4 t_5^3 t_6^5 +  y_{s}^4 y_{o_1}^7 y_{o_2}^9 t_1^3 t_2^6 t_3^6 t_4 t_5^3 t_6^5 +  y_{s}^4 y_{o_1}^7 y_{o_2}^9 t_1^6 t_2^3 t_3^5 t_4^2 t_5^3 t_6^5 -  y_{s}^4 y_{o_1}^7 y_{o_2}^9 t_1^5 t_2^4 t_3^5 t_4^2 t_5^3 t_6^5 -  y_{s}^4 y_{o_1}^7 y_{o_2}^9 t_1^4 t_2^5 t_3^5 t_4^2 t_5^3 t_6^5 +  y_{s}^4 y_{o_1}^7 y_{o_2}^9 t_1^3 t_2^6 t_3^5 t_4^2 t_5^3 t_6^5 +  2 y_{s}^4 y_{o_1}^7 y_{o_2}^9 t_1^6 t_2^3 t_3^4 t_4^3 t_5^3 t_6^5 +  y_{s}^4 y_{o_1}^7 y_{o_2}^9 t_1^5 t_2^4 t_3^4 t_4^3 t_5^3 t_6^5 +  y_{s}^4 y_{o_1}^7 y_{o_2}^9 t_1^4 t_2^5 t_3^4 t_4^3 t_5^3 t_6^5 +  2 y_{s}^4 y_{o_1}^7 y_{o_2}^9 t_1^3 t_2^6 t_3^4 t_4^3 t_5^3 t_6^5 +  2 y_{s}^4 y_{o_1}^7 y_{o_2}^9 t_1^6 t_2^3 t_3^3 t_4^4 t_5^3 t_6^5 +  y_{s}^4 y_{o_1}^7 y_{o_2}^9 t_1^5 t_2^4 t_3^3 t_4^4 t_5^3 t_6^5 +  y_{s}^4 y_{o_1}^7 y_{o_2}^9 t_1^4 t_2^5 t_3^3 t_4^4 t_5^3 t_6^5 +  2 y_{s}^4 y_{o_1}^7 y_{o_2}^9 t_1^3 t_2^6 t_3^3 t_4^4 t_5^3 t_6^5 +  y_{s}^4 y_{o_1}^7 y_{o_2}^9 t_1^6 t_2^3 t_3^2 t_4^5 t_5^3 t_6^5 -  y_{s}^4 y_{o_1}^7 y_{o_2}^9 t_1^5 t_2^4 t_3^2 t_4^5 t_5^3 t_6^5 -  y_{s}^4 y_{o_1}^7 y_{o_2}^9 t_1^4 t_2^5 t_3^2 t_4^5 t_5^3 t_6^5 +  y_{s}^4 y_{o_1}^7 y_{o_2}^9 t_1^3 t_2^6 t_3^2 t_4^5 t_5^3 t_6^5 +  y_{s}^4 y_{o_1}^7 y_{o_2}^9 t_1^6 t_2^3 t_3 t_4^6 t_5^3 t_6^5 +  y_{s}^4 y_{o_1}^7 y_{o_2}^9 t_1^3 t_2^6 t_3 t_4^6 t_5^3 t_6^5 -  y_{s}^5 y_{o_1}^{10} y_{o_2}^{10} t_1^6 t_2^4 t_3^7 t_4^3 t_5^5 t_6^5 -  y_{s}^5 y_{o_1}^{10} y_{o_2}^{10} t_1^5 t_2^5 t_3^7 t_4^3 t_5^5 t_6^5 -  y_{s}^5 y_{o_1}^{10} y_{o_2}^{10} t_1^4 t_2^6 t_3^7 t_4^3 t_5^5 t_6^5 -  y_{s}^5 y_{o_1}^{10} y_{o_2}^{10} t_1^7 t_2^3 t_3^6 t_4^4 t_5^5 t_6^5 -  2 y_{s}^5 y_{o_1}^{10} y_{o_2}^{10} t_1^6 t_2^4 t_3^6 t_4^4 t_5^5 t_6^5 -  y_{s}^5 y_{o_1}^{10} y_{o_2}^{10} t_1^5 t_2^5 t_3^6 t_4^4 t_5^5 t_6^5 -  2 y_{s}^5 y_{o_1}^{10} y_{o_2}^{10} t_1^4 t_2^6 t_3^6 t_4^4 t_5^5 t_6^5 -  y_{s}^5 y_{o_1}^{10} y_{o_2}^{10} t_1^3 t_2^7 t_3^6 t_4^4 t_5^5 t_6^5 -  y_{s}^5 y_{o_1}^{10} y_{o_2}^{10} t_1^7 t_2^3 t_3^5 t_4^5 t_5^5 t_6^5 -  y_{s}^5 y_{o_1}^{10} y_{o_2}^{10} t_1^6 t_2^4 t_3^5 t_4^5 t_5^5 t_6^5 +  y_{s}^5 y_{o_1}^{10} y_{o_2}^{10} t_1^5 t_2^5 t_3^5 t_4^5 t_5^5 t_6^5 -  y_{s}^5 y_{o_1}^{10} y_{o_2}^{10} t_1^4 t_2^6 t_3^5 t_4^5 t_5^5 t_6^5 -  y_{s}^5 y_{o_1}^{10} y_{o_2}^{10} t_1^3 t_2^7 t_3^5 t_4^5 t_5^5 t_6^5 -  y_{s}^5 y_{o_1}^{10} y_{o_2}^{10} t_1^7 t_2^3 t_3^4 t_4^6 t_5^5 t_6^5 -  2 y_{s}^5 y_{o_1}^{10} y_{o_2}^{10} t_1^6 t_2^4 t_3^4 t_4^6 t_5^5 t_6^5 -  y_{s}^5 y_{o_1}^{10} y_{o_2}^{10} t_1^5 t_2^5 t_3^4 t_4^6 t_5^5 t_6^5 -  2 y_{s}^5 y_{o_1}^{10} y_{o_2}^{10} t_1^4 t_2^6 t_3^4 t_4^6 t_5^5 t_6^5 -  y_{s}^5 y_{o_1}^{10} y_{o_2}^{10} t_1^3 t_2^7 t_3^4 t_4^6 t_5^5 t_6^5 -  y_{s}^5 y_{o_1}^{10} y_{o_2}^{10} t_1^6 t_2^4 t_3^3 t_4^7 t_5^5 t_6^5 -  y_{s}^5 y_{o_1}^{10} y_{o_2}^{10} t_1^5 t_2^5 t_3^3 t_4^7 t_5^5 t_6^5 -  y_{s}^5 y_{o_1}^{10} y_{o_2}^{10} t_1^4 t_2^6 t_3^3 t_4^7 t_5^5 t_6^5 +  y_{s}^6 y_{o_1}^{13} y_{o_2}^{11} t_1^6 t_2^5 t_3^7 t_4^6 t_5^7 t_6^5 +  y_{s}^6 y_{o_1}^{13} y_{o_2}^{11} t_1^5 t_2^6 t_3^7 t_4^6 t_5^7 t_6^5 +  y_{s}^6 y_{o_1}^{13} y_{o_2}^{11} t_1^6 t_2^5 t_3^6 t_4^7 t_5^7 t_6^5 +  y_{s}^6 y_{o_1}^{13} y_{o_2}^{11} t_1^5 t_2^6 t_3^6 t_4^7 t_5^7 t_6^5 +  y_{s}^4 y_{o_1}^6 y_{o_2}^{10} t_1^5 t_2^5 t_3^4 t_4^2 t_5^2 t_6^6 +  y_{s}^4 y_{o_1}^6 y_{o_2}^{10} t_1^5 t_2^5 t_3^3 t_4^3 t_5^2 t_6^6 +  y_{s}^4 y_{o_1}^6 y_{o_2}^{10} t_1^5 t_2^5 t_3^2 t_4^4 t_5^2 t_6^6 -  y_{s}^5 y_{o_1}^9 y_{o_2}^{11} t_1^6 t_2^5 t_3^7 t_4^2 t_5^4 t_6^6 -  y_{s}^5 y_{o_1}^9 y_{o_2}^{11} t_1^5 t_2^6 t_3^7 t_4^2 t_5^4 t_6^6 -  y_{s}^5 y_{o_1}^9 y_{o_2}^{11} t_1^6 t_2^5 t_3^6 t_4^3 t_5^4 t_6^6 -  y_{s}^5 y_{o_1}^9 y_{o_2}^{11} t_1^5 t_2^6 t_3^6 t_4^3 t_5^4 t_6^6 -  y_{s}^5 y_{o_1}^9 y_{o_2}^{11} t_1^6 t_2^5 t_3^5 t_4^4 t_5^4 t_6^6 -  y_{s}^5 y_{o_1}^9 y_{o_2}^{11} t_1^5 t_2^6 t_3^5 t_4^4 t_5^4 t_6^6 -  y_{s}^5 y_{o_1}^9 y_{o_2}^{11} t_1^6 t_2^5 t_3^4 t_4^5 t_5^4 t_6^6 -  y_{s}^5 y_{o_1}^9 y_{o_2}^{11} t_1^5 t_2^6 t_3^4 t_4^5 t_5^4 t_6^6 -  y_{s}^5 y_{o_1}^9 y_{o_2}^{11} t_1^6 t_2^5 t_3^3 t_4^6 t_5^4 t_6^6 -  y_{s}^5 y_{o_1}^9 y_{o_2}^{11} t_1^5 t_2^6 t_3^3 t_4^6 t_5^4 t_6^6 -  y_{s}^5 y_{o_1}^9 y_{o_2}^{11} t_1^6 t_2^5 t_3^2 t_4^7 t_5^4 t_6^6 -  y_{s}^5 y_{o_1}^9 y_{o_2}^{11} t_1^5 t_2^6 t_3^2 t_4^7 t_5^4 t_6^6 +  y_{s}^6 y_{o_1}^{12} y_{o_2}^{12} t_1^7 t_2^5 t_3^7 t_4^5 t_5^6 t_6^6 +  y_{s}^6 y_{o_1}^{12} y_{o_2}^{12} t_1^6 t_2^6 t_3^7 t_4^5 t_5^6 t_6^6 +  y_{s}^6 y_{o_1}^{12} y_{o_2}^{12} t_1^5 t_2^7 t_3^7 t_4^5 t_5^6 t_6^6 +  y_{s}^6 y_{o_1}^{12} y_{o_2}^{12} t_1^7 t_2^5 t_3^6 t_4^6 t_5^6 t_6^6 +  y_{s}^6 y_{o_1}^{12} y_{o_2}^{12} t_1^6 t_2^6 t_3^6 t_4^6 t_5^6 t_6^6 +  y_{s}^6 y_{o_1}^{12} y_{o_2}^{12} t_1^5 t_2^7 t_3^6 t_4^6 t_5^6 t_6^6 +  y_{s}^6 y_{o_1}^{12} y_{o_2}^{12} t_1^7 t_2^5 t_3^5 t_4^7 t_5^6 t_6^6 +  y_{s}^6 y_{o_1}^{12} y_{o_2}^{12} t_1^6 t_2^6 t_3^5 t_4^7 t_5^6 t_6^6 +  y_{s}^6 y_{o_1}^{12} y_{o_2}^{12} t_1^5 t_2^7 t_3^5 t_4^7 t_5^6 t_6^6 -  y_{s}^5 y_{o_1}^8 y_{o_2}^{12} t_1^6 t_2^6 t_3^6 t_4^2 t_5^3 t_6^7 -  y_{s}^5 y_{o_1}^8 y_{o_2}^{12} t_1^6 t_2^6 t_3^5 t_4^3 t_5^3 t_6^7 -  y_{s}^5 y_{o_1}^8 y_{o_2}^{12} t_1^6 t_2^6 t_3^4 t_4^4 t_5^3 t_6^7 -  y_{s}^5 y_{o_1}^8 y_{o_2}^{12} t_1^6 t_2^6 t_3^3 t_4^5 t_5^3 t_6^7 -  y_{s}^5 y_{o_1}^8 y_{o_2}^{12} t_1^6 t_2^6 t_3^2 t_4^6 t_5^3 t_6^7 +  y_{s}^6 y_{o_1}^{11} y_{o_2}^{13} t_1^7 t_2^6 t_3^6 t_4^5 t_5^5 t_6^7 +  y_{s}^6 y_{o_1}^{11} y_{o_2}^{13} t_1^6 t_2^7 t_3^6 t_4^5 t_5^5 t_6^7 +  y_{s}^6 y_{o_1}^{11} y_{o_2}^{13} t_1^7 t_2^6 t_3^5 t_4^6 t_5^5 t_6^7 +  y_{s}^6 y_{o_1}^{11} y_{o_2}^{13} t_1^6 t_2^7 t_3^5 t_4^6 t_5^5 t_6^7 +  y_{s}^7 y_{o_1}^{14} y_{o_2}^{14} t_1^7 t_2^7 t_3^7 t_4^7 t_5^7 t_6^7~,~$
\end{quote}
\endgroup
%
where $y_{s}$ counts $\prod_{a=1}^{11}s_a$, $y_{o_1}$ counts $o_1 o_2$ and $y_{o_2}$ counts $o_3 o_4$.

\subsection{Model 7 \label{app_num_07}}

\begingroup\makeatletter\def\f@size{7}\check@mathfonts
\begin{quote}\raggedright
$
P(t_i,y_s,y_{o_1},y_{o_2},y_{o_3}; \mathcal{M}_7) =
1 + y_{s} y_{o_1}^3 y_{o_2}^2 y_{o_3}^3 t_1^2 t_2 t_3^2 t_4^2 +  y_{s} y_{o_1}^4 y_{o_2}^3 y_{o_3}^3 t_1 t_2^2 t_3^2 t_4^3 +  y_{s} y_{o_1}^2 y_{o_2}^2 y_{o_3}^2 t_1^2 t_2 t_3 t_4 t_5 +  y_{s} y_{o_1}^3 y_{o_2}^3 y_{o_3}^2 t_1 t_2^2 t_3 t_4^2 t_5 -  y_{s}^2 y_{o_1}^4 y_{o_2}^3 y_{o_3}^5 t_1^5 t_2 t_3^3 t_4^2 t_5 +  y_{s} y_{o_1}^4 y_{o_2}^4 y_{o_3}^2 t_2^3 t_3 t_4^3 t_5 -  y_{s}^2 y_{o_1}^5 y_{o_2}^4 y_{o_3}^5 t_1^4 t_2^2 t_3^3 t_4^3 t_5 -  y_{s}^2 y_{o_1}^6 y_{o_2}^5 y_{o_3}^5 t_1^3 t_2^3 t_3^3 t_4^4 t_5 +  y_{s} y_{o_1}^2 y_{o_2}^3 y_{o_3} t_1 t_2^2 t_4 t_5^2 -  y_{s}^2 y_{o_1}^3 y_{o_2}^3 y_{o_3}^4 t_1^5 t_2 t_3^2 t_4 t_5^2 -  2 y_{s}^2 y_{o_1}^4 y_{o_2}^4 y_{o_3}^4 t_1^4 t_2^2 t_3^2 t_4^2 t_5^2 -  2 y_{s}^2 y_{o_1}^5 y_{o_2}^5 y_{o_3}^4 t_1^3 t_2^3 t_3^2 t_4^3 t_5^2 -  y_{s}^2 y_{o_1}^6 y_{o_2}^6 y_{o_3}^4 t_1^2 t_2^4 t_3^2 t_4^4 t_5^2 -  y_{s}^2 y_{o_1}^7 y_{o_2}^7 y_{o_3}^4 t_1 t_2^5 t_3^2 t_4^5 t_5^2 -  y_{s}^2 y_{o_1}^3 y_{o_2}^4 y_{o_3}^3 t_1^4 t_2^2 t_3 t_4 t_5^3 -  y_{s}^2 y_{o_1}^4 y_{o_2}^5 y_{o_3}^3 t_1^3 t_2^3 t_3 t_4^2 t_5^3 +  y_{s}^3 y_{o_1}^5 y_{o_2}^5 y_{o_3}^6 t_1^7 t_2^2 t_3^3 t_4^2 t_5^3 -  y_{s}^2 y_{o_1}^5 y_{o_2}^6 y_{o_3}^3 t_1^2 t_2^4 t_3 t_4^3 t_5^3 +  y_{s}^3 y_{o_1}^6 y_{o_2}^6 y_{o_3}^6 t_1^6 t_2^3 t_3^3 t_4^3 t_5^3 +  y_{s}^3 y_{o_1}^7 y_{o_2}^7 y_{o_3}^6 t_1^5 t_2^4 t_3^3 t_4^4 t_5^3 +  y_{s}^3 y_{o_1}^6 y_{o_2}^7 y_{o_3}^5 t_1^5 t_2^4 t_3^2 t_4^3 t_5^4 +  y_{s}^4 y_{o_1}^9 y_{o_2}^9 y_{o_3}^8 t_1^7 t_2^5 t_3^4 t_4^5 t_5^4 +  y_{s}^4 y_{o_1}^{10} y_{o_2}^{10} y_{o_3}^8 t_1^6 t_2^6 t_3^4 t_4^6 t_5^4 +  y_{s} y_{o_1}^2 y_{o_2} y_{o_3}^3 t_1^2 t_3^2 t_4 t_6 +  y_{s} y_{o_1}^3 y_{o_2}^2 y_{o_3}^3 t_1 t_2 t_3^2 t_4^2 t_6 +  y_{s} y_{o_1}^4 y_{o_2}^3 y_{o_3}^3 t_2^2 t_3^2 t_4^3 t_6 +  y_{s} y_{o_1} y_{o_2} y_{o_3}^2 t_1^2 t_3 t_5 t_6 +  y_{s} y_{o_1}^2 y_{o_2}^2 y_{o_3}^2 t_1 t_2 t_3 t_4 t_5 t_6 -  y_{s}^2 y_{o_1}^3 y_{o_2}^2 y_{o_3}^5 t_1^5 t_3^3 t_4 t_5 t_6 +  y_{s} y_{o_1}^3 y_{o_2}^3 y_{o_3}^2 t_2^2 t_3 t_4^2 t_5 t_6 -  y_{s}^2 y_{o_1}^4 y_{o_2}^3 y_{o_3}^5 t_1^4 t_2 t_3^3 t_4^2 t_5 t_6 -  y_{s}^2 y_{o_1}^5 y_{o_2}^4 y_{o_3}^5 t_1^3 t_2^2 t_3^3 t_4^3 t_5 t_6 +  y_{s} y_{o_1} y_{o_2}^2 y_{o_3} t_1 t_2 t_5^2 t_6 + y_{s} y_{o_1}^2 y_{o_2}^3 y_{o_3} t_2^2 t_4 t_5^2 t_6 -  2 y_{s}^2 y_{o_1}^3 y_{o_2}^3 y_{o_3}^4 t_1^4 t_2 t_3^2 t_4 t_5^2 t_6 -  2 y_{s}^2 y_{o_1}^4 y_{o_2}^4 y_{o_3}^4 t_1^3 t_2^2 t_3^2 t_4^2 t_5^2 t_6 -  y_{s}^2 y_{o_1}^5 y_{o_2}^5 y_{o_3}^4 t_1^2 t_2^3 t_3^2 t_4^3 t_5^2 t_6 -  y_{s}^2 y_{o_1}^6 y_{o_2}^6 y_{o_3}^4 t_1 t_2^4 t_3^2 t_4^4 t_5^2 t_6 -  y_{s}^3 y_{o_1}^7 y_{o_2}^6 y_{o_3}^7 t_1^5 t_2^3 t_3^4 t_4^4 t_5^2 t_6 -  y_{s}^2 y_{o_1}^7 y_{o_2}^7 y_{o_3}^4 t_2^5 t_3^2 t_4^5 t_5^2 t_6 -  y_{s}^2 y_{o_1}^2 y_{o_2}^3 y_{o_3}^3 t_1^4 t_2 t_3 t_5^3 t_6 -  y_{s}^2 y_{o_1}^3 y_{o_2}^4 y_{o_3}^3 t_1^3 t_2^2 t_3 t_4 t_5^3 t_6 +  y_{s}^3 y_{o_1}^4 y_{o_2}^4 y_{o_3}^6 t_1^7 t_2 t_3^3 t_4 t_5^3 t_6 -  y_{s}^2 y_{o_1}^4 y_{o_2}^5 y_{o_3}^3 t_1^2 t_2^3 t_3 t_4^2 t_5^3 t_6 +  y_{s}^3 y_{o_1}^5 y_{o_2}^5 y_{o_3}^6 t_1^6 t_2^2 t_3^3 t_4^2 t_5^3 t_6 +  y_{s}^3 y_{o_1}^6 y_{o_2}^6 y_{o_3}^6 t_1^5 t_2^3 t_3^3 t_4^3 t_5^3 t_6 +  y_{s}^4 y_{o_1}^8 y_{o_2}^8 y_{o_3}^8 t_1^7 t_2^4 t_3^4 t_4^4 t_5^4 t_6 +  y_{s}^4 y_{o_1}^9 y_{o_2}^9 y_{o_3}^8 t_1^6 t_2^5 t_3^4 t_4^5 t_5^4 t_6 +  y_{s}^4 y_{o_1}^{10} y_{o_2}^{10} y_{o_3}^8 t_1^5 t_2^6 t_3^4 t_4^6 t_5^4 t_6 +  y_{s} y_{o_1}^2 y_{o_2} y_{o_3}^3 t_1 t_3^2 t_4 t_6^2 +  y_{s} y_{o_1}^3 y_{o_2}^2 y_{o_3}^3 t_2 t_3^2 t_4^2 t_6^2 +  y_{s}^2 y_{o_1}^6 y_{o_2}^4 y_{o_3}^6 t_1^2 t_2^2 t_3^4 t_4^4 t_6^2 +  y_{s} y_{o_1} y_{o_2} y_{o_3}^2 t_1 t_3 t_5 t_6^2 +  y_{s} y_{o_1}^2 y_{o_2}^2 y_{o_3}^2 t_2 t_3 t_4 t_5 t_6^2 -  y_{s}^2 y_{o_1}^3 y_{o_2}^2 y_{o_3}^5 t_1^4 t_3^3 t_4 t_5 t_6^2 -  y_{s}^2 y_{o_1}^4 y_{o_2}^3 y_{o_3}^5 t_1^3 t_2 t_3^3 t_4^2 t_5 t_6^2 +  y_{s}^2 y_{o_1}^5 y_{o_2}^4 y_{o_3}^5 t_1^2 t_2^2 t_3^3 t_4^3 t_5 t_6^2 -  y_{s}^3 y_{o_1}^7 y_{o_2}^5 y_{o_3}^8 t_1^5 t_2^2 t_3^5 t_4^4 t_5 t_6^2 -  2 y_{s}^2 y_{o_1}^3 y_{o_2}^3 y_{o_3}^4 t_1^3 t_2 t_3^2 t_4 t_5^2 t_6^2 -  y_{s}^2 y_{o_1}^4 y_{o_2}^4 y_{o_3}^4 t_1^2 t_2^2 t_3^2 t_4^2 t_5^2 t_6^2 -  y_{s}^2 y_{o_1}^5 y_{o_2}^5 y_{o_3}^4 t_1 t_2^3 t_3^2 t_4^3 t_5^2 t_6^2 -  y_{s}^3 y_{o_1}^6 y_{o_2}^5 y_{o_3}^7 t_1^5 t_2^2 t_3^4 t_4^3 t_5^2 t_6^2 -  y_{s}^2 y_{o_1}^6 y_{o_2}^6 y_{o_3}^4 t_2^4 t_3^2 t_4^4 t_5^2 t_6^2 -  2 y_{s}^3 y_{o_1}^7 y_{o_2}^6 y_{o_3}^7 t_1^4 t_2^3 t_3^4 t_4^4 t_5^2 t_6^2 -  y_{s}^3 y_{o_1}^9 y_{o_2}^8 y_{o_3}^7 t_1^2 t_2^5 t_3^4 t_4^6 t_5^2 t_6^2 -  y_{s}^2 y_{o_1}^2 y_{o_2}^3 y_{o_3}^3 t_1^3 t_2 t_3 t_5^3 t_6^2 -  y_{s}^2 y_{o_1}^3 y_{o_2}^4 y_{o_3}^3 t_1^2 t_2^2 t_3 t_4 t_5^3 t_6^2 +  y_{s}^3 y_{o_1}^4 y_{o_2}^4 y_{o_3}^6 t_1^6 t_2 t_3^3 t_4 t_5^3 t_6^2 -  y_{s}^2 y_{o_1}^4 y_{o_2}^5 y_{o_3}^3 t_1 t_2^3 t_3 t_4^2 t_5^3 t_6^2 +  y_{s}^3 y_{o_1}^5 y_{o_2}^5 y_{o_3}^6 t_1^5 t_2^2 t_3^3 t_4^2 t_5^3 t_6^2 -  y_{s}^2 y_{o_1}^5 y_{o_2}^6 y_{o_3}^3 t_2^4 t_3 t_4^3 t_5^3 t_6^2 +  y_{s}^3 y_{o_1}^7 y_{o_2}^7 y_{o_3}^6 t_1^3 t_2^4 t_3^3 t_4^4 t_5^3 t_6^2 +  y_{s}^4 y_{o_1}^8 y_{o_2}^7 y_{o_3}^9 t_1^7 t_2^3 t_3^5 t_4^4 t_5^3 t_6^2 -  y_{s}^3 y_{o_1}^8 y_{o_2}^8 y_{o_3}^6 t_1^2 t_2^5 t_3^3 t_4^5 t_5^3 t_6^2 +  y_{s}^4 y_{o_1}^{10} y_{o_2}^9 y_{o_3}^9 t_1^5 t_2^5 t_3^5 t_4^6 t_5^3 t_6^2 +  y_{s}^3 y_{o_1}^4 y_{o_2}^5 y_{o_3}^5 t_1^5 t_2^2 t_3^2 t_4 t_5^4 t_6^2 +  2 y_{s}^3 y_{o_1}^6 y_{o_2}^7 y_{o_3}^5 t_1^3 t_2^4 t_3^2 t_4^3 t_5^4 t_6^2 +  y_{s}^4 y_{o_1}^7 y_{o_2}^7 y_{o_3}^8 t_1^7 t_2^3 t_3^4 t_4^3 t_5^4 t_6^2 +  y_{s}^3 y_{o_1}^7 y_{o_2}^8 y_{o_3}^5 t_1^2 t_2^5 t_3^2 t_4^4 t_5^4 t_6^2 +  y_{s}^4 y_{o_1}^8 y_{o_2}^8 y_{o_3}^8 t_1^6 t_2^4 t_3^4 t_4^4 t_5^4 t_6^2 +  y_{s}^4 y_{o_1}^9 y_{o_2}^9 y_{o_3}^8 t_1^5 t_2^5 t_3^4 t_4^5 t_5^4 t_6^2 +  2 y_{s}^4 y_{o_1}^{10} y_{o_2}^{10} y_{o_3}^8 t_1^4 t_2^6 t_3^4 t_4^6 t_5^4 t_6^2 +  y_{s}^3 y_{o_1}^5 y_{o_2}^7 y_{o_3}^4 t_1^3 t_2^4 t_3 t_4^2 t_5^5 t_6^2 +  y_{s}^3 y_{o_1}^6 y_{o_2}^8 y_{o_3}^4 t_1^2 t_2^5 t_3 t_4^3 t_5^5 t_6^2 -  y_{s}^4 y_{o_1}^7 y_{o_2}^8 y_{o_3}^7 t_1^6 t_2^4 t_3^3 t_4^3 t_5^5 t_6^2 -  y_{s}^4 y_{o_1}^8 y_{o_2}^9 y_{o_3}^7 t_1^5 t_2^5 t_3^3 t_4^4 t_5^5 t_6^2 +  y_{s}^4 y_{o_1}^9 y_{o_2}^{10} y_{o_3}^7 t_1^4 t_2^6 t_3^3 t_4^5 t_5^5 t_6^2 -  y_{s}^5 y_{o_1}^{11} y_{o_2}^{11} y_{o_3}^{10} t_1^7 t_2^6 t_3^5 t_4^6 t_5^5 t_6^2 -  y_{s}^4 y_{o_1}^7 y_{o_2}^9 y_{o_3}^6 t_1^5 t_2^5 t_3^2 t_4^3 t_5^6 t_6^2 -  y_{s}^5 y_{o_1}^{10} y_{o_2}^{11} y_{o_3}^9 t_1^7 t_2^6 t_3^4 t_4^5 t_5^6 t_6^2 -  y_{s}^5 y_{o_1}^{11} y_{o_2}^{12} y_{o_3}^9 t_1^6 t_2^7 t_3^4 t_4^6 t_5^6 t_6^2 -  y_{s}^2 y_{o_1}^3 y_{o_2}^2 y_{o_3}^5 t_1^3 t_3^3 t_4 t_5 t_6^3 -  y_{s}^2 y_{o_1}^4 y_{o_2}^3 y_{o_3}^5 t_1^2 t_2 t_3^3 t_4^2 t_5 t_6^3 -  y_{s}^2 y_{o_1}^5 y_{o_2}^4 y_{o_3}^5 t_1 t_2^2 t_3^3 t_4^3 t_5 t_6^3 -  y_{s}^2 y_{o_1}^6 y_{o_2}^5 y_{o_3}^5 t_2^3 t_3^3 t_4^4 t_5 t_6^3 +  y_{s}^3 y_{o_1}^8 y_{o_2}^6 y_{o_3}^8 t_1^3 t_2^3 t_3^5 t_4^5 t_5 t_6^3 -  2 y_{s}^2 y_{o_1}^3 y_{o_2}^3 y_{o_3}^4 t_1^2 t_2 t_3^2 t_4 t_5^2 t_6^3 -  2 y_{s}^2 y_{o_1}^4 y_{o_2}^4 y_{o_3}^4 t_1 t_2^2 t_3^2 t_4^2 t_5^2 t_6^3 +  y_{s}^3 y_{o_1}^5 y_{o_2}^4 y_{o_3}^7 t_1^5 t_2 t_3^4 t_4^2 t_5^2 t_6^3 -  2 y_{s}^2 y_{o_1}^5 y_{o_2}^5 y_{o_3}^4 t_2^3 t_3^2 t_4^3 t_5^2 t_6^3 +  y_{s}^3 y_{o_1}^6 y_{o_2}^5 y_{o_3}^7 t_1^4 t_2^2 t_3^4 t_4^3 t_5^2 t_6^3 +  y_{s}^3 y_{o_1}^7 y_{o_2}^6 y_{o_3}^7 t_1^3 t_2^3 t_3^4 t_4^4 t_5^2 t_6^3 -  y_{s}^2 y_{o_1}^2 y_{o_2}^3 y_{o_3}^3 t_1^2 t_2 t_3 t_5^3 t_6^3 -  y_{s}^2 y_{o_1}^3 y_{o_2}^4 y_{o_3}^3 t_1 t_2^2 t_3 t_4 t_5^3 t_6^3 +  2 y_{s}^3 y_{o_1}^4 y_{o_2}^4 y_{o_3}^6 t_1^5 t_2 t_3^3 t_4 t_5^3 t_6^3 -  y_{s}^2 y_{o_1}^4 y_{o_2}^5 y_{o_3}^3 t_2^3 t_3 t_4^2 t_5^3 t_6^3 +  2 y_{s}^3 y_{o_1}^5 y_{o_2}^5 y_{o_3}^6 t_1^4 t_2^2 t_3^3 t_4^2 t_5^3 t_6^3 +  3 y_{s}^3 y_{o_1}^6 y_{o_2}^6 y_{o_3}^6 t_1^3 t_2^3 t_3^3 t_4^3 t_5^3 t_6^3 +  y_{s}^3 y_{o_1}^7 y_{o_2}^7 y_{o_3}^6 t_1^2 t_2^4 t_3^3 t_4^4 t_5^3 t_6^3 +  y_{s}^4 y_{o_1}^{10} y_{o_2}^9 y_{o_3}^9 t_1^4 t_2^5 t_3^5 t_4^6 t_5^3 t_6^3 +  y_{s}^3 y_{o_1}^4 y_{o_2}^5 y_{o_3}^5 t_1^4 t_2^2 t_3^2 t_4 t_5^4 t_6^3 +  y_{s}^3 y_{o_1}^5 y_{o_2}^6 y_{o_3}^5 t_1^3 t_2^3 t_3^2 t_4^2 t_5^4 t_6^3 +  2 y_{s}^3 y_{o_1}^6 y_{o_2}^7 y_{o_3}^5 t_1^2 t_2^4 t_3^2 t_4^3 t_5^4 t_6^3 +  2 y_{s}^4 y_{o_1}^9 y_{o_2}^9 y_{o_3}^8 t_1^4 t_2^5 t_3^4 t_4^5 t_5^4 t_6^3 +  2 y_{s}^4 y_{o_1}^{10} y_{o_2}^{10} y_{o_3}^8 t_1^3 t_2^6 t_3^4 t_4^6 t_5^4 t_6^3 -  y_{s}^5 y_{o_1}^{11} y_{o_2}^{10} y_{o_3}^{11} t_1^7 t_2^5 t_3^6 t_4^6 t_5^4 t_6^3 -  y_{s}^5 y_{o_1}^{12} y_{o_2}^{11} y_{o_3}^{11} t_1^6 t_2^6 t_3^6 t_4^7 t_5^4 t_6^3 +  y_{s}^3 y_{o_1}^5 y_{o_2}^7 y_{o_3}^4 t_1^2 t_2^4 t_3 t_4^2 t_5^5 t_6^3 -  2 y_{s}^4 y_{o_1}^7 y_{o_2}^8 y_{o_3}^7 t_1^5 t_2^4 t_3^3 t_4^3 t_5^5 t_6^3 +  y_{s}^4 y_{o_1}^8 y_{o_2}^9 y_{o_3}^7 t_1^4 t_2^5 t_3^3 t_4^4 t_5^5 t_6^3 +  y_{s}^4 y_{o_1}^9 y_{o_2}^{10} y_{o_3}^7 t_1^3 t_2^6 t_3^3 t_4^5 t_5^5 t_6^3 -  2 y_{s}^5 y_{o_1}^{10} y_{o_2}^{10} y_{o_3}^{10} t_1^7 t_2^5 t_3^5 t_4^5 t_5^5 t_6^3 -  2 y_{s}^5 y_{o_1}^{11} y_{o_2}^{11} y_{o_3}^{10} t_1^6 t_2^6 t_3^5 t_4^6 t_5^5 t_6^3 -  y_{s}^5 y_{o_1}^9 y_{o_2}^{10} y_{o_3}^9 t_1^7 t_2^5 t_3^4 t_4^4 t_5^6 t_6^3 -  y_{s}^5 y_{o_1}^{10} y_{o_2}^{11} y_{o_3}^9 t_1^6 t_2^6 t_3^4 t_4^5 t_5^6 t_6^3 -  y_{s}^5 y_{o_1}^{11} y_{o_2}^{12} y_{o_3}^9 t_1^5 t_2^7 t_3^4 t_4^6 t_5^6 t_6^3 -  y_{s}^2 y_{o_1}^3 y_{o_2}^2 y_{o_3}^5 t_1^2 t_3^3 t_4 t_5 t_6^4 -  y_{s}^2 y_{o_1}^4 y_{o_2}^3 y_{o_3}^5 t_1 t_2 t_3^3 t_4^2 t_5 t_6^4 -  y_{s}^2 y_{o_1}^5 y_{o_2}^4 y_{o_3}^5 t_2^2 t_3^3 t_4^3 t_5 t_6^4 -  2 y_{s}^2 y_{o_1}^3 y_{o_2}^3 y_{o_3}^4 t_1 t_2 t_3^2 t_4 t_5^2 t_6^4 -  2 y_{s}^2 y_{o_1}^4 y_{o_2}^4 y_{o_3}^4 t_2^2 t_3^2 t_4^2 t_5^2 t_6^4 +  y_{s}^3 y_{o_1}^5 y_{o_2}^4 y_{o_3}^7 t_1^4 t_2 t_3^4 t_4^2 t_5^2 t_6^4 +  y_{s}^3 y_{o_1}^6 y_{o_2}^5 y_{o_3}^7 t_1^3 t_2^2 t_3^4 t_4^3 t_5^2 t_6^4 -  2 y_{s}^3 y_{o_1}^7 y_{o_2}^6 y_{o_3}^7 t_1^2 t_2^3 t_3^4 t_4^4 t_5^2 t_6^4 +  y_{s}^4 y_{o_1}^9 y_{o_2}^7 y_{o_3}^{10} t_1^5 t_2^3 t_3^6 t_4^5 t_5^2 t_6^4 -  y_{s}^2 y_{o_1}^2 y_{o_2}^3 y_{o_3}^3 t_1 t_2 t_3 t_5^3 t_6^4 -  y_{s}^2 y_{o_1}^3 y_{o_2}^4 y_{o_3}^3 t_2^2 t_3 t_4 t_5^3 t_6^4 +  2 y_{s}^3 y_{o_1}^4 y_{o_2}^4 y_{o_3}^6 t_1^4 t_2 t_3^3 t_4 t_5^3 t_6^4 +  2 y_{s}^3 y_{o_1}^5 y_{o_2}^5 y_{o_3}^6 t_1^3 t_2^2 t_3^3 t_4^2 t_5^3 t_6^4 +  2 y_{s}^4 y_{o_1}^8 y_{o_2}^7 y_{o_3}^9 t_1^5 t_2^3 t_3^5 t_4^4 t_5^3 t_6^4 +  y_{s}^4 y_{o_1}^9 y_{o_2}^8 y_{o_3}^9 t_1^4 t_2^4 t_3^5 t_4^5 t_5^3 t_6^4 +  y_{s}^4 y_{o_1}^{10} y_{o_2}^9 y_{o_3}^9 t_1^3 t_2^5 t_3^5 t_4^6 t_5^3 t_6^4 +  y_{s}^3 y_{o_1}^4 y_{o_2}^5 y_{o_3}^5 t_1^3 t_2^2 t_3^2 t_4 t_5^4 t_6^4 +  y_{s}^4 y_{o_1}^7 y_{o_2}^7 y_{o_3}^8 t_1^5 t_2^3 t_3^4 t_4^3 t_5^4 t_6^4 +  3 y_{s}^4 y_{o_1}^8 y_{o_2}^8 y_{o_3}^8 t_1^4 t_2^4 t_3^4 t_4^4 t_5^4 t_6^4 +  2 y_{s}^4 y_{o_1}^9 y_{o_2}^9 y_{o_3}^8 t_1^3 t_2^5 t_3^4 t_4^5 t_5^4 t_6^4 -  y_{s}^5 y_{o_1}^{10} y_{o_2}^9 y_{o_3}^{11} t_1^7 t_2^4 t_3^6 t_4^5 t_5^4 t_6^4 +  2 y_{s}^4 y_{o_1}^{10} y_{o_2}^{10} y_{o_3}^8 t_1^2 t_2^6 t_3^4 t_4^6 t_5^4 t_6^4 -  y_{s}^5 y_{o_1}^{11} y_{o_2}^{10} y_{o_3}^{11} t_1^6 t_2^5 t_3^6 t_4^6 t_5^4 t_6^4 -  y_{s}^5 y_{o_1}^{12} y_{o_2}^{11} y_{o_3}^{11} t_1^5 t_2^6 t_3^6 t_4^7 t_5^4 t_6^4 +  y_{s}^4 y_{o_1}^7 y_{o_2}^8 y_{o_3}^7 t_1^4 t_2^4 t_3^3 t_4^3 t_5^5 t_6^4 +  y_{s}^4 y_{o_1}^8 y_{o_2}^9 y_{o_3}^7 t_1^3 t_2^5 t_3^3 t_4^4 t_5^5 t_6^4 -  2 y_{s}^5 y_{o_1}^9 y_{o_2}^9 y_{o_3}^{10} t_1^7 t_2^4 t_3^5 t_4^4 t_5^5 t_6^4 +  y_{s}^4 y_{o_1}^9 y_{o_2}^{10} y_{o_3}^7 t_1^2 t_2^6 t_3^3 t_4^5 t_5^5 t_6^4 -  2 y_{s}^5 y_{o_1}^{10} y_{o_2}^{10} y_{o_3}^{10} t_1^6 t_2^5 t_3^5 t_4^5 t_5^5 t_6^4 -  2 y_{s}^5 y_{o_1}^{11} y_{o_2}^{11} y_{o_3}^{10} t_1^5 t_2^6 t_3^5 t_4^6 t_5^5 t_6^4 +  y_{s}^4 y_{o_1}^6 y_{o_2}^8 y_{o_3}^6 t_1^4 t_2^4 t_3^2 t_4^2 t_5^6 t_6^4 -  y_{s}^5 y_{o_1}^8 y_{o_2}^9 y_{o_3}^9 t_1^7 t_2^4 t_3^4 t_4^3 t_5^6 t_6^4 -  y_{s}^5 y_{o_1}^9 y_{o_2}^{10} y_{o_3}^9 t_1^6 t_2^5 t_3^4 t_4^4 t_5^6 t_6^4 -  y_{s}^5 y_{o_1}^{10} y_{o_2}^{11} y_{o_3}^9 t_1^5 t_2^6 t_3^4 t_4^5 t_5^6 t_6^4 -  y_{s}^5 y_{o_1}^{11} y_{o_2}^{12} y_{o_3}^9 t_1^4 t_2^7 t_3^4 t_4^6 t_5^6 t_6^4 -  y_{s}^2 y_{o_1}^3 y_{o_2}^2 y_{o_3}^5 t_1 t_3^3 t_4 t_5 t_6^5 -  y_{s}^2 y_{o_1}^4 y_{o_2}^3 y_{o_3}^5 t_2 t_3^3 t_4^2 t_5 t_6^5 -  y_{s}^3 y_{o_1}^7 y_{o_2}^5 y_{o_3}^8 t_1^2 t_2^2 t_3^5 t_4^4 t_5 t_6^5 -  y_{s}^2 y_{o_1}^3 y_{o_2}^3 y_{o_3}^4 t_2 t_3^2 t_4 t_5^2 t_6^5 +  y_{s}^3 y_{o_1}^5 y_{o_2}^4 y_{o_3}^7 t_1^3 t_2 t_3^4 t_4^2 t_5^2 t_6^5 -  y_{s}^3 y_{o_1}^6 y_{o_2}^5 y_{o_3}^7 t_1^2 t_2^2 t_3^4 t_4^3 t_5^2 t_6^5 -  y_{s}^3 y_{o_1}^7 y_{o_2}^6 y_{o_3}^7 t_1 t_2^3 t_3^4 t_4^4 t_5^2 t_6^5 +  y_{s}^4 y_{o_1}^8 y_{o_2}^6 y_{o_3}^{10} t_1^5 t_2^2 t_3^6 t_4^4 t_5^2 t_6^5 +  y_{s}^4 y_{o_1}^9 y_{o_2}^7 y_{o_3}^{10} t_1^4 t_2^3 t_3^6 t_4^5 t_5^2 t_6^5 +  2 y_{s}^3 y_{o_1}^4 y_{o_2}^4 y_{o_3}^6 t_1^3 t_2 t_3^3 t_4 t_5^3 t_6^5 +  y_{s}^3 y_{o_1}^5 y_{o_2}^5 y_{o_3}^6 t_1^2 t_2^2 t_3^3 t_4^2 t_5^3 t_6^5 +  y_{s}^3 y_{o_1}^6 y_{o_2}^6 y_{o_3}^6 t_1 t_2^3 t_3^3 t_4^3 t_5^3 t_6^5 +  y_{s}^4 y_{o_1}^7 y_{o_2}^6 y_{o_3}^9 t_1^5 t_2^2 t_3^5 t_4^3 t_5^3 t_6^5 +  y_{s}^3 y_{o_1}^7 y_{o_2}^7 y_{o_3}^6 t_2^4 t_3^3 t_4^4 t_5^3 t_6^5 +  2 y_{s}^4 y_{o_1}^8 y_{o_2}^7 y_{o_3}^9 t_1^4 t_2^3 t_3^5 t_4^4 t_5^3 t_6^5 +  y_{s}^4 y_{o_1}^{10} y_{o_2}^9 y_{o_3}^9 t_1^2 t_2^5 t_3^5 t_4^6 t_5^3 t_6^5 +  y_{s}^3 y_{o_1}^4 y_{o_2}^5 y_{o_3}^5 t_1^2 t_2^2 t_3^2 t_4 t_5^4 t_6^5 -  y_{s}^4 y_{o_1}^6 y_{o_2}^6 y_{o_3}^8 t_1^5 t_2^2 t_3^4 t_4^2 t_5^4 t_6^5 +  y_{s}^3 y_{o_1}^6 y_{o_2}^7 y_{o_3}^5 t_2^4 t_3^2 t_4^3 t_5^4 t_6^5 +  y_{s}^4 y_{o_1}^7 y_{o_2}^7 y_{o_3}^8 t_1^4 t_2^3 t_3^4 t_4^3 t_5^4 t_6^5 -  y_{s}^5 y_{o_1}^9 y_{o_2}^8 y_{o_3}^{11} t_1^7 t_2^3 t_3^6 t_4^4 t_5^4 t_6^5 +  y_{s}^4 y_{o_1}^9 y_{o_2}^9 y_{o_3}^8 t_1^2 t_2^5 t_3^4 t_4^5 t_5^4 t_6^5 -  y_{s}^5 y_{o_1}^{10} y_{o_2}^9 y_{o_3}^{11} t_1^6 t_2^4 t_3^6 t_4^5 t_5^4 t_6^5 +  y_{s}^4 y_{o_1}^{10} y_{o_2}^{10} y_{o_3}^8 t_1 t_2^6 t_3^4 t_4^6 t_5^4 t_6^5 -  y_{s}^5 y_{o_1}^{11} y_{o_2}^{10} y_{o_3}^{11} t_1^5 t_2^5 t_3^6 t_4^6 t_5^4 t_6^5 -  y_{s}^5 y_{o_1}^{12} y_{o_2}^{11} y_{o_3}^{11} t_1^4 t_2^6 t_3^6 t_4^7 t_5^4 t_6^5 -  y_{s}^4 y_{o_1}^5 y_{o_2}^6 y_{o_3}^7 t_1^5 t_2^2 t_3^3 t_4 t_5^5 t_6^5 -  2 y_{s}^4 y_{o_1}^7 y_{o_2}^8 y_{o_3}^7 t_1^3 t_2^4 t_3^3 t_4^3 t_5^5 t_6^5 -  y_{s}^5 y_{o_1}^8 y_{o_2}^8 y_{o_3}^{10} t_1^7 t_2^3 t_3^5 t_4^3 t_5^5 t_6^5 -  y_{s}^4 y_{o_1}^8 y_{o_2}^9 y_{o_3}^7 t_1^2 t_2^5 t_3^3 t_4^4 t_5^5 t_6^5 -  y_{s}^5 y_{o_1}^9 y_{o_2}^9 y_{o_3}^{10} t_1^6 t_2^4 t_3^5 t_4^4 t_5^5 t_6^5 -  y_{s}^5 y_{o_1}^{10} y_{o_2}^{10} y_{o_3}^{10} t_1^5 t_2^5 t_3^5 t_4^5 t_5^5 t_6^5 -  2 y_{s}^5 y_{o_1}^{11} y_{o_2}^{11} y_{o_3}^{10} t_1^4 t_2^6 t_3^5 t_4^6 t_5^5 t_6^5 -  y_{s}^4 y_{o_1}^7 y_{o_2}^9 y_{o_3}^6 t_1^2 t_2^5 t_3^2 t_4^3 t_5^6 t_6^5 +  y_{s}^5 y_{o_1}^9 y_{o_2}^{10} y_{o_3}^9 t_1^5 t_2^5 t_3^4 t_4^4 t_5^6 t_6^5 -  y_{s}^5 y_{o_1}^{10} y_{o_2}^{11} y_{o_3}^9 t_1^4 t_2^6 t_3^4 t_4^5 t_5^6 t_6^5 -  y_{s}^5 y_{o_1}^{11} y_{o_2}^{12} y_{o_3}^9 t_1^3 t_2^7 t_3^4 t_4^6 t_5^6 t_6^5 +  y_{s}^6 y_{o_1}^{12} y_{o_2}^{12} y_{o_3}^{12} t_1^7 t_2^6 t_3^6 t_4^6 t_5^6 t_6^5 +  y_{s}^6 y_{o_1}^{13} y_{o_2}^{13} y_{o_3}^{12} t_1^6 t_2^7 t_3^6 t_4^7 t_5^6 t_6^5 +  y_{s}^5 y_{o_1}^8 y_{o_2}^{10} y_{o_3}^8 t_1^5 t_2^5 t_3^3 t_4^3 t_5^7 t_6^5 +  y_{s}^6 y_{o_1}^{11} y_{o_2}^{12} y_{o_3}^{11} t_1^7 t_2^6 t_3^5 t_4^5 t_5^7 t_6^5 +  y_{s}^6 y_{o_1}^{12} y_{o_2}^{13} y_{o_3}^{11} t_1^6 t_2^7 t_3^5 t_4^6 t_5^7 t_6^5 +  y_{s}^3 y_{o_1}^4 y_{o_2}^4 y_{o_3}^6 t_1^2 t_2 t_3^3 t_4 t_5^3 t_6^6 +  y_{s}^3 y_{o_1}^5 y_{o_2}^5 y_{o_3}^6 t_1 t_2^2 t_3^3 t_4^2 t_5^3 t_6^6 +  y_{s}^3 y_{o_1}^6 y_{o_2}^6 y_{o_3}^6 t_2^3 t_3^3 t_4^3 t_5^3 t_6^6 +  y_{s}^4 y_{o_1}^8 y_{o_2}^8 y_{o_3}^8 t_1^2 t_2^4 t_3^4 t_4^4 t_5^4 t_6^6 +  y_{s}^4 y_{o_1}^9 y_{o_2}^9 y_{o_3}^8 t_1 t_2^5 t_3^4 t_4^5 t_5^4 t_6^6 -  y_{s}^5 y_{o_1}^{10} y_{o_2}^9 y_{o_3}^{11} t_1^5 t_2^4 t_3^6 t_4^5 t_5^4 t_6^6 +  y_{s}^4 y_{o_1}^{10} y_{o_2}^{10} y_{o_3}^8 t_2^6 t_3^4 t_4^6 t_5^4 t_6^6 -  y_{s}^5 y_{o_1}^{11} y_{o_2}^{10} y_{o_3}^{11} t_1^4 t_2^5 t_3^6 t_4^6 t_5^4 t_6^6 -  y_{s}^5 y_{o_1}^{12} y_{o_2}^{11} y_{o_3}^{11} t_1^3 t_2^6 t_3^6 t_4^7 t_5^4 t_6^6 -  y_{s}^5 y_{o_1}^7 y_{o_2}^7 y_{o_3}^{10} t_1^7 t_2^2 t_3^5 t_4^2 t_5^5 t_6^6 -  y_{s}^4 y_{o_1}^7 y_{o_2}^8 y_{o_3}^7 t_1^2 t_2^4 t_3^3 t_4^3 t_5^5 t_6^6 -  y_{s}^5 y_{o_1}^8 y_{o_2}^8 y_{o_3}^{10} t_1^6 t_2^3 t_3^5 t_4^3 t_5^5 t_6^6 -  y_{s}^5 y_{o_1}^9 y_{o_2}^9 y_{o_3}^{10} t_1^5 t_2^4 t_3^5 t_4^4 t_5^5 t_6^6 -  2 y_{s}^5 y_{o_1}^{10} y_{o_2}^{10} y_{o_3}^{10} t_1^4 t_2^5 t_3^5 t_4^5 t_5^5 t_6^6 -  2 y_{s}^5 y_{o_1}^{11} y_{o_2}^{11} y_{o_3}^{10} t_1^3 t_2^6 t_3^5 t_4^6 t_5^5 t_6^6 +  y_{s}^6 y_{o_1}^{12} y_{o_2}^{11} y_{o_3}^{13} t_1^7 t_2^5 t_3^7 t_4^6 t_5^5 t_6^6 +  y_{s}^6 y_{o_1}^{13} y_{o_2}^{12} y_{o_3}^{13} t_1^6 t_2^6 t_3^7 t_4^7 t_5^5 t_6^6 -  y_{s}^5 y_{o_1}^9 y_{o_2}^{10} y_{o_3}^9 t_1^4 t_2^5 t_3^4 t_4^4 t_5^6 t_6^6 -  y_{s}^5 y_{o_1}^{10} y_{o_2}^{11} y_{o_3}^9 t_1^3 t_2^6 t_3^4 t_4^5 t_5^6 t_6^6 +  y_{s}^6 y_{o_1}^{11} y_{o_2}^{11} y_{o_3}^{12} t_1^7 t_2^5 t_3^6 t_4^5 t_5^6 t_6^6 -  y_{s}^5 y_{o_1}^{11} y_{o_2}^{12} y_{o_3}^9 t_1^2 t_2^7 t_3^4 t_4^6 t_5^6 t_6^6 +  y_{s}^6 y_{o_1}^{12} y_{o_2}^{12} y_{o_3}^{12} t_1^6 t_2^6 t_3^6 t_4^6 t_5^6 t_6^6 +  y_{s}^6 y_{o_1}^{13} y_{o_2}^{13} y_{o_3}^{12} t_1^5 t_2^7 t_3^6 t_4^7 t_5^6 t_6^6 +  y_{s}^6 y_{o_1}^{10} y_{o_2}^{11} y_{o_3}^{11} t_1^7 t_2^5 t_3^5 t_4^4 t_5^7 t_6^6 +  y_{s}^6 y_{o_1}^{11} y_{o_2}^{12} y_{o_3}^{11} t_1^6 t_2^6 t_3^5 t_4^5 t_5^7 t_6^6 +  y_{s}^6 y_{o_1}^{12} y_{o_2}^{13} y_{o_3}^{11} t_1^5 t_2^7 t_3^5 t_4^6 t_5^7 t_6^6 +  y_{s}^3 y_{o_1}^4 y_{o_2}^4 y_{o_3}^6 t_1 t_2 t_3^3 t_4 t_5^3 t_6^7 +  y_{s}^3 y_{o_1}^5 y_{o_2}^5 y_{o_3}^6 t_2^2 t_3^3 t_4^2 t_5^3 t_6^7 +  y_{s}^4 y_{o_1}^8 y_{o_2}^7 y_{o_3}^9 t_1^2 t_2^3 t_3^5 t_4^4 t_5^3 t_6^7 +  y_{s}^4 y_{o_1}^7 y_{o_2}^7 y_{o_3}^8 t_1^2 t_2^3 t_3^4 t_4^3 t_5^4 t_6^7 +  y_{s}^4 y_{o_1}^8 y_{o_2}^8 y_{o_3}^8 t_1 t_2^4 t_3^4 t_4^4 t_5^4 t_6^7 -  y_{s}^5 y_{o_1}^9 y_{o_2}^8 y_{o_3}^{11} t_1^5 t_2^3 t_3^6 t_4^4 t_5^4 t_6^7 +  y_{s}^4 y_{o_1}^9 y_{o_2}^9 y_{o_3}^8 t_2^5 t_3^4 t_4^5 t_5^4 t_6^7 -  y_{s}^5 y_{o_1}^{10} y_{o_2}^9 y_{o_3}^{11} t_1^4 t_2^4 t_3^6 t_4^5 t_5^4 t_6^7 -  y_{s}^5 y_{o_1}^{11} y_{o_2}^{10} y_{o_3}^{11} t_1^3 t_2^5 t_3^6 t_4^6 t_5^4 t_6^7 -  y_{s}^5 y_{o_1}^7 y_{o_2}^7 y_{o_3}^{10} t_1^6 t_2^2 t_3^5 t_4^2 t_5^5 t_6^7 -  y_{s}^5 y_{o_1}^8 y_{o_2}^8 y_{o_3}^{10} t_1^5 t_2^3 t_3^5 t_4^3 t_5^5 t_6^7 -  2 y_{s}^5 y_{o_1}^9 y_{o_2}^9 y_{o_3}^{10} t_1^4 t_2^4 t_3^5 t_4^4 t_5^5 t_6^7 -  2 y_{s}^5 y_{o_1}^{10} y_{o_2}^{10} y_{o_3}^{10} t_1^3 t_2^5 t_3^5 t_4^5 t_5^5 t_6^7 -  y_{s}^5 y_{o_1}^{11} y_{o_2}^{11} y_{o_3}^{10} t_1^2 t_2^6 t_3^5 t_4^6 t_5^5 t_6^7 +  y_{s}^6 y_{o_1}^{12} y_{o_2}^{11} y_{o_3}^{13} t_1^6 t_2^5 t_3^7 t_4^6 t_5^5 t_6^7 -  y_{s}^5 y_{o_1}^8 y_{o_2}^9 y_{o_3}^9 t_1^4 t_2^4 t_3^4 t_4^3 t_5^6 t_6^7 -  y_{s}^5 y_{o_1}^9 y_{o_2}^{10} y_{o_3}^9 t_1^3 t_2^5 t_3^4 t_4^4 t_5^6 t_6^7 +  y_{s}^6 y_{o_1}^{10} y_{o_2}^{10} y_{o_3}^{12} t_1^7 t_2^4 t_3^6 t_4^4 t_5^6 t_6^7 -  y_{s}^5 y_{o_1}^{10} y_{o_2}^{11} y_{o_3}^9 t_1^2 t_2^6 t_3^4 t_4^5 t_5^6 t_6^7 +  y_{s}^6 y_{o_1}^{11} y_{o_2}^{11} y_{o_3}^{12} t_1^6 t_2^5 t_3^6 t_4^5 t_5^6 t_6^7 +  y_{s}^6 y_{o_1}^{12} y_{o_2}^{12} y_{o_3}^{12} t_1^5 t_2^6 t_3^6 t_4^6 t_5^6 t_6^7 +  y_{s}^6 y_{o_1}^{10} y_{o_2}^{11} y_{o_3}^{11} t_1^6 t_2^5 t_3^5 t_4^4 t_5^7 t_6^7 +  y_{s}^6 y_{o_1}^{11} y_{o_2}^{12} y_{o_3}^{11} t_1^5 t_2^6 t_3^5 t_4^5 t_5^7 t_6^7 +  y_{s}^7 y_{o_1}^{14} y_{o_2}^{14} y_{o_3}^{14} t_1^7 t_2^7 t_3^7 t_4^7 t_5^7 t_6^7
~,~
$
\end{quote}
\endgroup

\subsection{Model 8 \label{app_num_08}}

\begingroup\makeatletter\def\f@size{7}\check@mathfonts
\begin{quote}\raggedright
$P(t_i,y_s,y_{o_1},y_{o_2},y_{o_3}; \mathcal{M}_8) = 1 + y_{s} y_{o_1} y_{o_2} y_{o_3}^2 t_1^2 t_2 t_4 t_6 + y_{s} y_{o_1} y_{o_2} y_{o_3}^2 t_1 t_2^2 t_4 t_6 +  y_{s} y_{o_1}^2 y_{o_2}^2 y_{o_3}^2 t_1^2 t_3 t_4 t_5 t_6 +  y_{s} y_{o_1}^2 y_{o_2}^2 y_{o_3}^2 t_1 t_2 t_3 t_4 t_5 t_6 +  y_{s} y_{o_1}^2 y_{o_2}^2 y_{o_3}^2 t_2^2 t_3 t_4 t_5 t_6 -  y_{s}^2 y_{o_1}^3 y_{o_2}^2 y_{o_3}^3 t_1^3 t_2 t_4^3 t_5 t_6 -  y_{s}^2 y_{o_1}^3 y_{o_2}^2 y_{o_3}^3 t_1^2 t_2^2 t_4^3 t_5 t_6 -  y_{s}^2 y_{o_1}^3 y_{o_2}^2 y_{o_3}^3 t_1 t_2^3 t_4^3 t_5 t_6 +  y_{s} y_{o_1}^3 y_{o_2}^3 y_{o_3}^2 t_1 t_3^2 t_4 t_5^2 t_6 +  y_{s} y_{o_1}^3 y_{o_2}^3 y_{o_3}^2 t_2 t_3^2 t_4 t_5^2 t_6 -  y_{s}^2 y_{o_1}^4 y_{o_2}^3 y_{o_3}^3 t_1^3 t_3 t_4^3 t_5^2 t_6 -  2 y_{s}^2 y_{o_1}^4 y_{o_2}^3 y_{o_3}^3 t_1^2 t_2 t_3 t_4^3 t_5^2 t_6 -  2 y_{s}^2 y_{o_1}^4 y_{o_2}^3 y_{o_3}^3 t_1 t_2^2 t_3 t_4^3 t_5^2 t_6 -  y_{s}^2 y_{o_1}^4 y_{o_2}^3 y_{o_3}^3 t_2^3 t_3 t_4^3 t_5^2 t_6 +  y_{s} y_{o_1}^4 y_{o_2}^4 y_{o_3}^2 t_3^3 t_4 t_5^3 t_6 -  y_{s}^2 y_{o_1}^5 y_{o_2}^4 y_{o_3}^3 t_1^2 t_3^2 t_4^3 t_5^3 t_6 -  2 y_{s}^2 y_{o_1}^5 y_{o_2}^4 y_{o_3}^3 t_1 t_2 t_3^2 t_4^3 t_5^3 t_6 -  y_{s}^2 y_{o_1}^5 y_{o_2}^4 y_{o_3}^3 t_2^2 t_3^2 t_4^3 t_5^3 t_6 +  y_{s}^3 y_{o_1}^6 y_{o_2}^4 y_{o_3}^4 t_1^3 t_2 t_3 t_4^5 t_5^3 t_6 +  y_{s}^3 y_{o_1}^6 y_{o_2}^4 y_{o_3}^4 t_1^2 t_2^2 t_3 t_4^5 t_5^3 t_6 +  y_{s}^3 y_{o_1}^6 y_{o_2}^4 y_{o_3}^4 t_1 t_2^3 t_3 t_4^5 t_5^3 t_6 -  y_{s}^2 y_{o_1}^6 y_{o_2}^5 y_{o_3}^3 t_1 t_3^3 t_4^3 t_5^4 t_6 -  y_{s}^2 y_{o_1}^6 y_{o_2}^5 y_{o_3}^3 t_2 t_3^3 t_4^3 t_5^4 t_6 +  y_{s}^3 y_{o_1}^7 y_{o_2}^5 y_{o_3}^4 t_1^2 t_2 t_3^2 t_4^5 t_5^4 t_6 +  y_{s}^3 y_{o_1}^7 y_{o_2}^5 y_{o_3}^4 t_1 t_2^2 t_3^2 t_4^5 t_5^4 t_6 +  y_{s}^3 y_{o_1}^8 y_{o_2}^6 y_{o_3}^4 t_1 t_2 t_3^3 t_4^5 t_5^5 t_6 +  y_{s} y_{o_1} y_{o_2}^2 y_{o_3}^3 t_1^3 t_2 t_3 t_6^2 +  y_{s} y_{o_1} y_{o_2}^2 y_{o_3}^3 t_1^2 t_2^2 t_3 t_6^2 +  y_{s} y_{o_1} y_{o_2}^2 y_{o_3}^3 t_1 t_2^3 t_3 t_6^2 +  y_{s} y_{o_1}^2 y_{o_2}^3 y_{o_3}^3 t_1^3 t_3^2 t_5 t_6^2 +  y_{s} y_{o_1}^2 y_{o_2}^3 y_{o_3}^3 t_1^2 t_2 t_3^2 t_5 t_6^2 +  y_{s} y_{o_1}^2 y_{o_2}^3 y_{o_3}^3 t_1 t_2^2 t_3^2 t_5 t_6^2 +  y_{s} y_{o_1}^2 y_{o_2}^3 y_{o_3}^3 t_2^3 t_3^2 t_5 t_6^2 -  y_{s}^2 y_{o_1}^3 y_{o_2}^3 y_{o_3}^4 t_1^5 t_3 t_4^2 t_5 t_6^2 -  2 y_{s}^2 y_{o_1}^3 y_{o_2}^3 y_{o_3}^4 t_1^4 t_2 t_3 t_4^2 t_5 t_6^2 -  2 y_{s}^2 y_{o_1}^3 y_{o_2}^3 y_{o_3}^4 t_1^3 t_2^2 t_3 t_4^2 t_5 t_6^2 -  2 y_{s}^2 y_{o_1}^3 y_{o_2}^3 y_{o_3}^4 t_1^2 t_2^3 t_3 t_4^2 t_5 t_6^2 -  2 y_{s}^2 y_{o_1}^3 y_{o_2}^3 y_{o_3}^4 t_1 t_2^4 t_3 t_4^2 t_5 t_6^2 -  y_{s}^2 y_{o_1}^3 y_{o_2}^3 y_{o_3}^4 t_2^5 t_3 t_4^2 t_5 t_6^2 +  y_{s} y_{o_1}^3 y_{o_2}^4 y_{o_3}^3 t_1^2 t_3^3 t_5^2 t_6^2 +  y_{s} y_{o_1}^3 y_{o_2}^4 y_{o_3}^3 t_1 t_2 t_3^3 t_5^2 t_6^2 +  y_{s} y_{o_1}^3 y_{o_2}^4 y_{o_3}^3 t_2^2 t_3^3 t_5^2 t_6^2 -  2 y_{s}^2 y_{o_1}^4 y_{o_2}^4 y_{o_3}^4 t_1^4 t_3^2 t_4^2 t_5^2 t_6^2 -  3 y_{s}^2 y_{o_1}^4 y_{o_2}^4 y_{o_3}^4 t_1^3 t_2 t_3^2 t_4^2 t_5^2 t_6^2 -  2 y_{s}^2 y_{o_1}^4 y_{o_2}^4 y_{o_3}^4 t_1^2 t_2^2 t_3^2 t_4^2 t_5^2 t_6^2 -  3 y_{s}^2 y_{o_1}^4 y_{o_2}^4 y_{o_3}^4 t_1 t_2^3 t_3^2 t_4^2 t_5^2 t_6^2 -  2 y_{s}^2 y_{o_1}^4 y_{o_2}^4 y_{o_3}^4 t_2^4 t_3^2 t_4^2 t_5^2 t_6^2 +  y_{s}^3 y_{o_1}^5 y_{o_2}^4 y_{o_3}^5 t_1^5 t_2 t_3 t_4^4 t_5^2 t_6^2 +  y_{s}^3 y_{o_1}^5 y_{o_2}^4 y_{o_3}^5 t_1^4 t_2^2 t_3 t_4^4 t_5^2 t_6^2 +  y_{s}^3 y_{o_1}^5 y_{o_2}^4 y_{o_3}^5 t_1^3 t_2^3 t_3 t_4^4 t_5^2 t_6^2 +  y_{s}^3 y_{o_1}^5 y_{o_2}^4 y_{o_3}^5 t_1^2 t_2^4 t_3 t_4^4 t_5^2 t_6^2 +  y_{s}^3 y_{o_1}^5 y_{o_2}^4 y_{o_3}^5 t_1 t_2^5 t_3 t_4^4 t_5^2 t_6^2 +  y_{s} y_{o_1}^4 y_{o_2}^5 y_{o_3}^3 t_1 t_3^4 t_5^3 t_6^2 +  y_{s} y_{o_1}^4 y_{o_2}^5 y_{o_3}^3 t_2 t_3^4 t_5^3 t_6^2 -  2 y_{s}^2 y_{o_1}^5 y_{o_2}^5 y_{o_3}^4 t_1^3 t_3^3 t_4^2 t_5^3 t_6^2 -  2 y_{s}^2 y_{o_1}^5 y_{o_2}^5 y_{o_3}^4 t_1^2 t_2 t_3^3 t_4^2 t_5^3 t_6^2 -  2 y_{s}^2 y_{o_1}^5 y_{o_2}^5 y_{o_3}^4 t_1 t_2^2 t_3^3 t_4^2 t_5^3 t_6^2 -  2 y_{s}^2 y_{o_1}^5 y_{o_2}^5 y_{o_3}^4 t_2^3 t_3^3 t_4^2 t_5^3 t_6^2 +  y_{s}^3 y_{o_1}^6 y_{o_2}^5 y_{o_3}^5 t_1^5 t_3^2 t_4^4 t_5^3 t_6^2 +  3 y_{s}^3 y_{o_1}^6 y_{o_2}^5 y_{o_3}^5 t_1^4 t_2 t_3^2 t_4^4 t_5^3 t_6^2 +  2 y_{s}^3 y_{o_1}^6 y_{o_2}^5 y_{o_3}^5 t_1^3 t_2^2 t_3^2 t_4^4 t_5^3 t_6^2 +  2 y_{s}^3 y_{o_1}^6 y_{o_2}^5 y_{o_3}^5 t_1^2 t_2^3 t_3^2 t_4^4 t_5^3 t_6^2 +  3 y_{s}^3 y_{o_1}^6 y_{o_2}^5 y_{o_3}^5 t_1 t_2^4 t_3^2 t_4^4 t_5^3 t_6^2 +  y_{s}^3 y_{o_1}^6 y_{o_2}^5 y_{o_3}^5 t_2^5 t_3^2 t_4^4 t_5^3 t_6^2 -  y_{s}^2 y_{o_1}^6 y_{o_2}^6 y_{o_3}^4 t_1^2 t_3^4 t_4^2 t_5^4 t_6^2 -  2 y_{s}^2 y_{o_1}^6 y_{o_2}^6 y_{o_3}^4 t_1 t_2 t_3^4 t_4^2 t_5^4 t_6^2 -  y_{s}^2 y_{o_1}^6 y_{o_2}^6 y_{o_3}^4 t_2^2 t_3^4 t_4^2 t_5^4 t_6^2 +  y_{s}^3 y_{o_1}^7 y_{o_2}^6 y_{o_3}^5 t_1^4 t_3^3 t_4^4 t_5^4 t_6^2 +  2 y_{s}^3 y_{o_1}^7 y_{o_2}^6 y_{o_3}^5 t_1^3 t_2 t_3^3 t_4^4 t_5^4 t_6^2 +  y_{s}^3 y_{o_1}^7 y_{o_2}^6 y_{o_3}^5 t_1^2 t_2^2 t_3^3 t_4^4 t_5^4 t_6^2 +  2 y_{s}^3 y_{o_1}^7 y_{o_2}^6 y_{o_3}^5 t_1 t_2^3 t_3^3 t_4^4 t_5^4 t_6^2 +  y_{s}^3 y_{o_1}^7 y_{o_2}^6 y_{o_3}^5 t_2^4 t_3^3 t_4^4 t_5^4 t_6^2 -  y_{s}^4 y_{o_1}^8 y_{o_2}^6 y_{o_3}^6 t_1^5 t_2 t_3^2 t_4^6 t_5^4 t_6^2 -  y_{s}^4 y_{o_1}^8 y_{o_2}^6 y_{o_3}^6 t_1^4 t_2^2 t_3^2 t_4^6 t_5^4 t_6^2 -  y_{s}^4 y_{o_1}^8 y_{o_2}^6 y_{o_3}^6 t_1^2 t_2^4 t_3^2 t_4^6 t_5^4 t_6^2 -  y_{s}^4 y_{o_1}^8 y_{o_2}^6 y_{o_3}^6 t_1 t_2^5 t_3^2 t_4^6 t_5^4 t_6^2 -  y_{s}^2 y_{o_1}^7 y_{o_2}^7 y_{o_3}^4 t_1 t_3^5 t_4^2 t_5^5 t_6^2 -  y_{s}^2 y_{o_1}^7 y_{o_2}^7 y_{o_3}^4 t_2 t_3^5 t_4^2 t_5^5 t_6^2 +  y_{s}^3 y_{o_1}^8 y_{o_2}^7 y_{o_3}^5 t_1^2 t_2 t_3^4 t_4^4 t_5^5 t_6^2 +  y_{s}^3 y_{o_1}^8 y_{o_2}^7 y_{o_3}^5 t_1 t_2^2 t_3^4 t_4^4 t_5^5 t_6^2 -  y_{s}^4 y_{o_1}^9 y_{o_2}^7 y_{o_3}^6 t_1^4 t_2 t_3^3 t_4^6 t_5^5 t_6^2 -  y_{s}^4 y_{o_1}^9 y_{o_2}^7 y_{o_3}^6 t_1 t_2^4 t_3^3 t_4^6 t_5^5 t_6^2 +  y_{s}^3 y_{o_1}^9 y_{o_2}^8 y_{o_3}^5 t_1 t_2 t_3^5 t_4^4 t_5^6 t_6^2 -  y_{s}^2 y_{o_1}^2 y_{o_2}^3 y_{o_3}^5 t_1^6 t_2 t_3 t_4 t_6^3 -  y_{s}^2 y_{o_1}^2 y_{o_2}^3 y_{o_3}^5 t_1^5 t_2^2 t_3 t_4 t_6^3 -  y_{s}^2 y_{o_1}^2 y_{o_2}^3 y_{o_3}^5 t_1^4 t_2^3 t_3 t_4 t_6^3 -  y_{s}^2 y_{o_1}^2 y_{o_2}^3 y_{o_3}^5 t_1^3 t_2^4 t_3 t_4 t_6^3 -  y_{s}^2 y_{o_1}^2 y_{o_2}^3 y_{o_3}^5 t_1^2 t_2^5 t_3 t_4 t_6^3 -  y_{s}^2 y_{o_1}^2 y_{o_2}^3 y_{o_3}^5 t_1 t_2^6 t_3 t_4 t_6^3 -  y_{s}^2 y_{o_1}^3 y_{o_2}^4 y_{o_3}^5 t_1^6 t_3^2 t_4 t_5 t_6^3 -  y_{s}^2 y_{o_1}^3 y_{o_2}^4 y_{o_3}^5 t_1^5 t_2 t_3^2 t_4 t_5 t_6^3 -  y_{s}^2 y_{o_1}^3 y_{o_2}^4 y_{o_3}^5 t_1^4 t_2^2 t_3^2 t_4 t_5 t_6^3 -  y_{s}^2 y_{o_1}^3 y_{o_2}^4 y_{o_3}^5 t_1^3 t_2^3 t_3^2 t_4 t_5 t_6^3 -  y_{s}^2 y_{o_1}^3 y_{o_2}^4 y_{o_3}^5 t_1^2 t_2^4 t_3^2 t_4 t_5 t_6^3 -  y_{s}^2 y_{o_1}^3 y_{o_2}^4 y_{o_3}^5 t_1 t_2^5 t_3^2 t_4 t_5 t_6^3 -  y_{s}^2 y_{o_1}^3 y_{o_2}^4 y_{o_3}^5 t_2^6 t_3^2 t_4 t_5 t_6^3 +  y_{s}^3 y_{o_1}^4 y_{o_2}^4 y_{o_3}^6 t_1^7 t_2 t_3 t_4^3 t_5 t_6^3 +  y_{s}^3 y_{o_1}^4 y_{o_2}^4 y_{o_3}^6 t_1^6 t_2^2 t_3 t_4^3 t_5 t_6^3 +  2 y_{s}^3 y_{o_1}^4 y_{o_2}^4 y_{o_3}^6 t_1^5 t_2^3 t_3 t_4^3 t_5 t_6^3 +  2 y_{s}^3 y_{o_1}^4 y_{o_2}^4 y_{o_3}^6 t_1^4 t_2^4 t_3 t_4^3 t_5 t_6^3 +  2 y_{s}^3 y_{o_1}^4 y_{o_2}^4 y_{o_3}^6 t_1^3 t_2^5 t_3 t_4^3 t_5 t_6^3 +  y_{s}^3 y_{o_1}^4 y_{o_2}^4 y_{o_3}^6 t_1^2 t_2^6 t_3 t_4^3 t_5 t_6^3 +  y_{s}^3 y_{o_1}^4 y_{o_2}^4 y_{o_3}^6 t_1 t_2^7 t_3 t_4^3 t_5 t_6^3 -  y_{s}^2 y_{o_1}^4 y_{o_2}^5 y_{o_3}^5 t_1^5 t_3^3 t_4 t_5^2 t_6^3 -  y_{s}^2 y_{o_1}^4 y_{o_2}^5 y_{o_3}^5 t_1^4 t_2 t_3^3 t_4 t_5^2 t_6^3 -  y_{s}^2 y_{o_1}^4 y_{o_2}^5 y_{o_3}^5 t_1 t_2^4 t_3^3 t_4 t_5^2 t_6^3 -  y_{s}^2 y_{o_1}^4 y_{o_2}^5 y_{o_3}^5 t_2^5 t_3^3 t_4 t_5^2 t_6^3 +  y_{s}^3 y_{o_1}^5 y_{o_2}^5 y_{o_3}^6 t_1^7 t_3^2 t_4^3 t_5^2 t_6^3 +  2 y_{s}^3 y_{o_1}^5 y_{o_2}^5 y_{o_3}^6 t_1^6 t_2 t_3^2 t_4^3 t_5^2 t_6^3 +  2 y_{s}^3 y_{o_1}^5 y_{o_2}^5 y_{o_3}^6 t_1^5 t_2^2 t_3^2 t_4^3 t_5^2 t_6^3 +  3 y_{s}^3 y_{o_1}^5 y_{o_2}^5 y_{o_3}^6 t_1^4 t_2^3 t_3^2 t_4^3 t_5^2 t_6^3 +  3 y_{s}^3 y_{o_1}^5 y_{o_2}^5 y_{o_3}^6 t_1^3 t_2^4 t_3^2 t_4^3 t_5^2 t_6^3 +  2 y_{s}^3 y_{o_1}^5 y_{o_2}^5 y_{o_3}^6 t_1^2 t_2^5 t_3^2 t_4^3 t_5^2 t_6^3 +  2 y_{s}^3 y_{o_1}^5 y_{o_2}^5 y_{o_3}^6 t_1 t_2^6 t_3^2 t_4^3 t_5^2 t_6^3 +  y_{s}^3 y_{o_1}^5 y_{o_2}^5 y_{o_3}^6 t_2^7 t_3^2 t_4^3 t_5^2 t_6^3 -  y_{s}^4 y_{o_1}^6 y_{o_2}^5 y_{o_3}^7 t_1^5 t_2^4 t_3 t_4^5 t_5^2 t_6^3 -  y_{s}^4 y_{o_1}^6 y_{o_2}^5 y_{o_3}^7 t_1^4 t_2^5 t_3 t_4^5 t_5^2 t_6^3 -  y_{s}^2 y_{o_1}^5 y_{o_2}^6 y_{o_3}^5 t_1^4 t_3^4 t_4 t_5^3 t_6^3 +  y_{s}^2 y_{o_1}^5 y_{o_2}^6 y_{o_3}^5 t_1^2 t_2^2 t_3^4 t_4 t_5^3 t_6^3 -  y_{s}^2 y_{o_1}^5 y_{o_2}^6 y_{o_3}^5 t_2^4 t_3^4 t_4 t_5^3 t_6^3 +  y_{s}^3 y_{o_1}^6 y_{o_2}^6 y_{o_3}^6 t_1^6 t_3^3 t_4^3 t_5^3 t_6^3 +  2 y_{s}^3 y_{o_1}^6 y_{o_2}^6 y_{o_3}^6 t_1^5 t_2 t_3^3 t_4^3 t_5^3 t_6^3 +  y_{s}^3 y_{o_1}^6 y_{o_2}^6 y_{o_3}^6 t_1^4 t_2^2 t_3^3 t_4^3 t_5^3 t_6^3 +  y_{s}^3 y_{o_1}^6 y_{o_2}^6 y_{o_3}^6 t_1^3 t_2^3 t_3^3 t_4^3 t_5^3 t_6^3 +  y_{s}^3 y_{o_1}^6 y_{o_2}^6 y_{o_3}^6 t_1^2 t_2^4 t_3^3 t_4^3 t_5^3 t_6^3 +  2 y_{s}^3 y_{o_1}^6 y_{o_2}^6 y_{o_3}^6 t_1 t_2^5 t_3^3 t_4^3 t_5^3 t_6^3 +  y_{s}^3 y_{o_1}^6 y_{o_2}^6 y_{o_3}^6 t_2^6 t_3^3 t_4^3 t_5^3 t_6^3 -  y_{s}^4 y_{o_1}^7 y_{o_2}^6 y_{o_3}^7 t_1^7 t_2 t_3^2 t_4^5 t_5^3 t_6^3 -  y_{s}^4 y_{o_1}^7 y_{o_2}^6 y_{o_3}^7 t_1^6 t_2^2 t_3^2 t_4^5 t_5^3 t_6^3 -  2 y_{s}^4 y_{o_1}^7 y_{o_2}^6 y_{o_3}^7 t_1^5 t_2^3 t_3^2 t_4^5 t_5^3 t_6^3 -  3 y_{s}^4 y_{o_1}^7 y_{o_2}^6 y_{o_3}^7 t_1^4 t_2^4 t_3^2 t_4^5 t_5^3 t_6^3 -  2 y_{s}^4 y_{o_1}^7 y_{o_2}^6 y_{o_3}^7 t_1^3 t_2^5 t_3^2 t_4^5 t_5^3 t_6^3 -  y_{s}^4 y_{o_1}^7 y_{o_2}^6 y_{o_3}^7 t_1^2 t_2^6 t_3^2 t_4^5 t_5^3 t_6^3 -  y_{s}^4 y_{o_1}^7 y_{o_2}^6 y_{o_3}^7 t_1 t_2^7 t_3^2 t_4^5 t_5^3 t_6^3 +  y_{s}^3 y_{o_1}^7 y_{o_2}^7 y_{o_3}^6 t_1^5 t_3^4 t_4^3 t_5^4 t_6^3 +  y_{s}^3 y_{o_1}^7 y_{o_2}^7 y_{o_3}^6 t_1^4 t_2 t_3^4 t_4^3 t_5^4 t_6^3 -  y_{s}^3 y_{o_1}^7 y_{o_2}^7 y_{o_3}^6 t_1^3 t_2^2 t_3^4 t_4^3 t_5^4 t_6^3 -  y_{s}^3 y_{o_1}^7 y_{o_2}^7 y_{o_3}^6 t_1^2 t_2^3 t_3^4 t_4^3 t_5^4 t_6^3 +  y_{s}^3 y_{o_1}^7 y_{o_2}^7 y_{o_3}^6 t_1 t_2^4 t_3^4 t_4^3 t_5^4 t_6^3 +  y_{s}^3 y_{o_1}^7 y_{o_2}^7 y_{o_3}^6 t_2^5 t_3^4 t_4^3 t_5^4 t_6^3 -  y_{s}^4 y_{o_1}^8 y_{o_2}^7 y_{o_3}^7 t_1^6 t_2 t_3^3 t_4^5 t_5^4 t_6^3 -  y_{s}^4 y_{o_1}^8 y_{o_2}^7 y_{o_3}^7 t_1^5 t_2^2 t_3^3 t_4^5 t_5^4 t_6^3 -  y_{s}^4 y_{o_1}^8 y_{o_2}^7 y_{o_3}^7 t_1^4 t_2^3 t_3^3 t_4^5 t_5^4 t_6^3 -  y_{s}^4 y_{o_1}^8 y_{o_2}^7 y_{o_3}^7 t_1^3 t_2^4 t_3^3 t_4^5 t_5^4 t_6^3 -  y_{s}^4 y_{o_1}^8 y_{o_2}^7 y_{o_3}^7 t_1^2 t_2^5 t_3^3 t_4^5 t_5^4 t_6^3 -  y_{s}^4 y_{o_1}^8 y_{o_2}^7 y_{o_3}^7 t_1 t_2^6 t_3^3 t_4^5 t_5^4 t_6^3 +  y_{s}^5 y_{o_1}^9 y_{o_2}^7 y_{o_3}^8 t_1^5 t_2^4 t_3^2 t_4^7 t_5^4 t_6^3 +  y_{s}^5 y_{o_1}^9 y_{o_2}^7 y_{o_3}^8 t_1^4 t_2^5 t_3^2 t_4^7 t_5^4 t_6^3 -  y_{s}^3 y_{o_1}^8 y_{o_2}^8 y_{o_3}^6 t_1^2 t_2^2 t_3^5 t_4^3 t_5^5 t_6^3 -  y_{s}^4 y_{o_1}^9 y_{o_2}^8 y_{o_3}^7 t_1^5 t_2 t_3^4 t_4^5 t_5^5 t_6^3 +  y_{s}^4 y_{o_1}^9 y_{o_2}^8 y_{o_3}^7 t_1^3 t_2^3 t_3^4 t_4^5 t_5^5 t_6^3 -  y_{s}^4 y_{o_1}^9 y_{o_2}^8 y_{o_3}^7 t_1 t_2^5 t_3^4 t_4^5 t_5^5 t_6^3 +  y_{s}^5 y_{o_1}^{10} y_{o_2}^8 y_{o_3}^8 t_1^4 t_2^4 t_3^3 t_4^7 t_5^5 t_6^3 +  y_{s}^4 y_{o_1}^{10} y_{o_2}^9 y_{o_3}^7 t_1^3 t_2^2 t_3^5 t_4^5 t_5^6 t_6^3 +  y_{s}^4 y_{o_1}^{10} y_{o_2}^9 y_{o_3}^7 t_1^2 t_2^3 t_3^5 t_4^5 t_5^6 t_6^3 +  y_{s}^3 y_{o_1}^4 y_{o_2}^5 y_{o_3}^7 t_1^5 t_2^4 t_3^2 t_4^2 t_5 t_6^4 +  y_{s}^3 y_{o_1}^4 y_{o_2}^5 y_{o_3}^7 t_1^4 t_2^5 t_3^2 t_4^2 t_5 t_6^4 +  y_{s}^2 y_{o_1}^4 y_{o_2}^6 y_{o_3}^6 t_1^3 t_2^3 t_3^4 t_5^2 t_6^4 -  y_{s}^3 y_{o_1}^5 y_{o_2}^6 y_{o_3}^7 t_1^6 t_2^2 t_3^3 t_4^2 t_5^2 t_6^4 +  y_{s}^3 y_{o_1}^5 y_{o_2}^6 y_{o_3}^7 t_1^4 t_2^4 t_3^3 t_4^2 t_5^2 t_6^4 -  y_{s}^3 y_{o_1}^5 y_{o_2}^6 y_{o_3}^7 t_1^2 t_2^6 t_3^3 t_4^2 t_5^2 t_6^4 -  y_{s}^4 y_{o_1}^6 y_{o_2}^6 y_{o_3}^8 t_1^5 t_2^5 t_3^2 t_4^4 t_5^2 t_6^4 +  y_{s}^2 y_{o_1}^5 y_{o_2}^7 y_{o_3}^6 t_1^3 t_2^2 t_3^5 t_5^3 t_6^4 +  y_{s}^2 y_{o_1}^5 y_{o_2}^7 y_{o_3}^6 t_1^2 t_2^3 t_3^5 t_5^3 t_6^4 -  y_{s}^3 y_{o_1}^6 y_{o_2}^7 y_{o_3}^7 t_1^6 t_2 t_3^4 t_4^2 t_5^3 t_6^4 -  y_{s}^3 y_{o_1}^6 y_{o_2}^7 y_{o_3}^7 t_1^5 t_2^2 t_3^4 t_4^2 t_5^3 t_6^4 -  y_{s}^3 y_{o_1}^6 y_{o_2}^7 y_{o_3}^7 t_1^4 t_2^3 t_3^4 t_4^2 t_5^3 t_6^4 -  y_{s}^3 y_{o_1}^6 y_{o_2}^7 y_{o_3}^7 t_1^3 t_2^4 t_3^4 t_4^2 t_5^3 t_6^4 -  y_{s}^3 y_{o_1}^6 y_{o_2}^7 y_{o_3}^7 t_1^2 t_2^5 t_3^4 t_4^2 t_5^3 t_6^4 -  y_{s}^3 y_{o_1}^6 y_{o_2}^7 y_{o_3}^7 t_1 t_2^6 t_3^4 t_4^2 t_5^3 t_6^4 +  y_{s}^4 y_{o_1}^7 y_{o_2}^7 y_{o_3}^8 t_1^7 t_2^2 t_3^3 t_4^4 t_5^3 t_6^4 +  y_{s}^4 y_{o_1}^7 y_{o_2}^7 y_{o_3}^8 t_1^6 t_2^3 t_3^3 t_4^4 t_5^3 t_6^4 -  y_{s}^4 y_{o_1}^7 y_{o_2}^7 y_{o_3}^8 t_1^5 t_2^4 t_3^3 t_4^4 t_5^3 t_6^4 -  y_{s}^4 y_{o_1}^7 y_{o_2}^7 y_{o_3}^8 t_1^4 t_2^5 t_3^3 t_4^4 t_5^3 t_6^4 +  y_{s}^4 y_{o_1}^7 y_{o_2}^7 y_{o_3}^8 t_1^3 t_2^6 t_3^3 t_4^4 t_5^3 t_6^4 +  y_{s}^4 y_{o_1}^7 y_{o_2}^7 y_{o_3}^8 t_1^2 t_2^7 t_3^3 t_4^4 t_5^3 t_6^4 -  y_{s}^3 y_{o_1}^7 y_{o_2}^8 y_{o_3}^7 t_1^6 t_3^5 t_4^2 t_5^4 t_6^4 -  y_{s}^3 y_{o_1}^7 y_{o_2}^8 y_{o_3}^7 t_1^5 t_2 t_3^5 t_4^2 t_5^4 t_6^4 -  2 y_{s}^3 y_{o_1}^7 y_{o_2}^8 y_{o_3}^7 t_1^4 t_2^2 t_3^5 t_4^2 t_5^4 t_6^4 -  3 y_{s}^3 y_{o_1}^7 y_{o_2}^8 y_{o_3}^7 t_1^3 t_2^3 t_3^5 t_4^2 t_5^4 t_6^4 -  2 y_{s}^3 y_{o_1}^7 y_{o_2}^8 y_{o_3}^7 t_1^2 t_2^4 t_3^5 t_4^2 t_5^4 t_6^4 -  y_{s}^3 y_{o_1}^7 y_{o_2}^8 y_{o_3}^7 t_1 t_2^5 t_3^5 t_4^2 t_5^4 t_6^4 -  y_{s}^3 y_{o_1}^7 y_{o_2}^8 y_{o_3}^7 t_2^6 t_3^5 t_4^2 t_5^4 t_6^4 +  y_{s}^4 y_{o_1}^8 y_{o_2}^8 y_{o_3}^8 t_1^7 t_2 t_3^4 t_4^4 t_5^4 t_6^4 +  2 y_{s}^4 y_{o_1}^8 y_{o_2}^8 y_{o_3}^8 t_1^6 t_2^2 t_3^4 t_4^4 t_5^4 t_6^4 +  y_{s}^4 y_{o_1}^8 y_{o_2}^8 y_{o_3}^8 t_1^5 t_2^3 t_3^4 t_4^4 t_5^4 t_6^4 +  y_{s}^4 y_{o_1}^8 y_{o_2}^8 y_{o_3}^8 t_1^4 t_2^4 t_3^4 t_4^4 t_5^4 t_6^4 +  y_{s}^4 y_{o_1}^8 y_{o_2}^8 y_{o_3}^8 t_1^3 t_2^5 t_3^4 t_4^4 t_5^4 t_6^4 +  2 y_{s}^4 y_{o_1}^8 y_{o_2}^8 y_{o_3}^8 t_1^2 t_2^6 t_3^4 t_4^4 t_5^4 t_6^4 +  y_{s}^4 y_{o_1}^8 y_{o_2}^8 y_{o_3}^8 t_1 t_2^7 t_3^4 t_4^4 t_5^4 t_6^4 -  y_{s}^5 y_{o_1}^9 y_{o_2}^8 y_{o_3}^9 t_1^7 t_2^3 t_3^3 t_4^6 t_5^4 t_6^4 +  y_{s}^5 y_{o_1}^9 y_{o_2}^8 y_{o_3}^9 t_1^5 t_2^5 t_3^3 t_4^6 t_5^4 t_6^4 -  y_{s}^5 y_{o_1}^9 y_{o_2}^8 y_{o_3}^9 t_1^3 t_2^7 t_3^3 t_4^6 t_5^4 t_6^4 -  y_{s}^3 y_{o_1}^8 y_{o_2}^9 y_{o_3}^7 t_1^3 t_2^2 t_3^6 t_4^2 t_5^5 t_6^4 -  y_{s}^3 y_{o_1}^8 y_{o_2}^9 y_{o_3}^7 t_1^2 t_2^3 t_3^6 t_4^2 t_5^5 t_6^4 +  y_{s}^4 y_{o_1}^9 y_{o_2}^9 y_{o_3}^8 t_1^7 t_3^5 t_4^4 t_5^5 t_6^4 +  2 y_{s}^4 y_{o_1}^9 y_{o_2}^9 y_{o_3}^8 t_1^6 t_2 t_3^5 t_4^4 t_5^5 t_6^4 +  2 y_{s}^4 y_{o_1}^9 y_{o_2}^9 y_{o_3}^8 t_1^5 t_2^2 t_3^5 t_4^4 t_5^5 t_6^4 +  3 y_{s}^4 y_{o_1}^9 y_{o_2}^9 y_{o_3}^8 t_1^4 t_2^3 t_3^5 t_4^4 t_5^5 t_6^4 +  3 y_{s}^4 y_{o_1}^9 y_{o_2}^9 y_{o_3}^8 t_1^3 t_2^4 t_3^5 t_4^4 t_5^5 t_6^4 +  2 y_{s}^4 y_{o_1}^9 y_{o_2}^9 y_{o_3}^8 t_1^2 t_2^5 t_3^5 t_4^4 t_5^5 t_6^4 +  2 y_{s}^4 y_{o_1}^9 y_{o_2}^9 y_{o_3}^8 t_1 t_2^6 t_3^5 t_4^4 t_5^5 t_6^4 +  y_{s}^4 y_{o_1}^9 y_{o_2}^9 y_{o_3}^8 t_2^7 t_3^5 t_4^4 t_5^5 t_6^4 -  y_{s}^5 y_{o_1}^{10} y_{o_2}^9 y_{o_3}^9 t_1^7 t_2^2 t_3^4 t_4^6 t_5^5 t_6^4 -  y_{s}^5 y_{o_1}^{10} y_{o_2}^9 y_{o_3}^9 t_1^6 t_2^3 t_3^4 t_4^6 t_5^5 t_6^4 -  y_{s}^5 y_{o_1}^{10} y_{o_2}^9 y_{o_3}^9 t_1^3 t_2^6 t_3^4 t_4^6 t_5^5 t_6^4 -  y_{s}^5 y_{o_1}^{10} y_{o_2}^9 y_{o_3}^9 t_1^2 t_2^7 t_3^4 t_4^6 t_5^5 t_6^4 +  y_{s}^4 y_{o_1}^{10} y_{o_2}^{10} y_{o_3}^8 t_1^6 t_3^6 t_4^4 t_5^6 t_6^4 +  y_{s}^4 y_{o_1}^{10} y_{o_2}^{10} y_{o_3}^8 t_1^5 t_2 t_3^6 t_4^4 t_5^6 t_6^4 +  2 y_{s}^4 y_{o_1}^{10} y_{o_2}^{10} y_{o_3}^8 t_1^4 t_2^2 t_3^6 t_4^4 t_5^6 t_6^4 +  2 y_{s}^4 y_{o_1}^{10} y_{o_2}^{10} y_{o_3}^8 t_1^3 t_2^3 t_3^6 t_4^4 t_5^6 t_6^4 +  2 y_{s}^4 y_{o_1}^{10} y_{o_2}^{10} y_{o_3}^8 t_1^2 t_2^4 t_3^6 t_4^4 t_5^6 t_6^4 +  y_{s}^4 y_{o_1}^{10} y_{o_2}^{10} y_{o_3}^8 t_1 t_2^5 t_3^6 t_4^4 t_5^6 t_6^4 +  y_{s}^4 y_{o_1}^{10} y_{o_2}^{10} y_{o_3}^8 t_2^6 t_3^6 t_4^4 t_5^6 t_6^4 -  y_{s}^5 y_{o_1}^{11} y_{o_2}^{10} y_{o_3}^9 t_1^7 t_2 t_3^5 t_4^6 t_5^6 t_6^4 -  y_{s}^5 y_{o_1}^{11} y_{o_2}^{10} y_{o_3}^9 t_1^6 t_2^2 t_3^5 t_4^6 t_5^6 t_6^4 -  y_{s}^5 y_{o_1}^{11} y_{o_2}^{10} y_{o_3}^9 t_1^5 t_2^3 t_3^5 t_4^6 t_5^6 t_6^4 -  y_{s}^5 y_{o_1}^{11} y_{o_2}^{10} y_{o_3}^9 t_1^4 t_2^4 t_3^5 t_4^6 t_5^6 t_6^4 -  y_{s}^5 y_{o_1}^{11} y_{o_2}^{10} y_{o_3}^9 t_1^3 t_2^5 t_3^5 t_4^6 t_5^6 t_6^4 -  y_{s}^5 y_{o_1}^{11} y_{o_2}^{10} y_{o_3}^9 t_1^2 t_2^6 t_3^5 t_4^6 t_5^6 t_6^4 -  y_{s}^5 y_{o_1}^{11} y_{o_2}^{10} y_{o_3}^9 t_1 t_2^7 t_3^5 t_4^6 t_5^6 t_6^4 -  y_{s}^5 y_{o_1}^{12} y_{o_2}^{11} y_{o_3}^9 t_1^6 t_2 t_3^6 t_4^6 t_5^7 t_6^4 -  y_{s}^5 y_{o_1}^{12} y_{o_2}^{11} y_{o_3}^9 t_1^5 t_2^2 t_3^6 t_4^6 t_5^7 t_6^4 -  y_{s}^5 y_{o_1}^{12} y_{o_2}^{11} y_{o_3}^9 t_1^4 t_2^3 t_3^6 t_4^6 t_5^7 t_6^4 -  y_{s}^5 y_{o_1}^{12} y_{o_2}^{11} y_{o_3}^9 t_1^3 t_2^4 t_3^6 t_4^6 t_5^7 t_6^4 -  y_{s}^5 y_{o_1}^{12} y_{o_2}^{11} y_{o_3}^9 t_1^2 t_2^5 t_3^6 t_4^6 t_5^7 t_6^4 -  y_{s}^5 y_{o_1}^{12} y_{o_2}^{11} y_{o_3}^9 t_1 t_2^6 t_3^6 t_4^6 t_5^7 t_6^4 +  y_{s}^4 y_{o_1}^5 y_{o_2}^6 y_{o_3}^9 t_1^6 t_2^6 t_3^2 t_4^3 t_5 t_6^5 -  y_{s}^3 y_{o_1}^5 y_{o_2}^7 y_{o_3}^8 t_1^6 t_2^3 t_3^4 t_4 t_5^2 t_6^5 -  y_{s}^3 y_{o_1}^5 y_{o_2}^7 y_{o_3}^8 t_1^3 t_2^6 t_3^4 t_4 t_5^2 t_6^5 +  y_{s}^4 y_{o_1}^6 y_{o_2}^7 y_{o_3}^9 t_1^6 t_2^5 t_3^3 t_4^3 t_5^2 t_6^5 +  y_{s}^4 y_{o_1}^6 y_{o_2}^7 y_{o_3}^9 t_1^5 t_2^6 t_3^3 t_4^3 t_5^2 t_6^5 -  y_{s}^5 y_{o_1}^7 y_{o_2}^7 y_{o_3}^{10} t_1^7 t_2^6 t_3^2 t_4^5 t_5^2 t_6^5 -  y_{s}^5 y_{o_1}^7 y_{o_2}^7 y_{o_3}^{10} t_1^6 t_2^7 t_3^2 t_4^5 t_5^2 t_6^5 -  y_{s}^3 y_{o_1}^6 y_{o_2}^8 y_{o_3}^8 t_1^6 t_2^2 t_3^5 t_4 t_5^3 t_6^5 -  y_{s}^3 y_{o_1}^6 y_{o_2}^8 y_{o_3}^8 t_1^5 t_2^3 t_3^5 t_4 t_5^3 t_6^5 -  y_{s}^3 y_{o_1}^6 y_{o_2}^8 y_{o_3}^8 t_1^3 t_2^5 t_3^5 t_4 t_5^3 t_6^5 -  y_{s}^3 y_{o_1}^6 y_{o_2}^8 y_{o_3}^8 t_1^2 t_2^6 t_3^5 t_4 t_5^3 t_6^5 +  y_{s}^4 y_{o_1}^7 y_{o_2}^8 y_{o_3}^9 t_1^7 t_2^3 t_3^4 t_4^3 t_5^3 t_6^5 +  2 y_{s}^4 y_{o_1}^7 y_{o_2}^8 y_{o_3}^9 t_1^6 t_2^4 t_3^4 t_4^3 t_5^3 t_6^5 +  y_{s}^4 y_{o_1}^7 y_{o_2}^8 y_{o_3}^9 t_1^5 t_2^5 t_3^4 t_4^3 t_5^3 t_6^5 +  2 y_{s}^4 y_{o_1}^7 y_{o_2}^8 y_{o_3}^9 t_1^4 t_2^6 t_3^4 t_4^3 t_5^3 t_6^5 +  y_{s}^4 y_{o_1}^7 y_{o_2}^8 y_{o_3}^9 t_1^3 t_2^7 t_3^4 t_4^3 t_5^3 t_6^5 -  y_{s}^5 y_{o_1}^8 y_{o_2}^8 y_{o_3}^{10} t_1^7 t_2^5 t_3^3 t_4^5 t_5^3 t_6^5 -  2 y_{s}^5 y_{o_1}^8 y_{o_2}^8 y_{o_3}^{10} t_1^6 t_2^6 t_3^3 t_4^5 t_5^3 t_6^5 -  y_{s}^5 y_{o_1}^8 y_{o_2}^8 y_{o_3}^{10} t_1^5 t_2^7 t_3^3 t_4^5 t_5^3 t_6^5 +  y_{s}^4 y_{o_1}^8 y_{o_2}^9 y_{o_3}^9 t_1^7 t_2^2 t_3^5 t_4^3 t_5^4 t_6^5 +  3 y_{s}^4 y_{o_1}^8 y_{o_2}^9 y_{o_3}^9 t_1^6 t_2^3 t_3^5 t_4^3 t_5^4 t_6^5 +  2 y_{s}^4 y_{o_1}^8 y_{o_2}^9 y_{o_3}^9 t_1^5 t_2^4 t_3^5 t_4^3 t_5^4 t_6^5 +  2 y_{s}^4 y_{o_1}^8 y_{o_2}^9 y_{o_3}^9 t_1^4 t_2^5 t_3^5 t_4^3 t_5^4 t_6^5 +  3 y_{s}^4 y_{o_1}^8 y_{o_2}^9 y_{o_3}^9 t_1^3 t_2^6 t_3^5 t_4^3 t_5^4 t_6^5 +  y_{s}^4 y_{o_1}^8 y_{o_2}^9 y_{o_3}^9 t_1^2 t_2^7 t_3^5 t_4^3 t_5^4 t_6^5 -  2 y_{s}^5 y_{o_1}^9 y_{o_2}^9 y_{o_3}^{10} t_1^7 t_2^4 t_3^4 t_4^5 t_5^4 t_6^5 -  2 y_{s}^5 y_{o_1}^9 y_{o_2}^9 y_{o_3}^{10} t_1^6 t_2^5 t_3^4 t_4^5 t_5^4 t_6^5 -  2 y_{s}^5 y_{o_1}^9 y_{o_2}^9 y_{o_3}^{10} t_1^5 t_2^6 t_3^4 t_4^5 t_5^4 t_6^5 -  2 y_{s}^5 y_{o_1}^9 y_{o_2}^9 y_{o_3}^{10} t_1^4 t_2^7 t_3^4 t_4^5 t_5^4 t_6^5 +  y_{s}^6 y_{o_1}^{10} y_{o_2}^9 y_{o_3}^{11} t_1^7 t_2^6 t_3^3 t_4^7 t_5^4 t_6^5 +  y_{s}^6 y_{o_1}^{10} y_{o_2}^9 y_{o_3}^{11} t_1^6 t_2^7 t_3^3 t_4^7 t_5^4 t_6^5 +  y_{s}^4 y_{o_1}^9 y_{o_2}^{10} y_{o_3}^9 t_1^6 t_2^2 t_3^6 t_4^3 t_5^5 t_6^5 +  y_{s}^4 y_{o_1}^9 y_{o_2}^{10} y_{o_3}^9 t_1^5 t_2^3 t_3^6 t_4^3 t_5^5 t_6^5 +  y_{s}^4 y_{o_1}^9 y_{o_2}^{10} y_{o_3}^9 t_1^4 t_2^4 t_3^6 t_4^3 t_5^5 t_6^5 +  y_{s}^4 y_{o_1}^9 y_{o_2}^{10} y_{o_3}^9 t_1^3 t_2^5 t_3^6 t_4^3 t_5^5 t_6^5 +  y_{s}^4 y_{o_1}^9 y_{o_2}^{10} y_{o_3}^9 t_1^2 t_2^6 t_3^6 t_4^3 t_5^5 t_6^5 -  2 y_{s}^5 y_{o_1}^{10} y_{o_2}^{10} y_{o_3}^{10} t_1^7 t_2^3 t_3^5 t_4^5 t_5^5 t_6^5 -  3 y_{s}^5 y_{o_1}^{10} y_{o_2}^{10} y_{o_3}^{10} t_1^6 t_2^4 t_3^5 t_4^5 t_5^5 t_6^5 -  2 y_{s}^5 y_{o_1}^{10} y_{o_2}^{10} y_{o_3}^{10} t_1^5 t_2^5 t_3^5 t_4^5 t_5^5 t_6^5 -  3 y_{s}^5 y_{o_1}^{10} y_{o_2}^{10} y_{o_3}^{10} t_1^4 t_2^6 t_3^5 t_4^5 t_5^5 t_6^5 -  2 y_{s}^5 y_{o_1}^{10} y_{o_2}^{10} y_{o_3}^{10} t_1^3 t_2^7 t_3^5 t_4^5 t_5^5 t_6^5 +  y_{s}^6 y_{o_1}^{11} y_{o_2}^{10} y_{o_3}^{11} t_1^7 t_2^5 t_3^4 t_4^7 t_5^5 t_6^5 +  y_{s}^6 y_{o_1}^{11} y_{o_2}^{10} y_{o_3}^{11} t_1^6 t_2^6 t_3^4 t_4^7 t_5^5 t_6^5 +  y_{s}^6 y_{o_1}^{11} y_{o_2}^{10} y_{o_3}^{11} t_1^5 t_2^7 t_3^4 t_4^7 t_5^5 t_6^5 -  y_{s}^5 y_{o_1}^{11} y_{o_2}^{11} y_{o_3}^{10} t_1^7 t_2^2 t_3^6 t_4^5 t_5^6 t_6^5 -  2 y_{s}^5 y_{o_1}^{11} y_{o_2}^{11} y_{o_3}^{10} t_1^6 t_2^3 t_3^6 t_4^5 t_5^6 t_6^5 -  2 y_{s}^5 y_{o_1}^{11} y_{o_2}^{11} y_{o_3}^{10} t_1^5 t_2^4 t_3^6 t_4^5 t_5^6 t_6^5 -  2 y_{s}^5 y_{o_1}^{11} y_{o_2}^{11} y_{o_3}^{10} t_1^4 t_2^5 t_3^6 t_4^5 t_5^6 t_6^5 -  2 y_{s}^5 y_{o_1}^{11} y_{o_2}^{11} y_{o_3}^{10} t_1^3 t_2^6 t_3^6 t_4^5 t_5^6 t_6^5 -  y_{s}^5 y_{o_1}^{11} y_{o_2}^{11} y_{o_3}^{10} t_1^2 t_2^7 t_3^6 t_4^5 t_5^6 t_6^5 +  y_{s}^6 y_{o_1}^{12} y_{o_2}^{11} y_{o_3}^{11} t_1^7 t_2^4 t_3^5 t_4^7 t_5^6 t_6^5 +  y_{s}^6 y_{o_1}^{12} y_{o_2}^{11} y_{o_3}^{11} t_1^6 t_2^5 t_3^5 t_4^7 t_5^6 t_6^5 +  y_{s}^6 y_{o_1}^{12} y_{o_2}^{11} y_{o_3}^{11} t_1^5 t_2^6 t_3^5 t_4^7 t_5^6 t_6^5 +  y_{s}^6 y_{o_1}^{12} y_{o_2}^{11} y_{o_3}^{11} t_1^4 t_2^7 t_3^5 t_4^7 t_5^6 t_6^5 +  y_{s}^6 y_{o_1}^{13} y_{o_2}^{12} y_{o_3}^{11} t_1^6 t_2^4 t_3^6 t_4^7 t_5^7 t_6^5 +  y_{s}^6 y_{o_1}^{13} y_{o_2}^{12} y_{o_3}^{11} t_1^5 t_2^5 t_3^6 t_4^7 t_5^7 t_6^5 +  y_{s}^6 y_{o_1}^{13} y_{o_2}^{12} y_{o_3}^{11} t_1^4 t_2^6 t_3^6 t_4^7 t_5^7 t_6^5 +  y_{s}^4 y_{o_1}^6 y_{o_2}^8 y_{o_3}^{10} t_1^6 t_2^6 t_3^4 t_4^2 t_5^2 t_6^6 +  y_{s}^4 y_{o_1}^7 y_{o_2}^9 y_{o_3}^{10} t_1^6 t_2^5 t_3^5 t_4^2 t_5^3 t_6^6 +  y_{s}^4 y_{o_1}^7 y_{o_2}^9 y_{o_3}^{10} t_1^5 t_2^6 t_3^5 t_4^2 t_5^3 t_6^6 -  y_{s}^5 y_{o_1}^8 y_{o_2}^9 y_{o_3}^{11} t_1^7 t_2^6 t_3^4 t_4^4 t_5^3 t_6^6 -  y_{s}^5 y_{o_1}^8 y_{o_2}^9 y_{o_3}^{11} t_1^6 t_2^7 t_3^4 t_4^4 t_5^3 t_6^6 +  y_{s}^4 y_{o_1}^8 y_{o_2}^{10} y_{o_3}^{10} t_1^6 t_2^4 t_3^6 t_4^2 t_5^4 t_6^6 +  y_{s}^4 y_{o_1}^8 y_{o_2}^{10} y_{o_3}^{10} t_1^5 t_2^5 t_3^6 t_4^2 t_5^4 t_6^6 +  y_{s}^4 y_{o_1}^8 y_{o_2}^{10} y_{o_3}^{10} t_1^4 t_2^6 t_3^6 t_4^2 t_5^4 t_6^6 -  y_{s}^5 y_{o_1}^9 y_{o_2}^{10} y_{o_3}^{11} t_1^7 t_2^5 t_3^5 t_4^4 t_5^4 t_6^6 -  2 y_{s}^5 y_{o_1}^9 y_{o_2}^{10} y_{o_3}^{11} t_1^6 t_2^6 t_3^5 t_4^4 t_5^4 t_6^6 -  y_{s}^5 y_{o_1}^9 y_{o_2}^{10} y_{o_3}^{11} t_1^5 t_2^7 t_3^5 t_4^4 t_5^4 t_6^6 +  y_{s}^6 y_{o_1}^{10} y_{o_2}^{10} y_{o_3}^{12} t_1^7 t_2^7 t_3^4 t_4^6 t_5^4 t_6^6 -  y_{s}^5 y_{o_1}^{10} y_{o_2}^{11} y_{o_3}^{11} t_1^7 t_2^4 t_3^6 t_4^4 t_5^5 t_6^6 -  2 y_{s}^5 y_{o_1}^{10} y_{o_2}^{11} y_{o_3}^{11} t_1^6 t_2^5 t_3^6 t_4^4 t_5^5 t_6^6 -  2 y_{s}^5 y_{o_1}^{10} y_{o_2}^{11} y_{o_3}^{11} t_1^5 t_2^6 t_3^6 t_4^4 t_5^5 t_6^6 -  y_{s}^5 y_{o_1}^{10} y_{o_2}^{11} y_{o_3}^{11} t_1^4 t_2^7 t_3^6 t_4^4 t_5^5 t_6^6 +  y_{s}^6 y_{o_1}^{11} y_{o_2}^{11} y_{o_3}^{12} t_1^7 t_2^6 t_3^5 t_4^6 t_5^5 t_6^6 +  y_{s}^6 y_{o_1}^{11} y_{o_2}^{11} y_{o_3}^{12} t_1^6 t_2^7 t_3^5 t_4^6 t_5^5 t_6^6 -  y_{s}^5 y_{o_1}^{11} y_{o_2}^{12} y_{o_3}^{11} t_1^6 t_2^4 t_3^7 t_4^4 t_5^6 t_6^6 -  y_{s}^5 y_{o_1}^{11} y_{o_2}^{12} y_{o_3}^{11} t_1^5 t_2^5 t_3^7 t_4^4 t_5^6 t_6^6 -  y_{s}^5 y_{o_1}^{11} y_{o_2}^{12} y_{o_3}^{11} t_1^4 t_2^6 t_3^7 t_4^4 t_5^6 t_6^6 +  y_{s}^6 y_{o_1}^{12} y_{o_2}^{12} y_{o_3}^{12} t_1^7 t_2^5 t_3^6 t_4^6 t_5^6 t_6^6 +  y_{s}^6 y_{o_1}^{12} y_{o_2}^{12} y_{o_3}^{12} t_1^6 t_2^6 t_3^6 t_4^6 t_5^6 t_6^6 +  y_{s}^6 y_{o_1}^{12} y_{o_2}^{12} y_{o_3}^{12} t_1^5 t_2^7 t_3^6 t_4^6 t_5^6 t_6^6 +  y_{s}^6 y_{o_1}^{13} y_{o_2}^{13} y_{o_3}^{12} t_1^6 t_2^5 t_3^7 t_4^6 t_5^7 t_6^6 +  y_{s}^6 y_{o_1}^{13} y_{o_2}^{13} y_{o_3}^{12} t_1^5 t_2^6 t_3^7 t_4^6 t_5^7 t_6^6 +  y_{s}^7 y_{o_1}^{14} y_{o_2}^{14} y_{o_3}^{14} t_1^7 t_2^7 t_3^7 t_4^7 t_5^7 t_6^7~,~$
\end{quote}
\endgroup

\subsection{Model 9 \label{app_num_09}}

\begingroup\makeatletter\def\f@size{7}\check@mathfonts
\begin{quote}\raggedright
$
P(t_i,y_s,y_{o_1},y_{o_2}; \mathcal{M}_9) =
1 + y_{s} y_{o_1}^3 y_{o_2}^4 t_1^2 t_2 t_3^3 t_5^2 +  y_{s} y_{o_1}^3 y_{o_2}^4 t_1 t_2^2 t_3^3 t_5^2 + y_{s} y_{o_1}^3 y_{o_2}^4 t_1^3 t_3^2 t_4 t_5^2 +  y_{s} y_{o_1}^3 y_{o_2}^4 t_1^2 t_2 t_3^2 t_4 t_5^2 +  y_{s} y_{o_1}^3 y_{o_2}^4 t_1 t_2^2 t_3^2 t_4 t_5^2 +  y_{s} y_{o_1}^3 y_{o_2}^4 t_2^3 t_3^2 t_4 t_5^2 + y_{s} y_{o_1}^3 y_{o_2}^4 t_1^3 t_3 t_4^2 t_5^2 +  y_{s} y_{o_1}^3 y_{o_2}^4 t_1^2 t_2 t_3 t_4^2 t_5^2 +  y_{s} y_{o_1}^3 y_{o_2}^4 t_1 t_2^2 t_3 t_4^2 t_5^2 +  y_{s} y_{o_1}^3 y_{o_2}^4 t_2^3 t_3 t_4^2 t_5^2 + y_{s} y_{o_1}^3 y_{o_2}^4 t_1^2 t_2 t_4^3 t_5^2 +  y_{s} y_{o_1}^3 y_{o_2}^4 t_1 t_2^2 t_4^3 t_5^2 -  y_{s}^2 y_{o_1}^6 y_{o_2}^8 t_1^3 t_2^3 t_3^5 t_4 t_5^4 -  y_{s}^2 y_{o_1}^6 y_{o_2}^8 t_1^3 t_2^3 t_3^4 t_4^2 t_5^4 -  y_{s}^2 y_{o_1}^6 y_{o_2}^8 t_1^5 t_2 t_3^3 t_4^3 t_5^4 -  y_{s}^2 y_{o_1}^6 y_{o_2}^8 t_1^4 t_2^2 t_3^3 t_4^3 t_5^4 -  y_{s}^2 y_{o_1}^6 y_{o_2}^8 t_1^3 t_2^3 t_3^3 t_4^3 t_5^4 -  y_{s}^2 y_{o_1}^6 y_{o_2}^8 t_1^2 t_2^4 t_3^3 t_4^3 t_5^4 -  y_{s}^2 y_{o_1}^6 y_{o_2}^8 t_1 t_2^5 t_3^3 t_4^3 t_5^4 -  y_{s}^2 y_{o_1}^6 y_{o_2}^8 t_1^3 t_2^3 t_3^2 t_4^4 t_5^4 -  y_{s}^2 y_{o_1}^6 y_{o_2}^8 t_1^3 t_2^3 t_3 t_4^5 t_5^4 -  y_{s}^3 y_{o_1}^9 y_{o_2}^{12} t_1^5 t_2^4 t_3^5 t_4^4 t_5^6 -  y_{s}^3 y_{o_1}^9 y_{o_2}^{12} t_1^4 t_2^5 t_3^5 t_4^4 t_5^6 -  y_{s}^3 y_{o_1}^9 y_{o_2}^{12} t_1^5 t_2^4 t_3^4 t_4^5 t_5^6 -  y_{s}^3 y_{o_1}^9 y_{o_2}^{12} t_1^4 t_2^5 t_3^4 t_4^5 t_5^6 +  y_{s} y_{o_1}^2 y_{o_2}^3 t_1^2 t_3^2 t_5 t_6 + y_{s} y_{o_1}^2 y_{o_2}^3 t_1 t_2 t_3^2 t_5 t_6 +  y_{s} y_{o_1}^2 y_{o_2}^3 t_2^2 t_3^2 t_5 t_6 + y_{s} y_{o_1}^2 y_{o_2}^3 t_1^2 t_3 t_4 t_5 t_6 +  y_{s} y_{o_1}^2 y_{o_2}^3 t_1 t_2 t_3 t_4 t_5 t_6 + y_{s} y_{o_1}^2 y_{o_2}^3 t_2^2 t_3 t_4 t_5 t_6 +  y_{s} y_{o_1}^2 y_{o_2}^3 t_1^2 t_4^2 t_5 t_6 + y_{s} y_{o_1}^2 y_{o_2}^3 t_1 t_2 t_4^2 t_5 t_6 +  y_{s} y_{o_1}^2 y_{o_2}^3 t_2^2 t_4^2 t_5 t_6 -  y_{s}^3 y_{o_1}^8 y_{o_2}^{11} t_1^5 t_2^3 t_3^5 t_4^3 t_5^5 t_6 -  y_{s}^3 y_{o_1}^8 y_{o_2}^{11} t_1^4 t_2^4 t_3^5 t_4^3 t_5^5 t_6 -  y_{s}^3 y_{o_1}^8 y_{o_2}^{11} t_1^3 t_2^5 t_3^5 t_4^3 t_5^5 t_6 -  y_{s}^3 y_{o_1}^8 y_{o_2}^{11} t_1^5 t_2^3 t_3^4 t_4^4 t_5^5 t_6 -  y_{s}^3 y_{o_1}^8 y_{o_2}^{11} t_1^4 t_2^4 t_3^4 t_4^4 t_5^5 t_6 -  y_{s}^3 y_{o_1}^8 y_{o_2}^{11} t_1^3 t_2^5 t_3^4 t_4^4 t_5^5 t_6 -  y_{s}^3 y_{o_1}^8 y_{o_2}^{11} t_1^5 t_2^3 t_3^3 t_4^5 t_5^5 t_6 -  y_{s}^3 y_{o_1}^8 y_{o_2}^{11} t_1^4 t_2^4 t_3^3 t_4^5 t_5^5 t_6 -  y_{s}^3 y_{o_1}^8 y_{o_2}^{11} t_1^3 t_2^5 t_3^3 t_4^5 t_5^5 t_6 -  y_{s}^2 y_{o_1}^4 y_{o_2}^6 t_1^3 t_2 t_3^4 t_5^2 t_6^2 -  y_{s}^2 y_{o_1}^4 y_{o_2}^6 t_1^2 t_2^2 t_3^4 t_5^2 t_6^2 -  y_{s}^2 y_{o_1}^4 y_{o_2}^6 t_1 t_2^3 t_3^4 t_5^2 t_6^2 -  y_{s}^2 y_{o_1}^4 y_{o_2}^6 t_1^4 t_3^3 t_4 t_5^2 t_6^2 -  3 y_{s}^2 y_{o_1}^4 y_{o_2}^6 t_1^3 t_2 t_3^3 t_4 t_5^2 t_6^2 -  3 y_{s}^2 y_{o_1}^4 y_{o_2}^6 t_1^2 t_2^2 t_3^3 t_4 t_5^2 t_6^2 -  3 y_{s}^2 y_{o_1}^4 y_{o_2}^6 t_1 t_2^3 t_3^3 t_4 t_5^2 t_6^2 -  y_{s}^2 y_{o_1}^4 y_{o_2}^6 t_2^4 t_3^3 t_4 t_5^2 t_6^2 -  y_{s}^2 y_{o_1}^4 y_{o_2}^6 t_1^4 t_3^2 t_4^2 t_5^2 t_6^2 -  3 y_{s}^2 y_{o_1}^4 y_{o_2}^6 t_1^3 t_2 t_3^2 t_4^2 t_5^2 t_6^2 -  3 y_{s}^2 y_{o_1}^4 y_{o_2}^6 t_1^2 t_2^2 t_3^2 t_4^2 t_5^2 t_6^2 -  3 y_{s}^2 y_{o_1}^4 y_{o_2}^6 t_1 t_2^3 t_3^2 t_4^2 t_5^2 t_6^2 -  y_{s}^2 y_{o_1}^4 y_{o_2}^6 t_2^4 t_3^2 t_4^2 t_5^2 t_6^2 -  y_{s}^2 y_{o_1}^4 y_{o_2}^6 t_1^4 t_3 t_4^3 t_5^2 t_6^2 -  3 y_{s}^2 y_{o_1}^4 y_{o_2}^6 t_1^3 t_2 t_3 t_4^3 t_5^2 t_6^2 -  3 y_{s}^2 y_{o_1}^4 y_{o_2}^6 t_1^2 t_2^2 t_3 t_4^3 t_5^2 t_6^2 -  3 y_{s}^2 y_{o_1}^4 y_{o_2}^6 t_1 t_2^3 t_3 t_4^3 t_5^2 t_6^2 -  y_{s}^2 y_{o_1}^4 y_{o_2}^6 t_2^4 t_3 t_4^3 t_5^2 t_6^2 -  y_{s}^2 y_{o_1}^4 y_{o_2}^6 t_1^3 t_2 t_4^4 t_5^2 t_6^2 -  y_{s}^2 y_{o_1}^4 y_{o_2}^6 t_1^2 t_2^2 t_4^4 t_5^2 t_6^2 -  y_{s}^2 y_{o_1}^4 y_{o_2}^6 t_1 t_2^3 t_4^4 t_5^2 t_6^2 +  y_{s}^3 y_{o_1}^7 y_{o_2}^{10} t_1^4 t_2^3 t_3^6 t_4 t_5^4 t_6^2 +  y_{s}^3 y_{o_1}^7 y_{o_2}^{10} t_1^3 t_2^4 t_3^6 t_4 t_5^4 t_6^2 +  y_{s}^3 y_{o_1}^7 y_{o_2}^{10} t_1^4 t_2^3 t_3^5 t_4^2 t_5^4 t_6^2 +  y_{s}^3 y_{o_1}^7 y_{o_2}^{10} t_1^3 t_2^4 t_3^5 t_4^2 t_5^4 t_6^2 +  y_{s}^3 y_{o_1}^7 y_{o_2}^{10} t_1^6 t_2 t_3^4 t_4^3 t_5^4 t_6^2 +  y_{s}^3 y_{o_1}^7 y_{o_2}^{10} t_1^5 t_2^2 t_3^4 t_4^3 t_5^4 t_6^2 +  2 y_{s}^3 y_{o_1}^7 y_{o_2}^{10} t_1^4 t_2^3 t_3^4 t_4^3 t_5^4 t_6^2 +  2 y_{s}^3 y_{o_1}^7 y_{o_2}^{10} t_1^3 t_2^4 t_3^4 t_4^3 t_5^4 t_6^2 +  y_{s}^3 y_{o_1}^7 y_{o_2}^{10} t_1^2 t_2^5 t_3^4 t_4^3 t_5^4 t_6^2 +  y_{s}^3 y_{o_1}^7 y_{o_2}^{10} t_1 t_2^6 t_3^4 t_4^3 t_5^4 t_6^2 +  y_{s}^3 y_{o_1}^7 y_{o_2}^{10} t_1^6 t_2 t_3^3 t_4^4 t_5^4 t_6^2 +  y_{s}^3 y_{o_1}^7 y_{o_2}^{10} t_1^5 t_2^2 t_3^3 t_4^4 t_5^4 t_6^2 +  2 y_{s}^3 y_{o_1}^7 y_{o_2}^{10} t_1^4 t_2^3 t_3^3 t_4^4 t_5^4 t_6^2 +  2 y_{s}^3 y_{o_1}^7 y_{o_2}^{10} t_1^3 t_2^4 t_3^3 t_4^4 t_5^4 t_6^2 +  y_{s}^3 y_{o_1}^7 y_{o_2}^{10} t_1^2 t_2^5 t_3^3 t_4^4 t_5^4 t_6^2 +  y_{s}^3 y_{o_1}^7 y_{o_2}^{10} t_1 t_2^6 t_3^3 t_4^4 t_5^4 t_6^2 +  y_{s}^3 y_{o_1}^7 y_{o_2}^{10} t_1^4 t_2^3 t_3^2 t_4^5 t_5^4 t_6^2 +  y_{s}^3 y_{o_1}^7 y_{o_2}^{10} t_1^3 t_2^4 t_3^2 t_4^5 t_5^4 t_6^2 +  y_{s}^3 y_{o_1}^7 y_{o_2}^{10} t_1^4 t_2^3 t_3 t_4^6 t_5^4 t_6^2 +  y_{s}^3 y_{o_1}^7 y_{o_2}^{10} t_1^3 t_2^4 t_3 t_4^6 t_5^4 t_6^2 +  y_{s}^4 y_{o_1}^{10} y_{o_2}^{14} t_1^6 t_2^4 t_3^6 t_4^4 t_5^6 t_6^2 +  2 y_{s}^4 y_{o_1}^{10} y_{o_2}^{14} t_1^5 t_2^5 t_3^6 t_4^4 t_5^6 t_6^2 +  y_{s}^4 y_{o_1}^{10} y_{o_2}^{14} t_1^4 t_2^6 t_3^6 t_4^4 t_5^6 t_6^2 +  2 y_{s}^4 y_{o_1}^{10} y_{o_2}^{14} t_1^6 t_2^4 t_3^5 t_4^5 t_5^6 t_6^2 +  3 y_{s}^4 y_{o_1}^{10} y_{o_2}^{14} t_1^5 t_2^5 t_3^5 t_4^5 t_5^6 t_6^2 +  2 y_{s}^4 y_{o_1}^{10} y_{o_2}^{14} t_1^4 t_2^6 t_3^5 t_4^5 t_5^6 t_6^2 +  y_{s}^4 y_{o_1}^{10} y_{o_2}^{14} t_1^6 t_2^4 t_3^4 t_4^6 t_5^6 t_6^2 +  2 y_{s}^4 y_{o_1}^{10} y_{o_2}^{14} t_1^5 t_2^5 t_3^4 t_4^6 t_5^6 t_6^2 +  y_{s}^4 y_{o_1}^{10} y_{o_2}^{14} t_1^4 t_2^6 t_3^4 t_4^6 t_5^6 t_6^2 -  y_{s}^2 y_{o_1}^3 y_{o_2}^5 t_1^2 t_2 t_3^3 t_5 t_6^3 -  y_{s}^2 y_{o_1}^3 y_{o_2}^5 t_1 t_2^2 t_3^3 t_5 t_6^3 -  y_{s}^2 y_{o_1}^3 y_{o_2}^5 t_1^3 t_3^2 t_4 t_5 t_6^3 -  3 y_{s}^2 y_{o_1}^3 y_{o_2}^5 t_1^2 t_2 t_3^2 t_4 t_5 t_6^3 -  3 y_{s}^2 y_{o_1}^3 y_{o_2}^5 t_1 t_2^2 t_3^2 t_4 t_5 t_6^3 -  y_{s}^2 y_{o_1}^3 y_{o_2}^5 t_2^3 t_3^2 t_4 t_5 t_6^3 -  y_{s}^2 y_{o_1}^3 y_{o_2}^5 t_1^3 t_3 t_4^2 t_5 t_6^3 -  3 y_{s}^2 y_{o_1}^3 y_{o_2}^5 t_1^2 t_2 t_3 t_4^2 t_5 t_6^3 -  3 y_{s}^2 y_{o_1}^3 y_{o_2}^5 t_1 t_2^2 t_3 t_4^2 t_5 t_6^3 -  y_{s}^2 y_{o_1}^3 y_{o_2}^5 t_2^3 t_3 t_4^2 t_5 t_6^3 -  y_{s}^2 y_{o_1}^3 y_{o_2}^5 t_1^2 t_2 t_4^3 t_5 t_6^3 -  y_{s}^2 y_{o_1}^3 y_{o_2}^5 t_1 t_2^2 t_4^3 t_5 t_6^3 -  y_{s}^3 y_{o_1}^6 y_{o_2}^9 t_1^3 t_2^3 t_3^6 t_5^3 t_6^3 -  y_{s}^3 y_{o_1}^6 y_{o_2}^9 t_1^3 t_2^3 t_3^5 t_4 t_5^3 t_6^3 -  y_{s}^3 y_{o_1}^6 y_{o_2}^9 t_1^3 t_2^3 t_3^4 t_4^2 t_5^3 t_6^3 -  y_{s}^3 y_{o_1}^6 y_{o_2}^9 t_1^6 t_3^3 t_4^3 t_5^3 t_6^3 -  y_{s}^3 y_{o_1}^6 y_{o_2}^9 t_1^5 t_2 t_3^3 t_4^3 t_5^3 t_6^3 -  y_{s}^3 y_{o_1}^6 y_{o_2}^9 t_1^4 t_2^2 t_3^3 t_4^3 t_5^3 t_6^3 -  3 y_{s}^3 y_{o_1}^6 y_{o_2}^9 t_1^3 t_2^3 t_3^3 t_4^3 t_5^3 t_6^3 -  y_{s}^3 y_{o_1}^6 y_{o_2}^9 t_1^2 t_2^4 t_3^3 t_4^3 t_5^3 t_6^3 -  y_{s}^3 y_{o_1}^6 y_{o_2}^9 t_1 t_2^5 t_3^3 t_4^3 t_5^3 t_6^3 -  y_{s}^3 y_{o_1}^6 y_{o_2}^9 t_2^6 t_3^3 t_4^3 t_5^3 t_6^3 -  y_{s}^3 y_{o_1}^6 y_{o_2}^9 t_1^3 t_2^3 t_3^2 t_4^4 t_5^3 t_6^3 -  y_{s}^3 y_{o_1}^6 y_{o_2}^9 t_1^3 t_2^3 t_3 t_4^5 t_5^3 t_6^3 -  y_{s}^3 y_{o_1}^6 y_{o_2}^9 t_1^3 t_2^3 t_4^6 t_5^3 t_6^3 +  2 y_{s}^4 y_{o_1}^9 y_{o_2}^{13} t_1^6 t_2^3 t_3^6 t_4^3 t_5^5 t_6^3 +  2 y_{s}^4 y_{o_1}^9 y_{o_2}^{13} t_1^5 t_2^4 t_3^6 t_4^3 t_5^5 t_6^3 +  2 y_{s}^4 y_{o_1}^9 y_{o_2}^{13} t_1^4 t_2^5 t_3^6 t_4^3 t_5^5 t_6^3 +  2 y_{s}^4 y_{o_1}^9 y_{o_2}^{13} t_1^3 t_2^6 t_3^6 t_4^3 t_5^5 t_6^3 +  2 y_{s}^4 y_{o_1}^9 y_{o_2}^{13} t_1^6 t_2^3 t_3^5 t_4^4 t_5^5 t_6^3 +  3 y_{s}^4 y_{o_1}^9 y_{o_2}^{13} t_1^5 t_2^4 t_3^5 t_4^4 t_5^5 t_6^3 +  3 y_{s}^4 y_{o_1}^9 y_{o_2}^{13} t_1^4 t_2^5 t_3^5 t_4^4 t_5^5 t_6^3 +  2 y_{s}^4 y_{o_1}^9 y_{o_2}^{13} t_1^3 t_2^6 t_3^5 t_4^4 t_5^5 t_6^3 +  2 y_{s}^4 y_{o_1}^9 y_{o_2}^{13} t_1^6 t_2^3 t_3^4 t_4^5 t_5^5 t_6^3 +  3 y_{s}^4 y_{o_1}^9 y_{o_2}^{13} t_1^5 t_2^4 t_3^4 t_4^5 t_5^5 t_6^3 +  3 y_{s}^4 y_{o_1}^9 y_{o_2}^{13} t_1^4 t_2^5 t_3^4 t_4^5 t_5^5 t_6^3 +  2 y_{s}^4 y_{o_1}^9 y_{o_2}^{13} t_1^3 t_2^6 t_3^4 t_4^5 t_5^5 t_6^3 +  2 y_{s}^4 y_{o_1}^9 y_{o_2}^{13} t_1^6 t_2^3 t_3^3 t_4^6 t_5^5 t_6^3 +  2 y_{s}^4 y_{o_1}^9 y_{o_2}^{13} t_1^5 t_2^4 t_3^3 t_4^6 t_5^5 t_6^3 +  2 y_{s}^4 y_{o_1}^9 y_{o_2}^{13} t_1^4 t_2^5 t_3^3 t_4^6 t_5^5 t_6^3 +  2 y_{s}^4 y_{o_1}^9 y_{o_2}^{13} t_1^3 t_2^6 t_3^3 t_4^6 t_5^5 t_6^3 -  y_{s}^5 y_{o_1}^{12} y_{o_2}^{17} t_1^6 t_2^6 t_3^6 t_4^6 t_5^7 t_6^3 -  y_{s}^2 y_{o_1}^2 y_{o_2}^4 t_1 t_2 t_3 t_4 t_6^4 +  2 y_{s}^3 y_{o_1}^5 y_{o_2}^8 t_1^4 t_2 t_3^4 t_4 t_5^2 t_6^4 +  2 y_{s}^3 y_{o_1}^5 y_{o_2}^8 t_1^3 t_2^2 t_3^4 t_4 t_5^2 t_6^4 +  2 y_{s}^3 y_{o_1}^5 y_{o_2}^8 t_1^2 t_2^3 t_3^4 t_4 t_5^2 t_6^4 +  2 y_{s}^3 y_{o_1}^5 y_{o_2}^8 t_1 t_2^4 t_3^4 t_4 t_5^2 t_6^4 +  2 y_{s}^3 y_{o_1}^5 y_{o_2}^8 t_1^4 t_2 t_3^3 t_4^2 t_5^2 t_6^4 +  3 y_{s}^3 y_{o_1}^5 y_{o_2}^8 t_1^3 t_2^2 t_3^3 t_4^2 t_5^2 t_6^4 +  3 y_{s}^3 y_{o_1}^5 y_{o_2}^8 t_1^2 t_2^3 t_3^3 t_4^2 t_5^2 t_6^4 +  2 y_{s}^3 y_{o_1}^5 y_{o_2}^8 t_1 t_2^4 t_3^3 t_4^2 t_5^2 t_6^4 +  2 y_{s}^3 y_{o_1}^5 y_{o_2}^8 t_1^4 t_2 t_3^2 t_4^3 t_5^2 t_6^4 +  3 y_{s}^3 y_{o_1}^5 y_{o_2}^8 t_1^3 t_2^2 t_3^2 t_4^3 t_5^2 t_6^4 +  3 y_{s}^3 y_{o_1}^5 y_{o_2}^8 t_1^2 t_2^3 t_3^2 t_4^3 t_5^2 t_6^4 +  2 y_{s}^3 y_{o_1}^5 y_{o_2}^8 t_1 t_2^4 t_3^2 t_4^3 t_5^2 t_6^4 +  2 y_{s}^3 y_{o_1}^5 y_{o_2}^8 t_1^4 t_2 t_3 t_4^4 t_5^2 t_6^4 +  2 y_{s}^3 y_{o_1}^5 y_{o_2}^8 t_1^3 t_2^2 t_3 t_4^4 t_5^2 t_6^4 +  2 y_{s}^3 y_{o_1}^5 y_{o_2}^8 t_1^2 t_2^3 t_3 t_4^4 t_5^2 t_6^4 +  2 y_{s}^3 y_{o_1}^5 y_{o_2}^8 t_1 t_2^4 t_3 t_4^4 t_5^2 t_6^4 -  y_{s}^4 y_{o_1}^8 y_{o_2}^{12} t_1^4 t_2^4 t_3^7 t_4 t_5^4 t_6^4 -  y_{s}^4 y_{o_1}^8 y_{o_2}^{12} t_1^4 t_2^4 t_3^6 t_4^2 t_5^4 t_6^4 -  y_{s}^4 y_{o_1}^8 y_{o_2}^{12} t_1^4 t_2^4 t_3^5 t_4^3 t_5^4 t_6^4 -  y_{s}^4 y_{o_1}^8 y_{o_2}^{12} t_1^7 t_2 t_3^4 t_4^4 t_5^4 t_6^4 -  y_{s}^4 y_{o_1}^8 y_{o_2}^{12} t_1^6 t_2^2 t_3^4 t_4^4 t_5^4 t_6^4 -  y_{s}^4 y_{o_1}^8 y_{o_2}^{12} t_1^5 t_2^3 t_3^4 t_4^4 t_5^4 t_6^4 -  3 y_{s}^4 y_{o_1}^8 y_{o_2}^{12} t_1^4 t_2^4 t_3^4 t_4^4 t_5^4 t_6^4 -  y_{s}^4 y_{o_1}^8 y_{o_2}^{12} t_1^3 t_2^5 t_3^4 t_4^4 t_5^4 t_6^4 -  y_{s}^4 y_{o_1}^8 y_{o_2}^{12} t_1^2 t_2^6 t_3^4 t_4^4 t_5^4 t_6^4 -  y_{s}^4 y_{o_1}^8 y_{o_2}^{12} t_1 t_2^7 t_3^4 t_4^4 t_5^4 t_6^4 -  y_{s}^4 y_{o_1}^8 y_{o_2}^{12} t_1^4 t_2^4 t_3^3 t_4^5 t_5^4 t_6^4 -  y_{s}^4 y_{o_1}^8 y_{o_2}^{12} t_1^4 t_2^4 t_3^2 t_4^6 t_5^4 t_6^4 -  y_{s}^4 y_{o_1}^8 y_{o_2}^{12} t_1^4 t_2^4 t_3 t_4^7 t_5^4 t_6^4 -  y_{s}^5 y_{o_1}^{11} y_{o_2}^{16} t_1^6 t_2^5 t_3^7 t_4^4 t_5^6 t_6^4 -  y_{s}^5 y_{o_1}^{11} y_{o_2}^{16} t_1^5 t_2^6 t_3^7 t_4^4 t_5^6 t_6^4 -  y_{s}^5 y_{o_1}^{11} y_{o_2}^{16} t_1^7 t_2^4 t_3^6 t_4^5 t_5^6 t_6^4 -  3 y_{s}^5 y_{o_1}^{11} y_{o_2}^{16} t_1^6 t_2^5 t_3^6 t_4^5 t_5^6 t_6^4 -  3 y_{s}^5 y_{o_1}^{11} y_{o_2}^{16} t_1^5 t_2^6 t_3^6 t_4^5 t_5^6 t_6^4 -  y_{s}^5 y_{o_1}^{11} y_{o_2}^{16} t_1^4 t_2^7 t_3^6 t_4^5 t_5^6 t_6^4 -  y_{s}^5 y_{o_1}^{11} y_{o_2}^{16} t_1^7 t_2^4 t_3^5 t_4^6 t_5^6 t_6^4 -  3 y_{s}^5 y_{o_1}^{11} y_{o_2}^{16} t_1^6 t_2^5 t_3^5 t_4^6 t_5^6 t_6^4 -  3 y_{s}^5 y_{o_1}^{11} y_{o_2}^{16} t_1^5 t_2^6 t_3^5 t_4^6 t_5^6 t_6^4 -  y_{s}^5 y_{o_1}^{11} y_{o_2}^{16} t_1^4 t_2^7 t_3^5 t_4^6 t_5^6 t_6^4 -  y_{s}^5 y_{o_1}^{11} y_{o_2}^{16} t_1^6 t_2^5 t_3^4 t_4^7 t_5^6 t_6^4 -  y_{s}^5 y_{o_1}^{11} y_{o_2}^{16} t_1^5 t_2^6 t_3^4 t_4^7 t_5^6 t_6^4 +  y_{s}^3 y_{o_1}^4 y_{o_2}^7 t_1^3 t_2 t_3^3 t_4 t_5 t_6^5 +  2 y_{s}^3 y_{o_1}^4 y_{o_2}^7 t_1^2 t_2^2 t_3^3 t_4 t_5 t_6^5 +  y_{s}^3 y_{o_1}^4 y_{o_2}^7 t_1 t_2^3 t_3^3 t_4 t_5 t_6^5 +  2 y_{s}^3 y_{o_1}^4 y_{o_2}^7 t_1^3 t_2 t_3^2 t_4^2 t_5 t_6^5 +  3 y_{s}^3 y_{o_1}^4 y_{o_2}^7 t_1^2 t_2^2 t_3^2 t_4^2 t_5 t_6^5 +  2 y_{s}^3 y_{o_1}^4 y_{o_2}^7 t_1 t_2^3 t_3^2 t_4^2 t_5 t_6^5 +  y_{s}^3 y_{o_1}^4 y_{o_2}^7 t_1^3 t_2 t_3 t_4^3 t_5 t_6^5 +  2 y_{s}^3 y_{o_1}^4 y_{o_2}^7 t_1^2 t_2^2 t_3 t_4^3 t_5 t_6^5 +  y_{s}^3 y_{o_1}^4 y_{o_2}^7 t_1 t_2^3 t_3 t_4^3 t_5 t_6^5 +  y_{s}^4 y_{o_1}^7 y_{o_2}^{11} t_1^4 t_2^3 t_3^6 t_4 t_5^3 t_6^5 +  y_{s}^4 y_{o_1}^7 y_{o_2}^{11} t_1^3 t_2^4 t_3^6 t_4 t_5^3 t_6^5 +  y_{s}^4 y_{o_1}^7 y_{o_2}^{11} t_1^4 t_2^3 t_3^5 t_4^2 t_5^3 t_6^5 +  y_{s}^4 y_{o_1}^7 y_{o_2}^{11} t_1^3 t_2^4 t_3^5 t_4^2 t_5^3 t_6^5 +  y_{s}^4 y_{o_1}^7 y_{o_2}^{11} t_1^6 t_2 t_3^4 t_4^3 t_5^3 t_6^5 +  y_{s}^4 y_{o_1}^7 y_{o_2}^{11} t_1^5 t_2^2 t_3^4 t_4^3 t_5^3 t_6^5 +  2 y_{s}^4 y_{o_1}^7 y_{o_2}^{11} t_1^4 t_2^3 t_3^4 t_4^3 t_5^3 t_6^5 +  2 y_{s}^4 y_{o_1}^7 y_{o_2}^{11} t_1^3 t_2^4 t_3^4 t_4^3 t_5^3 t_6^5 +  y_{s}^4 y_{o_1}^7 y_{o_2}^{11} t_1^2 t_2^5 t_3^4 t_4^3 t_5^3 t_6^5 +  y_{s}^4 y_{o_1}^7 y_{o_2}^{11} t_1 t_2^6 t_3^4 t_4^3 t_5^3 t_6^5 +  y_{s}^4 y_{o_1}^7 y_{o_2}^{11} t_1^6 t_2 t_3^3 t_4^4 t_5^3 t_6^5 +  y_{s}^4 y_{o_1}^7 y_{o_2}^{11} t_1^5 t_2^2 t_3^3 t_4^4 t_5^3 t_6^5 +  2 y_{s}^4 y_{o_1}^7 y_{o_2}^{11} t_1^4 t_2^3 t_3^3 t_4^4 t_5^3 t_6^5 +  2 y_{s}^4 y_{o_1}^7 y_{o_2}^{11} t_1^3 t_2^4 t_3^3 t_4^4 t_5^3 t_6^5 +  y_{s}^4 y_{o_1}^7 y_{o_2}^{11} t_1^2 t_2^5 t_3^3 t_4^4 t_5^3 t_6^5 +  y_{s}^4 y_{o_1}^7 y_{o_2}^{11} t_1 t_2^6 t_3^3 t_4^4 t_5^3 t_6^5 +  y_{s}^4 y_{o_1}^7 y_{o_2}^{11} t_1^4 t_2^3 t_3^2 t_4^5 t_5^3 t_6^5 +  y_{s}^4 y_{o_1}^7 y_{o_2}^{11} t_1^3 t_2^4 t_3^2 t_4^5 t_5^3 t_6^5 +  y_{s}^4 y_{o_1}^7 y_{o_2}^{11} t_1^4 t_2^3 t_3 t_4^6 t_5^3 t_6^5 +  y_{s}^4 y_{o_1}^7 y_{o_2}^{11} t_1^3 t_2^4 t_3 t_4^6 t_5^3 t_6^5 -  y_{s}^5 y_{o_1}^{10} y_{o_2}^{15} t_1^6 t_2^4 t_3^7 t_4^3 t_5^5 t_6^5 -  y_{s}^5 y_{o_1}^{10} y_{o_2}^{15} t_1^5 t_2^5 t_3^7 t_4^3 t_5^5 t_6^5 -  y_{s}^5 y_{o_1}^{10} y_{o_2}^{15} t_1^4 t_2^6 t_3^7 t_4^3 t_5^5 t_6^5 -  y_{s}^5 y_{o_1}^{10} y_{o_2}^{15} t_1^7 t_2^3 t_3^6 t_4^4 t_5^5 t_6^5 -  3 y_{s}^5 y_{o_1}^{10} y_{o_2}^{15} t_1^6 t_2^4 t_3^6 t_4^4 t_5^5 t_6^5 -  3 y_{s}^5 y_{o_1}^{10} y_{o_2}^{15} t_1^5 t_2^5 t_3^6 t_4^4 t_5^5 t_6^5 -  3 y_{s}^5 y_{o_1}^{10} y_{o_2}^{15} t_1^4 t_2^6 t_3^6 t_4^4 t_5^5 t_6^5 -  y_{s}^5 y_{o_1}^{10} y_{o_2}^{15} t_1^3 t_2^7 t_3^6 t_4^4 t_5^5 t_6^5 -  y_{s}^5 y_{o_1}^{10} y_{o_2}^{15} t_1^7 t_2^3 t_3^5 t_4^5 t_5^5 t_6^5 -  3 y_{s}^5 y_{o_1}^{10} y_{o_2}^{15} t_1^6 t_2^4 t_3^5 t_4^5 t_5^5 t_6^5 -  3 y_{s}^5 y_{o_1}^{10} y_{o_2}^{15} t_1^5 t_2^5 t_3^5 t_4^5 t_5^5 t_6^5 -  3 y_{s}^5 y_{o_1}^{10} y_{o_2}^{15} t_1^4 t_2^6 t_3^5 t_4^5 t_5^5 t_6^5 -  y_{s}^5 y_{o_1}^{10} y_{o_2}^{15} t_1^3 t_2^7 t_3^5 t_4^5 t_5^5 t_6^5 -  y_{s}^5 y_{o_1}^{10} y_{o_2}^{15} t_1^7 t_2^3 t_3^4 t_4^6 t_5^5 t_6^5 -  3 y_{s}^5 y_{o_1}^{10} y_{o_2}^{15} t_1^6 t_2^4 t_3^4 t_4^6 t_5^5 t_6^5 -  3 y_{s}^5 y_{o_1}^{10} y_{o_2}^{15} t_1^5 t_2^5 t_3^4 t_4^6 t_5^5 t_6^5 -  3 y_{s}^5 y_{o_1}^{10} y_{o_2}^{15} t_1^4 t_2^6 t_3^4 t_4^6 t_5^5 t_6^5 -  y_{s}^5 y_{o_1}^{10} y_{o_2}^{15} t_1^3 t_2^7 t_3^4 t_4^6 t_5^5 t_6^5 -  y_{s}^5 y_{o_1}^{10} y_{o_2}^{15} t_1^6 t_2^4 t_3^3 t_4^7 t_5^5 t_6^5 -  y_{s}^5 y_{o_1}^{10} y_{o_2}^{15} t_1^5 t_2^5 t_3^3 t_4^7 t_5^5 t_6^5 -  y_{s}^5 y_{o_1}^{10} y_{o_2}^{15} t_1^4 t_2^6 t_3^3 t_4^7 t_5^5 t_6^5 -  y_{s}^4 y_{o_1}^6 y_{o_2}^{10} t_1^4 t_2^2 t_3^4 t_4^2 t_5^2 t_6^6 -  y_{s}^4 y_{o_1}^6 y_{o_2}^{10} t_1^3 t_2^3 t_3^4 t_4^2 t_5^2 t_6^6 -  y_{s}^4 y_{o_1}^6 y_{o_2}^{10} t_1^2 t_2^4 t_3^4 t_4^2 t_5^2 t_6^6 -  y_{s}^4 y_{o_1}^6 y_{o_2}^{10} t_1^4 t_2^2 t_3^3 t_4^3 t_5^2 t_6^6 -  y_{s}^4 y_{o_1}^6 y_{o_2}^{10} t_1^3 t_2^3 t_3^3 t_4^3 t_5^2 t_6^6 -  y_{s}^4 y_{o_1}^6 y_{o_2}^{10} t_1^2 t_2^4 t_3^3 t_4^3 t_5^2 t_6^6 -  y_{s}^4 y_{o_1}^6 y_{o_2}^{10} t_1^4 t_2^2 t_3^2 t_4^4 t_5^2 t_6^6 -  y_{s}^4 y_{o_1}^6 y_{o_2}^{10} t_1^3 t_2^3 t_3^2 t_4^4 t_5^2 t_6^6 -  y_{s}^4 y_{o_1}^6 y_{o_2}^{10} t_1^2 t_2^4 t_3^2 t_4^4 t_5^2 t_6^6 +  y_{s}^6 y_{o_1}^{12} y_{o_2}^{18} t_1^7 t_2^5 t_3^7 t_4^5 t_5^6 t_6^6 +  y_{s}^6 y_{o_1}^{12} y_{o_2}^{18} t_1^6 t_2^6 t_3^7 t_4^5 t_5^6 t_6^6 +  y_{s}^6 y_{o_1}^{12} y_{o_2}^{18} t_1^5 t_2^7 t_3^7 t_4^5 t_5^6 t_6^6 +  y_{s}^6 y_{o_1}^{12} y_{o_2}^{18} t_1^7 t_2^5 t_3^6 t_4^6 t_5^6 t_6^6 +  y_{s}^6 y_{o_1}^{12} y_{o_2}^{18} t_1^6 t_2^6 t_3^6 t_4^6 t_5^6 t_6^6 +  y_{s}^6 y_{o_1}^{12} y_{o_2}^{18} t_1^5 t_2^7 t_3^6 t_4^6 t_5^6 t_6^6 +  y_{s}^6 y_{o_1}^{12} y_{o_2}^{18} t_1^7 t_2^5 t_3^5 t_4^7 t_5^6 t_6^6 +  y_{s}^6 y_{o_1}^{12} y_{o_2}^{18} t_1^6 t_2^6 t_3^5 t_4^7 t_5^6 t_6^6 +  y_{s}^6 y_{o_1}^{12} y_{o_2}^{18} t_1^5 t_2^7 t_3^5 t_4^7 t_5^6 t_6^6 -  y_{s}^4 y_{o_1}^5 y_{o_2}^9 t_1^3 t_2^2 t_3^3 t_4^2 t_5 t_6^7 -  y_{s}^4 y_{o_1}^5 y_{o_2}^9 t_1^2 t_2^3 t_3^3 t_4^2 t_5 t_6^7 -  y_{s}^4 y_{o_1}^5 y_{o_2}^9 t_1^3 t_2^2 t_3^2 t_4^3 t_5 t_6^7 -  y_{s}^4 y_{o_1}^5 y_{o_2}^9 t_1^2 t_2^3 t_3^2 t_4^3 t_5 t_6^7 -  y_{s}^5 y_{o_1}^8 y_{o_2}^{13} t_1^4 t_2^4 t_3^6 t_4^2 t_5^3 t_6^7 -  y_{s}^5 y_{o_1}^8 y_{o_2}^{13} t_1^4 t_2^4 t_3^5 t_4^3 t_5^3 t_6^7 -  y_{s}^5 y_{o_1}^8 y_{o_2}^{13} t_1^6 t_2^2 t_3^4 t_4^4 t_5^3 t_6^7 -  y_{s}^5 y_{o_1}^8 y_{o_2}^{13} t_1^5 t_2^3 t_3^4 t_4^4 t_5^3 t_6^7 -  y_{s}^5 y_{o_1}^8 y_{o_2}^{13} t_1^4 t_2^4 t_3^4 t_4^4 t_5^3 t_6^7 -  y_{s}^5 y_{o_1}^8 y_{o_2}^{13} t_1^3 t_2^5 t_3^4 t_4^4 t_5^3 t_6^7 -  y_{s}^5 y_{o_1}^8 y_{o_2}^{13} t_1^2 t_2^6 t_3^4 t_4^4 t_5^3 t_6^7 -  y_{s}^5 y_{o_1}^8 y_{o_2}^{13} t_1^4 t_2^4 t_3^3 t_4^5 t_5^3 t_6^7 -  y_{s}^5 y_{o_1}^8 y_{o_2}^{13} t_1^4 t_2^4 t_3^2 t_4^6 t_5^3 t_6^7 +  y_{s}^6 y_{o_1}^{11} y_{o_2}^{17} t_1^6 t_2^5 t_3^7 t_4^4 t_5^5 t_6^7 +  y_{s}^6 y_{o_1}^{11} y_{o_2}^{17} t_1^5 t_2^6 t_3^7 t_4^4 t_5^5 t_6^7 +  y_{s}^6 y_{o_1}^{11} y_{o_2}^{17} t_1^7 t_2^4 t_3^6 t_4^5 t_5^5 t_6^7 +  y_{s}^6 y_{o_1}^{11} y_{o_2}^{17} t_1^6 t_2^5 t_3^6 t_4^5 t_5^5 t_6^7 +  y_{s}^6 y_{o_1}^{11} y_{o_2}^{17} t_1^5 t_2^6 t_3^6 t_4^5 t_5^5 t_6^7 +  y_{s}^6 y_{o_1}^{11} y_{o_2}^{17} t_1^4 t_2^7 t_3^6 t_4^5 t_5^5 t_6^7 +  y_{s}^6 y_{o_1}^{11} y_{o_2}^{17} t_1^7 t_2^4 t_3^5 t_4^6 t_5^5 t_6^7 +  y_{s}^6 y_{o_1}^{11} y_{o_2}^{17} t_1^6 t_2^5 t_3^5 t_4^6 t_5^5 t_6^7 +  y_{s}^6 y_{o_1}^{11} y_{o_2}^{17} t_1^5 t_2^6 t_3^5 t_4^6 t_5^5 t_6^7 +  y_{s}^6 y_{o_1}^{11} y_{o_2}^{17} t_1^4 t_2^7 t_3^5 t_4^6 t_5^5 t_6^7 +  y_{s}^6 y_{o_1}^{11} y_{o_2}^{17} t_1^6 t_2^5 t_3^4 t_4^7 t_5^5 t_6^7 +  y_{s}^6 y_{o_1}^{11} y_{o_2}^{17} t_1^5 t_2^6 t_3^4 t_4^7 t_5^5 t_6^7 +  y_{s}^7 y_{o_1}^{14} y_{o_2}^{21} t_1^7 t_2^7 t_3^7 t_4^7 t_5^7 t_6^7
~,~
$
\end{quote}
\endgroup

\subsection{Model 10 \label{app_num_10}}

\begingroup\makeatletter\def\f@size{7}\check@mathfonts
\begin{quote}\raggedright
$
P(t_i,y_s,y_{o_1},y_{o_2}; \mathcal{M}_{10}) =
1 + y_{s} y_{o_1}^2 y_{o_2} t_1 t_2 t_3 t_4^2 + y_{s} y_{o_1}^3 y_{o_2}^2 t_1^3 t_3^2 t_4 t_5 +  y_{s} y_{o_1}^3 y_{o_2}^2 t_1^2 t_2 t_3^2 t_4 t_5 + y_{s} y_{o_1}^3 y_{o_2}^2 t_1 t_2^2 t_3^2 t_4 t_5 +  y_{s} y_{o_1}^3 y_{o_2}^2 t_2^3 t_3^2 t_4 t_5 -  y_{s}^2 y_{o_1}^5 y_{o_2}^3 t_1^3 t_2^2 t_3^3 t_4^3 t_5 -  y_{s}^2 y_{o_1}^5 y_{o_2}^3 t_1^2 t_2^3 t_3^3 t_4^3 t_5 +  y_{s} y_{o_1}^4 y_{o_2}^3 t_1^3 t_2 t_3^3 t_5^2 + y_{s} y_{o_1}^4 y_{o_2}^3 t_1^2 t_2^2 t_3^3 t_5^2 +  y_{s} y_{o_1}^4 y_{o_2}^3 t_1 t_2^3 t_3^3 t_5^2 -  y_{s}^2 y_{o_1}^6 y_{o_2}^4 t_1^5 t_2 t_3^4 t_4^2 t_5^2 -  y_{s}^2 y_{o_1}^6 y_{o_2}^4 t_1^4 t_2^2 t_3^4 t_4^2 t_5^2 -  y_{s}^2 y_{o_1}^6 y_{o_2}^4 t_1^3 t_2^3 t_3^4 t_4^2 t_5^2 -  y_{s}^2 y_{o_1}^6 y_{o_2}^4 t_1^2 t_2^4 t_3^4 t_4^2 t_5^2 -  y_{s}^2 y_{o_1}^6 y_{o_2}^4 t_1 t_2^5 t_3^4 t_4^2 t_5^2 -  y_{s}^3 y_{o_1}^9 y_{o_2}^6 t_1^5 t_2^4 t_3^6 t_4^3 t_5^3 -  y_{s}^3 y_{o_1}^9 y_{o_2}^6 t_1^4 t_2^5 t_3^6 t_4^3 t_5^3 -  y_{s}^2 y_{o_1}^3 y_{o_2}^2 t_1^2 t_2 t_3 t_4^4 t_6 -  y_{s}^2 y_{o_1}^3 y_{o_2}^2 t_1 t_2^2 t_3 t_4^4 t_6 + y_{s} y_{o_1}^2 y_{o_2}^2 t_1^2 t_3 t_4 t_5 t_6 +  y_{s} y_{o_1}^2 y_{o_2}^2 t_1 t_2 t_3 t_4 t_5 t_6 + y_{s} y_{o_1}^2 y_{o_2}^2 t_2^2 t_3 t_4 t_5 t_6 -  y_{s}^2 y_{o_1}^4 y_{o_2}^3 t_1^4 t_3^2 t_4^3 t_5 t_6 -  2 y_{s}^2 y_{o_1}^4 y_{o_2}^3 t_1^3 t_2 t_3^2 t_4^3 t_5 t_6 -  3 y_{s}^2 y_{o_1}^4 y_{o_2}^3 t_1^2 t_2^2 t_3^2 t_4^3 t_5 t_6 -  2 y_{s}^2 y_{o_1}^4 y_{o_2}^3 t_1 t_2^3 t_3^2 t_4^3 t_5 t_6 -  y_{s}^2 y_{o_1}^4 y_{o_2}^3 t_2^4 t_3^2 t_4^3 t_5 t_6 +  y_{s}^3 y_{o_1}^6 y_{o_2}^4 t_1^4 t_2^2 t_3^3 t_4^5 t_5 t_6 +  2 y_{s}^3 y_{o_1}^6 y_{o_2}^4 t_1^3 t_2^3 t_3^3 t_4^5 t_5 t_6 +  y_{s}^3 y_{o_1}^6 y_{o_2}^4 t_1^2 t_2^4 t_3^3 t_4^5 t_5 t_6 +  y_{s} y_{o_1}^3 y_{o_2}^3 t_1^3 t_3^2 t_5^2 t_6 +  y_{s} y_{o_1}^3 y_{o_2}^3 t_1^2 t_2 t_3^2 t_5^2 t_6 +  y_{s} y_{o_1}^3 y_{o_2}^3 t_1 t_2^2 t_3^2 t_5^2 t_6 +  y_{s} y_{o_1}^3 y_{o_2}^3 t_2^3 t_3^2 t_5^2 t_6 -  y_{s}^2 y_{o_1}^5 y_{o_2}^4 t_1^5 t_3^3 t_4^2 t_5^2 t_6 -  2 y_{s}^2 y_{o_1}^5 y_{o_2}^4 t_1^4 t_2 t_3^3 t_4^2 t_5^2 t_6 -  3 y_{s}^2 y_{o_1}^5 y_{o_2}^4 t_1^3 t_2^2 t_3^3 t_4^2 t_5^2 t_6 -  3 y_{s}^2 y_{o_1}^5 y_{o_2}^4 t_1^2 t_2^3 t_3^3 t_4^2 t_5^2 t_6 -  2 y_{s}^2 y_{o_1}^5 y_{o_2}^4 t_1 t_2^4 t_3^3 t_4^2 t_5^2 t_6 -  y_{s}^2 y_{o_1}^5 y_{o_2}^4 t_2^5 t_3^3 t_4^2 t_5^2 t_6 +  y_{s}^3 y_{o_1}^7 y_{o_2}^5 t_1^6 t_2 t_3^4 t_4^4 t_5^2 t_6 +  2 y_{s}^3 y_{o_1}^7 y_{o_2}^5 t_1^5 t_2^2 t_3^4 t_4^4 t_5^2 t_6 +  2 y_{s}^3 y_{o_1}^7 y_{o_2}^5 t_1^4 t_2^3 t_3^4 t_4^4 t_5^2 t_6 +  2 y_{s}^3 y_{o_1}^7 y_{o_2}^5 t_1^3 t_2^4 t_3^4 t_4^4 t_5^2 t_6 +  2 y_{s}^3 y_{o_1}^7 y_{o_2}^5 t_1^2 t_2^5 t_3^4 t_4^4 t_5^2 t_6 +  y_{s}^3 y_{o_1}^7 y_{o_2}^5 t_1 t_2^6 t_3^4 t_4^4 t_5^2 t_6 +  y_{s}^2 y_{o_1}^6 y_{o_2}^5 t_1^3 t_2^3 t_3^4 t_4 t_5^3 t_6 -  y_{s}^3 y_{o_1}^8 y_{o_2}^6 t_1^5 t_2^3 t_3^5 t_4^3 t_5^3 t_6 -  y_{s}^3 y_{o_1}^8 y_{o_2}^6 t_1^4 t_2^4 t_3^5 t_4^3 t_5^3 t_6 -  y_{s}^3 y_{o_1}^8 y_{o_2}^6 t_1^3 t_2^5 t_3^5 t_4^3 t_5^3 t_6 +  y_{s}^4 y_{o_1}^{10} y_{o_2}^7 t_1^6 t_2^4 t_3^6 t_4^5 t_5^3 t_6 +  2 y_{s}^4 y_{o_1}^{10} y_{o_2}^7 t_1^5 t_2^5 t_3^6 t_4^5 t_5^3 t_6 +  y_{s}^4 y_{o_1}^{10} y_{o_2}^7 t_1^4 t_2^6 t_3^6 t_4^5 t_5^3 t_6 +  y_{s} y_{o_1} y_{o_2}^2 t_1 t_4 t_5 t_6^2 + y_{s} y_{o_1} y_{o_2}^2 t_2 t_4 t_5 t_6^2 -  y_{s}^2 y_{o_1}^3 y_{o_2}^3 t_1^3 t_3 t_4^3 t_5 t_6^2 -  2 y_{s}^2 y_{o_1}^3 y_{o_2}^3 t_1^2 t_2 t_3 t_4^3 t_5 t_6^2 -  2 y_{s}^2 y_{o_1}^3 y_{o_2}^3 t_1 t_2^2 t_3 t_4^3 t_5 t_6^2 -  y_{s}^2 y_{o_1}^3 y_{o_2}^3 t_2^3 t_3 t_4^3 t_5 t_6^2 +  y_{s}^3 y_{o_1}^5 y_{o_2}^4 t_1^4 t_2 t_3^2 t_4^5 t_5 t_6^2 +  2 y_{s}^3 y_{o_1}^5 y_{o_2}^4 t_1^3 t_2^2 t_3^2 t_4^5 t_5 t_6^2 +  2 y_{s}^3 y_{o_1}^5 y_{o_2}^4 t_1^2 t_2^3 t_3^2 t_4^5 t_5 t_6^2 +  y_{s}^3 y_{o_1}^5 y_{o_2}^4 t_1 t_2^4 t_3^2 t_4^5 t_5 t_6^2 -  y_{s}^4 y_{o_1}^7 y_{o_2}^5 t_1^4 t_2^3 t_3^3 t_4^7 t_5 t_6^2 -  y_{s}^4 y_{o_1}^7 y_{o_2}^5 t_1^3 t_2^4 t_3^3 t_4^7 t_5 t_6^2 +  y_{s} y_{o_1}^2 y_{o_2}^3 t_1^2 t_3 t_5^2 t_6^2 + y_{s} y_{o_1}^2 y_{o_2}^3 t_1 t_2 t_3 t_5^2 t_6^2 +  y_{s} y_{o_1}^2 y_{o_2}^3 t_2^2 t_3 t_5^2 t_6^2 -  y_{s}^2 y_{o_1}^4 y_{o_2}^4 t_1^4 t_3^2 t_4^2 t_5^2 t_6^2 -  2 y_{s}^2 y_{o_1}^4 y_{o_2}^4 t_1^3 t_2 t_3^2 t_4^2 t_5^2 t_6^2 -  3 y_{s}^2 y_{o_1}^4 y_{o_2}^4 t_1^2 t_2^2 t_3^2 t_4^2 t_5^2 t_6^2 -  2 y_{s}^2 y_{o_1}^4 y_{o_2}^4 t_1 t_2^3 t_3^2 t_4^2 t_5^2 t_6^2 -  y_{s}^2 y_{o_1}^4 y_{o_2}^4 t_2^4 t_3^2 t_4^2 t_5^2 t_6^2 +  y_{s}^3 y_{o_1}^6 y_{o_2}^5 t_1^5 t_2 t_3^3 t_4^4 t_5^2 t_6^2 +  2 y_{s}^3 y_{o_1}^6 y_{o_2}^5 t_1^4 t_2^2 t_3^3 t_4^4 t_5^2 t_6^2 +  3 y_{s}^3 y_{o_1}^6 y_{o_2}^5 t_1^3 t_2^3 t_3^3 t_4^4 t_5^2 t_6^2 +  2 y_{s}^3 y_{o_1}^6 y_{o_2}^5 t_1^2 t_2^4 t_3^3 t_4^4 t_5^2 t_6^2 +  y_{s}^3 y_{o_1}^6 y_{o_2}^5 t_1 t_2^5 t_3^3 t_4^4 t_5^2 t_6^2 -  y_{s}^4 y_{o_1}^8 y_{o_2}^6 t_1^5 t_2^3 t_3^4 t_4^6 t_5^2 t_6^2 -  y_{s}^4 y_{o_1}^8 y_{o_2}^6 t_1^4 t_2^4 t_3^4 t_4^6 t_5^2 t_6^2 -  y_{s}^4 y_{o_1}^8 y_{o_2}^6 t_1^3 t_2^5 t_3^4 t_4^6 t_5^2 t_6^2 +  y_{s}^2 y_{o_1}^5 y_{o_2}^5 t_1^3 t_2^2 t_3^3 t_4 t_5^3 t_6^2 +  y_{s}^2 y_{o_1}^5 y_{o_2}^5 t_1^2 t_2^3 t_3^3 t_4 t_5^3 t_6^2 -  y_{s}^3 y_{o_1}^7 y_{o_2}^6 t_1^5 t_2^2 t_3^4 t_4^3 t_5^3 t_6^2 -  2 y_{s}^3 y_{o_1}^7 y_{o_2}^6 t_1^4 t_2^3 t_3^4 t_4^3 t_5^3 t_6^2 -  2 y_{s}^3 y_{o_1}^7 y_{o_2}^6 t_1^3 t_2^4 t_3^4 t_4^3 t_5^3 t_6^2 -  y_{s}^3 y_{o_1}^7 y_{o_2}^6 t_1^2 t_2^5 t_3^4 t_4^3 t_5^3 t_6^2 +  y_{s}^4 y_{o_1}^9 y_{o_2}^7 t_1^6 t_2^3 t_3^5 t_4^5 t_5^3 t_6^2 +  2 y_{s}^4 y_{o_1}^9 y_{o_2}^7 t_1^5 t_2^4 t_3^5 t_4^5 t_5^3 t_6^2 +  2 y_{s}^4 y_{o_1}^9 y_{o_2}^7 t_1^4 t_2^5 t_3^5 t_4^5 t_5^3 t_6^2 +  y_{s}^4 y_{o_1}^9 y_{o_2}^7 t_1^3 t_2^6 t_3^5 t_4^5 t_5^3 t_6^2 -  y_{s}^5 y_{o_1}^{11} y_{o_2}^8 t_1^6 t_2^5 t_3^6 t_4^7 t_5^3 t_6^2 -  y_{s}^5 y_{o_1}^{11} y_{o_2}^8 t_1^5 t_2^6 t_3^6 t_4^7 t_5^3 t_6^2 +  y_{s}^2 y_{o_1}^6 y_{o_2}^6 t_1^3 t_2^3 t_3^4 t_5^4 t_6^2 -  y_{s}^3 y_{o_1}^8 y_{o_2}^7 t_1^5 t_2^3 t_3^5 t_4^2 t_5^4 t_6^2 -  y_{s}^3 y_{o_1}^8 y_{o_2}^7 t_1^4 t_2^4 t_3^5 t_4^2 t_5^4 t_6^2 -  y_{s}^3 y_{o_1}^8 y_{o_2}^7 t_1^3 t_2^5 t_3^5 t_4^2 t_5^4 t_6^2 +  y_{s}^4 y_{o_1}^{10} y_{o_2}^8 t_1^6 t_2^4 t_3^6 t_4^4 t_5^4 t_6^2 +  y_{s}^4 y_{o_1}^{10} y_{o_2}^8 t_1^5 t_2^5 t_3^6 t_4^4 t_5^4 t_6^2 +  y_{s}^4 y_{o_1}^{10} y_{o_2}^8 t_1^4 t_2^6 t_3^6 t_4^4 t_5^4 t_6^2 -  y_{s}^5 y_{o_1}^{12} y_{o_2}^9 t_1^6 t_2^6 t_3^7 t_4^6 t_5^4 t_6^2 -  y_{s}^2 y_{o_1}^2 y_{o_2}^3 t_1 t_2 t_4^3 t_5 t_6^3 +  y_{s}^3 y_{o_1}^4 y_{o_2}^4 t_1^3 t_2 t_3 t_4^5 t_5 t_6^3 +  y_{s}^3 y_{o_1}^4 y_{o_2}^4 t_1 t_2^3 t_3 t_4^5 t_5 t_6^3 -  y_{s}^4 y_{o_1}^6 y_{o_2}^5 t_1^3 t_2^3 t_3^2 t_4^7 t_5 t_6^3 -  y_{s}^2 y_{o_1}^3 y_{o_2}^4 t_1^3 t_3 t_4^2 t_5^2 t_6^3 -  3 y_{s}^2 y_{o_1}^3 y_{o_2}^4 t_1^2 t_2 t_3 t_4^2 t_5^2 t_6^3 -  3 y_{s}^2 y_{o_1}^3 y_{o_2}^4 t_1 t_2^2 t_3 t_4^2 t_5^2 t_6^3 -  y_{s}^2 y_{o_1}^3 y_{o_2}^4 t_2^3 t_3 t_4^2 t_5^2 t_6^3 +  y_{s}^3 y_{o_1}^5 y_{o_2}^5 t_1^4 t_2 t_3^2 t_4^4 t_5^2 t_6^3 +  2 y_{s}^3 y_{o_1}^5 y_{o_2}^5 t_1^3 t_2^2 t_3^2 t_4^4 t_5^2 t_6^3 +  2 y_{s}^3 y_{o_1}^5 y_{o_2}^5 t_1^2 t_2^3 t_3^2 t_4^4 t_5^2 t_6^3 +  y_{s}^3 y_{o_1}^5 y_{o_2}^5 t_1 t_2^4 t_3^2 t_4^4 t_5^2 t_6^3 -  y_{s}^4 y_{o_1}^7 y_{o_2}^6 t_1^4 t_2^3 t_3^3 t_4^6 t_5^2 t_6^3 -  y_{s}^4 y_{o_1}^7 y_{o_2}^6 t_1^3 t_2^4 t_3^3 t_4^6 t_5^2 t_6^3 -  y_{s}^2 y_{o_1}^4 y_{o_2}^5 t_1^3 t_2 t_3^2 t_4 t_5^3 t_6^3 -  y_{s}^2 y_{o_1}^4 y_{o_2}^5 t_1^2 t_2^2 t_3^2 t_4 t_5^3 t_6^3 -  y_{s}^2 y_{o_1}^4 y_{o_2}^5 t_1 t_2^3 t_3^2 t_4 t_5^3 t_6^3 -  y_{s}^3 y_{o_1}^6 y_{o_2}^6 t_1^6 t_3^3 t_4^3 t_5^3 t_6^3 -  y_{s}^3 y_{o_1}^6 y_{o_2}^6 t_1^5 t_2 t_3^3 t_4^3 t_5^3 t_6^3 -  2 y_{s}^3 y_{o_1}^6 y_{o_2}^6 t_1^4 t_2^2 t_3^3 t_4^3 t_5^3 t_6^3 -  y_{s}^3 y_{o_1}^6 y_{o_2}^6 t_1^3 t_2^3 t_3^3 t_4^3 t_5^3 t_6^3 -  2 y_{s}^3 y_{o_1}^6 y_{o_2}^6 t_1^2 t_2^4 t_3^3 t_4^3 t_5^3 t_6^3 -  y_{s}^3 y_{o_1}^6 y_{o_2}^6 t_1 t_2^5 t_3^3 t_4^3 t_5^3 t_6^3 -  y_{s}^3 y_{o_1}^6 y_{o_2}^6 t_2^6 t_3^3 t_4^3 t_5^3 t_6^3 +  2 y_{s}^4 y_{o_1}^8 y_{o_2}^7 t_1^6 t_2^2 t_3^4 t_4^5 t_5^3 t_6^3 +  2 y_{s}^4 y_{o_1}^8 y_{o_2}^7 t_1^5 t_2^3 t_3^4 t_4^5 t_5^3 t_6^3 +  3 y_{s}^4 y_{o_1}^8 y_{o_2}^7 t_1^4 t_2^4 t_3^4 t_4^5 t_5^3 t_6^3 +  2 y_{s}^4 y_{o_1}^8 y_{o_2}^7 t_1^3 t_2^5 t_3^4 t_4^5 t_5^3 t_6^3 +  2 y_{s}^4 y_{o_1}^8 y_{o_2}^7 t_1^2 t_2^6 t_3^4 t_4^5 t_5^3 t_6^3 -  y_{s}^5 y_{o_1}^{10} y_{o_2}^8 t_1^6 t_2^4 t_3^5 t_4^7 t_5^3 t_6^3 -  y_{s}^5 y_{o_1}^{10} y_{o_2}^8 t_1^5 t_2^5 t_3^5 t_4^7 t_5^3 t_6^3 -  y_{s}^5 y_{o_1}^{10} y_{o_2}^8 t_1^4 t_2^6 t_3^5 t_4^7 t_5^3 t_6^3 -  y_{s}^2 y_{o_1}^5 y_{o_2}^6 t_1^4 t_2 t_3^3 t_5^4 t_6^3 -  y_{s}^2 y_{o_1}^5 y_{o_2}^6 t_1^3 t_2^2 t_3^3 t_5^4 t_6^3 -  y_{s}^2 y_{o_1}^5 y_{o_2}^6 t_1^2 t_2^3 t_3^3 t_5^4 t_6^3 -  y_{s}^2 y_{o_1}^5 y_{o_2}^6 t_1 t_2^4 t_3^3 t_5^4 t_6^3 +  y_{s}^3 y_{o_1}^7 y_{o_2}^7 t_1^6 t_2 t_3^4 t_4^2 t_5^4 t_6^3 +  y_{s}^3 y_{o_1}^7 y_{o_2}^7 t_1^5 t_2^2 t_3^4 t_4^2 t_5^4 t_6^3 +  y_{s}^3 y_{o_1}^7 y_{o_2}^7 t_1^2 t_2^5 t_3^4 t_4^2 t_5^4 t_6^3 +  y_{s}^3 y_{o_1}^7 y_{o_2}^7 t_1 t_2^6 t_3^4 t_4^2 t_5^4 t_6^3 +  y_{s}^4 y_{o_1}^9 y_{o_2}^8 t_1^6 t_2^3 t_3^5 t_4^4 t_5^4 t_6^3 +  2 y_{s}^4 y_{o_1}^9 y_{o_2}^8 t_1^5 t_2^4 t_3^5 t_4^4 t_5^4 t_6^3 +  2 y_{s}^4 y_{o_1}^9 y_{o_2}^8 t_1^4 t_2^5 t_3^5 t_4^4 t_5^4 t_6^3 +  y_{s}^4 y_{o_1}^9 y_{o_2}^8 t_1^3 t_2^6 t_3^5 t_4^4 t_5^4 t_6^3 -  y_{s}^5 y_{o_1}^{11} y_{o_2}^9 t_1^6 t_2^5 t_3^6 t_4^6 t_5^4 t_6^3 -  y_{s}^5 y_{o_1}^{11} y_{o_2}^9 t_1^5 t_2^6 t_3^6 t_4^6 t_5^4 t_6^3 -  y_{s}^3 y_{o_1}^8 y_{o_2}^8 t_1^4 t_2^4 t_3^5 t_4 t_5^5 t_6^3 +  2 y_{s}^4 y_{o_1}^{10} y_{o_2}^9 t_1^6 t_2^4 t_3^6 t_4^3 t_5^5 t_6^3 +  2 y_{s}^4 y_{o_1}^{10} y_{o_2}^9 t_1^5 t_2^5 t_3^6 t_4^3 t_5^5 t_6^3 +  2 y_{s}^4 y_{o_1}^{10} y_{o_2}^9 t_1^4 t_2^6 t_3^6 t_4^3 t_5^5 t_6^3 -  y_{s}^5 y_{o_1}^{12} y_{o_2}^{10} t_1^6 t_2^6 t_3^7 t_4^5 t_5^5 t_6^3 -  y_{s}^2 y_{o_1}^2 y_{o_2}^4 t_1 t_2 t_4^2 t_5^2 t_6^4 +  2 y_{s}^3 y_{o_1}^4 y_{o_2}^5 t_1^3 t_2 t_3 t_4^4 t_5^2 t_6^4 +  2 y_{s}^3 y_{o_1}^4 y_{o_2}^5 t_1^2 t_2^2 t_3 t_4^4 t_5^2 t_6^4 +  2 y_{s}^3 y_{o_1}^4 y_{o_2}^5 t_1 t_2^3 t_3 t_4^4 t_5^2 t_6^4 -  y_{s}^4 y_{o_1}^6 y_{o_2}^6 t_1^3 t_2^3 t_3^2 t_4^6 t_5^2 t_6^4 -  y_{s}^2 y_{o_1}^3 y_{o_2}^5 t_1^2 t_2 t_3 t_4 t_5^3 t_6^4 -  y_{s}^2 y_{o_1}^3 y_{o_2}^5 t_1 t_2^2 t_3 t_4 t_5^3 t_6^4 +  y_{s}^3 y_{o_1}^5 y_{o_2}^6 t_1^4 t_2 t_3^2 t_4^3 t_5^3 t_6^4 +  2 y_{s}^3 y_{o_1}^5 y_{o_2}^6 t_1^3 t_2^2 t_3^2 t_4^3 t_5^3 t_6^4 +  2 y_{s}^3 y_{o_1}^5 y_{o_2}^6 t_1^2 t_2^3 t_3^2 t_4^3 t_5^3 t_6^4 +  y_{s}^3 y_{o_1}^5 y_{o_2}^6 t_1 t_2^4 t_3^2 t_4^3 t_5^3 t_6^4 +  y_{s}^4 y_{o_1}^7 y_{o_2}^7 t_1^6 t_2 t_3^3 t_4^5 t_5^3 t_6^4 +  y_{s}^4 y_{o_1}^7 y_{o_2}^7 t_1^5 t_2^2 t_3^3 t_4^5 t_5^3 t_6^4 +  y_{s}^4 y_{o_1}^7 y_{o_2}^7 t_1^2 t_2^5 t_3^3 t_4^5 t_5^3 t_6^4 +  y_{s}^4 y_{o_1}^7 y_{o_2}^7 t_1 t_2^6 t_3^3 t_4^5 t_5^3 t_6^4 -  y_{s}^5 y_{o_1}^9 y_{o_2}^8 t_1^6 t_2^3 t_3^4 t_4^7 t_5^3 t_6^4 -  y_{s}^5 y_{o_1}^9 y_{o_2}^8 t_1^5 t_2^4 t_3^4 t_4^7 t_5^3 t_6^4 -  y_{s}^5 y_{o_1}^9 y_{o_2}^8 t_1^4 t_2^5 t_3^4 t_4^7 t_5^3 t_6^4 -  y_{s}^5 y_{o_1}^9 y_{o_2}^8 t_1^3 t_2^6 t_3^4 t_4^7 t_5^3 t_6^4 -  y_{s}^2 y_{o_1}^4 y_{o_2}^6 t_1^3 t_2 t_3^2 t_5^4 t_6^4 -  y_{s}^2 y_{o_1}^4 y_{o_2}^6 t_1^2 t_2^2 t_3^2 t_5^4 t_6^4 -  y_{s}^2 y_{o_1}^4 y_{o_2}^6 t_1 t_2^3 t_3^2 t_5^4 t_6^4 +  2 y_{s}^3 y_{o_1}^6 y_{o_2}^7 t_1^5 t_2 t_3^3 t_4^2 t_5^4 t_6^4 +  2 y_{s}^3 y_{o_1}^6 y_{o_2}^7 t_1^4 t_2^2 t_3^3 t_4^2 t_5^4 t_6^4 +  3 y_{s}^3 y_{o_1}^6 y_{o_2}^7 t_1^3 t_2^3 t_3^3 t_4^2 t_5^4 t_6^4 +  2 y_{s}^3 y_{o_1}^6 y_{o_2}^7 t_1^2 t_2^4 t_3^3 t_4^2 t_5^4 t_6^4 +  2 y_{s}^3 y_{o_1}^6 y_{o_2}^7 t_1 t_2^5 t_3^3 t_4^2 t_5^4 t_6^4 -  y_{s}^4 y_{o_1}^8 y_{o_2}^8 t_1^7 t_2 t_3^4 t_4^4 t_5^4 t_6^4 -  y_{s}^4 y_{o_1}^8 y_{o_2}^8 t_1^6 t_2^2 t_3^4 t_4^4 t_5^4 t_6^4 -  2 y_{s}^4 y_{o_1}^8 y_{o_2}^8 t_1^5 t_2^3 t_3^4 t_4^4 t_5^4 t_6^4 -  y_{s}^4 y_{o_1}^8 y_{o_2}^8 t_1^4 t_2^4 t_3^4 t_4^4 t_5^4 t_6^4 -  2 y_{s}^4 y_{o_1}^8 y_{o_2}^8 t_1^3 t_2^5 t_3^4 t_4^4 t_5^4 t_6^4 -  y_{s}^4 y_{o_1}^8 y_{o_2}^8 t_1^2 t_2^6 t_3^4 t_4^4 t_5^4 t_6^4 -  y_{s}^4 y_{o_1}^8 y_{o_2}^8 t_1 t_2^7 t_3^4 t_4^4 t_5^4 t_6^4 -  y_{s}^5 y_{o_1}^{10} y_{o_2}^9 t_1^6 t_2^4 t_3^5 t_4^6 t_5^4 t_6^4 -  y_{s}^5 y_{o_1}^{10} y_{o_2}^9 t_1^5 t_2^5 t_3^5 t_4^6 t_5^4 t_6^4 -  y_{s}^5 y_{o_1}^{10} y_{o_2}^9 t_1^4 t_2^6 t_3^5 t_4^6 t_5^4 t_6^4 -  y_{s}^3 y_{o_1}^7 y_{o_2}^8 t_1^4 t_2^3 t_3^4 t_4 t_5^5 t_6^4 -  y_{s}^3 y_{o_1}^7 y_{o_2}^8 t_1^3 t_2^4 t_3^4 t_4 t_5^5 t_6^4 +  y_{s}^4 y_{o_1}^9 y_{o_2}^9 t_1^6 t_2^3 t_3^5 t_4^3 t_5^5 t_6^4 +  2 y_{s}^4 y_{o_1}^9 y_{o_2}^9 t_1^5 t_2^4 t_3^5 t_4^3 t_5^5 t_6^4 +  2 y_{s}^4 y_{o_1}^9 y_{o_2}^9 t_1^4 t_2^5 t_3^5 t_4^3 t_5^5 t_6^4 +  y_{s}^4 y_{o_1}^9 y_{o_2}^9 t_1^3 t_2^6 t_3^5 t_4^3 t_5^5 t_6^4 -  y_{s}^5 y_{o_1}^{11} y_{o_2}^{10} t_1^7 t_2^4 t_3^6 t_4^5 t_5^5 t_6^4 -  3 y_{s}^5 y_{o_1}^{11} y_{o_2}^{10} t_1^6 t_2^5 t_3^6 t_4^5 t_5^5 t_6^4 -  3 y_{s}^5 y_{o_1}^{11} y_{o_2}^{10} t_1^5 t_2^6 t_3^6 t_4^5 t_5^5 t_6^4 -  y_{s}^5 y_{o_1}^{11} y_{o_2}^{10} t_1^4 t_2^7 t_3^6 t_4^5 t_5^5 t_6^4 -  y_{s}^3 y_{o_1}^8 y_{o_2}^9 t_1^4 t_2^4 t_3^5 t_5^6 t_6^4 +  y_{s}^4 y_{o_1}^{10} y_{o_2}^{10} t_1^6 t_2^4 t_3^6 t_4^2 t_5^6 t_6^4 +  y_{s}^4 y_{o_1}^{10} y_{o_2}^{10} t_1^4 t_2^6 t_3^6 t_4^2 t_5^6 t_6^4 -  y_{s}^5 y_{o_1}^{12} y_{o_2}^{11} t_1^6 t_2^6 t_3^7 t_4^4 t_5^6 t_6^4 -  y_{s}^2 y_{o_1}^2 y_{o_2}^5 t_1 t_2 t_4 t_5^3 t_6^5 +  y_{s}^3 y_{o_1}^4 y_{o_2}^6 t_1^3 t_2 t_3 t_4^3 t_5^3 t_6^5 +  y_{s}^3 y_{o_1}^4 y_{o_2}^6 t_1^2 t_2^2 t_3 t_4^3 t_5^3 t_6^5 +  y_{s}^3 y_{o_1}^4 y_{o_2}^6 t_1 t_2^3 t_3 t_4^3 t_5^3 t_6^5 -  y_{s}^4 y_{o_1}^6 y_{o_2}^7 t_1^4 t_2^2 t_3^2 t_4^5 t_5^3 t_6^5 -  y_{s}^4 y_{o_1}^6 y_{o_2}^7 t_1^3 t_2^3 t_3^2 t_4^5 t_5^3 t_6^5 -  y_{s}^4 y_{o_1}^6 y_{o_2}^7 t_1^2 t_2^4 t_3^2 t_4^5 t_5^3 t_6^5 +  y_{s}^5 y_{o_1}^8 y_{o_2}^8 t_1^4 t_2^4 t_3^3 t_4^7 t_5^3 t_6^5 -  y_{s}^2 y_{o_1}^3 y_{o_2}^6 t_1^2 t_2 t_3 t_5^4 t_6^5 -  y_{s}^2 y_{o_1}^3 y_{o_2}^6 t_1 t_2^2 t_3 t_5^4 t_6^5 +  y_{s}^3 y_{o_1}^5 y_{o_2}^7 t_1^4 t_2 t_3^2 t_4^2 t_5^4 t_6^5 +  2 y_{s}^3 y_{o_1}^5 y_{o_2}^7 t_1^3 t_2^2 t_3^2 t_4^2 t_5^4 t_6^5 +  2 y_{s}^3 y_{o_1}^5 y_{o_2}^7 t_1^2 t_2^3 t_3^2 t_4^2 t_5^4 t_6^5 +  y_{s}^3 y_{o_1}^5 y_{o_2}^7 t_1 t_2^4 t_3^2 t_4^2 t_5^4 t_6^5 -  y_{s}^4 y_{o_1}^7 y_{o_2}^8 t_1^5 t_2^2 t_3^3 t_4^4 t_5^4 t_6^5 -  2 y_{s}^4 y_{o_1}^7 y_{o_2}^8 t_1^4 t_2^3 t_3^3 t_4^4 t_5^4 t_6^5 -  2 y_{s}^4 y_{o_1}^7 y_{o_2}^8 t_1^3 t_2^4 t_3^3 t_4^4 t_5^4 t_6^5 -  y_{s}^4 y_{o_1}^7 y_{o_2}^8 t_1^2 t_2^5 t_3^3 t_4^4 t_5^4 t_6^5 +  y_{s}^5 y_{o_1}^9 y_{o_2}^9 t_1^5 t_2^4 t_3^4 t_4^6 t_5^4 t_6^5 +  y_{s}^5 y_{o_1}^9 y_{o_2}^9 t_1^4 t_2^5 t_3^4 t_4^6 t_5^4 t_6^5 -  y_{s}^3 y_{o_1}^6 y_{o_2}^8 t_1^4 t_2^2 t_3^3 t_4 t_5^5 t_6^5 -  y_{s}^3 y_{o_1}^6 y_{o_2}^8 t_1^3 t_2^3 t_3^3 t_4 t_5^5 t_6^5 -  y_{s}^3 y_{o_1}^6 y_{o_2}^8 t_1^2 t_2^4 t_3^3 t_4 t_5^5 t_6^5 +  y_{s}^4 y_{o_1}^8 y_{o_2}^9 t_1^6 t_2^2 t_3^4 t_4^3 t_5^5 t_6^5 +  2 y_{s}^4 y_{o_1}^8 y_{o_2}^9 t_1^5 t_2^3 t_3^4 t_4^3 t_5^5 t_6^5 +  3 y_{s}^4 y_{o_1}^8 y_{o_2}^9 t_1^4 t_2^4 t_3^4 t_4^3 t_5^5 t_6^5 +  2 y_{s}^4 y_{o_1}^8 y_{o_2}^9 t_1^3 t_2^5 t_3^4 t_4^3 t_5^5 t_6^5 +  y_{s}^4 y_{o_1}^8 y_{o_2}^9 t_1^2 t_2^6 t_3^4 t_4^3 t_5^5 t_6^5 -  y_{s}^5 y_{o_1}^{10} y_{o_2}^{10} t_1^7 t_2^3 t_3^5 t_4^5 t_5^5 t_6^5 -  2 y_{s}^5 y_{o_1}^{10} y_{o_2}^{10} t_1^6 t_2^4 t_3^5 t_4^5 t_5^5 t_6^5 -  3 y_{s}^5 y_{o_1}^{10} y_{o_2}^{10} t_1^5 t_2^5 t_3^5 t_4^5 t_5^5 t_6^5 -  2 y_{s}^5 y_{o_1}^{10} y_{o_2}^{10} t_1^4 t_2^6 t_3^5 t_4^5 t_5^5 t_6^5 -  y_{s}^5 y_{o_1}^{10} y_{o_2}^{10} t_1^3 t_2^7 t_3^5 t_4^5 t_5^5 t_6^5 +  y_{s}^6 y_{o_1}^{12} y_{o_2}^{11} t_1^7 t_2^5 t_3^6 t_4^7 t_5^5 t_6^5 +  y_{s}^6 y_{o_1}^{12} y_{o_2}^{11} t_1^6 t_2^6 t_3^6 t_4^7 t_5^5 t_6^5 +  y_{s}^6 y_{o_1}^{12} y_{o_2}^{11} t_1^5 t_2^7 t_3^6 t_4^7 t_5^5 t_6^5 -  y_{s}^3 y_{o_1}^7 y_{o_2}^9 t_1^4 t_2^3 t_3^4 t_5^6 t_6^5 -  y_{s}^3 y_{o_1}^7 y_{o_2}^9 t_1^3 t_2^4 t_3^4 t_5^6 t_6^5 +  y_{s}^4 y_{o_1}^9 y_{o_2}^{10} t_1^6 t_2^3 t_3^5 t_4^2 t_5^6 t_6^5 +  2 y_{s}^4 y_{o_1}^9 y_{o_2}^{10} t_1^5 t_2^4 t_3^5 t_4^2 t_5^6 t_6^5 +  2 y_{s}^4 y_{o_1}^9 y_{o_2}^{10} t_1^4 t_2^5 t_3^5 t_4^2 t_5^6 t_6^5 +  y_{s}^4 y_{o_1}^9 y_{o_2}^{10} t_1^3 t_2^6 t_3^5 t_4^2 t_5^6 t_6^5 -  y_{s}^5 y_{o_1}^{11} y_{o_2}^{11} t_1^7 t_2^4 t_3^6 t_4^4 t_5^6 t_6^5 -  2 y_{s}^5 y_{o_1}^{11} y_{o_2}^{11} t_1^6 t_2^5 t_3^6 t_4^4 t_5^6 t_6^5 -  2 y_{s}^5 y_{o_1}^{11} y_{o_2}^{11} t_1^5 t_2^6 t_3^6 t_4^4 t_5^6 t_6^5 -  y_{s}^5 y_{o_1}^{11} y_{o_2}^{11} t_1^4 t_2^7 t_3^6 t_4^4 t_5^6 t_6^5 +  y_{s}^6 y_{o_1}^{13} y_{o_2}^{12} t_1^7 t_2^6 t_3^7 t_4^6 t_5^6 t_6^5 +  y_{s}^6 y_{o_1}^{13} y_{o_2}^{12} t_1^6 t_2^7 t_3^7 t_4^6 t_5^6 t_6^5 +  y_{s}^3 y_{o_1}^4 y_{o_2}^7 t_1^3 t_2 t_3 t_4^2 t_5^4 t_6^6 +  2 y_{s}^3 y_{o_1}^4 y_{o_2}^7 t_1^2 t_2^2 t_3 t_4^2 t_5^4 t_6^6 +  y_{s}^3 y_{o_1}^4 y_{o_2}^7 t_1 t_2^3 t_3 t_4^2 t_5^4 t_6^6 -  y_{s}^4 y_{o_1}^6 y_{o_2}^8 t_1^4 t_2^2 t_3^2 t_4^4 t_5^4 t_6^6 -  y_{s}^4 y_{o_1}^6 y_{o_2}^8 t_1^3 t_2^3 t_3^2 t_4^4 t_5^4 t_6^6 -  y_{s}^4 y_{o_1}^6 y_{o_2}^8 t_1^2 t_2^4 t_3^2 t_4^4 t_5^4 t_6^6 +  y_{s}^5 y_{o_1}^8 y_{o_2}^9 t_1^4 t_2^4 t_3^3 t_4^6 t_5^4 t_6^6 +  y_{s}^4 y_{o_1}^7 y_{o_2}^9 t_1^6 t_2 t_3^3 t_4^3 t_5^5 t_6^6 +  2 y_{s}^4 y_{o_1}^7 y_{o_2}^9 t_1^5 t_2^2 t_3^3 t_4^3 t_5^5 t_6^6 +  2 y_{s}^4 y_{o_1}^7 y_{o_2}^9 t_1^4 t_2^3 t_3^3 t_4^3 t_5^5 t_6^6 +  2 y_{s}^4 y_{o_1}^7 y_{o_2}^9 t_1^3 t_2^4 t_3^3 t_4^3 t_5^5 t_6^6 +  2 y_{s}^4 y_{o_1}^7 y_{o_2}^9 t_1^2 t_2^5 t_3^3 t_4^3 t_5^5 t_6^6 +  y_{s}^4 y_{o_1}^7 y_{o_2}^9 t_1 t_2^6 t_3^3 t_4^3 t_5^5 t_6^6 -  y_{s}^5 y_{o_1}^9 y_{o_2}^{10} t_1^7 t_2^2 t_3^4 t_4^5 t_5^5 t_6^6 -  2 y_{s}^5 y_{o_1}^9 y_{o_2}^{10} t_1^6 t_2^3 t_3^4 t_4^5 t_5^5 t_6^6 -  3 y_{s}^5 y_{o_1}^9 y_{o_2}^{10} t_1^5 t_2^4 t_3^4 t_4^5 t_5^5 t_6^6 -  3 y_{s}^5 y_{o_1}^9 y_{o_2}^{10} t_1^4 t_2^5 t_3^4 t_4^5 t_5^5 t_6^6 -  2 y_{s}^5 y_{o_1}^9 y_{o_2}^{10} t_1^3 t_2^6 t_3^4 t_4^5 t_5^5 t_6^6 -  y_{s}^5 y_{o_1}^9 y_{o_2}^{10} t_1^2 t_2^7 t_3^4 t_4^5 t_5^5 t_6^6 +  y_{s}^6 y_{o_1}^{11} y_{o_2}^{11} t_1^7 t_2^4 t_3^5 t_4^7 t_5^5 t_6^6 +  y_{s}^6 y_{o_1}^{11} y_{o_2}^{11} t_1^6 t_2^5 t_3^5 t_4^7 t_5^5 t_6^6 +  y_{s}^6 y_{o_1}^{11} y_{o_2}^{11} t_1^5 t_2^6 t_3^5 t_4^7 t_5^5 t_6^6 +  y_{s}^6 y_{o_1}^{11} y_{o_2}^{11} t_1^4 t_2^7 t_3^5 t_4^7 t_5^5 t_6^6 +  y_{s}^4 y_{o_1}^8 y_{o_2}^{10} t_1^5 t_2^3 t_3^4 t_4^2 t_5^6 t_6^6 +  2 y_{s}^4 y_{o_1}^8 y_{o_2}^{10} t_1^4 t_2^4 t_3^4 t_4^2 t_5^6 t_6^6 +  y_{s}^4 y_{o_1}^8 y_{o_2}^{10} t_1^3 t_2^5 t_3^4 t_4^2 t_5^6 t_6^6 -  y_{s}^5 y_{o_1}^{10} y_{o_2}^{11} t_1^7 t_2^3 t_3^5 t_4^4 t_5^6 t_6^6 -  2 y_{s}^5 y_{o_1}^{10} y_{o_2}^{11} t_1^6 t_2^4 t_3^5 t_4^4 t_5^6 t_6^6 -  3 y_{s}^5 y_{o_1}^{10} y_{o_2}^{11} t_1^5 t_2^5 t_3^5 t_4^4 t_5^6 t_6^6 -  2 y_{s}^5 y_{o_1}^{10} y_{o_2}^{11} t_1^4 t_2^6 t_3^5 t_4^4 t_5^6 t_6^6 -  y_{s}^5 y_{o_1}^{10} y_{o_2}^{11} t_1^3 t_2^7 t_3^5 t_4^4 t_5^6 t_6^6 +  y_{s}^6 y_{o_1}^{12} y_{o_2}^{12} t_1^7 t_2^5 t_3^6 t_4^6 t_5^6 t_6^6 +  y_{s}^6 y_{o_1}^{12} y_{o_2}^{12} t_1^6 t_2^6 t_3^6 t_4^6 t_5^6 t_6^6 +  y_{s}^6 y_{o_1}^{12} y_{o_2}^{12} t_1^5 t_2^7 t_3^6 t_4^6 t_5^6 t_6^6 -  y_{s}^5 y_{o_1}^{11} y_{o_2}^{12} t_1^6 t_2^5 t_3^6 t_4^3 t_5^7 t_6^6 -  y_{s}^5 y_{o_1}^{11} y_{o_2}^{12} t_1^5 t_2^6 t_3^6 t_4^3 t_5^7 t_6^6 -  y_{s}^4 y_{o_1}^5 y_{o_2}^8 t_1^3 t_2^2 t_3 t_4^4 t_5^4 t_6^7 -  y_{s}^4 y_{o_1}^5 y_{o_2}^8 t_1^2 t_2^3 t_3 t_4^4 t_5^4 t_6^7 -  y_{s}^5 y_{o_1}^8 y_{o_2}^{10} t_1^6 t_2^2 t_3^3 t_4^5 t_5^5 t_6^7 -  y_{s}^5 y_{o_1}^8 y_{o_2}^{10} t_1^5 t_2^3 t_3^3 t_4^5 t_5^5 t_6^7 -  y_{s}^5 y_{o_1}^8 y_{o_2}^{10} t_1^4 t_2^4 t_3^3 t_4^5 t_5^5 t_6^7 -  y_{s}^5 y_{o_1}^8 y_{o_2}^{10} t_1^3 t_2^5 t_3^3 t_4^5 t_5^5 t_6^7 -  y_{s}^5 y_{o_1}^8 y_{o_2}^{10} t_1^2 t_2^6 t_3^3 t_4^5 t_5^5 t_6^7 +  y_{s}^6 y_{o_1}^{10} y_{o_2}^{11} t_1^6 t_2^4 t_3^4 t_4^7 t_5^5 t_6^7 +  y_{s}^6 y_{o_1}^{10} y_{o_2}^{11} t_1^5 t_2^5 t_3^4 t_4^7 t_5^5 t_6^7 +  y_{s}^6 y_{o_1}^{10} y_{o_2}^{11} t_1^4 t_2^6 t_3^4 t_4^7 t_5^5 t_6^7 -  y_{s}^5 y_{o_1}^9 y_{o_2}^{11} t_1^5 t_2^4 t_3^4 t_4^4 t_5^6 t_6^7 -  y_{s}^5 y_{o_1}^9 y_{o_2}^{11} t_1^4 t_2^5 t_3^4 t_4^4 t_5^6 t_6^7 +  y_{s}^6 y_{o_1}^{11} y_{o_2}^{12} t_1^7 t_2^4 t_3^5 t_4^6 t_5^6 t_6^7 +  y_{s}^6 y_{o_1}^{11} y_{o_2}^{12} t_1^6 t_2^5 t_3^5 t_4^6 t_5^6 t_6^7 +  y_{s}^6 y_{o_1}^{11} y_{o_2}^{12} t_1^5 t_2^6 t_3^5 t_4^6 t_5^6 t_6^7 +  y_{s}^6 y_{o_1}^{11} y_{o_2}^{12} t_1^4 t_2^7 t_3^5 t_4^6 t_5^6 t_6^7 +  y_{s}^6 y_{o_1}^{12} y_{o_2}^{13} t_1^6 t_2^6 t_3^6 t_4^5 t_5^7 t_6^7 +  y_{s}^7 y_{o_1}^{14} y_{o_2}^{14} t_1^7 t_2^7 t_3^7 t_4^7 t_5^7 t_6^7
~,~
$
\end{quote}
\endgroup

\subsection{Model 11 \label{app_num_11}}

\begingroup\makeatletter\def\f@size{7}\check@mathfonts
\begin{quote}\raggedright
$
P(t_i,y_s,y_{o}; \mathcal{M}_{11}) =
1 + y_{s} y_{o}^3 t_1 t_2 t_3^3 t_5^2 + y_{s} y_{o}^3 t_1^2 t_3^2 t_4 t_5^2 +  y_{s} y_{o}^3 t_1 t_2 t_3^2 t_4 t_5^2 + y_{s} y_{o}^3 t_2^2 t_3^2 t_4 t_5^2 +  y_{s} y_{o}^3 t_1^2 t_3 t_4^2 t_5^2 + y_{s} y_{o}^3 t_1 t_2 t_3 t_4^2 t_5^2 +  y_{s} y_{o}^3 t_2^2 t_3 t_4^2 t_5^2 + y_{s} y_{o}^3 t_1 t_2 t_4^3 t_5^2 -  y_{s}^2 y_{o}^6 t_1^2 t_2^2 t_3^5 t_4 t_5^4 - y_{s}^2 y_{o}^6 t_1^2 t_2^2 t_3^4 t_4^2 t_5^4 -  y_{s}^2 y_{o}^6 t_1^3 t_2 t_3^3 t_4^3 t_5^4 -  y_{s}^2 y_{o}^6 t_1^2 t_2^2 t_3^3 t_4^3 t_5^4 - y_{s}^2 y_{o}^6 t_1 t_2^3 t_3^3 t_4^3 t_5^4 -  y_{s}^2 y_{o}^6 t_1^2 t_2^2 t_3^2 t_4^4 t_5^4 -  y_{s}^2 y_{o}^6 t_1^2 t_2^2 t_3 t_4^5 t_5^4 - y_{s}^3 y_{o}^9 t_1^3 t_2^3 t_3^5 t_4^4 t_5^6 -  y_{s}^3 y_{o}^9 t_1^3 t_2^3 t_3^4 t_4^5 t_5^6 + y_{s} y_{o}^2 t_1^2 t_3^2 t_5 t_6 +  y_{s} y_{o}^2 t_1 t_2 t_3^2 t_5 t_6 + y_{s} y_{o}^2 t_2^2 t_3^2 t_5 t_6 +  y_{s} y_{o}^2 t_1^2 t_3 t_4 t_5 t_6 + y_{s} y_{o}^2 t_1 t_2 t_3 t_4 t_5 t_6 +  y_{s} y_{o}^2 t_2^2 t_3 t_4 t_5 t_6 + y_{s} y_{o}^2 t_1^2 t_4^2 t_5 t_6 +  y_{s} y_{o}^2 t_1 t_2 t_4^2 t_5 t_6 + y_{s} y_{o}^2 t_2^2 t_4^2 t_5 t_6 -  y_{s}^2 y_{o}^5 t_1^2 t_2^2 t_3^5 t_5^3 t_6 -  y_{s}^2 y_{o}^5 t_1^2 t_2^2 t_3^4 t_4 t_5^3 t_6 -  y_{s}^2 y_{o}^5 t_1^2 t_2^2 t_3^3 t_4^2 t_5^3 t_6 -  y_{s}^2 y_{o}^5 t_1^2 t_2^2 t_3^2 t_4^3 t_5^3 t_6 -  y_{s}^2 y_{o}^5 t_1^2 t_2^2 t_3 t_4^4 t_5^3 t_6 -  y_{s}^2 y_{o}^5 t_1^2 t_2^2 t_4^5 t_5^3 t_6 -  y_{s}^3 y_{o}^8 t_1^3 t_2^3 t_3^5 t_4^3 t_5^5 t_6 -  y_{s}^3 y_{o}^8 t_1^3 t_2^3 t_3^4 t_4^4 t_5^5 t_6 -  y_{s}^3 y_{o}^8 t_1^3 t_2^3 t_3^3 t_4^5 t_5^5 t_6 + y_{s} y_{o} t_1 t_2 t_3 t_6^2 +  y_{s} y_{o} t_1 t_2 t_4 t_6^2 - y_{s}^2 y_{o}^4 t_1^3 t_2 t_3^4 t_5^2 t_6^2 -  y_{s}^2 y_{o}^4 t_1^2 t_2^2 t_3^4 t_5^2 t_6^2 - y_{s}^2 y_{o}^4 t_1 t_2^3 t_3^4 t_5^2 t_6^2 -  y_{s}^2 y_{o}^4 t_1^4 t_3^3 t_4 t_5^2 t_6^2 -  2 y_{s}^2 y_{o}^4 t_1^3 t_2 t_3^3 t_4 t_5^2 t_6^2 -  3 y_{s}^2 y_{o}^4 t_1^2 t_2^2 t_3^3 t_4 t_5^2 t_6^2 -  2 y_{s}^2 y_{o}^4 t_1 t_2^3 t_3^3 t_4 t_5^2 t_6^2 -  y_{s}^2 y_{o}^4 t_2^4 t_3^3 t_4 t_5^2 t_6^2 - y_{s}^2 y_{o}^4 t_1^4 t_3^2 t_4^2 t_5^2 t_6^2 -  y_{s}^2 y_{o}^4 t_1^3 t_2 t_3^2 t_4^2 t_5^2 t_6^2 -  3 y_{s}^2 y_{o}^4 t_1^2 t_2^2 t_3^2 t_4^2 t_5^2 t_6^2 -  y_{s}^2 y_{o}^4 t_1 t_2^3 t_3^2 t_4^2 t_5^2 t_6^2 -  y_{s}^2 y_{o}^4 t_2^4 t_3^2 t_4^2 t_5^2 t_6^2 -  y_{s}^2 y_{o}^4 t_1^4 t_3 t_4^3 t_5^2 t_6^2 -  2 y_{s}^2 y_{o}^4 t_1^3 t_2 t_3 t_4^3 t_5^2 t_6^2 -  3 y_{s}^2 y_{o}^4 t_1^2 t_2^2 t_3 t_4^3 t_5^2 t_6^2 -  2 y_{s}^2 y_{o}^4 t_1 t_2^3 t_3 t_4^3 t_5^2 t_6^2 -  y_{s}^2 y_{o}^4 t_2^4 t_3 t_4^3 t_5^2 t_6^2 - y_{s}^2 y_{o}^4 t_1^3 t_2 t_4^4 t_5^2 t_6^2 -  y_{s}^2 y_{o}^4 t_1^2 t_2^2 t_4^4 t_5^2 t_6^2 - y_{s}^2 y_{o}^4 t_1 t_2^3 t_4^4 t_5^2 t_6^2 +  y_{s}^3 y_{o}^7 t_1^4 t_2^2 t_3^6 t_4 t_5^4 t_6^2 +  y_{s}^3 y_{o}^7 t_1^2 t_2^4 t_3^6 t_4 t_5^4 t_6^2 +  y_{s}^3 y_{o}^7 t_1^4 t_2^2 t_3^5 t_4^2 t_5^4 t_6^2 -  y_{s}^3 y_{o}^7 t_1^3 t_2^3 t_3^5 t_4^2 t_5^4 t_6^2 +  y_{s}^3 y_{o}^7 t_1^2 t_2^4 t_3^5 t_4^2 t_5^4 t_6^2 +  y_{s}^3 y_{o}^7 t_1^5 t_2 t_3^4 t_4^3 t_5^4 t_6^2 +  2 y_{s}^3 y_{o}^7 t_1^4 t_2^2 t_3^4 t_4^3 t_5^4 t_6^2 +  y_{s}^3 y_{o}^7 t_1^3 t_2^3 t_3^4 t_4^3 t_5^4 t_6^2 +  2 y_{s}^3 y_{o}^7 t_1^2 t_2^4 t_3^4 t_4^3 t_5^4 t_6^2 +  y_{s}^3 y_{o}^7 t_1 t_2^5 t_3^4 t_4^3 t_5^4 t_6^2 +  y_{s}^3 y_{o}^7 t_1^5 t_2 t_3^3 t_4^4 t_5^4 t_6^2 +  2 y_{s}^3 y_{o}^7 t_1^4 t_2^2 t_3^3 t_4^4 t_5^4 t_6^2 +  y_{s}^3 y_{o}^7 t_1^3 t_2^3 t_3^3 t_4^4 t_5^4 t_6^2 +  2 y_{s}^3 y_{o}^7 t_1^2 t_2^4 t_3^3 t_4^4 t_5^4 t_6^2 +  y_{s}^3 y_{o}^7 t_1 t_2^5 t_3^3 t_4^4 t_5^4 t_6^2 +  y_{s}^3 y_{o}^7 t_1^4 t_2^2 t_3^2 t_4^5 t_5^4 t_6^2 -  y_{s}^3 y_{o}^7 t_1^3 t_2^3 t_3^2 t_4^5 t_5^4 t_6^2 +  y_{s}^3 y_{o}^7 t_1^2 t_2^4 t_3^2 t_4^5 t_5^4 t_6^2 +  y_{s}^3 y_{o}^7 t_1^4 t_2^2 t_3 t_4^6 t_5^4 t_6^2 +  y_{s}^3 y_{o}^7 t_1^2 t_2^4 t_3 t_4^6 t_5^4 t_6^2 +  y_{s}^4 y_{o}^{10} t_1^5 t_2^3 t_3^6 t_4^4 t_5^6 t_6^2 +  y_{s}^4 y_{o}^{10} t_1^3 t_2^5 t_3^6 t_4^4 t_5^6 t_6^2 +  2 y_{s}^4 y_{o}^{10} t_1^5 t_2^3 t_3^5 t_4^5 t_5^6 t_6^2 +  y_{s}^4 y_{o}^{10} t_1^4 t_2^4 t_3^5 t_4^5 t_5^6 t_6^2 +  2 y_{s}^4 y_{o}^{10} t_1^3 t_2^5 t_3^5 t_4^5 t_5^6 t_6^2 +  y_{s}^4 y_{o}^{10} t_1^5 t_2^3 t_3^4 t_4^6 t_5^6 t_6^2 +  y_{s}^4 y_{o}^{10} t_1^3 t_2^5 t_3^4 t_4^6 t_5^6 t_6^2 -  y_{s}^2 y_{o}^3 t_1^2 t_2^2 t_3^3 t_5 t_6^3 - y_{s}^2 y_{o}^3 t_1^4 t_3^2 t_4 t_5 t_6^3 -  y_{s}^2 y_{o}^3 t_1^3 t_2 t_3^2 t_4 t_5 t_6^3 -  3 y_{s}^2 y_{o}^3 t_1^2 t_2^2 t_3^2 t_4 t_5 t_6^3 -  y_{s}^2 y_{o}^3 t_1 t_2^3 t_3^2 t_4 t_5 t_6^3 - y_{s}^2 y_{o}^3 t_2^4 t_3^2 t_4 t_5 t_6^3 -  y_{s}^2 y_{o}^3 t_1^4 t_3 t_4^2 t_5 t_6^3 - y_{s}^2 y_{o}^3 t_1^3 t_2 t_3 t_4^2 t_5 t_6^3 -  3 y_{s}^2 y_{o}^3 t_1^2 t_2^2 t_3 t_4^2 t_5 t_6^3 -  y_{s}^2 y_{o}^3 t_1 t_2^3 t_3 t_4^2 t_5 t_6^3 - y_{s}^2 y_{o}^3 t_2^4 t_3 t_4^2 t_5 t_6^3 -  y_{s}^2 y_{o}^3 t_1^2 t_2^2 t_4^3 t_5 t_6^3 - y_{s}^3 y_{o}^6 t_1^3 t_2^3 t_3^6 t_5^3 t_6^3 +  y_{s}^3 y_{o}^6 t_1^4 t_2^2 t_3^5 t_4 t_5^3 t_6^3 -  y_{s}^3 y_{o}^6 t_1^3 t_2^3 t_3^5 t_4 t_5^3 t_6^3 +  y_{s}^3 y_{o}^6 t_1^2 t_2^4 t_3^5 t_4 t_5^3 t_6^3 +  y_{s}^3 y_{o}^6 t_1^4 t_2^2 t_3^4 t_4^2 t_5^3 t_6^3 -  y_{s}^3 y_{o}^6 t_1^3 t_2^3 t_3^4 t_4^2 t_5^3 t_6^3 +  y_{s}^3 y_{o}^6 t_1^2 t_2^4 t_3^4 t_4^2 t_5^3 t_6^3 -  y_{s}^3 y_{o}^6 t_1^6 t_3^3 t_4^3 t_5^3 t_6^3 -  y_{s}^3 y_{o}^6 t_1^5 t_2 t_3^3 t_4^3 t_5^3 t_6^3 -  y_{s}^3 y_{o}^6 t_1^4 t_2^2 t_3^3 t_4^3 t_5^3 t_6^3 -  3 y_{s}^3 y_{o}^6 t_1^3 t_2^3 t_3^3 t_4^3 t_5^3 t_6^3 -  y_{s}^3 y_{o}^6 t_1^2 t_2^4 t_3^3 t_4^3 t_5^3 t_6^3 -  y_{s}^3 y_{o}^6 t_1 t_2^5 t_3^3 t_4^3 t_5^3 t_6^3 -  y_{s}^3 y_{o}^6 t_2^6 t_3^3 t_4^3 t_5^3 t_6^3 +  y_{s}^3 y_{o}^6 t_1^4 t_2^2 t_3^2 t_4^4 t_5^3 t_6^3 -  y_{s}^3 y_{o}^6 t_1^3 t_2^3 t_3^2 t_4^4 t_5^3 t_6^3 +  y_{s}^3 y_{o}^6 t_1^2 t_2^4 t_3^2 t_4^4 t_5^3 t_6^3 +  y_{s}^3 y_{o}^6 t_1^4 t_2^2 t_3 t_4^5 t_5^3 t_6^3 -  y_{s}^3 y_{o}^6 t_1^3 t_2^3 t_3 t_4^5 t_5^3 t_6^3 +  y_{s}^3 y_{o}^6 t_1^2 t_2^4 t_3 t_4^5 t_5^3 t_6^3 -  y_{s}^3 y_{o}^6 t_1^3 t_2^3 t_4^6 t_5^3 t_6^3 +  y_{s}^4 y_{o}^9 t_1^6 t_2^2 t_3^6 t_4^3 t_5^5 t_6^3 +  2 y_{s}^4 y_{o}^9 t_1^5 t_2^3 t_3^6 t_4^3 t_5^5 t_6^3 +  2 y_{s}^4 y_{o}^9 t_1^4 t_2^4 t_3^6 t_4^3 t_5^5 t_6^3 +  2 y_{s}^4 y_{o}^9 t_1^3 t_2^5 t_3^6 t_4^3 t_5^5 t_6^3 +  y_{s}^4 y_{o}^9 t_1^2 t_2^6 t_3^6 t_4^3 t_5^5 t_6^3 +  2 y_{s}^4 y_{o}^9 t_1^5 t_2^3 t_3^5 t_4^4 t_5^5 t_6^3 +  y_{s}^4 y_{o}^9 t_1^4 t_2^4 t_3^5 t_4^4 t_5^5 t_6^3 +  2 y_{s}^4 y_{o}^9 t_1^3 t_2^5 t_3^5 t_4^4 t_5^5 t_6^3 +  2 y_{s}^4 y_{o}^9 t_1^5 t_2^3 t_3^4 t_4^5 t_5^5 t_6^3 +  y_{s}^4 y_{o}^9 t_1^4 t_2^4 t_3^4 t_4^5 t_5^5 t_6^3 +  2 y_{s}^4 y_{o}^9 t_1^3 t_2^5 t_3^4 t_4^5 t_5^5 t_6^3 +  y_{s}^4 y_{o}^9 t_1^6 t_2^2 t_3^3 t_4^6 t_5^5 t_6^3 +  2 y_{s}^4 y_{o}^9 t_1^5 t_2^3 t_3^3 t_4^6 t_5^5 t_6^3 +  2 y_{s}^4 y_{o}^9 t_1^4 t_2^4 t_3^3 t_4^6 t_5^5 t_6^3 +  2 y_{s}^4 y_{o}^9 t_1^3 t_2^5 t_3^3 t_4^6 t_5^5 t_6^3 +  y_{s}^4 y_{o}^9 t_1^2 t_2^6 t_3^3 t_4^6 t_5^5 t_6^3 -  y_{s}^5 y_{o}^{12} t_1^6 t_2^4 t_3^6 t_4^6 t_5^7 t_6^3 -  y_{s}^5 y_{o}^{12} t_1^5 t_2^5 t_3^6 t_4^6 t_5^7 t_6^3 -  y_{s}^5 y_{o}^{12} t_1^4 t_2^6 t_3^6 t_4^6 t_5^7 t_6^3 -  y_{s}^2 y_{o}^2 t_1^3 t_2 t_3 t_4 t_6^4 - y_{s}^2 y_{o}^2 t_1^2 t_2^2 t_3 t_4 t_6^4 -  y_{s}^2 y_{o}^2 t_1 t_2^3 t_3 t_4 t_6^4 + y_{s}^3 y_{o}^5 t_1^5 t_2 t_3^4 t_4 t_5^2 t_6^4 +  2 y_{s}^3 y_{o}^5 t_1^4 t_2^2 t_3^4 t_4 t_5^2 t_6^4 +  2 y_{s}^3 y_{o}^5 t_1^3 t_2^3 t_3^4 t_4 t_5^2 t_6^4 +  2 y_{s}^3 y_{o}^5 t_1^2 t_2^4 t_3^4 t_4 t_5^2 t_6^4 +  y_{s}^3 y_{o}^5 t_1 t_2^5 t_3^4 t_4 t_5^2 t_6^4 +  2 y_{s}^3 y_{o}^5 t_1^4 t_2^2 t_3^3 t_4^2 t_5^2 t_6^4 +  y_{s}^3 y_{o}^5 t_1^3 t_2^3 t_3^3 t_4^2 t_5^2 t_6^4 +  2 y_{s}^3 y_{o}^5 t_1^2 t_2^4 t_3^3 t_4^2 t_5^2 t_6^4 +  2 y_{s}^3 y_{o}^5 t_1^4 t_2^2 t_3^2 t_4^3 t_5^2 t_6^4 +  y_{s}^3 y_{o}^5 t_1^3 t_2^3 t_3^2 t_4^3 t_5^2 t_6^4 +  2 y_{s}^3 y_{o}^5 t_1^2 t_2^4 t_3^2 t_4^3 t_5^2 t_6^4 +  y_{s}^3 y_{o}^5 t_1^5 t_2 t_3 t_4^4 t_5^2 t_6^4 +  2 y_{s}^3 y_{o}^5 t_1^4 t_2^2 t_3 t_4^4 t_5^2 t_6^4 +  2 y_{s}^3 y_{o}^5 t_1^3 t_2^3 t_3 t_4^4 t_5^2 t_6^4 +  2 y_{s}^3 y_{o}^5 t_1^2 t_2^4 t_3 t_4^4 t_5^2 t_6^4 +  y_{s}^3 y_{o}^5 t_1 t_2^5 t_3 t_4^4 t_5^2 t_6^4 -  y_{s}^4 y_{o}^8 t_1^4 t_2^4 t_3^7 t_4 t_5^4 t_6^4 +  y_{s}^4 y_{o}^8 t_1^5 t_2^3 t_3^6 t_4^2 t_5^4 t_6^4 -  y_{s}^4 y_{o}^8 t_1^4 t_2^4 t_3^6 t_4^2 t_5^4 t_6^4 +  y_{s}^4 y_{o}^8 t_1^3 t_2^5 t_3^6 t_4^2 t_5^4 t_6^4 +  y_{s}^4 y_{o}^8 t_1^5 t_2^3 t_3^5 t_4^3 t_5^4 t_6^4 -  y_{s}^4 y_{o}^8 t_1^4 t_2^4 t_3^5 t_4^3 t_5^4 t_6^4 +  y_{s}^4 y_{o}^8 t_1^3 t_2^5 t_3^5 t_4^3 t_5^4 t_6^4 -  y_{s}^4 y_{o}^8 t_1^7 t_2 t_3^4 t_4^4 t_5^4 t_6^4 -  y_{s}^4 y_{o}^8 t_1^6 t_2^2 t_3^4 t_4^4 t_5^4 t_6^4 -  y_{s}^4 y_{o}^8 t_1^5 t_2^3 t_3^4 t_4^4 t_5^4 t_6^4 -  3 y_{s}^4 y_{o}^8 t_1^4 t_2^4 t_3^4 t_4^4 t_5^4 t_6^4 -  y_{s}^4 y_{o}^8 t_1^3 t_2^5 t_3^4 t_4^4 t_5^4 t_6^4 -  y_{s}^4 y_{o}^8 t_1^2 t_2^6 t_3^4 t_4^4 t_5^4 t_6^4 -  y_{s}^4 y_{o}^8 t_1 t_2^7 t_3^4 t_4^4 t_5^4 t_6^4 +  y_{s}^4 y_{o}^8 t_1^5 t_2^3 t_3^3 t_4^5 t_5^4 t_6^4 -  y_{s}^4 y_{o}^8 t_1^4 t_2^4 t_3^3 t_4^5 t_5^4 t_6^4 +  y_{s}^4 y_{o}^8 t_1^3 t_2^5 t_3^3 t_4^5 t_5^4 t_6^4 +  y_{s}^4 y_{o}^8 t_1^5 t_2^3 t_3^2 t_4^6 t_5^4 t_6^4 -  y_{s}^4 y_{o}^8 t_1^4 t_2^4 t_3^2 t_4^6 t_5^4 t_6^4 +  y_{s}^4 y_{o}^8 t_1^3 t_2^5 t_3^2 t_4^6 t_5^4 t_6^4 -  y_{s}^4 y_{o}^8 t_1^4 t_2^4 t_3 t_4^7 t_5^4 t_6^4 -  y_{s}^5 y_{o}^{11} t_1^5 t_2^5 t_3^7 t_4^4 t_5^6 t_6^4 -  y_{s}^5 y_{o}^{11} t_1^7 t_2^3 t_3^6 t_4^5 t_5^6 t_6^4 -  y_{s}^5 y_{o}^{11} t_1^6 t_2^4 t_3^6 t_4^5 t_5^6 t_6^4 -  3 y_{s}^5 y_{o}^{11} t_1^5 t_2^5 t_3^6 t_4^5 t_5^6 t_6^4 -  y_{s}^5 y_{o}^{11} t_1^4 t_2^6 t_3^6 t_4^5 t_5^6 t_6^4 -  y_{s}^5 y_{o}^{11} t_1^3 t_2^7 t_3^6 t_4^5 t_5^6 t_6^4 -  y_{s}^5 y_{o}^{11} t_1^7 t_2^3 t_3^5 t_4^6 t_5^6 t_6^4 -  y_{s}^5 y_{o}^{11} t_1^6 t_2^4 t_3^5 t_4^6 t_5^6 t_6^4 -  3 y_{s}^5 y_{o}^{11} t_1^5 t_2^5 t_3^5 t_4^6 t_5^6 t_6^4 -  y_{s}^5 y_{o}^{11} t_1^4 t_2^6 t_3^5 t_4^6 t_5^6 t_6^4 -  y_{s}^5 y_{o}^{11} t_1^3 t_2^7 t_3^5 t_4^6 t_5^6 t_6^4 -  y_{s}^5 y_{o}^{11} t_1^5 t_2^5 t_3^4 t_4^7 t_5^6 t_6^4 +  y_{s}^3 y_{o}^4 t_1^4 t_2^2 t_3^3 t_4 t_5 t_6^5 +  y_{s}^3 y_{o}^4 t_1^2 t_2^4 t_3^3 t_4 t_5 t_6^5 +  2 y_{s}^3 y_{o}^4 t_1^4 t_2^2 t_3^2 t_4^2 t_5 t_6^5 +  y_{s}^3 y_{o}^4 t_1^3 t_2^3 t_3^2 t_4^2 t_5 t_6^5 +  2 y_{s}^3 y_{o}^4 t_1^2 t_2^4 t_3^2 t_4^2 t_5 t_6^5 +  y_{s}^3 y_{o}^4 t_1^4 t_2^2 t_3 t_4^3 t_5 t_6^5 +  y_{s}^3 y_{o}^4 t_1^2 t_2^4 t_3 t_4^3 t_5 t_6^5 +  y_{s}^4 y_{o}^7 t_1^5 t_2^3 t_3^6 t_4 t_5^3 t_6^5 +  y_{s}^4 y_{o}^7 t_1^3 t_2^5 t_3^6 t_4 t_5^3 t_6^5 +  y_{s}^4 y_{o}^7 t_1^5 t_2^3 t_3^5 t_4^2 t_5^3 t_6^5 -  y_{s}^4 y_{o}^7 t_1^4 t_2^4 t_3^5 t_4^2 t_5^3 t_6^5 +  y_{s}^4 y_{o}^7 t_1^3 t_2^5 t_3^5 t_4^2 t_5^3 t_6^5 +  y_{s}^4 y_{o}^7 t_1^6 t_2^2 t_3^4 t_4^3 t_5^3 t_6^5 +  2 y_{s}^4 y_{o}^7 t_1^5 t_2^3 t_3^4 t_4^3 t_5^3 t_6^5 +  y_{s}^4 y_{o}^7 t_1^4 t_2^4 t_3^4 t_4^3 t_5^3 t_6^5 +  2 y_{s}^4 y_{o}^7 t_1^3 t_2^5 t_3^4 t_4^3 t_5^3 t_6^5 +  y_{s}^4 y_{o}^7 t_1^2 t_2^6 t_3^4 t_4^3 t_5^3 t_6^5 +  y_{s}^4 y_{o}^7 t_1^6 t_2^2 t_3^3 t_4^4 t_5^3 t_6^5 +  2 y_{s}^4 y_{o}^7 t_1^5 t_2^3 t_3^3 t_4^4 t_5^3 t_6^5 +  y_{s}^4 y_{o}^7 t_1^4 t_2^4 t_3^3 t_4^4 t_5^3 t_6^5 +  2 y_{s}^4 y_{o}^7 t_1^3 t_2^5 t_3^3 t_4^4 t_5^3 t_6^5 +  y_{s}^4 y_{o}^7 t_1^2 t_2^6 t_3^3 t_4^4 t_5^3 t_6^5 +  y_{s}^4 y_{o}^7 t_1^5 t_2^3 t_3^2 t_4^5 t_5^3 t_6^5 -  y_{s}^4 y_{o}^7 t_1^4 t_2^4 t_3^2 t_4^5 t_5^3 t_6^5 +  y_{s}^4 y_{o}^7 t_1^3 t_2^5 t_3^2 t_4^5 t_5^3 t_6^5 +  y_{s}^4 y_{o}^7 t_1^5 t_2^3 t_3 t_4^6 t_5^3 t_6^5 +  y_{s}^4 y_{o}^7 t_1^3 t_2^5 t_3 t_4^6 t_5^3 t_6^5 -  y_{s}^5 y_{o}^{10} t_1^6 t_2^4 t_3^7 t_4^3 t_5^5 t_6^5 -  y_{s}^5 y_{o}^{10} t_1^5 t_2^5 t_3^7 t_4^3 t_5^5 t_6^5 -  y_{s}^5 y_{o}^{10} t_1^4 t_2^6 t_3^7 t_4^3 t_5^5 t_6^5 -  y_{s}^5 y_{o}^{10} t_1^7 t_2^3 t_3^6 t_4^4 t_5^5 t_6^5 -  2 y_{s}^5 y_{o}^{10} t_1^6 t_2^4 t_3^6 t_4^4 t_5^5 t_6^5 -  3 y_{s}^5 y_{o}^{10} t_1^5 t_2^5 t_3^6 t_4^4 t_5^5 t_6^5 -  2 y_{s}^5 y_{o}^{10} t_1^4 t_2^6 t_3^6 t_4^4 t_5^5 t_6^5 -  y_{s}^5 y_{o}^{10} t_1^3 t_2^7 t_3^6 t_4^4 t_5^5 t_6^5 -  y_{s}^5 y_{o}^{10} t_1^7 t_2^3 t_3^5 t_4^5 t_5^5 t_6^5 -  y_{s}^5 y_{o}^{10} t_1^6 t_2^4 t_3^5 t_4^5 t_5^5 t_6^5 -  3 y_{s}^5 y_{o}^{10} t_1^5 t_2^5 t_3^5 t_4^5 t_5^5 t_6^5 -  y_{s}^5 y_{o}^{10} t_1^4 t_2^6 t_3^5 t_4^5 t_5^5 t_6^5 -  y_{s}^5 y_{o}^{10} t_1^3 t_2^7 t_3^5 t_4^5 t_5^5 t_6^5 -  y_{s}^5 y_{o}^{10} t_1^7 t_2^3 t_3^4 t_4^6 t_5^5 t_6^5 -  2 y_{s}^5 y_{o}^{10} t_1^6 t_2^4 t_3^4 t_4^6 t_5^5 t_6^5 -  3 y_{s}^5 y_{o}^{10} t_1^5 t_2^5 t_3^4 t_4^6 t_5^5 t_6^5 -  2 y_{s}^5 y_{o}^{10} t_1^4 t_2^6 t_3^4 t_4^6 t_5^5 t_6^5 -  y_{s}^5 y_{o}^{10} t_1^3 t_2^7 t_3^4 t_4^6 t_5^5 t_6^5 -  y_{s}^5 y_{o}^{10} t_1^6 t_2^4 t_3^3 t_4^7 t_5^5 t_6^5 -  y_{s}^5 y_{o}^{10} t_1^5 t_2^5 t_3^3 t_4^7 t_5^5 t_6^5 -  y_{s}^5 y_{o}^{10} t_1^4 t_2^6 t_3^3 t_4^7 t_5^5 t_6^5 +  y_{s}^6 y_{o}^{13} t_1^6 t_2^6 t_3^7 t_4^6 t_5^7 t_6^5 +  y_{s}^6 y_{o}^{13} t_1^6 t_2^6 t_3^6 t_4^7 t_5^7 t_6^5 -  y_{s}^4 y_{o}^6 t_1^4 t_2^4 t_3^4 t_4^2 t_5^2 t_6^6 -  y_{s}^4 y_{o}^6 t_1^4 t_2^4 t_3^3 t_4^3 t_5^2 t_6^6 -  y_{s}^4 y_{o}^6 t_1^4 t_2^4 t_3^2 t_4^4 t_5^2 t_6^6 -  y_{s}^5 y_{o}^9 t_1^5 t_2^5 t_3^7 t_4^2 t_5^4 t_6^6 -  y_{s}^5 y_{o}^9 t_1^5 t_2^5 t_3^6 t_4^3 t_5^4 t_6^6 -  y_{s}^5 y_{o}^9 t_1^5 t_2^5 t_3^5 t_4^4 t_5^4 t_6^6 -  y_{s}^5 y_{o}^9 t_1^5 t_2^5 t_3^4 t_4^5 t_5^4 t_6^6 -  y_{s}^5 y_{o}^9 t_1^5 t_2^5 t_3^3 t_4^6 t_5^4 t_6^6 -  y_{s}^5 y_{o}^9 t_1^5 t_2^5 t_3^2 t_4^7 t_5^4 t_6^6 +  y_{s}^6 y_{o}^{12} t_1^7 t_2^5 t_3^7 t_4^5 t_5^6 t_6^6 +  y_{s}^6 y_{o}^{12} t_1^6 t_2^6 t_3^7 t_4^5 t_5^6 t_6^6 +  y_{s}^6 y_{o}^{12} t_1^5 t_2^7 t_3^7 t_4^5 t_5^6 t_6^6 +  y_{s}^6 y_{o}^{12} t_1^7 t_2^5 t_3^6 t_4^6 t_5^6 t_6^6 +  y_{s}^6 y_{o}^{12} t_1^6 t_2^6 t_3^6 t_4^6 t_5^6 t_6^6 +  y_{s}^6 y_{o}^{12} t_1^5 t_2^7 t_3^6 t_4^6 t_5^6 t_6^6 +  y_{s}^6 y_{o}^{12} t_1^7 t_2^5 t_3^5 t_4^7 t_5^6 t_6^6 +  y_{s}^6 y_{o}^{12} t_1^6 t_2^6 t_3^5 t_4^7 t_5^6 t_6^6 +  y_{s}^6 y_{o}^{12} t_1^5 t_2^7 t_3^5 t_4^7 t_5^6 t_6^6 -  y_{s}^4 y_{o}^5 t_1^4 t_2^4 t_3^3 t_4^2 t_5 t_6^7 -  y_{s}^4 y_{o}^5 t_1^4 t_2^4 t_3^2 t_4^3 t_5 t_6^7 -  y_{s}^5 y_{o}^8 t_1^5 t_2^5 t_3^6 t_4^2 t_5^3 t_6^7 -  y_{s}^5 y_{o}^8 t_1^5 t_2^5 t_3^5 t_4^3 t_5^3 t_6^7 -  y_{s}^5 y_{o}^8 t_1^6 t_2^4 t_3^4 t_4^4 t_5^3 t_6^7 -  y_{s}^5 y_{o}^8 t_1^5 t_2^5 t_3^4 t_4^4 t_5^3 t_6^7 -  y_{s}^5 y_{o}^8 t_1^4 t_2^6 t_3^4 t_4^4 t_5^3 t_6^7 -  y_{s}^5 y_{o}^8 t_1^5 t_2^5 t_3^3 t_4^5 t_5^3 t_6^7 -  y_{s}^5 y_{o}^8 t_1^5 t_2^5 t_3^2 t_4^6 t_5^3 t_6^7 +  y_{s}^6 y_{o}^{11} t_1^6 t_2^6 t_3^7 t_4^4 t_5^5 t_6^7 +  y_{s}^6 y_{o}^{11} t_1^7 t_2^5 t_3^6 t_4^5 t_5^5 t_6^7 +  y_{s}^6 y_{o}^{11} t_1^6 t_2^6 t_3^6 t_4^5 t_5^5 t_6^7 +  y_{s}^6 y_{o}^{11} t_1^5 t_2^7 t_3^6 t_4^5 t_5^5 t_6^7 +  y_{s}^6 y_{o}^{11} t_1^7 t_2^5 t_3^5 t_4^6 t_5^5 t_6^7 +  y_{s}^6 y_{o}^{11} t_1^6 t_2^6 t_3^5 t_4^6 t_5^5 t_6^7 +  y_{s}^6 y_{o}^{11} t_1^5 t_2^7 t_3^5 t_4^6 t_5^5 t_6^7 +  y_{s}^6 y_{o}^{11} t_1^6 t_2^6 t_3^4 t_4^7 t_5^5 t_6^7 +  y_{s}^7 y_{o}^{14} t_1^7 t_2^7 t_3^7 t_4^7 t_5^7 t_6^7
~,~
$
\end{quote}
\endgroup

\subsection{Model 12 \label{app_num_12}}

\begingroup\makeatletter\def\f@size{7}\check@mathfonts
\begin{quote}\raggedright
$
P(t_i,y_s; \mathcal{M}_{12}) =
1 + y_{s} t_1 t_2 t_3^2 t_5^2 + y_{s} t_1^2 t_3 t_4 t_5^2 + y_{s} t_1 t_2 t_3 t_4 t_5^2 +  y_{s} t_2^2 t_3 t_4 t_5^2 + y_{s} t_1 t_2 t_4^2 t_5^2 - y_{s}^2 t_1^2 t_2^2 t_3^3 t_4 t_5^4 -  y_{s}^2 t_1^3 t_2 t_3^2 t_4^2 t_5^4 - y_{s}^2 t_1^2 t_2^2 t_3^2 t_4^2 t_5^4 -  y_{s}^2 t_1 t_2^3 t_3^2 t_4^2 t_5^4 - y_{s}^2 t_1^2 t_2^2 t_3 t_4^3 t_5^4 -  y_{s}^3 t_1^3 t_2^3 t_3^3 t_4^3 t_5^6 + y_{s} t_1^2 t_3^2 t_5 t_6 +  y_{s} t_1 t_2 t_3^2 t_5 t_6 + y_{s} t_2^2 t_3^2 t_5 t_6 + y_{s} t_1^2 t_3 t_4 t_5 t_6 +  y_{s} t_1 t_2 t_3 t_4 t_5 t_6 + y_{s} t_2^2 t_3 t_4 t_5 t_6 + y_{s} t_1^2 t_4^2 t_5 t_6 +  y_{s} t_1 t_2 t_4^2 t_5 t_6 + y_{s} t_2^2 t_4^2 t_5 t_6 - y_{s}^2 t_1^2 t_2^2 t_3^4 t_5^3 t_6 -  y_{s}^2 t_1^2 t_2^2 t_3^3 t_4 t_5^3 t_6 - y_{s}^2 t_1^4 t_3^2 t_4^2 t_5^3 t_6 -  y_{s}^2 t_1^3 t_2 t_3^2 t_4^2 t_5^3 t_6 - 3 y_{s}^2 t_1^2 t_2^2 t_3^2 t_4^2 t_5^3 t_6 -  y_{s}^2 t_1 t_2^3 t_3^2 t_4^2 t_5^3 t_6 - y_{s}^2 t_2^4 t_3^2 t_4^2 t_5^3 t_6 -  y_{s}^2 t_1^2 t_2^2 t_3 t_4^3 t_5^3 t_6 - y_{s}^2 t_1^2 t_2^2 t_4^4 t_5^3 t_6 +  y_{s}^3 t_1^4 t_2^2 t_3^4 t_4^2 t_5^5 t_6 + y_{s}^3 t_1^2 t_2^4 t_3^4 t_4^2 t_5^5 t_6 -  y_{s}^3 t_1^3 t_2^3 t_3^3 t_4^3 t_5^5 t_6 + y_{s}^3 t_1^4 t_2^2 t_3^2 t_4^4 t_5^5 t_6 +  y_{s}^3 t_1^2 t_2^4 t_3^2 t_4^4 t_5^5 t_6 - y_{s}^4 t_1^4 t_2^4 t_3^4 t_4^4 t_5^7 t_6 +  y_{s} t_1 t_2 t_3^2 t_6^2 + y_{s} t_1^2 t_3 t_4 t_6^2 + y_{s} t_1 t_2 t_3 t_4 t_6^2 +  y_{s} t_2^2 t_3 t_4 t_6^2 + y_{s} t_1 t_2 t_4^2 t_6^2 - y_{s}^2 t_1^3 t_2 t_3^4 t_5^2 t_6^2 -  y_{s}^2 t_1^2 t_2^2 t_3^4 t_5^2 t_6^2 - y_{s}^2 t_1 t_2^3 t_3^4 t_5^2 t_6^2 -  y_{s}^2 t_1^4 t_3^3 t_4 t_5^2 t_6^2 - y_{s}^2 t_1^3 t_2 t_3^3 t_4 t_5^2 t_6^2 -  3 y_{s}^2 t_1^2 t_2^2 t_3^3 t_4 t_5^2 t_6^2 - y_{s}^2 t_1 t_2^3 t_3^3 t_4 t_5^2 t_6^2 -  y_{s}^2 t_2^4 t_3^3 t_4 t_5^2 t_6^2 - y_{s}^2 t_1^4 t_3^2 t_4^2 t_5^2 t_6^2 -  3 y_{s}^2 t_1^3 t_2 t_3^2 t_4^2 t_5^2 t_6^2 -  3 y_{s}^2 t_1^2 t_2^2 t_3^2 t_4^2 t_5^2 t_6^2 -  3 y_{s}^2 t_1 t_2^3 t_3^2 t_4^2 t_5^2 t_6^2 - y_{s}^2 t_2^4 t_3^2 t_4^2 t_5^2 t_6^2 -  y_{s}^2 t_1^4 t_3 t_4^3 t_5^2 t_6^2 - y_{s}^2 t_1^3 t_2 t_3 t_4^3 t_5^2 t_6^2 -  3 y_{s}^2 t_1^2 t_2^2 t_3 t_4^3 t_5^2 t_6^2 - y_{s}^2 t_1 t_2^3 t_3 t_4^3 t_5^2 t_6^2 -  y_{s}^2 t_2^4 t_3 t_4^3 t_5^2 t_6^2 - y_{s}^2 t_1^3 t_2 t_4^4 t_5^2 t_6^2 -  y_{s}^2 t_1^2 t_2^2 t_4^4 t_5^2 t_6^2 - y_{s}^2 t_1 t_2^3 t_4^4 t_5^2 t_6^2 +  y_{s}^3 t_1^4 t_2^2 t_3^5 t_4 t_5^4 t_6^2 + y_{s}^3 t_1^2 t_2^4 t_3^5 t_4 t_5^4 t_6^2 +  y_{s}^3 t_1^5 t_2 t_3^4 t_4^2 t_5^4 t_6^2 +  2 y_{s}^3 t_1^4 t_2^2 t_3^4 t_4^2 t_5^4 t_6^2 +  2 y_{s}^3 t_1^3 t_2^3 t_3^4 t_4^2 t_5^4 t_6^2 +  2 y_{s}^3 t_1^2 t_2^4 t_3^4 t_4^2 t_5^4 t_6^2 +  y_{s}^3 t_1 t_2^5 t_3^4 t_4^2 t_5^4 t_6^2 +  2 y_{s}^3 t_1^4 t_2^2 t_3^3 t_4^3 t_5^4 t_6^2 -  y_{s}^3 t_1^3 t_2^3 t_3^3 t_4^3 t_5^4 t_6^2 +  2 y_{s}^3 t_1^2 t_2^4 t_3^3 t_4^3 t_5^4 t_6^2 +  y_{s}^3 t_1^5 t_2 t_3^2 t_4^4 t_5^4 t_6^2 +  2 y_{s}^3 t_1^4 t_2^2 t_3^2 t_4^4 t_5^4 t_6^2 +  2 y_{s}^3 t_1^3 t_2^3 t_3^2 t_4^4 t_5^4 t_6^2 +  2 y_{s}^3 t_1^2 t_2^4 t_3^2 t_4^4 t_5^4 t_6^2 +  y_{s}^3 t_1 t_2^5 t_3^2 t_4^4 t_5^4 t_6^2 + y_{s}^3 t_1^4 t_2^2 t_3 t_4^5 t_5^4 t_6^2 +  y_{s}^3 t_1^2 t_2^4 t_3 t_4^5 t_5^4 t_6^2 + y_{s}^4 t_1^5 t_2^3 t_3^5 t_4^3 t_5^6 t_6^2 +  y_{s}^4 t_1^3 t_2^5 t_3^5 t_4^3 t_5^6 t_6^2 -  y_{s}^4 t_1^4 t_2^4 t_3^4 t_4^4 t_5^6 t_6^2 +  y_{s}^4 t_1^5 t_2^3 t_3^3 t_4^5 t_5^6 t_6^2 +  y_{s}^4 t_1^3 t_2^5 t_3^3 t_4^5 t_5^6 t_6^2 - y_{s}^2 t_1^2 t_2^2 t_3^4 t_5 t_6^3 -  y_{s}^2 t_1^2 t_2^2 t_3^3 t_4 t_5 t_6^3 - y_{s}^2 t_1^4 t_3^2 t_4^2 t_5 t_6^3 -  y_{s}^2 t_1^3 t_2 t_3^2 t_4^2 t_5 t_6^3 - 3 y_{s}^2 t_1^2 t_2^2 t_3^2 t_4^2 t_5 t_6^3 -  y_{s}^2 t_1 t_2^3 t_3^2 t_4^2 t_5 t_6^3 - y_{s}^2 t_2^4 t_3^2 t_4^2 t_5 t_6^3 -  y_{s}^2 t_1^2 t_2^2 t_3 t_4^3 t_5 t_6^3 - y_{s}^2 t_1^2 t_2^2 t_4^4 t_5 t_6^3 -  y_{s}^3 t_1^3 t_2^3 t_3^6 t_5^3 t_6^3 - y_{s}^3 t_1^3 t_2^3 t_3^5 t_4 t_5^3 t_6^3 +  2 y_{s}^3 t_1^4 t_2^2 t_3^4 t_4^2 t_5^3 t_6^3 -  y_{s}^3 t_1^3 t_2^3 t_3^4 t_4^2 t_5^3 t_6^3 +  2 y_{s}^3 t_1^2 t_2^4 t_3^4 t_4^2 t_5^3 t_6^3 - y_{s}^3 t_1^6 t_3^3 t_4^3 t_5^3 t_6^3 -  y_{s}^3 t_1^5 t_2 t_3^3 t_4^3 t_5^3 t_6^3 -  y_{s}^3 t_1^4 t_2^2 t_3^3 t_4^3 t_5^3 t_6^3 -  3 y_{s}^3 t_1^3 t_2^3 t_3^3 t_4^3 t_5^3 t_6^3 -  y_{s}^3 t_1^2 t_2^4 t_3^3 t_4^3 t_5^3 t_6^3 - y_{s}^3 t_1 t_2^5 t_3^3 t_4^3 t_5^3 t_6^3 -  y_{s}^3 t_2^6 t_3^3 t_4^3 t_5^3 t_6^3 + 2 y_{s}^3 t_1^4 t_2^2 t_3^2 t_4^4 t_5^3 t_6^3 -  y_{s}^3 t_1^3 t_2^3 t_3^2 t_4^4 t_5^3 t_6^3 +  2 y_{s}^3 t_1^2 t_2^4 t_3^2 t_4^4 t_5^3 t_6^3 -  y_{s}^3 t_1^3 t_2^3 t_3 t_4^5 t_5^3 t_6^3 - y_{s}^3 t_1^3 t_2^3 t_4^6 t_5^3 t_6^3 +  y_{s}^4 t_1^5 t_2^3 t_3^6 t_4^2 t_5^5 t_6^3 +  y_{s}^4 t_1^3 t_2^5 t_3^6 t_4^2 t_5^5 t_6^3 +  y_{s}^4 t_1^6 t_2^2 t_3^5 t_4^3 t_5^5 t_6^3 +  2 y_{s}^4 t_1^5 t_2^3 t_3^5 t_4^3 t_5^5 t_6^3 +  2 y_{s}^4 t_1^4 t_2^4 t_3^5 t_4^3 t_5^5 t_6^3 +  2 y_{s}^4 t_1^3 t_2^5 t_3^5 t_4^3 t_5^5 t_6^3 +  y_{s}^4 t_1^2 t_2^6 t_3^5 t_4^3 t_5^5 t_6^3 +  2 y_{s}^4 t_1^5 t_2^3 t_3^4 t_4^4 t_5^5 t_6^3 -  y_{s}^4 t_1^4 t_2^4 t_3^4 t_4^4 t_5^5 t_6^3 +  2 y_{s}^4 t_1^3 t_2^5 t_3^4 t_4^4 t_5^5 t_6^3 +  y_{s}^4 t_1^6 t_2^2 t_3^3 t_4^5 t_5^5 t_6^3 +  2 y_{s}^4 t_1^5 t_2^3 t_3^3 t_4^5 t_5^5 t_6^3 +  2 y_{s}^4 t_1^4 t_2^4 t_3^3 t_4^5 t_5^5 t_6^3 +  2 y_{s}^4 t_1^3 t_2^5 t_3^3 t_4^5 t_5^5 t_6^3 +  y_{s}^4 t_1^2 t_2^6 t_3^3 t_4^5 t_5^5 t_6^3 +  y_{s}^4 t_1^5 t_2^3 t_3^2 t_4^6 t_5^5 t_6^3 +  y_{s}^4 t_1^3 t_2^5 t_3^2 t_4^6 t_5^5 t_6^3 -  y_{s}^5 t_1^5 t_2^5 t_3^6 t_4^4 t_5^7 t_6^3 -  y_{s}^5 t_1^6 t_2^4 t_3^5 t_4^5 t_5^7 t_6^3 -  y_{s}^5 t_1^5 t_2^5 t_3^5 t_4^5 t_5^7 t_6^3 -  y_{s}^5 t_1^4 t_2^6 t_3^5 t_4^5 t_5^7 t_6^3 -  y_{s}^5 t_1^5 t_2^5 t_3^4 t_4^6 t_5^7 t_6^3 - y_{s}^2 t_1^2 t_2^2 t_3^3 t_4 t_6^4 -  y_{s}^2 t_1^3 t_2 t_3^2 t_4^2 t_6^4 - y_{s}^2 t_1^2 t_2^2 t_3^2 t_4^2 t_6^4 -  y_{s}^2 t_1 t_2^3 t_3^2 t_4^2 t_6^4 - y_{s}^2 t_1^2 t_2^2 t_3 t_4^3 t_6^4 +  y_{s}^3 t_1^4 t_2^2 t_3^5 t_4 t_5^2 t_6^4 + y_{s}^3 t_1^2 t_2^4 t_3^5 t_4 t_5^2 t_6^4 +  y_{s}^3 t_1^5 t_2 t_3^4 t_4^2 t_5^2 t_6^4 +  2 y_{s}^3 t_1^4 t_2^2 t_3^4 t_4^2 t_5^2 t_6^4 +  2 y_{s}^3 t_1^3 t_2^3 t_3^4 t_4^2 t_5^2 t_6^4 +  2 y_{s}^3 t_1^2 t_2^4 t_3^4 t_4^2 t_5^2 t_6^4 +  y_{s}^3 t_1 t_2^5 t_3^4 t_4^2 t_5^2 t_6^4 +  2 y_{s}^3 t_1^4 t_2^2 t_3^3 t_4^3 t_5^2 t_6^4 -  y_{s}^3 t_1^3 t_2^3 t_3^3 t_4^3 t_5^2 t_6^4 +  2 y_{s}^3 t_1^2 t_2^4 t_3^3 t_4^3 t_5^2 t_6^4 +  y_{s}^3 t_1^5 t_2 t_3^2 t_4^4 t_5^2 t_6^4 +  2 y_{s}^3 t_1^4 t_2^2 t_3^2 t_4^4 t_5^2 t_6^4 +  2 y_{s}^3 t_1^3 t_2^3 t_3^2 t_4^4 t_5^2 t_6^4 +  2 y_{s}^3 t_1^2 t_2^4 t_3^2 t_4^4 t_5^2 t_6^4 +  y_{s}^3 t_1 t_2^5 t_3^2 t_4^4 t_5^2 t_6^4 + y_{s}^3 t_1^4 t_2^2 t_3 t_4^5 t_5^2 t_6^4 +  y_{s}^3 t_1^2 t_2^4 t_3 t_4^5 t_5^2 t_6^4 - y_{s}^4 t_1^4 t_2^4 t_3^7 t_4 t_5^4 t_6^4 -  y_{s}^4 t_1^4 t_2^4 t_3^6 t_4^2 t_5^4 t_6^4 +  2 y_{s}^4 t_1^5 t_2^3 t_3^5 t_4^3 t_5^4 t_6^4 -  y_{s}^4 t_1^4 t_2^4 t_3^5 t_4^3 t_5^4 t_6^4 +  2 y_{s}^4 t_1^3 t_2^5 t_3^5 t_4^3 t_5^4 t_6^4 -  y_{s}^4 t_1^7 t_2 t_3^4 t_4^4 t_5^4 t_6^4 - y_{s}^4 t_1^6 t_2^2 t_3^4 t_4^4 t_5^4 t_6^4 -  y_{s}^4 t_1^5 t_2^3 t_3^4 t_4^4 t_5^4 t_6^4 -  3 y_{s}^4 t_1^4 t_2^4 t_3^4 t_4^4 t_5^4 t_6^4 -  y_{s}^4 t_1^3 t_2^5 t_3^4 t_4^4 t_5^4 t_6^4 -  y_{s}^4 t_1^2 t_2^6 t_3^4 t_4^4 t_5^4 t_6^4 -  y_{s}^4 t_1 t_2^7 t_3^4 t_4^4 t_5^4 t_6^4 +  2 y_{s}^4 t_1^5 t_2^3 t_3^3 t_4^5 t_5^4 t_6^4 -  y_{s}^4 t_1^4 t_2^4 t_3^3 t_4^5 t_5^4 t_6^4 +  2 y_{s}^4 t_1^3 t_2^5 t_3^3 t_4^5 t_5^4 t_6^4 -  y_{s}^4 t_1^4 t_2^4 t_3^2 t_4^6 t_5^4 t_6^4 - y_{s}^4 t_1^4 t_2^4 t_3 t_4^7 t_5^4 t_6^4 -  y_{s}^5 t_1^5 t_2^5 t_3^7 t_4^3 t_5^6 t_6^4 -  y_{s}^5 t_1^5 t_2^5 t_3^6 t_4^4 t_5^6 t_6^4 -  y_{s}^5 t_1^7 t_2^3 t_3^5 t_4^5 t_5^6 t_6^4 -  y_{s}^5 t_1^6 t_2^4 t_3^5 t_4^5 t_5^6 t_6^4 -  3 y_{s}^5 t_1^5 t_2^5 t_3^5 t_4^5 t_5^6 t_6^4 -  y_{s}^5 t_1^4 t_2^6 t_3^5 t_4^5 t_5^6 t_6^4 -  y_{s}^5 t_1^3 t_2^7 t_3^5 t_4^5 t_5^6 t_6^4 -  y_{s}^5 t_1^5 t_2^5 t_3^4 t_4^6 t_5^6 t_6^4 -  y_{s}^5 t_1^5 t_2^5 t_3^3 t_4^7 t_5^6 t_6^4 + y_{s}^3 t_1^4 t_2^2 t_3^4 t_4^2 t_5 t_6^5 +  y_{s}^3 t_1^2 t_2^4 t_3^4 t_4^2 t_5 t_6^5 - y_{s}^3 t_1^3 t_2^3 t_3^3 t_4^3 t_5 t_6^5 +  y_{s}^3 t_1^4 t_2^2 t_3^2 t_4^4 t_5 t_6^5 + y_{s}^3 t_1^2 t_2^4 t_3^2 t_4^4 t_5 t_6^5 +  y_{s}^4 t_1^5 t_2^3 t_3^6 t_4^2 t_5^3 t_6^5 +  y_{s}^4 t_1^3 t_2^5 t_3^6 t_4^2 t_5^3 t_6^5 +  y_{s}^4 t_1^6 t_2^2 t_3^5 t_4^3 t_5^3 t_6^5 +  2 y_{s}^4 t_1^5 t_2^3 t_3^5 t_4^3 t_5^3 t_6^5 +  2 y_{s}^4 t_1^4 t_2^4 t_3^5 t_4^3 t_5^3 t_6^5 +  2 y_{s}^4 t_1^3 t_2^5 t_3^5 t_4^3 t_5^3 t_6^5 +  y_{s}^4 t_1^2 t_2^6 t_3^5 t_4^3 t_5^3 t_6^5 +  2 y_{s}^4 t_1^5 t_2^3 t_3^4 t_4^4 t_5^3 t_6^5 -  y_{s}^4 t_1^4 t_2^4 t_3^4 t_4^4 t_5^3 t_6^5 +  2 y_{s}^4 t_1^3 t_2^5 t_3^4 t_4^4 t_5^3 t_6^5 +  y_{s}^4 t_1^6 t_2^2 t_3^3 t_4^5 t_5^3 t_6^5 +  2 y_{s}^4 t_1^5 t_2^3 t_3^3 t_4^5 t_5^3 t_6^5 +  2 y_{s}^4 t_1^4 t_2^4 t_3^3 t_4^5 t_5^3 t_6^5 +  2 y_{s}^4 t_1^3 t_2^5 t_3^3 t_4^5 t_5^3 t_6^5 +  y_{s}^4 t_1^2 t_2^6 t_3^3 t_4^5 t_5^3 t_6^5 +  y_{s}^4 t_1^5 t_2^3 t_3^2 t_4^6 t_5^3 t_6^5 +  y_{s}^4 t_1^3 t_2^5 t_3^2 t_4^6 t_5^3 t_6^5 -  y_{s}^5 t_1^6 t_2^4 t_3^7 t_4^3 t_5^5 t_6^5 -  y_{s}^5 t_1^5 t_2^5 t_3^7 t_4^3 t_5^5 t_6^5 -  y_{s}^5 t_1^4 t_2^6 t_3^7 t_4^3 t_5^5 t_6^5 -  y_{s}^5 t_1^7 t_2^3 t_3^6 t_4^4 t_5^5 t_6^5 -  y_{s}^5 t_1^6 t_2^4 t_3^6 t_4^4 t_5^5 t_6^5 -  3 y_{s}^5 t_1^5 t_2^5 t_3^6 t_4^4 t_5^5 t_6^5 -  y_{s}^5 t_1^4 t_2^6 t_3^6 t_4^4 t_5^5 t_6^5 -  y_{s}^5 t_1^3 t_2^7 t_3^6 t_4^4 t_5^5 t_6^5 -  y_{s}^5 t_1^7 t_2^3 t_3^5 t_4^5 t_5^5 t_6^5 -  3 y_{s}^5 t_1^6 t_2^4 t_3^5 t_4^5 t_5^5 t_6^5 -  3 y_{s}^5 t_1^5 t_2^5 t_3^5 t_4^5 t_5^5 t_6^5 -  3 y_{s}^5 t_1^4 t_2^6 t_3^5 t_4^5 t_5^5 t_6^5 -  y_{s}^5 t_1^3 t_2^7 t_3^5 t_4^5 t_5^5 t_6^5 -  y_{s}^5 t_1^7 t_2^3 t_3^4 t_4^6 t_5^5 t_6^5 -  y_{s}^5 t_1^6 t_2^4 t_3^4 t_4^6 t_5^5 t_6^5 -  3 y_{s}^5 t_1^5 t_2^5 t_3^4 t_4^6 t_5^5 t_6^5 -  y_{s}^5 t_1^4 t_2^6 t_3^4 t_4^6 t_5^5 t_6^5 -  y_{s}^5 t_1^3 t_2^7 t_3^4 t_4^6 t_5^5 t_6^5 -  y_{s}^5 t_1^6 t_2^4 t_3^3 t_4^7 t_5^5 t_6^5 -  y_{s}^5 t_1^5 t_2^5 t_3^3 t_4^7 t_5^5 t_6^5 -  y_{s}^5 t_1^4 t_2^6 t_3^3 t_4^7 t_5^5 t_6^5 +  y_{s}^6 t_1^6 t_2^6 t_3^7 t_4^5 t_5^7 t_6^5 +  y_{s}^6 t_1^7 t_2^5 t_3^6 t_4^6 t_5^7 t_6^5 +  y_{s}^6 t_1^6 t_2^6 t_3^6 t_4^6 t_5^7 t_6^5 +  y_{s}^6 t_1^5 t_2^7 t_3^6 t_4^6 t_5^7 t_6^5 +  y_{s}^6 t_1^6 t_2^6 t_3^5 t_4^7 t_5^7 t_6^5 - y_{s}^3 t_1^3 t_2^3 t_3^3 t_4^3 t_6^6 +  y_{s}^4 t_1^5 t_2^3 t_3^5 t_4^3 t_5^2 t_6^6 +  y_{s}^4 t_1^3 t_2^5 t_3^5 t_4^3 t_5^2 t_6^6 -  y_{s}^4 t_1^4 t_2^4 t_3^4 t_4^4 t_5^2 t_6^6 +  y_{s}^4 t_1^5 t_2^3 t_3^3 t_4^5 t_5^2 t_6^6 +  y_{s}^4 t_1^3 t_2^5 t_3^3 t_4^5 t_5^2 t_6^6 -  y_{s}^5 t_1^5 t_2^5 t_3^7 t_4^3 t_5^4 t_6^6 -  y_{s}^5 t_1^5 t_2^5 t_3^6 t_4^4 t_5^4 t_6^6 -  y_{s}^5 t_1^7 t_2^3 t_3^5 t_4^5 t_5^4 t_6^6 -  y_{s}^5 t_1^6 t_2^4 t_3^5 t_4^5 t_5^4 t_6^6 -  3 y_{s}^5 t_1^5 t_2^5 t_3^5 t_4^5 t_5^4 t_6^6 -  y_{s}^5 t_1^4 t_2^6 t_3^5 t_4^5 t_5^4 t_6^6 -  y_{s}^5 t_1^3 t_2^7 t_3^5 t_4^5 t_5^4 t_6^6 -  y_{s}^5 t_1^5 t_2^5 t_3^4 t_4^6 t_5^4 t_6^6 -  y_{s}^5 t_1^5 t_2^5 t_3^3 t_4^7 t_5^4 t_6^6 +  y_{s}^6 t_1^7 t_2^5 t_3^7 t_4^5 t_5^6 t_6^6 +  y_{s}^6 t_1^6 t_2^6 t_3^7 t_4^5 t_5^6 t_6^6 +  y_{s}^6 t_1^5 t_2^7 t_3^7 t_4^5 t_5^6 t_6^6 +  y_{s}^6 t_1^7 t_2^5 t_3^6 t_4^6 t_5^6 t_6^6 +  y_{s}^6 t_1^6 t_2^6 t_3^6 t_4^6 t_5^6 t_6^6 +  y_{s}^6 t_1^5 t_2^7 t_3^6 t_4^6 t_5^6 t_6^6 +  y_{s}^6 t_1^7 t_2^5 t_3^5 t_4^7 t_5^6 t_6^6 +  y_{s}^6 t_1^6 t_2^6 t_3^5 t_4^7 t_5^6 t_6^6 +  y_{s}^6 t_1^5 t_2^7 t_3^5 t_4^7 t_5^6 t_6^6 - y_{s}^4 t_1^4 t_2^4 t_3^4 t_4^4 t_5 t_6^7 -  y_{s}^5 t_1^5 t_2^5 t_3^6 t_4^4 t_5^3 t_6^7 -  y_{s}^5 t_1^6 t_2^4 t_3^5 t_4^5 t_5^3 t_6^7 -  y_{s}^5 t_1^5 t_2^5 t_3^5 t_4^5 t_5^3 t_6^7 -  y_{s}^5 t_1^4 t_2^6 t_3^5 t_4^5 t_5^3 t_6^7 -  y_{s}^5 t_1^5 t_2^5 t_3^4 t_4^6 t_5^3 t_6^7 +  y_{s}^6 t_1^6 t_2^6 t_3^7 t_4^5 t_5^5 t_6^7 +  y_{s}^6 t_1^7 t_2^5 t_3^6 t_4^6 t_5^5 t_6^7 +  y_{s}^6 t_1^6 t_2^6 t_3^6 t_4^6 t_5^5 t_6^7 +  y_{s}^6 t_1^5 t_2^7 t_3^6 t_4^6 t_5^5 t_6^7 +  y_{s}^6 t_1^6 t_2^6 t_3^5 t_4^7 t_5^5 t_6^7 + y_{s}^7 t_1^7 t_2^7 t_3^7 t_4^7 t_5^7 t_6^7
~,~
$
\end{quote}
\endgroup

\subsection{Model 13 \label{app_num_13}}

\begingroup\makeatletter\def\f@size{7}\check@mathfonts
\begin{quote}\raggedright
$
P(t_i,y_s,y_{o_1},y_{o_2},y_{o_3},y_{o_4}; \mathcal{M}_{13}) =
1 + y_{s} y_{o_1}^2 y_{o_2} y_{o_3}^3 y_{o_4}^3 t_1^2 t_2 t_3 t_4^2 t_6^2 +  y_{s} y_{o_1}^2 y_{o_2} y_{o_3}^3 y_{o_4}^3 t_1 t_2^2 t_3 t_4^2 t_6^2 +  y_{s} y_{o_1}^3 y_{o_2}^2 y_{o_3}^3 y_{o_4}^2 t_1^2 t_3^2 t_4 t_5 t_6^2 +  y_{s} y_{o_1}^3 y_{o_2}^2 y_{o_3}^3 y_{o_4}^2 t_1 t_2 t_3^2 t_4 t_5 t_6^2 +  y_{s} y_{o_1}^3 y_{o_2}^2 y_{o_3}^3 y_{o_4}^2 t_2^2 t_3^2 t_4 t_5 t_6^2 -  y_{s}^2 y_{o_1}^6 y_{o_2}^4 y_{o_3}^6 y_{o_4}^4 t_1^3 t_2 t_3^4 t_4^2 t_5^2 t_6^4 -  y_{s}^2 y_{o_1}^6 y_{o_2}^4 y_{o_3}^6 y_{o_4}^4 t_1^2 t_2^2 t_3^4 t_4^2 t_5^2 t_6^4 -  y_{s}^2 y_{o_1}^6 y_{o_2}^4 y_{o_3}^6 y_{o_4}^4 t_1 t_2^3 t_3^4 t_4^2 t_5^2 t_6^4 -  y_{s}^2 y_{o_1}^7 y_{o_2}^5 y_{o_3}^6 y_{o_4}^3 t_1^2 t_2 t_3^5 t_4 t_5^3 t_6^4 -  y_{s}^2 y_{o_1}^7 y_{o_2}^5 y_{o_3}^6 y_{o_4}^3 t_1 t_2^2 t_3^5 t_4 t_5^3 t_6^4 -  y_{s}^3 y_{o_1}^9 y_{o_2}^6 y_{o_3}^9 y_{o_4}^6 t_1^3 t_2^3 t_3^6 t_4^3 t_5^3 t_6^6 +  y_{s} y_{o_1} y_{o_2} y_{o_3}^2 y_{o_4}^3 t_1^2 t_2 t_4^2 t_6 t_7 +  y_{s} y_{o_1} y_{o_2} y_{o_3}^2 y_{o_4}^3 t_1 t_2^2 t_4^2 t_6 t_7 +  y_{s} y_{o_1}^2 y_{o_2}^2 y_{o_3}^2 y_{o_4}^2 t_1^2 t_3 t_4 t_5 t_6 t_7 +  y_{s} y_{o_1}^2 y_{o_2}^2 y_{o_3}^2 y_{o_4}^2 t_1 t_2 t_3 t_4 t_5 t_6 t_7 +  y_{s} y_{o_1}^2 y_{o_2}^2 y_{o_3}^2 y_{o_4}^2 t_2^2 t_3 t_4 t_5 t_6 t_7 +  y_{s} y_{o_1}^3 y_{o_2}^3 y_{o_3}^2 y_{o_4} t_1 t_3^2 t_5^2 t_6 t_7 +  y_{s} y_{o_1}^3 y_{o_2}^3 y_{o_3}^2 y_{o_4} t_2 t_3^2 t_5^2 t_6 t_7 -  y_{s}^2 y_{o_1}^3 y_{o_2}^2 y_{o_3}^5 y_{o_4}^6 t_1^5 t_2 t_3 t_4^4 t_6^3 t_7 -  y_{s}^2 y_{o_1}^3 y_{o_2}^2 y_{o_3}^5 y_{o_4}^6 t_1^4 t_2^2 t_3 t_4^4 t_6^3 t_7 -  y_{s}^2 y_{o_1}^3 y_{o_2}^2 y_{o_3}^5 y_{o_4}^6 t_1^3 t_2^3 t_3 t_4^4 t_6^3 t_7 -  y_{s}^2 y_{o_1}^3 y_{o_2}^2 y_{o_3}^5 y_{o_4}^6 t_1^2 t_2^4 t_3 t_4^4 t_6^3 t_7 -  y_{s}^2 y_{o_1}^3 y_{o_2}^2 y_{o_3}^5 y_{o_4}^6 t_1 t_2^5 t_3 t_4^4 t_6^3 t_7 -  y_{s}^2 y_{o_1}^4 y_{o_2}^3 y_{o_3}^5 y_{o_4}^5 t_1^5 t_3^2 t_4^3 t_5 t_6^3 t_7 -  y_{s}^2 y_{o_1}^4 y_{o_2}^3 y_{o_3}^5 y_{o_4}^5 t_1^4 t_2 t_3^2 t_4^3 t_5 t_6^3 t_7 -  y_{s}^2 y_{o_1}^4 y_{o_2}^3 y_{o_3}^5 y_{o_4}^5 t_1^3 t_2^2 t_3^2 t_4^3 t_5 t_6^3 t_7 -  y_{s}^2 y_{o_1}^4 y_{o_2}^3 y_{o_3}^5 y_{o_4}^5 t_1^2 t_2^3 t_3^2 t_4^3 t_5 t_6^3 t_7 -  y_{s}^2 y_{o_1}^4 y_{o_2}^3 y_{o_3}^5 y_{o_4}^5 t_1 t_2^4 t_3^2 t_4^3 t_5 t_6^3 t_7 -  y_{s}^2 y_{o_1}^4 y_{o_2}^3 y_{o_3}^5 y_{o_4}^5 t_2^5 t_3^2 t_4^3 t_5 t_6^3 t_7 -  y_{s}^2 y_{o_1}^5 y_{o_2}^4 y_{o_3}^5 y_{o_4}^4 t_1^4 t_3^3 t_4^2 t_5^2 t_6^3 t_7 -  2 y_{s}^2 y_{o_1}^5 y_{o_2}^4 y_{o_3}^5 y_{o_4}^4 t_1^3 t_2 t_3^3 t_4^2 t_5^2 t_6^3 t_7 -  y_{s}^2 y_{o_1}^5 y_{o_2}^4 y_{o_3}^5 y_{o_4}^4 t_1^2 t_2^2 t_3^3 t_4^2 t_5^2 t_6^3 t_7 -  2 y_{s}^2 y_{o_1}^5 y_{o_2}^4 y_{o_3}^5 y_{o_4}^4 t_1 t_2^3 t_3^3 t_4^2 t_5^2 t_6^3 t_7 -  y_{s}^2 y_{o_1}^5 y_{o_2}^4 y_{o_3}^5 y_{o_4}^4 t_2^4 t_3^3 t_4^2 t_5^2 t_6^3 t_7 -  y_{s}^2 y_{o_1}^6 y_{o_2}^5 y_{o_3}^5 y_{o_4}^3 t_1^2 t_2 t_3^4 t_4 t_5^3 t_6^3 t_7 -  y_{s}^2 y_{o_1}^6 y_{o_2}^5 y_{o_3}^5 y_{o_4}^3 t_1 t_2^2 t_3^4 t_4 t_5^3 t_6^3 t_7 -  y_{s}^2 y_{o_1}^7 y_{o_2}^6 y_{o_3}^5 y_{o_4}^2 t_1 t_2 t_3^5 t_5^4 t_6^3 t_7 +  y_{s}^3 y_{o_1}^7 y_{o_2}^5 y_{o_3}^8 y_{o_4}^7 t_1^6 t_2 t_3^4 t_4^4 t_5^2 t_6^5 t_7 +  y_{s}^3 y_{o_1}^7 y_{o_2}^5 y_{o_3}^8 y_{o_4}^7 t_1^5 t_2^2 t_3^4 t_4^4 t_5^2 t_6^5 t_7 +  2 y_{s}^3 y_{o_1}^7 y_{o_2}^5 y_{o_3}^8 y_{o_4}^7 t_1^4 t_2^3 t_3^4 t_4^4 t_5^2 t_6^5 t_7 +  2 y_{s}^3 y_{o_1}^7 y_{o_2}^5 y_{o_3}^8 y_{o_4}^7 t_1^3 t_2^4 t_3^4 t_4^4 t_5^2 t_6^5 t_7 +  y_{s}^3 y_{o_1}^7 y_{o_2}^5 y_{o_3}^8 y_{o_4}^7 t_1^2 t_2^5 t_3^4 t_4^4 t_5^2 t_6^5 t_7 +  y_{s}^3 y_{o_1}^7 y_{o_2}^5 y_{o_3}^8 y_{o_4}^7 t_1 t_2^6 t_3^4 t_4^4 t_5^2 t_6^5 t_7 +  y_{s}^3 y_{o_1}^8 y_{o_2}^6 y_{o_3}^8 y_{o_4}^6 t_1^5 t_2 t_3^5 t_4^3 t_5^3 t_6^5 t_7 +  y_{s}^3 y_{o_1}^8 y_{o_2}^6 y_{o_3}^8 y_{o_4}^6 t_1^4 t_2^2 t_3^5 t_4^3 t_5^3 t_6^5 t_7 -  y_{s}^3 y_{o_1}^8 y_{o_2}^6 y_{o_3}^8 y_{o_4}^6 t_1^3 t_2^3 t_3^5 t_4^3 t_5^3 t_6^5 t_7 +  y_{s}^3 y_{o_1}^8 y_{o_2}^6 y_{o_3}^8 y_{o_4}^6 t_1^2 t_2^4 t_3^5 t_4^3 t_5^3 t_6^5 t_7 +  y_{s}^3 y_{o_1}^8 y_{o_2}^6 y_{o_3}^8 y_{o_4}^6 t_1 t_2^5 t_3^5 t_4^3 t_5^3 t_6^5 t_7 +  y_{s}^3 y_{o_1}^9 y_{o_2}^7 y_{o_3}^8 y_{o_4}^5 t_1^4 t_2 t_3^6 t_4^2 t_5^4 t_6^5 t_7 +  y_{s}^3 y_{o_1}^9 y_{o_2}^7 y_{o_3}^8 y_{o_4}^5 t_1 t_2^4 t_3^6 t_4^2 t_5^4 t_6^5 t_7 +  y_{s}^4 y_{o_1}^{10} y_{o_2}^7 y_{o_3}^{11} y_{o_4}^9 t_1^6 t_2^3 t_3^6 t_4^5 t_5^3 t_6^7 t_7 +  y_{s}^4 y_{o_1}^{10} y_{o_2}^7 y_{o_3}^{11} y_{o_4}^9 t_1^3 t_2^6 t_3^6 t_4^5 t_5^3 t_6^7 t_7 -  y_{s}^4 y_{o_1}^{11} y_{o_2}^8 y_{o_3}^{11} y_{o_4}^8 t_1^4 t_2^4 t_3^7 t_4^4 t_5^4 t_6^7 t_7 +  y_{s} y_{o_1} y_{o_2}^2 y_{o_3} y_{o_4}^2 t_1 t_2 t_4 t_5 t_7^2 -  y_{s}^2 y_{o_1}^3 y_{o_2}^3 y_{o_3}^4 y_{o_4}^5 t_1^5 t_3 t_4^3 t_5 t_6^2 t_7^2 -  2 y_{s}^2 y_{o_1}^3 y_{o_2}^3 y_{o_3}^4 y_{o_4}^5 t_1^4 t_2 t_3 t_4^3 t_5 t_6^2 t_7^2 -  2 y_{s}^2 y_{o_1}^3 y_{o_2}^3 y_{o_3}^4 y_{o_4}^5 t_1^3 t_2^2 t_3 t_4^3 t_5 t_6^2 t_7^2 -  2 y_{s}^2 y_{o_1}^3 y_{o_2}^3 y_{o_3}^4 y_{o_4}^5 t_1^2 t_2^3 t_3 t_4^3 t_5 t_6^2 t_7^2 -  2 y_{s}^2 y_{o_1}^3 y_{o_2}^3 y_{o_3}^4 y_{o_4}^5 t_1 t_2^4 t_3 t_4^3 t_5 t_6^2 t_7^2 -  y_{s}^2 y_{o_1}^3 y_{o_2}^3 y_{o_3}^4 y_{o_4}^5 t_2^5 t_3 t_4^3 t_5 t_6^2 t_7^2 -  2 y_{s}^2 y_{o_1}^4 y_{o_2}^4 y_{o_3}^4 y_{o_4}^4 t_1^4 t_3^2 t_4^2 t_5^2 t_6^2 t_7^2 -  3 y_{s}^2 y_{o_1}^4 y_{o_2}^4 y_{o_3}^4 y_{o_4}^4 t_1^3 t_2 t_3^2 t_4^2 t_5^2 t_6^2 t_7^2 -  3 y_{s}^2 y_{o_1}^4 y_{o_2}^4 y_{o_3}^4 y_{o_4}^4 t_1^2 t_2^2 t_3^2 t_4^2 t_5^2 t_6^2 t_7^2 -  3 y_{s}^2 y_{o_1}^4 y_{o_2}^4 y_{o_3}^4 y_{o_4}^4 t_1 t_2^3 t_3^2 t_4^2 t_5^2 t_6^2 t_7^2 -  2 y_{s}^2 y_{o_1}^4 y_{o_2}^4 y_{o_3}^4 y_{o_4}^4 t_2^4 t_3^2 t_4^2 t_5^2 t_6^2 t_7^2 -  y_{s}^2 y_{o_1}^5 y_{o_2}^5 y_{o_3}^4 y_{o_4}^3 t_1^3 t_3^3 t_4 t_5^3 t_6^2 t_7^2 -  3 y_{s}^2 y_{o_1}^5 y_{o_2}^5 y_{o_3}^4 y_{o_4}^3 t_1^2 t_2 t_3^3 t_4 t_5^3 t_6^2 t_7^2 -  3 y_{s}^2 y_{o_1}^5 y_{o_2}^5 y_{o_3}^4 y_{o_4}^3 t_1 t_2^2 t_3^3 t_4 t_5^3 t_6^2 t_7^2 -  y_{s}^2 y_{o_1}^5 y_{o_2}^5 y_{o_3}^4 y_{o_4}^3 t_2^3 t_3^3 t_4 t_5^3 t_6^2 t_7^2 -  y_{s}^2 y_{o_1}^6 y_{o_2}^6 y_{o_3}^4 y_{o_4}^2 t_1 t_2 t_3^4 t_5^4 t_6^2 t_7^2 +  y_{s}^3 y_{o_1}^5 y_{o_2}^4 y_{o_3}^7 y_{o_4}^8 t_1^5 t_2^3 t_3^2 t_4^5 t_5 t_6^4 t_7^2 +  y_{s}^3 y_{o_1}^5 y_{o_2}^4 y_{o_3}^7 y_{o_4}^8 t_1^4 t_2^4 t_3^2 t_4^5 t_5 t_6^4 t_7^2 +  y_{s}^3 y_{o_1}^5 y_{o_2}^4 y_{o_3}^7 y_{o_4}^8 t_1^3 t_2^5 t_3^2 t_4^5 t_5 t_6^4 t_7^2 -  y_{s}^3 y_{o_1}^6 y_{o_2}^5 y_{o_3}^7 y_{o_4}^7 t_1^5 t_2^2 t_3^3 t_4^4 t_5^2 t_6^4 t_7^2 +  y_{s}^3 y_{o_1}^6 y_{o_2}^5 y_{o_3}^7 y_{o_4}^7 t_1^4 t_2^3 t_3^3 t_4^4 t_5^2 t_6^4 t_7^2 +  y_{s}^3 y_{o_1}^6 y_{o_2}^5 y_{o_3}^7 y_{o_4}^7 t_1^3 t_2^4 t_3^3 t_4^4 t_5^2 t_6^4 t_7^2 -  y_{s}^3 y_{o_1}^6 y_{o_2}^5 y_{o_3}^7 y_{o_4}^7 t_1^2 t_2^5 t_3^3 t_4^4 t_5^2 t_6^4 t_7^2 +  2 y_{s}^3 y_{o_1}^7 y_{o_2}^6 y_{o_3}^7 y_{o_4}^6 t_1^5 t_2 t_3^4 t_4^3 t_5^3 t_6^4 t_7^2 +  2 y_{s}^3 y_{o_1}^7 y_{o_2}^6 y_{o_3}^7 y_{o_4}^6 t_1^4 t_2^2 t_3^4 t_4^3 t_5^3 t_6^4 t_7^2 +  y_{s}^3 y_{o_1}^7 y_{o_2}^6 y_{o_3}^7 y_{o_4}^6 t_1^3 t_2^3 t_3^4 t_4^3 t_5^3 t_6^4 t_7^2 +  2 y_{s}^3 y_{o_1}^7 y_{o_2}^6 y_{o_3}^7 y_{o_4}^6 t_1^2 t_2^4 t_3^4 t_4^3 t_5^3 t_6^4 t_7^2 +  2 y_{s}^3 y_{o_1}^7 y_{o_2}^6 y_{o_3}^7 y_{o_4}^6 t_1 t_2^5 t_3^4 t_4^3 t_5^3 t_6^4 t_7^2 +  3 y_{s}^3 y_{o_1}^8 y_{o_2}^7 y_{o_3}^7 y_{o_4}^5 t_1^4 t_2 t_3^5 t_4^2 t_5^4 t_6^4 t_7^2 +  2 y_{s}^3 y_{o_1}^8 y_{o_2}^7 y_{o_3}^7 y_{o_4}^5 t_1^3 t_2^2 t_3^5 t_4^2 t_5^4 t_6^4 t_7^2 +  2 y_{s}^3 y_{o_1}^8 y_{o_2}^7 y_{o_3}^7 y_{o_4}^5 t_1^2 t_2^3 t_3^5 t_4^2 t_5^4 t_6^4 t_7^2 +  3 y_{s}^3 y_{o_1}^8 y_{o_2}^7 y_{o_3}^7 y_{o_4}^5 t_1 t_2^4 t_3^5 t_4^2 t_5^4 t_6^4 t_7^2 +  y_{s}^3 y_{o_1}^9 y_{o_2}^8 y_{o_3}^7 y_{o_4}^4 t_1^3 t_2 t_3^6 t_4 t_5^5 t_6^4 t_7^2 +  2 y_{s}^3 y_{o_1}^9 y_{o_2}^8 y_{o_3}^7 y_{o_4}^4 t_1^2 t_2^2 t_3^6 t_4 t_5^5 t_6^4 t_7^2 +  y_{s}^3 y_{o_1}^9 y_{o_2}^8 y_{o_3}^7 y_{o_4}^4 t_1 t_2^3 t_3^6 t_4 t_5^5 t_6^4 t_7^2 +  y_{s}^4 y_{o_1}^8 y_{o_2}^6 y_{o_3}^{10} y_{o_4}^{10} t_1^5 t_2^5 t_3^4 t_4^6 t_5^2 t_6^6 t_7^2 +  y_{s}^4 y_{o_1}^9 y_{o_2}^7 y_{o_3}^{10} y_{o_4}^9 t_1^6 t_2^3 t_3^5 t_4^5 t_5^3 t_6^6 t_7^2 -  y_{s}^4 y_{o_1}^9 y_{o_2}^7 y_{o_3}^{10} y_{o_4}^9 t_1^5 t_2^4 t_3^5 t_4^5 t_5^3 t_6^6 t_7^2 -  y_{s}^4 y_{o_1}^9 y_{o_2}^7 y_{o_3}^{10} y_{o_4}^9 t_1^4 t_2^5 t_3^5 t_4^5 t_5^3 t_6^6 t_7^2 +  y_{s}^4 y_{o_1}^9 y_{o_2}^7 y_{o_3}^{10} y_{o_4}^9 t_1^3 t_2^6 t_3^5 t_4^5 t_5^3 t_6^6 t_7^2 +  y_{s}^4 y_{o_1}^{10} y_{o_2}^8 y_{o_3}^{10} y_{o_4}^8 t_1^6 t_2^2 t_3^6 t_4^4 t_5^4 t_6^6 t_7^2 +  2 y_{s}^4 y_{o_1}^{10} y_{o_2}^8 y_{o_3}^{10} y_{o_4}^8 t_1^5 t_2^3 t_3^6 t_4^4 t_5^4 t_6^6 t_7^2 -  y_{s}^4 y_{o_1}^{10} y_{o_2}^8 y_{o_3}^{10} y_{o_4}^8 t_1^4 t_2^4 t_3^6 t_4^4 t_5^4 t_6^6 t_7^2 +  2 y_{s}^4 y_{o_1}^{10} y_{o_2}^8 y_{o_3}^{10} y_{o_4}^8 t_1^3 t_2^5 t_3^6 t_4^4 t_5^4 t_6^6 t_7^2 +  y_{s}^4 y_{o_1}^{10} y_{o_2}^8 y_{o_3}^{10} y_{o_4}^8 t_1^2 t_2^6 t_3^6 t_4^4 t_5^4 t_6^6 t_7^2 +  y_{s}^4 y_{o_1}^{11} y_{o_2}^9 y_{o_3}^{10} y_{o_4}^7 t_1^4 t_2^3 t_3^7 t_4^3 t_5^5 t_6^6 t_7^2 +  y_{s}^4 y_{o_1}^{11} y_{o_2}^9 y_{o_3}^{10} y_{o_4}^7 t_1^3 t_2^4 t_3^7 t_4^3 t_5^5 t_6^6 t_7^2 -  y_{s}^5 y_{o_1}^{11} y_{o_2}^8 y_{o_3}^{13} y_{o_4}^{12} t_1^6 t_2^6 t_3^6 t_4^7 t_5^3 t_6^8 t_7^2 -  y_{s}^5 y_{o_1}^{12} y_{o_2}^9 y_{o_3}^{13} y_{o_4}^{11} t_1^6 t_2^5 t_3^7 t_4^6 t_5^4 t_6^8 t_7^2 -  y_{s}^5 y_{o_1}^{12} y_{o_2}^9 y_{o_3}^{13} y_{o_4}^{11} t_1^5 t_2^6 t_3^7 t_4^6 t_5^4 t_6^8 t_7^2 -  y_{s}^2 y_{o_1}^2 y_{o_2}^3 y_{o_3}^3 y_{o_4}^5 t_1^4 t_2 t_4^3 t_5 t_6 t_7^3 -  y_{s}^2 y_{o_1}^2 y_{o_2}^3 y_{o_3}^3 y_{o_4}^5 t_1^3 t_2^2 t_4^3 t_5 t_6 t_7^3 -  y_{s}^2 y_{o_1}^2 y_{o_2}^3 y_{o_3}^3 y_{o_4}^5 t_1^2 t_2^3 t_4^3 t_5 t_6 t_7^3 -  y_{s}^2 y_{o_1}^2 y_{o_2}^3 y_{o_3}^3 y_{o_4}^5 t_1 t_2^4 t_4^3 t_5 t_6 t_7^3 -  y_{s}^2 y_{o_1}^3 y_{o_2}^4 y_{o_3}^3 y_{o_4}^4 t_1^4 t_3 t_4^2 t_5^2 t_6 t_7^3 -  2 y_{s}^2 y_{o_1}^3 y_{o_2}^4 y_{o_3}^3 y_{o_4}^4 t_1^3 t_2 t_3 t_4^2 t_5^2 t_6 t_7^3 -  3 y_{s}^2 y_{o_1}^3 y_{o_2}^4 y_{o_3}^3 y_{o_4}^4 t_1^2 t_2^2 t_3 t_4^2 t_5^2 t_6 t_7^3 -  2 y_{s}^2 y_{o_1}^3 y_{o_2}^4 y_{o_3}^3 y_{o_4}^4 t_1 t_2^3 t_3 t_4^2 t_5^2 t_6 t_7^3 -  y_{s}^2 y_{o_1}^3 y_{o_2}^4 y_{o_3}^3 y_{o_4}^4 t_2^4 t_3 t_4^2 t_5^2 t_6 t_7^3 -  y_{s}^2 y_{o_1}^4 y_{o_2}^5 y_{o_3}^3 y_{o_4}^3 t_1^3 t_3^2 t_4 t_5^3 t_6 t_7^3 -  2 y_{s}^2 y_{o_1}^4 y_{o_2}^5 y_{o_3}^3 y_{o_4}^3 t_1^2 t_2 t_3^2 t_4 t_5^3 t_6 t_7^3 -  2 y_{s}^2 y_{o_1}^4 y_{o_2}^5 y_{o_3}^3 y_{o_4}^3 t_1 t_2^2 t_3^2 t_4 t_5^3 t_6 t_7^3 -  y_{s}^2 y_{o_1}^4 y_{o_2}^5 y_{o_3}^3 y_{o_4}^3 t_2^3 t_3^2 t_4 t_5^3 t_6 t_7^3 -  y_{s}^2 y_{o_1}^5 y_{o_2}^6 y_{o_3}^3 y_{o_4}^2 t_1 t_2 t_3^3 t_5^4 t_6 t_7^3 +  y_{s}^3 y_{o_1}^4 y_{o_2}^4 y_{o_3}^6 y_{o_4}^8 t_1^7 t_2 t_3 t_4^5 t_5 t_6^3 t_7^3 +  y_{s}^3 y_{o_1}^4 y_{o_2}^4 y_{o_3}^6 y_{o_4}^8 t_1^6 t_2^2 t_3 t_4^5 t_5 t_6^3 t_7^3 +  2 y_{s}^3 y_{o_1}^4 y_{o_2}^4 y_{o_3}^6 y_{o_4}^8 t_1^5 t_2^3 t_3 t_4^5 t_5 t_6^3 t_7^3 +  2 y_{s}^3 y_{o_1}^4 y_{o_2}^4 y_{o_3}^6 y_{o_4}^8 t_1^4 t_2^4 t_3 t_4^5 t_5 t_6^3 t_7^3 +  2 y_{s}^3 y_{o_1}^4 y_{o_2}^4 y_{o_3}^6 y_{o_4}^8 t_1^3 t_2^5 t_3 t_4^5 t_5 t_6^3 t_7^3 +  y_{s}^3 y_{o_1}^4 y_{o_2}^4 y_{o_3}^6 y_{o_4}^8 t_1^2 t_2^6 t_3 t_4^5 t_5 t_6^3 t_7^3 +  y_{s}^3 y_{o_1}^4 y_{o_2}^4 y_{o_3}^6 y_{o_4}^8 t_1 t_2^7 t_3 t_4^5 t_5 t_6^3 t_7^3 +  y_{s}^3 y_{o_1}^5 y_{o_2}^5 y_{o_3}^6 y_{o_4}^7 t_1^7 t_3^2 t_4^4 t_5^2 t_6^3 t_7^3 +  2 y_{s}^3 y_{o_1}^5 y_{o_2}^5 y_{o_3}^6 y_{o_4}^7 t_1^6 t_2 t_3^2 t_4^4 t_5^2 t_6^3 t_7^3 +  3 y_{s}^3 y_{o_1}^5 y_{o_2}^5 y_{o_3}^6 y_{o_4}^7 t_1^5 t_2^2 t_3^2 t_4^4 t_5^2 t_6^3 t_7^3 +  4 y_{s}^3 y_{o_1}^5 y_{o_2}^5 y_{o_3}^6 y_{o_4}^7 t_1^4 t_2^3 t_3^2 t_4^4 t_5^2 t_6^3 t_7^3 +  4 y_{s}^3 y_{o_1}^5 y_{o_2}^5 y_{o_3}^6 y_{o_4}^7 t_1^3 t_2^4 t_3^2 t_4^4 t_5^2 t_6^3 t_7^3 +  3 y_{s}^3 y_{o_1}^5 y_{o_2}^5 y_{o_3}^6 y_{o_4}^7 t_1^2 t_2^5 t_3^2 t_4^4 t_5^2 t_6^3 t_7^3 +  2 y_{s}^3 y_{o_1}^5 y_{o_2}^5 y_{o_3}^6 y_{o_4}^7 t_1 t_2^6 t_3^2 t_4^4 t_5^2 t_6^3 t_7^3 +  y_{s}^3 y_{o_1}^5 y_{o_2}^5 y_{o_3}^6 y_{o_4}^7 t_2^7 t_3^2 t_4^4 t_5^2 t_6^3 t_7^3 +  y_{s}^3 y_{o_1}^6 y_{o_2}^6 y_{o_3}^6 y_{o_4}^6 t_1^6 t_3^3 t_4^3 t_5^3 t_6^3 t_7^3 +  3 y_{s}^3 y_{o_1}^6 y_{o_2}^6 y_{o_3}^6 y_{o_4}^6 t_1^5 t_2 t_3^3 t_4^3 t_5^3 t_6^3 t_7^3 +  3 y_{s}^3 y_{o_1}^6 y_{o_2}^6 y_{o_3}^6 y_{o_4}^6 t_1^4 t_2^2 t_3^3 t_4^3 t_5^3 t_6^3 t_7^3 +  3 y_{s}^3 y_{o_1}^6 y_{o_2}^6 y_{o_3}^6 y_{o_4}^6 t_1^3 t_2^3 t_3^3 t_4^3 t_5^3 t_6^3 t_7^3 +  3 y_{s}^3 y_{o_1}^6 y_{o_2}^6 y_{o_3}^6 y_{o_4}^6 t_1^2 t_2^4 t_3^3 t_4^3 t_5^3 t_6^3 t_7^3 +  3 y_{s}^3 y_{o_1}^6 y_{o_2}^6 y_{o_3}^6 y_{o_4}^6 t_1 t_2^5 t_3^3 t_4^3 t_5^3 t_6^3 t_7^3 +  y_{s}^3 y_{o_1}^6 y_{o_2}^6 y_{o_3}^6 y_{o_4}^6 t_2^6 t_3^3 t_4^3 t_5^3 t_6^3 t_7^3 +  3 y_{s}^3 y_{o_1}^7 y_{o_2}^7 y_{o_3}^6 y_{o_4}^5 t_1^4 t_2 t_3^4 t_4^2 t_5^4 t_6^3 t_7^3 +  2 y_{s}^3 y_{o_1}^7 y_{o_2}^7 y_{o_3}^6 y_{o_4}^5 t_1^3 t_2^2 t_3^4 t_4^2 t_5^4 t_6^3 t_7^3 +  2 y_{s}^3 y_{o_1}^7 y_{o_2}^7 y_{o_3}^6 y_{o_4}^5 t_1^2 t_2^3 t_3^4 t_4^2 t_5^4 t_6^3 t_7^3 +  3 y_{s}^3 y_{o_1}^7 y_{o_2}^7 y_{o_3}^6 y_{o_4}^5 t_1 t_2^4 t_3^4 t_4^2 t_5^4 t_6^3 t_7^3 +  y_{s}^3 y_{o_1}^8 y_{o_2}^8 y_{o_3}^6 y_{o_4}^4 t_1^3 t_2 t_3^5 t_4 t_5^5 t_6^3 t_7^3 +  y_{s}^3 y_{o_1}^8 y_{o_2}^8 y_{o_3}^6 y_{o_4}^4 t_1^2 t_2^2 t_3^5 t_4 t_5^5 t_6^3 t_7^3 +  y_{s}^3 y_{o_1}^8 y_{o_2}^8 y_{o_3}^6 y_{o_4}^4 t_1 t_2^3 t_3^5 t_4 t_5^5 t_6^3 t_7^3 +  y_{s}^4 y_{o_1}^7 y_{o_2}^6 y_{o_3}^9 y_{o_4}^{10} t_1^5 t_2^5 t_3^3 t_4^6 t_5^2 t_6^5 t_7^3 -  y_{s}^4 y_{o_1}^8 y_{o_2}^7 y_{o_3}^9 y_{o_4}^9 t_1^8 t_2 t_3^4 t_4^5 t_5^3 t_6^5 t_7^3 -  y_{s}^4 y_{o_1}^8 y_{o_2}^7 y_{o_3}^9 y_{o_4}^9 t_1^7 t_2^2 t_3^4 t_4^5 t_5^3 t_6^5 t_7^3 -  y_{s}^4 y_{o_1}^8 y_{o_2}^7 y_{o_3}^9 y_{o_4}^9 t_1^6 t_2^3 t_3^4 t_4^5 t_5^3 t_6^5 t_7^3 -  2 y_{s}^4 y_{o_1}^8 y_{o_2}^7 y_{o_3}^9 y_{o_4}^9 t_1^5 t_2^4 t_3^4 t_4^5 t_5^3 t_6^5 t_7^3 -  2 y_{s}^4 y_{o_1}^8 y_{o_2}^7 y_{o_3}^9 y_{o_4}^9 t_1^4 t_2^5 t_3^4 t_4^5 t_5^3 t_6^5 t_7^3 -  y_{s}^4 y_{o_1}^8 y_{o_2}^7 y_{o_3}^9 y_{o_4}^9 t_1^3 t_2^6 t_3^4 t_4^5 t_5^3 t_6^5 t_7^3 -  y_{s}^4 y_{o_1}^8 y_{o_2}^7 y_{o_3}^9 y_{o_4}^9 t_1^2 t_2^7 t_3^4 t_4^5 t_5^3 t_6^5 t_7^3 -  y_{s}^4 y_{o_1}^8 y_{o_2}^7 y_{o_3}^9 y_{o_4}^9 t_1 t_2^8 t_3^4 t_4^5 t_5^3 t_6^5 t_7^3 -  2 y_{s}^4 y_{o_1}^9 y_{o_2}^8 y_{o_3}^9 y_{o_4}^8 t_1^7 t_2 t_3^5 t_4^4 t_5^4 t_6^5 t_7^3 -  2 y_{s}^4 y_{o_1}^9 y_{o_2}^8 y_{o_3}^9 y_{o_4}^8 t_1^6 t_2^2 t_3^5 t_4^4 t_5^4 t_6^5 t_7^3 -  y_{s}^4 y_{o_1}^9 y_{o_2}^8 y_{o_3}^9 y_{o_4}^8 t_1^5 t_2^3 t_3^5 t_4^4 t_5^4 t_6^5 t_7^3 -  5 y_{s}^4 y_{o_1}^9 y_{o_2}^8 y_{o_3}^9 y_{o_4}^8 t_1^4 t_2^4 t_3^5 t_4^4 t_5^4 t_6^5 t_7^3 -  y_{s}^4 y_{o_1}^9 y_{o_2}^8 y_{o_3}^9 y_{o_4}^8 t_1^3 t_2^5 t_3^5 t_4^4 t_5^4 t_6^5 t_7^3 -  2 y_{s}^4 y_{o_1}^9 y_{o_2}^8 y_{o_3}^9 y_{o_4}^8 t_1^2 t_2^6 t_3^5 t_4^4 t_5^4 t_6^5 t_7^3 -  2 y_{s}^4 y_{o_1}^9 y_{o_2}^8 y_{o_3}^9 y_{o_4}^8 t_1 t_2^7 t_3^5 t_4^4 t_5^4 t_6^5 t_7^3 -  y_{s}^4 y_{o_1}^{10} y_{o_2}^9 y_{o_3}^9 y_{o_4}^7 t_1^6 t_2 t_3^6 t_4^3 t_5^5 t_6^5 t_7^3 -  y_{s}^4 y_{o_1}^{10} y_{o_2}^9 y_{o_3}^9 y_{o_4}^7 t_1^5 t_2^2 t_3^6 t_4^3 t_5^5 t_6^5 t_7^3 -  y_{s}^4 y_{o_1}^{10} y_{o_2}^9 y_{o_3}^9 y_{o_4}^7 t_1^2 t_2^5 t_3^6 t_4^3 t_5^5 t_6^5 t_7^3 -  y_{s}^4 y_{o_1}^{10} y_{o_2}^9 y_{o_3}^9 y_{o_4}^7 t_1 t_2^6 t_3^6 t_4^3 t_5^5 t_6^5 t_7^3 -  y_{s}^5 y_{o_1}^{10} y_{o_2}^8 y_{o_3}^{12} y_{o_4}^{12} t_1^6 t_2^6 t_3^5 t_4^7 t_5^3 t_6^7 t_7^3 -  y_{s}^5 y_{o_1}^{11} y_{o_2}^9 y_{o_3}^{12} y_{o_4}^{11} t_1^8 t_2^3 t_3^6 t_4^6 t_5^4 t_6^7 t_7^3 -  2 y_{s}^5 y_{o_1}^{11} y_{o_2}^9 y_{o_3}^{12} y_{o_4}^{11} t_1^6 t_2^5 t_3^6 t_4^6 t_5^4 t_6^7 t_7^3 - 2 y_{s}^5 y_{o_1}^{11} y_{o_2}^9 y_{o_3}^{12} y_{o_4}^{11} t_1^5 t_2^6 t_3^6 t_4^6 t_5^4 t_6^7 t_7^3 - y_{s}^5 y_{o_1}^{11} y_{o_2}^9 y_{o_3}^{12} y_{o_4}^{11} t_1^3 t_2^8 t_3^6 t_4^6 t_5^4 t_6^7 t_7^3 -  y_{s}^5 y_{o_1}^{12} y_{o_2}^{10} y_{o_3}^{12} y_{o_4}^{10} t_1^7 t_2^3 t_3^7 t_4^5 t_5^5 t_6^7 t_7^3 -  y_{s}^5 y_{o_1}^{12} y_{o_2}^{10} y_{o_3}^{12} y_{o_4}^{10} t_1^6 t_2^4 t_3^7 t_4^5 t_5^5 t_6^7 t_7^3 -  y_{s}^5 y_{o_1}^{12} y_{o_2}^{10} y_{o_3}^{12} y_{o_4}^{10} t_1^5 t_2^5 t_3^7 t_4^5 t_5^5 t_6^7 t_7^3 -  y_{s}^5 y_{o_1}^{12} y_{o_2}^{10} y_{o_3}^{12} y_{o_4}^{10} t_1^4 t_2^6 t_3^7 t_4^5 t_5^5 t_6^7 t_7^3 -  y_{s}^5 y_{o_1}^{12} y_{o_2}^{10} y_{o_3}^{12} y_{o_4}^{10} t_1^3 t_2^7 t_3^7 t_4^5 t_5^5 t_6^7 t_7^3 -  y_{s}^2 y_{o_1}^3 y_{o_2}^5 y_{o_3}^2 y_{o_4}^3 t_1^2 t_2 t_3 t_4 t_5^3 t_7^4 -  y_{s}^2 y_{o_1}^3 y_{o_2}^5 y_{o_3}^2 y_{o_4}^3 t_1 t_2^2 t_3 t_4 t_5^3 t_7^4 +  y_{s}^3 y_{o_1}^4 y_{o_2}^5 y_{o_3}^5 y_{o_4}^7 t_1^5 t_2^2 t_3 t_4^4 t_5^2 t_6^2 t_7^4 +  y_{s}^3 y_{o_1}^4 y_{o_2}^5 y_{o_3}^5 y_{o_4}^7 t_1^4 t_2^3 t_3 t_4^4 t_5^2 t_6^2 t_7^4 +  y_{s}^3 y_{o_1}^4 y_{o_2}^5 y_{o_3}^5 y_{o_4}^7 t_1^3 t_2^4 t_3 t_4^4 t_5^2 t_6^2 t_7^4 +  y_{s}^3 y_{o_1}^4 y_{o_2}^5 y_{o_3}^5 y_{o_4}^7 t_1^2 t_2^5 t_3 t_4^4 t_5^2 t_6^2 t_7^4 +  y_{s}^3 y_{o_1}^5 y_{o_2}^6 y_{o_3}^5 y_{o_4}^6 t_1^6 t_3^2 t_4^3 t_5^3 t_6^2 t_7^4 +  3 y_{s}^3 y_{o_1}^5 y_{o_2}^6 y_{o_3}^5 y_{o_4}^6 t_1^5 t_2 t_3^2 t_4^3 t_5^3 t_6^2 t_7^4 +  4 y_{s}^3 y_{o_1}^5 y_{o_2}^6 y_{o_3}^5 y_{o_4}^6 t_1^4 t_2^2 t_3^2 t_4^3 t_5^3 t_6^2 t_7^4 +  4 y_{s}^3 y_{o_1}^5 y_{o_2}^6 y_{o_3}^5 y_{o_4}^6 t_1^3 t_2^3 t_3^2 t_4^3 t_5^3 t_6^2 t_7^4 +  4 y_{s}^3 y_{o_1}^5 y_{o_2}^6 y_{o_3}^5 y_{o_4}^6 t_1^2 t_2^4 t_3^2 t_4^3 t_5^3 t_6^2 t_7^4 +  3 y_{s}^3 y_{o_1}^5 y_{o_2}^6 y_{o_3}^5 y_{o_4}^6 t_1 t_2^5 t_3^2 t_4^3 t_5^3 t_6^2 t_7^4 +  y_{s}^3 y_{o_1}^5 y_{o_2}^6 y_{o_3}^5 y_{o_4}^6 t_2^6 t_3^2 t_4^3 t_5^3 t_6^2 t_7^4 +  2 y_{s}^3 y_{o_1}^6 y_{o_2}^7 y_{o_3}^5 y_{o_4}^5 t_1^4 t_2 t_3^3 t_4^2 t_5^4 t_6^2 t_7^4 +  2 y_{s}^3 y_{o_1}^6 y_{o_2}^7 y_{o_3}^5 y_{o_4}^5 t_1^3 t_2^2 t_3^3 t_4^2 t_5^4 t_6^2 t_7^4 +  2 y_{s}^3 y_{o_1}^6 y_{o_2}^7 y_{o_3}^5 y_{o_4}^5 t_1^2 t_2^3 t_3^3 t_4^2 t_5^4 t_6^2 t_7^4 +  2 y_{s}^3 y_{o_1}^6 y_{o_2}^7 y_{o_3}^5 y_{o_4}^5 t_1 t_2^4 t_3^3 t_4^2 t_5^4 t_6^2 t_7^4 +  2 y_{s}^3 y_{o_1}^7 y_{o_2}^8 y_{o_3}^5 y_{o_4}^4 t_1^3 t_2 t_3^4 t_4 t_5^5 t_6^2 t_7^4 +  2 y_{s}^3 y_{o_1}^7 y_{o_2}^8 y_{o_3}^5 y_{o_4}^4 t_1^2 t_2^2 t_3^4 t_4 t_5^5 t_6^2 t_7^4 +  2 y_{s}^3 y_{o_1}^7 y_{o_2}^8 y_{o_3}^5 y_{o_4}^4 t_1 t_2^3 t_3^4 t_4 t_5^5 t_6^2 t_7^4 -  y_{s}^4 y_{o_1}^6 y_{o_2}^6 y_{o_3}^8 y_{o_4}^{10} t_1^5 t_2^5 t_3^2 t_4^6 t_5^2 t_6^4 t_7^4 +  y_{s}^4 y_{o_1}^7 y_{o_2}^7 y_{o_3}^8 y_{o_4}^9 t_1^7 t_2^2 t_3^3 t_4^5 t_5^3 t_6^4 t_7^4 -  y_{s}^4 y_{o_1}^7 y_{o_2}^7 y_{o_3}^8 y_{o_4}^9 t_1^5 t_2^4 t_3^3 t_4^5 t_5^3 t_6^4 t_7^4 -  y_{s}^4 y_{o_1}^7 y_{o_2}^7 y_{o_3}^8 y_{o_4}^9 t_1^4 t_2^5 t_3^3 t_4^5 t_5^3 t_6^4 t_7^4 +  y_{s}^4 y_{o_1}^7 y_{o_2}^7 y_{o_3}^8 y_{o_4}^9 t_1^2 t_2^7 t_3^3 t_4^5 t_5^3 t_6^4 t_7^4 +  y_{s}^4 y_{o_1}^8 y_{o_2}^8 y_{o_3}^8 y_{o_4}^8 t_1^8 t_3^4 t_4^4 t_5^4 t_6^4 t_7^4 +  y_{s}^4 y_{o_1}^8 y_{o_2}^8 y_{o_3}^8 y_{o_4}^8 t_1^7 t_2 t_3^4 t_4^4 t_5^4 t_6^4 t_7^4 +  2 y_{s}^4 y_{o_1}^8 y_{o_2}^8 y_{o_3}^8 y_{o_4}^8 t_1^6 t_2^2 t_3^4 t_4^4 t_5^4 t_6^4 t_7^4 +  3 y_{s}^4 y_{o_1}^8 y_{o_2}^8 y_{o_3}^8 y_{o_4}^8 t_1^5 t_2^3 t_3^4 t_4^4 t_5^4 t_6^4 t_7^4 +  3 y_{s}^4 y_{o_1}^8 y_{o_2}^8 y_{o_3}^8 y_{o_4}^8 t_1^3 t_2^5 t_3^4 t_4^4 t_5^4 t_6^4 t_7^4 +  2 y_{s}^4 y_{o_1}^8 y_{o_2}^8 y_{o_3}^8 y_{o_4}^8 t_1^2 t_2^6 t_3^4 t_4^4 t_5^4 t_6^4 t_7^4 +  y_{s}^4 y_{o_1}^8 y_{o_2}^8 y_{o_3}^8 y_{o_4}^8 t_1 t_2^7 t_3^4 t_4^4 t_5^4 t_6^4 t_7^4 +  y_{s}^4 y_{o_1}^8 y_{o_2}^8 y_{o_3}^8 y_{o_4}^8 t_2^8 t_3^4 t_4^4 t_5^4 t_6^4 t_7^4 -  2 y_{s}^4 y_{o_1}^9 y_{o_2}^9 y_{o_3}^8 y_{o_4}^7 t_1^6 t_2 t_3^5 t_4^3 t_5^5 t_6^4 t_7^4 -  2 y_{s}^4 y_{o_1}^9 y_{o_2}^9 y_{o_3}^8 y_{o_4}^7 t_1^5 t_2^2 t_3^5 t_4^3 t_5^5 t_6^4 t_7^4 -  3 y_{s}^4 y_{o_1}^9 y_{o_2}^9 y_{o_3}^8 y_{o_4}^7 t_1^4 t_2^3 t_3^5 t_4^3 t_5^5 t_6^4 t_7^4 -  3 y_{s}^4 y_{o_1}^9 y_{o_2}^9 y_{o_3}^8 y_{o_4}^7 t_1^3 t_2^4 t_3^5 t_4^3 t_5^5 t_6^4 t_7^4 -  2 y_{s}^4 y_{o_1}^9 y_{o_2}^9 y_{o_3}^8 y_{o_4}^7 t_1^2 t_2^5 t_3^5 t_4^3 t_5^5 t_6^4 t_7^4 -  2 y_{s}^4 y_{o_1}^9 y_{o_2}^9 y_{o_3}^8 y_{o_4}^7 t_1 t_2^6 t_3^5 t_4^3 t_5^5 t_6^4 t_7^4 -  y_{s}^4 y_{o_1}^{10} y_{o_2}^{10} y_{o_3}^8 y_{o_4}^6 t_1^4 t_2^2 t_3^6 t_4^2 t_5^6 t_6^4 t_7^4 -  y_{s}^4 y_{o_1}^{10} y_{o_2}^{10} y_{o_3}^8 y_{o_4}^6 t_1^3 t_2^3 t_3^6 t_4^2 t_5^6 t_6^4 t_7^4 -  y_{s}^4 y_{o_1}^{10} y_{o_2}^{10} y_{o_3}^8 y_{o_4}^6 t_1^2 t_2^4 t_3^6 t_4^2 t_5^6 t_6^4 t_7^4 -  y_{s}^4 y_{o_1}^{11} y_{o_2}^{11} y_{o_3}^8 y_{o_4}^5 t_1^3 t_2^2 t_3^7 t_4 t_5^7 t_6^4 t_7^4 -  y_{s}^4 y_{o_1}^{11} y_{o_2}^{11} y_{o_3}^8 y_{o_4}^5 t_1^2 t_2^3 t_3^7 t_4 t_5^7 t_6^4 t_7^4 -  y_{s}^5 y_{o_1}^9 y_{o_2}^8 y_{o_3}^{11} y_{o_4}^{12} t_1^7 t_2^5 t_3^4 t_4^7 t_5^3 t_6^6 t_7^4 -  y_{s}^5 y_{o_1}^9 y_{o_2}^8 y_{o_3}^{11} y_{o_4}^{12} t_1^6 t_2^6 t_3^4 t_4^7 t_5^3 t_6^6 t_7^4 -  y_{s}^5 y_{o_1}^9 y_{o_2}^8 y_{o_3}^{11} y_{o_4}^{12} t_1^5 t_2^7 t_3^4 t_4^7 t_5^3 t_6^6 t_7^4 -  2 y_{s}^5 y_{o_1}^{10} y_{o_2}^9 y_{o_3}^{11} y_{o_4}^{11} t_1^8 t_2^3 t_3^5 t_4^6 t_5^4 t_6^6 t_7^4 -  y_{s}^5 y_{o_1}^{10} y_{o_2}^9 y_{o_3}^{11} y_{o_4}^{11} t_1^7 t_2^4 t_3^5 t_4^6 t_5^4 t_6^6 t_7^4 -  3 y_{s}^5 y_{o_1}^{10} y_{o_2}^9 y_{o_3}^{11} y_{o_4}^{11} t_1^6 t_2^5 t_3^5 t_4^6 t_5^4 t_6^6 t_7^4 - 3 y_{s}^5 y_{o_1}^{10} y_{o_2}^9 y_{o_3}^{11} y_{o_4}^{11} t_1^5 t_2^6 t_3^5 t_4^6 t_5^4 t_6^6 t_7^4 - y_{s}^5 y_{o_1}^{10} y_{o_2}^9 y_{o_3}^{11} y_{o_4}^{11} t_1^4 t_2^7 t_3^5 t_4^6 t_5^4 t_6^6 t_7^4 -  2 y_{s}^5 y_{o_1}^{10} y_{o_2}^9 y_{o_3}^{11} y_{o_4}^{11} t_1^3 t_2^8 t_3^5 t_4^6 t_5^4 t_6^6 t_7^4 -  y_{s}^5 y_{o_1}^{11} y_{o_2}^{10} y_{o_3}^{11} y_{o_4}^{10} t_1^8 t_2^2 t_3^6 t_4^5 t_5^5 t_6^6 t_7^4 - 2 y_{s}^5 y_{o_1}^{11} y_{o_2}^{10} y_{o_3}^{11} y_{o_4}^{10} t_1^7 t_2^3 t_3^6 t_4^5 t_5^5 t_6^6 t_7^4 - y_{s}^5 y_{o_1}^{11} y_{o_2}^{10} y_{o_3}^{11} y_{o_4}^{10} t_1^6 t_2^4 t_3^6 t_4^5 t_5^5 t_6^6 t_7^4 - 2 y_{s}^5 y_{o_1}^{11} y_{o_2}^{10} y_{o_3}^{11} y_{o_4}^{10} t_1^5 t_2^5 t_3^6 t_4^5 t_5^5 t_6^6 t_7^4 -  y_{s}^5 y_{o_1}^{11} y_{o_2}^{10} y_{o_3}^{11} y_{o_4}^{10} t_1^4 t_2^6 t_3^6 t_4^5 t_5^5 t_6^6 t_7^4 -  2 y_{s}^5 y_{o_1}^{11} y_{o_2}^{10} y_{o_3}^{11} y_{o_4}^{10} t_1^3 t_2^7 t_3^6 t_4^5 t_5^5 t_6^6 t_7^4 - y_{s}^5 y_{o_1}^{11} y_{o_2}^{10} y_{o_3}^{11} y_{o_4}^{10} t_1^2 t_2^8 t_3^6 t_4^5 t_5^5 t_6^6 t_7^4 -  y_{s}^5 y_{o_1}^{12} y_{o_2}^{11} y_{o_3}^{11} y_{o_4}^9 t_1^8 t_2 t_3^7 t_4^4 t_5^6 t_6^6 t_7^4 -  2 y_{s}^5 y_{o_1}^{12} y_{o_2}^{11} y_{o_3}^{11} y_{o_4}^9 t_1^7 t_2^2 t_3^7 t_4^4 t_5^6 t_6^6 t_7^4 - 3 y_{s}^5 y_{o_1}^{12} y_{o_2}^{11} y_{o_3}^{11} y_{o_4}^9 t_1^6 t_2^3 t_3^7 t_4^4 t_5^6 t_6^6 t_7^4 - 4 y_{s}^5 y_{o_1}^{12} y_{o_2}^{11} y_{o_3}^{11} y_{o_4}^9 t_1^5 t_2^4 t_3^7 t_4^4 t_5^6 t_6^6 t_7^4 - 4 y_{s}^5 y_{o_1}^{12} y_{o_2}^{11} y_{o_3}^{11} y_{o_4}^9 t_1^4 t_2^5 t_3^7 t_4^4 t_5^6 t_6^6 t_7^4 - 3 y_{s}^5 y_{o_1}^{12} y_{o_2}^{11} y_{o_3}^{11} y_{o_4}^9 t_1^3 t_2^6 t_3^7 t_4^4 t_5^6 t_6^6 t_7^4 - 2 y_{s}^5 y_{o_1}^{12} y_{o_2}^{11} y_{o_3}^{11} y_{o_4}^9 t_1^2 t_2^7 t_3^7 t_4^4 t_5^6 t_6^6 t_7^4 - y_{s}^5 y_{o_1}^{12} y_{o_2}^{11} y_{o_3}^{11} y_{o_4}^9 t_1 t_2^8 t_3^7 t_4^4 t_5^6 t_6^6 t_7^4 -  y_{s}^5 y_{o_1}^{13} y_{o_2}^{12} y_{o_3}^{11} y_{o_4}^8 t_1^4 t_2^4 t_3^8 t_4^3 t_5^7 t_6^6 t_7^4 +  y_{s}^6 y_{o_1}^{12} y_{o_2}^{10} y_{o_3}^{14} y_{o_4}^{14} t_1^8 t_2^6 t_3^6 t_4^8 t_5^4 t_6^8 t_7^4 +  y_{s}^6 y_{o_1}^{12} y_{o_2}^{10} y_{o_3}^{14} y_{o_4}^{14} t_1^6 t_2^8 t_3^6 t_4^8 t_5^4 t_6^8 t_7^4 +  y_{s}^6 y_{o_1}^{13} y_{o_2}^{11} y_{o_3}^{14} y_{o_4}^{13} t_1^8 t_2^5 t_3^7 t_4^7 t_5^5 t_6^8 t_7^4 +  2 y_{s}^6 y_{o_1}^{13} y_{o_2}^{11} y_{o_3}^{14} y_{o_4}^{13} t_1^7 t_2^6 t_3^7 t_4^7 t_5^5 t_6^8 t_7^4 + 2 y_{s}^6 y_{o_1}^{13} y_{o_2}^{11} y_{o_3}^{14} y_{o_4}^{13} t_1^6 t_2^7 t_3^7 t_4^7 t_5^5 t_6^8 t_7^4 + y_{s}^6 y_{o_1}^{13} y_{o_2}^{11} y_{o_3}^{14} y_{o_4}^{13} t_1^5 t_2^8 t_3^7 t_4^7 t_5^5 t_6^8 t_7^4 + y_{s}^6 y_{o_1}^{14} y_{o_2}^{12} y_{o_3}^{14} y_{o_4}^{12} t_1^8 t_2^4 t_3^8 t_4^6 t_5^6 t_6^8 t_7^4 + 2 y_{s}^6 y_{o_1}^{14} y_{o_2}^{12} y_{o_3}^{14} y_{o_4}^{12} t_1^7 t_2^5 t_3^8 t_4^6 t_5^6 t_6^8 t_7^4 +  2 y_{s}^6 y_{o_1}^{14} y_{o_2}^{12} y_{o_3}^{14} y_{o_4}^{12} t_1^6 t_2^6 t_3^8 t_4^6 t_5^6 t_6^8 t_7^4 + 2 y_{s}^6 y_{o_1}^{14} y_{o_2}^{12} y_{o_3}^{14} y_{o_4}^{12} t_1^5 t_2^7 t_3^8 t_4^6 t_5^6 t_6^8 t_7^4 + y_{s}^6 y_{o_1}^{14} y_{o_2}^{12} y_{o_3}^{14} y_{o_4}^{12} t_1^4 t_2^8 t_3^8 t_4^6 t_5^6 t_6^8 t_7^4 + y_{s}^3 y_{o_1}^4 y_{o_2}^6 y_{o_3}^4 y_{o_4}^6 t_1^5 t_2 t_3 t_4^3 t_5^3 t_6 t_7^5 +  2 y_{s}^3 y_{o_1}^4 y_{o_2}^6 y_{o_3}^4 y_{o_4}^6 t_1^4 t_2^2 t_3 t_4^3 t_5^3 t_6 t_7^5 +  2 y_{s}^3 y_{o_1}^4 y_{o_2}^6 y_{o_3}^4 y_{o_4}^6 t_1^3 t_2^3 t_3 t_4^3 t_5^3 t_6 t_7^5 +  2 y_{s}^3 y_{o_1}^4 y_{o_2}^6 y_{o_3}^4 y_{o_4}^6 t_1^2 t_2^4 t_3 t_4^3 t_5^3 t_6 t_7^5 +  y_{s}^3 y_{o_1}^4 y_{o_2}^6 y_{o_3}^4 y_{o_4}^6 t_1 t_2^5 t_3 t_4^3 t_5^3 t_6 t_7^5 +  y_{s}^3 y_{o_1}^5 y_{o_2}^7 y_{o_3}^4 y_{o_4}^5 t_1^4 t_2 t_3^2 t_4^2 t_5^4 t_6 t_7^5 +  2 y_{s}^3 y_{o_1}^5 y_{o_2}^7 y_{o_3}^4 y_{o_4}^5 t_1^3 t_2^2 t_3^2 t_4^2 t_5^4 t_6 t_7^5 +  2 y_{s}^3 y_{o_1}^5 y_{o_2}^7 y_{o_3}^4 y_{o_4}^5 t_1^2 t_2^3 t_3^2 t_4^2 t_5^4 t_6 t_7^5 +  y_{s}^3 y_{o_1}^5 y_{o_2}^7 y_{o_3}^4 y_{o_4}^5 t_1 t_2^4 t_3^2 t_4^2 t_5^4 t_6 t_7^5 +  y_{s}^3 y_{o_1}^6 y_{o_2}^8 y_{o_3}^4 y_{o_4}^4 t_1^3 t_2 t_3^3 t_4 t_5^5 t_6 t_7^5 +  y_{s}^3 y_{o_1}^6 y_{o_2}^8 y_{o_3}^4 y_{o_4}^4 t_1 t_2^3 t_3^3 t_4 t_5^5 t_6 t_7^5 -  y_{s}^4 y_{o_1}^5 y_{o_2}^6 y_{o_3}^7 y_{o_4}^{10} t_1^5 t_2^5 t_3 t_4^6 t_5^2 t_6^3 t_7^5 -  y_{s}^4 y_{o_1}^6 y_{o_2}^7 y_{o_3}^7 y_{o_4}^9 t_1^8 t_2 t_3^2 t_4^5 t_5^3 t_6^3 t_7^5 -  2 y_{s}^4 y_{o_1}^6 y_{o_2}^7 y_{o_3}^7 y_{o_4}^9 t_1^7 t_2^2 t_3^2 t_4^5 t_5^3 t_6^3 t_7^5 -  3 y_{s}^4 y_{o_1}^6 y_{o_2}^7 y_{o_3}^7 y_{o_4}^9 t_1^6 t_2^3 t_3^2 t_4^5 t_5^3 t_6^3 t_7^5 -  4 y_{s}^4 y_{o_1}^6 y_{o_2}^7 y_{o_3}^7 y_{o_4}^9 t_1^5 t_2^4 t_3^2 t_4^5 t_5^3 t_6^3 t_7^5 -  4 y_{s}^4 y_{o_1}^6 y_{o_2}^7 y_{o_3}^7 y_{o_4}^9 t_1^4 t_2^5 t_3^2 t_4^5 t_5^3 t_6^3 t_7^5 -  3 y_{s}^4 y_{o_1}^6 y_{o_2}^7 y_{o_3}^7 y_{o_4}^9 t_1^3 t_2^6 t_3^2 t_4^5 t_5^3 t_6^3 t_7^5 -  2 y_{s}^4 y_{o_1}^6 y_{o_2}^7 y_{o_3}^7 y_{o_4}^9 t_1^2 t_2^7 t_3^2 t_4^5 t_5^3 t_6^3 t_7^5 -  y_{s}^4 y_{o_1}^6 y_{o_2}^7 y_{o_3}^7 y_{o_4}^9 t_1 t_2^8 t_3^2 t_4^5 t_5^3 t_6^3 t_7^5 -  y_{s}^4 y_{o_1}^7 y_{o_2}^8 y_{o_3}^7 y_{o_4}^8 t_1^7 t_2 t_3^3 t_4^4 t_5^4 t_6^3 t_7^5 -  2 y_{s}^4 y_{o_1}^7 y_{o_2}^8 y_{o_3}^7 y_{o_4}^8 t_1^6 t_2^2 t_3^3 t_4^4 t_5^4 t_6^3 t_7^5 -  y_{s}^4 y_{o_1}^7 y_{o_2}^8 y_{o_3}^7 y_{o_4}^8 t_1^5 t_2^3 t_3^3 t_4^4 t_5^4 t_6^3 t_7^5 -  2 y_{s}^4 y_{o_1}^7 y_{o_2}^8 y_{o_3}^7 y_{o_4}^8 t_1^4 t_2^4 t_3^3 t_4^4 t_5^4 t_6^3 t_7^5 -  y_{s}^4 y_{o_1}^7 y_{o_2}^8 y_{o_3}^7 y_{o_4}^8 t_1^3 t_2^5 t_3^3 t_4^4 t_5^4 t_6^3 t_7^5 -  2 y_{s}^4 y_{o_1}^7 y_{o_2}^8 y_{o_3}^7 y_{o_4}^8 t_1^2 t_2^6 t_3^3 t_4^4 t_5^4 t_6^3 t_7^5 -  y_{s}^4 y_{o_1}^7 y_{o_2}^8 y_{o_3}^7 y_{o_4}^8 t_1 t_2^7 t_3^3 t_4^4 t_5^4 t_6^3 t_7^5 -  2 y_{s}^4 y_{o_1}^8 y_{o_2}^9 y_{o_3}^7 y_{o_4}^7 t_1^6 t_2 t_3^4 t_4^3 t_5^5 t_6^3 t_7^5 -  y_{s}^4 y_{o_1}^8 y_{o_2}^9 y_{o_3}^7 y_{o_4}^7 t_1^5 t_2^2 t_3^4 t_4^3 t_5^5 t_6^3 t_7^5 -  3 y_{s}^4 y_{o_1}^8 y_{o_2}^9 y_{o_3}^7 y_{o_4}^7 t_1^4 t_2^3 t_3^4 t_4^3 t_5^5 t_6^3 t_7^5 -  3 y_{s}^4 y_{o_1}^8 y_{o_2}^9 y_{o_3}^7 y_{o_4}^7 t_1^3 t_2^4 t_3^4 t_4^3 t_5^5 t_6^3 t_7^5 -  y_{s}^4 y_{o_1}^8 y_{o_2}^9 y_{o_3}^7 y_{o_4}^7 t_1^2 t_2^5 t_3^4 t_4^3 t_5^5 t_6^3 t_7^5 -  2 y_{s}^4 y_{o_1}^8 y_{o_2}^9 y_{o_3}^7 y_{o_4}^7 t_1 t_2^6 t_3^4 t_4^3 t_5^5 t_6^3 t_7^5 -  y_{s}^4 y_{o_1}^9 y_{o_2}^{10} y_{o_3}^7 y_{o_4}^6 t_1^4 t_2^2 t_3^5 t_4^2 t_5^6 t_6^3 t_7^5 -  y_{s}^4 y_{o_1}^9 y_{o_2}^{10} y_{o_3}^7 y_{o_4}^6 t_1^3 t_2^3 t_3^5 t_4^2 t_5^6 t_6^3 t_7^5 -  y_{s}^4 y_{o_1}^9 y_{o_2}^{10} y_{o_3}^7 y_{o_4}^6 t_1^2 t_2^4 t_3^5 t_4^2 t_5^6 t_6^3 t_7^5 -  y_{s}^5 y_{o_1}^7 y_{o_2}^7 y_{o_3}^{10} y_{o_4}^{13} t_1^7 t_2^6 t_3^2 t_4^8 t_5^2 t_6^5 t_7^5 -  y_{s}^5 y_{o_1}^7 y_{o_2}^7 y_{o_3}^{10} y_{o_4}^{13} t_1^6 t_2^7 t_3^2 t_4^8 t_5^2 t_6^5 t_7^5 -  y_{s}^5 y_{o_1}^8 y_{o_2}^8 y_{o_3}^{10} y_{o_4}^{12} t_1^7 t_2^5 t_3^3 t_4^7 t_5^3 t_6^5 t_7^5 -  y_{s}^5 y_{o_1}^8 y_{o_2}^8 y_{o_3}^{10} y_{o_4}^{12} t_1^6 t_2^6 t_3^3 t_4^7 t_5^3 t_6^5 t_7^5 -  y_{s}^5 y_{o_1}^8 y_{o_2}^8 y_{o_3}^{10} y_{o_4}^{12} t_1^5 t_2^7 t_3^3 t_4^7 t_5^3 t_6^5 t_7^5 -  2 y_{s}^5 y_{o_1}^9 y_{o_2}^9 y_{o_3}^{10} y_{o_4}^{11} t_1^8 t_2^3 t_3^4 t_4^6 t_5^4 t_6^5 t_7^5 -  2 y_{s}^5 y_{o_1}^9 y_{o_2}^9 y_{o_3}^{10} y_{o_4}^{11} t_1^7 t_2^4 t_3^4 t_4^6 t_5^4 t_6^5 t_7^5 -  3 y_{s}^5 y_{o_1}^9 y_{o_2}^9 y_{o_3}^{10} y_{o_4}^{11} t_1^6 t_2^5 t_3^4 t_4^6 t_5^4 t_6^5 t_7^5 -  3 y_{s}^5 y_{o_1}^9 y_{o_2}^9 y_{o_3}^{10} y_{o_4}^{11} t_1^5 t_2^6 t_3^4 t_4^6 t_5^4 t_6^5 t_7^5 -  2 y_{s}^5 y_{o_1}^9 y_{o_2}^9 y_{o_3}^{10} y_{o_4}^{11} t_1^4 t_2^7 t_3^4 t_4^6 t_5^4 t_6^5 t_7^5 -  2 y_{s}^5 y_{o_1}^9 y_{o_2}^9 y_{o_3}^{10} y_{o_4}^{11} t_1^3 t_2^8 t_3^4 t_4^6 t_5^4 t_6^5 t_7^5 +  y_{s}^5 y_{o_1}^{10} y_{o_2}^{10} y_{o_3}^{10} y_{o_4}^{10} t_1^9 t_2 t_3^5 t_4^5 t_5^5 t_6^5 t_7^5 +  y_{s}^5 y_{o_1}^{10} y_{o_2}^{10} y_{o_3}^{10} y_{o_4}^{10} t_1^8 t_2^2 t_3^5 t_4^5 t_5^5 t_6^5 t_7^5 +  2 y_{s}^5 y_{o_1}^{10} y_{o_2}^{10} y_{o_3}^{10} y_{o_4}^{10} t_1^7 t_2^3 t_3^5 t_4^5 t_5^5 t_6^5 t_7^5 + 3 y_{s}^5 y_{o_1}^{10} y_{o_2}^{10} y_{o_3}^{10} y_{o_4}^{10} t_1^6 t_2^4 t_3^5 t_4^5 t_5^5 t_6^5 t_7^5 + 3 y_{s}^5 y_{o_1}^{10} y_{o_2}^{10} y_{o_3}^{10} y_{o_4}^{10} t_1^4 t_2^6 t_3^5 t_4^5 t_5^5 t_6^5 t_7^5 +  2 y_{s}^5 y_{o_1}^{10} y_{o_2}^{10} y_{o_3}^{10} y_{o_4}^{10} t_1^3 t_2^7 t_3^5 t_4^5 t_5^5 t_6^5 t_7^5 + y_{s}^5 y_{o_1}^{10} y_{o_2}^{10} y_{o_3}^{10} y_{o_4}^{10} t_1^2 t_2^8 t_3^5 t_4^5 t_5^5 t_6^5 t_7^5 +  y_{s}^5 y_{o_1}^{10} y_{o_2}^{10} y_{o_3}^{10} y_{o_4}^{10} t_1 t_2^9 t_3^5 t_4^5 t_5^5 t_6^5 t_7^5 +  y_{s}^5 y_{o_1}^{11} y_{o_2}^{11} y_{o_3}^{10} y_{o_4}^9 t_1^7 t_2^2 t_3^6 t_4^4 t_5^6 t_6^5 t_7^5 -  y_{s}^5 y_{o_1}^{11} y_{o_2}^{11} y_{o_3}^{10} y_{o_4}^9 t_1^5 t_2^4 t_3^6 t_4^4 t_5^6 t_6^5 t_7^5 -  y_{s}^5 y_{o_1}^{11} y_{o_2}^{11} y_{o_3}^{10} y_{o_4}^9 t_1^4 t_2^5 t_3^6 t_4^4 t_5^6 t_6^5 t_7^5 +  y_{s}^5 y_{o_1}^{11} y_{o_2}^{11} y_{o_3}^{10} y_{o_4}^9 t_1^2 t_2^7 t_3^6 t_4^4 t_5^6 t_6^5 t_7^5 -  y_{s}^5 y_{o_1}^{12} y_{o_2}^{12} y_{o_3}^{10} y_{o_4}^8 t_1^4 t_2^4 t_3^7 t_4^3 t_5^7 t_6^5 t_7^5 +  2 y_{s}^6 y_{o_1}^{11} y_{o_2}^{10} y_{o_3}^{13} y_{o_4}^{14} t_1^8 t_2^6 t_3^5 t_4^8 t_5^4 t_6^7 t_7^5 + 2 y_{s}^6 y_{o_1}^{11} y_{o_2}^{10} y_{o_3}^{13} y_{o_4}^{14} t_1^7 t_2^7 t_3^5 t_4^8 t_5^4 t_6^7 t_7^5 + 2 y_{s}^6 y_{o_1}^{11} y_{o_2}^{10} y_{o_3}^{13} y_{o_4}^{14} t_1^6 t_2^8 t_3^5 t_4^8 t_5^4 t_6^7 t_7^5 +  2 y_{s}^6 y_{o_1}^{12} y_{o_2}^{11} y_{o_3}^{13} y_{o_4}^{13} t_1^8 t_2^5 t_3^6 t_4^7 t_5^5 t_6^7 t_7^5 + 2 y_{s}^6 y_{o_1}^{12} y_{o_2}^{11} y_{o_3}^{13} y_{o_4}^{13} t_1^7 t_2^6 t_3^6 t_4^7 t_5^5 t_6^7 t_7^5 + 2 y_{s}^6 y_{o_1}^{12} y_{o_2}^{11} y_{o_3}^{13} y_{o_4}^{13} t_1^6 t_2^7 t_3^6 t_4^7 t_5^5 t_6^7 t_7^5 +  2 y_{s}^6 y_{o_1}^{12} y_{o_2}^{11} y_{o_3}^{13} y_{o_4}^{13} t_1^5 t_2^8 t_3^6 t_4^7 t_5^5 t_6^7 t_7^5 + y_{s}^6 y_{o_1}^{13} y_{o_2}^{12} y_{o_3}^{13} y_{o_4}^{12} t_1^9 t_2^3 t_3^7 t_4^6 t_5^6 t_6^7 t_7^5 + 3 y_{s}^6 y_{o_1}^{13} y_{o_2}^{12} y_{o_3}^{13} y_{o_4}^{12} t_1^8 t_2^4 t_3^7 t_4^6 t_5^6 t_6^7 t_7^5 + 4 y_{s}^6 y_{o_1}^{13} y_{o_2}^{12} y_{o_3}^{13} y_{o_4}^{12} t_1^7 t_2^5 t_3^7 t_4^6 t_5^6 t_6^7 t_7^5 +  4 y_{s}^6 y_{o_1}^{13} y_{o_2}^{12} y_{o_3}^{13} y_{o_4}^{12} t_1^6 t_2^6 t_3^7 t_4^6 t_5^6 t_6^7 t_7^5 + 4 y_{s}^6 y_{o_1}^{13} y_{o_2}^{12} y_{o_3}^{13} y_{o_4}^{12} t_1^5 t_2^7 t_3^7 t_4^6 t_5^6 t_6^7 t_7^5 + 3 y_{s}^6 y_{o_1}^{13} y_{o_2}^{12} y_{o_3}^{13} y_{o_4}^{12} t_1^4 t_2^8 t_3^7 t_4^6 t_5^6 t_6^7 t_7^5 +  y_{s}^6 y_{o_1}^{13} y_{o_2}^{12} y_{o_3}^{13} y_{o_4}^{12} t_1^3 t_2^9 t_3^7 t_4^6 t_5^6 t_6^7 t_7^5 +  y_{s}^6 y_{o_1}^{14} y_{o_2}^{13} y_{o_3}^{13} y_{o_4}^{11} t_1^7 t_2^4 t_3^8 t_4^5 t_5^7 t_6^7 t_7^5 +  y_{s}^6 y_{o_1}^{14} y_{o_2}^{13} y_{o_3}^{13} y_{o_4}^{11} t_1^6 t_2^5 t_3^8 t_4^5 t_5^7 t_6^7 t_7^5 +  y_{s}^6 y_{o_1}^{14} y_{o_2}^{13} y_{o_3}^{13} y_{o_4}^{11} t_1^5 t_2^6 t_3^8 t_4^5 t_5^7 t_6^7 t_7^5 +  y_{s}^6 y_{o_1}^{14} y_{o_2}^{13} y_{o_3}^{13} y_{o_4}^{11} t_1^4 t_2^7 t_3^8 t_4^5 t_5^7 t_6^7 t_7^5 -  y_{s}^7 y_{o_1}^{15} y_{o_2}^{13} y_{o_3}^{16} y_{o_4}^{15} t_1^8 t_2^7 t_3^8 t_4^8 t_5^6 t_6^9 t_7^5 -  y_{s}^7 y_{o_1}^{15} y_{o_2}^{13} y_{o_3}^{16} y_{o_4}^{15} t_1^7 t_2^8 t_3^8 t_4^8 t_5^6 t_6^9 t_7^5 -  y_{s}^4 y_{o_1}^6 y_{o_2}^8 y_{o_3}^6 y_{o_4}^8 t_1^6 t_2^2 t_3^2 t_4^4 t_5^4 t_6^2 t_7^6 -  y_{s}^4 y_{o_1}^6 y_{o_2}^8 y_{o_3}^6 y_{o_4}^8 t_1^5 t_2^3 t_3^2 t_4^4 t_5^4 t_6^2 t_7^6 -  y_{s}^4 y_{o_1}^6 y_{o_2}^8 y_{o_3}^6 y_{o_4}^8 t_1^4 t_2^4 t_3^2 t_4^4 t_5^4 t_6^2 t_7^6 -  y_{s}^4 y_{o_1}^6 y_{o_2}^8 y_{o_3}^6 y_{o_4}^8 t_1^3 t_2^5 t_3^2 t_4^4 t_5^4 t_6^2 t_7^6 -  y_{s}^4 y_{o_1}^6 y_{o_2}^8 y_{o_3}^6 y_{o_4}^8 t_1^2 t_2^6 t_3^2 t_4^4 t_5^4 t_6^2 t_7^6 -  y_{s}^4 y_{o_1}^7 y_{o_2}^9 y_{o_3}^6 y_{o_4}^7 t_1^6 t_2 t_3^3 t_4^3 t_5^5 t_6^2 t_7^6 -  2 y_{s}^4 y_{o_1}^7 y_{o_2}^9 y_{o_3}^6 y_{o_4}^7 t_1^4 t_2^3 t_3^3 t_4^3 t_5^5 t_6^2 t_7^6 -  2 y_{s}^4 y_{o_1}^7 y_{o_2}^9 y_{o_3}^6 y_{o_4}^7 t_1^3 t_2^4 t_3^3 t_4^3 t_5^5 t_6^2 t_7^6 -  y_{s}^4 y_{o_1}^7 y_{o_2}^9 y_{o_3}^6 y_{o_4}^7 t_1 t_2^6 t_3^3 t_4^3 t_5^5 t_6^2 t_7^6 -  y_{s}^4 y_{o_1}^8 y_{o_2}^{10} y_{o_3}^6 y_{o_4}^6 t_1^3 t_2^3 t_3^4 t_4^2 t_5^6 t_6^2 t_7^6 -  y_{s}^5 y_{o_1}^8 y_{o_2}^9 y_{o_3}^9 y_{o_4}^{11} t_1^8 t_2^3 t_3^3 t_4^6 t_5^4 t_6^4 t_7^6 -  y_{s}^5 y_{o_1}^8 y_{o_2}^9 y_{o_3}^9 y_{o_4}^{11} t_1^7 t_2^4 t_3^3 t_4^6 t_5^4 t_6^4 t_7^6 -  y_{s}^5 y_{o_1}^8 y_{o_2}^9 y_{o_3}^9 y_{o_4}^{11} t_1^4 t_2^7 t_3^3 t_4^6 t_5^4 t_6^4 t_7^6 -  y_{s}^5 y_{o_1}^8 y_{o_2}^9 y_{o_3}^9 y_{o_4}^{11} t_1^3 t_2^8 t_3^3 t_4^6 t_5^4 t_6^4 t_7^6 -  2 y_{s}^5 y_{o_1}^9 y_{o_2}^{10} y_{o_3}^9 y_{o_4}^{10} t_1^8 t_2^2 t_3^4 t_4^5 t_5^5 t_6^4 t_7^6 -  2 y_{s}^5 y_{o_1}^9 y_{o_2}^{10} y_{o_3}^9 y_{o_4}^{10} t_1^7 t_2^3 t_3^4 t_4^5 t_5^5 t_6^4 t_7^6 -  y_{s}^5 y_{o_1}^9 y_{o_2}^{10} y_{o_3}^9 y_{o_4}^{10} t_1^6 t_2^4 t_3^4 t_4^5 t_5^5 t_6^4 t_7^6 -  5 y_{s}^5 y_{o_1}^9 y_{o_2}^{10} y_{o_3}^9 y_{o_4}^{10} t_1^5 t_2^5 t_3^4 t_4^5 t_5^5 t_6^4 t_7^6 -  y_{s}^5 y_{o_1}^9 y_{o_2}^{10} y_{o_3}^9 y_{o_4}^{10} t_1^4 t_2^6 t_3^4 t_4^5 t_5^5 t_6^4 t_7^6 -  2 y_{s}^5 y_{o_1}^9 y_{o_2}^{10} y_{o_3}^9 y_{o_4}^{10} t_1^3 t_2^7 t_3^4 t_4^5 t_5^5 t_6^4 t_7^6 -  2 y_{s}^5 y_{o_1}^9 y_{o_2}^{10} y_{o_3}^9 y_{o_4}^{10} t_1^2 t_2^8 t_3^4 t_4^5 t_5^5 t_6^4 t_7^6 -  y_{s}^5 y_{o_1}^{10} y_{o_2}^{11} y_{o_3}^9 y_{o_4}^9 t_1^8 t_2 t_3^5 t_4^4 t_5^6 t_6^4 t_7^6 -  y_{s}^5 y_{o_1}^{10} y_{o_2}^{11} y_{o_3}^9 y_{o_4}^9 t_1^7 t_2^2 t_3^5 t_4^4 t_5^6 t_6^4 t_7^6 -  y_{s}^5 y_{o_1}^{10} y_{o_2}^{11} y_{o_3}^9 y_{o_4}^9 t_1^6 t_2^3 t_3^5 t_4^4 t_5^6 t_6^4 t_7^6 -  2 y_{s}^5 y_{o_1}^{10} y_{o_2}^{11} y_{o_3}^9 y_{o_4}^9 t_1^5 t_2^4 t_3^5 t_4^4 t_5^6 t_6^4 t_7^6 -  2 y_{s}^5 y_{o_1}^{10} y_{o_2}^{11} y_{o_3}^9 y_{o_4}^9 t_1^4 t_2^5 t_3^5 t_4^4 t_5^6 t_6^4 t_7^6 -  y_{s}^5 y_{o_1}^{10} y_{o_2}^{11} y_{o_3}^9 y_{o_4}^9 t_1^3 t_2^6 t_3^5 t_4^4 t_5^6 t_6^4 t_7^6 -  y_{s}^5 y_{o_1}^{10} y_{o_2}^{11} y_{o_3}^9 y_{o_4}^9 t_1^2 t_2^7 t_3^5 t_4^4 t_5^6 t_6^4 t_7^6 -  y_{s}^5 y_{o_1}^{10} y_{o_2}^{11} y_{o_3}^9 y_{o_4}^9 t_1 t_2^8 t_3^5 t_4^4 t_5^6 t_6^4 t_7^6 +  y_{s}^5 y_{o_1}^{11} y_{o_2}^{12} y_{o_3}^9 y_{o_4}^8 t_1^4 t_2^4 t_3^6 t_4^3 t_5^7 t_6^4 t_7^6 +  y_{s}^6 y_{o_1}^{10} y_{o_2}^{10} y_{o_3}^{12} y_{o_4}^{14} t_1^8 t_2^6 t_3^4 t_4^8 t_5^4 t_6^6 t_7^6 +  y_{s}^6 y_{o_1}^{10} y_{o_2}^{10} y_{o_3}^{12} y_{o_4}^{14} t_1^7 t_2^7 t_3^4 t_4^8 t_5^4 t_6^6 t_7^6 +  y_{s}^6 y_{o_1}^{10} y_{o_2}^{10} y_{o_3}^{12} y_{o_4}^{14} t_1^6 t_2^8 t_3^4 t_4^8 t_5^4 t_6^6 t_7^6 +  3 y_{s}^6 y_{o_1}^{11} y_{o_2}^{11} y_{o_3}^{12} y_{o_4}^{13} t_1^8 t_2^5 t_3^5 t_4^7 t_5^5 t_6^6 t_7^6 + 2 y_{s}^6 y_{o_1}^{11} y_{o_2}^{11} y_{o_3}^{12} y_{o_4}^{13} t_1^7 t_2^6 t_3^5 t_4^7 t_5^5 t_6^6 t_7^6 + 2 y_{s}^6 y_{o_1}^{11} y_{o_2}^{11} y_{o_3}^{12} y_{o_4}^{13} t_1^6 t_2^7 t_3^5 t_4^7 t_5^5 t_6^6 t_7^6 +  3 y_{s}^6 y_{o_1}^{11} y_{o_2}^{11} y_{o_3}^{12} y_{o_4}^{13} t_1^5 t_2^8 t_3^5 t_4^7 t_5^5 t_6^6 t_7^6 + y_{s}^6 y_{o_1}^{12} y_{o_2}^{12} y_{o_3}^{12} y_{o_4}^{12} t_1^9 t_2^3 t_3^6 t_4^6 t_5^6 t_6^6 t_7^6 + 3 y_{s}^6 y_{o_1}^{12} y_{o_2}^{12} y_{o_3}^{12} y_{o_4}^{12} t_1^8 t_2^4 t_3^6 t_4^6 t_5^6 t_6^6 t_7^6 + 3 y_{s}^6 y_{o_1}^{12} y_{o_2}^{12} y_{o_3}^{12} y_{o_4}^{12} t_1^7 t_2^5 t_3^6 t_4^6 t_5^6 t_6^6 t_7^6 +  3 y_{s}^6 y_{o_1}^{12} y_{o_2}^{12} y_{o_3}^{12} y_{o_4}^{12} t_1^6 t_2^6 t_3^6 t_4^6 t_5^6 t_6^6 t_7^6 + 3 y_{s}^6 y_{o_1}^{12} y_{o_2}^{12} y_{o_3}^{12} y_{o_4}^{12} t_1^5 t_2^7 t_3^6 t_4^6 t_5^6 t_6^6 t_7^6 + 3 y_{s}^6 y_{o_1}^{12} y_{o_2}^{12} y_{o_3}^{12} y_{o_4}^{12} t_1^4 t_2^8 t_3^6 t_4^6 t_5^6 t_6^6 t_7^6 +  y_{s}^6 y_{o_1}^{12} y_{o_2}^{12} y_{o_3}^{12} y_{o_4}^{12} t_1^3 t_2^9 t_3^6 t_4^6 t_5^6 t_6^6 t_7^6 +  y_{s}^6 y_{o_1}^{13} y_{o_2}^{13} y_{o_3}^{12} y_{o_4}^{11} t_1^9 t_2^2 t_3^7 t_4^5 t_5^7 t_6^6 t_7^6 +  2 y_{s}^6 y_{o_1}^{13} y_{o_2}^{13} y_{o_3}^{12} y_{o_4}^{11} t_1^8 t_2^3 t_3^7 t_4^5 t_5^7 t_6^6 t_7^6 + 3 y_{s}^6 y_{o_1}^{13} y_{o_2}^{13} y_{o_3}^{12} y_{o_4}^{11} t_1^7 t_2^4 t_3^7 t_4^5 t_5^7 t_6^6 t_7^6 + 4 y_{s}^6 y_{o_1}^{13} y_{o_2}^{13} y_{o_3}^{12} y_{o_4}^{11} t_1^6 t_2^5 t_3^7 t_4^5 t_5^7 t_6^6 t_7^6 +  4 y_{s}^6 y_{o_1}^{13} y_{o_2}^{13} y_{o_3}^{12} y_{o_4}^{11} t_1^5 t_2^6 t_3^7 t_4^5 t_5^7 t_6^6 t_7^6 + 3 y_{s}^6 y_{o_1}^{13} y_{o_2}^{13} y_{o_3}^{12} y_{o_4}^{11} t_1^4 t_2^7 t_3^7 t_4^5 t_5^7 t_6^6 t_7^6 + 2 y_{s}^6 y_{o_1}^{13} y_{o_2}^{13} y_{o_3}^{12} y_{o_4}^{11} t_1^3 t_2^8 t_3^7 t_4^5 t_5^7 t_6^6 t_7^6 +  y_{s}^6 y_{o_1}^{13} y_{o_2}^{13} y_{o_3}^{12} y_{o_4}^{11} t_1^2 t_2^9 t_3^7 t_4^5 t_5^7 t_6^6 t_7^6 +  y_{s}^6 y_{o_1}^{14} y_{o_2}^{14} y_{o_3}^{12} y_{o_4}^{10} t_1^8 t_2^2 t_3^8 t_4^4 t_5^8 t_6^6 t_7^6 +  y_{s}^6 y_{o_1}^{14} y_{o_2}^{14} y_{o_3}^{12} y_{o_4}^{10} t_1^7 t_2^3 t_3^8 t_4^4 t_5^8 t_6^6 t_7^6 +  2 y_{s}^6 y_{o_1}^{14} y_{o_2}^{14} y_{o_3}^{12} y_{o_4}^{10} t_1^6 t_2^4 t_3^8 t_4^4 t_5^8 t_6^6 t_7^6 + 2 y_{s}^6 y_{o_1}^{14} y_{o_2}^{14} y_{o_3}^{12} y_{o_4}^{10} t_1^5 t_2^5 t_3^8 t_4^4 t_5^8 t_6^6 t_7^6 + 2 y_{s}^6 y_{o_1}^{14} y_{o_2}^{14} y_{o_3}^{12} y_{o_4}^{10} t_1^4 t_2^6 t_3^8 t_4^4 t_5^8 t_6^6 t_7^6 +  y_{s}^6 y_{o_1}^{14} y_{o_2}^{14} y_{o_3}^{12} y_{o_4}^{10} t_1^3 t_2^7 t_3^8 t_4^4 t_5^8 t_6^6 t_7^6 +  y_{s}^6 y_{o_1}^{14} y_{o_2}^{14} y_{o_3}^{12} y_{o_4}^{10} t_1^2 t_2^8 t_3^8 t_4^4 t_5^8 t_6^6 t_7^6 -  y_{s}^7 y_{o_1}^{13} y_{o_2}^{12} y_{o_3}^{15} y_{o_4}^{16} t_1^8 t_2^8 t_3^6 t_4^9 t_5^5 t_6^8 t_7^6 -  y_{s}^7 y_{o_1}^{14} y_{o_2}^{13} y_{o_3}^{15} y_{o_4}^{15} t_1^9 t_2^6 t_3^7 t_4^8 t_5^6 t_6^8 t_7^6 -  2 y_{s}^7 y_{o_1}^{14} y_{o_2}^{13} y_{o_3}^{15} y_{o_4}^{15} t_1^8 t_2^7 t_3^7 t_4^8 t_5^6 t_6^8 t_7^6 - 2 y_{s}^7 y_{o_1}^{14} y_{o_2}^{13} y_{o_3}^{15} y_{o_4}^{15} t_1^7 t_2^8 t_3^7 t_4^8 t_5^6 t_6^8 t_7^6 - y_{s}^7 y_{o_1}^{14} y_{o_2}^{13} y_{o_3}^{15} y_{o_4}^{15} t_1^6 t_2^9 t_3^7 t_4^8 t_5^6 t_6^8 t_7^6 - y_{s}^7 y_{o_1}^{15} y_{o_2}^{14} y_{o_3}^{15} y_{o_4}^{14} t_1^9 t_2^5 t_3^8 t_4^7 t_5^7 t_6^8 t_7^6 - 2 y_{s}^7 y_{o_1}^{15} y_{o_2}^{14} y_{o_3}^{15} y_{o_4}^{14} t_1^8 t_2^6 t_3^8 t_4^7 t_5^7 t_6^8 t_7^6 -  3 y_{s}^7 y_{o_1}^{15} y_{o_2}^{14} y_{o_3}^{15} y_{o_4}^{14} t_1^7 t_2^7 t_3^8 t_4^7 t_5^7 t_6^8 t_7^6 - 2 y_{s}^7 y_{o_1}^{15} y_{o_2}^{14} y_{o_3}^{15} y_{o_4}^{14} t_1^6 t_2^8 t_3^8 t_4^7 t_5^7 t_6^8 t_7^6 - y_{s}^7 y_{o_1}^{15} y_{o_2}^{14} y_{o_3}^{15} y_{o_4}^{14} t_1^5 t_2^9 t_3^8 t_4^7 t_5^7 t_6^8 t_7^6 - y_{s}^7 y_{o_1}^{16} y_{o_2}^{15} y_{o_3}^{15} y_{o_4}^{13} t_1^8 t_2^5 t_3^9 t_4^6 t_5^8 t_6^8 t_7^6 - y_{s}^7 y_{o_1}^{16} y_{o_2}^{15} y_{o_3}^{15} y_{o_4}^{13} t_1^7 t_2^6 t_3^9 t_4^6 t_5^8 t_6^8 t_7^6 - y_{s}^7 y_{o_1}^{16} y_{o_2}^{15} y_{o_3}^{15} y_{o_4}^{13} t_1^6 t_2^7 t_3^9 t_4^6 t_5^8 t_6^8 t_7^6 - y_{s}^7 y_{o_1}^{16} y_{o_2}^{15} y_{o_3}^{15} y_{o_4}^{13} t_1^5 t_2^8 t_3^9 t_4^6 t_5^8 t_6^8 t_7^6 - y_{s}^4 y_{o_1}^6 y_{o_2}^9 y_{o_3}^5 y_{o_4}^7 t_1^4 t_2^3 t_3^2 t_4^3 t_5^5 t_6 t_7^7 -  y_{s}^4 y_{o_1}^6 y_{o_2}^9 y_{o_3}^5 y_{o_4}^7 t_1^3 t_2^4 t_3^2 t_4^3 t_5^5 t_6 t_7^7 -  y_{s}^4 y_{o_1}^7 y_{o_2}^{10} y_{o_3}^5 y_{o_4}^6 t_1^3 t_2^3 t_3^3 t_4^2 t_5^6 t_6 t_7^7 +  y_{s}^5 y_{o_1}^7 y_{o_2}^9 y_{o_3}^8 y_{o_4}^{11} t_1^6 t_2^5 t_3^2 t_4^6 t_5^4 t_6^3 t_7^7 +  y_{s}^5 y_{o_1}^7 y_{o_2}^9 y_{o_3}^8 y_{o_4}^{11} t_1^5 t_2^6 t_3^2 t_4^6 t_5^4 t_6^3 t_7^7 +  y_{s}^5 y_{o_1}^8 y_{o_2}^{10} y_{o_3}^8 y_{o_4}^{10} t_1^7 t_2^3 t_3^3 t_4^5 t_5^5 t_6^3 t_7^7 +  2 y_{s}^5 y_{o_1}^8 y_{o_2}^{10} y_{o_3}^8 y_{o_4}^{10} t_1^6 t_2^4 t_3^3 t_4^5 t_5^5 t_6^3 t_7^7 -  y_{s}^5 y_{o_1}^8 y_{o_2}^{10} y_{o_3}^8 y_{o_4}^{10} t_1^5 t_2^5 t_3^3 t_4^5 t_5^5 t_6^3 t_7^7 +  2 y_{s}^5 y_{o_1}^8 y_{o_2}^{10} y_{o_3}^8 y_{o_4}^{10} t_1^4 t_2^6 t_3^3 t_4^5 t_5^5 t_6^3 t_7^7 +  y_{s}^5 y_{o_1}^8 y_{o_2}^{10} y_{o_3}^8 y_{o_4}^{10} t_1^3 t_2^7 t_3^3 t_4^5 t_5^5 t_6^3 t_7^7 +  y_{s}^5 y_{o_1}^9 y_{o_2}^{11} y_{o_3}^8 y_{o_4}^9 t_1^6 t_2^3 t_3^4 t_4^4 t_5^6 t_6^3 t_7^7 -  y_{s}^5 y_{o_1}^9 y_{o_2}^{11} y_{o_3}^8 y_{o_4}^9 t_1^5 t_2^4 t_3^4 t_4^4 t_5^6 t_6^3 t_7^7 -  y_{s}^5 y_{o_1}^9 y_{o_2}^{11} y_{o_3}^8 y_{o_4}^9 t_1^4 t_2^5 t_3^4 t_4^4 t_5^6 t_6^3 t_7^7 +  y_{s}^5 y_{o_1}^9 y_{o_2}^{11} y_{o_3}^8 y_{o_4}^9 t_1^3 t_2^6 t_3^4 t_4^4 t_5^6 t_6^3 t_7^7 +  y_{s}^5 y_{o_1}^{10} y_{o_2}^{12} y_{o_3}^8 y_{o_4}^8 t_1^4 t_2^4 t_3^5 t_4^3 t_5^7 t_6^3 t_7^7 +  y_{s}^6 y_{o_1}^9 y_{o_2}^{10} y_{o_3}^{11} y_{o_4}^{14} t_1^8 t_2^6 t_3^3 t_4^8 t_5^4 t_6^5 t_7^7 +  2 y_{s}^6 y_{o_1}^9 y_{o_2}^{10} y_{o_3}^{11} y_{o_4}^{14} t_1^7 t_2^7 t_3^3 t_4^8 t_5^4 t_6^5 t_7^7 +  y_{s}^6 y_{o_1}^9 y_{o_2}^{10} y_{o_3}^{11} y_{o_4}^{14} t_1^6 t_2^8 t_3^3 t_4^8 t_5^4 t_6^5 t_7^7 +  3 y_{s}^6 y_{o_1}^{10} y_{o_2}^{11} y_{o_3}^{11} y_{o_4}^{13} t_1^8 t_2^5 t_3^4 t_4^7 t_5^5 t_6^5 t_7^7 + 2 y_{s}^6 y_{o_1}^{10} y_{o_2}^{11} y_{o_3}^{11} y_{o_4}^{13} t_1^7 t_2^6 t_3^4 t_4^7 t_5^5 t_6^5 t_7^7 + 2 y_{s}^6 y_{o_1}^{10} y_{o_2}^{11} y_{o_3}^{11} y_{o_4}^{13} t_1^6 t_2^7 t_3^4 t_4^7 t_5^5 t_6^5 t_7^7 +  3 y_{s}^6 y_{o_1}^{10} y_{o_2}^{11} y_{o_3}^{11} y_{o_4}^{13} t_1^5 t_2^8 t_3^4 t_4^7 t_5^5 t_6^5 t_7^7 + 2 y_{s}^6 y_{o_1}^{11} y_{o_2}^{12} y_{o_3}^{11} y_{o_4}^{12} t_1^8 t_2^4 t_3^5 t_4^6 t_5^6 t_6^5 t_7^7 + 2 y_{s}^6 y_{o_1}^{11} y_{o_2}^{12} y_{o_3}^{11} y_{o_4}^{12} t_1^7 t_2^5 t_3^5 t_4^6 t_5^6 t_6^5 t_7^7 +  y_{s}^6 y_{o_1}^{11} y_{o_2}^{12} y_{o_3}^{11} y_{o_4}^{12} t_1^6 t_2^6 t_3^5 t_4^6 t_5^6 t_6^5 t_7^7 +  2 y_{s}^6 y_{o_1}^{11} y_{o_2}^{12} y_{o_3}^{11} y_{o_4}^{12} t_1^5 t_2^7 t_3^5 t_4^6 t_5^6 t_6^5 t_7^7 + 2 y_{s}^6 y_{o_1}^{11} y_{o_2}^{12} y_{o_3}^{11} y_{o_4}^{12} t_1^4 t_2^8 t_3^5 t_4^6 t_5^6 t_6^5 t_7^7 - y_{s}^6 y_{o_1}^{12} y_{o_2}^{13} y_{o_3}^{11} y_{o_4}^{11} t_1^7 t_2^4 t_3^6 t_4^5 t_5^7 t_6^5 t_7^7 + y_{s}^6 y_{o_1}^{12} y_{o_2}^{13} y_{o_3}^{11} y_{o_4}^{11} t_1^6 t_2^5 t_3^6 t_4^5 t_5^7 t_6^5 t_7^7 + y_{s}^6 y_{o_1}^{12} y_{o_2}^{13} y_{o_3}^{11} y_{o_4}^{11} t_1^5 t_2^6 t_3^6 t_4^5 t_5^7 t_6^5 t_7^7 - y_{s}^6 y_{o_1}^{12} y_{o_2}^{13} y_{o_3}^{11} y_{o_4}^{11} t_1^4 t_2^7 t_3^6 t_4^5 t_5^7 t_6^5 t_7^7 + y_{s}^6 y_{o_1}^{13} y_{o_2}^{14} y_{o_3}^{11} y_{o_4}^{10} t_1^6 t_2^4 t_3^7 t_4^4 t_5^8 t_6^5 t_7^7 + y_{s}^6 y_{o_1}^{13} y_{o_2}^{14} y_{o_3}^{11} y_{o_4}^{10} t_1^5 t_2^5 t_3^7 t_4^4 t_5^8 t_6^5 t_7^7 + y_{s}^6 y_{o_1}^{13} y_{o_2}^{14} y_{o_3}^{11} y_{o_4}^{10} t_1^4 t_2^6 t_3^7 t_4^4 t_5^8 t_6^5 t_7^7 - y_{s}^7 y_{o_1}^{12} y_{o_2}^{12} y_{o_3}^{14} y_{o_4}^{16} t_1^8 t_2^8 t_3^5 t_4^9 t_5^5 t_6^7 t_7^7 - y_{s}^7 y_{o_1}^{13} y_{o_2}^{13} y_{o_3}^{14} y_{o_4}^{15} t_1^9 t_2^6 t_3^6 t_4^8 t_5^6 t_6^7 t_7^7 - 3 y_{s}^7 y_{o_1}^{13} y_{o_2}^{13} y_{o_3}^{14} y_{o_4}^{15} t_1^8 t_2^7 t_3^6 t_4^8 t_5^6 t_6^7 t_7^7 -  3 y_{s}^7 y_{o_1}^{13} y_{o_2}^{13} y_{o_3}^{14} y_{o_4}^{15} t_1^7 t_2^8 t_3^6 t_4^8 t_5^6 t_6^7 t_7^7 - y_{s}^7 y_{o_1}^{13} y_{o_2}^{13} y_{o_3}^{14} y_{o_4}^{15} t_1^6 t_2^9 t_3^6 t_4^8 t_5^6 t_6^7 t_7^7 - 2 y_{s}^7 y_{o_1}^{14} y_{o_2}^{14} y_{o_3}^{14} y_{o_4}^{14} t_1^9 t_2^5 t_3^7 t_4^7 t_5^7 t_6^7 t_7^7 - 3 y_{s}^7 y_{o_1}^{14} y_{o_2}^{14} y_{o_3}^{14} y_{o_4}^{14} t_1^8 t_2^6 t_3^7 t_4^7 t_5^7 t_6^7 t_7^7 -  3 y_{s}^7 y_{o_1}^{14} y_{o_2}^{14} y_{o_3}^{14} y_{o_4}^{14} t_1^7 t_2^7 t_3^7 t_4^7 t_5^7 t_6^7 t_7^7 - 3 y_{s}^7 y_{o_1}^{14} y_{o_2}^{14} y_{o_3}^{14} y_{o_4}^{14} t_1^6 t_2^8 t_3^7 t_4^7 t_5^7 t_6^7 t_7^7 - 2 y_{s}^7 y_{o_1}^{14} y_{o_2}^{14} y_{o_3}^{14} y_{o_4}^{14} t_1^5 t_2^9 t_3^7 t_4^7 t_5^7 t_6^7 t_7^7 -  y_{s}^7 y_{o_1}^{15} y_{o_2}^{15} y_{o_3}^{14} y_{o_4}^{13} t_1^9 t_2^4 t_3^8 t_4^6 t_5^8 t_6^7 t_7^7 -  2 y_{s}^7 y_{o_1}^{15} y_{o_2}^{15} y_{o_3}^{14} y_{o_4}^{13} t_1^8 t_2^5 t_3^8 t_4^6 t_5^8 t_6^7 t_7^7 - 2 y_{s}^7 y_{o_1}^{15} y_{o_2}^{15} y_{o_3}^{14} y_{o_4}^{13} t_1^7 t_2^6 t_3^8 t_4^6 t_5^8 t_6^7 t_7^7 - 2 y_{s}^7 y_{o_1}^{15} y_{o_2}^{15} y_{o_3}^{14} y_{o_4}^{13} t_1^6 t_2^7 t_3^8 t_4^6 t_5^8 t_6^7 t_7^7 -  2 y_{s}^7 y_{o_1}^{15} y_{o_2}^{15} y_{o_3}^{14} y_{o_4}^{13} t_1^5 t_2^8 t_3^8 t_4^6 t_5^8 t_6^7 t_7^7 - y_{s}^7 y_{o_1}^{15} y_{o_2}^{15} y_{o_3}^{14} y_{o_4}^{13} t_1^4 t_2^9 t_3^8 t_4^6 t_5^8 t_6^7 t_7^7 +  y_{s}^8 y_{o_1}^{17} y_{o_2}^{16} y_{o_3}^{17} y_{o_4}^{16} t_1^8 t_2^8 t_3^9 t_4^8 t_5^8 t_6^9 t_7^7 -  y_{s}^5 y_{o_1}^7 y_{o_2}^{10} y_{o_3}^7 y_{o_4}^{10} t_1^5 t_2^5 t_3^2 t_4^5 t_5^5 t_6^2 t_7^8 +  y_{s}^5 y_{o_1}^8 y_{o_2}^{11} y_{o_3}^7 y_{o_4}^9 t_1^6 t_2^3 t_3^3 t_4^4 t_5^6 t_6^2 t_7^8 +  y_{s}^5 y_{o_1}^8 y_{o_2}^{11} y_{o_3}^7 y_{o_4}^9 t_1^3 t_2^6 t_3^3 t_4^4 t_5^6 t_6^2 t_7^8 +  y_{s}^6 y_{o_1}^9 y_{o_2}^{11} y_{o_3}^{10} y_{o_4}^{13} t_1^8 t_2^5 t_3^3 t_4^7 t_5^5 t_6^4 t_7^8 +  y_{s}^6 y_{o_1}^9 y_{o_2}^{11} y_{o_3}^{10} y_{o_4}^{13} t_1^5 t_2^8 t_3^3 t_4^7 t_5^5 t_6^4 t_7^8 +  y_{s}^6 y_{o_1}^{10} y_{o_2}^{12} y_{o_3}^{10} y_{o_4}^{12} t_1^8 t_2^4 t_3^4 t_4^6 t_5^6 t_6^4 t_7^8 +  y_{s}^6 y_{o_1}^{10} y_{o_2}^{12} y_{o_3}^{10} y_{o_4}^{12} t_1^7 t_2^5 t_3^4 t_4^6 t_5^6 t_6^4 t_7^8 -  y_{s}^6 y_{o_1}^{10} y_{o_2}^{12} y_{o_3}^{10} y_{o_4}^{12} t_1^6 t_2^6 t_3^4 t_4^6 t_5^6 t_6^4 t_7^8 +  y_{s}^6 y_{o_1}^{10} y_{o_2}^{12} y_{o_3}^{10} y_{o_4}^{12} t_1^5 t_2^7 t_3^4 t_4^6 t_5^6 t_6^4 t_7^8 +  y_{s}^6 y_{o_1}^{10} y_{o_2}^{12} y_{o_3}^{10} y_{o_4}^{12} t_1^4 t_2^8 t_3^4 t_4^6 t_5^6 t_6^4 t_7^8 +  y_{s}^6 y_{o_1}^{11} y_{o_2}^{13} y_{o_3}^{10} y_{o_4}^{11} t_1^8 t_2^3 t_3^5 t_4^5 t_5^7 t_6^4 t_7^8 +  y_{s}^6 y_{o_1}^{11} y_{o_2}^{13} y_{o_3}^{10} y_{o_4}^{11} t_1^7 t_2^4 t_3^5 t_4^5 t_5^7 t_6^4 t_7^8 +  2 y_{s}^6 y_{o_1}^{11} y_{o_2}^{13} y_{o_3}^{10} y_{o_4}^{11} t_1^6 t_2^5 t_3^5 t_4^5 t_5^7 t_6^4 t_7^8 + 2 y_{s}^6 y_{o_1}^{11} y_{o_2}^{13} y_{o_3}^{10} y_{o_4}^{11} t_1^5 t_2^6 t_3^5 t_4^5 t_5^7 t_6^4 t_7^8 + y_{s}^6 y_{o_1}^{11} y_{o_2}^{13} y_{o_3}^{10} y_{o_4}^{11} t_1^4 t_2^7 t_3^5 t_4^5 t_5^7 t_6^4 t_7^8 + y_{s}^6 y_{o_1}^{11} y_{o_2}^{13} y_{o_3}^{10} y_{o_4}^{11} t_1^3 t_2^8 t_3^5 t_4^5 t_5^7 t_6^4 t_7^8 - y_{s}^7 y_{o_1}^{11} y_{o_2}^{12} y_{o_3}^{13} y_{o_4}^{16} t_1^8 t_2^8 t_3^4 t_4^9 t_5^5 t_6^6 t_7^8 - y_{s}^7 y_{o_1}^{12} y_{o_2}^{13} y_{o_3}^{13} y_{o_4}^{15} t_1^8 t_2^7 t_3^5 t_4^8 t_5^6 t_6^6 t_7^8 - y_{s}^7 y_{o_1}^{12} y_{o_2}^{13} y_{o_3}^{13} y_{o_4}^{15} t_1^7 t_2^8 t_3^5 t_4^8 t_5^6 t_6^6 t_7^8 - y_{s}^7 y_{o_1}^{13} y_{o_2}^{14} y_{o_3}^{13} y_{o_4}^{14} t_1^9 t_2^5 t_3^6 t_4^7 t_5^7 t_6^6 t_7^8 - 2 y_{s}^7 y_{o_1}^{13} y_{o_2}^{14} y_{o_3}^{13} y_{o_4}^{14} t_1^8 t_2^6 t_3^6 t_4^7 t_5^7 t_6^6 t_7^8 -  y_{s}^7 y_{o_1}^{13} y_{o_2}^{14} y_{o_3}^{13} y_{o_4}^{14} t_1^7 t_2^7 t_3^6 t_4^7 t_5^7 t_6^6 t_7^8 -  2 y_{s}^7 y_{o_1}^{13} y_{o_2}^{14} y_{o_3}^{13} y_{o_4}^{14} t_1^6 t_2^8 t_3^6 t_4^7 t_5^7 t_6^6 t_7^8 - y_{s}^7 y_{o_1}^{13} y_{o_2}^{14} y_{o_3}^{13} y_{o_4}^{14} t_1^5 t_2^9 t_3^6 t_4^7 t_5^7 t_6^6 t_7^8 -  y_{s}^7 y_{o_1}^{14} y_{o_2}^{15} y_{o_3}^{13} y_{o_4}^{13} t_1^9 t_2^4 t_3^7 t_4^6 t_5^8 t_6^6 t_7^8 -  y_{s}^7 y_{o_1}^{14} y_{o_2}^{15} y_{o_3}^{13} y_{o_4}^{13} t_1^8 t_2^5 t_3^7 t_4^6 t_5^8 t_6^6 t_7^8 -  y_{s}^7 y_{o_1}^{14} y_{o_2}^{15} y_{o_3}^{13} y_{o_4}^{13} t_1^7 t_2^6 t_3^7 t_4^6 t_5^8 t_6^6 t_7^8 -  y_{s}^7 y_{o_1}^{14} y_{o_2}^{15} y_{o_3}^{13} y_{o_4}^{13} t_1^6 t_2^7 t_3^7 t_4^6 t_5^8 t_6^6 t_7^8 -  y_{s}^7 y_{o_1}^{14} y_{o_2}^{15} y_{o_3}^{13} y_{o_4}^{13} t_1^5 t_2^8 t_3^7 t_4^6 t_5^8 t_6^6 t_7^8 -  y_{s}^7 y_{o_1}^{14} y_{o_2}^{15} y_{o_3}^{13} y_{o_4}^{13} t_1^4 t_2^9 t_3^7 t_4^6 t_5^8 t_6^6 t_7^8 -  y_{s}^7 y_{o_1}^{15} y_{o_2}^{16} y_{o_3}^{13} y_{o_4}^{12} t_1^8 t_2^4 t_3^8 t_4^5 t_5^9 t_6^6 t_7^8 -  y_{s}^7 y_{o_1}^{15} y_{o_2}^{16} y_{o_3}^{13} y_{o_4}^{12} t_1^7 t_2^5 t_3^8 t_4^5 t_5^9 t_6^6 t_7^8 -  y_{s}^7 y_{o_1}^{15} y_{o_2}^{16} y_{o_3}^{13} y_{o_4}^{12} t_1^6 t_2^6 t_3^8 t_4^5 t_5^9 t_6^6 t_7^8 -  y_{s}^7 y_{o_1}^{15} y_{o_2}^{16} y_{o_3}^{13} y_{o_4}^{12} t_1^5 t_2^7 t_3^8 t_4^5 t_5^9 t_6^6 t_7^8 -  y_{s}^7 y_{o_1}^{15} y_{o_2}^{16} y_{o_3}^{13} y_{o_4}^{12} t_1^4 t_2^8 t_3^8 t_4^5 t_5^9 t_6^6 t_7^8 +  y_{s}^8 y_{o_1}^{15} y_{o_2}^{15} y_{o_3}^{16} y_{o_4}^{17} t_1^9 t_2^8 t_3^7 t_4^9 t_5^7 t_6^8 t_7^8 +  y_{s}^8 y_{o_1}^{15} y_{o_2}^{15} y_{o_3}^{16} y_{o_4}^{17} t_1^8 t_2^9 t_3^7 t_4^9 t_5^7 t_6^8 t_7^8 +  y_{s}^8 y_{o_1}^{16} y_{o_2}^{16} y_{o_3}^{16} y_{o_4}^{16} t_1^9 t_2^7 t_3^8 t_4^8 t_5^8 t_6^8 t_7^8 +  y_{s}^8 y_{o_1}^{16} y_{o_2}^{16} y_{o_3}^{16} y_{o_4}^{16} t_1^8 t_2^8 t_3^8 t_4^8 t_5^8 t_6^8 t_7^8 +  y_{s}^8 y_{o_1}^{16} y_{o_2}^{16} y_{o_3}^{16} y_{o_4}^{16} t_1^7 t_2^9 t_3^8 t_4^8 t_5^8 t_6^8 t_7^8 +  y_{s}^8 y_{o_1}^{17} y_{o_2}^{17} y_{o_3}^{16} y_{o_4}^{15} t_1^8 t_2^7 t_3^9 t_4^7 t_5^9 t_6^8 t_7^8 +  y_{s}^8 y_{o_1}^{17} y_{o_2}^{17} y_{o_3}^{16} y_{o_4}^{15} t_1^7 t_2^8 t_3^9 t_4^7 t_5^9 t_6^8 t_7^8 -  y_{s}^6 y_{o_1}^9 y_{o_2}^{12} y_{o_3}^9 y_{o_4}^{12} t_1^6 t_2^6 t_3^3 t_4^6 t_5^6 t_6^3 t_7^9 -  y_{s}^7 y_{o_1}^{11} y_{o_2}^{13} y_{o_3}^{12} y_{o_4}^{15} t_1^8 t_2^7 t_3^4 t_4^8 t_5^6 t_6^5 t_7^9 -  y_{s}^7 y_{o_1}^{11} y_{o_2}^{13} y_{o_3}^{12} y_{o_4}^{15} t_1^7 t_2^8 t_3^4 t_4^8 t_5^6 t_6^5 t_7^9 -  y_{s}^7 y_{o_1}^{12} y_{o_2}^{14} y_{o_3}^{12} y_{o_4}^{14} t_1^8 t_2^6 t_3^5 t_4^7 t_5^7 t_6^5 t_7^9 -  y_{s}^7 y_{o_1}^{12} y_{o_2}^{14} y_{o_3}^{12} y_{o_4}^{14} t_1^7 t_2^7 t_3^5 t_4^7 t_5^7 t_6^5 t_7^9 -  y_{s}^7 y_{o_1}^{12} y_{o_2}^{14} y_{o_3}^{12} y_{o_4}^{14} t_1^6 t_2^8 t_3^5 t_4^7 t_5^7 t_6^5 t_7^9 +  y_{s}^8 y_{o_1}^{15} y_{o_2}^{16} y_{o_3}^{15} y_{o_4}^{16} t_1^9 t_2^7 t_3^7 t_4^8 t_5^8 t_6^7 t_7^9 +  y_{s}^8 y_{o_1}^{15} y_{o_2}^{16} y_{o_3}^{15} y_{o_4}^{16} t_1^8 t_2^8 t_3^7 t_4^8 t_5^8 t_6^7 t_7^9 +  y_{s}^8 y_{o_1}^{15} y_{o_2}^{16} y_{o_3}^{15} y_{o_4}^{16} t_1^7 t_2^9 t_3^7 t_4^8 t_5^8 t_6^7 t_7^9 +  y_{s}^8 y_{o_1}^{16} y_{o_2}^{17} y_{o_3}^{15} y_{o_4}^{15} t_1^8 t_2^7 t_3^8 t_4^7 t_5^9 t_6^7 t_7^9 +  y_{s}^8 y_{o_1}^{16} y_{o_2}^{17} y_{o_3}^{15} y_{o_4}^{15} t_1^7 t_2^8 t_3^8 t_4^7 t_5^9 t_6^7 t_7^9 +  y_{s}^9 y_{o_1}^{18} y_{o_2}^{18} y_{o_3}^{18} y_{o_4}^{18} t_1^9 t_2^9 t_3^9 t_4^9 t_5^9 t_6^9 t_7^9
~,~
$
\end{quote}
\endgroup

\subsection{Model 14 \label{app_num_14}}

\begingroup\makeatletter\def\f@size{7}\check@mathfonts
\begin{quote}\raggedright
$
P(t_i,y_s,y_{o_1},y_{o_2},y_{o_3},y_{o_4}; \mathcal{M}_{14}) =
1 + y_{s} y_{o_1} y_{o_2} y_{o_3}^2 y_{o_4}^3 t_1 t_2 t_3 t_4 t_6^2 -  y_{s}^2 y_{o_1}^2 y_{o_2}^3 y_{o_3}^3 y_{o_4}^5 t_1^2 t_2 t_3 t_4^3 t_5 t_6^3 -  y_{s}^2 y_{o_1}^2 y_{o_2}^3 y_{o_3}^3 y_{o_4}^5 t_1 t_2^2 t_3 t_4^3 t_5 t_6^3 +  y_{s} y_{o_1}^2 y_{o_2}^2 y_{o_3}^2 y_{o_4}^3 t_1^2 t_3 t_4 t_5 t_6 t_7 +  y_{s} y_{o_1}^2 y_{o_2}^2 y_{o_3}^2 y_{o_4}^3 t_1 t_2 t_3 t_4 t_5 t_6 t_7 +  y_{s} y_{o_1}^2 y_{o_2}^2 y_{o_3}^2 y_{o_4}^3 t_2^2 t_3 t_4 t_5 t_6 t_7 -  y_{s}^2 y_{o_1}^3 y_{o_2}^5 y_{o_3}^2 y_{o_4}^4 t_1 t_2 t_4^4 t_5^3 t_6 t_7 +  y_{s} y_{o_1}^2 y_{o_2} y_{o_3}^3 y_{o_4}^4 t_1^2 t_2 t_3^2 t_6^2 t_7 +  y_{s} y_{o_1}^2 y_{o_2} y_{o_3}^3 y_{o_4}^4 t_1 t_2^2 t_3^2 t_6^2 t_7 -  y_{s}^2 y_{o_1}^3 y_{o_2}^4 y_{o_3}^3 y_{o_4}^5 t_1^3 t_3 t_4^3 t_5^2 t_6^2 t_7 -  3 y_{s}^2 y_{o_1}^3 y_{o_2}^4 y_{o_3}^3 y_{o_4}^5 t_1^2 t_2 t_3 t_4^3 t_5^2 t_6^2 t_7 -  3 y_{s}^2 y_{o_1}^3 y_{o_2}^4 y_{o_3}^3 y_{o_4}^5 t_1 t_2^2 t_3 t_4^3 t_5^2 t_6^2 t_7 -  y_{s}^2 y_{o_1}^3 y_{o_2}^4 y_{o_3}^3 y_{o_4}^5 t_2^3 t_3 t_4^3 t_5^2 t_6^2 t_7 -  y_{s}^2 y_{o_1}^3 y_{o_2}^3 y_{o_3}^4 y_{o_4}^6 t_1^4 t_3^2 t_4^2 t_5 t_6^3 t_7 -  2 y_{s}^2 y_{o_1}^3 y_{o_2}^3 y_{o_3}^4 y_{o_4}^6 t_1^3 t_2 t_3^2 t_4^2 t_5 t_6^3 t_7 -  3 y_{s}^2 y_{o_1}^3 y_{o_2}^3 y_{o_3}^4 y_{o_4}^6 t_1^2 t_2^2 t_3^2 t_4^2 t_5 t_6^3 t_7 -  2 y_{s}^2 y_{o_1}^3 y_{o_2}^3 y_{o_3}^4 y_{o_4}^6 t_1 t_2^3 t_3^2 t_4^2 t_5 t_6^3 t_7 -  y_{s}^2 y_{o_1}^3 y_{o_2}^3 y_{o_3}^4 y_{o_4}^6 t_2^4 t_3^2 t_4^2 t_5 t_6^3 t_7 +  2 y_{s}^3 y_{o_1}^4 y_{o_2}^6 y_{o_3}^4 y_{o_4}^7 t_1^3 t_2 t_3 t_4^5 t_5^3 t_6^3 t_7 +  2 y_{s}^3 y_{o_1}^4 y_{o_2}^6 y_{o_3}^4 y_{o_4}^7 t_1^2 t_2^2 t_3 t_4^5 t_5^3 t_6^3 t_7 +  2 y_{s}^3 y_{o_1}^4 y_{o_2}^6 y_{o_3}^4 y_{o_4}^7 t_1 t_2^3 t_3 t_4^5 t_5^3 t_6^3 t_7 -  y_{s}^2 y_{o_1}^3 y_{o_2}^2 y_{o_3}^5 y_{o_4}^7 t_1^4 t_2 t_3^3 t_4 t_6^4 t_7 -  y_{s}^2 y_{o_1}^3 y_{o_2}^2 y_{o_3}^5 y_{o_4}^7 t_1^3 t_2^2 t_3^3 t_4 t_6^4 t_7 -  y_{s}^2 y_{o_1}^3 y_{o_2}^2 y_{o_3}^5 y_{o_4}^7 t_1^2 t_2^3 t_3^3 t_4 t_6^4 t_7 -  y_{s}^2 y_{o_1}^3 y_{o_2}^2 y_{o_3}^5 y_{o_4}^7 t_1 t_2^4 t_3^3 t_4 t_6^4 t_7 +  y_{s}^3 y_{o_1}^4 y_{o_2}^5 y_{o_3}^5 y_{o_4}^8 t_1^4 t_2 t_3^2 t_4^4 t_5^2 t_6^4 t_7 +  2 y_{s}^3 y_{o_1}^4 y_{o_2}^5 y_{o_3}^5 y_{o_4}^8 t_1^3 t_2^2 t_3^2 t_4^4 t_5^2 t_6^4 t_7 +  2 y_{s}^3 y_{o_1}^4 y_{o_2}^5 y_{o_3}^5 y_{o_4}^8 t_1^2 t_2^3 t_3^2 t_4^4 t_5^2 t_6^4 t_7 +  y_{s}^3 y_{o_1}^4 y_{o_2}^5 y_{o_3}^5 y_{o_4}^8 t_1 t_2^4 t_3^2 t_4^4 t_5^2 t_6^4 t_7 +  y_{s}^3 y_{o_1}^4 y_{o_2}^4 y_{o_3}^6 y_{o_4}^9 t_1^5 t_2 t_3^3 t_4^3 t_5 t_6^5 t_7 +  2 y_{s}^3 y_{o_1}^4 y_{o_2}^4 y_{o_3}^6 y_{o_4}^9 t_1^4 t_2^2 t_3^3 t_4^3 t_5 t_6^5 t_7 +  2 y_{s}^3 y_{o_1}^4 y_{o_2}^4 y_{o_3}^6 y_{o_4}^9 t_1^3 t_2^3 t_3^3 t_4^3 t_5 t_6^5 t_7 +  2 y_{s}^3 y_{o_1}^4 y_{o_2}^4 y_{o_3}^6 y_{o_4}^9 t_1^2 t_2^4 t_3^3 t_4^3 t_5 t_6^5 t_7 +  y_{s}^3 y_{o_1}^4 y_{o_2}^4 y_{o_3}^6 y_{o_4}^9 t_1 t_2^5 t_3^3 t_4^3 t_5 t_6^5 t_7 -  y_{s}^4 y_{o_1}^5 y_{o_2}^7 y_{o_3}^6 y_{o_4}^{10} t_1^3 t_2^3 t_3^2 t_4^6 t_5^3 t_6^5 t_7 -  y_{s}^4 y_{o_1}^5 y_{o_2}^6 y_{o_3}^7 y_{o_4}^{11} t_1^4 t_2^3 t_3^3 t_4^5 t_5^2 t_6^6 t_7 -  y_{s}^4 y_{o_1}^5 y_{o_2}^6 y_{o_3}^7 y_{o_4}^{11} t_1^3 t_2^4 t_3^3 t_4^5 t_5^2 t_6^6 t_7 +  y_{s} y_{o_1}^3 y_{o_2}^3 y_{o_3}^2 y_{o_4}^3 t_1^2 t_3 t_4 t_5^2 t_7^2 +  y_{s} y_{o_1}^3 y_{o_2}^3 y_{o_3}^2 y_{o_4}^3 t_1 t_2 t_3 t_4 t_5^2 t_7^2 +  y_{s} y_{o_1}^3 y_{o_2}^3 y_{o_3}^2 y_{o_4}^3 t_2^2 t_3 t_4 t_5^2 t_7^2 +  y_{s} y_{o_1}^3 y_{o_2}^2 y_{o_3}^3 y_{o_4}^4 t_1^3 t_3^2 t_5 t_6 t_7^2 +  y_{s} y_{o_1}^3 y_{o_2}^2 y_{o_3}^3 y_{o_4}^4 t_1^2 t_2 t_3^2 t_5 t_6 t_7^2 +  y_{s} y_{o_1}^3 y_{o_2}^2 y_{o_3}^3 y_{o_4}^4 t_1 t_2^2 t_3^2 t_5 t_6 t_7^2 +  y_{s} y_{o_1}^3 y_{o_2}^2 y_{o_3}^3 y_{o_4}^4 t_2^3 t_3^2 t_5 t_6 t_7^2 -  y_{s}^2 y_{o_1}^4 y_{o_2}^5 y_{o_3}^3 y_{o_4}^5 t_1^3 t_3 t_4^3 t_5^3 t_6 t_7^2 -  3 y_{s}^2 y_{o_1}^4 y_{o_2}^5 y_{o_3}^3 y_{o_4}^5 t_1^2 t_2 t_3 t_4^3 t_5^3 t_6 t_7^2 -  3 y_{s}^2 y_{o_1}^4 y_{o_2}^5 y_{o_3}^3 y_{o_4}^5 t_1 t_2^2 t_3 t_4^3 t_5^3 t_6 t_7^2 -  y_{s}^2 y_{o_1}^4 y_{o_2}^5 y_{o_3}^3 y_{o_4}^5 t_2^3 t_3 t_4^3 t_5^3 t_6 t_7^2 -  2 y_{s}^2 y_{o_1}^4 y_{o_2}^4 y_{o_3}^4 y_{o_4}^6 t_1^4 t_3^2 t_4^2 t_5^2 t_6^2 t_7^2 -  4 y_{s}^2 y_{o_1}^4 y_{o_2}^4 y_{o_3}^4 y_{o_4}^6 t_1^3 t_2 t_3^2 t_4^2 t_5^2 t_6^2 t_7^2 -  5 y_{s}^2 y_{o_1}^4 y_{o_2}^4 y_{o_3}^4 y_{o_4}^6 t_1^2 t_2^2 t_3^2 t_4^2 t_5^2 t_6^2 t_7^2 -  4 y_{s}^2 y_{o_1}^4 y_{o_2}^4 y_{o_3}^4 y_{o_4}^6 t_1 t_2^3 t_3^2 t_4^2 t_5^2 t_6^2 t_7^2 -  2 y_{s}^2 y_{o_1}^4 y_{o_2}^4 y_{o_3}^4 y_{o_4}^6 t_2^4 t_3^2 t_4^2 t_5^2 t_6^2 t_7^2 +  2 y_{s}^3 y_{o_1}^5 y_{o_2}^7 y_{o_3}^4 y_{o_4}^7 t_1^3 t_2 t_3 t_4^5 t_5^4 t_6^2 t_7^2 +  3 y_{s}^3 y_{o_1}^5 y_{o_2}^7 y_{o_3}^4 y_{o_4}^7 t_1^2 t_2^2 t_3 t_4^5 t_5^4 t_6^2 t_7^2 +  2 y_{s}^3 y_{o_1}^5 y_{o_2}^7 y_{o_3}^4 y_{o_4}^7 t_1 t_2^3 t_3 t_4^5 t_5^4 t_6^2 t_7^2 -  y_{s}^2 y_{o_1}^4 y_{o_2}^3 y_{o_3}^5 y_{o_4}^7 t_1^5 t_3^3 t_4 t_5 t_6^3 t_7^2 -  y_{s}^2 y_{o_1}^4 y_{o_2}^3 y_{o_3}^5 y_{o_4}^7 t_1^4 t_2 t_3^3 t_4 t_5 t_6^3 t_7^2 -  2 y_{s}^2 y_{o_1}^4 y_{o_2}^3 y_{o_3}^5 y_{o_4}^7 t_1^3 t_2^2 t_3^3 t_4 t_5 t_6^3 t_7^2 -  2 y_{s}^2 y_{o_1}^4 y_{o_2}^3 y_{o_3}^5 y_{o_4}^7 t_1^2 t_2^3 t_3^3 t_4 t_5 t_6^3 t_7^2 -  y_{s}^2 y_{o_1}^4 y_{o_2}^3 y_{o_3}^5 y_{o_4}^7 t_1 t_2^4 t_3^3 t_4 t_5 t_6^3 t_7^2 -  y_{s}^2 y_{o_1}^4 y_{o_2}^3 y_{o_3}^5 y_{o_4}^7 t_2^5 t_3^3 t_4 t_5 t_6^3 t_7^2 +  y_{s}^3 y_{o_1}^5 y_{o_2}^6 y_{o_3}^5 y_{o_4}^8 t_1^5 t_3^2 t_4^4 t_5^3 t_6^3 t_7^2 +  5 y_{s}^3 y_{o_1}^5 y_{o_2}^6 y_{o_3}^5 y_{o_4}^8 t_1^4 t_2 t_3^2 t_4^4 t_5^3 t_6^3 t_7^2 +  6 y_{s}^3 y_{o_1}^5 y_{o_2}^6 y_{o_3}^5 y_{o_4}^8 t_1^3 t_2^2 t_3^2 t_4^4 t_5^3 t_6^3 t_7^2 +  6 y_{s}^3 y_{o_1}^5 y_{o_2}^6 y_{o_3}^5 y_{o_4}^8 t_1^2 t_2^3 t_3^2 t_4^4 t_5^3 t_6^3 t_7^2 +  5 y_{s}^3 y_{o_1}^5 y_{o_2}^6 y_{o_3}^5 y_{o_4}^8 t_1 t_2^4 t_3^2 t_4^4 t_5^3 t_6^3 t_7^2 +  y_{s}^3 y_{o_1}^5 y_{o_2}^6 y_{o_3}^5 y_{o_4}^8 t_2^5 t_3^2 t_4^4 t_5^3 t_6^3 t_7^2 -  y_{s}^4 y_{o_1}^6 y_{o_2}^9 y_{o_3}^5 y_{o_4}^9 t_1^3 t_2^2 t_3 t_4^7 t_5^5 t_6^3 t_7^2 -  y_{s}^4 y_{o_1}^6 y_{o_2}^9 y_{o_3}^5 y_{o_4}^9 t_1^2 t_2^3 t_3 t_4^7 t_5^5 t_6^3 t_7^2 +  y_{s}^3 y_{o_1}^5 y_{o_2}^5 y_{o_3}^6 y_{o_4}^9 t_1^6 t_3^3 t_4^3 t_5^2 t_6^4 t_7^2 +  3 y_{s}^3 y_{o_1}^5 y_{o_2}^5 y_{o_3}^6 y_{o_4}^9 t_1^5 t_2 t_3^3 t_4^3 t_5^2 t_6^4 t_7^2 +  5 y_{s}^3 y_{o_1}^5 y_{o_2}^5 y_{o_3}^6 y_{o_4}^9 t_1^4 t_2^2 t_3^3 t_4^3 t_5^2 t_6^4 t_7^2 +  6 y_{s}^3 y_{o_1}^5 y_{o_2}^5 y_{o_3}^6 y_{o_4}^9 t_1^3 t_2^3 t_3^3 t_4^3 t_5^2 t_6^4 t_7^2 +  5 y_{s}^3 y_{o_1}^5 y_{o_2}^5 y_{o_3}^6 y_{o_4}^9 t_1^2 t_2^4 t_3^3 t_4^3 t_5^2 t_6^4 t_7^2 +  3 y_{s}^3 y_{o_1}^5 y_{o_2}^5 y_{o_3}^6 y_{o_4}^9 t_1 t_2^5 t_3^3 t_4^3 t_5^2 t_6^4 t_7^2 +  y_{s}^3 y_{o_1}^5 y_{o_2}^5 y_{o_3}^6 y_{o_4}^9 t_2^6 t_3^3 t_4^3 t_5^2 t_6^4 t_7^2 -  y_{s}^4 y_{o_1}^6 y_{o_2}^8 y_{o_3}^6 y_{o_4}^{10} t_1^5 t_2 t_3^2 t_4^6 t_5^4 t_6^4 t_7^2 -  2 y_{s}^4 y_{o_1}^6 y_{o_2}^8 y_{o_3}^6 y_{o_4}^{10} t_1^4 t_2^2 t_3^2 t_4^6 t_5^4 t_6^4 t_7^2 -  3 y_{s}^4 y_{o_1}^6 y_{o_2}^8 y_{o_3}^6 y_{o_4}^{10} t_1^3 t_2^3 t_3^2 t_4^6 t_5^4 t_6^4 t_7^2 -  2 y_{s}^4 y_{o_1}^6 y_{o_2}^8 y_{o_3}^6 y_{o_4}^{10} t_1^2 t_2^4 t_3^2 t_4^6 t_5^4 t_6^4 t_7^2 -  y_{s}^4 y_{o_1}^6 y_{o_2}^8 y_{o_3}^6 y_{o_4}^{10} t_1 t_2^5 t_3^2 t_4^6 t_5^4 t_6^4 t_7^2 +  y_{s}^3 y_{o_1}^5 y_{o_2}^4 y_{o_3}^7 y_{o_4}^{10} t_1^5 t_2^2 t_3^4 t_4^2 t_5 t_6^5 t_7^2 +  y_{s}^3 y_{o_1}^5 y_{o_2}^4 y_{o_3}^7 y_{o_4}^{10} t_1^4 t_2^3 t_3^4 t_4^2 t_5 t_6^5 t_7^2 +  y_{s}^3 y_{o_1}^5 y_{o_2}^4 y_{o_3}^7 y_{o_4}^{10} t_1^3 t_2^4 t_3^4 t_4^2 t_5 t_6^5 t_7^2 +  y_{s}^3 y_{o_1}^5 y_{o_2}^4 y_{o_3}^7 y_{o_4}^{10} t_1^2 t_2^5 t_3^4 t_4^2 t_5 t_6^5 t_7^2 -  2 y_{s}^4 y_{o_1}^6 y_{o_2}^7 y_{o_3}^7 y_{o_4}^{11} t_1^6 t_2 t_3^3 t_4^5 t_5^3 t_6^5 t_7^2 -  3 y_{s}^4 y_{o_1}^6 y_{o_2}^7 y_{o_3}^7 y_{o_4}^{11} t_1^5 t_2^2 t_3^3 t_4^5 t_5^3 t_6^5 t_7^2 -  6 y_{s}^4 y_{o_1}^6 y_{o_2}^7 y_{o_3}^7 y_{o_4}^{11} t_1^4 t_2^3 t_3^3 t_4^5 t_5^3 t_6^5 t_7^2 -  6 y_{s}^4 y_{o_1}^6 y_{o_2}^7 y_{o_3}^7 y_{o_4}^{11} t_1^3 t_2^4 t_3^3 t_4^5 t_5^3 t_6^5 t_7^2 -  3 y_{s}^4 y_{o_1}^6 y_{o_2}^7 y_{o_3}^7 y_{o_4}^{11} t_1^2 t_2^5 t_3^3 t_4^5 t_5^3 t_6^5 t_7^2 -  2 y_{s}^4 y_{o_1}^6 y_{o_2}^7 y_{o_3}^7 y_{o_4}^{11} t_1 t_2^6 t_3^3 t_4^5 t_5^3 t_6^5 t_7^2 -  y_{s}^4 y_{o_1}^6 y_{o_2}^6 y_{o_3}^8 y_{o_4}^{12} t_1^6 t_2^2 t_3^4 t_4^4 t_5^2 t_6^6 t_7^2 -  y_{s}^4 y_{o_1}^6 y_{o_2}^6 y_{o_3}^8 y_{o_4}^{12} t_1^5 t_2^3 t_3^4 t_4^4 t_5^2 t_6^6 t_7^2 -  y_{s}^4 y_{o_1}^6 y_{o_2}^6 y_{o_3}^8 y_{o_4}^{12} t_1^4 t_2^4 t_3^4 t_4^4 t_5^2 t_6^6 t_7^2 -  y_{s}^4 y_{o_1}^6 y_{o_2}^6 y_{o_3}^8 y_{o_4}^{12} t_1^3 t_2^5 t_3^4 t_4^4 t_5^2 t_6^6 t_7^2 -  y_{s}^4 y_{o_1}^6 y_{o_2}^6 y_{o_3}^8 y_{o_4}^{12} t_1^2 t_2^6 t_3^4 t_4^4 t_5^2 t_6^6 t_7^2 +  y_{s}^5 y_{o_1}^7 y_{o_2}^9 y_{o_3}^8 y_{o_4}^{13} t_1^5 t_2^3 t_3^3 t_4^7 t_5^4 t_6^6 t_7^2 +  2 y_{s}^5 y_{o_1}^7 y_{o_2}^9 y_{o_3}^8 y_{o_4}^{13} t_1^4 t_2^4 t_3^3 t_4^7 t_5^4 t_6^6 t_7^2 +  y_{s}^5 y_{o_1}^7 y_{o_2}^9 y_{o_3}^8 y_{o_4}^{13} t_1^3 t_2^5 t_3^3 t_4^7 t_5^4 t_6^6 t_7^2 +  y_{s}^5 y_{o_1}^7 y_{o_2}^8 y_{o_3}^9 y_{o_4}^{14} t_1^6 t_2^3 t_3^4 t_4^6 t_5^3 t_6^7 t_7^2 +  y_{s}^5 y_{o_1}^7 y_{o_2}^8 y_{o_3}^9 y_{o_4}^{14} t_1^3 t_2^6 t_3^4 t_4^6 t_5^3 t_6^7 t_7^2 -  y_{s}^5 y_{o_1}^7 y_{o_2}^7 y_{o_3}^{10} y_{o_4}^{15} t_1^5 t_2^5 t_3^5 t_4^5 t_5^2 t_6^8 t_7^2 +  y_{s} y_{o_1}^4 y_{o_2}^3 y_{o_3}^3 y_{o_4}^4 t_1^2 t_2 t_3^2 t_5^2 t_7^3 +  y_{s} y_{o_1}^4 y_{o_2}^3 y_{o_3}^3 y_{o_4}^4 t_1 t_2^2 t_3^2 t_5^2 t_7^3 -  y_{s}^2 y_{o_1}^5 y_{o_2}^6 y_{o_3}^3 y_{o_4}^5 t_1^2 t_2 t_3 t_4^3 t_5^4 t_7^3 -  y_{s}^2 y_{o_1}^5 y_{o_2}^6 y_{o_3}^3 y_{o_4}^5 t_1 t_2^2 t_3 t_4^3 t_5^4 t_7^3 -  y_{s}^2 y_{o_1}^5 y_{o_2}^5 y_{o_3}^4 y_{o_4}^6 t_1^4 t_3^2 t_4^2 t_5^3 t_6 t_7^3 -  3 y_{s}^2 y_{o_1}^5 y_{o_2}^5 y_{o_3}^4 y_{o_4}^6 t_1^3 t_2 t_3^2 t_4^2 t_5^3 t_6 t_7^3 -  3 y_{s}^2 y_{o_1}^5 y_{o_2}^5 y_{o_3}^4 y_{o_4}^6 t_1^2 t_2^2 t_3^2 t_4^2 t_5^3 t_6 t_7^3 -  3 y_{s}^2 y_{o_1}^5 y_{o_2}^5 y_{o_3}^4 y_{o_4}^6 t_1 t_2^3 t_3^2 t_4^2 t_5^3 t_6 t_7^3 -  y_{s}^2 y_{o_1}^5 y_{o_2}^5 y_{o_3}^4 y_{o_4}^6 t_2^4 t_3^2 t_4^2 t_5^3 t_6 t_7^3 +  y_{s}^3 y_{o_1}^6 y_{o_2}^8 y_{o_3}^4 y_{o_4}^7 t_1^3 t_2 t_3 t_4^5 t_5^5 t_6 t_7^3 +  2 y_{s}^3 y_{o_1}^6 y_{o_2}^8 y_{o_3}^4 y_{o_4}^7 t_1^2 t_2^2 t_3 t_4^5 t_5^5 t_6 t_7^3 +  y_{s}^3 y_{o_1}^6 y_{o_2}^8 y_{o_3}^4 y_{o_4}^7 t_1 t_2^3 t_3 t_4^5 t_5^5 t_6 t_7^3 -  y_{s}^2 y_{o_1}^5 y_{o_2}^4 y_{o_3}^5 y_{o_4}^7 t_1^5 t_3^3 t_4 t_5^2 t_6^2 t_7^3 -  2 y_{s}^2 y_{o_1}^5 y_{o_2}^4 y_{o_3}^5 y_{o_4}^7 t_1^4 t_2 t_3^3 t_4 t_5^2 t_6^2 t_7^3 -  2 y_{s}^2 y_{o_1}^5 y_{o_2}^4 y_{o_3}^5 y_{o_4}^7 t_1^3 t_2^2 t_3^3 t_4 t_5^2 t_6^2 t_7^3 -  2 y_{s}^2 y_{o_1}^5 y_{o_2}^4 y_{o_3}^5 y_{o_4}^7 t_1^2 t_2^3 t_3^3 t_4 t_5^2 t_6^2 t_7^3 -  2 y_{s}^2 y_{o_1}^5 y_{o_2}^4 y_{o_3}^5 y_{o_4}^7 t_1 t_2^4 t_3^3 t_4 t_5^2 t_6^2 t_7^3 -  y_{s}^2 y_{o_1}^5 y_{o_2}^4 y_{o_3}^5 y_{o_4}^7 t_2^5 t_3^3 t_4 t_5^2 t_6^2 t_7^3 +  3 y_{s}^3 y_{o_1}^6 y_{o_2}^7 y_{o_3}^5 y_{o_4}^8 t_1^4 t_2 t_3^2 t_4^4 t_5^4 t_6^2 t_7^3 +  4 y_{s}^3 y_{o_1}^6 y_{o_2}^7 y_{o_3}^5 y_{o_4}^8 t_1^3 t_2^2 t_3^2 t_4^4 t_5^4 t_6^2 t_7^3 +  4 y_{s}^3 y_{o_1}^6 y_{o_2}^7 y_{o_3}^5 y_{o_4}^8 t_1^2 t_2^3 t_3^2 t_4^4 t_5^4 t_6^2 t_7^3 +  3 y_{s}^3 y_{o_1}^6 y_{o_2}^7 y_{o_3}^5 y_{o_4}^8 t_1 t_2^4 t_3^2 t_4^4 t_5^4 t_6^2 t_7^3 -  y_{s}^4 y_{o_1}^7 y_{o_2}^{10} y_{o_3}^5 y_{o_4}^9 t_1^3 t_2^2 t_3 t_4^7 t_5^6 t_6^2 t_7^3 -  y_{s}^4 y_{o_1}^7 y_{o_2}^{10} y_{o_3}^5 y_{o_4}^9 t_1^2 t_2^3 t_3 t_4^7 t_5^6 t_6^2 t_7^3 -  y_{s}^2 y_{o_1}^5 y_{o_2}^3 y_{o_3}^6 y_{o_4}^8 t_1^3 t_2^3 t_3^4 t_5 t_6^3 t_7^3 +  y_{s}^3 y_{o_1}^6 y_{o_2}^6 y_{o_3}^6 y_{o_4}^9 t_1^6 t_3^3 t_4^3 t_5^3 t_6^3 t_7^3 +  4 y_{s}^3 y_{o_1}^6 y_{o_2}^6 y_{o_3}^6 y_{o_4}^9 t_1^5 t_2 t_3^3 t_4^3 t_5^3 t_6^3 t_7^3 +  5 y_{s}^3 y_{o_1}^6 y_{o_2}^6 y_{o_3}^6 y_{o_4}^9 t_1^4 t_2^2 t_3^3 t_4^3 t_5^3 t_6^3 t_7^3 +  5 y_{s}^3 y_{o_1}^6 y_{o_2}^6 y_{o_3}^6 y_{o_4}^9 t_1^3 t_2^3 t_3^3 t_4^3 t_5^3 t_6^3 t_7^3 +  5 y_{s}^3 y_{o_1}^6 y_{o_2}^6 y_{o_3}^6 y_{o_4}^9 t_1^2 t_2^4 t_3^3 t_4^3 t_5^3 t_6^3 t_7^3 +  4 y_{s}^3 y_{o_1}^6 y_{o_2}^6 y_{o_3}^6 y_{o_4}^9 t_1 t_2^5 t_3^3 t_4^3 t_5^3 t_6^3 t_7^3 +  y_{s}^3 y_{o_1}^6 y_{o_2}^6 y_{o_3}^6 y_{o_4}^9 t_2^6 t_3^3 t_4^3 t_5^3 t_6^3 t_7^3 -  y_{s}^4 y_{o_1}^7 y_{o_2}^9 y_{o_3}^6 y_{o_4}^{10} t_1^5 t_2 t_3^2 t_4^6 t_5^5 t_6^3 t_7^3 -  3 y_{s}^4 y_{o_1}^7 y_{o_2}^9 y_{o_3}^6 y_{o_4}^{10} t_1^4 t_2^2 t_3^2 t_4^6 t_5^5 t_6^3 t_7^3 -  3 y_{s}^4 y_{o_1}^7 y_{o_2}^9 y_{o_3}^6 y_{o_4}^{10} t_1^3 t_2^3 t_3^2 t_4^6 t_5^5 t_6^3 t_7^3 -  3 y_{s}^4 y_{o_1}^7 y_{o_2}^9 y_{o_3}^6 y_{o_4}^{10} t_1^2 t_2^4 t_3^2 t_4^6 t_5^5 t_6^3 t_7^3 -  y_{s}^4 y_{o_1}^7 y_{o_2}^9 y_{o_3}^6 y_{o_4}^{10} t_1 t_2^5 t_3^2 t_4^6 t_5^5 t_6^3 t_7^3 +  y_{s}^3 y_{o_1}^6 y_{o_2}^5 y_{o_3}^7 y_{o_4}^{10} t_1^5 t_2^2 t_3^4 t_4^2 t_5^2 t_6^4 t_7^3 +  2 y_{s}^3 y_{o_1}^6 y_{o_2}^5 y_{o_3}^7 y_{o_4}^{10} t_1^4 t_2^3 t_3^4 t_4^2 t_5^2 t_6^4 t_7^3 +  2 y_{s}^3 y_{o_1}^6 y_{o_2}^5 y_{o_3}^7 y_{o_4}^{10} t_1^3 t_2^4 t_3^4 t_4^2 t_5^2 t_6^4 t_7^3 +  y_{s}^3 y_{o_1}^6 y_{o_2}^5 y_{o_3}^7 y_{o_4}^{10} t_1^2 t_2^5 t_3^4 t_4^2 t_5^2 t_6^4 t_7^3 -  2 y_{s}^4 y_{o_1}^7 y_{o_2}^8 y_{o_3}^7 y_{o_4}^{11} t_1^6 t_2 t_3^3 t_4^5 t_5^4 t_6^4 t_7^3 -  3 y_{s}^4 y_{o_1}^7 y_{o_2}^8 y_{o_3}^7 y_{o_4}^{11} t_1^5 t_2^2 t_3^3 t_4^5 t_5^4 t_6^4 t_7^3 -  5 y_{s}^4 y_{o_1}^7 y_{o_2}^8 y_{o_3}^7 y_{o_4}^{11} t_1^4 t_2^3 t_3^3 t_4^5 t_5^4 t_6^4 t_7^3 -  5 y_{s}^4 y_{o_1}^7 y_{o_2}^8 y_{o_3}^7 y_{o_4}^{11} t_1^3 t_2^4 t_3^3 t_4^5 t_5^4 t_6^4 t_7^3 -  3 y_{s}^4 y_{o_1}^7 y_{o_2}^8 y_{o_3}^7 y_{o_4}^{11} t_1^2 t_2^5 t_3^3 t_4^5 t_5^4 t_6^4 t_7^3 -  2 y_{s}^4 y_{o_1}^7 y_{o_2}^8 y_{o_3}^7 y_{o_4}^{11} t_1 t_2^6 t_3^3 t_4^5 t_5^4 t_6^4 t_7^3 +  y_{s}^5 y_{o_1}^8 y_{o_2}^{11} y_{o_3}^7 y_{o_4}^{12} t_1^5 t_2^2 t_3^2 t_4^8 t_5^6 t_6^4 t_7^3 +  y_{s}^5 y_{o_1}^8 y_{o_2}^{11} y_{o_3}^7 y_{o_4}^{12} t_1^4 t_2^3 t_3^2 t_4^8 t_5^6 t_6^4 t_7^3 +  y_{s}^5 y_{o_1}^8 y_{o_2}^{11} y_{o_3}^7 y_{o_4}^{12} t_1^3 t_2^4 t_3^2 t_4^8 t_5^6 t_6^4 t_7^3 +  y_{s}^5 y_{o_1}^8 y_{o_2}^{11} y_{o_3}^7 y_{o_4}^{12} t_1^2 t_2^5 t_3^2 t_4^8 t_5^6 t_6^4 t_7^3 +  y_{s}^3 y_{o_1}^6 y_{o_2}^4 y_{o_3}^8 y_{o_4}^{11} t_1^5 t_2^3 t_3^5 t_4 t_5 t_6^5 t_7^3 +  y_{s}^3 y_{o_1}^6 y_{o_2}^4 y_{o_3}^8 y_{o_4}^{11} t_1^3 t_2^5 t_3^5 t_4 t_5 t_6^5 t_7^3 -  y_{s}^4 y_{o_1}^7 y_{o_2}^7 y_{o_3}^8 y_{o_4}^{12} t_1^6 t_2^2 t_3^4 t_4^4 t_5^3 t_6^5 t_7^3 -  2 y_{s}^4 y_{o_1}^7 y_{o_2}^7 y_{o_3}^8 y_{o_4}^{12} t_1^5 t_2^3 t_3^4 t_4^4 t_5^3 t_6^5 t_7^3 -  4 y_{s}^4 y_{o_1}^7 y_{o_2}^7 y_{o_3}^8 y_{o_4}^{12} t_1^4 t_2^4 t_3^4 t_4^4 t_5^3 t_6^5 t_7^3 -  2 y_{s}^4 y_{o_1}^7 y_{o_2}^7 y_{o_3}^8 y_{o_4}^{12} t_1^3 t_2^5 t_3^4 t_4^4 t_5^3 t_6^5 t_7^3 -  y_{s}^4 y_{o_1}^7 y_{o_2}^7 y_{o_3}^8 y_{o_4}^{12} t_1^2 t_2^6 t_3^4 t_4^4 t_5^3 t_6^5 t_7^3 +  y_{s}^5 y_{o_1}^8 y_{o_2}^{10} y_{o_3}^8 y_{o_4}^{13} t_1^6 t_2^2 t_3^3 t_4^7 t_5^5 t_6^5 t_7^3 +  2 y_{s}^5 y_{o_1}^8 y_{o_2}^{10} y_{o_3}^8 y_{o_4}^{13} t_1^5 t_2^3 t_3^3 t_4^7 t_5^5 t_6^5 t_7^3 +  3 y_{s}^5 y_{o_1}^8 y_{o_2}^{10} y_{o_3}^8 y_{o_4}^{13} t_1^4 t_2^4 t_3^3 t_4^7 t_5^5 t_6^5 t_7^3 +  2 y_{s}^5 y_{o_1}^8 y_{o_2}^{10} y_{o_3}^8 y_{o_4}^{13} t_1^3 t_2^5 t_3^3 t_4^7 t_5^5 t_6^5 t_7^3 +  y_{s}^5 y_{o_1}^8 y_{o_2}^{10} y_{o_3}^8 y_{o_4}^{13} t_1^2 t_2^6 t_3^3 t_4^7 t_5^5 t_6^5 t_7^3 -  y_{s}^4 y_{o_1}^7 y_{o_2}^6 y_{o_3}^9 y_{o_4}^{13} t_1^6 t_2^3 t_3^5 t_4^3 t_5^2 t_6^6 t_7^3 -  y_{s}^4 y_{o_1}^7 y_{o_2}^6 y_{o_3}^9 y_{o_4}^{13} t_1^5 t_2^4 t_3^5 t_4^3 t_5^2 t_6^6 t_7^3 -  y_{s}^4 y_{o_1}^7 y_{o_2}^6 y_{o_3}^9 y_{o_4}^{13} t_1^4 t_2^5 t_3^5 t_4^3 t_5^2 t_6^6 t_7^3 -  y_{s}^4 y_{o_1}^7 y_{o_2}^6 y_{o_3}^9 y_{o_4}^{13} t_1^3 t_2^6 t_3^5 t_4^3 t_5^2 t_6^6 t_7^3 +  y_{s}^5 y_{o_1}^8 y_{o_2}^9 y_{o_3}^9 y_{o_4}^{14} t_1^6 t_2^3 t_3^4 t_4^6 t_5^4 t_6^6 t_7^3 +  y_{s}^5 y_{o_1}^8 y_{o_2}^9 y_{o_3}^9 y_{o_4}^{14} t_1^5 t_2^4 t_3^4 t_4^6 t_5^4 t_6^6 t_7^3 +  y_{s}^5 y_{o_1}^8 y_{o_2}^9 y_{o_3}^9 y_{o_4}^{14} t_1^4 t_2^5 t_3^4 t_4^6 t_5^4 t_6^6 t_7^3 +  y_{s}^5 y_{o_1}^8 y_{o_2}^9 y_{o_3}^9 y_{o_4}^{14} t_1^3 t_2^6 t_3^4 t_4^6 t_5^4 t_6^6 t_7^3 -  y_{s}^6 y_{o_1}^9 y_{o_2}^{12} y_{o_3}^9 y_{o_4}^{15} t_1^5 t_2^4 t_3^3 t_4^9 t_5^6 t_6^6 t_7^3 -  y_{s}^6 y_{o_1}^9 y_{o_2}^{12} y_{o_3}^9 y_{o_4}^{15} t_1^4 t_2^5 t_3^3 t_4^9 t_5^6 t_6^6 t_7^3 -  y_{s}^4 y_{o_1}^7 y_{o_2}^5 y_{o_3}^{10} y_{o_4}^{14} t_1^5 t_2^5 t_3^6 t_4^2 t_5 t_6^7 t_7^3 +  y_{s}^5 y_{o_1}^8 y_{o_2}^8 y_{o_3}^{10} y_{o_4}^{15} t_1^6 t_2^4 t_3^5 t_4^5 t_5^3 t_6^7 t_7^3 -  y_{s}^5 y_{o_1}^8 y_{o_2}^8 y_{o_3}^{10} y_{o_4}^{15} t_1^5 t_2^5 t_3^5 t_4^5 t_5^3 t_6^7 t_7^3 +  y_{s}^5 y_{o_1}^8 y_{o_2}^8 y_{o_3}^{10} y_{o_4}^{15} t_1^4 t_2^6 t_3^5 t_4^5 t_5^3 t_6^7 t_7^3 +  y_{s}^5 y_{o_1}^8 y_{o_2}^7 y_{o_3}^{11} y_{o_4}^{16} t_1^6 t_2^5 t_3^6 t_4^4 t_5^2 t_6^8 t_7^3 +  y_{s}^5 y_{o_1}^8 y_{o_2}^7 y_{o_3}^{11} y_{o_4}^{16} t_1^5 t_2^6 t_3^6 t_4^4 t_5^2 t_6^8 t_7^3 +  y_{s}^6 y_{o_1}^9 y_{o_2}^{10} y_{o_3}^{11} y_{o_4}^{17} t_1^6 t_2^5 t_3^5 t_4^7 t_5^4 t_6^8 t_7^3 +  y_{s}^6 y_{o_1}^9 y_{o_2}^{10} y_{o_3}^{11} y_{o_4}^{17} t_1^5 t_2^6 t_3^5 t_4^7 t_5^4 t_6^8 t_7^3 -  y_{s}^6 y_{o_1}^9 y_{o_2}^9 y_{o_3}^{12} y_{o_4}^{18} t_1^6 t_2^6 t_3^6 t_4^6 t_5^3 t_6^9 t_7^3 -  y_{s}^2 y_{o_1}^6 y_{o_2}^6 y_{o_3}^4 y_{o_4}^6 t_1^3 t_2 t_3^2 t_4^2 t_5^4 t_7^4 -  y_{s}^2 y_{o_1}^6 y_{o_2}^6 y_{o_3}^4 y_{o_4}^6 t_1^2 t_2^2 t_3^2 t_4^2 t_5^4 t_7^4 -  y_{s}^2 y_{o_1}^6 y_{o_2}^6 y_{o_3}^4 y_{o_4}^6 t_1 t_2^3 t_3^2 t_4^2 t_5^4 t_7^4 +  2 y_{s}^3 y_{o_1}^7 y_{o_2}^8 y_{o_3}^5 y_{o_4}^8 t_1^4 t_2 t_3^2 t_4^4 t_5^5 t_6 t_7^4 +  2 y_{s}^3 y_{o_1}^7 y_{o_2}^8 y_{o_3}^5 y_{o_4}^8 t_1^3 t_2^2 t_3^2 t_4^4 t_5^5 t_6 t_7^4 +  2 y_{s}^3 y_{o_1}^7 y_{o_2}^8 y_{o_3}^5 y_{o_4}^8 t_1^2 t_2^3 t_3^2 t_4^4 t_5^5 t_6 t_7^4 +  2 y_{s}^3 y_{o_1}^7 y_{o_2}^8 y_{o_3}^5 y_{o_4}^8 t_1 t_2^4 t_3^2 t_4^4 t_5^5 t_6 t_7^4 -  y_{s}^2 y_{o_1}^6 y_{o_2}^4 y_{o_3}^6 y_{o_4}^8 t_1^5 t_2 t_3^4 t_5^2 t_6^2 t_7^4 -  y_{s}^2 y_{o_1}^6 y_{o_2}^4 y_{o_3}^6 y_{o_4}^8 t_1^4 t_2^2 t_3^4 t_5^2 t_6^2 t_7^4 -  y_{s}^2 y_{o_1}^6 y_{o_2}^4 y_{o_3}^6 y_{o_4}^8 t_1^3 t_2^3 t_3^4 t_5^2 t_6^2 t_7^4 -  y_{s}^2 y_{o_1}^6 y_{o_2}^4 y_{o_3}^6 y_{o_4}^8 t_1^2 t_2^4 t_3^4 t_5^2 t_6^2 t_7^4 -  y_{s}^2 y_{o_1}^6 y_{o_2}^4 y_{o_3}^6 y_{o_4}^8 t_1 t_2^5 t_3^4 t_5^2 t_6^2 t_7^4 +  2 y_{s}^3 y_{o_1}^7 y_{o_2}^7 y_{o_3}^6 y_{o_4}^9 t_1^5 t_2 t_3^3 t_4^3 t_5^4 t_6^2 t_7^4 +  2 y_{s}^3 y_{o_1}^7 y_{o_2}^7 y_{o_3}^6 y_{o_4}^9 t_1^4 t_2^2 t_3^3 t_4^3 t_5^4 t_6^2 t_7^4 +  y_{s}^3 y_{o_1}^7 y_{o_2}^7 y_{o_3}^6 y_{o_4}^9 t_1^3 t_2^3 t_3^3 t_4^3 t_5^4 t_6^2 t_7^4 +  2 y_{s}^3 y_{o_1}^7 y_{o_2}^7 y_{o_3}^6 y_{o_4}^9 t_1^2 t_2^4 t_3^3 t_4^3 t_5^4 t_6^2 t_7^4 +  2 y_{s}^3 y_{o_1}^7 y_{o_2}^7 y_{o_3}^6 y_{o_4}^9 t_1 t_2^5 t_3^3 t_4^3 t_5^4 t_6^2 t_7^4 -  y_{s}^4 y_{o_1}^8 y_{o_2}^{10} y_{o_3}^6 y_{o_4}^{10} t_1^4 t_2^2 t_3^2 t_4^6 t_5^6 t_6^2 t_7^4 -  y_{s}^4 y_{o_1}^8 y_{o_2}^{10} y_{o_3}^6 y_{o_4}^{10} t_1^3 t_2^3 t_3^2 t_4^6 t_5^6 t_6^2 t_7^4 -  y_{s}^4 y_{o_1}^8 y_{o_2}^{10} y_{o_3}^6 y_{o_4}^{10} t_1^2 t_2^4 t_3^2 t_4^6 t_5^6 t_6^2 t_7^4 +  y_{s}^3 y_{o_1}^7 y_{o_2}^6 y_{o_3}^7 y_{o_4}^{10} t_1^6 t_2 t_3^4 t_4^2 t_5^3 t_6^3 t_7^4 +  y_{s}^3 y_{o_1}^7 y_{o_2}^6 y_{o_3}^7 y_{o_4}^{10} t_1^5 t_2^2 t_3^4 t_4^2 t_5^3 t_6^3 t_7^4 +  2 y_{s}^3 y_{o_1}^7 y_{o_2}^6 y_{o_3}^7 y_{o_4}^{10} t_1^4 t_2^3 t_3^4 t_4^2 t_5^3 t_6^3 t_7^4 +  2 y_{s}^3 y_{o_1}^7 y_{o_2}^6 y_{o_3}^7 y_{o_4}^{10} t_1^3 t_2^4 t_3^4 t_4^2 t_5^3 t_6^3 t_7^4 +  y_{s}^3 y_{o_1}^7 y_{o_2}^6 y_{o_3}^7 y_{o_4}^{10} t_1^2 t_2^5 t_3^4 t_4^2 t_5^3 t_6^3 t_7^4 +  y_{s}^3 y_{o_1}^7 y_{o_2}^6 y_{o_3}^7 y_{o_4}^{10} t_1 t_2^6 t_3^4 t_4^2 t_5^3 t_6^3 t_7^4 -  2 y_{s}^4 y_{o_1}^8 y_{o_2}^9 y_{o_3}^7 y_{o_4}^{11} t_1^6 t_2 t_3^3 t_4^5 t_5^5 t_6^3 t_7^4 -  3 y_{s}^4 y_{o_1}^8 y_{o_2}^9 y_{o_3}^7 y_{o_4}^{11} t_1^5 t_2^2 t_3^3 t_4^5 t_5^5 t_6^3 t_7^4 -  3 y_{s}^4 y_{o_1}^8 y_{o_2}^9 y_{o_3}^7 y_{o_4}^{11} t_1^4 t_2^3 t_3^3 t_4^5 t_5^5 t_6^3 t_7^4 -  3 y_{s}^4 y_{o_1}^8 y_{o_2}^9 y_{o_3}^7 y_{o_4}^{11} t_1^3 t_2^4 t_3^3 t_4^5 t_5^5 t_6^3 t_7^4 -  3 y_{s}^4 y_{o_1}^8 y_{o_2}^9 y_{o_3}^7 y_{o_4}^{11} t_1^2 t_2^5 t_3^3 t_4^5 t_5^5 t_6^3 t_7^4 -  2 y_{s}^4 y_{o_1}^8 y_{o_2}^9 y_{o_3}^7 y_{o_4}^{11} t_1 t_2^6 t_3^3 t_4^5 t_5^5 t_6^3 t_7^4 +  y_{s}^3 y_{o_1}^7 y_{o_2}^5 y_{o_3}^8 y_{o_4}^{11} t_1^7 t_2 t_3^5 t_4 t_5^2 t_6^4 t_7^4 +  y_{s}^3 y_{o_1}^7 y_{o_2}^5 y_{o_3}^8 y_{o_4}^{11} t_1^6 t_2^2 t_3^5 t_4 t_5^2 t_6^4 t_7^4 +  2 y_{s}^3 y_{o_1}^7 y_{o_2}^5 y_{o_3}^8 y_{o_4}^{11} t_1^5 t_2^3 t_3^5 t_4 t_5^2 t_6^4 t_7^4 +  2 y_{s}^3 y_{o_1}^7 y_{o_2}^5 y_{o_3}^8 y_{o_4}^{11} t_1^4 t_2^4 t_3^5 t_4 t_5^2 t_6^4 t_7^4 +  2 y_{s}^3 y_{o_1}^7 y_{o_2}^5 y_{o_3}^8 y_{o_4}^{11} t_1^3 t_2^5 t_3^5 t_4 t_5^2 t_6^4 t_7^4 +  y_{s}^3 y_{o_1}^7 y_{o_2}^5 y_{o_3}^8 y_{o_4}^{11} t_1^2 t_2^6 t_3^5 t_4 t_5^2 t_6^4 t_7^4 +  y_{s}^3 y_{o_1}^7 y_{o_2}^5 y_{o_3}^8 y_{o_4}^{11} t_1 t_2^7 t_3^5 t_4 t_5^2 t_6^4 t_7^4 +  y_{s}^4 y_{o_1}^8 y_{o_2}^8 y_{o_3}^8 y_{o_4}^{12} t_1^8 t_3^4 t_4^4 t_5^4 t_6^4 t_7^4 +  y_{s}^4 y_{o_1}^8 y_{o_2}^8 y_{o_3}^8 y_{o_4}^{12} t_1^7 t_2 t_3^4 t_4^4 t_5^4 t_6^4 t_7^4 +  y_{s}^4 y_{o_1}^8 y_{o_2}^8 y_{o_3}^8 y_{o_4}^{12} t_1^6 t_2^2 t_3^4 t_4^4 t_5^4 t_6^4 t_7^4 +  2 y_{s}^4 y_{o_1}^8 y_{o_2}^8 y_{o_3}^8 y_{o_4}^{12} t_1^5 t_2^3 t_3^4 t_4^4 t_5^4 t_6^4 t_7^4 +  2 y_{s}^4 y_{o_1}^8 y_{o_2}^8 y_{o_3}^8 y_{o_4}^{12} t_1^3 t_2^5 t_3^4 t_4^4 t_5^4 t_6^4 t_7^4 +  y_{s}^4 y_{o_1}^8 y_{o_2}^8 y_{o_3}^8 y_{o_4}^{12} t_1^2 t_2^6 t_3^4 t_4^4 t_5^4 t_6^4 t_7^4 +  y_{s}^4 y_{o_1}^8 y_{o_2}^8 y_{o_3}^8 y_{o_4}^{12} t_1 t_2^7 t_3^4 t_4^4 t_5^4 t_6^4 t_7^4 +  y_{s}^4 y_{o_1}^8 y_{o_2}^8 y_{o_3}^8 y_{o_4}^{12} t_2^8 t_3^4 t_4^4 t_5^4 t_6^4 t_7^4 +  y_{s}^5 y_{o_1}^9 y_{o_2}^{11} y_{o_3}^8 y_{o_4}^{13} t_1^6 t_2^2 t_3^3 t_4^7 t_5^6 t_6^4 t_7^4 +  y_{s}^5 y_{o_1}^9 y_{o_2}^{11} y_{o_3}^8 y_{o_4}^{13} t_1^5 t_2^3 t_3^3 t_4^7 t_5^6 t_6^4 t_7^4 +  y_{s}^5 y_{o_1}^9 y_{o_2}^{11} y_{o_3}^8 y_{o_4}^{13} t_1^4 t_2^4 t_3^3 t_4^7 t_5^6 t_6^4 t_7^4 +  y_{s}^5 y_{o_1}^9 y_{o_2}^{11} y_{o_3}^8 y_{o_4}^{13} t_1^3 t_2^5 t_3^3 t_4^7 t_5^6 t_6^4 t_7^4 +  y_{s}^5 y_{o_1}^9 y_{o_2}^{11} y_{o_3}^8 y_{o_4}^{13} t_1^2 t_2^6 t_3^3 t_4^7 t_5^6 t_6^4 t_7^4 -  y_{s}^4 y_{o_1}^8 y_{o_2}^7 y_{o_3}^9 y_{o_4}^{13} t_1^8 t_2 t_3^5 t_4^3 t_5^3 t_6^5 t_7^4 -  y_{s}^4 y_{o_1}^8 y_{o_2}^7 y_{o_3}^9 y_{o_4}^{13} t_1^7 t_2^2 t_3^5 t_4^3 t_5^3 t_6^5 t_7^4 -  2 y_{s}^4 y_{o_1}^8 y_{o_2}^7 y_{o_3}^9 y_{o_4}^{13} t_1^6 t_2^3 t_3^5 t_4^3 t_5^3 t_6^5 t_7^4 -  3 y_{s}^4 y_{o_1}^8 y_{o_2}^7 y_{o_3}^9 y_{o_4}^{13} t_1^5 t_2^4 t_3^5 t_4^3 t_5^3 t_6^5 t_7^4 -  3 y_{s}^4 y_{o_1}^8 y_{o_2}^7 y_{o_3}^9 y_{o_4}^{13} t_1^4 t_2^5 t_3^5 t_4^3 t_5^3 t_6^5 t_7^4 -  2 y_{s}^4 y_{o_1}^8 y_{o_2}^7 y_{o_3}^9 y_{o_4}^{13} t_1^3 t_2^6 t_3^5 t_4^3 t_5^3 t_6^5 t_7^4 -  y_{s}^4 y_{o_1}^8 y_{o_2}^7 y_{o_3}^9 y_{o_4}^{13} t_1^2 t_2^7 t_3^5 t_4^3 t_5^3 t_6^5 t_7^4 -  y_{s}^4 y_{o_1}^8 y_{o_2}^7 y_{o_3}^9 y_{o_4}^{13} t_1 t_2^8 t_3^5 t_4^3 t_5^3 t_6^5 t_7^4 -  y_{s}^5 y_{o_1}^9 y_{o_2}^{10} y_{o_3}^9 y_{o_4}^{14} t_1^8 t_2 t_3^4 t_4^6 t_5^5 t_6^5 t_7^4 -  y_{s}^5 y_{o_1}^9 y_{o_2}^{10} y_{o_3}^9 y_{o_4}^{14} t_1^7 t_2^2 t_3^4 t_4^6 t_5^5 t_6^5 t_7^4 -  y_{s}^5 y_{o_1}^9 y_{o_2}^{10} y_{o_3}^9 y_{o_4}^{14} t_1^6 t_2^3 t_3^4 t_4^6 t_5^5 t_6^5 t_7^4 -  y_{s}^5 y_{o_1}^9 y_{o_2}^{10} y_{o_3}^9 y_{o_4}^{14} t_1^5 t_2^4 t_3^4 t_4^6 t_5^5 t_6^5 t_7^4 -  y_{s}^5 y_{o_1}^9 y_{o_2}^{10} y_{o_3}^9 y_{o_4}^{14} t_1^4 t_2^5 t_3^4 t_4^6 t_5^5 t_6^5 t_7^4 -  y_{s}^5 y_{o_1}^9 y_{o_2}^{10} y_{o_3}^9 y_{o_4}^{14} t_1^3 t_2^6 t_3^4 t_4^6 t_5^5 t_6^5 t_7^4 -  y_{s}^5 y_{o_1}^9 y_{o_2}^{10} y_{o_3}^9 y_{o_4}^{14} t_1^2 t_2^7 t_3^4 t_4^6 t_5^5 t_6^5 t_7^4 -  y_{s}^5 y_{o_1}^9 y_{o_2}^{10} y_{o_3}^9 y_{o_4}^{14} t_1 t_2^8 t_3^4 t_4^6 t_5^5 t_6^5 t_7^4 -  y_{s}^4 y_{o_1}^8 y_{o_2}^6 y_{o_3}^{10} y_{o_4}^{14} t_1^5 t_2^5 t_3^6 t_4^2 t_5^2 t_6^6 t_7^4 -  y_{s}^5 y_{o_1}^9 y_{o_2}^9 y_{o_3}^{10} y_{o_4}^{15} t_1^8 t_2^2 t_3^5 t_4^5 t_5^4 t_6^6 t_7^4 -  y_{s}^5 y_{o_1}^9 y_{o_2}^9 y_{o_3}^{10} y_{o_4}^{15} t_1^7 t_2^3 t_3^5 t_4^5 t_5^4 t_6^6 t_7^4 -  3 y_{s}^5 y_{o_1}^9 y_{o_2}^9 y_{o_3}^{10} y_{o_4}^{15} t_1^5 t_2^5 t_3^5 t_4^5 t_5^4 t_6^6 t_7^4 -  y_{s}^5 y_{o_1}^9 y_{o_2}^9 y_{o_3}^{10} y_{o_4}^{15} t_1^3 t_2^7 t_3^5 t_4^5 t_5^4 t_6^6 t_7^4 -  y_{s}^5 y_{o_1}^9 y_{o_2}^9 y_{o_3}^{10} y_{o_4}^{15} t_1^2 t_2^8 t_3^5 t_4^5 t_5^4 t_6^6 t_7^4 -  y_{s}^6 y_{o_1}^{10} y_{o_2}^{12} y_{o_3}^{10} y_{o_4}^{16} t_1^5 t_2^5 t_3^4 t_4^8 t_5^6 t_6^6 t_7^4 +  y_{s}^5 y_{o_1}^9 y_{o_2}^8 y_{o_3}^{11} y_{o_4}^{16} t_1^6 t_2^5 t_3^6 t_4^4 t_5^3 t_6^7 t_7^4 +  y_{s}^5 y_{o_1}^9 y_{o_2}^8 y_{o_3}^{11} y_{o_4}^{16} t_1^5 t_2^6 t_3^6 t_4^4 t_5^3 t_6^7 t_7^4 +  y_{s}^6 y_{o_1}^{10} y_{o_2}^{11} y_{o_3}^{11} y_{o_4}^{17} t_1^8 t_2^3 t_3^5 t_4^7 t_5^5 t_6^7 t_7^4 +  y_{s}^6 y_{o_1}^{10} y_{o_2}^{11} y_{o_3}^{11} y_{o_4}^{17} t_1^7 t_2^4 t_3^5 t_4^7 t_5^5 t_6^7 t_7^4 +  2 y_{s}^6 y_{o_1}^{10} y_{o_2}^{11} y_{o_3}^{11} y_{o_4}^{17} t_1^6 t_2^5 t_3^5 t_4^7 t_5^5 t_6^7 t_7^4 + 2 y_{s}^6 y_{o_1}^{10} y_{o_2}^{11} y_{o_3}^{11} y_{o_4}^{17} t_1^5 t_2^6 t_3^5 t_4^7 t_5^5 t_6^7 t_7^4 + y_{s}^6 y_{o_1}^{10} y_{o_2}^{11} y_{o_3}^{11} y_{o_4}^{17} t_1^4 t_2^7 t_3^5 t_4^7 t_5^5 t_6^7 t_7^4 + y_{s}^6 y_{o_1}^{10} y_{o_2}^{11} y_{o_3}^{11} y_{o_4}^{17} t_1^3 t_2^8 t_3^5 t_4^7 t_5^5 t_6^7 t_7^4 - y_{s}^6 y_{o_1}^{10} y_{o_2}^{10} y_{o_3}^{12} y_{o_4}^{18} t_1^6 t_2^6 t_3^6 t_4^6 t_5^4 t_6^8 t_7^4 - y_{s}^7 y_{o_1}^{11} y_{o_2}^{13} y_{o_3}^{12} y_{o_4}^{19} t_1^6 t_2^6 t_3^5 t_4^9 t_5^6 t_6^8 t_7^4 - y_{s}^2 y_{o_1}^7 y_{o_2}^5 y_{o_3}^6 y_{o_4}^8 t_1^3 t_2^3 t_3^4 t_5^3 t_6 t_7^5 -  y_{s}^3 y_{o_1}^8 y_{o_2}^8 y_{o_3}^6 y_{o_4}^9 t_1^3 t_2^3 t_3^3 t_4^3 t_5^5 t_6 t_7^5 +  y_{s}^3 y_{o_1}^8 y_{o_2}^7 y_{o_3}^7 y_{o_4}^{10} t_1^6 t_2 t_3^4 t_4^2 t_5^4 t_6^2 t_7^5 +  y_{s}^3 y_{o_1}^8 y_{o_2}^7 y_{o_3}^7 y_{o_4}^{10} t_1^5 t_2^2 t_3^4 t_4^2 t_5^4 t_6^2 t_7^5 +  2 y_{s}^3 y_{o_1}^8 y_{o_2}^7 y_{o_3}^7 y_{o_4}^{10} t_1^4 t_2^3 t_3^4 t_4^2 t_5^4 t_6^2 t_7^5 +  2 y_{s}^3 y_{o_1}^8 y_{o_2}^7 y_{o_3}^7 y_{o_4}^{10} t_1^3 t_2^4 t_3^4 t_4^2 t_5^4 t_6^2 t_7^5 +  y_{s}^3 y_{o_1}^8 y_{o_2}^7 y_{o_3}^7 y_{o_4}^{10} t_1^2 t_2^5 t_3^4 t_4^2 t_5^4 t_6^2 t_7^5 +  y_{s}^3 y_{o_1}^8 y_{o_2}^7 y_{o_3}^7 y_{o_4}^{10} t_1 t_2^6 t_3^4 t_4^2 t_5^4 t_6^2 t_7^5 +  y_{s}^4 y_{o_1}^9 y_{o_2}^{10} y_{o_3}^7 y_{o_4}^{11} t_1^4 t_2^3 t_3^3 t_4^5 t_5^6 t_6^2 t_7^5 +  y_{s}^4 y_{o_1}^9 y_{o_2}^{10} y_{o_3}^7 y_{o_4}^{11} t_1^3 t_2^4 t_3^3 t_4^5 t_5^6 t_6^2 t_7^5 -  y_{s}^3 y_{o_1}^8 y_{o_2}^6 y_{o_3}^8 y_{o_4}^{11} t_1^4 t_2^4 t_3^5 t_4 t_5^3 t_6^3 t_7^5 -  y_{s}^4 y_{o_1}^9 y_{o_2}^9 y_{o_3}^8 y_{o_4}^{12} t_1^7 t_2 t_3^4 t_4^4 t_5^5 t_6^3 t_7^5 -  y_{s}^4 y_{o_1}^9 y_{o_2}^9 y_{o_3}^8 y_{o_4}^{12} t_1^6 t_2^2 t_3^4 t_4^4 t_5^5 t_6^3 t_7^5 -  3 y_{s}^4 y_{o_1}^9 y_{o_2}^9 y_{o_3}^8 y_{o_4}^{12} t_1^4 t_2^4 t_3^4 t_4^4 t_5^5 t_6^3 t_7^5 -  y_{s}^4 y_{o_1}^9 y_{o_2}^9 y_{o_3}^8 y_{o_4}^{12} t_1^2 t_2^6 t_3^4 t_4^4 t_5^5 t_6^3 t_7^5 -  y_{s}^4 y_{o_1}^9 y_{o_2}^9 y_{o_3}^8 y_{o_4}^{12} t_1 t_2^7 t_3^4 t_4^4 t_5^5 t_6^3 t_7^5 -  y_{s}^5 y_{o_1}^{10} y_{o_2}^{12} y_{o_3}^8 y_{o_4}^{13} t_1^4 t_2^4 t_3^3 t_4^7 t_5^7 t_6^3 t_7^5 -  y_{s}^4 y_{o_1}^9 y_{o_2}^8 y_{o_3}^9 y_{o_4}^{13} t_1^8 t_2 t_3^5 t_4^3 t_5^4 t_6^4 t_7^5 -  y_{s}^4 y_{o_1}^9 y_{o_2}^8 y_{o_3}^9 y_{o_4}^{13} t_1^7 t_2^2 t_3^5 t_4^3 t_5^4 t_6^4 t_7^5 -  y_{s}^4 y_{o_1}^9 y_{o_2}^8 y_{o_3}^9 y_{o_4}^{13} t_1^6 t_2^3 t_3^5 t_4^3 t_5^4 t_6^4 t_7^5 -  y_{s}^4 y_{o_1}^9 y_{o_2}^8 y_{o_3}^9 y_{o_4}^{13} t_1^5 t_2^4 t_3^5 t_4^3 t_5^4 t_6^4 t_7^5 -  y_{s}^4 y_{o_1}^9 y_{o_2}^8 y_{o_3}^9 y_{o_4}^{13} t_1^4 t_2^5 t_3^5 t_4^3 t_5^4 t_6^4 t_7^5 -  y_{s}^4 y_{o_1}^9 y_{o_2}^8 y_{o_3}^9 y_{o_4}^{13} t_1^3 t_2^6 t_3^5 t_4^3 t_5^4 t_6^4 t_7^5 -  y_{s}^4 y_{o_1}^9 y_{o_2}^8 y_{o_3}^9 y_{o_4}^{13} t_1^2 t_2^7 t_3^5 t_4^3 t_5^4 t_6^4 t_7^5 -  y_{s}^4 y_{o_1}^9 y_{o_2}^8 y_{o_3}^9 y_{o_4}^{13} t_1 t_2^8 t_3^5 t_4^3 t_5^4 t_6^4 t_7^5 -  y_{s}^5 y_{o_1}^{10} y_{o_2}^{11} y_{o_3}^9 y_{o_4}^{14} t_1^8 t_2 t_3^4 t_4^6 t_5^6 t_6^4 t_7^5 -  y_{s}^5 y_{o_1}^{10} y_{o_2}^{11} y_{o_3}^9 y_{o_4}^{14} t_1^7 t_2^2 t_3^4 t_4^6 t_5^6 t_6^4 t_7^5 -  2 y_{s}^5 y_{o_1}^{10} y_{o_2}^{11} y_{o_3}^9 y_{o_4}^{14} t_1^6 t_2^3 t_3^4 t_4^6 t_5^6 t_6^4 t_7^5 - 3 y_{s}^5 y_{o_1}^{10} y_{o_2}^{11} y_{o_3}^9 y_{o_4}^{14} t_1^5 t_2^4 t_3^4 t_4^6 t_5^6 t_6^4 t_7^5 - 3 y_{s}^5 y_{o_1}^{10} y_{o_2}^{11} y_{o_3}^9 y_{o_4}^{14} t_1^4 t_2^5 t_3^4 t_4^6 t_5^6 t_6^4 t_7^5 - 2 y_{s}^5 y_{o_1}^{10} y_{o_2}^{11} y_{o_3}^9 y_{o_4}^{14} t_1^3 t_2^6 t_3^4 t_4^6 t_5^6 t_6^4 t_7^5 - y_{s}^5 y_{o_1}^{10} y_{o_2}^{11} y_{o_3}^9 y_{o_4}^{14} t_1^2 t_2^7 t_3^4 t_4^6 t_5^6 t_6^4 t_7^5 -  y_{s}^5 y_{o_1}^{10} y_{o_2}^{11} y_{o_3}^9 y_{o_4}^{14} t_1 t_2^8 t_3^4 t_4^6 t_5^6 t_6^4 t_7^5 +  y_{s}^4 y_{o_1}^9 y_{o_2}^7 y_{o_3}^{10} y_{o_4}^{14} t_1^7 t_2^3 t_3^6 t_4^2 t_5^3 t_6^5 t_7^5 +  y_{s}^4 y_{o_1}^9 y_{o_2}^7 y_{o_3}^{10} y_{o_4}^{14} t_1^6 t_2^4 t_3^6 t_4^2 t_5^3 t_6^5 t_7^5 +  y_{s}^4 y_{o_1}^9 y_{o_2}^7 y_{o_3}^{10} y_{o_4}^{14} t_1^5 t_2^5 t_3^6 t_4^2 t_5^3 t_6^5 t_7^5 +  y_{s}^4 y_{o_1}^9 y_{o_2}^7 y_{o_3}^{10} y_{o_4}^{14} t_1^4 t_2^6 t_3^6 t_4^2 t_5^3 t_6^5 t_7^5 +  y_{s}^4 y_{o_1}^9 y_{o_2}^7 y_{o_3}^{10} y_{o_4}^{14} t_1^3 t_2^7 t_3^6 t_4^2 t_5^3 t_6^5 t_7^5 +  y_{s}^5 y_{o_1}^{10} y_{o_2}^{10} y_{o_3}^{10} y_{o_4}^{15} t_1^9 t_2 t_3^5 t_4^5 t_5^5 t_6^5 t_7^5 +  y_{s}^5 y_{o_1}^{10} y_{o_2}^{10} y_{o_3}^{10} y_{o_4}^{15} t_1^8 t_2^2 t_3^5 t_4^5 t_5^5 t_6^5 t_7^5 +  y_{s}^5 y_{o_1}^{10} y_{o_2}^{10} y_{o_3}^{10} y_{o_4}^{15} t_1^7 t_2^3 t_3^5 t_4^5 t_5^5 t_6^5 t_7^5 +  2 y_{s}^5 y_{o_1}^{10} y_{o_2}^{10} y_{o_3}^{10} y_{o_4}^{15} t_1^6 t_2^4 t_3^5 t_4^5 t_5^5 t_6^5 t_7^5 + 2 y_{s}^5 y_{o_1}^{10} y_{o_2}^{10} y_{o_3}^{10} y_{o_4}^{15} t_1^4 t_2^6 t_3^5 t_4^5 t_5^5 t_6^5 t_7^5 + y_{s}^5 y_{o_1}^{10} y_{o_2}^{10} y_{o_3}^{10} y_{o_4}^{15} t_1^3 t_2^7 t_3^5 t_4^5 t_5^5 t_6^5 t_7^5 + y_{s}^5 y_{o_1}^{10} y_{o_2}^{10} y_{o_3}^{10} y_{o_4}^{15} t_1^2 t_2^8 t_3^5 t_4^5 t_5^5 t_6^5 t_7^5 + y_{s}^5 y_{o_1}^{10} y_{o_2}^{10} y_{o_3}^{10} y_{o_4}^{15} t_1 t_2^9 t_3^5 t_4^5 t_5^5 t_6^5 t_7^5 + y_{s}^6 y_{o_1}^{11} y_{o_2}^{13} y_{o_3}^{10} y_{o_4}^{16} t_1^8 t_2^2 t_3^4 t_4^8 t_5^7 t_6^5 t_7^5 + y_{s}^6 y_{o_1}^{11} y_{o_2}^{13} y_{o_3}^{10} y_{o_4}^{16} t_1^7 t_2^3 t_3^4 t_4^8 t_5^7 t_6^5 t_7^5 + 2 y_{s}^6 y_{o_1}^{11} y_{o_2}^{13} y_{o_3}^{10} y_{o_4}^{16} t_1^6 t_2^4 t_3^4 t_4^8 t_5^7 t_6^5 t_7^5 +  2 y_{s}^6 y_{o_1}^{11} y_{o_2}^{13} y_{o_3}^{10} y_{o_4}^{16} t_1^5 t_2^5 t_3^4 t_4^8 t_5^7 t_6^5 t_7^5 + 2 y_{s}^6 y_{o_1}^{11} y_{o_2}^{13} y_{o_3}^{10} y_{o_4}^{16} t_1^4 t_2^6 t_3^4 t_4^8 t_5^7 t_6^5 t_7^5 + y_{s}^6 y_{o_1}^{11} y_{o_2}^{13} y_{o_3}^{10} y_{o_4}^{16} t_1^3 t_2^7 t_3^4 t_4^8 t_5^7 t_6^5 t_7^5 + y_{s}^6 y_{o_1}^{11} y_{o_2}^{13} y_{o_3}^{10} y_{o_4}^{16} t_1^2 t_2^8 t_3^4 t_4^8 t_5^7 t_6^5 t_7^5 - 2 y_{s}^5 y_{o_1}^{10} y_{o_2}^9 y_{o_3}^{11} y_{o_4}^{16} t_1^8 t_2^3 t_3^6 t_4^4 t_5^4 t_6^6 t_7^5 - 3 y_{s}^5 y_{o_1}^{10} y_{o_2}^9 y_{o_3}^{11} y_{o_4}^{16} t_1^7 t_2^4 t_3^6 t_4^4 t_5^4 t_6^6 t_7^5 - 3 y_{s}^5 y_{o_1}^{10} y_{o_2}^9 y_{o_3}^{11} y_{o_4}^{16} t_1^6 t_2^5 t_3^6 t_4^4 t_5^4 t_6^6 t_7^5 - 3 y_{s}^5 y_{o_1}^{10} y_{o_2}^9 y_{o_3}^{11} y_{o_4}^{16} t_1^5 t_2^6 t_3^6 t_4^4 t_5^4 t_6^6 t_7^5 - 3 y_{s}^5 y_{o_1}^{10} y_{o_2}^9 y_{o_3}^{11} y_{o_4}^{16} t_1^4 t_2^7 t_3^6 t_4^4 t_5^4 t_6^6 t_7^5 - 2 y_{s}^5 y_{o_1}^{10} y_{o_2}^9 y_{o_3}^{11} y_{o_4}^{16} t_1^3 t_2^8 t_3^6 t_4^4 t_5^4 t_6^6 t_7^5 + y_{s}^6 y_{o_1}^{11} y_{o_2}^{12} y_{o_3}^{11} y_{o_4}^{17} t_1^8 t_2^3 t_3^5 t_4^7 t_5^6 t_6^6 t_7^5 + y_{s}^6 y_{o_1}^{11} y_{o_2}^{12} y_{o_3}^{11} y_{o_4}^{17} t_1^7 t_2^4 t_3^5 t_4^7 t_5^6 t_6^6 t_7^5 + 2 y_{s}^6 y_{o_1}^{11} y_{o_2}^{12} y_{o_3}^{11} y_{o_4}^{17} t_1^6 t_2^5 t_3^5 t_4^7 t_5^6 t_6^6 t_7^5 +  2 y_{s}^6 y_{o_1}^{11} y_{o_2}^{12} y_{o_3}^{11} y_{o_4}^{17} t_1^5 t_2^6 t_3^5 t_4^7 t_5^6 t_6^6 t_7^5 + y_{s}^6 y_{o_1}^{11} y_{o_2}^{12} y_{o_3}^{11} y_{o_4}^{17} t_1^4 t_2^7 t_3^5 t_4^7 t_5^6 t_6^6 t_7^5 +  y_{s}^6 y_{o_1}^{11} y_{o_2}^{12} y_{o_3}^{11} y_{o_4}^{17} t_1^3 t_2^8 t_3^5 t_4^7 t_5^6 t_6^6 t_7^5 -  y_{s}^5 y_{o_1}^{10} y_{o_2}^8 y_{o_3}^{12} y_{o_4}^{17} t_1^7 t_2^5 t_3^7 t_4^3 t_5^3 t_6^7 t_7^5 -  y_{s}^5 y_{o_1}^{10} y_{o_2}^8 y_{o_3}^{12} y_{o_4}^{17} t_1^6 t_2^6 t_3^7 t_4^3 t_5^3 t_6^7 t_7^5 -  y_{s}^5 y_{o_1}^{10} y_{o_2}^8 y_{o_3}^{12} y_{o_4}^{17} t_1^5 t_2^7 t_3^7 t_4^3 t_5^3 t_6^7 t_7^5 +  2 y_{s}^6 y_{o_1}^{11} y_{o_2}^{11} y_{o_3}^{12} y_{o_4}^{18} t_1^8 t_2^4 t_3^6 t_4^6 t_5^5 t_6^7 t_7^5 + 2 y_{s}^6 y_{o_1}^{11} y_{o_2}^{11} y_{o_3}^{12} y_{o_4}^{18} t_1^7 t_2^5 t_3^6 t_4^6 t_5^5 t_6^7 t_7^5 + y_{s}^6 y_{o_1}^{11} y_{o_2}^{11} y_{o_3}^{12} y_{o_4}^{18} t_1^6 t_2^6 t_3^6 t_4^6 t_5^5 t_6^7 t_7^5 + 2 y_{s}^6 y_{o_1}^{11} y_{o_2}^{11} y_{o_3}^{12} y_{o_4}^{18} t_1^5 t_2^7 t_3^6 t_4^6 t_5^5 t_6^7 t_7^5 +  2 y_{s}^6 y_{o_1}^{11} y_{o_2}^{11} y_{o_3}^{12} y_{o_4}^{18} t_1^4 t_2^8 t_3^6 t_4^6 t_5^5 t_6^7 t_7^5 - y_{s}^7 y_{o_1}^{12} y_{o_2}^{14} y_{o_3}^{12} y_{o_4}^{19} t_1^8 t_2^4 t_3^5 t_4^9 t_5^7 t_6^7 t_7^5 -  y_{s}^7 y_{o_1}^{12} y_{o_2}^{14} y_{o_3}^{12} y_{o_4}^{19} t_1^7 t_2^5 t_3^5 t_4^9 t_5^7 t_6^7 t_7^5 -  y_{s}^7 y_{o_1}^{12} y_{o_2}^{14} y_{o_3}^{12} y_{o_4}^{19} t_1^6 t_2^6 t_3^5 t_4^9 t_5^7 t_6^7 t_7^5 -  y_{s}^7 y_{o_1}^{12} y_{o_2}^{14} y_{o_3}^{12} y_{o_4}^{19} t_1^5 t_2^7 t_3^5 t_4^9 t_5^7 t_6^7 t_7^5 -  y_{s}^7 y_{o_1}^{12} y_{o_2}^{14} y_{o_3}^{12} y_{o_4}^{19} t_1^4 t_2^8 t_3^5 t_4^9 t_5^7 t_6^7 t_7^5 + 2 y_{s}^6 y_{o_1}^{11} y_{o_2}^{10} y_{o_3}^{13} y_{o_4}^{19} t_1^8 t_2^5 t_3^7 t_4^5 t_5^4 t_6^8 t_7^5 + 2 y_{s}^6 y_{o_1}^{11} y_{o_2}^{10} y_{o_3}^{13} y_{o_4}^{19} t_1^7 t_2^6 t_3^7 t_4^5 t_5^4 t_6^8 t_7^5 +  2 y_{s}^6 y_{o_1}^{11} y_{o_2}^{10} y_{o_3}^{13} y_{o_4}^{19} t_1^6 t_2^7 t_3^7 t_4^5 t_5^4 t_6^8 t_7^5 + 2 y_{s}^6 y_{o_1}^{11} y_{o_2}^{10} y_{o_3}^{13} y_{o_4}^{19} t_1^5 t_2^8 t_3^7 t_4^5 t_5^4 t_6^8 t_7^5 - y_{s}^7 y_{o_1}^{12} y_{o_2}^{12} y_{o_3}^{14} y_{o_4}^{21} t_1^8 t_2^6 t_3^7 t_4^7 t_5^5 t_6^9 t_7^5 - y_{s}^7 y_{o_1}^{12} y_{o_2}^{12} y_{o_3}^{14} y_{o_4}^{21} t_1^7 t_2^7 t_3^7 t_4^7 t_5^5 t_6^9 t_7^5 - y_{s}^7 y_{o_1}^{12} y_{o_2}^{12} y_{o_3}^{14} y_{o_4}^{21} t_1^6 t_2^8 t_3^7 t_4^7 t_5^5 t_6^9 t_7^5 - y_{s}^3 y_{o_1}^9 y_{o_2}^9 y_{o_3}^6 y_{o_4}^9 t_1^3 t_2^3 t_3^3 t_4^3 t_5^6 t_7^6 +  y_{s}^3 y_{o_1}^9 y_{o_2}^8 y_{o_3}^7 y_{o_4}^{10} t_1^4 t_2^3 t_3^4 t_4^2 t_5^5 t_6 t_7^6 +  y_{s}^3 y_{o_1}^9 y_{o_2}^8 y_{o_3}^7 y_{o_4}^{10} t_1^3 t_2^4 t_3^4 t_4^2 t_5^5 t_6 t_7^6 +  y_{s}^4 y_{o_1}^{10} y_{o_2}^{11} y_{o_3}^7 y_{o_4}^{11} t_1^4 t_2^3 t_3^3 t_4^5 t_5^7 t_6 t_7^6 +  y_{s}^4 y_{o_1}^{10} y_{o_2}^{11} y_{o_3}^7 y_{o_4}^{11} t_1^3 t_2^4 t_3^3 t_4^5 t_5^7 t_6 t_7^6 +  y_{s}^4 y_{o_1}^{10} y_{o_2}^{10} y_{o_3}^8 y_{o_4}^{12} t_1^5 t_2^3 t_3^4 t_4^4 t_5^6 t_6^2 t_7^6 -  y_{s}^4 y_{o_1}^{10} y_{o_2}^{10} y_{o_3}^8 y_{o_4}^{12} t_1^4 t_2^4 t_3^4 t_4^4 t_5^6 t_6^2 t_7^6 +  y_{s}^4 y_{o_1}^{10} y_{o_2}^{10} y_{o_3}^8 y_{o_4}^{12} t_1^3 t_2^5 t_3^4 t_4^4 t_5^6 t_6^2 t_7^6 -  y_{s}^5 y_{o_1}^{11} y_{o_2}^{13} y_{o_3}^8 y_{o_4}^{13} t_1^4 t_2^4 t_3^3 t_4^7 t_5^8 t_6^2 t_7^6 -  y_{s}^3 y_{o_1}^9 y_{o_2}^6 y_{o_3}^9 y_{o_4}^{12} t_1^5 t_2^4 t_3^6 t_5^3 t_6^3 t_7^6 -  y_{s}^3 y_{o_1}^9 y_{o_2}^6 y_{o_3}^9 y_{o_4}^{12} t_1^4 t_2^5 t_3^6 t_5^3 t_6^3 t_7^6 +  y_{s}^4 y_{o_1}^{10} y_{o_2}^9 y_{o_3}^9 y_{o_4}^{13} t_1^6 t_2^3 t_3^5 t_4^3 t_5^5 t_6^3 t_7^6 +  y_{s}^4 y_{o_1}^{10} y_{o_2}^9 y_{o_3}^9 y_{o_4}^{13} t_1^5 t_2^4 t_3^5 t_4^3 t_5^5 t_6^3 t_7^6 +  y_{s}^4 y_{o_1}^{10} y_{o_2}^9 y_{o_3}^9 y_{o_4}^{13} t_1^4 t_2^5 t_3^5 t_4^3 t_5^5 t_6^3 t_7^6 +  y_{s}^4 y_{o_1}^{10} y_{o_2}^9 y_{o_3}^9 y_{o_4}^{13} t_1^3 t_2^6 t_3^5 t_4^3 t_5^5 t_6^3 t_7^6 -  y_{s}^5 y_{o_1}^{11} y_{o_2}^{12} y_{o_3}^9 y_{o_4}^{14} t_1^6 t_2^3 t_3^4 t_4^6 t_5^7 t_6^3 t_7^6 -  y_{s}^5 y_{o_1}^{11} y_{o_2}^{12} y_{o_3}^9 y_{o_4}^{14} t_1^5 t_2^4 t_3^4 t_4^6 t_5^7 t_6^3 t_7^6 -  y_{s}^5 y_{o_1}^{11} y_{o_2}^{12} y_{o_3}^9 y_{o_4}^{14} t_1^4 t_2^5 t_3^4 t_4^6 t_5^7 t_6^3 t_7^6 -  y_{s}^5 y_{o_1}^{11} y_{o_2}^{12} y_{o_3}^9 y_{o_4}^{14} t_1^3 t_2^6 t_3^4 t_4^6 t_5^7 t_6^3 t_7^6 +  y_{s}^4 y_{o_1}^{10} y_{o_2}^8 y_{o_3}^{10} y_{o_4}^{14} t_1^7 t_2^3 t_3^6 t_4^2 t_5^4 t_6^4 t_7^6 +  2 y_{s}^4 y_{o_1}^{10} y_{o_2}^8 y_{o_3}^{10} y_{o_4}^{14} t_1^6 t_2^4 t_3^6 t_4^2 t_5^4 t_6^4 t_7^6 + 3 y_{s}^4 y_{o_1}^{10} y_{o_2}^8 y_{o_3}^{10} y_{o_4}^{14} t_1^5 t_2^5 t_3^6 t_4^2 t_5^4 t_6^4 t_7^6 + 2 y_{s}^4 y_{o_1}^{10} y_{o_2}^8 y_{o_3}^{10} y_{o_4}^{14} t_1^4 t_2^6 t_3^6 t_4^2 t_5^4 t_6^4 t_7^6 + y_{s}^4 y_{o_1}^{10} y_{o_2}^8 y_{o_3}^{10} y_{o_4}^{14} t_1^3 t_2^7 t_3^6 t_4^2 t_5^4 t_6^4 t_7^6 -  y_{s}^5 y_{o_1}^{11} y_{o_2}^{11} y_{o_3}^{10} y_{o_4}^{15} t_1^7 t_2^3 t_3^5 t_4^5 t_5^6 t_6^4 t_7^6 -  2 y_{s}^5 y_{o_1}^{11} y_{o_2}^{11} y_{o_3}^{10} y_{o_4}^{15} t_1^6 t_2^4 t_3^5 t_4^5 t_5^6 t_6^4 t_7^6 - 4 y_{s}^5 y_{o_1}^{11} y_{o_2}^{11} y_{o_3}^{10} y_{o_4}^{15} t_1^5 t_2^5 t_3^5 t_4^5 t_5^6 t_6^4 t_7^6 - 2 y_{s}^5 y_{o_1}^{11} y_{o_2}^{11} y_{o_3}^{10} y_{o_4}^{15} t_1^4 t_2^6 t_3^5 t_4^5 t_5^6 t_6^4 t_7^6 -  y_{s}^5 y_{o_1}^{11} y_{o_2}^{11} y_{o_3}^{10} y_{o_4}^{15} t_1^3 t_2^7 t_3^5 t_4^5 t_5^6 t_6^4 t_7^6 +  y_{s}^6 y_{o_1}^{12} y_{o_2}^{14} y_{o_3}^{10} y_{o_4}^{16} t_1^6 t_2^4 t_3^4 t_4^8 t_5^8 t_6^4 t_7^6 +  y_{s}^6 y_{o_1}^{12} y_{o_2}^{14} y_{o_3}^{10} y_{o_4}^{16} t_1^4 t_2^6 t_3^4 t_4^8 t_5^8 t_6^4 t_7^6 +  y_{s}^4 y_{o_1}^{10} y_{o_2}^7 y_{o_3}^{11} y_{o_4}^{15} t_1^7 t_2^4 t_3^7 t_4 t_5^3 t_6^5 t_7^6 +  y_{s}^4 y_{o_1}^{10} y_{o_2}^7 y_{o_3}^{11} y_{o_4}^{15} t_1^6 t_2^5 t_3^7 t_4 t_5^3 t_6^5 t_7^6 +  y_{s}^4 y_{o_1}^{10} y_{o_2}^7 y_{o_3}^{11} y_{o_4}^{15} t_1^5 t_2^6 t_3^7 t_4 t_5^3 t_6^5 t_7^6 +  y_{s}^4 y_{o_1}^{10} y_{o_2}^7 y_{o_3}^{11} y_{o_4}^{15} t_1^4 t_2^7 t_3^7 t_4 t_5^3 t_6^5 t_7^6 -  2 y_{s}^5 y_{o_1}^{11} y_{o_2}^{10} y_{o_3}^{11} y_{o_4}^{16} t_1^8 t_2^3 t_3^6 t_4^4 t_5^5 t_6^5 t_7^6 - 3 y_{s}^5 y_{o_1}^{11} y_{o_2}^{10} y_{o_3}^{11} y_{o_4}^{16} t_1^7 t_2^4 t_3^6 t_4^4 t_5^5 t_6^5 t_7^6 - 5 y_{s}^5 y_{o_1}^{11} y_{o_2}^{10} y_{o_3}^{11} y_{o_4}^{16} t_1^6 t_2^5 t_3^6 t_4^4 t_5^5 t_6^5 t_7^6 -  5 y_{s}^5 y_{o_1}^{11} y_{o_2}^{10} y_{o_3}^{11} y_{o_4}^{16} t_1^5 t_2^6 t_3^6 t_4^4 t_5^5 t_6^5 t_7^6 - 3 y_{s}^5 y_{o_1}^{11} y_{o_2}^{10} y_{o_3}^{11} y_{o_4}^{16} t_1^4 t_2^7 t_3^6 t_4^4 t_5^5 t_6^5 t_7^6 - 2 y_{s}^5 y_{o_1}^{11} y_{o_2}^{10} y_{o_3}^{11} y_{o_4}^{16} t_1^3 t_2^8 t_3^6 t_4^4 t_5^5 t_6^5 t_7^6 +  y_{s}^6 y_{o_1}^{12} y_{o_2}^{13} y_{o_3}^{11} y_{o_4}^{17} t_1^7 t_2^4 t_3^5 t_4^7 t_5^7 t_6^5 t_7^6 +  2 y_{s}^6 y_{o_1}^{12} y_{o_2}^{13} y_{o_3}^{11} y_{o_4}^{17} t_1^6 t_2^5 t_3^5 t_4^7 t_5^7 t_6^5 t_7^6 + 2 y_{s}^6 y_{o_1}^{12} y_{o_2}^{13} y_{o_3}^{11} y_{o_4}^{17} t_1^5 t_2^6 t_3^5 t_4^7 t_5^7 t_6^5 t_7^6 + y_{s}^6 y_{o_1}^{12} y_{o_2}^{13} y_{o_3}^{11} y_{o_4}^{17} t_1^4 t_2^7 t_3^5 t_4^7 t_5^7 t_6^5 t_7^6 - y_{s}^5 y_{o_1}^{11} y_{o_2}^9 y_{o_3}^{12} y_{o_4}^{17} t_1^8 t_2^4 t_3^7 t_4^3 t_5^4 t_6^6 t_7^6 - 3 y_{s}^5 y_{o_1}^{11} y_{o_2}^9 y_{o_3}^{12} y_{o_4}^{17} t_1^7 t_2^5 t_3^7 t_4^3 t_5^4 t_6^6 t_7^6 - 3 y_{s}^5 y_{o_1}^{11} y_{o_2}^9 y_{o_3}^{12} y_{o_4}^{17} t_1^6 t_2^6 t_3^7 t_4^3 t_5^4 t_6^6 t_7^6 - 3 y_{s}^5 y_{o_1}^{11} y_{o_2}^9 y_{o_3}^{12} y_{o_4}^{17} t_1^5 t_2^7 t_3^7 t_4^3 t_5^4 t_6^6 t_7^6 - y_{s}^5 y_{o_1}^{11} y_{o_2}^9 y_{o_3}^{12} y_{o_4}^{17} t_1^4 t_2^8 t_3^7 t_4^3 t_5^4 t_6^6 t_7^6 + y_{s}^6 y_{o_1}^{12} y_{o_2}^{12} y_{o_3}^{12} y_{o_4}^{18} t_1^9 t_2^3 t_3^6 t_4^6 t_5^6 t_6^6 t_7^6 + 4 y_{s}^6 y_{o_1}^{12} y_{o_2}^{12} y_{o_3}^{12} y_{o_4}^{18} t_1^8 t_2^4 t_3^6 t_4^6 t_5^6 t_6^6 t_7^6 +  5 y_{s}^6 y_{o_1}^{12} y_{o_2}^{12} y_{o_3}^{12} y_{o_4}^{18} t_1^7 t_2^5 t_3^6 t_4^6 t_5^6 t_6^6 t_7^6 + 5 y_{s}^6 y_{o_1}^{12} y_{o_2}^{12} y_{o_3}^{12} y_{o_4}^{18} t_1^6 t_2^6 t_3^6 t_4^6 t_5^6 t_6^6 t_7^6 + 5 y_{s}^6 y_{o_1}^{12} y_{o_2}^{12} y_{o_3}^{12} y_{o_4}^{18} t_1^5 t_2^7 t_3^6 t_4^6 t_5^6 t_6^6 t_7^6 +  4 y_{s}^6 y_{o_1}^{12} y_{o_2}^{12} y_{o_3}^{12} y_{o_4}^{18} t_1^4 t_2^8 t_3^6 t_4^6 t_5^6 t_6^6 t_7^6 + y_{s}^6 y_{o_1}^{12} y_{o_2}^{12} y_{o_3}^{12} y_{o_4}^{18} t_1^3 t_2^9 t_3^6 t_4^6 t_5^6 t_6^6 t_7^6 -  y_{s}^7 y_{o_1}^{13} y_{o_2}^{15} y_{o_3}^{12} y_{o_4}^{19} t_1^6 t_2^6 t_3^5 t_4^9 t_5^8 t_6^6 t_7^6 -  y_{s}^5 y_{o_1}^{11} y_{o_2}^8 y_{o_3}^{13} y_{o_4}^{18} t_1^7 t_2^6 t_3^8 t_4^2 t_5^3 t_6^7 t_7^6 -  y_{s}^5 y_{o_1}^{11} y_{o_2}^8 y_{o_3}^{13} y_{o_4}^{18} t_1^6 t_2^7 t_3^8 t_4^2 t_5^3 t_6^7 t_7^6 +  3 y_{s}^6 y_{o_1}^{12} y_{o_2}^{11} y_{o_3}^{13} y_{o_4}^{19} t_1^8 t_2^5 t_3^7 t_4^5 t_5^5 t_6^7 t_7^6 + 4 y_{s}^6 y_{o_1}^{12} y_{o_2}^{11} y_{o_3}^{13} y_{o_4}^{19} t_1^7 t_2^6 t_3^7 t_4^5 t_5^5 t_6^7 t_7^6 + 4 y_{s}^6 y_{o_1}^{12} y_{o_2}^{11} y_{o_3}^{13} y_{o_4}^{19} t_1^6 t_2^7 t_3^7 t_4^5 t_5^5 t_6^7 t_7^6 +  3 y_{s}^6 y_{o_1}^{12} y_{o_2}^{11} y_{o_3}^{13} y_{o_4}^{19} t_1^5 t_2^8 t_3^7 t_4^5 t_5^5 t_6^7 t_7^6 - y_{s}^7 y_{o_1}^{13} y_{o_2}^{14} y_{o_3}^{13} y_{o_4}^{20} t_1^9 t_2^4 t_3^6 t_4^8 t_5^7 t_6^7 t_7^6 - 2 y_{s}^7 y_{o_1}^{13} y_{o_2}^{14} y_{o_3}^{13} y_{o_4}^{20} t_1^8 t_2^5 t_3^6 t_4^8 t_5^7 t_6^7 t_7^6 - 2 y_{s}^7 y_{o_1}^{13} y_{o_2}^{14} y_{o_3}^{13} y_{o_4}^{20} t_1^7 t_2^6 t_3^6 t_4^8 t_5^7 t_6^7 t_7^6 -  2 y_{s}^7 y_{o_1}^{13} y_{o_2}^{14} y_{o_3}^{13} y_{o_4}^{20} t_1^6 t_2^7 t_3^6 t_4^8 t_5^7 t_6^7 t_7^6 - 2 y_{s}^7 y_{o_1}^{13} y_{o_2}^{14} y_{o_3}^{13} y_{o_4}^{20} t_1^5 t_2^8 t_3^6 t_4^8 t_5^7 t_6^7 t_7^6 - y_{s}^7 y_{o_1}^{13} y_{o_2}^{14} y_{o_3}^{13} y_{o_4}^{20} t_1^4 t_2^9 t_3^6 t_4^8 t_5^7 t_6^7 t_7^6 + y_{s}^6 y_{o_1}^{12} y_{o_2}^{10} y_{o_3}^{14} y_{o_4}^{20} t_1^8 t_2^6 t_3^8 t_4^4 t_5^4 t_6^8 t_7^6 + 2 y_{s}^6 y_{o_1}^{12} y_{o_2}^{10} y_{o_3}^{14} y_{o_4}^{20} t_1^7 t_2^7 t_3^8 t_4^4 t_5^4 t_6^8 t_7^6 +  y_{s}^6 y_{o_1}^{12} y_{o_2}^{10} y_{o_3}^{14} y_{o_4}^{20} t_1^6 t_2^8 t_3^8 t_4^4 t_5^4 t_6^8 t_7^6 -  y_{s}^7 y_{o_1}^{13} y_{o_2}^{13} y_{o_3}^{14} y_{o_4}^{21} t_1^9 t_2^5 t_3^7 t_4^7 t_5^6 t_6^8 t_7^6 -  3 y_{s}^7 y_{o_1}^{13} y_{o_2}^{13} y_{o_3}^{14} y_{o_4}^{21} t_1^8 t_2^6 t_3^7 t_4^7 t_5^6 t_6^8 t_7^6 - 3 y_{s}^7 y_{o_1}^{13} y_{o_2}^{13} y_{o_3}^{14} y_{o_4}^{21} t_1^7 t_2^7 t_3^7 t_4^7 t_5^6 t_6^8 t_7^6 - 3 y_{s}^7 y_{o_1}^{13} y_{o_2}^{13} y_{o_3}^{14} y_{o_4}^{21} t_1^6 t_2^8 t_3^7 t_4^7 t_5^6 t_6^8 t_7^6 -  y_{s}^7 y_{o_1}^{13} y_{o_2}^{13} y_{o_3}^{14} y_{o_4}^{21} t_1^5 t_2^9 t_3^7 t_4^7 t_5^6 t_6^8 t_7^6 -  y_{s}^7 y_{o_1}^{13} y_{o_2}^{12} y_{o_3}^{15} y_{o_4}^{22} t_1^8 t_2^7 t_3^8 t_4^6 t_5^5 t_6^9 t_7^6 -  y_{s}^7 y_{o_1}^{13} y_{o_2}^{12} y_{o_3}^{15} y_{o_4}^{22} t_1^7 t_2^8 t_3^8 t_4^6 t_5^5 t_6^9 t_7^6 +  y_{s}^8 y_{o_1}^{14} y_{o_2}^{15} y_{o_3}^{15} y_{o_4}^{23} t_1^8 t_2^7 t_3^7 t_4^9 t_5^7 t_6^9 t_7^6 +  y_{s}^8 y_{o_1}^{14} y_{o_2}^{15} y_{o_3}^{15} y_{o_4}^{23} t_1^7 t_2^8 t_3^7 t_4^9 t_5^7 t_6^9 t_7^6 -  y_{s}^4 y_{o_1}^{11} y_{o_2}^{11} y_{o_3}^8 y_{o_4}^{12} t_1^4 t_2^4 t_3^4 t_4^4 t_5^7 t_6 t_7^7 +  y_{s}^4 y_{o_1}^{11} y_{o_2}^{10} y_{o_3}^9 y_{o_4}^{13} t_1^6 t_2^3 t_3^5 t_4^3 t_5^6 t_6^2 t_7^7 +  y_{s}^4 y_{o_1}^{11} y_{o_2}^{10} y_{o_3}^9 y_{o_4}^{13} t_1^3 t_2^6 t_3^5 t_4^3 t_5^6 t_6^2 t_7^7 +  y_{s}^4 y_{o_1}^{11} y_{o_2}^9 y_{o_3}^{10} y_{o_4}^{14} t_1^6 t_2^4 t_3^6 t_4^2 t_5^5 t_6^3 t_7^7 +  2 y_{s}^4 y_{o_1}^{11} y_{o_2}^9 y_{o_3}^{10} y_{o_4}^{14} t_1^5 t_2^5 t_3^6 t_4^2 t_5^5 t_6^3 t_7^7 +  y_{s}^4 y_{o_1}^{11} y_{o_2}^9 y_{o_3}^{10} y_{o_4}^{14} t_1^4 t_2^6 t_3^6 t_4^2 t_5^5 t_6^3 t_7^7 -  y_{s}^5 y_{o_1}^{12} y_{o_2}^{12} y_{o_3}^{10} y_{o_4}^{15} t_1^7 t_2^3 t_3^5 t_4^5 t_5^7 t_6^3 t_7^7 -  y_{s}^5 y_{o_1}^{12} y_{o_2}^{12} y_{o_3}^{10} y_{o_4}^{15} t_1^6 t_2^4 t_3^5 t_4^5 t_5^7 t_6^3 t_7^7 -  y_{s}^5 y_{o_1}^{12} y_{o_2}^{12} y_{o_3}^{10} y_{o_4}^{15} t_1^5 t_2^5 t_3^5 t_4^5 t_5^7 t_6^3 t_7^7 -  y_{s}^5 y_{o_1}^{12} y_{o_2}^{12} y_{o_3}^{10} y_{o_4}^{15} t_1^4 t_2^6 t_3^5 t_4^5 t_5^7 t_6^3 t_7^7 -  y_{s}^5 y_{o_1}^{12} y_{o_2}^{12} y_{o_3}^{10} y_{o_4}^{15} t_1^3 t_2^7 t_3^5 t_4^5 t_5^7 t_6^3 t_7^7 -  2 y_{s}^5 y_{o_1}^{12} y_{o_2}^{11} y_{o_3}^{11} y_{o_4}^{16} t_1^8 t_2^3 t_3^6 t_4^4 t_5^6 t_6^4 t_7^7 - 3 y_{s}^5 y_{o_1}^{12} y_{o_2}^{11} y_{o_3}^{11} y_{o_4}^{16} t_1^7 t_2^4 t_3^6 t_4^4 t_5^6 t_6^4 t_7^7 - 6 y_{s}^5 y_{o_1}^{12} y_{o_2}^{11} y_{o_3}^{11} y_{o_4}^{16} t_1^6 t_2^5 t_3^6 t_4^4 t_5^6 t_6^4 t_7^7 -  6 y_{s}^5 y_{o_1}^{12} y_{o_2}^{11} y_{o_3}^{11} y_{o_4}^{16} t_1^5 t_2^6 t_3^6 t_4^4 t_5^6 t_6^4 t_7^7 - 3 y_{s}^5 y_{o_1}^{12} y_{o_2}^{11} y_{o_3}^{11} y_{o_4}^{16} t_1^4 t_2^7 t_3^6 t_4^4 t_5^6 t_6^4 t_7^7 - 2 y_{s}^5 y_{o_1}^{12} y_{o_2}^{11} y_{o_3}^{11} y_{o_4}^{16} t_1^3 t_2^8 t_3^6 t_4^4 t_5^6 t_6^4 t_7^7 +  y_{s}^6 y_{o_1}^{13} y_{o_2}^{14} y_{o_3}^{11} y_{o_4}^{17} t_1^7 t_2^4 t_3^5 t_4^7 t_5^8 t_6^4 t_7^7 +  y_{s}^6 y_{o_1}^{13} y_{o_2}^{14} y_{o_3}^{11} y_{o_4}^{17} t_1^6 t_2^5 t_3^5 t_4^7 t_5^8 t_6^4 t_7^7 +  y_{s}^6 y_{o_1}^{13} y_{o_2}^{14} y_{o_3}^{11} y_{o_4}^{17} t_1^5 t_2^6 t_3^5 t_4^7 t_5^8 t_6^4 t_7^7 +  y_{s}^6 y_{o_1}^{13} y_{o_2}^{14} y_{o_3}^{11} y_{o_4}^{17} t_1^4 t_2^7 t_3^5 t_4^7 t_5^8 t_6^4 t_7^7 -  y_{s}^5 y_{o_1}^{12} y_{o_2}^{10} y_{o_3}^{12} y_{o_4}^{17} t_1^8 t_2^4 t_3^7 t_4^3 t_5^5 t_6^5 t_7^7 -  2 y_{s}^5 y_{o_1}^{12} y_{o_2}^{10} y_{o_3}^{12} y_{o_4}^{17} t_1^7 t_2^5 t_3^7 t_4^3 t_5^5 t_6^5 t_7^7 - 3 y_{s}^5 y_{o_1}^{12} y_{o_2}^{10} y_{o_3}^{12} y_{o_4}^{17} t_1^6 t_2^6 t_3^7 t_4^3 t_5^5 t_6^5 t_7^7 - 2 y_{s}^5 y_{o_1}^{12} y_{o_2}^{10} y_{o_3}^{12} y_{o_4}^{17} t_1^5 t_2^7 t_3^7 t_4^3 t_5^5 t_6^5 t_7^7 -  y_{s}^5 y_{o_1}^{12} y_{o_2}^{10} y_{o_3}^{12} y_{o_4}^{17} t_1^4 t_2^8 t_3^7 t_4^3 t_5^5 t_6^5 t_7^7 +  y_{s}^6 y_{o_1}^{13} y_{o_2}^{13} y_{o_3}^{12} y_{o_4}^{18} t_1^9 t_2^3 t_3^6 t_4^6 t_5^7 t_6^5 t_7^7 +  3 y_{s}^6 y_{o_1}^{13} y_{o_2}^{13} y_{o_3}^{12} y_{o_4}^{18} t_1^8 t_2^4 t_3^6 t_4^6 t_5^7 t_6^5 t_7^7 + 5 y_{s}^6 y_{o_1}^{13} y_{o_2}^{13} y_{o_3}^{12} y_{o_4}^{18} t_1^7 t_2^5 t_3^6 t_4^6 t_5^7 t_6^5 t_7^7 + 6 y_{s}^6 y_{o_1}^{13} y_{o_2}^{13} y_{o_3}^{12} y_{o_4}^{18} t_1^6 t_2^6 t_3^6 t_4^6 t_5^7 t_6^5 t_7^7 +  5 y_{s}^6 y_{o_1}^{13} y_{o_2}^{13} y_{o_3}^{12} y_{o_4}^{18} t_1^5 t_2^7 t_3^6 t_4^6 t_5^7 t_6^5 t_7^7 + 3 y_{s}^6 y_{o_1}^{13} y_{o_2}^{13} y_{o_3}^{12} y_{o_4}^{18} t_1^4 t_2^8 t_3^6 t_4^6 t_5^7 t_6^5 t_7^7 + y_{s}^6 y_{o_1}^{13} y_{o_2}^{13} y_{o_3}^{12} y_{o_4}^{18} t_1^3 t_2^9 t_3^6 t_4^6 t_5^7 t_6^5 t_7^7 - y_{s}^5 y_{o_1}^{12} y_{o_2}^9 y_{o_3}^{13} y_{o_4}^{18} t_1^7 t_2^6 t_3^8 t_4^2 t_5^4 t_6^6 t_7^7 - y_{s}^5 y_{o_1}^{12} y_{o_2}^9 y_{o_3}^{13} y_{o_4}^{18} t_1^6 t_2^7 t_3^8 t_4^2 t_5^4 t_6^6 t_7^7 + y_{s}^6 y_{o_1}^{13} y_{o_2}^{12} y_{o_3}^{13} y_{o_4}^{19} t_1^9 t_2^4 t_3^7 t_4^5 t_5^6 t_6^6 t_7^7 + 5 y_{s}^6 y_{o_1}^{13} y_{o_2}^{12} y_{o_3}^{13} y_{o_4}^{19} t_1^8 t_2^5 t_3^7 t_4^5 t_5^6 t_6^6 t_7^7 +  6 y_{s}^6 y_{o_1}^{13} y_{o_2}^{12} y_{o_3}^{13} y_{o_4}^{19} t_1^7 t_2^6 t_3^7 t_4^5 t_5^6 t_6^6 t_7^7 + 6 y_{s}^6 y_{o_1}^{13} y_{o_2}^{12} y_{o_3}^{13} y_{o_4}^{19} t_1^6 t_2^7 t_3^7 t_4^5 t_5^6 t_6^6 t_7^7 + 5 y_{s}^6 y_{o_1}^{13} y_{o_2}^{12} y_{o_3}^{13} y_{o_4}^{19} t_1^5 t_2^8 t_3^7 t_4^5 t_5^6 t_6^6 t_7^7 +  y_{s}^6 y_{o_1}^{13} y_{o_2}^{12} y_{o_3}^{13} y_{o_4}^{19} t_1^4 t_2^9 t_3^7 t_4^5 t_5^6 t_6^6 t_7^7 -  y_{s}^7 y_{o_1}^{14} y_{o_2}^{15} y_{o_3}^{13} y_{o_4}^{20} t_1^9 t_2^4 t_3^6 t_4^8 t_5^8 t_6^6 t_7^7 -  y_{s}^7 y_{o_1}^{14} y_{o_2}^{15} y_{o_3}^{13} y_{o_4}^{20} t_1^8 t_2^5 t_3^6 t_4^8 t_5^8 t_6^6 t_7^7 -  2 y_{s}^7 y_{o_1}^{14} y_{o_2}^{15} y_{o_3}^{13} y_{o_4}^{20} t_1^7 t_2^6 t_3^6 t_4^8 t_5^8 t_6^6 t_7^7 - 2 y_{s}^7 y_{o_1}^{14} y_{o_2}^{15} y_{o_3}^{13} y_{o_4}^{20} t_1^6 t_2^7 t_3^6 t_4^8 t_5^8 t_6^6 t_7^7 - y_{s}^7 y_{o_1}^{14} y_{o_2}^{15} y_{o_3}^{13} y_{o_4}^{20} t_1^5 t_2^8 t_3^6 t_4^8 t_5^8 t_6^6 t_7^7 - y_{s}^7 y_{o_1}^{14} y_{o_2}^{15} y_{o_3}^{13} y_{o_4}^{20} t_1^4 t_2^9 t_3^6 t_4^8 t_5^8 t_6^6 t_7^7 + 2 y_{s}^6 y_{o_1}^{13} y_{o_2}^{11} y_{o_3}^{14} y_{o_4}^{20} t_1^8 t_2^6 t_3^8 t_4^4 t_5^5 t_6^7 t_7^7 +  3 y_{s}^6 y_{o_1}^{13} y_{o_2}^{11} y_{o_3}^{14} y_{o_4}^{20} t_1^7 t_2^7 t_3^8 t_4^4 t_5^5 t_6^7 t_7^7 + 2 y_{s}^6 y_{o_1}^{13} y_{o_2}^{11} y_{o_3}^{14} y_{o_4}^{20} t_1^6 t_2^8 t_3^8 t_4^4 t_5^5 t_6^7 t_7^7 - 2 y_{s}^7 y_{o_1}^{14} y_{o_2}^{14} y_{o_3}^{14} y_{o_4}^{21} t_1^9 t_2^5 t_3^7 t_4^7 t_5^7 t_6^7 t_7^7 -  4 y_{s}^7 y_{o_1}^{14} y_{o_2}^{14} y_{o_3}^{14} y_{o_4}^{21} t_1^8 t_2^6 t_3^7 t_4^7 t_5^7 t_6^7 t_7^7 - 5 y_{s}^7 y_{o_1}^{14} y_{o_2}^{14} y_{o_3}^{14} y_{o_4}^{21} t_1^7 t_2^7 t_3^7 t_4^7 t_5^7 t_6^7 t_7^7 - 4 y_{s}^7 y_{o_1}^{14} y_{o_2}^{14} y_{o_3}^{14} y_{o_4}^{21} t_1^6 t_2^8 t_3^7 t_4^7 t_5^7 t_6^7 t_7^7 -  2 y_{s}^7 y_{o_1}^{14} y_{o_2}^{14} y_{o_3}^{14} y_{o_4}^{21} t_1^5 t_2^9 t_3^7 t_4^7 t_5^7 t_6^7 t_7^7 - y_{s}^7 y_{o_1}^{14} y_{o_2}^{13} y_{o_3}^{15} y_{o_4}^{22} t_1^9 t_2^6 t_3^8 t_4^6 t_5^6 t_6^8 t_7^7 - 3 y_{s}^7 y_{o_1}^{14} y_{o_2}^{13} y_{o_3}^{15} y_{o_4}^{22} t_1^8 t_2^7 t_3^8 t_4^6 t_5^6 t_6^8 t_7^7 - 3 y_{s}^7 y_{o_1}^{14} y_{o_2}^{13} y_{o_3}^{15} y_{o_4}^{22} t_1^7 t_2^8 t_3^8 t_4^6 t_5^6 t_6^8 t_7^7 -  y_{s}^7 y_{o_1}^{14} y_{o_2}^{13} y_{o_3}^{15} y_{o_4}^{22} t_1^6 t_2^9 t_3^8 t_4^6 t_5^6 t_6^8 t_7^7 +  y_{s}^8 y_{o_1}^{15} y_{o_2}^{16} y_{o_3}^{15} y_{o_4}^{23} t_1^9 t_2^6 t_3^7 t_4^9 t_5^8 t_6^8 t_7^7 +  y_{s}^8 y_{o_1}^{15} y_{o_2}^{16} y_{o_3}^{15} y_{o_4}^{23} t_1^8 t_2^7 t_3^7 t_4^9 t_5^8 t_6^8 t_7^7 +  y_{s}^8 y_{o_1}^{15} y_{o_2}^{16} y_{o_3}^{15} y_{o_4}^{23} t_1^7 t_2^8 t_3^7 t_4^9 t_5^8 t_6^8 t_7^7 +  y_{s}^8 y_{o_1}^{15} y_{o_2}^{16} y_{o_3}^{15} y_{o_4}^{23} t_1^6 t_2^9 t_3^7 t_4^9 t_5^8 t_6^8 t_7^7 +  y_{s}^8 y_{o_1}^{15} y_{o_2}^{15} y_{o_3}^{16} y_{o_4}^{24} t_1^9 t_2^7 t_3^8 t_4^8 t_5^7 t_6^9 t_7^7 +  y_{s}^8 y_{o_1}^{15} y_{o_2}^{15} y_{o_3}^{16} y_{o_4}^{24} t_1^8 t_2^8 t_3^8 t_4^8 t_5^7 t_6^9 t_7^7 +  y_{s}^8 y_{o_1}^{15} y_{o_2}^{15} y_{o_3}^{16} y_{o_4}^{24} t_1^7 t_2^9 t_3^8 t_4^8 t_5^7 t_6^9 t_7^7 -  y_{s}^5 y_{o_1}^{13} y_{o_2}^{12} y_{o_3}^{11} y_{o_4}^{16} t_1^6 t_2^5 t_3^6 t_4^4 t_5^7 t_6^3 t_7^8 -  y_{s}^5 y_{o_1}^{13} y_{o_2}^{12} y_{o_3}^{11} y_{o_4}^{16} t_1^5 t_2^6 t_3^6 t_4^4 t_5^7 t_6^3 t_7^8 -  y_{s}^5 y_{o_1}^{13} y_{o_2}^{11} y_{o_3}^{12} y_{o_4}^{17} t_1^6 t_2^6 t_3^7 t_4^3 t_5^6 t_6^4 t_7^8 +  y_{s}^6 y_{o_1}^{14} y_{o_2}^{14} y_{o_3}^{12} y_{o_4}^{18} t_1^8 t_2^4 t_3^6 t_4^6 t_5^8 t_6^4 t_7^8 +  2 y_{s}^6 y_{o_1}^{14} y_{o_2}^{14} y_{o_3}^{12} y_{o_4}^{18} t_1^7 t_2^5 t_3^6 t_4^6 t_5^8 t_6^4 t_7^8 + 2 y_{s}^6 y_{o_1}^{14} y_{o_2}^{14} y_{o_3}^{12} y_{o_4}^{18} t_1^6 t_2^6 t_3^6 t_4^6 t_5^8 t_6^4 t_7^8 + 2 y_{s}^6 y_{o_1}^{14} y_{o_2}^{14} y_{o_3}^{12} y_{o_4}^{18} t_1^5 t_2^7 t_3^6 t_4^6 t_5^8 t_6^4 t_7^8 +  y_{s}^6 y_{o_1}^{14} y_{o_2}^{14} y_{o_3}^{12} y_{o_4}^{18} t_1^4 t_2^8 t_3^6 t_4^6 t_5^8 t_6^4 t_7^8 +  y_{s}^6 y_{o_1}^{14} y_{o_2}^{13} y_{o_3}^{13} y_{o_4}^{19} t_1^8 t_2^5 t_3^7 t_4^5 t_5^7 t_6^5 t_7^8 +  2 y_{s}^6 y_{o_1}^{14} y_{o_2}^{13} y_{o_3}^{13} y_{o_4}^{19} t_1^7 t_2^6 t_3^7 t_4^5 t_5^7 t_6^5 t_7^8 + 2 y_{s}^6 y_{o_1}^{14} y_{o_2}^{13} y_{o_3}^{13} y_{o_4}^{19} t_1^6 t_2^7 t_3^7 t_4^5 t_5^7 t_6^5 t_7^8 + y_{s}^6 y_{o_1}^{14} y_{o_2}^{13} y_{o_3}^{13} y_{o_4}^{19} t_1^5 t_2^8 t_3^7 t_4^5 t_5^7 t_6^5 t_7^8 - y_{s}^7 y_{o_1}^{15} y_{o_2}^{16} y_{o_3}^{13} y_{o_4}^{20} t_1^8 t_2^5 t_3^6 t_4^8 t_5^9 t_6^5 t_7^8 - y_{s}^7 y_{o_1}^{15} y_{o_2}^{16} y_{o_3}^{13} y_{o_4}^{20} t_1^7 t_2^6 t_3^6 t_4^8 t_5^9 t_6^5 t_7^8 - y_{s}^7 y_{o_1}^{15} y_{o_2}^{16} y_{o_3}^{13} y_{o_4}^{20} t_1^6 t_2^7 t_3^6 t_4^8 t_5^9 t_6^5 t_7^8 - y_{s}^7 y_{o_1}^{15} y_{o_2}^{16} y_{o_3}^{13} y_{o_4}^{20} t_1^5 t_2^8 t_3^6 t_4^8 t_5^9 t_6^5 t_7^8 + 2 y_{s}^6 y_{o_1}^{14} y_{o_2}^{12} y_{o_3}^{14} y_{o_4}^{20} t_1^8 t_2^6 t_3^8 t_4^4 t_5^6 t_6^6 t_7^8 +  2 y_{s}^6 y_{o_1}^{14} y_{o_2}^{12} y_{o_3}^{14} y_{o_4}^{20} t_1^7 t_2^7 t_3^8 t_4^4 t_5^6 t_6^6 t_7^8 + 2 y_{s}^6 y_{o_1}^{14} y_{o_2}^{12} y_{o_3}^{14} y_{o_4}^{20} t_1^6 t_2^8 t_3^8 t_4^4 t_5^6 t_6^6 t_7^8 - y_{s}^7 y_{o_1}^{15} y_{o_2}^{15} y_{o_3}^{14} y_{o_4}^{21} t_1^9 t_2^5 t_3^7 t_4^7 t_5^8 t_6^6 t_7^8 - 2 y_{s}^7 y_{o_1}^{15} y_{o_2}^{15} y_{o_3}^{14} y_{o_4}^{21} t_1^8 t_2^6 t_3^7 t_4^7 t_5^8 t_6^6 t_7^8 -  3 y_{s}^7 y_{o_1}^{15} y_{o_2}^{15} y_{o_3}^{14} y_{o_4}^{21} t_1^7 t_2^7 t_3^7 t_4^7 t_5^8 t_6^6 t_7^8 - 2 y_{s}^7 y_{o_1}^{15} y_{o_2}^{15} y_{o_3}^{14} y_{o_4}^{21} t_1^6 t_2^8 t_3^7 t_4^7 t_5^8 t_6^6 t_7^8 - y_{s}^7 y_{o_1}^{15} y_{o_2}^{15} y_{o_3}^{14} y_{o_4}^{21} t_1^5 t_2^9 t_3^7 t_4^7 t_5^8 t_6^6 t_7^8 - y_{s}^7 y_{o_1}^{15} y_{o_2}^{14} y_{o_3}^{15} y_{o_4}^{22} t_1^9 t_2^6 t_3^8 t_4^6 t_5^7 t_6^7 t_7^8 - 3 y_{s}^7 y_{o_1}^{15} y_{o_2}^{14} y_{o_3}^{15} y_{o_4}^{22} t_1^8 t_2^7 t_3^8 t_4^6 t_5^7 t_6^7 t_7^8 -  3 y_{s}^7 y_{o_1}^{15} y_{o_2}^{14} y_{o_3}^{15} y_{o_4}^{22} t_1^7 t_2^8 t_3^8 t_4^6 t_5^7 t_6^7 t_7^8 - y_{s}^7 y_{o_1}^{15} y_{o_2}^{14} y_{o_3}^{15} y_{o_4}^{22} t_1^6 t_2^9 t_3^8 t_4^6 t_5^7 t_6^7 t_7^8 +  y_{s}^8 y_{o_1}^{16} y_{o_2}^{17} y_{o_3}^{15} y_{o_4}^{23} t_1^8 t_2^7 t_3^7 t_4^9 t_5^9 t_6^7 t_7^8 +  y_{s}^8 y_{o_1}^{16} y_{o_2}^{17} y_{o_3}^{15} y_{o_4}^{23} t_1^7 t_2^8 t_3^7 t_4^9 t_5^9 t_6^7 t_7^8 -  y_{s}^7 y_{o_1}^{15} y_{o_2}^{13} y_{o_3}^{16} y_{o_4}^{23} t_1^8 t_2^8 t_3^9 t_4^5 t_5^6 t_6^8 t_7^8 +  y_{s}^8 y_{o_1}^{16} y_{o_2}^{16} y_{o_3}^{16} y_{o_4}^{24} t_1^9 t_2^7 t_3^8 t_4^8 t_5^8 t_6^8 t_7^8 +  y_{s}^8 y_{o_1}^{16} y_{o_2}^{16} y_{o_3}^{16} y_{o_4}^{24} t_1^8 t_2^8 t_3^8 t_4^8 t_5^8 t_6^8 t_7^8 +  y_{s}^8 y_{o_1}^{16} y_{o_2}^{16} y_{o_3}^{16} y_{o_4}^{24} t_1^7 t_2^9 t_3^8 t_4^8 t_5^8 t_6^8 t_7^8 -  y_{s}^7 y_{o_1}^{16} y_{o_2}^{15} y_{o_3}^{15} y_{o_4}^{22} t_1^8 t_2^7 t_3^8 t_4^6 t_5^8 t_6^6 t_7^9 -  y_{s}^7 y_{o_1}^{16} y_{o_2}^{15} y_{o_3}^{15} y_{o_4}^{22} t_1^7 t_2^8 t_3^8 t_4^6 t_5^8 t_6^6 t_7^9 +  y_{s}^8 y_{o_1}^{17} y_{o_2}^{17} y_{o_3}^{16} y_{o_4}^{24} t_1^8 t_2^8 t_3^8 t_4^8 t_5^9 t_6^7 t_7^9 +  y_{s}^9 y_{o_1}^{18} y_{o_2}^{18} y_{o_3}^{18} y_{o_4}^{27} t_1^9 t_2^9 t_3^9 t_4^9 t_5^9 t_6^9 t_7^9
~,~
$
\end{quote}
\endgroup

\subsection{Model 15 \label{app_num_15}}

\begingroup\makeatletter\def\f@size{7}\check@mathfonts
\begin{quote}\raggedright
$
P(t_i,y_s,y_{o_1},y_{o_2},y_{o_3},y_{o_4}; \mathcal{M}_{15}) =
1 - y_{s}^2 y_{o_1}^2 y_{o_2}^3 y_{o_3}^3 y_{o_4}^4 t_1 t_2 t_3 t_4^3 t_5 t_6^3 +  y_{s} y_{o_1}^2 y_{o_2} y_{o_3}^3 y_{o_4}^3 t_1 t_2 t_4^2 t_5^2 t_7 +  y_{s} y_{o_1}^2 y_{o_2}^2 y_{o_3}^2 y_{o_4}^3 t_1^2 t_3 t_4 t_5 t_6 t_7 +  y_{s} y_{o_1}^2 y_{o_2}^2 y_{o_3}^2 y_{o_4}^3 t_1 t_2 t_3 t_4 t_5 t_6 t_7 +  y_{s} y_{o_1}^2 y_{o_2}^2 y_{o_3}^2 y_{o_4}^3 t_2^2 t_3 t_4 t_5 t_6 t_7 -  y_{s}^2 y_{o_1}^3 y_{o_2}^2 y_{o_3}^5 y_{o_4}^5 t_1^2 t_2 t_4^4 t_5^3 t_6 t_7 -  y_{s}^2 y_{o_1}^3 y_{o_2}^2 y_{o_3}^5 y_{o_4}^5 t_1 t_2^2 t_4^4 t_5^3 t_6 t_7 +  y_{s} y_{o_1}^2 y_{o_2}^3 y_{o_3} y_{o_4}^3 t_1 t_2 t_3^2 t_6^2 t_7 -  y_{s}^2 y_{o_1}^3 y_{o_2}^3 y_{o_3}^4 y_{o_4}^5 t_1^3 t_3 t_4^3 t_5^2 t_6^2 t_7 -  3 y_{s}^2 y_{o_1}^3 y_{o_2}^3 y_{o_3}^4 y_{o_4}^5 t_1^2 t_2 t_3 t_4^3 t_5^2 t_6^2 t_7 -  3 y_{s}^2 y_{o_1}^3 y_{o_2}^3 y_{o_3}^4 y_{o_4}^5 t_1 t_2^2 t_3 t_4^3 t_5^2 t_6^2 t_7 -  y_{s}^2 y_{o_1}^3 y_{o_2}^3 y_{o_3}^4 y_{o_4}^5 t_2^3 t_3 t_4^3 t_5^2 t_6^2 t_7 -  y_{s}^2 y_{o_1}^3 y_{o_2}^4 y_{o_3}^3 y_{o_4}^5 t_1^3 t_3^2 t_4^2 t_5 t_6^3 t_7 -  3 y_{s}^2 y_{o_1}^3 y_{o_2}^4 y_{o_3}^3 y_{o_4}^5 t_1^2 t_2 t_3^2 t_4^2 t_5 t_6^3 t_7 -  3 y_{s}^2 y_{o_1}^3 y_{o_2}^4 y_{o_3}^3 y_{o_4}^5 t_1 t_2^2 t_3^2 t_4^2 t_5 t_6^3 t_7 -  y_{s}^2 y_{o_1}^3 y_{o_2}^4 y_{o_3}^3 y_{o_4}^5 t_2^3 t_3^2 t_4^2 t_5 t_6^3 t_7 +  2 y_{s}^3 y_{o_1}^4 y_{o_2}^4 y_{o_3}^6 y_{o_4}^7 t_1^3 t_2 t_3 t_4^5 t_5^3 t_6^3 t_7 +  2 y_{s}^3 y_{o_1}^4 y_{o_2}^4 y_{o_3}^6 y_{o_4}^7 t_1^2 t_2^2 t_3 t_4^5 t_5^3 t_6^3 t_7 +  2 y_{s}^3 y_{o_1}^4 y_{o_2}^4 y_{o_3}^6 y_{o_4}^7 t_1 t_2^3 t_3 t_4^5 t_5^3 t_6^3 t_7 -  y_{s}^2 y_{o_1}^3 y_{o_2}^5 y_{o_3}^2 y_{o_4}^5 t_1^2 t_2 t_3^3 t_4 t_6^4 t_7 -  y_{s}^2 y_{o_1}^3 y_{o_2}^5 y_{o_3}^2 y_{o_4}^5 t_1 t_2^2 t_3^3 t_4 t_6^4 t_7 +  2 y_{s}^3 y_{o_1}^4 y_{o_2}^5 y_{o_3}^5 y_{o_4}^7 t_1^3 t_2 t_3^2 t_4^4 t_5^2 t_6^4 t_7 +  3 y_{s}^3 y_{o_1}^4 y_{o_2}^5 y_{o_3}^5 y_{o_4}^7 t_1^2 t_2^2 t_3^2 t_4^4 t_5^2 t_6^4 t_7 +  2 y_{s}^3 y_{o_1}^4 y_{o_2}^5 y_{o_3}^5 y_{o_4}^7 t_1 t_2^3 t_3^2 t_4^4 t_5^2 t_6^4 t_7 +  2 y_{s}^3 y_{o_1}^4 y_{o_2}^6 y_{o_3}^4 y_{o_4}^7 t_1^3 t_2 t_3^3 t_4^3 t_5 t_6^5 t_7 +  2 y_{s}^3 y_{o_1}^4 y_{o_2}^6 y_{o_3}^4 y_{o_4}^7 t_1^2 t_2^2 t_3^3 t_4^3 t_5 t_6^5 t_7 +  2 y_{s}^3 y_{o_1}^4 y_{o_2}^6 y_{o_3}^4 y_{o_4}^7 t_1 t_2^3 t_3^3 t_4^3 t_5 t_6^5 t_7 -  y_{s}^4 y_{o_1}^5 y_{o_2}^6 y_{o_3}^7 y_{o_4}^9 t_1^3 t_2^2 t_3^2 t_4^6 t_5^3 t_6^5 t_7 -  y_{s}^4 y_{o_1}^5 y_{o_2}^6 y_{o_3}^7 y_{o_4}^9 t_1^2 t_2^3 t_3^2 t_4^6 t_5^3 t_6^5 t_7 -  y_{s}^4 y_{o_1}^5 y_{o_2}^7 y_{o_3}^6 y_{o_4}^9 t_1^3 t_2^2 t_3^3 t_4^5 t_5^2 t_6^6 t_7 -  y_{s}^4 y_{o_1}^5 y_{o_2}^7 y_{o_3}^6 y_{o_4}^9 t_1^2 t_2^3 t_3^3 t_4^5 t_5^2 t_6^6 t_7 +  y_{s} y_{o_1}^3 y_{o_2}^2 y_{o_3}^3 y_{o_4}^4 t_1^3 t_3 t_4 t_5^2 t_7^2 +  y_{s} y_{o_1}^3 y_{o_2}^2 y_{o_3}^3 y_{o_4}^4 t_1^2 t_2 t_3 t_4 t_5^2 t_7^2 +  y_{s} y_{o_1}^3 y_{o_2}^2 y_{o_3}^3 y_{o_4}^4 t_1 t_2^2 t_3 t_4 t_5^2 t_7^2 +  y_{s} y_{o_1}^3 y_{o_2}^2 y_{o_3}^3 y_{o_4}^4 t_2^3 t_3 t_4 t_5^2 t_7^2 +  y_{s} y_{o_1}^3 y_{o_2}^3 y_{o_3}^2 y_{o_4}^4 t_1^3 t_3^2 t_5 t_6 t_7^2 +  y_{s} y_{o_1}^3 y_{o_2}^3 y_{o_3}^2 y_{o_4}^4 t_1^2 t_2 t_3^2 t_5 t_6 t_7^2 +  y_{s} y_{o_1}^3 y_{o_2}^3 y_{o_3}^2 y_{o_4}^4 t_1 t_2^2 t_3^2 t_5 t_6 t_7^2 +  y_{s} y_{o_1}^3 y_{o_2}^3 y_{o_3}^2 y_{o_4}^4 t_2^3 t_3^2 t_5 t_6 t_7^2 -  y_{s}^2 y_{o_1}^4 y_{o_2}^3 y_{o_3}^5 y_{o_4}^6 t_1^4 t_3 t_4^3 t_5^3 t_6 t_7^2 -  2 y_{s}^2 y_{o_1}^4 y_{o_2}^3 y_{o_3}^5 y_{o_4}^6 t_1^3 t_2 t_3 t_4^3 t_5^3 t_6 t_7^2 -  3 y_{s}^2 y_{o_1}^4 y_{o_2}^3 y_{o_3}^5 y_{o_4}^6 t_1^2 t_2^2 t_3 t_4^3 t_5^3 t_6 t_7^2 -  2 y_{s}^2 y_{o_1}^4 y_{o_2}^3 y_{o_3}^5 y_{o_4}^6 t_1 t_2^3 t_3 t_4^3 t_5^3 t_6 t_7^2 -  y_{s}^2 y_{o_1}^4 y_{o_2}^3 y_{o_3}^5 y_{o_4}^6 t_2^4 t_3 t_4^3 t_5^3 t_6 t_7^2 -  2 y_{s}^2 y_{o_1}^4 y_{o_2}^4 y_{o_3}^4 y_{o_4}^6 t_1^4 t_3^2 t_4^2 t_5^2 t_6^2 t_7^2 -  5 y_{s}^2 y_{o_1}^4 y_{o_2}^4 y_{o_3}^4 y_{o_4}^6 t_1^3 t_2 t_3^2 t_4^2 t_5^2 t_6^2 t_7^2 -  5 y_{s}^2 y_{o_1}^4 y_{o_2}^4 y_{o_3}^4 y_{o_4}^6 t_1^2 t_2^2 t_3^2 t_4^2 t_5^2 t_6^2 t_7^2 -  5 y_{s}^2 y_{o_1}^4 y_{o_2}^4 y_{o_3}^4 y_{o_4}^6 t_1 t_2^3 t_3^2 t_4^2 t_5^2 t_6^2 t_7^2 -  2 y_{s}^2 y_{o_1}^4 y_{o_2}^4 y_{o_3}^4 y_{o_4}^6 t_2^4 t_3^2 t_4^2 t_5^2 t_6^2 t_7^2 +  y_{s}^3 y_{o_1}^5 y_{o_2}^4 y_{o_3}^7 y_{o_4}^8 t_1^4 t_2 t_3 t_4^5 t_5^4 t_6^2 t_7^2 +  2 y_{s}^3 y_{o_1}^5 y_{o_2}^4 y_{o_3}^7 y_{o_4}^8 t_1^3 t_2^2 t_3 t_4^5 t_5^4 t_6^2 t_7^2 +  2 y_{s}^3 y_{o_1}^5 y_{o_2}^4 y_{o_3}^7 y_{o_4}^8 t_1^2 t_2^3 t_3 t_4^5 t_5^4 t_6^2 t_7^2 +  y_{s}^3 y_{o_1}^5 y_{o_2}^4 y_{o_3}^7 y_{o_4}^8 t_1 t_2^4 t_3 t_4^5 t_5^4 t_6^2 t_7^2 -  y_{s}^2 y_{o_1}^4 y_{o_2}^5 y_{o_3}^3 y_{o_4}^6 t_1^4 t_3^3 t_4 t_5 t_6^3 t_7^2 -  2 y_{s}^2 y_{o_1}^4 y_{o_2}^5 y_{o_3}^3 y_{o_4}^6 t_1^3 t_2 t_3^3 t_4 t_5 t_6^3 t_7^2 -  3 y_{s}^2 y_{o_1}^4 y_{o_2}^5 y_{o_3}^3 y_{o_4}^6 t_1^2 t_2^2 t_3^3 t_4 t_5 t_6^3 t_7^2 -  2 y_{s}^2 y_{o_1}^4 y_{o_2}^5 y_{o_3}^3 y_{o_4}^6 t_1 t_2^3 t_3^3 t_4 t_5 t_6^3 t_7^2 -  y_{s}^2 y_{o_1}^4 y_{o_2}^5 y_{o_3}^3 y_{o_4}^6 t_2^4 t_3^3 t_4 t_5 t_6^3 t_7^2 +  y_{s}^3 y_{o_1}^5 y_{o_2}^5 y_{o_3}^6 y_{o_4}^8 t_1^5 t_3^2 t_4^4 t_5^3 t_6^3 t_7^2 +  5 y_{s}^3 y_{o_1}^5 y_{o_2}^5 y_{o_3}^6 y_{o_4}^8 t_1^4 t_2 t_3^2 t_4^4 t_5^3 t_6^3 t_7^2 +  7 y_{s}^3 y_{o_1}^5 y_{o_2}^5 y_{o_3}^6 y_{o_4}^8 t_1^3 t_2^2 t_3^2 t_4^4 t_5^3 t_6^3 t_7^2 +  7 y_{s}^3 y_{o_1}^5 y_{o_2}^5 y_{o_3}^6 y_{o_4}^8 t_1^2 t_2^3 t_3^2 t_4^4 t_5^3 t_6^3 t_7^2 +  5 y_{s}^3 y_{o_1}^5 y_{o_2}^5 y_{o_3}^6 y_{o_4}^8 t_1 t_2^4 t_3^2 t_4^4 t_5^3 t_6^3 t_7^2 +  y_{s}^3 y_{o_1}^5 y_{o_2}^5 y_{o_3}^6 y_{o_4}^8 t_2^5 t_3^2 t_4^4 t_5^3 t_6^3 t_7^2 -  y_{s}^4 y_{o_1}^6 y_{o_2}^5 y_{o_3}^9 y_{o_4}^{10} t_1^3 t_2^3 t_3 t_4^7 t_5^5 t_6^3 t_7^2 +  y_{s}^3 y_{o_1}^5 y_{o_2}^6 y_{o_3}^5 y_{o_4}^8 t_1^5 t_3^3 t_4^3 t_5^2 t_6^4 t_7^2 +  5 y_{s}^3 y_{o_1}^5 y_{o_2}^6 y_{o_3}^5 y_{o_4}^8 t_1^4 t_2 t_3^3 t_4^3 t_5^2 t_6^4 t_7^2 +  7 y_{s}^3 y_{o_1}^5 y_{o_2}^6 y_{o_3}^5 y_{o_4}^8 t_1^3 t_2^2 t_3^3 t_4^3 t_5^2 t_6^4 t_7^2 +  7 y_{s}^3 y_{o_1}^5 y_{o_2}^6 y_{o_3}^5 y_{o_4}^8 t_1^2 t_2^3 t_3^3 t_4^3 t_5^2 t_6^4 t_7^2 +  5 y_{s}^3 y_{o_1}^5 y_{o_2}^6 y_{o_3}^5 y_{o_4}^8 t_1 t_2^4 t_3^3 t_4^3 t_5^2 t_6^4 t_7^2 +  y_{s}^3 y_{o_1}^5 y_{o_2}^6 y_{o_3}^5 y_{o_4}^8 t_2^5 t_3^3 t_4^3 t_5^2 t_6^4 t_7^2 -  y_{s}^4 y_{o_1}^6 y_{o_2}^6 y_{o_3}^8 y_{o_4}^{10} t_1^5 t_2 t_3^2 t_4^6 t_5^4 t_6^4 t_7^2 -  2 y_{s}^4 y_{o_1}^6 y_{o_2}^6 y_{o_3}^8 y_{o_4}^{10} t_1^4 t_2^2 t_3^2 t_4^6 t_5^4 t_6^4 t_7^2 -  3 y_{s}^4 y_{o_1}^6 y_{o_2}^6 y_{o_3}^8 y_{o_4}^{10} t_1^3 t_2^3 t_3^2 t_4^6 t_5^4 t_6^4 t_7^2 -  2 y_{s}^4 y_{o_1}^6 y_{o_2}^6 y_{o_3}^8 y_{o_4}^{10} t_1^2 t_2^4 t_3^2 t_4^6 t_5^4 t_6^4 t_7^2 -  y_{s}^4 y_{o_1}^6 y_{o_2}^6 y_{o_3}^8 y_{o_4}^{10} t_1 t_2^5 t_3^2 t_4^6 t_5^4 t_6^4 t_7^2 +  y_{s}^3 y_{o_1}^5 y_{o_2}^7 y_{o_3}^4 y_{o_4}^8 t_1^4 t_2 t_3^4 t_4^2 t_5 t_6^5 t_7^2 +  2 y_{s}^3 y_{o_1}^5 y_{o_2}^7 y_{o_3}^4 y_{o_4}^8 t_1^3 t_2^2 t_3^4 t_4^2 t_5 t_6^5 t_7^2 +  2 y_{s}^3 y_{o_1}^5 y_{o_2}^7 y_{o_3}^4 y_{o_4}^8 t_1^2 t_2^3 t_3^4 t_4^2 t_5 t_6^5 t_7^2 +  y_{s}^3 y_{o_1}^5 y_{o_2}^7 y_{o_3}^4 y_{o_4}^8 t_1 t_2^4 t_3^4 t_4^2 t_5 t_6^5 t_7^2 -  3 y_{s}^4 y_{o_1}^6 y_{o_2}^7 y_{o_3}^7 y_{o_4}^{10} t_1^5 t_2 t_3^3 t_4^5 t_5^3 t_6^5 t_7^2 -  5 y_{s}^4 y_{o_1}^6 y_{o_2}^7 y_{o_3}^7 y_{o_4}^{10} t_1^4 t_2^2 t_3^3 t_4^5 t_5^3 t_6^5 t_7^2 -  7 y_{s}^4 y_{o_1}^6 y_{o_2}^7 y_{o_3}^7 y_{o_4}^{10} t_1^3 t_2^3 t_3^3 t_4^5 t_5^3 t_6^5 t_7^2 -  5 y_{s}^4 y_{o_1}^6 y_{o_2}^7 y_{o_3}^7 y_{o_4}^{10} t_1^2 t_2^4 t_3^3 t_4^5 t_5^3 t_6^5 t_7^2 -  3 y_{s}^4 y_{o_1}^6 y_{o_2}^7 y_{o_3}^7 y_{o_4}^{10} t_1 t_2^5 t_3^3 t_4^5 t_5^3 t_6^5 t_7^2 -  y_{s}^4 y_{o_1}^6 y_{o_2}^8 y_{o_3}^6 y_{o_4}^{10} t_1^5 t_2 t_3^4 t_4^4 t_5^2 t_6^6 t_7^2 -  2 y_{s}^4 y_{o_1}^6 y_{o_2}^8 y_{o_3}^6 y_{o_4}^{10} t_1^4 t_2^2 t_3^4 t_4^4 t_5^2 t_6^6 t_7^2 -  3 y_{s}^4 y_{o_1}^6 y_{o_2}^8 y_{o_3}^6 y_{o_4}^{10} t_1^3 t_2^3 t_3^4 t_4^4 t_5^2 t_6^6 t_7^2 -  2 y_{s}^4 y_{o_1}^6 y_{o_2}^8 y_{o_3}^6 y_{o_4}^{10} t_1^2 t_2^4 t_3^4 t_4^4 t_5^2 t_6^6 t_7^2 -  y_{s}^4 y_{o_1}^6 y_{o_2}^8 y_{o_3}^6 y_{o_4}^{10} t_1 t_2^5 t_3^4 t_4^4 t_5^2 t_6^6 t_7^2 +  y_{s}^5 y_{o_1}^7 y_{o_2}^8 y_{o_3}^9 y_{o_4}^{12} t_1^5 t_2^2 t_3^3 t_4^7 t_5^4 t_6^6 t_7^2 +  y_{s}^5 y_{o_1}^7 y_{o_2}^8 y_{o_3}^9 y_{o_4}^{12} t_1^4 t_2^3 t_3^3 t_4^7 t_5^4 t_6^6 t_7^2 +  y_{s}^5 y_{o_1}^7 y_{o_2}^8 y_{o_3}^9 y_{o_4}^{12} t_1^3 t_2^4 t_3^3 t_4^7 t_5^4 t_6^6 t_7^2 +  y_{s}^5 y_{o_1}^7 y_{o_2}^8 y_{o_3}^9 y_{o_4}^{12} t_1^2 t_2^5 t_3^3 t_4^7 t_5^4 t_6^6 t_7^2 -  y_{s}^4 y_{o_1}^6 y_{o_2}^9 y_{o_3}^5 y_{o_4}^{10} t_1^3 t_2^3 t_3^5 t_4^3 t_5 t_6^7 t_7^2 +  y_{s}^5 y_{o_1}^7 y_{o_2}^9 y_{o_3}^8 y_{o_4}^{12} t_1^5 t_2^2 t_3^4 t_4^6 t_5^3 t_6^7 t_7^2 +  y_{s}^5 y_{o_1}^7 y_{o_2}^9 y_{o_3}^8 y_{o_4}^{12} t_1^4 t_2^3 t_3^4 t_4^6 t_5^3 t_6^7 t_7^2 +  y_{s}^5 y_{o_1}^7 y_{o_2}^9 y_{o_3}^8 y_{o_4}^{12} t_1^3 t_2^4 t_3^4 t_4^6 t_5^3 t_6^7 t_7^2 +  y_{s}^5 y_{o_1}^7 y_{o_2}^9 y_{o_3}^8 y_{o_4}^{12} t_1^2 t_2^5 t_3^4 t_4^6 t_5^3 t_6^7 t_7^2 +  y_{s} y_{o_1}^4 y_{o_2}^3 y_{o_3}^3 y_{o_4}^5 t_1^3 t_2 t_3^2 t_5^2 t_7^3 +  y_{s} y_{o_1}^4 y_{o_2}^3 y_{o_3}^3 y_{o_4}^5 t_1^2 t_2^2 t_3^2 t_5^2 t_7^3 +  y_{s} y_{o_1}^4 y_{o_2}^3 y_{o_3}^3 y_{o_4}^5 t_1 t_2^3 t_3^2 t_5^2 t_7^3 -  y_{s}^2 y_{o_1}^5 y_{o_2}^3 y_{o_3}^6 y_{o_4}^7 t_1^3 t_2^2 t_3 t_4^3 t_5^4 t_7^3 -  y_{s}^2 y_{o_1}^5 y_{o_2}^3 y_{o_3}^6 y_{o_4}^7 t_1^2 t_2^3 t_3 t_4^3 t_5^4 t_7^3 -  y_{s}^2 y_{o_1}^5 y_{o_2}^4 y_{o_3}^5 y_{o_4}^7 t_1^5 t_3^2 t_4^2 t_5^3 t_6 t_7^3 -  2 y_{s}^2 y_{o_1}^5 y_{o_2}^4 y_{o_3}^5 y_{o_4}^7 t_1^4 t_2 t_3^2 t_4^2 t_5^3 t_6 t_7^3 -  3 y_{s}^2 y_{o_1}^5 y_{o_2}^4 y_{o_3}^5 y_{o_4}^7 t_1^3 t_2^2 t_3^2 t_4^2 t_5^3 t_6 t_7^3 -  3 y_{s}^2 y_{o_1}^5 y_{o_2}^4 y_{o_3}^5 y_{o_4}^7 t_1^2 t_2^3 t_3^2 t_4^2 t_5^3 t_6 t_7^3 -  2 y_{s}^2 y_{o_1}^5 y_{o_2}^4 y_{o_3}^5 y_{o_4}^7 t_1 t_2^4 t_3^2 t_4^2 t_5^3 t_6 t_7^3 -  y_{s}^2 y_{o_1}^5 y_{o_2}^4 y_{o_3}^5 y_{o_4}^7 t_2^5 t_3^2 t_4^2 t_5^3 t_6 t_7^3 +  y_{s}^3 y_{o_1}^6 y_{o_2}^4 y_{o_3}^8 y_{o_4}^9 t_1^4 t_2^2 t_3 t_4^5 t_5^5 t_6 t_7^3 +  2 y_{s}^3 y_{o_1}^6 y_{o_2}^4 y_{o_3}^8 y_{o_4}^9 t_1^3 t_2^3 t_3 t_4^5 t_5^5 t_6 t_7^3 +  y_{s}^3 y_{o_1}^6 y_{o_2}^4 y_{o_3}^8 y_{o_4}^9 t_1^2 t_2^4 t_3 t_4^5 t_5^5 t_6 t_7^3 -  y_{s}^2 y_{o_1}^5 y_{o_2}^5 y_{o_3}^4 y_{o_4}^7 t_1^5 t_3^3 t_4 t_5^2 t_6^2 t_7^3 -  2 y_{s}^2 y_{o_1}^5 y_{o_2}^5 y_{o_3}^4 y_{o_4}^7 t_1^4 t_2 t_3^3 t_4 t_5^2 t_6^2 t_7^3 -  3 y_{s}^2 y_{o_1}^5 y_{o_2}^5 y_{o_3}^4 y_{o_4}^7 t_1^3 t_2^2 t_3^3 t_4 t_5^2 t_6^2 t_7^3 -  3 y_{s}^2 y_{o_1}^5 y_{o_2}^5 y_{o_3}^4 y_{o_4}^7 t_1^2 t_2^3 t_3^3 t_4 t_5^2 t_6^2 t_7^3 -  2 y_{s}^2 y_{o_1}^5 y_{o_2}^5 y_{o_3}^4 y_{o_4}^7 t_1 t_2^4 t_3^3 t_4 t_5^2 t_6^2 t_7^3 -  y_{s}^2 y_{o_1}^5 y_{o_2}^5 y_{o_3}^4 y_{o_4}^7 t_2^5 t_3^3 t_4 t_5^2 t_6^2 t_7^3 +  y_{s}^3 y_{o_1}^6 y_{o_2}^5 y_{o_3}^7 y_{o_4}^9 t_1^5 t_2 t_3^2 t_4^4 t_5^4 t_6^2 t_7^3 +  3 y_{s}^3 y_{o_1}^6 y_{o_2}^5 y_{o_3}^7 y_{o_4}^9 t_1^4 t_2^2 t_3^2 t_4^4 t_5^4 t_6^2 t_7^3 +  5 y_{s}^3 y_{o_1}^6 y_{o_2}^5 y_{o_3}^7 y_{o_4}^9 t_1^3 t_2^3 t_3^2 t_4^4 t_5^4 t_6^2 t_7^3 +  3 y_{s}^3 y_{o_1}^6 y_{o_2}^5 y_{o_3}^7 y_{o_4}^9 t_1^2 t_2^4 t_3^2 t_4^4 t_5^4 t_6^2 t_7^3 +  y_{s}^3 y_{o_1}^6 y_{o_2}^5 y_{o_3}^7 y_{o_4}^9 t_1 t_2^5 t_3^2 t_4^4 t_5^4 t_6^2 t_7^3 -  y_{s}^4 y_{o_1}^7 y_{o_2}^5 y_{o_3}^{10} y_{o_4}^{11} t_1^4 t_2^3 t_3 t_4^7 t_5^6 t_6^2 t_7^3 -  y_{s}^4 y_{o_1}^7 y_{o_2}^5 y_{o_3}^{10} y_{o_4}^{11} t_1^3 t_2^4 t_3 t_4^7 t_5^6 t_6^2 t_7^3 -  y_{s}^2 y_{o_1}^5 y_{o_2}^6 y_{o_3}^3 y_{o_4}^7 t_1^3 t_2^2 t_3^4 t_5 t_6^3 t_7^3 -  y_{s}^2 y_{o_1}^5 y_{o_2}^6 y_{o_3}^3 y_{o_4}^7 t_1^2 t_2^3 t_3^4 t_5 t_6^3 t_7^3 +  y_{s}^3 y_{o_1}^6 y_{o_2}^6 y_{o_3}^6 y_{o_4}^9 t_1^6 t_3^3 t_4^3 t_5^3 t_6^3 t_7^3 +  4 y_{s}^3 y_{o_1}^6 y_{o_2}^6 y_{o_3}^6 y_{o_4}^9 t_1^5 t_2 t_3^3 t_4^3 t_5^3 t_6^3 t_7^3 +  6 y_{s}^3 y_{o_1}^6 y_{o_2}^6 y_{o_3}^6 y_{o_4}^9 t_1^4 t_2^2 t_3^3 t_4^3 t_5^3 t_6^3 t_7^3 +  7 y_{s}^3 y_{o_1}^6 y_{o_2}^6 y_{o_3}^6 y_{o_4}^9 t_1^3 t_2^3 t_3^3 t_4^3 t_5^3 t_6^3 t_7^3 +  6 y_{s}^3 y_{o_1}^6 y_{o_2}^6 y_{o_3}^6 y_{o_4}^9 t_1^2 t_2^4 t_3^3 t_4^3 t_5^3 t_6^3 t_7^3 +  4 y_{s}^3 y_{o_1}^6 y_{o_2}^6 y_{o_3}^6 y_{o_4}^9 t_1 t_2^5 t_3^3 t_4^3 t_5^3 t_6^3 t_7^3 +  y_{s}^3 y_{o_1}^6 y_{o_2}^6 y_{o_3}^6 y_{o_4}^9 t_2^6 t_3^3 t_4^3 t_5^3 t_6^3 t_7^3 -  y_{s}^4 y_{o_1}^7 y_{o_2}^6 y_{o_3}^9 y_{o_4}^{11} t_1^5 t_2^2 t_3^2 t_4^6 t_5^5 t_6^3 t_7^3 -  4 y_{s}^4 y_{o_1}^7 y_{o_2}^6 y_{o_3}^9 y_{o_4}^{11} t_1^4 t_2^3 t_3^2 t_4^6 t_5^5 t_6^3 t_7^3 -  4 y_{s}^4 y_{o_1}^7 y_{o_2}^6 y_{o_3}^9 y_{o_4}^{11} t_1^3 t_2^4 t_3^2 t_4^6 t_5^5 t_6^3 t_7^3 -  y_{s}^4 y_{o_1}^7 y_{o_2}^6 y_{o_3}^9 y_{o_4}^{11} t_1^2 t_2^5 t_3^2 t_4^6 t_5^5 t_6^3 t_7^3 +  y_{s}^3 y_{o_1}^6 y_{o_2}^7 y_{o_3}^5 y_{o_4}^9 t_1^5 t_2 t_3^4 t_4^2 t_5^2 t_6^4 t_7^3 +  3 y_{s}^3 y_{o_1}^6 y_{o_2}^7 y_{o_3}^5 y_{o_4}^9 t_1^4 t_2^2 t_3^4 t_4^2 t_5^2 t_6^4 t_7^3 +  5 y_{s}^3 y_{o_1}^6 y_{o_2}^7 y_{o_3}^5 y_{o_4}^9 t_1^3 t_2^3 t_3^4 t_4^2 t_5^2 t_6^4 t_7^3 +  3 y_{s}^3 y_{o_1}^6 y_{o_2}^7 y_{o_3}^5 y_{o_4}^9 t_1^2 t_2^4 t_3^4 t_4^2 t_5^2 t_6^4 t_7^3 +  y_{s}^3 y_{o_1}^6 y_{o_2}^7 y_{o_3}^5 y_{o_4}^9 t_1 t_2^5 t_3^4 t_4^2 t_5^2 t_6^4 t_7^3 -  y_{s}^4 y_{o_1}^7 y_{o_2}^7 y_{o_3}^8 y_{o_4}^{11} t_1^6 t_2 t_3^3 t_4^5 t_5^4 t_6^4 t_7^3 -  3 y_{s}^4 y_{o_1}^7 y_{o_2}^7 y_{o_3}^8 y_{o_4}^{11} t_1^5 t_2^2 t_3^3 t_4^5 t_5^4 t_6^4 t_7^3 -  6 y_{s}^4 y_{o_1}^7 y_{o_2}^7 y_{o_3}^8 y_{o_4}^{11} t_1^4 t_2^3 t_3^3 t_4^5 t_5^4 t_6^4 t_7^3 -  6 y_{s}^4 y_{o_1}^7 y_{o_2}^7 y_{o_3}^8 y_{o_4}^{11} t_1^3 t_2^4 t_3^3 t_4^5 t_5^4 t_6^4 t_7^3 -  3 y_{s}^4 y_{o_1}^7 y_{o_2}^7 y_{o_3}^8 y_{o_4}^{11} t_1^2 t_2^5 t_3^3 t_4^5 t_5^4 t_6^4 t_7^3 -  y_{s}^4 y_{o_1}^7 y_{o_2}^7 y_{o_3}^8 y_{o_4}^{11} t_1 t_2^6 t_3^3 t_4^5 t_5^4 t_6^4 t_7^3 +  y_{s}^5 y_{o_1}^8 y_{o_2}^7 y_{o_3}^{11} y_{o_4}^{13} t_1^5 t_2^3 t_3^2 t_4^8 t_5^6 t_6^4 t_7^3 +  2 y_{s}^5 y_{o_1}^8 y_{o_2}^7 y_{o_3}^{11} y_{o_4}^{13} t_1^4 t_2^4 t_3^2 t_4^8 t_5^6 t_6^4 t_7^3 +  y_{s}^5 y_{o_1}^8 y_{o_2}^7 y_{o_3}^{11} y_{o_4}^{13} t_1^3 t_2^5 t_3^2 t_4^8 t_5^6 t_6^4 t_7^3 +  y_{s}^3 y_{o_1}^6 y_{o_2}^8 y_{o_3}^4 y_{o_4}^9 t_1^4 t_2^2 t_3^5 t_4 t_5 t_6^5 t_7^3 +  2 y_{s}^3 y_{o_1}^6 y_{o_2}^8 y_{o_3}^4 y_{o_4}^9 t_1^3 t_2^3 t_3^5 t_4 t_5 t_6^5 t_7^3 +  y_{s}^3 y_{o_1}^6 y_{o_2}^8 y_{o_3}^4 y_{o_4}^9 t_1^2 t_2^4 t_3^5 t_4 t_5 t_6^5 t_7^3 -  y_{s}^4 y_{o_1}^7 y_{o_2}^8 y_{o_3}^7 y_{o_4}^{11} t_1^6 t_2 t_3^4 t_4^4 t_5^3 t_6^5 t_7^3 -  3 y_{s}^4 y_{o_1}^7 y_{o_2}^8 y_{o_3}^7 y_{o_4}^{11} t_1^5 t_2^2 t_3^4 t_4^4 t_5^3 t_6^5 t_7^3 -  6 y_{s}^4 y_{o_1}^7 y_{o_2}^8 y_{o_3}^7 y_{o_4}^{11} t_1^4 t_2^3 t_3^4 t_4^4 t_5^3 t_6^5 t_7^3 -  6 y_{s}^4 y_{o_1}^7 y_{o_2}^8 y_{o_3}^7 y_{o_4}^{11} t_1^3 t_2^4 t_3^4 t_4^4 t_5^3 t_6^5 t_7^3 -  3 y_{s}^4 y_{o_1}^7 y_{o_2}^8 y_{o_3}^7 y_{o_4}^{11} t_1^2 t_2^5 t_3^4 t_4^4 t_5^3 t_6^5 t_7^3 -  y_{s}^4 y_{o_1}^7 y_{o_2}^8 y_{o_3}^7 y_{o_4}^{11} t_1 t_2^6 t_3^4 t_4^4 t_5^3 t_6^5 t_7^3 +  2 y_{s}^5 y_{o_1}^8 y_{o_2}^8 y_{o_3}^{10} y_{o_4}^{13} t_1^5 t_2^3 t_3^3 t_4^7 t_5^5 t_6^5 t_7^3 +  3 y_{s}^5 y_{o_1}^8 y_{o_2}^8 y_{o_3}^{10} y_{o_4}^{13} t_1^4 t_2^4 t_3^3 t_4^7 t_5^5 t_6^5 t_7^3 +  2 y_{s}^5 y_{o_1}^8 y_{o_2}^8 y_{o_3}^{10} y_{o_4}^{13} t_1^3 t_2^5 t_3^3 t_4^7 t_5^5 t_6^5 t_7^3 -  y_{s}^4 y_{o_1}^7 y_{o_2}^9 y_{o_3}^6 y_{o_4}^{11} t_1^5 t_2^2 t_3^5 t_4^3 t_5^2 t_6^6 t_7^3 -  4 y_{s}^4 y_{o_1}^7 y_{o_2}^9 y_{o_3}^6 y_{o_4}^{11} t_1^4 t_2^3 t_3^5 t_4^3 t_5^2 t_6^6 t_7^3 -  4 y_{s}^4 y_{o_1}^7 y_{o_2}^9 y_{o_3}^6 y_{o_4}^{11} t_1^3 t_2^4 t_3^5 t_4^3 t_5^2 t_6^6 t_7^3 -  y_{s}^4 y_{o_1}^7 y_{o_2}^9 y_{o_3}^6 y_{o_4}^{11} t_1^2 t_2^5 t_3^5 t_4^3 t_5^2 t_6^6 t_7^3 +  y_{s}^5 y_{o_1}^8 y_{o_2}^9 y_{o_3}^9 y_{o_4}^{13} t_1^6 t_2^2 t_3^4 t_4^6 t_5^4 t_6^6 t_7^3 +  2 y_{s}^5 y_{o_1}^8 y_{o_2}^9 y_{o_3}^9 y_{o_4}^{13} t_1^5 t_2^3 t_3^4 t_4^6 t_5^4 t_6^6 t_7^3 +  2 y_{s}^5 y_{o_1}^8 y_{o_2}^9 y_{o_3}^9 y_{o_4}^{13} t_1^4 t_2^4 t_3^4 t_4^6 t_5^4 t_6^6 t_7^3 +  2 y_{s}^5 y_{o_1}^8 y_{o_2}^9 y_{o_3}^9 y_{o_4}^{13} t_1^3 t_2^5 t_3^4 t_4^6 t_5^4 t_6^6 t_7^3 +  y_{s}^5 y_{o_1}^8 y_{o_2}^9 y_{o_3}^9 y_{o_4}^{13} t_1^2 t_2^6 t_3^4 t_4^6 t_5^4 t_6^6 t_7^3 -  y_{s}^6 y_{o_1}^9 y_{o_2}^9 y_{o_3}^{12} y_{o_4}^{15} t_1^5 t_2^4 t_3^3 t_4^9 t_5^6 t_6^6 t_7^3 -  y_{s}^6 y_{o_1}^9 y_{o_2}^9 y_{o_3}^{12} y_{o_4}^{15} t_1^4 t_2^5 t_3^3 t_4^9 t_5^6 t_6^6 t_7^3 -  y_{s}^4 y_{o_1}^7 y_{o_2}^{10} y_{o_3}^5 y_{o_4}^{11} t_1^4 t_2^3 t_3^6 t_4^2 t_5 t_6^7 t_7^3 -  y_{s}^4 y_{o_1}^7 y_{o_2}^{10} y_{o_3}^5 y_{o_4}^{11} t_1^3 t_2^4 t_3^6 t_4^2 t_5 t_6^7 t_7^3 +  2 y_{s}^5 y_{o_1}^8 y_{o_2}^{10} y_{o_3}^8 y_{o_4}^{13} t_1^5 t_2^3 t_3^5 t_4^5 t_5^3 t_6^7 t_7^3 +  3 y_{s}^5 y_{o_1}^8 y_{o_2}^{10} y_{o_3}^8 y_{o_4}^{13} t_1^4 t_2^4 t_3^5 t_4^5 t_5^3 t_6^7 t_7^3 +  2 y_{s}^5 y_{o_1}^8 y_{o_2}^{10} y_{o_3}^8 y_{o_4}^{13} t_1^3 t_2^5 t_3^5 t_4^5 t_5^3 t_6^7 t_7^3 +  y_{s}^5 y_{o_1}^8 y_{o_2}^{11} y_{o_3}^7 y_{o_4}^{13} t_1^5 t_2^3 t_3^6 t_4^4 t_5^2 t_6^8 t_7^3 +  2 y_{s}^5 y_{o_1}^8 y_{o_2}^{11} y_{o_3}^7 y_{o_4}^{13} t_1^4 t_2^4 t_3^6 t_4^4 t_5^2 t_6^8 t_7^3 +  y_{s}^5 y_{o_1}^8 y_{o_2}^{11} y_{o_3}^7 y_{o_4}^{13} t_1^3 t_2^5 t_3^6 t_4^4 t_5^2 t_6^8 t_7^3 -  y_{s}^6 y_{o_1}^9 y_{o_2}^{12} y_{o_3}^9 y_{o_4}^{15} t_1^5 t_2^4 t_3^6 t_4^6 t_5^3 t_6^9 t_7^3 -  y_{s}^6 y_{o_1}^9 y_{o_2}^{12} y_{o_3}^9 y_{o_4}^{15} t_1^4 t_2^5 t_3^6 t_4^6 t_5^3 t_6^9 t_7^3 -  y_{s}^2 y_{o_1}^6 y_{o_2}^4 y_{o_3}^6 y_{o_4}^8 t_1^5 t_2 t_3^2 t_4^2 t_5^4 t_7^4 -  y_{s}^2 y_{o_1}^6 y_{o_2}^4 y_{o_3}^6 y_{o_4}^8 t_1^4 t_2^2 t_3^2 t_4^2 t_5^4 t_7^4 -  y_{s}^2 y_{o_1}^6 y_{o_2}^4 y_{o_3}^6 y_{o_4}^8 t_1^3 t_2^3 t_3^2 t_4^2 t_5^4 t_7^4 -  y_{s}^2 y_{o_1}^6 y_{o_2}^4 y_{o_3}^6 y_{o_4}^8 t_1^2 t_2^4 t_3^2 t_4^2 t_5^4 t_7^4 -  y_{s}^2 y_{o_1}^6 y_{o_2}^4 y_{o_3}^6 y_{o_4}^8 t_1 t_2^5 t_3^2 t_4^2 t_5^4 t_7^4 +  y_{s}^2 y_{o_1}^6 y_{o_2}^5 y_{o_3}^5 y_{o_4}^8 t_1^3 t_2^3 t_3^3 t_4 t_5^3 t_6 t_7^4 +  y_{s}^3 y_{o_1}^7 y_{o_2}^5 y_{o_3}^8 y_{o_4}^{10} t_1^6 t_2 t_3^2 t_4^4 t_5^5 t_6 t_7^4 +  2 y_{s}^3 y_{o_1}^7 y_{o_2}^5 y_{o_3}^8 y_{o_4}^{10} t_1^5 t_2^2 t_3^2 t_4^4 t_5^5 t_6 t_7^4 +  2 y_{s}^3 y_{o_1}^7 y_{o_2}^5 y_{o_3}^8 y_{o_4}^{10} t_1^4 t_2^3 t_3^2 t_4^4 t_5^5 t_6 t_7^4 +  2 y_{s}^3 y_{o_1}^7 y_{o_2}^5 y_{o_3}^8 y_{o_4}^{10} t_1^3 t_2^4 t_3^2 t_4^4 t_5^5 t_6 t_7^4 +  2 y_{s}^3 y_{o_1}^7 y_{o_2}^5 y_{o_3}^8 y_{o_4}^{10} t_1^2 t_2^5 t_3^2 t_4^4 t_5^5 t_6 t_7^4 +  y_{s}^3 y_{o_1}^7 y_{o_2}^5 y_{o_3}^8 y_{o_4}^{10} t_1 t_2^6 t_3^2 t_4^4 t_5^5 t_6 t_7^4 -  y_{s}^2 y_{o_1}^6 y_{o_2}^6 y_{o_3}^4 y_{o_4}^8 t_1^5 t_2 t_3^4 t_5^2 t_6^2 t_7^4 -  y_{s}^2 y_{o_1}^6 y_{o_2}^6 y_{o_3}^4 y_{o_4}^8 t_1^4 t_2^2 t_3^4 t_5^2 t_6^2 t_7^4 -  y_{s}^2 y_{o_1}^6 y_{o_2}^6 y_{o_3}^4 y_{o_4}^8 t_1^3 t_2^3 t_3^4 t_5^2 t_6^2 t_7^4 -  y_{s}^2 y_{o_1}^6 y_{o_2}^6 y_{o_3}^4 y_{o_4}^8 t_1^2 t_2^4 t_3^4 t_5^2 t_6^2 t_7^4 -  y_{s}^2 y_{o_1}^6 y_{o_2}^6 y_{o_3}^4 y_{o_4}^8 t_1 t_2^5 t_3^4 t_5^2 t_6^2 t_7^4 +  y_{s}^3 y_{o_1}^7 y_{o_2}^6 y_{o_3}^7 y_{o_4}^{10} t_1^6 t_2 t_3^3 t_4^3 t_5^4 t_6^2 t_7^4 +  2 y_{s}^3 y_{o_1}^7 y_{o_2}^6 y_{o_3}^7 y_{o_4}^{10} t_1^5 t_2^2 t_3^3 t_4^3 t_5^4 t_6^2 t_7^4 +  y_{s}^3 y_{o_1}^7 y_{o_2}^6 y_{o_3}^7 y_{o_4}^{10} t_1^4 t_2^3 t_3^3 t_4^3 t_5^4 t_6^2 t_7^4 +  y_{s}^3 y_{o_1}^7 y_{o_2}^6 y_{o_3}^7 y_{o_4}^{10} t_1^3 t_2^4 t_3^3 t_4^3 t_5^4 t_6^2 t_7^4 +  2 y_{s}^3 y_{o_1}^7 y_{o_2}^6 y_{o_3}^7 y_{o_4}^{10} t_1^2 t_2^5 t_3^3 t_4^3 t_5^4 t_6^2 t_7^4 +  y_{s}^3 y_{o_1}^7 y_{o_2}^6 y_{o_3}^7 y_{o_4}^{10} t_1 t_2^6 t_3^3 t_4^3 t_5^4 t_6^2 t_7^4 -  y_{s}^4 y_{o_1}^8 y_{o_2}^6 y_{o_3}^{10} y_{o_4}^{12} t_1^5 t_2^3 t_3^2 t_4^6 t_5^6 t_6^2 t_7^4 -  y_{s}^4 y_{o_1}^8 y_{o_2}^6 y_{o_3}^{10} y_{o_4}^{12} t_1^4 t_2^4 t_3^2 t_4^6 t_5^6 t_6^2 t_7^4 -  y_{s}^4 y_{o_1}^8 y_{o_2}^6 y_{o_3}^{10} y_{o_4}^{12} t_1^3 t_2^5 t_3^2 t_4^6 t_5^6 t_6^2 t_7^4 +  y_{s}^3 y_{o_1}^7 y_{o_2}^7 y_{o_3}^6 y_{o_4}^{10} t_1^6 t_2 t_3^4 t_4^2 t_5^3 t_6^3 t_7^4 +  2 y_{s}^3 y_{o_1}^7 y_{o_2}^7 y_{o_3}^6 y_{o_4}^{10} t_1^5 t_2^2 t_3^4 t_4^2 t_5^3 t_6^3 t_7^4 +  y_{s}^3 y_{o_1}^7 y_{o_2}^7 y_{o_3}^6 y_{o_4}^{10} t_1^4 t_2^3 t_3^4 t_4^2 t_5^3 t_6^3 t_7^4 +  y_{s}^3 y_{o_1}^7 y_{o_2}^7 y_{o_3}^6 y_{o_4}^{10} t_1^3 t_2^4 t_3^4 t_4^2 t_5^3 t_6^3 t_7^4 +  2 y_{s}^3 y_{o_1}^7 y_{o_2}^7 y_{o_3}^6 y_{o_4}^{10} t_1^2 t_2^5 t_3^4 t_4^2 t_5^3 t_6^3 t_7^4 +  y_{s}^3 y_{o_1}^7 y_{o_2}^7 y_{o_3}^6 y_{o_4}^{10} t_1 t_2^6 t_3^4 t_4^2 t_5^3 t_6^3 t_7^4 -  y_{s}^4 y_{o_1}^8 y_{o_2}^7 y_{o_3}^9 y_{o_4}^{12} t_1^7 t_2 t_3^3 t_4^5 t_5^5 t_6^3 t_7^4 -  2 y_{s}^4 y_{o_1}^8 y_{o_2}^7 y_{o_3}^9 y_{o_4}^{12} t_1^6 t_2^2 t_3^3 t_4^5 t_5^5 t_6^3 t_7^4 -  4 y_{s}^4 y_{o_1}^8 y_{o_2}^7 y_{o_3}^9 y_{o_4}^{12} t_1^5 t_2^3 t_3^3 t_4^5 t_5^5 t_6^3 t_7^4 -  3 y_{s}^4 y_{o_1}^8 y_{o_2}^7 y_{o_3}^9 y_{o_4}^{12} t_1^4 t_2^4 t_3^3 t_4^5 t_5^5 t_6^3 t_7^4 -  4 y_{s}^4 y_{o_1}^8 y_{o_2}^7 y_{o_3}^9 y_{o_4}^{12} t_1^3 t_2^5 t_3^3 t_4^5 t_5^5 t_6^3 t_7^4 -  2 y_{s}^4 y_{o_1}^8 y_{o_2}^7 y_{o_3}^9 y_{o_4}^{12} t_1^2 t_2^6 t_3^3 t_4^5 t_5^5 t_6^3 t_7^4 -  y_{s}^4 y_{o_1}^8 y_{o_2}^7 y_{o_3}^9 y_{o_4}^{12} t_1 t_2^7 t_3^3 t_4^5 t_5^5 t_6^3 t_7^4 +  y_{s}^3 y_{o_1}^7 y_{o_2}^8 y_{o_3}^5 y_{o_4}^{10} t_1^6 t_2 t_3^5 t_4 t_5^2 t_6^4 t_7^4 +  2 y_{s}^3 y_{o_1}^7 y_{o_2}^8 y_{o_3}^5 y_{o_4}^{10} t_1^5 t_2^2 t_3^5 t_4 t_5^2 t_6^4 t_7^4 +  2 y_{s}^3 y_{o_1}^7 y_{o_2}^8 y_{o_3}^5 y_{o_4}^{10} t_1^4 t_2^3 t_3^5 t_4 t_5^2 t_6^4 t_7^4 +  2 y_{s}^3 y_{o_1}^7 y_{o_2}^8 y_{o_3}^5 y_{o_4}^{10} t_1^3 t_2^4 t_3^5 t_4 t_5^2 t_6^4 t_7^4 +  2 y_{s}^3 y_{o_1}^7 y_{o_2}^8 y_{o_3}^5 y_{o_4}^{10} t_1^2 t_2^5 t_3^5 t_4 t_5^2 t_6^4 t_7^4 +  y_{s}^3 y_{o_1}^7 y_{o_2}^8 y_{o_3}^5 y_{o_4}^{10} t_1 t_2^6 t_3^5 t_4 t_5^2 t_6^4 t_7^4 +  y_{s}^4 y_{o_1}^8 y_{o_2}^8 y_{o_3}^8 y_{o_4}^{12} t_1^8 t_3^4 t_4^4 t_5^4 t_6^4 t_7^4 +  y_{s}^4 y_{o_1}^8 y_{o_2}^8 y_{o_3}^8 y_{o_4}^{12} t_1^7 t_2 t_3^4 t_4^4 t_5^4 t_6^4 t_7^4 +  y_{s}^4 y_{o_1}^8 y_{o_2}^8 y_{o_3}^8 y_{o_4}^{12} t_1^6 t_2^2 t_3^4 t_4^4 t_5^4 t_6^4 t_7^4 +  2 y_{s}^4 y_{o_1}^8 y_{o_2}^8 y_{o_3}^8 y_{o_4}^{12} t_1^4 t_2^4 t_3^4 t_4^4 t_5^4 t_6^4 t_7^4 +  y_{s}^4 y_{o_1}^8 y_{o_2}^8 y_{o_3}^8 y_{o_4}^{12} t_1^2 t_2^6 t_3^4 t_4^4 t_5^4 t_6^4 t_7^4 +  y_{s}^4 y_{o_1}^8 y_{o_2}^8 y_{o_3}^8 y_{o_4}^{12} t_1 t_2^7 t_3^4 t_4^4 t_5^4 t_6^4 t_7^4 +  y_{s}^4 y_{o_1}^8 y_{o_2}^8 y_{o_3}^8 y_{o_4}^{12} t_2^8 t_3^4 t_4^4 t_5^4 t_6^4 t_7^4 +  y_{s}^5 y_{o_1}^9 y_{o_2}^8 y_{o_3}^{11} y_{o_4}^{14} t_1^6 t_2^3 t_3^3 t_4^7 t_5^6 t_6^4 t_7^4 +  2 y_{s}^5 y_{o_1}^9 y_{o_2}^8 y_{o_3}^{11} y_{o_4}^{14} t_1^5 t_2^4 t_3^3 t_4^7 t_5^6 t_6^4 t_7^4 +  2 y_{s}^5 y_{o_1}^9 y_{o_2}^8 y_{o_3}^{11} y_{o_4}^{14} t_1^4 t_2^5 t_3^3 t_4^7 t_5^6 t_6^4 t_7^4 +  y_{s}^5 y_{o_1}^9 y_{o_2}^8 y_{o_3}^{11} y_{o_4}^{14} t_1^3 t_2^6 t_3^3 t_4^7 t_5^6 t_6^4 t_7^4 -  y_{s}^4 y_{o_1}^8 y_{o_2}^9 y_{o_3}^7 y_{o_4}^{12} t_1^7 t_2 t_3^5 t_4^3 t_5^3 t_6^5 t_7^4 -  2 y_{s}^4 y_{o_1}^8 y_{o_2}^9 y_{o_3}^7 y_{o_4}^{12} t_1^6 t_2^2 t_3^5 t_4^3 t_5^3 t_6^5 t_7^4 -  4 y_{s}^4 y_{o_1}^8 y_{o_2}^9 y_{o_3}^7 y_{o_4}^{12} t_1^5 t_2^3 t_3^5 t_4^3 t_5^3 t_6^5 t_7^4 -  3 y_{s}^4 y_{o_1}^8 y_{o_2}^9 y_{o_3}^7 y_{o_4}^{12} t_1^4 t_2^4 t_3^5 t_4^3 t_5^3 t_6^5 t_7^4 -  4 y_{s}^4 y_{o_1}^8 y_{o_2}^9 y_{o_3}^7 y_{o_4}^{12} t_1^3 t_2^5 t_3^5 t_4^3 t_5^3 t_6^5 t_7^4 -  2 y_{s}^4 y_{o_1}^8 y_{o_2}^9 y_{o_3}^7 y_{o_4}^{12} t_1^2 t_2^6 t_3^5 t_4^3 t_5^3 t_6^5 t_7^4 -  y_{s}^4 y_{o_1}^8 y_{o_2}^9 y_{o_3}^7 y_{o_4}^{12} t_1 t_2^7 t_3^5 t_4^3 t_5^3 t_6^5 t_7^4 -  y_{s}^5 y_{o_1}^9 y_{o_2}^9 y_{o_3}^{10} y_{o_4}^{14} t_1^8 t_2 t_3^4 t_4^6 t_5^5 t_6^5 t_7^4 -  y_{s}^5 y_{o_1}^9 y_{o_2}^9 y_{o_3}^{10} y_{o_4}^{14} t_1^7 t_2^2 t_3^4 t_4^6 t_5^5 t_6^5 t_7^4 -  y_{s}^5 y_{o_1}^9 y_{o_2}^9 y_{o_3}^{10} y_{o_4}^{14} t_1^2 t_2^7 t_3^4 t_4^6 t_5^5 t_6^5 t_7^4 -  y_{s}^5 y_{o_1}^9 y_{o_2}^9 y_{o_3}^{10} y_{o_4}^{14} t_1 t_2^8 t_3^4 t_4^6 t_5^5 t_6^5 t_7^4 -  y_{s}^4 y_{o_1}^8 y_{o_2}^{10} y_{o_3}^6 y_{o_4}^{12} t_1^5 t_2^3 t_3^6 t_4^2 t_5^2 t_6^6 t_7^4 -  y_{s}^4 y_{o_1}^8 y_{o_2}^{10} y_{o_3}^6 y_{o_4}^{12} t_1^4 t_2^4 t_3^6 t_4^2 t_5^2 t_6^6 t_7^4 -  y_{s}^4 y_{o_1}^8 y_{o_2}^{10} y_{o_3}^6 y_{o_4}^{12} t_1^3 t_2^5 t_3^6 t_4^2 t_5^2 t_6^6 t_7^4 -  y_{s}^5 y_{o_1}^9 y_{o_2}^{10} y_{o_3}^9 y_{o_4}^{14} t_1^8 t_2 t_3^5 t_4^5 t_5^4 t_6^6 t_7^4 -  y_{s}^5 y_{o_1}^9 y_{o_2}^{10} y_{o_3}^9 y_{o_4}^{14} t_1^7 t_2^2 t_3^5 t_4^5 t_5^4 t_6^6 t_7^4 -  y_{s}^5 y_{o_1}^9 y_{o_2}^{10} y_{o_3}^9 y_{o_4}^{14} t_1^2 t_2^7 t_3^5 t_4^5 t_5^4 t_6^6 t_7^4 -  y_{s}^5 y_{o_1}^9 y_{o_2}^{10} y_{o_3}^9 y_{o_4}^{14} t_1 t_2^8 t_3^5 t_4^5 t_5^4 t_6^6 t_7^4 -  y_{s}^6 y_{o_1}^{10} y_{o_2}^{10} y_{o_3}^{12} y_{o_4}^{16} t_1^6 t_2^4 t_3^4 t_4^8 t_5^6 t_6^6 t_7^4 -  y_{s}^6 y_{o_1}^{10} y_{o_2}^{10} y_{o_3}^{12} y_{o_4}^{16} t_1^5 t_2^5 t_3^4 t_4^8 t_5^6 t_6^6 t_7^4 -  y_{s}^6 y_{o_1}^{10} y_{o_2}^{10} y_{o_3}^{12} y_{o_4}^{16} t_1^4 t_2^6 t_3^4 t_4^8 t_5^6 t_6^6 t_7^4 +  y_{s}^5 y_{o_1}^9 y_{o_2}^{11} y_{o_3}^8 y_{o_4}^{14} t_1^6 t_2^3 t_3^6 t_4^4 t_5^3 t_6^7 t_7^4 +  2 y_{s}^5 y_{o_1}^9 y_{o_2}^{11} y_{o_3}^8 y_{o_4}^{14} t_1^5 t_2^4 t_3^6 t_4^4 t_5^3 t_6^7 t_7^4 +  2 y_{s}^5 y_{o_1}^9 y_{o_2}^{11} y_{o_3}^8 y_{o_4}^{14} t_1^4 t_2^5 t_3^6 t_4^4 t_5^3 t_6^7 t_7^4 +  y_{s}^5 y_{o_1}^9 y_{o_2}^{11} y_{o_3}^8 y_{o_4}^{14} t_1^3 t_2^6 t_3^6 t_4^4 t_5^3 t_6^7 t_7^4 +  y_{s}^6 y_{o_1}^{10} y_{o_2}^{11} y_{o_3}^{11} y_{o_4}^{16} t_1^8 t_2^2 t_3^5 t_4^7 t_5^5 t_6^7 t_7^4 +  y_{s}^6 y_{o_1}^{10} y_{o_2}^{11} y_{o_3}^{11} y_{o_4}^{16} t_1^7 t_2^3 t_3^5 t_4^7 t_5^5 t_6^7 t_7^4 +  y_{s}^6 y_{o_1}^{10} y_{o_2}^{11} y_{o_3}^{11} y_{o_4}^{16} t_1^6 t_2^4 t_3^5 t_4^7 t_5^5 t_6^7 t_7^4 +  y_{s}^6 y_{o_1}^{10} y_{o_2}^{11} y_{o_3}^{11} y_{o_4}^{16} t_1^5 t_2^5 t_3^5 t_4^7 t_5^5 t_6^7 t_7^4 +  y_{s}^6 y_{o_1}^{10} y_{o_2}^{11} y_{o_3}^{11} y_{o_4}^{16} t_1^4 t_2^6 t_3^5 t_4^7 t_5^5 t_6^7 t_7^4 +  y_{s}^6 y_{o_1}^{10} y_{o_2}^{11} y_{o_3}^{11} y_{o_4}^{16} t_1^3 t_2^7 t_3^5 t_4^7 t_5^5 t_6^7 t_7^4 +  y_{s}^6 y_{o_1}^{10} y_{o_2}^{11} y_{o_3}^{11} y_{o_4}^{16} t_1^2 t_2^8 t_3^5 t_4^7 t_5^5 t_6^7 t_7^4 -  y_{s}^6 y_{o_1}^{10} y_{o_2}^{12} y_{o_3}^{10} y_{o_4}^{16} t_1^6 t_2^4 t_3^6 t_4^6 t_5^4 t_6^8 t_7^4 -  y_{s}^6 y_{o_1}^{10} y_{o_2}^{12} y_{o_3}^{10} y_{o_4}^{16} t_1^5 t_2^5 t_3^6 t_4^6 t_5^4 t_6^8 t_7^4 -  y_{s}^6 y_{o_1}^{10} y_{o_2}^{12} y_{o_3}^{10} y_{o_4}^{16} t_1^4 t_2^6 t_3^6 t_4^6 t_5^4 t_6^8 t_7^4 -  y_{s}^3 y_{o_1}^8 y_{o_2}^6 y_{o_3}^8 y_{o_4}^{11} t_1^5 t_2^3 t_3^3 t_4^3 t_5^5 t_6 t_7^5 -  y_{s}^3 y_{o_1}^8 y_{o_2}^6 y_{o_3}^8 y_{o_4}^{11} t_1^4 t_2^4 t_3^3 t_4^3 t_5^5 t_6 t_7^5 -  y_{s}^3 y_{o_1}^8 y_{o_2}^6 y_{o_3}^8 y_{o_4}^{11} t_1^3 t_2^5 t_3^3 t_4^3 t_5^5 t_6 t_7^5 +  y_{s}^3 y_{o_1}^8 y_{o_2}^7 y_{o_3}^7 y_{o_4}^{11} t_1^7 t_2 t_3^4 t_4^2 t_5^4 t_6^2 t_7^5 +  y_{s}^3 y_{o_1}^8 y_{o_2}^7 y_{o_3}^7 y_{o_4}^{11} t_1^6 t_2^2 t_3^4 t_4^2 t_5^4 t_6^2 t_7^5 +  y_{s}^3 y_{o_1}^8 y_{o_2}^7 y_{o_3}^7 y_{o_4}^{11} t_1^5 t_2^3 t_3^4 t_4^2 t_5^4 t_6^2 t_7^5 +  y_{s}^3 y_{o_1}^8 y_{o_2}^7 y_{o_3}^7 y_{o_4}^{11} t_1^4 t_2^4 t_3^4 t_4^2 t_5^4 t_6^2 t_7^5 +  y_{s}^3 y_{o_1}^8 y_{o_2}^7 y_{o_3}^7 y_{o_4}^{11} t_1^3 t_2^5 t_3^4 t_4^2 t_5^4 t_6^2 t_7^5 +  y_{s}^3 y_{o_1}^8 y_{o_2}^7 y_{o_3}^7 y_{o_4}^{11} t_1^2 t_2^6 t_3^4 t_4^2 t_5^4 t_6^2 t_7^5 +  y_{s}^3 y_{o_1}^8 y_{o_2}^7 y_{o_3}^7 y_{o_4}^{11} t_1 t_2^7 t_3^4 t_4^2 t_5^4 t_6^2 t_7^5 +  y_{s}^4 y_{o_1}^9 y_{o_2}^7 y_{o_3}^{10} y_{o_4}^{13} t_1^6 t_2^3 t_3^3 t_4^5 t_5^6 t_6^2 t_7^5 +  2 y_{s}^4 y_{o_1}^9 y_{o_2}^7 y_{o_3}^{10} y_{o_4}^{13} t_1^5 t_2^4 t_3^3 t_4^5 t_5^6 t_6^2 t_7^5 +  2 y_{s}^4 y_{o_1}^9 y_{o_2}^7 y_{o_3}^{10} y_{o_4}^{13} t_1^4 t_2^5 t_3^3 t_4^5 t_5^6 t_6^2 t_7^5 +  y_{s}^4 y_{o_1}^9 y_{o_2}^7 y_{o_3}^{10} y_{o_4}^{13} t_1^3 t_2^6 t_3^3 t_4^5 t_5^6 t_6^2 t_7^5 -  y_{s}^3 y_{o_1}^8 y_{o_2}^8 y_{o_3}^6 y_{o_4}^{11} t_1^5 t_2^3 t_3^5 t_4 t_5^3 t_6^3 t_7^5 -  y_{s}^3 y_{o_1}^8 y_{o_2}^8 y_{o_3}^6 y_{o_4}^{11} t_1^4 t_2^4 t_3^5 t_4 t_5^3 t_6^3 t_7^5 -  y_{s}^3 y_{o_1}^8 y_{o_2}^8 y_{o_3}^6 y_{o_4}^{11} t_1^3 t_2^5 t_3^5 t_4 t_5^3 t_6^3 t_7^5 -  y_{s}^4 y_{o_1}^9 y_{o_2}^8 y_{o_3}^9 y_{o_4}^{13} t_1^8 t_2 t_3^4 t_4^4 t_5^5 t_6^3 t_7^5 -  y_{s}^4 y_{o_1}^9 y_{o_2}^8 y_{o_3}^9 y_{o_4}^{13} t_1^7 t_2^2 t_3^4 t_4^4 t_5^5 t_6^3 t_7^5 -  y_{s}^4 y_{o_1}^9 y_{o_2}^8 y_{o_3}^9 y_{o_4}^{13} t_1^2 t_2^7 t_3^4 t_4^4 t_5^5 t_6^3 t_7^5 -  y_{s}^4 y_{o_1}^9 y_{o_2}^8 y_{o_3}^9 y_{o_4}^{13} t_1 t_2^8 t_3^4 t_4^4 t_5^5 t_6^3 t_7^5 -  y_{s}^5 y_{o_1}^{10} y_{o_2}^8 y_{o_3}^{12} y_{o_4}^{15} t_1^6 t_2^4 t_3^3 t_4^7 t_5^7 t_6^3 t_7^5 -  y_{s}^5 y_{o_1}^{10} y_{o_2}^8 y_{o_3}^{12} y_{o_4}^{15} t_1^5 t_2^5 t_3^3 t_4^7 t_5^7 t_6^3 t_7^5 -  y_{s}^5 y_{o_1}^{10} y_{o_2}^8 y_{o_3}^{12} y_{o_4}^{15} t_1^4 t_2^6 t_3^3 t_4^7 t_5^7 t_6^3 t_7^5 -  y_{s}^4 y_{o_1}^9 y_{o_2}^9 y_{o_3}^8 y_{o_4}^{13} t_1^8 t_2 t_3^5 t_4^3 t_5^4 t_6^4 t_7^5 -  y_{s}^4 y_{o_1}^9 y_{o_2}^9 y_{o_3}^8 y_{o_4}^{13} t_1^7 t_2^2 t_3^5 t_4^3 t_5^4 t_6^4 t_7^5 -  y_{s}^4 y_{o_1}^9 y_{o_2}^9 y_{o_3}^8 y_{o_4}^{13} t_1^2 t_2^7 t_3^5 t_4^3 t_5^4 t_6^4 t_7^5 -  y_{s}^4 y_{o_1}^9 y_{o_2}^9 y_{o_3}^8 y_{o_4}^{13} t_1 t_2^8 t_3^5 t_4^3 t_5^4 t_6^4 t_7^5 -  y_{s}^5 y_{o_1}^{10} y_{o_2}^9 y_{o_3}^{11} y_{o_4}^{15} t_1^8 t_2^2 t_3^4 t_4^6 t_5^6 t_6^4 t_7^5 -  2 y_{s}^5 y_{o_1}^{10} y_{o_2}^9 y_{o_3}^{11} y_{o_4}^{15} t_1^7 t_2^3 t_3^4 t_4^6 t_5^6 t_6^4 t_7^5 - 4 y_{s}^5 y_{o_1}^{10} y_{o_2}^9 y_{o_3}^{11} y_{o_4}^{15} t_1^6 t_2^4 t_3^4 t_4^6 t_5^6 t_6^4 t_7^5 - 3 y_{s}^5 y_{o_1}^{10} y_{o_2}^9 y_{o_3}^{11} y_{o_4}^{15} t_1^5 t_2^5 t_3^4 t_4^6 t_5^6 t_6^4 t_7^5 - 4 y_{s}^5 y_{o_1}^{10} y_{o_2}^9 y_{o_3}^{11} y_{o_4}^{15} t_1^4 t_2^6 t_3^4 t_4^6 t_5^6 t_6^4 t_7^5 - 2 y_{s}^5 y_{o_1}^{10} y_{o_2}^9 y_{o_3}^{11} y_{o_4}^{15} t_1^3 t_2^7 t_3^4 t_4^6 t_5^6 t_6^4 t_7^5 - y_{s}^5 y_{o_1}^{10} y_{o_2}^9 y_{o_3}^{11} y_{o_4}^{15} t_1^2 t_2^8 t_3^4 t_4^6 t_5^6 t_6^4 t_7^5 +  y_{s}^4 y_{o_1}^9 y_{o_2}^{10} y_{o_3}^7 y_{o_4}^{13} t_1^6 t_2^3 t_3^6 t_4^2 t_5^3 t_6^5 t_7^5 +  2 y_{s}^4 y_{o_1}^9 y_{o_2}^{10} y_{o_3}^7 y_{o_4}^{13} t_1^5 t_2^4 t_3^6 t_4^2 t_5^3 t_6^5 t_7^5 +  2 y_{s}^4 y_{o_1}^9 y_{o_2}^{10} y_{o_3}^7 y_{o_4}^{13} t_1^4 t_2^5 t_3^6 t_4^2 t_5^3 t_6^5 t_7^5 +  y_{s}^4 y_{o_1}^9 y_{o_2}^{10} y_{o_3}^7 y_{o_4}^{13} t_1^3 t_2^6 t_3^6 t_4^2 t_5^3 t_6^5 t_7^5 +  y_{s}^5 y_{o_1}^{10} y_{o_2}^{10} y_{o_3}^{10} y_{o_4}^{15} t_1^9 t_2 t_3^5 t_4^5 t_5^5 t_6^5 t_7^5 +  y_{s}^5 y_{o_1}^{10} y_{o_2}^{10} y_{o_3}^{10} y_{o_4}^{15} t_1^8 t_2^2 t_3^5 t_4^5 t_5^5 t_6^5 t_7^5 +  y_{s}^5 y_{o_1}^{10} y_{o_2}^{10} y_{o_3}^{10} y_{o_4}^{15} t_1^7 t_2^3 t_3^5 t_4^5 t_5^5 t_6^5 t_7^5 +  2 y_{s}^5 y_{o_1}^{10} y_{o_2}^{10} y_{o_3}^{10} y_{o_4}^{15} t_1^5 t_2^5 t_3^5 t_4^5 t_5^5 t_6^5 t_7^5 + y_{s}^5 y_{o_1}^{10} y_{o_2}^{10} y_{o_3}^{10} y_{o_4}^{15} t_1^3 t_2^7 t_3^5 t_4^5 t_5^5 t_6^5 t_7^5 +  y_{s}^5 y_{o_1}^{10} y_{o_2}^{10} y_{o_3}^{10} y_{o_4}^{15} t_1^2 t_2^8 t_3^5 t_4^5 t_5^5 t_6^5 t_7^5 +  y_{s}^5 y_{o_1}^{10} y_{o_2}^{10} y_{o_3}^{10} y_{o_4}^{15} t_1 t_2^9 t_3^5 t_4^5 t_5^5 t_6^5 t_7^5 +  y_{s}^6 y_{o_1}^{11} y_{o_2}^{10} y_{o_3}^{13} y_{o_4}^{17} t_1^8 t_2^3 t_3^4 t_4^8 t_5^7 t_6^5 t_7^5 +  2 y_{s}^6 y_{o_1}^{11} y_{o_2}^{10} y_{o_3}^{13} y_{o_4}^{17} t_1^7 t_2^4 t_3^4 t_4^8 t_5^7 t_6^5 t_7^5 + 2 y_{s}^6 y_{o_1}^{11} y_{o_2}^{10} y_{o_3}^{13} y_{o_4}^{17} t_1^6 t_2^5 t_3^4 t_4^8 t_5^7 t_6^5 t_7^5 + 2 y_{s}^6 y_{o_1}^{11} y_{o_2}^{10} y_{o_3}^{13} y_{o_4}^{17} t_1^5 t_2^6 t_3^4 t_4^8 t_5^7 t_6^5 t_7^5 +  2 y_{s}^6 y_{o_1}^{11} y_{o_2}^{10} y_{o_3}^{13} y_{o_4}^{17} t_1^4 t_2^7 t_3^4 t_4^8 t_5^7 t_6^5 t_7^5 + y_{s}^6 y_{o_1}^{11} y_{o_2}^{10} y_{o_3}^{13} y_{o_4}^{17} t_1^3 t_2^8 t_3^4 t_4^8 t_5^7 t_6^5 t_7^5 -  y_{s}^5 y_{o_1}^{10} y_{o_2}^{11} y_{o_3}^9 y_{o_4}^{15} t_1^8 t_2^2 t_3^6 t_4^4 t_5^4 t_6^6 t_7^5 -  2 y_{s}^5 y_{o_1}^{10} y_{o_2}^{11} y_{o_3}^9 y_{o_4}^{15} t_1^7 t_2^3 t_3^6 t_4^4 t_5^4 t_6^6 t_7^5 - 4 y_{s}^5 y_{o_1}^{10} y_{o_2}^{11} y_{o_3}^9 y_{o_4}^{15} t_1^6 t_2^4 t_3^6 t_4^4 t_5^4 t_6^6 t_7^5 - 3 y_{s}^5 y_{o_1}^{10} y_{o_2}^{11} y_{o_3}^9 y_{o_4}^{15} t_1^5 t_2^5 t_3^6 t_4^4 t_5^4 t_6^6 t_7^5 - 4 y_{s}^5 y_{o_1}^{10} y_{o_2}^{11} y_{o_3}^9 y_{o_4}^{15} t_1^4 t_2^6 t_3^6 t_4^4 t_5^4 t_6^6 t_7^5 - 2 y_{s}^5 y_{o_1}^{10} y_{o_2}^{11} y_{o_3}^9 y_{o_4}^{15} t_1^3 t_2^7 t_3^6 t_4^4 t_5^4 t_6^6 t_7^5 - y_{s}^5 y_{o_1}^{10} y_{o_2}^{11} y_{o_3}^9 y_{o_4}^{15} t_1^2 t_2^8 t_3^6 t_4^4 t_5^4 t_6^6 t_7^5 +  y_{s}^6 y_{o_1}^{11} y_{o_2}^{11} y_{o_3}^{12} y_{o_4}^{17} t_1^8 t_2^3 t_3^5 t_4^7 t_5^6 t_6^6 t_7^5 +  2 y_{s}^6 y_{o_1}^{11} y_{o_2}^{11} y_{o_3}^{12} y_{o_4}^{17} t_1^7 t_2^4 t_3^5 t_4^7 t_5^6 t_6^6 t_7^5 + y_{s}^6 y_{o_1}^{11} y_{o_2}^{11} y_{o_3}^{12} y_{o_4}^{17} t_1^6 t_2^5 t_3^5 t_4^7 t_5^6 t_6^6 t_7^5 +  y_{s}^6 y_{o_1}^{11} y_{o_2}^{11} y_{o_3}^{12} y_{o_4}^{17} t_1^5 t_2^6 t_3^5 t_4^7 t_5^6 t_6^6 t_7^5 + 2 y_{s}^6 y_{o_1}^{11} y_{o_2}^{11} y_{o_3}^{12} y_{o_4}^{17} t_1^4 t_2^7 t_3^5 t_4^7 t_5^6 t_6^6 t_7^5 + y_{s}^6 y_{o_1}^{11} y_{o_2}^{11} y_{o_3}^{12} y_{o_4}^{17} t_1^3 t_2^8 t_3^5 t_4^7 t_5^6 t_6^6 t_7^5 - y_{s}^5 y_{o_1}^{10} y_{o_2}^{12} y_{o_3}^8 y_{o_4}^{15} t_1^6 t_2^4 t_3^7 t_4^3 t_5^3 t_6^7 t_7^5 - y_{s}^5 y_{o_1}^{10} y_{o_2}^{12} y_{o_3}^8 y_{o_4}^{15} t_1^5 t_2^5 t_3^7 t_4^3 t_5^3 t_6^7 t_7^5 - y_{s}^5 y_{o_1}^{10} y_{o_2}^{12} y_{o_3}^8 y_{o_4}^{15} t_1^4 t_2^6 t_3^7 t_4^3 t_5^3 t_6^7 t_7^5 + y_{s}^6 y_{o_1}^{11} y_{o_2}^{12} y_{o_3}^{11} y_{o_4}^{17} t_1^8 t_2^3 t_3^6 t_4^6 t_5^5 t_6^7 t_7^5 + 2 y_{s}^6 y_{o_1}^{11} y_{o_2}^{12} y_{o_3}^{11} y_{o_4}^{17} t_1^7 t_2^4 t_3^6 t_4^6 t_5^5 t_6^7 t_7^5 +  y_{s}^6 y_{o_1}^{11} y_{o_2}^{12} y_{o_3}^{11} y_{o_4}^{17} t_1^6 t_2^5 t_3^6 t_4^6 t_5^5 t_6^7 t_7^5 +  y_{s}^6 y_{o_1}^{11} y_{o_2}^{12} y_{o_3}^{11} y_{o_4}^{17} t_1^5 t_2^6 t_3^6 t_4^6 t_5^5 t_6^7 t_7^5 +  2 y_{s}^6 y_{o_1}^{11} y_{o_2}^{12} y_{o_3}^{11} y_{o_4}^{17} t_1^4 t_2^7 t_3^6 t_4^6 t_5^5 t_6^7 t_7^5 + y_{s}^6 y_{o_1}^{11} y_{o_2}^{12} y_{o_3}^{11} y_{o_4}^{17} t_1^3 t_2^8 t_3^6 t_4^6 t_5^5 t_6^7 t_7^5 -  y_{s}^7 y_{o_1}^{12} y_{o_2}^{12} y_{o_3}^{14} y_{o_4}^{19} t_1^8 t_2^4 t_3^5 t_4^9 t_5^7 t_6^7 t_7^5 -  y_{s}^7 y_{o_1}^{12} y_{o_2}^{12} y_{o_3}^{14} y_{o_4}^{19} t_1^7 t_2^5 t_3^5 t_4^9 t_5^7 t_6^7 t_7^5 -  y_{s}^7 y_{o_1}^{12} y_{o_2}^{12} y_{o_3}^{14} y_{o_4}^{19} t_1^6 t_2^6 t_3^5 t_4^9 t_5^7 t_6^7 t_7^5 -  y_{s}^7 y_{o_1}^{12} y_{o_2}^{12} y_{o_3}^{14} y_{o_4}^{19} t_1^5 t_2^7 t_3^5 t_4^9 t_5^7 t_6^7 t_7^5 -  y_{s}^7 y_{o_1}^{12} y_{o_2}^{12} y_{o_3}^{14} y_{o_4}^{19} t_1^4 t_2^8 t_3^5 t_4^9 t_5^7 t_6^7 t_7^5 +  y_{s}^6 y_{o_1}^{11} y_{o_2}^{13} y_{o_3}^{10} y_{o_4}^{17} t_1^8 t_2^3 t_3^7 t_4^5 t_5^4 t_6^8 t_7^5 + 2 y_{s}^6 y_{o_1}^{11} y_{o_2}^{13} y_{o_3}^{10} y_{o_4}^{17} t_1^7 t_2^4 t_3^7 t_4^5 t_5^4 t_6^8 t_7^5 + 2 y_{s}^6 y_{o_1}^{11} y_{o_2}^{13} y_{o_3}^{10} y_{o_4}^{17} t_1^6 t_2^5 t_3^7 t_4^5 t_5^4 t_6^8 t_7^5 +  2 y_{s}^6 y_{o_1}^{11} y_{o_2}^{13} y_{o_3}^{10} y_{o_4}^{17} t_1^5 t_2^6 t_3^7 t_4^5 t_5^4 t_6^8 t_7^5 + 2 y_{s}^6 y_{o_1}^{11} y_{o_2}^{13} y_{o_3}^{10} y_{o_4}^{17} t_1^4 t_2^7 t_3^7 t_4^5 t_5^4 t_6^8 t_7^5 + y_{s}^6 y_{o_1}^{11} y_{o_2}^{13} y_{o_3}^{10} y_{o_4}^{17} t_1^3 t_2^8 t_3^7 t_4^5 t_5^4 t_6^8 t_7^5 + y_{s}^7 y_{o_1}^{12} y_{o_2}^{13} y_{o_3}^{13} y_{o_4}^{19} t_1^6 t_2^6 t_3^6 t_4^8 t_5^6 t_6^8 t_7^5 - y_{s}^7 y_{o_1}^{12} y_{o_2}^{14} y_{o_3}^{12} y_{o_4}^{19} t_1^8 t_2^4 t_3^7 t_4^7 t_5^5 t_6^9 t_7^5 - y_{s}^7 y_{o_1}^{12} y_{o_2}^{14} y_{o_3}^{12} y_{o_4}^{19} t_1^7 t_2^5 t_3^7 t_4^7 t_5^5 t_6^9 t_7^5 - y_{s}^7 y_{o_1}^{12} y_{o_2}^{14} y_{o_3}^{12} y_{o_4}^{19} t_1^6 t_2^6 t_3^7 t_4^7 t_5^5 t_6^9 t_7^5 - y_{s}^7 y_{o_1}^{12} y_{o_2}^{14} y_{o_3}^{12} y_{o_4}^{19} t_1^5 t_2^7 t_3^7 t_4^7 t_5^5 t_6^9 t_7^5 - y_{s}^7 y_{o_1}^{12} y_{o_2}^{14} y_{o_3}^{12} y_{o_4}^{19} t_1^4 t_2^8 t_3^7 t_4^7 t_5^5 t_6^9 t_7^5 - y_{s}^3 y_{o_1}^9 y_{o_2}^6 y_{o_3}^9 y_{o_4}^{12} t_1^5 t_2^4 t_3^3 t_4^3 t_5^6 t_7^6 -  y_{s}^3 y_{o_1}^9 y_{o_2}^6 y_{o_3}^9 y_{o_4}^{12} t_1^4 t_2^5 t_3^3 t_4^3 t_5^6 t_7^6 +  y_{s}^4 y_{o_1}^{10} y_{o_2}^7 y_{o_3}^{11} y_{o_4}^{14} t_1^6 t_2^4 t_3^3 t_4^5 t_5^7 t_6 t_7^6 +  2 y_{s}^4 y_{o_1}^{10} y_{o_2}^7 y_{o_3}^{11} y_{o_4}^{14} t_1^5 t_2^5 t_3^3 t_4^5 t_5^7 t_6 t_7^6 +  y_{s}^4 y_{o_1}^{10} y_{o_2}^7 y_{o_3}^{11} y_{o_4}^{14} t_1^4 t_2^6 t_3^3 t_4^5 t_5^7 t_6 t_7^6 +  2 y_{s}^4 y_{o_1}^{10} y_{o_2}^8 y_{o_3}^{10} y_{o_4}^{14} t_1^6 t_2^4 t_3^4 t_4^4 t_5^6 t_6^2 t_7^6 + 3 y_{s}^4 y_{o_1}^{10} y_{o_2}^8 y_{o_3}^{10} y_{o_4}^{14} t_1^5 t_2^5 t_3^4 t_4^4 t_5^6 t_6^2 t_7^6 + 2 y_{s}^4 y_{o_1}^{10} y_{o_2}^8 y_{o_3}^{10} y_{o_4}^{14} t_1^4 t_2^6 t_3^4 t_4^4 t_5^6 t_6^2 t_7^6 - y_{s}^5 y_{o_1}^{11} y_{o_2}^8 y_{o_3}^{13} y_{o_4}^{16} t_1^6 t_2^5 t_3^3 t_4^7 t_5^8 t_6^2 t_7^6 -  y_{s}^5 y_{o_1}^{11} y_{o_2}^8 y_{o_3}^{13} y_{o_4}^{16} t_1^5 t_2^6 t_3^3 t_4^7 t_5^8 t_6^2 t_7^6 -  y_{s}^3 y_{o_1}^9 y_{o_2}^9 y_{o_3}^6 y_{o_4}^{12} t_1^5 t_2^4 t_3^6 t_5^3 t_6^3 t_7^6 -  y_{s}^3 y_{o_1}^9 y_{o_2}^9 y_{o_3}^6 y_{o_4}^{12} t_1^4 t_2^5 t_3^6 t_5^3 t_6^3 t_7^6 +  y_{s}^4 y_{o_1}^{10} y_{o_2}^9 y_{o_3}^9 y_{o_4}^{14} t_1^7 t_2^3 t_3^5 t_4^3 t_5^5 t_6^3 t_7^6 +  2 y_{s}^4 y_{o_1}^{10} y_{o_2}^9 y_{o_3}^9 y_{o_4}^{14} t_1^6 t_2^4 t_3^5 t_4^3 t_5^5 t_6^3 t_7^6 +  2 y_{s}^4 y_{o_1}^{10} y_{o_2}^9 y_{o_3}^9 y_{o_4}^{14} t_1^5 t_2^5 t_3^5 t_4^3 t_5^5 t_6^3 t_7^6 +  2 y_{s}^4 y_{o_1}^{10} y_{o_2}^9 y_{o_3}^9 y_{o_4}^{14} t_1^4 t_2^6 t_3^5 t_4^3 t_5^5 t_6^3 t_7^6 +  y_{s}^4 y_{o_1}^{10} y_{o_2}^9 y_{o_3}^9 y_{o_4}^{14} t_1^3 t_2^7 t_3^5 t_4^3 t_5^5 t_6^3 t_7^6 -  y_{s}^5 y_{o_1}^{11} y_{o_2}^9 y_{o_3}^{12} y_{o_4}^{16} t_1^7 t_2^4 t_3^4 t_4^6 t_5^7 t_6^3 t_7^6 -  4 y_{s}^5 y_{o_1}^{11} y_{o_2}^9 y_{o_3}^{12} y_{o_4}^{16} t_1^6 t_2^5 t_3^4 t_4^6 t_5^7 t_6^3 t_7^6 - 4 y_{s}^5 y_{o_1}^{11} y_{o_2}^9 y_{o_3}^{12} y_{o_4}^{16} t_1^5 t_2^6 t_3^4 t_4^6 t_5^7 t_6^3 t_7^6 - y_{s}^5 y_{o_1}^{11} y_{o_2}^9 y_{o_3}^{12} y_{o_4}^{16} t_1^4 t_2^7 t_3^4 t_4^6 t_5^7 t_6^3 t_7^6 +  2 y_{s}^4 y_{o_1}^{10} y_{o_2}^{10} y_{o_3}^8 y_{o_4}^{14} t_1^6 t_2^4 t_3^6 t_4^2 t_5^4 t_6^4 t_7^6 + 3 y_{s}^4 y_{o_1}^{10} y_{o_2}^{10} y_{o_3}^8 y_{o_4}^{14} t_1^5 t_2^5 t_3^6 t_4^2 t_5^4 t_6^4 t_7^6 + 2 y_{s}^4 y_{o_1}^{10} y_{o_2}^{10} y_{o_3}^8 y_{o_4}^{14} t_1^4 t_2^6 t_3^6 t_4^2 t_5^4 t_6^4 t_7^6 - y_{s}^5 y_{o_1}^{11} y_{o_2}^{10} y_{o_3}^{11} y_{o_4}^{16} t_1^8 t_2^3 t_3^5 t_4^5 t_5^6 t_6^4 t_7^6 - 3 y_{s}^5 y_{o_1}^{11} y_{o_2}^{10} y_{o_3}^{11} y_{o_4}^{16} t_1^7 t_2^4 t_3^5 t_4^5 t_5^6 t_6^4 t_7^6 - 6 y_{s}^5 y_{o_1}^{11} y_{o_2}^{10} y_{o_3}^{11} y_{o_4}^{16} t_1^6 t_2^5 t_3^5 t_4^5 t_5^6 t_6^4 t_7^6 -  6 y_{s}^5 y_{o_1}^{11} y_{o_2}^{10} y_{o_3}^{11} y_{o_4}^{16} t_1^5 t_2^6 t_3^5 t_4^5 t_5^6 t_6^4 t_7^6 - 3 y_{s}^5 y_{o_1}^{11} y_{o_2}^{10} y_{o_3}^{11} y_{o_4}^{16} t_1^4 t_2^7 t_3^5 t_4^5 t_5^6 t_6^4 t_7^6 - y_{s}^5 y_{o_1}^{11} y_{o_2}^{10} y_{o_3}^{11} y_{o_4}^{16} t_1^3 t_2^8 t_3^5 t_4^5 t_5^6 t_6^4 t_7^6 + y_{s}^6 y_{o_1}^{12} y_{o_2}^{10} y_{o_3}^{14} y_{o_4}^{18} t_1^7 t_2^5 t_3^4 t_4^8 t_5^8 t_6^4 t_7^6 + 2 y_{s}^6 y_{o_1}^{12} y_{o_2}^{10} y_{o_3}^{14} y_{o_4}^{18} t_1^6 t_2^6 t_3^4 t_4^8 t_5^8 t_6^4 t_7^6 +  y_{s}^6 y_{o_1}^{12} y_{o_2}^{10} y_{o_3}^{14} y_{o_4}^{18} t_1^5 t_2^7 t_3^4 t_4^8 t_5^8 t_6^4 t_7^6 +  y_{s}^4 y_{o_1}^{10} y_{o_2}^{11} y_{o_3}^7 y_{o_4}^{14} t_1^6 t_2^4 t_3^7 t_4 t_5^3 t_6^5 t_7^6 +  2 y_{s}^4 y_{o_1}^{10} y_{o_2}^{11} y_{o_3}^7 y_{o_4}^{14} t_1^5 t_2^5 t_3^7 t_4 t_5^3 t_6^5 t_7^6 +  y_{s}^4 y_{o_1}^{10} y_{o_2}^{11} y_{o_3}^7 y_{o_4}^{14} t_1^4 t_2^6 t_3^7 t_4 t_5^3 t_6^5 t_7^6 -  y_{s}^5 y_{o_1}^{11} y_{o_2}^{11} y_{o_3}^{10} y_{o_4}^{16} t_1^8 t_2^3 t_3^6 t_4^4 t_5^5 t_6^5 t_7^6 -  3 y_{s}^5 y_{o_1}^{11} y_{o_2}^{11} y_{o_3}^{10} y_{o_4}^{16} t_1^7 t_2^4 t_3^6 t_4^4 t_5^5 t_6^5 t_7^6 - 6 y_{s}^5 y_{o_1}^{11} y_{o_2}^{11} y_{o_3}^{10} y_{o_4}^{16} t_1^6 t_2^5 t_3^6 t_4^4 t_5^5 t_6^5 t_7^6 - 6 y_{s}^5 y_{o_1}^{11} y_{o_2}^{11} y_{o_3}^{10} y_{o_4}^{16} t_1^5 t_2^6 t_3^6 t_4^4 t_5^5 t_6^5 t_7^6 -  3 y_{s}^5 y_{o_1}^{11} y_{o_2}^{11} y_{o_3}^{10} y_{o_4}^{16} t_1^4 t_2^7 t_3^6 t_4^4 t_5^5 t_6^5 t_7^6 - y_{s}^5 y_{o_1}^{11} y_{o_2}^{11} y_{o_3}^{10} y_{o_4}^{16} t_1^3 t_2^8 t_3^6 t_4^4 t_5^5 t_6^5 t_7^6 +  y_{s}^6 y_{o_1}^{12} y_{o_2}^{11} y_{o_3}^{13} y_{o_4}^{18} t_1^8 t_2^4 t_3^5 t_4^7 t_5^7 t_6^5 t_7^6 + 3 y_{s}^6 y_{o_1}^{12} y_{o_2}^{11} y_{o_3}^{13} y_{o_4}^{18} t_1^7 t_2^5 t_3^5 t_4^7 t_5^7 t_6^5 t_7^6 + 5 y_{s}^6 y_{o_1}^{12} y_{o_2}^{11} y_{o_3}^{13} y_{o_4}^{18} t_1^6 t_2^6 t_3^5 t_4^7 t_5^7 t_6^5 t_7^6 +  3 y_{s}^6 y_{o_1}^{12} y_{o_2}^{11} y_{o_3}^{13} y_{o_4}^{18} t_1^5 t_2^7 t_3^5 t_4^7 t_5^7 t_6^5 t_7^6 + y_{s}^6 y_{o_1}^{12} y_{o_2}^{11} y_{o_3}^{13} y_{o_4}^{18} t_1^4 t_2^8 t_3^5 t_4^7 t_5^7 t_6^5 t_7^6 -  y_{s}^5 y_{o_1}^{11} y_{o_2}^{12} y_{o_3}^9 y_{o_4}^{16} t_1^7 t_2^4 t_3^7 t_4^3 t_5^4 t_6^6 t_7^6 -  4 y_{s}^5 y_{o_1}^{11} y_{o_2}^{12} y_{o_3}^9 y_{o_4}^{16} t_1^6 t_2^5 t_3^7 t_4^3 t_5^4 t_6^6 t_7^6 - 4 y_{s}^5 y_{o_1}^{11} y_{o_2}^{12} y_{o_3}^9 y_{o_4}^{16} t_1^5 t_2^6 t_3^7 t_4^3 t_5^4 t_6^6 t_7^6 - y_{s}^5 y_{o_1}^{11} y_{o_2}^{12} y_{o_3}^9 y_{o_4}^{16} t_1^4 t_2^7 t_3^7 t_4^3 t_5^4 t_6^6 t_7^6 +  y_{s}^6 y_{o_1}^{12} y_{o_2}^{12} y_{o_3}^{12} y_{o_4}^{18} t_1^9 t_2^3 t_3^6 t_4^6 t_5^6 t_6^6 t_7^6 +  4 y_{s}^6 y_{o_1}^{12} y_{o_2}^{12} y_{o_3}^{12} y_{o_4}^{18} t_1^8 t_2^4 t_3^6 t_4^6 t_5^6 t_6^6 t_7^6 + 6 y_{s}^6 y_{o_1}^{12} y_{o_2}^{12} y_{o_3}^{12} y_{o_4}^{18} t_1^7 t_2^5 t_3^6 t_4^6 t_5^6 t_6^6 t_7^6 + 7 y_{s}^6 y_{o_1}^{12} y_{o_2}^{12} y_{o_3}^{12} y_{o_4}^{18} t_1^6 t_2^6 t_3^6 t_4^6 t_5^6 t_6^6 t_7^6 +  6 y_{s}^6 y_{o_1}^{12} y_{o_2}^{12} y_{o_3}^{12} y_{o_4}^{18} t_1^5 t_2^7 t_3^6 t_4^6 t_5^6 t_6^6 t_7^6 + 4 y_{s}^6 y_{o_1}^{12} y_{o_2}^{12} y_{o_3}^{12} y_{o_4}^{18} t_1^4 t_2^8 t_3^6 t_4^6 t_5^6 t_6^6 t_7^6 + y_{s}^6 y_{o_1}^{12} y_{o_2}^{12} y_{o_3}^{12} y_{o_4}^{18} t_1^3 t_2^9 t_3^6 t_4^6 t_5^6 t_6^6 t_7^6 - y_{s}^7 y_{o_1}^{13} y_{o_2}^{12} y_{o_3}^{15} y_{o_4}^{20} t_1^7 t_2^6 t_3^5 t_4^9 t_5^8 t_6^6 t_7^6 - y_{s}^7 y_{o_1}^{13} y_{o_2}^{12} y_{o_3}^{15} y_{o_4}^{20} t_1^6 t_2^7 t_3^5 t_4^9 t_5^8 t_6^6 t_7^6 - y_{s}^5 y_{o_1}^{11} y_{o_2}^{13} y_{o_3}^8 y_{o_4}^{16} t_1^6 t_2^5 t_3^8 t_4^2 t_5^3 t_6^7 t_7^6 - y_{s}^5 y_{o_1}^{11} y_{o_2}^{13} y_{o_3}^8 y_{o_4}^{16} t_1^5 t_2^6 t_3^8 t_4^2 t_5^3 t_6^7 t_7^6 + y_{s}^6 y_{o_1}^{12} y_{o_2}^{13} y_{o_3}^{11} y_{o_4}^{18} t_1^8 t_2^4 t_3^7 t_4^5 t_5^5 t_6^7 t_7^6 + 3 y_{s}^6 y_{o_1}^{12} y_{o_2}^{13} y_{o_3}^{11} y_{o_4}^{18} t_1^7 t_2^5 t_3^7 t_4^5 t_5^5 t_6^7 t_7^6 +  5 y_{s}^6 y_{o_1}^{12} y_{o_2}^{13} y_{o_3}^{11} y_{o_4}^{18} t_1^6 t_2^6 t_3^7 t_4^5 t_5^5 t_6^7 t_7^6 + 3 y_{s}^6 y_{o_1}^{12} y_{o_2}^{13} y_{o_3}^{11} y_{o_4}^{18} t_1^5 t_2^7 t_3^7 t_4^5 t_5^5 t_6^7 t_7^6 + y_{s}^6 y_{o_1}^{12} y_{o_2}^{13} y_{o_3}^{11} y_{o_4}^{18} t_1^4 t_2^8 t_3^7 t_4^5 t_5^5 t_6^7 t_7^6 - y_{s}^7 y_{o_1}^{13} y_{o_2}^{13} y_{o_3}^{14} y_{o_4}^{20} t_1^9 t_2^4 t_3^6 t_4^8 t_5^7 t_6^7 t_7^6 - 2 y_{s}^7 y_{o_1}^{13} y_{o_2}^{13} y_{o_3}^{14} y_{o_4}^{20} t_1^8 t_2^5 t_3^6 t_4^8 t_5^7 t_6^7 t_7^6 -  3 y_{s}^7 y_{o_1}^{13} y_{o_2}^{13} y_{o_3}^{14} y_{o_4}^{20} t_1^7 t_2^6 t_3^6 t_4^8 t_5^7 t_6^7 t_7^6 - 3 y_{s}^7 y_{o_1}^{13} y_{o_2}^{13} y_{o_3}^{14} y_{o_4}^{20} t_1^6 t_2^7 t_3^6 t_4^8 t_5^7 t_6^7 t_7^6 - 2 y_{s}^7 y_{o_1}^{13} y_{o_2}^{13} y_{o_3}^{14} y_{o_4}^{20} t_1^5 t_2^8 t_3^6 t_4^8 t_5^7 t_6^7 t_7^6 -  y_{s}^7 y_{o_1}^{13} y_{o_2}^{13} y_{o_3}^{14} y_{o_4}^{20} t_1^4 t_2^9 t_3^6 t_4^8 t_5^7 t_6^7 t_7^6 +  y_{s}^6 y_{o_1}^{12} y_{o_2}^{14} y_{o_3}^{10} y_{o_4}^{18} t_1^7 t_2^5 t_3^8 t_4^4 t_5^4 t_6^8 t_7^6 +  2 y_{s}^6 y_{o_1}^{12} y_{o_2}^{14} y_{o_3}^{10} y_{o_4}^{18} t_1^6 t_2^6 t_3^8 t_4^4 t_5^4 t_6^8 t_7^6 + y_{s}^6 y_{o_1}^{12} y_{o_2}^{14} y_{o_3}^{10} y_{o_4}^{18} t_1^5 t_2^7 t_3^8 t_4^4 t_5^4 t_6^8 t_7^6 -  y_{s}^7 y_{o_1}^{13} y_{o_2}^{14} y_{o_3}^{13} y_{o_4}^{20} t_1^9 t_2^4 t_3^7 t_4^7 t_5^6 t_6^8 t_7^6 - 2 y_{s}^7 y_{o_1}^{13} y_{o_2}^{14} y_{o_3}^{13} y_{o_4}^{20} t_1^8 t_2^5 t_3^7 t_4^7 t_5^6 t_6^8 t_7^6 - 3 y_{s}^7 y_{o_1}^{13} y_{o_2}^{14} y_{o_3}^{13} y_{o_4}^{20} t_1^7 t_2^6 t_3^7 t_4^7 t_5^6 t_6^8 t_7^6 -  3 y_{s}^7 y_{o_1}^{13} y_{o_2}^{14} y_{o_3}^{13} y_{o_4}^{20} t_1^6 t_2^7 t_3^7 t_4^7 t_5^6 t_6^8 t_7^6 - 2 y_{s}^7 y_{o_1}^{13} y_{o_2}^{14} y_{o_3}^{13} y_{o_4}^{20} t_1^5 t_2^8 t_3^7 t_4^7 t_5^6 t_6^8 t_7^6 - y_{s}^7 y_{o_1}^{13} y_{o_2}^{14} y_{o_3}^{13} y_{o_4}^{20} t_1^4 t_2^9 t_3^7 t_4^7 t_5^6 t_6^8 t_7^6 - y_{s}^7 y_{o_1}^{13} y_{o_2}^{15} y_{o_3}^{12} y_{o_4}^{20} t_1^7 t_2^6 t_3^8 t_4^6 t_5^5 t_6^9 t_7^6 - y_{s}^7 y_{o_1}^{13} y_{o_2}^{15} y_{o_3}^{12} y_{o_4}^{20} t_1^6 t_2^7 t_3^8 t_4^6 t_5^5 t_6^9 t_7^6 + y_{s}^8 y_{o_1}^{14} y_{o_2}^{15} y_{o_3}^{15} y_{o_4}^{22} t_1^8 t_2^6 t_3^7 t_4^9 t_5^7 t_6^9 t_7^6 + y_{s}^8 y_{o_1}^{14} y_{o_2}^{15} y_{o_3}^{15} y_{o_4}^{22} t_1^7 t_2^7 t_3^7 t_4^9 t_5^7 t_6^9 t_7^6 + y_{s}^8 y_{o_1}^{14} y_{o_2}^{15} y_{o_3}^{15} y_{o_4}^{22} t_1^6 t_2^8 t_3^7 t_4^9 t_5^7 t_6^9 t_7^6 + y_{s}^4 y_{o_1}^{11} y_{o_2}^9 y_{o_3}^{10} y_{o_4}^{15} t_1^7 t_2^4 t_3^5 t_4^3 t_5^6 t_6^2 t_7^7 + y_{s}^4 y_{o_1}^{11} y_{o_2}^9 y_{o_3}^{10} y_{o_4}^{15} t_1^6 t_2^5 t_3^5 t_4^3 t_5^6 t_6^2 t_7^7 + y_{s}^4 y_{o_1}^{11} y_{o_2}^9 y_{o_3}^{10} y_{o_4}^{15} t_1^5 t_2^6 t_3^5 t_4^3 t_5^6 t_6^2 t_7^7 + y_{s}^4 y_{o_1}^{11} y_{o_2}^9 y_{o_3}^{10} y_{o_4}^{15} t_1^4 t_2^7 t_3^5 t_4^3 t_5^6 t_6^2 t_7^7 - y_{s}^5 y_{o_1}^{12} y_{o_2}^9 y_{o_3}^{13} y_{o_4}^{17} t_1^6 t_2^6 t_3^4 t_4^6 t_5^8 t_6^2 t_7^7 + y_{s}^4 y_{o_1}^{11} y_{o_2}^{10} y_{o_3}^9 y_{o_4}^{15} t_1^7 t_2^4 t_3^6 t_4^2 t_5^5 t_6^3 t_7^7 + y_{s}^4 y_{o_1}^{11} y_{o_2}^{10} y_{o_3}^9 y_{o_4}^{15} t_1^6 t_2^5 t_3^6 t_4^2 t_5^5 t_6^3 t_7^7 + y_{s}^4 y_{o_1}^{11} y_{o_2}^{10} y_{o_3}^9 y_{o_4}^{15} t_1^5 t_2^6 t_3^6 t_4^2 t_5^5 t_6^3 t_7^7 + y_{s}^4 y_{o_1}^{11} y_{o_2}^{10} y_{o_3}^9 y_{o_4}^{15} t_1^4 t_2^7 t_3^6 t_4^2 t_5^5 t_6^3 t_7^7 - y_{s}^5 y_{o_1}^{12} y_{o_2}^{10} y_{o_3}^{12} y_{o_4}^{17} t_1^8 t_2^4 t_3^5 t_4^5 t_5^7 t_6^3 t_7^7 - 2 y_{s}^5 y_{o_1}^{12} y_{o_2}^{10} y_{o_3}^{12} y_{o_4}^{17} t_1^7 t_2^5 t_3^5 t_4^5 t_5^7 t_6^3 t_7^7 -  3 y_{s}^5 y_{o_1}^{12} y_{o_2}^{10} y_{o_3}^{12} y_{o_4}^{17} t_1^6 t_2^6 t_3^5 t_4^5 t_5^7 t_6^3 t_7^7 - 2 y_{s}^5 y_{o_1}^{12} y_{o_2}^{10} y_{o_3}^{12} y_{o_4}^{17} t_1^5 t_2^7 t_3^5 t_4^5 t_5^7 t_6^3 t_7^7 - y_{s}^5 y_{o_1}^{12} y_{o_2}^{10} y_{o_3}^{12} y_{o_4}^{17} t_1^4 t_2^8 t_3^5 t_4^5 t_5^7 t_6^3 t_7^7 - 3 y_{s}^5 y_{o_1}^{12} y_{o_2}^{11} y_{o_3}^{11} y_{o_4}^{17} t_1^8 t_2^4 t_3^6 t_4^4 t_5^6 t_6^4 t_7^7 -  5 y_{s}^5 y_{o_1}^{12} y_{o_2}^{11} y_{o_3}^{11} y_{o_4}^{17} t_1^7 t_2^5 t_3^6 t_4^4 t_5^6 t_6^4 t_7^7 - 7 y_{s}^5 y_{o_1}^{12} y_{o_2}^{11} y_{o_3}^{11} y_{o_4}^{17} t_1^6 t_2^6 t_3^6 t_4^4 t_5^6 t_6^4 t_7^7 - 5 y_{s}^5 y_{o_1}^{12} y_{o_2}^{11} y_{o_3}^{11} y_{o_4}^{17} t_1^5 t_2^7 t_3^6 t_4^4 t_5^6 t_6^4 t_7^7 -  3 y_{s}^5 y_{o_1}^{12} y_{o_2}^{11} y_{o_3}^{11} y_{o_4}^{17} t_1^4 t_2^8 t_3^6 t_4^4 t_5^6 t_6^4 t_7^7 + y_{s}^6 y_{o_1}^{13} y_{o_2}^{11} y_{o_3}^{14} y_{o_4}^{19} t_1^8 t_2^5 t_3^5 t_4^7 t_5^8 t_6^4 t_7^7 + 2 y_{s}^6 y_{o_1}^{13} y_{o_2}^{11} y_{o_3}^{14} y_{o_4}^{19} t_1^7 t_2^6 t_3^5 t_4^7 t_5^8 t_6^4 t_7^7 + 2 y_{s}^6 y_{o_1}^{13} y_{o_2}^{11} y_{o_3}^{14} y_{o_4}^{19} t_1^6 t_2^7 t_3^5 t_4^7 t_5^8 t_6^4 t_7^7 +  y_{s}^6 y_{o_1}^{13} y_{o_2}^{11} y_{o_3}^{14} y_{o_4}^{19} t_1^5 t_2^8 t_3^5 t_4^7 t_5^8 t_6^4 t_7^7 -  y_{s}^5 y_{o_1}^{12} y_{o_2}^{12} y_{o_3}^{10} y_{o_4}^{17} t_1^8 t_2^4 t_3^7 t_4^3 t_5^5 t_6^5 t_7^7 -  2 y_{s}^5 y_{o_1}^{12} y_{o_2}^{12} y_{o_3}^{10} y_{o_4}^{17} t_1^7 t_2^5 t_3^7 t_4^3 t_5^5 t_6^5 t_7^7 - 3 y_{s}^5 y_{o_1}^{12} y_{o_2}^{12} y_{o_3}^{10} y_{o_4}^{17} t_1^6 t_2^6 t_3^7 t_4^3 t_5^5 t_6^5 t_7^7 - 2 y_{s}^5 y_{o_1}^{12} y_{o_2}^{12} y_{o_3}^{10} y_{o_4}^{17} t_1^5 t_2^7 t_3^7 t_4^3 t_5^5 t_6^5 t_7^7 -  y_{s}^5 y_{o_1}^{12} y_{o_2}^{12} y_{o_3}^{10} y_{o_4}^{17} t_1^4 t_2^8 t_3^7 t_4^3 t_5^5 t_6^5 t_7^7 +  y_{s}^6 y_{o_1}^{13} y_{o_2}^{12} y_{o_3}^{13} y_{o_4}^{19} t_1^9 t_2^4 t_3^6 t_4^6 t_5^7 t_6^5 t_7^7 +  5 y_{s}^6 y_{o_1}^{13} y_{o_2}^{12} y_{o_3}^{13} y_{o_4}^{19} t_1^8 t_2^5 t_3^6 t_4^6 t_5^7 t_6^5 t_7^7 + 7 y_{s}^6 y_{o_1}^{13} y_{o_2}^{12} y_{o_3}^{13} y_{o_4}^{19} t_1^7 t_2^6 t_3^6 t_4^6 t_5^7 t_6^5 t_7^7 + 7 y_{s}^6 y_{o_1}^{13} y_{o_2}^{12} y_{o_3}^{13} y_{o_4}^{19} t_1^6 t_2^7 t_3^6 t_4^6 t_5^7 t_6^5 t_7^7 +  5 y_{s}^6 y_{o_1}^{13} y_{o_2}^{12} y_{o_3}^{13} y_{o_4}^{19} t_1^5 t_2^8 t_3^6 t_4^6 t_5^7 t_6^5 t_7^7 + y_{s}^6 y_{o_1}^{13} y_{o_2}^{12} y_{o_3}^{13} y_{o_4}^{19} t_1^4 t_2^9 t_3^6 t_4^6 t_5^7 t_6^5 t_7^7 -  y_{s}^5 y_{o_1}^{12} y_{o_2}^{13} y_{o_3}^9 y_{o_4}^{17} t_1^6 t_2^6 t_3^8 t_4^2 t_5^4 t_6^6 t_7^7 +  y_{s}^6 y_{o_1}^{13} y_{o_2}^{13} y_{o_3}^{12} y_{o_4}^{19} t_1^9 t_2^4 t_3^7 t_4^5 t_5^6 t_6^6 t_7^7 +  5 y_{s}^6 y_{o_1}^{13} y_{o_2}^{13} y_{o_3}^{12} y_{o_4}^{19} t_1^8 t_2^5 t_3^7 t_4^5 t_5^6 t_6^6 t_7^7 + 7 y_{s}^6 y_{o_1}^{13} y_{o_2}^{13} y_{o_3}^{12} y_{o_4}^{19} t_1^7 t_2^6 t_3^7 t_4^5 t_5^6 t_6^6 t_7^7 + 7 y_{s}^6 y_{o_1}^{13} y_{o_2}^{13} y_{o_3}^{12} y_{o_4}^{19} t_1^6 t_2^7 t_3^7 t_4^5 t_5^6 t_6^6 t_7^7 +  5 y_{s}^6 y_{o_1}^{13} y_{o_2}^{13} y_{o_3}^{12} y_{o_4}^{19} t_1^5 t_2^8 t_3^7 t_4^5 t_5^6 t_6^6 t_7^7 + y_{s}^6 y_{o_1}^{13} y_{o_2}^{13} y_{o_3}^{12} y_{o_4}^{19} t_1^4 t_2^9 t_3^7 t_4^5 t_5^6 t_6^6 t_7^7 -  y_{s}^7 y_{o_1}^{14} y_{o_2}^{13} y_{o_3}^{15} y_{o_4}^{21} t_1^9 t_2^5 t_3^6 t_4^8 t_5^8 t_6^6 t_7^7 - 2 y_{s}^7 y_{o_1}^{14} y_{o_2}^{13} y_{o_3}^{15} y_{o_4}^{21} t_1^8 t_2^6 t_3^6 t_4^8 t_5^8 t_6^6 t_7^7 - 3 y_{s}^7 y_{o_1}^{14} y_{o_2}^{13} y_{o_3}^{15} y_{o_4}^{21} t_1^7 t_2^7 t_3^6 t_4^8 t_5^8 t_6^6 t_7^7 -  2 y_{s}^7 y_{o_1}^{14} y_{o_2}^{13} y_{o_3}^{15} y_{o_4}^{21} t_1^6 t_2^8 t_3^6 t_4^8 t_5^8 t_6^6 t_7^7 - y_{s}^7 y_{o_1}^{14} y_{o_2}^{13} y_{o_3}^{15} y_{o_4}^{21} t_1^5 t_2^9 t_3^6 t_4^8 t_5^8 t_6^6 t_7^7 +  y_{s}^6 y_{o_1}^{13} y_{o_2}^{14} y_{o_3}^{11} y_{o_4}^{19} t_1^8 t_2^5 t_3^8 t_4^4 t_5^5 t_6^7 t_7^7 + 2 y_{s}^6 y_{o_1}^{13} y_{o_2}^{14} y_{o_3}^{11} y_{o_4}^{19} t_1^7 t_2^6 t_3^8 t_4^4 t_5^5 t_6^7 t_7^7 + 2 y_{s}^6 y_{o_1}^{13} y_{o_2}^{14} y_{o_3}^{11} y_{o_4}^{19} t_1^6 t_2^7 t_3^8 t_4^4 t_5^5 t_6^7 t_7^7 +  y_{s}^6 y_{o_1}^{13} y_{o_2}^{14} y_{o_3}^{11} y_{o_4}^{19} t_1^5 t_2^8 t_3^8 t_4^4 t_5^5 t_6^7 t_7^7 -  2 y_{s}^7 y_{o_1}^{14} y_{o_2}^{14} y_{o_3}^{14} y_{o_4}^{21} t_1^9 t_2^5 t_3^7 t_4^7 t_5^7 t_6^7 t_7^7 - 5 y_{s}^7 y_{o_1}^{14} y_{o_2}^{14} y_{o_3}^{14} y_{o_4}^{21} t_1^8 t_2^6 t_3^7 t_4^7 t_5^7 t_6^7 t_7^7 - 5 y_{s}^7 y_{o_1}^{14} y_{o_2}^{14} y_{o_3}^{14} y_{o_4}^{21} t_1^7 t_2^7 t_3^7 t_4^7 t_5^7 t_6^7 t_7^7 -  5 y_{s}^7 y_{o_1}^{14} y_{o_2}^{14} y_{o_3}^{14} y_{o_4}^{21} t_1^6 t_2^8 t_3^7 t_4^7 t_5^7 t_6^7 t_7^7 - 2 y_{s}^7 y_{o_1}^{14} y_{o_2}^{14} y_{o_3}^{14} y_{o_4}^{21} t_1^5 t_2^9 t_3^7 t_4^7 t_5^7 t_6^7 t_7^7 - y_{s}^7 y_{o_1}^{14} y_{o_2}^{15} y_{o_3}^{13} y_{o_4}^{21} t_1^9 t_2^5 t_3^8 t_4^6 t_5^6 t_6^8 t_7^7 - 2 y_{s}^7 y_{o_1}^{14} y_{o_2}^{15} y_{o_3}^{13} y_{o_4}^{21} t_1^8 t_2^6 t_3^8 t_4^6 t_5^6 t_6^8 t_7^7 -  3 y_{s}^7 y_{o_1}^{14} y_{o_2}^{15} y_{o_3}^{13} y_{o_4}^{21} t_1^7 t_2^7 t_3^8 t_4^6 t_5^6 t_6^8 t_7^7 - 2 y_{s}^7 y_{o_1}^{14} y_{o_2}^{15} y_{o_3}^{13} y_{o_4}^{21} t_1^6 t_2^8 t_3^8 t_4^6 t_5^6 t_6^8 t_7^7 - y_{s}^7 y_{o_1}^{14} y_{o_2}^{15} y_{o_3}^{13} y_{o_4}^{21} t_1^5 t_2^9 t_3^8 t_4^6 t_5^6 t_6^8 t_7^7 + y_{s}^8 y_{o_1}^{15} y_{o_2}^{15} y_{o_3}^{16} y_{o_4}^{23} t_1^9 t_2^6 t_3^7 t_4^9 t_5^8 t_6^8 t_7^7 + y_{s}^8 y_{o_1}^{15} y_{o_2}^{15} y_{o_3}^{16} y_{o_4}^{23} t_1^8 t_2^7 t_3^7 t_4^9 t_5^8 t_6^8 t_7^7 + y_{s}^8 y_{o_1}^{15} y_{o_2}^{15} y_{o_3}^{16} y_{o_4}^{23} t_1^7 t_2^8 t_3^7 t_4^9 t_5^8 t_6^8 t_7^7 + y_{s}^8 y_{o_1}^{15} y_{o_2}^{15} y_{o_3}^{16} y_{o_4}^{23} t_1^6 t_2^9 t_3^7 t_4^9 t_5^8 t_6^8 t_7^7 + y_{s}^8 y_{o_1}^{15} y_{o_2}^{16} y_{o_3}^{15} y_{o_4}^{23} t_1^9 t_2^6 t_3^8 t_4^8 t_5^7 t_6^9 t_7^7 + y_{s}^8 y_{o_1}^{15} y_{o_2}^{16} y_{o_3}^{15} y_{o_4}^{23} t_1^8 t_2^7 t_3^8 t_4^8 t_5^7 t_6^9 t_7^7 + y_{s}^8 y_{o_1}^{15} y_{o_2}^{16} y_{o_3}^{15} y_{o_4}^{23} t_1^7 t_2^8 t_3^8 t_4^8 t_5^7 t_6^9 t_7^7 + y_{s}^8 y_{o_1}^{15} y_{o_2}^{16} y_{o_3}^{15} y_{o_4}^{23} t_1^6 t_2^9 t_3^8 t_4^8 t_5^7 t_6^9 t_7^7 - y_{s}^5 y_{o_1}^{13} y_{o_2}^{11} y_{o_3}^{12} y_{o_4}^{18} t_1^7 t_2^6 t_3^6 t_4^4 t_5^7 t_6^3 t_7^8 - y_{s}^5 y_{o_1}^{13} y_{o_2}^{11} y_{o_3}^{12} y_{o_4}^{18} t_1^6 t_2^7 t_3^6 t_4^4 t_5^7 t_6^3 t_7^8 - y_{s}^5 y_{o_1}^{13} y_{o_2}^{12} y_{o_3}^{11} y_{o_4}^{18} t_1^7 t_2^6 t_3^7 t_4^3 t_5^6 t_6^4 t_7^8 - y_{s}^5 y_{o_1}^{13} y_{o_2}^{12} y_{o_3}^{11} y_{o_4}^{18} t_1^6 t_2^7 t_3^7 t_4^3 t_5^6 t_6^4 t_7^8 + 2 y_{s}^6 y_{o_1}^{14} y_{o_2}^{12} y_{o_3}^{14} y_{o_4}^{20} t_1^8 t_2^6 t_3^6 t_4^6 t_5^8 t_6^4 t_7^8 +  2 y_{s}^6 y_{o_1}^{14} y_{o_2}^{12} y_{o_3}^{14} y_{o_4}^{20} t_1^7 t_2^7 t_3^6 t_4^6 t_5^8 t_6^4 t_7^8 + 2 y_{s}^6 y_{o_1}^{14} y_{o_2}^{12} y_{o_3}^{14} y_{o_4}^{20} t_1^6 t_2^8 t_3^6 t_4^6 t_5^8 t_6^4 t_7^8 + 2 y_{s}^6 y_{o_1}^{14} y_{o_2}^{13} y_{o_3}^{13} y_{o_4}^{20} t_1^8 t_2^6 t_3^7 t_4^5 t_5^7 t_6^5 t_7^8 +  3 y_{s}^6 y_{o_1}^{14} y_{o_2}^{13} y_{o_3}^{13} y_{o_4}^{20} t_1^7 t_2^7 t_3^7 t_4^5 t_5^7 t_6^5 t_7^8 + 2 y_{s}^6 y_{o_1}^{14} y_{o_2}^{13} y_{o_3}^{13} y_{o_4}^{20} t_1^6 t_2^8 t_3^7 t_4^5 t_5^7 t_6^5 t_7^8 - y_{s}^7 y_{o_1}^{15} y_{o_2}^{13} y_{o_3}^{16} y_{o_4}^{22} t_1^8 t_2^7 t_3^6 t_4^8 t_5^9 t_6^5 t_7^8 - y_{s}^7 y_{o_1}^{15} y_{o_2}^{13} y_{o_3}^{16} y_{o_4}^{22} t_1^7 t_2^8 t_3^6 t_4^8 t_5^9 t_6^5 t_7^8 + 2 y_{s}^6 y_{o_1}^{14} y_{o_2}^{14} y_{o_3}^{12} y_{o_4}^{20} t_1^8 t_2^6 t_3^8 t_4^4 t_5^6 t_6^6 t_7^8 +  2 y_{s}^6 y_{o_1}^{14} y_{o_2}^{14} y_{o_3}^{12} y_{o_4}^{20} t_1^7 t_2^7 t_3^8 t_4^4 t_5^6 t_6^6 t_7^8 + 2 y_{s}^6 y_{o_1}^{14} y_{o_2}^{14} y_{o_3}^{12} y_{o_4}^{20} t_1^6 t_2^8 t_3^8 t_4^4 t_5^6 t_6^6 t_7^8 - y_{s}^7 y_{o_1}^{15} y_{o_2}^{14} y_{o_3}^{15} y_{o_4}^{22} t_1^9 t_2^6 t_3^7 t_4^7 t_5^8 t_6^6 t_7^8 - 3 y_{s}^7 y_{o_1}^{15} y_{o_2}^{14} y_{o_3}^{15} y_{o_4}^{22} t_1^8 t_2^7 t_3^7 t_4^7 t_5^8 t_6^6 t_7^8 -  3 y_{s}^7 y_{o_1}^{15} y_{o_2}^{14} y_{o_3}^{15} y_{o_4}^{22} t_1^7 t_2^8 t_3^7 t_4^7 t_5^8 t_6^6 t_7^8 - y_{s}^7 y_{o_1}^{15} y_{o_2}^{14} y_{o_3}^{15} y_{o_4}^{22} t_1^6 t_2^9 t_3^7 t_4^7 t_5^8 t_6^6 t_7^8 -  y_{s}^7 y_{o_1}^{15} y_{o_2}^{15} y_{o_3}^{14} y_{o_4}^{22} t_1^9 t_2^6 t_3^8 t_4^6 t_5^7 t_6^7 t_7^8 - 3 y_{s}^7 y_{o_1}^{15} y_{o_2}^{15} y_{o_3}^{14} y_{o_4}^{22} t_1^8 t_2^7 t_3^8 t_4^6 t_5^7 t_6^7 t_7^8 - 3 y_{s}^7 y_{o_1}^{15} y_{o_2}^{15} y_{o_3}^{14} y_{o_4}^{22} t_1^7 t_2^8 t_3^8 t_4^6 t_5^7 t_6^7 t_7^8 -  y_{s}^7 y_{o_1}^{15} y_{o_2}^{15} y_{o_3}^{14} y_{o_4}^{22} t_1^6 t_2^9 t_3^8 t_4^6 t_5^7 t_6^7 t_7^8 +  y_{s}^8 y_{o_1}^{16} y_{o_2}^{15} y_{o_3}^{17} y_{o_4}^{24} t_1^8 t_2^8 t_3^7 t_4^9 t_5^9 t_6^7 t_7^8 -  y_{s}^7 y_{o_1}^{15} y_{o_2}^{16} y_{o_3}^{13} y_{o_4}^{22} t_1^8 t_2^7 t_3^9 t_4^5 t_5^6 t_6^8 t_7^8 -  y_{s}^7 y_{o_1}^{15} y_{o_2}^{16} y_{o_3}^{13} y_{o_4}^{22} t_1^7 t_2^8 t_3^9 t_4^5 t_5^6 t_6^8 t_7^8 +  y_{s}^8 y_{o_1}^{16} y_{o_2}^{16} y_{o_3}^{16} y_{o_4}^{24} t_1^9 t_2^7 t_3^8 t_4^8 t_5^8 t_6^8 t_7^8 +  y_{s}^8 y_{o_1}^{16} y_{o_2}^{16} y_{o_3}^{16} y_{o_4}^{24} t_1^8 t_2^8 t_3^8 t_4^8 t_5^8 t_6^8 t_7^8 +  y_{s}^8 y_{o_1}^{16} y_{o_2}^{16} y_{o_3}^{16} y_{o_4}^{24} t_1^7 t_2^9 t_3^8 t_4^8 t_5^8 t_6^8 t_7^8 +  y_{s}^8 y_{o_1}^{16} y_{o_2}^{17} y_{o_3}^{15} y_{o_4}^{24} t_1^8 t_2^8 t_3^9 t_4^7 t_5^7 t_6^9 t_7^8 -  y_{s}^7 y_{o_1}^{16} y_{o_2}^{15} y_{o_3}^{15} y_{o_4}^{23} t_1^8 t_2^8 t_3^8 t_4^6 t_5^8 t_6^6 t_7^9 +  y_{s}^9 y_{o_1}^{18} y_{o_2}^{18} y_{o_3}^{18} y_{o_4}^{27} t_1^9 t_2^9 t_3^9 t_4^9 t_5^9 t_6^9 t_7^9
~,~
$
\end{quote}
\endgroup

\subsection{Model 16 \label{app_num_16}}

\begingroup\makeatletter\def\f@size{7}\check@mathfonts
\begin{quote}\raggedright
$
P(t_i,y_s,y_{o_1},y_{o_2},y_{o_3}; \mathcal{M}_{16}) =
1 + y_{s} y_{o_1} y_{o_2} y_{o_3}^2 t_1 t_2 t_4^2 t_5 t_6 +  y_{s} y_{o_1}^2 y_{o_2} y_{o_3} t_1 t_2 t_3 t_4 t_6^2 -  y_{s}^2 y_{o_1}^3 y_{o_2}^2 y_{o_3}^3 t_1^3 t_2 t_3 t_4^3 t_5 t_6^3 -  y_{s}^2 y_{o_1}^3 y_{o_2}^2 y_{o_3}^3 t_1^2 t_2^2 t_3 t_4^3 t_5 t_6^3 -  y_{s}^2 y_{o_1}^3 y_{o_2}^2 y_{o_3}^3 t_1 t_2^3 t_3 t_4^3 t_5 t_6^3 +  y_{s} y_{o_1} y_{o_2}^2 y_{o_3}^3 t_1 t_2 t_4^2 t_5^2 t_7 +  y_{s} y_{o_1}^2 y_{o_2}^2 y_{o_3}^2 t_1^2 t_3 t_4 t_5 t_6 t_7 +  y_{s} y_{o_1}^2 y_{o_2}^2 y_{o_3}^2 t_1 t_2 t_3 t_4 t_5 t_6 t_7 +  y_{s} y_{o_1}^2 y_{o_2}^2 y_{o_3}^2 t_2^2 t_3 t_4 t_5 t_6 t_7 -  y_{s}^2 y_{o_1}^2 y_{o_2}^3 y_{o_3}^5 t_1^3 t_2 t_4^4 t_5^3 t_6 t_7 -  y_{s}^2 y_{o_1}^2 y_{o_2}^3 y_{o_3}^5 t_1^2 t_2^2 t_4^4 t_5^3 t_6 t_7 -  y_{s}^2 y_{o_1}^2 y_{o_2}^3 y_{o_3}^5 t_1 t_2^3 t_4^4 t_5^3 t_6 t_7 +  y_{s} y_{o_1}^3 y_{o_2}^2 y_{o_3} t_1 t_2 t_3^2 t_6^2 t_7 -  y_{s}^2 y_{o_1}^3 y_{o_2}^3 y_{o_3}^4 t_1^4 t_3 t_4^3 t_5^2 t_6^2 t_7 -  2 y_{s}^2 y_{o_1}^3 y_{o_2}^3 y_{o_3}^4 t_1^3 t_2 t_3 t_4^3 t_5^2 t_6^2 t_7 -  3 y_{s}^2 y_{o_1}^3 y_{o_2}^3 y_{o_3}^4 t_1^2 t_2^2 t_3 t_4^3 t_5^2 t_6^2 t_7 -  2 y_{s}^2 y_{o_1}^3 y_{o_2}^3 y_{o_3}^4 t_1 t_2^3 t_3 t_4^3 t_5^2 t_6^2 t_7 -  y_{s}^2 y_{o_1}^3 y_{o_2}^3 y_{o_3}^4 t_2^4 t_3 t_4^3 t_5^2 t_6^2 t_7 -  y_{s}^2 y_{o_1}^4 y_{o_2}^3 y_{o_3}^3 t_1^4 t_3^2 t_4^2 t_5 t_6^3 t_7 -  2 y_{s}^2 y_{o_1}^4 y_{o_2}^3 y_{o_3}^3 t_1^3 t_2 t_3^2 t_4^2 t_5 t_6^3 t_7 -  3 y_{s}^2 y_{o_1}^4 y_{o_2}^3 y_{o_3}^3 t_1^2 t_2^2 t_3^2 t_4^2 t_5 t_6^3 t_7 -  2 y_{s}^2 y_{o_1}^4 y_{o_2}^3 y_{o_3}^3 t_1 t_2^3 t_3^2 t_4^2 t_5 t_6^3 t_7 -  y_{s}^2 y_{o_1}^4 y_{o_2}^3 y_{o_3}^3 t_2^4 t_3^2 t_4^2 t_5 t_6^3 t_7 +  y_{s}^3 y_{o_1}^4 y_{o_2}^4 y_{o_3}^6 t_1^5 t_2 t_3 t_4^5 t_5^3 t_6^3 t_7 +  2 y_{s}^3 y_{o_1}^4 y_{o_2}^4 y_{o_3}^6 t_1^4 t_2^2 t_3 t_4^5 t_5^3 t_6^3 t_7 +  2 y_{s}^3 y_{o_1}^4 y_{o_2}^4 y_{o_3}^6 t_1^3 t_2^3 t_3 t_4^5 t_5^3 t_6^3 t_7 +  2 y_{s}^3 y_{o_1}^4 y_{o_2}^4 y_{o_3}^6 t_1^2 t_2^4 t_3 t_4^5 t_5^3 t_6^3 t_7 +  y_{s}^3 y_{o_1}^4 y_{o_2}^4 y_{o_3}^6 t_1 t_2^5 t_3 t_4^5 t_5^3 t_6^3 t_7 -  y_{s}^2 y_{o_1}^5 y_{o_2}^3 y_{o_3}^2 t_1^3 t_2 t_3^3 t_4 t_6^4 t_7 -  y_{s}^2 y_{o_1}^5 y_{o_2}^3 y_{o_3}^2 t_1^2 t_2^2 t_3^3 t_4 t_6^4 t_7 -  y_{s}^2 y_{o_1}^5 y_{o_2}^3 y_{o_3}^2 t_1 t_2^3 t_3^3 t_4 t_6^4 t_7 +  2 y_{s}^3 y_{o_1}^5 y_{o_2}^4 y_{o_3}^5 t_1^4 t_2^2 t_3^2 t_4^4 t_5^2 t_6^4 t_7 +  y_{s}^3 y_{o_1}^5 y_{o_2}^4 y_{o_3}^5 t_1^3 t_2^3 t_3^2 t_4^4 t_5^2 t_6^4 t_7 +  2 y_{s}^3 y_{o_1}^5 y_{o_2}^4 y_{o_3}^5 t_1^2 t_2^4 t_3^2 t_4^4 t_5^2 t_6^4 t_7 +  y_{s}^3 y_{o_1}^6 y_{o_2}^4 y_{o_3}^4 t_1^5 t_2 t_3^3 t_4^3 t_5 t_6^5 t_7 +  2 y_{s}^3 y_{o_1}^6 y_{o_2}^4 y_{o_3}^4 t_1^4 t_2^2 t_3^3 t_4^3 t_5 t_6^5 t_7 +  2 y_{s}^3 y_{o_1}^6 y_{o_2}^4 y_{o_3}^4 t_1^3 t_2^3 t_3^3 t_4^3 t_5 t_6^5 t_7 +  2 y_{s}^3 y_{o_1}^6 y_{o_2}^4 y_{o_3}^4 t_1^2 t_2^4 t_3^3 t_4^3 t_5 t_6^5 t_7 +  y_{s}^3 y_{o_1}^6 y_{o_2}^4 y_{o_3}^4 t_1 t_2^5 t_3^3 t_4^3 t_5 t_6^5 t_7 -  y_{s}^4 y_{o_1}^6 y_{o_2}^5 y_{o_3}^7 t_1^4 t_2^4 t_3^2 t_4^6 t_5^3 t_6^5 t_7 -  y_{s}^4 y_{o_1}^7 y_{o_2}^5 y_{o_3}^6 t_1^4 t_2^4 t_3^3 t_4^5 t_5^2 t_6^6 t_7 +  y_{s} y_{o_1}^2 y_{o_2}^3 y_{o_3}^3 t_1^2 t_3 t_4 t_5^2 t_7^2 +  y_{s} y_{o_1}^2 y_{o_2}^3 y_{o_3}^3 t_1 t_2 t_3 t_4 t_5^2 t_7^2 +  y_{s} y_{o_1}^2 y_{o_2}^3 y_{o_3}^3 t_2^2 t_3 t_4 t_5^2 t_7^2 +  y_{s} y_{o_1}^3 y_{o_2}^3 y_{o_3}^2 t_1^2 t_3^2 t_5 t_6 t_7^2 +  y_{s} y_{o_1}^3 y_{o_2}^3 y_{o_3}^2 t_1 t_2 t_3^2 t_5 t_6 t_7^2 +  y_{s} y_{o_1}^3 y_{o_2}^3 y_{o_3}^2 t_2^2 t_3^2 t_5 t_6 t_7^2 -  y_{s}^2 y_{o_1}^3 y_{o_2}^4 y_{o_3}^5 t_1^4 t_3 t_4^3 t_5^3 t_6 t_7^2 -  y_{s}^2 y_{o_1}^3 y_{o_2}^4 y_{o_3}^5 t_1^3 t_2 t_3 t_4^3 t_5^3 t_6 t_7^2 -  3 y_{s}^2 y_{o_1}^3 y_{o_2}^4 y_{o_3}^5 t_1^2 t_2^2 t_3 t_4^3 t_5^3 t_6 t_7^2 -  y_{s}^2 y_{o_1}^3 y_{o_2}^4 y_{o_3}^5 t_1 t_2^3 t_3 t_4^3 t_5^3 t_6 t_7^2 -  y_{s}^2 y_{o_1}^3 y_{o_2}^4 y_{o_3}^5 t_2^4 t_3 t_4^3 t_5^3 t_6 t_7^2 -  2 y_{s}^2 y_{o_1}^4 y_{o_2}^4 y_{o_3}^4 t_1^4 t_3^2 t_4^2 t_5^2 t_6^2 t_7^2 -  3 y_{s}^2 y_{o_1}^4 y_{o_2}^4 y_{o_3}^4 t_1^3 t_2 t_3^2 t_4^2 t_5^2 t_6^2 t_7^2 -  5 y_{s}^2 y_{o_1}^4 y_{o_2}^4 y_{o_3}^4 t_1^2 t_2^2 t_3^2 t_4^2 t_5^2 t_6^2 t_7^2 -  3 y_{s}^2 y_{o_1}^4 y_{o_2}^4 y_{o_3}^4 t_1 t_2^3 t_3^2 t_4^2 t_5^2 t_6^2 t_7^2 -  2 y_{s}^2 y_{o_1}^4 y_{o_2}^4 y_{o_3}^4 t_2^4 t_3^2 t_4^2 t_5^2 t_6^2 t_7^2 +  2 y_{s}^3 y_{o_1}^4 y_{o_2}^5 y_{o_3}^7 t_1^4 t_2^2 t_3 t_4^5 t_5^4 t_6^2 t_7^2 +  y_{s}^3 y_{o_1}^4 y_{o_2}^5 y_{o_3}^7 t_1^3 t_2^3 t_3 t_4^5 t_5^4 t_6^2 t_7^2 +  2 y_{s}^3 y_{o_1}^4 y_{o_2}^5 y_{o_3}^7 t_1^2 t_2^4 t_3 t_4^5 t_5^4 t_6^2 t_7^2 -  y_{s}^2 y_{o_1}^5 y_{o_2}^4 y_{o_3}^3 t_1^4 t_3^3 t_4 t_5 t_6^3 t_7^2 -  y_{s}^2 y_{o_1}^5 y_{o_2}^4 y_{o_3}^3 t_1^3 t_2 t_3^3 t_4 t_5 t_6^3 t_7^2 -  3 y_{s}^2 y_{o_1}^5 y_{o_2}^4 y_{o_3}^3 t_1^2 t_2^2 t_3^3 t_4 t_5 t_6^3 t_7^2 -  y_{s}^2 y_{o_1}^5 y_{o_2}^4 y_{o_3}^3 t_1 t_2^3 t_3^3 t_4 t_5 t_6^3 t_7^2 -  y_{s}^2 y_{o_1}^5 y_{o_2}^4 y_{o_3}^3 t_2^4 t_3^3 t_4 t_5 t_6^3 t_7^2 +  y_{s}^3 y_{o_1}^5 y_{o_2}^5 y_{o_3}^6 t_1^6 t_3^2 t_4^4 t_5^3 t_6^3 t_7^2 +  2 y_{s}^3 y_{o_1}^5 y_{o_2}^5 y_{o_3}^6 t_1^5 t_2 t_3^2 t_4^4 t_5^3 t_6^3 t_7^2 +  6 y_{s}^3 y_{o_1}^5 y_{o_2}^5 y_{o_3}^6 t_1^4 t_2^2 t_3^2 t_4^4 t_5^3 t_6^3 t_7^2 +  4 y_{s}^3 y_{o_1}^5 y_{o_2}^5 y_{o_3}^6 t_1^3 t_2^3 t_3^2 t_4^4 t_5^3 t_6^3 t_7^2 +  6 y_{s}^3 y_{o_1}^5 y_{o_2}^5 y_{o_3}^6 t_1^2 t_2^4 t_3^2 t_4^4 t_5^3 t_6^3 t_7^2 +  2 y_{s}^3 y_{o_1}^5 y_{o_2}^5 y_{o_3}^6 t_1 t_2^5 t_3^2 t_4^4 t_5^3 t_6^3 t_7^2 +  y_{s}^3 y_{o_1}^5 y_{o_2}^5 y_{o_3}^6 t_2^6 t_3^2 t_4^4 t_5^3 t_6^3 t_7^2 -  y_{s}^4 y_{o_1}^5 y_{o_2}^6 y_{o_3}^9 t_1^4 t_2^4 t_3 t_4^7 t_5^5 t_6^3 t_7^2 +  y_{s}^3 y_{o_1}^6 y_{o_2}^5 y_{o_3}^5 t_1^6 t_3^3 t_4^3 t_5^2 t_6^4 t_7^2 +  2 y_{s}^3 y_{o_1}^6 y_{o_2}^5 y_{o_3}^5 t_1^5 t_2 t_3^3 t_4^3 t_5^2 t_6^4 t_7^2 +  6 y_{s}^3 y_{o_1}^6 y_{o_2}^5 y_{o_3}^5 t_1^4 t_2^2 t_3^3 t_4^3 t_5^2 t_6^4 t_7^2 +  4 y_{s}^3 y_{o_1}^6 y_{o_2}^5 y_{o_3}^5 t_1^3 t_2^3 t_3^3 t_4^3 t_5^2 t_6^4 t_7^2 +  6 y_{s}^3 y_{o_1}^6 y_{o_2}^5 y_{o_3}^5 t_1^2 t_2^4 t_3^3 t_4^3 t_5^2 t_6^4 t_7^2 +  2 y_{s}^3 y_{o_1}^6 y_{o_2}^5 y_{o_3}^5 t_1 t_2^5 t_3^3 t_4^3 t_5^2 t_6^4 t_7^2 +  y_{s}^3 y_{o_1}^6 y_{o_2}^5 y_{o_3}^5 t_2^6 t_3^3 t_4^3 t_5^2 t_6^4 t_7^2 -  y_{s}^4 y_{o_1}^6 y_{o_2}^6 y_{o_3}^8 t_1^6 t_2^2 t_3^2 t_4^6 t_5^4 t_6^4 t_7^2 -  3 y_{s}^4 y_{o_1}^6 y_{o_2}^6 y_{o_3}^8 t_1^4 t_2^4 t_3^2 t_4^6 t_5^4 t_6^4 t_7^2 -  y_{s}^4 y_{o_1}^6 y_{o_2}^6 y_{o_3}^8 t_1^2 t_2^6 t_3^2 t_4^6 t_5^4 t_6^4 t_7^2 +  2 y_{s}^3 y_{o_1}^7 y_{o_2}^5 y_{o_3}^4 t_1^4 t_2^2 t_3^4 t_4^2 t_5 t_6^5 t_7^2 +  y_{s}^3 y_{o_1}^7 y_{o_2}^5 y_{o_3}^4 t_1^3 t_2^3 t_3^4 t_4^2 t_5 t_6^5 t_7^2 +  2 y_{s}^3 y_{o_1}^7 y_{o_2}^5 y_{o_3}^4 t_1^2 t_2^4 t_3^4 t_4^2 t_5 t_6^5 t_7^2 -  y_{s}^4 y_{o_1}^7 y_{o_2}^6 y_{o_3}^7 t_1^7 t_2 t_3^3 t_4^5 t_5^3 t_6^5 t_7^2 -  3 y_{s}^4 y_{o_1}^7 y_{o_2}^6 y_{o_3}^7 t_1^6 t_2^2 t_3^3 t_4^5 t_5^3 t_6^5 t_7^2 -  3 y_{s}^4 y_{o_1}^7 y_{o_2}^6 y_{o_3}^7 t_1^5 t_2^3 t_3^3 t_4^5 t_5^3 t_6^5 t_7^2 -  7 y_{s}^4 y_{o_1}^7 y_{o_2}^6 y_{o_3}^7 t_1^4 t_2^4 t_3^3 t_4^5 t_5^3 t_6^5 t_7^2 -  3 y_{s}^4 y_{o_1}^7 y_{o_2}^6 y_{o_3}^7 t_1^3 t_2^5 t_3^3 t_4^5 t_5^3 t_6^5 t_7^2 -  3 y_{s}^4 y_{o_1}^7 y_{o_2}^6 y_{o_3}^7 t_1^2 t_2^6 t_3^3 t_4^5 t_5^3 t_6^5 t_7^2 -  y_{s}^4 y_{o_1}^7 y_{o_2}^6 y_{o_3}^7 t_1 t_2^7 t_3^3 t_4^5 t_5^3 t_6^5 t_7^2 -  y_{s}^5 y_{o_1}^7 y_{o_2}^7 y_{o_3}^{10} t_1^5 t_2^5 t_3^2 t_4^8 t_5^5 t_6^5 t_7^2 -  y_{s}^4 y_{o_1}^8 y_{o_2}^6 y_{o_3}^6 t_1^6 t_2^2 t_3^4 t_4^4 t_5^2 t_6^6 t_7^2 -  3 y_{s}^4 y_{o_1}^8 y_{o_2}^6 y_{o_3}^6 t_1^4 t_2^4 t_3^4 t_4^4 t_5^2 t_6^6 t_7^2 -  y_{s}^4 y_{o_1}^8 y_{o_2}^6 y_{o_3}^6 t_1^2 t_2^6 t_3^4 t_4^4 t_5^2 t_6^6 t_7^2 +  y_{s}^5 y_{o_1}^8 y_{o_2}^7 y_{o_3}^9 t_1^6 t_2^4 t_3^3 t_4^7 t_5^4 t_6^6 t_7^2 +  y_{s}^5 y_{o_1}^8 y_{o_2}^7 y_{o_3}^9 t_1^4 t_2^6 t_3^3 t_4^7 t_5^4 t_6^6 t_7^2 -  y_{s}^4 y_{o_1}^9 y_{o_2}^6 y_{o_3}^5 t_1^4 t_2^4 t_3^5 t_4^3 t_5 t_6^7 t_7^2 +  y_{s}^5 y_{o_1}^9 y_{o_2}^7 y_{o_3}^8 t_1^6 t_2^4 t_3^4 t_4^6 t_5^3 t_6^7 t_7^2 +  y_{s}^5 y_{o_1}^9 y_{o_2}^7 y_{o_3}^8 t_1^4 t_2^6 t_3^4 t_4^6 t_5^3 t_6^7 t_7^2 -  y_{s}^5 y_{o_1}^{10} y_{o_2}^7 y_{o_3}^7 t_1^5 t_2^5 t_3^5 t_4^5 t_5^2 t_6^8 t_7^2 +  y_{s} y_{o_1}^3 y_{o_2}^4 y_{o_3}^3 t_1 t_2 t_3^2 t_5^2 t_7^3 -  y_{s}^2 y_{o_1}^3 y_{o_2}^5 y_{o_3}^6 t_1^2 t_2^2 t_3 t_4^3 t_5^4 t_7^3 -  y_{s}^2 y_{o_1}^4 y_{o_2}^5 y_{o_3}^5 t_1^4 t_3^2 t_4^2 t_5^3 t_6 t_7^3 -  2 y_{s}^2 y_{o_1}^4 y_{o_2}^5 y_{o_3}^5 t_1^3 t_2 t_3^2 t_4^2 t_5^3 t_6 t_7^3 -  3 y_{s}^2 y_{o_1}^4 y_{o_2}^5 y_{o_3}^5 t_1^2 t_2^2 t_3^2 t_4^2 t_5^3 t_6 t_7^3 -  2 y_{s}^2 y_{o_1}^4 y_{o_2}^5 y_{o_3}^5 t_1 t_2^3 t_3^2 t_4^2 t_5^3 t_6 t_7^3 -  y_{s}^2 y_{o_1}^4 y_{o_2}^5 y_{o_3}^5 t_2^4 t_3^2 t_4^2 t_5^3 t_6 t_7^3 +  y_{s}^3 y_{o_1}^4 y_{o_2}^6 y_{o_3}^8 t_1^4 t_2^2 t_3 t_4^5 t_5^5 t_6 t_7^3 +  y_{s}^3 y_{o_1}^4 y_{o_2}^6 y_{o_3}^8 t_1^2 t_2^4 t_3 t_4^5 t_5^5 t_6 t_7^3 -  y_{s}^2 y_{o_1}^5 y_{o_2}^5 y_{o_3}^4 t_1^4 t_3^3 t_4 t_5^2 t_6^2 t_7^3 -  2 y_{s}^2 y_{o_1}^5 y_{o_2}^5 y_{o_3}^4 t_1^3 t_2 t_3^3 t_4 t_5^2 t_6^2 t_7^3 -  3 y_{s}^2 y_{o_1}^5 y_{o_2}^5 y_{o_3}^4 t_1^2 t_2^2 t_3^3 t_4 t_5^2 t_6^2 t_7^3 -  2 y_{s}^2 y_{o_1}^5 y_{o_2}^5 y_{o_3}^4 t_1 t_2^3 t_3^3 t_4 t_5^2 t_6^2 t_7^3 -  y_{s}^2 y_{o_1}^5 y_{o_2}^5 y_{o_3}^4 t_2^4 t_3^3 t_4 t_5^2 t_6^2 t_7^3 +  3 y_{s}^3 y_{o_1}^5 y_{o_2}^6 y_{o_3}^7 t_1^4 t_2^2 t_3^2 t_4^4 t_5^4 t_6^2 t_7^3 +  y_{s}^3 y_{o_1}^5 y_{o_2}^6 y_{o_3}^7 t_1^3 t_2^3 t_3^2 t_4^4 t_5^4 t_6^2 t_7^3 +  3 y_{s}^3 y_{o_1}^5 y_{o_2}^6 y_{o_3}^7 t_1^2 t_2^4 t_3^2 t_4^4 t_5^4 t_6^2 t_7^3 -  y_{s}^4 y_{o_1}^5 y_{o_2}^7 y_{o_3}^{10} t_1^4 t_2^4 t_3 t_4^7 t_5^6 t_6^2 t_7^3 -  y_{s}^2 y_{o_1}^6 y_{o_2}^5 y_{o_3}^3 t_1^2 t_2^2 t_3^4 t_5 t_6^3 t_7^3 +  y_{s}^3 y_{o_1}^6 y_{o_2}^6 y_{o_3}^6 t_1^6 t_3^3 t_4^3 t_5^3 t_6^3 t_7^3 +  2 y_{s}^3 y_{o_1}^6 y_{o_2}^6 y_{o_3}^6 t_1^5 t_2 t_3^3 t_4^3 t_5^3 t_6^3 t_7^3 +  6 y_{s}^3 y_{o_1}^6 y_{o_2}^6 y_{o_3}^6 t_1^4 t_2^2 t_3^3 t_4^3 t_5^3 t_6^3 t_7^3 +  3 y_{s}^3 y_{o_1}^6 y_{o_2}^6 y_{o_3}^6 t_1^3 t_2^3 t_3^3 t_4^3 t_5^3 t_6^3 t_7^3 +  6 y_{s}^3 y_{o_1}^6 y_{o_2}^6 y_{o_3}^6 t_1^2 t_2^4 t_3^3 t_4^3 t_5^3 t_6^3 t_7^3 +  2 y_{s}^3 y_{o_1}^6 y_{o_2}^6 y_{o_3}^6 t_1 t_2^5 t_3^3 t_4^3 t_5^3 t_6^3 t_7^3 +  y_{s}^3 y_{o_1}^6 y_{o_2}^6 y_{o_3}^6 t_2^6 t_3^3 t_4^3 t_5^3 t_6^3 t_7^3 -  y_{s}^4 y_{o_1}^6 y_{o_2}^7 y_{o_3}^9 t_1^6 t_2^2 t_3^2 t_4^6 t_5^5 t_6^3 t_7^3 -  3 y_{s}^4 y_{o_1}^6 y_{o_2}^7 y_{o_3}^9 t_1^4 t_2^4 t_3^2 t_4^6 t_5^5 t_6^3 t_7^3 -  y_{s}^4 y_{o_1}^6 y_{o_2}^7 y_{o_3}^9 t_1^2 t_2^6 t_3^2 t_4^6 t_5^5 t_6^3 t_7^3 +  3 y_{s}^3 y_{o_1}^7 y_{o_2}^6 y_{o_3}^5 t_1^4 t_2^2 t_3^4 t_4^2 t_5^2 t_6^4 t_7^3 +  y_{s}^3 y_{o_1}^7 y_{o_2}^6 y_{o_3}^5 t_1^3 t_2^3 t_3^4 t_4^2 t_5^2 t_6^4 t_7^3 +  3 y_{s}^3 y_{o_1}^7 y_{o_2}^6 y_{o_3}^5 t_1^2 t_2^4 t_3^4 t_4^2 t_5^2 t_6^4 t_7^3 -  2 y_{s}^4 y_{o_1}^7 y_{o_2}^7 y_{o_3}^8 t_1^6 t_2^2 t_3^3 t_4^5 t_5^4 t_6^4 t_7^3 -  6 y_{s}^4 y_{o_1}^7 y_{o_2}^7 y_{o_3}^8 t_1^4 t_2^4 t_3^3 t_4^5 t_5^4 t_6^4 t_7^3 -  2 y_{s}^4 y_{o_1}^7 y_{o_2}^7 y_{o_3}^8 t_1^2 t_2^6 t_3^3 t_4^5 t_5^4 t_6^4 t_7^3 +  y_{s}^5 y_{o_1}^7 y_{o_2}^8 y_{o_3}^{11} t_1^6 t_2^4 t_3^2 t_4^8 t_5^6 t_6^4 t_7^3 +  y_{s}^5 y_{o_1}^7 y_{o_2}^8 y_{o_3}^{11} t_1^4 t_2^6 t_3^2 t_4^8 t_5^6 t_6^4 t_7^3 +  y_{s}^3 y_{o_1}^8 y_{o_2}^6 y_{o_3}^4 t_1^4 t_2^2 t_3^5 t_4 t_5 t_6^5 t_7^3 +  y_{s}^3 y_{o_1}^8 y_{o_2}^6 y_{o_3}^4 t_1^2 t_2^4 t_3^5 t_4 t_5 t_6^5 t_7^3 -  2 y_{s}^4 y_{o_1}^8 y_{o_2}^7 y_{o_3}^7 t_1^6 t_2^2 t_3^4 t_4^4 t_5^3 t_6^5 t_7^3 -  6 y_{s}^4 y_{o_1}^8 y_{o_2}^7 y_{o_3}^7 t_1^4 t_2^4 t_3^4 t_4^4 t_5^3 t_6^5 t_7^3 -  2 y_{s}^4 y_{o_1}^8 y_{o_2}^7 y_{o_3}^7 t_1^2 t_2^6 t_3^4 t_4^4 t_5^3 t_6^5 t_7^3 +  2 y_{s}^5 y_{o_1}^8 y_{o_2}^8 y_{o_3}^{10} t_1^6 t_2^4 t_3^3 t_4^7 t_5^5 t_6^5 t_7^3 -  y_{s}^5 y_{o_1}^8 y_{o_2}^8 y_{o_3}^{10} t_1^5 t_2^5 t_3^3 t_4^7 t_5^5 t_6^5 t_7^3 +  2 y_{s}^5 y_{o_1}^8 y_{o_2}^8 y_{o_3}^{10} t_1^4 t_2^6 t_3^3 t_4^7 t_5^5 t_6^5 t_7^3 -  y_{s}^4 y_{o_1}^9 y_{o_2}^7 y_{o_3}^6 t_1^6 t_2^2 t_3^5 t_4^3 t_5^2 t_6^6 t_7^3 -  3 y_{s}^4 y_{o_1}^9 y_{o_2}^7 y_{o_3}^6 t_1^4 t_2^4 t_3^5 t_4^3 t_5^2 t_6^6 t_7^3 -  y_{s}^4 y_{o_1}^9 y_{o_2}^7 y_{o_3}^6 t_1^2 t_2^6 t_3^5 t_4^3 t_5^2 t_6^6 t_7^3 -  y_{s}^5 y_{o_1}^9 y_{o_2}^8 y_{o_3}^9 t_1^7 t_2^3 t_3^4 t_4^6 t_5^4 t_6^6 t_7^3 +  2 y_{s}^5 y_{o_1}^9 y_{o_2}^8 y_{o_3}^9 t_1^6 t_2^4 t_3^4 t_4^6 t_5^4 t_6^6 t_7^3 -  2 y_{s}^5 y_{o_1}^9 y_{o_2}^8 y_{o_3}^9 t_1^5 t_2^5 t_3^4 t_4^6 t_5^4 t_6^6 t_7^3 +  2 y_{s}^5 y_{o_1}^9 y_{o_2}^8 y_{o_3}^9 t_1^4 t_2^6 t_3^4 t_4^6 t_5^4 t_6^6 t_7^3 -  y_{s}^5 y_{o_1}^9 y_{o_2}^8 y_{o_3}^9 t_1^3 t_2^7 t_3^4 t_4^6 t_5^4 t_6^6 t_7^3 -  y_{s}^6 y_{o_1}^9 y_{o_2}^9 y_{o_3}^{12} t_1^6 t_2^6 t_3^3 t_4^9 t_5^6 t_6^6 t_7^3 -  y_{s}^4 y_{o_1}^{10} y_{o_2}^7 y_{o_3}^5 t_1^4 t_2^4 t_3^6 t_4^2 t_5 t_6^7 t_7^3 +  2 y_{s}^5 y_{o_1}^{10} y_{o_2}^8 y_{o_3}^8 t_1^6 t_2^4 t_3^5 t_4^5 t_5^3 t_6^7 t_7^3 -  y_{s}^5 y_{o_1}^{10} y_{o_2}^8 y_{o_3}^8 t_1^5 t_2^5 t_3^5 t_4^5 t_5^3 t_6^7 t_7^3 +  2 y_{s}^5 y_{o_1}^{10} y_{o_2}^8 y_{o_3}^8 t_1^4 t_2^6 t_3^5 t_4^5 t_5^3 t_6^7 t_7^3 +  y_{s}^6 y_{o_1}^{10} y_{o_2}^9 y_{o_3}^{11} t_1^7 t_2^5 t_3^4 t_4^8 t_5^5 t_6^7 t_7^3 +  y_{s}^6 y_{o_1}^{10} y_{o_2}^9 y_{o_3}^{11} t_1^5 t_2^7 t_3^4 t_4^8 t_5^5 t_6^7 t_7^3 +  y_{s}^5 y_{o_1}^{11} y_{o_2}^8 y_{o_3}^7 t_1^6 t_2^4 t_3^6 t_4^4 t_5^2 t_6^8 t_7^3 +  y_{s}^5 y_{o_1}^{11} y_{o_2}^8 y_{o_3}^7 t_1^4 t_2^6 t_3^6 t_4^4 t_5^2 t_6^8 t_7^3 +  y_{s}^6 y_{o_1}^{11} y_{o_2}^9 y_{o_3}^{10} t_1^7 t_2^5 t_3^5 t_4^7 t_5^4 t_6^8 t_7^3 +  y_{s}^6 y_{o_1}^{11} y_{o_2}^9 y_{o_3}^{10} t_1^5 t_2^7 t_3^5 t_4^7 t_5^4 t_6^8 t_7^3 -  y_{s}^6 y_{o_1}^{12} y_{o_2}^9 y_{o_3}^9 t_1^6 t_2^6 t_3^6 t_4^6 t_5^3 t_6^9 t_7^3 -  y_{s}^2 y_{o_1}^4 y_{o_2}^6 y_{o_3}^6 t_1^3 t_2 t_3^2 t_4^2 t_5^4 t_7^4 -  y_{s}^2 y_{o_1}^4 y_{o_2}^6 y_{o_3}^6 t_1^2 t_2^2 t_3^2 t_4^2 t_5^4 t_7^4 -  y_{s}^2 y_{o_1}^4 y_{o_2}^6 y_{o_3}^6 t_1 t_2^3 t_3^2 t_4^2 t_5^4 t_7^4 -  y_{s}^2 y_{o_1}^5 y_{o_2}^6 y_{o_3}^5 t_1^2 t_2^2 t_3^3 t_4 t_5^3 t_6 t_7^4 +  y_{s}^3 y_{o_1}^5 y_{o_2}^7 y_{o_3}^8 t_1^5 t_2 t_3^2 t_4^4 t_5^5 t_6 t_7^4 +  2 y_{s}^3 y_{o_1}^5 y_{o_2}^7 y_{o_3}^8 t_1^4 t_2^2 t_3^2 t_4^4 t_5^5 t_6 t_7^4 +  2 y_{s}^3 y_{o_1}^5 y_{o_2}^7 y_{o_3}^8 t_1^3 t_2^3 t_3^2 t_4^4 t_5^5 t_6 t_7^4 +  2 y_{s}^3 y_{o_1}^5 y_{o_2}^7 y_{o_3}^8 t_1^2 t_2^4 t_3^2 t_4^4 t_5^5 t_6 t_7^4 +  y_{s}^3 y_{o_1}^5 y_{o_2}^7 y_{o_3}^8 t_1 t_2^5 t_3^2 t_4^4 t_5^5 t_6 t_7^4 -  y_{s}^2 y_{o_1}^6 y_{o_2}^6 y_{o_3}^4 t_1^3 t_2 t_3^4 t_5^2 t_6^2 t_7^4 -  y_{s}^2 y_{o_1}^6 y_{o_2}^6 y_{o_3}^4 t_1^2 t_2^2 t_3^4 t_5^2 t_6^2 t_7^4 -  y_{s}^2 y_{o_1}^6 y_{o_2}^6 y_{o_3}^4 t_1 t_2^3 t_3^4 t_5^2 t_6^2 t_7^4 +  y_{s}^3 y_{o_1}^6 y_{o_2}^7 y_{o_3}^7 t_1^5 t_2 t_3^3 t_4^3 t_5^4 t_6^2 t_7^4 +  3 y_{s}^3 y_{o_1}^6 y_{o_2}^7 y_{o_3}^7 t_1^4 t_2^2 t_3^3 t_4^3 t_5^4 t_6^2 t_7^4 +  y_{s}^3 y_{o_1}^6 y_{o_2}^7 y_{o_3}^7 t_1^3 t_2^3 t_3^3 t_4^3 t_5^4 t_6^2 t_7^4 +  3 y_{s}^3 y_{o_1}^6 y_{o_2}^7 y_{o_3}^7 t_1^2 t_2^4 t_3^3 t_4^3 t_5^4 t_6^2 t_7^4 +  y_{s}^3 y_{o_1}^6 y_{o_2}^7 y_{o_3}^7 t_1 t_2^5 t_3^3 t_4^3 t_5^4 t_6^2 t_7^4 -  y_{s}^4 y_{o_1}^6 y_{o_2}^8 y_{o_3}^{10} t_1^4 t_2^4 t_3^2 t_4^6 t_5^6 t_6^2 t_7^4 +  y_{s}^3 y_{o_1}^7 y_{o_2}^7 y_{o_3}^6 t_1^5 t_2 t_3^4 t_4^2 t_5^3 t_6^3 t_7^4 +  3 y_{s}^3 y_{o_1}^7 y_{o_2}^7 y_{o_3}^6 t_1^4 t_2^2 t_3^4 t_4^2 t_5^3 t_6^3 t_7^4 +  y_{s}^3 y_{o_1}^7 y_{o_2}^7 y_{o_3}^6 t_1^3 t_2^3 t_3^4 t_4^2 t_5^3 t_6^3 t_7^4 +  3 y_{s}^3 y_{o_1}^7 y_{o_2}^7 y_{o_3}^6 t_1^2 t_2^4 t_3^4 t_4^2 t_5^3 t_6^3 t_7^4 +  y_{s}^3 y_{o_1}^7 y_{o_2}^7 y_{o_3}^6 t_1 t_2^5 t_3^4 t_4^2 t_5^3 t_6^3 t_7^4 -  y_{s}^4 y_{o_1}^7 y_{o_2}^8 y_{o_3}^9 t_1^7 t_2 t_3^3 t_4^5 t_5^5 t_6^3 t_7^4 -  2 y_{s}^4 y_{o_1}^7 y_{o_2}^8 y_{o_3}^9 t_1^6 t_2^2 t_3^3 t_4^5 t_5^5 t_6^3 t_7^4 -  y_{s}^4 y_{o_1}^7 y_{o_2}^8 y_{o_3}^9 t_1^5 t_2^3 t_3^3 t_4^5 t_5^5 t_6^3 t_7^4 -  5 y_{s}^4 y_{o_1}^7 y_{o_2}^8 y_{o_3}^9 t_1^4 t_2^4 t_3^3 t_4^5 t_5^5 t_6^3 t_7^4 -  y_{s}^4 y_{o_1}^7 y_{o_2}^8 y_{o_3}^9 t_1^3 t_2^5 t_3^3 t_4^5 t_5^5 t_6^3 t_7^4 -  2 y_{s}^4 y_{o_1}^7 y_{o_2}^8 y_{o_3}^9 t_1^2 t_2^6 t_3^3 t_4^5 t_5^5 t_6^3 t_7^4 -  y_{s}^4 y_{o_1}^7 y_{o_2}^8 y_{o_3}^9 t_1 t_2^7 t_3^3 t_4^5 t_5^5 t_6^3 t_7^4 +  y_{s}^3 y_{o_1}^8 y_{o_2}^7 y_{o_3}^5 t_1^5 t_2 t_3^5 t_4 t_5^2 t_6^4 t_7^4 +  2 y_{s}^3 y_{o_1}^8 y_{o_2}^7 y_{o_3}^5 t_1^4 t_2^2 t_3^5 t_4 t_5^2 t_6^4 t_7^4 +  2 y_{s}^3 y_{o_1}^8 y_{o_2}^7 y_{o_3}^5 t_1^3 t_2^3 t_3^5 t_4 t_5^2 t_6^4 t_7^4 +  2 y_{s}^3 y_{o_1}^8 y_{o_2}^7 y_{o_3}^5 t_1^2 t_2^4 t_3^5 t_4 t_5^2 t_6^4 t_7^4 +  y_{s}^3 y_{o_1}^8 y_{o_2}^7 y_{o_3}^5 t_1 t_2^5 t_3^5 t_4 t_5^2 t_6^4 t_7^4 +  y_{s}^4 y_{o_1}^8 y_{o_2}^8 y_{o_3}^8 t_1^8 t_3^4 t_4^4 t_5^4 t_6^4 t_7^4 +  y_{s}^4 y_{o_1}^8 y_{o_2}^8 y_{o_3}^8 t_1^7 t_2 t_3^4 t_4^4 t_5^4 t_6^4 t_7^4 +  y_{s}^4 y_{o_1}^8 y_{o_2}^8 y_{o_3}^8 t_1^6 t_2^2 t_3^4 t_4^4 t_5^4 t_6^4 t_7^4 +  4 y_{s}^4 y_{o_1}^8 y_{o_2}^8 y_{o_3}^8 t_1^5 t_2^3 t_3^4 t_4^4 t_5^4 t_6^4 t_7^4 -  2 y_{s}^4 y_{o_1}^8 y_{o_2}^8 y_{o_3}^8 t_1^4 t_2^4 t_3^4 t_4^4 t_5^4 t_6^4 t_7^4 +  4 y_{s}^4 y_{o_1}^8 y_{o_2}^8 y_{o_3}^8 t_1^3 t_2^5 t_3^4 t_4^4 t_5^4 t_6^4 t_7^4 +  y_{s}^4 y_{o_1}^8 y_{o_2}^8 y_{o_3}^8 t_1^2 t_2^6 t_3^4 t_4^4 t_5^4 t_6^4 t_7^4 +  y_{s}^4 y_{o_1}^8 y_{o_2}^8 y_{o_3}^8 t_1 t_2^7 t_3^4 t_4^4 t_5^4 t_6^4 t_7^4 +  y_{s}^4 y_{o_1}^8 y_{o_2}^8 y_{o_3}^8 t_2^8 t_3^4 t_4^4 t_5^4 t_6^4 t_7^4 +  y_{s}^5 y_{o_1}^8 y_{o_2}^9 y_{o_3}^{11} t_1^6 t_2^4 t_3^3 t_4^7 t_5^6 t_6^4 t_7^4 -  y_{s}^5 y_{o_1}^8 y_{o_2}^9 y_{o_3}^{11} t_1^5 t_2^5 t_3^3 t_4^7 t_5^6 t_6^4 t_7^4 +  y_{s}^5 y_{o_1}^8 y_{o_2}^9 y_{o_3}^{11} t_1^4 t_2^6 t_3^3 t_4^7 t_5^6 t_6^4 t_7^4 -  y_{s}^4 y_{o_1}^9 y_{o_2}^8 y_{o_3}^7 t_1^7 t_2 t_3^5 t_4^3 t_5^3 t_6^5 t_7^4 -  2 y_{s}^4 y_{o_1}^9 y_{o_2}^8 y_{o_3}^7 t_1^6 t_2^2 t_3^5 t_4^3 t_5^3 t_6^5 t_7^4 -  y_{s}^4 y_{o_1}^9 y_{o_2}^8 y_{o_3}^7 t_1^5 t_2^3 t_3^5 t_4^3 t_5^3 t_6^5 t_7^4 -  5 y_{s}^4 y_{o_1}^9 y_{o_2}^8 y_{o_3}^7 t_1^4 t_2^4 t_3^5 t_4^3 t_5^3 t_6^5 t_7^4 -  y_{s}^4 y_{o_1}^9 y_{o_2}^8 y_{o_3}^7 t_1^3 t_2^5 t_3^5 t_4^3 t_5^3 t_6^5 t_7^4 -  2 y_{s}^4 y_{o_1}^9 y_{o_2}^8 y_{o_3}^7 t_1^2 t_2^6 t_3^5 t_4^3 t_5^3 t_6^5 t_7^4 -  y_{s}^4 y_{o_1}^9 y_{o_2}^8 y_{o_3}^7 t_1 t_2^7 t_3^5 t_4^3 t_5^3 t_6^5 t_7^4 -  y_{s}^5 y_{o_1}^9 y_{o_2}^9 y_{o_3}^{10} t_1^8 t_2^2 t_3^4 t_4^6 t_5^5 t_6^5 t_7^4 -  2 y_{s}^5 y_{o_1}^9 y_{o_2}^9 y_{o_3}^{10} t_1^7 t_2^3 t_3^4 t_4^6 t_5^5 t_6^5 t_7^4 -  5 y_{s}^5 y_{o_1}^9 y_{o_2}^9 y_{o_3}^{10} t_1^5 t_2^5 t_3^4 t_4^6 t_5^5 t_6^5 t_7^4 -  2 y_{s}^5 y_{o_1}^9 y_{o_2}^9 y_{o_3}^{10} t_1^3 t_2^7 t_3^4 t_4^6 t_5^5 t_6^5 t_7^4 -  y_{s}^5 y_{o_1}^9 y_{o_2}^9 y_{o_3}^{10} t_1^2 t_2^8 t_3^4 t_4^6 t_5^5 t_6^5 t_7^4 -  y_{s}^4 y_{o_1}^{10} y_{o_2}^8 y_{o_3}^6 t_1^4 t_2^4 t_3^6 t_4^2 t_5^2 t_6^6 t_7^4 -  y_{s}^5 y_{o_1}^{10} y_{o_2}^9 y_{o_3}^9 t_1^8 t_2^2 t_3^5 t_4^5 t_5^4 t_6^6 t_7^4 -  2 y_{s}^5 y_{o_1}^{10} y_{o_2}^9 y_{o_3}^9 t_1^7 t_2^3 t_3^5 t_4^5 t_5^4 t_6^6 t_7^4 -  5 y_{s}^5 y_{o_1}^{10} y_{o_2}^9 y_{o_3}^9 t_1^5 t_2^5 t_3^5 t_4^5 t_5^4 t_6^6 t_7^4 -  2 y_{s}^5 y_{o_1}^{10} y_{o_2}^9 y_{o_3}^9 t_1^3 t_2^7 t_3^5 t_4^5 t_5^4 t_6^6 t_7^4 -  y_{s}^5 y_{o_1}^{10} y_{o_2}^9 y_{o_3}^9 t_1^2 t_2^8 t_3^5 t_4^5 t_5^4 t_6^6 t_7^4 +  y_{s}^6 y_{o_1}^{10} y_{o_2}^{10} y_{o_3}^{12} t_1^7 t_2^5 t_3^4 t_4^8 t_5^6 t_6^6 t_7^4 -  y_{s}^6 y_{o_1}^{10} y_{o_2}^{10} y_{o_3}^{12} t_1^6 t_2^6 t_3^4 t_4^8 t_5^6 t_6^6 t_7^4 +  y_{s}^6 y_{o_1}^{10} y_{o_2}^{10} y_{o_3}^{12} t_1^5 t_2^7 t_3^4 t_4^8 t_5^6 t_6^6 t_7^4 +  y_{s}^5 y_{o_1}^{11} y_{o_2}^9 y_{o_3}^8 t_1^6 t_2^4 t_3^6 t_4^4 t_5^3 t_6^7 t_7^4 -  y_{s}^5 y_{o_1}^{11} y_{o_2}^9 y_{o_3}^8 t_1^5 t_2^5 t_3^6 t_4^4 t_5^3 t_6^7 t_7^4 +  y_{s}^5 y_{o_1}^{11} y_{o_2}^9 y_{o_3}^8 t_1^4 t_2^6 t_3^6 t_4^4 t_5^3 t_6^7 t_7^4 +  y_{s}^6 y_{o_1}^{11} y_{o_2}^{10} y_{o_3}^{11} t_1^8 t_2^4 t_3^5 t_4^7 t_5^5 t_6^7 t_7^4 +  3 y_{s}^6 y_{o_1}^{11} y_{o_2}^{10} y_{o_3}^{11} t_1^7 t_2^5 t_3^5 t_4^7 t_5^5 t_6^7 t_7^4 +  y_{s}^6 y_{o_1}^{11} y_{o_2}^{10} y_{o_3}^{11} t_1^6 t_2^6 t_3^5 t_4^7 t_5^5 t_6^7 t_7^4 +  3 y_{s}^6 y_{o_1}^{11} y_{o_2}^{10} y_{o_3}^{11} t_1^5 t_2^7 t_3^5 t_4^7 t_5^5 t_6^7 t_7^4 +  y_{s}^6 y_{o_1}^{11} y_{o_2}^{10} y_{o_3}^{11} t_1^4 t_2^8 t_3^5 t_4^7 t_5^5 t_6^7 t_7^4 +  y_{s}^6 y_{o_1}^{12} y_{o_2}^{10} y_{o_3}^{10} t_1^7 t_2^5 t_3^6 t_4^6 t_5^4 t_6^8 t_7^4 -  y_{s}^6 y_{o_1}^{12} y_{o_2}^{10} y_{o_3}^{10} t_1^6 t_2^6 t_3^6 t_4^6 t_5^4 t_6^8 t_7^4 +  y_{s}^6 y_{o_1}^{12} y_{o_2}^{10} y_{o_3}^{10} t_1^5 t_2^7 t_3^6 t_4^6 t_5^4 t_6^8 t_7^4 -  y_{s}^7 y_{o_1}^{12} y_{o_2}^{11} y_{o_3}^{13} t_1^7 t_2^7 t_3^5 t_4^9 t_5^6 t_6^8 t_7^4 -  y_{s}^7 y_{o_1}^{13} y_{o_2}^{11} y_{o_3}^{12} t_1^7 t_2^7 t_3^6 t_4^8 t_5^5 t_6^9 t_7^4 -  y_{s}^2 y_{o_1}^5 y_{o_2}^7 y_{o_3}^6 t_1^2 t_2^2 t_3^3 t_4 t_5^4 t_7^5 -  y_{s}^2 y_{o_1}^6 y_{o_2}^7 y_{o_3}^5 t_1^2 t_2^2 t_3^4 t_5^3 t_6 t_7^5 +  y_{s}^3 y_{o_1}^6 y_{o_2}^8 y_{o_3}^8 t_1^4 t_2^2 t_3^3 t_4^3 t_5^5 t_6 t_7^5 -  y_{s}^3 y_{o_1}^6 y_{o_2}^8 y_{o_3}^8 t_1^3 t_2^3 t_3^3 t_4^3 t_5^5 t_6 t_7^5 +  y_{s}^3 y_{o_1}^6 y_{o_2}^8 y_{o_3}^8 t_1^2 t_2^4 t_3^3 t_4^3 t_5^5 t_6 t_7^5 +  y_{s}^3 y_{o_1}^7 y_{o_2}^8 y_{o_3}^7 t_1^5 t_2 t_3^4 t_4^2 t_5^4 t_6^2 t_7^5 +  3 y_{s}^3 y_{o_1}^7 y_{o_2}^8 y_{o_3}^7 t_1^4 t_2^2 t_3^4 t_4^2 t_5^4 t_6^2 t_7^5 +  y_{s}^3 y_{o_1}^7 y_{o_2}^8 y_{o_3}^7 t_1^3 t_2^3 t_3^4 t_4^2 t_5^4 t_6^2 t_7^5 +  3 y_{s}^3 y_{o_1}^7 y_{o_2}^8 y_{o_3}^7 t_1^2 t_2^4 t_3^4 t_4^2 t_5^4 t_6^2 t_7^5 +  y_{s}^3 y_{o_1}^7 y_{o_2}^8 y_{o_3}^7 t_1 t_2^5 t_3^4 t_4^2 t_5^4 t_6^2 t_7^5 +  y_{s}^4 y_{o_1}^7 y_{o_2}^9 y_{o_3}^{10} t_1^5 t_2^3 t_3^3 t_4^5 t_5^6 t_6^2 t_7^5 -  y_{s}^4 y_{o_1}^7 y_{o_2}^9 y_{o_3}^{10} t_1^4 t_2^4 t_3^3 t_4^5 t_5^6 t_6^2 t_7^5 +  y_{s}^4 y_{o_1}^7 y_{o_2}^9 y_{o_3}^{10} t_1^3 t_2^5 t_3^3 t_4^5 t_5^6 t_6^2 t_7^5 +  y_{s}^3 y_{o_1}^8 y_{o_2}^8 y_{o_3}^6 t_1^4 t_2^2 t_3^5 t_4 t_5^3 t_6^3 t_7^5 -  y_{s}^3 y_{o_1}^8 y_{o_2}^8 y_{o_3}^6 t_1^3 t_2^3 t_3^5 t_4 t_5^3 t_6^3 t_7^5 +  y_{s}^3 y_{o_1}^8 y_{o_2}^8 y_{o_3}^6 t_1^2 t_2^4 t_3^5 t_4 t_5^3 t_6^3 t_7^5 -  y_{s}^4 y_{o_1}^8 y_{o_2}^9 y_{o_3}^9 t_1^7 t_2 t_3^4 t_4^4 t_5^5 t_6^3 t_7^5 -  2 y_{s}^4 y_{o_1}^8 y_{o_2}^9 y_{o_3}^9 t_1^6 t_2^2 t_3^4 t_4^4 t_5^5 t_6^3 t_7^5 -  5 y_{s}^4 y_{o_1}^8 y_{o_2}^9 y_{o_3}^9 t_1^4 t_2^4 t_3^4 t_4^4 t_5^5 t_6^3 t_7^5 -  2 y_{s}^4 y_{o_1}^8 y_{o_2}^9 y_{o_3}^9 t_1^2 t_2^6 t_3^4 t_4^4 t_5^5 t_6^3 t_7^5 -  y_{s}^4 y_{o_1}^8 y_{o_2}^9 y_{o_3}^9 t_1 t_2^7 t_3^4 t_4^4 t_5^5 t_6^3 t_7^5 -  y_{s}^5 y_{o_1}^8 y_{o_2}^{10} y_{o_3}^{12} t_1^5 t_2^5 t_3^3 t_4^7 t_5^7 t_6^3 t_7^5 -  y_{s}^4 y_{o_1}^9 y_{o_2}^9 y_{o_3}^8 t_1^7 t_2 t_3^5 t_4^3 t_5^4 t_6^4 t_7^5 -  2 y_{s}^4 y_{o_1}^9 y_{o_2}^9 y_{o_3}^8 t_1^6 t_2^2 t_3^5 t_4^3 t_5^4 t_6^4 t_7^5 -  5 y_{s}^4 y_{o_1}^9 y_{o_2}^9 y_{o_3}^8 t_1^4 t_2^4 t_3^5 t_4^3 t_5^4 t_6^4 t_7^5 -  2 y_{s}^4 y_{o_1}^9 y_{o_2}^9 y_{o_3}^8 t_1^2 t_2^6 t_3^5 t_4^3 t_5^4 t_6^4 t_7^5 -  y_{s}^4 y_{o_1}^9 y_{o_2}^9 y_{o_3}^8 t_1 t_2^7 t_3^5 t_4^3 t_5^4 t_6^4 t_7^5 -  y_{s}^5 y_{o_1}^9 y_{o_2}^{10} y_{o_3}^{11} t_1^8 t_2^2 t_3^4 t_4^6 t_5^6 t_6^4 t_7^5 -  2 y_{s}^5 y_{o_1}^9 y_{o_2}^{10} y_{o_3}^{11} t_1^7 t_2^3 t_3^4 t_4^6 t_5^6 t_6^4 t_7^5 -  y_{s}^5 y_{o_1}^9 y_{o_2}^{10} y_{o_3}^{11} t_1^6 t_2^4 t_3^4 t_4^6 t_5^6 t_6^4 t_7^5 -  5 y_{s}^5 y_{o_1}^9 y_{o_2}^{10} y_{o_3}^{11} t_1^5 t_2^5 t_3^4 t_4^6 t_5^6 t_6^4 t_7^5 -  y_{s}^5 y_{o_1}^9 y_{o_2}^{10} y_{o_3}^{11} t_1^4 t_2^6 t_3^4 t_4^6 t_5^6 t_6^4 t_7^5 -  2 y_{s}^5 y_{o_1}^9 y_{o_2}^{10} y_{o_3}^{11} t_1^3 t_2^7 t_3^4 t_4^6 t_5^6 t_6^4 t_7^5 -  y_{s}^5 y_{o_1}^9 y_{o_2}^{10} y_{o_3}^{11} t_1^2 t_2^8 t_3^4 t_4^6 t_5^6 t_6^4 t_7^5 +  y_{s}^4 y_{o_1}^{10} y_{o_2}^9 y_{o_3}^7 t_1^5 t_2^3 t_3^6 t_4^2 t_5^3 t_6^5 t_7^5 -  y_{s}^4 y_{o_1}^{10} y_{o_2}^9 y_{o_3}^7 t_1^4 t_2^4 t_3^6 t_4^2 t_5^3 t_6^5 t_7^5 +  y_{s}^4 y_{o_1}^{10} y_{o_2}^9 y_{o_3}^7 t_1^3 t_2^5 t_3^6 t_4^2 t_5^3 t_6^5 t_7^5 +  y_{s}^5 y_{o_1}^{10} y_{o_2}^{10} y_{o_3}^{10} t_1^9 t_2 t_3^5 t_4^5 t_5^5 t_6^5 t_7^5 +  y_{s}^5 y_{o_1}^{10} y_{o_2}^{10} y_{o_3}^{10} t_1^8 t_2^2 t_3^5 t_4^5 t_5^5 t_6^5 t_7^5 +  y_{s}^5 y_{o_1}^{10} y_{o_2}^{10} y_{o_3}^{10} t_1^7 t_2^3 t_3^5 t_4^5 t_5^5 t_6^5 t_7^5 +  4 y_{s}^5 y_{o_1}^{10} y_{o_2}^{10} y_{o_3}^{10} t_1^6 t_2^4 t_3^5 t_4^5 t_5^5 t_6^5 t_7^5 -  2 y_{s}^5 y_{o_1}^{10} y_{o_2}^{10} y_{o_3}^{10} t_1^5 t_2^5 t_3^5 t_4^5 t_5^5 t_6^5 t_7^5 +  4 y_{s}^5 y_{o_1}^{10} y_{o_2}^{10} y_{o_3}^{10} t_1^4 t_2^6 t_3^5 t_4^5 t_5^5 t_6^5 t_7^5 +  y_{s}^5 y_{o_1}^{10} y_{o_2}^{10} y_{o_3}^{10} t_1^3 t_2^7 t_3^5 t_4^5 t_5^5 t_6^5 t_7^5 +  y_{s}^5 y_{o_1}^{10} y_{o_2}^{10} y_{o_3}^{10} t_1^2 t_2^8 t_3^5 t_4^5 t_5^5 t_6^5 t_7^5 +  y_{s}^5 y_{o_1}^{10} y_{o_2}^{10} y_{o_3}^{10} t_1 t_2^9 t_3^5 t_4^5 t_5^5 t_6^5 t_7^5 +  y_{s}^6 y_{o_1}^{10} y_{o_2}^{11} y_{o_3}^{13} t_1^8 t_2^4 t_3^4 t_4^8 t_5^7 t_6^5 t_7^5 +  2 y_{s}^6 y_{o_1}^{10} y_{o_2}^{11} y_{o_3}^{13} t_1^7 t_2^5 t_3^4 t_4^8 t_5^7 t_6^5 t_7^5 +  2 y_{s}^6 y_{o_1}^{10} y_{o_2}^{11} y_{o_3}^{13} t_1^6 t_2^6 t_3^4 t_4^8 t_5^7 t_6^5 t_7^5 +  2 y_{s}^6 y_{o_1}^{10} y_{o_2}^{11} y_{o_3}^{13} t_1^5 t_2^7 t_3^4 t_4^8 t_5^7 t_6^5 t_7^5 +  y_{s}^6 y_{o_1}^{10} y_{o_2}^{11} y_{o_3}^{13} t_1^4 t_2^8 t_3^4 t_4^8 t_5^7 t_6^5 t_7^5 -  y_{s}^5 y_{o_1}^{11} y_{o_2}^{10} y_{o_3}^9 t_1^8 t_2^2 t_3^6 t_4^4 t_5^4 t_6^6 t_7^5 -  2 y_{s}^5 y_{o_1}^{11} y_{o_2}^{10} y_{o_3}^9 t_1^7 t_2^3 t_3^6 t_4^4 t_5^4 t_6^6 t_7^5 -  y_{s}^5 y_{o_1}^{11} y_{o_2}^{10} y_{o_3}^9 t_1^6 t_2^4 t_3^6 t_4^4 t_5^4 t_6^6 t_7^5 -  5 y_{s}^5 y_{o_1}^{11} y_{o_2}^{10} y_{o_3}^9 t_1^5 t_2^5 t_3^6 t_4^4 t_5^4 t_6^6 t_7^5 -  y_{s}^5 y_{o_1}^{11} y_{o_2}^{10} y_{o_3}^9 t_1^4 t_2^6 t_3^6 t_4^4 t_5^4 t_6^6 t_7^5 -  2 y_{s}^5 y_{o_1}^{11} y_{o_2}^{10} y_{o_3}^9 t_1^3 t_2^7 t_3^6 t_4^4 t_5^4 t_6^6 t_7^5 -  y_{s}^5 y_{o_1}^{11} y_{o_2}^{10} y_{o_3}^9 t_1^2 t_2^8 t_3^6 t_4^4 t_5^4 t_6^6 t_7^5 +  y_{s}^6 y_{o_1}^{11} y_{o_2}^{11} y_{o_3}^{12} t_1^8 t_2^4 t_3^5 t_4^7 t_5^6 t_6^6 t_7^5 +  3 y_{s}^6 y_{o_1}^{11} y_{o_2}^{11} y_{o_3}^{12} t_1^7 t_2^5 t_3^5 t_4^7 t_5^6 t_6^6 t_7^5 +  y_{s}^6 y_{o_1}^{11} y_{o_2}^{11} y_{o_3}^{12} t_1^6 t_2^6 t_3^5 t_4^7 t_5^6 t_6^6 t_7^5 +  3 y_{s}^6 y_{o_1}^{11} y_{o_2}^{11} y_{o_3}^{12} t_1^5 t_2^7 t_3^5 t_4^7 t_5^6 t_6^6 t_7^5 +  y_{s}^6 y_{o_1}^{11} y_{o_2}^{11} y_{o_3}^{12} t_1^4 t_2^8 t_3^5 t_4^7 t_5^6 t_6^6 t_7^5 -  y_{s}^5 y_{o_1}^{12} y_{o_2}^{10} y_{o_3}^8 t_1^5 t_2^5 t_3^7 t_4^3 t_5^3 t_6^7 t_7^5 +  y_{s}^6 y_{o_1}^{12} y_{o_2}^{11} y_{o_3}^{11} t_1^8 t_2^4 t_3^6 t_4^6 t_5^5 t_6^7 t_7^5 +  3 y_{s}^6 y_{o_1}^{12} y_{o_2}^{11} y_{o_3}^{11} t_1^7 t_2^5 t_3^6 t_4^6 t_5^5 t_6^7 t_7^5 +  y_{s}^6 y_{o_1}^{12} y_{o_2}^{11} y_{o_3}^{11} t_1^6 t_2^6 t_3^6 t_4^6 t_5^5 t_6^7 t_7^5 +  3 y_{s}^6 y_{o_1}^{12} y_{o_2}^{11} y_{o_3}^{11} t_1^5 t_2^7 t_3^6 t_4^6 t_5^5 t_6^7 t_7^5 +  y_{s}^6 y_{o_1}^{12} y_{o_2}^{11} y_{o_3}^{11} t_1^4 t_2^8 t_3^6 t_4^6 t_5^5 t_6^7 t_7^5 -  y_{s}^7 y_{o_1}^{12} y_{o_2}^{12} y_{o_3}^{14} t_1^8 t_2^6 t_3^5 t_4^9 t_5^7 t_6^7 t_7^5 -  y_{s}^7 y_{o_1}^{12} y_{o_2}^{12} y_{o_3}^{14} t_1^7 t_2^7 t_3^5 t_4^9 t_5^7 t_6^7 t_7^5 -  y_{s}^7 y_{o_1}^{12} y_{o_2}^{12} y_{o_3}^{14} t_1^6 t_2^8 t_3^5 t_4^9 t_5^7 t_6^7 t_7^5 +  y_{s}^6 y_{o_1}^{13} y_{o_2}^{11} y_{o_3}^{10} t_1^8 t_2^4 t_3^7 t_4^5 t_5^4 t_6^8 t_7^5 +  2 y_{s}^6 y_{o_1}^{13} y_{o_2}^{11} y_{o_3}^{10} t_1^7 t_2^5 t_3^7 t_4^5 t_5^4 t_6^8 t_7^5 +  2 y_{s}^6 y_{o_1}^{13} y_{o_2}^{11} y_{o_3}^{10} t_1^6 t_2^6 t_3^7 t_4^5 t_5^4 t_6^8 t_7^5 +  2 y_{s}^6 y_{o_1}^{13} y_{o_2}^{11} y_{o_3}^{10} t_1^5 t_2^7 t_3^7 t_4^5 t_5^4 t_6^8 t_7^5 +  y_{s}^6 y_{o_1}^{13} y_{o_2}^{11} y_{o_3}^{10} t_1^4 t_2^8 t_3^7 t_4^5 t_5^4 t_6^8 t_7^5 -  y_{s}^7 y_{o_1}^{13} y_{o_2}^{12} y_{o_3}^{13} t_1^7 t_2^7 t_3^6 t_4^8 t_5^6 t_6^8 t_7^5 -  y_{s}^7 y_{o_1}^{14} y_{o_2}^{12} y_{o_3}^{12} t_1^8 t_2^6 t_3^7 t_4^7 t_5^5 t_6^9 t_7^5 -  y_{s}^7 y_{o_1}^{14} y_{o_2}^{12} y_{o_3}^{12} t_1^7 t_2^7 t_3^7 t_4^7 t_5^5 t_6^9 t_7^5 -  y_{s}^7 y_{o_1}^{14} y_{o_2}^{12} y_{o_3}^{12} t_1^6 t_2^8 t_3^7 t_4^7 t_5^5 t_6^9 t_7^5 -  y_{s}^3 y_{o_1}^6 y_{o_2}^9 y_{o_3}^9 t_1^3 t_2^3 t_3^3 t_4^3 t_5^6 t_7^6 +  y_{s}^3 y_{o_1}^7 y_{o_2}^9 y_{o_3}^8 t_1^4 t_2^2 t_3^4 t_4^2 t_5^5 t_6 t_7^6 +  y_{s}^3 y_{o_1}^7 y_{o_2}^9 y_{o_3}^8 t_1^2 t_2^4 t_3^4 t_4^2 t_5^5 t_6 t_7^6 +  y_{s}^4 y_{o_1}^7 y_{o_2}^{10} y_{o_3}^{11} t_1^5 t_2^3 t_3^3 t_4^5 t_5^7 t_6 t_7^6 +  y_{s}^4 y_{o_1}^7 y_{o_2}^{10} y_{o_3}^{11} t_1^3 t_2^5 t_3^3 t_4^5 t_5^7 t_6 t_7^6 +  y_{s}^3 y_{o_1}^8 y_{o_2}^9 y_{o_3}^7 t_1^4 t_2^2 t_3^5 t_4 t_5^4 t_6^2 t_7^6 +  y_{s}^3 y_{o_1}^8 y_{o_2}^9 y_{o_3}^7 t_1^2 t_2^4 t_3^5 t_4 t_5^4 t_6^2 t_7^6 +  2 y_{s}^4 y_{o_1}^8 y_{o_2}^{10} y_{o_3}^{10} t_1^5 t_2^3 t_3^4 t_4^4 t_5^6 t_6^2 t_7^6 -  y_{s}^4 y_{o_1}^8 y_{o_2}^{10} y_{o_3}^{10} t_1^4 t_2^4 t_3^4 t_4^4 t_5^6 t_6^2 t_7^6 +  2 y_{s}^4 y_{o_1}^8 y_{o_2}^{10} y_{o_3}^{10} t_1^3 t_2^5 t_3^4 t_4^4 t_5^6 t_6^2 t_7^6 -  y_{s}^5 y_{o_1}^8 y_{o_2}^{11} y_{o_3}^{13} t_1^5 t_2^5 t_3^3 t_4^7 t_5^8 t_6^2 t_7^6 -  y_{s}^3 y_{o_1}^9 y_{o_2}^9 y_{o_3}^6 t_1^3 t_2^3 t_3^6 t_5^3 t_6^3 t_7^6 -  y_{s}^4 y_{o_1}^9 y_{o_2}^{10} y_{o_3}^9 t_1^6 t_2^2 t_3^5 t_4^3 t_5^5 t_6^3 t_7^6 +  2 y_{s}^4 y_{o_1}^9 y_{o_2}^{10} y_{o_3}^9 t_1^5 t_2^3 t_3^5 t_4^3 t_5^5 t_6^3 t_7^6 -  2 y_{s}^4 y_{o_1}^9 y_{o_2}^{10} y_{o_3}^9 t_1^4 t_2^4 t_3^5 t_4^3 t_5^5 t_6^3 t_7^6 +  2 y_{s}^4 y_{o_1}^9 y_{o_2}^{10} y_{o_3}^9 t_1^3 t_2^5 t_3^5 t_4^3 t_5^5 t_6^3 t_7^6 -  y_{s}^4 y_{o_1}^9 y_{o_2}^{10} y_{o_3}^9 t_1^2 t_2^6 t_3^5 t_4^3 t_5^5 t_6^3 t_7^6 -  y_{s}^5 y_{o_1}^9 y_{o_2}^{11} y_{o_3}^{12} t_1^7 t_2^3 t_3^4 t_4^6 t_5^7 t_6^3 t_7^6 -  3 y_{s}^5 y_{o_1}^9 y_{o_2}^{11} y_{o_3}^{12} t_1^5 t_2^5 t_3^4 t_4^6 t_5^7 t_6^3 t_7^6 -  y_{s}^5 y_{o_1}^9 y_{o_2}^{11} y_{o_3}^{12} t_1^3 t_2^7 t_3^4 t_4^6 t_5^7 t_6^3 t_7^6 +  2 y_{s}^4 y_{o_1}^{10} y_{o_2}^{10} y_{o_3}^8 t_1^5 t_2^3 t_3^6 t_4^2 t_5^4 t_6^4 t_7^6 -  y_{s}^4 y_{o_1}^{10} y_{o_2}^{10} y_{o_3}^8 t_1^4 t_2^4 t_3^6 t_4^2 t_5^4 t_6^4 t_7^6 +  2 y_{s}^4 y_{o_1}^{10} y_{o_2}^{10} y_{o_3}^8 t_1^3 t_2^5 t_3^6 t_4^2 t_5^4 t_6^4 t_7^6 -  2 y_{s}^5 y_{o_1}^{10} y_{o_2}^{11} y_{o_3}^{11} t_1^7 t_2^3 t_3^5 t_4^5 t_5^6 t_6^4 t_7^6 -  6 y_{s}^5 y_{o_1}^{10} y_{o_2}^{11} y_{o_3}^{11} t_1^5 t_2^5 t_3^5 t_4^5 t_5^6 t_6^4 t_7^6 -  2 y_{s}^5 y_{o_1}^{10} y_{o_2}^{11} y_{o_3}^{11} t_1^3 t_2^7 t_3^5 t_4^5 t_5^6 t_6^4 t_7^6 +  y_{s}^6 y_{o_1}^{10} y_{o_2}^{12} y_{o_3}^{14} t_1^7 t_2^5 t_3^4 t_4^8 t_5^8 t_6^4 t_7^6 +  y_{s}^6 y_{o_1}^{10} y_{o_2}^{12} y_{o_3}^{14} t_1^5 t_2^7 t_3^4 t_4^8 t_5^8 t_6^4 t_7^6 +  y_{s}^4 y_{o_1}^{11} y_{o_2}^{10} y_{o_3}^7 t_1^5 t_2^3 t_3^7 t_4 t_5^3 t_6^5 t_7^6 +  y_{s}^4 y_{o_1}^{11} y_{o_2}^{10} y_{o_3}^7 t_1^3 t_2^5 t_3^7 t_4 t_5^3 t_6^5 t_7^6 -  2 y_{s}^5 y_{o_1}^{11} y_{o_2}^{11} y_{o_3}^{10} t_1^7 t_2^3 t_3^6 t_4^4 t_5^5 t_6^5 t_7^6 -  6 y_{s}^5 y_{o_1}^{11} y_{o_2}^{11} y_{o_3}^{10} t_1^5 t_2^5 t_3^6 t_4^4 t_5^5 t_6^5 t_7^6 -  2 y_{s}^5 y_{o_1}^{11} y_{o_2}^{11} y_{o_3}^{10} t_1^3 t_2^7 t_3^6 t_4^4 t_5^5 t_6^5 t_7^6 +  3 y_{s}^6 y_{o_1}^{11} y_{o_2}^{12} y_{o_3}^{13} t_1^7 t_2^5 t_3^5 t_4^7 t_5^7 t_6^5 t_7^6 +  y_{s}^6 y_{o_1}^{11} y_{o_2}^{12} y_{o_3}^{13} t_1^6 t_2^6 t_3^5 t_4^7 t_5^7 t_6^5 t_7^6 +  3 y_{s}^6 y_{o_1}^{11} y_{o_2}^{12} y_{o_3}^{13} t_1^5 t_2^7 t_3^5 t_4^7 t_5^7 t_6^5 t_7^6 -  y_{s}^5 y_{o_1}^{12} y_{o_2}^{11} y_{o_3}^9 t_1^7 t_2^3 t_3^7 t_4^3 t_5^4 t_6^6 t_7^6 -  3 y_{s}^5 y_{o_1}^{12} y_{o_2}^{11} y_{o_3}^9 t_1^5 t_2^5 t_3^7 t_4^3 t_5^4 t_6^6 t_7^6 -  y_{s}^5 y_{o_1}^{12} y_{o_2}^{11} y_{o_3}^9 t_1^3 t_2^7 t_3^7 t_4^3 t_5^4 t_6^6 t_7^6 +  y_{s}^6 y_{o_1}^{12} y_{o_2}^{12} y_{o_3}^{12} t_1^9 t_2^3 t_3^6 t_4^6 t_5^6 t_6^6 t_7^6 +  2 y_{s}^6 y_{o_1}^{12} y_{o_2}^{12} y_{o_3}^{12} t_1^8 t_2^4 t_3^6 t_4^6 t_5^6 t_6^6 t_7^6 +  6 y_{s}^6 y_{o_1}^{12} y_{o_2}^{12} y_{o_3}^{12} t_1^7 t_2^5 t_3^6 t_4^6 t_5^6 t_6^6 t_7^6 +  3 y_{s}^6 y_{o_1}^{12} y_{o_2}^{12} y_{o_3}^{12} t_1^6 t_2^6 t_3^6 t_4^6 t_5^6 t_6^6 t_7^6 +  6 y_{s}^6 y_{o_1}^{12} y_{o_2}^{12} y_{o_3}^{12} t_1^5 t_2^7 t_3^6 t_4^6 t_5^6 t_6^6 t_7^6 +  2 y_{s}^6 y_{o_1}^{12} y_{o_2}^{12} y_{o_3}^{12} t_1^4 t_2^8 t_3^6 t_4^6 t_5^6 t_6^6 t_7^6 +  y_{s}^6 y_{o_1}^{12} y_{o_2}^{12} y_{o_3}^{12} t_1^3 t_2^9 t_3^6 t_4^6 t_5^6 t_6^6 t_7^6 -  y_{s}^7 y_{o_1}^{12} y_{o_2}^{13} y_{o_3}^{15} t_1^7 t_2^7 t_3^5 t_4^9 t_5^8 t_6^6 t_7^6 -  y_{s}^5 y_{o_1}^{13} y_{o_2}^{11} y_{o_3}^8 t_1^5 t_2^5 t_3^8 t_4^2 t_5^3 t_6^7 t_7^6 +  3 y_{s}^6 y_{o_1}^{13} y_{o_2}^{12} y_{o_3}^{11} t_1^7 t_2^5 t_3^7 t_4^5 t_5^5 t_6^7 t_7^6 +  y_{s}^6 y_{o_1}^{13} y_{o_2}^{12} y_{o_3}^{11} t_1^6 t_2^6 t_3^7 t_4^5 t_5^5 t_6^7 t_7^6 +  3 y_{s}^6 y_{o_1}^{13} y_{o_2}^{12} y_{o_3}^{11} t_1^5 t_2^7 t_3^7 t_4^5 t_5^5 t_6^7 t_7^6 -  y_{s}^7 y_{o_1}^{13} y_{o_2}^{13} y_{o_3}^{14} t_1^9 t_2^5 t_3^6 t_4^8 t_5^7 t_6^7 t_7^6 -  2 y_{s}^7 y_{o_1}^{13} y_{o_2}^{13} y_{o_3}^{14} t_1^8 t_2^6 t_3^6 t_4^8 t_5^7 t_6^7 t_7^6 -  3 y_{s}^7 y_{o_1}^{13} y_{o_2}^{13} y_{o_3}^{14} t_1^7 t_2^7 t_3^6 t_4^8 t_5^7 t_6^7 t_7^6 -  2 y_{s}^7 y_{o_1}^{13} y_{o_2}^{13} y_{o_3}^{14} t_1^6 t_2^8 t_3^6 t_4^8 t_5^7 t_6^7 t_7^6 -  y_{s}^7 y_{o_1}^{13} y_{o_2}^{13} y_{o_3}^{14} t_1^5 t_2^9 t_3^6 t_4^8 t_5^7 t_6^7 t_7^6 +  y_{s}^6 y_{o_1}^{14} y_{o_2}^{12} y_{o_3}^{10} t_1^7 t_2^5 t_3^8 t_4^4 t_5^4 t_6^8 t_7^6 +  y_{s}^6 y_{o_1}^{14} y_{o_2}^{12} y_{o_3}^{10} t_1^5 t_2^7 t_3^8 t_4^4 t_5^4 t_6^8 t_7^6 -  y_{s}^7 y_{o_1}^{14} y_{o_2}^{13} y_{o_3}^{13} t_1^9 t_2^5 t_3^7 t_4^7 t_5^6 t_6^8 t_7^6 -  2 y_{s}^7 y_{o_1}^{14} y_{o_2}^{13} y_{o_3}^{13} t_1^8 t_2^6 t_3^7 t_4^7 t_5^6 t_6^8 t_7^6 -  3 y_{s}^7 y_{o_1}^{14} y_{o_2}^{13} y_{o_3}^{13} t_1^7 t_2^7 t_3^7 t_4^7 t_5^6 t_6^8 t_7^6 -  2 y_{s}^7 y_{o_1}^{14} y_{o_2}^{13} y_{o_3}^{13} t_1^6 t_2^8 t_3^7 t_4^7 t_5^6 t_6^8 t_7^6 -  y_{s}^7 y_{o_1}^{14} y_{o_2}^{13} y_{o_3}^{13} t_1^5 t_2^9 t_3^7 t_4^7 t_5^6 t_6^8 t_7^6 -  y_{s}^7 y_{o_1}^{15} y_{o_2}^{13} y_{o_3}^{12} t_1^7 t_2^7 t_3^8 t_4^6 t_5^5 t_6^9 t_7^6 +  y_{s}^8 y_{o_1}^{15} y_{o_2}^{14} y_{o_3}^{15} t_1^8 t_2^8 t_3^7 t_4^9 t_5^7 t_6^9 t_7^6 -  y_{s}^4 y_{o_1}^8 y_{o_2}^{11} y_{o_3}^{11} t_1^4 t_2^4 t_3^4 t_4^4 t_5^7 t_6 t_7^7 +  y_{s}^4 y_{o_1}^9 y_{o_2}^{11} y_{o_3}^{10} t_1^5 t_2^3 t_3^5 t_4^3 t_5^6 t_6^2 t_7^7 +  y_{s}^4 y_{o_1}^9 y_{o_2}^{11} y_{o_3}^{10} t_1^3 t_2^5 t_3^5 t_4^3 t_5^6 t_6^2 t_7^7 -  y_{s}^5 y_{o_1}^9 y_{o_2}^{12} y_{o_3}^{13} t_1^5 t_2^5 t_3^4 t_4^6 t_5^8 t_6^2 t_7^7 +  y_{s}^4 y_{o_1}^{10} y_{o_2}^{11} y_{o_3}^9 t_1^5 t_2^3 t_3^6 t_4^2 t_5^5 t_6^3 t_7^7 +  y_{s}^4 y_{o_1}^{10} y_{o_2}^{11} y_{o_3}^9 t_1^3 t_2^5 t_3^6 t_4^2 t_5^5 t_6^3 t_7^7 -  y_{s}^5 y_{o_1}^{10} y_{o_2}^{12} y_{o_3}^{12} t_1^7 t_2^3 t_3^5 t_4^5 t_5^7 t_6^3 t_7^7 -  3 y_{s}^5 y_{o_1}^{10} y_{o_2}^{12} y_{o_3}^{12} t_1^5 t_2^5 t_3^5 t_4^5 t_5^7 t_6^3 t_7^7 -  y_{s}^5 y_{o_1}^{10} y_{o_2}^{12} y_{o_3}^{12} t_1^3 t_2^7 t_3^5 t_4^5 t_5^7 t_6^3 t_7^7 -  y_{s}^4 y_{o_1}^{11} y_{o_2}^{11} y_{o_3}^8 t_1^4 t_2^4 t_3^7 t_4 t_5^4 t_6^4 t_7^7 -  y_{s}^5 y_{o_1}^{11} y_{o_2}^{12} y_{o_3}^{11} t_1^8 t_2^2 t_3^6 t_4^4 t_5^6 t_6^4 t_7^7 -  3 y_{s}^5 y_{o_1}^{11} y_{o_2}^{12} y_{o_3}^{11} t_1^7 t_2^3 t_3^6 t_4^4 t_5^6 t_6^4 t_7^7 -  3 y_{s}^5 y_{o_1}^{11} y_{o_2}^{12} y_{o_3}^{11} t_1^6 t_2^4 t_3^6 t_4^4 t_5^6 t_6^4 t_7^7 -  7 y_{s}^5 y_{o_1}^{11} y_{o_2}^{12} y_{o_3}^{11} t_1^5 t_2^5 t_3^6 t_4^4 t_5^6 t_6^4 t_7^7 -  3 y_{s}^5 y_{o_1}^{11} y_{o_2}^{12} y_{o_3}^{11} t_1^4 t_2^6 t_3^6 t_4^4 t_5^6 t_6^4 t_7^7 -  3 y_{s}^5 y_{o_1}^{11} y_{o_2}^{12} y_{o_3}^{11} t_1^3 t_2^7 t_3^6 t_4^4 t_5^6 t_6^4 t_7^7 -  y_{s}^5 y_{o_1}^{11} y_{o_2}^{12} y_{o_3}^{11} t_1^2 t_2^8 t_3^6 t_4^4 t_5^6 t_6^4 t_7^7 +  2 y_{s}^6 y_{o_1}^{11} y_{o_2}^{13} y_{o_3}^{14} t_1^7 t_2^5 t_3^5 t_4^7 t_5^8 t_6^4 t_7^7 +  y_{s}^6 y_{o_1}^{11} y_{o_2}^{13} y_{o_3}^{14} t_1^6 t_2^6 t_3^5 t_4^7 t_5^8 t_6^4 t_7^7 +  2 y_{s}^6 y_{o_1}^{11} y_{o_2}^{13} y_{o_3}^{14} t_1^5 t_2^7 t_3^5 t_4^7 t_5^8 t_6^4 t_7^7 -  y_{s}^5 y_{o_1}^{12} y_{o_2}^{12} y_{o_3}^{10} t_1^7 t_2^3 t_3^7 t_4^3 t_5^5 t_6^5 t_7^7 -  3 y_{s}^5 y_{o_1}^{12} y_{o_2}^{12} y_{o_3}^{10} t_1^5 t_2^5 t_3^7 t_4^3 t_5^5 t_6^5 t_7^7 -  y_{s}^5 y_{o_1}^{12} y_{o_2}^{12} y_{o_3}^{10} t_1^3 t_2^7 t_3^7 t_4^3 t_5^5 t_6^5 t_7^7 +  y_{s}^6 y_{o_1}^{12} y_{o_2}^{13} y_{o_3}^{13} t_1^9 t_2^3 t_3^6 t_4^6 t_5^7 t_6^5 t_7^7 +  2 y_{s}^6 y_{o_1}^{12} y_{o_2}^{13} y_{o_3}^{13} t_1^8 t_2^4 t_3^6 t_4^6 t_5^7 t_6^5 t_7^7 +  6 y_{s}^6 y_{o_1}^{12} y_{o_2}^{13} y_{o_3}^{13} t_1^7 t_2^5 t_3^6 t_4^6 t_5^7 t_6^5 t_7^7 +  4 y_{s}^6 y_{o_1}^{12} y_{o_2}^{13} y_{o_3}^{13} t_1^6 t_2^6 t_3^6 t_4^6 t_5^7 t_6^5 t_7^7 +  6 y_{s}^6 y_{o_1}^{12} y_{o_2}^{13} y_{o_3}^{13} t_1^5 t_2^7 t_3^6 t_4^6 t_5^7 t_6^5 t_7^7 +  2 y_{s}^6 y_{o_1}^{12} y_{o_2}^{13} y_{o_3}^{13} t_1^4 t_2^8 t_3^6 t_4^6 t_5^7 t_6^5 t_7^7 +  y_{s}^6 y_{o_1}^{12} y_{o_2}^{13} y_{o_3}^{13} t_1^3 t_2^9 t_3^6 t_4^6 t_5^7 t_6^5 t_7^7 -  y_{s}^5 y_{o_1}^{13} y_{o_2}^{12} y_{o_3}^9 t_1^5 t_2^5 t_3^8 t_4^2 t_5^4 t_6^6 t_7^7 +  y_{s}^6 y_{o_1}^{13} y_{o_2}^{13} y_{o_3}^{12} t_1^9 t_2^3 t_3^7 t_4^5 t_5^6 t_6^6 t_7^7 +  2 y_{s}^6 y_{o_1}^{13} y_{o_2}^{13} y_{o_3}^{12} t_1^8 t_2^4 t_3^7 t_4^5 t_5^6 t_6^6 t_7^7 +  6 y_{s}^6 y_{o_1}^{13} y_{o_2}^{13} y_{o_3}^{12} t_1^7 t_2^5 t_3^7 t_4^5 t_5^6 t_6^6 t_7^7 +  4 y_{s}^6 y_{o_1}^{13} y_{o_2}^{13} y_{o_3}^{12} t_1^6 t_2^6 t_3^7 t_4^5 t_5^6 t_6^6 t_7^7 +  6 y_{s}^6 y_{o_1}^{13} y_{o_2}^{13} y_{o_3}^{12} t_1^5 t_2^7 t_3^7 t_4^5 t_5^6 t_6^6 t_7^7 +  2 y_{s}^6 y_{o_1}^{13} y_{o_2}^{13} y_{o_3}^{12} t_1^4 t_2^8 t_3^7 t_4^5 t_5^6 t_6^6 t_7^7 +  y_{s}^6 y_{o_1}^{13} y_{o_2}^{13} y_{o_3}^{12} t_1^3 t_2^9 t_3^7 t_4^5 t_5^6 t_6^6 t_7^7 -  y_{s}^7 y_{o_1}^{13} y_{o_2}^{14} y_{o_3}^{15} t_1^9 t_2^5 t_3^6 t_4^8 t_5^8 t_6^6 t_7^7 -  y_{s}^7 y_{o_1}^{13} y_{o_2}^{14} y_{o_3}^{15} t_1^8 t_2^6 t_3^6 t_4^8 t_5^8 t_6^6 t_7^7 -  3 y_{s}^7 y_{o_1}^{13} y_{o_2}^{14} y_{o_3}^{15} t_1^7 t_2^7 t_3^6 t_4^8 t_5^8 t_6^6 t_7^7 -  y_{s}^7 y_{o_1}^{13} y_{o_2}^{14} y_{o_3}^{15} t_1^6 t_2^8 t_3^6 t_4^8 t_5^8 t_6^6 t_7^7 -  y_{s}^7 y_{o_1}^{13} y_{o_2}^{14} y_{o_3}^{15} t_1^5 t_2^9 t_3^6 t_4^8 t_5^8 t_6^6 t_7^7 +  2 y_{s}^6 y_{o_1}^{14} y_{o_2}^{13} y_{o_3}^{11} t_1^7 t_2^5 t_3^8 t_4^4 t_5^5 t_6^7 t_7^7 +  y_{s}^6 y_{o_1}^{14} y_{o_2}^{13} y_{o_3}^{11} t_1^6 t_2^6 t_3^8 t_4^4 t_5^5 t_6^7 t_7^7 +  2 y_{s}^6 y_{o_1}^{14} y_{o_2}^{13} y_{o_3}^{11} t_1^5 t_2^7 t_3^8 t_4^4 t_5^5 t_6^7 t_7^7 -  2 y_{s}^7 y_{o_1}^{14} y_{o_2}^{14} y_{o_3}^{14} t_1^9 t_2^5 t_3^7 t_4^7 t_5^7 t_6^7 t_7^7 -  3 y_{s}^7 y_{o_1}^{14} y_{o_2}^{14} y_{o_3}^{14} t_1^8 t_2^6 t_3^7 t_4^7 t_5^7 t_6^7 t_7^7 -  5 y_{s}^7 y_{o_1}^{14} y_{o_2}^{14} y_{o_3}^{14} t_1^7 t_2^7 t_3^7 t_4^7 t_5^7 t_6^7 t_7^7 -  3 y_{s}^7 y_{o_1}^{14} y_{o_2}^{14} y_{o_3}^{14} t_1^6 t_2^8 t_3^7 t_4^7 t_5^7 t_6^7 t_7^7 -  2 y_{s}^7 y_{o_1}^{14} y_{o_2}^{14} y_{o_3}^{14} t_1^5 t_2^9 t_3^7 t_4^7 t_5^7 t_6^7 t_7^7 -  y_{s}^7 y_{o_1}^{15} y_{o_2}^{14} y_{o_3}^{13} t_1^9 t_2^5 t_3^8 t_4^6 t_5^6 t_6^8 t_7^7 -  y_{s}^7 y_{o_1}^{15} y_{o_2}^{14} y_{o_3}^{13} t_1^8 t_2^6 t_3^8 t_4^6 t_5^6 t_6^8 t_7^7 -  3 y_{s}^7 y_{o_1}^{15} y_{o_2}^{14} y_{o_3}^{13} t_1^7 t_2^7 t_3^8 t_4^6 t_5^6 t_6^8 t_7^7 -  y_{s}^7 y_{o_1}^{15} y_{o_2}^{14} y_{o_3}^{13} t_1^6 t_2^8 t_3^8 t_4^6 t_5^6 t_6^8 t_7^7 -  y_{s}^7 y_{o_1}^{15} y_{o_2}^{14} y_{o_3}^{13} t_1^5 t_2^9 t_3^8 t_4^6 t_5^6 t_6^8 t_7^7 +  y_{s}^8 y_{o_1}^{15} y_{o_2}^{15} y_{o_3}^{16} t_1^9 t_2^7 t_3^7 t_4^9 t_5^8 t_6^8 t_7^7 +  y_{s}^8 y_{o_1}^{15} y_{o_2}^{15} y_{o_3}^{16} t_1^8 t_2^8 t_3^7 t_4^9 t_5^8 t_6^8 t_7^7 +  y_{s}^8 y_{o_1}^{15} y_{o_2}^{15} y_{o_3}^{16} t_1^7 t_2^9 t_3^7 t_4^9 t_5^8 t_6^8 t_7^7 +  y_{s}^8 y_{o_1}^{16} y_{o_2}^{15} y_{o_3}^{15} t_1^9 t_2^7 t_3^8 t_4^8 t_5^7 t_6^9 t_7^7 +  y_{s}^8 y_{o_1}^{16} y_{o_2}^{15} y_{o_3}^{15} t_1^8 t_2^8 t_3^8 t_4^8 t_5^7 t_6^9 t_7^7 +  y_{s}^8 y_{o_1}^{16} y_{o_2}^{15} y_{o_3}^{15} t_1^7 t_2^9 t_3^8 t_4^8 t_5^7 t_6^9 t_7^7 -  y_{s}^5 y_{o_1}^{11} y_{o_2}^{13} y_{o_3}^{12} t_1^5 t_2^5 t_3^6 t_4^4 t_5^7 t_6^3 t_7^8 -  y_{s}^5 y_{o_1}^{12} y_{o_2}^{13} y_{o_3}^{11} t_1^5 t_2^5 t_3^7 t_4^3 t_5^6 t_6^4 t_7^8 +  y_{s}^6 y_{o_1}^{12} y_{o_2}^{14} y_{o_3}^{14} t_1^8 t_2^4 t_3^6 t_4^6 t_5^8 t_6^4 t_7^8 +  2 y_{s}^6 y_{o_1}^{12} y_{o_2}^{14} y_{o_3}^{14} t_1^7 t_2^5 t_3^6 t_4^6 t_5^8 t_6^4 t_7^8 +  2 y_{s}^6 y_{o_1}^{12} y_{o_2}^{14} y_{o_3}^{14} t_1^6 t_2^6 t_3^6 t_4^6 t_5^8 t_6^4 t_7^8 +  2 y_{s}^6 y_{o_1}^{12} y_{o_2}^{14} y_{o_3}^{14} t_1^5 t_2^7 t_3^6 t_4^6 t_5^8 t_6^4 t_7^8 +  y_{s}^6 y_{o_1}^{12} y_{o_2}^{14} y_{o_3}^{14} t_1^4 t_2^8 t_3^6 t_4^6 t_5^8 t_6^4 t_7^8 +  2 y_{s}^6 y_{o_1}^{13} y_{o_2}^{14} y_{o_3}^{13} t_1^7 t_2^5 t_3^7 t_4^5 t_5^7 t_6^5 t_7^8 +  y_{s}^6 y_{o_1}^{13} y_{o_2}^{14} y_{o_3}^{13} t_1^6 t_2^6 t_3^7 t_4^5 t_5^7 t_6^5 t_7^8 +  2 y_{s}^6 y_{o_1}^{13} y_{o_2}^{14} y_{o_3}^{13} t_1^5 t_2^7 t_3^7 t_4^5 t_5^7 t_6^5 t_7^8 -  y_{s}^7 y_{o_1}^{13} y_{o_2}^{15} y_{o_3}^{16} t_1^8 t_2^6 t_3^6 t_4^8 t_5^9 t_6^5 t_7^8 -  y_{s}^7 y_{o_1}^{13} y_{o_2}^{15} y_{o_3}^{16} t_1^7 t_2^7 t_3^6 t_4^8 t_5^9 t_6^5 t_7^8 -  y_{s}^7 y_{o_1}^{13} y_{o_2}^{15} y_{o_3}^{16} t_1^6 t_2^8 t_3^6 t_4^8 t_5^9 t_6^5 t_7^8 +  y_{s}^6 y_{o_1}^{14} y_{o_2}^{14} y_{o_3}^{12} t_1^8 t_2^4 t_3^8 t_4^4 t_5^6 t_6^6 t_7^8 +  2 y_{s}^6 y_{o_1}^{14} y_{o_2}^{14} y_{o_3}^{12} t_1^7 t_2^5 t_3^8 t_4^4 t_5^6 t_6^6 t_7^8 +  2 y_{s}^6 y_{o_1}^{14} y_{o_2}^{14} y_{o_3}^{12} t_1^6 t_2^6 t_3^8 t_4^4 t_5^6 t_6^6 t_7^8 +  2 y_{s}^6 y_{o_1}^{14} y_{o_2}^{14} y_{o_3}^{12} t_1^5 t_2^7 t_3^8 t_4^4 t_5^6 t_6^6 t_7^8 +  y_{s}^6 y_{o_1}^{14} y_{o_2}^{14} y_{o_3}^{12} t_1^4 t_2^8 t_3^8 t_4^4 t_5^6 t_6^6 t_7^8 -  y_{s}^7 y_{o_1}^{14} y_{o_2}^{15} y_{o_3}^{15} t_1^9 t_2^5 t_3^7 t_4^7 t_5^8 t_6^6 t_7^8 -  2 y_{s}^7 y_{o_1}^{14} y_{o_2}^{15} y_{o_3}^{15} t_1^8 t_2^6 t_3^7 t_4^7 t_5^8 t_6^6 t_7^8 -  3 y_{s}^7 y_{o_1}^{14} y_{o_2}^{15} y_{o_3}^{15} t_1^7 t_2^7 t_3^7 t_4^7 t_5^8 t_6^6 t_7^8 -  2 y_{s}^7 y_{o_1}^{14} y_{o_2}^{15} y_{o_3}^{15} t_1^6 t_2^8 t_3^7 t_4^7 t_5^8 t_6^6 t_7^8 -  y_{s}^7 y_{o_1}^{14} y_{o_2}^{15} y_{o_3}^{15} t_1^5 t_2^9 t_3^7 t_4^7 t_5^8 t_6^6 t_7^8 -  y_{s}^7 y_{o_1}^{15} y_{o_2}^{15} y_{o_3}^{14} t_1^9 t_2^5 t_3^8 t_4^6 t_5^7 t_6^7 t_7^8 -  2 y_{s}^7 y_{o_1}^{15} y_{o_2}^{15} y_{o_3}^{14} t_1^8 t_2^6 t_3^8 t_4^6 t_5^7 t_6^7 t_7^8 -  3 y_{s}^7 y_{o_1}^{15} y_{o_2}^{15} y_{o_3}^{14} t_1^7 t_2^7 t_3^8 t_4^6 t_5^7 t_6^7 t_7^8 -  2 y_{s}^7 y_{o_1}^{15} y_{o_2}^{15} y_{o_3}^{14} t_1^6 t_2^8 t_3^8 t_4^6 t_5^7 t_6^7 t_7^8 -  y_{s}^7 y_{o_1}^{15} y_{o_2}^{15} y_{o_3}^{14} t_1^5 t_2^9 t_3^8 t_4^6 t_5^7 t_6^7 t_7^8 +  y_{s}^8 y_{o_1}^{15} y_{o_2}^{16} y_{o_3}^{17} t_1^8 t_2^8 t_3^7 t_4^9 t_5^9 t_6^7 t_7^8 -  y_{s}^7 y_{o_1}^{16} y_{o_2}^{15} y_{o_3}^{13} t_1^8 t_2^6 t_3^9 t_4^5 t_5^6 t_6^8 t_7^8 -  y_{s}^7 y_{o_1}^{16} y_{o_2}^{15} y_{o_3}^{13} t_1^7 t_2^7 t_3^9 t_4^5 t_5^6 t_6^8 t_7^8 -  y_{s}^7 y_{o_1}^{16} y_{o_2}^{15} y_{o_3}^{13} t_1^6 t_2^8 t_3^9 t_4^5 t_5^6 t_6^8 t_7^8 +  y_{s}^8 y_{o_1}^{16} y_{o_2}^{16} y_{o_3}^{16} t_1^9 t_2^7 t_3^8 t_4^8 t_5^8 t_6^8 t_7^8 +  y_{s}^8 y_{o_1}^{16} y_{o_2}^{16} y_{o_3}^{16} t_1^8 t_2^8 t_3^8 t_4^8 t_5^8 t_6^8 t_7^8 +  y_{s}^8 y_{o_1}^{16} y_{o_2}^{16} y_{o_3}^{16} t_1^7 t_2^9 t_3^8 t_4^8 t_5^8 t_6^8 t_7^8 +  y_{s}^8 y_{o_1}^{17} y_{o_2}^{16} y_{o_3}^{15} t_1^8 t_2^8 t_3^9 t_4^7 t_5^7 t_6^9 t_7^8 -  y_{s}^7 y_{o_1}^{15} y_{o_2}^{16} y_{o_3}^{15} t_1^8 t_2^6 t_3^8 t_4^6 t_5^8 t_6^6 t_7^9 -  y_{s}^7 y_{o_1}^{15} y_{o_2}^{16} y_{o_3}^{15} t_1^7 t_2^7 t_3^8 t_4^6 t_5^8 t_6^6 t_7^9 -  y_{s}^7 y_{o_1}^{15} y_{o_2}^{16} y_{o_3}^{15} t_1^6 t_2^8 t_3^8 t_4^6 t_5^8 t_6^6 t_7^9 +  y_{s}^8 y_{o_1}^{16} y_{o_2}^{17} y_{o_3}^{17} t_1^8 t_2^8 t_3^8 t_4^8 t_5^9 t_6^7 t_7^9 +  y_{s}^8 y_{o_1}^{17} y_{o_2}^{17} y_{o_3}^{16} t_1^8 t_2^8 t_3^9 t_4^7 t_5^8 t_6^8 t_7^9 +  y_{s}^9 y_{o_1}^{18} y_{o_2}^{18} y_{o_3}^{18} t_1^9 t_2^9 t_3^9 t_4^9 t_5^9 t_6^9 t_7^9
~,~
$
\end{quote}
\endgroup

\subsection{Model 17 \label{app_num_17}}

\begingroup\makeatletter\def\f@size{7}\check@mathfonts
\begin{quote}\raggedright
$
P(t_i,y_s,y_{o_1},y_{o_2},y_{o_3},y_{o_4},y_{o_5},y_{o_6}; \mathcal{M}_{17}) =
1 + y_{s} y_{o_1}^3 y_{o_2} y_{o_3}^3 y_{o_4} y_{o_5}^2 y_{o_6}^2 t_1 t_2 t_3 t_4 t_5^2 t_7^2 +  y_{s} y_{o_1}^3 y_{o_2} y_{o_3}^2 y_{o_4}^2 y_{o_5}^3 y_{o_6} t_1 t_2 t_4^2 t_5 t_6 t_7^2 -  y_{s}^2 y_{o_1}^6 y_{o_2}^2 y_{o_3}^5 y_{o_4}^3 y_{o_5}^5 y_{o_6}^3 t_1^3 t_2 t_3 t_4^3 t_5^3 t_6 t_7^4 - y_{s}^2 y_{o_1}^6 y_{o_2}^2 y_{o_3}^5 y_{o_4}^3 y_{o_5}^5 y_{o_6}^3 t_1^2 t_2^2 t_3 t_4^3 t_5^3 t_6 t_7^4 - y_{s}^2 y_{o_1}^6 y_{o_2}^2 y_{o_3}^5 y_{o_4}^3 y_{o_5}^5 y_{o_6}^3 t_1 t_2^3 t_3 t_4^3 t_5^3 t_6 t_7^4 + y_{s} y_{o_1}^2 y_{o_2}^2 y_{o_3}^3 y_{o_4} y_{o_5} y_{o_6}^3 t_1 t_2 t_3^2 t_5^2 t_7 t_8 +  y_{s} y_{o_1}^2 y_{o_2}^2 y_{o_3}^2 y_{o_4}^2 y_{o_5}^2 y_{o_6}^2 t_1^2 t_3 t_4 t_5 t_6 t_7 t_8 +  y_{s} y_{o_1}^2 y_{o_2}^2 y_{o_3}^2 y_{o_4}^2 y_{o_5}^2 y_{o_6}^2 t_1 t_2 t_3 t_4 t_5 t_6 t_7 t_8 +  y_{s} y_{o_1}^2 y_{o_2}^2 y_{o_3}^2 y_{o_4}^2 y_{o_5}^2 y_{o_6}^2 t_2^2 t_3 t_4 t_5 t_6 t_7 t_8 +  y_{s} y_{o_1}^2 y_{o_2}^2 y_{o_3} y_{o_4}^3 y_{o_5}^3 y_{o_6} t_1 t_2 t_4^2 t_6^2 t_7 t_8 -  y_{s}^2 y_{o_1}^5 y_{o_2}^3 y_{o_3}^6 y_{o_4}^2 y_{o_5}^3 y_{o_6}^5 t_1^3 t_2 t_3^3 t_4 t_5^4 t_7^3 t_8 - y_{s}^2 y_{o_1}^5 y_{o_2}^3 y_{o_3}^6 y_{o_4}^2 y_{o_5}^3 y_{o_6}^5 t_1^2 t_2^2 t_3^3 t_4 t_5^4 t_7^3 t_8 - y_{s}^2 y_{o_1}^5 y_{o_2}^3 y_{o_3}^6 y_{o_4}^2 y_{o_5}^3 y_{o_6}^5 t_1 t_2^3 t_3^3 t_4 t_5^4 t_7^3 t_8 -  y_{s}^2 y_{o_1}^5 y_{o_2}^3 y_{o_3}^5 y_{o_4}^3 y_{o_5}^4 y_{o_6}^4 t_1^4 t_3^2 t_4^2 t_5^3 t_6 t_7^3 t_8 - 2 y_{s}^2 y_{o_1}^5 y_{o_2}^3 y_{o_3}^5 y_{o_4}^3 y_{o_5}^4 y_{o_6}^4 t_1^3 t_2 t_3^2 t_4^2 t_5^3 t_6 t_7^3 t_8 -  3 y_{s}^2 y_{o_1}^5 y_{o_2}^3 y_{o_3}^5 y_{o_4}^3 y_{o_5}^4 y_{o_6}^4 t_1^2 t_2^2 t_3^2 t_4^2 t_5^3 t_6 t_7^3 t_8 -  2 y_{s}^2 y_{o_1}^5 y_{o_2}^3 y_{o_3}^5 y_{o_4}^3 y_{o_5}^4 y_{o_6}^4 t_1 t_2^3 t_3^2 t_4^2 t_5^3 t_6 t_7^3 t_8 -  y_{s}^2 y_{o_1}^5 y_{o_2}^3 y_{o_3}^5 y_{o_4}^3 y_{o_5}^4 y_{o_6}^4 t_2^4 t_3^2 t_4^2 t_5^3 t_6 t_7^3 t_8 - y_{s}^2 y_{o_1}^5 y_{o_2}^3 y_{o_3}^4 y_{o_4}^4 y_{o_5}^5 y_{o_6}^3 t_1^4 t_3 t_4^3 t_5^2 t_6^2 t_7^3 t_8 -  2 y_{s}^2 y_{o_1}^5 y_{o_2}^3 y_{o_3}^4 y_{o_4}^4 y_{o_5}^5 y_{o_6}^3 t_1^3 t_2 t_3 t_4^3 t_5^2 t_6^2 t_7^3 t_8 -  3 y_{s}^2 y_{o_1}^5 y_{o_2}^3 y_{o_3}^4 y_{o_4}^4 y_{o_5}^5 y_{o_6}^3 t_1^2 t_2^2 t_3 t_4^3 t_5^2 t_6^2 t_7^3 t_8 -  2 y_{s}^2 y_{o_1}^5 y_{o_2}^3 y_{o_3}^4 y_{o_4}^4 y_{o_5}^5 y_{o_6}^3 t_1 t_2^3 t_3 t_4^3 t_5^2 t_6^2 t_7^3 t_8 -  y_{s}^2 y_{o_1}^5 y_{o_2}^3 y_{o_3}^4 y_{o_4}^4 y_{o_5}^5 y_{o_6}^3 t_2^4 t_3 t_4^3 t_5^2 t_6^2 t_7^3 t_8 - y_{s}^2 y_{o_1}^5 y_{o_2}^3 y_{o_3}^3 y_{o_4}^5 y_{o_5}^6 y_{o_6}^2 t_1^3 t_2 t_4^4 t_5 t_6^3 t_7^3 t_8 -  y_{s}^2 y_{o_1}^5 y_{o_2}^3 y_{o_3}^3 y_{o_4}^5 y_{o_5}^6 y_{o_6}^2 t_1^2 t_2^2 t_4^4 t_5 t_6^3 t_7^3 t_8 - y_{s}^2 y_{o_1}^5 y_{o_2}^3 y_{o_3}^3 y_{o_4}^5 y_{o_5}^6 y_{o_6}^2 t_1 t_2^3 t_4^4 t_5 t_6^3 t_7^3 t_8 +  y_{s}^3 y_{o_1}^8 y_{o_2}^4 y_{o_3}^8 y_{o_4}^4 y_{o_5}^6 y_{o_6}^6 t_1^5 t_2 t_3^3 t_4^3 t_5^5 t_6 t_7^5 t_8 +  2 y_{s}^3 y_{o_1}^8 y_{o_2}^4 y_{o_3}^8 y_{o_4}^4 y_{o_5}^6 y_{o_6}^6 t_1^4 t_2^2 t_3^3 t_4^3 t_5^5 t_6 t_7^5 t_8 +  2 y_{s}^3 y_{o_1}^8 y_{o_2}^4 y_{o_3}^8 y_{o_4}^4 y_{o_5}^6 y_{o_6}^6 t_1^3 t_2^3 t_3^3 t_4^3 t_5^5 t_6 t_7^5 t_8 +  2 y_{s}^3 y_{o_1}^8 y_{o_2}^4 y_{o_3}^8 y_{o_4}^4 y_{o_5}^6 y_{o_6}^6 t_1^2 t_2^4 t_3^3 t_4^3 t_5^5 t_6 t_7^5 t_8 +  y_{s}^3 y_{o_1}^8 y_{o_2}^4 y_{o_3}^8 y_{o_4}^4 y_{o_5}^6 y_{o_6}^6 t_1 t_2^5 t_3^3 t_4^3 t_5^5 t_6 t_7^5 t_8 +  2 y_{s}^3 y_{o_1}^8 y_{o_2}^4 y_{o_3}^7 y_{o_4}^5 y_{o_5}^7 y_{o_6}^5 t_1^4 t_2^2 t_3^2 t_4^4 t_5^4 t_6^2 t_7^5 t_8 +  y_{s}^3 y_{o_1}^8 y_{o_2}^4 y_{o_3}^7 y_{o_4}^5 y_{o_5}^7 y_{o_6}^5 t_1^3 t_2^3 t_3^2 t_4^4 t_5^4 t_6^2 t_7^5 t_8 +  2 y_{s}^3 y_{o_1}^8 y_{o_2}^4 y_{o_3}^7 y_{o_4}^5 y_{o_5}^7 y_{o_6}^5 t_1^2 t_2^4 t_3^2 t_4^4 t_5^4 t_6^2 t_7^5 t_8 +  y_{s}^3 y_{o_1}^8 y_{o_2}^4 y_{o_3}^6 y_{o_4}^6 y_{o_5}^8 y_{o_6}^4 t_1^5 t_2 t_3 t_4^5 t_5^3 t_6^3 t_7^5 t_8 +  2 y_{s}^3 y_{o_1}^8 y_{o_2}^4 y_{o_3}^6 y_{o_4}^6 y_{o_5}^8 y_{o_6}^4 t_1^4 t_2^2 t_3 t_4^5 t_5^3 t_6^3 t_7^5 t_8 +  2 y_{s}^3 y_{o_1}^8 y_{o_2}^4 y_{o_3}^6 y_{o_4}^6 y_{o_5}^8 y_{o_6}^4 t_1^3 t_2^3 t_3 t_4^5 t_5^3 t_6^3 t_7^5 t_8 +  2 y_{s}^3 y_{o_1}^8 y_{o_2}^4 y_{o_3}^6 y_{o_4}^6 y_{o_5}^8 y_{o_6}^4 t_1^2 t_2^4 t_3 t_4^5 t_5^3 t_6^3 t_7^5 t_8 +  y_{s}^3 y_{o_1}^8 y_{o_2}^4 y_{o_3}^6 y_{o_4}^6 y_{o_5}^8 y_{o_6}^4 t_1 t_2^5 t_3 t_4^5 t_5^3 t_6^3 t_7^5 t_8 -  y_{s}^4 y_{o_1}^{11} y_{o_2}^5 y_{o_3}^{10} y_{o_4}^6 y_{o_5}^9 y_{o_6}^7 t_1^4 t_2^4 t_3^3 t_4^5 t_5^6 t_6^2 t_7^7 t_8 -  y_{s}^4 y_{o_1}^{11} y_{o_2}^5 y_{o_3}^9 y_{o_4}^7 y_{o_5}^{10} y_{o_6}^6 t_1^4 t_2^4 t_3^2 t_4^6 t_5^5 t_6^3 t_7^7 t_8 +  y_{s} y_{o_1} y_{o_2}^3 y_{o_3}^2 y_{o_4}^2 y_{o_5} y_{o_6}^3 t_1 t_2 t_3^2 t_5 t_6 t_8^2 +  y_{s} y_{o_1} y_{o_2}^3 y_{o_3} y_{o_4}^3 y_{o_5}^2 y_{o_6}^2 t_1 t_2 t_3 t_4 t_6^2 t_8^2 -  y_{s}^2 y_{o_1}^4 y_{o_2}^4 y_{o_3}^5 y_{o_4}^3 y_{o_5}^3 y_{o_6}^5 t_1^4 t_3^3 t_4 t_5^3 t_6 t_7^2 t_8^2 - 2 y_{s}^2 y_{o_1}^4 y_{o_2}^4 y_{o_3}^5 y_{o_4}^3 y_{o_5}^3 y_{o_6}^5 t_1^3 t_2 t_3^3 t_4 t_5^3 t_6 t_7^2 t_8^2 -  3 y_{s}^2 y_{o_1}^4 y_{o_2}^4 y_{o_3}^5 y_{o_4}^3 y_{o_5}^3 y_{o_6}^5 t_1^2 t_2^2 t_3^3 t_4 t_5^3 t_6 t_7^2 t_8^2 -  2 y_{s}^2 y_{o_1}^4 y_{o_2}^4 y_{o_3}^5 y_{o_4}^3 y_{o_5}^3 y_{o_6}^5 t_1 t_2^3 t_3^3 t_4 t_5^3 t_6 t_7^2 t_8^2 -  y_{s}^2 y_{o_1}^4 y_{o_2}^4 y_{o_3}^5 y_{o_4}^3 y_{o_5}^3 y_{o_6}^5 t_2^4 t_3^3 t_4 t_5^3 t_6 t_7^2 t_8^2 - 2 y_{s}^2 y_{o_1}^4 y_{o_2}^4 y_{o_3}^4 y_{o_4}^4 y_{o_5}^4 y_{o_6}^4 t_1^4 t_3^2 t_4^2 t_5^2 t_6^2 t_7^2 t_8^2 -  5 y_{s}^2 y_{o_1}^4 y_{o_2}^4 y_{o_3}^4 y_{o_4}^4 y_{o_5}^4 y_{o_6}^4 t_1^3 t_2 t_3^2 t_4^2 t_5^2 t_6^2 t_7^2 t_8^2 -  5 y_{s}^2 y_{o_1}^4 y_{o_2}^4 y_{o_3}^4 y_{o_4}^4 y_{o_5}^4 y_{o_6}^4 t_1^2 t_2^2 t_3^2 t_4^2 t_5^2 t_6^2 t_7^2 t_8^2 -  5 y_{s}^2 y_{o_1}^4 y_{o_2}^4 y_{o_3}^4 y_{o_4}^4 y_{o_5}^4 y_{o_6}^4 t_1 t_2^3 t_3^2 t_4^2 t_5^2 t_6^2 t_7^2 t_8^2 -  2 y_{s}^2 y_{o_1}^4 y_{o_2}^4 y_{o_3}^4 y_{o_4}^4 y_{o_5}^4 y_{o_6}^4 t_2^4 t_3^2 t_4^2 t_5^2 t_6^2 t_7^2 t_8^2 -  y_{s}^2 y_{o_1}^4 y_{o_2}^4 y_{o_3}^3 y_{o_4}^5 y_{o_5}^5 y_{o_6}^3 t_1^4 t_3 t_4^3 t_5 t_6^3 t_7^2 t_8^2 - 2 y_{s}^2 y_{o_1}^4 y_{o_2}^4 y_{o_3}^3 y_{o_4}^5 y_{o_5}^5 y_{o_6}^3 t_1^3 t_2 t_3 t_4^3 t_5 t_6^3 t_7^2 t_8^2 -  3 y_{s}^2 y_{o_1}^4 y_{o_2}^4 y_{o_3}^3 y_{o_4}^5 y_{o_5}^5 y_{o_6}^3 t_1^2 t_2^2 t_3 t_4^3 t_5 t_6^3 t_7^2 t_8^2 -  2 y_{s}^2 y_{o_1}^4 y_{o_2}^4 y_{o_3}^3 y_{o_4}^5 y_{o_5}^5 y_{o_6}^3 t_1 t_2^3 t_3 t_4^3 t_5 t_6^3 t_7^2 t_8^2 -  y_{s}^2 y_{o_1}^4 y_{o_2}^4 y_{o_3}^3 y_{o_4}^5 y_{o_5}^5 y_{o_6}^3 t_2^4 t_3 t_4^3 t_5 t_6^3 t_7^2 t_8^2 + 2 y_{s}^3 y_{o_1}^7 y_{o_2}^5 y_{o_3}^8 y_{o_4}^4 y_{o_5}^5 y_{o_6}^7 t_1^4 t_2^2 t_3^4 t_4^2 t_5^5 t_6 t_7^4 t_8^2 +  y_{s}^3 y_{o_1}^7 y_{o_2}^5 y_{o_3}^8 y_{o_4}^4 y_{o_5}^5 y_{o_6}^7 t_1^3 t_2^3 t_3^4 t_4^2 t_5^5 t_6 t_7^4 t_8^2 +  2 y_{s}^3 y_{o_1}^7 y_{o_2}^5 y_{o_3}^8 y_{o_4}^4 y_{o_5}^5 y_{o_6}^7 t_1^2 t_2^4 t_3^4 t_4^2 t_5^5 t_6 t_7^4 t_8^2 +  y_{s}^3 y_{o_1}^7 y_{o_2}^5 y_{o_3}^7 y_{o_4}^5 y_{o_5}^6 y_{o_6}^6 t_1^6 t_3^3 t_4^3 t_5^4 t_6^2 t_7^4 t_8^2 +  3 y_{s}^3 y_{o_1}^7 y_{o_2}^5 y_{o_3}^7 y_{o_4}^5 y_{o_5}^6 y_{o_6}^6 t_1^5 t_2 t_3^3 t_4^3 t_5^4 t_6^2 t_7^4 t_8^2 +  7 y_{s}^3 y_{o_1}^7 y_{o_2}^5 y_{o_3}^7 y_{o_4}^5 y_{o_5}^6 y_{o_6}^6 t_1^4 t_2^2 t_3^3 t_4^3 t_5^4 t_6^2 t_7^4 t_8^2 +  6 y_{s}^3 y_{o_1}^7 y_{o_2}^5 y_{o_3}^7 y_{o_4}^5 y_{o_5}^6 y_{o_6}^6 t_1^3 t_2^3 t_3^3 t_4^3 t_5^4 t_6^2 t_7^4 t_8^2 +  7 y_{s}^3 y_{o_1}^7 y_{o_2}^5 y_{o_3}^7 y_{o_4}^5 y_{o_5}^6 y_{o_6}^6 t_1^2 t_2^4 t_3^3 t_4^3 t_5^4 t_6^2 t_7^4 t_8^2 +  3 y_{s}^3 y_{o_1}^7 y_{o_2}^5 y_{o_3}^7 y_{o_4}^5 y_{o_5}^6 y_{o_6}^6 t_1 t_2^5 t_3^3 t_4^3 t_5^4 t_6^2 t_7^4 t_8^2 +  y_{s}^3 y_{o_1}^7 y_{o_2}^5 y_{o_3}^7 y_{o_4}^5 y_{o_5}^6 y_{o_6}^6 t_2^6 t_3^3 t_4^3 t_5^4 t_6^2 t_7^4 t_8^2 +  y_{s}^3 y_{o_1}^7 y_{o_2}^5 y_{o_3}^6 y_{o_4}^6 y_{o_5}^7 y_{o_6}^5 t_1^6 t_3^2 t_4^4 t_5^3 t_6^3 t_7^4 t_8^2 +  3 y_{s}^3 y_{o_1}^7 y_{o_2}^5 y_{o_3}^6 y_{o_4}^6 y_{o_5}^7 y_{o_6}^5 t_1^5 t_2 t_3^2 t_4^4 t_5^3 t_6^3 t_7^4 t_8^2 +  7 y_{s}^3 y_{o_1}^7 y_{o_2}^5 y_{o_3}^6 y_{o_4}^6 y_{o_5}^7 y_{o_6}^5 t_1^4 t_2^2 t_3^2 t_4^4 t_5^3 t_6^3 t_7^4 t_8^2 +  6 y_{s}^3 y_{o_1}^7 y_{o_2}^5 y_{o_3}^6 y_{o_4}^6 y_{o_5}^7 y_{o_6}^5 t_1^3 t_2^3 t_3^2 t_4^4 t_5^3 t_6^3 t_7^4 t_8^2 +  7 y_{s}^3 y_{o_1}^7 y_{o_2}^5 y_{o_3}^6 y_{o_4}^6 y_{o_5}^7 y_{o_6}^5 t_1^2 t_2^4 t_3^2 t_4^4 t_5^3 t_6^3 t_7^4 t_8^2 +  3 y_{s}^3 y_{o_1}^7 y_{o_2}^5 y_{o_3}^6 y_{o_4}^6 y_{o_5}^7 y_{o_6}^5 t_1 t_2^5 t_3^2 t_4^4 t_5^3 t_6^3 t_7^4 t_8^2 +  y_{s}^3 y_{o_1}^7 y_{o_2}^5 y_{o_3}^6 y_{o_4}^6 y_{o_5}^7 y_{o_6}^5 t_2^6 t_3^2 t_4^4 t_5^3 t_6^3 t_7^4 t_8^2 +  2 y_{s}^3 y_{o_1}^7 y_{o_2}^5 y_{o_3}^5 y_{o_4}^7 y_{o_5}^8 y_{o_6}^4 t_1^4 t_2^2 t_3 t_4^5 t_5^2 t_6^4 t_7^4 t_8^2 +  y_{s}^3 y_{o_1}^7 y_{o_2}^5 y_{o_3}^5 y_{o_4}^7 y_{o_5}^8 y_{o_6}^4 t_1^3 t_2^3 t_3 t_4^5 t_5^2 t_6^4 t_7^4 t_8^2 +  2 y_{s}^3 y_{o_1}^7 y_{o_2}^5 y_{o_3}^5 y_{o_4}^7 y_{o_5}^8 y_{o_6}^4 t_1^2 t_2^4 t_3 t_4^5 t_5^2 t_6^4 t_7^4 t_8^2 -  y_{s}^4 y_{o_1}^{10} y_{o_2}^6 y_{o_3}^{11} y_{o_4}^5 y_{o_5}^7 y_{o_6}^9 t_1^4 t_2^4 t_3^5 t_4^3 t_5^7 t_6 t_7^6 t_8^2 -  y_{s}^4 y_{o_1}^{10} y_{o_2}^6 y_{o_3}^{10} y_{o_4}^6 y_{o_5}^8 y_{o_6}^8 t_1^6 t_2^2 t_3^4 t_4^4 t_5^6 t_6^2 t_7^6 t_8^2 -  3 y_{s}^4 y_{o_1}^{10} y_{o_2}^6 y_{o_3}^{10} y_{o_4}^6 y_{o_5}^8 y_{o_6}^8 t_1^4 t_2^4 t_3^4 t_4^4 t_5^6 t_6^2 t_7^6 t_8^2 -  y_{s}^4 y_{o_1}^{10} y_{o_2}^6 y_{o_3}^{10} y_{o_4}^6 y_{o_5}^8 y_{o_6}^8 t_1^2 t_2^6 t_3^4 t_4^4 t_5^6 t_6^2 t_7^6 t_8^2 -  y_{s}^4 y_{o_1}^{10} y_{o_2}^6 y_{o_3}^9 y_{o_4}^7 y_{o_5}^9 y_{o_6}^7 t_1^7 t_2 t_3^3 t_4^5 t_5^5 t_6^3 t_7^6 t_8^2 -  3 y_{s}^4 y_{o_1}^{10} y_{o_2}^6 y_{o_3}^9 y_{o_4}^7 y_{o_5}^9 y_{o_6}^7 t_1^6 t_2^2 t_3^3 t_4^5 t_5^5 t_6^3 t_7^6 t_8^2 -  3 y_{s}^4 y_{o_1}^{10} y_{o_2}^6 y_{o_3}^9 y_{o_4}^7 y_{o_5}^9 y_{o_6}^7 t_1^5 t_2^3 t_3^3 t_4^5 t_5^5 t_6^3 t_7^6 t_8^2 -  7 y_{s}^4 y_{o_1}^{10} y_{o_2}^6 y_{o_3}^9 y_{o_4}^7 y_{o_5}^9 y_{o_6}^7 t_1^4 t_2^4 t_3^3 t_4^5 t_5^5 t_6^3 t_7^6 t_8^2 -  3 y_{s}^4 y_{o_1}^{10} y_{o_2}^6 y_{o_3}^9 y_{o_4}^7 y_{o_5}^9 y_{o_6}^7 t_1^3 t_2^5 t_3^3 t_4^5 t_5^5 t_6^3 t_7^6 t_8^2 -  3 y_{s}^4 y_{o_1}^{10} y_{o_2}^6 y_{o_3}^9 y_{o_4}^7 y_{o_5}^9 y_{o_6}^7 t_1^2 t_2^6 t_3^3 t_4^5 t_5^5 t_6^3 t_7^6 t_8^2 -  y_{s}^4 y_{o_1}^{10} y_{o_2}^6 y_{o_3}^9 y_{o_4}^7 y_{o_5}^9 y_{o_6}^7 t_1 t_2^7 t_3^3 t_4^5 t_5^5 t_6^3 t_7^6 t_8^2 -  y_{s}^4 y_{o_1}^{10} y_{o_2}^6 y_{o_3}^8 y_{o_4}^8 y_{o_5}^{10} y_{o_6}^6 t_1^6 t_2^2 t_3^2 t_4^6 t_5^4 t_6^4 t_7^6 t_8^2 -  3 y_{s}^4 y_{o_1}^{10} y_{o_2}^6 y_{o_3}^8 y_{o_4}^8 y_{o_5}^{10} y_{o_6}^6 t_1^4 t_2^4 t_3^2 t_4^6 t_5^4 t_6^4 t_7^6 t_8^2 -  y_{s}^4 y_{o_1}^{10} y_{o_2}^6 y_{o_3}^8 y_{o_4}^8 y_{o_5}^{10} y_{o_6}^6 t_1^2 t_2^6 t_3^2 t_4^6 t_5^4 t_6^4 t_7^6 t_8^2 -  y_{s}^4 y_{o_1}^{10} y_{o_2}^6 y_{o_3}^7 y_{o_4}^9 y_{o_5}^{11} y_{o_6}^5 t_1^4 t_2^4 t_3 t_4^7 t_5^3 t_6^5 t_7^6 t_8^2 -  y_{s}^5 y_{o_1}^{13} y_{o_2}^7 y_{o_3}^{13} y_{o_4}^7 y_{o_5}^{10} y_{o_6}^{10} t_1^5 t_2^5 t_3^5 t_4^5 t_5^8 t_6^2 t_7^8 t_8^2 +  y_{s}^5 y_{o_1}^{13} y_{o_2}^7 y_{o_3}^{12} y_{o_4}^8 y_{o_5}^{11} y_{o_6}^9 t_1^6 t_2^4 t_3^4 t_4^6 t_5^7 t_6^3 t_7^8 t_8^2 +  y_{s}^5 y_{o_1}^{13} y_{o_2}^7 y_{o_3}^{12} y_{o_4}^8 y_{o_5}^{11} y_{o_6}^9 t_1^4 t_2^6 t_3^4 t_4^6 t_5^7 t_6^3 t_7^8 t_8^2 +  y_{s}^5 y_{o_1}^{13} y_{o_2}^7 y_{o_3}^{11} y_{o_4}^9 y_{o_5}^{12} y_{o_6}^8 t_1^6 t_2^4 t_3^3 t_4^7 t_5^6 t_6^4 t_7^8 t_8^2 +  y_{s}^5 y_{o_1}^{13} y_{o_2}^7 y_{o_3}^{11} y_{o_4}^9 y_{o_5}^{12} y_{o_6}^8 t_1^4 t_2^6 t_3^3 t_4^7 t_5^6 t_6^4 t_7^8 t_8^2 -  y_{s}^5 y_{o_1}^{13} y_{o_2}^7 y_{o_3}^{10} y_{o_4}^{10} y_{o_5}^{13} y_{o_6}^7 t_1^5 t_2^5 t_3^2 t_4^8 t_5^5 t_6^5 t_7^8 t_8^2 -  y_{s}^2 y_{o_1}^3 y_{o_2}^5 y_{o_3}^5 y_{o_4}^3 y_{o_5}^2 y_{o_6}^6 t_1^3 t_2 t_3^4 t_5^3 t_6 t_7 t_8^3 - y_{s}^2 y_{o_1}^3 y_{o_2}^5 y_{o_3}^5 y_{o_4}^3 y_{o_5}^2 y_{o_6}^6 t_1^2 t_2^2 t_3^4 t_5^3 t_6 t_7 t_8^3 - y_{s}^2 y_{o_1}^3 y_{o_2}^5 y_{o_3}^5 y_{o_4}^3 y_{o_5}^2 y_{o_6}^6 t_1 t_2^3 t_3^4 t_5^3 t_6 t_7 t_8^3 -  y_{s}^2 y_{o_1}^3 y_{o_2}^5 y_{o_3}^4 y_{o_4}^4 y_{o_5}^3 y_{o_6}^5 t_1^4 t_3^3 t_4 t_5^2 t_6^2 t_7 t_8^3 - 2 y_{s}^2 y_{o_1}^3 y_{o_2}^5 y_{o_3}^4 y_{o_4}^4 y_{o_5}^3 y_{o_6}^5 t_1^3 t_2 t_3^3 t_4 t_5^2 t_6^2 t_7 t_8^3 -  3 y_{s}^2 y_{o_1}^3 y_{o_2}^5 y_{o_3}^4 y_{o_4}^4 y_{o_5}^3 y_{o_6}^5 t_1^2 t_2^2 t_3^3 t_4 t_5^2 t_6^2 t_7 t_8^3 -  2 y_{s}^2 y_{o_1}^3 y_{o_2}^5 y_{o_3}^4 y_{o_4}^4 y_{o_5}^3 y_{o_6}^5 t_1 t_2^3 t_3^3 t_4 t_5^2 t_6^2 t_7 t_8^3 -  y_{s}^2 y_{o_1}^3 y_{o_2}^5 y_{o_3}^4 y_{o_4}^4 y_{o_5}^3 y_{o_6}^5 t_2^4 t_3^3 t_4 t_5^2 t_6^2 t_7 t_8^3 - y_{s}^2 y_{o_1}^3 y_{o_2}^5 y_{o_3}^3 y_{o_4}^5 y_{o_5}^4 y_{o_6}^4 t_1^4 t_3^2 t_4^2 t_5 t_6^3 t_7 t_8^3 -  2 y_{s}^2 y_{o_1}^3 y_{o_2}^5 y_{o_3}^3 y_{o_4}^5 y_{o_5}^4 y_{o_6}^4 t_1^3 t_2 t_3^2 t_4^2 t_5 t_6^3 t_7 t_8^3 -  3 y_{s}^2 y_{o_1}^3 y_{o_2}^5 y_{o_3}^3 y_{o_4}^5 y_{o_5}^4 y_{o_6}^4 t_1^2 t_2^2 t_3^2 t_4^2 t_5 t_6^3 t_7 t_8^3 -  2 y_{s}^2 y_{o_1}^3 y_{o_2}^5 y_{o_3}^3 y_{o_4}^5 y_{o_5}^4 y_{o_6}^4 t_1 t_2^3 t_3^2 t_4^2 t_5 t_6^3 t_7 t_8^3 -  y_{s}^2 y_{o_1}^3 y_{o_2}^5 y_{o_3}^3 y_{o_4}^5 y_{o_5}^4 y_{o_6}^4 t_2^4 t_3^2 t_4^2 t_5 t_6^3 t_7 t_8^3 - y_{s}^2 y_{o_1}^3 y_{o_2}^5 y_{o_3}^2 y_{o_4}^6 y_{o_5}^5 y_{o_6}^3 t_1^3 t_2 t_3 t_4^3 t_6^4 t_7 t_8^3 -  y_{s}^2 y_{o_1}^3 y_{o_2}^5 y_{o_3}^2 y_{o_4}^6 y_{o_5}^5 y_{o_6}^3 t_1^2 t_2^2 t_3 t_4^3 t_6^4 t_7 t_8^3 - y_{s}^2 y_{o_1}^3 y_{o_2}^5 y_{o_3}^2 y_{o_4}^6 y_{o_5}^5 y_{o_6}^3 t_1 t_2^3 t_3 t_4^3 t_6^4 t_7 t_8^3 +  y_{s}^3 y_{o_1}^6 y_{o_2}^6 y_{o_3}^8 y_{o_4}^4 y_{o_5}^4 y_{o_6}^8 t_1^5 t_2 t_3^5 t_4 t_5^5 t_6 t_7^3 t_8^3 + 2 y_{s}^3 y_{o_1}^6 y_{o_2}^6 y_{o_3}^8 y_{o_4}^4 y_{o_5}^4 y_{o_6}^8 t_1^4 t_2^2 t_3^5 t_4 t_5^5 t_6 t_7^3 t_8^3 +  2 y_{s}^3 y_{o_1}^6 y_{o_2}^6 y_{o_3}^8 y_{o_4}^4 y_{o_5}^4 y_{o_6}^8 t_1^3 t_2^3 t_3^5 t_4 t_5^5 t_6 t_7^3 t_8^3 +  2 y_{s}^3 y_{o_1}^6 y_{o_2}^6 y_{o_3}^8 y_{o_4}^4 y_{o_5}^4 y_{o_6}^8 t_1^2 t_2^4 t_3^5 t_4 t_5^5 t_6 t_7^3 t_8^3 +  y_{s}^3 y_{o_1}^6 y_{o_2}^6 y_{o_3}^8 y_{o_4}^4 y_{o_5}^4 y_{o_6}^8 t_1 t_2^5 t_3^5 t_4 t_5^5 t_6 t_7^3 t_8^3 + y_{s}^3 y_{o_1}^6 y_{o_2}^6 y_{o_3}^7 y_{o_4}^5 y_{o_5}^5 y_{o_6}^7 t_1^6 t_3^4 t_4^2 t_5^4 t_6^2 t_7^3 t_8^3 +  3 y_{s}^3 y_{o_1}^6 y_{o_2}^6 y_{o_3}^7 y_{o_4}^5 y_{o_5}^5 y_{o_6}^7 t_1^5 t_2 t_3^4 t_4^2 t_5^4 t_6^2 t_7^3 t_8^3 +  7 y_{s}^3 y_{o_1}^6 y_{o_2}^6 y_{o_3}^7 y_{o_4}^5 y_{o_5}^5 y_{o_6}^7 t_1^4 t_2^2 t_3^4 t_4^2 t_5^4 t_6^2 t_7^3 t_8^3 +  6 y_{s}^3 y_{o_1}^6 y_{o_2}^6 y_{o_3}^7 y_{o_4}^5 y_{o_5}^5 y_{o_6}^7 t_1^3 t_2^3 t_3^4 t_4^2 t_5^4 t_6^2 t_7^3 t_8^3 +  7 y_{s}^3 y_{o_1}^6 y_{o_2}^6 y_{o_3}^7 y_{o_4}^5 y_{o_5}^5 y_{o_6}^7 t_1^2 t_2^4 t_3^4 t_4^2 t_5^4 t_6^2 t_7^3 t_8^3 +  3 y_{s}^3 y_{o_1}^6 y_{o_2}^6 y_{o_3}^7 y_{o_4}^5 y_{o_5}^5 y_{o_6}^7 t_1 t_2^5 t_3^4 t_4^2 t_5^4 t_6^2 t_7^3 t_8^3 +  y_{s}^3 y_{o_1}^6 y_{o_2}^6 y_{o_3}^7 y_{o_4}^5 y_{o_5}^5 y_{o_6}^7 t_2^6 t_3^4 t_4^2 t_5^4 t_6^2 t_7^3 t_8^3 +  2 y_{s}^3 y_{o_1}^6 y_{o_2}^6 y_{o_3}^6 y_{o_4}^6 y_{o_5}^6 y_{o_6}^6 t_1^6 t_3^3 t_4^3 t_5^3 t_6^3 t_7^3 t_8^3 +  4 y_{s}^3 y_{o_1}^6 y_{o_2}^6 y_{o_3}^6 y_{o_4}^6 y_{o_5}^6 y_{o_6}^6 t_1^5 t_2 t_3^3 t_4^3 t_5^3 t_6^3 t_7^3 t_8^3 +  {10} y_{s}^3 y_{o_1}^6 y_{o_2}^6 y_{o_3}^6 y_{o_4}^6 y_{o_5}^6 y_{o_6}^6 t_1^4 t_2^2 t_3^3 t_4^3 t_5^3 t_6^3 t_7^3 t_8^3 +  7 y_{s}^3 y_{o_1}^6 y_{o_2}^6 y_{o_3}^6 y_{o_4}^6 y_{o_5}^6 y_{o_6}^6 t_1^3 t_2^3 t_3^3 t_4^3 t_5^3 t_6^3 t_7^3 t_8^3 +  {10} y_{s}^3 y_{o_1}^6 y_{o_2}^6 y_{o_3}^6 y_{o_4}^6 y_{o_5}^6 y_{o_6}^6 t_1^2 t_2^4 t_3^3 t_4^3 t_5^3 t_6^3 t_7^3 t_8^3 +  4 y_{s}^3 y_{o_1}^6 y_{o_2}^6 y_{o_3}^6 y_{o_4}^6 y_{o_5}^6 y_{o_6}^6 t_1 t_2^5 t_3^3 t_4^3 t_5^3 t_6^3 t_7^3 t_8^3 +  2 y_{s}^3 y_{o_1}^6 y_{o_2}^6 y_{o_3}^6 y_{o_4}^6 y_{o_5}^6 y_{o_6}^6 t_2^6 t_3^3 t_4^3 t_5^3 t_6^3 t_7^3 t_8^3 +  y_{s}^3 y_{o_1}^6 y_{o_2}^6 y_{o_3}^5 y_{o_4}^7 y_{o_5}^7 y_{o_6}^5 t_1^6 t_3^2 t_4^4 t_5^2 t_6^4 t_7^3 t_8^3 +  3 y_{s}^3 y_{o_1}^6 y_{o_2}^6 y_{o_3}^5 y_{o_4}^7 y_{o_5}^7 y_{o_6}^5 t_1^5 t_2 t_3^2 t_4^4 t_5^2 t_6^4 t_7^3 t_8^3 +  7 y_{s}^3 y_{o_1}^6 y_{o_2}^6 y_{o_3}^5 y_{o_4}^7 y_{o_5}^7 y_{o_6}^5 t_1^4 t_2^2 t_3^2 t_4^4 t_5^2 t_6^4 t_7^3 t_8^3 +  6 y_{s}^3 y_{o_1}^6 y_{o_2}^6 y_{o_3}^5 y_{o_4}^7 y_{o_5}^7 y_{o_6}^5 t_1^3 t_2^3 t_3^2 t_4^4 t_5^2 t_6^4 t_7^3 t_8^3 +  7 y_{s}^3 y_{o_1}^6 y_{o_2}^6 y_{o_3}^5 y_{o_4}^7 y_{o_5}^7 y_{o_6}^5 t_1^2 t_2^4 t_3^2 t_4^4 t_5^2 t_6^4 t_7^3 t_8^3 +  3 y_{s}^3 y_{o_1}^6 y_{o_2}^6 y_{o_3}^5 y_{o_4}^7 y_{o_5}^7 y_{o_6}^5 t_1 t_2^5 t_3^2 t_4^4 t_5^2 t_6^4 t_7^3 t_8^3 +  y_{s}^3 y_{o_1}^6 y_{o_2}^6 y_{o_3}^5 y_{o_4}^7 y_{o_5}^7 y_{o_6}^5 t_2^6 t_3^2 t_4^4 t_5^2 t_6^4 t_7^3 t_8^3 +  y_{s}^3 y_{o_1}^6 y_{o_2}^6 y_{o_3}^4 y_{o_4}^8 y_{o_5}^8 y_{o_6}^4 t_1^5 t_2 t_3 t_4^5 t_5 t_6^5 t_7^3 t_8^3 + 2 y_{s}^3 y_{o_1}^6 y_{o_2}^6 y_{o_3}^4 y_{o_4}^8 y_{o_5}^8 y_{o_6}^4 t_1^4 t_2^2 t_3 t_4^5 t_5 t_6^5 t_7^3 t_8^3 +  2 y_{s}^3 y_{o_1}^6 y_{o_2}^6 y_{o_3}^4 y_{o_4}^8 y_{o_5}^8 y_{o_6}^4 t_1^3 t_2^3 t_3 t_4^5 t_5 t_6^5 t_7^3 t_8^3 +  2 y_{s}^3 y_{o_1}^6 y_{o_2}^6 y_{o_3}^4 y_{o_4}^8 y_{o_5}^8 y_{o_6}^4 t_1^2 t_2^4 t_3 t_4^5 t_5 t_6^5 t_7^3 t_8^3 +  y_{s}^3 y_{o_1}^6 y_{o_2}^6 y_{o_3}^4 y_{o_4}^8 y_{o_5}^8 y_{o_6}^4 t_1 t_2^5 t_3 t_4^5 t_5 t_6^5 t_7^3 t_8^3 - y_{s}^4 y_{o_1}^9 y_{o_2}^7 y_{o_3}^{11} y_{o_4}^5 y_{o_5}^6 y_{o_6}^{10} t_1^4 t_2^4 t_3^6 t_4^2 t_5^7 t_6 t_7^5 t_8^3 -  y_{s}^4 y_{o_1}^9 y_{o_2}^7 y_{o_3}^{10} y_{o_4}^6 y_{o_5}^7 y_{o_6}^9 t_1^7 t_2 t_3^5 t_4^3 t_5^6 t_6^2 t_7^5 t_8^3 -  3 y_{s}^4 y_{o_1}^9 y_{o_2}^7 y_{o_3}^{10} y_{o_4}^6 y_{o_5}^7 y_{o_6}^9 t_1^6 t_2^2 t_3^5 t_4^3 t_5^6 t_6^2 t_7^5 t_8^3 -  3 y_{s}^4 y_{o_1}^9 y_{o_2}^7 y_{o_3}^{10} y_{o_4}^6 y_{o_5}^7 y_{o_6}^9 t_1^5 t_2^3 t_3^5 t_4^3 t_5^6 t_6^2 t_7^5 t_8^3 -  7 y_{s}^4 y_{o_1}^9 y_{o_2}^7 y_{o_3}^{10} y_{o_4}^6 y_{o_5}^7 y_{o_6}^9 t_1^4 t_2^4 t_3^5 t_4^3 t_5^6 t_6^2 t_7^5 t_8^3 -  3 y_{s}^4 y_{o_1}^9 y_{o_2}^7 y_{o_3}^{10} y_{o_4}^6 y_{o_5}^7 y_{o_6}^9 t_1^3 t_2^5 t_3^5 t_4^3 t_5^6 t_6^2 t_7^5 t_8^3 -  3 y_{s}^4 y_{o_1}^9 y_{o_2}^7 y_{o_3}^{10} y_{o_4}^6 y_{o_5}^7 y_{o_6}^9 t_1^2 t_2^6 t_3^5 t_4^3 t_5^6 t_6^2 t_7^5 t_8^3 -  y_{s}^4 y_{o_1}^9 y_{o_2}^7 y_{o_3}^{10} y_{o_4}^6 y_{o_5}^7 y_{o_6}^9 t_1 t_2^7 t_3^5 t_4^3 t_5^6 t_6^2 t_7^5 t_8^3 -  y_{s}^4 y_{o_1}^9 y_{o_2}^7 y_{o_3}^9 y_{o_4}^7 y_{o_5}^8 y_{o_6}^8 t_1^7 t_2 t_3^4 t_4^4 t_5^5 t_6^3 t_7^5 t_8^3 -  5 y_{s}^4 y_{o_1}^9 y_{o_2}^7 y_{o_3}^9 y_{o_4}^7 y_{o_5}^8 y_{o_6}^8 t_1^6 t_2^2 t_3^4 t_4^4 t_5^5 t_6^3 t_7^5 t_8^3 -  3 y_{s}^4 y_{o_1}^9 y_{o_2}^7 y_{o_3}^9 y_{o_4}^7 y_{o_5}^8 y_{o_6}^8 t_1^5 t_2^3 t_3^4 t_4^4 t_5^5 t_6^3 t_7^5 t_8^3 -  {12} y_{s}^4 y_{o_1}^9 y_{o_2}^7 y_{o_3}^9 y_{o_4}^7 y_{o_5}^8 y_{o_6}^8 t_1^4 t_2^4 t_3^4 t_4^4 t_5^5 t_6^3 t_7^5 t_8^3 -  3 y_{s}^4 y_{o_1}^9 y_{o_2}^7 y_{o_3}^9 y_{o_4}^7 y_{o_5}^8 y_{o_6}^8 t_1^3 t_2^5 t_3^4 t_4^4 t_5^5 t_6^3 t_7^5 t_8^3 -  5 y_{s}^4 y_{o_1}^9 y_{o_2}^7 y_{o_3}^9 y_{o_4}^7 y_{o_5}^8 y_{o_6}^8 t_1^2 t_2^6 t_3^4 t_4^4 t_5^5 t_6^3 t_7^5 t_8^3 -  y_{s}^4 y_{o_1}^9 y_{o_2}^7 y_{o_3}^9 y_{o_4}^7 y_{o_5}^8 y_{o_6}^8 t_1 t_2^7 t_3^4 t_4^4 t_5^5 t_6^3 t_7^5 t_8^3 -  y_{s}^4 y_{o_1}^9 y_{o_2}^7 y_{o_3}^8 y_{o_4}^8 y_{o_5}^9 y_{o_6}^7 t_1^7 t_2 t_3^3 t_4^5 t_5^4 t_6^4 t_7^5 t_8^3 -  5 y_{s}^4 y_{o_1}^9 y_{o_2}^7 y_{o_3}^8 y_{o_4}^8 y_{o_5}^9 y_{o_6}^7 t_1^6 t_2^2 t_3^3 t_4^5 t_5^4 t_6^4 t_7^5 t_8^3 -  3 y_{s}^4 y_{o_1}^9 y_{o_2}^7 y_{o_3}^8 y_{o_4}^8 y_{o_5}^9 y_{o_6}^7 t_1^5 t_2^3 t_3^3 t_4^5 t_5^4 t_6^4 t_7^5 t_8^3 -  {12} y_{s}^4 y_{o_1}^9 y_{o_2}^7 y_{o_3}^8 y_{o_4}^8 y_{o_5}^9 y_{o_6}^7 t_1^4 t_2^4 t_3^3 t_4^5 t_5^4 t_6^4 t_7^5 t_8^3 -  3 y_{s}^4 y_{o_1}^9 y_{o_2}^7 y_{o_3}^8 y_{o_4}^8 y_{o_5}^9 y_{o_6}^7 t_1^3 t_2^5 t_3^3 t_4^5 t_5^4 t_6^4 t_7^5 t_8^3 -  5 y_{s}^4 y_{o_1}^9 y_{o_2}^7 y_{o_3}^8 y_{o_4}^8 y_{o_5}^9 y_{o_6}^7 t_1^2 t_2^6 t_3^3 t_4^5 t_5^4 t_6^4 t_7^5 t_8^3 -  y_{s}^4 y_{o_1}^9 y_{o_2}^7 y_{o_3}^8 y_{o_4}^8 y_{o_5}^9 y_{o_6}^7 t_1 t_2^7 t_3^3 t_4^5 t_5^4 t_6^4 t_7^5 t_8^3 -  y_{s}^4 y_{o_1}^9 y_{o_2}^7 y_{o_3}^7 y_{o_4}^9 y_{o_5}^{10} y_{o_6}^6 t_1^7 t_2 t_3^2 t_4^6 t_5^3 t_6^5 t_7^5 t_8^3 -  3 y_{s}^4 y_{o_1}^9 y_{o_2}^7 y_{o_3}^7 y_{o_4}^9 y_{o_5}^{10} y_{o_6}^6 t_1^6 t_2^2 t_3^2 t_4^6 t_5^3 t_6^5 t_7^5 t_8^3 -  3 y_{s}^4 y_{o_1}^9 y_{o_2}^7 y_{o_3}^7 y_{o_4}^9 y_{o_5}^{10} y_{o_6}^6 t_1^5 t_2^3 t_3^2 t_4^6 t_5^3 t_6^5 t_7^5 t_8^3 -  7 y_{s}^4 y_{o_1}^9 y_{o_2}^7 y_{o_3}^7 y_{o_4}^9 y_{o_5}^{10} y_{o_6}^6 t_1^4 t_2^4 t_3^2 t_4^6 t_5^3 t_6^5 t_7^5 t_8^3 -  3 y_{s}^4 y_{o_1}^9 y_{o_2}^7 y_{o_3}^7 y_{o_4}^9 y_{o_5}^{10} y_{o_6}^6 t_1^3 t_2^5 t_3^2 t_4^6 t_5^3 t_6^5 t_7^5 t_8^3 -  3 y_{s}^4 y_{o_1}^9 y_{o_2}^7 y_{o_3}^7 y_{o_4}^9 y_{o_5}^{10} y_{o_6}^6 t_1^2 t_2^6 t_3^2 t_4^6 t_5^3 t_6^5 t_7^5 t_8^3 -  y_{s}^4 y_{o_1}^9 y_{o_2}^7 y_{o_3}^7 y_{o_4}^9 y_{o_5}^{10} y_{o_6}^6 t_1 t_2^7 t_3^2 t_4^6 t_5^3 t_6^5 t_7^5 t_8^3 -  y_{s}^4 y_{o_1}^9 y_{o_2}^7 y_{o_3}^6 y_{o_4}^{10} y_{o_5}^{11} y_{o_6}^5 t_1^4 t_2^4 t_3 t_4^7 t_5^2 t_6^6 t_7^5 t_8^3 +  y_{s}^5 y_{o_1}^{12} y_{o_2}^8 y_{o_3}^{13} y_{o_4}^7 y_{o_5}^9 y_{o_6}^{11} t_1^6 t_2^4 t_3^6 t_4^4 t_5^8 t_6^2 t_7^7 t_8^3 +  y_{s}^5 y_{o_1}^{12} y_{o_2}^8 y_{o_3}^{13} y_{o_4}^7 y_{o_5}^9 y_{o_6}^{11} t_1^4 t_2^6 t_3^6 t_4^4 t_5^8 t_6^2 t_7^7 t_8^3 +  3 y_{s}^5 y_{o_1}^{12} y_{o_2}^8 y_{o_3}^{12} y_{o_4}^8 y_{o_5}^{10} y_{o_6}^{10} t_1^6 t_2^4 t_3^5 t_4^5 t_5^7 t_6^3 t_7^7 t_8^3 -  y_{s}^5 y_{o_1}^{12} y_{o_2}^8 y_{o_3}^{12} y_{o_4}^8 y_{o_5}^{10} y_{o_6}^{10} t_1^5 t_2^5 t_3^5 t_4^5 t_5^7 t_6^3 t_7^7 t_8^3 +  3 y_{s}^5 y_{o_1}^{12} y_{o_2}^8 y_{o_3}^{12} y_{o_4}^8 y_{o_5}^{10} y_{o_6}^{10} t_1^4 t_2^6 t_3^5 t_4^5 t_5^7 t_6^3 t_7^7 t_8^3 -  y_{s}^5 y_{o_1}^{12} y_{o_2}^8 y_{o_3}^{11} y_{o_4}^9 y_{o_5}^{11} y_{o_6}^9 t_1^7 t_2^3 t_3^4 t_4^6 t_5^6 t_6^4 t_7^7 t_8^3 +  3 y_{s}^5 y_{o_1}^{12} y_{o_2}^8 y_{o_3}^{11} y_{o_4}^9 y_{o_5}^{11} y_{o_6}^9 t_1^6 t_2^4 t_3^4 t_4^6 t_5^6 t_6^4 t_7^7 t_8^3 -  3 y_{s}^5 y_{o_1}^{12} y_{o_2}^8 y_{o_3}^{11} y_{o_4}^9 y_{o_5}^{11} y_{o_6}^9 t_1^5 t_2^5 t_3^4 t_4^6 t_5^6 t_6^4 t_7^7 t_8^3 +  3 y_{s}^5 y_{o_1}^{12} y_{o_2}^8 y_{o_3}^{11} y_{o_4}^9 y_{o_5}^{11} y_{o_6}^9 t_1^4 t_2^6 t_3^4 t_4^6 t_5^6 t_6^4 t_7^7 t_8^3 -  y_{s}^5 y_{o_1}^{12} y_{o_2}^8 y_{o_3}^{11} y_{o_4}^9 y_{o_5}^{11} y_{o_6}^9 t_1^3 t_2^7 t_3^4 t_4^6 t_5^6 t_6^4 t_7^7 t_8^3 +  3 y_{s}^5 y_{o_1}^{12} y_{o_2}^8 y_{o_3}^{10} y_{o_4}^{10} y_{o_5}^{12} y_{o_6}^8 t_1^6 t_2^4 t_3^3 t_4^7 t_5^5 t_6^5 t_7^7 t_8^3 -  y_{s}^5 y_{o_1}^{12} y_{o_2}^8 y_{o_3}^{10} y_{o_4}^{10} y_{o_5}^{12} y_{o_6}^8 t_1^5 t_2^5 t_3^3 t_4^7 t_5^5 t_6^5 t_7^7 t_8^3 +  3 y_{s}^5 y_{o_1}^{12} y_{o_2}^8 y_{o_3}^{10} y_{o_4}^{10} y_{o_5}^{12} y_{o_6}^8 t_1^4 t_2^6 t_3^3 t_4^7 t_5^5 t_6^5 t_7^7 t_8^3 +  y_{s}^5 y_{o_1}^{12} y_{o_2}^8 y_{o_3}^9 y_{o_4}^{11} y_{o_5}^{13} y_{o_6}^7 t_1^6 t_2^4 t_3^2 t_4^8 t_5^4 t_6^6 t_7^7 t_8^3 +  y_{s}^5 y_{o_1}^{12} y_{o_2}^8 y_{o_3}^9 y_{o_4}^{11} y_{o_5}^{13} y_{o_6}^7 t_1^4 t_2^6 t_3^2 t_4^8 t_5^4 t_6^6 t_7^7 t_8^3 -  y_{s}^6 y_{o_1}^{15} y_{o_2}^9 y_{o_3}^{15} y_{o_4}^9 y_{o_5}^{12} y_{o_6}^{12} t_1^6 t_2^6 t_3^6 t_4^6 t_5^9 t_6^3 t_7^9 t_8^3 +  y_{s}^6 y_{o_1}^{15} y_{o_2}^9 y_{o_3}^{14} y_{o_4}^{10} y_{o_5}^{13} y_{o_6}^{11} t_1^7 t_2^5 t_3^5 t_4^7 t_5^8 t_6^4 t_7^9 t_8^3 +  y_{s}^6 y_{o_1}^{15} y_{o_2}^9 y_{o_3}^{14} y_{o_4}^{10} y_{o_5}^{13} y_{o_6}^{11} t_1^5 t_2^7 t_3^5 t_4^7 t_5^8 t_6^4 t_7^9 t_8^3 +  y_{s}^6 y_{o_1}^{15} y_{o_2}^9 y_{o_3}^{13} y_{o_4}^{11} y_{o_5}^{14} y_{o_6}^{10} t_1^7 t_2^5 t_3^4 t_4^8 t_5^7 t_6^5 t_7^9 t_8^3 +  y_{s}^6 y_{o_1}^{15} y_{o_2}^9 y_{o_3}^{13} y_{o_4}^{11} y_{o_5}^{14} y_{o_6}^{10} t_1^5 t_2^7 t_3^4 t_4^8 t_5^7 t_6^5 t_7^9 t_8^3 -  y_{s}^6 y_{o_1}^{15} y_{o_2}^9 y_{o_3}^{12} y_{o_4}^{12} y_{o_5}^{15} y_{o_6}^9 t_1^6 t_2^6 t_3^3 t_4^9 t_5^6 t_6^6 t_7^9 t_8^3 -  y_{s}^2 y_{o_1}^2 y_{o_2}^6 y_{o_3}^3 y_{o_4}^5 y_{o_5}^3 y_{o_6}^5 t_1^3 t_2 t_3^3 t_4 t_5 t_6^3 t_8^4 - y_{s}^2 y_{o_1}^2 y_{o_2}^6 y_{o_3}^3 y_{o_4}^5 y_{o_5}^3 y_{o_6}^5 t_1^2 t_2^2 t_3^3 t_4 t_5 t_6^3 t_8^4 - y_{s}^2 y_{o_1}^2 y_{o_2}^6 y_{o_3}^3 y_{o_4}^5 y_{o_5}^3 y_{o_6}^5 t_1 t_2^3 t_3^3 t_4 t_5 t_6^3 t_8^4 +  2 y_{s}^3 y_{o_1}^5 y_{o_2}^7 y_{o_3}^7 y_{o_4}^5 y_{o_5}^4 y_{o_6}^8 t_1^4 t_2^2 t_3^5 t_4 t_5^4 t_6^2 t_7^2 t_8^4 +  y_{s}^3 y_{o_1}^5 y_{o_2}^7 y_{o_3}^7 y_{o_4}^5 y_{o_5}^4 y_{o_6}^8 t_1^3 t_2^3 t_3^5 t_4 t_5^4 t_6^2 t_7^2 t_8^4 +  2 y_{s}^3 y_{o_1}^5 y_{o_2}^7 y_{o_3}^7 y_{o_4}^5 y_{o_5}^4 y_{o_6}^8 t_1^2 t_2^4 t_3^5 t_4 t_5^4 t_6^2 t_7^2 t_8^4 +  y_{s}^3 y_{o_1}^5 y_{o_2}^7 y_{o_3}^6 y_{o_4}^6 y_{o_5}^5 y_{o_6}^7 t_1^6 t_3^4 t_4^2 t_5^3 t_6^3 t_7^2 t_8^4 +  3 y_{s}^3 y_{o_1}^5 y_{o_2}^7 y_{o_3}^6 y_{o_4}^6 y_{o_5}^5 y_{o_6}^7 t_1^5 t_2 t_3^4 t_4^2 t_5^3 t_6^3 t_7^2 t_8^4 +  7 y_{s}^3 y_{o_1}^5 y_{o_2}^7 y_{o_3}^6 y_{o_4}^6 y_{o_5}^5 y_{o_6}^7 t_1^4 t_2^2 t_3^4 t_4^2 t_5^3 t_6^3 t_7^2 t_8^4 +  6 y_{s}^3 y_{o_1}^5 y_{o_2}^7 y_{o_3}^6 y_{o_4}^6 y_{o_5}^5 y_{o_6}^7 t_1^3 t_2^3 t_3^4 t_4^2 t_5^3 t_6^3 t_7^2 t_8^4 +  7 y_{s}^3 y_{o_1}^5 y_{o_2}^7 y_{o_3}^6 y_{o_4}^6 y_{o_5}^5 y_{o_6}^7 t_1^2 t_2^4 t_3^4 t_4^2 t_5^3 t_6^3 t_7^2 t_8^4 +  3 y_{s}^3 y_{o_1}^5 y_{o_2}^7 y_{o_3}^6 y_{o_4}^6 y_{o_5}^5 y_{o_6}^7 t_1 t_2^5 t_3^4 t_4^2 t_5^3 t_6^3 t_7^2 t_8^4 +  y_{s}^3 y_{o_1}^5 y_{o_2}^7 y_{o_3}^6 y_{o_4}^6 y_{o_5}^5 y_{o_6}^7 t_2^6 t_3^4 t_4^2 t_5^3 t_6^3 t_7^2 t_8^4 +  y_{s}^3 y_{o_1}^5 y_{o_2}^7 y_{o_3}^5 y_{o_4}^7 y_{o_5}^6 y_{o_6}^6 t_1^6 t_3^3 t_4^3 t_5^2 t_6^4 t_7^2 t_8^4 +  3 y_{s}^3 y_{o_1}^5 y_{o_2}^7 y_{o_3}^5 y_{o_4}^7 y_{o_5}^6 y_{o_6}^6 t_1^5 t_2 t_3^3 t_4^3 t_5^2 t_6^4 t_7^2 t_8^4 +  7 y_{s}^3 y_{o_1}^5 y_{o_2}^7 y_{o_3}^5 y_{o_4}^7 y_{o_5}^6 y_{o_6}^6 t_1^4 t_2^2 t_3^3 t_4^3 t_5^2 t_6^4 t_7^2 t_8^4 +  6 y_{s}^3 y_{o_1}^5 y_{o_2}^7 y_{o_3}^5 y_{o_4}^7 y_{o_5}^6 y_{o_6}^6 t_1^3 t_2^3 t_3^3 t_4^3 t_5^2 t_6^4 t_7^2 t_8^4 +  7 y_{s}^3 y_{o_1}^5 y_{o_2}^7 y_{o_3}^5 y_{o_4}^7 y_{o_5}^6 y_{o_6}^6 t_1^2 t_2^4 t_3^3 t_4^3 t_5^2 t_6^4 t_7^2 t_8^4 +  3 y_{s}^3 y_{o_1}^5 y_{o_2}^7 y_{o_3}^5 y_{o_4}^7 y_{o_5}^6 y_{o_6}^6 t_1 t_2^5 t_3^3 t_4^3 t_5^2 t_6^4 t_7^2 t_8^4 +  y_{s}^3 y_{o_1}^5 y_{o_2}^7 y_{o_3}^5 y_{o_4}^7 y_{o_5}^6 y_{o_6}^6 t_2^6 t_3^3 t_4^3 t_5^2 t_6^4 t_7^2 t_8^4 +  2 y_{s}^3 y_{o_1}^5 y_{o_2}^7 y_{o_3}^4 y_{o_4}^8 y_{o_5}^7 y_{o_6}^5 t_1^4 t_2^2 t_3^2 t_4^4 t_5 t_6^5 t_7^2 t_8^4 +  y_{s}^3 y_{o_1}^5 y_{o_2}^7 y_{o_3}^4 y_{o_4}^8 y_{o_5}^7 y_{o_6}^5 t_1^3 t_2^3 t_3^2 t_4^4 t_5 t_6^5 t_7^2 t_8^4 +  2 y_{s}^3 y_{o_1}^5 y_{o_2}^7 y_{o_3}^4 y_{o_4}^8 y_{o_5}^7 y_{o_6}^5 t_1^2 t_2^4 t_3^2 t_4^4 t_5 t_6^5 t_7^2 t_8^4 -  y_{s}^4 y_{o_1}^8 y_{o_2}^8 y_{o_3}^{10} y_{o_4}^6 y_{o_5}^6 y_{o_6}^{10} t_1^6 t_2^2 t_3^6 t_4^2 t_5^6 t_6^2 t_7^4 t_8^4 -  3 y_{s}^4 y_{o_1}^8 y_{o_2}^8 y_{o_3}^{10} y_{o_4}^6 y_{o_5}^6 y_{o_6}^{10} t_1^4 t_2^4 t_3^6 t_4^2 t_5^6 t_6^2 t_7^4 t_8^4 -  y_{s}^4 y_{o_1}^8 y_{o_2}^8 y_{o_3}^{10} y_{o_4}^6 y_{o_5}^6 y_{o_6}^{10} t_1^2 t_2^6 t_3^6 t_4^2 t_5^6 t_6^2 t_7^4 t_8^4 -  y_{s}^4 y_{o_1}^8 y_{o_2}^8 y_{o_3}^9 y_{o_4}^7 y_{o_5}^7 y_{o_6}^9 t_1^7 t_2 t_3^5 t_4^3 t_5^5 t_6^3 t_7^4 t_8^4 -  5 y_{s}^4 y_{o_1}^8 y_{o_2}^8 y_{o_3}^9 y_{o_4}^7 y_{o_5}^7 y_{o_6}^9 t_1^6 t_2^2 t_3^5 t_4^3 t_5^5 t_6^3 t_7^4 t_8^4 -  3 y_{s}^4 y_{o_1}^8 y_{o_2}^8 y_{o_3}^9 y_{o_4}^7 y_{o_5}^7 y_{o_6}^9 t_1^5 t_2^3 t_3^5 t_4^3 t_5^5 t_6^3 t_7^4 t_8^4 -  {12} y_{s}^4 y_{o_1}^8 y_{o_2}^8 y_{o_3}^9 y_{o_4}^7 y_{o_5}^7 y_{o_6}^9 t_1^4 t_2^4 t_3^5 t_4^3 t_5^5 t_6^3 t_7^4 t_8^4 -  3 y_{s}^4 y_{o_1}^8 y_{o_2}^8 y_{o_3}^9 y_{o_4}^7 y_{o_5}^7 y_{o_6}^9 t_1^3 t_2^5 t_3^5 t_4^3 t_5^5 t_6^3 t_7^4 t_8^4 -  5 y_{s}^4 y_{o_1}^8 y_{o_2}^8 y_{o_3}^9 y_{o_4}^7 y_{o_5}^7 y_{o_6}^9 t_1^2 t_2^6 t_3^5 t_4^3 t_5^5 t_6^3 t_7^4 t_8^4 -  y_{s}^4 y_{o_1}^8 y_{o_2}^8 y_{o_3}^9 y_{o_4}^7 y_{o_5}^7 y_{o_6}^9 t_1 t_2^7 t_3^5 t_4^3 t_5^5 t_6^3 t_7^4 t_8^4 -  y_{s}^4 y_{o_1}^8 y_{o_2}^8 y_{o_3}^8 y_{o_4}^8 y_{o_5}^8 y_{o_6}^8 t_1^8 t_3^4 t_4^4 t_5^4 t_6^4 t_7^4 t_8^4 -  4 y_{s}^4 y_{o_1}^8 y_{o_2}^8 y_{o_3}^8 y_{o_4}^8 y_{o_5}^8 y_{o_6}^8 t_1^7 t_2 t_3^4 t_4^4 t_5^4 t_6^4 t_7^4 t_8^4 -  {10} y_{s}^4 y_{o_1}^8 y_{o_2}^8 y_{o_3}^8 y_{o_4}^8 y_{o_5}^8 y_{o_6}^8 t_1^6 t_2^2 t_3^4 t_4^4 t_5^4 t_6^4 t_7^4 t_8^4 -  8 y_{s}^4 y_{o_1}^8 y_{o_2}^8 y_{o_3}^8 y_{o_4}^8 y_{o_5}^8 y_{o_6}^8 t_1^5 t_2^3 t_3^4 t_4^4 t_5^4 t_6^4 t_7^4 t_8^4 -  {20} y_{s}^4 y_{o_1}^8 y_{o_2}^8 y_{o_3}^8 y_{o_4}^8 y_{o_5}^8 y_{o_6}^8 t_1^4 t_2^4 t_3^4 t_4^4 t_5^4 t_6^4 t_7^4 t_8^4 -  8 y_{s}^4 y_{o_1}^8 y_{o_2}^8 y_{o_3}^8 y_{o_4}^8 y_{o_5}^8 y_{o_6}^8 t_1^3 t_2^5 t_3^4 t_4^4 t_5^4 t_6^4 t_7^4 t_8^4 -  {10} y_{s}^4 y_{o_1}^8 y_{o_2}^8 y_{o_3}^8 y_{o_4}^8 y_{o_5}^8 y_{o_6}^8 t_1^2 t_2^6 t_3^4 t_4^4 t_5^4 t_6^4 t_7^4 t_8^4 -  4 y_{s}^4 y_{o_1}^8 y_{o_2}^8 y_{o_3}^8 y_{o_4}^8 y_{o_5}^8 y_{o_6}^8 t_1 t_2^7 t_3^4 t_4^4 t_5^4 t_6^4 t_7^4 t_8^4 -  y_{s}^4 y_{o_1}^8 y_{o_2}^8 y_{o_3}^8 y_{o_4}^8 y_{o_5}^8 y_{o_6}^8 t_2^8 t_3^4 t_4^4 t_5^4 t_6^4 t_7^4 t_8^4 -  y_{s}^4 y_{o_1}^8 y_{o_2}^8 y_{o_3}^7 y_{o_4}^9 y_{o_5}^9 y_{o_6}^7 t_1^7 t_2 t_3^3 t_4^5 t_5^3 t_6^5 t_7^4 t_8^4 -  5 y_{s}^4 y_{o_1}^8 y_{o_2}^8 y_{o_3}^7 y_{o_4}^9 y_{o_5}^9 y_{o_6}^7 t_1^6 t_2^2 t_3^3 t_4^5 t_5^3 t_6^5 t_7^4 t_8^4 -  3 y_{s}^4 y_{o_1}^8 y_{o_2}^8 y_{o_3}^7 y_{o_4}^9 y_{o_5}^9 y_{o_6}^7 t_1^5 t_2^3 t_3^3 t_4^5 t_5^3 t_6^5 t_7^4 t_8^4 -  {12} y_{s}^4 y_{o_1}^8 y_{o_2}^8 y_{o_3}^7 y_{o_4}^9 y_{o_5}^9 y_{o_6}^7 t_1^4 t_2^4 t_3^3 t_4^5 t_5^3 t_6^5 t_7^4 t_8^4 -  3 y_{s}^4 y_{o_1}^8 y_{o_2}^8 y_{o_3}^7 y_{o_4}^9 y_{o_5}^9 y_{o_6}^7 t_1^3 t_2^5 t_3^3 t_4^5 t_5^3 t_6^5 t_7^4 t_8^4 -  5 y_{s}^4 y_{o_1}^8 y_{o_2}^8 y_{o_3}^7 y_{o_4}^9 y_{o_5}^9 y_{o_6}^7 t_1^2 t_2^6 t_3^3 t_4^5 t_5^3 t_6^5 t_7^4 t_8^4 -  y_{s}^4 y_{o_1}^8 y_{o_2}^8 y_{o_3}^7 y_{o_4}^9 y_{o_5}^9 y_{o_6}^7 t_1 t_2^7 t_3^3 t_4^5 t_5^3 t_6^5 t_7^4 t_8^4 -  y_{s}^4 y_{o_1}^8 y_{o_2}^8 y_{o_3}^6 y_{o_4}^{10} y_{o_5}^{10} y_{o_6}^6 t_1^6 t_2^2 t_3^2 t_4^6 t_5^2 t_6^6 t_7^4 t_8^4 -  3 y_{s}^4 y_{o_1}^8 y_{o_2}^8 y_{o_3}^6 y_{o_4}^{10} y_{o_5}^{10} y_{o_6}^6 t_1^4 t_2^4 t_3^2 t_4^6 t_5^2 t_6^6 t_7^4 t_8^4 -  y_{s}^4 y_{o_1}^8 y_{o_2}^8 y_{o_3}^6 y_{o_4}^{10} y_{o_5}^{10} y_{o_6}^6 t_1^2 t_2^6 t_3^2 t_4^6 t_5^2 t_6^6 t_7^4 t_8^4 +  y_{s}^5 y_{o_1}^{11} y_{o_2}^9 y_{o_3}^{13} y_{o_4}^7 y_{o_5}^8 y_{o_6}^{12} t_1^6 t_2^4 t_3^7 t_4^3 t_5^8 t_6^2 t_7^6 t_8^4 +  y_{s}^5 y_{o_1}^{11} y_{o_2}^9 y_{o_3}^{13} y_{o_4}^7 y_{o_5}^8 y_{o_6}^{12} t_1^4 t_2^6 t_3^7 t_4^3 t_5^8 t_6^2 t_7^6 t_8^4 -  y_{s}^5 y_{o_1}^{11} y_{o_2}^9 y_{o_3}^{12} y_{o_4}^8 y_{o_5}^9 y_{o_6}^{11} t_1^7 t_2^3 t_3^6 t_4^4 t_5^7 t_6^3 t_7^6 t_8^4 +  3 y_{s}^5 y_{o_1}^{11} y_{o_2}^9 y_{o_3}^{12} y_{o_4}^8 y_{o_5}^9 y_{o_6}^{11} t_1^6 t_2^4 t_3^6 t_4^4 t_5^7 t_6^3 t_7^6 t_8^4 -  3 y_{s}^5 y_{o_1}^{11} y_{o_2}^9 y_{o_3}^{12} y_{o_4}^8 y_{o_5}^9 y_{o_6}^{11} t_1^5 t_2^5 t_3^6 t_4^4 t_5^7 t_6^3 t_7^6 t_8^4 +  3 y_{s}^5 y_{o_1}^{11} y_{o_2}^9 y_{o_3}^{12} y_{o_4}^8 y_{o_5}^9 y_{o_6}^{11} t_1^4 t_2^6 t_3^6 t_4^4 t_5^7 t_6^3 t_7^6 t_8^4 -  y_{s}^5 y_{o_1}^{11} y_{o_2}^9 y_{o_3}^{12} y_{o_4}^8 y_{o_5}^9 y_{o_6}^{11} t_1^3 t_2^7 t_3^6 t_4^4 t_5^7 t_6^3 t_7^6 t_8^4 +  y_{s}^5 y_{o_1}^{11} y_{o_2}^9 y_{o_3}^{11} y_{o_4}^9 y_{o_5}^{10} y_{o_6}^{10} t_1^9 t_2 t_3^5 t_4^5 t_5^6 t_6^4 t_7^6 t_8^4 +  2 y_{s}^5 y_{o_1}^{11} y_{o_2}^9 y_{o_3}^{11} y_{o_4}^9 y_{o_5}^{10} y_{o_6}^{10} t_1^8 t_2^2 t_3^5 t_4^5 t_5^6 t_6^4 t_7^6 t_8^4 +  y_{s}^5 y_{o_1}^{11} y_{o_2}^9 y_{o_3}^{11} y_{o_4}^9 y_{o_5}^{10} y_{o_6}^{10} t_1^7 t_2^3 t_3^5 t_4^5 t_5^6 t_6^4 t_7^6 t_8^4 +  {10} y_{s}^5 y_{o_1}^{11} y_{o_2}^9 y_{o_3}^{11} y_{o_4}^9 y_{o_5}^{10} y_{o_6}^{10} t_1^6 t_2^4 t_3^5 t_4^5 t_5^6 t_6^4 t_7^6 t_8^4 -  y_{s}^5 y_{o_1}^{11} y_{o_2}^9 y_{o_3}^{11} y_{o_4}^9 y_{o_5}^{10} y_{o_6}^{10} t_1^5 t_2^5 t_3^5 t_4^5 t_5^6 t_6^4 t_7^6 t_8^4 +  {10} y_{s}^5 y_{o_1}^{11} y_{o_2}^9 y_{o_3}^{11} y_{o_4}^9 y_{o_5}^{10} y_{o_6}^{10} t_1^4 t_2^6 t_3^5 t_4^5 t_5^6 t_6^4 t_7^6 t_8^4 +  y_{s}^5 y_{o_1}^{11} y_{o_2}^9 y_{o_3}^{11} y_{o_4}^9 y_{o_5}^{10} y_{o_6}^{10} t_1^3 t_2^7 t_3^5 t_4^5 t_5^6 t_6^4 t_7^6 t_8^4 +  2 y_{s}^5 y_{o_1}^{11} y_{o_2}^9 y_{o_3}^{11} y_{o_4}^9 y_{o_5}^{10} y_{o_6}^{10} t_1^2 t_2^8 t_3^5 t_4^5 t_5^6 t_6^4 t_7^6 t_8^4 +  y_{s}^5 y_{o_1}^{11} y_{o_2}^9 y_{o_3}^{11} y_{o_4}^9 y_{o_5}^{10} y_{o_6}^{10} t_1 t_2^9 t_3^5 t_4^5 t_5^6 t_6^4 t_7^6 t_8^4 +  y_{s}^5 y_{o_1}^{11} y_{o_2}^9 y_{o_3}^{10} y_{o_4}^{10} y_{o_5}^{11} y_{o_6}^9 t_1^9 t_2 t_3^4 t_4^6 t_5^5 t_6^5 t_7^6 t_8^4 +  2 y_{s}^5 y_{o_1}^{11} y_{o_2}^9 y_{o_3}^{10} y_{o_4}^{10} y_{o_5}^{11} y_{o_6}^9 t_1^8 t_2^2 t_3^4 t_4^6 t_5^5 t_6^5 t_7^6 t_8^4 +  y_{s}^5 y_{o_1}^{11} y_{o_2}^9 y_{o_3}^{10} y_{o_4}^{10} y_{o_5}^{11} y_{o_6}^9 t_1^7 t_2^3 t_3^4 t_4^6 t_5^5 t_6^5 t_7^6 t_8^4 +  {10} y_{s}^5 y_{o_1}^{11} y_{o_2}^9 y_{o_3}^{10} y_{o_4}^{10} y_{o_5}^{11} y_{o_6}^9 t_1^6 t_2^4 t_3^4 t_4^6 t_5^5 t_6^5 t_7^6 t_8^4 -  y_{s}^5 y_{o_1}^{11} y_{o_2}^9 y_{o_3}^{10} y_{o_4}^{10} y_{o_5}^{11} y_{o_6}^9 t_1^5 t_2^5 t_3^4 t_4^6 t_5^5 t_6^5 t_7^6 t_8^4 +  {10} y_{s}^5 y_{o_1}^{11} y_{o_2}^9 y_{o_3}^{10} y_{o_4}^{10} y_{o_5}^{11} y_{o_6}^9 t_1^4 t_2^6 t_3^4 t_4^6 t_5^5 t_6^5 t_7^6 t_8^4 +  y_{s}^5 y_{o_1}^{11} y_{o_2}^9 y_{o_3}^{10} y_{o_4}^{10} y_{o_5}^{11} y_{o_6}^9 t_1^3 t_2^7 t_3^4 t_4^6 t_5^5 t_6^5 t_7^6 t_8^4 +  2 y_{s}^5 y_{o_1}^{11} y_{o_2}^9 y_{o_3}^{10} y_{o_4}^{10} y_{o_5}^{11} y_{o_6}^9 t_1^2 t_2^8 t_3^4 t_4^6 t_5^5 t_6^5 t_7^6 t_8^4 +  y_{s}^5 y_{o_1}^{11} y_{o_2}^9 y_{o_3}^{10} y_{o_4}^{10} y_{o_5}^{11} y_{o_6}^9 t_1 t_2^9 t_3^4 t_4^6 t_5^5 t_6^5 t_7^6 t_8^4 -  y_{s}^5 y_{o_1}^{11} y_{o_2}^9 y_{o_3}^9 y_{o_4}^{11} y_{o_5}^{12} y_{o_6}^8 t_1^7 t_2^3 t_3^3 t_4^7 t_5^4 t_6^6 t_7^6 t_8^4 +  3 y_{s}^5 y_{o_1}^{11} y_{o_2}^9 y_{o_3}^9 y_{o_4}^{11} y_{o_5}^{12} y_{o_6}^8 t_1^6 t_2^4 t_3^3 t_4^7 t_5^4 t_6^6 t_7^6 t_8^4 -  3 y_{s}^5 y_{o_1}^{11} y_{o_2}^9 y_{o_3}^9 y_{o_4}^{11} y_{o_5}^{12} y_{o_6}^8 t_1^5 t_2^5 t_3^3 t_4^7 t_5^4 t_6^6 t_7^6 t_8^4 +  3 y_{s}^5 y_{o_1}^{11} y_{o_2}^9 y_{o_3}^9 y_{o_4}^{11} y_{o_5}^{12} y_{o_6}^8 t_1^4 t_2^6 t_3^3 t_4^7 t_5^4 t_6^6 t_7^6 t_8^4 -  y_{s}^5 y_{o_1}^{11} y_{o_2}^9 y_{o_3}^9 y_{o_4}^{11} y_{o_5}^{12} y_{o_6}^8 t_1^3 t_2^7 t_3^3 t_4^7 t_5^4 t_6^6 t_7^6 t_8^4 +  y_{s}^5 y_{o_1}^{11} y_{o_2}^9 y_{o_3}^8 y_{o_4}^{12} y_{o_5}^{13} y_{o_6}^7 t_1^6 t_2^4 t_3^2 t_4^8 t_5^3 t_6^7 t_7^6 t_8^4 +  y_{s}^5 y_{o_1}^{11} y_{o_2}^9 y_{o_3}^8 y_{o_4}^{12} y_{o_5}^{13} y_{o_6}^7 t_1^4 t_2^6 t_3^2 t_4^8 t_5^3 t_6^7 t_7^6 t_8^4 +  y_{s}^6 y_{o_1}^{14} y_{o_2}^{10} y_{o_3}^{15} y_{o_4}^9 y_{o_5}^{11} y_{o_6}^{13} t_1^7 t_2^5 t_3^7 t_4^5 t_5^9 t_6^3 t_7^8 t_8^4 +  y_{s}^6 y_{o_1}^{14} y_{o_2}^{10} y_{o_3}^{15} y_{o_4}^9 y_{o_5}^{11} y_{o_6}^{13} t_1^5 t_2^7 t_3^7 t_4^5 t_5^9 t_6^3 t_7^8 t_8^4 +  3 y_{s}^6 y_{o_1}^{14} y_{o_2}^{10} y_{o_3}^{14} y_{o_4}^{10} y_{o_5}^{12} y_{o_6}^{12} t_1^7 t_2^5 t_3^6 t_4^6 t_5^8 t_6^4 t_7^8 t_8^4 -  y_{s}^6 y_{o_1}^{14} y_{o_2}^{10} y_{o_3}^{14} y_{o_4}^{10} y_{o_5}^{12} y_{o_6}^{12} t_1^6 t_2^6 t_3^6 t_4^6 t_5^8 t_6^4 t_7^8 t_8^4 +  3 y_{s}^6 y_{o_1}^{14} y_{o_2}^{10} y_{o_3}^{14} y_{o_4}^{10} y_{o_5}^{12} y_{o_6}^{12} t_1^5 t_2^7 t_3^6 t_4^6 t_5^8 t_6^4 t_7^8 t_8^4 -  y_{s}^6 y_{o_1}^{14} y_{o_2}^{10} y_{o_3}^{13} y_{o_4}^{11} y_{o_5}^{13} y_{o_6}^{11} t_1^8 t_2^4 t_3^5 t_4^7 t_5^7 t_6^5 t_7^8 t_8^4 +  3 y_{s}^6 y_{o_1}^{14} y_{o_2}^{10} y_{o_3}^{13} y_{o_4}^{11} y_{o_5}^{13} y_{o_6}^{11} t_1^7 t_2^5 t_3^5 t_4^7 t_5^7 t_6^5 t_7^8 t_8^4 -  3 y_{s}^6 y_{o_1}^{14} y_{o_2}^{10} y_{o_3}^{13} y_{o_4}^{11} y_{o_5}^{13} y_{o_6}^{11} t_1^6 t_2^6 t_3^5 t_4^7 t_5^7 t_6^5 t_7^8 t_8^4 +  3 y_{s}^6 y_{o_1}^{14} y_{o_2}^{10} y_{o_3}^{13} y_{o_4}^{11} y_{o_5}^{13} y_{o_6}^{11} t_1^5 t_2^7 t_3^5 t_4^7 t_5^7 t_6^5 t_7^8 t_8^4 -  y_{s}^6 y_{o_1}^{14} y_{o_2}^{10} y_{o_3}^{13} y_{o_4}^{11} y_{o_5}^{13} y_{o_6}^{11} t_1^4 t_2^8 t_3^5 t_4^7 t_5^7 t_6^5 t_7^8 t_8^4 +  3 y_{s}^6 y_{o_1}^{14} y_{o_2}^{10} y_{o_3}^{12} y_{o_4}^{12} y_{o_5}^{14} y_{o_6}^{10} t_1^7 t_2^5 t_3^4 t_4^8 t_5^6 t_6^6 t_7^8 t_8^4 -  y_{s}^6 y_{o_1}^{14} y_{o_2}^{10} y_{o_3}^{12} y_{o_4}^{12} y_{o_5}^{14} y_{o_6}^{10} t_1^6 t_2^6 t_3^4 t_4^8 t_5^6 t_6^6 t_7^8 t_8^4 +  3 y_{s}^6 y_{o_1}^{14} y_{o_2}^{10} y_{o_3}^{12} y_{o_4}^{12} y_{o_5}^{14} y_{o_6}^{10} t_1^5 t_2^7 t_3^4 t_4^8 t_5^6 t_6^6 t_7^8 t_8^4 +  y_{s}^6 y_{o_1}^{14} y_{o_2}^{10} y_{o_3}^{11} y_{o_4}^{13} y_{o_5}^{15} y_{o_6}^9 t_1^7 t_2^5 t_3^3 t_4^9 t_5^5 t_6^7 t_7^8 t_8^4 +  y_{s}^6 y_{o_1}^{14} y_{o_2}^{10} y_{o_3}^{11} y_{o_4}^{13} y_{o_5}^{15} y_{o_6}^9 t_1^5 t_2^7 t_3^3 t_4^9 t_5^5 t_6^7 t_7^8 t_8^4 -  y_{s}^7 y_{o_1}^{17} y_{o_2}^{11} y_{o_3}^{16} y_{o_4}^{12} y_{o_5}^{15} y_{o_6}^{13} t_1^7 t_2^7 t_3^6 t_4^8 t_5^9 t_6^5 t_7^{10} t_8^4 -  y_{s}^7 y_{o_1}^{17} y_{o_2}^{11} y_{o_3}^{15} y_{o_4}^{13} y_{o_5}^{16} y_{o_6}^{12} t_1^7 t_2^7 t_3^5 t_4^9 t_5^8 t_6^6 t_7^{10} t_8^4 +  y_{s}^3 y_{o_1}^4 y_{o_2}^8 y_{o_3}^6 y_{o_4}^6 y_{o_5}^4 y_{o_6}^8 t_1^5 t_2 t_3^5 t_4 t_5^3 t_6^3 t_7 t_8^5 + 2 y_{s}^3 y_{o_1}^4 y_{o_2}^8 y_{o_3}^6 y_{o_4}^6 y_{o_5}^4 y_{o_6}^8 t_1^4 t_2^2 t_3^5 t_4 t_5^3 t_6^3 t_7 t_8^5 +  2 y_{s}^3 y_{o_1}^4 y_{o_2}^8 y_{o_3}^6 y_{o_4}^6 y_{o_5}^4 y_{o_6}^8 t_1^3 t_2^3 t_3^5 t_4 t_5^3 t_6^3 t_7 t_8^5 +  2 y_{s}^3 y_{o_1}^4 y_{o_2}^8 y_{o_3}^6 y_{o_4}^6 y_{o_5}^4 y_{o_6}^8 t_1^2 t_2^4 t_3^5 t_4 t_5^3 t_6^3 t_7 t_8^5 +  y_{s}^3 y_{o_1}^4 y_{o_2}^8 y_{o_3}^6 y_{o_4}^6 y_{o_5}^4 y_{o_6}^8 t_1 t_2^5 t_3^5 t_4 t_5^3 t_6^3 t_7 t_8^5 + 2 y_{s}^3 y_{o_1}^4 y_{o_2}^8 y_{o_3}^5 y_{o_4}^7 y_{o_5}^5 y_{o_6}^7 t_1^4 t_2^2 t_3^4 t_4^2 t_5^2 t_6^4 t_7 t_8^5 +  y_{s}^3 y_{o_1}^4 y_{o_2}^8 y_{o_3}^5 y_{o_4}^7 y_{o_5}^5 y_{o_6}^7 t_1^3 t_2^3 t_3^4 t_4^2 t_5^2 t_6^4 t_7 t_8^5 +  2 y_{s}^3 y_{o_1}^4 y_{o_2}^8 y_{o_3}^5 y_{o_4}^7 y_{o_5}^5 y_{o_6}^7 t_1^2 t_2^4 t_3^4 t_4^2 t_5^2 t_6^4 t_7 t_8^5 +  y_{s}^3 y_{o_1}^4 y_{o_2}^8 y_{o_3}^4 y_{o_4}^8 y_{o_5}^6 y_{o_6}^6 t_1^5 t_2 t_3^3 t_4^3 t_5 t_6^5 t_7 t_8^5 + 2 y_{s}^3 y_{o_1}^4 y_{o_2}^8 y_{o_3}^4 y_{o_4}^8 y_{o_5}^6 y_{o_6}^6 t_1^4 t_2^2 t_3^3 t_4^3 t_5 t_6^5 t_7 t_8^5 +  2 y_{s}^3 y_{o_1}^4 y_{o_2}^8 y_{o_3}^4 y_{o_4}^8 y_{o_5}^6 y_{o_6}^6 t_1^3 t_2^3 t_3^3 t_4^3 t_5 t_6^5 t_7 t_8^5 +  2 y_{s}^3 y_{o_1}^4 y_{o_2}^8 y_{o_3}^4 y_{o_4}^8 y_{o_5}^6 y_{o_6}^6 t_1^2 t_2^4 t_3^3 t_4^3 t_5 t_6^5 t_7 t_8^5 +  y_{s}^3 y_{o_1}^4 y_{o_2}^8 y_{o_3}^4 y_{o_4}^8 y_{o_5}^6 y_{o_6}^6 t_1 t_2^5 t_3^3 t_4^3 t_5 t_6^5 t_7 t_8^5 - y_{s}^4 y_{o_1}^7 y_{o_2}^9 y_{o_3}^{10} y_{o_4}^6 y_{o_5}^5 y_{o_6}^{11} t_1^4 t_2^4 t_3^7 t_4 t_5^6 t_6^2 t_7^3 t_8^5 -  y_{s}^4 y_{o_1}^7 y_{o_2}^9 y_{o_3}^9 y_{o_4}^7 y_{o_5}^6 y_{o_6}^{10} t_1^7 t_2 t_3^6 t_4^2 t_5^5 t_6^3 t_7^3 t_8^5 -  3 y_{s}^4 y_{o_1}^7 y_{o_2}^9 y_{o_3}^9 y_{o_4}^7 y_{o_5}^6 y_{o_6}^{10} t_1^6 t_2^2 t_3^6 t_4^2 t_5^5 t_6^3 t_7^3 t_8^5 -  3 y_{s}^4 y_{o_1}^7 y_{o_2}^9 y_{o_3}^9 y_{o_4}^7 y_{o_5}^6 y_{o_6}^{10} t_1^5 t_2^3 t_3^6 t_4^2 t_5^5 t_6^3 t_7^3 t_8^5 -  7 y_{s}^4 y_{o_1}^7 y_{o_2}^9 y_{o_3}^9 y_{o_4}^7 y_{o_5}^6 y_{o_6}^{10} t_1^4 t_2^4 t_3^6 t_4^2 t_5^5 t_6^3 t_7^3 t_8^5 -  3 y_{s}^4 y_{o_1}^7 y_{o_2}^9 y_{o_3}^9 y_{o_4}^7 y_{o_5}^6 y_{o_6}^{10} t_1^3 t_2^5 t_3^6 t_4^2 t_5^5 t_6^3 t_7^3 t_8^5 -  3 y_{s}^4 y_{o_1}^7 y_{o_2}^9 y_{o_3}^9 y_{o_4}^7 y_{o_5}^6 y_{o_6}^{10} t_1^2 t_2^6 t_3^6 t_4^2 t_5^5 t_6^3 t_7^3 t_8^5 -  y_{s}^4 y_{o_1}^7 y_{o_2}^9 y_{o_3}^9 y_{o_4}^7 y_{o_5}^6 y_{o_6}^{10} t_1 t_2^7 t_3^6 t_4^2 t_5^5 t_6^3 t_7^3 t_8^5 -  y_{s}^4 y_{o_1}^7 y_{o_2}^9 y_{o_3}^8 y_{o_4}^8 y_{o_5}^7 y_{o_6}^9 t_1^7 t_2 t_3^5 t_4^3 t_5^4 t_6^4 t_7^3 t_8^5 -  5 y_{s}^4 y_{o_1}^7 y_{o_2}^9 y_{o_3}^8 y_{o_4}^8 y_{o_5}^7 y_{o_6}^9 t_1^6 t_2^2 t_3^5 t_4^3 t_5^4 t_6^4 t_7^3 t_8^5 -  3 y_{s}^4 y_{o_1}^7 y_{o_2}^9 y_{o_3}^8 y_{o_4}^8 y_{o_5}^7 y_{o_6}^9 t_1^5 t_2^3 t_3^5 t_4^3 t_5^4 t_6^4 t_7^3 t_8^5 -  {12} y_{s}^4 y_{o_1}^7 y_{o_2}^9 y_{o_3}^8 y_{o_4}^8 y_{o_5}^7 y_{o_6}^9 t_1^4 t_2^4 t_3^5 t_4^3 t_5^4 t_6^4 t_7^3 t_8^5 -  3 y_{s}^4 y_{o_1}^7 y_{o_2}^9 y_{o_3}^8 y_{o_4}^8 y_{o_5}^7 y_{o_6}^9 t_1^3 t_2^5 t_3^5 t_4^3 t_5^4 t_6^4 t_7^3 t_8^5 -  5 y_{s}^4 y_{o_1}^7 y_{o_2}^9 y_{o_3}^8 y_{o_4}^8 y_{o_5}^7 y_{o_6}^9 t_1^2 t_2^6 t_3^5 t_4^3 t_5^4 t_6^4 t_7^3 t_8^5 -  y_{s}^4 y_{o_1}^7 y_{o_2}^9 y_{o_3}^8 y_{o_4}^8 y_{o_5}^7 y_{o_6}^9 t_1 t_2^7 t_3^5 t_4^3 t_5^4 t_6^4 t_7^3 t_8^5 -  y_{s}^4 y_{o_1}^7 y_{o_2}^9 y_{o_3}^7 y_{o_4}^9 y_{o_5}^8 y_{o_6}^8 t_1^7 t_2 t_3^4 t_4^4 t_5^3 t_6^5 t_7^3 t_8^5 -  5 y_{s}^4 y_{o_1}^7 y_{o_2}^9 y_{o_3}^7 y_{o_4}^9 y_{o_5}^8 y_{o_6}^8 t_1^6 t_2^2 t_3^4 t_4^4 t_5^3 t_6^5 t_7^3 t_8^5 -  3 y_{s}^4 y_{o_1}^7 y_{o_2}^9 y_{o_3}^7 y_{o_4}^9 y_{o_5}^8 y_{o_6}^8 t_1^5 t_2^3 t_3^4 t_4^4 t_5^3 t_6^5 t_7^3 t_8^5 -  {12} y_{s}^4 y_{o_1}^7 y_{o_2}^9 y_{o_3}^7 y_{o_4}^9 y_{o_5}^8 y_{o_6}^8 t_1^4 t_2^4 t_3^4 t_4^4 t_5^3 t_6^5 t_7^3 t_8^5 -  3 y_{s}^4 y_{o_1}^7 y_{o_2}^9 y_{o_3}^7 y_{o_4}^9 y_{o_5}^8 y_{o_6}^8 t_1^3 t_2^5 t_3^4 t_4^4 t_5^3 t_6^5 t_7^3 t_8^5 -  5 y_{s}^4 y_{o_1}^7 y_{o_2}^9 y_{o_3}^7 y_{o_4}^9 y_{o_5}^8 y_{o_6}^8 t_1^2 t_2^6 t_3^4 t_4^4 t_5^3 t_6^5 t_7^3 t_8^5 -  y_{s}^4 y_{o_1}^7 y_{o_2}^9 y_{o_3}^7 y_{o_4}^9 y_{o_5}^8 y_{o_6}^8 t_1 t_2^7 t_3^4 t_4^4 t_5^3 t_6^5 t_7^3 t_8^5 -  y_{s}^4 y_{o_1}^7 y_{o_2}^9 y_{o_3}^6 y_{o_4}^{10} y_{o_5}^9 y_{o_6}^7 t_1^7 t_2 t_3^3 t_4^5 t_5^2 t_6^6 t_7^3 t_8^5 -  3 y_{s}^4 y_{o_1}^7 y_{o_2}^9 y_{o_3}^6 y_{o_4}^{10} y_{o_5}^9 y_{o_6}^7 t_1^6 t_2^2 t_3^3 t_4^5 t_5^2 t_6^6 t_7^3 t_8^5 -  3 y_{s}^4 y_{o_1}^7 y_{o_2}^9 y_{o_3}^6 y_{o_4}^{10} y_{o_5}^9 y_{o_6}^7 t_1^5 t_2^3 t_3^3 t_4^5 t_5^2 t_6^6 t_7^3 t_8^5 -  7 y_{s}^4 y_{o_1}^7 y_{o_2}^9 y_{o_3}^6 y_{o_4}^{10} y_{o_5}^9 y_{o_6}^7 t_1^4 t_2^4 t_3^3 t_4^5 t_5^2 t_6^6 t_7^3 t_8^5 -  3 y_{s}^4 y_{o_1}^7 y_{o_2}^9 y_{o_3}^6 y_{o_4}^{10} y_{o_5}^9 y_{o_6}^7 t_1^3 t_2^5 t_3^3 t_4^5 t_5^2 t_6^6 t_7^3 t_8^5 -  3 y_{s}^4 y_{o_1}^7 y_{o_2}^9 y_{o_3}^6 y_{o_4}^{10} y_{o_5}^9 y_{o_6}^7 t_1^2 t_2^6 t_3^3 t_4^5 t_5^2 t_6^6 t_7^3 t_8^5 -  y_{s}^4 y_{o_1}^7 y_{o_2}^9 y_{o_3}^6 y_{o_4}^{10} y_{o_5}^9 y_{o_6}^7 t_1 t_2^7 t_3^3 t_4^5 t_5^2 t_6^6 t_7^3 t_8^5 -  y_{s}^4 y_{o_1}^7 y_{o_2}^9 y_{o_3}^5 y_{o_4}^{11} y_{o_5}^{10} y_{o_6}^6 t_1^4 t_2^4 t_3^2 t_4^6 t_5 t_6^7 t_7^3 t_8^5 -  y_{s}^5 y_{o_1}^{10} y_{o_2}^{10} y_{o_3}^{13} y_{o_4}^7 y_{o_5}^7 y_{o_6}^{13} t_1^5 t_2^5 t_3^8 t_4^2 t_5^8 t_6^2 t_7^5 t_8^5 +  3 y_{s}^5 y_{o_1}^{10} y_{o_2}^{10} y_{o_3}^{12} y_{o_4}^8 y_{o_5}^8 y_{o_6}^{12} t_1^6 t_2^4 t_3^7 t_4^3 t_5^7 t_6^3 t_7^5 t_8^5 -  y_{s}^5 y_{o_1}^{10} y_{o_2}^{10} y_{o_3}^{12} y_{o_4}^8 y_{o_5}^8 y_{o_6}^{12} t_1^5 t_2^5 t_3^7 t_4^3 t_5^7 t_6^3 t_7^5 t_8^5 +  3 y_{s}^5 y_{o_1}^{10} y_{o_2}^{10} y_{o_3}^{12} y_{o_4}^8 y_{o_5}^8 y_{o_6}^{12} t_1^4 t_2^6 t_3^7 t_4^3 t_5^7 t_6^3 t_7^5 t_8^5 +  y_{s}^5 y_{o_1}^{10} y_{o_2}^{10} y_{o_3}^{11} y_{o_4}^9 y_{o_5}^9 y_{o_6}^{11} t_1^9 t_2 t_3^6 t_4^4 t_5^6 t_6^4 t_7^5 t_8^5 +  2 y_{s}^5 y_{o_1}^{10} y_{o_2}^{10} y_{o_3}^{11} y_{o_4}^9 y_{o_5}^9 y_{o_6}^{11} t_1^8 t_2^2 t_3^6 t_4^4 t_5^6 t_6^4 t_7^5 t_8^5 +  y_{s}^5 y_{o_1}^{10} y_{o_2}^{10} y_{o_3}^{11} y_{o_4}^9 y_{o_5}^9 y_{o_6}^{11} t_1^7 t_2^3 t_3^6 t_4^4 t_5^6 t_6^4 t_7^5 t_8^5 +  {10} y_{s}^5 y_{o_1}^{10} y_{o_2}^{10} y_{o_3}^{11} y_{o_4}^9 y_{o_5}^9 y_{o_6}^{11} t_1^6 t_2^4 t_3^6 t_4^4 t_5^6 t_6^4 t_7^5 t_8^5 -  y_{s}^5 y_{o_1}^{10} y_{o_2}^{10} y_{o_3}^{11} y_{o_4}^9 y_{o_5}^9 y_{o_6}^{11} t_1^5 t_2^5 t_3^6 t_4^4 t_5^6 t_6^4 t_7^5 t_8^5 +  {10} y_{s}^5 y_{o_1}^{10} y_{o_2}^{10} y_{o_3}^{11} y_{o_4}^9 y_{o_5}^9 y_{o_6}^{11} t_1^4 t_2^6 t_3^6 t_4^4 t_5^6 t_6^4 t_7^5 t_8^5 +  y_{s}^5 y_{o_1}^{10} y_{o_2}^{10} y_{o_3}^{11} y_{o_4}^9 y_{o_5}^9 y_{o_6}^{11} t_1^3 t_2^7 t_3^6 t_4^4 t_5^6 t_6^4 t_7^5 t_8^5 +  2 y_{s}^5 y_{o_1}^{10} y_{o_2}^{10} y_{o_3}^{11} y_{o_4}^9 y_{o_5}^9 y_{o_6}^{11} t_1^2 t_2^8 t_3^6 t_4^4 t_5^6 t_6^4 t_7^5 t_8^5 +  y_{s}^5 y_{o_1}^{10} y_{o_2}^{10} y_{o_3}^{11} y_{o_4}^9 y_{o_5}^9 y_{o_6}^{11} t_1 t_2^9 t_3^6 t_4^4 t_5^6 t_6^4 t_7^5 t_8^5 -  y_{s}^5 y_{o_1}^{10} y_{o_2}^{10} y_{o_3}^{10} y_{o_4}^{10} y_{o_5}^{10} y_{o_6}^{10} t_1^{10} t_3^5 t_4^5 t_5^5 t_6^5 t_7^5 t_8^5 -  y_{s}^5 y_{o_1}^{10} y_{o_2}^{10} y_{o_3}^{10} y_{o_4}^{10} y_{o_5}^{10} y_{o_6}^{10} t_1^9 t_2 t_3^5 t_4^5 t_5^5 t_6^5 t_7^5 t_8^5 -  y_{s}^5 y_{o_1}^{10} y_{o_2}^{10} y_{o_3}^{10} y_{o_4}^{10} y_{o_5}^{10} y_{o_6}^{10} t_1^8 t_2^2 t_3^5 t_4^5 t_5^5 t_6^5 t_7^5 t_8^5 -  4 y_{s}^5 y_{o_1}^{10} y_{o_2}^{10} y_{o_3}^{10} y_{o_4}^{10} y_{o_5}^{10} y_{o_6}^{10} t_1^7 t_2^3 t_3^5 t_4^5 t_5^5 t_6^5 t_7^5 t_8^5 +  8 y_{s}^5 y_{o_1}^{10} y_{o_2}^{10} y_{o_3}^{10} y_{o_4}^{10} y_{o_5}^{10} y_{o_6}^{10} t_1^6 t_2^4 t_3^5 t_4^5 t_5^5 t_6^5 t_7^5 t_8^5 -  8 y_{s}^5 y_{o_1}^{10} y_{o_2}^{10} y_{o_3}^{10} y_{o_4}^{10} y_{o_5}^{10} y_{o_6}^{10} t_1^5 t_2^5 t_3^5 t_4^5 t_5^5 t_6^5 t_7^5 t_8^5 +  8 y_{s}^5 y_{o_1}^{10} y_{o_2}^{10} y_{o_3}^{10} y_{o_4}^{10} y_{o_5}^{10} y_{o_6}^{10} t_1^4 t_2^6 t_3^5 t_4^5 t_5^5 t_6^5 t_7^5 t_8^5 -  4 y_{s}^5 y_{o_1}^{10} y_{o_2}^{10} y_{o_3}^{10} y_{o_4}^{10} y_{o_5}^{10} y_{o_6}^{10} t_1^3 t_2^7 t_3^5 t_4^5 t_5^5 t_6^5 t_7^5 t_8^5 -  y_{s}^5 y_{o_1}^{10} y_{o_2}^{10} y_{o_3}^{10} y_{o_4}^{10} y_{o_5}^{10} y_{o_6}^{10} t_1^2 t_2^8 t_3^5 t_4^5 t_5^5 t_6^5 t_7^5 t_8^5 -  y_{s}^5 y_{o_1}^{10} y_{o_2}^{10} y_{o_3}^{10} y_{o_4}^{10} y_{o_5}^{10} y_{o_6}^{10} t_1 t_2^9 t_3^5 t_4^5 t_5^5 t_6^5 t_7^5 t_8^5 -  y_{s}^5 y_{o_1}^{10} y_{o_2}^{10} y_{o_3}^{10} y_{o_4}^{10} y_{o_5}^{10} y_{o_6}^{10} t_2^{10} t_3^5 t_4^5 t_5^5 t_6^5 t_7^5 t_8^5 +  y_{s}^5 y_{o_1}^{10} y_{o_2}^{10} y_{o_3}^9 y_{o_4}^{11} y_{o_5}^{11} y_{o_6}^9 t_1^9 t_2 t_3^4 t_4^6 t_5^4 t_6^6 t_7^5 t_8^5 +  2 y_{s}^5 y_{o_1}^{10} y_{o_2}^{10} y_{o_3}^9 y_{o_4}^{11} y_{o_5}^{11} y_{o_6}^9 t_1^8 t_2^2 t_3^4 t_4^6 t_5^4 t_6^6 t_7^5 t_8^5 +  y_{s}^5 y_{o_1}^{10} y_{o_2}^{10} y_{o_3}^9 y_{o_4}^{11} y_{o_5}^{11} y_{o_6}^9 t_1^7 t_2^3 t_3^4 t_4^6 t_5^4 t_6^6 t_7^5 t_8^5 +  {10} y_{s}^5 y_{o_1}^{10} y_{o_2}^{10} y_{o_3}^9 y_{o_4}^{11} y_{o_5}^{11} y_{o_6}^9 t_1^6 t_2^4 t_3^4 t_4^6 t_5^4 t_6^6 t_7^5 t_8^5 -  y_{s}^5 y_{o_1}^{10} y_{o_2}^{10} y_{o_3}^9 y_{o_4}^{11} y_{o_5}^{11} y_{o_6}^9 t_1^5 t_2^5 t_3^4 t_4^6 t_5^4 t_6^6 t_7^5 t_8^5 +  {10} y_{s}^5 y_{o_1}^{10} y_{o_2}^{10} y_{o_3}^9 y_{o_4}^{11} y_{o_5}^{11} y_{o_6}^9 t_1^4 t_2^6 t_3^4 t_4^6 t_5^4 t_6^6 t_7^5 t_8^5 +  y_{s}^5 y_{o_1}^{10} y_{o_2}^{10} y_{o_3}^9 y_{o_4}^{11} y_{o_5}^{11} y_{o_6}^9 t_1^3 t_2^7 t_3^4 t_4^6 t_5^4 t_6^6 t_7^5 t_8^5 +  2 y_{s}^5 y_{o_1}^{10} y_{o_2}^{10} y_{o_3}^9 y_{o_4}^{11} y_{o_5}^{11} y_{o_6}^9 t_1^2 t_2^8 t_3^4 t_4^6 t_5^4 t_6^6 t_7^5 t_8^5 +  y_{s}^5 y_{o_1}^{10} y_{o_2}^{10} y_{o_3}^9 y_{o_4}^{11} y_{o_5}^{11} y_{o_6}^9 t_1 t_2^9 t_3^4 t_4^6 t_5^4 t_6^6 t_7^5 t_8^5 +  3 y_{s}^5 y_{o_1}^{10} y_{o_2}^{10} y_{o_3}^8 y_{o_4}^{12} y_{o_5}^{12} y_{o_6}^8 t_1^6 t_2^4 t_3^3 t_4^7 t_5^3 t_6^7 t_7^5 t_8^5 -  y_{s}^5 y_{o_1}^{10} y_{o_2}^{10} y_{o_3}^8 y_{o_4}^{12} y_{o_5}^{12} y_{o_6}^8 t_1^5 t_2^5 t_3^3 t_4^7 t_5^3 t_6^7 t_7^5 t_8^5 +  3 y_{s}^5 y_{o_1}^{10} y_{o_2}^{10} y_{o_3}^8 y_{o_4}^{12} y_{o_5}^{12} y_{o_6}^8 t_1^4 t_2^6 t_3^3 t_4^7 t_5^3 t_6^7 t_7^5 t_8^5 -  y_{s}^5 y_{o_1}^{10} y_{o_2}^{10} y_{o_3}^7 y_{o_4}^{13} y_{o_5}^{13} y_{o_6}^7 t_1^5 t_2^5 t_3^2 t_4^8 t_5^2 t_6^8 t_7^5 t_8^5 +  y_{s}^6 y_{o_1}^{13} y_{o_2}^{11} y_{o_3}^{15} y_{o_4}^9 y_{o_5}^{10} y_{o_6}^{14} t_1^7 t_2^5 t_3^8 t_4^4 t_5^9 t_6^3 t_7^7 t_8^5 +  y_{s}^6 y_{o_1}^{13} y_{o_2}^{11} y_{o_3}^{15} y_{o_4}^9 y_{o_5}^{10} y_{o_6}^{14} t_1^5 t_2^7 t_3^8 t_4^4 t_5^9 t_6^3 t_7^7 t_8^5 -  y_{s}^6 y_{o_1}^{13} y_{o_2}^{11} y_{o_3}^{14} y_{o_4}^{10} y_{o_5}^{11} y_{o_6}^{13} t_1^8 t_2^4 t_3^7 t_4^5 t_5^8 t_6^4 t_7^7 t_8^5 +  3 y_{s}^6 y_{o_1}^{13} y_{o_2}^{11} y_{o_3}^{14} y_{o_4}^{10} y_{o_5}^{11} y_{o_6}^{13} t_1^7 t_2^5 t_3^7 t_4^5 t_5^8 t_6^4 t_7^7 t_8^5 -  3 y_{s}^6 y_{o_1}^{13} y_{o_2}^{11} y_{o_3}^{14} y_{o_4}^{10} y_{o_5}^{11} y_{o_6}^{13} t_1^6 t_2^6 t_3^7 t_4^5 t_5^8 t_6^4 t_7^7 t_8^5 +  3 y_{s}^6 y_{o_1}^{13} y_{o_2}^{11} y_{o_3}^{14} y_{o_4}^{10} y_{o_5}^{11} y_{o_6}^{13} t_1^5 t_2^7 t_3^7 t_4^5 t_5^8 t_6^4 t_7^7 t_8^5 -  y_{s}^6 y_{o_1}^{13} y_{o_2}^{11} y_{o_3}^{14} y_{o_4}^{10} y_{o_5}^{11} y_{o_6}^{13} t_1^4 t_2^8 t_3^7 t_4^5 t_5^8 t_6^4 t_7^7 t_8^5 +  y_{s}^6 y_{o_1}^{13} y_{o_2}^{11} y_{o_3}^{13} y_{o_4}^{11} y_{o_5}^{12} y_{o_6}^{12} t_1^{10} t_2^2 t_3^6 t_4^6 t_5^7 t_6^5 t_7^7 t_8^5 +  2 y_{s}^6 y_{o_1}^{13} y_{o_2}^{11} y_{o_3}^{13} y_{o_4}^{11} y_{o_5}^{12} y_{o_6}^{12} t_1^9 t_2^3 t_3^6 t_4^6 t_5^7 t_6^5 t_7^7 t_8^5 +  y_{s}^6 y_{o_1}^{13} y_{o_2}^{11} y_{o_3}^{13} y_{o_4}^{11} y_{o_5}^{12} y_{o_6}^{12} t_1^8 t_2^4 t_3^6 t_4^6 t_5^7 t_6^5 t_7^7 t_8^5 +  {10} y_{s}^6 y_{o_1}^{13} y_{o_2}^{11} y_{o_3}^{13} y_{o_4}^{11} y_{o_5}^{12} y_{o_6}^{12} t_1^7 t_2^5 t_3^6 t_4^6 t_5^7 t_6^5 t_7^7 t_8^5 -  y_{s}^6 y_{o_1}^{13} y_{o_2}^{11} y_{o_3}^{13} y_{o_4}^{11} y_{o_5}^{12} y_{o_6}^{12} t_1^6 t_2^6 t_3^6 t_4^6 t_5^7 t_6^5 t_7^7 t_8^5 +  {10} y_{s}^6 y_{o_1}^{13} y_{o_2}^{11} y_{o_3}^{13} y_{o_4}^{11} y_{o_5}^{12} y_{o_6}^{12} t_1^5 t_2^7 t_3^6 t_4^6 t_5^7 t_6^5 t_7^7 t_8^5 +  y_{s}^6 y_{o_1}^{13} y_{o_2}^{11} y_{o_3}^{13} y_{o_4}^{11} y_{o_5}^{12} y_{o_6}^{12} t_1^4 t_2^8 t_3^6 t_4^6 t_5^7 t_6^5 t_7^7 t_8^5 +  2 y_{s}^6 y_{o_1}^{13} y_{o_2}^{11} y_{o_3}^{13} y_{o_4}^{11} y_{o_5}^{12} y_{o_6}^{12} t_1^3 t_2^9 t_3^6 t_4^6 t_5^7 t_6^5 t_7^7 t_8^5 +  y_{s}^6 y_{o_1}^{13} y_{o_2}^{11} y_{o_3}^{13} y_{o_4}^{11} y_{o_5}^{12} y_{o_6}^{12} t_1^2 t_2^{10} t_3^6 t_4^6 t_5^7 t_6^5 t_7^7 t_8^5 +  y_{s}^6 y_{o_1}^{13} y_{o_2}^{11} y_{o_3}^{12} y_{o_4}^{12} y_{o_5}^{13} y_{o_6}^{11} t_1^{10} t_2^2 t_3^5 t_4^7 t_5^6 t_6^6 t_7^7 t_8^5 +  2 y_{s}^6 y_{o_1}^{13} y_{o_2}^{11} y_{o_3}^{12} y_{o_4}^{12} y_{o_5}^{13} y_{o_6}^{11} t_1^9 t_2^3 t_3^5 t_4^7 t_5^6 t_6^6 t_7^7 t_8^5 +  y_{s}^6 y_{o_1}^{13} y_{o_2}^{11} y_{o_3}^{12} y_{o_4}^{12} y_{o_5}^{13} y_{o_6}^{11} t_1^8 t_2^4 t_3^5 t_4^7 t_5^6 t_6^6 t_7^7 t_8^5 +  {10} y_{s}^6 y_{o_1}^{13} y_{o_2}^{11} y_{o_3}^{12} y_{o_4}^{12} y_{o_5}^{13} y_{o_6}^{11} t_1^7 t_2^5 t_3^5 t_4^7 t_5^6 t_6^6 t_7^7 t_8^5 -  y_{s}^6 y_{o_1}^{13} y_{o_2}^{11} y_{o_3}^{12} y_{o_4}^{12} y_{o_5}^{13} y_{o_6}^{11} t_1^6 t_2^6 t_3^5 t_4^7 t_5^6 t_6^6 t_7^7 t_8^5 +  {10} y_{s}^6 y_{o_1}^{13} y_{o_2}^{11} y_{o_3}^{12} y_{o_4}^{12} y_{o_5}^{13} y_{o_6}^{11} t_1^5 t_2^7 t_3^5 t_4^7 t_5^6 t_6^6 t_7^7 t_8^5 +  y_{s}^6 y_{o_1}^{13} y_{o_2}^{11} y_{o_3}^{12} y_{o_4}^{12} y_{o_5}^{13} y_{o_6}^{11} t_1^4 t_2^8 t_3^5 t_4^7 t_5^6 t_6^6 t_7^7 t_8^5 +  2 y_{s}^6 y_{o_1}^{13} y_{o_2}^{11} y_{o_3}^{12} y_{o_4}^{12} y_{o_5}^{13} y_{o_6}^{11} t_1^3 t_2^9 t_3^5 t_4^7 t_5^6 t_6^6 t_7^7 t_8^5 +  y_{s}^6 y_{o_1}^{13} y_{o_2}^{11} y_{o_3}^{12} y_{o_4}^{12} y_{o_5}^{13} y_{o_6}^{11} t_1^2 t_2^{10} t_3^5 t_4^7 t_5^6 t_6^6 t_7^7 t_8^5 -  y_{s}^6 y_{o_1}^{13} y_{o_2}^{11} y_{o_3}^{11} y_{o_4}^{13} y_{o_5}^{14} y_{o_6}^{10} t_1^8 t_2^4 t_3^4 t_4^8 t_5^5 t_6^7 t_7^7 t_8^5 +  3 y_{s}^6 y_{o_1}^{13} y_{o_2}^{11} y_{o_3}^{11} y_{o_4}^{13} y_{o_5}^{14} y_{o_6}^{10} t_1^7 t_2^5 t_3^4 t_4^8 t_5^5 t_6^7 t_7^7 t_8^5 -  3 y_{s}^6 y_{o_1}^{13} y_{o_2}^{11} y_{o_3}^{11} y_{o_4}^{13} y_{o_5}^{14} y_{o_6}^{10} t_1^6 t_2^6 t_3^4 t_4^8 t_5^5 t_6^7 t_7^7 t_8^5 +  3 y_{s}^6 y_{o_1}^{13} y_{o_2}^{11} y_{o_3}^{11} y_{o_4}^{13} y_{o_5}^{14} y_{o_6}^{10} t_1^5 t_2^7 t_3^4 t_4^8 t_5^5 t_6^7 t_7^7 t_8^5 -  y_{s}^6 y_{o_1}^{13} y_{o_2}^{11} y_{o_3}^{11} y_{o_4}^{13} y_{o_5}^{14} y_{o_6}^{10} t_1^4 t_2^8 t_3^4 t_4^8 t_5^5 t_6^7 t_7^7 t_8^5 +  y_{s}^6 y_{o_1}^{13} y_{o_2}^{11} y_{o_3}^{10} y_{o_4}^{14} y_{o_5}^{15} y_{o_6}^9 t_1^7 t_2^5 t_3^3 t_4^9 t_5^4 t_6^8 t_7^7 t_8^5 +  y_{s}^6 y_{o_1}^{13} y_{o_2}^{11} y_{o_3}^{10} y_{o_4}^{14} y_{o_5}^{15} y_{o_6}^9 t_1^5 t_2^7 t_3^3 t_4^9 t_5^4 t_6^8 t_7^7 t_8^5 -  y_{s}^7 y_{o_1}^{16} y_{o_2}^{12} y_{o_3}^{17} y_{o_4}^{11} y_{o_5}^{13} y_{o_6}^{15} t_1^7 t_2^7 t_3^8 t_4^6 t_5^{10} t_6^4 t_7^9 t_8^5 -  y_{s}^7 y_{o_1}^{16} y_{o_2}^{12} y_{o_3}^{16} y_{o_4}^{12} y_{o_5}^{14} y_{o_6}^{14} t_1^9 t_2^5 t_3^7 t_4^7 t_5^9 t_6^5 t_7^9 t_8^5 -  3 y_{s}^7 y_{o_1}^{16} y_{o_2}^{12} y_{o_3}^{16} y_{o_4}^{12} y_{o_5}^{14} y_{o_6}^{14} t_1^7 t_2^7 t_3^7 t_4^7 t_5^9 t_6^5 t_7^9 t_8^5 -  y_{s}^7 y_{o_1}^{16} y_{o_2}^{12} y_{o_3}^{16} y_{o_4}^{12} y_{o_5}^{14} y_{o_6}^{14} t_1^5 t_2^9 t_3^7 t_4^7 t_5^9 t_6^5 t_7^9 t_8^5 -  y_{s}^7 y_{o_1}^{16} y_{o_2}^{12} y_{o_3}^{15} y_{o_4}^{13} y_{o_5}^{15} y_{o_6}^{13} t_1^{10} t_2^4 t_3^6 t_4^8 t_5^8 t_6^6 t_7^9 t_8^5 -  3 y_{s}^7 y_{o_1}^{16} y_{o_2}^{12} y_{o_3}^{15} y_{o_4}^{13} y_{o_5}^{15} y_{o_6}^{13} t_1^9 t_2^5 t_3^6 t_4^8 t_5^8 t_6^6 t_7^9 t_8^5 -  3 y_{s}^7 y_{o_1}^{16} y_{o_2}^{12} y_{o_3}^{15} y_{o_4}^{13} y_{o_5}^{15} y_{o_6}^{13} t_1^8 t_2^6 t_3^6 t_4^8 t_5^8 t_6^6 t_7^9 t_8^5 -  7 y_{s}^7 y_{o_1}^{16} y_{o_2}^{12} y_{o_3}^{15} y_{o_4}^{13} y_{o_5}^{15} y_{o_6}^{13} t_1^7 t_2^7 t_3^6 t_4^8 t_5^8 t_6^6 t_7^9 t_8^5 -  3 y_{s}^7 y_{o_1}^{16} y_{o_2}^{12} y_{o_3}^{15} y_{o_4}^{13} y_{o_5}^{15} y_{o_6}^{13} t_1^6 t_2^8 t_3^6 t_4^8 t_5^8 t_6^6 t_7^9 t_8^5 -  3 y_{s}^7 y_{o_1}^{16} y_{o_2}^{12} y_{o_3}^{15} y_{o_4}^{13} y_{o_5}^{15} y_{o_6}^{13} t_1^5 t_2^9 t_3^6 t_4^8 t_5^8 t_6^6 t_7^9 t_8^5 -  y_{s}^7 y_{o_1}^{16} y_{o_2}^{12} y_{o_3}^{15} y_{o_4}^{13} y_{o_5}^{15} y_{o_6}^{13} t_1^4 t_2^{10} t_3^6 t_4^8 t_5^8 t_6^6 t_7^9 t_8^5 -  y_{s}^7 y_{o_1}^{16} y_{o_2}^{12} y_{o_3}^{14} y_{o_4}^{14} y_{o_5}^{16} y_{o_6}^{12} t_1^9 t_2^5 t_3^5 t_4^9 t_5^7 t_6^7 t_7^9 t_8^5 -  3 y_{s}^7 y_{o_1}^{16} y_{o_2}^{12} y_{o_3}^{14} y_{o_4}^{14} y_{o_5}^{16} y_{o_6}^{12} t_1^7 t_2^7 t_3^5 t_4^9 t_5^7 t_6^7 t_7^9 t_8^5 -  y_{s}^7 y_{o_1}^{16} y_{o_2}^{12} y_{o_3}^{14} y_{o_4}^{14} y_{o_5}^{16} y_{o_6}^{12} t_1^5 t_2^9 t_3^5 t_4^9 t_5^7 t_6^7 t_7^9 t_8^5 -  y_{s}^7 y_{o_1}^{16} y_{o_2}^{12} y_{o_3}^{13} y_{o_4}^{15} y_{o_5}^{17} y_{o_6}^{11} t_1^7 t_2^7 t_3^4 t_4^{10} t_5^6 t_6^8 t_7^9 t_8^5 -  y_{s}^4 y_{o_1}^6 y_{o_2}^{10} y_{o_3}^9 y_{o_4}^7 y_{o_5}^5 y_{o_6}^{11} t_1^4 t_2^4 t_3^7 t_4 t_5^5 t_6^3 t_7^2 t_8^6 -  y_{s}^4 y_{o_1}^6 y_{o_2}^{10} y_{o_3}^8 y_{o_4}^8 y_{o_5}^6 y_{o_6}^{10} t_1^6 t_2^2 t_3^6 t_4^2 t_5^4 t_6^4 t_7^2 t_8^6 -  3 y_{s}^4 y_{o_1}^6 y_{o_2}^{10} y_{o_3}^8 y_{o_4}^8 y_{o_5}^6 y_{o_6}^{10} t_1^4 t_2^4 t_3^6 t_4^2 t_5^4 t_6^4 t_7^2 t_8^6 -  y_{s}^4 y_{o_1}^6 y_{o_2}^{10} y_{o_3}^8 y_{o_4}^8 y_{o_5}^6 y_{o_6}^{10} t_1^2 t_2^6 t_3^6 t_4^2 t_5^4 t_6^4 t_7^2 t_8^6 -  y_{s}^4 y_{o_1}^6 y_{o_2}^{10} y_{o_3}^7 y_{o_4}^9 y_{o_5}^7 y_{o_6}^9 t_1^7 t_2 t_3^5 t_4^3 t_5^3 t_6^5 t_7^2 t_8^6 -  3 y_{s}^4 y_{o_1}^6 y_{o_2}^{10} y_{o_3}^7 y_{o_4}^9 y_{o_5}^7 y_{o_6}^9 t_1^6 t_2^2 t_3^5 t_4^3 t_5^3 t_6^5 t_7^2 t_8^6 -  3 y_{s}^4 y_{o_1}^6 y_{o_2}^{10} y_{o_3}^7 y_{o_4}^9 y_{o_5}^7 y_{o_6}^9 t_1^5 t_2^3 t_3^5 t_4^3 t_5^3 t_6^5 t_7^2 t_8^6 -  7 y_{s}^4 y_{o_1}^6 y_{o_2}^{10} y_{o_3}^7 y_{o_4}^9 y_{o_5}^7 y_{o_6}^9 t_1^4 t_2^4 t_3^5 t_4^3 t_5^3 t_6^5 t_7^2 t_8^6 -  3 y_{s}^4 y_{o_1}^6 y_{o_2}^{10} y_{o_3}^7 y_{o_4}^9 y_{o_5}^7 y_{o_6}^9 t_1^3 t_2^5 t_3^5 t_4^3 t_5^3 t_6^5 t_7^2 t_8^6 -  3 y_{s}^4 y_{o_1}^6 y_{o_2}^{10} y_{o_3}^7 y_{o_4}^9 y_{o_5}^7 y_{o_6}^9 t_1^2 t_2^6 t_3^5 t_4^3 t_5^3 t_6^5 t_7^2 t_8^6 -  y_{s}^4 y_{o_1}^6 y_{o_2}^{10} y_{o_3}^7 y_{o_4}^9 y_{o_5}^7 y_{o_6}^9 t_1 t_2^7 t_3^5 t_4^3 t_5^3 t_6^5 t_7^2 t_8^6 -  y_{s}^4 y_{o_1}^6 y_{o_2}^{10} y_{o_3}^6 y_{o_4}^{10} y_{o_5}^8 y_{o_6}^8 t_1^6 t_2^2 t_3^4 t_4^4 t_5^2 t_6^6 t_7^2 t_8^6 -  3 y_{s}^4 y_{o_1}^6 y_{o_2}^{10} y_{o_3}^6 y_{o_4}^{10} y_{o_5}^8 y_{o_6}^8 t_1^4 t_2^4 t_3^4 t_4^4 t_5^2 t_6^6 t_7^2 t_8^6 -  y_{s}^4 y_{o_1}^6 y_{o_2}^{10} y_{o_3}^6 y_{o_4}^{10} y_{o_5}^8 y_{o_6}^8 t_1^2 t_2^6 t_3^4 t_4^4 t_5^2 t_6^6 t_7^2 t_8^6 -  y_{s}^4 y_{o_1}^6 y_{o_2}^{10} y_{o_3}^5 y_{o_4}^{11} y_{o_5}^9 y_{o_6}^7 t_1^4 t_2^4 t_3^3 t_4^5 t_5 t_6^7 t_7^2 t_8^6 +  y_{s}^5 y_{o_1}^9 y_{o_2}^{11} y_{o_3}^{12} y_{o_4}^8 y_{o_5}^7 y_{o_6}^{13} t_1^6 t_2^4 t_3^8 t_4^2 t_5^7 t_6^3 t_7^4 t_8^6 +  y_{s}^5 y_{o_1}^9 y_{o_2}^{11} y_{o_3}^{12} y_{o_4}^8 y_{o_5}^7 y_{o_6}^{13} t_1^4 t_2^6 t_3^8 t_4^2 t_5^7 t_6^3 t_7^4 t_8^6 -  y_{s}^5 y_{o_1}^9 y_{o_2}^{11} y_{o_3}^{11} y_{o_4}^9 y_{o_5}^8 y_{o_6}^{12} t_1^7 t_2^3 t_3^7 t_4^3 t_5^6 t_6^4 t_7^4 t_8^6 +  3 y_{s}^5 y_{o_1}^9 y_{o_2}^{11} y_{o_3}^{11} y_{o_4}^9 y_{o_5}^8 y_{o_6}^{12} t_1^6 t_2^4 t_3^7 t_4^3 t_5^6 t_6^4 t_7^4 t_8^6 -  3 y_{s}^5 y_{o_1}^9 y_{o_2}^{11} y_{o_3}^{11} y_{o_4}^9 y_{o_5}^8 y_{o_6}^{12} t_1^5 t_2^5 t_3^7 t_4^3 t_5^6 t_6^4 t_7^4 t_8^6 +  3 y_{s}^5 y_{o_1}^9 y_{o_2}^{11} y_{o_3}^{11} y_{o_4}^9 y_{o_5}^8 y_{o_6}^{12} t_1^4 t_2^6 t_3^7 t_4^3 t_5^6 t_6^4 t_7^4 t_8^6 -  y_{s}^5 y_{o_1}^9 y_{o_2}^{11} y_{o_3}^{11} y_{o_4}^9 y_{o_5}^8 y_{o_6}^{12} t_1^3 t_2^7 t_3^7 t_4^3 t_5^6 t_6^4 t_7^4 t_8^6 +  y_{s}^5 y_{o_1}^9 y_{o_2}^{11} y_{o_3}^{10} y_{o_4}^{10} y_{o_5}^9 y_{o_6}^{11} t_1^9 t_2 t_3^6 t_4^4 t_5^5 t_6^5 t_7^4 t_8^6 +  2 y_{s}^5 y_{o_1}^9 y_{o_2}^{11} y_{o_3}^{10} y_{o_4}^{10} y_{o_5}^9 y_{o_6}^{11} t_1^8 t_2^2 t_3^6 t_4^4 t_5^5 t_6^5 t_7^4 t_8^6 +  y_{s}^5 y_{o_1}^9 y_{o_2}^{11} y_{o_3}^{10} y_{o_4}^{10} y_{o_5}^9 y_{o_6}^{11} t_1^7 t_2^3 t_3^6 t_4^4 t_5^5 t_6^5 t_7^4 t_8^6 +  {10} y_{s}^5 y_{o_1}^9 y_{o_2}^{11} y_{o_3}^{10} y_{o_4}^{10} y_{o_5}^9 y_{o_6}^{11} t_1^6 t_2^4 t_3^6 t_4^4 t_5^5 t_6^5 t_7^4 t_8^6 -  y_{s}^5 y_{o_1}^9 y_{o_2}^{11} y_{o_3}^{10} y_{o_4}^{10} y_{o_5}^9 y_{o_6}^{11} t_1^5 t_2^5 t_3^6 t_4^4 t_5^5 t_6^5 t_7^4 t_8^6 +  {10} y_{s}^5 y_{o_1}^9 y_{o_2}^{11} y_{o_3}^{10} y_{o_4}^{10} y_{o_5}^9 y_{o_6}^{11} t_1^4 t_2^6 t_3^6 t_4^4 t_5^5 t_6^5 t_7^4 t_8^6 +  y_{s}^5 y_{o_1}^9 y_{o_2}^{11} y_{o_3}^{10} y_{o_4}^{10} y_{o_5}^9 y_{o_6}^{11} t_1^3 t_2^7 t_3^6 t_4^4 t_5^5 t_6^5 t_7^4 t_8^6 +  2 y_{s}^5 y_{o_1}^9 y_{o_2}^{11} y_{o_3}^{10} y_{o_4}^{10} y_{o_5}^9 y_{o_6}^{11} t_1^2 t_2^8 t_3^6 t_4^4 t_5^5 t_6^5 t_7^4 t_8^6 +  y_{s}^5 y_{o_1}^9 y_{o_2}^{11} y_{o_3}^{10} y_{o_4}^{10} y_{o_5}^9 y_{o_6}^{11} t_1 t_2^9 t_3^6 t_4^4 t_5^5 t_6^5 t_7^4 t_8^6 +  y_{s}^5 y_{o_1}^9 y_{o_2}^{11} y_{o_3}^9 y_{o_4}^{11} y_{o_5}^{10} y_{o_6}^{10} t_1^9 t_2 t_3^5 t_4^5 t_5^4 t_6^6 t_7^4 t_8^6 +  2 y_{s}^5 y_{o_1}^9 y_{o_2}^{11} y_{o_3}^9 y_{o_4}^{11} y_{o_5}^{10} y_{o_6}^{10} t_1^8 t_2^2 t_3^5 t_4^5 t_5^4 t_6^6 t_7^4 t_8^6 +  y_{s}^5 y_{o_1}^9 y_{o_2}^{11} y_{o_3}^9 y_{o_4}^{11} y_{o_5}^{10} y_{o_6}^{10} t_1^7 t_2^3 t_3^5 t_4^5 t_5^4 t_6^6 t_7^4 t_8^6 +  {10} y_{s}^5 y_{o_1}^9 y_{o_2}^{11} y_{o_3}^9 y_{o_4}^{11} y_{o_5}^{10} y_{o_6}^{10} t_1^6 t_2^4 t_3^5 t_4^5 t_5^4 t_6^6 t_7^4 t_8^6 -  y_{s}^5 y_{o_1}^9 y_{o_2}^{11} y_{o_3}^9 y_{o_4}^{11} y_{o_5}^{10} y_{o_6}^{10} t_1^5 t_2^5 t_3^5 t_4^5 t_5^4 t_6^6 t_7^4 t_8^6 +  {10} y_{s}^5 y_{o_1}^9 y_{o_2}^{11} y_{o_3}^9 y_{o_4}^{11} y_{o_5}^{10} y_{o_6}^{10} t_1^4 t_2^6 t_3^5 t_4^5 t_5^4 t_6^6 t_7^4 t_8^6 +  y_{s}^5 y_{o_1}^9 y_{o_2}^{11} y_{o_3}^9 y_{o_4}^{11} y_{o_5}^{10} y_{o_6}^{10} t_1^3 t_2^7 t_3^5 t_4^5 t_5^4 t_6^6 t_7^4 t_8^6 +  2 y_{s}^5 y_{o_1}^9 y_{o_2}^{11} y_{o_3}^9 y_{o_4}^{11} y_{o_5}^{10} y_{o_6}^{10} t_1^2 t_2^8 t_3^5 t_4^5 t_5^4 t_6^6 t_7^4 t_8^6 +  y_{s}^5 y_{o_1}^9 y_{o_2}^{11} y_{o_3}^9 y_{o_4}^{11} y_{o_5}^{10} y_{o_6}^{10} t_1 t_2^9 t_3^5 t_4^5 t_5^4 t_6^6 t_7^4 t_8^6 -  y_{s}^5 y_{o_1}^9 y_{o_2}^{11} y_{o_3}^8 y_{o_4}^{12} y_{o_5}^{11} y_{o_6}^9 t_1^7 t_2^3 t_3^4 t_4^6 t_5^3 t_6^7 t_7^4 t_8^6 +  3 y_{s}^5 y_{o_1}^9 y_{o_2}^{11} y_{o_3}^8 y_{o_4}^{12} y_{o_5}^{11} y_{o_6}^9 t_1^6 t_2^4 t_3^4 t_4^6 t_5^3 t_6^7 t_7^4 t_8^6 -  3 y_{s}^5 y_{o_1}^9 y_{o_2}^{11} y_{o_3}^8 y_{o_4}^{12} y_{o_5}^{11} y_{o_6}^9 t_1^5 t_2^5 t_3^4 t_4^6 t_5^3 t_6^7 t_7^4 t_8^6 +  3 y_{s}^5 y_{o_1}^9 y_{o_2}^{11} y_{o_3}^8 y_{o_4}^{12} y_{o_5}^{11} y_{o_6}^9 t_1^4 t_2^6 t_3^4 t_4^6 t_5^3 t_6^7 t_7^4 t_8^6 -  y_{s}^5 y_{o_1}^9 y_{o_2}^{11} y_{o_3}^8 y_{o_4}^{12} y_{o_5}^{11} y_{o_6}^9 t_1^3 t_2^7 t_3^4 t_4^6 t_5^3 t_6^7 t_7^4 t_8^6 +  y_{s}^5 y_{o_1}^9 y_{o_2}^{11} y_{o_3}^7 y_{o_4}^{13} y_{o_5}^{12} y_{o_6}^8 t_1^6 t_2^4 t_3^3 t_4^7 t_5^2 t_6^8 t_7^4 t_8^6 +  y_{s}^5 y_{o_1}^9 y_{o_2}^{11} y_{o_3}^7 y_{o_4}^{13} y_{o_5}^{12} y_{o_6}^8 t_1^4 t_2^6 t_3^3 t_4^7 t_5^2 t_6^8 t_7^4 t_8^6 -  y_{s}^6 y_{o_1}^{12} y_{o_2}^{12} y_{o_3}^{15} y_{o_4}^9 y_{o_5}^9 y_{o_6}^{15} t_1^6 t_2^6 t_3^9 t_4^3 t_5^9 t_6^3 t_7^6 t_8^6 +  3 y_{s}^6 y_{o_1}^{12} y_{o_2}^{12} y_{o_3}^{14} y_{o_4}^{10} y_{o_5}^{10} y_{o_6}^{14} t_1^7 t_2^5 t_3^8 t_4^4 t_5^8 t_6^4 t_7^6 t_8^6 -  y_{s}^6 y_{o_1}^{12} y_{o_2}^{12} y_{o_3}^{14} y_{o_4}^{10} y_{o_5}^{10} y_{o_6}^{14} t_1^6 t_2^6 t_3^8 t_4^4 t_5^8 t_6^4 t_7^6 t_8^6 +  3 y_{s}^6 y_{o_1}^{12} y_{o_2}^{12} y_{o_3}^{14} y_{o_4}^{10} y_{o_5}^{10} y_{o_6}^{14} t_1^5 t_2^7 t_3^8 t_4^4 t_5^8 t_6^4 t_7^6 t_8^6 +  y_{s}^6 y_{o_1}^{12} y_{o_2}^{12} y_{o_3}^{13} y_{o_4}^{11} y_{o_5}^{11} y_{o_6}^{13} t_1^{10} t_2^2 t_3^7 t_4^5 t_5^7 t_6^5 t_7^6 t_8^6 +  2 y_{s}^6 y_{o_1}^{12} y_{o_2}^{12} y_{o_3}^{13} y_{o_4}^{11} y_{o_5}^{11} y_{o_6}^{13} t_1^9 t_2^3 t_3^7 t_4^5 t_5^7 t_6^5 t_7^6 t_8^6 +  y_{s}^6 y_{o_1}^{12} y_{o_2}^{12} y_{o_3}^{13} y_{o_4}^{11} y_{o_5}^{11} y_{o_6}^{13} t_1^8 t_2^4 t_3^7 t_4^5 t_5^7 t_6^5 t_7^6 t_8^6 +  {10} y_{s}^6 y_{o_1}^{12} y_{o_2}^{12} y_{o_3}^{13} y_{o_4}^{11} y_{o_5}^{11} y_{o_6}^{13} t_1^7 t_2^5 t_3^7 t_4^5 t_5^7 t_6^5 t_7^6 t_8^6 -  y_{s}^6 y_{o_1}^{12} y_{o_2}^{12} y_{o_3}^{13} y_{o_4}^{11} y_{o_5}^{11} y_{o_6}^{13} t_1^6 t_2^6 t_3^7 t_4^5 t_5^7 t_6^5 t_7^6 t_8^6 +  {10} y_{s}^6 y_{o_1}^{12} y_{o_2}^{12} y_{o_3}^{13} y_{o_4}^{11} y_{o_5}^{11} y_{o_6}^{13} t_1^5 t_2^7 t_3^7 t_4^5 t_5^7 t_6^5 t_7^6 t_8^6 +  y_{s}^6 y_{o_1}^{12} y_{o_2}^{12} y_{o_3}^{13} y_{o_4}^{11} y_{o_5}^{11} y_{o_6}^{13} t_1^4 t_2^8 t_3^7 t_4^5 t_5^7 t_6^5 t_7^6 t_8^6 +  2 y_{s}^6 y_{o_1}^{12} y_{o_2}^{12} y_{o_3}^{13} y_{o_4}^{11} y_{o_5}^{11} y_{o_6}^{13} t_1^3 t_2^9 t_3^7 t_4^5 t_5^7 t_6^5 t_7^6 t_8^6 +  y_{s}^6 y_{o_1}^{12} y_{o_2}^{12} y_{o_3}^{13} y_{o_4}^{11} y_{o_5}^{11} y_{o_6}^{13} t_1^2 t_2^{10} t_3^7 t_4^5 t_5^7 t_6^5 t_7^6 t_8^6 -  y_{s}^6 y_{o_1}^{12} y_{o_2}^{12} y_{o_3}^{12} y_{o_4}^{12} y_{o_5}^{12} y_{o_6}^{12} t_1^{11} t_2 t_3^6 t_4^6 t_5^6 t_6^6 t_7^6 t_8^6 -  y_{s}^6 y_{o_1}^{12} y_{o_2}^{12} y_{o_3}^{12} y_{o_4}^{12} y_{o_5}^{12} y_{o_6}^{12} t_1^{10} t_2^2 t_3^6 t_4^6 t_5^6 t_6^6 t_7^6 t_8^6 -  y_{s}^6 y_{o_1}^{12} y_{o_2}^{12} y_{o_3}^{12} y_{o_4}^{12} y_{o_5}^{12} y_{o_6}^{12} t_1^9 t_2^3 t_3^6 t_4^6 t_5^6 t_6^6 t_7^6 t_8^6 -  4 y_{s}^6 y_{o_1}^{12} y_{o_2}^{12} y_{o_3}^{12} y_{o_4}^{12} y_{o_5}^{12} y_{o_6}^{12} t_1^8 t_2^4 t_3^6 t_4^6 t_5^6 t_6^6 t_7^6 t_8^6 +  8 y_{s}^6 y_{o_1}^{12} y_{o_2}^{12} y_{o_3}^{12} y_{o_4}^{12} y_{o_5}^{12} y_{o_6}^{12} t_1^7 t_2^5 t_3^6 t_4^6 t_5^6 t_6^6 t_7^6 t_8^6 -  8 y_{s}^6 y_{o_1}^{12} y_{o_2}^{12} y_{o_3}^{12} y_{o_4}^{12} y_{o_5}^{12} y_{o_6}^{12} t_1^6 t_2^6 t_3^6 t_4^6 t_5^6 t_6^6 t_7^6 t_8^6 +  8 y_{s}^6 y_{o_1}^{12} y_{o_2}^{12} y_{o_3}^{12} y_{o_4}^{12} y_{o_5}^{12} y_{o_6}^{12} t_1^5 t_2^7 t_3^6 t_4^6 t_5^6 t_6^6 t_7^6 t_8^6 -  4 y_{s}^6 y_{o_1}^{12} y_{o_2}^{12} y_{o_3}^{12} y_{o_4}^{12} y_{o_5}^{12} y_{o_6}^{12} t_1^4 t_2^8 t_3^6 t_4^6 t_5^6 t_6^6 t_7^6 t_8^6 -  y_{s}^6 y_{o_1}^{12} y_{o_2}^{12} y_{o_3}^{12} y_{o_4}^{12} y_{o_5}^{12} y_{o_6}^{12} t_1^3 t_2^9 t_3^6 t_4^6 t_5^6 t_6^6 t_7^6 t_8^6 -  y_{s}^6 y_{o_1}^{12} y_{o_2}^{12} y_{o_3}^{12} y_{o_4}^{12} y_{o_5}^{12} y_{o_6}^{12} t_1^2 t_2^{10} t_3^6 t_4^6 t_5^6 t_6^6 t_7^6 t_8^6 -  y_{s}^6 y_{o_1}^{12} y_{o_2}^{12} y_{o_3}^{12} y_{o_4}^{12} y_{o_5}^{12} y_{o_6}^{12} t_1 t_2^{11} t_3^6 t_4^6 t_5^6 t_6^6 t_7^6 t_8^6 +  y_{s}^6 y_{o_1}^{12} y_{o_2}^{12} y_{o_3}^{11} y_{o_4}^{13} y_{o_5}^{13} y_{o_6}^{11} t_1^{10} t_2^2 t_3^5 t_4^7 t_5^5 t_6^7 t_7^6 t_8^6 +  2 y_{s}^6 y_{o_1}^{12} y_{o_2}^{12} y_{o_3}^{11} y_{o_4}^{13} y_{o_5}^{13} y_{o_6}^{11} t_1^9 t_2^3 t_3^5 t_4^7 t_5^5 t_6^7 t_7^6 t_8^6 +  y_{s}^6 y_{o_1}^{12} y_{o_2}^{12} y_{o_3}^{11} y_{o_4}^{13} y_{o_5}^{13} y_{o_6}^{11} t_1^8 t_2^4 t_3^5 t_4^7 t_5^5 t_6^7 t_7^6 t_8^6 +  {10} y_{s}^6 y_{o_1}^{12} y_{o_2}^{12} y_{o_3}^{11} y_{o_4}^{13} y_{o_5}^{13} y_{o_6}^{11} t_1^7 t_2^5 t_3^5 t_4^7 t_5^5 t_6^7 t_7^6 t_8^6 -  y_{s}^6 y_{o_1}^{12} y_{o_2}^{12} y_{o_3}^{11} y_{o_4}^{13} y_{o_5}^{13} y_{o_6}^{11} t_1^6 t_2^6 t_3^5 t_4^7 t_5^5 t_6^7 t_7^6 t_8^6 +  {10} y_{s}^6 y_{o_1}^{12} y_{o_2}^{12} y_{o_3}^{11} y_{o_4}^{13} y_{o_5}^{13} y_{o_6}^{11} t_1^5 t_2^7 t_3^5 t_4^7 t_5^5 t_6^7 t_7^6 t_8^6 +  y_{s}^6 y_{o_1}^{12} y_{o_2}^{12} y_{o_3}^{11} y_{o_4}^{13} y_{o_5}^{13} y_{o_6}^{11} t_1^4 t_2^8 t_3^5 t_4^7 t_5^5 t_6^7 t_7^6 t_8^6 +  2 y_{s}^6 y_{o_1}^{12} y_{o_2}^{12} y_{o_3}^{11} y_{o_4}^{13} y_{o_5}^{13} y_{o_6}^{11} t_1^3 t_2^9 t_3^5 t_4^7 t_5^5 t_6^7 t_7^6 t_8^6 +  y_{s}^6 y_{o_1}^{12} y_{o_2}^{12} y_{o_3}^{11} y_{o_4}^{13} y_{o_5}^{13} y_{o_6}^{11} t_1^2 t_2^{10} t_3^5 t_4^7 t_5^5 t_6^7 t_7^6 t_8^6 +  3 y_{s}^6 y_{o_1}^{12} y_{o_2}^{12} y_{o_3}^{10} y_{o_4}^{14} y_{o_5}^{14} y_{o_6}^{10} t_1^7 t_2^5 t_3^4 t_4^8 t_5^4 t_6^8 t_7^6 t_8^6 -  y_{s}^6 y_{o_1}^{12} y_{o_2}^{12} y_{o_3}^{10} y_{o_4}^{14} y_{o_5}^{14} y_{o_6}^{10} t_1^6 t_2^6 t_3^4 t_4^8 t_5^4 t_6^8 t_7^6 t_8^6 +  3 y_{s}^6 y_{o_1}^{12} y_{o_2}^{12} y_{o_3}^{10} y_{o_4}^{14} y_{o_5}^{14} y_{o_6}^{10} t_1^5 t_2^7 t_3^4 t_4^8 t_5^4 t_6^8 t_7^6 t_8^6 -  y_{s}^6 y_{o_1}^{12} y_{o_2}^{12} y_{o_3}^9 y_{o_4}^{15} y_{o_5}^{15} y_{o_6}^9 t_1^6 t_2^6 t_3^3 t_4^9 t_5^3 t_6^9 t_7^6 t_8^6 -  y_{s}^7 y_{o_1}^{15} y_{o_2}^{13} y_{o_3}^{17} y_{o_4}^{11} y_{o_5}^{12} y_{o_6}^{16} t_1^7 t_2^7 t_3^9 t_4^5 t_5^{10} t_6^4 t_7^8 t_8^6 -  y_{s}^7 y_{o_1}^{15} y_{o_2}^{13} y_{o_3}^{16} y_{o_4}^{12} y_{o_5}^{13} y_{o_6}^{15} t_1^{10} t_2^4 t_3^8 t_4^6 t_5^9 t_6^5 t_7^8 t_8^6 -  3 y_{s}^7 y_{o_1}^{15} y_{o_2}^{13} y_{o_3}^{16} y_{o_4}^{12} y_{o_5}^{13} y_{o_6}^{15} t_1^9 t_2^5 t_3^8 t_4^6 t_5^9 t_6^5 t_7^8 t_8^6 -  3 y_{s}^7 y_{o_1}^{15} y_{o_2}^{13} y_{o_3}^{16} y_{o_4}^{12} y_{o_5}^{13} y_{o_6}^{15} t_1^8 t_2^6 t_3^8 t_4^6 t_5^9 t_6^5 t_7^8 t_8^6 -  7 y_{s}^7 y_{o_1}^{15} y_{o_2}^{13} y_{o_3}^{16} y_{o_4}^{12} y_{o_5}^{13} y_{o_6}^{15} t_1^7 t_2^7 t_3^8 t_4^6 t_5^9 t_6^5 t_7^8 t_8^6 -  3 y_{s}^7 y_{o_1}^{15} y_{o_2}^{13} y_{o_3}^{16} y_{o_4}^{12} y_{o_5}^{13} y_{o_6}^{15} t_1^6 t_2^8 t_3^8 t_4^6 t_5^9 t_6^5 t_7^8 t_8^6 -  3 y_{s}^7 y_{o_1}^{15} y_{o_2}^{13} y_{o_3}^{16} y_{o_4}^{12} y_{o_5}^{13} y_{o_6}^{15} t_1^5 t_2^9 t_3^8 t_4^6 t_5^9 t_6^5 t_7^8 t_8^6 -  y_{s}^7 y_{o_1}^{15} y_{o_2}^{13} y_{o_3}^{16} y_{o_4}^{12} y_{o_5}^{13} y_{o_6}^{15} t_1^4 t_2^{10} t_3^8 t_4^6 t_5^9 t_6^5 t_7^8 t_8^6 -  y_{s}^7 y_{o_1}^{15} y_{o_2}^{13} y_{o_3}^{15} y_{o_4}^{13} y_{o_5}^{14} y_{o_6}^{14} t_1^{10} t_2^4 t_3^7 t_4^7 t_5^8 t_6^6 t_7^8 t_8^6 -  5 y_{s}^7 y_{o_1}^{15} y_{o_2}^{13} y_{o_3}^{15} y_{o_4}^{13} y_{o_5}^{14} y_{o_6}^{14} t_1^9 t_2^5 t_3^7 t_4^7 t_5^8 t_6^6 t_7^8 t_8^6 -  3 y_{s}^7 y_{o_1}^{15} y_{o_2}^{13} y_{o_3}^{15} y_{o_4}^{13} y_{o_5}^{14} y_{o_6}^{14} t_1^8 t_2^6 t_3^7 t_4^7 t_5^8 t_6^6 t_7^8 t_8^6 -  {12} y_{s}^7 y_{o_1}^{15} y_{o_2}^{13} y_{o_3}^{15} y_{o_4}^{13} y_{o_5}^{14} y_{o_6}^{14} t_1^7 t_2^7 t_3^7 t_4^7 t_5^8 t_6^6 t_7^8 t_8^6 -  3 y_{s}^7 y_{o_1}^{15} y_{o_2}^{13} y_{o_3}^{15} y_{o_4}^{13} y_{o_5}^{14} y_{o_6}^{14} t_1^6 t_2^8 t_3^7 t_4^7 t_5^8 t_6^6 t_7^8 t_8^6 -  5 y_{s}^7 y_{o_1}^{15} y_{o_2}^{13} y_{o_3}^{15} y_{o_4}^{13} y_{o_5}^{14} y_{o_6}^{14} t_1^5 t_2^9 t_3^7 t_4^7 t_5^8 t_6^6 t_7^8 t_8^6 -  y_{s}^7 y_{o_1}^{15} y_{o_2}^{13} y_{o_3}^{15} y_{o_4}^{13} y_{o_5}^{14} y_{o_6}^{14} t_1^4 t_2^{10} t_3^7 t_4^7 t_5^8 t_6^6 t_7^8 t_8^6 -  y_{s}^7 y_{o_1}^{15} y_{o_2}^{13} y_{o_3}^{14} y_{o_4}^{14} y_{o_5}^{15} y_{o_6}^{13} t_1^{10} t_2^4 t_3^6 t_4^8 t_5^7 t_6^7 t_7^8 t_8^6 -  5 y_{s}^7 y_{o_1}^{15} y_{o_2}^{13} y_{o_3}^{14} y_{o_4}^{14} y_{o_5}^{15} y_{o_6}^{13} t_1^9 t_2^5 t_3^6 t_4^8 t_5^7 t_6^7 t_7^8 t_8^6 -  3 y_{s}^7 y_{o_1}^{15} y_{o_2}^{13} y_{o_3}^{14} y_{o_4}^{14} y_{o_5}^{15} y_{o_6}^{13} t_1^8 t_2^6 t_3^6 t_4^8 t_5^7 t_6^7 t_7^8 t_8^6 -  {12} y_{s}^7 y_{o_1}^{15} y_{o_2}^{13} y_{o_3}^{14} y_{o_4}^{14} y_{o_5}^{15} y_{o_6}^{13} t_1^7 t_2^7 t_3^6 t_4^8 t_5^7 t_6^7 t_7^8 t_8^6 -  3 y_{s}^7 y_{o_1}^{15} y_{o_2}^{13} y_{o_3}^{14} y_{o_4}^{14} y_{o_5}^{15} y_{o_6}^{13} t_1^6 t_2^8 t_3^6 t_4^8 t_5^7 t_6^7 t_7^8 t_8^6 -  5 y_{s}^7 y_{o_1}^{15} y_{o_2}^{13} y_{o_3}^{14} y_{o_4}^{14} y_{o_5}^{15} y_{o_6}^{13} t_1^5 t_2^9 t_3^6 t_4^8 t_5^7 t_6^7 t_7^8 t_8^6 -  y_{s}^7 y_{o_1}^{15} y_{o_2}^{13} y_{o_3}^{14} y_{o_4}^{14} y_{o_5}^{15} y_{o_6}^{13} t_1^4 t_2^{10} t_3^6 t_4^8 t_5^7 t_6^7 t_7^8 t_8^6 -  y_{s}^7 y_{o_1}^{15} y_{o_2}^{13} y_{o_3}^{13} y_{o_4}^{15} y_{o_5}^{16} y_{o_6}^{12} t_1^{10} t_2^4 t_3^5 t_4^9 t_5^6 t_6^8 t_7^8 t_8^6 -  3 y_{s}^7 y_{o_1}^{15} y_{o_2}^{13} y_{o_3}^{13} y_{o_4}^{15} y_{o_5}^{16} y_{o_6}^{12} t_1^9 t_2^5 t_3^5 t_4^9 t_5^6 t_6^8 t_7^8 t_8^6 -  3 y_{s}^7 y_{o_1}^{15} y_{o_2}^{13} y_{o_3}^{13} y_{o_4}^{15} y_{o_5}^{16} y_{o_6}^{12} t_1^8 t_2^6 t_3^5 t_4^9 t_5^6 t_6^8 t_7^8 t_8^6 -  7 y_{s}^7 y_{o_1}^{15} y_{o_2}^{13} y_{o_3}^{13} y_{o_4}^{15} y_{o_5}^{16} y_{o_6}^{12} t_1^7 t_2^7 t_3^5 t_4^9 t_5^6 t_6^8 t_7^8 t_8^6 -  3 y_{s}^7 y_{o_1}^{15} y_{o_2}^{13} y_{o_3}^{13} y_{o_4}^{15} y_{o_5}^{16} y_{o_6}^{12} t_1^6 t_2^8 t_3^5 t_4^9 t_5^6 t_6^8 t_7^8 t_8^6 -  3 y_{s}^7 y_{o_1}^{15} y_{o_2}^{13} y_{o_3}^{13} y_{o_4}^{15} y_{o_5}^{16} y_{o_6}^{12} t_1^5 t_2^9 t_3^5 t_4^9 t_5^6 t_6^8 t_7^8 t_8^6 -  y_{s}^7 y_{o_1}^{15} y_{o_2}^{13} y_{o_3}^{13} y_{o_4}^{15} y_{o_5}^{16} y_{o_6}^{12} t_1^4 t_2^{10} t_3^5 t_4^9 t_5^6 t_6^8 t_7^8 t_8^6 -  y_{s}^7 y_{o_1}^{15} y_{o_2}^{13} y_{o_3}^{12} y_{o_4}^{16} y_{o_5}^{17} y_{o_6}^{11} t_1^7 t_2^7 t_3^4 t_4^{10} t_5^5 t_6^9 t_7^8 t_8^6 +  y_{s}^8 y_{o_1}^{18} y_{o_2}^{14} y_{o_3}^{18} y_{o_4}^{14} y_{o_5}^{16} y_{o_6}^{16} t_1^{10} t_2^6 t_3^8 t_4^8 t_5^{10} t_6^6 t_7^{10} t_8^6 +  2 y_{s}^8 y_{o_1}^{18} y_{o_2}^{14} y_{o_3}^{18} y_{o_4}^{14} y_{o_5}^{16} y_{o_6}^{16} t_1^9 t_2^7 t_3^8 t_4^8 t_5^{10} t_6^6 t_7^{10} t_8^6 +  2 y_{s}^8 y_{o_1}^{18} y_{o_2}^{14} y_{o_3}^{18} y_{o_4}^{14} y_{o_5}^{16} y_{o_6}^{16} t_1^8 t_2^8 t_3^8 t_4^8 t_5^{10} t_6^6 t_7^{10} t_8^6 +  2 y_{s}^8 y_{o_1}^{18} y_{o_2}^{14} y_{o_3}^{18} y_{o_4}^{14} y_{o_5}^{16} y_{o_6}^{16} t_1^7 t_2^9 t_3^8 t_4^8 t_5^{10} t_6^6 t_7^{10} t_8^6 +  y_{s}^8 y_{o_1}^{18} y_{o_2}^{14} y_{o_3}^{18} y_{o_4}^{14} y_{o_5}^{16} y_{o_6}^{16} t_1^6 t_2^{10} t_3^8 t_4^8 t_5^{10} t_6^6 t_7^{10} t_8^6 +  2 y_{s}^8 y_{o_1}^{18} y_{o_2}^{14} y_{o_3}^{17} y_{o_4}^{15} y_{o_5}^{17} y_{o_6}^{15} t_1^9 t_2^7 t_3^7 t_4^9 t_5^9 t_6^7 t_7^{10} t_8^6 +  y_{s}^8 y_{o_1}^{18} y_{o_2}^{14} y_{o_3}^{17} y_{o_4}^{15} y_{o_5}^{17} y_{o_6}^{15} t_1^8 t_2^8 t_3^7 t_4^9 t_5^9 t_6^7 t_7^{10} t_8^6 +  2 y_{s}^8 y_{o_1}^{18} y_{o_2}^{14} y_{o_3}^{17} y_{o_4}^{15} y_{o_5}^{17} y_{o_6}^{15} t_1^7 t_2^9 t_3^7 t_4^9 t_5^9 t_6^7 t_7^{10} t_8^6 +  y_{s}^8 y_{o_1}^{18} y_{o_2}^{14} y_{o_3}^{16} y_{o_4}^{16} y_{o_5}^{18} y_{o_6}^{14} t_1^{10} t_2^6 t_3^6 t_4^{10} t_5^8 t_6^8 t_7^{10} t_8^6 +  2 y_{s}^8 y_{o_1}^{18} y_{o_2}^{14} y_{o_3}^{16} y_{o_4}^{16} y_{o_5}^{18} y_{o_6}^{14} t_1^9 t_2^7 t_3^6 t_4^{10} t_5^8 t_6^8 t_7^{10} t_8^6 +  2 y_{s}^8 y_{o_1}^{18} y_{o_2}^{14} y_{o_3}^{16} y_{o_4}^{16} y_{o_5}^{18} y_{o_6}^{14} t_1^8 t_2^8 t_3^6 t_4^{10} t_5^8 t_6^8 t_7^{10} t_8^6 +  2 y_{s}^8 y_{o_1}^{18} y_{o_2}^{14} y_{o_3}^{16} y_{o_4}^{16} y_{o_5}^{18} y_{o_6}^{14} t_1^7 t_2^9 t_3^6 t_4^{10} t_5^8 t_6^8 t_7^{10} t_8^6 +  y_{s}^8 y_{o_1}^{18} y_{o_2}^{14} y_{o_3}^{16} y_{o_4}^{16} y_{o_5}^{18} y_{o_6}^{14} t_1^6 t_2^{10} t_3^6 t_4^{10} t_5^8 t_6^8 t_7^{10} t_8^6 -  y_{s}^4 y_{o_1}^5 y_{o_2}^{11} y_{o_3}^7 y_{o_4}^9 y_{o_5}^6 y_{o_6}^{10} t_1^4 t_2^4 t_3^6 t_4^2 t_5^3 t_6^5 t_7 t_8^7 -  y_{s}^4 y_{o_1}^5 y_{o_2}^{11} y_{o_3}^6 y_{o_4}^{10} y_{o_5}^7 y_{o_6}^9 t_1^4 t_2^4 t_3^5 t_4^3 t_5^2 t_6^6 t_7 t_8^7 +  y_{s}^5 y_{o_1}^8 y_{o_2}^{12} y_{o_3}^{11} y_{o_4}^9 y_{o_5}^7 y_{o_6}^{13} t_1^6 t_2^4 t_3^8 t_4^2 t_5^6 t_6^4 t_7^3 t_8^7 +  y_{s}^5 y_{o_1}^8 y_{o_2}^{12} y_{o_3}^{11} y_{o_4}^9 y_{o_5}^7 y_{o_6}^{13} t_1^4 t_2^6 t_3^8 t_4^2 t_5^6 t_6^4 t_7^3 t_8^7 +  3 y_{s}^5 y_{o_1}^8 y_{o_2}^{12} y_{o_3}^{10} y_{o_4}^{10} y_{o_5}^8 y_{o_6}^{12} t_1^6 t_2^4 t_3^7 t_4^3 t_5^5 t_6^5 t_7^3 t_8^7 -  y_{s}^5 y_{o_1}^8 y_{o_2}^{12} y_{o_3}^{10} y_{o_4}^{10} y_{o_5}^8 y_{o_6}^{12} t_1^5 t_2^5 t_3^7 t_4^3 t_5^5 t_6^5 t_7^3 t_8^7 +  3 y_{s}^5 y_{o_1}^8 y_{o_2}^{12} y_{o_3}^{10} y_{o_4}^{10} y_{o_5}^8 y_{o_6}^{12} t_1^4 t_2^6 t_3^7 t_4^3 t_5^5 t_6^5 t_7^3 t_8^7 -  y_{s}^5 y_{o_1}^8 y_{o_2}^{12} y_{o_3}^9 y_{o_4}^{11} y_{o_5}^9 y_{o_6}^{11} t_1^7 t_2^3 t_3^6 t_4^4 t_5^4 t_6^6 t_7^3 t_8^7 +  3 y_{s}^5 y_{o_1}^8 y_{o_2}^{12} y_{o_3}^9 y_{o_4}^{11} y_{o_5}^9 y_{o_6}^{11} t_1^6 t_2^4 t_3^6 t_4^4 t_5^4 t_6^6 t_7^3 t_8^7 -  3 y_{s}^5 y_{o_1}^8 y_{o_2}^{12} y_{o_3}^9 y_{o_4}^{11} y_{o_5}^9 y_{o_6}^{11} t_1^5 t_2^5 t_3^6 t_4^4 t_5^4 t_6^6 t_7^3 t_8^7 +  3 y_{s}^5 y_{o_1}^8 y_{o_2}^{12} y_{o_3}^9 y_{o_4}^{11} y_{o_5}^9 y_{o_6}^{11} t_1^4 t_2^6 t_3^6 t_4^4 t_5^4 t_6^6 t_7^3 t_8^7 -  y_{s}^5 y_{o_1}^8 y_{o_2}^{12} y_{o_3}^9 y_{o_4}^{11} y_{o_5}^9 y_{o_6}^{11} t_1^3 t_2^7 t_3^6 t_4^4 t_5^4 t_6^6 t_7^3 t_8^7 +  3 y_{s}^5 y_{o_1}^8 y_{o_2}^{12} y_{o_3}^8 y_{o_4}^{12} y_{o_5}^{10} y_{o_6}^{10} t_1^6 t_2^4 t_3^5 t_4^5 t_5^3 t_6^7 t_7^3 t_8^7 -  y_{s}^5 y_{o_1}^8 y_{o_2}^{12} y_{o_3}^8 y_{o_4}^{12} y_{o_5}^{10} y_{o_6}^{10} t_1^5 t_2^5 t_3^5 t_4^5 t_5^3 t_6^7 t_7^3 t_8^7 +  3 y_{s}^5 y_{o_1}^8 y_{o_2}^{12} y_{o_3}^8 y_{o_4}^{12} y_{o_5}^{10} y_{o_6}^{10} t_1^4 t_2^6 t_3^5 t_4^5 t_5^3 t_6^7 t_7^3 t_8^7 +  y_{s}^5 y_{o_1}^8 y_{o_2}^{12} y_{o_3}^7 y_{o_4}^{13} y_{o_5}^{11} y_{o_6}^9 t_1^6 t_2^4 t_3^4 t_4^6 t_5^2 t_6^8 t_7^3 t_8^7 +  y_{s}^5 y_{o_1}^8 y_{o_2}^{12} y_{o_3}^7 y_{o_4}^{13} y_{o_5}^{11} y_{o_6}^9 t_1^4 t_2^6 t_3^4 t_4^6 t_5^2 t_6^8 t_7^3 t_8^7 +  y_{s}^6 y_{o_1}^{11} y_{o_2}^{13} y_{o_3}^{14} y_{o_4}^{10} y_{o_5}^9 y_{o_6}^{15} t_1^7 t_2^5 t_3^9 t_4^3 t_5^8 t_6^4 t_7^5 t_8^7 +  y_{s}^6 y_{o_1}^{11} y_{o_2}^{13} y_{o_3}^{14} y_{o_4}^{10} y_{o_5}^9 y_{o_6}^{15} t_1^5 t_2^7 t_3^9 t_4^3 t_5^8 t_6^4 t_7^5 t_8^7 -  y_{s}^6 y_{o_1}^{11} y_{o_2}^{13} y_{o_3}^{13} y_{o_4}^{11} y_{o_5}^{10} y_{o_6}^{14} t_1^8 t_2^4 t_3^8 t_4^4 t_5^7 t_6^5 t_7^5 t_8^7 +  3 y_{s}^6 y_{o_1}^{11} y_{o_2}^{13} y_{o_3}^{13} y_{o_4}^{11} y_{o_5}^{10} y_{o_6}^{14} t_1^7 t_2^5 t_3^8 t_4^4 t_5^7 t_6^5 t_7^5 t_8^7 -  3 y_{s}^6 y_{o_1}^{11} y_{o_2}^{13} y_{o_3}^{13} y_{o_4}^{11} y_{o_5}^{10} y_{o_6}^{14} t_1^6 t_2^6 t_3^8 t_4^4 t_5^7 t_6^5 t_7^5 t_8^7 +  3 y_{s}^6 y_{o_1}^{11} y_{o_2}^{13} y_{o_3}^{13} y_{o_4}^{11} y_{o_5}^{10} y_{o_6}^{14} t_1^5 t_2^7 t_3^8 t_4^4 t_5^7 t_6^5 t_7^5 t_8^7 -  y_{s}^6 y_{o_1}^{11} y_{o_2}^{13} y_{o_3}^{13} y_{o_4}^{11} y_{o_5}^{10} y_{o_6}^{14} t_1^4 t_2^8 t_3^8 t_4^4 t_5^7 t_6^5 t_7^5 t_8^7 +  y_{s}^6 y_{o_1}^{11} y_{o_2}^{13} y_{o_3}^{12} y_{o_4}^{12} y_{o_5}^{11} y_{o_6}^{13} t_1^{10} t_2^2 t_3^7 t_4^5 t_5^6 t_6^6 t_7^5 t_8^7 +  2 y_{s}^6 y_{o_1}^{11} y_{o_2}^{13} y_{o_3}^{12} y_{o_4}^{12} y_{o_5}^{11} y_{o_6}^{13} t_1^9 t_2^3 t_3^7 t_4^5 t_5^6 t_6^6 t_7^5 t_8^7 +  y_{s}^6 y_{o_1}^{11} y_{o_2}^{13} y_{o_3}^{12} y_{o_4}^{12} y_{o_5}^{11} y_{o_6}^{13} t_1^8 t_2^4 t_3^7 t_4^5 t_5^6 t_6^6 t_7^5 t_8^7 +  {10} y_{s}^6 y_{o_1}^{11} y_{o_2}^{13} y_{o_3}^{12} y_{o_4}^{12} y_{o_5}^{11} y_{o_6}^{13} t_1^7 t_2^5 t_3^7 t_4^5 t_5^6 t_6^6 t_7^5 t_8^7 -  y_{s}^6 y_{o_1}^{11} y_{o_2}^{13} y_{o_3}^{12} y_{o_4}^{12} y_{o_5}^{11} y_{o_6}^{13} t_1^6 t_2^6 t_3^7 t_4^5 t_5^6 t_6^6 t_7^5 t_8^7 +  {10} y_{s}^6 y_{o_1}^{11} y_{o_2}^{13} y_{o_3}^{12} y_{o_4}^{12} y_{o_5}^{11} y_{o_6}^{13} t_1^5 t_2^7 t_3^7 t_4^5 t_5^6 t_6^6 t_7^5 t_8^7 +  y_{s}^6 y_{o_1}^{11} y_{o_2}^{13} y_{o_3}^{12} y_{o_4}^{12} y_{o_5}^{11} y_{o_6}^{13} t_1^4 t_2^8 t_3^7 t_4^5 t_5^6 t_6^6 t_7^5 t_8^7 +  2 y_{s}^6 y_{o_1}^{11} y_{o_2}^{13} y_{o_3}^{12} y_{o_4}^{12} y_{o_5}^{11} y_{o_6}^{13} t_1^3 t_2^9 t_3^7 t_4^5 t_5^6 t_6^6 t_7^5 t_8^7 +  y_{s}^6 y_{o_1}^{11} y_{o_2}^{13} y_{o_3}^{12} y_{o_4}^{12} y_{o_5}^{11} y_{o_6}^{13} t_1^2 t_2^{10} t_3^7 t_4^5 t_5^6 t_6^6 t_7^5 t_8^7 +  y_{s}^6 y_{o_1}^{11} y_{o_2}^{13} y_{o_3}^{11} y_{o_4}^{13} y_{o_5}^{12} y_{o_6}^{12} t_1^{10} t_2^2 t_3^6 t_4^6 t_5^5 t_6^7 t_7^5 t_8^7 +  2 y_{s}^6 y_{o_1}^{11} y_{o_2}^{13} y_{o_3}^{11} y_{o_4}^{13} y_{o_5}^{12} y_{o_6}^{12} t_1^9 t_2^3 t_3^6 t_4^6 t_5^5 t_6^7 t_7^5 t_8^7 +  y_{s}^6 y_{o_1}^{11} y_{o_2}^{13} y_{o_3}^{11} y_{o_4}^{13} y_{o_5}^{12} y_{o_6}^{12} t_1^8 t_2^4 t_3^6 t_4^6 t_5^5 t_6^7 t_7^5 t_8^7 +  {10} y_{s}^6 y_{o_1}^{11} y_{o_2}^{13} y_{o_3}^{11} y_{o_4}^{13} y_{o_5}^{12} y_{o_6}^{12} t_1^7 t_2^5 t_3^6 t_4^6 t_5^5 t_6^7 t_7^5 t_8^7 -  y_{s}^6 y_{o_1}^{11} y_{o_2}^{13} y_{o_3}^{11} y_{o_4}^{13} y_{o_5}^{12} y_{o_6}^{12} t_1^6 t_2^6 t_3^6 t_4^6 t_5^5 t_6^7 t_7^5 t_8^7 +  {10} y_{s}^6 y_{o_1}^{11} y_{o_2}^{13} y_{o_3}^{11} y_{o_4}^{13} y_{o_5}^{12} y_{o_6}^{12} t_1^5 t_2^7 t_3^6 t_4^6 t_5^5 t_6^7 t_7^5 t_8^7 +  y_{s}^6 y_{o_1}^{11} y_{o_2}^{13} y_{o_3}^{11} y_{o_4}^{13} y_{o_5}^{12} y_{o_6}^{12} t_1^4 t_2^8 t_3^6 t_4^6 t_5^5 t_6^7 t_7^5 t_8^7 +  2 y_{s}^6 y_{o_1}^{11} y_{o_2}^{13} y_{o_3}^{11} y_{o_4}^{13} y_{o_5}^{12} y_{o_6}^{12} t_1^3 t_2^9 t_3^6 t_4^6 t_5^5 t_6^7 t_7^5 t_8^7 +  y_{s}^6 y_{o_1}^{11} y_{o_2}^{13} y_{o_3}^{11} y_{o_4}^{13} y_{o_5}^{12} y_{o_6}^{12} t_1^2 t_2^{10} t_3^6 t_4^6 t_5^5 t_6^7 t_7^5 t_8^7 -  y_{s}^6 y_{o_1}^{11} y_{o_2}^{13} y_{o_3}^{10} y_{o_4}^{14} y_{o_5}^{13} y_{o_6}^{11} t_1^8 t_2^4 t_3^5 t_4^7 t_5^4 t_6^8 t_7^5 t_8^7 +  3 y_{s}^6 y_{o_1}^{11} y_{o_2}^{13} y_{o_3}^{10} y_{o_4}^{14} y_{o_5}^{13} y_{o_6}^{11} t_1^7 t_2^5 t_3^5 t_4^7 t_5^4 t_6^8 t_7^5 t_8^7 -  3 y_{s}^6 y_{o_1}^{11} y_{o_2}^{13} y_{o_3}^{10} y_{o_4}^{14} y_{o_5}^{13} y_{o_6}^{11} t_1^6 t_2^6 t_3^5 t_4^7 t_5^4 t_6^8 t_7^5 t_8^7 +  3 y_{s}^6 y_{o_1}^{11} y_{o_2}^{13} y_{o_3}^{10} y_{o_4}^{14} y_{o_5}^{13} y_{o_6}^{11} t_1^5 t_2^7 t_3^5 t_4^7 t_5^4 t_6^8 t_7^5 t_8^7 -  y_{s}^6 y_{o_1}^{11} y_{o_2}^{13} y_{o_3}^{10} y_{o_4}^{14} y_{o_5}^{13} y_{o_6}^{11} t_1^4 t_2^8 t_3^5 t_4^7 t_5^4 t_6^8 t_7^5 t_8^7 +  y_{s}^6 y_{o_1}^{11} y_{o_2}^{13} y_{o_3}^9 y_{o_4}^{15} y_{o_5}^{14} y_{o_6}^{10} t_1^7 t_2^5 t_3^4 t_4^8 t_5^3 t_6^9 t_7^5 t_8^7 +  y_{s}^6 y_{o_1}^{11} y_{o_2}^{13} y_{o_3}^9 y_{o_4}^{15} y_{o_5}^{14} y_{o_6}^{10} t_1^5 t_2^7 t_3^4 t_4^8 t_5^3 t_6^9 t_7^5 t_8^7 -  y_{s}^7 y_{o_1}^{14} y_{o_2}^{14} y_{o_3}^{16} y_{o_4}^{12} y_{o_5}^{12} y_{o_6}^{16} t_1^9 t_2^5 t_3^9 t_4^5 t_5^9 t_6^5 t_7^7 t_8^7 -  3 y_{s}^7 y_{o_1}^{14} y_{o_2}^{14} y_{o_3}^{16} y_{o_4}^{12} y_{o_5}^{12} y_{o_6}^{16} t_1^7 t_2^7 t_3^9 t_4^5 t_5^9 t_6^5 t_7^7 t_8^7 -  y_{s}^7 y_{o_1}^{14} y_{o_2}^{14} y_{o_3}^{16} y_{o_4}^{12} y_{o_5}^{12} y_{o_6}^{16} t_1^5 t_2^9 t_3^9 t_4^5 t_5^9 t_6^5 t_7^7 t_8^7 -  y_{s}^7 y_{o_1}^{14} y_{o_2}^{14} y_{o_3}^{15} y_{o_4}^{13} y_{o_5}^{13} y_{o_6}^{15} t_1^{10} t_2^4 t_3^8 t_4^6 t_5^8 t_6^6 t_7^7 t_8^7 -  5 y_{s}^7 y_{o_1}^{14} y_{o_2}^{14} y_{o_3}^{15} y_{o_4}^{13} y_{o_5}^{13} y_{o_6}^{15} t_1^9 t_2^5 t_3^8 t_4^6 t_5^8 t_6^6 t_7^7 t_8^7 -  3 y_{s}^7 y_{o_1}^{14} y_{o_2}^{14} y_{o_3}^{15} y_{o_4}^{13} y_{o_5}^{13} y_{o_6}^{15} t_1^8 t_2^6 t_3^8 t_4^6 t_5^8 t_6^6 t_7^7 t_8^7 -  {12} y_{s}^7 y_{o_1}^{14} y_{o_2}^{14} y_{o_3}^{15} y_{o_4}^{13} y_{o_5}^{13} y_{o_6}^{15} t_1^7 t_2^7 t_3^8 t_4^6 t_5^8 t_6^6 t_7^7 t_8^7 -  3 y_{s}^7 y_{o_1}^{14} y_{o_2}^{14} y_{o_3}^{15} y_{o_4}^{13} y_{o_5}^{13} y_{o_6}^{15} t_1^6 t_2^8 t_3^8 t_4^6 t_5^8 t_6^6 t_7^7 t_8^7 -  5 y_{s}^7 y_{o_1}^{14} y_{o_2}^{14} y_{o_3}^{15} y_{o_4}^{13} y_{o_5}^{13} y_{o_6}^{15} t_1^5 t_2^9 t_3^8 t_4^6 t_5^8 t_6^6 t_7^7 t_8^7 -  y_{s}^7 y_{o_1}^{14} y_{o_2}^{14} y_{o_3}^{15} y_{o_4}^{13} y_{o_5}^{13} y_{o_6}^{15} t_1^4 t_2^{10} t_3^8 t_4^6 t_5^8 t_6^6 t_7^7 t_8^7 -  y_{s}^7 y_{o_1}^{14} y_{o_2}^{14} y_{o_3}^{14} y_{o_4}^{14} y_{o_5}^{14} y_{o_6}^{14} t_1^{11} t_2^3 t_3^7 t_4^7 t_5^7 t_6^7 t_7^7 t_8^7 -  4 y_{s}^7 y_{o_1}^{14} y_{o_2}^{14} y_{o_3}^{14} y_{o_4}^{14} y_{o_5}^{14} y_{o_6}^{14} t_1^{10} t_2^4 t_3^7 t_4^7 t_5^7 t_6^7 t_7^7 t_8^7 -  {10} y_{s}^7 y_{o_1}^{14} y_{o_2}^{14} y_{o_3}^{14} y_{o_4}^{14} y_{o_5}^{14} y_{o_6}^{14} t_1^9 t_2^5 t_3^7 t_4^7 t_5^7 t_6^7 t_7^7 t_8^7 -  8 y_{s}^7 y_{o_1}^{14} y_{o_2}^{14} y_{o_3}^{14} y_{o_4}^{14} y_{o_5}^{14} y_{o_6}^{14} t_1^8 t_2^6 t_3^7 t_4^7 t_5^7 t_6^7 t_7^7 t_8^7 -  {20} y_{s}^7 y_{o_1}^{14} y_{o_2}^{14} y_{o_3}^{14} y_{o_4}^{14} y_{o_5}^{14} y_{o_6}^{14} t_1^7 t_2^7 t_3^7 t_4^7 t_5^7 t_6^7 t_7^7 t_8^7 -  8 y_{s}^7 y_{o_1}^{14} y_{o_2}^{14} y_{o_3}^{14} y_{o_4}^{14} y_{o_5}^{14} y_{o_6}^{14} t_1^6 t_2^8 t_3^7 t_4^7 t_5^7 t_6^7 t_7^7 t_8^7 -  {10} y_{s}^7 y_{o_1}^{14} y_{o_2}^{14} y_{o_3}^{14} y_{o_4}^{14} y_{o_5}^{14} y_{o_6}^{14} t_1^5 t_2^9 t_3^7 t_4^7 t_5^7 t_6^7 t_7^7 t_8^7 -  4 y_{s}^7 y_{o_1}^{14} y_{o_2}^{14} y_{o_3}^{14} y_{o_4}^{14} y_{o_5}^{14} y_{o_6}^{14} t_1^4 t_2^{10} t_3^7 t_4^7 t_5^7 t_6^7 t_7^7 t_8^7 -  y_{s}^7 y_{o_1}^{14} y_{o_2}^{14} y_{o_3}^{14} y_{o_4}^{14} y_{o_5}^{14} y_{o_6}^{14} t_1^3 t_2^{11} t_3^7 t_4^7 t_5^7 t_6^7 t_7^7 t_8^7 -  y_{s}^7 y_{o_1}^{14} y_{o_2}^{14} y_{o_3}^{13} y_{o_4}^{15} y_{o_5}^{15} y_{o_6}^{13} t_1^{10} t_2^4 t_3^6 t_4^8 t_5^6 t_6^8 t_7^7 t_8^7 -  5 y_{s}^7 y_{o_1}^{14} y_{o_2}^{14} y_{o_3}^{13} y_{o_4}^{15} y_{o_5}^{15} y_{o_6}^{13} t_1^9 t_2^5 t_3^6 t_4^8 t_5^6 t_6^8 t_7^7 t_8^7 -  3 y_{s}^7 y_{o_1}^{14} y_{o_2}^{14} y_{o_3}^{13} y_{o_4}^{15} y_{o_5}^{15} y_{o_6}^{13} t_1^8 t_2^6 t_3^6 t_4^8 t_5^6 t_6^8 t_7^7 t_8^7 -  {12} y_{s}^7 y_{o_1}^{14} y_{o_2}^{14} y_{o_3}^{13} y_{o_4}^{15} y_{o_5}^{15} y_{o_6}^{13} t_1^7 t_2^7 t_3^6 t_4^8 t_5^6 t_6^8 t_7^7 t_8^7 -  3 y_{s}^7 y_{o_1}^{14} y_{o_2}^{14} y_{o_3}^{13} y_{o_4}^{15} y_{o_5}^{15} y_{o_6}^{13} t_1^6 t_2^8 t_3^6 t_4^8 t_5^6 t_6^8 t_7^7 t_8^7 -  5 y_{s}^7 y_{o_1}^{14} y_{o_2}^{14} y_{o_3}^{13} y_{o_4}^{15} y_{o_5}^{15} y_{o_6}^{13} t_1^5 t_2^9 t_3^6 t_4^8 t_5^6 t_6^8 t_7^7 t_8^7 -  y_{s}^7 y_{o_1}^{14} y_{o_2}^{14} y_{o_3}^{13} y_{o_4}^{15} y_{o_5}^{15} y_{o_6}^{13} t_1^4 t_2^{10} t_3^6 t_4^8 t_5^6 t_6^8 t_7^7 t_8^7 -  y_{s}^7 y_{o_1}^{14} y_{o_2}^{14} y_{o_3}^{12} y_{o_4}^{16} y_{o_5}^{16} y_{o_6}^{12} t_1^9 t_2^5 t_3^5 t_4^9 t_5^5 t_6^9 t_7^7 t_8^7 -  3 y_{s}^7 y_{o_1}^{14} y_{o_2}^{14} y_{o_3}^{12} y_{o_4}^{16} y_{o_5}^{16} y_{o_6}^{12} t_1^7 t_2^7 t_3^5 t_4^9 t_5^5 t_6^9 t_7^7 t_8^7 -  y_{s}^7 y_{o_1}^{14} y_{o_2}^{14} y_{o_3}^{12} y_{o_4}^{16} y_{o_5}^{16} y_{o_6}^{12} t_1^5 t_2^9 t_3^5 t_4^9 t_5^5 t_6^9 t_7^7 t_8^7 +  2 y_{s}^8 y_{o_1}^{17} y_{o_2}^{15} y_{o_3}^{18} y_{o_4}^{14} y_{o_5}^{15} y_{o_6}^{17} t_1^9 t_2^7 t_3^9 t_4^7 t_5^{10} t_6^6 t_7^9 t_8^7 +  y_{s}^8 y_{o_1}^{17} y_{o_2}^{15} y_{o_3}^{18} y_{o_4}^{14} y_{o_5}^{15} y_{o_6}^{17} t_1^8 t_2^8 t_3^9 t_4^7 t_5^{10} t_6^6 t_7^9 t_8^7 +  2 y_{s}^8 y_{o_1}^{17} y_{o_2}^{15} y_{o_3}^{18} y_{o_4}^{14} y_{o_5}^{15} y_{o_6}^{17} t_1^7 t_2^9 t_3^9 t_4^7 t_5^{10} t_6^6 t_7^9 t_8^7 +  y_{s}^8 y_{o_1}^{17} y_{o_2}^{15} y_{o_3}^{17} y_{o_4}^{15} y_{o_5}^{16} y_{o_6}^{16} t_1^{11} t_2^5 t_3^8 t_4^8 t_5^9 t_6^7 t_7^9 t_8^7 +  3 y_{s}^8 y_{o_1}^{17} y_{o_2}^{15} y_{o_3}^{17} y_{o_4}^{15} y_{o_5}^{16} y_{o_6}^{16} t_1^{10} t_2^6 t_3^8 t_4^8 t_5^9 t_6^7 t_7^9 t_8^7 +  7 y_{s}^8 y_{o_1}^{17} y_{o_2}^{15} y_{o_3}^{17} y_{o_4}^{15} y_{o_5}^{16} y_{o_6}^{16} t_1^9 t_2^7 t_3^8 t_4^8 t_5^9 t_6^7 t_7^9 t_8^7 +  6 y_{s}^8 y_{o_1}^{17} y_{o_2}^{15} y_{o_3}^{17} y_{o_4}^{15} y_{o_5}^{16} y_{o_6}^{16} t_1^8 t_2^8 t_3^8 t_4^8 t_5^9 t_6^7 t_7^9 t_8^7 +  7 y_{s}^8 y_{o_1}^{17} y_{o_2}^{15} y_{o_3}^{17} y_{o_4}^{15} y_{o_5}^{16} y_{o_6}^{16} t_1^7 t_2^9 t_3^8 t_4^8 t_5^9 t_6^7 t_7^9 t_8^7 +  3 y_{s}^8 y_{o_1}^{17} y_{o_2}^{15} y_{o_3}^{17} y_{o_4}^{15} y_{o_5}^{16} y_{o_6}^{16} t_1^6 t_2^{10} t_3^8 t_4^8 t_5^9 t_6^7 t_7^9 t_8^7 +  y_{s}^8 y_{o_1}^{17} y_{o_2}^{15} y_{o_3}^{17} y_{o_4}^{15} y_{o_5}^{16} y_{o_6}^{16} t_1^5 t_2^{11} t_3^8 t_4^8 t_5^9 t_6^7 t_7^9 t_8^7 +  y_{s}^8 y_{o_1}^{17} y_{o_2}^{15} y_{o_3}^{16} y_{o_4}^{16} y_{o_5}^{17} y_{o_6}^{15} t_1^{11} t_2^5 t_3^7 t_4^9 t_5^8 t_6^8 t_7^9 t_8^7 +  3 y_{s}^8 y_{o_1}^{17} y_{o_2}^{15} y_{o_3}^{16} y_{o_4}^{16} y_{o_5}^{17} y_{o_6}^{15} t_1^{10} t_2^6 t_3^7 t_4^9 t_5^8 t_6^8 t_7^9 t_8^7 +  7 y_{s}^8 y_{o_1}^{17} y_{o_2}^{15} y_{o_3}^{16} y_{o_4}^{16} y_{o_5}^{17} y_{o_6}^{15} t_1^9 t_2^7 t_3^7 t_4^9 t_5^8 t_6^8 t_7^9 t_8^7 +  6 y_{s}^8 y_{o_1}^{17} y_{o_2}^{15} y_{o_3}^{16} y_{o_4}^{16} y_{o_5}^{17} y_{o_6}^{15} t_1^8 t_2^8 t_3^7 t_4^9 t_5^8 t_6^8 t_7^9 t_8^7 +  7 y_{s}^8 y_{o_1}^{17} y_{o_2}^{15} y_{o_3}^{16} y_{o_4}^{16} y_{o_5}^{17} y_{o_6}^{15} t_1^7 t_2^9 t_3^7 t_4^9 t_5^8 t_6^8 t_7^9 t_8^7 +  3 y_{s}^8 y_{o_1}^{17} y_{o_2}^{15} y_{o_3}^{16} y_{o_4}^{16} y_{o_5}^{17} y_{o_6}^{15} t_1^6 t_2^{10} t_3^7 t_4^9 t_5^8 t_6^8 t_7^9 t_8^7 +  y_{s}^8 y_{o_1}^{17} y_{o_2}^{15} y_{o_3}^{16} y_{o_4}^{16} y_{o_5}^{17} y_{o_6}^{15} t_1^5 t_2^{11} t_3^7 t_4^9 t_5^8 t_6^8 t_7^9 t_8^7 +  2 y_{s}^8 y_{o_1}^{17} y_{o_2}^{15} y_{o_3}^{15} y_{o_4}^{17} y_{o_5}^{18} y_{o_6}^{14} t_1^9 t_2^7 t_3^6 t_4^{10} t_5^7 t_6^9 t_7^9 t_8^7 +  y_{s}^8 y_{o_1}^{17} y_{o_2}^{15} y_{o_3}^{15} y_{o_4}^{17} y_{o_5}^{18} y_{o_6}^{14} t_1^8 t_2^8 t_3^6 t_4^{10} t_5^7 t_6^9 t_7^9 t_8^7 +  2 y_{s}^8 y_{o_1}^{17} y_{o_2}^{15} y_{o_3}^{15} y_{o_4}^{17} y_{o_5}^{18} y_{o_6}^{14} t_1^7 t_2^9 t_3^6 t_4^{10} t_5^7 t_6^9 t_7^9 t_8^7 -  y_{s}^9 y_{o_1}^{20} y_{o_2}^{16} y_{o_3}^{19} y_{o_4}^{17} y_{o_5}^{19} y_{o_6}^{17} t_1^{10} t_2^8 t_3^8 t_4^{10} t_5^{10} t_6^8 t_7^{11} t_8^7 -  y_{s}^9 y_{o_1}^{20} y_{o_2}^{16} y_{o_3}^{19} y_{o_4}^{17} y_{o_5}^{19} y_{o_6}^{17} t_1^9 t_2^9 t_3^8 t_4^{10} t_5^{10} t_6^8 t_7^{11} t_8^7 -  y_{s}^9 y_{o_1}^{20} y_{o_2}^{16} y_{o_3}^{19} y_{o_4}^{17} y_{o_5}^{19} y_{o_6}^{17} t_1^8 t_2^{10} t_3^8 t_4^{10} t_5^{10} t_6^8 t_7^{11} t_8^7 -  y_{s}^5 y_{o_1}^7 y_{o_2}^{13} y_{o_3}^{10} y_{o_4}^{10} y_{o_5}^7 y_{o_6}^{13} t_1^5 t_2^5 t_3^8 t_4^2 t_5^5 t_6^5 t_7^2 t_8^8 +  y_{s}^5 y_{o_1}^7 y_{o_2}^{13} y_{o_3}^9 y_{o_4}^{11} y_{o_5}^8 y_{o_6}^{12} t_1^6 t_2^4 t_3^7 t_4^3 t_5^4 t_6^6 t_7^2 t_8^8 +  y_{s}^5 y_{o_1}^7 y_{o_2}^{13} y_{o_3}^9 y_{o_4}^{11} y_{o_5}^8 y_{o_6}^{12} t_1^4 t_2^6 t_3^7 t_4^3 t_5^4 t_6^6 t_7^2 t_8^8 +  y_{s}^5 y_{o_1}^7 y_{o_2}^{13} y_{o_3}^8 y_{o_4}^{12} y_{o_5}^9 y_{o_6}^{11} t_1^6 t_2^4 t_3^6 t_4^4 t_5^3 t_6^7 t_7^2 t_8^8 +  y_{s}^5 y_{o_1}^7 y_{o_2}^{13} y_{o_3}^8 y_{o_4}^{12} y_{o_5}^9 y_{o_6}^{11} t_1^4 t_2^6 t_3^6 t_4^4 t_5^3 t_6^7 t_7^2 t_8^8 -  y_{s}^5 y_{o_1}^7 y_{o_2}^{13} y_{o_3}^7 y_{o_4}^{13} y_{o_5}^{10} y_{o_6}^{10} t_1^5 t_2^5 t_3^5 t_4^5 t_5^2 t_6^8 t_7^2 t_8^8 +  y_{s}^6 y_{o_1}^{10} y_{o_2}^{14} y_{o_3}^{13} y_{o_4}^{11} y_{o_5}^9 y_{o_6}^{15} t_1^7 t_2^5 t_3^9 t_4^3 t_5^7 t_6^5 t_7^4 t_8^8 +  y_{s}^6 y_{o_1}^{10} y_{o_2}^{14} y_{o_3}^{13} y_{o_4}^{11} y_{o_5}^9 y_{o_6}^{15} t_1^5 t_2^7 t_3^9 t_4^3 t_5^7 t_6^5 t_7^4 t_8^8 +  3 y_{s}^6 y_{o_1}^{10} y_{o_2}^{14} y_{o_3}^{12} y_{o_4}^{12} y_{o_5}^{10} y_{o_6}^{14} t_1^7 t_2^5 t_3^8 t_4^4 t_5^6 t_6^6 t_7^4 t_8^8 -  y_{s}^6 y_{o_1}^{10} y_{o_2}^{14} y_{o_3}^{12} y_{o_4}^{12} y_{o_5}^{10} y_{o_6}^{14} t_1^6 t_2^6 t_3^8 t_4^4 t_5^6 t_6^6 t_7^4 t_8^8 +  3 y_{s}^6 y_{o_1}^{10} y_{o_2}^{14} y_{o_3}^{12} y_{o_4}^{12} y_{o_5}^{10} y_{o_6}^{14} t_1^5 t_2^7 t_3^8 t_4^4 t_5^6 t_6^6 t_7^4 t_8^8 -  y_{s}^6 y_{o_1}^{10} y_{o_2}^{14} y_{o_3}^{11} y_{o_4}^{13} y_{o_5}^{11} y_{o_6}^{13} t_1^8 t_2^4 t_3^7 t_4^5 t_5^5 t_6^7 t_7^4 t_8^8 +  3 y_{s}^6 y_{o_1}^{10} y_{o_2}^{14} y_{o_3}^{11} y_{o_4}^{13} y_{o_5}^{11} y_{o_6}^{13} t_1^7 t_2^5 t_3^7 t_4^5 t_5^5 t_6^7 t_7^4 t_8^8 -  3 y_{s}^6 y_{o_1}^{10} y_{o_2}^{14} y_{o_3}^{11} y_{o_4}^{13} y_{o_5}^{11} y_{o_6}^{13} t_1^6 t_2^6 t_3^7 t_4^5 t_5^5 t_6^7 t_7^4 t_8^8 +  3 y_{s}^6 y_{o_1}^{10} y_{o_2}^{14} y_{o_3}^{11} y_{o_4}^{13} y_{o_5}^{11} y_{o_6}^{13} t_1^5 t_2^7 t_3^7 t_4^5 t_5^5 t_6^7 t_7^4 t_8^8 -  y_{s}^6 y_{o_1}^{10} y_{o_2}^{14} y_{o_3}^{11} y_{o_4}^{13} y_{o_5}^{11} y_{o_6}^{13} t_1^4 t_2^8 t_3^7 t_4^5 t_5^5 t_6^7 t_7^4 t_8^8 +  3 y_{s}^6 y_{o_1}^{10} y_{o_2}^{14} y_{o_3}^{10} y_{o_4}^{14} y_{o_5}^{12} y_{o_6}^{12} t_1^7 t_2^5 t_3^6 t_4^6 t_5^4 t_6^8 t_7^4 t_8^8 -  y_{s}^6 y_{o_1}^{10} y_{o_2}^{14} y_{o_3}^{10} y_{o_4}^{14} y_{o_5}^{12} y_{o_6}^{12} t_1^6 t_2^6 t_3^6 t_4^6 t_5^4 t_6^8 t_7^4 t_8^8 +  3 y_{s}^6 y_{o_1}^{10} y_{o_2}^{14} y_{o_3}^{10} y_{o_4}^{14} y_{o_5}^{12} y_{o_6}^{12} t_1^5 t_2^7 t_3^6 t_4^6 t_5^4 t_6^8 t_7^4 t_8^8 +  y_{s}^6 y_{o_1}^{10} y_{o_2}^{14} y_{o_3}^9 y_{o_4}^{15} y_{o_5}^{13} y_{o_6}^{11} t_1^7 t_2^5 t_3^5 t_4^7 t_5^3 t_6^9 t_7^4 t_8^8 +  y_{s}^6 y_{o_1}^{10} y_{o_2}^{14} y_{o_3}^9 y_{o_4}^{15} y_{o_5}^{13} y_{o_6}^{11} t_1^5 t_2^7 t_3^5 t_4^7 t_5^3 t_6^9 t_7^4 t_8^8 -  y_{s}^7 y_{o_1}^{13} y_{o_2}^{15} y_{o_3}^{16} y_{o_4}^{12} y_{o_5}^{11} y_{o_6}^{17} t_1^7 t_2^7 t_3^{10} t_4^4 t_5^9 t_6^5 t_7^6 t_8^8 -  y_{s}^7 y_{o_1}^{13} y_{o_2}^{15} y_{o_3}^{15} y_{o_4}^{13} y_{o_5}^{12} y_{o_6}^{16} t_1^{10} t_2^4 t_3^9 t_4^5 t_5^8 t_6^6 t_7^6 t_8^8 -  3 y_{s}^7 y_{o_1}^{13} y_{o_2}^{15} y_{o_3}^{15} y_{o_4}^{13} y_{o_5}^{12} y_{o_6}^{16} t_1^9 t_2^5 t_3^9 t_4^5 t_5^8 t_6^6 t_7^6 t_8^8 -  3 y_{s}^7 y_{o_1}^{13} y_{o_2}^{15} y_{o_3}^{15} y_{o_4}^{13} y_{o_5}^{12} y_{o_6}^{16} t_1^8 t_2^6 t_3^9 t_4^5 t_5^8 t_6^6 t_7^6 t_8^8 -  7 y_{s}^7 y_{o_1}^{13} y_{o_2}^{15} y_{o_3}^{15} y_{o_4}^{13} y_{o_5}^{12} y_{o_6}^{16} t_1^7 t_2^7 t_3^9 t_4^5 t_5^8 t_6^6 t_7^6 t_8^8 -  3 y_{s}^7 y_{o_1}^{13} y_{o_2}^{15} y_{o_3}^{15} y_{o_4}^{13} y_{o_5}^{12} y_{o_6}^{16} t_1^6 t_2^8 t_3^9 t_4^5 t_5^8 t_6^6 t_7^6 t_8^8 -  3 y_{s}^7 y_{o_1}^{13} y_{o_2}^{15} y_{o_3}^{15} y_{o_4}^{13} y_{o_5}^{12} y_{o_6}^{16} t_1^5 t_2^9 t_3^9 t_4^5 t_5^8 t_6^6 t_7^6 t_8^8 -  y_{s}^7 y_{o_1}^{13} y_{o_2}^{15} y_{o_3}^{15} y_{o_4}^{13} y_{o_5}^{12} y_{o_6}^{16} t_1^4 t_2^{10} t_3^9 t_4^5 t_5^8 t_6^6 t_7^6 t_8^8 -  y_{s}^7 y_{o_1}^{13} y_{o_2}^{15} y_{o_3}^{14} y_{o_4}^{14} y_{o_5}^{13} y_{o_6}^{15} t_1^{10} t_2^4 t_3^8 t_4^6 t_5^7 t_6^7 t_7^6 t_8^8 -  5 y_{s}^7 y_{o_1}^{13} y_{o_2}^{15} y_{o_3}^{14} y_{o_4}^{14} y_{o_5}^{13} y_{o_6}^{15} t_1^9 t_2^5 t_3^8 t_4^6 t_5^7 t_6^7 t_7^6 t_8^8 -  3 y_{s}^7 y_{o_1}^{13} y_{o_2}^{15} y_{o_3}^{14} y_{o_4}^{14} y_{o_5}^{13} y_{o_6}^{15} t_1^8 t_2^6 t_3^8 t_4^6 t_5^7 t_6^7 t_7^6 t_8^8 -  {12} y_{s}^7 y_{o_1}^{13} y_{o_2}^{15} y_{o_3}^{14} y_{o_4}^{14} y_{o_5}^{13} y_{o_6}^{15} t_1^7 t_2^7 t_3^8 t_4^6 t_5^7 t_6^7 t_7^6 t_8^8 -  3 y_{s}^7 y_{o_1}^{13} y_{o_2}^{15} y_{o_3}^{14} y_{o_4}^{14} y_{o_5}^{13} y_{o_6}^{15} t_1^6 t_2^8 t_3^8 t_4^6 t_5^7 t_6^7 t_7^6 t_8^8 -  5 y_{s}^7 y_{o_1}^{13} y_{o_2}^{15} y_{o_3}^{14} y_{o_4}^{14} y_{o_5}^{13} y_{o_6}^{15} t_1^5 t_2^9 t_3^8 t_4^6 t_5^7 t_6^7 t_7^6 t_8^8 -  y_{s}^7 y_{o_1}^{13} y_{o_2}^{15} y_{o_3}^{14} y_{o_4}^{14} y_{o_5}^{13} y_{o_6}^{15} t_1^4 t_2^{10} t_3^8 t_4^6 t_5^7 t_6^7 t_7^6 t_8^8 -  y_{s}^7 y_{o_1}^{13} y_{o_2}^{15} y_{o_3}^{13} y_{o_4}^{15} y_{o_5}^{14} y_{o_6}^{14} t_1^{10} t_2^4 t_3^7 t_4^7 t_5^6 t_6^8 t_7^6 t_8^8 -  5 y_{s}^7 y_{o_1}^{13} y_{o_2}^{15} y_{o_3}^{13} y_{o_4}^{15} y_{o_5}^{14} y_{o_6}^{14} t_1^9 t_2^5 t_3^7 t_4^7 t_5^6 t_6^8 t_7^6 t_8^8 -  3 y_{s}^7 y_{o_1}^{13} y_{o_2}^{15} y_{o_3}^{13} y_{o_4}^{15} y_{o_5}^{14} y_{o_6}^{14} t_1^8 t_2^6 t_3^7 t_4^7 t_5^6 t_6^8 t_7^6 t_8^8 -  {12} y_{s}^7 y_{o_1}^{13} y_{o_2}^{15} y_{o_3}^{13} y_{o_4}^{15} y_{o_5}^{14} y_{o_6}^{14} t_1^7 t_2^7 t_3^7 t_4^7 t_5^6 t_6^8 t_7^6 t_8^8 -  3 y_{s}^7 y_{o_1}^{13} y_{o_2}^{15} y_{o_3}^{13} y_{o_4}^{15} y_{o_5}^{14} y_{o_6}^{14} t_1^6 t_2^8 t_3^7 t_4^7 t_5^6 t_6^8 t_7^6 t_8^8 -  5 y_{s}^7 y_{o_1}^{13} y_{o_2}^{15} y_{o_3}^{13} y_{o_4}^{15} y_{o_5}^{14} y_{o_6}^{14} t_1^5 t_2^9 t_3^7 t_4^7 t_5^6 t_6^8 t_7^6 t_8^8 -  y_{s}^7 y_{o_1}^{13} y_{o_2}^{15} y_{o_3}^{13} y_{o_4}^{15} y_{o_5}^{14} y_{o_6}^{14} t_1^4 t_2^{10} t_3^7 t_4^7 t_5^6 t_6^8 t_7^6 t_8^8 -  y_{s}^7 y_{o_1}^{13} y_{o_2}^{15} y_{o_3}^{12} y_{o_4}^{16} y_{o_5}^{15} y_{o_6}^{13} t_1^{10} t_2^4 t_3^6 t_4^8 t_5^5 t_6^9 t_7^6 t_8^8 -  3 y_{s}^7 y_{o_1}^{13} y_{o_2}^{15} y_{o_3}^{12} y_{o_4}^{16} y_{o_5}^{15} y_{o_6}^{13} t_1^9 t_2^5 t_3^6 t_4^8 t_5^5 t_6^9 t_7^6 t_8^8 -  3 y_{s}^7 y_{o_1}^{13} y_{o_2}^{15} y_{o_3}^{12} y_{o_4}^{16} y_{o_5}^{15} y_{o_6}^{13} t_1^8 t_2^6 t_3^6 t_4^8 t_5^5 t_6^9 t_7^6 t_8^8 -  7 y_{s}^7 y_{o_1}^{13} y_{o_2}^{15} y_{o_3}^{12} y_{o_4}^{16} y_{o_5}^{15} y_{o_6}^{13} t_1^7 t_2^7 t_3^6 t_4^8 t_5^5 t_6^9 t_7^6 t_8^8 -  3 y_{s}^7 y_{o_1}^{13} y_{o_2}^{15} y_{o_3}^{12} y_{o_4}^{16} y_{o_5}^{15} y_{o_6}^{13} t_1^6 t_2^8 t_3^6 t_4^8 t_5^5 t_6^9 t_7^6 t_8^8 -  3 y_{s}^7 y_{o_1}^{13} y_{o_2}^{15} y_{o_3}^{12} y_{o_4}^{16} y_{o_5}^{15} y_{o_6}^{13} t_1^5 t_2^9 t_3^6 t_4^8 t_5^5 t_6^9 t_7^6 t_8^8 -  y_{s}^7 y_{o_1}^{13} y_{o_2}^{15} y_{o_3}^{12} y_{o_4}^{16} y_{o_5}^{15} y_{o_6}^{13} t_1^4 t_2^{10} t_3^6 t_4^8 t_5^5 t_6^9 t_7^6 t_8^8 -  y_{s}^7 y_{o_1}^{13} y_{o_2}^{15} y_{o_3}^{11} y_{o_4}^{17} y_{o_5}^{16} y_{o_6}^{12} t_1^7 t_2^7 t_3^5 t_4^9 t_5^4 t_6^{10} t_7^6 t_8^8 +  y_{s}^8 y_{o_1}^{16} y_{o_2}^{16} y_{o_3}^{18} y_{o_4}^{14} y_{o_5}^{14} y_{o_6}^{18} t_1^{10} t_2^6 t_3^{10} t_4^6 t_5^{10} t_6^6 t_7^8 t_8^8 +  2 y_{s}^8 y_{o_1}^{16} y_{o_2}^{16} y_{o_3}^{18} y_{o_4}^{14} y_{o_5}^{14} y_{o_6}^{18} t_1^9 t_2^7 t_3^{10} t_4^6 t_5^{10} t_6^6 t_7^8 t_8^8 +  2 y_{s}^8 y_{o_1}^{16} y_{o_2}^{16} y_{o_3}^{18} y_{o_4}^{14} y_{o_5}^{14} y_{o_6}^{18} t_1^8 t_2^8 t_3^{10} t_4^6 t_5^{10} t_6^6 t_7^8 t_8^8 +  2 y_{s}^8 y_{o_1}^{16} y_{o_2}^{16} y_{o_3}^{18} y_{o_4}^{14} y_{o_5}^{14} y_{o_6}^{18} t_1^7 t_2^9 t_3^{10} t_4^6 t_5^{10} t_6^6 t_7^8 t_8^8 +  y_{s}^8 y_{o_1}^{16} y_{o_2}^{16} y_{o_3}^{18} y_{o_4}^{14} y_{o_5}^{14} y_{o_6}^{18} t_1^6 t_2^{10} t_3^{10} t_4^6 t_5^{10} t_6^6 t_7^8 t_8^8 +  y_{s}^8 y_{o_1}^{16} y_{o_2}^{16} y_{o_3}^{17} y_{o_4}^{15} y_{o_5}^{15} y_{o_6}^{17} t_1^{11} t_2^5 t_3^9 t_4^7 t_5^9 t_6^7 t_7^8 t_8^8 +  3 y_{s}^8 y_{o_1}^{16} y_{o_2}^{16} y_{o_3}^{17} y_{o_4}^{15} y_{o_5}^{15} y_{o_6}^{17} t_1^{10} t_2^6 t_3^9 t_4^7 t_5^9 t_6^7 t_7^8 t_8^8 +  7 y_{s}^8 y_{o_1}^{16} y_{o_2}^{16} y_{o_3}^{17} y_{o_4}^{15} y_{o_5}^{15} y_{o_6}^{17} t_1^9 t_2^7 t_3^9 t_4^7 t_5^9 t_6^7 t_7^8 t_8^8 +  6 y_{s}^8 y_{o_1}^{16} y_{o_2}^{16} y_{o_3}^{17} y_{o_4}^{15} y_{o_5}^{15} y_{o_6}^{17} t_1^8 t_2^8 t_3^9 t_4^7 t_5^9 t_6^7 t_7^8 t_8^8 +  7 y_{s}^8 y_{o_1}^{16} y_{o_2}^{16} y_{o_3}^{17} y_{o_4}^{15} y_{o_5}^{15} y_{o_6}^{17} t_1^7 t_2^9 t_3^9 t_4^7 t_5^9 t_6^7 t_7^8 t_8^8 +  3 y_{s}^8 y_{o_1}^{16} y_{o_2}^{16} y_{o_3}^{17} y_{o_4}^{15} y_{o_5}^{15} y_{o_6}^{17} t_1^6 t_2^{10} t_3^9 t_4^7 t_5^9 t_6^7 t_7^8 t_8^8 +  y_{s}^8 y_{o_1}^{16} y_{o_2}^{16} y_{o_3}^{17} y_{o_4}^{15} y_{o_5}^{15} y_{o_6}^{17} t_1^5 t_2^{11} t_3^9 t_4^7 t_5^9 t_6^7 t_7^8 t_8^8 +  2 y_{s}^8 y_{o_1}^{16} y_{o_2}^{16} y_{o_3}^{16} y_{o_4}^{16} y_{o_5}^{16} y_{o_6}^{16} t_1^{11} t_2^5 t_3^8 t_4^8 t_5^8 t_6^8 t_7^8 t_8^8 +  4 y_{s}^8 y_{o_1}^{16} y_{o_2}^{16} y_{o_3}^{16} y_{o_4}^{16} y_{o_5}^{16} y_{o_6}^{16} t_1^{10} t_2^6 t_3^8 t_4^8 t_5^8 t_6^8 t_7^8 t_8^8 +  {10} y_{s}^8 y_{o_1}^{16} y_{o_2}^{16} y_{o_3}^{16} y_{o_4}^{16} y_{o_5}^{16} y_{o_6}^{16} t_1^9 t_2^7 t_3^8 t_4^8 t_5^8 t_6^8 t_7^8 t_8^8 +  7 y_{s}^8 y_{o_1}^{16} y_{o_2}^{16} y_{o_3}^{16} y_{o_4}^{16} y_{o_5}^{16} y_{o_6}^{16} t_1^8 t_2^8 t_3^8 t_4^8 t_5^8 t_6^8 t_7^8 t_8^8 +  {10} y_{s}^8 y_{o_1}^{16} y_{o_2}^{16} y_{o_3}^{16} y_{o_4}^{16} y_{o_5}^{16} y_{o_6}^{16} t_1^7 t_2^9 t_3^8 t_4^8 t_5^8 t_6^8 t_7^8 t_8^8 +  4 y_{s}^8 y_{o_1}^{16} y_{o_2}^{16} y_{o_3}^{16} y_{o_4}^{16} y_{o_5}^{16} y_{o_6}^{16} t_1^6 t_2^{10} t_3^8 t_4^8 t_5^8 t_6^8 t_7^8 t_8^8 +  2 y_{s}^8 y_{o_1}^{16} y_{o_2}^{16} y_{o_3}^{16} y_{o_4}^{16} y_{o_5}^{16} y_{o_6}^{16} t_1^5 t_2^{11} t_3^8 t_4^8 t_5^8 t_6^8 t_7^8 t_8^8 +  y_{s}^8 y_{o_1}^{16} y_{o_2}^{16} y_{o_3}^{15} y_{o_4}^{17} y_{o_5}^{17} y_{o_6}^{15} t_1^{11} t_2^5 t_3^7 t_4^9 t_5^7 t_6^9 t_7^8 t_8^8 +  3 y_{s}^8 y_{o_1}^{16} y_{o_2}^{16} y_{o_3}^{15} y_{o_4}^{17} y_{o_5}^{17} y_{o_6}^{15} t_1^{10} t_2^6 t_3^7 t_4^9 t_5^7 t_6^9 t_7^8 t_8^8 +  7 y_{s}^8 y_{o_1}^{16} y_{o_2}^{16} y_{o_3}^{15} y_{o_4}^{17} y_{o_5}^{17} y_{o_6}^{15} t_1^9 t_2^7 t_3^7 t_4^9 t_5^7 t_6^9 t_7^8 t_8^8 +  6 y_{s}^8 y_{o_1}^{16} y_{o_2}^{16} y_{o_3}^{15} y_{o_4}^{17} y_{o_5}^{17} y_{o_6}^{15} t_1^8 t_2^8 t_3^7 t_4^9 t_5^7 t_6^9 t_7^8 t_8^8 +  7 y_{s}^8 y_{o_1}^{16} y_{o_2}^{16} y_{o_3}^{15} y_{o_4}^{17} y_{o_5}^{17} y_{o_6}^{15} t_1^7 t_2^9 t_3^7 t_4^9 t_5^7 t_6^9 t_7^8 t_8^8 +  3 y_{s}^8 y_{o_1}^{16} y_{o_2}^{16} y_{o_3}^{15} y_{o_4}^{17} y_{o_5}^{17} y_{o_6}^{15} t_1^6 t_2^{10} t_3^7 t_4^9 t_5^7 t_6^9 t_7^8 t_8^8 +  y_{s}^8 y_{o_1}^{16} y_{o_2}^{16} y_{o_3}^{15} y_{o_4}^{17} y_{o_5}^{17} y_{o_6}^{15} t_1^5 t_2^{11} t_3^7 t_4^9 t_5^7 t_6^9 t_7^8 t_8^8 +  y_{s}^8 y_{o_1}^{16} y_{o_2}^{16} y_{o_3}^{14} y_{o_4}^{18} y_{o_5}^{18} y_{o_6}^{14} t_1^{10} t_2^6 t_3^6 t_4^{10} t_5^6 t_6^{10} t_7^8 t_8^8 +  2 y_{s}^8 y_{o_1}^{16} y_{o_2}^{16} y_{o_3}^{14} y_{o_4}^{18} y_{o_5}^{18} y_{o_6}^{14} t_1^9 t_2^7 t_3^6 t_4^{10} t_5^6 t_6^{10} t_7^8 t_8^8 +  2 y_{s}^8 y_{o_1}^{16} y_{o_2}^{16} y_{o_3}^{14} y_{o_4}^{18} y_{o_5}^{18} y_{o_6}^{14} t_1^8 t_2^8 t_3^6 t_4^{10} t_5^6 t_6^{10} t_7^8 t_8^8 +  2 y_{s}^8 y_{o_1}^{16} y_{o_2}^{16} y_{o_3}^{14} y_{o_4}^{18} y_{o_5}^{18} y_{o_6}^{14} t_1^7 t_2^9 t_3^6 t_4^{10} t_5^6 t_6^{10} t_7^8 t_8^8 +  y_{s}^8 y_{o_1}^{16} y_{o_2}^{16} y_{o_3}^{14} y_{o_4}^{18} y_{o_5}^{18} y_{o_6}^{14} t_1^6 t_2^{10} t_3^6 t_4^{10} t_5^6 t_6^{10} t_7^8 t_8^8 -  y_{s}^9 y_{o_1}^{19} y_{o_2}^{17} y_{o_3}^{20} y_{o_4}^{16} y_{o_5}^{17} y_{o_6}^{19} t_1^{10} t_2^8 t_3^{10} t_4^8 t_5^{11} t_6^7 t_7^{10} t_8^8 -  y_{s}^9 y_{o_1}^{19} y_{o_2}^{17} y_{o_3}^{20} y_{o_4}^{16} y_{o_5}^{17} y_{o_6}^{19} t_1^9 t_2^9 t_3^{10} t_4^8 t_5^{11} t_6^7 t_7^{10} t_8^8 -  y_{s}^9 y_{o_1}^{19} y_{o_2}^{17} y_{o_3}^{20} y_{o_4}^{16} y_{o_5}^{17} y_{o_6}^{19} t_1^8 t_2^{10} t_3^{10} t_4^8 t_5^{11} t_6^7 t_7^{10} t_8^8 -  y_{s}^9 y_{o_1}^{19} y_{o_2}^{17} y_{o_3}^{19} y_{o_4}^{17} y_{o_5}^{18} y_{o_6}^{18} t_1^{11} t_2^7 t_3^9 t_4^9 t_5^{10} t_6^8 t_7^{10} t_8^8 -  2 y_{s}^9 y_{o_1}^{19} y_{o_2}^{17} y_{o_3}^{19} y_{o_4}^{17} y_{o_5}^{18} y_{o_6}^{18} t_1^{10} t_2^8 t_3^9 t_4^9 t_5^{10} t_6^8 t_7^{10} t_8^8 -  3 y_{s}^9 y_{o_1}^{19} y_{o_2}^{17} y_{o_3}^{19} y_{o_4}^{17} y_{o_5}^{18} y_{o_6}^{18} t_1^9 t_2^9 t_3^9 t_4^9 t_5^{10} t_6^8 t_7^{10} t_8^8 -  2 y_{s}^9 y_{o_1}^{19} y_{o_2}^{17} y_{o_3}^{19} y_{o_4}^{17} y_{o_5}^{18} y_{o_6}^{18} t_1^8 t_2^{10} t_3^9 t_4^9 t_5^{10} t_6^8 t_7^{10} t_8^8 -  y_{s}^9 y_{o_1}^{19} y_{o_2}^{17} y_{o_3}^{19} y_{o_4}^{17} y_{o_5}^{18} y_{o_6}^{18} t_1^7 t_2^{11} t_3^9 t_4^9 t_5^{10} t_6^8 t_7^{10} t_8^8 -  y_{s}^9 y_{o_1}^{19} y_{o_2}^{17} y_{o_3}^{18} y_{o_4}^{18} y_{o_5}^{19} y_{o_6}^{17} t_1^{11} t_2^7 t_3^8 t_4^{10} t_5^9 t_6^9 t_7^{10} t_8^8 -  2 y_{s}^9 y_{o_1}^{19} y_{o_2}^{17} y_{o_3}^{18} y_{o_4}^{18} y_{o_5}^{19} y_{o_6}^{17} t_1^{10} t_2^8 t_3^8 t_4^{10} t_5^9 t_6^9 t_7^{10} t_8^8 -  3 y_{s}^9 y_{o_1}^{19} y_{o_2}^{17} y_{o_3}^{18} y_{o_4}^{18} y_{o_5}^{19} y_{o_6}^{17} t_1^9 t_2^9 t_3^8 t_4^{10} t_5^9 t_6^9 t_7^{10} t_8^8 -  2 y_{s}^9 y_{o_1}^{19} y_{o_2}^{17} y_{o_3}^{18} y_{o_4}^{18} y_{o_5}^{19} y_{o_6}^{17} t_1^8 t_2^{10} t_3^8 t_4^{10} t_5^9 t_6^9 t_7^{10} t_8^8 -  y_{s}^9 y_{o_1}^{19} y_{o_2}^{17} y_{o_3}^{18} y_{o_4}^{18} y_{o_5}^{19} y_{o_6}^{17} t_1^7 t_2^{11} t_3^8 t_4^{10} t_5^9 t_6^9 t_7^{10} t_8^8 -  y_{s}^9 y_{o_1}^{19} y_{o_2}^{17} y_{o_3}^{17} y_{o_4}^{19} y_{o_5}^{20} y_{o_6}^{16} t_1^{10} t_2^8 t_3^7 t_4^{11} t_5^8 t_6^{10} t_7^{10} t_8^8 -  y_{s}^9 y_{o_1}^{19} y_{o_2}^{17} y_{o_3}^{17} y_{o_4}^{19} y_{o_5}^{20} y_{o_6}^{16} t_1^9 t_2^9 t_3^7 t_4^{11} t_5^8 t_6^{10} t_7^{10} t_8^8 -  y_{s}^9 y_{o_1}^{19} y_{o_2}^{17} y_{o_3}^{17} y_{o_4}^{19} y_{o_5}^{20} y_{o_6}^{16} t_1^8 t_2^{10} t_3^7 t_4^{11} t_5^8 t_6^{10} t_7^{10} t_8^8 -  y_{s}^6 y_{o_1}^9 y_{o_2}^{15} y_{o_3}^{12} y_{o_4}^{12} y_{o_5}^9 y_{o_6}^{15} t_1^6 t_2^6 t_3^9 t_4^3 t_5^6 t_6^6 t_7^3 t_8^9 +  y_{s}^6 y_{o_1}^9 y_{o_2}^{15} y_{o_3}^{11} y_{o_4}^{13} y_{o_5}^{10} y_{o_6}^{14} t_1^7 t_2^5 t_3^8 t_4^4 t_5^5 t_6^7 t_7^3 t_8^9 +  y_{s}^6 y_{o_1}^9 y_{o_2}^{15} y_{o_3}^{11} y_{o_4}^{13} y_{o_5}^{10} y_{o_6}^{14} t_1^5 t_2^7 t_3^8 t_4^4 t_5^5 t_6^7 t_7^3 t_8^9 +  y_{s}^6 y_{o_1}^9 y_{o_2}^{15} y_{o_3}^{10} y_{o_4}^{14} y_{o_5}^{11} y_{o_6}^{13} t_1^7 t_2^5 t_3^7 t_4^5 t_5^4 t_6^8 t_7^3 t_8^9 +  y_{s}^6 y_{o_1}^9 y_{o_2}^{15} y_{o_3}^{10} y_{o_4}^{14} y_{o_5}^{11} y_{o_6}^{13} t_1^5 t_2^7 t_3^7 t_4^5 t_5^4 t_6^8 t_7^3 t_8^9 -  y_{s}^6 y_{o_1}^9 y_{o_2}^{15} y_{o_3}^9 y_{o_4}^{15} y_{o_5}^{12} y_{o_6}^{12} t_1^6 t_2^6 t_3^6 t_4^6 t_5^3 t_6^9 t_7^3 t_8^9 -  y_{s}^7 y_{o_1}^{12} y_{o_2}^{16} y_{o_3}^{15} y_{o_4}^{13} y_{o_5}^{11} y_{o_6}^{17} t_1^7 t_2^7 t_3^{10} t_4^4 t_5^8 t_6^6 t_7^5 t_8^9 -  y_{s}^7 y_{o_1}^{12} y_{o_2}^{16} y_{o_3}^{14} y_{o_4}^{14} y_{o_5}^{12} y_{o_6}^{16} t_1^9 t_2^5 t_3^9 t_4^5 t_5^7 t_6^7 t_7^5 t_8^9 -  3 y_{s}^7 y_{o_1}^{12} y_{o_2}^{16} y_{o_3}^{14} y_{o_4}^{14} y_{o_5}^{12} y_{o_6}^{16} t_1^7 t_2^7 t_3^9 t_4^5 t_5^7 t_6^7 t_7^5 t_8^9 -  y_{s}^7 y_{o_1}^{12} y_{o_2}^{16} y_{o_3}^{14} y_{o_4}^{14} y_{o_5}^{12} y_{o_6}^{16} t_1^5 t_2^9 t_3^9 t_4^5 t_5^7 t_6^7 t_7^5 t_8^9 -  y_{s}^7 y_{o_1}^{12} y_{o_2}^{16} y_{o_3}^{13} y_{o_4}^{15} y_{o_5}^{13} y_{o_6}^{15} t_1^{10} t_2^4 t_3^8 t_4^6 t_5^6 t_6^8 t_7^5 t_8^9 -  3 y_{s}^7 y_{o_1}^{12} y_{o_2}^{16} y_{o_3}^{13} y_{o_4}^{15} y_{o_5}^{13} y_{o_6}^{15} t_1^9 t_2^5 t_3^8 t_4^6 t_5^6 t_6^8 t_7^5 t_8^9 -  3 y_{s}^7 y_{o_1}^{12} y_{o_2}^{16} y_{o_3}^{13} y_{o_4}^{15} y_{o_5}^{13} y_{o_6}^{15} t_1^8 t_2^6 t_3^8 t_4^6 t_5^6 t_6^8 t_7^5 t_8^9 -  7 y_{s}^7 y_{o_1}^{12} y_{o_2}^{16} y_{o_3}^{13} y_{o_4}^{15} y_{o_5}^{13} y_{o_6}^{15} t_1^7 t_2^7 t_3^8 t_4^6 t_5^6 t_6^8 t_7^5 t_8^9 -  3 y_{s}^7 y_{o_1}^{12} y_{o_2}^{16} y_{o_3}^{13} y_{o_4}^{15} y_{o_5}^{13} y_{o_6}^{15} t_1^6 t_2^8 t_3^8 t_4^6 t_5^6 t_6^8 t_7^5 t_8^9 -  3 y_{s}^7 y_{o_1}^{12} y_{o_2}^{16} y_{o_3}^{13} y_{o_4}^{15} y_{o_5}^{13} y_{o_6}^{15} t_1^5 t_2^9 t_3^8 t_4^6 t_5^6 t_6^8 t_7^5 t_8^9 -  y_{s}^7 y_{o_1}^{12} y_{o_2}^{16} y_{o_3}^{13} y_{o_4}^{15} y_{o_5}^{13} y_{o_6}^{15} t_1^4 t_2^{10} t_3^8 t_4^6 t_5^6 t_6^8 t_7^5 t_8^9 -  y_{s}^7 y_{o_1}^{12} y_{o_2}^{16} y_{o_3}^{12} y_{o_4}^{16} y_{o_5}^{14} y_{o_6}^{14} t_1^9 t_2^5 t_3^7 t_4^7 t_5^5 t_6^9 t_7^5 t_8^9 -  3 y_{s}^7 y_{o_1}^{12} y_{o_2}^{16} y_{o_3}^{12} y_{o_4}^{16} y_{o_5}^{14} y_{o_6}^{14} t_1^7 t_2^7 t_3^7 t_4^7 t_5^5 t_6^9 t_7^5 t_8^9 -  y_{s}^7 y_{o_1}^{12} y_{o_2}^{16} y_{o_3}^{12} y_{o_4}^{16} y_{o_5}^{14} y_{o_6}^{14} t_1^5 t_2^9 t_3^7 t_4^7 t_5^5 t_6^9 t_7^5 t_8^9 -  y_{s}^7 y_{o_1}^{12} y_{o_2}^{16} y_{o_3}^{11} y_{o_4}^{17} y_{o_5}^{15} y_{o_6}^{13} t_1^7 t_2^7 t_3^6 t_4^8 t_5^4 t_6^{10} t_7^5 t_8^9 +  2 y_{s}^8 y_{o_1}^{15} y_{o_2}^{17} y_{o_3}^{17} y_{o_4}^{15} y_{o_5}^{14} y_{o_6}^{18} t_1^9 t_2^7 t_3^{10} t_4^6 t_5^9 t_6^7 t_7^7 t_8^9 +  y_{s}^8 y_{o_1}^{15} y_{o_2}^{17} y_{o_3}^{17} y_{o_4}^{15} y_{o_5}^{14} y_{o_6}^{18} t_1^8 t_2^8 t_3^{10} t_4^6 t_5^9 t_6^7 t_7^7 t_8^9 +  2 y_{s}^8 y_{o_1}^{15} y_{o_2}^{17} y_{o_3}^{17} y_{o_4}^{15} y_{o_5}^{14} y_{o_6}^{18} t_1^7 t_2^9 t_3^{10} t_4^6 t_5^9 t_6^7 t_7^7 t_8^9 +  y_{s}^8 y_{o_1}^{15} y_{o_2}^{17} y_{o_3}^{16} y_{o_4}^{16} y_{o_5}^{15} y_{o_6}^{17} t_1^{11} t_2^5 t_3^9 t_4^7 t_5^8 t_6^8 t_7^7 t_8^9 +  3 y_{s}^8 y_{o_1}^{15} y_{o_2}^{17} y_{o_3}^{16} y_{o_4}^{16} y_{o_5}^{15} y_{o_6}^{17} t_1^{10} t_2^6 t_3^9 t_4^7 t_5^8 t_6^8 t_7^7 t_8^9 +  7 y_{s}^8 y_{o_1}^{15} y_{o_2}^{17} y_{o_3}^{16} y_{o_4}^{16} y_{o_5}^{15} y_{o_6}^{17} t_1^9 t_2^7 t_3^9 t_4^7 t_5^8 t_6^8 t_7^7 t_8^9 +  6 y_{s}^8 y_{o_1}^{15} y_{o_2}^{17} y_{o_3}^{16} y_{o_4}^{16} y_{o_5}^{15} y_{o_6}^{17} t_1^8 t_2^8 t_3^9 t_4^7 t_5^8 t_6^8 t_7^7 t_8^9 +  7 y_{s}^8 y_{o_1}^{15} y_{o_2}^{17} y_{o_3}^{16} y_{o_4}^{16} y_{o_5}^{15} y_{o_6}^{17} t_1^7 t_2^9 t_3^9 t_4^7 t_5^8 t_6^8 t_7^7 t_8^9 +  3 y_{s}^8 y_{o_1}^{15} y_{o_2}^{17} y_{o_3}^{16} y_{o_4}^{16} y_{o_5}^{15} y_{o_6}^{17} t_1^6 t_2^{10} t_3^9 t_4^7 t_5^8 t_6^8 t_7^7 t_8^9 +  y_{s}^8 y_{o_1}^{15} y_{o_2}^{17} y_{o_3}^{16} y_{o_4}^{16} y_{o_5}^{15} y_{o_6}^{17} t_1^5 t_2^{11} t_3^9 t_4^7 t_5^8 t_6^8 t_7^7 t_8^9 +  y_{s}^8 y_{o_1}^{15} y_{o_2}^{17} y_{o_3}^{15} y_{o_4}^{17} y_{o_5}^{16} y_{o_6}^{16} t_1^{11} t_2^5 t_3^8 t_4^8 t_5^7 t_6^9 t_7^7 t_8^9 +  3 y_{s}^8 y_{o_1}^{15} y_{o_2}^{17} y_{o_3}^{15} y_{o_4}^{17} y_{o_5}^{16} y_{o_6}^{16} t_1^{10} t_2^6 t_3^8 t_4^8 t_5^7 t_6^9 t_7^7 t_8^9 +  7 y_{s}^8 y_{o_1}^{15} y_{o_2}^{17} y_{o_3}^{15} y_{o_4}^{17} y_{o_5}^{16} y_{o_6}^{16} t_1^9 t_2^7 t_3^8 t_4^8 t_5^7 t_6^9 t_7^7 t_8^9 +  6 y_{s}^8 y_{o_1}^{15} y_{o_2}^{17} y_{o_3}^{15} y_{o_4}^{17} y_{o_5}^{16} y_{o_6}^{16} t_1^8 t_2^8 t_3^8 t_4^8 t_5^7 t_6^9 t_7^7 t_8^9 +  7 y_{s}^8 y_{o_1}^{15} y_{o_2}^{17} y_{o_3}^{15} y_{o_4}^{17} y_{o_5}^{16} y_{o_6}^{16} t_1^7 t_2^9 t_3^8 t_4^8 t_5^7 t_6^9 t_7^7 t_8^9 +  3 y_{s}^8 y_{o_1}^{15} y_{o_2}^{17} y_{o_3}^{15} y_{o_4}^{17} y_{o_5}^{16} y_{o_6}^{16} t_1^6 t_2^{10} t_3^8 t_4^8 t_5^7 t_6^9 t_7^7 t_8^9 +  y_{s}^8 y_{o_1}^{15} y_{o_2}^{17} y_{o_3}^{15} y_{o_4}^{17} y_{o_5}^{16} y_{o_6}^{16} t_1^5 t_2^{11} t_3^8 t_4^8 t_5^7 t_6^9 t_7^7 t_8^9 +  2 y_{s}^8 y_{o_1}^{15} y_{o_2}^{17} y_{o_3}^{14} y_{o_4}^{18} y_{o_5}^{17} y_{o_6}^{15} t_1^9 t_2^7 t_3^7 t_4^9 t_5^6 t_6^{10} t_7^7 t_8^9 +  y_{s}^8 y_{o_1}^{15} y_{o_2}^{17} y_{o_3}^{14} y_{o_4}^{18} y_{o_5}^{17} y_{o_6}^{15} t_1^8 t_2^8 t_3^7 t_4^9 t_5^6 t_6^{10} t_7^7 t_8^9 +  2 y_{s}^8 y_{o_1}^{15} y_{o_2}^{17} y_{o_3}^{14} y_{o_4}^{18} y_{o_5}^{17} y_{o_6}^{15} t_1^7 t_2^9 t_3^7 t_4^9 t_5^6 t_6^{10} t_7^7 t_8^9 -  y_{s}^9 y_{o_1}^{18} y_{o_2}^{18} y_{o_3}^{19} y_{o_4}^{17} y_{o_5}^{17} y_{o_6}^{19} t_1^{11} t_2^7 t_3^{10} t_4^8 t_5^{10} t_6^8 t_7^9 t_8^9 -  2 y_{s}^9 y_{o_1}^{18} y_{o_2}^{18} y_{o_3}^{19} y_{o_4}^{17} y_{o_5}^{17} y_{o_6}^{19} t_1^{10} t_2^8 t_3^{10} t_4^8 t_5^{10} t_6^8 t_7^9 t_8^9 -  3 y_{s}^9 y_{o_1}^{18} y_{o_2}^{18} y_{o_3}^{19} y_{o_4}^{17} y_{o_5}^{17} y_{o_6}^{19} t_1^9 t_2^9 t_3^{10} t_4^8 t_5^{10} t_6^8 t_7^9 t_8^9 -  2 y_{s}^9 y_{o_1}^{18} y_{o_2}^{18} y_{o_3}^{19} y_{o_4}^{17} y_{o_5}^{17} y_{o_6}^{19} t_1^8 t_2^{10} t_3^{10} t_4^8 t_5^{10} t_6^8 t_7^9 t_8^9 -  y_{s}^9 y_{o_1}^{18} y_{o_2}^{18} y_{o_3}^{19} y_{o_4}^{17} y_{o_5}^{17} y_{o_6}^{19} t_1^7 t_2^{11} t_3^{10} t_4^8 t_5^{10} t_6^8 t_7^9 t_8^9 -  2 y_{s}^9 y_{o_1}^{18} y_{o_2}^{18} y_{o_3}^{18} y_{o_4}^{18} y_{o_5}^{18} y_{o_6}^{18} t_1^{11} t_2^7 t_3^9 t_4^9 t_5^9 t_6^9 t_7^9 t_8^9 -  5 y_{s}^9 y_{o_1}^{18} y_{o_2}^{18} y_{o_3}^{18} y_{o_4}^{18} y_{o_5}^{18} y_{o_6}^{18} t_1^{10} t_2^8 t_3^9 t_4^9 t_5^9 t_6^9 t_7^9 t_8^9 -  5 y_{s}^9 y_{o_1}^{18} y_{o_2}^{18} y_{o_3}^{18} y_{o_4}^{18} y_{o_5}^{18} y_{o_6}^{18} t_1^9 t_2^9 t_3^9 t_4^9 t_5^9 t_6^9 t_7^9 t_8^9 -  5 y_{s}^9 y_{o_1}^{18} y_{o_2}^{18} y_{o_3}^{18} y_{o_4}^{18} y_{o_5}^{18} y_{o_6}^{18} t_1^8 t_2^{10} t_3^9 t_4^9 t_5^9 t_6^9 t_7^9 t_8^9 -  2 y_{s}^9 y_{o_1}^{18} y_{o_2}^{18} y_{o_3}^{18} y_{o_4}^{18} y_{o_5}^{18} y_{o_6}^{18} t_1^7 t_2^{11} t_3^9 t_4^9 t_5^9 t_6^9 t_7^9 t_8^9 -  y_{s}^9 y_{o_1}^{18} y_{o_2}^{18} y_{o_3}^{17} y_{o_4}^{19} y_{o_5}^{19} y_{o_6}^{17} t_1^{11} t_2^7 t_3^8 t_4^{10} t_5^8 t_6^{10} t_7^9 t_8^9 -  2 y_{s}^9 y_{o_1}^{18} y_{o_2}^{18} y_{o_3}^{17} y_{o_4}^{19} y_{o_5}^{19} y_{o_6}^{17} t_1^{10} t_2^8 t_3^8 t_4^{10} t_5^8 t_6^{10} t_7^9 t_8^9 -  3 y_{s}^9 y_{o_1}^{18} y_{o_2}^{18} y_{o_3}^{17} y_{o_4}^{19} y_{o_5}^{19} y_{o_6}^{17} t_1^9 t_2^9 t_3^8 t_4^{10} t_5^8 t_6^{10} t_7^9 t_8^9 -  2 y_{s}^9 y_{o_1}^{18} y_{o_2}^{18} y_{o_3}^{17} y_{o_4}^{19} y_{o_5}^{19} y_{o_6}^{17} t_1^8 t_2^{10} t_3^8 t_4^{10} t_5^8 t_6^{10} t_7^9 t_8^9 -  y_{s}^9 y_{o_1}^{18} y_{o_2}^{18} y_{o_3}^{17} y_{o_4}^{19} y_{o_5}^{19} y_{o_6}^{17} t_1^7 t_2^{11} t_3^8 t_4^{10} t_5^8 t_6^{10} t_7^9 t_8^9 +  y_{s}^{10} y_{o_1}^{21} y_{o_2}^{19} y_{o_3}^{21} y_{o_4}^{19} y_{o_5}^{20} y_{o_6}^{20} t_1^{10} t_2^{10} t_3^{10} t_4^{10} t_5^{11} t_6^9 t_7^{11} t_8^9 +  y_{s}^{10} y_{o_1}^{21} y_{o_2}^{19} y_{o_3}^{20} y_{o_4}^{20} y_{o_5}^{21} y_{o_6}^{19} t_1^{10} t_2^{10} t_3^9 t_4^{11} t_5^{10} t_6^{10} t_7^{11} t_8^9 -  y_{s}^7 y_{o_1}^{11} y_{o_2}^{17} y_{o_3}^{13} y_{o_4}^{15} y_{o_5}^{12} y_{o_6}^{16} t_1^7 t_2^7 t_3^9 t_4^5 t_5^6 t_6^8 t_7^4 t_8^{10} -  y_{s}^7 y_{o_1}^{11} y_{o_2}^{17} y_{o_3}^{12} y_{o_4}^{16} y_{o_5}^{13} y_{o_6}^{15} t_1^7 t_2^7 t_3^8 t_4^6 t_5^5 t_6^9 t_7^4 t_8^{10} +  y_{s}^8 y_{o_1}^{14} y_{o_2}^{18} y_{o_3}^{16} y_{o_4}^{16} y_{o_5}^{14} y_{o_6}^{18} t_1^{10} t_2^6 t_3^{10} t_4^6 t_5^8 t_6^8 t_7^6 t_8^{10} +  2 y_{s}^8 y_{o_1}^{14} y_{o_2}^{18} y_{o_3}^{16} y_{o_4}^{16} y_{o_5}^{14} y_{o_6}^{18} t_1^9 t_2^7 t_3^{10} t_4^6 t_5^8 t_6^8 t_7^6 t_8^{10} +  2 y_{s}^8 y_{o_1}^{14} y_{o_2}^{18} y_{o_3}^{16} y_{o_4}^{16} y_{o_5}^{14} y_{o_6}^{18} t_1^8 t_2^8 t_3^{10} t_4^6 t_5^8 t_6^8 t_7^6 t_8^{10} +  2 y_{s}^8 y_{o_1}^{14} y_{o_2}^{18} y_{o_3}^{16} y_{o_4}^{16} y_{o_5}^{14} y_{o_6}^{18} t_1^7 t_2^9 t_3^{10} t_4^6 t_5^8 t_6^8 t_7^6 t_8^{10} +  y_{s}^8 y_{o_1}^{14} y_{o_2}^{18} y_{o_3}^{16} y_{o_4}^{16} y_{o_5}^{14} y_{o_6}^{18} t_1^6 t_2^{10} t_3^{10} t_4^6 t_5^8 t_6^8 t_7^6 t_8^{10} +  2 y_{s}^8 y_{o_1}^{14} y_{o_2}^{18} y_{o_3}^{15} y_{o_4}^{17} y_{o_5}^{15} y_{o_6}^{17} t_1^9 t_2^7 t_3^9 t_4^7 t_5^7 t_6^9 t_7^6 t_8^{10} +  y_{s}^8 y_{o_1}^{14} y_{o_2}^{18} y_{o_3}^{15} y_{o_4}^{17} y_{o_5}^{15} y_{o_6}^{17} t_1^8 t_2^8 t_3^9 t_4^7 t_5^7 t_6^9 t_7^6 t_8^{10} +  2 y_{s}^8 y_{o_1}^{14} y_{o_2}^{18} y_{o_3}^{15} y_{o_4}^{17} y_{o_5}^{15} y_{o_6}^{17} t_1^7 t_2^9 t_3^9 t_4^7 t_5^7 t_6^9 t_7^6 t_8^{10} +  y_{s}^8 y_{o_1}^{14} y_{o_2}^{18} y_{o_3}^{14} y_{o_4}^{18} y_{o_5}^{16} y_{o_6}^{16} t_1^{10} t_2^6 t_3^8 t_4^8 t_5^6 t_6^{10} t_7^6 t_8^{10} +  2 y_{s}^8 y_{o_1}^{14} y_{o_2}^{18} y_{o_3}^{14} y_{o_4}^{18} y_{o_5}^{16} y_{o_6}^{16} t_1^9 t_2^7 t_3^8 t_4^8 t_5^6 t_6^{10} t_7^6 t_8^{10} +  2 y_{s}^8 y_{o_1}^{14} y_{o_2}^{18} y_{o_3}^{14} y_{o_4}^{18} y_{o_5}^{16} y_{o_6}^{16} t_1^8 t_2^8 t_3^8 t_4^8 t_5^6 t_6^{10} t_7^6 t_8^{10} +  2 y_{s}^8 y_{o_1}^{14} y_{o_2}^{18} y_{o_3}^{14} y_{o_4}^{18} y_{o_5}^{16} y_{o_6}^{16} t_1^7 t_2^9 t_3^8 t_4^8 t_5^6 t_6^{10} t_7^6 t_8^{10} +  y_{s}^8 y_{o_1}^{14} y_{o_2}^{18} y_{o_3}^{14} y_{o_4}^{18} y_{o_5}^{16} y_{o_6}^{16} t_1^6 t_2^{10} t_3^8 t_4^8 t_5^6 t_6^{10} t_7^6 t_8^{10} -  y_{s}^9 y_{o_1}^{17} y_{o_2}^{19} y_{o_3}^{19} y_{o_4}^{17} y_{o_5}^{16} y_{o_6}^{20} t_1^{10} t_2^8 t_3^{11} t_4^7 t_5^{10} t_6^8 t_7^8 t_8^{10} -  y_{s}^9 y_{o_1}^{17} y_{o_2}^{19} y_{o_3}^{19} y_{o_4}^{17} y_{o_5}^{16} y_{o_6}^{20} t_1^9 t_2^9 t_3^{11} t_4^7 t_5^{10} t_6^8 t_7^8 t_8^{10} -  y_{s}^9 y_{o_1}^{17} y_{o_2}^{19} y_{o_3}^{19} y_{o_4}^{17} y_{o_5}^{16} y_{o_6}^{20} t_1^8 t_2^{10} t_3^{11} t_4^7 t_5^{10} t_6^8 t_7^8 t_8^{10} -  y_{s}^9 y_{o_1}^{17} y_{o_2}^{19} y_{o_3}^{18} y_{o_4}^{18} y_{o_5}^{17} y_{o_6}^{19} t_1^{11} t_2^7 t_3^{10} t_4^8 t_5^9 t_6^9 t_7^8 t_8^{10} -  2 y_{s}^9 y_{o_1}^{17} y_{o_2}^{19} y_{o_3}^{18} y_{o_4}^{18} y_{o_5}^{17} y_{o_6}^{19} t_1^{10} t_2^8 t_3^{10} t_4^8 t_5^9 t_6^9 t_7^8 t_8^{10} -  3 y_{s}^9 y_{o_1}^{17} y_{o_2}^{19} y_{o_3}^{18} y_{o_4}^{18} y_{o_5}^{17} y_{o_6}^{19} t_1^9 t_2^9 t_3^{10} t_4^8 t_5^9 t_6^9 t_7^8 t_8^{10} -  2 y_{s}^9 y_{o_1}^{17} y_{o_2}^{19} y_{o_3}^{18} y_{o_4}^{18} y_{o_5}^{17} y_{o_6}^{19} t_1^8 t_2^{10} t_3^{10} t_4^8 t_5^9 t_6^9 t_7^8 t_8^{10} -  y_{s}^9 y_{o_1}^{17} y_{o_2}^{19} y_{o_3}^{18} y_{o_4}^{18} y_{o_5}^{17} y_{o_6}^{19} t_1^7 t_2^{11} t_3^{10} t_4^8 t_5^9 t_6^9 t_7^8 t_8^{10} -  y_{s}^9 y_{o_1}^{17} y_{o_2}^{19} y_{o_3}^{17} y_{o_4}^{19} y_{o_5}^{18} y_{o_6}^{18} t_1^{11} t_2^7 t_3^9 t_4^9 t_5^8 t_6^{10} t_7^8 t_8^{10} -  2 y_{s}^9 y_{o_1}^{17} y_{o_2}^{19} y_{o_3}^{17} y_{o_4}^{19} y_{o_5}^{18} y_{o_6}^{18} t_1^{10} t_2^8 t_3^9 t_4^9 t_5^8 t_6^{10} t_7^8 t_8^{10} -  3 y_{s}^9 y_{o_1}^{17} y_{o_2}^{19} y_{o_3}^{17} y_{o_4}^{19} y_{o_5}^{18} y_{o_6}^{18} t_1^9 t_2^9 t_3^9 t_4^9 t_5^8 t_6^{10} t_7^8 t_8^{10} -  2 y_{s}^9 y_{o_1}^{17} y_{o_2}^{19} y_{o_3}^{17} y_{o_4}^{19} y_{o_5}^{18} y_{o_6}^{18} t_1^8 t_2^{10} t_3^9 t_4^9 t_5^8 t_6^{10} t_7^8 t_8^{10} -  y_{s}^9 y_{o_1}^{17} y_{o_2}^{19} y_{o_3}^{17} y_{o_4}^{19} y_{o_5}^{18} y_{o_6}^{18} t_1^7 t_2^{11} t_3^9 t_4^9 t_5^8 t_6^{10} t_7^8 t_8^{10} -  y_{s}^9 y_{o_1}^{17} y_{o_2}^{19} y_{o_3}^{16} y_{o_4}^{20} y_{o_5}^{19} y_{o_6}^{17} t_1^{10} t_2^8 t_3^8 t_4^{10} t_5^7 t_6^{11} t_7^8 t_8^{10} -  y_{s}^9 y_{o_1}^{17} y_{o_2}^{19} y_{o_3}^{16} y_{o_4}^{20} y_{o_5}^{19} y_{o_6}^{17} t_1^9 t_2^9 t_3^8 t_4^{10} t_5^7 t_6^{11} t_7^8 t_8^{10} -  y_{s}^9 y_{o_1}^{17} y_{o_2}^{19} y_{o_3}^{16} y_{o_4}^{20} y_{o_5}^{19} y_{o_6}^{17} t_1^8 t_2^{10} t_3^8 t_4^{10} t_5^7 t_6^{11} t_7^8 t_8^{10} +  y_{s}^{10} y_{o_1}^{20} y_{o_2}^{20} y_{o_3}^{21} y_{o_4}^{19} y_{o_5}^{19} y_{o_6}^{21} t_1^{10} t_2^{10} t_3^{11} t_4^9 t_5^{11} t_6^9 t_7^{10} t_8^{10} +  y_{s}^{10} y_{o_1}^{20} y_{o_2}^{20} y_{o_3}^{20} y_{o_4}^{20} y_{o_5}^{20} y_{o_6}^{20} t_1^{11} t_2^9 t_3^{10} t_4^{10} t_5^{10} t_6^{10} t_7^{10} t_8^{10} +  y_{s}^{10} y_{o_1}^{20} y_{o_2}^{20} y_{o_3}^{20} y_{o_4}^{20} y_{o_5}^{20} y_{o_6}^{20} t_1^{10} t_2^{10} t_3^{10} t_4^{10} t_5^{10} t_6^{10} t_7^{10} t_8^{10} +  y_{s}^{10} y_{o_1}^{20} y_{o_2}^{20} y_{o_3}^{20} y_{o_4}^{20} y_{o_5}^{20} y_{o_6}^{20} t_1^9 t_2^{11} t_3^{10} t_4^{10} t_5^{10} t_6^{10} t_7^{10} t_8^{10} +  y_{s}^{10} y_{o_1}^{20} y_{o_2}^{20} y_{o_3}^{19} y_{o_4}^{21} y_{o_5}^{21} y_{o_6}^{19} t_1^{10} t_2^{10} t_3^9 t_4^{11} t_5^9 t_6^{11} t_7^{10} t_8^{10} -  y_{s}^9 y_{o_1}^{16} y_{o_2}^{20} y_{o_3}^{17} y_{o_4}^{19} y_{o_5}^{17} y_{o_6}^{19} t_1^{10} t_2^8 t_3^{10} t_4^8 t_5^8 t_6^{10} t_7^7 t_8^{11} -  y_{s}^9 y_{o_1}^{16} y_{o_2}^{20} y_{o_3}^{17} y_{o_4}^{19} y_{o_5}^{17} y_{o_6}^{19} t_1^9 t_2^9 t_3^{10} t_4^8 t_5^8 t_6^{10} t_7^7 t_8^{11} -  y_{s}^9 y_{o_1}^{16} y_{o_2}^{20} y_{o_3}^{17} y_{o_4}^{19} y_{o_5}^{17} y_{o_6}^{19} t_1^8 t_2^{10} t_3^{10} t_4^8 t_5^8 t_6^{10} t_7^7 t_8^{11} +  y_{s}^{10} y_{o_1}^{19} y_{o_2}^{21} y_{o_3}^{20} y_{o_4}^{20} y_{o_5}^{19} y_{o_6}^{21} t_1^{10} t_2^{10} t_3^{11} t_4^9 t_5^{10} t_6^{10} t_7^9 t_8^{11} +  y_{s}^{10} y_{o_1}^{19} y_{o_2}^{21} y_{o_3}^{19} y_{o_4}^{21} y_{o_5}^{20} y_{o_6}^{20} t_1^{10} t_2^{10} t_3^{10} t_4^{10} t_5^9 t_6^{11} t_7^9 t_8^{11} +  y_{s}^{11} y_{o_1}^{22} y_{o_2}^{22} y_{o_3}^{22} y_{o_4}^{22} y_{o_5}^{22} y_{o_6}^{22} t_1^{11} t_2^{11} t_3^{11} t_4^{11} t_5^{11} t_6^{11} t_7^{11} t_8^{11}
~,~
$
\end{quote}
\endgroup

\subsection{Model 18 \label{app_num_18}}

\begingroup\makeatletter\def\f@size{7}\check@mathfonts
\begin{quote}\raggedright
$
P(t_i,y_s,y_{o_1},y_{o_2},y_{o_3},y_{o_4},y_{o_5},y_{o_6},y_{o_7}; \mathcal{M}_{18}) =
1 - y_{s}^2 y_{o_1}^3 y_{o_2}^5 y_{o_3}^5 y_{o_4}^3 y_{o_5}^2 y_{o_6}^6 y_{o_7}^4 t_1 t_2 t_3^3 t_4^3 t_5^4 t_6 t_7 +  y_{s} y_{o_1} y_{o_2}^3 y_{o_3} y_{o_4}^3 y_{o_5}^2 y_{o_6}^2 y_{o_7}^3 t_1 t_2 t_4^2 t_5 t_6^2 t_8 -  y_{s}^2 y_{o_1}^2 y_{o_2}^6 y_{o_3}^3 y_{o_4}^5 y_{o_5}^3 y_{o_6}^5 y_{o_7}^5 t_1^2 t_2 t_3 t_4^4 t_5^3 t_6^3 t_8 -  y_{s}^2 y_{o_1}^2 y_{o_2}^6 y_{o_3}^3 y_{o_4}^5 y_{o_5}^3 y_{o_6}^5 y_{o_7}^5 t_1 t_2^2 t_3 t_4^4 t_5^3 t_6^3 t_8 +  y_{s} y_{o_1}^2 y_{o_2}^2 y_{o_3}^2 y_{o_4}^2 y_{o_5}^2 y_{o_6}^2 y_{o_7}^3 t_1^2 t_3 t_4 t_5 t_6 t_7 t_8 +  y_{s} y_{o_1}^2 y_{o_2}^2 y_{o_3}^2 y_{o_4}^2 y_{o_5}^2 y_{o_6}^2 y_{o_7}^3 t_1 t_2 t_3 t_4 t_5 t_6 t_7 t_8 +  y_{s} y_{o_1}^2 y_{o_2}^2 y_{o_3}^2 y_{o_4}^2 y_{o_5}^2 y_{o_6}^2 y_{o_7}^3 t_2^2 t_3 t_4 t_5 t_6 t_7 t_8 -  y_{s}^2 y_{o_1}^3 y_{o_2}^5 y_{o_3}^4 y_{o_4}^4 y_{o_5}^3 y_{o_6}^5 y_{o_7}^5 t_1^3 t_3^2 t_4^3 t_5^3 t_6^2 t_7 t_8 -  3 y_{s}^2 y_{o_1}^3 y_{o_2}^5 y_{o_3}^4 y_{o_4}^4 y_{o_5}^3 y_{o_6}^5 y_{o_7}^5 t_1^2 t_2 t_3^2 t_4^3 t_5^3 t_6^2 t_7 t_8 -  3 y_{s}^2 y_{o_1}^3 y_{o_2}^5 y_{o_3}^4 y_{o_4}^4 y_{o_5}^3 y_{o_6}^5 y_{o_7}^5 t_1 t_2^2 t_3^2 t_4^3 t_5^3 t_6^2 t_7 t_8 -  y_{s}^2 y_{o_1}^3 y_{o_2}^5 y_{o_3}^4 y_{o_4}^4 y_{o_5}^3 y_{o_6}^5 y_{o_7}^5 t_2^3 t_3^2 t_4^3 t_5^3 t_6^2 t_7 t_8 +  2 y_{s}^3 y_{o_1}^4 y_{o_2}^8 y_{o_3}^6 y_{o_4}^6 y_{o_5}^4 y_{o_6}^8 y_{o_7}^7 t_1^3 t_2 t_3^3 t_4^5 t_5^5 t_6^3 t_7 t_8 +  2 y_{s}^3 y_{o_1}^4 y_{o_2}^8 y_{o_3}^6 y_{o_4}^6 y_{o_5}^4 y_{o_6}^8 y_{o_7}^7 t_1^2 t_2^2 t_3^3 t_4^5 t_5^5 t_6^3 t_7 t_8 +  2 y_{s}^3 y_{o_1}^4 y_{o_2}^8 y_{o_3}^6 y_{o_4}^6 y_{o_5}^4 y_{o_6}^8 y_{o_7}^7 t_1 t_2^3 t_3^3 t_4^5 t_5^5 t_6^3 t_7 t_8 +  y_{s} y_{o_1}^3 y_{o_2} y_{o_3}^3 y_{o_4} y_{o_5}^2 y_{o_6}^2 y_{o_7}^3 t_1 t_2 t_3^2 t_5 t_7^2 t_8 -  y_{s}^2 y_{o_1}^4 y_{o_2}^4 y_{o_3}^5 y_{o_4}^3 y_{o_5}^3 y_{o_6}^5 y_{o_7}^5 t_1^3 t_3^3 t_4^2 t_5^3 t_6 t_7^2 t_8 -  3 y_{s}^2 y_{o_1}^4 y_{o_2}^4 y_{o_3}^5 y_{o_4}^3 y_{o_5}^3 y_{o_6}^5 y_{o_7}^5 t_1^2 t_2 t_3^3 t_4^2 t_5^3 t_6 t_7^2 t_8 -  3 y_{s}^2 y_{o_1}^4 y_{o_2}^4 y_{o_3}^5 y_{o_4}^3 y_{o_5}^3 y_{o_6}^5 y_{o_7}^5 t_1 t_2^2 t_3^3 t_4^2 t_5^3 t_6 t_7^2 t_8 -  y_{s}^2 y_{o_1}^4 y_{o_2}^4 y_{o_3}^5 y_{o_4}^3 y_{o_5}^3 y_{o_6}^5 y_{o_7}^5 t_2^3 t_3^3 t_4^2 t_5^3 t_6 t_7^2 t_8 +  2 y_{s}^3 y_{o_1}^5 y_{o_2}^7 y_{o_3}^7 y_{o_4}^5 y_{o_5}^4 y_{o_6}^8 y_{o_7}^7 t_1^3 t_2 t_3^4 t_4^4 t_5^5 t_6^2 t_7^2 t_8 +  3 y_{s}^3 y_{o_1}^5 y_{o_2}^7 y_{o_3}^7 y_{o_4}^5 y_{o_5}^4 y_{o_6}^8 y_{o_7}^7 t_1^2 t_2^2 t_3^4 t_4^4 t_5^5 t_6^2 t_7^2 t_8 +  2 y_{s}^3 y_{o_1}^5 y_{o_2}^7 y_{o_3}^7 y_{o_4}^5 y_{o_5}^4 y_{o_6}^8 y_{o_7}^7 t_1 t_2^3 t_3^4 t_4^4 t_5^5 t_6^2 t_7^2 t_8 -  y_{s}^4 y_{o_1}^6 y_{o_2}^{10} y_{o_3}^9 y_{o_4}^7 y_{o_5}^5 y_{o_6}^{11} y_{o_7}^9 t_1^3 t_2^2 t_3^5 t_4^6 t_5^7 t_6^3 t_7^2 t_8 -  y_{s}^4 y_{o_1}^6 y_{o_2}^{10} y_{o_3}^9 y_{o_4}^7 y_{o_5}^5 y_{o_6}^{11} y_{o_7}^9 t_1^2 t_2^3 t_3^5 t_4^6 t_5^7 t_6^3 t_7^2 t_8 -  y_{s}^2 y_{o_1}^5 y_{o_2}^3 y_{o_3}^6 y_{o_4}^2 y_{o_5}^3 y_{o_6}^5 y_{o_7}^5 t_1^2 t_2 t_3^4 t_4 t_5^3 t_7^3 t_8 -  y_{s}^2 y_{o_1}^5 y_{o_2}^3 y_{o_3}^6 y_{o_4}^2 y_{o_5}^3 y_{o_6}^5 y_{o_7}^5 t_1 t_2^2 t_3^4 t_4 t_5^3 t_7^3 t_8 +  2 y_{s}^3 y_{o_1}^6 y_{o_2}^6 y_{o_3}^8 y_{o_4}^4 y_{o_5}^4 y_{o_6}^8 y_{o_7}^7 t_1^3 t_2 t_3^5 t_4^3 t_5^5 t_6 t_7^3 t_8 +  2 y_{s}^3 y_{o_1}^6 y_{o_2}^6 y_{o_3}^8 y_{o_4}^4 y_{o_5}^4 y_{o_6}^8 y_{o_7}^7 t_1^2 t_2^2 t_3^5 t_4^3 t_5^5 t_6 t_7^3 t_8 +  2 y_{s}^3 y_{o_1}^6 y_{o_2}^6 y_{o_3}^8 y_{o_4}^4 y_{o_5}^4 y_{o_6}^8 y_{o_7}^7 t_1 t_2^3 t_3^5 t_4^3 t_5^5 t_6 t_7^3 t_8 -  y_{s}^4 y_{o_1}^7 y_{o_2}^9 y_{o_3}^{10} y_{o_4}^6 y_{o_5}^5 y_{o_6}^{11} y_{o_7}^9 t_1^3 t_2^2 t_3^6 t_4^5 t_5^7 t_6^2 t_7^3 t_8 -  y_{s}^4 y_{o_1}^7 y_{o_2}^9 y_{o_3}^{10} y_{o_4}^6 y_{o_5}^5 y_{o_6}^{11} y_{o_7}^9 t_1^2 t_2^3 t_3^6 t_4^5 t_5^7 t_6^2 t_7^3 t_8 +  y_{s} y_{o_1}^2 y_{o_2}^2 y_{o_3} y_{o_4}^3 y_{o_5}^3 y_{o_6} y_{o_7}^4 t_1^2 t_2 t_4 t_6^2 t_7 t_8^2 +  y_{s} y_{o_1}^2 y_{o_2}^2 y_{o_3} y_{o_4}^3 y_{o_5}^3 y_{o_6} y_{o_7}^4 t_1 t_2^2 t_4 t_6^2 t_7 t_8^2 -  y_{s}^2 y_{o_1}^3 y_{o_2}^5 y_{o_3}^3 y_{o_4}^5 y_{o_5}^4 y_{o_6}^4 y_{o_7}^6 t_1^4 t_3 t_4^3 t_5^2 t_6^3 t_7 t_8^2 -  2 y_{s}^2 y_{o_1}^3 y_{o_2}^5 y_{o_3}^3 y_{o_4}^5 y_{o_5}^4 y_{o_6}^4 y_{o_7}^6 t_1^3 t_2 t_3 t_4^3 t_5^2 t_6^3 t_7 t_8^2 -  3 y_{s}^2 y_{o_1}^3 y_{o_2}^5 y_{o_3}^3 y_{o_4}^5 y_{o_5}^4 y_{o_6}^4 y_{o_7}^6 t_1^2 t_2^2 t_3 t_4^3 t_5^2 t_6^3 t_7 t_8^2 -  2 y_{s}^2 y_{o_1}^3 y_{o_2}^5 y_{o_3}^3 y_{o_4}^5 y_{o_5}^4 y_{o_6}^4 y_{o_7}^6 t_1 t_2^3 t_3 t_4^3 t_5^2 t_6^3 t_7 t_8^2 -  y_{s}^2 y_{o_1}^3 y_{o_2}^5 y_{o_3}^3 y_{o_4}^5 y_{o_5}^4 y_{o_6}^4 y_{o_7}^6 t_2^4 t_3 t_4^3 t_5^2 t_6^3 t_7 t_8^2 +  y_{s}^3 y_{o_1}^4 y_{o_2}^8 y_{o_3}^5 y_{o_4}^7 y_{o_5}^5 y_{o_6}^7 y_{o_7}^8 t_1^4 t_2 t_3^2 t_4^5 t_5^4 t_6^4 t_7 t_8^2 +  2 y_{s}^3 y_{o_1}^4 y_{o_2}^8 y_{o_3}^5 y_{o_4}^7 y_{o_5}^5 y_{o_6}^7 y_{o_7}^8 t_1^3 t_2^2 t_3^2 t_4^5 t_5^4 t_6^4 t_7 t_8^2 +  2 y_{s}^3 y_{o_1}^4 y_{o_2}^8 y_{o_3}^5 y_{o_4}^7 y_{o_5}^5 y_{o_6}^7 y_{o_7}^8 t_1^2 t_2^3 t_3^2 t_4^5 t_5^4 t_6^4 t_7 t_8^2 +  y_{s}^3 y_{o_1}^4 y_{o_2}^8 y_{o_3}^5 y_{o_4}^7 y_{o_5}^5 y_{o_6}^7 y_{o_7}^8 t_1 t_2^4 t_3^2 t_4^5 t_5^4 t_6^4 t_7 t_8^2 -  y_{s}^4 y_{o_1}^5 y_{o_2}^{11} y_{o_3}^7 y_{o_4}^9 y_{o_5}^6 y_{o_6}^{10} y_{o_7}^{10} t_1^3 t_2^3 t_3^3 t_4^7 t_5^6 t_6^5 t_7 t_8^2 +  y_{s} y_{o_1}^3 y_{o_2} y_{o_3}^2 y_{o_4}^2 y_{o_5}^3 y_{o_6} y_{o_7}^4 t_1^2 t_2 t_3 t_6 t_7^2 t_8^2 +  y_{s} y_{o_1}^3 y_{o_2} y_{o_3}^2 y_{o_4}^2 y_{o_5}^3 y_{o_6} y_{o_7}^4 t_1 t_2^2 t_3 t_6 t_7^2 t_8^2 -  2 y_{s}^2 y_{o_1}^4 y_{o_2}^4 y_{o_3}^4 y_{o_4}^4 y_{o_5}^4 y_{o_6}^4 y_{o_7}^6 t_1^4 t_3^2 t_4^2 t_5^2 t_6^2 t_7^2 t_8^2 -  5 y_{s}^2 y_{o_1}^4 y_{o_2}^4 y_{o_3}^4 y_{o_4}^4 y_{o_5}^4 y_{o_6}^4 y_{o_7}^6 t_1^3 t_2 t_3^2 t_4^2 t_5^2 t_6^2 t_7^2 t_8^2 -  5 y_{s}^2 y_{o_1}^4 y_{o_2}^4 y_{o_3}^4 y_{o_4}^4 y_{o_5}^4 y_{o_6}^4 y_{o_7}^6 t_1^2 t_2^2 t_3^2 t_4^2 t_5^2 t_6^2 t_7^2 t_8^2 -  5 y_{s}^2 y_{o_1}^4 y_{o_2}^4 y_{o_3}^4 y_{o_4}^4 y_{o_5}^4 y_{o_6}^4 y_{o_7}^6 t_1 t_2^3 t_3^2 t_4^2 t_5^2 t_6^2 t_7^2 t_8^2 -  2 y_{s}^2 y_{o_1}^4 y_{o_2}^4 y_{o_3}^4 y_{o_4}^4 y_{o_5}^4 y_{o_6}^4 y_{o_7}^6 t_2^4 t_3^2 t_4^2 t_5^2 t_6^2 t_7^2 t_8^2 +  y_{s}^3 y_{o_1}^5 y_{o_2}^7 y_{o_3}^6 y_{o_4}^6 y_{o_5}^5 y_{o_6}^7 y_{o_7}^8 t_1^5 t_3^3 t_4^4 t_5^4 t_6^3 t_7^2 t_8^2 +  6 y_{s}^3 y_{o_1}^5 y_{o_2}^7 y_{o_3}^6 y_{o_4}^6 y_{o_5}^5 y_{o_6}^7 y_{o_7}^8 t_1^4 t_2 t_3^3 t_4^4 t_5^4 t_6^3 t_7^2 t_8^2 +  7 y_{s}^3 y_{o_1}^5 y_{o_2}^7 y_{o_3}^6 y_{o_4}^6 y_{o_5}^5 y_{o_6}^7 y_{o_7}^8 t_1^3 t_2^2 t_3^3 t_4^4 t_5^4 t_6^3 t_7^2 t_8^2 +  7 y_{s}^3 y_{o_1}^5 y_{o_2}^7 y_{o_3}^6 y_{o_4}^6 y_{o_5}^5 y_{o_6}^7 y_{o_7}^8 t_1^2 t_2^3 t_3^3 t_4^4 t_5^4 t_6^3 t_7^2 t_8^2 +  6 y_{s}^3 y_{o_1}^5 y_{o_2}^7 y_{o_3}^6 y_{o_4}^6 y_{o_5}^5 y_{o_6}^7 y_{o_7}^8 t_1 t_2^4 t_3^3 t_4^4 t_5^4 t_6^3 t_7^2 t_8^2 +  y_{s}^3 y_{o_1}^5 y_{o_2}^7 y_{o_3}^6 y_{o_4}^6 y_{o_5}^5 y_{o_6}^7 y_{o_7}^8 t_2^5 t_3^3 t_4^4 t_5^4 t_6^3 t_7^2 t_8^2 -  y_{s}^4 y_{o_1}^6 y_{o_2}^{10} y_{o_3}^8 y_{o_4}^8 y_{o_5}^6 y_{o_6}^{10} y_{o_7}^{10} t_1^5 t_2 t_3^4 t_4^6 t_5^6 t_6^4 t_7^2 t_8^2 -  2 y_{s}^4 y_{o_1}^6 y_{o_2}^{10} y_{o_3}^8 y_{o_4}^8 y_{o_5}^6 y_{o_6}^{10} y_{o_7}^{10} t_1^4 t_2^2 t_3^4 t_4^6 t_5^6 t_6^4 t_7^2 t_8^2 -  3 y_{s}^4 y_{o_1}^6 y_{o_2}^{10} y_{o_3}^8 y_{o_4}^8 y_{o_5}^6 y_{o_6}^{10} y_{o_7}^{10} t_1^3 t_2^3 t_3^4 t_4^6 t_5^6 t_6^4 t_7^2 t_8^2 -  2 y_{s}^4 y_{o_1}^6 y_{o_2}^{10} y_{o_3}^8 y_{o_4}^8 y_{o_5}^6 y_{o_6}^{10} y_{o_7}^{10} t_1^2 t_2^4 t_3^4 t_4^6 t_5^6 t_6^4 t_7^2 t_8^2 -  y_{s}^4 y_{o_1}^6 y_{o_2}^{10} y_{o_3}^8 y_{o_4}^8 y_{o_5}^6 y_{o_6}^{10} y_{o_7}^{10} t_1 t_2^5 t_3^4 t_4^6 t_5^6 t_6^4 t_7^2 t_8^2 -  y_{s}^2 y_{o_1}^5 y_{o_2}^3 y_{o_3}^5 y_{o_4}^3 y_{o_5}^4 y_{o_6}^4 y_{o_7}^6 t_1^4 t_3^3 t_4 t_5^2 t_6 t_7^3 t_8^2 -  2 y_{s}^2 y_{o_1}^5 y_{o_2}^3 y_{o_3}^5 y_{o_4}^3 y_{o_5}^4 y_{o_6}^4 y_{o_7}^6 t_1^3 t_2 t_3^3 t_4 t_5^2 t_6 t_7^3 t_8^2 -  3 y_{s}^2 y_{o_1}^5 y_{o_2}^3 y_{o_3}^5 y_{o_4}^3 y_{o_5}^4 y_{o_6}^4 y_{o_7}^6 t_1^2 t_2^2 t_3^3 t_4 t_5^2 t_6 t_7^3 t_8^2 -  2 y_{s}^2 y_{o_1}^5 y_{o_2}^3 y_{o_3}^5 y_{o_4}^3 y_{o_5}^4 y_{o_6}^4 y_{o_7}^6 t_1 t_2^3 t_3^3 t_4 t_5^2 t_6 t_7^3 t_8^2 -  y_{s}^2 y_{o_1}^5 y_{o_2}^3 y_{o_3}^5 y_{o_4}^3 y_{o_5}^4 y_{o_6}^4 y_{o_7}^6 t_2^4 t_3^3 t_4 t_5^2 t_6 t_7^3 t_8^2 +  y_{s}^3 y_{o_1}^6 y_{o_2}^6 y_{o_3}^7 y_{o_4}^5 y_{o_5}^5 y_{o_6}^7 y_{o_7}^8 t_1^5 t_3^4 t_4^3 t_5^4 t_6^2 t_7^3 t_8^2 +  6 y_{s}^3 y_{o_1}^6 y_{o_2}^6 y_{o_3}^7 y_{o_4}^5 y_{o_5}^5 y_{o_6}^7 y_{o_7}^8 t_1^4 t_2 t_3^4 t_4^3 t_5^4 t_6^2 t_7^3 t_8^2 +  7 y_{s}^3 y_{o_1}^6 y_{o_2}^6 y_{o_3}^7 y_{o_4}^5 y_{o_5}^5 y_{o_6}^7 y_{o_7}^8 t_1^3 t_2^2 t_3^4 t_4^3 t_5^4 t_6^2 t_7^3 t_8^2 +  7 y_{s}^3 y_{o_1}^6 y_{o_2}^6 y_{o_3}^7 y_{o_4}^5 y_{o_5}^5 y_{o_6}^7 y_{o_7}^8 t_1^2 t_2^3 t_3^4 t_4^3 t_5^4 t_6^2 t_7^3 t_8^2 +  6 y_{s}^3 y_{o_1}^6 y_{o_2}^6 y_{o_3}^7 y_{o_4}^5 y_{o_5}^5 y_{o_6}^7 y_{o_7}^8 t_1 t_2^4 t_3^4 t_4^3 t_5^4 t_6^2 t_7^3 t_8^2 +  y_{s}^3 y_{o_1}^6 y_{o_2}^6 y_{o_3}^7 y_{o_4}^5 y_{o_5}^5 y_{o_6}^7 y_{o_7}^8 t_2^5 t_3^4 t_4^3 t_5^4 t_6^2 t_7^3 t_8^2 -  3 y_{s}^4 y_{o_1}^7 y_{o_2}^9 y_{o_3}^9 y_{o_4}^7 y_{o_5}^6 y_{o_6}^{10} y_{o_7}^{10} t_1^5 t_2 t_3^5 t_4^5 t_5^6 t_6^3 t_7^3 t_8^2 -  5 y_{s}^4 y_{o_1}^7 y_{o_2}^9 y_{o_3}^9 y_{o_4}^7 y_{o_5}^6 y_{o_6}^{10} y_{o_7}^{10} t_1^4 t_2^2 t_3^5 t_4^5 t_5^6 t_6^3 t_7^3 t_8^2 -  7 y_{s}^4 y_{o_1}^7 y_{o_2}^9 y_{o_3}^9 y_{o_4}^7 y_{o_5}^6 y_{o_6}^{10} y_{o_7}^{10} t_1^3 t_2^3 t_3^5 t_4^5 t_5^6 t_6^3 t_7^3 t_8^2 -  5 y_{s}^4 y_{o_1}^7 y_{o_2}^9 y_{o_3}^9 y_{o_4}^7 y_{o_5}^6 y_{o_6}^{10} y_{o_7}^{10} t_1^2 t_2^4 t_3^5 t_4^5 t_5^6 t_6^3 t_7^3 t_8^2 -  3 y_{s}^4 y_{o_1}^7 y_{o_2}^9 y_{o_3}^9 y_{o_4}^7 y_{o_5}^6 y_{o_6}^{10} y_{o_7}^{10} t_1 t_2^5 t_3^5 t_4^5 t_5^6 t_6^3 t_7^3 t_8^2 +  y_{s}^5 y_{o_1}^8 y_{o_2}^{12} y_{o_3}^{11} y_{o_4}^9 y_{o_5}^7 y_{o_6}^{13} y_{o_7}^{12} t_1^5 t_2^2 t_3^6 t_4^7 t_5^8 t_6^4 t_7^3 t_8^2 +  y_{s}^5 y_{o_1}^8 y_{o_2}^{12} y_{o_3}^{11} y_{o_4}^9 y_{o_5}^7 y_{o_6}^{13} y_{o_7}^{12} t_1^4 t_2^3 t_3^6 t_4^7 t_5^8 t_6^4 t_7^3 t_8^2 +  y_{s}^5 y_{o_1}^8 y_{o_2}^{12} y_{o_3}^{11} y_{o_4}^9 y_{o_5}^7 y_{o_6}^{13} y_{o_7}^{12} t_1^3 t_2^4 t_3^6 t_4^7 t_5^8 t_6^4 t_7^3 t_8^2 +  y_{s}^5 y_{o_1}^8 y_{o_2}^{12} y_{o_3}^{11} y_{o_4}^9 y_{o_5}^7 y_{o_6}^{13} y_{o_7}^{12} t_1^2 t_2^5 t_3^6 t_4^7 t_5^8 t_6^4 t_7^3 t_8^2 +  y_{s}^3 y_{o_1}^7 y_{o_2}^5 y_{o_3}^8 y_{o_4}^4 y_{o_5}^5 y_{o_6}^7 y_{o_7}^8 t_1^4 t_2 t_3^5 t_4^2 t_5^4 t_6 t_7^4 t_8^2 +  2 y_{s}^3 y_{o_1}^7 y_{o_2}^5 y_{o_3}^8 y_{o_4}^4 y_{o_5}^5 y_{o_6}^7 y_{o_7}^8 t_1^3 t_2^2 t_3^5 t_4^2 t_5^4 t_6 t_7^4 t_8^2 +  2 y_{s}^3 y_{o_1}^7 y_{o_2}^5 y_{o_3}^8 y_{o_4}^4 y_{o_5}^5 y_{o_6}^7 y_{o_7}^8 t_1^2 t_2^3 t_3^5 t_4^2 t_5^4 t_6 t_7^4 t_8^2 +  y_{s}^3 y_{o_1}^7 y_{o_2}^5 y_{o_3}^8 y_{o_4}^4 y_{o_5}^5 y_{o_6}^7 y_{o_7}^8 t_1 t_2^4 t_3^5 t_4^2 t_5^4 t_6 t_7^4 t_8^2 -  y_{s}^4 y_{o_1}^8 y_{o_2}^8 y_{o_3}^{10} y_{o_4}^6 y_{o_5}^6 y_{o_6}^{10} y_{o_7}^{10} t_1^5 t_2 t_3^6 t_4^4 t_5^6 t_6^2 t_7^4 t_8^2 -  2 y_{s}^4 y_{o_1}^8 y_{o_2}^8 y_{o_3}^{10} y_{o_4}^6 y_{o_5}^6 y_{o_6}^{10} y_{o_7}^{10} t_1^4 t_2^2 t_3^6 t_4^4 t_5^6 t_6^2 t_7^4 t_8^2 -  3 y_{s}^4 y_{o_1}^8 y_{o_2}^8 y_{o_3}^{10} y_{o_4}^6 y_{o_5}^6 y_{o_6}^{10} y_{o_7}^{10} t_1^3 t_2^3 t_3^6 t_4^4 t_5^6 t_6^2 t_7^4 t_8^2 -  2 y_{s}^4 y_{o_1}^8 y_{o_2}^8 y_{o_3}^{10} y_{o_4}^6 y_{o_5}^6 y_{o_6}^{10} y_{o_7}^{10} t_1^2 t_2^4 t_3^6 t_4^4 t_5^6 t_6^2 t_7^4 t_8^2 -  y_{s}^4 y_{o_1}^8 y_{o_2}^8 y_{o_3}^{10} y_{o_4}^6 y_{o_5}^6 y_{o_6}^{10} y_{o_7}^{10} t_1 t_2^5 t_3^6 t_4^4 t_5^6 t_6^2 t_7^4 t_8^2 +  y_{s}^5 y_{o_1}^9 y_{o_2}^{11} y_{o_3}^{12} y_{o_4}^8 y_{o_5}^7 y_{o_6}^{13} y_{o_7}^{12} t_1^5 t_2^2 t_3^7 t_4^6 t_5^8 t_6^3 t_7^4 t_8^2 +  y_{s}^5 y_{o_1}^9 y_{o_2}^{11} y_{o_3}^{12} y_{o_4}^8 y_{o_5}^7 y_{o_6}^{13} y_{o_7}^{12} t_1^4 t_2^3 t_3^7 t_4^6 t_5^8 t_6^3 t_7^4 t_8^2 +  y_{s}^5 y_{o_1}^9 y_{o_2}^{11} y_{o_3}^{12} y_{o_4}^8 y_{o_5}^7 y_{o_6}^{13} y_{o_7}^{12} t_1^3 t_2^4 t_3^7 t_4^6 t_5^8 t_6^3 t_7^4 t_8^2 +  y_{s}^5 y_{o_1}^9 y_{o_2}^{11} y_{o_3}^{12} y_{o_4}^8 y_{o_5}^7 y_{o_6}^{13} y_{o_7}^{12} t_1^2 t_2^5 t_3^7 t_4^6 t_5^8 t_6^3 t_7^4 t_8^2 -  y_{s}^4 y_{o_1}^9 y_{o_2}^7 y_{o_3}^{11} y_{o_4}^5 y_{o_5}^6 y_{o_6}^{10} y_{o_7}^{10} t_1^3 t_2^3 t_3^7 t_4^3 t_5^6 t_6 t_7^5 t_8^2 -  y_{s}^2 y_{o_1}^3 y_{o_2}^5 y_{o_3}^2 y_{o_4}^6 y_{o_5}^5 y_{o_6}^3 y_{o_7}^7 t_1^4 t_2 t_4^3 t_5 t_6^4 t_7 t_8^3 -  y_{s}^2 y_{o_1}^3 y_{o_2}^5 y_{o_3}^2 y_{o_4}^6 y_{o_5}^5 y_{o_6}^3 y_{o_7}^7 t_1^3 t_2^2 t_4^3 t_5 t_6^4 t_7 t_8^3 -  y_{s}^2 y_{o_1}^3 y_{o_2}^5 y_{o_3}^2 y_{o_4}^6 y_{o_5}^5 y_{o_6}^3 y_{o_7}^7 t_1^2 t_2^3 t_4^3 t_5 t_6^4 t_7 t_8^3 -  y_{s}^2 y_{o_1}^3 y_{o_2}^5 y_{o_3}^2 y_{o_4}^6 y_{o_5}^5 y_{o_6}^3 y_{o_7}^7 t_1 t_2^4 t_4^3 t_5 t_6^4 t_7 t_8^3 +  y_{s}^3 y_{o_1}^4 y_{o_2}^8 y_{o_3}^4 y_{o_4}^8 y_{o_5}^6 y_{o_6}^6 y_{o_7}^9 t_1^5 t_2 t_3 t_4^5 t_5^3 t_6^5 t_7 t_8^3 +  2 y_{s}^3 y_{o_1}^4 y_{o_2}^8 y_{o_3}^4 y_{o_4}^8 y_{o_5}^6 y_{o_6}^6 y_{o_7}^9 t_1^4 t_2^2 t_3 t_4^5 t_5^3 t_6^5 t_7 t_8^3 +  2 y_{s}^3 y_{o_1}^4 y_{o_2}^8 y_{o_3}^4 y_{o_4}^8 y_{o_5}^6 y_{o_6}^6 y_{o_7}^9 t_1^3 t_2^3 t_3 t_4^5 t_5^3 t_6^5 t_7 t_8^3 +  2 y_{s}^3 y_{o_1}^4 y_{o_2}^8 y_{o_3}^4 y_{o_4}^8 y_{o_5}^6 y_{o_6}^6 y_{o_7}^9 t_1^2 t_2^4 t_3 t_4^5 t_5^3 t_6^5 t_7 t_8^3 +  y_{s}^3 y_{o_1}^4 y_{o_2}^8 y_{o_3}^4 y_{o_4}^8 y_{o_5}^6 y_{o_6}^6 y_{o_7}^9 t_1 t_2^5 t_3 t_4^5 t_5^3 t_6^5 t_7 t_8^3 -  y_{s}^4 y_{o_1}^5 y_{o_2}^{11} y_{o_3}^6 y_{o_4}^{10} y_{o_5}^7 y_{o_6}^9 y_{o_7}^{11} t_1^4 t_2^3 t_3^2 t_4^7 t_5^5 t_6^6 t_7 t_8^3 -  y_{s}^4 y_{o_1}^5 y_{o_2}^{11} y_{o_3}^6 y_{o_4}^{10} y_{o_5}^7 y_{o_6}^9 y_{o_7}^{11} t_1^3 t_2^4 t_3^2 t_4^7 t_5^5 t_6^6 t_7 t_8^3 -  y_{s}^2 y_{o_1}^4 y_{o_2}^4 y_{o_3}^3 y_{o_4}^5 y_{o_5}^5 y_{o_6}^3 y_{o_7}^7 t_1^5 t_3 t_4^2 t_5 t_6^3 t_7^2 t_8^3 -  2 y_{s}^2 y_{o_1}^4 y_{o_2}^4 y_{o_3}^3 y_{o_4}^5 y_{o_5}^5 y_{o_6}^3 y_{o_7}^7 t_1^4 t_2 t_3 t_4^2 t_5 t_6^3 t_7^2 t_8^3 -  2 y_{s}^2 y_{o_1}^4 y_{o_2}^4 y_{o_3}^3 y_{o_4}^5 y_{o_5}^5 y_{o_6}^3 y_{o_7}^7 t_1^3 t_2^2 t_3 t_4^2 t_5 t_6^3 t_7^2 t_8^3 -  2 y_{s}^2 y_{o_1}^4 y_{o_2}^4 y_{o_3}^3 y_{o_4}^5 y_{o_5}^5 y_{o_6}^3 y_{o_7}^7 t_1^2 t_2^3 t_3 t_4^2 t_5 t_6^3 t_7^2 t_8^3 -  2 y_{s}^2 y_{o_1}^4 y_{o_2}^4 y_{o_3}^3 y_{o_4}^5 y_{o_5}^5 y_{o_6}^3 y_{o_7}^7 t_1 t_2^4 t_3 t_4^2 t_5 t_6^3 t_7^2 t_8^3 -  y_{s}^2 y_{o_1}^4 y_{o_2}^4 y_{o_3}^3 y_{o_4}^5 y_{o_5}^5 y_{o_6}^3 y_{o_7}^7 t_2^5 t_3 t_4^2 t_5 t_6^3 t_7^2 t_8^3 +  y_{s}^3 y_{o_1}^5 y_{o_2}^7 y_{o_3}^5 y_{o_4}^7 y_{o_5}^6 y_{o_6}^6 y_{o_7}^9 t_1^6 t_3^2 t_4^4 t_5^3 t_6^4 t_7^2 t_8^3 +  4 y_{s}^3 y_{o_1}^5 y_{o_2}^7 y_{o_3}^5 y_{o_4}^7 y_{o_5}^6 y_{o_6}^6 y_{o_7}^9 t_1^5 t_2 t_3^2 t_4^4 t_5^3 t_6^4 t_7^2 t_8^3 +  6 y_{s}^3 y_{o_1}^5 y_{o_2}^7 y_{o_3}^5 y_{o_4}^7 y_{o_5}^6 y_{o_6}^6 y_{o_7}^9 t_1^4 t_2^2 t_3^2 t_4^4 t_5^3 t_6^4 t_7^2 t_8^3 +  6 y_{s}^3 y_{o_1}^5 y_{o_2}^7 y_{o_3}^5 y_{o_4}^7 y_{o_5}^6 y_{o_6}^6 y_{o_7}^9 t_1^3 t_2^3 t_3^2 t_4^4 t_5^3 t_6^4 t_7^2 t_8^3 +  6 y_{s}^3 y_{o_1}^5 y_{o_2}^7 y_{o_3}^5 y_{o_4}^7 y_{o_5}^6 y_{o_6}^6 y_{o_7}^9 t_1^2 t_2^4 t_3^2 t_4^4 t_5^3 t_6^4 t_7^2 t_8^3 +  4 y_{s}^3 y_{o_1}^5 y_{o_2}^7 y_{o_3}^5 y_{o_4}^7 y_{o_5}^6 y_{o_6}^6 y_{o_7}^9 t_1 t_2^5 t_3^2 t_4^4 t_5^3 t_6^4 t_7^2 t_8^3 +  y_{s}^3 y_{o_1}^5 y_{o_2}^7 y_{o_3}^5 y_{o_4}^7 y_{o_5}^6 y_{o_6}^6 y_{o_7}^9 t_2^6 t_3^2 t_4^4 t_5^3 t_6^4 t_7^2 t_8^3 -  2 y_{s}^4 y_{o_1}^6 y_{o_2}^{10} y_{o_3}^7 y_{o_4}^9 y_{o_5}^7 y_{o_6}^9 y_{o_7}^{11} t_1^6 t_2 t_3^3 t_4^6 t_5^5 t_6^5 t_7^2 t_8^3 -  3 y_{s}^4 y_{o_1}^6 y_{o_2}^{10} y_{o_3}^7 y_{o_4}^9 y_{o_5}^7 y_{o_6}^9 y_{o_7}^{11} t_1^5 t_2^2 t_3^3 t_4^6 t_5^5 t_6^5 t_7^2 t_8^3 -  6 y_{s}^4 y_{o_1}^6 y_{o_2}^{10} y_{o_3}^7 y_{o_4}^9 y_{o_5}^7 y_{o_6}^9 y_{o_7}^{11} t_1^4 t_2^3 t_3^3 t_4^6 t_5^5 t_6^5 t_7^2 t_8^3 -  6 y_{s}^4 y_{o_1}^6 y_{o_2}^{10} y_{o_3}^7 y_{o_4}^9 y_{o_5}^7 y_{o_6}^9 y_{o_7}^{11} t_1^3 t_2^4 t_3^3 t_4^6 t_5^5 t_6^5 t_7^2 t_8^3 -  3 y_{s}^4 y_{o_1}^6 y_{o_2}^{10} y_{o_3}^7 y_{o_4}^9 y_{o_5}^7 y_{o_6}^9 y_{o_7}^{11} t_1^2 t_2^5 t_3^3 t_4^6 t_5^5 t_6^5 t_7^2 t_8^3 -  2 y_{s}^4 y_{o_1}^6 y_{o_2}^{10} y_{o_3}^7 y_{o_4}^9 y_{o_5}^7 y_{o_6}^9 y_{o_7}^{11} t_1 t_2^6 t_3^3 t_4^6 t_5^5 t_6^5 t_7^2 t_8^3 +  y_{s}^5 y_{o_1}^7 y_{o_2}^{13} y_{o_3}^9 y_{o_4}^{11} y_{o_5}^8 y_{o_6}^{12} y_{o_7}^{13} t_1^5 t_2^3 t_3^4 t_4^8 t_5^7 t_6^6 t_7^2 t_8^3 +  2 y_{s}^5 y_{o_1}^7 y_{o_2}^{13} y_{o_3}^9 y_{o_4}^{11} y_{o_5}^8 y_{o_6}^{12} y_{o_7}^{13} t_1^4 t_2^4 t_3^4 t_4^8 t_5^7 t_6^6 t_7^2 t_8^3 +  y_{s}^5 y_{o_1}^7 y_{o_2}^{13} y_{o_3}^9 y_{o_4}^{11} y_{o_5}^8 y_{o_6}^{12} y_{o_7}^{13} t_1^3 t_2^5 t_3^4 t_4^8 t_5^7 t_6^6 t_7^2 t_8^3 -  y_{s}^2 y_{o_1}^5 y_{o_2}^3 y_{o_3}^4 y_{o_4}^4 y_{o_5}^5 y_{o_6}^3 y_{o_7}^7 t_1^5 t_3^2 t_4 t_5 t_6^2 t_7^3 t_8^3 -  2 y_{s}^2 y_{o_1}^5 y_{o_2}^3 y_{o_3}^4 y_{o_4}^4 y_{o_5}^5 y_{o_6}^3 y_{o_7}^7 t_1^4 t_2 t_3^2 t_4 t_5 t_6^2 t_7^3 t_8^3 -  2 y_{s}^2 y_{o_1}^5 y_{o_2}^3 y_{o_3}^4 y_{o_4}^4 y_{o_5}^5 y_{o_6}^3 y_{o_7}^7 t_1^3 t_2^2 t_3^2 t_4 t_5 t_6^2 t_7^3 t_8^3 -  2 y_{s}^2 y_{o_1}^5 y_{o_2}^3 y_{o_3}^4 y_{o_4}^4 y_{o_5}^5 y_{o_6}^3 y_{o_7}^7 t_1^2 t_2^3 t_3^2 t_4 t_5 t_6^2 t_7^3 t_8^3 -  2 y_{s}^2 y_{o_1}^5 y_{o_2}^3 y_{o_3}^4 y_{o_4}^4 y_{o_5}^5 y_{o_6}^3 y_{o_7}^7 t_1 t_2^4 t_3^2 t_4 t_5 t_6^2 t_7^3 t_8^3 -  y_{s}^2 y_{o_1}^5 y_{o_2}^3 y_{o_3}^4 y_{o_4}^4 y_{o_5}^5 y_{o_6}^3 y_{o_7}^7 t_2^5 t_3^2 t_4 t_5 t_6^2 t_7^3 t_8^3 +  2 y_{s}^3 y_{o_1}^6 y_{o_2}^6 y_{o_3}^6 y_{o_4}^6 y_{o_5}^6 y_{o_6}^6 y_{o_7}^9 t_1^6 t_3^3 t_4^3 t_5^3 t_6^3 t_7^3 t_8^3 +  6 y_{s}^3 y_{o_1}^6 y_{o_2}^6 y_{o_3}^6 y_{o_4}^6 y_{o_5}^6 y_{o_6}^6 y_{o_7}^9 t_1^5 t_2 t_3^3 t_4^3 t_5^3 t_6^3 t_7^3 t_8^3 +  8 y_{s}^3 y_{o_1}^6 y_{o_2}^6 y_{o_3}^6 y_{o_4}^6 y_{o_5}^6 y_{o_6}^6 y_{o_7}^9 t_1^4 t_2^2 t_3^3 t_4^3 t_5^3 t_6^3 t_7^3 t_8^3 +  7 y_{s}^3 y_{o_1}^6 y_{o_2}^6 y_{o_3}^6 y_{o_4}^6 y_{o_5}^6 y_{o_6}^6 y_{o_7}^9 t_1^3 t_2^3 t_3^3 t_4^3 t_5^3 t_6^3 t_7^3 t_8^3 +  8 y_{s}^3 y_{o_1}^6 y_{o_2}^6 y_{o_3}^6 y_{o_4}^6 y_{o_5}^6 y_{o_6}^6 y_{o_7}^9 t_1^2 t_2^4 t_3^3 t_4^3 t_5^3 t_6^3 t_7^3 t_8^3 +  6 y_{s}^3 y_{o_1}^6 y_{o_2}^6 y_{o_3}^6 y_{o_4}^6 y_{o_5}^6 y_{o_6}^6 y_{o_7}^9 t_1 t_2^5 t_3^3 t_4^3 t_5^3 t_6^3 t_7^3 t_8^3 +  2 y_{s}^3 y_{o_1}^6 y_{o_2}^6 y_{o_3}^6 y_{o_4}^6 y_{o_5}^6 y_{o_6}^6 y_{o_7}^9 t_2^6 t_3^3 t_4^3 t_5^3 t_6^3 t_7^3 t_8^3 -  4 y_{s}^4 y_{o_1}^7 y_{o_2}^9 y_{o_3}^8 y_{o_4}^8 y_{o_5}^7 y_{o_6}^9 y_{o_7}^{11} t_1^6 t_2 t_3^4 t_4^5 t_5^5 t_6^4 t_7^3 t_8^3 -  6 y_{s}^4 y_{o_1}^7 y_{o_2}^9 y_{o_3}^8 y_{o_4}^8 y_{o_5}^7 y_{o_6}^9 y_{o_7}^{11} t_1^5 t_2^2 t_3^4 t_4^5 t_5^5 t_6^4 t_7^3 t_8^3 -  8 y_{s}^4 y_{o_1}^7 y_{o_2}^9 y_{o_3}^8 y_{o_4}^8 y_{o_5}^7 y_{o_6}^9 y_{o_7}^{11} t_1^4 t_2^3 t_3^4 t_4^5 t_5^5 t_6^4 t_7^3 t_8^3 -  8 y_{s}^4 y_{o_1}^7 y_{o_2}^9 y_{o_3}^8 y_{o_4}^8 y_{o_5}^7 y_{o_6}^9 y_{o_7}^{11} t_1^3 t_2^4 t_3^4 t_4^5 t_5^5 t_6^4 t_7^3 t_8^3 -  6 y_{s}^4 y_{o_1}^7 y_{o_2}^9 y_{o_3}^8 y_{o_4}^8 y_{o_5}^7 y_{o_6}^9 y_{o_7}^{11} t_1^2 t_2^5 t_3^4 t_4^5 t_5^5 t_6^4 t_7^3 t_8^3 -  4 y_{s}^4 y_{o_1}^7 y_{o_2}^9 y_{o_3}^8 y_{o_4}^8 y_{o_5}^7 y_{o_6}^9 y_{o_7}^{11} t_1 t_2^6 t_3^4 t_4^5 t_5^5 t_6^4 t_7^3 t_8^3 +  y_{s}^5 y_{o_1}^8 y_{o_2}^{12} y_{o_3}^{10} y_{o_4}^{10} y_{o_5}^8 y_{o_6}^{12} y_{o_7}^{13} t_1^6 t_2^2 t_3^5 t_4^7 t_5^7 t_6^5 t_7^3 t_8^3 +  3 y_{s}^5 y_{o_1}^8 y_{o_2}^{12} y_{o_3}^{10} y_{o_4}^{10} y_{o_5}^8 y_{o_6}^{12} y_{o_7}^{13} t_1^5 t_2^3 t_3^5 t_4^7 t_5^7 t_6^5 t_7^3 t_8^3 +  3 y_{s}^5 y_{o_1}^8 y_{o_2}^{12} y_{o_3}^{10} y_{o_4}^{10} y_{o_5}^8 y_{o_6}^{12} y_{o_7}^{13} t_1^4 t_2^4 t_3^5 t_4^7 t_5^7 t_6^5 t_7^3 t_8^3 +  3 y_{s}^5 y_{o_1}^8 y_{o_2}^{12} y_{o_3}^{10} y_{o_4}^{10} y_{o_5}^8 y_{o_6}^{12} y_{o_7}^{13} t_1^3 t_2^5 t_3^5 t_4^7 t_5^7 t_6^5 t_7^3 t_8^3 +  y_{s}^5 y_{o_1}^8 y_{o_2}^{12} y_{o_3}^{10} y_{o_4}^{10} y_{o_5}^8 y_{o_6}^{12} y_{o_7}^{13} t_1^2 t_2^6 t_3^5 t_4^7 t_5^7 t_6^5 t_7^3 t_8^3 -  y_{s}^6 y_{o_1}^9 y_{o_2}^{15} y_{o_3}^{12} y_{o_4}^{12} y_{o_5}^9 y_{o_6}^{15} y_{o_7}^{15} t_1^5 t_2^4 t_3^6 t_4^9 t_5^9 t_6^6 t_7^3 t_8^3 -  y_{s}^6 y_{o_1}^9 y_{o_2}^{15} y_{o_3}^{12} y_{o_4}^{12} y_{o_5}^9 y_{o_6}^{15} y_{o_7}^{15} t_1^4 t_2^5 t_3^6 t_4^9 t_5^9 t_6^6 t_7^3 t_8^3 -  y_{s}^2 y_{o_1}^6 y_{o_2}^2 y_{o_3}^5 y_{o_4}^3 y_{o_5}^5 y_{o_6}^3 y_{o_7}^7 t_1^4 t_2 t_3^3 t_5 t_6 t_7^4 t_8^3 -  y_{s}^2 y_{o_1}^6 y_{o_2}^2 y_{o_3}^5 y_{o_4}^3 y_{o_5}^5 y_{o_6}^3 y_{o_7}^7 t_1^3 t_2^2 t_3^3 t_5 t_6 t_7^4 t_8^3 -  y_{s}^2 y_{o_1}^6 y_{o_2}^2 y_{o_3}^5 y_{o_4}^3 y_{o_5}^5 y_{o_6}^3 y_{o_7}^7 t_1^2 t_2^3 t_3^3 t_5 t_6 t_7^4 t_8^3 -  y_{s}^2 y_{o_1}^6 y_{o_2}^2 y_{o_3}^5 y_{o_4}^3 y_{o_5}^5 y_{o_6}^3 y_{o_7}^7 t_1 t_2^4 t_3^3 t_5 t_6 t_7^4 t_8^3 +  y_{s}^3 y_{o_1}^7 y_{o_2}^5 y_{o_3}^7 y_{o_4}^5 y_{o_5}^6 y_{o_6}^6 y_{o_7}^9 t_1^6 t_3^4 t_4^2 t_5^3 t_6^2 t_7^4 t_8^3 +  4 y_{s}^3 y_{o_1}^7 y_{o_2}^5 y_{o_3}^7 y_{o_4}^5 y_{o_5}^6 y_{o_6}^6 y_{o_7}^9 t_1^5 t_2 t_3^4 t_4^2 t_5^3 t_6^2 t_7^4 t_8^3 +  6 y_{s}^3 y_{o_1}^7 y_{o_2}^5 y_{o_3}^7 y_{o_4}^5 y_{o_5}^6 y_{o_6}^6 y_{o_7}^9 t_1^4 t_2^2 t_3^4 t_4^2 t_5^3 t_6^2 t_7^4 t_8^3 +  6 y_{s}^3 y_{o_1}^7 y_{o_2}^5 y_{o_3}^7 y_{o_4}^5 y_{o_5}^6 y_{o_6}^6 y_{o_7}^9 t_1^3 t_2^3 t_3^4 t_4^2 t_5^3 t_6^2 t_7^4 t_8^3 +  6 y_{s}^3 y_{o_1}^7 y_{o_2}^5 y_{o_3}^7 y_{o_4}^5 y_{o_5}^6 y_{o_6}^6 y_{o_7}^9 t_1^2 t_2^4 t_3^4 t_4^2 t_5^3 t_6^2 t_7^4 t_8^3 +  4 y_{s}^3 y_{o_1}^7 y_{o_2}^5 y_{o_3}^7 y_{o_4}^5 y_{o_5}^6 y_{o_6}^6 y_{o_7}^9 t_1 t_2^5 t_3^4 t_4^2 t_5^3 t_6^2 t_7^4 t_8^3 +  y_{s}^3 y_{o_1}^7 y_{o_2}^5 y_{o_3}^7 y_{o_4}^5 y_{o_5}^6 y_{o_6}^6 y_{o_7}^9 t_2^6 t_3^4 t_4^2 t_5^3 t_6^2 t_7^4 t_8^3 -  4 y_{s}^4 y_{o_1}^8 y_{o_2}^8 y_{o_3}^9 y_{o_4}^7 y_{o_5}^7 y_{o_6}^9 y_{o_7}^{11} t_1^6 t_2 t_3^5 t_4^4 t_5^5 t_6^3 t_7^4 t_8^3 -  6 y_{s}^4 y_{o_1}^8 y_{o_2}^8 y_{o_3}^9 y_{o_4}^7 y_{o_5}^7 y_{o_6}^9 y_{o_7}^{11} t_1^5 t_2^2 t_3^5 t_4^4 t_5^5 t_6^3 t_7^4 t_8^3 -  8 y_{s}^4 y_{o_1}^8 y_{o_2}^8 y_{o_3}^9 y_{o_4}^7 y_{o_5}^7 y_{o_6}^9 y_{o_7}^{11} t_1^4 t_2^3 t_3^5 t_4^4 t_5^5 t_6^3 t_7^4 t_8^3 -  8 y_{s}^4 y_{o_1}^8 y_{o_2}^8 y_{o_3}^9 y_{o_4}^7 y_{o_5}^7 y_{o_6}^9 y_{o_7}^{11} t_1^3 t_2^4 t_3^5 t_4^4 t_5^5 t_6^3 t_7^4 t_8^3 -  6 y_{s}^4 y_{o_1}^8 y_{o_2}^8 y_{o_3}^9 y_{o_4}^7 y_{o_5}^7 y_{o_6}^9 y_{o_7}^{11} t_1^2 t_2^5 t_3^5 t_4^4 t_5^5 t_6^3 t_7^4 t_8^3 -  4 y_{s}^4 y_{o_1}^8 y_{o_2}^8 y_{o_3}^9 y_{o_4}^7 y_{o_5}^7 y_{o_6}^9 y_{o_7}^{11} t_1 t_2^6 t_3^5 t_4^4 t_5^5 t_6^3 t_7^4 t_8^3 +  2 y_{s}^5 y_{o_1}^9 y_{o_2}^{11} y_{o_3}^{11} y_{o_4}^9 y_{o_5}^8 y_{o_6}^{12} y_{o_7}^{13} t_1^6 t_2^2 t_3^6 t_4^6 t_5^7 t_6^4 t_7^4 t_8^3 +  3 y_{s}^5 y_{o_1}^9 y_{o_2}^{11} y_{o_3}^{11} y_{o_4}^9 y_{o_5}^8 y_{o_6}^{12} y_{o_7}^{13} t_1^5 t_2^3 t_3^6 t_4^6 t_5^7 t_6^4 t_7^4 t_8^3 +  y_{s}^5 y_{o_1}^9 y_{o_2}^{11} y_{o_3}^{11} y_{o_4}^9 y_{o_5}^8 y_{o_6}^{12} y_{o_7}^{13} t_1^4 t_2^4 t_3^6 t_4^6 t_5^7 t_6^4 t_7^4 t_8^3 +  3 y_{s}^5 y_{o_1}^9 y_{o_2}^{11} y_{o_3}^{11} y_{o_4}^9 y_{o_5}^8 y_{o_6}^{12} y_{o_7}^{13} t_1^3 t_2^5 t_3^6 t_4^6 t_5^7 t_6^4 t_7^4 t_8^3 +  2 y_{s}^5 y_{o_1}^9 y_{o_2}^{11} y_{o_3}^{11} y_{o_4}^9 y_{o_5}^8 y_{o_6}^{12} y_{o_7}^{13} t_1^2 t_2^6 t_3^6 t_4^6 t_5^7 t_6^4 t_7^4 t_8^3 +  y_{s}^3 y_{o_1}^8 y_{o_2}^4 y_{o_3}^8 y_{o_4}^4 y_{o_5}^6 y_{o_6}^6 y_{o_7}^9 t_1^5 t_2 t_3^5 t_4 t_5^3 t_6 t_7^5 t_8^3 +  2 y_{s}^3 y_{o_1}^8 y_{o_2}^4 y_{o_3}^8 y_{o_4}^4 y_{o_5}^6 y_{o_6}^6 y_{o_7}^9 t_1^4 t_2^2 t_3^5 t_4 t_5^3 t_6 t_7^5 t_8^3 +  2 y_{s}^3 y_{o_1}^8 y_{o_2}^4 y_{o_3}^8 y_{o_4}^4 y_{o_5}^6 y_{o_6}^6 y_{o_7}^9 t_1^3 t_2^3 t_3^5 t_4 t_5^3 t_6 t_7^5 t_8^3 +  2 y_{s}^3 y_{o_1}^8 y_{o_2}^4 y_{o_3}^8 y_{o_4}^4 y_{o_5}^6 y_{o_6}^6 y_{o_7}^9 t_1^2 t_2^4 t_3^5 t_4 t_5^3 t_6 t_7^5 t_8^3 +  y_{s}^3 y_{o_1}^8 y_{o_2}^4 y_{o_3}^8 y_{o_4}^4 y_{o_5}^6 y_{o_6}^6 y_{o_7}^9 t_1 t_2^5 t_3^5 t_4 t_5^3 t_6 t_7^5 t_8^3 -  2 y_{s}^4 y_{o_1}^9 y_{o_2}^7 y_{o_3}^{10} y_{o_4}^6 y_{o_5}^7 y_{o_6}^9 y_{o_7}^{11} t_1^6 t_2 t_3^6 t_4^3 t_5^5 t_6^2 t_7^5 t_8^3 -  3 y_{s}^4 y_{o_1}^9 y_{o_2}^7 y_{o_3}^{10} y_{o_4}^6 y_{o_5}^7 y_{o_6}^9 y_{o_7}^{11} t_1^5 t_2^2 t_3^6 t_4^3 t_5^5 t_6^2 t_7^5 t_8^3 -  6 y_{s}^4 y_{o_1}^9 y_{o_2}^7 y_{o_3}^{10} y_{o_4}^6 y_{o_5}^7 y_{o_6}^9 y_{o_7}^{11} t_1^4 t_2^3 t_3^6 t_4^3 t_5^5 t_6^2 t_7^5 t_8^3 -  6 y_{s}^4 y_{o_1}^9 y_{o_2}^7 y_{o_3}^{10} y_{o_4}^6 y_{o_5}^7 y_{o_6}^9 y_{o_7}^{11} t_1^3 t_2^4 t_3^6 t_4^3 t_5^5 t_6^2 t_7^5 t_8^3 -  3 y_{s}^4 y_{o_1}^9 y_{o_2}^7 y_{o_3}^{10} y_{o_4}^6 y_{o_5}^7 y_{o_6}^9 y_{o_7}^{11} t_1^2 t_2^5 t_3^6 t_4^3 t_5^5 t_6^2 t_7^5 t_8^3 -  2 y_{s}^4 y_{o_1}^9 y_{o_2}^7 y_{o_3}^{10} y_{o_4}^6 y_{o_5}^7 y_{o_6}^9 y_{o_7}^{11} t_1 t_2^6 t_3^6 t_4^3 t_5^5 t_6^2 t_7^5 t_8^3 +  y_{s}^5 y_{o_1}^{10} y_{o_2}^{10} y_{o_3}^{12} y_{o_4}^8 y_{o_5}^8 y_{o_6}^{12} y_{o_7}^{13} t_1^6 t_2^2 t_3^7 t_4^5 t_5^7 t_6^3 t_7^5 t_8^3 +  3 y_{s}^5 y_{o_1}^{10} y_{o_2}^{10} y_{o_3}^{12} y_{o_4}^8 y_{o_5}^8 y_{o_6}^{12} y_{o_7}^{13} t_1^5 t_2^3 t_3^7 t_4^5 t_5^7 t_6^3 t_7^5 t_8^3 +  3 y_{s}^5 y_{o_1}^{10} y_{o_2}^{10} y_{o_3}^{12} y_{o_4}^8 y_{o_5}^8 y_{o_6}^{12} y_{o_7}^{13} t_1^4 t_2^4 t_3^7 t_4^5 t_5^7 t_6^3 t_7^5 t_8^3 +  3 y_{s}^5 y_{o_1}^{10} y_{o_2}^{10} y_{o_3}^{12} y_{o_4}^8 y_{o_5}^8 y_{o_6}^{12} y_{o_7}^{13} t_1^3 t_2^5 t_3^7 t_4^5 t_5^7 t_6^3 t_7^5 t_8^3 +  y_{s}^5 y_{o_1}^{10} y_{o_2}^{10} y_{o_3}^{12} y_{o_4}^8 y_{o_5}^8 y_{o_6}^{12} y_{o_7}^{13} t_1^2 t_2^6 t_3^7 t_4^5 t_5^7 t_6^3 t_7^5 t_8^3 -  y_{s}^4 y_{o_1}^{10} y_{o_2}^6 y_{o_3}^{11} y_{o_4}^5 y_{o_5}^7 y_{o_6}^9 y_{o_7}^{11} t_1^4 t_2^3 t_3^7 t_4^2 t_5^5 t_6 t_7^6 t_8^3 -  y_{s}^4 y_{o_1}^{10} y_{o_2}^6 y_{o_3}^{11} y_{o_4}^5 y_{o_5}^7 y_{o_6}^9 y_{o_7}^{11} t_1^3 t_2^4 t_3^7 t_4^2 t_5^5 t_6 t_7^6 t_8^3 +  y_{s}^5 y_{o_1}^{11} y_{o_2}^9 y_{o_3}^{13} y_{o_4}^7 y_{o_5}^8 y_{o_6}^{12} y_{o_7}^{13} t_1^5 t_2^3 t_3^8 t_4^4 t_5^7 t_6^2 t_7^6 t_8^3 +  2 y_{s}^5 y_{o_1}^{11} y_{o_2}^9 y_{o_3}^{13} y_{o_4}^7 y_{o_5}^8 y_{o_6}^{12} y_{o_7}^{13} t_1^4 t_2^4 t_3^8 t_4^4 t_5^7 t_6^2 t_7^6 t_8^3 +  y_{s}^5 y_{o_1}^{11} y_{o_2}^9 y_{o_3}^{13} y_{o_4}^7 y_{o_5}^8 y_{o_6}^{12} y_{o_7}^{13} t_1^3 t_2^5 t_3^8 t_4^4 t_5^7 t_6^2 t_7^6 t_8^3 -  y_{s}^6 y_{o_1}^{12} y_{o_2}^{12} y_{o_3}^{15} y_{o_4}^9 y_{o_5}^9 y_{o_6}^{15} y_{o_7}^{15} t_1^5 t_2^4 t_3^9 t_4^6 t_5^9 t_6^3 t_7^6 t_8^3 -  y_{s}^6 y_{o_1}^{12} y_{o_2}^{12} y_{o_3}^{15} y_{o_4}^9 y_{o_5}^9 y_{o_6}^{15} y_{o_7}^{15} t_1^4 t_2^5 t_3^9 t_4^6 t_5^9 t_6^3 t_7^6 t_8^3 +  y_{s}^3 y_{o_1}^5 y_{o_2}^7 y_{o_3}^4 y_{o_4}^8 y_{o_5}^7 y_{o_6}^5 y_{o_7}^{10} t_1^5 t_2^2 t_3 t_4^4 t_5^2 t_6^5 t_7^2 t_8^4 +  y_{s}^3 y_{o_1}^5 y_{o_2}^7 y_{o_3}^4 y_{o_4}^8 y_{o_5}^7 y_{o_6}^5 y_{o_7}^{10} t_1^4 t_2^3 t_3 t_4^4 t_5^2 t_6^5 t_7^2 t_8^4 +  y_{s}^3 y_{o_1}^5 y_{o_2}^7 y_{o_3}^4 y_{o_4}^8 y_{o_5}^7 y_{o_6}^5 y_{o_7}^{10} t_1^3 t_2^4 t_3 t_4^4 t_5^2 t_6^5 t_7^2 t_8^4 +  y_{s}^3 y_{o_1}^5 y_{o_2}^7 y_{o_3}^4 y_{o_4}^8 y_{o_5}^7 y_{o_6}^5 y_{o_7}^{10} t_1^2 t_2^5 t_3 t_4^4 t_5^2 t_6^5 t_7^2 t_8^4 -  y_{s}^4 y_{o_1}^6 y_{o_2}^{10} y_{o_3}^6 y_{o_4}^{10} y_{o_5}^8 y_{o_6}^8 y_{o_7}^{12} t_1^6 t_2^2 t_3^2 t_4^6 t_5^4 t_6^6 t_7^2 t_8^4 -  y_{s}^4 y_{o_1}^6 y_{o_2}^{10} y_{o_3}^6 y_{o_4}^{10} y_{o_5}^8 y_{o_6}^8 y_{o_7}^{12} t_1^5 t_2^3 t_3^2 t_4^6 t_5^4 t_6^6 t_7^2 t_8^4 -  y_{s}^4 y_{o_1}^6 y_{o_2}^{10} y_{o_3}^6 y_{o_4}^{10} y_{o_5}^8 y_{o_6}^8 y_{o_7}^{12} t_1^4 t_2^4 t_3^2 t_4^6 t_5^4 t_6^6 t_7^2 t_8^4 -  y_{s}^4 y_{o_1}^6 y_{o_2}^{10} y_{o_3}^6 y_{o_4}^{10} y_{o_5}^8 y_{o_6}^8 y_{o_7}^{12} t_1^3 t_2^5 t_3^2 t_4^6 t_5^4 t_6^6 t_7^2 t_8^4 -  y_{s}^4 y_{o_1}^6 y_{o_2}^{10} y_{o_3}^6 y_{o_4}^{10} y_{o_5}^8 y_{o_6}^8 y_{o_7}^{12} t_1^2 t_2^6 t_3^2 t_4^6 t_5^4 t_6^6 t_7^2 t_8^4 +  y_{s}^5 y_{o_1}^7 y_{o_2}^{13} y_{o_3}^8 y_{o_4}^{12} y_{o_5}^9 y_{o_6}^{11} y_{o_7}^{14} t_1^6 t_2^3 t_3^3 t_4^8 t_5^6 t_6^7 t_7^2 t_8^4 +  y_{s}^5 y_{o_1}^7 y_{o_2}^{13} y_{o_3}^8 y_{o_4}^{12} y_{o_5}^9 y_{o_6}^{11} y_{o_7}^{14} t_1^3 t_2^6 t_3^3 t_4^8 t_5^6 t_6^7 t_7^2 t_8^4 -  y_{s}^2 y_{o_1}^5 y_{o_2}^3 y_{o_3}^3 y_{o_4}^5 y_{o_5}^6 y_{o_6}^2 y_{o_7}^8 t_1^5 t_2 t_3 t_4 t_6^3 t_7^3 t_8^4 -  y_{s}^2 y_{o_1}^5 y_{o_2}^3 y_{o_3}^3 y_{o_4}^5 y_{o_5}^6 y_{o_6}^2 y_{o_7}^8 t_1^4 t_2^2 t_3 t_4 t_6^3 t_7^3 t_8^4 -  y_{s}^2 y_{o_1}^5 y_{o_2}^3 y_{o_3}^3 y_{o_4}^5 y_{o_5}^6 y_{o_6}^2 y_{o_7}^8 t_1^3 t_2^3 t_3 t_4 t_6^3 t_7^3 t_8^4 -  y_{s}^2 y_{o_1}^5 y_{o_2}^3 y_{o_3}^3 y_{o_4}^5 y_{o_5}^6 y_{o_6}^2 y_{o_7}^8 t_1^2 t_2^4 t_3 t_4 t_6^3 t_7^3 t_8^4 -  y_{s}^2 y_{o_1}^5 y_{o_2}^3 y_{o_3}^3 y_{o_4}^5 y_{o_5}^6 y_{o_6}^2 y_{o_7}^8 t_1 t_2^5 t_3 t_4 t_6^3 t_7^3 t_8^4 +  y_{s}^3 y_{o_1}^6 y_{o_2}^6 y_{o_3}^5 y_{o_4}^7 y_{o_5}^7 y_{o_6}^5 y_{o_7}^{10} t_1^7 t_3^2 t_4^3 t_5^2 t_6^4 t_7^3 t_8^4 +  3 y_{s}^3 y_{o_1}^6 y_{o_2}^6 y_{o_3}^5 y_{o_4}^7 y_{o_5}^7 y_{o_6}^5 y_{o_7}^{10} t_1^6 t_2 t_3^2 t_4^3 t_5^2 t_6^4 t_7^3 t_8^4 +  4 y_{s}^3 y_{o_1}^6 y_{o_2}^6 y_{o_3}^5 y_{o_4}^7 y_{o_5}^7 y_{o_6}^5 y_{o_7}^{10} t_1^5 t_2^2 t_3^2 t_4^3 t_5^2 t_6^4 t_7^3 t_8^4 +  6 y_{s}^3 y_{o_1}^6 y_{o_2}^6 y_{o_3}^5 y_{o_4}^7 y_{o_5}^7 y_{o_6}^5 y_{o_7}^{10} t_1^4 t_2^3 t_3^2 t_4^3 t_5^2 t_6^4 t_7^3 t_8^4 +  6 y_{s}^3 y_{o_1}^6 y_{o_2}^6 y_{o_3}^5 y_{o_4}^7 y_{o_5}^7 y_{o_6}^5 y_{o_7}^{10} t_1^3 t_2^4 t_3^2 t_4^3 t_5^2 t_6^4 t_7^3 t_8^4 +  4 y_{s}^3 y_{o_1}^6 y_{o_2}^6 y_{o_3}^5 y_{o_4}^7 y_{o_5}^7 y_{o_6}^5 y_{o_7}^{10} t_1^2 t_2^5 t_3^2 t_4^3 t_5^2 t_6^4 t_7^3 t_8^4 +  3 y_{s}^3 y_{o_1}^6 y_{o_2}^6 y_{o_3}^5 y_{o_4}^7 y_{o_5}^7 y_{o_6}^5 y_{o_7}^{10} t_1 t_2^6 t_3^2 t_4^3 t_5^2 t_6^4 t_7^3 t_8^4 +  y_{s}^3 y_{o_1}^6 y_{o_2}^6 y_{o_3}^5 y_{o_4}^7 y_{o_5}^7 y_{o_6}^5 y_{o_7}^{10} t_2^7 t_3^2 t_4^3 t_5^2 t_6^4 t_7^3 t_8^4 -  2 y_{s}^4 y_{o_1}^7 y_{o_2}^9 y_{o_3}^7 y_{o_4}^9 y_{o_5}^8 y_{o_6}^8 y_{o_7}^{12} t_1^7 t_2 t_3^3 t_4^5 t_5^4 t_6^5 t_7^3 t_8^4 -  4 y_{s}^4 y_{o_1}^7 y_{o_2}^9 y_{o_3}^7 y_{o_4}^9 y_{o_5}^8 y_{o_6}^8 y_{o_7}^{12} t_1^6 t_2^2 t_3^3 t_4^5 t_5^4 t_6^5 t_7^3 t_8^4 -  5 y_{s}^4 y_{o_1}^7 y_{o_2}^9 y_{o_3}^7 y_{o_4}^9 y_{o_5}^8 y_{o_6}^8 y_{o_7}^{12} t_1^5 t_2^3 t_3^3 t_4^5 t_5^4 t_6^5 t_7^3 t_8^4 -  8 y_{s}^4 y_{o_1}^7 y_{o_2}^9 y_{o_3}^7 y_{o_4}^9 y_{o_5}^8 y_{o_6}^8 y_{o_7}^{12} t_1^4 t_2^4 t_3^3 t_4^5 t_5^4 t_6^5 t_7^3 t_8^4 -  5 y_{s}^4 y_{o_1}^7 y_{o_2}^9 y_{o_3}^7 y_{o_4}^9 y_{o_5}^8 y_{o_6}^8 y_{o_7}^{12} t_1^3 t_2^5 t_3^3 t_4^5 t_5^4 t_6^5 t_7^3 t_8^4 -  4 y_{s}^4 y_{o_1}^7 y_{o_2}^9 y_{o_3}^7 y_{o_4}^9 y_{o_5}^8 y_{o_6}^8 y_{o_7}^{12} t_1^2 t_2^6 t_3^3 t_4^5 t_5^4 t_6^5 t_7^3 t_8^4 -  2 y_{s}^4 y_{o_1}^7 y_{o_2}^9 y_{o_3}^7 y_{o_4}^9 y_{o_5}^8 y_{o_6}^8 y_{o_7}^{12} t_1 t_2^7 t_3^3 t_4^5 t_5^4 t_6^5 t_7^3 t_8^4 +  2 y_{s}^5 y_{o_1}^8 y_{o_2}^{12} y_{o_3}^9 y_{o_4}^{11} y_{o_5}^9 y_{o_6}^{11} y_{o_7}^{14} t_1^6 t_2^3 t_3^4 t_4^7 t_5^6 t_6^6 t_7^3 t_8^4 +  y_{s}^5 y_{o_1}^8 y_{o_2}^{12} y_{o_3}^9 y_{o_4}^{11} y_{o_5}^9 y_{o_6}^{11} y_{o_7}^{14} t_1^5 t_2^4 t_3^4 t_4^7 t_5^6 t_6^6 t_7^3 t_8^4 +  y_{s}^5 y_{o_1}^8 y_{o_2}^{12} y_{o_3}^9 y_{o_4}^{11} y_{o_5}^9 y_{o_6}^{11} y_{o_7}^{14} t_1^4 t_2^5 t_3^4 t_4^7 t_5^6 t_6^6 t_7^3 t_8^4 +  2 y_{s}^5 y_{o_1}^8 y_{o_2}^{12} y_{o_3}^9 y_{o_4}^{11} y_{o_5}^9 y_{o_6}^{11} y_{o_7}^{14} t_1^3 t_2^6 t_3^4 t_4^7 t_5^6 t_6^6 t_7^3 t_8^4 +  y_{s}^3 y_{o_1}^7 y_{o_2}^5 y_{o_3}^6 y_{o_4}^6 y_{o_5}^7 y_{o_6}^5 y_{o_7}^{10} t_1^7 t_3^3 t_4^2 t_5^2 t_6^3 t_7^4 t_8^4 +  3 y_{s}^3 y_{o_1}^7 y_{o_2}^5 y_{o_3}^6 y_{o_4}^6 y_{o_5}^7 y_{o_6}^5 y_{o_7}^{10} t_1^6 t_2 t_3^3 t_4^2 t_5^2 t_6^3 t_7^4 t_8^4 +  4 y_{s}^3 y_{o_1}^7 y_{o_2}^5 y_{o_3}^6 y_{o_4}^6 y_{o_5}^7 y_{o_6}^5 y_{o_7}^{10} t_1^5 t_2^2 t_3^3 t_4^2 t_5^2 t_6^3 t_7^4 t_8^4 +  6 y_{s}^3 y_{o_1}^7 y_{o_2}^5 y_{o_3}^6 y_{o_4}^6 y_{o_5}^7 y_{o_6}^5 y_{o_7}^{10} t_1^4 t_2^3 t_3^3 t_4^2 t_5^2 t_6^3 t_7^4 t_8^4 +  6 y_{s}^3 y_{o_1}^7 y_{o_2}^5 y_{o_3}^6 y_{o_4}^6 y_{o_5}^7 y_{o_6}^5 y_{o_7}^{10} t_1^3 t_2^4 t_3^3 t_4^2 t_5^2 t_6^3 t_7^4 t_8^4 +  4 y_{s}^3 y_{o_1}^7 y_{o_2}^5 y_{o_3}^6 y_{o_4}^6 y_{o_5}^7 y_{o_6}^5 y_{o_7}^{10} t_1^2 t_2^5 t_3^3 t_4^2 t_5^2 t_6^3 t_7^4 t_8^4 +  3 y_{s}^3 y_{o_1}^7 y_{o_2}^5 y_{o_3}^6 y_{o_4}^6 y_{o_5}^7 y_{o_6}^5 y_{o_7}^{10} t_1 t_2^6 t_3^3 t_4^2 t_5^2 t_6^3 t_7^4 t_8^4 +  y_{s}^3 y_{o_1}^7 y_{o_2}^5 y_{o_3}^6 y_{o_4}^6 y_{o_5}^7 y_{o_6}^5 y_{o_7}^{10} t_2^7 t_3^3 t_4^2 t_5^2 t_6^3 t_7^4 t_8^4 -  y_{s}^4 y_{o_1}^8 y_{o_2}^8 y_{o_3}^8 y_{o_4}^8 y_{o_5}^8 y_{o_6}^8 y_{o_7}^{12} t_1^8 t_3^4 t_4^4 t_5^4 t_6^4 t_7^4 t_8^4 -  6 y_{s}^4 y_{o_1}^8 y_{o_2}^8 y_{o_3}^8 y_{o_4}^8 y_{o_5}^8 y_{o_6}^8 y_{o_7}^{12} t_1^7 t_2 t_3^4 t_4^4 t_5^4 t_6^4 t_7^4 t_8^4 -  8 y_{s}^4 y_{o_1}^8 y_{o_2}^8 y_{o_3}^8 y_{o_4}^8 y_{o_5}^8 y_{o_6}^8 y_{o_7}^{12} t_1^6 t_2^2 t_3^4 t_4^4 t_5^4 t_6^4 t_7^4 t_8^4 -  {10} y_{s}^4 y_{o_1}^8 y_{o_2}^8 y_{o_3}^8 y_{o_4}^8 y_{o_5}^8 y_{o_6}^8 y_{o_7}^{12} t_1^5 t_2^3 t_3^4 t_4^4 t_5^4 t_6^4 t_7^4 t_8^4 -  {16} y_{s}^4 y_{o_1}^8 y_{o_2}^8 y_{o_3}^8 y_{o_4}^8 y_{o_5}^8 y_{o_6}^8 y_{o_7}^{12} t_1^4 t_2^4 t_3^4 t_4^4 t_5^4 t_6^4 t_7^4 t_8^4 -  {10} y_{s}^4 y_{o_1}^8 y_{o_2}^8 y_{o_3}^8 y_{o_4}^8 y_{o_5}^8 y_{o_6}^8 y_{o_7}^{12} t_1^3 t_2^5 t_3^4 t_4^4 t_5^4 t_6^4 t_7^4 t_8^4 -  8 y_{s}^4 y_{o_1}^8 y_{o_2}^8 y_{o_3}^8 y_{o_4}^8 y_{o_5}^8 y_{o_6}^8 y_{o_7}^{12} t_1^2 t_2^6 t_3^4 t_4^4 t_5^4 t_6^4 t_7^4 t_8^4 -  6 y_{s}^4 y_{o_1}^8 y_{o_2}^8 y_{o_3}^8 y_{o_4}^8 y_{o_5}^8 y_{o_6}^8 y_{o_7}^{12} t_1 t_2^7 t_3^4 t_4^4 t_5^4 t_6^4 t_7^4 t_8^4 -  y_{s}^4 y_{o_1}^8 y_{o_2}^8 y_{o_3}^8 y_{o_4}^8 y_{o_5}^8 y_{o_6}^8 y_{o_7}^{12} t_2^8 t_3^4 t_4^4 t_5^4 t_6^4 t_7^4 t_8^4 +  2 y_{s}^5 y_{o_1}^9 y_{o_2}^{11} y_{o_3}^{10} y_{o_4}^{10} y_{o_5}^9 y_{o_6}^{11} y_{o_7}^{14} t_1^8 t_2 t_3^5 t_4^6 t_5^6 t_6^5 t_7^4 t_8^4 +  3 y_{s}^5 y_{o_1}^9 y_{o_2}^{11} y_{o_3}^{10} y_{o_4}^{10} y_{o_5}^9 y_{o_6}^{11} y_{o_7}^{14} t_1^7 t_2^2 t_3^5 t_4^6 t_5^6 t_6^5 t_7^4 t_8^4 +  6 y_{s}^5 y_{o_1}^9 y_{o_2}^{11} y_{o_3}^{10} y_{o_4}^{10} y_{o_5}^9 y_{o_6}^{11} y_{o_7}^{14} t_1^6 t_2^3 t_3^5 t_4^6 t_5^6 t_6^5 t_7^4 t_8^4 +  7 y_{s}^5 y_{o_1}^9 y_{o_2}^{11} y_{o_3}^{10} y_{o_4}^{10} y_{o_5}^9 y_{o_6}^{11} y_{o_7}^{14} t_1^5 t_2^4 t_3^5 t_4^6 t_5^6 t_6^5 t_7^4 t_8^4 +  7 y_{s}^5 y_{o_1}^9 y_{o_2}^{11} y_{o_3}^{10} y_{o_4}^{10} y_{o_5}^9 y_{o_6}^{11} y_{o_7}^{14} t_1^4 t_2^5 t_3^5 t_4^6 t_5^6 t_6^5 t_7^4 t_8^4 +  6 y_{s}^5 y_{o_1}^9 y_{o_2}^{11} y_{o_3}^{10} y_{o_4}^{10} y_{o_5}^9 y_{o_6}^{11} y_{o_7}^{14} t_1^3 t_2^6 t_3^5 t_4^6 t_5^6 t_6^5 t_7^4 t_8^4 +  3 y_{s}^5 y_{o_1}^9 y_{o_2}^{11} y_{o_3}^{10} y_{o_4}^{10} y_{o_5}^9 y_{o_6}^{11} y_{o_7}^{14} t_1^2 t_2^7 t_3^5 t_4^6 t_5^6 t_6^5 t_7^4 t_8^4 +  2 y_{s}^5 y_{o_1}^9 y_{o_2}^{11} y_{o_3}^{10} y_{o_4}^{10} y_{o_5}^9 y_{o_6}^{11} y_{o_7}^{14} t_1 t_2^8 t_3^5 t_4^6 t_5^6 t_6^5 t_7^4 t_8^4 -  y_{s}^6 y_{o_1}^{10} y_{o_2}^{14} y_{o_3}^{12} y_{o_4}^{12} y_{o_5}^{10} y_{o_6}^{14} y_{o_7}^{16} t_1^5 t_2^5 t_3^6 t_4^8 t_5^8 t_6^6 t_7^4 t_8^4 +  y_{s}^3 y_{o_1}^8 y_{o_2}^4 y_{o_3}^7 y_{o_4}^5 y_{o_5}^7 y_{o_6}^5 y_{o_7}^{10} t_1^5 t_2^2 t_3^4 t_4 t_5^2 t_6^2 t_7^5 t_8^4 +  y_{s}^3 y_{o_1}^8 y_{o_2}^4 y_{o_3}^7 y_{o_4}^5 y_{o_5}^7 y_{o_6}^5 y_{o_7}^{10} t_1^4 t_2^3 t_3^4 t_4 t_5^2 t_6^2 t_7^5 t_8^4 +  y_{s}^3 y_{o_1}^8 y_{o_2}^4 y_{o_3}^7 y_{o_4}^5 y_{o_5}^7 y_{o_6}^5 y_{o_7}^{10} t_1^3 t_2^4 t_3^4 t_4 t_5^2 t_6^2 t_7^5 t_8^4 +  y_{s}^3 y_{o_1}^8 y_{o_2}^4 y_{o_3}^7 y_{o_4}^5 y_{o_5}^7 y_{o_6}^5 y_{o_7}^{10} t_1^2 t_2^5 t_3^4 t_4 t_5^2 t_6^2 t_7^5 t_8^4 -  2 y_{s}^4 y_{o_1}^9 y_{o_2}^7 y_{o_3}^9 y_{o_4}^7 y_{o_5}^8 y_{o_6}^8 y_{o_7}^{12} t_1^7 t_2 t_3^5 t_4^3 t_5^4 t_6^3 t_7^5 t_8^4 -  4 y_{s}^4 y_{o_1}^9 y_{o_2}^7 y_{o_3}^9 y_{o_4}^7 y_{o_5}^8 y_{o_6}^8 y_{o_7}^{12} t_1^6 t_2^2 t_3^5 t_4^3 t_5^4 t_6^3 t_7^5 t_8^4 -  5 y_{s}^4 y_{o_1}^9 y_{o_2}^7 y_{o_3}^9 y_{o_4}^7 y_{o_5}^8 y_{o_6}^8 y_{o_7}^{12} t_1^5 t_2^3 t_3^5 t_4^3 t_5^4 t_6^3 t_7^5 t_8^4 -  8 y_{s}^4 y_{o_1}^9 y_{o_2}^7 y_{o_3}^9 y_{o_4}^7 y_{o_5}^8 y_{o_6}^8 y_{o_7}^{12} t_1^4 t_2^4 t_3^5 t_4^3 t_5^4 t_6^3 t_7^5 t_8^4 -  5 y_{s}^4 y_{o_1}^9 y_{o_2}^7 y_{o_3}^9 y_{o_4}^7 y_{o_5}^8 y_{o_6}^8 y_{o_7}^{12} t_1^3 t_2^5 t_3^5 t_4^3 t_5^4 t_6^3 t_7^5 t_8^4 -  4 y_{s}^4 y_{o_1}^9 y_{o_2}^7 y_{o_3}^9 y_{o_4}^7 y_{o_5}^8 y_{o_6}^8 y_{o_7}^{12} t_1^2 t_2^6 t_3^5 t_4^3 t_5^4 t_6^3 t_7^5 t_8^4 -  2 y_{s}^4 y_{o_1}^9 y_{o_2}^7 y_{o_3}^9 y_{o_4}^7 y_{o_5}^8 y_{o_6}^8 y_{o_7}^{12} t_1 t_2^7 t_3^5 t_4^3 t_5^4 t_6^3 t_7^5 t_8^4 +  2 y_{s}^5 y_{o_1}^{10} y_{o_2}^{10} y_{o_3}^{11} y_{o_4}^9 y_{o_5}^9 y_{o_6}^{11} y_{o_7}^{14} t_1^8 t_2 t_3^6 t_4^5 t_5^6 t_6^4 t_7^5 t_8^4 +  3 y_{s}^5 y_{o_1}^{10} y_{o_2}^{10} y_{o_3}^{11} y_{o_4}^9 y_{o_5}^9 y_{o_6}^{11} y_{o_7}^{14} t_1^7 t_2^2 t_3^6 t_4^5 t_5^6 t_6^4 t_7^5 t_8^4 +  6 y_{s}^5 y_{o_1}^{10} y_{o_2}^{10} y_{o_3}^{11} y_{o_4}^9 y_{o_5}^9 y_{o_6}^{11} y_{o_7}^{14} t_1^6 t_2^3 t_3^6 t_4^5 t_5^6 t_6^4 t_7^5 t_8^4 +  7 y_{s}^5 y_{o_1}^{10} y_{o_2}^{10} y_{o_3}^{11} y_{o_4}^9 y_{o_5}^9 y_{o_6}^{11} y_{o_7}^{14} t_1^5 t_2^4 t_3^6 t_4^5 t_5^6 t_6^4 t_7^5 t_8^4 +  7 y_{s}^5 y_{o_1}^{10} y_{o_2}^{10} y_{o_3}^{11} y_{o_4}^9 y_{o_5}^9 y_{o_6}^{11} y_{o_7}^{14} t_1^4 t_2^5 t_3^6 t_4^5 t_5^6 t_6^4 t_7^5 t_8^4 +  6 y_{s}^5 y_{o_1}^{10} y_{o_2}^{10} y_{o_3}^{11} y_{o_4}^9 y_{o_5}^9 y_{o_6}^{11} y_{o_7}^{14} t_1^3 t_2^6 t_3^6 t_4^5 t_5^6 t_6^4 t_7^5 t_8^4 +  3 y_{s}^5 y_{o_1}^{10} y_{o_2}^{10} y_{o_3}^{11} y_{o_4}^9 y_{o_5}^9 y_{o_6}^{11} y_{o_7}^{14} t_1^2 t_2^7 t_3^6 t_4^5 t_5^6 t_6^4 t_7^5 t_8^4 +  2 y_{s}^5 y_{o_1}^{10} y_{o_2}^{10} y_{o_3}^{11} y_{o_4}^9 y_{o_5}^9 y_{o_6}^{11} y_{o_7}^{14} t_1 t_2^8 t_3^6 t_4^5 t_5^6 t_6^4 t_7^5 t_8^4 -  y_{s}^6 y_{o_1}^{11} y_{o_2}^{13} y_{o_3}^{13} y_{o_4}^{11} y_{o_5}^{10} y_{o_6}^{14} y_{o_7}^{16} t_1^8 t_2^2 t_3^7 t_4^7 t_5^8 t_6^5 t_7^5 t_8^4 -  y_{s}^6 y_{o_1}^{11} y_{o_2}^{13} y_{o_3}^{13} y_{o_4}^{11} y_{o_5}^{10} y_{o_6}^{14} y_{o_7}^{16} t_1^7 t_2^3 t_3^7 t_4^7 t_5^8 t_6^5 t_7^5 t_8^4 -  y_{s}^6 y_{o_1}^{11} y_{o_2}^{13} y_{o_3}^{13} y_{o_4}^{11} y_{o_5}^{10} y_{o_6}^{14} y_{o_7}^{16} t_1^6 t_2^4 t_3^7 t_4^7 t_5^8 t_6^5 t_7^5 t_8^4 -  3 y_{s}^6 y_{o_1}^{11} y_{o_2}^{13} y_{o_3}^{13} y_{o_4}^{11} y_{o_5}^{10} y_{o_6}^{14} y_{o_7}^{16} t_1^5 t_2^5 t_3^7 t_4^7 t_5^8 t_6^5 t_7^5 t_8^4 -  y_{s}^6 y_{o_1}^{11} y_{o_2}^{13} y_{o_3}^{13} y_{o_4}^{11} y_{o_5}^{10} y_{o_6}^{14} y_{o_7}^{16} t_1^4 t_2^6 t_3^7 t_4^7 t_5^8 t_6^5 t_7^5 t_8^4 -  y_{s}^6 y_{o_1}^{11} y_{o_2}^{13} y_{o_3}^{13} y_{o_4}^{11} y_{o_5}^{10} y_{o_6}^{14} y_{o_7}^{16} t_1^3 t_2^7 t_3^7 t_4^7 t_5^8 t_6^5 t_7^5 t_8^4 -  y_{s}^6 y_{o_1}^{11} y_{o_2}^{13} y_{o_3}^{13} y_{o_4}^{11} y_{o_5}^{10} y_{o_6}^{14} y_{o_7}^{16} t_1^2 t_2^8 t_3^7 t_4^7 t_5^8 t_6^5 t_7^5 t_8^4 -  y_{s}^4 y_{o_1}^{10} y_{o_2}^6 y_{o_3}^{10} y_{o_4}^6 y_{o_5}^8 y_{o_6}^8 y_{o_7}^{12} t_1^6 t_2^2 t_3^6 t_4^2 t_5^4 t_6^2 t_7^6 t_8^4 -  y_{s}^4 y_{o_1}^{10} y_{o_2}^6 y_{o_3}^{10} y_{o_4}^6 y_{o_5}^8 y_{o_6}^8 y_{o_7}^{12} t_1^5 t_2^3 t_3^6 t_4^2 t_5^4 t_6^2 t_7^6 t_8^4 -  y_{s}^4 y_{o_1}^{10} y_{o_2}^6 y_{o_3}^{10} y_{o_4}^6 y_{o_5}^8 y_{o_6}^8 y_{o_7}^{12} t_1^4 t_2^4 t_3^6 t_4^2 t_5^4 t_6^2 t_7^6 t_8^4 -  y_{s}^4 y_{o_1}^{10} y_{o_2}^6 y_{o_3}^{10} y_{o_4}^6 y_{o_5}^8 y_{o_6}^8 y_{o_7}^{12} t_1^3 t_2^5 t_3^6 t_4^2 t_5^4 t_6^2 t_7^6 t_8^4 -  y_{s}^4 y_{o_1}^{10} y_{o_2}^6 y_{o_3}^{10} y_{o_4}^6 y_{o_5}^8 y_{o_6}^8 y_{o_7}^{12} t_1^2 t_2^6 t_3^6 t_4^2 t_5^4 t_6^2 t_7^6 t_8^4 +  2 y_{s}^5 y_{o_1}^{11} y_{o_2}^9 y_{o_3}^{12} y_{o_4}^8 y_{o_5}^9 y_{o_6}^{11} y_{o_7}^{14} t_1^6 t_2^3 t_3^7 t_4^4 t_5^6 t_6^3 t_7^6 t_8^4 +  y_{s}^5 y_{o_1}^{11} y_{o_2}^9 y_{o_3}^{12} y_{o_4}^8 y_{o_5}^9 y_{o_6}^{11} y_{o_7}^{14} t_1^5 t_2^4 t_3^7 t_4^4 t_5^6 t_6^3 t_7^6 t_8^4 +  y_{s}^5 y_{o_1}^{11} y_{o_2}^9 y_{o_3}^{12} y_{o_4}^8 y_{o_5}^9 y_{o_6}^{11} y_{o_7}^{14} t_1^4 t_2^5 t_3^7 t_4^4 t_5^6 t_6^3 t_7^6 t_8^4 +  2 y_{s}^5 y_{o_1}^{11} y_{o_2}^9 y_{o_3}^{12} y_{o_4}^8 y_{o_5}^9 y_{o_6}^{11} y_{o_7}^{14} t_1^3 t_2^6 t_3^7 t_4^4 t_5^6 t_6^3 t_7^6 t_8^4 -  y_{s}^6 y_{o_1}^{12} y_{o_2}^{12} y_{o_3}^{14} y_{o_4}^{10} y_{o_5}^{10} y_{o_6}^{14} y_{o_7}^{16} t_1^5 t_2^5 t_3^8 t_4^6 t_5^8 t_6^4 t_7^6 t_8^4 +  y_{s}^5 y_{o_1}^{12} y_{o_2}^8 y_{o_3}^{13} y_{o_4}^7 y_{o_5}^9 y_{o_6}^{11} y_{o_7}^{14} t_1^6 t_2^3 t_3^8 t_4^3 t_5^6 t_6^2 t_7^7 t_8^4 +  y_{s}^5 y_{o_1}^{12} y_{o_2}^8 y_{o_3}^{13} y_{o_4}^7 y_{o_5}^9 y_{o_6}^{11} y_{o_7}^{14} t_1^3 t_2^6 t_3^8 t_4^3 t_5^6 t_6^2 t_7^7 t_8^4 -  y_{s}^5 y_{o_1}^7 y_{o_2}^{13} y_{o_3}^7 y_{o_4}^{13} y_{o_5}^{10} y_{o_6}^{10} y_{o_7}^{15} t_1^5 t_2^5 t_3^2 t_4^8 t_5^5 t_6^8 t_7^2 t_8^5 +  y_{s}^3 y_{o_1}^6 y_{o_2}^6 y_{o_3}^4 y_{o_4}^8 y_{o_5}^8 y_{o_6}^4 y_{o_7}^{11} t_1^7 t_2 t_3 t_4^3 t_5 t_6^5 t_7^3 t_8^5 +  y_{s}^3 y_{o_1}^6 y_{o_2}^6 y_{o_3}^4 y_{o_4}^8 y_{o_5}^8 y_{o_6}^4 y_{o_7}^{11} t_1^6 t_2^2 t_3 t_4^3 t_5 t_6^5 t_7^3 t_8^5 +  2 y_{s}^3 y_{o_1}^6 y_{o_2}^6 y_{o_3}^4 y_{o_4}^8 y_{o_5}^8 y_{o_6}^4 y_{o_7}^{11} t_1^5 t_2^3 t_3 t_4^3 t_5 t_6^5 t_7^3 t_8^5 +  2 y_{s}^3 y_{o_1}^6 y_{o_2}^6 y_{o_3}^4 y_{o_4}^8 y_{o_5}^8 y_{o_6}^4 y_{o_7}^{11} t_1^4 t_2^4 t_3 t_4^3 t_5 t_6^5 t_7^3 t_8^5 +  2 y_{s}^3 y_{o_1}^6 y_{o_2}^6 y_{o_3}^4 y_{o_4}^8 y_{o_5}^8 y_{o_6}^4 y_{o_7}^{11} t_1^3 t_2^5 t_3 t_4^3 t_5 t_6^5 t_7^3 t_8^5 +  y_{s}^3 y_{o_1}^6 y_{o_2}^6 y_{o_3}^4 y_{o_4}^8 y_{o_5}^8 y_{o_6}^4 y_{o_7}^{11} t_1^2 t_2^6 t_3 t_4^3 t_5 t_6^5 t_7^3 t_8^5 +  y_{s}^3 y_{o_1}^6 y_{o_2}^6 y_{o_3}^4 y_{o_4}^8 y_{o_5}^8 y_{o_6}^4 y_{o_7}^{11} t_1 t_2^7 t_3 t_4^3 t_5 t_6^5 t_7^3 t_8^5 -  y_{s}^4 y_{o_1}^7 y_{o_2}^9 y_{o_3}^6 y_{o_4}^{10} y_{o_5}^9 y_{o_6}^7 y_{o_7}^{13} t_1^8 t_2 t_3^2 t_4^5 t_5^3 t_6^6 t_7^3 t_8^5 -  2 y_{s}^4 y_{o_1}^7 y_{o_2}^9 y_{o_3}^6 y_{o_4}^{10} y_{o_5}^9 y_{o_6}^7 y_{o_7}^{13} t_1^7 t_2^2 t_3^2 t_4^5 t_5^3 t_6^6 t_7^3 t_8^5 -  3 y_{s}^4 y_{o_1}^7 y_{o_2}^9 y_{o_3}^6 y_{o_4}^{10} y_{o_5}^9 y_{o_6}^7 y_{o_7}^{13} t_1^6 t_2^3 t_3^2 t_4^5 t_5^3 t_6^6 t_7^3 t_8^5 -  4 y_{s}^4 y_{o_1}^7 y_{o_2}^9 y_{o_3}^6 y_{o_4}^{10} y_{o_5}^9 y_{o_6}^7 y_{o_7}^{13} t_1^5 t_2^4 t_3^2 t_4^5 t_5^3 t_6^6 t_7^3 t_8^5 -  4 y_{s}^4 y_{o_1}^7 y_{o_2}^9 y_{o_3}^6 y_{o_4}^{10} y_{o_5}^9 y_{o_6}^7 y_{o_7}^{13} t_1^4 t_2^5 t_3^2 t_4^5 t_5^3 t_6^6 t_7^3 t_8^5 -  3 y_{s}^4 y_{o_1}^7 y_{o_2}^9 y_{o_3}^6 y_{o_4}^{10} y_{o_5}^9 y_{o_6}^7 y_{o_7}^{13} t_1^3 t_2^6 t_3^2 t_4^5 t_5^3 t_6^6 t_7^3 t_8^5 -  2 y_{s}^4 y_{o_1}^7 y_{o_2}^9 y_{o_3}^6 y_{o_4}^{10} y_{o_5}^9 y_{o_6}^7 y_{o_7}^{13} t_1^2 t_2^7 t_3^2 t_4^5 t_5^3 t_6^6 t_7^3 t_8^5 -  y_{s}^4 y_{o_1}^7 y_{o_2}^9 y_{o_3}^6 y_{o_4}^{10} y_{o_5}^9 y_{o_6}^7 y_{o_7}^{13} t_1 t_2^8 t_3^2 t_4^5 t_5^3 t_6^6 t_7^3 t_8^5 +  y_{s}^5 y_{o_1}^8 y_{o_2}^{12} y_{o_3}^8 y_{o_4}^{12} y_{o_5}^{10} y_{o_6}^{10} y_{o_7}^{15} t_1^7 t_2^3 t_3^3 t_4^7 t_5^5 t_6^7 t_7^3 t_8^5 +  2 y_{s}^5 y_{o_1}^8 y_{o_2}^{12} y_{o_3}^8 y_{o_4}^{12} y_{o_5}^{10} y_{o_6}^{10} y_{o_7}^{15} t_1^6 t_2^4 t_3^3 t_4^7 t_5^5 t_6^7 t_7^3 t_8^5 -  y_{s}^5 y_{o_1}^8 y_{o_2}^{12} y_{o_3}^8 y_{o_4}^{12} y_{o_5}^{10} y_{o_6}^{10} y_{o_7}^{15} t_1^5 t_2^5 t_3^3 t_4^7 t_5^5 t_6^7 t_7^3 t_8^5 +  2 y_{s}^5 y_{o_1}^8 y_{o_2}^{12} y_{o_3}^8 y_{o_4}^{12} y_{o_5}^{10} y_{o_6}^{10} y_{o_7}^{15} t_1^4 t_2^6 t_3^3 t_4^7 t_5^5 t_6^7 t_7^3 t_8^5 +  y_{s}^5 y_{o_1}^8 y_{o_2}^{12} y_{o_3}^8 y_{o_4}^{12} y_{o_5}^{10} y_{o_6}^{10} y_{o_7}^{15} t_1^3 t_2^7 t_3^3 t_4^7 t_5^5 t_6^7 t_7^3 t_8^5 +  y_{s}^6 y_{o_1}^9 y_{o_2}^{15} y_{o_3}^{10} y_{o_4}^{14} y_{o_5}^{11} y_{o_6}^{13} y_{o_7}^{17} t_1^6 t_2^5 t_3^4 t_4^9 t_5^7 t_6^8 t_7^3 t_8^5 +  y_{s}^6 y_{o_1}^9 y_{o_2}^{15} y_{o_3}^{10} y_{o_4}^{14} y_{o_5}^{11} y_{o_6}^{13} y_{o_7}^{17} t_1^5 t_2^6 t_3^4 t_4^9 t_5^7 t_6^8 t_7^3 t_8^5 +  y_{s}^3 y_{o_1}^7 y_{o_2}^5 y_{o_3}^5 y_{o_4}^7 y_{o_5}^8 y_{o_6}^4 y_{o_7}^{11} t_1^5 t_2^3 t_3^2 t_4^2 t_5 t_6^4 t_7^4 t_8^5 +  y_{s}^3 y_{o_1}^7 y_{o_2}^5 y_{o_3}^5 y_{o_4}^7 y_{o_5}^8 y_{o_6}^4 y_{o_7}^{11} t_1^4 t_2^4 t_3^2 t_4^2 t_5 t_6^4 t_7^4 t_8^5 +  y_{s}^3 y_{o_1}^7 y_{o_2}^5 y_{o_3}^5 y_{o_4}^7 y_{o_5}^8 y_{o_6}^4 y_{o_7}^{11} t_1^3 t_2^5 t_3^2 t_4^2 t_5 t_6^4 t_7^4 t_8^5 -  y_{s}^4 y_{o_1}^8 y_{o_2}^8 y_{o_3}^7 y_{o_4}^9 y_{o_5}^9 y_{o_6}^7 y_{o_7}^{13} t_1^8 t_2 t_3^3 t_4^4 t_5^3 t_6^5 t_7^4 t_8^5 -  2 y_{s}^4 y_{o_1}^8 y_{o_2}^8 y_{o_3}^7 y_{o_4}^9 y_{o_5}^9 y_{o_6}^7 y_{o_7}^{13} t_1^7 t_2^2 t_3^3 t_4^4 t_5^3 t_6^5 t_7^4 t_8^5 -  3 y_{s}^4 y_{o_1}^8 y_{o_2}^8 y_{o_3}^7 y_{o_4}^9 y_{o_5}^9 y_{o_6}^7 y_{o_7}^{13} t_1^6 t_2^3 t_3^3 t_4^4 t_5^3 t_6^5 t_7^4 t_8^5 -  6 y_{s}^4 y_{o_1}^8 y_{o_2}^8 y_{o_3}^7 y_{o_4}^9 y_{o_5}^9 y_{o_6}^7 y_{o_7}^{13} t_1^5 t_2^4 t_3^3 t_4^4 t_5^3 t_6^5 t_7^4 t_8^5 -  6 y_{s}^4 y_{o_1}^8 y_{o_2}^8 y_{o_3}^7 y_{o_4}^9 y_{o_5}^9 y_{o_6}^7 y_{o_7}^{13} t_1^4 t_2^5 t_3^3 t_4^4 t_5^3 t_6^5 t_7^4 t_8^5 -  3 y_{s}^4 y_{o_1}^8 y_{o_2}^8 y_{o_3}^7 y_{o_4}^9 y_{o_5}^9 y_{o_6}^7 y_{o_7}^{13} t_1^3 t_2^6 t_3^3 t_4^4 t_5^3 t_6^5 t_7^4 t_8^5 -  2 y_{s}^4 y_{o_1}^8 y_{o_2}^8 y_{o_3}^7 y_{o_4}^9 y_{o_5}^9 y_{o_6}^7 y_{o_7}^{13} t_1^2 t_2^7 t_3^3 t_4^4 t_5^3 t_6^5 t_7^4 t_8^5 -  y_{s}^4 y_{o_1}^8 y_{o_2}^8 y_{o_3}^7 y_{o_4}^9 y_{o_5}^9 y_{o_6}^7 y_{o_7}^{13} t_1 t_2^8 t_3^3 t_4^4 t_5^3 t_6^5 t_7^4 t_8^5 +  y_{s}^5 y_{o_1}^9 y_{o_2}^{11} y_{o_3}^9 y_{o_4}^{11} y_{o_5}^{10} y_{o_6}^{10} y_{o_7}^{15} t_1^9 t_2 t_3^4 t_4^6 t_5^5 t_6^6 t_7^4 t_8^5 +  2 y_{s}^5 y_{o_1}^9 y_{o_2}^{11} y_{o_3}^9 y_{o_4}^{11} y_{o_5}^{10} y_{o_6}^{10} y_{o_7}^{15} t_1^8 t_2^2 t_3^4 t_4^6 t_5^5 t_6^6 t_7^4 t_8^5 +  3 y_{s}^5 y_{o_1}^9 y_{o_2}^{11} y_{o_3}^9 y_{o_4}^{11} y_{o_5}^{10} y_{o_6}^{10} y_{o_7}^{15} t_1^7 t_2^3 t_3^4 t_4^6 t_5^5 t_6^6 t_7^4 t_8^5 +  6 y_{s}^5 y_{o_1}^9 y_{o_2}^{11} y_{o_3}^9 y_{o_4}^{11} y_{o_5}^{10} y_{o_6}^{10} y_{o_7}^{15} t_1^6 t_2^4 t_3^4 t_4^6 t_5^5 t_6^6 t_7^4 t_8^5 +  3 y_{s}^5 y_{o_1}^9 y_{o_2}^{11} y_{o_3}^9 y_{o_4}^{11} y_{o_5}^{10} y_{o_6}^{10} y_{o_7}^{15} t_1^5 t_2^5 t_3^4 t_4^6 t_5^5 t_6^6 t_7^4 t_8^5 +  6 y_{s}^5 y_{o_1}^9 y_{o_2}^{11} y_{o_3}^9 y_{o_4}^{11} y_{o_5}^{10} y_{o_6}^{10} y_{o_7}^{15} t_1^4 t_2^6 t_3^4 t_4^6 t_5^5 t_6^6 t_7^4 t_8^5 +  3 y_{s}^5 y_{o_1}^9 y_{o_2}^{11} y_{o_3}^9 y_{o_4}^{11} y_{o_5}^{10} y_{o_6}^{10} y_{o_7}^{15} t_1^3 t_2^7 t_3^4 t_4^6 t_5^5 t_6^6 t_7^4 t_8^5 +  2 y_{s}^5 y_{o_1}^9 y_{o_2}^{11} y_{o_3}^9 y_{o_4}^{11} y_{o_5}^{10} y_{o_6}^{10} y_{o_7}^{15} t_1^2 t_2^8 t_3^4 t_4^6 t_5^5 t_6^6 t_7^4 t_8^5 +  y_{s}^5 y_{o_1}^9 y_{o_2}^{11} y_{o_3}^9 y_{o_4}^{11} y_{o_5}^{10} y_{o_6}^{10} y_{o_7}^{15} t_1 t_2^9 t_3^4 t_4^6 t_5^5 t_6^6 t_7^4 t_8^5 -  y_{s}^6 y_{o_1}^{10} y_{o_2}^{14} y_{o_3}^{11} y_{o_4}^{13} y_{o_5}^{11} y_{o_6}^{13} y_{o_7}^{17} t_1^8 t_2^3 t_3^5 t_4^8 t_5^7 t_6^7 t_7^4 t_8^5 -  y_{s}^6 y_{o_1}^{10} y_{o_2}^{14} y_{o_3}^{11} y_{o_4}^{13} y_{o_5}^{11} y_{o_6}^{13} y_{o_7}^{17} t_1^7 t_2^4 t_3^5 t_4^8 t_5^7 t_6^7 t_7^4 t_8^5 -  y_{s}^6 y_{o_1}^{10} y_{o_2}^{14} y_{o_3}^{11} y_{o_4}^{13} y_{o_5}^{11} y_{o_6}^{13} y_{o_7}^{17} t_1^4 t_2^7 t_3^5 t_4^8 t_5^7 t_6^7 t_7^4 t_8^5 -  y_{s}^6 y_{o_1}^{10} y_{o_2}^{14} y_{o_3}^{11} y_{o_4}^{13} y_{o_5}^{11} y_{o_6}^{13} y_{o_7}^{17} t_1^3 t_2^8 t_3^5 t_4^8 t_5^7 t_6^7 t_7^4 t_8^5 -  y_{s}^7 y_{o_1}^{11} y_{o_2}^{17} y_{o_3}^{13} y_{o_4}^{15} y_{o_5}^{12} y_{o_6}^{16} y_{o_7}^{19} t_1^6 t_2^6 t_3^6 t_4^{10} t_5^9 t_6^8 t_7^4 t_8^5 +  y_{s}^3 y_{o_1}^8 y_{o_2}^4 y_{o_3}^6 y_{o_4}^6 y_{o_5}^8 y_{o_6}^4 y_{o_7}^{11} t_1^7 t_2 t_3^3 t_4 t_5 t_6^3 t_7^5 t_8^5 +  y_{s}^3 y_{o_1}^8 y_{o_2}^4 y_{o_3}^6 y_{o_4}^6 y_{o_5}^8 y_{o_6}^4 y_{o_7}^{11} t_1^6 t_2^2 t_3^3 t_4 t_5 t_6^3 t_7^5 t_8^5 +  2 y_{s}^3 y_{o_1}^8 y_{o_2}^4 y_{o_3}^6 y_{o_4}^6 y_{o_5}^8 y_{o_6}^4 y_{o_7}^{11} t_1^5 t_2^3 t_3^3 t_4 t_5 t_6^3 t_7^5 t_8^5 +  2 y_{s}^3 y_{o_1}^8 y_{o_2}^4 y_{o_3}^6 y_{o_4}^6 y_{o_5}^8 y_{o_6}^4 y_{o_7}^{11} t_1^4 t_2^4 t_3^3 t_4 t_5 t_6^3 t_7^5 t_8^5 +  2 y_{s}^3 y_{o_1}^8 y_{o_2}^4 y_{o_3}^6 y_{o_4}^6 y_{o_5}^8 y_{o_6}^4 y_{o_7}^{11} t_1^3 t_2^5 t_3^3 t_4 t_5 t_6^3 t_7^5 t_8^5 +  y_{s}^3 y_{o_1}^8 y_{o_2}^4 y_{o_3}^6 y_{o_4}^6 y_{o_5}^8 y_{o_6}^4 y_{o_7}^{11} t_1^2 t_2^6 t_3^3 t_4 t_5 t_6^3 t_7^5 t_8^5 +  y_{s}^3 y_{o_1}^8 y_{o_2}^4 y_{o_3}^6 y_{o_4}^6 y_{o_5}^8 y_{o_6}^4 y_{o_7}^{11} t_1 t_2^7 t_3^3 t_4 t_5 t_6^3 t_7^5 t_8^5 -  y_{s}^4 y_{o_1}^9 y_{o_2}^7 y_{o_3}^8 y_{o_4}^8 y_{o_5}^9 y_{o_6}^7 y_{o_7}^{13} t_1^8 t_2 t_3^4 t_4^3 t_5^3 t_6^4 t_7^5 t_8^5 -  2 y_{s}^4 y_{o_1}^9 y_{o_2}^7 y_{o_3}^8 y_{o_4}^8 y_{o_5}^9 y_{o_6}^7 y_{o_7}^{13} t_1^7 t_2^2 t_3^4 t_4^3 t_5^3 t_6^4 t_7^5 t_8^5 -  3 y_{s}^4 y_{o_1}^9 y_{o_2}^7 y_{o_3}^8 y_{o_4}^8 y_{o_5}^9 y_{o_6}^7 y_{o_7}^{13} t_1^6 t_2^3 t_3^4 t_4^3 t_5^3 t_6^4 t_7^5 t_8^5 -  6 y_{s}^4 y_{o_1}^9 y_{o_2}^7 y_{o_3}^8 y_{o_4}^8 y_{o_5}^9 y_{o_6}^7 y_{o_7}^{13} t_1^5 t_2^4 t_3^4 t_4^3 t_5^3 t_6^4 t_7^5 t_8^5 -  6 y_{s}^4 y_{o_1}^9 y_{o_2}^7 y_{o_3}^8 y_{o_4}^8 y_{o_5}^9 y_{o_6}^7 y_{o_7}^{13} t_1^4 t_2^5 t_3^4 t_4^3 t_5^3 t_6^4 t_7^5 t_8^5 -  3 y_{s}^4 y_{o_1}^9 y_{o_2}^7 y_{o_3}^8 y_{o_4}^8 y_{o_5}^9 y_{o_6}^7 y_{o_7}^{13} t_1^3 t_2^6 t_3^4 t_4^3 t_5^3 t_6^4 t_7^5 t_8^5 -  2 y_{s}^4 y_{o_1}^9 y_{o_2}^7 y_{o_3}^8 y_{o_4}^8 y_{o_5}^9 y_{o_6}^7 y_{o_7}^{13} t_1^2 t_2^7 t_3^4 t_4^3 t_5^3 t_6^4 t_7^5 t_8^5 -  y_{s}^4 y_{o_1}^9 y_{o_2}^7 y_{o_3}^8 y_{o_4}^8 y_{o_5}^9 y_{o_6}^7 y_{o_7}^{13} t_1 t_2^8 t_3^4 t_4^3 t_5^3 t_6^4 t_7^5 t_8^5 -  y_{s}^5 y_{o_1}^{10} y_{o_2}^{10} y_{o_3}^{10} y_{o_4}^{10} y_{o_5}^{10} y_{o_6}^{10} y_{o_7}^{15} t_1^{10} t_3^5 t_4^5 t_5^5 t_6^5 t_7^5 t_8^5 -  y_{s}^5 y_{o_1}^{10} y_{o_2}^{10} y_{o_3}^{10} y_{o_4}^{10} y_{o_5}^{10} y_{o_6}^{10} y_{o_7}^{15} t_1^9 t_2 t_3^5 t_4^5 t_5^5 t_6^5 t_7^5 t_8^5 -  y_{s}^5 y_{o_1}^{10} y_{o_2}^{10} y_{o_3}^{10} y_{o_4}^{10} y_{o_5}^{10} y_{o_6}^{10} y_{o_7}^{15} t_1^8 t_2^2 t_3^5 t_4^5 t_5^5 t_6^5 t_7^5 t_8^5 -  2 y_{s}^5 y_{o_1}^{10} y_{o_2}^{10} y_{o_3}^{10} y_{o_4}^{10} y_{o_5}^{10} y_{o_6}^{10} y_{o_7}^{15} t_1^7 t_2^3 t_3^5 t_4^5 t_5^5 t_6^5 t_7^5 t_8^5 +  2 y_{s}^5 y_{o_1}^{10} y_{o_2}^{10} y_{o_3}^{10} y_{o_4}^{10} y_{o_5}^{10} y_{o_6}^{10} y_{o_7}^{15} t_1^6 t_2^4 t_3^5 t_4^5 t_5^5 t_6^5 t_7^5 t_8^5 +  2 y_{s}^5 y_{o_1}^{10} y_{o_2}^{10} y_{o_3}^{10} y_{o_4}^{10} y_{o_5}^{10} y_{o_6}^{10} y_{o_7}^{15} t_1^4 t_2^6 t_3^5 t_4^5 t_5^5 t_6^5 t_7^5 t_8^5 -  2 y_{s}^5 y_{o_1}^{10} y_{o_2}^{10} y_{o_3}^{10} y_{o_4}^{10} y_{o_5}^{10} y_{o_6}^{10} y_{o_7}^{15} t_1^3 t_2^7 t_3^5 t_4^5 t_5^5 t_6^5 t_7^5 t_8^5 -  y_{s}^5 y_{o_1}^{10} y_{o_2}^{10} y_{o_3}^{10} y_{o_4}^{10} y_{o_5}^{10} y_{o_6}^{10} y_{o_7}^{15} t_1^2 t_2^8 t_3^5 t_4^5 t_5^5 t_6^5 t_7^5 t_8^5 -  y_{s}^5 y_{o_1}^{10} y_{o_2}^{10} y_{o_3}^{10} y_{o_4}^{10} y_{o_5}^{10} y_{o_6}^{10} y_{o_7}^{15} t_1 t_2^9 t_3^5 t_4^5 t_5^5 t_6^5 t_7^5 t_8^5 -  y_{s}^5 y_{o_1}^{10} y_{o_2}^{10} y_{o_3}^{10} y_{o_4}^{10} y_{o_5}^{10} y_{o_6}^{10} y_{o_7}^{15} t_2^{10} t_3^5 t_4^5 t_5^5 t_6^5 t_7^5 t_8^5 +  y_{s}^6 y_{o_1}^{11} y_{o_2}^{13} y_{o_3}^{12} y_{o_4}^{12} y_{o_5}^{11} y_{o_6}^{13} y_{o_7}^{17} t_1^{10} t_2 t_3^6 t_4^7 t_5^7 t_6^6 t_7^5 t_8^5 +  y_{s}^6 y_{o_1}^{11} y_{o_2}^{13} y_{o_3}^{12} y_{o_4}^{12} y_{o_5}^{11} y_{o_6}^{13} y_{o_7}^{17} t_1^9 t_2^2 t_3^6 t_4^7 t_5^7 t_6^6 t_7^5 t_8^5 +  y_{s}^6 y_{o_1}^{11} y_{o_2}^{13} y_{o_3}^{12} y_{o_4}^{12} y_{o_5}^{11} y_{o_6}^{13} y_{o_7}^{17} t_1^8 t_2^3 t_3^6 t_4^7 t_5^7 t_6^6 t_7^5 t_8^5 +  4 y_{s}^6 y_{o_1}^{11} y_{o_2}^{13} y_{o_3}^{12} y_{o_4}^{12} y_{o_5}^{11} y_{o_6}^{13} y_{o_7}^{17} t_1^7 t_2^4 t_3^6 t_4^7 t_5^7 t_6^6 t_7^5 t_8^5 +  2 y_{s}^6 y_{o_1}^{11} y_{o_2}^{13} y_{o_3}^{12} y_{o_4}^{12} y_{o_5}^{11} y_{o_6}^{13} y_{o_7}^{17} t_1^6 t_2^5 t_3^6 t_4^7 t_5^7 t_6^6 t_7^5 t_8^5 +  2 y_{s}^6 y_{o_1}^{11} y_{o_2}^{13} y_{o_3}^{12} y_{o_4}^{12} y_{o_5}^{11} y_{o_6}^{13} y_{o_7}^{17} t_1^5 t_2^6 t_3^6 t_4^7 t_5^7 t_6^6 t_7^5 t_8^5 +  4 y_{s}^6 y_{o_1}^{11} y_{o_2}^{13} y_{o_3}^{12} y_{o_4}^{12} y_{o_5}^{11} y_{o_6}^{13} y_{o_7}^{17} t_1^4 t_2^7 t_3^6 t_4^7 t_5^7 t_6^6 t_7^5 t_8^5 +  y_{s}^6 y_{o_1}^{11} y_{o_2}^{13} y_{o_3}^{12} y_{o_4}^{12} y_{o_5}^{11} y_{o_6}^{13} y_{o_7}^{17} t_1^3 t_2^8 t_3^6 t_4^7 t_5^7 t_6^6 t_7^5 t_8^5 +  y_{s}^6 y_{o_1}^{11} y_{o_2}^{13} y_{o_3}^{12} y_{o_4}^{12} y_{o_5}^{11} y_{o_6}^{13} y_{o_7}^{17} t_1^2 t_2^9 t_3^6 t_4^7 t_5^7 t_6^6 t_7^5 t_8^5 +  y_{s}^6 y_{o_1}^{11} y_{o_2}^{13} y_{o_3}^{12} y_{o_4}^{12} y_{o_5}^{11} y_{o_6}^{13} y_{o_7}^{17} t_1 t_2^{10} t_3^6 t_4^7 t_5^7 t_6^6 t_7^5 t_8^5 -  y_{s}^7 y_{o_1}^{12} y_{o_2}^{16} y_{o_3}^{14} y_{o_4}^{14} y_{o_5}^{12} y_{o_6}^{16} y_{o_7}^{19} t_1^6 t_2^6 t_3^7 t_4^9 t_5^9 t_6^7 t_7^5 t_8^5 -  y_{s}^4 y_{o_1}^{10} y_{o_2}^6 y_{o_3}^9 y_{o_4}^7 y_{o_5}^9 y_{o_6}^7 y_{o_7}^{13} t_1^8 t_2 t_3^5 t_4^2 t_5^3 t_6^3 t_7^6 t_8^5 -  2 y_{s}^4 y_{o_1}^{10} y_{o_2}^6 y_{o_3}^9 y_{o_4}^7 y_{o_5}^9 y_{o_6}^7 y_{o_7}^{13} t_1^7 t_2^2 t_3^5 t_4^2 t_5^3 t_6^3 t_7^6 t_8^5 -  3 y_{s}^4 y_{o_1}^{10} y_{o_2}^6 y_{o_3}^9 y_{o_4}^7 y_{o_5}^9 y_{o_6}^7 y_{o_7}^{13} t_1^6 t_2^3 t_3^5 t_4^2 t_5^3 t_6^3 t_7^6 t_8^5 -  4 y_{s}^4 y_{o_1}^{10} y_{o_2}^6 y_{o_3}^9 y_{o_4}^7 y_{o_5}^9 y_{o_6}^7 y_{o_7}^{13} t_1^5 t_2^4 t_3^5 t_4^2 t_5^3 t_6^3 t_7^6 t_8^5 -  4 y_{s}^4 y_{o_1}^{10} y_{o_2}^6 y_{o_3}^9 y_{o_4}^7 y_{o_5}^9 y_{o_6}^7 y_{o_7}^{13} t_1^4 t_2^5 t_3^5 t_4^2 t_5^3 t_6^3 t_7^6 t_8^5 -  3 y_{s}^4 y_{o_1}^{10} y_{o_2}^6 y_{o_3}^9 y_{o_4}^7 y_{o_5}^9 y_{o_6}^7 y_{o_7}^{13} t_1^3 t_2^6 t_3^5 t_4^2 t_5^3 t_6^3 t_7^6 t_8^5 -  2 y_{s}^4 y_{o_1}^{10} y_{o_2}^6 y_{o_3}^9 y_{o_4}^7 y_{o_5}^9 y_{o_6}^7 y_{o_7}^{13} t_1^2 t_2^7 t_3^5 t_4^2 t_5^3 t_6^3 t_7^6 t_8^5 -  y_{s}^4 y_{o_1}^{10} y_{o_2}^6 y_{o_3}^9 y_{o_4}^7 y_{o_5}^9 y_{o_6}^7 y_{o_7}^{13} t_1 t_2^8 t_3^5 t_4^2 t_5^3 t_6^3 t_7^6 t_8^5 +  y_{s}^5 y_{o_1}^{11} y_{o_2}^9 y_{o_3}^{11} y_{o_4}^9 y_{o_5}^{10} y_{o_6}^{10} y_{o_7}^{15} t_1^9 t_2 t_3^6 t_4^4 t_5^5 t_6^4 t_7^6 t_8^5 +  2 y_{s}^5 y_{o_1}^{11} y_{o_2}^9 y_{o_3}^{11} y_{o_4}^9 y_{o_5}^{10} y_{o_6}^{10} y_{o_7}^{15} t_1^8 t_2^2 t_3^6 t_4^4 t_5^5 t_6^4 t_7^6 t_8^5 +  3 y_{s}^5 y_{o_1}^{11} y_{o_2}^9 y_{o_3}^{11} y_{o_4}^9 y_{o_5}^{10} y_{o_6}^{10} y_{o_7}^{15} t_1^7 t_2^3 t_3^6 t_4^4 t_5^5 t_6^4 t_7^6 t_8^5 +  6 y_{s}^5 y_{o_1}^{11} y_{o_2}^9 y_{o_3}^{11} y_{o_4}^9 y_{o_5}^{10} y_{o_6}^{10} y_{o_7}^{15} t_1^6 t_2^4 t_3^6 t_4^4 t_5^5 t_6^4 t_7^6 t_8^5 +  3 y_{s}^5 y_{o_1}^{11} y_{o_2}^9 y_{o_3}^{11} y_{o_4}^9 y_{o_5}^{10} y_{o_6}^{10} y_{o_7}^{15} t_1^5 t_2^5 t_3^6 t_4^4 t_5^5 t_6^4 t_7^6 t_8^5 +  6 y_{s}^5 y_{o_1}^{11} y_{o_2}^9 y_{o_3}^{11} y_{o_4}^9 y_{o_5}^{10} y_{o_6}^{10} y_{o_7}^{15} t_1^4 t_2^6 t_3^6 t_4^4 t_5^5 t_6^4 t_7^6 t_8^5 +  3 y_{s}^5 y_{o_1}^{11} y_{o_2}^9 y_{o_3}^{11} y_{o_4}^9 y_{o_5}^{10} y_{o_6}^{10} y_{o_7}^{15} t_1^3 t_2^7 t_3^6 t_4^4 t_5^5 t_6^4 t_7^6 t_8^5 +  2 y_{s}^5 y_{o_1}^{11} y_{o_2}^9 y_{o_3}^{11} y_{o_4}^9 y_{o_5}^{10} y_{o_6}^{10} y_{o_7}^{15} t_1^2 t_2^8 t_3^6 t_4^4 t_5^5 t_6^4 t_7^6 t_8^5 +  y_{s}^5 y_{o_1}^{11} y_{o_2}^9 y_{o_3}^{11} y_{o_4}^9 y_{o_5}^{10} y_{o_6}^{10} y_{o_7}^{15} t_1 t_2^9 t_3^6 t_4^4 t_5^5 t_6^4 t_7^6 t_8^5 +  y_{s}^6 y_{o_1}^{12} y_{o_2}^{12} y_{o_3}^{13} y_{o_4}^{11} y_{o_5}^{11} y_{o_6}^{13} y_{o_7}^{17} t_1^{10} t_2 t_3^7 t_4^6 t_5^7 t_6^5 t_7^6 t_8^5 +  y_{s}^6 y_{o_1}^{12} y_{o_2}^{12} y_{o_3}^{13} y_{o_4}^{11} y_{o_5}^{11} y_{o_6}^{13} y_{o_7}^{17} t_1^9 t_2^2 t_3^7 t_4^6 t_5^7 t_6^5 t_7^6 t_8^5 +  y_{s}^6 y_{o_1}^{12} y_{o_2}^{12} y_{o_3}^{13} y_{o_4}^{11} y_{o_5}^{11} y_{o_6}^{13} y_{o_7}^{17} t_1^8 t_2^3 t_3^7 t_4^6 t_5^7 t_6^5 t_7^6 t_8^5 +  4 y_{s}^6 y_{o_1}^{12} y_{o_2}^{12} y_{o_3}^{13} y_{o_4}^{11} y_{o_5}^{11} y_{o_6}^{13} y_{o_7}^{17} t_1^7 t_2^4 t_3^7 t_4^6 t_5^7 t_6^5 t_7^6 t_8^5 +  2 y_{s}^6 y_{o_1}^{12} y_{o_2}^{12} y_{o_3}^{13} y_{o_4}^{11} y_{o_5}^{11} y_{o_6}^{13} y_{o_7}^{17} t_1^6 t_2^5 t_3^7 t_4^6 t_5^7 t_6^5 t_7^6 t_8^5 +  2 y_{s}^6 y_{o_1}^{12} y_{o_2}^{12} y_{o_3}^{13} y_{o_4}^{11} y_{o_5}^{11} y_{o_6}^{13} y_{o_7}^{17} t_1^5 t_2^6 t_3^7 t_4^6 t_5^7 t_6^5 t_7^6 t_8^5 +  4 y_{s}^6 y_{o_1}^{12} y_{o_2}^{12} y_{o_3}^{13} y_{o_4}^{11} y_{o_5}^{11} y_{o_6}^{13} y_{o_7}^{17} t_1^4 t_2^7 t_3^7 t_4^6 t_5^7 t_6^5 t_7^6 t_8^5 +  y_{s}^6 y_{o_1}^{12} y_{o_2}^{12} y_{o_3}^{13} y_{o_4}^{11} y_{o_5}^{11} y_{o_6}^{13} y_{o_7}^{17} t_1^3 t_2^8 t_3^7 t_4^6 t_5^7 t_6^5 t_7^6 t_8^5 +  y_{s}^6 y_{o_1}^{12} y_{o_2}^{12} y_{o_3}^{13} y_{o_4}^{11} y_{o_5}^{11} y_{o_6}^{13} y_{o_7}^{17} t_1^2 t_2^9 t_3^7 t_4^6 t_5^7 t_6^5 t_7^6 t_8^5 +  y_{s}^6 y_{o_1}^{12} y_{o_2}^{12} y_{o_3}^{13} y_{o_4}^{11} y_{o_5}^{11} y_{o_6}^{13} y_{o_7}^{17} t_1 t_2^{10} t_3^7 t_4^6 t_5^7 t_6^5 t_7^6 t_8^5 -  y_{s}^7 y_{o_1}^{13} y_{o_2}^{15} y_{o_3}^{15} y_{o_4}^{13} y_{o_5}^{12} y_{o_6}^{16} y_{o_7}^{19} t_1^{10} t_2^2 t_3^8 t_4^8 t_5^9 t_6^6 t_7^6 t_8^5 -  y_{s}^7 y_{o_1}^{13} y_{o_2}^{15} y_{o_3}^{15} y_{o_4}^{13} y_{o_5}^{12} y_{o_6}^{16} y_{o_7}^{19} t_1^9 t_2^3 t_3^8 t_4^8 t_5^9 t_6^6 t_7^6 t_8^5 -  3 y_{s}^7 y_{o_1}^{13} y_{o_2}^{15} y_{o_3}^{15} y_{o_4}^{13} y_{o_5}^{12} y_{o_6}^{16} y_{o_7}^{19} t_1^8 t_2^4 t_3^8 t_4^8 t_5^9 t_6^6 t_7^6 t_8^5 -  3 y_{s}^7 y_{o_1}^{13} y_{o_2}^{15} y_{o_3}^{15} y_{o_4}^{13} y_{o_5}^{12} y_{o_6}^{16} y_{o_7}^{19} t_1^7 t_2^5 t_3^8 t_4^8 t_5^9 t_6^6 t_7^6 t_8^5 -  3 y_{s}^7 y_{o_1}^{13} y_{o_2}^{15} y_{o_3}^{15} y_{o_4}^{13} y_{o_5}^{12} y_{o_6}^{16} y_{o_7}^{19} t_1^6 t_2^6 t_3^8 t_4^8 t_5^9 t_6^6 t_7^6 t_8^5 -  3 y_{s}^7 y_{o_1}^{13} y_{o_2}^{15} y_{o_3}^{15} y_{o_4}^{13} y_{o_5}^{12} y_{o_6}^{16} y_{o_7}^{19} t_1^5 t_2^7 t_3^8 t_4^8 t_5^9 t_6^6 t_7^6 t_8^5 -  3 y_{s}^7 y_{o_1}^{13} y_{o_2}^{15} y_{o_3}^{15} y_{o_4}^{13} y_{o_5}^{12} y_{o_6}^{16} y_{o_7}^{19} t_1^4 t_2^8 t_3^8 t_4^8 t_5^9 t_6^6 t_7^6 t_8^5 -  y_{s}^7 y_{o_1}^{13} y_{o_2}^{15} y_{o_3}^{15} y_{o_4}^{13} y_{o_5}^{12} y_{o_6}^{16} y_{o_7}^{19} t_1^3 t_2^9 t_3^8 t_4^8 t_5^9 t_6^6 t_7^6 t_8^5 -  y_{s}^7 y_{o_1}^{13} y_{o_2}^{15} y_{o_3}^{15} y_{o_4}^{13} y_{o_5}^{12} y_{o_6}^{16} y_{o_7}^{19} t_1^2 t_2^{10} t_3^8 t_4^8 t_5^9 t_6^6 t_7^6 t_8^5 +  y_{s}^5 y_{o_1}^{12} y_{o_2}^8 y_{o_3}^{12} y_{o_4}^8 y_{o_5}^{10} y_{o_6}^{10} y_{o_7}^{15} t_1^7 t_2^3 t_3^7 t_4^3 t_5^5 t_6^3 t_7^7 t_8^5 +  2 y_{s}^5 y_{o_1}^{12} y_{o_2}^8 y_{o_3}^{12} y_{o_4}^8 y_{o_5}^{10} y_{o_6}^{10} y_{o_7}^{15} t_1^6 t_2^4 t_3^7 t_4^3 t_5^5 t_6^3 t_7^7 t_8^5 -  y_{s}^5 y_{o_1}^{12} y_{o_2}^8 y_{o_3}^{12} y_{o_4}^8 y_{o_5}^{10} y_{o_6}^{10} y_{o_7}^{15} t_1^5 t_2^5 t_3^7 t_4^3 t_5^5 t_6^3 t_7^7 t_8^5 +  2 y_{s}^5 y_{o_1}^{12} y_{o_2}^8 y_{o_3}^{12} y_{o_4}^8 y_{o_5}^{10} y_{o_6}^{10} y_{o_7}^{15} t_1^4 t_2^6 t_3^7 t_4^3 t_5^5 t_6^3 t_7^7 t_8^5 +  y_{s}^5 y_{o_1}^{12} y_{o_2}^8 y_{o_3}^{12} y_{o_4}^8 y_{o_5}^{10} y_{o_6}^{10} y_{o_7}^{15} t_1^3 t_2^7 t_3^7 t_4^3 t_5^5 t_6^3 t_7^7 t_8^5 -  y_{s}^6 y_{o_1}^{13} y_{o_2}^{11} y_{o_3}^{14} y_{o_4}^{10} y_{o_5}^{11} y_{o_6}^{13} y_{o_7}^{17} t_1^8 t_2^3 t_3^8 t_4^5 t_5^7 t_6^4 t_7^7 t_8^5 -  y_{s}^6 y_{o_1}^{13} y_{o_2}^{11} y_{o_3}^{14} y_{o_4}^{10} y_{o_5}^{11} y_{o_6}^{13} y_{o_7}^{17} t_1^7 t_2^4 t_3^8 t_4^5 t_5^7 t_6^4 t_7^7 t_8^5 -  y_{s}^6 y_{o_1}^{13} y_{o_2}^{11} y_{o_3}^{14} y_{o_4}^{10} y_{o_5}^{11} y_{o_6}^{13} y_{o_7}^{17} t_1^4 t_2^7 t_3^8 t_4^5 t_5^7 t_6^4 t_7^7 t_8^5 -  y_{s}^6 y_{o_1}^{13} y_{o_2}^{11} y_{o_3}^{14} y_{o_4}^{10} y_{o_5}^{11} y_{o_6}^{13} y_{o_7}^{17} t_1^3 t_2^8 t_3^8 t_4^5 t_5^7 t_6^4 t_7^7 t_8^5 -  y_{s}^7 y_{o_1}^{14} y_{o_2}^{14} y_{o_3}^{16} y_{o_4}^{12} y_{o_5}^{12} y_{o_6}^{16} y_{o_7}^{19} t_1^6 t_2^6 t_3^9 t_4^7 t_5^9 t_6^5 t_7^7 t_8^5 -  y_{s}^5 y_{o_1}^{13} y_{o_2}^7 y_{o_3}^{13} y_{o_4}^7 y_{o_5}^{10} y_{o_6}^{10} y_{o_7}^{15} t_1^5 t_2^5 t_3^8 t_4^2 t_5^5 t_6^2 t_7^8 t_8^5 +  y_{s}^6 y_{o_1}^{14} y_{o_2}^{10} y_{o_3}^{15} y_{o_4}^9 y_{o_5}^{11} y_{o_6}^{13} y_{o_7}^{17} t_1^6 t_2^5 t_3^9 t_4^4 t_5^7 t_6^3 t_7^8 t_8^5 +  y_{s}^6 y_{o_1}^{14} y_{o_2}^{10} y_{o_3}^{15} y_{o_4}^9 y_{o_5}^{11} y_{o_6}^{13} y_{o_7}^{17} t_1^5 t_2^6 t_3^9 t_4^4 t_5^7 t_6^3 t_7^8 t_8^5 -  y_{s}^7 y_{o_1}^{15} y_{o_2}^{13} y_{o_3}^{17} y_{o_4}^{11} y_{o_5}^{12} y_{o_6}^{16} y_{o_7}^{19} t_1^6 t_2^6 t_3^{10} t_4^6 t_5^9 t_6^4 t_7^8 t_8^5 -  y_{s}^4 y_{o_1}^7 y_{o_2}^9 y_{o_3}^5 y_{o_4}^{11} y_{o_5}^{10} y_{o_6}^6 y_{o_7}^{14} t_1^5 t_2^5 t_3 t_4^5 t_5^2 t_6^7 t_7^3 t_8^6 +  y_{s}^5 y_{o_1}^8 y_{o_2}^{12} y_{o_3}^7 y_{o_4}^{13} y_{o_5}^{11} y_{o_6}^9 y_{o_7}^{16} t_1^6 t_2^5 t_3^2 t_4^7 t_5^4 t_6^8 t_7^3 t_8^6 +  y_{s}^5 y_{o_1}^8 y_{o_2}^{12} y_{o_3}^7 y_{o_4}^{13} y_{o_5}^{11} y_{o_6}^9 y_{o_7}^{16} t_1^5 t_2^6 t_3^2 t_4^7 t_5^4 t_6^8 t_7^3 t_8^6 -  y_{s}^6 y_{o_1}^9 y_{o_2}^{15} y_{o_3}^9 y_{o_4}^{15} y_{o_5}^{12} y_{o_6}^{12} y_{o_7}^{18} t_1^6 t_2^6 t_3^3 t_4^9 t_5^6 t_6^9 t_7^3 t_8^6 -  y_{s}^4 y_{o_1}^8 y_{o_2}^8 y_{o_3}^6 y_{o_4}^{10} y_{o_5}^{10} y_{o_6}^6 y_{o_7}^{14} t_1^5 t_2^5 t_3^2 t_4^4 t_5^2 t_6^6 t_7^4 t_8^6 -  y_{s}^5 y_{o_1}^9 y_{o_2}^{11} y_{o_3}^8 y_{o_4}^{12} y_{o_5}^{11} y_{o_6}^9 y_{o_7}^{16} t_1^8 t_2^3 t_3^3 t_4^6 t_5^4 t_6^7 t_7^4 t_8^6 -  y_{s}^5 y_{o_1}^9 y_{o_2}^{11} y_{o_3}^8 y_{o_4}^{12} y_{o_5}^{11} y_{o_6}^9 y_{o_7}^{16} t_1^7 t_2^4 t_3^3 t_4^6 t_5^4 t_6^7 t_7^4 t_8^6 -  y_{s}^5 y_{o_1}^9 y_{o_2}^{11} y_{o_3}^8 y_{o_4}^{12} y_{o_5}^{11} y_{o_6}^9 y_{o_7}^{16} t_1^4 t_2^7 t_3^3 t_4^6 t_5^4 t_6^7 t_7^4 t_8^6 -  y_{s}^5 y_{o_1}^9 y_{o_2}^{11} y_{o_3}^8 y_{o_4}^{12} y_{o_5}^{11} y_{o_6}^9 y_{o_7}^{16} t_1^3 t_2^8 t_3^3 t_4^6 t_5^4 t_6^7 t_7^4 t_8^6 +  y_{s}^6 y_{o_1}^{10} y_{o_2}^{14} y_{o_3}^{10} y_{o_4}^{14} y_{o_5}^{12} y_{o_6}^{12} y_{o_7}^{18} t_1^8 t_2^4 t_3^4 t_4^8 t_5^6 t_6^8 t_7^4 t_8^6 +  2 y_{s}^6 y_{o_1}^{10} y_{o_2}^{14} y_{o_3}^{10} y_{o_4}^{14} y_{o_5}^{12} y_{o_6}^{12} y_{o_7}^{18} t_1^7 t_2^5 t_3^4 t_4^8 t_5^6 t_6^8 t_7^4 t_8^6 -  y_{s}^6 y_{o_1}^{10} y_{o_2}^{14} y_{o_3}^{10} y_{o_4}^{14} y_{o_5}^{12} y_{o_6}^{12} y_{o_7}^{18} t_1^6 t_2^6 t_3^4 t_4^8 t_5^6 t_6^8 t_7^4 t_8^6 +  2 y_{s}^6 y_{o_1}^{10} y_{o_2}^{14} y_{o_3}^{10} y_{o_4}^{14} y_{o_5}^{12} y_{o_6}^{12} y_{o_7}^{18} t_1^5 t_2^7 t_3^4 t_4^8 t_5^6 t_6^8 t_7^4 t_8^6 +  y_{s}^6 y_{o_1}^{10} y_{o_2}^{14} y_{o_3}^{10} y_{o_4}^{14} y_{o_5}^{12} y_{o_6}^{12} y_{o_7}^{18} t_1^4 t_2^8 t_3^4 t_4^8 t_5^6 t_6^8 t_7^4 t_8^6 -  y_{s}^4 y_{o_1}^9 y_{o_2}^7 y_{o_3}^7 y_{o_4}^9 y_{o_5}^{10} y_{o_6}^6 y_{o_7}^{14} t_1^9 t_2 t_3^3 t_4^3 t_5^2 t_6^5 t_7^5 t_8^6 -  y_{s}^4 y_{o_1}^9 y_{o_2}^7 y_{o_3}^7 y_{o_4}^9 y_{o_5}^{10} y_{o_6}^6 y_{o_7}^{14} t_1^8 t_2^2 t_3^3 t_4^3 t_5^2 t_6^5 t_7^5 t_8^6 -  3 y_{s}^4 y_{o_1}^9 y_{o_2}^7 y_{o_3}^7 y_{o_4}^9 y_{o_5}^{10} y_{o_6}^6 y_{o_7}^{14} t_1^7 t_2^3 t_3^3 t_4^3 t_5^2 t_6^5 t_7^5 t_8^6 -  3 y_{s}^4 y_{o_1}^9 y_{o_2}^7 y_{o_3}^7 y_{o_4}^9 y_{o_5}^{10} y_{o_6}^6 y_{o_7}^{14} t_1^6 t_2^4 t_3^3 t_4^3 t_5^2 t_6^5 t_7^5 t_8^6 -  3 y_{s}^4 y_{o_1}^9 y_{o_2}^7 y_{o_3}^7 y_{o_4}^9 y_{o_5}^{10} y_{o_6}^6 y_{o_7}^{14} t_1^5 t_2^5 t_3^3 t_4^3 t_5^2 t_6^5 t_7^5 t_8^6 -  3 y_{s}^4 y_{o_1}^9 y_{o_2}^7 y_{o_3}^7 y_{o_4}^9 y_{o_5}^{10} y_{o_6}^6 y_{o_7}^{14} t_1^4 t_2^6 t_3^3 t_4^3 t_5^2 t_6^5 t_7^5 t_8^6 -  3 y_{s}^4 y_{o_1}^9 y_{o_2}^7 y_{o_3}^7 y_{o_4}^9 y_{o_5}^{10} y_{o_6}^6 y_{o_7}^{14} t_1^3 t_2^7 t_3^3 t_4^3 t_5^2 t_6^5 t_7^5 t_8^6 -  y_{s}^4 y_{o_1}^9 y_{o_2}^7 y_{o_3}^7 y_{o_4}^9 y_{o_5}^{10} y_{o_6}^6 y_{o_7}^{14} t_1^2 t_2^8 t_3^3 t_4^3 t_5^2 t_6^5 t_7^5 t_8^6 -  y_{s}^4 y_{o_1}^9 y_{o_2}^7 y_{o_3}^7 y_{o_4}^9 y_{o_5}^{10} y_{o_6}^6 y_{o_7}^{14} t_1 t_2^9 t_3^3 t_4^3 t_5^2 t_6^5 t_7^5 t_8^6 +  y_{s}^5 y_{o_1}^{10} y_{o_2}^{10} y_{o_3}^9 y_{o_4}^{11} y_{o_5}^{11} y_{o_6}^9 y_{o_7}^{16} t_1^{10} t_2 t_3^4 t_4^5 t_5^4 t_6^6 t_7^5 t_8^6 +  y_{s}^5 y_{o_1}^{10} y_{o_2}^{10} y_{o_3}^9 y_{o_4}^{11} y_{o_5}^{11} y_{o_6}^9 y_{o_7}^{16} t_1^9 t_2^2 t_3^4 t_4^5 t_5^4 t_6^6 t_7^5 t_8^6 +  y_{s}^5 y_{o_1}^{10} y_{o_2}^{10} y_{o_3}^9 y_{o_4}^{11} y_{o_5}^{11} y_{o_6}^9 y_{o_7}^{16} t_1^8 t_2^3 t_3^4 t_4^5 t_5^4 t_6^6 t_7^5 t_8^6 +  4 y_{s}^5 y_{o_1}^{10} y_{o_2}^{10} y_{o_3}^9 y_{o_4}^{11} y_{o_5}^{11} y_{o_6}^9 y_{o_7}^{16} t_1^7 t_2^4 t_3^4 t_4^5 t_5^4 t_6^6 t_7^5 t_8^6 +  2 y_{s}^5 y_{o_1}^{10} y_{o_2}^{10} y_{o_3}^9 y_{o_4}^{11} y_{o_5}^{11} y_{o_6}^9 y_{o_7}^{16} t_1^6 t_2^5 t_3^4 t_4^5 t_5^4 t_6^6 t_7^5 t_8^6 +  2 y_{s}^5 y_{o_1}^{10} y_{o_2}^{10} y_{o_3}^9 y_{o_4}^{11} y_{o_5}^{11} y_{o_6}^9 y_{o_7}^{16} t_1^5 t_2^6 t_3^4 t_4^5 t_5^4 t_6^6 t_7^5 t_8^6 +  4 y_{s}^5 y_{o_1}^{10} y_{o_2}^{10} y_{o_3}^9 y_{o_4}^{11} y_{o_5}^{11} y_{o_6}^9 y_{o_7}^{16} t_1^4 t_2^7 t_3^4 t_4^5 t_5^4 t_6^6 t_7^5 t_8^6 +  y_{s}^5 y_{o_1}^{10} y_{o_2}^{10} y_{o_3}^9 y_{o_4}^{11} y_{o_5}^{11} y_{o_6}^9 y_{o_7}^{16} t_1^3 t_2^8 t_3^4 t_4^5 t_5^4 t_6^6 t_7^5 t_8^6 +  y_{s}^5 y_{o_1}^{10} y_{o_2}^{10} y_{o_3}^9 y_{o_4}^{11} y_{o_5}^{11} y_{o_6}^9 y_{o_7}^{16} t_1^2 t_2^9 t_3^4 t_4^5 t_5^4 t_6^6 t_7^5 t_8^6 +  y_{s}^5 y_{o_1}^{10} y_{o_2}^{10} y_{o_3}^9 y_{o_4}^{11} y_{o_5}^{11} y_{o_6}^9 y_{o_7}^{16} t_1 t_2^{10} t_3^4 t_4^5 t_5^4 t_6^6 t_7^5 t_8^6 +  y_{s}^6 y_{o_1}^{11} y_{o_2}^{13} y_{o_3}^{11} y_{o_4}^{13} y_{o_5}^{12} y_{o_6}^{12} y_{o_7}^{18} t_1^{10} t_2^2 t_3^5 t_4^7 t_5^6 t_6^7 t_7^5 t_8^6 +  2 y_{s}^6 y_{o_1}^{11} y_{o_2}^{13} y_{o_3}^{11} y_{o_4}^{13} y_{o_5}^{12} y_{o_6}^{12} y_{o_7}^{18} t_1^9 t_2^3 t_3^5 t_4^7 t_5^6 t_6^7 t_7^5 t_8^6 +  3 y_{s}^6 y_{o_1}^{11} y_{o_2}^{13} y_{o_3}^{11} y_{o_4}^{13} y_{o_5}^{12} y_{o_6}^{12} y_{o_7}^{18} t_1^8 t_2^4 t_3^5 t_4^7 t_5^6 t_6^7 t_7^5 t_8^6 +  6 y_{s}^6 y_{o_1}^{11} y_{o_2}^{13} y_{o_3}^{11} y_{o_4}^{13} y_{o_5}^{12} y_{o_6}^{12} y_{o_7}^{18} t_1^7 t_2^5 t_3^5 t_4^7 t_5^6 t_6^7 t_7^5 t_8^6 +  3 y_{s}^6 y_{o_1}^{11} y_{o_2}^{13} y_{o_3}^{11} y_{o_4}^{13} y_{o_5}^{12} y_{o_6}^{12} y_{o_7}^{18} t_1^6 t_2^6 t_3^5 t_4^7 t_5^6 t_6^7 t_7^5 t_8^6 +  6 y_{s}^6 y_{o_1}^{11} y_{o_2}^{13} y_{o_3}^{11} y_{o_4}^{13} y_{o_5}^{12} y_{o_6}^{12} y_{o_7}^{18} t_1^5 t_2^7 t_3^5 t_4^7 t_5^6 t_6^7 t_7^5 t_8^6 +  3 y_{s}^6 y_{o_1}^{11} y_{o_2}^{13} y_{o_3}^{11} y_{o_4}^{13} y_{o_5}^{12} y_{o_6}^{12} y_{o_7}^{18} t_1^4 t_2^8 t_3^5 t_4^7 t_5^6 t_6^7 t_7^5 t_8^6 +  2 y_{s}^6 y_{o_1}^{11} y_{o_2}^{13} y_{o_3}^{11} y_{o_4}^{13} y_{o_5}^{12} y_{o_6}^{12} y_{o_7}^{18} t_1^3 t_2^9 t_3^5 t_4^7 t_5^6 t_6^7 t_7^5 t_8^6 +  y_{s}^6 y_{o_1}^{11} y_{o_2}^{13} y_{o_3}^{11} y_{o_4}^{13} y_{o_5}^{12} y_{o_6}^{12} y_{o_7}^{18} t_1^2 t_2^{10} t_3^5 t_4^7 t_5^6 t_6^7 t_7^5 t_8^6 -  y_{s}^7 y_{o_1}^{12} y_{o_2}^{16} y_{o_3}^{13} y_{o_4}^{15} y_{o_5}^{13} y_{o_6}^{15} y_{o_7}^{20} t_1^{10} t_2^3 t_3^6 t_4^9 t_5^8 t_6^8 t_7^5 t_8^6 -  2 y_{s}^7 y_{o_1}^{12} y_{o_2}^{16} y_{o_3}^{13} y_{o_4}^{15} y_{o_5}^{13} y_{o_6}^{15} y_{o_7}^{20} t_1^9 t_2^4 t_3^6 t_4^9 t_5^8 t_6^8 t_7^5 t_8^6 -  3 y_{s}^7 y_{o_1}^{12} y_{o_2}^{16} y_{o_3}^{13} y_{o_4}^{15} y_{o_5}^{13} y_{o_6}^{15} y_{o_7}^{20} t_1^8 t_2^5 t_3^6 t_4^9 t_5^8 t_6^8 t_7^5 t_8^6 -  4 y_{s}^7 y_{o_1}^{12} y_{o_2}^{16} y_{o_3}^{13} y_{o_4}^{15} y_{o_5}^{13} y_{o_6}^{15} y_{o_7}^{20} t_1^7 t_2^6 t_3^6 t_4^9 t_5^8 t_6^8 t_7^5 t_8^6 -  4 y_{s}^7 y_{o_1}^{12} y_{o_2}^{16} y_{o_3}^{13} y_{o_4}^{15} y_{o_5}^{13} y_{o_6}^{15} y_{o_7}^{20} t_1^6 t_2^7 t_3^6 t_4^9 t_5^8 t_6^8 t_7^5 t_8^6 -  3 y_{s}^7 y_{o_1}^{12} y_{o_2}^{16} y_{o_3}^{13} y_{o_4}^{15} y_{o_5}^{13} y_{o_6}^{15} y_{o_7}^{20} t_1^5 t_2^8 t_3^6 t_4^9 t_5^8 t_6^8 t_7^5 t_8^6 -  2 y_{s}^7 y_{o_1}^{12} y_{o_2}^{16} y_{o_3}^{13} y_{o_4}^{15} y_{o_5}^{13} y_{o_6}^{15} y_{o_7}^{20} t_1^4 t_2^9 t_3^6 t_4^9 t_5^8 t_6^8 t_7^5 t_8^6 -  y_{s}^7 y_{o_1}^{12} y_{o_2}^{16} y_{o_3}^{13} y_{o_4}^{15} y_{o_5}^{13} y_{o_6}^{15} y_{o_7}^{20} t_1^3 t_2^{10} t_3^6 t_4^9 t_5^8 t_6^8 t_7^5 t_8^6 -  y_{s}^4 y_{o_1}^{10} y_{o_2}^6 y_{o_3}^8 y_{o_4}^8 y_{o_5}^{10} y_{o_6}^6 y_{o_7}^{14} t_1^5 t_2^5 t_3^4 t_4^2 t_5^2 t_6^4 t_7^6 t_8^6 +  y_{s}^5 y_{o_1}^{11} y_{o_2}^9 y_{o_3}^{10} y_{o_4}^{10} y_{o_5}^{11} y_{o_6}^9 y_{o_7}^{16} t_1^{10} t_2 t_3^5 t_4^4 t_5^4 t_6^5 t_7^6 t_8^6 +  y_{s}^5 y_{o_1}^{11} y_{o_2}^9 y_{o_3}^{10} y_{o_4}^{10} y_{o_5}^{11} y_{o_6}^9 y_{o_7}^{16} t_1^9 t_2^2 t_3^5 t_4^4 t_5^4 t_6^5 t_7^6 t_8^6 +  y_{s}^5 y_{o_1}^{11} y_{o_2}^9 y_{o_3}^{10} y_{o_4}^{10} y_{o_5}^{11} y_{o_6}^9 y_{o_7}^{16} t_1^8 t_2^3 t_3^5 t_4^4 t_5^4 t_6^5 t_7^6 t_8^6 +  4 y_{s}^5 y_{o_1}^{11} y_{o_2}^9 y_{o_3}^{10} y_{o_4}^{10} y_{o_5}^{11} y_{o_6}^9 y_{o_7}^{16} t_1^7 t_2^4 t_3^5 t_4^4 t_5^4 t_6^5 t_7^6 t_8^6 +  2 y_{s}^5 y_{o_1}^{11} y_{o_2}^9 y_{o_3}^{10} y_{o_4}^{10} y_{o_5}^{11} y_{o_6}^9 y_{o_7}^{16} t_1^6 t_2^5 t_3^5 t_4^4 t_5^4 t_6^5 t_7^6 t_8^6 +  2 y_{s}^5 y_{o_1}^{11} y_{o_2}^9 y_{o_3}^{10} y_{o_4}^{10} y_{o_5}^{11} y_{o_6}^9 y_{o_7}^{16} t_1^5 t_2^6 t_3^5 t_4^4 t_5^4 t_6^5 t_7^6 t_8^6 +  4 y_{s}^5 y_{o_1}^{11} y_{o_2}^9 y_{o_3}^{10} y_{o_4}^{10} y_{o_5}^{11} y_{o_6}^9 y_{o_7}^{16} t_1^4 t_2^7 t_3^5 t_4^4 t_5^4 t_6^5 t_7^6 t_8^6 +  y_{s}^5 y_{o_1}^{11} y_{o_2}^9 y_{o_3}^{10} y_{o_4}^{10} y_{o_5}^{11} y_{o_6}^9 y_{o_7}^{16} t_1^3 t_2^8 t_3^5 t_4^4 t_5^4 t_6^5 t_7^6 t_8^6 +  y_{s}^5 y_{o_1}^{11} y_{o_2}^9 y_{o_3}^{10} y_{o_4}^{10} y_{o_5}^{11} y_{o_6}^9 y_{o_7}^{16} t_1^2 t_2^9 t_3^5 t_4^4 t_5^4 t_6^5 t_7^6 t_8^6 +  y_{s}^5 y_{o_1}^{11} y_{o_2}^9 y_{o_3}^{10} y_{o_4}^{10} y_{o_5}^{11} y_{o_6}^9 y_{o_7}^{16} t_1 t_2^{10} t_3^5 t_4^4 t_5^4 t_6^5 t_7^6 t_8^6 -  y_{s}^6 y_{o_1}^{12} y_{o_2}^{12} y_{o_3}^{12} y_{o_4}^{12} y_{o_5}^{12} y_{o_6}^{12} y_{o_7}^{18} t_1^{11} t_2 t_3^6 t_4^6 t_5^6 t_6^6 t_7^6 t_8^6 -  y_{s}^6 y_{o_1}^{12} y_{o_2}^{12} y_{o_3}^{12} y_{o_4}^{12} y_{o_5}^{12} y_{o_6}^{12} y_{o_7}^{18} t_1^{10} t_2^2 t_3^6 t_4^6 t_5^6 t_6^6 t_7^6 t_8^6 -  y_{s}^6 y_{o_1}^{12} y_{o_2}^{12} y_{o_3}^{12} y_{o_4}^{12} y_{o_5}^{12} y_{o_6}^{12} y_{o_7}^{18} t_1^9 t_2^3 t_3^6 t_4^6 t_5^6 t_6^6 t_7^6 t_8^6 -  2 y_{s}^6 y_{o_1}^{12} y_{o_2}^{12} y_{o_3}^{12} y_{o_4}^{12} y_{o_5}^{12} y_{o_6}^{12} y_{o_7}^{18} t_1^8 t_2^4 t_3^6 t_4^6 t_5^6 t_6^6 t_7^6 t_8^6 +  2 y_{s}^6 y_{o_1}^{12} y_{o_2}^{12} y_{o_3}^{12} y_{o_4}^{12} y_{o_5}^{12} y_{o_6}^{12} y_{o_7}^{18} t_1^7 t_2^5 t_3^6 t_4^6 t_5^6 t_6^6 t_7^6 t_8^6 +  2 y_{s}^6 y_{o_1}^{12} y_{o_2}^{12} y_{o_3}^{12} y_{o_4}^{12} y_{o_5}^{12} y_{o_6}^{12} y_{o_7}^{18} t_1^5 t_2^7 t_3^6 t_4^6 t_5^6 t_6^6 t_7^6 t_8^6 -  2 y_{s}^6 y_{o_1}^{12} y_{o_2}^{12} y_{o_3}^{12} y_{o_4}^{12} y_{o_5}^{12} y_{o_6}^{12} y_{o_7}^{18} t_1^4 t_2^8 t_3^6 t_4^6 t_5^6 t_6^6 t_7^6 t_8^6 -  y_{s}^6 y_{o_1}^{12} y_{o_2}^{12} y_{o_3}^{12} y_{o_4}^{12} y_{o_5}^{12} y_{o_6}^{12} y_{o_7}^{18} t_1^3 t_2^9 t_3^6 t_4^6 t_5^6 t_6^6 t_7^6 t_8^6 -  y_{s}^6 y_{o_1}^{12} y_{o_2}^{12} y_{o_3}^{12} y_{o_4}^{12} y_{o_5}^{12} y_{o_6}^{12} y_{o_7}^{18} t_1^2 t_2^{10} t_3^6 t_4^6 t_5^6 t_6^6 t_7^6 t_8^6 -  y_{s}^6 y_{o_1}^{12} y_{o_2}^{12} y_{o_3}^{12} y_{o_4}^{12} y_{o_5}^{12} y_{o_6}^{12} y_{o_7}^{18} t_1 t_2^{11} t_3^6 t_4^6 t_5^6 t_6^6 t_7^6 t_8^6 -  y_{s}^7 y_{o_1}^{13} y_{o_2}^{15} y_{o_3}^{14} y_{o_4}^{14} y_{o_5}^{13} y_{o_6}^{15} y_{o_7}^{20} t_1^{10} t_2^3 t_3^7 t_4^8 t_5^8 t_6^7 t_7^6 t_8^6 -  2 y_{s}^7 y_{o_1}^{13} y_{o_2}^{15} y_{o_3}^{14} y_{o_4}^{14} y_{o_5}^{13} y_{o_6}^{15} y_{o_7}^{20} t_1^9 t_2^4 t_3^7 t_4^8 t_5^8 t_6^7 t_7^6 t_8^6 -  3 y_{s}^7 y_{o_1}^{13} y_{o_2}^{15} y_{o_3}^{14} y_{o_4}^{14} y_{o_5}^{13} y_{o_6}^{15} y_{o_7}^{20} t_1^8 t_2^5 t_3^7 t_4^8 t_5^8 t_6^7 t_7^6 t_8^6 -  6 y_{s}^7 y_{o_1}^{13} y_{o_2}^{15} y_{o_3}^{14} y_{o_4}^{14} y_{o_5}^{13} y_{o_6}^{15} y_{o_7}^{20} t_1^7 t_2^6 t_3^7 t_4^8 t_5^8 t_6^7 t_7^6 t_8^6 -  6 y_{s}^7 y_{o_1}^{13} y_{o_2}^{15} y_{o_3}^{14} y_{o_4}^{14} y_{o_5}^{13} y_{o_6}^{15} y_{o_7}^{20} t_1^6 t_2^7 t_3^7 t_4^8 t_5^8 t_6^7 t_7^6 t_8^6 -  3 y_{s}^7 y_{o_1}^{13} y_{o_2}^{15} y_{o_3}^{14} y_{o_4}^{14} y_{o_5}^{13} y_{o_6}^{15} y_{o_7}^{20} t_1^5 t_2^8 t_3^7 t_4^8 t_5^8 t_6^7 t_7^6 t_8^6 -  2 y_{s}^7 y_{o_1}^{13} y_{o_2}^{15} y_{o_3}^{14} y_{o_4}^{14} y_{o_5}^{13} y_{o_6}^{15} y_{o_7}^{20} t_1^4 t_2^9 t_3^7 t_4^8 t_5^8 t_6^7 t_7^6 t_8^6 -  y_{s}^7 y_{o_1}^{13} y_{o_2}^{15} y_{o_3}^{14} y_{o_4}^{14} y_{o_5}^{13} y_{o_6}^{15} y_{o_7}^{20} t_1^3 t_2^{10} t_3^7 t_4^8 t_5^8 t_6^7 t_7^6 t_8^6 +  y_{s}^8 y_{o_1}^{14} y_{o_2}^{18} y_{o_3}^{16} y_{o_4}^{16} y_{o_5}^{14} y_{o_6}^{18} y_{o_7}^{22} t_1^{10} t_2^4 t_3^8 t_4^{10} t_5^{10} t_6^8 t_7^6 t_8^6 +  y_{s}^8 y_{o_1}^{14} y_{o_2}^{18} y_{o_3}^{16} y_{o_4}^{16} y_{o_5}^{14} y_{o_6}^{18} y_{o_7}^{22} t_1^9 t_2^5 t_3^8 t_4^{10} t_5^{10} t_6^8 t_7^6 t_8^6 +  2 y_{s}^8 y_{o_1}^{14} y_{o_2}^{18} y_{o_3}^{16} y_{o_4}^{16} y_{o_5}^{14} y_{o_6}^{18} y_{o_7}^{22} t_1^8 t_2^6 t_3^8 t_4^{10} t_5^{10} t_6^8 t_7^6 t_8^6 +  2 y_{s}^8 y_{o_1}^{14} y_{o_2}^{18} y_{o_3}^{16} y_{o_4}^{16} y_{o_5}^{14} y_{o_6}^{18} y_{o_7}^{22} t_1^7 t_2^7 t_3^8 t_4^{10} t_5^{10} t_6^8 t_7^6 t_8^6 +  2 y_{s}^8 y_{o_1}^{14} y_{o_2}^{18} y_{o_3}^{16} y_{o_4}^{16} y_{o_5}^{14} y_{o_6}^{18} y_{o_7}^{22} t_1^6 t_2^8 t_3^8 t_4^{10} t_5^{10} t_6^8 t_7^6 t_8^6 +  y_{s}^8 y_{o_1}^{14} y_{o_2}^{18} y_{o_3}^{16} y_{o_4}^{16} y_{o_5}^{14} y_{o_6}^{18} y_{o_7}^{22} t_1^5 t_2^9 t_3^8 t_4^{10} t_5^{10} t_6^8 t_7^6 t_8^6 +  y_{s}^8 y_{o_1}^{14} y_{o_2}^{18} y_{o_3}^{16} y_{o_4}^{16} y_{o_5}^{14} y_{o_6}^{18} y_{o_7}^{22} t_1^4 t_2^{10} t_3^8 t_4^{10} t_5^{10} t_6^8 t_7^6 t_8^6 -  y_{s}^4 y_{o_1}^{11} y_{o_2}^5 y_{o_3}^9 y_{o_4}^7 y_{o_5}^{10} y_{o_6}^6 y_{o_7}^{14} t_1^5 t_2^5 t_3^5 t_4 t_5^2 t_6^3 t_7^7 t_8^6 -  y_{s}^5 y_{o_1}^{12} y_{o_2}^8 y_{o_3}^{11} y_{o_4}^9 y_{o_5}^{11} y_{o_6}^9 y_{o_7}^{16} t_1^8 t_2^3 t_3^6 t_4^3 t_5^4 t_6^4 t_7^7 t_8^6 -  y_{s}^5 y_{o_1}^{12} y_{o_2}^8 y_{o_3}^{11} y_{o_4}^9 y_{o_5}^{11} y_{o_6}^9 y_{o_7}^{16} t_1^7 t_2^4 t_3^6 t_4^3 t_5^4 t_6^4 t_7^7 t_8^6 -  y_{s}^5 y_{o_1}^{12} y_{o_2}^8 y_{o_3}^{11} y_{o_4}^9 y_{o_5}^{11} y_{o_6}^9 y_{o_7}^{16} t_1^4 t_2^7 t_3^6 t_4^3 t_5^4 t_6^4 t_7^7 t_8^6 -  y_{s}^5 y_{o_1}^{12} y_{o_2}^8 y_{o_3}^{11} y_{o_4}^9 y_{o_5}^{11} y_{o_6}^9 y_{o_7}^{16} t_1^3 t_2^8 t_3^6 t_4^3 t_5^4 t_6^4 t_7^7 t_8^6 +  y_{s}^6 y_{o_1}^{13} y_{o_2}^{11} y_{o_3}^{13} y_{o_4}^{11} y_{o_5}^{12} y_{o_6}^{12} y_{o_7}^{18} t_1^{10} t_2^2 t_3^7 t_4^5 t_5^6 t_6^5 t_7^7 t_8^6 +  2 y_{s}^6 y_{o_1}^{13} y_{o_2}^{11} y_{o_3}^{13} y_{o_4}^{11} y_{o_5}^{12} y_{o_6}^{12} y_{o_7}^{18} t_1^9 t_2^3 t_3^7 t_4^5 t_5^6 t_6^5 t_7^7 t_8^6 +  3 y_{s}^6 y_{o_1}^{13} y_{o_2}^{11} y_{o_3}^{13} y_{o_4}^{11} y_{o_5}^{12} y_{o_6}^{12} y_{o_7}^{18} t_1^8 t_2^4 t_3^7 t_4^5 t_5^6 t_6^5 t_7^7 t_8^6 +  6 y_{s}^6 y_{o_1}^{13} y_{o_2}^{11} y_{o_3}^{13} y_{o_4}^{11} y_{o_5}^{12} y_{o_6}^{12} y_{o_7}^{18} t_1^7 t_2^5 t_3^7 t_4^5 t_5^6 t_6^5 t_7^7 t_8^6 +  3 y_{s}^6 y_{o_1}^{13} y_{o_2}^{11} y_{o_3}^{13} y_{o_4}^{11} y_{o_5}^{12} y_{o_6}^{12} y_{o_7}^{18} t_1^6 t_2^6 t_3^7 t_4^5 t_5^6 t_6^5 t_7^7 t_8^6 +  6 y_{s}^6 y_{o_1}^{13} y_{o_2}^{11} y_{o_3}^{13} y_{o_4}^{11} y_{o_5}^{12} y_{o_6}^{12} y_{o_7}^{18} t_1^5 t_2^7 t_3^7 t_4^5 t_5^6 t_6^5 t_7^7 t_8^6 +  3 y_{s}^6 y_{o_1}^{13} y_{o_2}^{11} y_{o_3}^{13} y_{o_4}^{11} y_{o_5}^{12} y_{o_6}^{12} y_{o_7}^{18} t_1^4 t_2^8 t_3^7 t_4^5 t_5^6 t_6^5 t_7^7 t_8^6 +  2 y_{s}^6 y_{o_1}^{13} y_{o_2}^{11} y_{o_3}^{13} y_{o_4}^{11} y_{o_5}^{12} y_{o_6}^{12} y_{o_7}^{18} t_1^3 t_2^9 t_3^7 t_4^5 t_5^6 t_6^5 t_7^7 t_8^6 +  y_{s}^6 y_{o_1}^{13} y_{o_2}^{11} y_{o_3}^{13} y_{o_4}^{11} y_{o_5}^{12} y_{o_6}^{12} y_{o_7}^{18} t_1^2 t_2^{10} t_3^7 t_4^5 t_5^6 t_6^5 t_7^7 t_8^6 -  y_{s}^7 y_{o_1}^{14} y_{o_2}^{14} y_{o_3}^{15} y_{o_4}^{13} y_{o_5}^{13} y_{o_6}^{15} y_{o_7}^{20} t_1^{10} t_2^3 t_3^8 t_4^7 t_5^8 t_6^6 t_7^7 t_8^6 -  2 y_{s}^7 y_{o_1}^{14} y_{o_2}^{14} y_{o_3}^{15} y_{o_4}^{13} y_{o_5}^{13} y_{o_6}^{15} y_{o_7}^{20} t_1^9 t_2^4 t_3^8 t_4^7 t_5^8 t_6^6 t_7^7 t_8^6 -  3 y_{s}^7 y_{o_1}^{14} y_{o_2}^{14} y_{o_3}^{15} y_{o_4}^{13} y_{o_5}^{13} y_{o_6}^{15} y_{o_7}^{20} t_1^8 t_2^5 t_3^8 t_4^7 t_5^8 t_6^6 t_7^7 t_8^6 -  6 y_{s}^7 y_{o_1}^{14} y_{o_2}^{14} y_{o_3}^{15} y_{o_4}^{13} y_{o_5}^{13} y_{o_6}^{15} y_{o_7}^{20} t_1^7 t_2^6 t_3^8 t_4^7 t_5^8 t_6^6 t_7^7 t_8^6 -  6 y_{s}^7 y_{o_1}^{14} y_{o_2}^{14} y_{o_3}^{15} y_{o_4}^{13} y_{o_5}^{13} y_{o_6}^{15} y_{o_7}^{20} t_1^6 t_2^7 t_3^8 t_4^7 t_5^8 t_6^6 t_7^7 t_8^6 -  3 y_{s}^7 y_{o_1}^{14} y_{o_2}^{14} y_{o_3}^{15} y_{o_4}^{13} y_{o_5}^{13} y_{o_6}^{15} y_{o_7}^{20} t_1^5 t_2^8 t_3^8 t_4^7 t_5^8 t_6^6 t_7^7 t_8^6 -  2 y_{s}^7 y_{o_1}^{14} y_{o_2}^{14} y_{o_3}^{15} y_{o_4}^{13} y_{o_5}^{13} y_{o_6}^{15} y_{o_7}^{20} t_1^4 t_2^9 t_3^8 t_4^7 t_5^8 t_6^6 t_7^7 t_8^6 -  y_{s}^7 y_{o_1}^{14} y_{o_2}^{14} y_{o_3}^{15} y_{o_4}^{13} y_{o_5}^{13} y_{o_6}^{15} y_{o_7}^{20} t_1^3 t_2^{10} t_3^8 t_4^7 t_5^8 t_6^6 t_7^7 t_8^6 +  y_{s}^8 y_{o_1}^{15} y_{o_2}^{17} y_{o_3}^{17} y_{o_4}^{15} y_{o_5}^{14} y_{o_6}^{18} y_{o_7}^{22} t_1^8 t_2^6 t_3^9 t_4^9 t_5^{10} t_6^7 t_7^7 t_8^6 +  y_{s}^8 y_{o_1}^{15} y_{o_2}^{17} y_{o_3}^{17} y_{o_4}^{15} y_{o_5}^{14} y_{o_6}^{18} y_{o_7}^{22} t_1^7 t_2^7 t_3^9 t_4^9 t_5^{10} t_6^7 t_7^7 t_8^6 +  y_{s}^8 y_{o_1}^{15} y_{o_2}^{17} y_{o_3}^{17} y_{o_4}^{15} y_{o_5}^{14} y_{o_6}^{18} y_{o_7}^{22} t_1^6 t_2^8 t_3^9 t_4^9 t_5^{10} t_6^7 t_7^7 t_8^6 +  y_{s}^5 y_{o_1}^{13} y_{o_2}^7 y_{o_3}^{12} y_{o_4}^8 y_{o_5}^{11} y_{o_6}^9 y_{o_7}^{16} t_1^6 t_2^5 t_3^7 t_4^2 t_5^4 t_6^3 t_7^8 t_8^6 +  y_{s}^5 y_{o_1}^{13} y_{o_2}^7 y_{o_3}^{12} y_{o_4}^8 y_{o_5}^{11} y_{o_6}^9 y_{o_7}^{16} t_1^5 t_2^6 t_3^7 t_4^2 t_5^4 t_6^3 t_7^8 t_8^6 +  y_{s}^6 y_{o_1}^{14} y_{o_2}^{10} y_{o_3}^{14} y_{o_4}^{10} y_{o_5}^{12} y_{o_6}^{12} y_{o_7}^{18} t_1^8 t_2^4 t_3^8 t_4^4 t_5^6 t_6^4 t_7^8 t_8^6 +  2 y_{s}^6 y_{o_1}^{14} y_{o_2}^{10} y_{o_3}^{14} y_{o_4}^{10} y_{o_5}^{12} y_{o_6}^{12} y_{o_7}^{18} t_1^7 t_2^5 t_3^8 t_4^4 t_5^6 t_6^4 t_7^8 t_8^6 -  y_{s}^6 y_{o_1}^{14} y_{o_2}^{10} y_{o_3}^{14} y_{o_4}^{10} y_{o_5}^{12} y_{o_6}^{12} y_{o_7}^{18} t_1^6 t_2^6 t_3^8 t_4^4 t_5^6 t_6^4 t_7^8 t_8^6 +  2 y_{s}^6 y_{o_1}^{14} y_{o_2}^{10} y_{o_3}^{14} y_{o_4}^{10} y_{o_5}^{12} y_{o_6}^{12} y_{o_7}^{18} t_1^5 t_2^7 t_3^8 t_4^4 t_5^6 t_6^4 t_7^8 t_8^6 +  y_{s}^6 y_{o_1}^{14} y_{o_2}^{10} y_{o_3}^{14} y_{o_4}^{10} y_{o_5}^{12} y_{o_6}^{12} y_{o_7}^{18} t_1^4 t_2^8 t_3^8 t_4^4 t_5^6 t_6^4 t_7^8 t_8^6 -  y_{s}^7 y_{o_1}^{15} y_{o_2}^{13} y_{o_3}^{16} y_{o_4}^{12} y_{o_5}^{13} y_{o_6}^{15} y_{o_7}^{20} t_1^{10} t_2^3 t_3^9 t_4^6 t_5^8 t_6^5 t_7^8 t_8^6 -  2 y_{s}^7 y_{o_1}^{15} y_{o_2}^{13} y_{o_3}^{16} y_{o_4}^{12} y_{o_5}^{13} y_{o_6}^{15} y_{o_7}^{20} t_1^9 t_2^4 t_3^9 t_4^6 t_5^8 t_6^5 t_7^8 t_8^6 -  3 y_{s}^7 y_{o_1}^{15} y_{o_2}^{13} y_{o_3}^{16} y_{o_4}^{12} y_{o_5}^{13} y_{o_6}^{15} y_{o_7}^{20} t_1^8 t_2^5 t_3^9 t_4^6 t_5^8 t_6^5 t_7^8 t_8^6 -  4 y_{s}^7 y_{o_1}^{15} y_{o_2}^{13} y_{o_3}^{16} y_{o_4}^{12} y_{o_5}^{13} y_{o_6}^{15} y_{o_7}^{20} t_1^7 t_2^6 t_3^9 t_4^6 t_5^8 t_6^5 t_7^8 t_8^6 -  4 y_{s}^7 y_{o_1}^{15} y_{o_2}^{13} y_{o_3}^{16} y_{o_4}^{12} y_{o_5}^{13} y_{o_6}^{15} y_{o_7}^{20} t_1^6 t_2^7 t_3^9 t_4^6 t_5^8 t_6^5 t_7^8 t_8^6 -  3 y_{s}^7 y_{o_1}^{15} y_{o_2}^{13} y_{o_3}^{16} y_{o_4}^{12} y_{o_5}^{13} y_{o_6}^{15} y_{o_7}^{20} t_1^5 t_2^8 t_3^9 t_4^6 t_5^8 t_6^5 t_7^8 t_8^6 -  2 y_{s}^7 y_{o_1}^{15} y_{o_2}^{13} y_{o_3}^{16} y_{o_4}^{12} y_{o_5}^{13} y_{o_6}^{15} y_{o_7}^{20} t_1^4 t_2^9 t_3^9 t_4^6 t_5^8 t_6^5 t_7^8 t_8^6 -  y_{s}^7 y_{o_1}^{15} y_{o_2}^{13} y_{o_3}^{16} y_{o_4}^{12} y_{o_5}^{13} y_{o_6}^{15} y_{o_7}^{20} t_1^3 t_2^{10} t_3^9 t_4^6 t_5^8 t_6^5 t_7^8 t_8^6 +  y_{s}^8 y_{o_1}^{16} y_{o_2}^{16} y_{o_3}^{18} y_{o_4}^{14} y_{o_5}^{14} y_{o_6}^{18} y_{o_7}^{22} t_1^{10} t_2^4 t_3^{10} t_4^8 t_5^{10} t_6^6 t_7^8 t_8^6 +  y_{s}^8 y_{o_1}^{16} y_{o_2}^{16} y_{o_3}^{18} y_{o_4}^{14} y_{o_5}^{14} y_{o_6}^{18} y_{o_7}^{22} t_1^9 t_2^5 t_3^{10} t_4^8 t_5^{10} t_6^6 t_7^8 t_8^6 +  2 y_{s}^8 y_{o_1}^{16} y_{o_2}^{16} y_{o_3}^{18} y_{o_4}^{14} y_{o_5}^{14} y_{o_6}^{18} y_{o_7}^{22} t_1^8 t_2^6 t_3^{10} t_4^8 t_5^{10} t_6^6 t_7^8 t_8^6 +  2 y_{s}^8 y_{o_1}^{16} y_{o_2}^{16} y_{o_3}^{18} y_{o_4}^{14} y_{o_5}^{14} y_{o_6}^{18} y_{o_7}^{22} t_1^7 t_2^7 t_3^{10} t_4^8 t_5^{10} t_6^6 t_7^8 t_8^6 +  2 y_{s}^8 y_{o_1}^{16} y_{o_2}^{16} y_{o_3}^{18} y_{o_4}^{14} y_{o_5}^{14} y_{o_6}^{18} y_{o_7}^{22} t_1^6 t_2^8 t_3^{10} t_4^8 t_5^{10} t_6^6 t_7^8 t_8^6 +  y_{s}^8 y_{o_1}^{16} y_{o_2}^{16} y_{o_3}^{18} y_{o_4}^{14} y_{o_5}^{14} y_{o_6}^{18} y_{o_7}^{22} t_1^5 t_2^9 t_3^{10} t_4^8 t_5^{10} t_6^6 t_7^8 t_8^6 +  y_{s}^8 y_{o_1}^{16} y_{o_2}^{16} y_{o_3}^{18} y_{o_4}^{14} y_{o_5}^{14} y_{o_6}^{18} y_{o_7}^{22} t_1^4 t_2^{10} t_3^{10} t_4^8 t_5^{10} t_6^6 t_7^8 t_8^6 -  y_{s}^6 y_{o_1}^{15} y_{o_2}^9 y_{o_3}^{15} y_{o_4}^9 y_{o_5}^{12} y_{o_6}^{12} y_{o_7}^{18} t_1^6 t_2^6 t_3^9 t_4^3 t_5^6 t_6^3 t_7^9 t_8^6 +  y_{s}^6 y_{o_1}^{10} y_{o_2}^{14} y_{o_3}^9 y_{o_4}^{15} y_{o_5}^{13} y_{o_6}^{11} y_{o_7}^{19} t_1^8 t_2^5 t_3^3 t_4^8 t_5^5 t_6^9 t_7^4 t_8^7 +  y_{s}^6 y_{o_1}^{10} y_{o_2}^{14} y_{o_3}^9 y_{o_4}^{15} y_{o_5}^{13} y_{o_6}^{11} y_{o_7}^{19} t_1^5 t_2^8 t_3^3 t_4^8 t_5^5 t_6^9 t_7^4 t_8^7 -  y_{s}^5 y_{o_1}^{10} y_{o_2}^{10} y_{o_3}^8 y_{o_4}^{12} y_{o_5}^{12} y_{o_6}^8 y_{o_7}^{17} t_1^6 t_2^6 t_3^3 t_4^5 t_5^3 t_6^7 t_7^5 t_8^7 +  2 y_{s}^6 y_{o_1}^{11} y_{o_2}^{13} y_{o_3}^{10} y_{o_4}^{14} y_{o_5}^{13} y_{o_6}^{11} y_{o_7}^{19} t_1^8 t_2^5 t_3^4 t_4^7 t_5^5 t_6^8 t_7^5 t_8^7 +  y_{s}^6 y_{o_1}^{11} y_{o_2}^{13} y_{o_3}^{10} y_{o_4}^{14} y_{o_5}^{13} y_{o_6}^{11} y_{o_7}^{19} t_1^7 t_2^6 t_3^4 t_4^7 t_5^5 t_6^8 t_7^5 t_8^7 +  y_{s}^6 y_{o_1}^{11} y_{o_2}^{13} y_{o_3}^{10} y_{o_4}^{14} y_{o_5}^{13} y_{o_6}^{11} y_{o_7}^{19} t_1^6 t_2^7 t_3^4 t_4^7 t_5^5 t_6^8 t_7^5 t_8^7 +  2 y_{s}^6 y_{o_1}^{11} y_{o_2}^{13} y_{o_3}^{10} y_{o_4}^{14} y_{o_5}^{13} y_{o_6}^{11} y_{o_7}^{19} t_1^5 t_2^8 t_3^4 t_4^7 t_5^5 t_6^8 t_7^5 t_8^7 -  y_{s}^7 y_{o_1}^{12} y_{o_2}^{16} y_{o_3}^{12} y_{o_4}^{16} y_{o_5}^{14} y_{o_6}^{14} y_{o_7}^{21} t_1^9 t_2^5 t_3^5 t_4^9 t_5^7 t_6^9 t_7^5 t_8^7 -  y_{s}^7 y_{o_1}^{12} y_{o_2}^{16} y_{o_3}^{12} y_{o_4}^{16} y_{o_5}^{14} y_{o_6}^{14} y_{o_7}^{21} t_1^8 t_2^6 t_3^5 t_4^9 t_5^7 t_6^9 t_7^5 t_8^7 -  y_{s}^7 y_{o_1}^{12} y_{o_2}^{16} y_{o_3}^{12} y_{o_4}^{16} y_{o_5}^{14} y_{o_6}^{14} y_{o_7}^{21} t_1^7 t_2^7 t_3^5 t_4^9 t_5^7 t_6^9 t_7^5 t_8^7 -  y_{s}^7 y_{o_1}^{12} y_{o_2}^{16} y_{o_3}^{12} y_{o_4}^{16} y_{o_5}^{14} y_{o_6}^{14} y_{o_7}^{21} t_1^6 t_2^8 t_3^5 t_4^9 t_5^7 t_6^9 t_7^5 t_8^7 -  y_{s}^7 y_{o_1}^{12} y_{o_2}^{16} y_{o_3}^{12} y_{o_4}^{16} y_{o_5}^{14} y_{o_6}^{14} y_{o_7}^{21} t_1^5 t_2^9 t_3^5 t_4^9 t_5^7 t_6^9 t_7^5 t_8^7 -  y_{s}^5 y_{o_1}^{11} y_{o_2}^9 y_{o_3}^9 y_{o_4}^{11} y_{o_5}^{12} y_{o_6}^8 y_{o_7}^{17} t_1^9 t_2^3 t_3^4 t_4^4 t_5^3 t_6^6 t_7^6 t_8^7 -  y_{s}^5 y_{o_1}^{11} y_{o_2}^9 y_{o_3}^9 y_{o_4}^{11} y_{o_5}^{12} y_{o_6}^8 y_{o_7}^{17} t_1^8 t_2^4 t_3^4 t_4^4 t_5^3 t_6^6 t_7^6 t_8^7 -  y_{s}^5 y_{o_1}^{11} y_{o_2}^9 y_{o_3}^9 y_{o_4}^{11} y_{o_5}^{12} y_{o_6}^8 y_{o_7}^{17} t_1^7 t_2^5 t_3^4 t_4^4 t_5^3 t_6^6 t_7^6 t_8^7 -  3 y_{s}^5 y_{o_1}^{11} y_{o_2}^9 y_{o_3}^9 y_{o_4}^{11} y_{o_5}^{12} y_{o_6}^8 y_{o_7}^{17} t_1^6 t_2^6 t_3^4 t_4^4 t_5^3 t_6^6 t_7^6 t_8^7 -  y_{s}^5 y_{o_1}^{11} y_{o_2}^9 y_{o_3}^9 y_{o_4}^{11} y_{o_5}^{12} y_{o_6}^8 y_{o_7}^{17} t_1^5 t_2^7 t_3^4 t_4^4 t_5^3 t_6^6 t_7^6 t_8^7 -  y_{s}^5 y_{o_1}^{11} y_{o_2}^9 y_{o_3}^9 y_{o_4}^{11} y_{o_5}^{12} y_{o_6}^8 y_{o_7}^{17} t_1^4 t_2^8 t_3^4 t_4^4 t_5^3 t_6^6 t_7^6 t_8^7 -  y_{s}^5 y_{o_1}^{11} y_{o_2}^9 y_{o_3}^9 y_{o_4}^{11} y_{o_5}^{12} y_{o_6}^8 y_{o_7}^{17} t_1^3 t_2^9 t_3^4 t_4^4 t_5^3 t_6^6 t_7^6 t_8^7 +  2 y_{s}^6 y_{o_1}^{12} y_{o_2}^{12} y_{o_3}^{11} y_{o_4}^{13} y_{o_5}^{13} y_{o_6}^{11} y_{o_7}^{19} t_1^{10} t_2^3 t_3^5 t_4^6 t_5^5 t_6^7 t_7^6 t_8^7 +  3 y_{s}^6 y_{o_1}^{12} y_{o_2}^{12} y_{o_3}^{11} y_{o_4}^{13} y_{o_5}^{13} y_{o_6}^{11} y_{o_7}^{19} t_1^9 t_2^4 t_3^5 t_4^6 t_5^5 t_6^7 t_7^6 t_8^7 +  6 y_{s}^6 y_{o_1}^{12} y_{o_2}^{12} y_{o_3}^{11} y_{o_4}^{13} y_{o_5}^{13} y_{o_6}^{11} y_{o_7}^{19} t_1^8 t_2^5 t_3^5 t_4^6 t_5^5 t_6^7 t_7^6 t_8^7 +  7 y_{s}^6 y_{o_1}^{12} y_{o_2}^{12} y_{o_3}^{11} y_{o_4}^{13} y_{o_5}^{13} y_{o_6}^{11} y_{o_7}^{19} t_1^7 t_2^6 t_3^5 t_4^6 t_5^5 t_6^7 t_7^6 t_8^7 +  7 y_{s}^6 y_{o_1}^{12} y_{o_2}^{12} y_{o_3}^{11} y_{o_4}^{13} y_{o_5}^{13} y_{o_6}^{11} y_{o_7}^{19} t_1^6 t_2^7 t_3^5 t_4^6 t_5^5 t_6^7 t_7^6 t_8^7 +  6 y_{s}^6 y_{o_1}^{12} y_{o_2}^{12} y_{o_3}^{11} y_{o_4}^{13} y_{o_5}^{13} y_{o_6}^{11} y_{o_7}^{19} t_1^5 t_2^8 t_3^5 t_4^6 t_5^5 t_6^7 t_7^6 t_8^7 +  3 y_{s}^6 y_{o_1}^{12} y_{o_2}^{12} y_{o_3}^{11} y_{o_4}^{13} y_{o_5}^{13} y_{o_6}^{11} y_{o_7}^{19} t_1^4 t_2^9 t_3^5 t_4^6 t_5^5 t_6^7 t_7^6 t_8^7 +  2 y_{s}^6 y_{o_1}^{12} y_{o_2}^{12} y_{o_3}^{11} y_{o_4}^{13} y_{o_5}^{13} y_{o_6}^{11} y_{o_7}^{19} t_1^3 t_2^{10} t_3^5 t_4^6 t_5^5 t_6^7 t_7^6 t_8^7 -  2 y_{s}^7 y_{o_1}^{13} y_{o_2}^{15} y_{o_3}^{13} y_{o_4}^{15} y_{o_5}^{14} y_{o_6}^{14} y_{o_7}^{21} t_1^{10} t_2^4 t_3^6 t_4^8 t_5^7 t_6^8 t_7^6 t_8^7 -  4 y_{s}^7 y_{o_1}^{13} y_{o_2}^{15} y_{o_3}^{13} y_{o_4}^{15} y_{o_5}^{14} y_{o_6}^{14} y_{o_7}^{21} t_1^9 t_2^5 t_3^6 t_4^8 t_5^7 t_6^8 t_7^6 t_8^7 -  5 y_{s}^7 y_{o_1}^{13} y_{o_2}^{15} y_{o_3}^{13} y_{o_4}^{15} y_{o_5}^{14} y_{o_6}^{14} y_{o_7}^{21} t_1^8 t_2^6 t_3^6 t_4^8 t_5^7 t_6^8 t_7^6 t_8^7 -  8 y_{s}^7 y_{o_1}^{13} y_{o_2}^{15} y_{o_3}^{13} y_{o_4}^{15} y_{o_5}^{14} y_{o_6}^{14} y_{o_7}^{21} t_1^7 t_2^7 t_3^6 t_4^8 t_5^7 t_6^8 t_7^6 t_8^7 -  5 y_{s}^7 y_{o_1}^{13} y_{o_2}^{15} y_{o_3}^{13} y_{o_4}^{15} y_{o_5}^{14} y_{o_6}^{14} y_{o_7}^{21} t_1^6 t_2^8 t_3^6 t_4^8 t_5^7 t_6^8 t_7^6 t_8^7 -  4 y_{s}^7 y_{o_1}^{13} y_{o_2}^{15} y_{o_3}^{13} y_{o_4}^{15} y_{o_5}^{14} y_{o_6}^{14} y_{o_7}^{21} t_1^5 t_2^9 t_3^6 t_4^8 t_5^7 t_6^8 t_7^6 t_8^7 -  2 y_{s}^7 y_{o_1}^{13} y_{o_2}^{15} y_{o_3}^{13} y_{o_4}^{15} y_{o_5}^{14} y_{o_6}^{14} y_{o_7}^{21} t_1^4 t_2^{10} t_3^6 t_4^8 t_5^7 t_6^8 t_7^6 t_8^7 +  y_{s}^8 y_{o_1}^{14} y_{o_2}^{18} y_{o_3}^{15} y_{o_4}^{17} y_{o_5}^{15} y_{o_6}^{17} y_{o_7}^{23} t_1^9 t_2^6 t_3^7 t_4^{10} t_5^9 t_6^9 t_7^6 t_8^7 +  y_{s}^8 y_{o_1}^{14} y_{o_2}^{18} y_{o_3}^{15} y_{o_4}^{17} y_{o_5}^{15} y_{o_6}^{17} y_{o_7}^{23} t_1^8 t_2^7 t_3^7 t_4^{10} t_5^9 t_6^9 t_7^6 t_8^7 +  y_{s}^8 y_{o_1}^{14} y_{o_2}^{18} y_{o_3}^{15} y_{o_4}^{17} y_{o_5}^{15} y_{o_6}^{17} y_{o_7}^{23} t_1^7 t_2^8 t_3^7 t_4^{10} t_5^9 t_6^9 t_7^6 t_8^7 +  y_{s}^8 y_{o_1}^{14} y_{o_2}^{18} y_{o_3}^{15} y_{o_4}^{17} y_{o_5}^{15} y_{o_6}^{17} y_{o_7}^{23} t_1^6 t_2^9 t_3^7 t_4^{10} t_5^9 t_6^9 t_7^6 t_8^7 -  y_{s}^5 y_{o_1}^{12} y_{o_2}^8 y_{o_3}^{10} y_{o_4}^{10} y_{o_5}^{12} y_{o_6}^8 y_{o_7}^{17} t_1^6 t_2^6 t_3^5 t_4^3 t_5^3 t_6^5 t_7^7 t_8^7 +  2 y_{s}^6 y_{o_1}^{13} y_{o_2}^{11} y_{o_3}^{12} y_{o_4}^{12} y_{o_5}^{13} y_{o_6}^{11} y_{o_7}^{19} t_1^{10} t_2^3 t_3^6 t_4^5 t_5^5 t_6^6 t_7^7 t_8^7 +  3 y_{s}^6 y_{o_1}^{13} y_{o_2}^{11} y_{o_3}^{12} y_{o_4}^{12} y_{o_5}^{13} y_{o_6}^{11} y_{o_7}^{19} t_1^9 t_2^4 t_3^6 t_4^5 t_5^5 t_6^6 t_7^7 t_8^7 +  6 y_{s}^6 y_{o_1}^{13} y_{o_2}^{11} y_{o_3}^{12} y_{o_4}^{12} y_{o_5}^{13} y_{o_6}^{11} y_{o_7}^{19} t_1^8 t_2^5 t_3^6 t_4^5 t_5^5 t_6^6 t_7^7 t_8^7 +  7 y_{s}^6 y_{o_1}^{13} y_{o_2}^{11} y_{o_3}^{12} y_{o_4}^{12} y_{o_5}^{13} y_{o_6}^{11} y_{o_7}^{19} t_1^7 t_2^6 t_3^6 t_4^5 t_5^5 t_6^6 t_7^7 t_8^7 +  7 y_{s}^6 y_{o_1}^{13} y_{o_2}^{11} y_{o_3}^{12} y_{o_4}^{12} y_{o_5}^{13} y_{o_6}^{11} y_{o_7}^{19} t_1^6 t_2^7 t_3^6 t_4^5 t_5^5 t_6^6 t_7^7 t_8^7 +  6 y_{s}^6 y_{o_1}^{13} y_{o_2}^{11} y_{o_3}^{12} y_{o_4}^{12} y_{o_5}^{13} y_{o_6}^{11} y_{o_7}^{19} t_1^5 t_2^8 t_3^6 t_4^5 t_5^5 t_6^6 t_7^7 t_8^7 +  3 y_{s}^6 y_{o_1}^{13} y_{o_2}^{11} y_{o_3}^{12} y_{o_4}^{12} y_{o_5}^{13} y_{o_6}^{11} y_{o_7}^{19} t_1^4 t_2^9 t_3^6 t_4^5 t_5^5 t_6^6 t_7^7 t_8^7 +  2 y_{s}^6 y_{o_1}^{13} y_{o_2}^{11} y_{o_3}^{12} y_{o_4}^{12} y_{o_5}^{13} y_{o_6}^{11} y_{o_7}^{19} t_1^3 t_2^{10} t_3^6 t_4^5 t_5^5 t_6^6 t_7^7 t_8^7 -  y_{s}^7 y_{o_1}^{14} y_{o_2}^{14} y_{o_3}^{14} y_{o_4}^{14} y_{o_5}^{14} y_{o_6}^{14} y_{o_7}^{21} t_1^{11} t_2^3 t_3^7 t_4^7 t_5^7 t_6^7 t_7^7 t_8^7 -  6 y_{s}^7 y_{o_1}^{14} y_{o_2}^{14} y_{o_3}^{14} y_{o_4}^{14} y_{o_5}^{14} y_{o_6}^{14} y_{o_7}^{21} t_1^{10} t_2^4 t_3^7 t_4^7 t_5^7 t_6^7 t_7^7 t_8^7 -  8 y_{s}^7 y_{o_1}^{14} y_{o_2}^{14} y_{o_3}^{14} y_{o_4}^{14} y_{o_5}^{14} y_{o_6}^{14} y_{o_7}^{21} t_1^9 t_2^5 t_3^7 t_4^7 t_5^7 t_6^7 t_7^7 t_8^7 -  {10} y_{s}^7 y_{o_1}^{14} y_{o_2}^{14} y_{o_3}^{14} y_{o_4}^{14} y_{o_5}^{14} y_{o_6}^{14} y_{o_7}^{21} t_1^8 t_2^6 t_3^7 t_4^7 t_5^7 t_6^7 t_7^7 t_8^7 -  {16} y_{s}^7 y_{o_1}^{14} y_{o_2}^{14} y_{o_3}^{14} y_{o_4}^{14} y_{o_5}^{14} y_{o_6}^{14} y_{o_7}^{21} t_1^7 t_2^7 t_3^7 t_4^7 t_5^7 t_6^7 t_7^7 t_8^7 -  {10} y_{s}^7 y_{o_1}^{14} y_{o_2}^{14} y_{o_3}^{14} y_{o_4}^{14} y_{o_5}^{14} y_{o_6}^{14} y_{o_7}^{21} t_1^6 t_2^8 t_3^7 t_4^7 t_5^7 t_6^7 t_7^7 t_8^7 -  8 y_{s}^7 y_{o_1}^{14} y_{o_2}^{14} y_{o_3}^{14} y_{o_4}^{14} y_{o_5}^{14} y_{o_6}^{14} y_{o_7}^{21} t_1^5 t_2^9 t_3^7 t_4^7 t_5^7 t_6^7 t_7^7 t_8^7 -  6 y_{s}^7 y_{o_1}^{14} y_{o_2}^{14} y_{o_3}^{14} y_{o_4}^{14} y_{o_5}^{14} y_{o_6}^{14} y_{o_7}^{21} t_1^4 t_2^{10} t_3^7 t_4^7 t_5^7 t_6^7 t_7^7 t_8^7 -  y_{s}^7 y_{o_1}^{14} y_{o_2}^{14} y_{o_3}^{14} y_{o_4}^{14} y_{o_5}^{14} y_{o_6}^{14} y_{o_7}^{21} t_1^3 t_2^{11} t_3^7 t_4^7 t_5^7 t_6^7 t_7^7 t_8^7 +  y_{s}^8 y_{o_1}^{15} y_{o_2}^{17} y_{o_3}^{16} y_{o_4}^{16} y_{o_5}^{15} y_{o_6}^{17} y_{o_7}^{23} t_1^{11} t_2^4 t_3^8 t_4^9 t_5^9 t_6^8 t_7^7 t_8^7 +  3 y_{s}^8 y_{o_1}^{15} y_{o_2}^{17} y_{o_3}^{16} y_{o_4}^{16} y_{o_5}^{15} y_{o_6}^{17} y_{o_7}^{23} t_1^{10} t_2^5 t_3^8 t_4^9 t_5^9 t_6^8 t_7^7 t_8^7 +  4 y_{s}^8 y_{o_1}^{15} y_{o_2}^{17} y_{o_3}^{16} y_{o_4}^{16} y_{o_5}^{15} y_{o_6}^{17} y_{o_7}^{23} t_1^9 t_2^6 t_3^8 t_4^9 t_5^9 t_6^8 t_7^7 t_8^7 +  6 y_{s}^8 y_{o_1}^{15} y_{o_2}^{17} y_{o_3}^{16} y_{o_4}^{16} y_{o_5}^{15} y_{o_6}^{17} y_{o_7}^{23} t_1^8 t_2^7 t_3^8 t_4^9 t_5^9 t_6^8 t_7^7 t_8^7 +  6 y_{s}^8 y_{o_1}^{15} y_{o_2}^{17} y_{o_3}^{16} y_{o_4}^{16} y_{o_5}^{15} y_{o_6}^{17} y_{o_7}^{23} t_1^7 t_2^8 t_3^8 t_4^9 t_5^9 t_6^8 t_7^7 t_8^7 +  4 y_{s}^8 y_{o_1}^{15} y_{o_2}^{17} y_{o_3}^{16} y_{o_4}^{16} y_{o_5}^{15} y_{o_6}^{17} y_{o_7}^{23} t_1^6 t_2^9 t_3^8 t_4^9 t_5^9 t_6^8 t_7^7 t_8^7 +  3 y_{s}^8 y_{o_1}^{15} y_{o_2}^{17} y_{o_3}^{16} y_{o_4}^{16} y_{o_5}^{15} y_{o_6}^{17} y_{o_7}^{23} t_1^5 t_2^{10} t_3^8 t_4^9 t_5^9 t_6^8 t_7^7 t_8^7 +  y_{s}^8 y_{o_1}^{15} y_{o_2}^{17} y_{o_3}^{16} y_{o_4}^{16} y_{o_5}^{15} y_{o_6}^{17} y_{o_7}^{23} t_1^4 t_2^{11} t_3^8 t_4^9 t_5^9 t_6^8 t_7^7 t_8^7 +  2 y_{s}^6 y_{o_1}^{14} y_{o_2}^{10} y_{o_3}^{13} y_{o_4}^{11} y_{o_5}^{13} y_{o_6}^{11} y_{o_7}^{19} t_1^8 t_2^5 t_3^7 t_4^4 t_5^5 t_6^5 t_7^8 t_8^7 +  y_{s}^6 y_{o_1}^{14} y_{o_2}^{10} y_{o_3}^{13} y_{o_4}^{11} y_{o_5}^{13} y_{o_6}^{11} y_{o_7}^{19} t_1^7 t_2^6 t_3^7 t_4^4 t_5^5 t_6^5 t_7^8 t_8^7 +  y_{s}^6 y_{o_1}^{14} y_{o_2}^{10} y_{o_3}^{13} y_{o_4}^{11} y_{o_5}^{13} y_{o_6}^{11} y_{o_7}^{19} t_1^6 t_2^7 t_3^7 t_4^4 t_5^5 t_6^5 t_7^8 t_8^7 +  2 y_{s}^6 y_{o_1}^{14} y_{o_2}^{10} y_{o_3}^{13} y_{o_4}^{11} y_{o_5}^{13} y_{o_6}^{11} y_{o_7}^{19} t_1^5 t_2^8 t_3^7 t_4^4 t_5^5 t_6^5 t_7^8 t_8^7 -  2 y_{s}^7 y_{o_1}^{15} y_{o_2}^{13} y_{o_3}^{15} y_{o_4}^{13} y_{o_5}^{14} y_{o_6}^{14} y_{o_7}^{21} t_1^{10} t_2^4 t_3^8 t_4^6 t_5^7 t_6^6 t_7^8 t_8^7 -  4 y_{s}^7 y_{o_1}^{15} y_{o_2}^{13} y_{o_3}^{15} y_{o_4}^{13} y_{o_5}^{14} y_{o_6}^{14} y_{o_7}^{21} t_1^9 t_2^5 t_3^8 t_4^6 t_5^7 t_6^6 t_7^8 t_8^7 -  5 y_{s}^7 y_{o_1}^{15} y_{o_2}^{13} y_{o_3}^{15} y_{o_4}^{13} y_{o_5}^{14} y_{o_6}^{14} y_{o_7}^{21} t_1^8 t_2^6 t_3^8 t_4^6 t_5^7 t_6^6 t_7^8 t_8^7 -  8 y_{s}^7 y_{o_1}^{15} y_{o_2}^{13} y_{o_3}^{15} y_{o_4}^{13} y_{o_5}^{14} y_{o_6}^{14} y_{o_7}^{21} t_1^7 t_2^7 t_3^8 t_4^6 t_5^7 t_6^6 t_7^8 t_8^7 -  5 y_{s}^7 y_{o_1}^{15} y_{o_2}^{13} y_{o_3}^{15} y_{o_4}^{13} y_{o_5}^{14} y_{o_6}^{14} y_{o_7}^{21} t_1^6 t_2^8 t_3^8 t_4^6 t_5^7 t_6^6 t_7^8 t_8^7 -  4 y_{s}^7 y_{o_1}^{15} y_{o_2}^{13} y_{o_3}^{15} y_{o_4}^{13} y_{o_5}^{14} y_{o_6}^{14} y_{o_7}^{21} t_1^5 t_2^9 t_3^8 t_4^6 t_5^7 t_6^6 t_7^8 t_8^7 -  2 y_{s}^7 y_{o_1}^{15} y_{o_2}^{13} y_{o_3}^{15} y_{o_4}^{13} y_{o_5}^{14} y_{o_6}^{14} y_{o_7}^{21} t_1^4 t_2^{10} t_3^8 t_4^6 t_5^7 t_6^6 t_7^8 t_8^7 +  y_{s}^8 y_{o_1}^{16} y_{o_2}^{16} y_{o_3}^{17} y_{o_4}^{15} y_{o_5}^{15} y_{o_6}^{17} y_{o_7}^{23} t_1^{11} t_2^4 t_3^9 t_4^8 t_5^9 t_6^7 t_7^8 t_8^7 +  3 y_{s}^8 y_{o_1}^{16} y_{o_2}^{16} y_{o_3}^{17} y_{o_4}^{15} y_{o_5}^{15} y_{o_6}^{17} y_{o_7}^{23} t_1^{10} t_2^5 t_3^9 t_4^8 t_5^9 t_6^7 t_7^8 t_8^7 +  4 y_{s}^8 y_{o_1}^{16} y_{o_2}^{16} y_{o_3}^{17} y_{o_4}^{15} y_{o_5}^{15} y_{o_6}^{17} y_{o_7}^{23} t_1^9 t_2^6 t_3^9 t_4^8 t_5^9 t_6^7 t_7^8 t_8^7 +  6 y_{s}^8 y_{o_1}^{16} y_{o_2}^{16} y_{o_3}^{17} y_{o_4}^{15} y_{o_5}^{15} y_{o_6}^{17} y_{o_7}^{23} t_1^8 t_2^7 t_3^9 t_4^8 t_5^9 t_6^7 t_7^8 t_8^7 +  6 y_{s}^8 y_{o_1}^{16} y_{o_2}^{16} y_{o_3}^{17} y_{o_4}^{15} y_{o_5}^{15} y_{o_6}^{17} y_{o_7}^{23} t_1^7 t_2^8 t_3^9 t_4^8 t_5^9 t_6^7 t_7^8 t_8^7 +  4 y_{s}^8 y_{o_1}^{16} y_{o_2}^{16} y_{o_3}^{17} y_{o_4}^{15} y_{o_5}^{15} y_{o_6}^{17} y_{o_7}^{23} t_1^6 t_2^9 t_3^9 t_4^8 t_5^9 t_6^7 t_7^8 t_8^7 +  3 y_{s}^8 y_{o_1}^{16} y_{o_2}^{16} y_{o_3}^{17} y_{o_4}^{15} y_{o_5}^{15} y_{o_6}^{17} y_{o_7}^{23} t_1^5 t_2^{10} t_3^9 t_4^8 t_5^9 t_6^7 t_7^8 t_8^7 +  y_{s}^8 y_{o_1}^{16} y_{o_2}^{16} y_{o_3}^{17} y_{o_4}^{15} y_{o_5}^{15} y_{o_6}^{17} y_{o_7}^{23} t_1^4 t_2^{11} t_3^9 t_4^8 t_5^9 t_6^7 t_7^8 t_8^7 -  y_{s}^9 y_{o_1}^{17} y_{o_2}^{19} y_{o_3}^{19} y_{o_4}^{17} y_{o_5}^{16} y_{o_6}^{20} y_{o_7}^{25} t_1^{10} t_2^6 t_3^{10} t_4^{10} t_5^{11} t_6^8 t_7^8 t_8^7 -  y_{s}^9 y_{o_1}^{17} y_{o_2}^{19} y_{o_3}^{19} y_{o_4}^{17} y_{o_5}^{16} y_{o_6}^{20} y_{o_7}^{25} t_1^9 t_2^7 t_3^{10} t_4^{10} t_5^{11} t_6^8 t_7^8 t_8^7 -  y_{s}^9 y_{o_1}^{17} y_{o_2}^{19} y_{o_3}^{19} y_{o_4}^{17} y_{o_5}^{16} y_{o_6}^{20} y_{o_7}^{25} t_1^8 t_2^8 t_3^{10} t_4^{10} t_5^{11} t_6^8 t_7^8 t_8^7 -  y_{s}^9 y_{o_1}^{17} y_{o_2}^{19} y_{o_3}^{19} y_{o_4}^{17} y_{o_5}^{16} y_{o_6}^{20} y_{o_7}^{25} t_1^7 t_2^9 t_3^{10} t_4^{10} t_5^{11} t_6^8 t_7^8 t_8^7 -  y_{s}^9 y_{o_1}^{17} y_{o_2}^{19} y_{o_3}^{19} y_{o_4}^{17} y_{o_5}^{16} y_{o_6}^{20} y_{o_7}^{25} t_1^6 t_2^{10} t_3^{10} t_4^{10} t_5^{11} t_6^8 t_7^8 t_8^7 +  y_{s}^6 y_{o_1}^{15} y_{o_2}^9 y_{o_3}^{14} y_{o_4}^{10} y_{o_5}^{13} y_{o_6}^{11} y_{o_7}^{19} t_1^8 t_2^5 t_3^8 t_4^3 t_5^5 t_6^4 t_7^9 t_8^7 +  y_{s}^6 y_{o_1}^{15} y_{o_2}^9 y_{o_3}^{14} y_{o_4}^{10} y_{o_5}^{13} y_{o_6}^{11} y_{o_7}^{19} t_1^5 t_2^8 t_3^8 t_4^3 t_5^5 t_6^4 t_7^9 t_8^7 -  y_{s}^7 y_{o_1}^{16} y_{o_2}^{12} y_{o_3}^{16} y_{o_4}^{12} y_{o_5}^{14} y_{o_6}^{14} y_{o_7}^{21} t_1^9 t_2^5 t_3^9 t_4^5 t_5^7 t_6^5 t_7^9 t_8^7 -  y_{s}^7 y_{o_1}^{16} y_{o_2}^{12} y_{o_3}^{16} y_{o_4}^{12} y_{o_5}^{14} y_{o_6}^{14} y_{o_7}^{21} t_1^8 t_2^6 t_3^9 t_4^5 t_5^7 t_6^5 t_7^9 t_8^7 -  y_{s}^7 y_{o_1}^{16} y_{o_2}^{12} y_{o_3}^{16} y_{o_4}^{12} y_{o_5}^{14} y_{o_6}^{14} y_{o_7}^{21} t_1^7 t_2^7 t_3^9 t_4^5 t_5^7 t_6^5 t_7^9 t_8^7 -  y_{s}^7 y_{o_1}^{16} y_{o_2}^{12} y_{o_3}^{16} y_{o_4}^{12} y_{o_5}^{14} y_{o_6}^{14} y_{o_7}^{21} t_1^6 t_2^8 t_3^9 t_4^5 t_5^7 t_6^5 t_7^9 t_8^7 -  y_{s}^7 y_{o_1}^{16} y_{o_2}^{12} y_{o_3}^{16} y_{o_4}^{12} y_{o_5}^{14} y_{o_6}^{14} y_{o_7}^{21} t_1^5 t_2^9 t_3^9 t_4^5 t_5^7 t_6^5 t_7^9 t_8^7 +  y_{s}^8 y_{o_1}^{17} y_{o_2}^{15} y_{o_3}^{18} y_{o_4}^{14} y_{o_5}^{15} y_{o_6}^{17} y_{o_7}^{23} t_1^9 t_2^6 t_3^{10} t_4^7 t_5^9 t_6^6 t_7^9 t_8^7 +  y_{s}^8 y_{o_1}^{17} y_{o_2}^{15} y_{o_3}^{18} y_{o_4}^{14} y_{o_5}^{15} y_{o_6}^{17} y_{o_7}^{23} t_1^8 t_2^7 t_3^{10} t_4^7 t_5^9 t_6^6 t_7^9 t_8^7 +  y_{s}^8 y_{o_1}^{17} y_{o_2}^{15} y_{o_3}^{18} y_{o_4}^{14} y_{o_5}^{15} y_{o_6}^{17} y_{o_7}^{23} t_1^7 t_2^8 t_3^{10} t_4^7 t_5^9 t_6^6 t_7^9 t_8^7 +  y_{s}^8 y_{o_1}^{17} y_{o_2}^{15} y_{o_3}^{18} y_{o_4}^{14} y_{o_5}^{15} y_{o_6}^{17} y_{o_7}^{23} t_1^6 t_2^9 t_3^{10} t_4^7 t_5^9 t_6^6 t_7^9 t_8^7 -  y_{s}^5 y_{o_1}^{10} y_{o_2}^{10} y_{o_3}^7 y_{o_4}^{13} y_{o_5}^{13} y_{o_6}^7 y_{o_7}^{18} t_1^7 t_2^6 t_3^2 t_4^5 t_5^2 t_6^8 t_7^5 t_8^8 -  y_{s}^5 y_{o_1}^{10} y_{o_2}^{10} y_{o_3}^7 y_{o_4}^{13} y_{o_5}^{13} y_{o_6}^7 y_{o_7}^{18} t_1^6 t_2^7 t_3^2 t_4^5 t_5^2 t_6^8 t_7^5 t_8^8 +  y_{s}^6 y_{o_1}^{11} y_{o_2}^{13} y_{o_3}^9 y_{o_4}^{15} y_{o_5}^{14} y_{o_6}^{10} y_{o_7}^{20} t_1^8 t_2^6 t_3^3 t_4^7 t_5^4 t_6^9 t_7^5 t_8^8 +  2 y_{s}^6 y_{o_1}^{11} y_{o_2}^{13} y_{o_3}^9 y_{o_4}^{15} y_{o_5}^{14} y_{o_6}^{10} y_{o_7}^{20} t_1^7 t_2^7 t_3^3 t_4^7 t_5^4 t_6^9 t_7^5 t_8^8 +  y_{s}^6 y_{o_1}^{11} y_{o_2}^{13} y_{o_3}^9 y_{o_4}^{15} y_{o_5}^{14} y_{o_6}^{10} y_{o_7}^{20} t_1^6 t_2^8 t_3^3 t_4^7 t_5^4 t_6^9 t_7^5 t_8^8 -  y_{s}^7 y_{o_1}^{12} y_{o_2}^{16} y_{o_3}^{11} y_{o_4}^{17} y_{o_5}^{15} y_{o_6}^{13} y_{o_7}^{22} t_1^8 t_2^7 t_3^4 t_4^9 t_5^6 t_6^{10} t_7^5 t_8^8 -  y_{s}^7 y_{o_1}^{12} y_{o_2}^{16} y_{o_3}^{11} y_{o_4}^{17} y_{o_5}^{15} y_{o_6}^{13} y_{o_7}^{22} t_1^7 t_2^8 t_3^4 t_4^9 t_5^6 t_6^{10} t_7^5 t_8^8 +  y_{s}^6 y_{o_1}^{12} y_{o_2}^{12} y_{o_3}^{10} y_{o_4}^{14} y_{o_5}^{14} y_{o_6}^{10} y_{o_7}^{20} t_1^9 t_2^5 t_3^4 t_4^6 t_5^4 t_6^8 t_7^6 t_8^8 +  3 y_{s}^6 y_{o_1}^{12} y_{o_2}^{12} y_{o_3}^{10} y_{o_4}^{14} y_{o_5}^{14} y_{o_6}^{10} y_{o_7}^{20} t_1^8 t_2^6 t_3^4 t_4^6 t_5^4 t_6^8 t_7^6 t_8^8 +  3 y_{s}^6 y_{o_1}^{12} y_{o_2}^{12} y_{o_3}^{10} y_{o_4}^{14} y_{o_5}^{14} y_{o_6}^{10} y_{o_7}^{20} t_1^7 t_2^7 t_3^4 t_4^6 t_5^4 t_6^8 t_7^6 t_8^8 +  3 y_{s}^6 y_{o_1}^{12} y_{o_2}^{12} y_{o_3}^{10} y_{o_4}^{14} y_{o_5}^{14} y_{o_6}^{10} y_{o_7}^{20} t_1^6 t_2^8 t_3^4 t_4^6 t_5^4 t_6^8 t_7^6 t_8^8 +  y_{s}^6 y_{o_1}^{12} y_{o_2}^{12} y_{o_3}^{10} y_{o_4}^{14} y_{o_5}^{14} y_{o_6}^{10} y_{o_7}^{20} t_1^5 t_2^9 t_3^4 t_4^6 t_5^4 t_6^8 t_7^6 t_8^8 -  2 y_{s}^7 y_{o_1}^{13} y_{o_2}^{15} y_{o_3}^{12} y_{o_4}^{16} y_{o_5}^{15} y_{o_6}^{13} y_{o_7}^{22} t_1^{10} t_2^5 t_3^5 t_4^8 t_5^6 t_6^9 t_7^6 t_8^8 -  3 y_{s}^7 y_{o_1}^{13} y_{o_2}^{15} y_{o_3}^{12} y_{o_4}^{16} y_{o_5}^{15} y_{o_6}^{13} y_{o_7}^{22} t_1^9 t_2^6 t_3^5 t_4^8 t_5^6 t_6^9 t_7^6 t_8^8 -  6 y_{s}^7 y_{o_1}^{13} y_{o_2}^{15} y_{o_3}^{12} y_{o_4}^{16} y_{o_5}^{15} y_{o_6}^{13} y_{o_7}^{22} t_1^8 t_2^7 t_3^5 t_4^8 t_5^6 t_6^9 t_7^6 t_8^8 -  6 y_{s}^7 y_{o_1}^{13} y_{o_2}^{15} y_{o_3}^{12} y_{o_4}^{16} y_{o_5}^{15} y_{o_6}^{13} y_{o_7}^{22} t_1^7 t_2^8 t_3^5 t_4^8 t_5^6 t_6^9 t_7^6 t_8^8 -  3 y_{s}^7 y_{o_1}^{13} y_{o_2}^{15} y_{o_3}^{12} y_{o_4}^{16} y_{o_5}^{15} y_{o_6}^{13} y_{o_7}^{22} t_1^6 t_2^9 t_3^5 t_4^8 t_5^6 t_6^9 t_7^6 t_8^8 -  2 y_{s}^7 y_{o_1}^{13} y_{o_2}^{15} y_{o_3}^{12} y_{o_4}^{16} y_{o_5}^{15} y_{o_6}^{13} y_{o_7}^{22} t_1^5 t_2^{10} t_3^5 t_4^8 t_5^6 t_6^9 t_7^6 t_8^8 +  y_{s}^8 y_{o_1}^{14} y_{o_2}^{18} y_{o_3}^{14} y_{o_4}^{18} y_{o_5}^{16} y_{o_6}^{16} y_{o_7}^{24} t_1^{10} t_2^6 t_3^6 t_4^{10} t_5^8 t_6^{10} t_7^6 t_8^8 +  2 y_{s}^8 y_{o_1}^{14} y_{o_2}^{18} y_{o_3}^{14} y_{o_4}^{18} y_{o_5}^{16} y_{o_6}^{16} y_{o_7}^{24} t_1^9 t_2^7 t_3^6 t_4^{10} t_5^8 t_6^{10} t_7^6 t_8^8 +  2 y_{s}^8 y_{o_1}^{14} y_{o_2}^{18} y_{o_3}^{14} y_{o_4}^{18} y_{o_5}^{16} y_{o_6}^{16} y_{o_7}^{24} t_1^8 t_2^8 t_3^6 t_4^{10} t_5^8 t_6^{10} t_7^6 t_8^8 +  2 y_{s}^8 y_{o_1}^{14} y_{o_2}^{18} y_{o_3}^{14} y_{o_4}^{18} y_{o_5}^{16} y_{o_6}^{16} y_{o_7}^{24} t_1^7 t_2^9 t_3^6 t_4^{10} t_5^8 t_6^{10} t_7^6 t_8^8 +  y_{s}^8 y_{o_1}^{14} y_{o_2}^{18} y_{o_3}^{14} y_{o_4}^{18} y_{o_5}^{16} y_{o_6}^{16} y_{o_7}^{24} t_1^6 t_2^{10} t_3^6 t_4^{10} t_5^8 t_6^{10} t_7^6 t_8^8 +  2 y_{s}^6 y_{o_1}^{13} y_{o_2}^{11} y_{o_3}^{11} y_{o_4}^{13} y_{o_5}^{14} y_{o_6}^{10} y_{o_7}^{20} t_1^9 t_2^5 t_3^5 t_4^5 t_5^4 t_6^7 t_7^7 t_8^8 +  3 y_{s}^6 y_{o_1}^{13} y_{o_2}^{11} y_{o_3}^{11} y_{o_4}^{13} y_{o_5}^{14} y_{o_6}^{10} y_{o_7}^{20} t_1^8 t_2^6 t_3^5 t_4^5 t_5^4 t_6^7 t_7^7 t_8^8 +  y_{s}^6 y_{o_1}^{13} y_{o_2}^{11} y_{o_3}^{11} y_{o_4}^{13} y_{o_5}^{14} y_{o_6}^{10} y_{o_7}^{20} t_1^7 t_2^7 t_3^5 t_4^5 t_5^4 t_6^7 t_7^7 t_8^8 +  3 y_{s}^6 y_{o_1}^{13} y_{o_2}^{11} y_{o_3}^{11} y_{o_4}^{13} y_{o_5}^{14} y_{o_6}^{10} y_{o_7}^{20} t_1^6 t_2^8 t_3^5 t_4^5 t_5^4 t_6^7 t_7^7 t_8^8 +  2 y_{s}^6 y_{o_1}^{13} y_{o_2}^{11} y_{o_3}^{11} y_{o_4}^{13} y_{o_5}^{14} y_{o_6}^{10} y_{o_7}^{20} t_1^5 t_2^9 t_3^5 t_4^5 t_5^4 t_6^7 t_7^7 t_8^8 -  4 y_{s}^7 y_{o_1}^{14} y_{o_2}^{14} y_{o_3}^{13} y_{o_4}^{15} y_{o_5}^{15} y_{o_6}^{13} y_{o_7}^{22} t_1^{10} t_2^5 t_3^6 t_4^7 t_5^6 t_6^8 t_7^7 t_8^8 -  6 y_{s}^7 y_{o_1}^{14} y_{o_2}^{14} y_{o_3}^{13} y_{o_4}^{15} y_{o_5}^{15} y_{o_6}^{13} y_{o_7}^{22} t_1^9 t_2^6 t_3^6 t_4^7 t_5^6 t_6^8 t_7^7 t_8^8 -  8 y_{s}^7 y_{o_1}^{14} y_{o_2}^{14} y_{o_3}^{13} y_{o_4}^{15} y_{o_5}^{15} y_{o_6}^{13} y_{o_7}^{22} t_1^8 t_2^7 t_3^6 t_4^7 t_5^6 t_6^8 t_7^7 t_8^8 -  8 y_{s}^7 y_{o_1}^{14} y_{o_2}^{14} y_{o_3}^{13} y_{o_4}^{15} y_{o_5}^{15} y_{o_6}^{13} y_{o_7}^{22} t_1^7 t_2^8 t_3^6 t_4^7 t_5^6 t_6^8 t_7^7 t_8^8 -  6 y_{s}^7 y_{o_1}^{14} y_{o_2}^{14} y_{o_3}^{13} y_{o_4}^{15} y_{o_5}^{15} y_{o_6}^{13} y_{o_7}^{22} t_1^6 t_2^9 t_3^6 t_4^7 t_5^6 t_6^8 t_7^7 t_8^8 -  4 y_{s}^7 y_{o_1}^{14} y_{o_2}^{14} y_{o_3}^{13} y_{o_4}^{15} y_{o_5}^{15} y_{o_6}^{13} y_{o_7}^{22} t_1^5 t_2^{10} t_3^6 t_4^7 t_5^6 t_6^8 t_7^7 t_8^8 +  y_{s}^8 y_{o_1}^{15} y_{o_2}^{17} y_{o_3}^{15} y_{o_4}^{17} y_{o_5}^{16} y_{o_6}^{16} y_{o_7}^{24} t_1^{11} t_2^5 t_3^7 t_4^9 t_5^8 t_6^9 t_7^7 t_8^8 +  4 y_{s}^8 y_{o_1}^{15} y_{o_2}^{17} y_{o_3}^{15} y_{o_4}^{17} y_{o_5}^{16} y_{o_6}^{16} y_{o_7}^{24} t_1^{10} t_2^6 t_3^7 t_4^9 t_5^8 t_6^9 t_7^7 t_8^8 +  6 y_{s}^8 y_{o_1}^{15} y_{o_2}^{17} y_{o_3}^{15} y_{o_4}^{17} y_{o_5}^{16} y_{o_6}^{16} y_{o_7}^{24} t_1^9 t_2^7 t_3^7 t_4^9 t_5^8 t_6^9 t_7^7 t_8^8 +  6 y_{s}^8 y_{o_1}^{15} y_{o_2}^{17} y_{o_3}^{15} y_{o_4}^{17} y_{o_5}^{16} y_{o_6}^{16} y_{o_7}^{24} t_1^8 t_2^8 t_3^7 t_4^9 t_5^8 t_6^9 t_7^7 t_8^8 +  6 y_{s}^8 y_{o_1}^{15} y_{o_2}^{17} y_{o_3}^{15} y_{o_4}^{17} y_{o_5}^{16} y_{o_6}^{16} y_{o_7}^{24} t_1^7 t_2^9 t_3^7 t_4^9 t_5^8 t_6^9 t_7^7 t_8^8 +  4 y_{s}^8 y_{o_1}^{15} y_{o_2}^{17} y_{o_3}^{15} y_{o_4}^{17} y_{o_5}^{16} y_{o_6}^{16} y_{o_7}^{24} t_1^6 t_2^{10} t_3^7 t_4^9 t_5^8 t_6^9 t_7^7 t_8^8 +  y_{s}^8 y_{o_1}^{15} y_{o_2}^{17} y_{o_3}^{15} y_{o_4}^{17} y_{o_5}^{16} y_{o_6}^{16} y_{o_7}^{24} t_1^5 t_2^{11} t_3^7 t_4^9 t_5^8 t_6^9 t_7^7 t_8^8 -  y_{s}^9 y_{o_1}^{16} y_{o_2}^{20} y_{o_3}^{17} y_{o_4}^{19} y_{o_5}^{17} y_{o_6}^{19} y_{o_7}^{26} t_1^{10} t_2^7 t_3^8 t_4^{11} t_5^{10} t_6^{10} t_7^7 t_8^8 -  y_{s}^9 y_{o_1}^{16} y_{o_2}^{20} y_{o_3}^{17} y_{o_4}^{19} y_{o_5}^{17} y_{o_6}^{19} y_{o_7}^{26} t_1^9 t_2^8 t_3^8 t_4^{11} t_5^{10} t_6^{10} t_7^7 t_8^8 -  y_{s}^9 y_{o_1}^{16} y_{o_2}^{20} y_{o_3}^{17} y_{o_4}^{19} y_{o_5}^{17} y_{o_6}^{19} y_{o_7}^{26} t_1^8 t_2^9 t_3^8 t_4^{11} t_5^{10} t_6^{10} t_7^7 t_8^8 -  y_{s}^9 y_{o_1}^{16} y_{o_2}^{20} y_{o_3}^{17} y_{o_4}^{19} y_{o_5}^{17} y_{o_6}^{19} y_{o_7}^{26} t_1^7 t_2^{10} t_3^8 t_4^{11} t_5^{10} t_6^{10} t_7^7 t_8^8 -  y_{s}^5 y_{o_1}^{13} y_{o_2}^7 y_{o_3}^{10} y_{o_4}^{10} y_{o_5}^{13} y_{o_6}^7 y_{o_7}^{18} t_1^7 t_2^6 t_3^5 t_4^2 t_5^2 t_6^5 t_7^8 t_8^8 -  y_{s}^5 y_{o_1}^{13} y_{o_2}^7 y_{o_3}^{10} y_{o_4}^{10} y_{o_5}^{13} y_{o_6}^7 y_{o_7}^{18} t_1^6 t_2^7 t_3^5 t_4^2 t_5^2 t_6^5 t_7^8 t_8^8 +  y_{s}^6 y_{o_1}^{14} y_{o_2}^{10} y_{o_3}^{12} y_{o_4}^{12} y_{o_5}^{14} y_{o_6}^{10} y_{o_7}^{20} t_1^9 t_2^5 t_3^6 t_4^4 t_5^4 t_6^6 t_7^8 t_8^8 +  3 y_{s}^6 y_{o_1}^{14} y_{o_2}^{10} y_{o_3}^{12} y_{o_4}^{12} y_{o_5}^{14} y_{o_6}^{10} y_{o_7}^{20} t_1^8 t_2^6 t_3^6 t_4^4 t_5^4 t_6^6 t_7^8 t_8^8 +  3 y_{s}^6 y_{o_1}^{14} y_{o_2}^{10} y_{o_3}^{12} y_{o_4}^{12} y_{o_5}^{14} y_{o_6}^{10} y_{o_7}^{20} t_1^7 t_2^7 t_3^6 t_4^4 t_5^4 t_6^6 t_7^8 t_8^8 +  3 y_{s}^6 y_{o_1}^{14} y_{o_2}^{10} y_{o_3}^{12} y_{o_4}^{12} y_{o_5}^{14} y_{o_6}^{10} y_{o_7}^{20} t_1^6 t_2^8 t_3^6 t_4^4 t_5^4 t_6^6 t_7^8 t_8^8 +  y_{s}^6 y_{o_1}^{14} y_{o_2}^{10} y_{o_3}^{12} y_{o_4}^{12} y_{o_5}^{14} y_{o_6}^{10} y_{o_7}^{20} t_1^5 t_2^9 t_3^6 t_4^4 t_5^4 t_6^6 t_7^8 t_8^8 -  4 y_{s}^7 y_{o_1}^{15} y_{o_2}^{13} y_{o_3}^{14} y_{o_4}^{14} y_{o_5}^{15} y_{o_6}^{13} y_{o_7}^{22} t_1^{10} t_2^5 t_3^7 t_4^6 t_5^6 t_6^7 t_7^8 t_8^8 -  6 y_{s}^7 y_{o_1}^{15} y_{o_2}^{13} y_{o_3}^{14} y_{o_4}^{14} y_{o_5}^{15} y_{o_6}^{13} y_{o_7}^{22} t_1^9 t_2^6 t_3^7 t_4^6 t_5^6 t_6^7 t_7^8 t_8^8 -  8 y_{s}^7 y_{o_1}^{15} y_{o_2}^{13} y_{o_3}^{14} y_{o_4}^{14} y_{o_5}^{15} y_{o_6}^{13} y_{o_7}^{22} t_1^8 t_2^7 t_3^7 t_4^6 t_5^6 t_6^7 t_7^8 t_8^8 -  8 y_{s}^7 y_{o_1}^{15} y_{o_2}^{13} y_{o_3}^{14} y_{o_4}^{14} y_{o_5}^{15} y_{o_6}^{13} y_{o_7}^{22} t_1^7 t_2^8 t_3^7 t_4^6 t_5^6 t_6^7 t_7^8 t_8^8 -  6 y_{s}^7 y_{o_1}^{15} y_{o_2}^{13} y_{o_3}^{14} y_{o_4}^{14} y_{o_5}^{15} y_{o_6}^{13} y_{o_7}^{22} t_1^6 t_2^9 t_3^7 t_4^6 t_5^6 t_6^7 t_7^8 t_8^8 -  4 y_{s}^7 y_{o_1}^{15} y_{o_2}^{13} y_{o_3}^{14} y_{o_4}^{14} y_{o_5}^{15} y_{o_6}^{13} y_{o_7}^{22} t_1^5 t_2^{10} t_3^7 t_4^6 t_5^6 t_6^7 t_7^8 t_8^8 +  2 y_{s}^8 y_{o_1}^{16} y_{o_2}^{16} y_{o_3}^{16} y_{o_4}^{16} y_{o_5}^{16} y_{o_6}^{16} y_{o_7}^{24} t_1^{11} t_2^5 t_3^8 t_4^8 t_5^8 t_6^8 t_7^8 t_8^8 +  6 y_{s}^8 y_{o_1}^{16} y_{o_2}^{16} y_{o_3}^{16} y_{o_4}^{16} y_{o_5}^{16} y_{o_6}^{16} y_{o_7}^{24} t_1^{10} t_2^6 t_3^8 t_4^8 t_5^8 t_6^8 t_7^8 t_8^8 +  8 y_{s}^8 y_{o_1}^{16} y_{o_2}^{16} y_{o_3}^{16} y_{o_4}^{16} y_{o_5}^{16} y_{o_6}^{16} y_{o_7}^{24} t_1^9 t_2^7 t_3^8 t_4^8 t_5^8 t_6^8 t_7^8 t_8^8 +  7 y_{s}^8 y_{o_1}^{16} y_{o_2}^{16} y_{o_3}^{16} y_{o_4}^{16} y_{o_5}^{16} y_{o_6}^{16} y_{o_7}^{24} t_1^8 t_2^8 t_3^8 t_4^8 t_5^8 t_6^8 t_7^8 t_8^8 +  8 y_{s}^8 y_{o_1}^{16} y_{o_2}^{16} y_{o_3}^{16} y_{o_4}^{16} y_{o_5}^{16} y_{o_6}^{16} y_{o_7}^{24} t_1^7 t_2^9 t_3^8 t_4^8 t_5^8 t_6^8 t_7^8 t_8^8 +  6 y_{s}^8 y_{o_1}^{16} y_{o_2}^{16} y_{o_3}^{16} y_{o_4}^{16} y_{o_5}^{16} y_{o_6}^{16} y_{o_7}^{24} t_1^6 t_2^{10} t_3^8 t_4^8 t_5^8 t_6^8 t_7^8 t_8^8 +  2 y_{s}^8 y_{o_1}^{16} y_{o_2}^{16} y_{o_3}^{16} y_{o_4}^{16} y_{o_5}^{16} y_{o_6}^{16} y_{o_7}^{24} t_1^5 t_2^{11} t_3^8 t_4^8 t_5^8 t_6^8 t_7^8 t_8^8 -  y_{s}^9 y_{o_1}^{17} y_{o_2}^{19} y_{o_3}^{18} y_{o_4}^{18} y_{o_5}^{17} y_{o_6}^{19} y_{o_7}^{26} t_1^{11} t_2^6 t_3^9 t_4^{10} t_5^{10} t_6^9 t_7^8 t_8^8 -  2 y_{s}^9 y_{o_1}^{17} y_{o_2}^{19} y_{o_3}^{18} y_{o_4}^{18} y_{o_5}^{17} y_{o_6}^{19} y_{o_7}^{26} t_1^{10} t_2^7 t_3^9 t_4^{10} t_5^{10} t_6^9 t_7^8 t_8^8 -  2 y_{s}^9 y_{o_1}^{17} y_{o_2}^{19} y_{o_3}^{18} y_{o_4}^{18} y_{o_5}^{17} y_{o_6}^{19} y_{o_7}^{26} t_1^9 t_2^8 t_3^9 t_4^{10} t_5^{10} t_6^9 t_7^8 t_8^8 -  2 y_{s}^9 y_{o_1}^{17} y_{o_2}^{19} y_{o_3}^{18} y_{o_4}^{18} y_{o_5}^{17} y_{o_6}^{19} y_{o_7}^{26} t_1^8 t_2^9 t_3^9 t_4^{10} t_5^{10} t_6^9 t_7^8 t_8^8 -  2 y_{s}^9 y_{o_1}^{17} y_{o_2}^{19} y_{o_3}^{18} y_{o_4}^{18} y_{o_5}^{17} y_{o_6}^{19} y_{o_7}^{26} t_1^7 t_2^{10} t_3^9 t_4^{10} t_5^{10} t_6^9 t_7^8 t_8^8 -  y_{s}^9 y_{o_1}^{17} y_{o_2}^{19} y_{o_3}^{18} y_{o_4}^{18} y_{o_5}^{17} y_{o_6}^{19} y_{o_7}^{26} t_1^6 t_2^{11} t_3^9 t_4^{10} t_5^{10} t_6^9 t_7^8 t_8^8 +  y_{s}^6 y_{o_1}^{15} y_{o_2}^9 y_{o_3}^{13} y_{o_4}^{11} y_{o_5}^{14} y_{o_6}^{10} y_{o_7}^{20} t_1^8 t_2^6 t_3^7 t_4^3 t_5^4 t_6^5 t_7^9 t_8^8 +  2 y_{s}^6 y_{o_1}^{15} y_{o_2}^9 y_{o_3}^{13} y_{o_4}^{11} y_{o_5}^{14} y_{o_6}^{10} y_{o_7}^{20} t_1^7 t_2^7 t_3^7 t_4^3 t_5^4 t_6^5 t_7^9 t_8^8 +  y_{s}^6 y_{o_1}^{15} y_{o_2}^9 y_{o_3}^{13} y_{o_4}^{11} y_{o_5}^{14} y_{o_6}^{10} y_{o_7}^{20} t_1^6 t_2^8 t_3^7 t_4^3 t_5^4 t_6^5 t_7^9 t_8^8 -  2 y_{s}^7 y_{o_1}^{16} y_{o_2}^{12} y_{o_3}^{15} y_{o_4}^{13} y_{o_5}^{15} y_{o_6}^{13} y_{o_7}^{22} t_1^{10} t_2^5 t_3^8 t_4^5 t_5^6 t_6^6 t_7^9 t_8^8 -  3 y_{s}^7 y_{o_1}^{16} y_{o_2}^{12} y_{o_3}^{15} y_{o_4}^{13} y_{o_5}^{15} y_{o_6}^{13} y_{o_7}^{22} t_1^9 t_2^6 t_3^8 t_4^5 t_5^6 t_6^6 t_7^9 t_8^8 -  6 y_{s}^7 y_{o_1}^{16} y_{o_2}^{12} y_{o_3}^{15} y_{o_4}^{13} y_{o_5}^{15} y_{o_6}^{13} y_{o_7}^{22} t_1^8 t_2^7 t_3^8 t_4^5 t_5^6 t_6^6 t_7^9 t_8^8 -  6 y_{s}^7 y_{o_1}^{16} y_{o_2}^{12} y_{o_3}^{15} y_{o_4}^{13} y_{o_5}^{15} y_{o_6}^{13} y_{o_7}^{22} t_1^7 t_2^8 t_3^8 t_4^5 t_5^6 t_6^6 t_7^9 t_8^8 -  3 y_{s}^7 y_{o_1}^{16} y_{o_2}^{12} y_{o_3}^{15} y_{o_4}^{13} y_{o_5}^{15} y_{o_6}^{13} y_{o_7}^{22} t_1^6 t_2^9 t_3^8 t_4^5 t_5^6 t_6^6 t_7^9 t_8^8 -  2 y_{s}^7 y_{o_1}^{16} y_{o_2}^{12} y_{o_3}^{15} y_{o_4}^{13} y_{o_5}^{15} y_{o_6}^{13} y_{o_7}^{22} t_1^5 t_2^{10} t_3^8 t_4^5 t_5^6 t_6^6 t_7^9 t_8^8 +  y_{s}^8 y_{o_1}^{17} y_{o_2}^{15} y_{o_3}^{17} y_{o_4}^{15} y_{o_5}^{16} y_{o_6}^{16} y_{o_7}^{24} t_1^{11} t_2^5 t_3^9 t_4^7 t_5^8 t_6^7 t_7^9 t_8^8 +  4 y_{s}^8 y_{o_1}^{17} y_{o_2}^{15} y_{o_3}^{17} y_{o_4}^{15} y_{o_5}^{16} y_{o_6}^{16} y_{o_7}^{24} t_1^{10} t_2^6 t_3^9 t_4^7 t_5^8 t_6^7 t_7^9 t_8^8 +  6 y_{s}^8 y_{o_1}^{17} y_{o_2}^{15} y_{o_3}^{17} y_{o_4}^{15} y_{o_5}^{16} y_{o_6}^{16} y_{o_7}^{24} t_1^9 t_2^7 t_3^9 t_4^7 t_5^8 t_6^7 t_7^9 t_8^8 +  6 y_{s}^8 y_{o_1}^{17} y_{o_2}^{15} y_{o_3}^{17} y_{o_4}^{15} y_{o_5}^{16} y_{o_6}^{16} y_{o_7}^{24} t_1^8 t_2^8 t_3^9 t_4^7 t_5^8 t_6^7 t_7^9 t_8^8 +  6 y_{s}^8 y_{o_1}^{17} y_{o_2}^{15} y_{o_3}^{17} y_{o_4}^{15} y_{o_5}^{16} y_{o_6}^{16} y_{o_7}^{24} t_1^7 t_2^9 t_3^9 t_4^7 t_5^8 t_6^7 t_7^9 t_8^8 +  4 y_{s}^8 y_{o_1}^{17} y_{o_2}^{15} y_{o_3}^{17} y_{o_4}^{15} y_{o_5}^{16} y_{o_6}^{16} y_{o_7}^{24} t_1^6 t_2^{10} t_3^9 t_4^7 t_5^8 t_6^7 t_7^9 t_8^8 +  y_{s}^8 y_{o_1}^{17} y_{o_2}^{15} y_{o_3}^{17} y_{o_4}^{15} y_{o_5}^{16} y_{o_6}^{16} y_{o_7}^{24} t_1^5 t_2^{11} t_3^9 t_4^7 t_5^8 t_6^7 t_7^9 t_8^8 -  y_{s}^9 y_{o_1}^{18} y_{o_2}^{18} y_{o_3}^{19} y_{o_4}^{17} y_{o_5}^{17} y_{o_6}^{19} y_{o_7}^{26} t_1^{11} t_2^6 t_3^{10} t_4^9 t_5^{10} t_6^8 t_7^9 t_8^8 -  2 y_{s}^9 y_{o_1}^{18} y_{o_2}^{18} y_{o_3}^{19} y_{o_4}^{17} y_{o_5}^{17} y_{o_6}^{19} y_{o_7}^{26} t_1^{10} t_2^7 t_3^{10} t_4^9 t_5^{10} t_6^8 t_7^9 t_8^8 -  2 y_{s}^9 y_{o_1}^{18} y_{o_2}^{18} y_{o_3}^{19} y_{o_4}^{17} y_{o_5}^{17} y_{o_6}^{19} y_{o_7}^{26} t_1^9 t_2^8 t_3^{10} t_4^9 t_5^{10} t_6^8 t_7^9 t_8^8 -  2 y_{s}^9 y_{o_1}^{18} y_{o_2}^{18} y_{o_3}^{19} y_{o_4}^{17} y_{o_5}^{17} y_{o_6}^{19} y_{o_7}^{26} t_1^8 t_2^9 t_3^{10} t_4^9 t_5^{10} t_6^8 t_7^9 t_8^8 -  2 y_{s}^9 y_{o_1}^{18} y_{o_2}^{18} y_{o_3}^{19} y_{o_4}^{17} y_{o_5}^{17} y_{o_6}^{19} y_{o_7}^{26} t_1^7 t_2^{10} t_3^{10} t_4^9 t_5^{10} t_6^8 t_7^9 t_8^8 -  y_{s}^9 y_{o_1}^{18} y_{o_2}^{18} y_{o_3}^{19} y_{o_4}^{17} y_{o_5}^{17} y_{o_6}^{19} y_{o_7}^{26} t_1^6 t_2^{11} t_3^{10} t_4^9 t_5^{10} t_6^8 t_7^9 t_8^8 -  y_{s}^7 y_{o_1}^{17} y_{o_2}^{11} y_{o_3}^{16} y_{o_4}^{12} y_{o_5}^{15} y_{o_6}^{13} y_{o_7}^{22} t_1^8 t_2^7 t_3^9 t_4^4 t_5^6 t_6^5 t_7^{10} t_8^8 -  y_{s}^7 y_{o_1}^{17} y_{o_2}^{11} y_{o_3}^{16} y_{o_4}^{12} y_{o_5}^{15} y_{o_6}^{13} y_{o_7}^{22} t_1^7 t_2^8 t_3^9 t_4^4 t_5^6 t_6^5 t_7^{10} t_8^8 +  y_{s}^8 y_{o_1}^{18} y_{o_2}^{14} y_{o_3}^{18} y_{o_4}^{14} y_{o_5}^{16} y_{o_6}^{16} y_{o_7}^{24} t_1^{10} t_2^6 t_3^{10} t_4^6 t_5^8 t_6^6 t_7^{10} t_8^8 +  2 y_{s}^8 y_{o_1}^{18} y_{o_2}^{14} y_{o_3}^{18} y_{o_4}^{14} y_{o_5}^{16} y_{o_6}^{16} y_{o_7}^{24} t_1^9 t_2^7 t_3^{10} t_4^6 t_5^8 t_6^6 t_7^{10} t_8^8 +  2 y_{s}^8 y_{o_1}^{18} y_{o_2}^{14} y_{o_3}^{18} y_{o_4}^{14} y_{o_5}^{16} y_{o_6}^{16} y_{o_7}^{24} t_1^8 t_2^8 t_3^{10} t_4^6 t_5^8 t_6^6 t_7^{10} t_8^8 +  2 y_{s}^8 y_{o_1}^{18} y_{o_2}^{14} y_{o_3}^{18} y_{o_4}^{14} y_{o_5}^{16} y_{o_6}^{16} y_{o_7}^{24} t_1^7 t_2^9 t_3^{10} t_4^6 t_5^8 t_6^6 t_7^{10} t_8^8 +  y_{s}^8 y_{o_1}^{18} y_{o_2}^{14} y_{o_3}^{18} y_{o_4}^{14} y_{o_5}^{16} y_{o_6}^{16} y_{o_7}^{24} t_1^6 t_2^{10} t_3^{10} t_4^6 t_5^8 t_6^6 t_7^{10} t_8^8 -  y_{s}^9 y_{o_1}^{19} y_{o_2}^{17} y_{o_3}^{20} y_{o_4}^{16} y_{o_5}^{17} y_{o_6}^{19} y_{o_7}^{26} t_1^{10} t_2^7 t_3^{11} t_4^8 t_5^{10} t_6^7 t_7^{10} t_8^8 -  y_{s}^9 y_{o_1}^{19} y_{o_2}^{17} y_{o_3}^{20} y_{o_4}^{16} y_{o_5}^{17} y_{o_6}^{19} y_{o_7}^{26} t_1^9 t_2^8 t_3^{11} t_4^8 t_5^{10} t_6^7 t_7^{10} t_8^8 -  y_{s}^9 y_{o_1}^{19} y_{o_2}^{17} y_{o_3}^{20} y_{o_4}^{16} y_{o_5}^{17} y_{o_6}^{19} y_{o_7}^{26} t_1^8 t_2^9 t_3^{11} t_4^8 t_5^{10} t_6^7 t_7^{10} t_8^8 -  y_{s}^9 y_{o_1}^{19} y_{o_2}^{17} y_{o_3}^{20} y_{o_4}^{16} y_{o_5}^{17} y_{o_6}^{19} y_{o_7}^{26} t_1^7 t_2^{10} t_3^{11} t_4^8 t_5^{10} t_6^7 t_7^{10} t_8^8 -  y_{s}^7 y_{o_1}^{13} y_{o_2}^{15} y_{o_3}^{11} y_{o_4}^{17} y_{o_5}^{16} y_{o_6}^{12} y_{o_7}^{23} t_1^8 t_2^8 t_3^4 t_4^8 t_5^5 t_6^{10} t_7^6 t_8^9 +  y_{s}^6 y_{o_1}^{13} y_{o_2}^{11} y_{o_3}^{10} y_{o_4}^{14} y_{o_5}^{15} y_{o_6}^9 y_{o_7}^{21} t_1^9 t_2^6 t_3^4 t_4^5 t_5^3 t_6^8 t_7^7 t_8^9 +  y_{s}^6 y_{o_1}^{13} y_{o_2}^{11} y_{o_3}^{10} y_{o_4}^{14} y_{o_5}^{15} y_{o_6}^9 y_{o_7}^{21} t_1^8 t_2^7 t_3^4 t_4^5 t_5^3 t_6^8 t_7^7 t_8^9 +  y_{s}^6 y_{o_1}^{13} y_{o_2}^{11} y_{o_3}^{10} y_{o_4}^{14} y_{o_5}^{15} y_{o_6}^9 y_{o_7}^{21} t_1^7 t_2^8 t_3^4 t_4^5 t_5^3 t_6^8 t_7^7 t_8^9 +  y_{s}^6 y_{o_1}^{13} y_{o_2}^{11} y_{o_3}^{10} y_{o_4}^{14} y_{o_5}^{15} y_{o_6}^9 y_{o_7}^{21} t_1^6 t_2^9 t_3^4 t_4^5 t_5^3 t_6^8 t_7^7 t_8^9 -  y_{s}^7 y_{o_1}^{14} y_{o_2}^{14} y_{o_3}^{12} y_{o_4}^{16} y_{o_5}^{16} y_{o_6}^{12} y_{o_7}^{23} t_1^{10} t_2^6 t_3^5 t_4^7 t_5^5 t_6^9 t_7^7 t_8^9 -  2 y_{s}^7 y_{o_1}^{14} y_{o_2}^{14} y_{o_3}^{12} y_{o_4}^{16} y_{o_5}^{16} y_{o_6}^{12} y_{o_7}^{23} t_1^9 t_2^7 t_3^5 t_4^7 t_5^5 t_6^9 t_7^7 t_8^9 -  3 y_{s}^7 y_{o_1}^{14} y_{o_2}^{14} y_{o_3}^{12} y_{o_4}^{16} y_{o_5}^{16} y_{o_6}^{12} y_{o_7}^{23} t_1^8 t_2^8 t_3^5 t_4^7 t_5^5 t_6^9 t_7^7 t_8^9 -  2 y_{s}^7 y_{o_1}^{14} y_{o_2}^{14} y_{o_3}^{12} y_{o_4}^{16} y_{o_5}^{16} y_{o_6}^{12} y_{o_7}^{23} t_1^7 t_2^9 t_3^5 t_4^7 t_5^5 t_6^9 t_7^7 t_8^9 -  y_{s}^7 y_{o_1}^{14} y_{o_2}^{14} y_{o_3}^{12} y_{o_4}^{16} y_{o_5}^{16} y_{o_6}^{12} y_{o_7}^{23} t_1^6 t_2^{10} t_3^5 t_4^7 t_5^5 t_6^9 t_7^7 t_8^9 +  y_{s}^8 y_{o_1}^{15} y_{o_2}^{17} y_{o_3}^{14} y_{o_4}^{18} y_{o_5}^{17} y_{o_6}^{15} y_{o_7}^{25} t_1^{10} t_2^7 t_3^6 t_4^9 t_5^7 t_6^{10} t_7^7 t_8^9 +  2 y_{s}^8 y_{o_1}^{15} y_{o_2}^{17} y_{o_3}^{14} y_{o_4}^{18} y_{o_5}^{17} y_{o_6}^{15} y_{o_7}^{25} t_1^9 t_2^8 t_3^6 t_4^9 t_5^7 t_6^{10} t_7^7 t_8^9 +  2 y_{s}^8 y_{o_1}^{15} y_{o_2}^{17} y_{o_3}^{14} y_{o_4}^{18} y_{o_5}^{17} y_{o_6}^{15} y_{o_7}^{25} t_1^8 t_2^9 t_3^6 t_4^9 t_5^7 t_6^{10} t_7^7 t_8^9 +  y_{s}^8 y_{o_1}^{15} y_{o_2}^{17} y_{o_3}^{14} y_{o_4}^{18} y_{o_5}^{17} y_{o_6}^{15} y_{o_7}^{25} t_1^7 t_2^{10} t_3^6 t_4^9 t_5^7 t_6^{10} t_7^7 t_8^9 +  y_{s}^6 y_{o_1}^{14} y_{o_2}^{10} y_{o_3}^{11} y_{o_4}^{13} y_{o_5}^{15} y_{o_6}^9 y_{o_7}^{21} t_1^9 t_2^6 t_3^5 t_4^4 t_5^3 t_6^7 t_7^8 t_8^9 +  y_{s}^6 y_{o_1}^{14} y_{o_2}^{10} y_{o_3}^{11} y_{o_4}^{13} y_{o_5}^{15} y_{o_6}^9 y_{o_7}^{21} t_1^8 t_2^7 t_3^5 t_4^4 t_5^3 t_6^7 t_7^8 t_8^9 +  y_{s}^6 y_{o_1}^{14} y_{o_2}^{10} y_{o_3}^{11} y_{o_4}^{13} y_{o_5}^{15} y_{o_6}^9 y_{o_7}^{21} t_1^7 t_2^8 t_3^5 t_4^4 t_5^3 t_6^7 t_7^8 t_8^9 +  y_{s}^6 y_{o_1}^{14} y_{o_2}^{10} y_{o_3}^{11} y_{o_4}^{13} y_{o_5}^{15} y_{o_6}^9 y_{o_7}^{21} t_1^6 t_2^9 t_3^5 t_4^4 t_5^3 t_6^7 t_7^8 t_8^9 -  3 y_{s}^7 y_{o_1}^{15} y_{o_2}^{13} y_{o_3}^{13} y_{o_4}^{15} y_{o_5}^{16} y_{o_6}^{12} y_{o_7}^{23} t_1^{10} t_2^6 t_3^6 t_4^6 t_5^5 t_6^8 t_7^8 t_8^9 -  5 y_{s}^7 y_{o_1}^{15} y_{o_2}^{13} y_{o_3}^{13} y_{o_4}^{15} y_{o_5}^{16} y_{o_6}^{12} y_{o_7}^{23} t_1^9 t_2^7 t_3^6 t_4^6 t_5^5 t_6^8 t_7^8 t_8^9 -  7 y_{s}^7 y_{o_1}^{15} y_{o_2}^{13} y_{o_3}^{13} y_{o_4}^{15} y_{o_5}^{16} y_{o_6}^{12} y_{o_7}^{23} t_1^8 t_2^8 t_3^6 t_4^6 t_5^5 t_6^8 t_7^8 t_8^9 -  5 y_{s}^7 y_{o_1}^{15} y_{o_2}^{13} y_{o_3}^{13} y_{o_4}^{15} y_{o_5}^{16} y_{o_6}^{12} y_{o_7}^{23} t_1^7 t_2^9 t_3^6 t_4^6 t_5^5 t_6^8 t_7^8 t_8^9 -  3 y_{s}^7 y_{o_1}^{15} y_{o_2}^{13} y_{o_3}^{13} y_{o_4}^{15} y_{o_5}^{16} y_{o_6}^{12} y_{o_7}^{23} t_1^6 t_2^{10} t_3^6 t_4^6 t_5^5 t_6^8 t_7^8 t_8^9 +  y_{s}^8 y_{o_1}^{16} y_{o_2}^{16} y_{o_3}^{15} y_{o_4}^{17} y_{o_5}^{17} y_{o_6}^{15} y_{o_7}^{25} t_1^{11} t_2^6 t_3^7 t_4^8 t_5^7 t_6^9 t_7^8 t_8^9 +  6 y_{s}^8 y_{o_1}^{16} y_{o_2}^{16} y_{o_3}^{15} y_{o_4}^{17} y_{o_5}^{17} y_{o_6}^{15} y_{o_7}^{25} t_1^{10} t_2^7 t_3^7 t_4^8 t_5^7 t_6^9 t_7^8 t_8^9 +  7 y_{s}^8 y_{o_1}^{16} y_{o_2}^{16} y_{o_3}^{15} y_{o_4}^{17} y_{o_5}^{17} y_{o_6}^{15} y_{o_7}^{25} t_1^9 t_2^8 t_3^7 t_4^8 t_5^7 t_6^9 t_7^8 t_8^9 +  7 y_{s}^8 y_{o_1}^{16} y_{o_2}^{16} y_{o_3}^{15} y_{o_4}^{17} y_{o_5}^{17} y_{o_6}^{15} y_{o_7}^{25} t_1^8 t_2^9 t_3^7 t_4^8 t_5^7 t_6^9 t_7^8 t_8^9 +  6 y_{s}^8 y_{o_1}^{16} y_{o_2}^{16} y_{o_3}^{15} y_{o_4}^{17} y_{o_5}^{17} y_{o_6}^{15} y_{o_7}^{25} t_1^7 t_2^{10} t_3^7 t_4^8 t_5^7 t_6^9 t_7^8 t_8^9 +  y_{s}^8 y_{o_1}^{16} y_{o_2}^{16} y_{o_3}^{15} y_{o_4}^{17} y_{o_5}^{17} y_{o_6}^{15} y_{o_7}^{25} t_1^6 t_2^{11} t_3^7 t_4^8 t_5^7 t_6^9 t_7^8 t_8^9 -  y_{s}^9 y_{o_1}^{17} y_{o_2}^{19} y_{o_3}^{17} y_{o_4}^{19} y_{o_5}^{18} y_{o_6}^{18} y_{o_7}^{27} t_1^{11} t_2^7 t_3^8 t_4^{10} t_5^9 t_6^{10} t_7^8 t_8^9 -  2 y_{s}^9 y_{o_1}^{17} y_{o_2}^{19} y_{o_3}^{17} y_{o_4}^{19} y_{o_5}^{18} y_{o_6}^{18} y_{o_7}^{27} t_1^{10} t_2^8 t_3^8 t_4^{10} t_5^9 t_6^{10} t_7^8 t_8^9 -  3 y_{s}^9 y_{o_1}^{17} y_{o_2}^{19} y_{o_3}^{17} y_{o_4}^{19} y_{o_5}^{18} y_{o_6}^{18} y_{o_7}^{27} t_1^9 t_2^9 t_3^8 t_4^{10} t_5^9 t_6^{10} t_7^8 t_8^9 -  2 y_{s}^9 y_{o_1}^{17} y_{o_2}^{19} y_{o_3}^{17} y_{o_4}^{19} y_{o_5}^{18} y_{o_6}^{18} y_{o_7}^{27} t_1^8 t_2^{10} t_3^8 t_4^{10} t_5^9 t_6^{10} t_7^8 t_8^9 -  y_{s}^9 y_{o_1}^{17} y_{o_2}^{19} y_{o_3}^{17} y_{o_4}^{19} y_{o_5}^{18} y_{o_6}^{18} y_{o_7}^{27} t_1^7 t_2^{11} t_3^8 t_4^{10} t_5^9 t_6^{10} t_7^8 t_8^9 -  y_{s}^7 y_{o_1}^{16} y_{o_2}^{12} y_{o_3}^{14} y_{o_4}^{14} y_{o_5}^{16} y_{o_6}^{12} y_{o_7}^{23} t_1^{10} t_2^6 t_3^7 t_4^5 t_5^5 t_6^7 t_7^9 t_8^9 -  2 y_{s}^7 y_{o_1}^{16} y_{o_2}^{12} y_{o_3}^{14} y_{o_4}^{14} y_{o_5}^{16} y_{o_6}^{12} y_{o_7}^{23} t_1^9 t_2^7 t_3^7 t_4^5 t_5^5 t_6^7 t_7^9 t_8^9 -  3 y_{s}^7 y_{o_1}^{16} y_{o_2}^{12} y_{o_3}^{14} y_{o_4}^{14} y_{o_5}^{16} y_{o_6}^{12} y_{o_7}^{23} t_1^8 t_2^8 t_3^7 t_4^5 t_5^5 t_6^7 t_7^9 t_8^9 -  2 y_{s}^7 y_{o_1}^{16} y_{o_2}^{12} y_{o_3}^{14} y_{o_4}^{14} y_{o_5}^{16} y_{o_6}^{12} y_{o_7}^{23} t_1^7 t_2^9 t_3^7 t_4^5 t_5^5 t_6^7 t_7^9 t_8^9 -  y_{s}^7 y_{o_1}^{16} y_{o_2}^{12} y_{o_3}^{14} y_{o_4}^{14} y_{o_5}^{16} y_{o_6}^{12} y_{o_7}^{23} t_1^6 t_2^{10} t_3^7 t_4^5 t_5^5 t_6^7 t_7^9 t_8^9 +  y_{s}^8 y_{o_1}^{17} y_{o_2}^{15} y_{o_3}^{16} y_{o_4}^{16} y_{o_5}^{17} y_{o_6}^{15} y_{o_7}^{25} t_1^{11} t_2^6 t_3^8 t_4^7 t_5^7 t_6^8 t_7^9 t_8^9 +  6 y_{s}^8 y_{o_1}^{17} y_{o_2}^{15} y_{o_3}^{16} y_{o_4}^{16} y_{o_5}^{17} y_{o_6}^{15} y_{o_7}^{25} t_1^{10} t_2^7 t_3^8 t_4^7 t_5^7 t_6^8 t_7^9 t_8^9 +  7 y_{s}^8 y_{o_1}^{17} y_{o_2}^{15} y_{o_3}^{16} y_{o_4}^{16} y_{o_5}^{17} y_{o_6}^{15} y_{o_7}^{25} t_1^9 t_2^8 t_3^8 t_4^7 t_5^7 t_6^8 t_7^9 t_8^9 +  7 y_{s}^8 y_{o_1}^{17} y_{o_2}^{15} y_{o_3}^{16} y_{o_4}^{16} y_{o_5}^{17} y_{o_6}^{15} y_{o_7}^{25} t_1^8 t_2^9 t_3^8 t_4^7 t_5^7 t_6^8 t_7^9 t_8^9 +  6 y_{s}^8 y_{o_1}^{17} y_{o_2}^{15} y_{o_3}^{16} y_{o_4}^{16} y_{o_5}^{17} y_{o_6}^{15} y_{o_7}^{25} t_1^7 t_2^{10} t_3^8 t_4^7 t_5^7 t_6^8 t_7^9 t_8^9 +  y_{s}^8 y_{o_1}^{17} y_{o_2}^{15} y_{o_3}^{16} y_{o_4}^{16} y_{o_5}^{17} y_{o_6}^{15} y_{o_7}^{25} t_1^6 t_2^{11} t_3^8 t_4^7 t_5^7 t_6^8 t_7^9 t_8^9 -  2 y_{s}^9 y_{o_1}^{18} y_{o_2}^{18} y_{o_3}^{18} y_{o_4}^{18} y_{o_5}^{18} y_{o_6}^{18} y_{o_7}^{27} t_1^{11} t_2^7 t_3^9 t_4^9 t_5^9 t_6^9 t_7^9 t_8^9 -  5 y_{s}^9 y_{o_1}^{18} y_{o_2}^{18} y_{o_3}^{18} y_{o_4}^{18} y_{o_5}^{18} y_{o_6}^{18} y_{o_7}^{27} t_1^{10} t_2^8 t_3^9 t_4^9 t_5^9 t_6^9 t_7^9 t_8^9 -  5 y_{s}^9 y_{o_1}^{18} y_{o_2}^{18} y_{o_3}^{18} y_{o_4}^{18} y_{o_5}^{18} y_{o_6}^{18} y_{o_7}^{27} t_1^9 t_2^9 t_3^9 t_4^9 t_5^9 t_6^9 t_7^9 t_8^9 -  5 y_{s}^9 y_{o_1}^{18} y_{o_2}^{18} y_{o_3}^{18} y_{o_4}^{18} y_{o_5}^{18} y_{o_6}^{18} y_{o_7}^{27} t_1^8 t_2^{10} t_3^9 t_4^9 t_5^9 t_6^9 t_7^9 t_8^9 -  2 y_{s}^9 y_{o_1}^{18} y_{o_2}^{18} y_{o_3}^{18} y_{o_4}^{18} y_{o_5}^{18} y_{o_6}^{18} y_{o_7}^{27} t_1^7 t_2^{11} t_3^9 t_4^9 t_5^9 t_6^9 t_7^9 t_8^9 +  y_{s}^{10} y_{o_1}^{19} y_{o_2}^{21} y_{o_3}^{20} y_{o_4}^{20} y_{o_5}^{19} y_{o_6}^{21} y_{o_7}^{29} t_1^{10} t_2^9 t_3^{10} t_4^{11} t_5^{11} t_6^{10} t_7^9 t_8^9 +  y_{s}^{10} y_{o_1}^{19} y_{o_2}^{21} y_{o_3}^{20} y_{o_4}^{20} y_{o_5}^{19} y_{o_6}^{21} y_{o_7}^{29} t_1^9 t_2^{10} t_3^{10} t_4^{11} t_5^{11} t_6^{10} t_7^9 t_8^9 -  y_{s}^7 y_{o_1}^{17} y_{o_2}^{11} y_{o_3}^{15} y_{o_4}^{13} y_{o_5}^{16} y_{o_6}^{12} y_{o_7}^{23} t_1^8 t_2^8 t_3^8 t_4^4 t_5^5 t_6^6 t_7^{10} t_8^9 +  y_{s}^8 y_{o_1}^{18} y_{o_2}^{14} y_{o_3}^{17} y_{o_4}^{15} y_{o_5}^{17} y_{o_6}^{15} y_{o_7}^{25} t_1^{10} t_2^7 t_3^9 t_4^6 t_5^7 t_6^7 t_7^{10} t_8^9 +  2 y_{s}^8 y_{o_1}^{18} y_{o_2}^{14} y_{o_3}^{17} y_{o_4}^{15} y_{o_5}^{17} y_{o_6}^{15} y_{o_7}^{25} t_1^9 t_2^8 t_3^9 t_4^6 t_5^7 t_6^7 t_7^{10} t_8^9 +  2 y_{s}^8 y_{o_1}^{18} y_{o_2}^{14} y_{o_3}^{17} y_{o_4}^{15} y_{o_5}^{17} y_{o_6}^{15} y_{o_7}^{25} t_1^8 t_2^9 t_3^9 t_4^6 t_5^7 t_6^7 t_7^{10} t_8^9 +  y_{s}^8 y_{o_1}^{18} y_{o_2}^{14} y_{o_3}^{17} y_{o_4}^{15} y_{o_5}^{17} y_{o_6}^{15} y_{o_7}^{25} t_1^7 t_2^{10} t_3^9 t_4^6 t_5^7 t_6^7 t_7^{10} t_8^9 -  y_{s}^9 y_{o_1}^{19} y_{o_2}^{17} y_{o_3}^{19} y_{o_4}^{17} y_{o_5}^{18} y_{o_6}^{18} y_{o_7}^{27} t_1^{11} t_2^7 t_3^{10} t_4^8 t_5^9 t_6^8 t_7^{10} t_8^9 -  2 y_{s}^9 y_{o_1}^{19} y_{o_2}^{17} y_{o_3}^{19} y_{o_4}^{17} y_{o_5}^{18} y_{o_6}^{18} y_{o_7}^{27} t_1^{10} t_2^8 t_3^{10} t_4^8 t_5^9 t_6^8 t_7^{10} t_8^9 -  3 y_{s}^9 y_{o_1}^{19} y_{o_2}^{17} y_{o_3}^{19} y_{o_4}^{17} y_{o_5}^{18} y_{o_6}^{18} y_{o_7}^{27} t_1^9 t_2^9 t_3^{10} t_4^8 t_5^9 t_6^8 t_7^{10} t_8^9 -  2 y_{s}^9 y_{o_1}^{19} y_{o_2}^{17} y_{o_3}^{19} y_{o_4}^{17} y_{o_5}^{18} y_{o_6}^{18} y_{o_7}^{27} t_1^8 t_2^{10} t_3^{10} t_4^8 t_5^9 t_6^8 t_7^{10} t_8^9 -  y_{s}^9 y_{o_1}^{19} y_{o_2}^{17} y_{o_3}^{19} y_{o_4}^{17} y_{o_5}^{18} y_{o_6}^{18} y_{o_7}^{27} t_1^7 t_2^{11} t_3^{10} t_4^8 t_5^9 t_6^8 t_7^{10} t_8^9 +  y_{s}^{10} y_{o_1}^{20} y_{o_2}^{20} y_{o_3}^{21} y_{o_4}^{19} y_{o_5}^{19} y_{o_6}^{21} y_{o_7}^{29} t_1^{10} t_2^9 t_3^{11} t_4^{10} t_5^{11} t_6^9 t_7^{10} t_8^9 +  y_{s}^{10} y_{o_1}^{20} y_{o_2}^{20} y_{o_3}^{21} y_{o_4}^{19} y_{o_5}^{19} y_{o_6}^{21} y_{o_7}^{29} t_1^9 t_2^{10} t_3^{11} t_4^{10} t_5^{11} t_6^9 t_7^{10} t_8^9 -  y_{s}^7 y_{o_1}^{15} y_{o_2}^{13} y_{o_3}^{12} y_{o_4}^{16} y_{o_5}^{17} y_{o_6}^{11} y_{o_7}^{24} t_1^9 t_2^8 t_3^5 t_4^6 t_5^4 t_6^9 t_7^8 t_8^{10} -  y_{s}^7 y_{o_1}^{15} y_{o_2}^{13} y_{o_3}^{12} y_{o_4}^{16} y_{o_5}^{17} y_{o_6}^{11} y_{o_7}^{24} t_1^8 t_2^9 t_3^5 t_4^6 t_5^4 t_6^9 t_7^8 t_8^{10} +  2 y_{s}^8 y_{o_1}^{16} y_{o_2}^{16} y_{o_3}^{14} y_{o_4}^{18} y_{o_5}^{18} y_{o_6}^{14} y_{o_7}^{26} t_1^{10} t_2^8 t_3^6 t_4^8 t_5^6 t_6^{10} t_7^8 t_8^{10} +  2 y_{s}^8 y_{o_1}^{16} y_{o_2}^{16} y_{o_3}^{14} y_{o_4}^{18} y_{o_5}^{18} y_{o_6}^{14} y_{o_7}^{26} t_1^9 t_2^9 t_3^6 t_4^8 t_5^6 t_6^{10} t_7^8 t_8^{10} +  2 y_{s}^8 y_{o_1}^{16} y_{o_2}^{16} y_{o_3}^{14} y_{o_4}^{18} y_{o_5}^{18} y_{o_6}^{14} y_{o_7}^{26} t_1^8 t_2^{10} t_3^6 t_4^8 t_5^6 t_6^{10} t_7^8 t_8^{10} -  y_{s}^9 y_{o_1}^{17} y_{o_2}^{19} y_{o_3}^{16} y_{o_4}^{20} y_{o_5}^{19} y_{o_6}^{17} y_{o_7}^{28} t_1^{10} t_2^9 t_3^7 t_4^{10} t_5^8 t_6^{11} t_7^8 t_8^{10} -  y_{s}^9 y_{o_1}^{17} y_{o_2}^{19} y_{o_3}^{16} y_{o_4}^{20} y_{o_5}^{19} y_{o_6}^{17} y_{o_7}^{28} t_1^9 t_2^{10} t_3^7 t_4^{10} t_5^8 t_6^{11} t_7^8 t_8^{10} -  y_{s}^7 y_{o_1}^{16} y_{o_2}^{12} y_{o_3}^{13} y_{o_4}^{15} y_{o_5}^{17} y_{o_6}^{11} y_{o_7}^{24} t_1^9 t_2^8 t_3^6 t_4^5 t_5^4 t_6^8 t_7^9 t_8^{10} -  y_{s}^7 y_{o_1}^{16} y_{o_2}^{12} y_{o_3}^{13} y_{o_4}^{15} y_{o_5}^{17} y_{o_6}^{11} y_{o_7}^{24} t_1^8 t_2^9 t_3^6 t_4^5 t_5^4 t_6^8 t_7^9 t_8^{10} +  2 y_{s}^8 y_{o_1}^{17} y_{o_2}^{15} y_{o_3}^{15} y_{o_4}^{17} y_{o_5}^{18} y_{o_6}^{14} y_{o_7}^{26} t_1^{10} t_2^8 t_3^7 t_4^7 t_5^6 t_6^9 t_7^9 t_8^{10} +  3 y_{s}^8 y_{o_1}^{17} y_{o_2}^{15} y_{o_3}^{15} y_{o_4}^{17} y_{o_5}^{18} y_{o_6}^{14} y_{o_7}^{26} t_1^9 t_2^9 t_3^7 t_4^7 t_5^6 t_6^9 t_7^9 t_8^{10} +  2 y_{s}^8 y_{o_1}^{17} y_{o_2}^{15} y_{o_3}^{15} y_{o_4}^{17} y_{o_5}^{18} y_{o_6}^{14} y_{o_7}^{26} t_1^8 t_2^{10} t_3^7 t_4^7 t_5^6 t_6^9 t_7^9 t_8^{10} -  y_{s}^9 y_{o_1}^{18} y_{o_2}^{18} y_{o_3}^{17} y_{o_4}^{19} y_{o_5}^{19} y_{o_6}^{17} y_{o_7}^{28} t_1^{11} t_2^8 t_3^8 t_4^9 t_5^8 t_6^{10} t_7^9 t_8^{10} -  3 y_{s}^9 y_{o_1}^{18} y_{o_2}^{18} y_{o_3}^{17} y_{o_4}^{19} y_{o_5}^{19} y_{o_6}^{17} y_{o_7}^{28} t_1^{10} t_2^9 t_3^8 t_4^9 t_5^8 t_6^{10} t_7^9 t_8^{10} -  3 y_{s}^9 y_{o_1}^{18} y_{o_2}^{18} y_{o_3}^{17} y_{o_4}^{19} y_{o_5}^{19} y_{o_6}^{17} y_{o_7}^{28} t_1^9 t_2^{10} t_3^8 t_4^9 t_5^8 t_6^{10} t_7^9 t_8^{10} -  y_{s}^9 y_{o_1}^{18} y_{o_2}^{18} y_{o_3}^{17} y_{o_4}^{19} y_{o_5}^{19} y_{o_6}^{17} y_{o_7}^{28} t_1^8 t_2^{11} t_3^8 t_4^9 t_5^8 t_6^{10} t_7^9 t_8^{10} +  y_{s}^{10} y_{o_1}^{19} y_{o_2}^{21} y_{o_3}^{19} y_{o_4}^{21} y_{o_5}^{20} y_{o_6}^{20} y_{o_7}^{30} t_1^{10} t_2^{10} t_3^9 t_4^{11} t_5^{10} t_6^{11} t_7^9 t_8^{10} +  2 y_{s}^8 y_{o_1}^{18} y_{o_2}^{14} y_{o_3}^{16} y_{o_4}^{16} y_{o_5}^{18} y_{o_6}^{14} y_{o_7}^{26} t_1^{10} t_2^8 t_3^8 t_4^6 t_5^6 t_6^8 t_7^{10} t_8^{10} +  2 y_{s}^8 y_{o_1}^{18} y_{o_2}^{14} y_{o_3}^{16} y_{o_4}^{16} y_{o_5}^{18} y_{o_6}^{14} y_{o_7}^{26} t_1^9 t_2^9 t_3^8 t_4^6 t_5^6 t_6^8 t_7^{10} t_8^{10} +  2 y_{s}^8 y_{o_1}^{18} y_{o_2}^{14} y_{o_3}^{16} y_{o_4}^{16} y_{o_5}^{18} y_{o_6}^{14} y_{o_7}^{26} t_1^8 t_2^{10} t_3^8 t_4^6 t_5^6 t_6^8 t_7^{10} t_8^{10} -  y_{s}^9 y_{o_1}^{19} y_{o_2}^{17} y_{o_3}^{18} y_{o_4}^{18} y_{o_5}^{19} y_{o_6}^{17} y_{o_7}^{28} t_1^{11} t_2^8 t_3^9 t_4^8 t_5^8 t_6^9 t_7^{10} t_8^{10} -  3 y_{s}^9 y_{o_1}^{19} y_{o_2}^{17} y_{o_3}^{18} y_{o_4}^{18} y_{o_5}^{19} y_{o_6}^{17} y_{o_7}^{28} t_1^{10} t_2^9 t_3^9 t_4^8 t_5^8 t_6^9 t_7^{10} t_8^{10} -  3 y_{s}^9 y_{o_1}^{19} y_{o_2}^{17} y_{o_3}^{18} y_{o_4}^{18} y_{o_5}^{19} y_{o_6}^{17} y_{o_7}^{28} t_1^9 t_2^{10} t_3^9 t_4^8 t_5^8 t_6^9 t_7^{10} t_8^{10} -  y_{s}^9 y_{o_1}^{19} y_{o_2}^{17} y_{o_3}^{18} y_{o_4}^{18} y_{o_5}^{19} y_{o_6}^{17} y_{o_7}^{28} t_1^8 t_2^{11} t_3^9 t_4^8 t_5^8 t_6^9 t_7^{10} t_8^{10} +  y_{s}^{10} y_{o_1}^{20} y_{o_2}^{20} y_{o_3}^{20} y_{o_4}^{20} y_{o_5}^{20} y_{o_6}^{20} y_{o_7}^{30} t_1^{11} t_2^9 t_3^{10} t_4^{10} t_5^{10} t_6^{10} t_7^{10} t_8^{10} +  y_{s}^{10} y_{o_1}^{20} y_{o_2}^{20} y_{o_3}^{20} y_{o_4}^{20} y_{o_5}^{20} y_{o_6}^{20} y_{o_7}^{30} t_1^{10} t_2^{10} t_3^{10} t_4^{10} t_5^{10} t_6^{10} t_7^{10} t_8^{10} +  y_{s}^{10} y_{o_1}^{20} y_{o_2}^{20} y_{o_3}^{20} y_{o_4}^{20} y_{o_5}^{20} y_{o_6}^{20} y_{o_7}^{30} t_1^9 t_2^{11} t_3^{10} t_4^{10} t_5^{10} t_6^{10} t_7^{10} t_8^{10} -  y_{s}^9 y_{o_1}^{20} y_{o_2}^{16} y_{o_3}^{19} y_{o_4}^{17} y_{o_5}^{19} y_{o_6}^{17} y_{o_7}^{28} t_1^{10} t_2^9 t_3^{10} t_4^7 t_5^8 t_6^8 t_7^{11} t_8^{10} -  y_{s}^9 y_{o_1}^{20} y_{o_2}^{16} y_{o_3}^{19} y_{o_4}^{17} y_{o_5}^{19} y_{o_6}^{17} y_{o_7}^{28} t_1^9 t_2^{10} t_3^{10} t_4^7 t_5^8 t_6^8 t_7^{11} t_8^{10} +  y_{s}^{10} y_{o_1}^{21} y_{o_2}^{19} y_{o_3}^{21} y_{o_4}^{19} y_{o_5}^{20} y_{o_6}^{20} y_{o_7}^{30} t_1^{10} t_2^{10} t_3^{11} t_4^9 t_5^{10} t_6^9 t_7^{11} t_8^{10} -  y_{s}^9 y_{o_1}^{19} y_{o_2}^{17} y_{o_3}^{17} y_{o_4}^{19} y_{o_5}^{20} y_{o_6}^{16} y_{o_7}^{29} t_1^{10} t_2^{10} t_3^8 t_4^8 t_5^7 t_6^{10} t_7^{10} t_8^{11} +  y_{s}^{11} y_{o_1}^{22} y_{o_2}^{22} y_{o_3}^{22} y_{o_4}^{22} y_{o_5}^{22} y_{o_6}^{22} y_{o_7}^{33} t_1^{11} t_2^{11} t_3^{11} t_4^{11} t_5^{11} t_6^{11} t_7^{11} t_8^{11}
~,~
$
\end{quote}
\endgroup

\bibliographystyle{JHEP}
\bibliography{mybib}

\end{document}